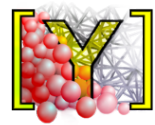

# Yade Documentation (3rd ed.)

## Authors

- **Václav Šmilauer** Freelance consultant (http://woodem.eu)

- **Vasileios Angelidakis** Newcastle University, UK

- **Emanuele Catalano** Univ. Grenoble Alpes, 3SR lab.

- **Robert Caulk** Univ. Grenoble Alpes, 3SR lab.

- **Bruno Chareyre** Univ. Grenoble Alpes, 3SR lab.

- **William Chèvremont** Univ. Grenoble Alpes, LRP

- **Sergei Dorofeenko** IPCP RAS, Chernogolovka

- **Jérôme Duriez** INRAE, Aix Marseille Univ, RECOVER, Aix-en-Provence, France

- **Nolan Dyck** Univ. of Western Ontario

- **Jan Eliáš** Brno University of Technology

- **Burak Er** Bursa Technical University

- **Alexander Eulitz** TU Berlin / Institute for Machine Tools and Factory Management

- **Anton Gladky** TU Bergakademie Freiberg

- **Ning Guo** Hong Kong Univ. of Science and Tech.

- **Christian Jakob** TU Bergakademie Freiberg

- **François Kneib** Univ. Grenoble Alpes, 3SR lab. / Irstea Grenoble

- **Janek Kozicki** Gdansk University of Technology

- **Donia Marzougui** Univ. Grenoble Alpes, 3SR lab.

- **Raphaël Maurin** Irstea Grenoble

- **Chiara Modenese** University of Oxford

- **Gerald Pekmezi** University of Alabama at Birmingham

- **Luc Scholtès** Univ. Grenoble Alpes, 3SR lab.

- **Luc Sibille** University of Nantes, lab. GeM

- **Jan Stránský** CVUT Prague

- **Thomas Sweijen** Utrecht University

- **Klaus Thoeni** The University of Newcastle (Australia)

- **Chao Yuan** Univ. Grenoble Alpes, 3SR lab.

## Citing this document

# Contents

# Chapter 1

# Guided tour

## 1.1 Introduction

### 1.1.1 Getting started

Before you start moving around in Yade, you should have some prior knowledge.

- Basics of command line in your Linux system are necessary for running yade. Look on the web for tutorials.

- Python language; we recommend the official Python tutorial. Reading further documents on the topic, such as Dive into Python will certainly not hurt either.

You are advised to try all commands described yourself. Don't be afraid to experiment.

---

**Hint:** Sometimes reading this documentation in a .pdf format can be more comfortable. For example in okular pdf viewer clicking links is faster than a page refresh in the web browser and to go back press the shortcut `Alt Shift ←`. To try it have a look at the inheritance graph of *PartialEngine* then go back.

---

#### Starting yade

Yade is being run primarily from terminal; the name of command is `yade`.[1] (In case you did not install from package, you might need to give specific path to the command[2]):

```
$ yade
Welcome to Yade
TCP python prompt on localhost:9001, auth cookie `sdksuy'
TCP info provider on localhost:21000
```

(continues on next page)

---

[1] The executable name can carry a suffix, such as version number (`yade-0.20`), depending on compilation options. Packaged versions on Debian systems always provide the plain `yade` alias, by default pointing to latest stable version (or latest snapshot, if no stable version is installed). You can use `update-alternatives` to change this.

[2] In general, Unix *shell* (command line) has environment variable `PATH` defined, which determines directories searched for executable files if you give name of the file without path. Typically, $PATH contains `/usr/bin/`, `/usr/local/bin`, `/bin` and others; you can inspect your `PATH` by typing `echo $PATH` in the shell (directories are separated by `:`).

If Yade executable is not in directory contained in `PATH`, you have to specify it by hand, i.e. by typing the path in front of the filename, such as in `/home/user/bin/yade` and similar. You can also navigate to the directory itself (`cd ~/bin/yade`, where `~` is replaced by your home directory automatically) and type `./yade` then (the `.` is the current directory, so `./` specifies that the file is to be found in the current directory).

To save typing, you can add the directory where Yade is installed to your `PATH`, typically by editing `~/.profile` (in normal cases automatically executed when shell starts up) file adding line like `export PATH=/home/user/bin:$PATH`. You can also define an *alias* by saying `alias yade="/home/users/bin/yade"` in that file.

Details depend on what shell you use (bash, zsh, tcsh, …) and you will find more information in introductory material on Linux/Unix.

```
[[ ^L clears screen, ^U kills line. F12 controller, F11 3d view, F10 both, F9 generator, F8
↪plot. ]]
Yade [1]:
```

These initial lines give you some information about

- some information for *Remote control*, which you are unlikely to need now;

- basic help for the command-line that just appeared (`Yade [1]:`).

Type `quit()`, `exit()` or simply press `^D` (`^` is a commonly used written shortcut for pressing the `Ctrl` key, so here `^D` means `Ctrl D`) to quit Yade.

The command-line is ipython, python shell with enhanced interactive capabilities; it features persistent history (remembers commands from your last sessions), searching and so on. See ipython's documentation for more details.

Typically, you will not type Yade commands by hand, but use *scripts*, python programs describing and running your simulations. Let us take the most simple script that will just print "Hello world!":

```python
print("Hello world!")
```

Saving such script as `hello.py`, it can be given as argument to Yade:

```
$ yade hello.py
Welcome to Yade
TCP python prompt on localhost:9001, auth cookie `askcsu'
TCP info provider on localhost:21000
Running script hello.py                               ## the script is being run
Hello world!                                          ## output from the script
[[ ^L clears screen, ^U kills line. F12 controller, F11 3d view, F10 both, F9 generator, F8
↪plot. ]]
Yade [1]:
```

Yade will run the script and then drop to the command-line again.[3] If you want Yade to quit immediately after running the script, use the `-x` switch:

```
$ yade -x script.py
```

There is more command-line options than just `-x`, run `yade -h` to see all of them.

**Options:**

| | |
|---|---|
| **-v, --version** | show program's version number and exit |
| **-h, --help** | show this help message and exit |
| **-j THREADS, --threads=THREADS** | Number of OpenMP threads to run; defaults to 1. Equivalent to setting OMP_-NUM_THREADS environment variable. |
| **--cores=CORES** | Set number of OpenMP threads (as –threads) and in addition set affinity of threads to the cores given. |
| **--update** | Update deprecated class names in given script(s) using text search & replace. Changed files will be backed up with ~ suffix. Exit when done without running any simulation. |
| **--nice=NICE** | Increase nice level (i.e. decrease priority) by given number. |

---

[3] Plain Python interpreter exits once it finishes running the script. The reason why Yade does the contrary is that most of the time script only sets up simulation and lets it run; since computation typically runs in background thread, the script is technically finished, but the computation is running.

| | |
|---|---|
| **-x** | Exit when the script finishes |
| **-f** | Set *logging verbosity*, default is -f3 (yade.log.WARN) for all classes |
| **-n** | Run without graphical interface (equivalent to unsetting the DISPLAY environment variable) |
| **--test** | Run regression test suite and exit; the exists status is 0 if all tests pass, 1 if a test fails and 2 for an unspecified exception. |
| **--check** | Run a series of user-defined check tests as described in scripts/checks-and-tests/checks/README and *Regression tests* |
| **--performance** | Starts a test to measure the productivity. |
| **--stdperformance** | Starts a standardized test to measure the productivity, which will keep retrying to run the benchmark until standard deviation of the performance is below 1%. A common type of simulation is done: the spheres fall down in a box and are given enough time to settle in there. Note: better to use this with argument *-j THREADS* (explained above). |
| **--quickperformance** | Starts a quick test to measure the productivity. Same as above, but only two short runs are performed, without the attempts to find the computer performance with small error. |
| **--no-gdb** | Do not show backtrace when yade crashes (only effective with --debug)[4]. |

### Quick inline help

All of functions callable from ipython shell have a quickly accessible help by appending `?` to the function name, or calling `help(…)` command on them:

```
Yade [1]: O.run?
Docstring:
run( (Omega)arg1 [, (int)nSteps=-1 [, (bool)wait=False]]) -> None :
    Run the simulation. *nSteps* how many steps to run, then stop (if positive); *wait* will↵
↪cause not returning to python until simulation will have stopped.
Type:       method

Yade [2]: help(O.pause)
Help on method pause:

pause(...) method of yade.wrapper.Omega instance
    pause( (Omega)arg1 ) -> None :
        Stop simulation execution. (May be called from within the loop, and it will stop after↵
↪the current step).
```

A quick way to discover available functions is by using the tab-completion mechanism, e.g. type `O.` then press tab.

### Creating simulation

To create simulation, one can either use a specialized class of type *FileGenerator* to create full scene, possibly receiving some parameters. Generators are written in C++ and their role is limited to well-

---

[4] On some linux systems stack trace will produce `Operation not permitted` error. See *debugging section* for solution.

defined scenarios. For instance, to create triaxial test scene:

```
Yade [3]: TriaxialTest(numberOfGrains=200).load()

Yade [4]: len(O.bodies)
Out[4]: 206
```

Generators are regular yade objects that support attribute access.

It is also possible to construct the scene by a python script; this gives much more flexibility and speed of development and is the recommended way to create simulation. Yade provides modules for streamlined body construction, import of geometries from files and reuse of common code. Since this topic is more involved, it is explained in the *User's manual*.

**Running simulation**

As explained below, the loop consists in running defined sequence of engines. Step number can be queried by `O.iter` and advancing by one step is done by `O.step()`. Every step advances *virtual time* by current timestep, `O.dt` that can be directly assigned or, which is usually better, automatically determined by a *GlobalStiffnessTimeStepper*, if present:

```
Yade [5]: O.iter
Out[5]: 0

Yade [6]: O.time
Out[6]: 0.0

Yade [7]: O.dt=1e-4

Yade [8]: O.dynDt=False #else it would be adjusted automaticaly during first iteration

Yade [9]: O.step()

Yade [10]: O.iter
Out[10]: 1

Yade [11]: O.time
Out[11]: 0.0001
```

Normal simulations, however, are run continuously. Starting/stopping the loop is done by `O.run()` and `O.pause()`; note that `O.run()` returns control to Python and the simulation runs in background; if you want to wait for it to finish, use `O.wait()`. Fixed number of steps can be run with `O.run(1000)`, `O.run(1000,True)` will run and wait. To stop at absolute step number, `O.stopAtIter` can be set and `O.run()` called normally.

```
Yade [12]: O.run()

Yade [13]: O.pause()

Yade [14]: O.iter
Out[14]: 1715

Yade [15]: O.run(100000,True)

Yade [16]: O.iter
Out[16]: 101715

Yade [17]: O.stopAtIter=500000

Yade [18]: O.run()
```

```
Yade [19]: O.wait()

Yade [20]: O.iter
Out[20]: 500000
```

### Saving and loading

Simulation can be saved at any point to a binary file (optionaly compressed if the filename has extensions such as ".gz" or ".bz2"). Saving to a XML file is also possible though resulting in larger files and slower save/load, it is used when the filename contains "xml". With some limitations, it is generally possible to load the scene later and resume the simulation as if it were not interrupted. Note that since the saved scene is a dump of Yade's internal objects, it might not (probably will not) open with different Yade version. This problem can be sometimes solved by migrating the saved file using ".xml" format.

```
Yade [21]: O.save('/tmp/a.yade.bz2')

Yade [22]: O.reload()

Yade [23]: O.load('/tmp/another.yade.bz2')
```

The principal use of saving the simulation to XML is to use it as temporary in-memory storage for checkpoints in simulation, e.g. for reloading the initial state and running again with different parameters (think tension/compression test, where each begins from the same virgin state). The functions `O.saveTmp()` and `O.loadTmp()` can be optionally given a slot name, under which they will be found in memory:

```
Yade [24]: O.saveTmp()

Yade [25]: O.loadTmp()

Yade [26]: O.saveTmp('init') ## named memory slot

Yade [27]: O.loadTmp('init')
```

Simulation can be reset to empty state by `O.reset()`.

It can be sometimes useful to run different simulation, while the original one is temporarily suspended, e.g. when dynamically creating packing. `O.switchWorld()` toggles between the primary and secondary simulation.

### Graphical interface

Yade can be optionally compiled with QT based graphical interface (qt4 and qt5 are supported). It can be started by pressing `F12` in the command-line, and also is started automatically when running a script.

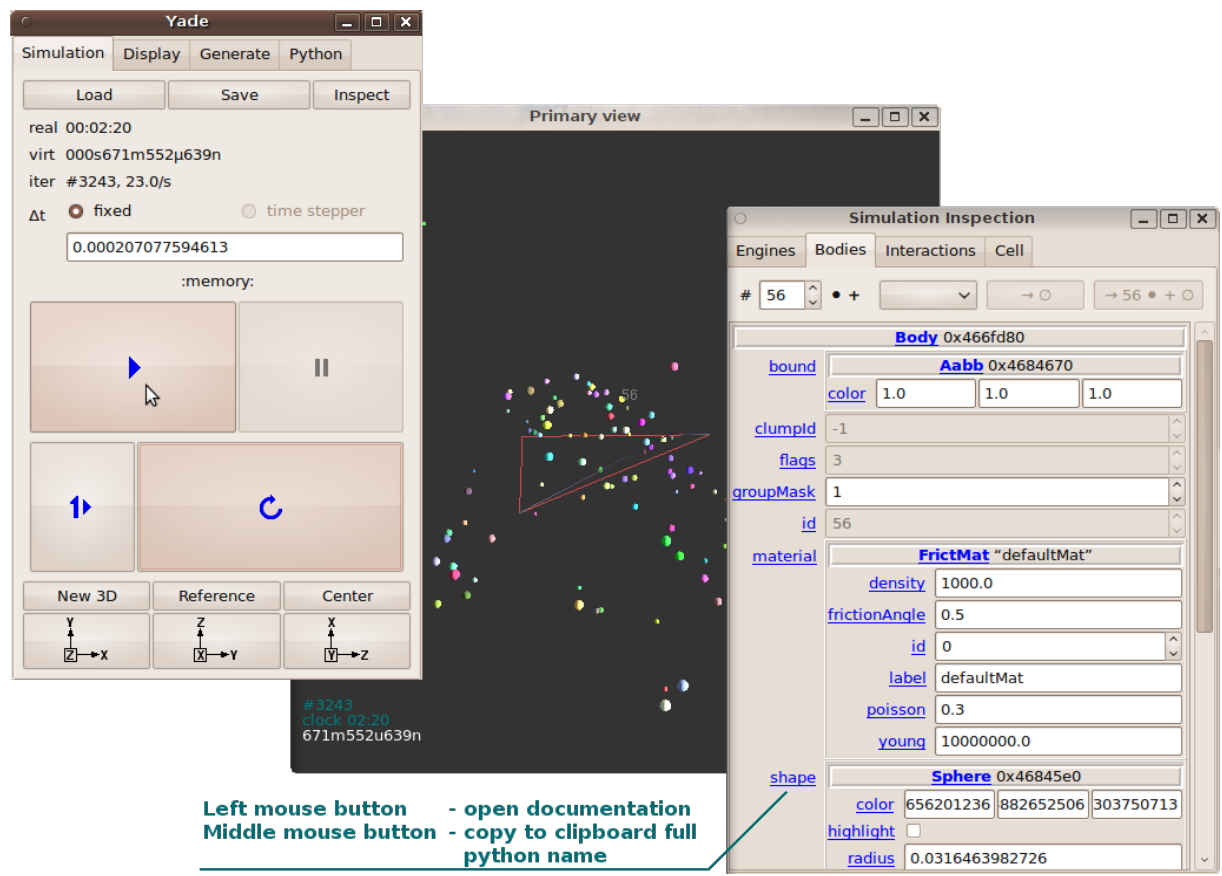

The control window on the left (fig. *imgQtGui*) is called `Controller` (can be invoked by `yade.qt.Controller()` from python or by pressing `F12` key in terminal):

1. The *Simulation* tab is mostly self-explanatory, and permits basic simulation control.

2. The *Display* tab has various rendering-related options, which apply to all opened views (they can be zero or more, new one is opened by the *New 3D* button).

3. The *Python* tab has only a simple text entry area; it can be useful to enter python commands while the command-line is blocked by running script, for instance.

Inside the *Inspect* window (on the right in fig. *imgQtGui*) all simulation data can be examined and modified in realtime.

1. Clicking left mouse button on any of the blue hyperlinks will open documentation.

2. Clicking middle mouse button will copy the fully qualified python name into clipboard, which can be pasted into terminal by clicking middle mouse button in the terminal (or pressing `Ctrl-V`).

3d views can be controlled using mouse and keyboard shortcuts; help is displayed if you press the `h` key while in the 3d view. Note that having the 3d view open can slow down running simulation significantly, it is meant only for quickly checking whether the simulation runs smoothly. Advanced post-processing is described in dedicated section *Data mining*.

### 1.1.2 Architecture overview

In the following, a high-level overview of Yade architecture will be given. As many of the features are directly represented in simulation scripts, which are written in Python, being familiar with this language will help you follow the examples. For the rest, this knowledge is not strictly necessary and you can ignore code examples.

**Data and functions**

To assure flexibility of software design, yade makes clear distinction of 2 families of classes: *data* components and *functional* components. The former only store data without providing functionality, while the latter define functions operating on the data. In programming, this is known as *visitor* pattern (as functional components "visit" the data, without being bound to them explicitly).

Entire simulation, i.e. both data and functions, are stored in a single `Scene` object. It is accessible through the *Omega* class in python (a singleton), which is by default stored in the `O` global variable:

```
Yade [28]: O.bodies          # some data components
Out[28]: <yade.wrapper.BodyContainer at 0x7f0cfbeadeb0>

Yade [29]: len(O.bodies)     # there are no bodies as of yet
Out[29]: 0

Yade [30]: O.engines         # functional components, empty at the moment
Out[30]: []
```

**Data components**

**Bodies**

Yade simulation (class *Scene*, but hidden inside *Omega* in Python) is represented by *Bodies*, their *Interactions* and resultant generalized *forces* (all stored internally in special containers).

Each *Body* comprises the following:

**Shape** represents particle's geometry (neutral with regards to its spatial orientation), such as *Sphere*, *Facet* or infinite *Wall*; it usually does not change during simulation.

**Material** stores characteristics pertaining to mechanical behavior, such as Young's modulus or density, which are independent on particle's shape and dimensions; usually constant, might be shared amongst multiple bodies.

**State** contains state variables, in particular spatial *position* and *orientation*, *linear* and *angular* velocity; it is updated by the *integrator* at every step. The derived classes would contain other information related to current state of this body, e.g. its temperature, *averaged damage* or *broken links* between components.

**Bound** is used for approximate ("pass 1") contact detection; updated as necessary following body's motion. Currently, *Aabb* is used most often as *Bound*. Some bodies may have no *Bound*, in which case they are exempt from contact detection.

(In addition to these 4 components, bodies have several more minor data associated, such as *Body::id* or *Body::mask*.)

All these four properties can be of different types, derived from their respective base types. Yade frequently makes decisions about computation based on those types: *Sphere* + *Sphere* collision has to be treated differently than *Facet* + *Sphere* collision. Objects making those decisions are called *Dispatchers* and are essential to understand Yade's functioning; they are discussed below.

Explicitly assigning all 4 properties to each particle by hand would be not practical; there are utility functions defined to create them with all necessary ingredients. For example, we can create sphere particle using *utils.sphere*:

```
Yade [31]: s=utils.sphere(center=[0,0,0],radius=1)

Yade [32]: s.shape, s.state, s.mat, s.bound
Out[32]:
(<Sphere instance at 0x38d9450>,
```

(continues on next page)

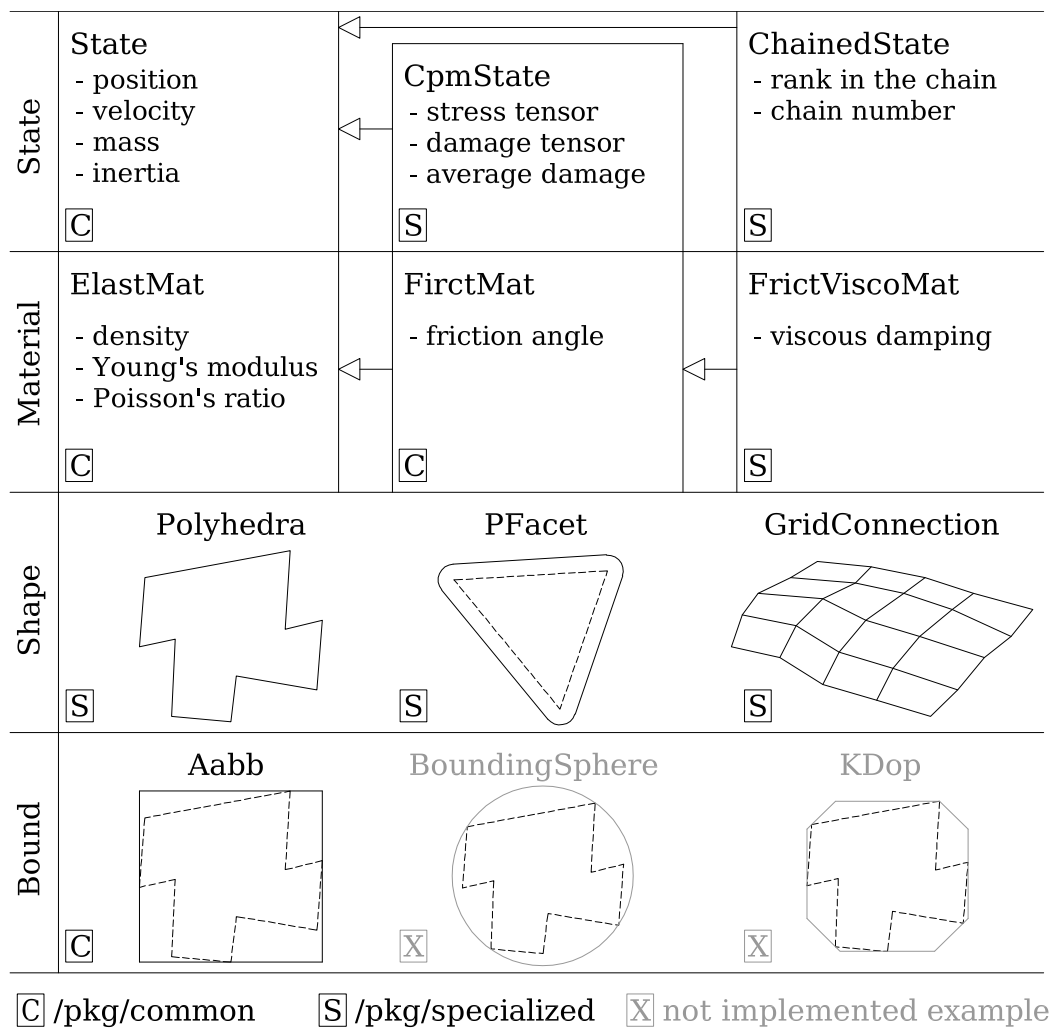

Fig. 1: Examples of concrete classes that might be used to describe a *Body*: *State*, *CpmState*, *Chained-State*, *Material*, *ElastMat*, *FrictMat*, *FrictViscoMat*, *Shape*, **Polyhedra**, *PFacet*, *GridConnection*, *Bound*, *Aabb*.

```
 <State instance at 0x3b2d060>,
 <FrictMat instance at 0x3d09ef0>,
 None)

Yade [33]: s.state.pos
Out[33]: Vector3(0,0,0)

Yade [34]: s.shape.radius
Out[34]: 1.0
```

We see that a sphere with material of type *FrictMat* (default, unless you provide another *Material*) and bounding volume of type *Aabb* (axis-aligned bounding box) was created. Its position is at the origin and its radius is 1.0. Finally, this object can be inserted into the simulation; and we can insert yet one sphere as well.

```
Yade [35]: O.bodies.append(s)
Out[35]: 0

Yade [36]: O.bodies.append(utils.sphere([0,0,2],.5))
Out[36]: 1
```

In each case, return value is *Body.id* of the body inserted.

Since till now the simulation was empty, its id is 0 for the first sphere and 1 for the second one. Saving the id value is not necessary, unless you want to access this particular body later; it is remembered internally in *Body* itself. You can address bodies by their id:

```
Yade [37]: O.bodies[1].state.pos
Out[37]: Vector3(0,0,2)

Yade [38]: O.bodies[100]          # error because there are only two bodies
---------------------------------------------------------------------
IndexError                                 Traceback (most recent call last)
~/yade/lib/x86_64-linux-gnu/yadeflip/py/yade/__init__.py in <module>
----> 1 O.bodies[100]    # error because there are only two bodies

IndexError: Body id out of range.
```

Adding the same body twice is, for reasons of the id uniqueness, not allowed:

```
Yade [39]: O.bodies.append(s)    # error because this sphere was already added
---------------------------------------------------------------------
IndexError                                 Traceback (most recent call last)
~/yade/lib/x86_64-linux-gnu/yadeflip/py/yade/__init__.py in <module>
----> 1 O.bodies.append(s)  # error because this sphere was already added

IndexError: Body already has id 0 set; appending such body (for the second time) is not␣
↪allowed.
```

Bodies can be iterated over using standard python iteration syntax:

```
Yade [40]: for b in O.bodies:
   ....:     print(b.id,b.shape.radius)
   ....:
0 1.0
1 0.5
```

### Interactions

*Interactions* are always between pair of bodies; usually, they are created by the collider based on spatial proximity; they can, however, be created explicitly and exist independently of distance. Each interaction has 2 components:

**IGeom** holding geometrical configuration of the two particles in collision; it is updated automatically as the particles in question move and can be queried for various geometrical characteristics, such as penetration distance or shear strain.

Based on combination of types of *Shapes* of the particles, there might be different storage requirements; for that reason, a number of derived classes exists, e.g. for representing geometry of contact between *Sphere+Sphere*, *Cylinder+Sphere* etc. Note, however, that it is possible to represent many type of contacts with the basic sphere-sphere geometry (for instance in *Ig2_Wall_Sphere_Sc-Geom*).

**IPhys** representing non-geometrical features of the interaction; some are computed from *Materials* of the particles in contact using some averaging algorithm (such as contact stiffness from Young's moduli of particles), others might be internal variables like damage.

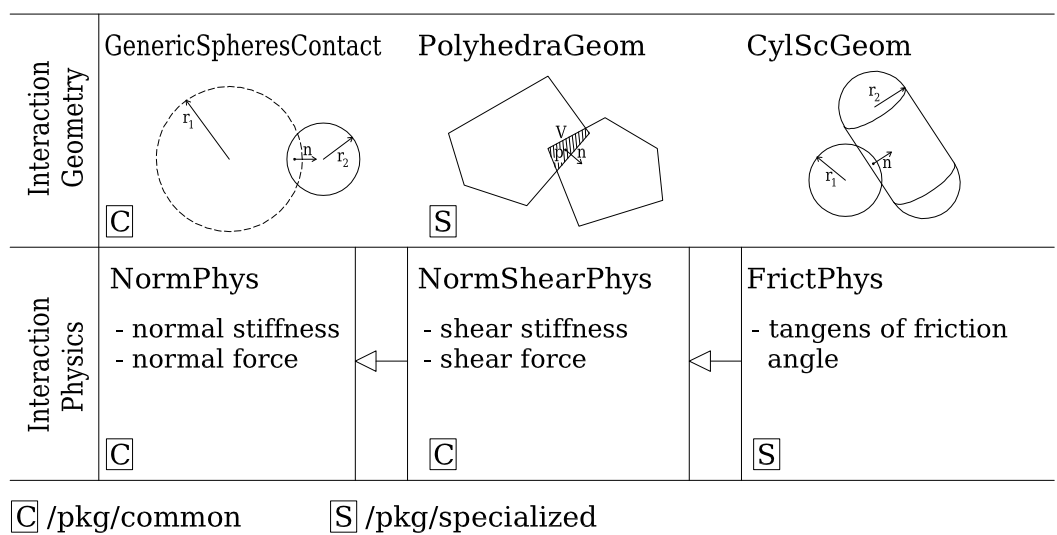

Fig. 2: Examples of concrete classes that might be used to describe an *Interaction*: *IGeom*, *Generic-SpheresContact*, *PolyhedraGeom*, *CylScGeom*, *IPhys*, *NormPhys*, *NormShearPhys*, *FrictPhys*.

Suppose now interactions have been already created. We can access them by the id pair:

```
Yade [41]: O.interactions[0,1]
Out[41]: <Interaction instance at 0x384c250>

Yade [42]: O.interactions[1,0]        # order of ids is not important
Out[42]: <Interaction instance at 0x384c250>

Yade [43]: i=O.interactions[0,1]

Yade [44]: i.id1,i.id2
Out[44]: (0, 1)

Yade [45]: i.geom
Out[45]: <ScGeom instance at 0x315d280>

Yade [46]: i.phys
Out[46]: <FrictPhys instance at 0x3d07c20>

Yade [47]: O.interactions[100,10111]    # asking for non existing interaction throws exception
---------------------------------------------------------------------------
IndexError                              Traceback (most recent call last)
~/yade/lib/x86_64-linux-gnu/yadeflip/py/yade/__init__.py in <module>
```

```
----> 1 O.interactions[100,10111]      # asking for non existing interaction throws exception

IndexError: No such interaction
```

### Generalized forces

Generalized forces include force, torque and forced displacement and rotation; they are stored only temporariliy, during one computation step, and reset to zero afterwards. For reasons of parallel computation, they work as accumulators, i.e. only can be added to, read and reset.

```
Yade [48]: O.forces.f(0)
Out[48]: Vector3(0,0,0)

Yade [49]: O.forces.addF(0,Vector3(1,2,3))

Yade [50]: O.forces.f(0)
Out[50]: Vector3(1,2,3)
```

You will only rarely modify forces from Python; it is usually done in c++ code and relevant documentation can be found in the Programmer's manual.

### Function components

In a typical DEM simulation, the following sequence is run repeatedly:

- reset forces on bodies from previous step
- approximate collision detection (pass 1)
- detect exact collisions of bodies, update interactions as necessary
- solve interactions, applying forces on bodies
- apply other external conditions (gravity, for instance).
- change position of bodies based on forces, by integrating motion equations.

Each of these actions is represented by an *Engine*, functional element of simulation. The sequence of engines is called *simulation loop*.

### Engines

Simulation loop, shown at fig. *img-yade-iter-loop*, can be described as follows in Python (details will be explained later); each of the `O.engines` items is instance of a type deriving from *Engine*:

```
O.engines=[
        # reset forces
        ForceResetter(),
        # approximate collision detection, create interactions
        InsertionSortCollider([Bo1_Sphere_Aabb(),Bo1_Facet_Aabb()]),
        # handle interactions
        InteractionLoop(
                [Ig2_Sphere_Sphere_ScGeom(),Ig2_Facet_Sphere_ScGeom()],
                [Ip2_FrictMat_FrictMat_FrictPhys()],
                [Law2_ScGeom_FrictPhys_CundallStrack()],
        ),
        # apply other conditions
        GravityEngine(gravity=(0,0,-9.81)),
```

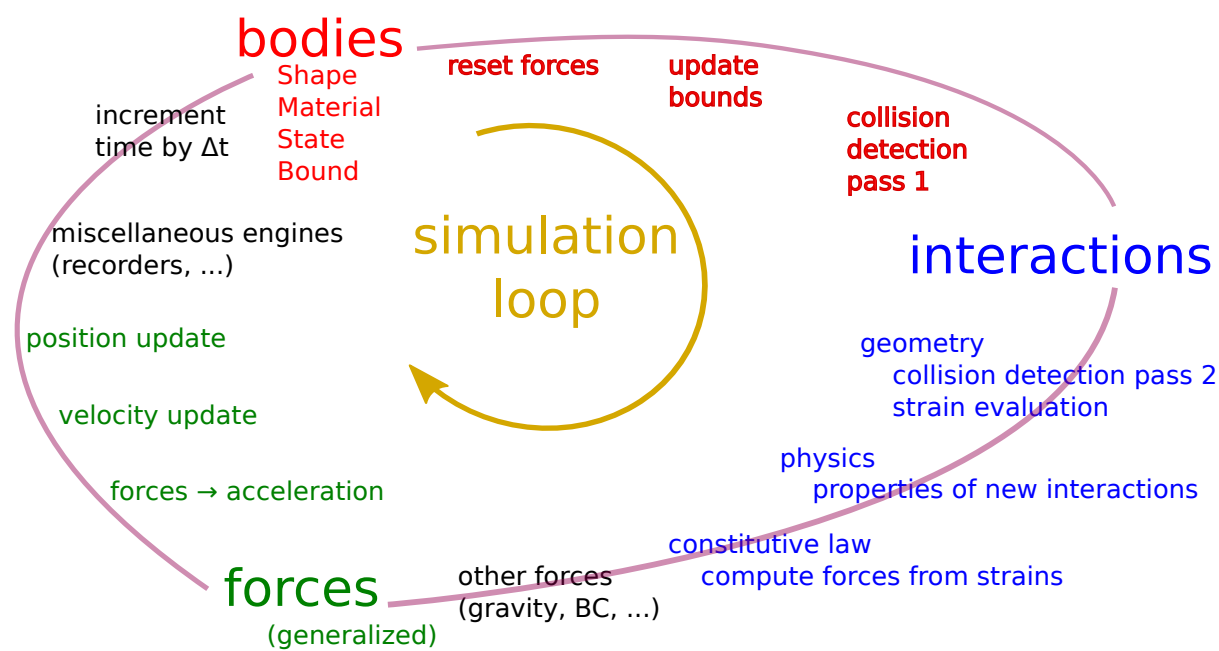

Fig. 3: Typical simulation loop; each step begins at body-centered bit at 11 o'clock, continues with interaction bit, force application bit, miscellanea and ends with time update.

```
        # update positions using Newton's equations
        NewtonIntegrator()
]
```

There are 3 fundamental types of Engines:

**GlobalEngines** operating on the whole simulation (e.g. *ForceResetter* which zeroes forces acting on bodies or *GravityEngine* looping over all bodies and applying force based on their mass)

**PartialEngine** operating only on some pre-selected bodies (e.g. *ForceEngine* applying constant force to some *selected* bodies)

**Dispatchers** do not perform any computation themselves; they merely call other functions, represented by function objects, *Functors*. Each functor is specialized, able to handle certain object types, and will be dispatched if such obejct is treated by the dispatcher.

**Dispatchers and functors**

For approximate collision detection (pass 1), we want to compute *bounds* for all *bodies* in the simulation; suppose we want bound of type *axis-aligned bounding box*. Since the exact algorithm is different depending on particular *shape*, we need to provide functors for handling all specific cases. In the `O.engines=[…]` declared above, the line:

```
InsertionSortCollider([Bo1_Sphere_Aabb(),Bo1_Facet_Aabb()])
```

creates *InsertionSortCollider* (it internally uses *BoundDispatcher*, but that is a detail). It traverses all bodies and will, based on *shape* type of each *body*, dispatch one of the functors to create/update *bound* for that particular body. In the case shown, it has 2 functors, one handling *spheres*, another *facets*.

The name is composed from several parts: `Bo` (functor creating *Bound*), which accepts `1` type *Sphere* and creates an *Aabb* (axis-aligned bounding box; it is derived from *Bound*). The *Aabb* objects are used by *InsertionSortCollider* itself. All `Bo1` functors derive from *BoundFunctor*.

The next part, reading

BoundFunctor

| Bo1_Sphere_Aabb | Bo1_Facet_Aabb | Bo1_Cylinder_Aabb |
|---|---|---|

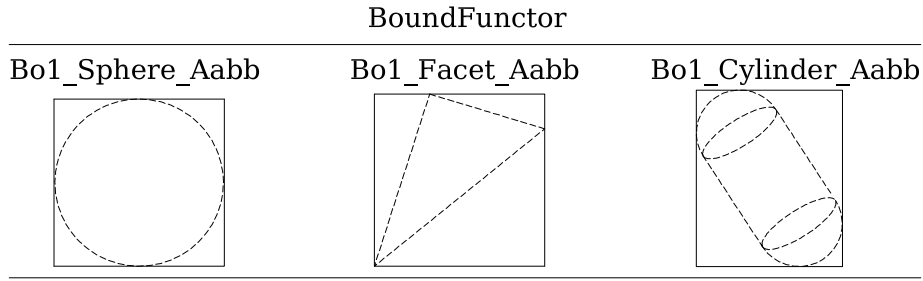

Fig. 4: Example *bound functors* producing *Aabb* accepting various different types, such as *Sphere*, *Facet* or *Cylinder*. In the case shown, the `Bo1` functors produce *Aabb* instances from single specific *Shape*, hence the number `1` in the functor name. Each of those functors uses specific geometry of the *Shape* i.e. position of nodes in *Facet* or *radius of sphere* to calculate the *Aabb*.

```
InteractionLoop(
        [Ig2_Sphere_Sphere_ScGeom(),Ig2_Facet_Sphere_ScGeom()],
        [Ip2_FrictMat_FrictMat_FrictPhys()],
        [Law2_ScGeom_FrictPhys_CundallStrack()],
),
```

hides 3 internal dispatchers within the *InteractionLoop* engine; they all operate on interactions and are, for performance reasons, put together:

**IGeomDispatcher which uses IGeomFunctor** uses the first set of functors (`Ig2`), which are dispatched based on combination of **2** *Shapes* objects. Dispatched functor resolves exact collision configuration and creates an Interaction Geometry *IGeom* (whence `Ig` in the name) associated with the interaction, if there is collision. The functor might as well determine that there is no real collision even if they did overlap in the approximate collision detection (e.g. the *Aabb* did overlap, but the shapes did not). In that case the attribute is set to false and interaction is scheduled for removal.

1. The first functor, *Ig2_Sphere_Sphere_ScGeom*, is called on interaction of 2 *Spheres* and creates *ScGeom* instance, if appropriate.

2. The second functor, *Ig2_Facet_Sphere_ScGeom*, is called for interaction of *Facet* with *Sphere* and might create (again) a *ScGeom* instance.

All `Ig2` functors derive from *IGeomFunctor* (they are documented at the same place).

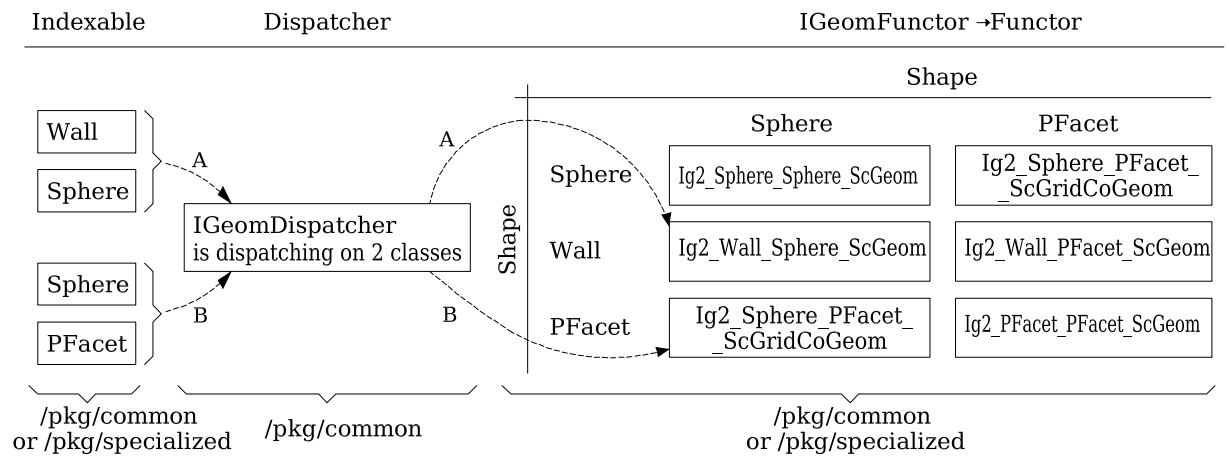

Fig. 5: Example *interaction geometry functors* producing *ScGeom* or *ScGridCoGeom* accepting two various different types (hence **2** in their name `Ig2`), such as *Sphere*, *Wall* or *PFacet*. Each of those functors uses specific geometry of the *Shape* i.e. position of nodes in *PFacet* or *radius of sphere* to calculate the *interaction geometry*.

**IPhysDispatcher** which uses **IPhysFunctor** dispatches to the second set of functors based on combination of 2 *Materials*; these functors return return *IPhys* instance (the Ip prefix). In our case, there is only 1 functor used, *Ip2_FrictMat_FrictMat_FrictPhys*, which create *FrictPhys* from 2 *FrictMat's*.

Ip2 functors are derived from *IPhysFunctor*.

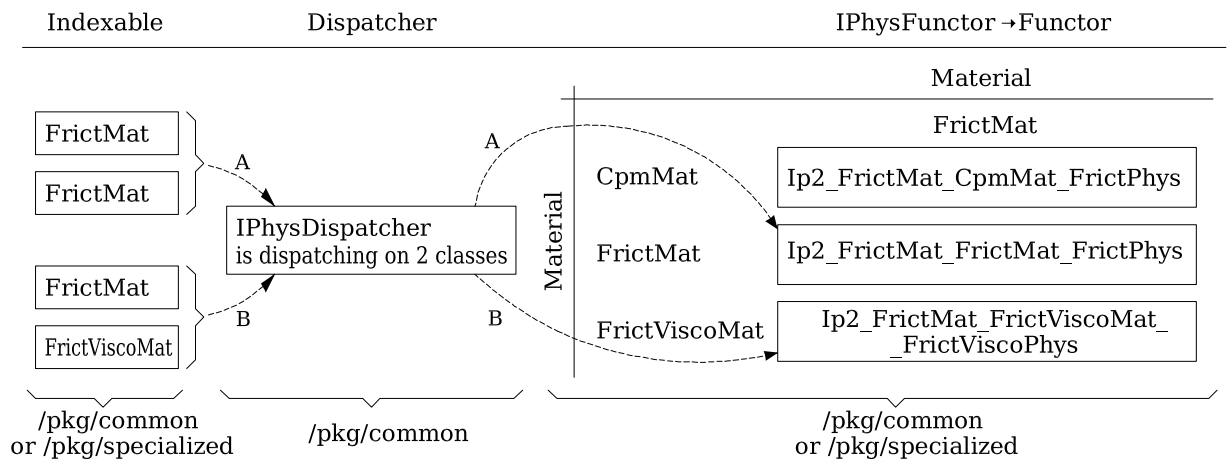

Fig. 6: Example *interaction physics functors* (*Ip2_FrictMat_CpmMat_FrictPhys*, *Ip2_FrictMat_Frict-Mat_FrictPhys* and *Ip2_FrictMat_FrictViscoMat_FrictViscoPhys*) producing *FrictPhys* or *FrictVisco-Phys* accepting two various different types of *Material* (hence Ip2), such as *CpmMat*, *FrictMat* or *FrictViscoMat*.

**LawDispatcher** which uses **LawFunctor** dispatches to the third set of functors, based on combinations of *IGeom* and *IPhys* (wherefore 2 in their name again) of each particular interaction, created by preceding functors. The Law2 functors represent constitutive law; they resolve the interaction by computing forces on the interacting bodies (repulsion, attraction, shear forces, …) or otherwise update interaction state variables.

Law2 functors all inherit from *LawFunctor*.

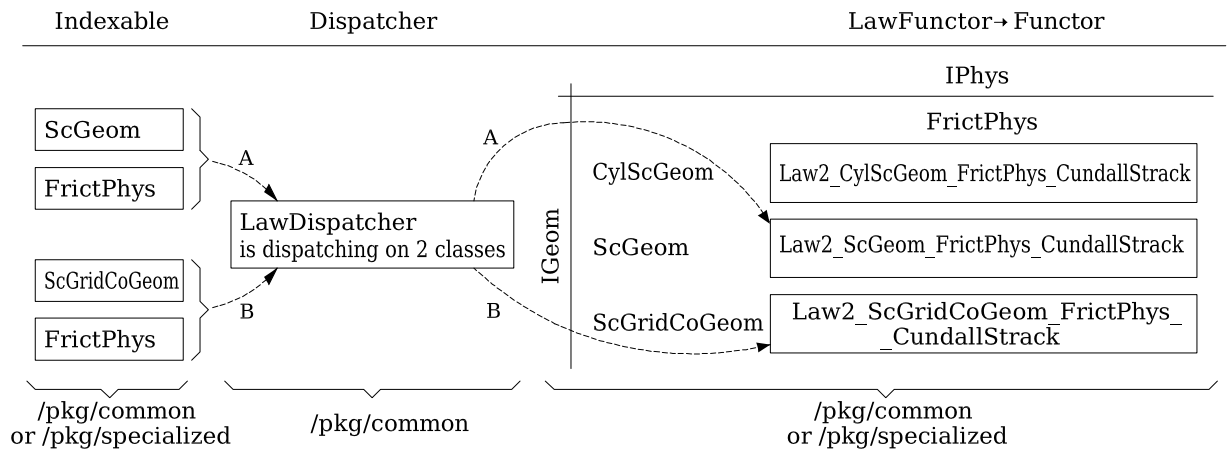

Fig. 7: Example *LawFunctors* (*Law2_CylScGeom_FrictPhys_CundallStrack*, *Law2_ScGeom_Frict-Phys_CundallStrack* and *Law2_ScGridCoGeom_FrictPhys_CundallStrack*) each of them performing calcuation of forces according to selected constitutive law.

There is chain of types produced by earlier functors and accepted by later ones; the user is responsible to satisfy type requirement (see img. *img-dispatch-loop*). An exception (with explanation) is raised in the contrary case.

---

**Note:** When Yade starts, O.engines is filled with a reasonable default list, so that it is not strictly

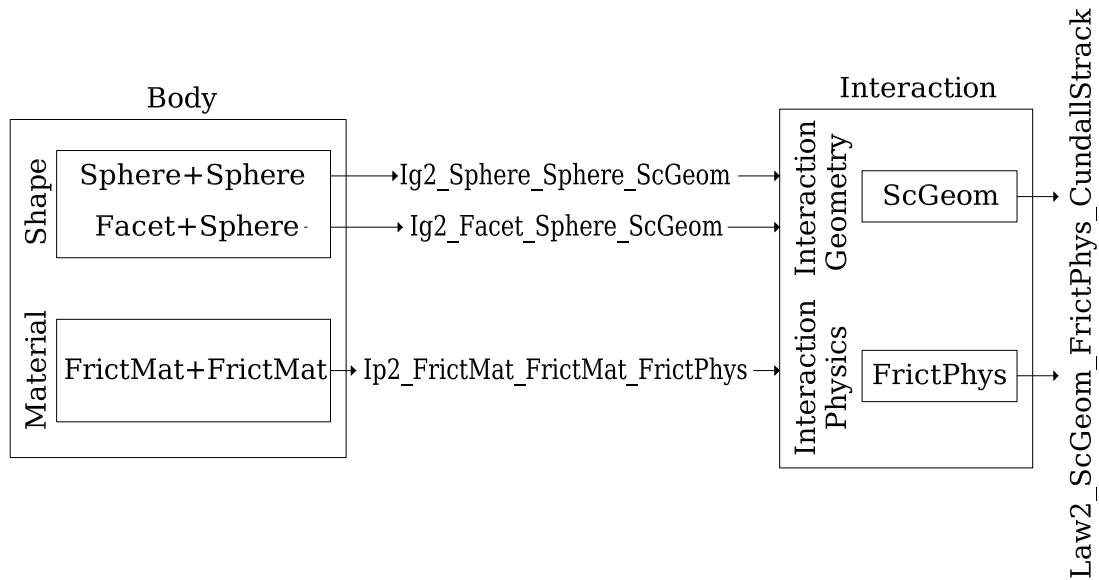

Fig. 8: Chain of functors producing and accepting certain types. In the case shown, the `Ig2` functors produce *ScGeom* instances from all handled *Shapes* combinations; the `Ig2` functor produces *FrictMat*. The constitutive law functor `Law2` accepts the combination of types produced. Note that the types are stated in the functor's class names.

necessary to redefine it when trying simple things. The default scene will handle spheres, boxes, and facets with *frictional* properties correctly, and adjusts the timestep dynamically. You can find an example in examples/simple-scene/simple-scene-default-engines.py.

## 1.2 Tutorial

This tutorial originated as handout for a course held at Technische Universität Dresden / Fakultät Bauingenieurwesen / Institut für Geotechnik in January 2011. The focus was to give quick and rather practical introduction to people without prior modeling experience, but with knowledge of mechanics. Some computer literacy was assumed, though basics are reviewed in the *Hands-on section*.

The course did not in reality follow this document, but was based on interactive writing and commenting simple *Examples*, which were mostly suggested by participants; many thanks to them for their ideas and suggestions.

### 1.2.1 Introduction

The chapter *Introduction* is summarized in following presentation Yade: past, present and future with some additional different examples. This presentation is from year 2011 and does not include latest additions. As of year 2019 it is factually correct.

### 1.2.2 Hands-on

#### Shell basics

#### Directory tree

Directory tree is hierarchical way to organize files in operating systems. A typical (reduced) tree in linux looks like this:

```
/           Root
  boot       System startup
  bin        Low-level programs
  lib        Low-level libraries
  dev        Hardware access
  sbin       Administration programs
  proc       System information
  var        Files modified by system services
  root       Root (administrator) home directory
  etc        Configuration files
  media      External drives
  tmp        Temporary files
  usr        Everything for normal operation (usr = UNIX system resources)
      bin       User programs
      sbin      Administration programs
      include   Header files for c/c++
      lib       Libraries
      local     Locally installed software
      doc       Documentation
  home       Contains the user's home directories
      user      Home directory for user
      user1     Home directory for user1
```

Note that there is a single root `/`; all other disks (such as USB sticks) attach to some point in the tree (e.g. in `/media`).

### Shell navigation

Shell is the UNIX command-line, interface for conversation with the machine. Don't be afraid.

### Moving around

The shell is always operated by some **user**, at some concrete `machine`; these two are constant. We can move in the directory structure, and the current place where we are is *current directory*. By default, it is the *home directory* which contains all files belonging to the respective user:

```
user@machine:~$                        # user operating at machine, in the directory ~ (= user
↪'s home directory)
user@machine:~$ ls .                   # list contents of the current directory
user@machine:~$ ls foo                 # list contents of directory foo, relative to the␣
↪current directory ~ (= ls ~/foo = ls /home/user/foo)
user@machine:~$ ls /tmp                # list contents of /tmp
user@machine:~$ cd foo                 # change directory to foo
user@machine:~/foo$ ls ~               # list home directory (= ls /home/user)
user@machine:~/foo$ cd bar             # change to bar (= cd ~/foo/bar)
user@machine:~/foo/bar$ cd ../../foo2  # go to the parent directory twice, then to foo2 (cd ~/
↪foo/bar/../../foo2 = cd ~/foo2 = cd /home/user/foo2)
user@machine:~/foo2$ cd                # go to the home directory (= ls ~ = ls /home/user)
user@machine:~$
```

Users typically have only permissions to write (i.e. modify files) only in their home directory (abbreviated ~, usually is `/home/user`) and `/tmp`, and permissions to read files in most other parts of the system:

```
user@machine:~$ ls /root    # see what files the administrator has
ls: cannot open directory /root: Permission denied
```

### Keys

Useful keys on the command-line are:

| | |
|---|---|
| \<tab\> | show possible completions of what is being typed (use abundantly!) |
| ^C (=Ctrl+C) | delete current line |
| ^D | exit the shell |
| ↑↓ | move up and down in the command history |
| ^C | interrupt currently running program |
| ^\ | kill currently running program |
| Shift-PgUp | scroll the screen up (show past output) |
| Shift-PgDown | scroll the screen down (show future output; works only on quantum computers) |

### Running programs

When a program is being run (without giving its full path), several directories are searched for program of that name; those directories are given by `$PATH`:

```
user@machine:~$ echo $PATH      # show the value of $PATH
/usr/local/sbin:/usr/local/bin:/usr/sbin:/usr/bin:/sbin:/bin:/usr/games
user@machine:~$ which ls        # say what is the real path of ls
```

The first part of the command-line is the program to be run (`which`), the remaining parts are *arguments* (`ls` in this case). It is up to the program which arguments it understands. Many programs can take special arguments called *options* starting with `-` (followed by a single letter) or `--` (followed by words); one of the common options is `-h` or `--help`, which displays how to use the program (try `ls --help`).

Full documentation for each program usually exists as *manual page* (or *man page*), which can be shown using e.g. `man ls` (`q` to exit)

### Starting yade

If yade is installed on the machine, it can be (roughly speaking) run as any other program; without any arguments, it runs in the "dialog mode", where a command-line is presented:

```
user@machine:~$ yade
Welcome to Yade 2019.01a
TCP python prompt on localhost:9002, auth cookie `adcusk'
XMLRPC info provider on http://localhost:21002
[[ ^L clears screen, ^U kills line. F12 controller, F11 3d view, F10 both, F9 generator, F8↵
↪plot. ]]
Yade [1]:                                ### hit ^D to exit
Do you really want to exit ([y]/n)?
Yade: normal exit.
```

The command-line is in fact `python`, enriched with some yade-specific features. (Pure python interpreter can be run with `python` or `ipython` commands).

Instead of typing commands on-by-one on the command line, they can be be written in a file (with the .py extension) and given as argument to Yade:

```
user@machine:~$ yade simulation.py
```

For a complete help, see `man yade`

**Exercises**

1. Open the terminal, navigate to your home directory
2. Create a new empty file and save it in `~/first.py`
3. Change directory to `/tmp`; delete the file `~/first.py`
4. Run program `xeyes`
5. Look at the help of Yade.
6. Look at the *manual page* of Yade
7. Run Yade, exit and run it again.

**Python basics**

We assume the reader is familar with Python tutorial and only briefly review some of the basic capabilities. The following will run in pure-python interpreter (`python` or `ipython`), but also inside Yade, which is a super-set of Python.

Numerical operations and modules:

```
Yade [1]: (1+3*4)**2         # usual rules for operator precedence, ** is exponentiation
Out[1]: 169

Yade [2]: import math        # gain access to "module" of functions

Yade [3]: math.sqrt(2)       # use a function from that module
Out[3]: 1.4142135623730951

Yade [4]: import math as m   # use the module under a different name

Yade [5]: m.cos(m.pi)
Out[5]: -1.0

Yade [6]: from math import * # import everything so that it can be used without module name

Yade [7]: cos(pi)
Out[7]: -1.0
```

Variables:

```
Yade [8]: a=1; b,c=2,3       # multiple commands separated with ;, multiple assignment

Yade [9]: a+b+c
Out[9]: 6
```

**Sequences**

**Lists**

Lists are variable-length sequences, which can be modified; they are written with braces `[...]`, and their elements are accessed with numerical indices:

```
Yade [10]: a=[1,2,3]         # list of numbers

Yade [11]: a[0]              # first element has index 0
Out[11]: 1
```

(continues on next page)

```
Yade [12]: a[-1]                # negative counts from the end
Out[12]: 3

Yade [13]: a[3]                 # error
---------------------------------------------------------------------------
IndexError                                Traceback (most recent call last)
~/yade/lib/x86_64-linux-gnu/yadeflip/py/yade/__init__.py in <module>
----> 1 a[3]                    # error

IndexError: list index out of range

Yade [14]: len(a)               # number of elements
Out[14]: 3

Yade [15]: a[1:]                # from second element to the end
Out[15]: [2, 3]

Yade [16]: a+=[4,5]             # extend the list

Yade [17]: a+=[6]; a.append(7)  # extend with single value, both have the same effect

Yade [18]: 9 in a               # test presence of an element
Out[18]: False
```

Lists can be created in various ways:

```
Yade [19]: range(10)
Out[19]: range(0, 10)

Yade [20]: range(10)[-1]
Out[20]: 9
```

List of squares of even number smaller than 20, i.e. $\left\{ a^2 \ \forall a \in \{0, \cdots, 19\} \ \big| \ 2 \| a \right\}$ (note the similarity):

```
Yade [21]: [a**2 for a in range(20) if a%2==0]
Out[21]: [0, 4, 16, 36, 64, 100, 144, 196, 256, 324]
```

## Tuples

Tuples are constant sequences:

```
Yade [22]: b=(1,2,3)

Yade [23]: b[0]
Out[23]: 1

Yade [24]: b[0]=4               # error
---------------------------------------------------------------------------
TypeError                                 Traceback (most recent call last)
~/yade/lib/x86_64-linux-gnu/yadeflip/py/yade/__init__.py in <module>
----> 1 b[0]=4                  # error

TypeError: 'tuple' object does not support item assignment
```

## Dictionaries

Mapping from keys to values:

```
Yade [25]: ende={'one':'ein' , 'two':'zwei' , 'three':'drei'}

Yade [26]: de={1:'ein' , 2:'zwei' , 3:'drei'}; en={1:'one' , 2:'two' , 3:'three'}

Yade [27]: ende['one']          ## access values
Out[27]: 'ein'

Yade [28]: de[1], en[2]
Out[28]: ('ein', 'two')
```

**Functions, conditionals**

```
Yade [29]: 4==5
Out[29]: False

Yade [30]: a=3.1

Yade [31]: if a<10:
   ....:      b=-2              # conditional statement
   ....: else:
   ....:      b=3
   ....:

Yade [32]: c=0 if a<1 else 1      # trenary conditional expression

Yade [33]: b,c
Out[33]: (-2, 1)

Yade [34]: def square(x): return x**2    # define a new function
   ....:

Yade [35]: square(2)                      # and call that function
Out[35]: 4
```

**Exercises**

1. Read the following code and say what wil be the values of **a** and **b**:

   ```
   a=range(5)
   b=[(aa**2 if aa%2==0 else -aa**2) for aa in a]
   ```

**Yade basics**

Yade objects are constructed in the following manner (this process is also called "instantiation", since we create concrete instances of abstract classes: one individual sphere is an instance of the abstract *Sphere*, like Socrates is an instance of "man"):

```
Yade [36]: Sphere              # try also Sphere?
Out[36]: yade.wrapper.Sphere

Yade [37]: s=Sphere()          # create a Sphere, without specifying any attributes

Yade [38]: s.radius            # 'nan' is a special value meaning "not a number" (i.e. not
   ↪defined)
Out[38]: nan
```

(continues on next page)

```
Yade [39]: s.radius=2          # set radius of an existing object

Yade [40]: s.radius
Out[40]: 2.0

Yade [41]: ss=Sphere(radius=3)    # create Sphere, giving radius directly

Yade [42]: s.radius, ss.radius      # also try typing s.<tab> to see defined attributes
Out[42]: (2.0, 3.0)
```

### Particles

Particles are the "data" component of simulation; they are the objects that will undergo some processes, though do not define those processes yet.

### Singles

There is a number of pre-defined functions to create particles of certain type; in order to create a sphere, one has to (see the source of *utils.sphere* for instance):

1. Create *Body*

2. Set *Body.shape* to be an instance of *Sphere* with some given radius

3. Set *Body.material* (last-defined material is used, otherwise a default material is created)

4. Set position and orientation in *Body.state*, compute mass and moment of inertia based on *Material* and *Shape*

In order to avoid such tasks, shorthand functions are defined in the *utils* module; to mention a few of them, they are *utils.sphere*, *utils.facet*, *utils.wall*.

```
Yade [43]: s=utils.sphere((0,0,0),radius=1)       # create sphere particle centered at (0,0,0)
↪with radius=1

Yade [44]: s.shape                          # s.shape describes the geometry of the particle
Out[44]: <Sphere instance at 0x3f1cd20>

Yade [45]: s.shape.radius                    # we already know the Sphere class
Out[45]: 1.0

Yade [46]: s.state.mass, s.state.inertia  # inertia is computed from density and geometry
Out[46]:
(4188.790204786391,
 Vector3(1675.516081914556253,1675.516081914556253,1675.516081914556253))

Yade [47]: s.state.pos                        # position is the one we prescribed
Out[47]: Vector3(0,0,0)

Yade [48]: s2=utils.sphere((-2,0,0),radius=1,fixed=True)      # explanation below
```

In the last example, the particle was fixed in space by the **fixed=True** parameter to *utils.sphere*; such a particle will not move, creating a primitive boundary condition.

A particle object is not yet part of the simulation; in order to do so, a special function *O.bodies.append* (also see *Omega::bodies* and *Scene*) is called:

```
Yade [49]: O.bodies.append(s)              # adds particle s to the simulation; returns id of
↪the particle(s) added
Out[49]: 24
```

## Packs

There are functions to generate a specific arrangement of particles in the *pack* module; for instance, cloud (random loose packing) of spheres can be generated with the *pack.SpherePack* class:

```
Yade [50]: from yade import pack

Yade [51]: sp=pack.SpherePack()                        # create an empty cloud; SpherePack contains
 ↪only geometrical information

Yade [52]: sp.makeCloud((1,1,1),(2,2,2),rMean=.2)  # put spheres with defined radius inside box
 ↪given by corners (1,1,1) and (2,2,2)
Out[52]: 6

Yade [53]: for c,r in sp: print(c,r)                   # print center and radius of all particles
 ↪(SpherePack is a sequence which can be iterated over)
   ....:
Vector3(1.274443445943540087,1.367880531586413095,1.275812677797829142) 0.2
Vector3(1.395144492365041344,1.682654071894964964,1.504498259316015663) 0.2
Vector3(1.691901039056097789,1.646021131505697621,1.775464815231047933) 0.2
Vector3(1.698276069049370784,1.239026483132536161,1.307143271925819583) 0.2
Vector3(1.546530002061732745,1.248592120112199888,1.780182798004842359) 0.2
Vector3(1.77882382544176032,1.662755981140189299,1.256899030310328236) 0.2

Yade [54]: sp.toSimulation()                           # create particles and add them to the
 ↪simulation
Out[54]: [25, 26, 27, 28, 29, 30]
```

## Boundaries

*utils.facet* (triangle *Facet*) and *utils.wall* (infinite axes-aligned plane *Wall*) geometries are typically used to define boundaries. For instance, a "floor" for the simulation can be created like this:

```
Yade [55]: O.bodies.append(utils.wall(-1,axis=2))
Out[55]: 31
```

There are other convenience functions (like *utils.facetBox* for creating closed or open rectangular box, or family of *ymport* functions)

## Look inside

The simulation can be inspected in several ways. All data can be accessed from python directly:

```
Yade [56]: len(O.bodies)
Out[56]: 32

Yade [57]: O.bodies[10].shape.radius    # radius of body #10 (will give error if not sphere,
 ↪since only spheres have radius defined)
Out[57]: 0.16

Yade [58]: O.bodies[12].state.pos        # position of body #12
Out[58]: Vector3(1.224408970802669305,1.614102010700998235,1.392374433964128411)
```

Besides that, Yade says this at startup (the line preceding the command-line):

```
[[ ^L clears screen, ^U kills line. F12 controller, F11 3d view, F10 both, F9 generator, F8
 ↪plot. ]]
```

***Controller*** Pressing F12 brings up a window for controlling the simulation. Although typically no human intervention is done in large simulations (which run "headless", without any graphical interaction), it can be handy in small examples. There are basic information on the simulation (will be used later).

***3d view*** The 3d view can be opened with F11 (or by clicking on button in the *Controller* – see below). There is a number of keyboard shortcuts to manipulate it (press h to get basic help), and it can be moved, rotated and zoomed using mouse. Display-related settings can be set in the "Display" tab of the controller (such as whether particles are drawn).

***Inspector*** *Inspector* is opened by clicking on the appropriate button in the *Controller*. It shows (and updates) internal data of the current simulation. In particular, one can have a look at engines, particles (*Bodies*) and interactions (*Interactions*). Clicking at each of the attribute names links to the appropriate section in the documentation.

### Exercises

1. What is this code going to do?

   ```
   Yade [59]: O.bodies.append([utils.sphere((2*i,0,0),1) for i in range(1,20)])
   Out[59]: [32, 33, 34, 35, 36, 37, 38, 39, 40, 41, 42, 43, 44, 45, 46, 47, 48, 49, 50]
   ```

2. Create a simple simulation with cloud of spheres enclosed in the box (0,0,0) and (1,1,1) with mean radius .1. (hint: *pack.SpherePack.makeCloud*)

3. Enclose the cloud created above in box with corners (0,0,0) and (1,1,1); keep the top of the box open. (hint: *utils.facetBox*; type utils.facetBox? or utils.facetBox?? to get help on the command line)

4. Open the 3D view, try zooming in/out; position axes so that z is upwards, y goes to the right and x towards you.

### Engines

Engines define processes undertaken by particles. As we know from the theoretical introduction, the sequence of engines is called *simulation loop*. Let us define a simple interaction loop:

```
Yade [60]: O.engines=[                              # newlines and indentations are not important until
     ↪the brace is closed
    ....:     ForceResetter(),
    ....:     InsertionSortCollider([Bo1_Sphere_Aabb(),Bo1_Wall_Aabb()]),
    ....:     InteractionLoop(                       # dtto for the parenthesis here
    ....:         [Ig2_Sphere_Sphere_ScGeom(),Ig2_Wall_Sphere_ScGeom()],
    ....:         [Ip2_FrictMat_FrictMat_FrictPhys()],
    ....:         [Law2_ScGeom_FrictPhys_CundallStrack()]
    ....:     ),
    ....:     NewtonIntegrator(damping=.2,label='newtonCustomLabel')    # define a label
     ↪newtonCustomLabel under which we can access this engine easily
    ....: ]
    ....:

Yade [61]: O.engines
Out[61]:
[<ForceResetter instance at 0x3cc9610>,
 <InsertionSortCollider instance at 0x2ac5490>,
 <InteractionLoop instance at 0x3c3c3d0>,
 <NewtonIntegrator instance at 0x37bfa50>]

Yade [62]: O.engines[-1]==newtonCustomLabel       # is it the same object?
Out[62]: True
```

(continues on next page)

```
Yade [63]: newtonCustomLabel.damping
Out[63]: 0.2
```

Instead of typing everything into the command-line, one can describe simulation in a file (*script*) and then run yade with that file as an argument. We will therefore no longer show the command-line unless necessary; instead, only the script part will be shown. Like this:

```
O.engines=[                        # newlines and indentations are not important until the brace is
↪closed
        ForceResetter(),
        InsertionSortCollider([Bo1_Sphere_Aabb(),Bo1_Wall_Aabb()]),
        InteractionLoop(               # dtto for the parenthesis here
                [Ig2_Sphere_Sphere_ScGeom(),Ig2_Wall_Sphere_ScGeom()],
                [Ip2_FrictMat_FrictMat_FrictPhys()],
                [Law2_ScGeom_FrictPhys_CundallStrack()]
        ),
        GravityEngine(gravity=(0,0,-9.81)),                    # 9.81 is the gravity
↪acceleration, and we say that
        NewtonIntegrator(damping=.2,label='newtonCustomLabel') # define a label under which we
↪can access this engine easily
]
```

Besides engines being run, it is likewise important to define how often they will run. Some engines can run only sometimes (we will see this later), while most of them will run always; the time between two successive runs of engines is *timestep* (Δt). There is a mathematical limit on the timestep value, called *critical timestep*, which is computed from properties of particles. Since there is a function for that, we can just set timestep using *utils.PWaveTimeStep*:

```
O.dt=utils.PWaveTimeStep()
```

Each time when the simulation loop finishes, time `O.time` is advanced by the timestep `O.dt`:

```
Yade [64]: O.dt=0.01

Yade [65]: O.time
Out[65]: 0.0

Yade [66]: O.step()

Yade [67]: O.time
Out[67]: 0.01
```

For experimenting with a single simulations, it is handy to save it to memory; this can be achieved, once everything is defined, with:

```
O.saveTmp()
```

### Exercises

1. Define *engines* as in the above example, run the *Inspector* and click through the engines to see their sequence.

2. Write a simple script which will

    1. define particles as in the previous exercise (cloud of spheres inside a box open from the top)

    2. define a simple simulation loop, as the one given above

    3. set Δt equal to the critical P-Wave Δt

    4. save the initial simulation state to memory

3. Run the previously-defined simulation multiple times, while changing the value of timestep (use the   button to reload the initial configuration).

    1. See what happens as you increase Δt above the P-Wave value.

    2. Try changing the *gravity* parameter, before running the simulation.

    3. Try changing *damping*

4. Reload the simulation, open the 3d view, open the *Inspector*, select a particle in the 3d view (shift-click). Then run the simulation and watch how forces on that particle change; pause the simulation somewhere in the middle, look at interactions of this particle.

5. At which point can we say that the deposition is done, so that the simulation can be stopped?

**See also:**

The *Bouncing sphere* example shows a basic simulation.

### 1.2.3 Data mining

**Read**

**Local data**

All data of the simulation are accessible from python; when you open the *Inspector*, blue labels of various data can be clicked – left button for getting to the documentation, middle click to copy the name of the object (use `Ctrl-V` or middle-click to paste elsewhere). The interesting objects are among others (see *Omega* for a full list):

1. *O.engines*

    Engines are accessed by their index (position) in the simulation loop:

```
O.engines[0]        # first engine
O.engines[-1]       # last engine
```

    **Note:** The index can change if *O.engines* is modified. *Labeling* introduced in the section below is a better solution for reliable access to a particular engine.

2. *O.bodies*

    Bodies are identified by their *id*, which is guaranteed to not change during the whole simulation:

```
O.bodies[0]                                                          # first body
[b.shape.radius for b in O.bodies if isinstance(b.shape,Sphere)]     # list of radii of
↪all spherical bodies
sum([b.state.mass for b in O.bodies])                                # sum of masses of
↪all bodies
numpy.average([b.state.vel[0] for b in O.bodies])                    # average velocity in
↪x direction
```

    **Note:** Uniqueness of *Body.id* is not guaranteed, since newly created bodies might recycle *ids* of *deleted* ones.

3. *O.forces*

Generalized forces (forces, torques) acting on each particle. They are (usually) reset at the beginning of each step with *ForceResetter*, subsequently forces from individual interactions are accumulated in *InteractionLoop*. To access the data, use:

```
O.forces.f(0)        # force on #0
O.forces.t(1)        # torque on #1
```

4. *O.interactions*

Interactions are identified by *ids* of the respective interacting particles (they are created and deleted automatically during the simulation):

```
O.interactions[0,1]       # interactions of #0 with #1
O.interactions[1,0]       # the same object
O.bodies[0].intrs()       # all interactions of body #0
for i in O.bodies[12].intrs(): print (i.isReal,i.id1,i.id2)        # get some info about
↪interactions of body #12
[(i.isReal,i.id1,i.id2) for i in O.bodies[12].intrs()]             # same thing, but make a
↪list
```

## Labels

*Engines* and *functors* can be *labeled*, which means that python variable of that name is automatically created.

```
Yade [1]: O.engines=[
   ...:      NewtonIntegrator(damping=.2,label='newtonCustomLabel')
   ...: ]
   ...:

Yade [2]: newtonCustomLabel.damping=.4

Yade [3]: O.engines[0].damping              # O.engines[0] and newtonCustomLabel are the same
↪objects
Out[3]: 0.4

Yade [4]: newtonCustomLabel==O.engines[0]   # O.engines[0] and newtonCustomLabel are the same
↪objects
Out[4]: True
```

## Exercises

1. Find meaning of this expression:

```
max([b.state.vel.norm() for b in O.bodies])
```

2. Run the *Gravity deposition* script, pause after a few seconds of simulation. Write expressions that compute

   1. kinetic energy $\sum \frac{1}{2} m_i |v_i|^2$
   2. average mass (hint: use numpy.average)
   3. maximum $z$-coordinate of all particles
   4. number of interactions of body #1

## Global data

Useful measures of what happens in the simulation globally:

**unbalanced force** ratio of maximum contact force and maximum per-body force; measure of staticity, computed with *unbalancedForce*.

**porosity** ratio of void volume and total volume; computed with *porosity*.

**coordination number** average number of interactions per particle, *avgNumInteractions*

**stress tensor (periodic boundary conditions)** averaged force in interactions, computed with *normalShearStressTensors*

**fabric tensor** distribution of contacts in space (not yet implemented); can be visualized with *plotDirections*

### Energies

Evaluating energy data for all components in the simulation (such as gravity work, kinetic energy, plastic dissipation, damping dissipation) can be enabled with

```
O.trackEnergy=True
```

Subsequently, energy values are accessible in the *O.energy*; it is a dictionary where its entries can be retrived with `keys()` and their values with `O.energy[key]`.

### Save

### PyRunner

To save data that we just learned to access, we need to call Python from within the *simulation loop*. *PyRunner* is created just for that; it inherits periodicy control from *PeriodicEngine* and takes the code to run as text (must be quoted, i.e. inside `'...'`) attribute called *command*. For instance, adding this to *O.engines* will print the current step number every one second wall clock time:

```
O.engines=O.engines+[ PyRunner(command='print(O.iter)',realPeriod=1) ]
```

Writing complicated code inside *command* is awkward; in such case, we define a function that will be called:

```
def myFunction():
        '''Print step number, and pause the simulation is unbalanced force is smaller than 0.
 ↪05.'''
        print(O.iter)
        if utils.unbalancedForce()<0.05:
                print('Unbalanced force is smaller than 0.05, pausing.')
                O.pause()
```

Now this function can be added to *O.engines*:

```
O.engines+=[PyRunner(command='myFunction()',iterPeriod=100)]
```

or, in general, like that:

```
O.engines=[
        # ...
        PyRunner(command='myFunction()',iterPeriod=100) # call myFunction every 100 steps
]
```

---

**Warning:** If a function was declared inside a *live* yade session (ipython) and PyRunner attribute *updateGlobals is set to False* then an error `NameError: name 'myFunction' is not defined` will occur unless python globals() are updated with command

---

```
globals().update(locals())
```

**Exercises**

1. Run the *Gravity deposition* simulation, but change it such that:

   1. *utils.unbalancedForce* is printed every 2 seconds.

   2. check every 1000 steps the value of unbalanced force

      - if smaller than 0.2, set *damping* to 0.8 (hint: use labels)

      - if smaller than 0.1, pause the simulation

**Keeping history**

Yade provides the *plot* module used for storing and plotting variables (plotting itself will be discussed later). Let us start by importing this module and declare variable names that will be plotted:

```
from yade import plot
plot.plots={'t':('coordNum','unForce',None,'Ek')}        # kinetic energy will have
↪legend on the right as indicated by None separator.
```

Periodic storing of data is done with *PyRunner* and the *plot.addData* function. Also let's enable energy tracking:

```
O.trackEnergy=True
def addPlotData():
        # this function adds current values to the history of data, under the names specified
        plot.addData(t=O.time,Ek=utils.kineticEnergy(),coordNum=utils.avgNumInteractions(),
↪unForce=utils.unbalancedForce())
```

Now this function can be added to *O.engines*:

```
O.engines+=[PyRunner(command='addPlotData()',iterPeriod=20)]
```

or, in general, like that:

```
O.engines=[  # ...,
        PyRunner(command='addPlotData()',iterPeriod=20)        # call the addPlotData
↪function every 20 iterations
]
```

History is stored in *plot.data*, and can be accessed using the variable name, e.g. `plot.data['Ek']`, and saved to text file (for post-processing outside yade) with *plot.saveDataTxt*.

**Plot**

*plot* provides facilities for plotting history saved with *plot.addData* as 2d plots. Data to be plotted are specified using dictionary *plot.plots*

```
plot.plots={'t':('coordNum','unForce',None,'Ek')}
```

History of all values is given as the name used for *plot.addData*; keys of the dictionary are x-axis values, and values are sequence of data on the `y` axis; the `None` separates data on the left and right axes (they are scaled independently). The plot itself is created with

```
plot.plot()              # on the command line, F8 can be used as shorthand
```

While the plot is open, it will be updated periodically, so that simulation evolution can be seen in real-time.

**Energy plots**

Plotting all energy contributions would be difficult, since names of all energies might not be known in advance. Fortunately, there is a way to handle that in Yade. It consists in two parts:

1. *plot.addData* is given all the energies that are currently defined:

   ```
   plot.addData(i=O.iter,total=O.energy.total(),**O.energy)
   ```

   The *O.energy.total* functions, which sums all energies together. The `**O.energy` is special python syntax for converting dictionary (remember that *O.energy* is a dictionary) to named functions arguments, so that the following two commands are identical:

   ```
   function(a=3,b=34)           # give arguments as arguments
   function(**{'a':3,'b':34})   # create arguments from dictionary
   ```

2. Data to plot are specified using a *function* that gives names of data to plot, rather than providing the data names directly:

   ```
   plot.plots={'i':['total']+O.energy.keys()}
   ```

   where `total` is the name we gave to `O.energy.total()` above, while `O.energy.keys()` will always return list of currently defined energies.

**Energy plot example**

Plotting energies inside a *live* yade session, for example by launching examples/test/triax-basic-without-plots.py would look following:

```
from yade import plot
O.trackEnergy=True
O.step()                     # performing a single simulation step is necessary to␣
↪populate O.energy.keys()
plot.plots={'t':O.energy.keys()+['total']}

def addPlotData():
        # this function adds current values to the history of data, under the names specified
        plot.addData( t=O.time , total=O.energy.total() , **O.energy )

O.engines+=[PyRunner(command='addPlotData()',iterPeriod=20)]

globals().update(locals())   # do this only because this is an example of a live yade␣
↪session
```

Press F8 to show plot window and F11 to show 3D view, then press   to start simulation.

**Using multiple plots**

It is also possible to make several separate plots, for example like this:

```
plot.plots={ 't':('total','kinetic') , 't ':['elastPotential','gravWork'] , 't  ':('nonviscDamp
↪') }
```

---

> **Warning:** There cannot be duplicate names declared in separate plots. This is why spaces were used above to indicate the same variable **t**.

---

With the caveat above, a following example inside a *live* yade session launched on examples/test/triax-basic-without-plots.py would look following:

```
from yade import plot
O.trackEnergy=True
plot.plots={ 't':('total','kinetic') , 't ':['elastPotential','gravWork'] , 't  ':('nonviscDamp
↪') }

def addPlotData():
        # assign value to all three: 't', 't ' and 't  ' with single t=... assignment
        plot.addData( t=O.time , total=O.energy.total() , **O.energy )

O.engines+=[PyRunner(command='addPlotData()',iterPeriod=20)]

globals().update(locals())       # do this only because this is an example of a live yade␣
↪session

plot.plot(subPlots=False)        # show plots in separate windows

plot.plot(subPlots=True)         # same as pressing F8: close current plot windows and reopen␣
↪a single new one
```

Press F8 to show plot window and F11 to show 3D view, then press   to start simulation, see video below:

**Exercises**

1. Calculate average momentum in y direction.

2. Run the *Gravity deposition* script, plotting unbalanced force and kinetic energy.

3. While the script is running, try changing the *NewtonIntegrator.damping* parameter (do it from both **Inspector** and from the command-line). What influence does it have on the evolution of unbalanced force and kinetic energy?

4. Think about and write down all energy sources (input); write down also all energy sinks (dissipation).

5. Simulate *Gravity deposition* and plot all energies as they evolve during the simulation.

**See also:**

Most *Examples with tutorial* use plotting facilities of Yade, some of them also track energy of the simulation.

## 1.2.4 Setting up a simulation

**See also:**

Examples *Gravity deposition*, *Oedometric test*, *Periodic simple shear*, *Periodic triaxial test* deal with topics discussed here.

**Parametric studies**

Input parameters of the simulation (such as size distribution, damping, various contact parameters, …) influence the results, but frequently an analytical relationship is not known. To study such influence, similar simulations differing only in a few parameters can be run and results compared. Yade can be run in *batch mode*, where one simulation script is used in conjunction with *parameter table*, which specifies

---

parameter values for each run of the script. Batch simulation are run non-interactively, i.e. without user intervention; the user must therefore start and stop the simulation explicitly.

Suppose we want to study the influence of *damping* on the evolution of kinetic energy. The script has to be adapted at several places:

1. We have to make sure the script reads relevant parameters from the *parameter table*. This is done using *utils.readParamsFromTable*; the parameters which are read are created as variables in the `yade.params.table` module:

```
readParamsFromTable(damping=.2)        # yade.params.table.damping variable will be created
from yade.params import table           # typing table.damping is easier than yade.
↪params.table.damping
```

Note that *utils.readParamsFromTable* takes default values of its parameters, which are used if the script is not run in non-batch mode.

2. Parameters from the table are used at appropriate places:

```
NewtonIntegrator(damping=table.damping),
```

3. The simulation is run non-interactively; we must therefore specify at which point it should stop:

```
O.engines+=[PyRunner(iterPeriod=1000,command='checkUnbalancedForce()')]     # call our
↪function defined below periodically

def checkUnbalancedForce():
    if unbalancedForce<0.05:                        # exit Yade if unbalanced force
↪drops below 0.05
        plot.saveDataTxt(O.tags['d.id']+'.data.bz2')      # save all data into a unique file
↪before exiting
        import sys
        sys.exit(0)                                # exit the program
```

4. Finally, we must start the simulation at the very end of the script:

```
O.run()                 # run forever, until stopped by checkUnbalancedForce()
waitIfBatch()           # do not finish the script until the simulation ends; does nothing
↪in non-batch mode
```

The *parameter table* is a simple text-file (e.g. `params.txt` ), where each line specifies a simulation to run:

```
# comments start with # as in python
damping      # first non-comment line is variable name
.2
.4
.6
```

Finally, the simulation is run using the special batch command:

```
user@machine:~$ yade-batch params.txt simulation.py
```

### Exercises

1. Run the *Gravity deposition* script in batch mode, varying *damping* to take values of `.2`, `.4`, `.6`.

2. See the http://localhost:9080 overview page while the batch is running (fig. *imgBatchExample*).

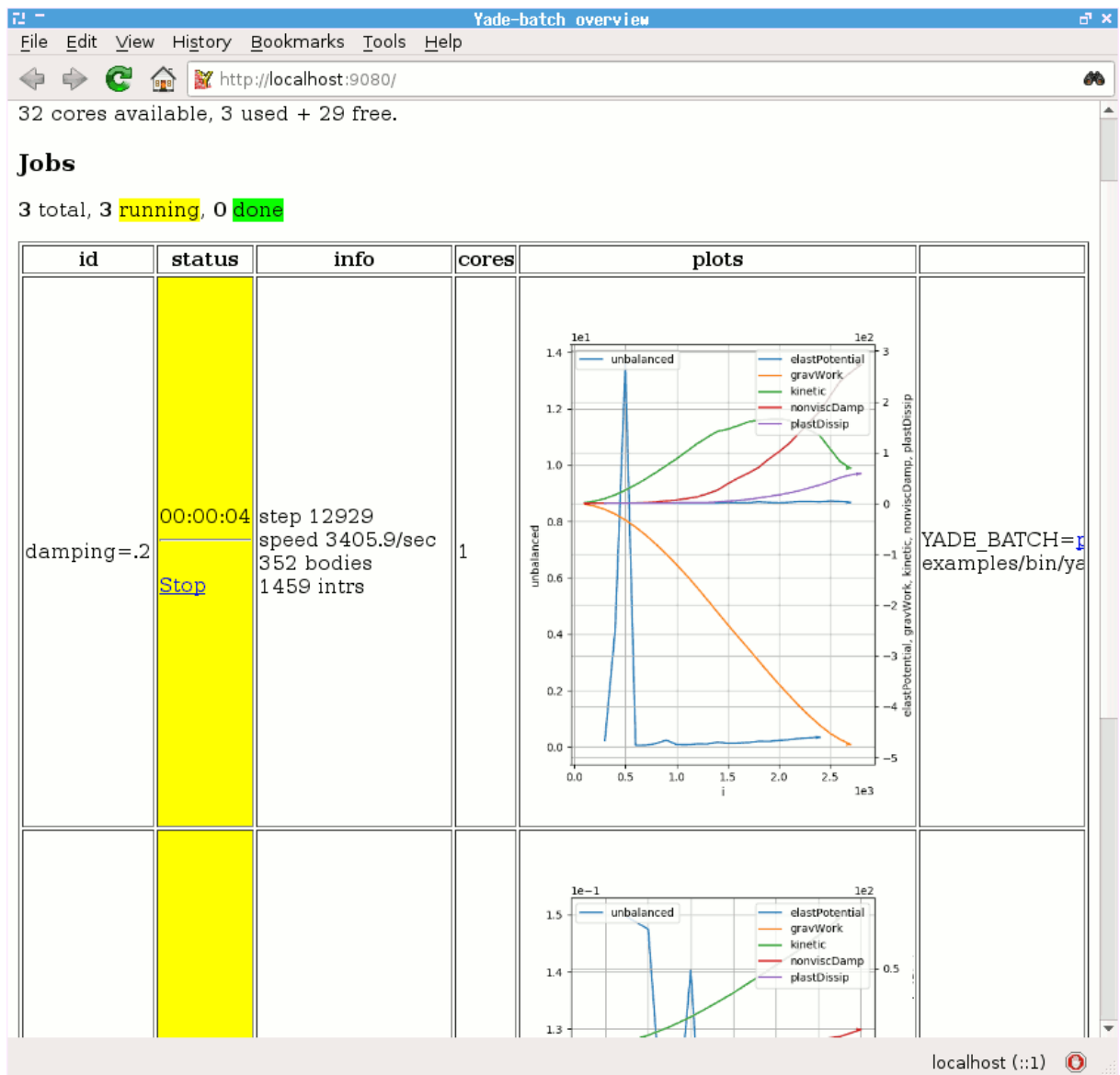

## Boundary

Particles moving in infinite space usually need some constraints to make the simulation meaningful.

## Supports

So far, supports (unmovable particles) were providing necessary boundary: in the *Gravity deposition* script the *geom.facetBox* is internally composed of *facets* (triangulation elements), which are `fixed` in space; facets are also used for arbitrary triangulated surfaces (see relevant sections of the *User's manual*). Another frequently used boundary is *utils.wall* (infinite axis-aligned plane).

## Periodic

Periodic boundary is a "boundary" created by using periodic (rather than infinite) space. Such boundary is activated by *O.periodic=True* , and the space configuration is decribed by *O.cell* . It is well suited for studying bulk material behavior, as boundary effects are avoided, leading to smaller number of particles. On the other hand, it might not be suitable for studying localization, as any cell-level effects (such as shear bands) have to satisfy periodicity as well.

The periodic cell is described by its *reference size* of box aligned with global axes, and *current transformation*, which can capture stretch, shear and rotation. Deformation is prescribed via *velocity gradient*, which updates the *transformation* before the next step. *Homothetic deformation* can smear *velocity gradient* across the cell, making the boundary dissolve in the whole cell.

Stress and strains can be controlled with *PeriTriaxController*; it is possible to prescribe mixed strain/stress *goal* state using *PeriTriaxController.stressMask*.

The following creates periodic cloud of spheres and compresses to achieve σ$_x$=-10 kPa, σ$_y$=-10kPa and ε$_z$=-0.1. Since stress is specified for y and z, *stressMask* is binary 0b011 (x→1, y→2, z→4, in decimal 1+2=3).

```
Yade [1]: sp=pack.SpherePack()

Yade [2]: sp.makeCloud((1,1,1),(2,2,2),rMean=.16,periodic=True)
Out[2]: 20

Yade [3]: sp.toSimulation()                    # implicitly sets O.periodic=True, and O.cell.refSize
 ↪to the packing period size
Out[3]: [4, 5, 6, 7, 8, 9, 10, 11, 12, 13, 14, 15, 16, 17, 18, 19, 20, 21, 22, 23]

Yade [4]: O.engines+=[PeriTriaxController(goal=(-1e4,-1e4,-.1),stressMask=0b011,maxUnbalanced=.
 ↪2,doneHook='functionToRunWhenFinished()')]
```

When the simulation *runs*, *PeriTriaxController* takes over the control and calls *doneHook* when *goal* is reached. A full simulation with PeriTriaxController might look like the following:

```python
from __future__ import print_function
from yade import pack, plot
sp = pack.SpherePack()
rMean = .05
sp.makeCloud((0, 0, 0), (1, 1, 1), rMean=rMean, periodic=True)
sp.toSimulation()
O.engines = [
        ForceResetter(),
        InsertionSortCollider([Bo1_Sphere_Aabb()], verletDist=.05 * rMean),
        InteractionLoop([Ig2_Sphere_Sphere_ScGeom()], [Ip2_FrictMat_FrictMat_FrictPhys()],
 ↪[Law2_ScGeom_FrictPhys_CundallStrack()]),
        NewtonIntegrator(damping=.6),
        PeriTriaxController(
                goal=(-1e6, -1e6, -.1), stressMask=0b011, maxUnbalanced=.2, doneHook=
 ↪'goalReached()', label='triax', maxStrainRate=(.1, .1, .1), dynCell=True
        ),
        PyRunner(iterPeriod=100, command='addPlotData()')
]
O.dt = .5 * utils.PWaveTimeStep()
O.trackEnergy = True

def goalReached():
        print('Goal reached, strain', triax.strain, ' stress', triax.stress)
        O.pause()

def addPlotData():
        plot.addData(
                sx=triax.stress[0],
                sy=triax.stress[1],
                sz=triax.stress[2],
                ex=triax.strain[0],
                ey=triax.strain[1],
                ez=triax.strain[2],
                i=O.iter,
```

```
                unbalanced=utils.unbalancedForce(),
                totalEnergy=O.energy.total(),
                **O.energy  # plot all energies
        )

plot.plots = {
        'i': (('unbalanced', 'go'), None, 'kinetic'),
        ' i': ('ex', 'ey', 'ez', None, 'sx', 'sy', 'sz'),
        'i ': (O.energy.keys, None, ('totalEnergy', 'bo'))
}
plot.plot()
O.saveTmp()
O.run()
```

## 1.2.5 Advanced & more

### Particle size distribution

See *Periodic triaxial test* and examples/test/psd.py

### Clumps

*Clump*; see *Periodic triaxial test*

### Testing laws

*LawTester*, scripts/checks-and-tests/law-test.py

### New law

### Visualization

See the example *3d-postprocessing and video recording*

- *VTKRecorder* & Paraview
- *makeVideo*
- *SnapshotEngine*
- doc/sphinx/tutorial/05-3d-postprocessing.py
- examples/test/force-network-video.py
- doc/sphinx/tutorial/make-simulation-video.py

### Convert python 2 scripts to python 3

Below is a non-exhaustive list of common things to do to convert your scripts to python 3.

**Mandatory:**

- `print ...` becomes `print(...)`,
- `myDict.iterkeys()`, `myDict.itervalues()`, `myDict.iteritems()` becomes `myDict.keys()`, `myDict.values()`, `myDict.items()`,
- `import cPickle` becomes `import pickle`,
- `''` and `<>` operators are no longer recognized,
- inconsistent use of tabs and spaces in indentation is prohibited, for this reason all scripts in yade use tabs for indentation.

**Should be checked, but not always mandatory:**

- (euclidian) division of two integers: `i1/i2` becomes `i1//i2`,
- `myDict.keys()`, `myDict.values()`, `myDict.items()` becomes sometimes `list(myDict.keys())`, `list(myDict.values())`, `list(myDict.items())` (depending on your usage),
- `map()`, `filter()`, `zip()` becomes sometimes `list(map())`, `list(filter())`, `list(zip())` (depending on your usage),
- string encoding is now UTF8 everywhere, it may cause problems on user inputs/outputs (keyboard, file...) with special chars.

**Optional:**

- `# encoding: utf-8` no longer needed

## 1.2.6 Examples with tutorial

The online version of this tutorial contains embedded videos.

**Bouncing sphere**

Following example is in file doc/sphinx/tutorial/01-bouncing-sphere.py.

```python
# basic simulation showing sphere falling ball gravity,
# bouncing against another sphere representing the support

# DATA COMPONENTS

# add 2 particles to the simulation
# they the default material (utils.defaultMat)
O.bodies.append(
        [
                # fixed: particle's position in space will not change (support)
                sphere(center=(0, 0, 0), radius=.5, fixed=True),
                # this particles is free, subject to dynamics
                sphere((0, 0, 2), .5)
        ]
)

# FUNCTIONAL COMPONENTS

# simulation loop -- see presentation for the explanation
O.engines = [
```

(continues on next page)

```
        ForceResetter(),
        InsertionSortCollider([Bo1_Sphere_Aabb()]),
        InteractionLoop(
                [Ig2_Sphere_Sphere_ScGeom()],  # collision geometry
                [Ip2_FrictMat_FrictMat_FrictPhys()],  # collision "physics"
                [Law2_ScGeom_FrictPhys_CundallStrack()]  # contact law -- apply forces
        ),
        # Apply gravity force to particles. damping: numerical dissipation of energy.
        NewtonIntegrator(gravity=(0, 0, -9.81), damping=0.1)
]

# set timestep to a fraction of the critical timestep
# the fraction is very small, so that the simulation is not too fast
# and the motion can be observed
O.dt = .5e-4 * PWaveTimeStep()

# save the simulation, so that it can be reloaded later, for experimentation
O.saveTmp()
```

## Gravity deposition

Following example is in file doc/sphinx/tutorial/02-gravity-deposition.py.

```
# gravity deposition in box, showing how to plot and save history of data,
# and how to control the simulation while it is running by calling
# python functions from within the simulation loop

# import yade modules that we will use below
from yade import pack, plot

# create rectangular box from facets
O.bodies.append(geom.facetBox((.5, .5, .5), (.5, .5, .5), wallMask=31))

# create empty sphere packing
# sphere packing is not equivalent to particles in simulation, it contains only the pure
# ↪geometry
sp = pack.SpherePack()
# generate randomly spheres with uniform radius distribution
sp.makeCloud((0, 0, 0), (1, 1, 1), rMean=.05, rRelFuzz=.5)
# add the sphere pack to the simulation
sp.toSimulation()

O.engines = [
        ForceResetter(),
        InsertionSortCollider([Bo1_Sphere_Aabb(), Bo1_Facet_Aabb()]),
        InteractionLoop(
                # handle sphere+sphere and facet+sphere collisions
                [Ig2_Sphere_Sphere_ScGeom(), Ig2_Facet_Sphere_ScGeom()],
                [Ip2_FrictMat_FrictMat_FrictPhys()],
                [Law2_ScGeom_FrictPhys_CundallStrack()]
        ),
        NewtonIntegrator(gravity=(0, 0, -9.81), damping=0.4),
        # call the checkUnbalanced function (defined below) every 2 seconds
        PyRunner(command='checkUnbalanced()', realPeriod=2),
        # call the addPlotData function every 200 steps
        PyRunner(command='addPlotData()', iterPeriod=100)
]
O.dt = .5 * PWaveTimeStep()

# enable energy tracking; any simulation parts supporting it
```

```python
# can create and update arbitrary energy types, which can be
# accessed as O.energy['energyName'] subsequently
O.trackEnergy = True

# if the unbalanced forces goes below .05, the packing
# is considered stabilized, therefore we stop collected
# data history and stop
def checkUnbalanced():
        if unbalancedForce() < .05:
                O.pause()
                plot.saveDataTxt('bbb.txt.bz2')
                # plot.saveGnuplot('bbb') is also possible

# collect history of data which will be plotted
def addPlotData():
        # each item is given a names, by which it can be the unsed in plot.plots
        # the **O.energy converts dictionary-like O.energy to plot.addData arguments
        plot.addData(i=O.iter, unbalanced=unbalancedForce(), **O.energy)

# define how to plot data: 'i' (step number) on the x-axis, unbalanced force
# on the left y-axis, all energies on the right y-axis
# (O.energy.keys is function which will be called to get all defined energies)
# None separates left and right y-axis
plot.plots = {'i': ('unbalanced', None, O.energy.keys)}

# show the plot on the screen, and update while the simulation runs
plot.plot()

O.saveTmp()
```

**Oedometric test**

Following example is in file doc/sphinx/tutorial/03-oedometric-test.py.

```python
# gravity deposition, continuing with oedometric test after stabilization
# shows also how to run parametric studies with yade-batch

# The components of the batch are:
# 1. table with parameters, one set of parameters per line (ccc.table)
# 2. readParamsFromTable which reads respective line from the parameter file
# 3. the simulation muse be run using yade-batch, not yade
#
# $ yade-batch --job-threads=1 03-oedometric-test.table 03-oedometric-test.py
#

# load parameters from file if run in batch
# default values are used if not run from batch
readParamsFromTable(rMean=.05, rRelFuzz=.3, maxLoad=1e6, minLoad=1e4)
# make rMean, rRelFuzz, maxLoad accessible directly as variables later
from yade.params.table import *

# create box with free top, and ceate loose packing inside the box
from yade import pack, plot
O.bodies.append(geom.facetBox((.5, .5, .5), (.5, .5, .5), wallMask=31))
sp = pack.SpherePack()
sp.makeCloud((0, 0, 0), (1, 1, 1), rMean=rMean, rRelFuzz=rRelFuzz)
sp.toSimulation()
```

```python
O.engines = [
        ForceResetter(),
        # sphere, facet, wall
        InsertionSortCollider([Bo1_Sphere_Aabb(), Bo1_Facet_Aabb(), Bo1_Wall_Aabb()]),
        InteractionLoop(
                # the loading plate is a wall, we need to handle sphere+sphere, sphere+facet,
                ↪sphere+wall
                [Ig2_Sphere_Sphere_ScGeom(), Ig2_Facet_Sphere_ScGeom(), Ig2_Wall_Sphere_
                ↪ScGeom()],
                [Ip2_FrictMat_FrictMat_FrictPhys()],
                [Law2_ScGeom_FrictPhys_CundallStrack()]
        ),
        NewtonIntegrator(gravity=(0, 0, -9.81), damping=0.5),
        # the label creates an automatic variable referring to this engine
        # we use it below to change its attributes from the functions called
        PyRunner(command='checkUnbalanced()', realPeriod=2, label='checker'),
]
O.dt = .5 * PWaveTimeStep()

# the following checkUnbalanced, unloadPlate and stopUnloading functions are all called by the
↪'checker'
# (the last engine) one after another; this sequence defines progression of different stages
↪of the
# simulation, as each of the functions, when the condition is satisfied, updates 'checker' to
↪call
# the next function when it is run from within the simulation next time

# check whether the gravity deposition has already finished
# if so, add wall on the top of the packing and start the oedometric test
def checkUnbalanced():
        # at the very start, unbalanced force can be low as there is only few contacts, but it
        ↪does not mean the packing is stable
        if O.iter < 5000:
                return
        # the rest will be run only if unbalanced is < .1 (stabilized packing)
        if unbalancedForce() > .1:
                return
        # add plate at the position on the top of the packing
        # the maximum finds the z-coordinate of the top of the topmost particle
        O.bodies.append(wall(max([b.state.pos[2] + b.shape.radius for b in O.bodies if
        ↪isinstance(b.shape, Sphere)]), axis=2, sense=-1))
        global plate  # without this line, the plate variable would only exist inside this
        ↪function
        plate = O.bodies[-1]  # the last particles is the plate
        # Wall objects are "fixed" by default, i.e. not subject to forces
        # prescribing a velocity will therefore make it move at constant velocity (downwards)
        plate.state.vel = (0, 0, -.1)
        # start plotting the data now, it was not interesting before
        O.engines = O.engines + [PyRunner(command='addPlotData()', iterPeriod=200)]
        # next time, do not call this function anymore, but the next one (unloadPlate) instead
        checker.command = 'unloadPlate()'

def unloadPlate():
        # if the force on plate exceeds maximum load, start unloading
        if abs(O.forces.f(plate.id)[2]) > maxLoad:
                plate.state.vel *= -1
                # next time, do not call this function anymore, but the next one
                ↪(stopUnloading) instead
```

```
                checker.command = 'stopUnloading()'

def stopUnloading():
        if abs(O.forces.f(plate.id)[2]) < minLoad:
                # O.tags can be used to retrieve unique identifiers of the simulation
                # if running in batch, subsequent simulation would overwrite each other's
↪output files otherwise
                # d (or description) is simulation description (composed of parameter values)
                # while the id is composed of time and process number
                plot.saveDataTxt(O.tags['d.id'] + '.txt')
                O.pause()

def addPlotData():
        if not isinstance(O.bodies[-1].shape, Wall):
                plot.addData()
                return
        Fz = O.forces.f(plate.id)[2]
        plot.addData(Fz=Fz, w=plate.state.pos[2] - plate.state.refPos[2],
↪unbalanced=unbalancedForce(), i=O.iter)

# besides unbalanced force evolution, also plot the displacement-force diagram
plot.plots = {'i': ('unbalanced',), 'w': ('Fz',)}
plot.plot()

O.run()
# when running with yade-batch, the script must not finish until the simulation is done fully
# this command will wait for that (has no influence in the non-batch mode)
waitIfBatch()
```

**Batch table**

To run the same script doc/sphinx/tutorial/03-oedometric-test.py in batch mode to test different parameters, execute command **yade-batch 03-oedometric-test.table 03-oedometric-test.py**, also visit page http://localhost:9080 to see the batch simulation progress.

```
rMean rRelFuzz maxLoad
.05 .1 1e6
.05 .2 1e6
.05 .3 1e6
```

**Periodic simple shear**

Following example is in file doc/sphinx/tutorial/04-periodic-simple-shear.py.

```
# encoding: utf-8

# script for periodic simple shear test, with periodic boundary
# first compresses to attain some isotropic stress (checkStress),
# then loads in shear (checkDistorsion)
#
# the initial packing is either regular (hexagonal), with empty bands along the boundary,
# or periodic random cloud of spheres
#
# material friction angle is initially set to zero, so that the resulting packing is dense
```

```python
# (sphere rearrangement is easier if there is no friction)
#

# setup the periodic boundary
from __future__ import print_function
O.periodic = True
O.cell.hSize = Matrix3(2, 0, 0, 0, 2, 0, 0, 0, 2)

from yade import pack, plot

# the "if 0:" block will be never executed, therefore the "else:" block will be
# to use cloud instead of regular packing, change to "if 1:" or something similar
if 0:
        # create cloud of spheres and insert them into the simulation
        # we give corners, mean radius, radius variation
        sp = pack.SpherePack()
        sp.makeCloud((0, 0, 0), (2, 2, 2), rMean=.1, rRelFuzz=.6, periodic=True)
        # insert the packing into the simulation
        sp.toSimulation(color=(0, 0, 1))  # pure blue
else:
        # in this case, add dense packing
        O.bodies.append(pack.regularHexa(pack.inAlignedBox((0, 0, 0), (2, 2, 2)), radius=.1,␣
↪gap=0, color=(0, 0, 1)))

# create "dense" packing by setting friction to zero initially
O.materials[0].frictionAngle = 0

# simulation loop (will be run at every step)
O.engines = [
        ForceResetter(),
        InsertionSortCollider([Bo1_Sphere_Aabb()]),
        InteractionLoop(
                # interaction loop
                [Ig2_Sphere_Sphere_ScGeom()],
                [Ip2_FrictMat_FrictMat_FrictPhys()],
                [Law2_ScGeom_FrictPhys_CundallStrack()]
        ),
        NewtonIntegrator(damping=.4),
        # run checkStress function (defined below) every second
        # the label is arbitrary, and is used later to refer to this engine
        PyRunner(command='checkStress()', realPeriod=1, label='checker'),
        # record data for plotting every 100 steps; addData function is defined below
        PyRunner(command='addData()', iterPeriod=100)
]

# set the integration timestep to be 1/2 of the "critical" timestep
O.dt = .5 * PWaveTimeStep()

# prescribe isotropic normal deformation (constant strain rate)
# of the periodic cell
O.cell.velGrad = Matrix3(-.1, 0, 0, 0, -.1, 0, 0, 0, -.1)

# when to stop the isotropic compression (used inside checkStress)
limitMeanStress = -5e5

# called every second by the PyRunner engine
def checkStress():
        # stress tensor as the sum of normal and shear contributions
        # Matrix3.Zero is the intial value for sum(...)
        stress = getStress().trace() / 3.
```

```
        print('mean stress', stress)
        # if mean stress is below (bigger in absolute value) limitMeanStress, start shearing
        if stress < limitMeanStress:
                # apply constant-rate distorsion on the periodic cell
                O.cell.velGrad = Matrix3(0, 0, .1, 0, 0, 0, 0, 0, 0)
                # change the function called by the checker engine
                # (checkStress will not be called anymore)
                checker.command = 'checkDistorsion()'
                # block rotations of particles to increase tanPhi, if desired
                # disabled by default
                if 0:
                        for b in O.bodies:
                                # block X,Y,Z rotations, translations are free
                                b.state.blockedDOFs = 'XYZ'
                                # stop rotations if any, as blockedDOFs block accelerations↵
↪really
                                b.state.angVel = (0, 0, 0)
                # set friction angle back to non-zero value
                # tangensOfFrictionAngle is computed by the Ip2_* functor from material
                # for future contacts change material (there is only one material for all↵
↪particles)
                O.materials[0].frictionAngle = .5  # radians
                # for existing contacts, set contact friction directly
                for i in O.interactions:
                        i.phys.tangensOfFrictionAngle = tan(.5)

# called from the 'checker' engine periodically, during the shear phase
def checkDistorsion():
        # if the distorsion value is >.3, exit; otherwise do nothing
        if abs(O.cell.trsf[0, 2]) > .5:
                # save data from addData(...) before exiting into file
                # use O.tags['id'] to distinguish individual runs of the same simulation
                plot.saveDataTxt(O.tags['id'] + '.txt')
                # exit the program
                #import sys
                #sys.exit(0) # no error (0)
                O.pause()

# called periodically to store data history
def addData():
        # get the stress tensor (as 3x3 matrix)
        stress = sum(normalShearStressTensors(), Matrix3.Zero)
        # give names to values we are interested in and save them
        plot.addData(exz=O.cell.trsf[0, 2], szz=stress[2, 2], sxz=stress[0, 2],↵
↪tanPhi=(stress[0, 2] / stress[2, 2]) if stress[2, 2] != 0 else 0, i=O.iter)
        # color particles based on rotation amount
        for b in O.bodies:
                # rot() gives rotation vector between reference and current position
                b.shape.color = scalarOnColorScale(b.state.rot().norm(), 0, pi / 2.)

# define what to plot (3 plots in total)
## exz(i), [left y axis, separate by None:] szz(i), sxz(i)
## szz(exz), sxz(exz)
## tanPhi(i)
# note the space in 'i ' so that it does not overwrite the 'i' entry
plot.plots = {'i': ('exz', None, 'szz', 'sxz'), 'exz': ('szz', 'sxz'), 'i ': ('tanPhi',)}

# better show rotation of particles
```

```
Gl1_Sphere.stripes = True

# open the plot on the screen
plot.plot()

O.saveTmp()
```

### 3d postprocessing

Following example is in file doc/sphinx/tutorial/05-3d-postprocessing.py. This example will run for 20000 iterations, saving *.png snapshots, then it will make a video **3d.mpeg** out of those snapshots.

```python
# demonstrate 3d postprocessing with yade
#
# 1. qt.SnapshotEngine saves images of the 3d view as it appears on the screen periodically
#    makeVideo is then used to make real movie from those images
# 2. VTKRecorder saves data in files which can be opened with Paraview
#    see the User's manual for an intro to Paraview

# generate loose packing
from yade import pack, qt
sp = pack.SpherePack()
sp.makeCloud((0, 0, 0), (2, 2, 2), rMean=.1, rRelFuzz=.6, periodic=True)
# add to scene, make it periodic
sp.toSimulation()

O.engines = [
        ForceResetter(),
        InsertionSortCollider([Bo1_Sphere_Aabb()]),
        InteractionLoop(
                # interaction loop
                [Ig2_Sphere_Sphere_ScGeom()],
                [Ip2_FrictMat_FrictMat_FrictPhys()],
                [Law2_ScGeom_FrictPhys_CundallStrack()]
        ),
        NewtonIntegrator(damping=.4),
        # save data for Paraview
        VTKRecorder(fileName='3d-vtk-', recorders=['all'], iterPeriod=1000),
        # save data from Yade's own 3d view
        qt.SnapshotEngine(fileBase='3d-', iterPeriod=200, label='snapshot'),
        # this engine will be called after 20000 steps, only once
        PyRunner(command='finish()', iterPeriod=20000)
]
O.dt = .5 * PWaveTimeStep()

# prescribe constant-strain deformation of the cell
O.cell.velGrad = Matrix3(-.1, 0, 0, 0, -.1, 0, 0, 0, -.1)

# we must open the view explicitly (limitation of the qt.SnapshotEngine)
qt.View()

# this function is called when the simulation is finished
def finish():
        # snapshot is label of qt.SnapshotEngine
        # the 'snapshots' attribute contains list of all saved files
        makeVideo(snapshot.snapshots, '3d.mpeg', fps=10, bps=10000)
        O.pause()
```

```
# set parameters of the renderer, to show network chains rather than particles
# these settings are accessible from the Controller window, on the second tab ("Display") as
↪well
rr = yade.qt.Renderer()
rr.shape = False
rr.intrPhys = True
```

**Periodic triaxial test**

Following example is in file doc/sphinx/tutorial/06-periodic-triaxial-test.py.

```python
# encoding: utf-8

# periodic triaxial test simulation
#
# The initial packing is either
#
# 1. random cloud with uniform distribution, or
# 2. cloud with specified granulometry (radii and percentages), or
# 3. cloud of clumps, i.e. rigid aggregates of several particles
#
# The triaxial consists of 2 stages:
#
# 1. isotropic compaction, until sigmaIso is reached in all directions;
#    this stage is ended by calling compactionFinished()
# 2. constant-strain deformation along the z-axis, while maintaining
#    constant stress (sigmaIso) laterally; this stage is ended by calling
#    triaxFinished()
#
# Controlling of strain and stresses is performed via PeriTriaxController,
# of which parameters determine type of control and also stability
# condition (maxUnbalanced) so that the packing is considered stabilized
# and the stage is done.
#

from __future__ import print_function
sigmaIso = -1e5

#import matplotlib
#matplotlib.use('Agg')

# generate loose packing
from yade import pack, qt, plot

O.periodic = True
sp = pack.SpherePack()
if 0:
        ## uniform distribution
        sp.makeCloud((0, 0, 0), (2, 2, 2), rMean=.1, rRelFuzz=.3, periodic=True)
else:
        ## create packing from clumps
        # configuration of one clump
        c1 = pack.SpherePack([((0, 0, 0), .03333), ((.03, 0, 0), .017), ((0, .03, 0), .017)])
        # make cloud using the configuration c1 (there could c2, c3, ...; selection between
↪them would be random)
        sp.makeClumpCloud((0, 0, 0), (2, 2, 2), [c1], periodic=True, num=500)

# setup periodic boundary, insert the packing
sp.toSimulation()
```

```python
O.engines = [
        ForceResetter(),
        InsertionSortCollider([Bo1_Sphere_Aabb()]),
        InteractionLoop([Ig2_Sphere_Sphere_ScGeom()], [Ip2_FrictMat_FrictMat_FrictPhys()],␣
↪[Law2_ScGeom_FrictPhys_CundallStrack()]),
        PeriTriaxController(
                label='triax',
                # specify target values and whether they are strains or stresses
                goal=(sigmaIso, sigmaIso, sigmaIso),
                stressMask=7,
                # type of servo-control
                dynCell=True,
                maxStrainRate=(10, 10, 10),
                # wait until the unbalanced force goes below this value
                maxUnbalanced=.1,
                relStressTol=1e-3,
                # call this function when goal is reached and the packing is stable
                doneHook='compactionFinished()'
        ),
        NewtonIntegrator(damping=.2),
        PyRunner(command='addPlotData()', iterPeriod=100),
]
O.dt = .5 * PWaveTimeStep()

def addPlotData():
        plot.addData(
                unbalanced=unbalancedForce(),
                i=O.iter,
                sxx=triax.stress[0],
                syy=triax.stress[1],
                szz=triax.stress[2],
                exx=triax.strain[0],
                eyy=triax.strain[1],
                ezz=triax.strain[2],
                # save all available energy data
                Etot=O.energy.total(),
                **O.energy
        )

# enable energy tracking in the code
O.trackEnergy = True

# define what to plot
plot.plots = {
        'i': ('unbalanced',),
        'i ': ('sxx', 'syy', 'szz'),
        ' i': ('exx', 'eyy', 'ezz'),
        # energy plot
        ' i ': (O.energy.keys, None, 'Etot'),
}
# show the plot
plot.plot()

def compactionFinished():
        # set the current cell configuration to be the reference one
        O.cell.trsf = Matrix3.Identity
        # change control type: keep constant confinement in x,y, 20% compression in z
```

```
        triax.goal = (sigmaIso, sigmaIso, -.2)
        triax.stressMask = 3
        # allow faster deformation along x,y to better maintain stresses
        triax.maxStrainRate = (1., 1., .1)
        # next time, call triaxFinished instead of compactionFinished
        triax.doneHook = 'triaxFinished()'
        # do not wait for stabilization before calling triaxFinished
        triax.maxUnbalanced = 10

def triaxFinished():
        print('Finished')
        O.pause()
```

## 1.2.7 More examples

The same list with embedded videos is available online, but not recommended for viewing on slow internet connection.

A full list of examples is in file examples/list__of__examples.txt. Videos of some of those examples are listed below.

### FluidCouplingLBM

- *refFastBuoyancy*, source file, video.

### FluidCouplingPFV

- *refFastOedometer*, source file, video.

### HydroForceEngine

- *refFastBuoyantParticles*, source file, video.
- *refFastFluidizedBed*, source file, video.
- *refFastSedimentTransportExample*, source file, video.
- *refFastLaminarShearFlow*, source file, video.
- *refFastPostProcessValidMaurin2015*, source file, video.
- *refFastValidMaurin2015*, source file, video.

### PeriodicBoundaries

- *refFastCellFlipping*, source file, video.
- *refFastPeri3dController-example1*, source file, video.
- *refFastPeri3dController-shear*, source file, video.
- *refFastPeri3dController-triaxialCompression*, source file, video.
- *refFastPeriodic-compress*, source file, video.
- *refFastPeriodic-shear*, source file, video.
- *refFastPeriodic-simple-shear*, source file, video.

- *refFastPeriodic-simple*, source file, video.
- *refFastPeriodic-triax-settingHsize*, source file, video.
- *refFastPeriodic-triax*, source file, video.
- *refFastPeriodicSandPile*, source file, video.

### PotentialBlocks

- *refFastWedgeYADE*, source file, video.
- *refFastCubePBscaled*, source file, video.

### PotentialParticles

- *refFastCubePPscaled*, source file, video.

### WireMatPM

- *refFastWirecontacttest*, source file, video.
- *refFastWirepackings*, source file, video.
- *refFastWiretensiltest*, source file, video.

### Adaptiveintegrator

- *refFastSimple-scene-plot-NewtonIntegrator*, source file, video.
- *refFastSimple-scene-plot-RungeKuttaCashKarp54*, source file, video.

### Agglomerate

- *refFastCompress*, source file, video.
- *refFastSimulation*, source file, video.

### Baraban

- *refFastBicyclePedalEngine*, source file, video.
- *refFastBaraban*, source file, video.
- *refFastRotating-cylinder*, source file, video.

### Bulldozer

- *refFastBulldozer*, source file, video.

### Capillary

- *refFastCapillar*, source file, video.

### CapillaryLaplaceYoung

- *refFastCapillaryPhys-example*, source file, video.
- *refFastCapillaryBridge*, source file, video.

### Chained-cylinders

- *refFastCohesiveCylinderSphere*, source file, video.
- *refFastChained-cylinder-roots*, source file, video.
- *refFastChained-cylinder-spring*, source file, video.

### Clumps

- *refFastAddToClump-example*, source file, video.
- *refFastApply-buoyancy-clumps*, source file, video.
- *refFastClump-hopper-test*, source file, video.
- *refFastClump-hopper-viscoelastic*, source file, video.
- *refFastClump-inbox-viscoelastic*, source file, video.
- *refFastClump-viscoelastic*, source file, video.
- *refFastReleaseFromClump-example*, source file, video.
- *refFastReplaceByClumps-example*, source file, video.
- *refFastTriax-basic-with-clumps*, source file, video.

### Concrete

- *refFastBrazilian*, source file, video.
- *refFastInteraction-histogram*, source file, video.
- *refFastPeriodic*, source file, video.
- *refFastTriax*, source file, video.
- *refFastUniax-post*, source file, video.
- *refFastUniax*, source file, video.

### Conveyor

- *refFastConveyor*, source file, video.

### Cylinders

- *refFastBendingbeams*, source file, video.
- *refFastCylinder-cylinder*, source file, video.
- *refFastCylinderconnection-roots*, source file, video.
- *refFastMikado*, source file, video.

**Deformableelem**

- *refFastMinimalTensileTest*, source file, video.
- *refFastTestDeformableBodies*, source file, video.
- *refFastTestDeformableBodies-pressure*, source file, video.

**Grids**

- *refFastCohesiveGridConnectionSphere*, source file, video.
- *refFastGridConnection-Spring*, source file, video.
- *refFastSimple-GridConnection-Falling*, source file, video.
- *refFastSimple-Grid-Falling*, source file, video.

**Gts-horse**

- *refFastGts-horse*, source file, video.
- *refFastGts-operators*, source file, video.
- *refFastGts-random-pack-obb*, source file, video.
- *refFastGts-random-pack*, source file, video.

**Hourglass**

- *refFastHourglass*, source file, video.

**Packs**

- *refFastPacks*, source file, video.

**Pfacet**

- *refFastGts-pfacet*, source file, video.
- *refFastMesh-pfacet*, source file, video.
- *refFastPFacets-grids-spheres-interacting*, source file, video.
- *refFastPfacetcreators*, source file, video.

**Polyhedra**

- *refFastBall*, source file, video.
- *refFastHorse*, source file, video.
- *refFastIrregular*, source file, video.
- *refFastSphere-interaction*, source file, video.
- *refFastSplitter*, source file, video.
- *refFastInteractinDetectionFactor*, source file, video.
- *refFastScGeom*, source file, video.
- *refFastTextExport*, source file, video.

**PolyhedraBreak**

- *refFastUniaxial-compression*, source file, video.

**Ring2d**

- *refFastRingCundallDamping*, source file, video.
- *refFastRingSimpleViscoelastic*, source file, video.

**Rod-penetration**

- *refFastModel*, source file, video.

**Simple-scene**

- *refFast2SpheresNormVisc*, source file, video.
- *refFastSave-then-reload*, source file, video.
- *refFastSimple-scene-default-engines*, source file, video.
- *refFastSimple-scene-energy-tracking*, source file, video.
- *refFastSimple-scene-plot*, source file, video.
- *refFastSimple-scene*, source file, video.

**Stl-gts**

- *refFastGts-stl*, source file, video.

**Tesselationwrapper**

- *refFastTesselationWrapper*, source file, video.

**Test**

- *refFastNet-2part-displ-unloading*, source file, video.
- *refFastNet-2part-displ*, source file, video.
- *refFastBeam-l6geom*, source file, video.
- *refFastClump-facet*, source file, video.
- *refFastClumpPack*, source file, video.
- *refFastCollider-stride-triax*, source file, video.
- *refFastCollider-stride*, source file, video.
- *refFastCombined-kinematic-engine*, source file, video.
- *refFastEnergy*, source file, video.
- *refFastFacet-box*, source file, video.
- *refFastFacet-sphere-ViscElBasic-peri*, source file, video.
- *refFastFacet-sphere-ViscElBasic*, source file, video.
- *refFastFacet-sphere*, source file, video.

- *refFastHelix*, source file, video.
- *refFastInterpolating-force*, source file, video.
- *refFastKinematic*, source file, video.
- *refFastMindlin*, source file, video.
- *refFastMulti*, source file, video.
- *refFastPack-cloud*, source file, video.
- *refFastPack-inConvexPolyhedron*, source file, video.
- *refFastPv-section*, source file, video.
- *refFastPeriodic-geom-compare*, source file, video.
- *refFastPsd*, source file, video.
- *refFastSphere-sphere-ViscElBasic-peri*, source file, video.
- *refFastSubdomain-balancer*, source file, video.
- *refFastTest-sphere-facet-corner*, source file, video.
- *refFastTest-sphere-facet*, source file, video.
- *refFastTriax-basic*, source file, video.
- *refFastTriax-basic-without-plots*, source file, video.
- *refFastUnvRead*, source file, video.

**Tetra**

- *refFastOneTetra*, source file, video.
- *refFastOneTetraPoly*, source file, video.
- *refFastTwoTetras*, source file, video.
- *refFastTwoTetrasPoly*, source file, video.

# Chapter 2

# Yade for users

## 2.1 DEM formulation

In this chapter, we mathematically describe general features of explicit DEM simulations, with some reference to Yade implementation of these algorithms. They are given roughly in the order as they appear in simulation; first, two particles might establish a new interaction, which consists in

1. detecting collision between particles;

2. creating new interaction and determining its properties (such as stiffness); they are either precomputed or derived from properties of both particles;

Then, for already existing interactions, the following is performed:

1. strain evaluation;

2. stress computation based on strains;

3. force application to particles in interaction.

This simplified description serves only to give meaning to the ordering of sections within this chapter. A more detailed description of this *simulation loop* is given later.

In this chapter we refer to kinematic variables of the contacts as "strains", although at this scale it is also common to speak of "displacements". Which semantic is more appropriate depends on the conceptual model one is starting from, and therefore it cannot be decided independently of specific problems. The reader familiar with displacements can mentaly replace normal strain and shear strain by normal displacement and shear displacement, respectively, without altering the meaning of what follows.

### 2.1.1 Collision detection

#### Generalities

Exact computation of collision configuration between two particles can be relatively expensive (for instance between *Sphere* and *Facet*). Taking a general pair of bodies $i$ and $j$ and their "exact" (In the sense of precision admissible by numerical implementation.) spatial predicates (called *Shape* in Yade) represented by point sets $P_i$, $P_j$ the detection generally proceeds in 2 passes:

1. fast collision detection using approximate predicate $\tilde{P}_i$ and $\tilde{P}_j$; they are pre-constructed in such a way as to abstract away individual features of $P_i$ and $P_j$ and satisfy the condition

$$\forall \mathbf{x} \in \mathsf{R}^3 : x \in P_i \Rightarrow x \in \tilde{P}_i \tag{2.1}$$

(likewise for $P_j$). The approximate predicate is called "bounding volume" (*Bound* in Yade) since it bounds any particle's volume from outside (by virtue of the implication). It follows that $(P_i \cap P_j) \neq$

$\emptyset \Rightarrow (\tilde{P}_i \cap \tilde{P}_j) \neq \emptyset$ and, by applying *modus tollens*,

$$(\tilde{P}_i \cap \tilde{P}_j) = \emptyset \Rightarrow (P_i \cap P_j) = \emptyset \qquad (2.2)$$

which is a candidate exclusion rule in the proper sense.

2. By filtering away impossible collisions in (2.2), a more expensive, exact collision detection algorithms can be run on possible interactions, filtering out remaining spurious couples $(\tilde{P}_i \cap \tilde{P}_j) \neq \emptyset \wedge (P_i \cap P_j) = \emptyset$. These algorithms operate on $P_i$ and $P_j$ and have to be able to handle all possible combinations of shape types.

It is only the first step we are concerned with here.

### Algorithms

Collision evaluation algorithms have been the subject of extensive research in fields such as robotics, computer graphics and simulations. They can be roughly divided in two groups:

**Hierarchical algorithms** which recursively subdivide space and restrict the number of approximate checks in the first pass, knowing that lower-level bounding volumes can intersect only if they are part of the same higher-level bounding volume. Hierarchy elements are bounding volumes of different kinds: octrees [Jung1997], bounding spheres [Hubbard1996], k-DOP's [Klosowski1998].

**Flat algorithms** work directly with bounding volumes without grouping them in hierarchies first; let us only mention two kinds commonly used in particle simulations:

    **Sweep and prune** algorithm operates on axis-aligned bounding boxes, which overlap if and only if they overlap along all axes. These algorithms have roughly $\mathcal{O}(n \log n)$ complexity, where $n$ is number of particles as long as they exploit *temporal coherence* of the simulation.

    **Grid algorithms** represent continuous $R^3$ space by a finite set of regularly spaced points, leading to very fast neighbor search; they can reach the $\mathcal{O}(n)$ complexity [Munjiza1998] and recent research suggests ways to overcome one of the major drawbacks of this method, which is the necessity to adjust grid cell size to the largest particle in the simulation ([Munjiza2006], the "multistep" extension).

**Temporal coherence** expresses the fact that motion of particles in simulation is not arbitrary but governed by physical laws. This knowledge can be exploited to optimize performance.

Numerical stability of integrating motion equations dictates an upper limit on $\Delta t$ (sect. *Stability considerations*) and, by consequence, on displacement of particles during one step. This consideration is taken into account in [Munjiza2006], implying that any particle may not move further than to a neighboring grid cell during one step allowing the $\mathcal{O}(n)$ complexity; it is also explored in the periodic variant of the sweep and prune algorithm described below.

On a finer level, it is common to enlarge $\tilde{P}_i$ predicates in such a way that they satisfy the (2.1) condition during *several* timesteps; the first collision detection pass might then be run with stride, speeding up the simulation considerably. The original publication of this optimization by Verlet [Verlet1967] used enlarged list of neighbors, giving this technique the name *Verlet list*. In general cases, however, where neighbor lists are not necessarily used, the term *Verlet distance* is employed.

### Sweep and prune

Let us describe in detail the sweep and prune algorithm used for collision detection in Yade (class *InsertionSortCollider*). Axis-aligned bounding boxes (*Aabb*) are used as $\tilde{P}_i$; each *Aabb* is given by lower and upper corner $\in R^3$ (in the following, $\tilde{P}_i^{x0}$, $\tilde{P}_i^{x1}$ are minimum/maximum coordinates of $\tilde{P}_i$ along the x-axis and so on). Construction of *Aabb* from various particle *Shape*'s (such as *Sphere*, *Facet*, *Wall*) is straightforward, handled by appropriate classes deriving form *BoundFunctor* (*Bo1_Sphere_Aabb*, *Bo1_Facet_Aabb*, …).

Presence of overlap of two *Aabb*'s can be determined from conjunction of separate overlaps of intervals along each axis (*fig-sweep-and-prune*):

$$\left(\tilde{P}_i \cap \tilde{P}_j\right) \neq \emptyset \Leftrightarrow \bigwedge_{w \in \{x,y,z\}} \left[\left(\left(\tilde{P}_i^{w0}, \tilde{P}_i^{w1}\right) \cap \left(\tilde{P}_j^{w0}, \tilde{P}_j^{w1}\right)\right) \neq \emptyset\right]$$

where $(a, b)$ denotes interval in $\mathbb{R}$.

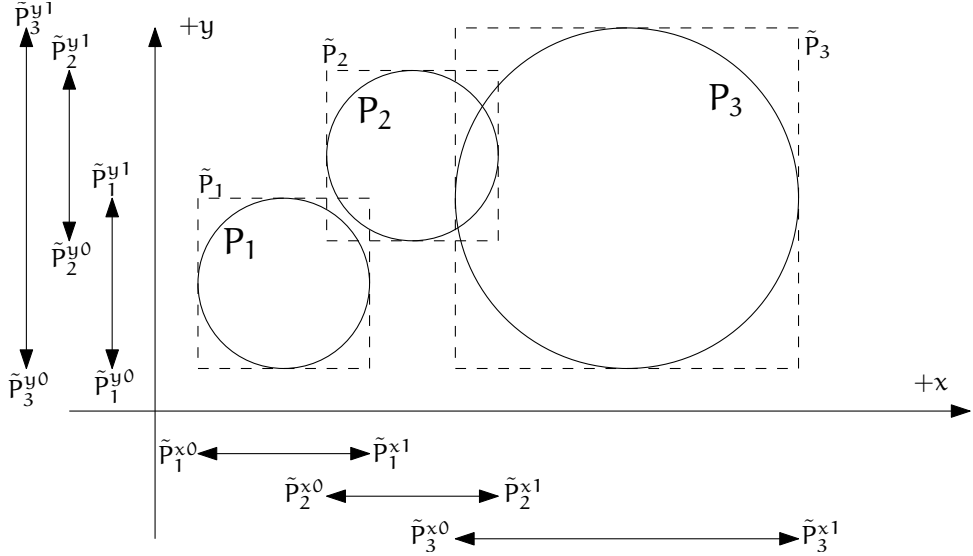

Fig. 1: Sweep and prune algorithm (shown in 2D), where *Aabb* of each sphere is represented by minimum and maximum value along each axis. Spatial overlap of *Aabb*'s is present if they overlap along all axes. In this case, $\tilde{P}_1 \cap \tilde{P}_2 \neq \emptyset$ (but note that $P_1 \cap P_2 = \emptyset$) and $\tilde{P}_2 \cap \tilde{P}_3 \neq \emptyset$.}

The collider keeps 3 separate lists (arrays) $L_w$ for each axis $w \in \{x, y, z\}$

$$L_w = \bigcup_i \left\{\tilde{P}_i^{w0}, \tilde{P}_i^{w1}\right\}$$

where $i$ traverses all particles. $L_w$ arrays (sorted sets) contain respective coordinates of minimum and maximum corners for each *Aabb* (we call these coordinates *bound* in the following); besides bound, each of list elements further carries `id` referring to particle it belongs to, and a flag whether it is lower or upper bound.

In the initial step, all lists are sorted (using quicksort, average $\mathcal{O}(n \log n)$) and one axis is used to create initial interactions: the range between lower and upper bound for each body is traversed, while bounds in-between indicate potential *Aabb* overlaps which must be checked on the remaining axes as well.

At each successive step, lists are already pre-sorted. Inversions occur where a particle's coordinate has just crossed another particle's coordinate; this number is limited by numerical stability of simulation and its physical meaning (giving spatio-temporal coherence to the algorithm). The insertion sort algorithm swaps neighboring elements if they are inverted, and has complexity between $\mathcal{O}(n)$ and $\mathcal{O}(n^2)$, for pre-sorted and unsorted lists respectively. For our purposes, we need only to handle inversions, which by nature of the sort algorithm are detected inside the sort loop. An inversion might signify:

- overlap along the current axis, if an upper bound inverts (swaps) with a lower bound (i.e. that the upper bound with a higher coordinate was out of order in coming before the lower bound with a lower coordinate). Overlap along the other 2 axes is checked and if there is overlap along all axes, a new potential interaction is created.

- End of overlap along the current axis, if lower bound inverts (swaps) with an upper bound. If there is only potential interaction between the two particles in question, it is deleted.

- Nothing if both bounds are upper or both lower.

**Aperiodic insertion sort**

Let us show the sort algorithm on a sample sequence of numbers:

$$\| \quad 3 \qquad 7 \qquad 2 \qquad 4 \quad \|$$

Elements are traversed from left to right; each of them keeps inverting (swapping) with neighbors to the left, moving left itself, until any of the following conditions is satisfied:

| ($\leq$) | the sorting order with the left neighbor is correct, or |
|---|---|
| ($\|$) | the element is at the beginning of the sequence. |

We start at the leftmost element (the current element is marked $\boxed{\text{i}}$)

$$\| \quad \boxed{3} \qquad 7 \qquad 2 \qquad 4 \quad \|.$$

It obviously immediately satisfies ($\|$), and we move to the next element:

$$\| \quad 3 \quad\underset{\leq}{\curvearrowleft}\quad \boxed{7} \qquad 2 \qquad 4 \quad \|.$$

Condition ($\leq$) holds, therefore we move to the right. The $\boxed{2}$ is not in order (violating ($\leq$)) and two inversions take place; after that, ($\|$) holds:

$$\| \quad 3 \qquad 7 \quad\underset{\nleq}{\curvearrowleft}\quad \boxed{2} \qquad 4 \quad \|,$$

$$\| \quad 3 \quad\underset{\nleq}{\curvearrowleft}\quad \boxed{2} \qquad 7 \qquad 4 \quad \|,$$

$$\| \quad \boxed{2} \qquad 3 \qquad 7 \qquad 4 \quad \|.$$

The last element $\boxed{4}$ first violates ($\leq$), but satisfies it after one inversion

$$\| \quad 2 \qquad 3 \qquad 7 \quad\underset{\nleq}{\curvearrowleft}\quad \boxed{4} \quad \|,$$

$$\| \quad 2 \qquad 3 \quad\underset{\leq}{\curvearrowleft}\quad \boxed{4} \qquad 7 \quad \|.$$

All elements having been traversed, the sequence is now sorted.

It is obvious that if the initial sequence were sorted, elements only would have to be traversed without any inversion to handle (that happens in $\mathcal{O}(n)$ time).

For each inversion during the sort in simulation, the function that investigates change in *Aabb* overlap is invoked, creating or deleting interactions.

The periodic variant of the sort algorithm is described in *Periodic insertion sort algorithm*, along with other periodic-boundary related topics.

### Optimization with Verlet distances

As noted above, [Verlet1967] explored the possibility of running the collision detection only sparsely by enlarging predicates $\bar{P_i}$.

In Yade, this is achieved by enlarging *Aabb* of particles by fixed relative length (or Verlet's distance) in all dimensions $\Delta L$ (*InsertionSortCollider.sweepLength*). Suppose the collider run last time at step $m$ and the current step is $n$. *NewtonIntegrator* tracks the cummulated distance traversed by each particle between $m$ and $n$ by comparing the current position with the reference position from time $n$ (*Bound::refPos*),

$$L_{mn} = |X^n - X^m| \tag{2.3}$$

triggering the collider re-run as soon as one particle gives:

$$L_{mn} > \Delta L. \tag{2.4}$$

$\Delta L$ is defined primarily by the parameter *InsertionSortCollider.verletDist*. It can be set directly by assigning a positive value, or indirectly by assigning negative value (which defines $\Delta L$ in proportion of the smallest particle radius). In addition, *InsertionSortCollider.targetInterv* can be used to adjust $\Delta L$ independently for each particle. Larger $\Delta L$ will be assigned to the fastest ones, so that all particles would ideally reach the edge of their bounds after this "target" number of iterations. Results of using Verlet distance depend highly on the nature of simulation and choice of *InsertionSortCollider.targetInterv*. Adjusting the sizes independently for each particle is especially efficient if some parts of a problem have high-speed particles will others are not moving. If it is not the case, no significant gain should be expected as compared to targetInterv=0 (assigning the same $\Delta L$ to all particles).

The number of particles and the number of available threads is also to be considered for choosing an appropriate Verlet's distance. A larger distance will result in less time spent in the collider (which runs single-threaded) and more time in computing interactions (multi-threaded). Typically, large $\Delta L$ will be used for large simulations with more than $10^5$ particles on multi-core computers. On the other hand simulations with less than $10^4$ particles on single processor will probably benefit from smaller $\Delta L$. Users benchmarks may be found on Yade's wiki (see e.g. https://yade-dem.org/wiki/Colliders_performace).

## 2.1.2 Creating interaction between particles

Collision detection described above is only approximate. Exact collision detection depends on the geometry of individual particles and is handled separately. In Yade terminology, the *Collider* creates only *potential* interactions; potential interactions are evaluated exactly using specialized algorithms for collision of two spheres or other combinations. Exact collision detection must be run at every timestep since it is at every step that particles can change their mutual position (the collider is only run sometimes if the Verlet distance optimization is in use). Some exact collision detection algorithms are described in *Kinematic variables*; in Yade, they are implemented in classes deriving from *IGeomFunctor* (prefixed with Ig2).

Besides detection of geometrical overlap (which corresponds to *IGeom* in Yade), there are also non-geometrical properties of the interaction to be determined (*IPhys*). In Yade, they are computed for every new interaction by calling a functor deriving from *IPhysFunctor* (prefixed with Ip2) which accepts the given combination of *Material* types of both particles.

### Stiffnesses

Basic DEM interaction defines two stiffnesses: normal stiffness $K_N$ and shear (tangent) stiffness $K_T$. It is desirable that $K_N$ be related to fictitious Young's modulus of the particles' material, while $K_T$ is typically determined as a given fraction of computed $K_N$. The $K_T/K_N$ ratio determines macroscopic Poisson's ratio of the arrangement, which can be shown by dimensional analysis: elastic continuum has two parameters ($E$ and $\nu$) and basic DEM model also has 2 parameters with the same dimensions $K_N$ and $K_T/K_N$; macroscopic Poisson's ratio is therefore determined solely by $K_T/K_N$ and macroscopic Young's modulus is then proportional to $K_N$ and affected by $K_T/K_N$.

Naturally, such analysis is highly simplifying and does not account for particle radius distribution, packing configuration and other possible parameters such as the interaction radius introduced later.

### Normal stiffness

The algorithm commonly used in Yade computes normal interaction stiffness as stiffness of two springs in serial configuration with lengths equal to the sphere radii (*fig-spheres-contact-stiffness*).

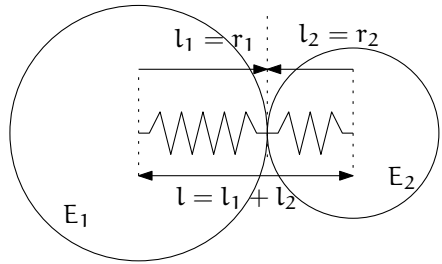

Fig. 2: Series of 2 springs representing normal stiffness of contact between 2 spheres.

Let us define distance $l = l_1 + l_2$, where $l_i$ are distances between contact point and sphere centers, which are initially (roughly speaking) equal to sphere radii. Change of distance between the sphere centers $\Delta l$ is distributed onto deformations of both spheres $\Delta l = \Delta l_1 + \Delta l_2$ proportionally to their compliances. Displacement change $\Delta l_i$ generates force $F_i = K_i \Delta l_i$, where $K_i$ assures proportionality and has physical meaning and dimension of stiffness; $K_i$ is related to the sphere material modulus $E_i$ and some length $\tilde{l}_i$ proportional to $r_i$.

$$\Delta l = \Delta l_1 + \Delta l_2$$

$$K_i = E_i \tilde{l}_i$$

$$K_N \Delta l = F = F_1 = F_2$$

$$K_N \left( \Delta l_1 + \Delta l_2 \right) = F$$

$$K_N \left( \frac{F}{K_1} + \frac{F}{K_2} \right) = F$$

$$K_1^{-1} + K_2^{-1} = K_N^{-1}$$

$$K_N = \frac{K_1 K_2}{K_1 + K_2}$$

$$K_N = \frac{E_1 \tilde{l}_1 E_2 \tilde{l}_2}{E_1 \tilde{l}_1 + E_2 \tilde{l}_2}$$

The most used class computing interaction properties *Ip2_FrictMat_FrictMat_FrictPhys* uses $\tilde{l}_i = 2r_i$.

Some formulations define an equivalent cross-section $A_{eq}$, which in that case appears in the $\tilde{l}_i$ term as $K_i = E_i \tilde{l}_i = E_i \frac{A_{eq}}{l_i}$. Such is the case for the concrete model (*Ip2_CpmMat_CpmMat_CpmPhys*), where $A_{eq} = \min(r_1, r_2)$.

For reasons given above, no pretense about equality of particle-level $E_i$ and macroscopic modulus $E$ should be made. Some formulations, such as [Hentz2003], introduce parameters to match them numerically. This is not appropriate, in our opinion, since it binds those values to particular features of the sphere arrangement that was used for calibration.

**Other parameters**

Non-elastic parameters differ for various material models. Usually, though, they are averaged from the particles' material properties, if it makes sense. For instance, *Ip2_CpmMat_CpmMat_CpmPhys* averages most quantities, while *Ip2_FrictMat_FrictMat_FrictPhys* computes internal friction angle as $\varphi = \min(\varphi_1, \varphi_2)$ to avoid friction with bodies that are frictionless.

### 2.1.3 Kinematic variables

In the general case, mutual configuration of two particles has 6 degrees of freedom (DoFs) just like a beam in 3D space: both particles have 6 DoFs each, but the interaction itself is free to move and rotate in space (with both spheres) having 6 DoFs itself; then $12 - 6 = 6$. They are shown at *fig-spheres-dofs*.

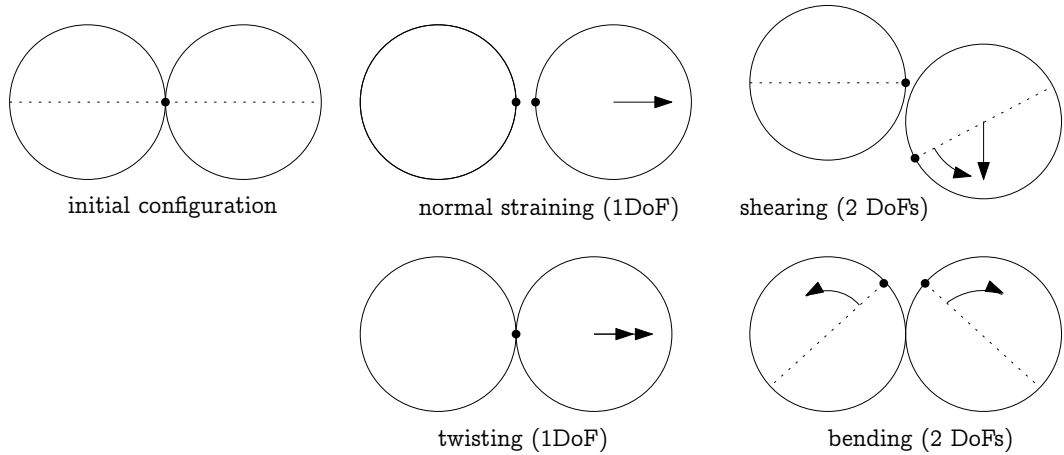

Fig. 3: Degrees of freedom of configuration of two spheres. Normal motion appears if there is a difference of linear velocity along the interaction axis ($\mathfrak{n}$); shearing originates from the difference of linear velocities perpendicular to $\mathfrak{n}$ *and* from the part of $\boldsymbol{\omega}_1 + \boldsymbol{\omega}_2$ perpendicular to $\mathfrak{n}$; twisting is caused by the part of $\boldsymbol{\omega}_1 - \boldsymbol{\omega}_2$ parallel with $\mathfrak{n}$; bending comes from the part of $\boldsymbol{\omega}_1 - \boldsymbol{\omega}_2$ perpendicular to $\mathfrak{n}$.

We will only describe normal and shear components of the relative movement in the following, leaving torsion and bending aside. The reason is that most constitutive laws for contacts do not use the latter two.

**Normal deformation**

**Constants**

Let us consider two spheres with *initial* centers $\bar{\mathbf{C}}_1$, $\bar{\mathbf{C}}_2$ and radii $r_1$, $r_2$ that enter into contact. The order of spheres within the contact is arbitrary and has no influence on the behavior. Then we define lengths

$$d_0 = |\bar{\mathbf{C}}_2 - \bar{\mathbf{C}}_1|$$

$$d_1 = r_1 + \frac{d_0 - r_1 - r_2}{2}, \qquad\qquad d_2 = d_0 - d_1.$$

These quantities are *constant* throughout the life of the interaction and are computed only once when the interaction is established. The distance $d_0$ is the *reference distance* and is used for the conversion of absolute displacements to dimensionless strain, for instance. It is also the distance where (for usual contact laws) there is neither repulsive nor attractive force between the spheres, whence the name *equilibrium distance*.

Distances $d_1$ and $d_2$ define reduced (or expanded) radii of spheres; geometrical radii $r_1$ and $r_2$ are used only for collision detection and may not be the same as $d_1$ and $d_2$, as shown in fig. *fig-sphere-sphere*.

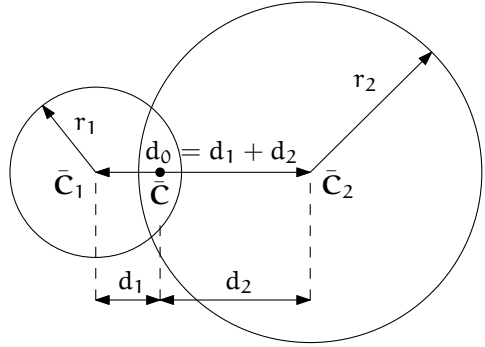

Fig. 4: Geometry of the initial contact of 2 spheres; this case pictures spheres which already overlap when the contact is created (which can be the case at the beginning of a simulation) for the sake of generality. The initial contact point $\bar{\mathbf{C}}$ is in the middle of the overlap zone.

This difference is exploited in cases where the average number of contacts between spheres should be increased, e.g. to influence the response in compression or to stabilize the packing. In such case, interactions will be created also for spheres that do not geometrically overlap based on the *interaction radius* $R_I$, a dimensionless parameter determining „non-locality" of contact detection. For $R_I = 1$, only spheres that touch are considered in contact; the general condition reads

$$d_0 \leq R_I(r_1 + r_2). \tag{2.5}$$

The value of $R_I$ directly influences the average number of interactions per sphere (percolation), which for some models is necessary in order to achieve realistic results. In such cases, *Aabb* (or $\bar{P}_i$ predicates in general) must be enlarged accordingly (*Bo1_Sphere_Aabb.aabbEnlargeFactor*).

### Contact cross-section

Some constitutive laws are formulated with strains and stresses (*Law2_ScGeom_CpmPhys_Cpm*, the concrete model described later, for instance); in that case, equivalent cross-section of the contact must be introduced for the sake of dimensionality. The exact definition is rather arbitrary; the CPM model (*Ip2_CpmMat_CpmMat_CpmPhys*) uses the relation

$$A_{eq} = \pi \min(r_1, r_2)^2 \tag{2.6}$$

which will be used to convert stresses to forces, if the constitutive law used is formulated in terms of stresses and strains. Note that other values than $\pi$ can be used; it will merely scale macroscopic packing stiffness; it is only for the intuitive notion of a truss-like element between the particle centers that we choose $A_{eq}$ representing the circle area. Besides that, another function than $\min(r_1, r_2)$ can be used, although the result should depend linearly on $r_1$ and $r_2$ so that the equation gives consistent results if the particle dimensions are scaled.

### Variables

The following state variables are updated as spheres undergo motion during the simulation (as $\mathbf{C}_1^\circ$ and $\mathbf{C}_2^\circ$ change):

$$\mathbf{n}^\circ = \frac{\mathbf{C}_2^\circ - \mathbf{C}_1^\circ}{|\mathbf{C}_2^\circ - \mathbf{C}_1^\circ|} \equiv \widehat{\mathbf{C}_2^\circ - \mathbf{C}_1^\circ} \tag{2.7}$$

and

$$\mathbf{C}^\circ = \mathbf{C}_1^\circ + \left(d_1 - \frac{d_0 - |\mathbf{C}_2^\circ - \mathbf{C}_1^\circ|}{2}\right)\mathbf{n}. \tag{2.8}$$

The contact point $\mathbf{C}^\circ$ is always in the middle of the spheres' overlap zone (even if the overlap is negative, when it is in the middle of the empty space between the spheres). The *contact plane* is always perpendicular to the contact plane normal $\mathbf{n}^\circ$ and passes through $\mathbf{C}^\circ$.

Normal displacement and strain can be defined as

$$u_N = |\mathbf{C}_2^\circ - \mathbf{C}_1^\circ| - d_0,$$
$$\varepsilon_N = \frac{u_N}{d_0} = \frac{|\mathbf{C}_2^\circ - \mathbf{C}_1^\circ|}{d_0} - 1.$$

Since $u_N$ is always aligned with $\mathbf{n}$, it can be stored as a scalar value multiplied by $\mathbf{n}$ if necessary.

For massively compressive simulations, it might be beneficial to use the logarithmic strain, such that the strain tends to $-\infty$ (rather than $-1$) as centers of both spheres approach. Otherwise, repulsive force would remain finite and the spheres could penetrate through each other. Therefore, we can adjust the definition of normal strain as follows:

$$\varepsilon_N = \begin{cases} \log\left(\frac{|\mathbf{C}_2^\circ - \mathbf{C}_1^\circ|}{d_0}\right) & \text{if } |\mathbf{C}_2^\circ - \mathbf{C}_1^\circ| < d_0 \\ \frac{|\mathbf{C}_2^\circ - \mathbf{C}_1^\circ|}{d_0} - 1 & \text{otherwise.} \end{cases}$$

Such definition, however, has the disadvantage of effectively increasing rigidity (up to infinity) of contacts, requiring $\Delta t$ to be adjusted, lest the simulation becomes unstable. Such dynamic adjustment is possible using a stiffness-based time-stepper (*GlobalStiffnessTimeStepper* in Yade).

### Shear deformation

In order to keep $\mathbf{u}_T$ consistent (e.g. that $\mathbf{u}_T$ must be constant if two spheres retain mutually constant configuration but move arbitrarily in space), then either $\mathbf{u}_T$ must track spheres' spatial motion or must (somehow) rely on sphere-local data exclusively.

Geometrical meaning of shear strain is shown in *fig-shear-2d*.

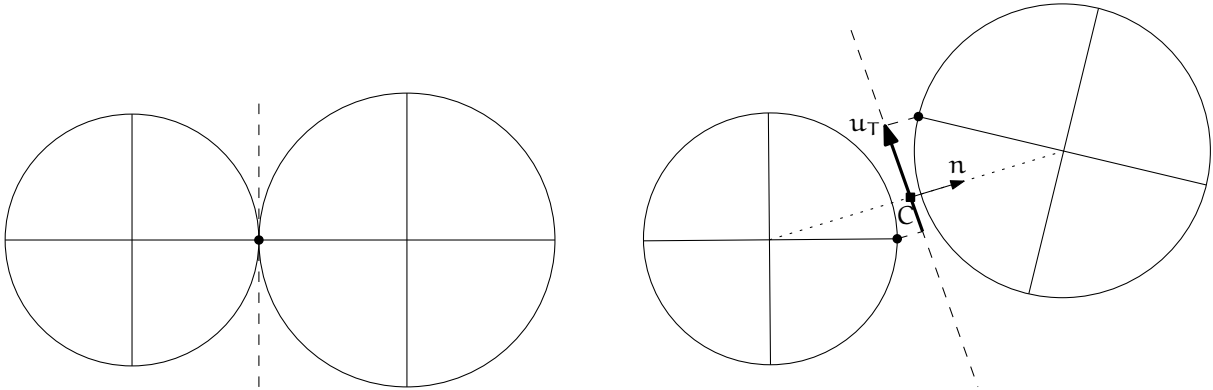

Fig. 5: Evolution of shear displacement $\mathbf{u}_T$ due to mutual motion of spheres, both linear and rotational. Left configuration is the initial contact, right configuration is after displacement and rotation of one particle.

The classical incremental algorithm is widely used in DEM codes and is described frequently ([Luding2008], [Alonso2004]). Yade implements this algorithm in the *ScGeom* class. At each step, shear displacement $\mathbf{u}_T$ is updated; the update increment can be decomposed in 2 parts: motion of the interaction (i.e. $\mathbf{C}$ and $\mathbf{n}$) in global space and mutual motion of spheres.

1. Contact moves dues to changes of the spheres' positions $\mathbf{C}_1$ and $\mathbf{C}_2$, which updates current $\mathbf{C}^\circ$ and $\mathbf{n}^\circ$ as per (2.8) and (2.7). $\mathbf{u}_T^-$ is perpendicular to the contact plane at the previous step $\mathbf{n}^-$ and must be updated so that $\mathbf{u}_T^- + (\Delta\mathbf{u}_T) = \mathbf{u}_T^\circ \perp \mathbf{n}^\circ$; this is done by perpendicular projection to the plane first (which might decrease $|\mathbf{u}_T|$) and adding what corresponds to spatial rotation of the

interaction instead:

$$(\Delta\mathbf{u}_T)_1 = -\mathbf{u}_T^- \times (\mathbf{n}^- \times \mathbf{n}^\circ)$$

$$(\Delta\mathbf{u}_T)_2 = -\mathbf{u}_T^- \times \left(\frac{\Delta t}{2}\mathbf{n}^\circ \cdot (\boldsymbol{\omega}_1^\ominus + \boldsymbol{\omega}_2^\ominus)\right)\mathbf{n}^\circ$$

2. Mutual movement of spheres, using only its part perpendicular to $\mathbf{n}^\circ$; $\mathbf{v}_{12}$ denotes mutual velocity of spheres at the contact point:

$$\mathbf{v}_{12} = \left(\mathbf{v}_2^\ominus + \boldsymbol{\omega}_2^\ominus \times (-d_2\mathbf{n}^\circ)\right) - \left(\mathbf{v}_1^\ominus + \boldsymbol{\omega}_1^\ominus \times (d_1\mathbf{n}^\circ)\right)$$

$$\mathbf{v}_{12}^\perp = \mathbf{v}_{12} - (\mathbf{n}^\circ \cdot \mathbf{v}_{12})\mathbf{n}^\circ$$

$$(\Delta\mathbf{u}_T)_3 = -\Delta t\mathbf{v}_{12}^\perp$$

Finally, we compute

$$\mathbf{u}_T^\circ = \mathbf{u}_T^- + (\Delta\mathbf{u}_T)_1 + (\Delta\mathbf{u}_T)_2 + (\Delta\mathbf{u}_T)_3.$$

## 2.1.4 Contact model (example)

The kinematic variables of an interaction are used to determine the forces acting on both spheres via a constitutive law. In DEM generally, some constitutive laws are expressed using strains and stresses while others prefer displacement/force formulation. The law described here falls in the latter category.

The constitutive law presented here is the most common in DEM, originally proposed by Cundall. While the kinematic variables are described in the previous section regardless of the contact model, the force evaluation depends on the nature of the material being modeled. The constitutive law presented here is the simplest non-cohesive elastic-frictional contact model, which Yade implements in *Law2_ScGeom_- FrictPhys_CundallStrack* (all constitutive laws derive from base class *LawFunctor*).

When new contact is established (discussed in *Engines*) it has its properties (*IPhys*) computed from *Materials* associated with both particles. In the simple case of frictional material *FrictMat*, *Ip2_- FrictMat_FrictMat_FrictPhys* creates a new *FrictPhys* instance, which defines normal stiffness $K_N$, shear stiffness $K_T$ and friction angle $\varphi$.

At each step, given normal and shear displacements $\mathbf{u}_N$, $\mathbf{u}_T$, normal and shear forces are computed (if $u_N > 0$, the contact is deleted without generating any forces):

$$\mathbf{F}_N = K_N u_N \mathbf{n},$$

$$\mathbf{F}_T^t = K_T \mathbf{u}_T$$

where $\mathbf{F}_N$ is normal force and $\mathbf{F}_T^t$ is trial shear force. A simple non-associated stress return algorithm is applied to compute final shear force

$$\mathbf{F}_T = \begin{cases} \mathbf{F}_T^t \frac{|\mathbf{F}_N|\tan\varphi}{|\mathbf{F}_T^t|} & \text{if } |\mathbf{F}_T^t| > |\mathbf{F}_N|\tan\varphi, \\ \mathbf{F}_T^t & \text{otherwise.} \end{cases}$$

Summary force $\mathbf{F} = \mathbf{F}_N + \mathbf{F}_T$ is then applied to both particles – each particle accumulates forces and torques acting on it in the course of each step. Because the force computed acts at contact point $\mathbf{C}$, which is difference from spheres' centers, torque generated by $\mathbf{F}$ must also be considered.

$$\mathbf{F}_1 += \mathbf{F} \qquad\qquad\qquad\qquad \mathbf{F}_2 += -\mathbf{F}$$

$$\mathbf{T}_1 += d_1(-\mathbf{n}) \times \mathbf{F} \qquad\qquad\qquad \mathbf{T}_2 += d_2\mathbf{n} \times \mathbf{F}.$$

## 2.1.5 Motion integration

Each particle accumulates generalized forces (forces and torques) from the contacts in which it participates. These generalized forces are then used to integrate motion equations for each particle separately; therefore, we omit $i$ indices denoting the $i$-th particle in this section.

The customary leapfrog scheme (also known as the Verlet scheme) is used, with some adjustments for rotation of non-spherical particles, as explained below. The "leapfrog" name comes from the fact that even derivatives of position/orientation are known at on-step points, whereas odd derivatives are known at mid-step points. Let us recall that we use $\mathfrak{a}^-$, $\mathfrak{a}^\circ$, $\mathfrak{a}^+$ for on-step values of $\mathfrak{a}$ at $t - \Delta t$, $t$ and $t + \Delta t$ respectively; and $\mathfrak{a}^\ominus$, $\mathfrak{a}^\oplus$ for mid-step values of $\mathfrak{a}$ at $t - \Delta t/2$, $t + \Delta t/2$.

Described integration algorithms are implemented in the *NewtonIntegrator* class in Yade.

### Position

Integrating motion consists in using current acceleration $\ddot{\mathbf{u}}^\circ$ on a particle to update its position from the current value $\mathbf{u}^\circ$ to its value at the next timestep $\mathbf{u}^+$. Computation of acceleration, knowing current forces $\mathbf{F}$ acting on the particle in question and its mass $\mathfrak{m}$, is simply

$$\ddot{\mathbf{u}}^\circ = \mathbf{F}/\mathfrak{m}.$$

Using the 2nd order finite difference with step $\Delta t$, we obtain

$$\ddot{\mathbf{u}}^\circ \cong \frac{\mathbf{u}^- - 2\mathbf{u}^\circ + \mathbf{u}^+}{\Delta t^2}$$

from which we express

$$\mathbf{u}^+ = 2\mathbf{u}^\circ - \mathbf{u}^- + \ddot{\mathbf{u}}^\circ \Delta t^2 =$$
$$= \mathbf{u}^\circ + \Delta t \underbrace{\left( \frac{\mathbf{u}^\circ - \mathbf{u}^-}{\Delta t} + \ddot{\mathbf{u}}^\circ \Delta t \right)}_{(\dagger)}.$$

Typically, $\mathbf{u}^-$ is already not known (only $\mathbf{u}^\circ$ is); we notice, however, that

$$\dot{\mathbf{u}}^\ominus \simeq \frac{\mathbf{u}^\circ - \mathbf{u}^-}{\Delta t},$$

i.e. the mean velocity during the previous step, which is known. Plugging this approximate into the $(\dagger)$ term, we also notice that mean velocity during the current step can be approximated as

$$\dot{\mathbf{u}}^\oplus \simeq \dot{\mathbf{u}}^\ominus + \ddot{\mathbf{u}}^\circ \Delta t,$$

which is $(\dagger)$; we arrive finally at

$$\mathbf{u}^+ = \mathbf{u}^\circ + \Delta t \left( \dot{\mathbf{u}}^\ominus + \ddot{\mathbf{u}}^\circ \Delta t \right).$$

The algorithm can then be written down by first computing current mean velocity $\dot{\mathbf{u}}^\oplus$ which we need to store for the next step (just as we use its old value $\dot{\mathbf{u}}^\ominus$ now), then computing the position for the next time step $\mathbf{u}^+$:

$$\dot{\mathbf{u}}^\oplus = \dot{\mathbf{u}}^\ominus + \ddot{\mathbf{u}}^\circ \Delta t$$
$$\mathbf{u}^+ = \mathbf{u}^\circ + \dot{\mathbf{u}}^\oplus \Delta t.$$

Positions are known at times $i\Delta t$ (if $\Delta t$ is constant) while velocities are known at $i\Delta t + \frac{\Delta t}{2}$. The fact that they interleave (jump over each other) in such way gave rise to the colloquial name "leapfrog" scheme.

### Orientation (spherical)

Updating particle orientation $\mathbf{q}^\circ$ proceeds in an analogous way to position update. First, we compute current angular acceleration $\dot{\boldsymbol{\omega}}^\circ$ from known current torque $\mathbf{T}$. For spherical particles where the inertia tensor is diagonal in any orientation (therefore also in current global orientation), satisfying $\mathbf{I}_{11} = \mathbf{I}_{22} = \mathbf{I}_{33}$, we can write

$$\dot{\boldsymbol{\omega}}_i^\circ = \mathbf{T}_i/\mathbf{I}_{11},$$

We use the same approximation scheme, obtaining an equation analogous to (2.1.5)

$$\boldsymbol{\omega}^{\oplus} = \boldsymbol{\omega}^{\ominus} + \Delta t \dot{\boldsymbol{\omega}}^{\circ}.$$

The quaternion $\Delta q$ representing rotation vector $\boldsymbol{\omega}^{\oplus} \Delta t$ is constructed, i.e. such that

$$(\Delta q)_{\vartheta} = |\boldsymbol{\omega}^{\oplus}|,$$
$$(\Delta q)_{\mathbf{u}} = \widehat{\boldsymbol{\omega}^{\oplus}}$$

Finally, we compute the next orientation $q^{+}$ by rotation composition

$$q^{+} = \Delta q q^{\circ}.$$

### Orientation (aspherical)

Integrating rotation of aspherical particles is considerably more complicated than their position, as their local reference frame is not inertial. Rotation of rigid body in the local frame, where inertia matrix $\mathbf{I}$ is diagonal, is described in the continuous form by Euler's equations ($i \in \{1, 2, 3\}$ and $i$, $j$, $k$ are subsequent indices):

$$\mathbf{T}_i = \mathbf{I}_{ii} \dot{\boldsymbol{\omega}}_i + (\mathbf{I}_{kk} - \mathbf{I}_{jj}) \boldsymbol{\omega}_j \boldsymbol{\omega}_k.$$

Due to the presence of the current values of both $\boldsymbol{\omega}$ and $\dot{\boldsymbol{\omega}}$, they cannot be solved using the standard leapfrog algorithm (that was the case for translational motion and also for the spherical bodies' rotation where this equation reduced to $\mathbf{T} = \mathbf{I} \dot{\boldsymbol{\omega}}$).

The algorithm presented here is described by [Allen1989] (pg. 84–89) and was designed by Fincham for molecular dynamics problems; it is based on extending the leapfrog algorithm by mid-step/on-step estimators of quantities known at on-step/mid-step points in the basic formulation. Although it has received criticism and more precise algorithms are known ([Omelyan1999], [Neto2006], [Johnson2008]), this one is currently implemented in Yade for its relative simplicity.

Each body has its local coordinate system based on the principal axes of inertia for that body. We use $\widetilde{\bullet}$ to denote vectors in local coordinates. The orientation of the local system is given by the current particle's orientation $q^{\circ}$ as a quaternion; this quaternion can be expressed as the (current) rotation matrix $\mathbf{A}$. Therefore, every vector $\mathbf{a}$ is transformed as $\widetilde{\mathbf{a}} = q \mathbf{a} q^{*} = \mathbf{A} \mathbf{a}$. Since $\mathbf{A}$ is a rotation (orthogonal) matrix, the inverse rotation $\mathbf{A}^{-1} = \mathbf{A}^{\mathsf{T}}$.

For given particle in question, we know

- $\widetilde{\mathbf{I}}^{\circ}$ (constant) inertia matrix; diagonal, since in local, principal coordinates,
- $\mathbf{T}^{\circ}$ external torque,
- $q^{\circ}$ current orientation (and its equivalent rotation matrix $\mathbf{A}$),
- $\boldsymbol{\omega}^{\ominus}$ mid-step angular velocity,
- $\mathbf{L}^{\ominus}$ mid-step angular momentum; this is an auxiliary variable that must be tracked in addition for use in this algorithm. It will be zero in the initial step.

Our goal is to compute new values of the latter three, that is $\mathbf{L}^{\oplus}$, $q^{+}$, $\boldsymbol{\omega}^{\oplus}$. We first estimate current angular momentum and compute current local angular velocity:

$$\mathbf{L}^{\circ} = \mathbf{L}^{\ominus} + \mathbf{T}^{\circ} \frac{\Delta t}{2}, \qquad\qquad \widetilde{\mathbf{L}}^{\circ} = \mathbf{A} \mathbf{L}^{\circ},$$
$$\mathbf{L}^{\oplus} = \mathbf{L}^{\ominus} + \mathbf{T}^{\circ} \Delta t, \qquad\qquad \widetilde{\mathbf{L}}^{\oplus} = \mathbf{A} \mathbf{L}^{\oplus},$$
$$\widetilde{\boldsymbol{\omega}}^{\circ} = \widetilde{\mathbf{I}}^{\circ -1} \widetilde{\mathbf{L}}^{\circ},$$
$$\widetilde{\boldsymbol{\omega}}^{\oplus} = \widetilde{\mathbf{I}}^{\circ -1} \widetilde{\mathbf{L}}^{\oplus}.$$

Then we compute $\dot{q}^\circ$, using $q^\circ$ and $\widetilde{\omega}^\circ$:

$$\begin{pmatrix} \dot{q}_w^\circ \\ \dot{q}_x^\circ \\ \dot{q}_y^\circ \\ \dot{q}_z^\circ \end{pmatrix} = \frac{1}{2} \begin{pmatrix} q_w^\circ & -q_x^\circ & -q_y^\circ & -q_z^\circ \\ q_x^\circ & q_w^\circ & -q_z^\circ & q_y^\circ \\ q_y^\circ & q_z^\circ & q_w^\circ & -q_x^\circ \\ q_z^\circ & -q_y^\circ & q_x^\circ & q_w^\circ \end{pmatrix} \begin{pmatrix} 0 \\ \widetilde{\omega}_x^\circ \\ \widetilde{\omega}_y^\circ \\ \widetilde{\omega}_z^\circ \end{pmatrix},$$

$$q^\oplus = q^\circ + \dot{q}^\circ \frac{\Delta t}{2}.$$

We evaluate $\dot{q}^\oplus$ from $q^\oplus$ and $\widetilde{\omega}^\oplus$ in the same way as in (2.1.5) but shifted by $\Delta t/2$ ahead. Then we can finally compute the desired values

$$q^+ = q^\circ + \dot{q}^\oplus \Delta t,$$
$$\omega^\oplus = \mathbf{A}^{-1} \widetilde{\omega}^\oplus$$

### Clumps (rigid aggregates)

DEM simulations frequently make use of rigid aggregates of particles to model complex shapes [Price2007] called *clumps*, typically composed of many spheres. Dynamic properties of clumps are computed from the properties of its members:

- For non-overlapping clump members the clump's mass $m_c$ is summed over members, the inertia tensor $\mathbf{I}_c$ is computed using the parallel axes theorem: $\mathbf{I}_c = \sum_i (m_i * d_i^2 + I_i)$, where $m_i$ is the mass of clump member $i$, $d_i$ is the distance from center of clump member $i$ to clump's centroid and $I_i$ is the inertia tensor of the clump member $i$.

- For overlapping clump members the clump's mass $m_c$ is summed over cells using a regular grid spacing inside axis-aligned bounding box (*Aabb*) of the clump, the inertia tensor is computed using the parallel axes theorem: $\mathbf{I}_c = \sum_j (m_j * d_j^2 + I_j)$, where $m_j$ is the mass of cell $j$, $d_j$ is the distance from cell center to clump's centroid and $I_j$ is the inertia tensor of the cell $j$.

Local axes are oriented such that they are principal and inertia tensor is diagonal and clump's orientation is changed to compensate rotation of the local system, as to not change the clump members' positions in global space. Initial positions and orientations of all clump members in local coordinate system are stored.

In Yade (class *Clump*), clump members behave as stand-alone particles during simulation for purposes of collision detection and contact resolution, except that they have no contacts created among themselves within one clump. It is at the stage of motion integration that they are treated specially. Instead of integrating each of them separately, forces/torques on those particles $F_i$, $T_i$ are converted to forces/torques on the clump itself. Let us denote $r_i$ relative position of each particle with regards to clump's centroid, in global orientation. Then summary force and torque on the clump are

$$F_c = \sum F_i,$$
$$T_c = \sum r_i \times F_i + T_i.$$

Motion of the clump is then integrated, using aspherical rotation integration. Afterwards, clump members are displaced in global space, to keep their initial positions and orientations in the clump's local coordinate system. In such a way, relative positions of clump members are always the same, resulting in the behavior of a rigid aggregate.

### Numerical damping

In simulations of quasi-static phenomena, it it desirable to dissipate kinetic energy of particles. Since most constitutive laws (including *Law_ScGeom_FrictPhys_Basic* shown above, *Contact model (example)*) do not include velocity-based damping (such as one in [Addetta2001]), it is possible to use artificial numerical damping. The formulation is described in [Pfc3dManual30], although our version is slightly adapted. The basic idea is to decrease forces which increase the particle velocities and vice versa by $(\Delta F)_d$, comparing the current acceleration sense and particle velocity sense. This is done by component, which makes the damping scheme clearly non-physical, as it is not invariant with respect to coordinate system rotation; on the other hand, it is very easy to compute. Cundall proposed the form (we omit particle indices $i$ since it applies to all of them separately):

$$\frac{(\Delta F)_{dw}}{F_w} = -\lambda_d \, \mathrm{sgn}(F_w \dot{u}_w^\ominus), \quad w \in \{x, y, z\}$$

where $\lambda_d$ is the damping coefficient. This formulation has several advantages [Hentz2003]:

- it acts on forces (accelerations), not constraining uniform motion;
- it is independent of eigenfrequencies of particles, they will be all damped equally;
- it needs only the dimensionless parameter $\lambda_d$ which does not have to be scaled.

In Yade, we use the adapted form

$$\frac{(\Delta F)_{dw}}{F_w} = -\lambda_d \, \mathrm{sgn} \, F_w \underbrace{\left( \dot{u}_w^\ominus + \frac{\ddot{u}_w^\circ \Delta t}{2} \right)}_{\simeq \dot{u}_w^\circ}, \tag{2.9}$$

where we replaced the previous mid-step velocity $\dot{u}^\ominus$ by its on-step estimate in parentheses. This is to avoid locked-in forces that appear if the velocity changes its sign due to force application at each step, i.e. when the particle in question oscillates around the position of equilibrium with $2\Delta t$ period.

In Yade, damping (2.9) is implemented in the *NewtonIntegrator* engine; the damping coefficient $\lambda_d$ is *NewtonIntegrator.damping*.

### Stability considerations

### Critical timestep

In order to ensure stability for the explicit integration sceheme, an upper limit is imposed on $\Delta t$:

$$\Delta t_{cr} = \frac{2}{\omega_{max}} \tag{2.10}$$

where $\omega_{max}$ is the highest eigenfrequency within the system.

### Single mass-spring system

Single 1D mass-spring system with mass $m$ and stiffness $K$ is governed by the equation

$$m\ddot{x} = -Kx$$

where $x$ is displacement from the mean (equilibrium) position. The solution of harmonic oscillation is $x(t) = A\cos(\omega t + \varphi)$ where phase $\varphi$ and amplitude $A$ are determined by initial conditions. The angular frequency

$$\omega^{(1)} = \sqrt{\frac{K}{m}} \tag{2.11}$$

does not depend on initial conditions. Since there is one single mass, $\omega_{max}^{(1)} = \omega^{(1)}$. Plugging (2.11) into (2.10), we obtain

$$\Delta t_{cr}^{(1)} = 2/\omega_{max}^{(1)} = 2\sqrt{m/K}$$

for a single oscillator.

**General mass-spring system**

In a general mass-spring system, the highest frequency occurs if two connected masses $\mathfrak{m}_i$, $\mathfrak{m}_j$ are in opposite motion; let us suppose they have equal velocities (which is conservative) and they are connected by a spring with stiffness $K_i$: displacement $\Delta x_i$ of $\mathfrak{m}_i$ will be accompanied by $\Delta x_j = -\Delta x_i$ of $\mathfrak{m}_j$, giving $\Delta F_i = -K_i(\Delta x_i - (-\Delta x_i)) = -2K_i\Delta x_i$. That results in apparent stiffness $K_i^{(2)} = 2K_i$, giving maximum eigenfrequency of the whole system

$$\omega_{\max} = \max_i \sqrt{K_i^{(2)}/\mathfrak{m}_i}.$$

The overall critical timestep is then

$$\Delta t_{cr} = \frac{2}{\omega_{\max}} = \min_i 2\sqrt{\frac{\mathfrak{m}_i}{K_i^{(2)}}} = \min_i 2\sqrt{\frac{\mathfrak{m}_i}{2K_i}} = \min_i \sqrt{2}\sqrt{\frac{\mathfrak{m}_i}{K_i}}. \tag{2.12}$$

This equation can be used for all 6 degrees of freedom (DOF) in translation and rotation, by considering generalized mass and stiffness matrices M and K, and replacing fractions $\frac{\mathfrak{m}_i}{K_i}$ by eigen values of $M.K^{-1}$. The critical timestep is then associated to the eigen mode with highest frequency :

$$\Delta t_{cr} = \min \Delta t_{crk}, \quad k \in \{1, ..., 6\}. \tag{2.13}$$

**DEM simulations**

In DEM simulations, per-particle stiffness $\mathbf{K}_{ij}$ is determined from the stiffnesses of contacts in which it participates. Suppose each contact has normal stiffness $K_{Nk}$, shear stiffness $K_{Tk} = \xi K_{Nk}$ and is oriented by normal $\mathbf{n}_k$. A translational stiffness matrix $\mathbf{K}_{ij}$ can be defined as the sum of contributions of all contacts in which it participates (indices k), as [Chareyre2005]:

$$\mathbf{K}_{ij} = \sum_k (K_{Nk} - K_{Tk})\mathbf{n}_i\mathbf{n}_j + K_{Tk} = \sum_j K_{Nk}\left((1-\xi)\mathbf{n}_i\mathbf{n}_j + \xi\right) \tag{2.14}$$

with i and j $\in \{x, y, z\}$. Equations (2.13) and (2.14) determine $\Delta t_{cr}$ in a simulation. A similar approach generalized to all 6 DOFs is implemented by the *GlobalStiffnessTimeStepper* engine in Yade. The derivation of generalized stiffness including rotational terms is very similar and can be found in [AboulHosn2017].

Note that for computation efficiency reasons, eigenvalues of the stiffness matrices are not computed. They are only approximated assuming than DOF's are uncoupled, and using the diagonal terms of $K.M^{-1}$. They give good approximates in typical mechanical systems.

There is one important condition that $\omega_{\max} > 0$: if there are no contacts between particles and $\omega_{\max} = 0$, we would obtain value $\Delta t_{cr} = \infty$. While formally correct, this value is numerically erroneous: we were silently supposing that stiffness remains constant during each timestep, which is not true if contacts are created as particles collide. In case of no contact, therefore, stiffness must be pre-estimated based on future interactions, as shown in the next section.

**Estimation of $\Delta t_{cr}$ by wave propagation speed**

Estimating timestep in absence of interactions is based on the connection between interaction stiffnesses and the particle's properties. Note that in this section, symbols E and $\rho$ refer exceptionally to Young's modulus and density of *particles*, not of macroscopic arrangement.

In Yade, particles have associated *Material* which defines density $\rho$ (*Material.density*), and also may define (in *ElastMat* and derived classes) particle's "Young's modulus" E (*ElastMat.young*). $\rho$ is used when particle's mass $\mathfrak{m}$ is initially computed from its $\rho$, while E is taken in account when creating new interaction between particles, affecting stiffness $K_N$. Knowing $\mathfrak{m}$ and $K_N$, we can estimate (2.14) for each particle; we obviously neglect

- number of interactions per particle $N_i$; for a "reasonable" radius distribution, however, there is a geometrically imposed upper limit (12 for a packing of spheres with equal radii, for instance);

- the exact relationship the between particles' rigidities $E_i$, $E_j$, supposing only that $K_N$ is somehow proportional to them.

By defining $E$ and $\rho$, particles have continuum-like quantities. Explicit integration schemes for continuum equations impose a critical timestep based on sonic speed $\sqrt{E/\rho}$; the elastic wave must not propagate farther than the minimum distance of integration points $l_{min}$ during one step. Since $E$, $\rho$ are parameters of the elastic continuum and $l_{min}$ is fixed beforehand, we obtain

$$\Delta t_{cr}^{(c)} = l_{min}\sqrt{\frac{\rho}{E}}.$$

For our purposes, we define $E$ and $\rho$ for each particle separately; $l_{min}$ can be replaced by the sphere's radius $R_i$; technically, $l_{min} = 2R_i$ could be used, but because of possible interactions of spheres and facets (which have zero thickness), we consider $l_{min} = R_i$ instead. Then

$$\Delta t_{cr}^{(p)} = \min_i R_i \sqrt{\frac{\rho_i}{E_i}}.$$

This algorithm is implemented in the *utils.PWaveTimeStep* function.

Let us compare this result to (2.12); this necessitates making several simplifying hypotheses:

- all particles are spherical and have the same radius $R$;

- the sphere's material has the same $E$ and $\rho$;

- the average number of contacts per sphere is $N$;

- the contacts have sufficiently uniform spatial distribution around each particle;

- the $\xi = K_N/K_T$ ratio is constant for all interactions;

- contact stiffness $K_N$ is computed from $E$ using a formula of the form

$$K_N = E\pi'R', \tag{2.15}$$

where $\pi'$ is some constant depending on the algorithm in usefootnote{For example, $\pi' = \pi/2$ in the concrete particle model (*Ip2_CpmMat_CpmMat_CpmPhys*), while $\pi' = 2$ in the classical DEM model (*Ip2_FrictMat_FrictMat_FrictPhys*) as implemented in Yade.} and $R'$ is half-distance between spheres in contact, equal to $R$ for the case of interaction radius $R_I = 1$. If $R_I = 1$ (and $R' \equiv R$ by consequence), all interactions will have the same stiffness $K_N$. In other cases, we will consider $K_N$ as the average stiffness computed from average $R'$ (see below).

As all particles have the same parameters, we drop the $i$ index in the following formulas.

We try to express the average per-particle stiffness from (2.14). It is a sum over all interactions where $K_N$ and $\xi$ are scalars that will not rotate with interaction, while $n_w$ is $w$-th component of unit interaction normal $\mathbf{n}$. Since we supposed uniform spatial distribution, we can replace $n_w^2$ by its average value $\overline{n}_w^2$. Recognizing components of $\mathbf{n}$ as direction cosines, the average values of $n_w^2$ is $1/3$. We find the average value by integrating over all possible orientations, which are uniformly distributed in space:

Moreover, since all directions are equal, we can write the per-body stiffness as $K = \mathbf{K}_w$ for all $w \in \{x, y, z\}$. We obtain

$$K = \sum K_N \left( (1 - \xi)\frac{1}{3} + \xi \right) = \sum K_N \frac{1 + 2\xi}{3}$$

and can put constant terms (everything) in front of the summation. $\sum 1$ equals the number of contacts per sphere, i.e. $N$. Arriving at

$$K = NK_N \frac{1 - 2\xi}{3},$$

we substitute K into (2.12) using (2.15):

$$\Delta t_{cr} = \sqrt{2}\sqrt{\frac{m}{K}} = \sqrt{2}\sqrt{\frac{\frac{4}{3}\pi R^3 \rho}{NE\pi'R\frac{1-2\xi}{3}}} = \underbrace{R\sqrt{\frac{\rho}{E}}}_{\Delta t_{cr}^{(p)}} 2\sqrt{\frac{\pi/\pi'}{N(1-2\xi)}}.$$

The ratio of timestep $\Delta t_{cr}^{(p)}$ predicted by the p-wave velocity and numerically stable timestep $\Delta t_{cr}$ is the inverse value of the last (dimensionless) term:

$$\frac{\Delta t_{cr}^{(p)}}{\Delta t_{cr}} = 2\sqrt{\frac{N(1+\xi)}{\pi/\pi'}}.$$

Actual values of this ratio depend on characteristics of packing $N$, $K_N/K_T = \xi$ ratio and the way of computing contact stiffness from particle rigidity. Let us show it for two models in Yade:

**Concrete particle model** computes contact stiffness from the equivalent area $A_{eq}$ first (2.6),

$$A_{eq} = \pi R^2 K_N \qquad\qquad = \frac{A_{eq}E}{d_0}.$$

$d_0$ is the initial contact length, which will be, for interaction radius (2.5) $R_I > 1$, in average larger than $2R$. For $R_I = 1.5$ , we can roughly estimate $\overline{d}_0 = 1.25 \cdot 2R = \frac{5}{2}R$, getting

$$K_N = E\left(\frac{2}{5}\pi\right) R$$

where $\frac{2}{5}\pi = \pi'$ by comparison with (2.15).

Interaction radius $R_I = 1.5$ leads to average $N \approx 12$ interactions per sphere for dense packing of spheres with the same radius $R$. $\xi = 0.2$ is calibrated to match the desired macroscopic Poisson's ratio $\nu = 0.2$.

Finally, we obtain the ratio

$$\frac{\Delta t_{cr}^{(p)}}{\Delta t_{cr}} = 2\sqrt{\frac{12(1-2\cdot 0.2)}{\frac{\pi}{(2/5)\pi}}} = 3.39,$$

showing significant overestimation by the p-wave algorithm.

**Non-cohesive dry friction model** is the basic model proposed by Cundall explained in *Contact model (example)*. Supposing almost-constant sphere radius $R$ and rather dense packing, each sphere will have $N = 6$ interactions on average (that corresponds to maximally dense packing of spheres with a constant radius). If we use the *Ip2_FrictMat_FrictMat_FrictPhys* class, we have $\pi' = 2$, as $K_N = E2R$; we again use $\xi = 0.2$ (for lack of a more significant value). In this case, we obtain the result

$$\frac{\Delta t_{cr}^{(p)}}{\Delta t_{cr}} = 2\sqrt{\frac{6(1-2\cdot 0.2)}{\pi/2}} = 3.02$$

which again overestimates the numerical critical timestep.

To conclude, p-wave timestep gives estimate proportional to the real $\Delta t_{cr}$, but in the cases shown, the value of about $\Delta t = 0.3\Delta t_{cr}^{(p)}$ should be used to guarantee stable simulation.

### Non-elastic $\Delta t$ constraints

Let us note at this place that not only $\Delta t_{cr}$ assuring numerical stability of motion integration is a constraint. In systems where particles move at relatively high velocities, position change during one timestep can lead to non-elastic irreversible effects such as damage. The $\Delta t$ needed for reasonable result can be lower $\Delta t_{cr}$. We have no rigorously derived rules for such cases.

## 2.1.6 Periodic boundary conditions

While most DEM simulations happen in $\mathbb{R}^3$ space, it is frequently useful to avoid boundary effects by using periodic space instead. In order to satisfy periodicity conditions, periodic space is created by repetition of parallelepiped-shaped cell. In Yade, periodic space is implemented in the *Cell* class. The geometry of the cell in the reference coordinates system is defined by three edges of the parallelepiped. The corresponding base vectors are stored in the columns of matrix **H** (*Cell.hSize*).

The initial **H** can be explicitly defined as a 3x3 matrix at the beginning of the simulation. There are no restrictions on the possible shapes: any parallelepiped is accepted as the initial cell. If the base vectors are axis-aligned, defining only their sizes can be more convenient than defining the full **H** matrix; in that case it is enough to define the norms of columns in **H** (see *Cell.size*).

After the definition of the initial cell's geometry, **H** should generally not be modified by direct assignment. Instead, its deformation rate will be defined via the velocity gradient *Cell.velGrad* described below. It is the only variable that let the period deformation be correctly accounted for in constitutive laws and Newton integrator (*NewtonIntegrator*).

### Deformations handling

The deformation of the cell over time is defined via a tensor representing the gradient of an homogeneous velocity field $\nabla\mathbf{v}$ (*Cell.velGrad*). This gradient represents arbitrary combinations of rotations and stretches. It can be imposed externaly or updated by *boundary controllers* (see *PeriTriaxController* or *Peri3dController*) in order to reach target strain values or to maintain some prescribed stress.

The velocity gradient is integrated automatically over time, and the cumulated transformation is reflected in the transformation matrix **F** (*Cell.trsf*) and the current shape of the cell **H**. The per-step transformation update reads (it is similar for **H**), with I the identity matrix:

$$\mathbf{F}^+ = (\mathbf{I} + \nabla\mathbf{v}\Delta\mathbf{t})\mathbf{F}^\circ.$$

**F** can be set back to identity at any point in simulations, in order to define the current state as reference for strains definition in boundary controllers. It will have no effect on **H**.

Along with the automatic integration of cell transformation, there is an option to homothetically displace all particles so that $\nabla\mathbf{v}$ is applied over the whole simulation (enabled via *Cell.homoDeform*). This avoids all boundary effects coming from change of the velocity gradient.

### Collision detection in periodic cell

In usual implementations, particle positions are forced to be inside the cell by wrapping their positions if they get over the boundary (so that they appear on the other side). As we wanted to avoid abrupt changes of position (it would make particle's velocity inconsistent with step displacement change), a different method was chosen.

### Approximate collision detection

Pass 1 collision detection (based on sweep and prune algorithm, sect. *Sweep and prune*) operates on axis-aligned bounding boxes (*Aabb*) of particles. During the collision detection phase, bounds of all *Aabb's* are wrapped inside the cell in the first step. At subsequent runs, every bound remembers by how many cells it was initially shifted from coordinate given by the *Aabb* and uses this offset repeatedly as it is being updated from *Aabb* during particle's motion. Bounds are sorted using the periodic insertion sort algorithm (sect. *Periodic insertion sort algorithm*), which tracks periodic cell boundary ∥.

Upon inversion of two *Aabb's*, their collision along all three axes is checked, wrapping real coordinates inside the cell for that purpose.

This algorithm detects collisions as if all particles were inside the cell but without the need of constructing "ghost particles" (to represent periodic image of a particle which enters the cell from the other side) or changing the particle's positions.

It is required by the implementation (and partly by the algorithm itself) that particles do not span more than half of the current cell size along any axis; the reason is that otherwise two (or more) contacts between both particles could appear, on each side. Since Yade identifies contacts by *Body.id* of both bodies, they would not be distinguishable.

In presence of shear, the sweep-and-prune collider could not sort bounds independently along three axes: collision along x axis depends on the mutual position of particles on the y axis. Therefore, bounding boxes *are expressed in transformed coordinates* which are perpendicular in the sense of collision detection. This requires some extra computation: *Aabb* of sphere in transformed coordinates will no longer be cube, but cuboid, as the sphere itself will appear as ellipsoid after transformation. Inversely, the sphere in simulation space will have a parallelepiped bounding "box", which is cuboid around the ellipsoid in transformed axes (the *Aabb* has axes aligned with transformed cell basis). This is shown in fig. *fig-cell-shear-aabb*.

The restriction of a single particle not spanning more than half of the transformed axis becomes stringent as *Aabb* is enlarged due to shear. Considering *Aabb* of a sphere with radius r in the cell where $x' \equiv x$, $z' \equiv z$, but $\angle(y, y') = \varphi$, the x-span of the *Aabb* will be multiplied by $1/\cos\varphi$. For the infinite shear $\varphi \to \pi/2$, which can be desirable to simulate, we have $1/\cos\varphi \to \infty$. Fortunately, this limitation can be easily circumvented by realizing the quasi-identity of all periodic cells which, if repeated in space, create the same grid with their corners: the periodic cell can be flipped, keeping all particle interactions intact, as shown in fig. *fig-cell-flip*. It only necessitates adjusting the *Interaction.cellDist* of interactions and re-initialization of the collider (`Collider::invalidatePersistentData`). Cell flipping is implemented in the *utils.flipCell* function.

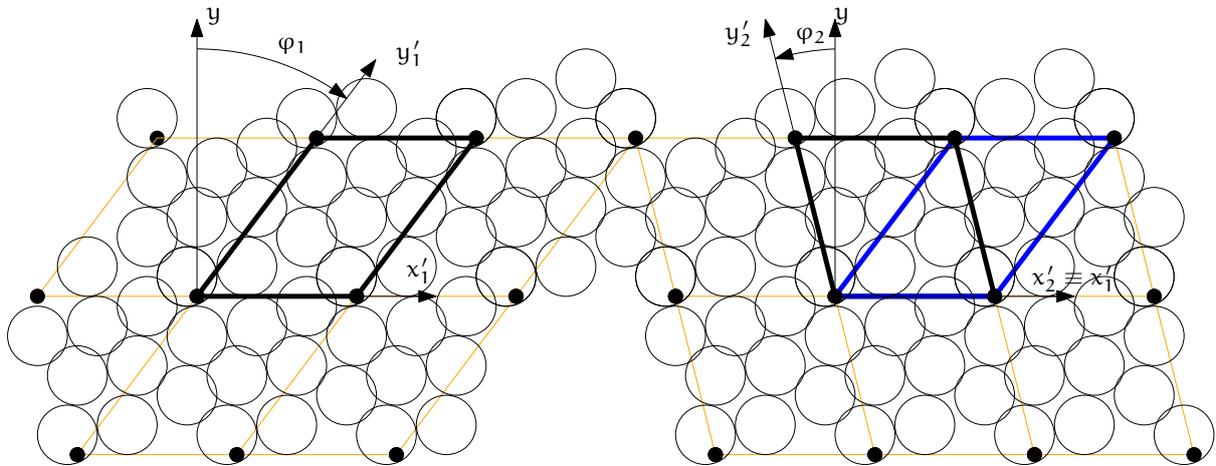

Fig. 6: Flipping cell (*utils.flipCell*) to avoid infinite stretch of the bounding boxes' spans with growing $\varphi$. Cell flip does not affect interactions from the point of view of the simulation. The periodic arrangement on the left is the same as the one on the right, only the cell is situated differently between identical grid points of repetition; at the same time $|\varphi_2| < |\varphi_1|$ and sphere bounding box's x-span stretched by $1/\cos\varphi$ becomes smaller. Flipping can be repeated, making effective infinite shear possible.

This algorithm is implemented in *InsertionSortCollider* and is used whenever simulation is periodic (*Omega.isPeriodic*); individual *BoundFunctor's* are responsible for computing sheared *Aabb's*; currently it is implemented for spheres and facets (in *Bo1_Sphere_Aabb* and *Bo1_Facet_Aabb* respectively).

**Exact collision detection**

When the collider detects approximate contact (on the *Aabb* level) and the contact does not yet exist, it creates *potential* contact, which is subsequently checked by exact collision algorithms (depending on the combination of *Shapes*). Since particles can interact over many periodic cells (recall we never change their positions in simulation space), the collider embeds the relative cell coordinate of particles in the interaction itself (*Interaction.cellDist*) as an *integer* vector `c`. Multiplying current cell size `Ts` by `c` component-wise, we obtain particle offset $\Delta x$ in aperiodic $\mathbb{R}^3$; this value is passed (from *InteractionLoop*) to the functor computing exact collision (*IGeomFunctor*), which adds it to the position of the particle *Interaction.id2*.

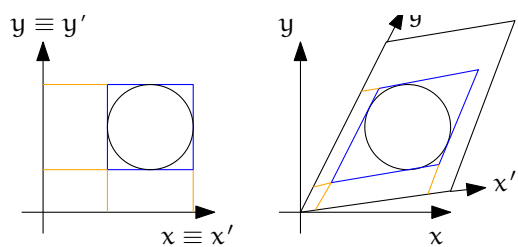

Fig. 7: Constructing axis-aligned bounding box (*Aabb*) of a sphere in simulation space coordinates (without periodic cell – left) and transformed cell coordinates (right), where collision detection axes $x'$, $y'$ are not identical with simulation space axes $x$, $y$. Bounds' projection to axes is shown by orange lines.

By storing the integral offset `c`, $\Delta x$ automatically updates as cell parameters change.

**Periodic insertion sort algorithm**

The extension of sweep and prune algorithm (described in *Sweep and prune*) to periodic boundary conditions is non-trivial. Its cornerstone is a periodic variant of the insertion sort algorithm, which involves keeping track of the "period" of each boundary; e.g. taking period $\langle 0, 10 \rangle$, then $\delta_1 \equiv -2_2 < 2_2$ (subscript indicating period). Doing so efficiently (without shuffling data in memory around as bound wraps from one period to another) requires moving period boundary rather than bounds themselves and making the comparison work transparently at the edge of the container.

This algorithm was also extended to handle non-orthogonal periodic *Cell* boundaries by working in transformed rather than Cartesian coordinates; this modifies computation of *Aabb* from Cartesian coordinates in which bodies are positioned (treated in detail in *Approximate collision detection*).

The sort algorithm is tracking *Aabb* extrema along all axes. At the collider's initialization, each value is assigned an integral period, i.e. its distance from the cell's interior expressed in the cell's dimension along its respective axis, and is wrapped to a value inside the cell. We put the period number in subscript.

Let us give an example of coordinate sequence along x axis (in a real case, the number of elements would be even, as there is maximum and minimum value couple for each particle; this demonstration only shows the sorting algorithm, however.)

$$4_1 \qquad 12_2 \quad \| \quad -1_2 \qquad -2_4 \qquad 5_0$$

with cell x-size $s_x = 10$. The $4_1$ value then means that the real coordinate $x_i$ of this extremum is $x_i + 1 \cdot 10 = 4$, i.e. $x_i = -4$. The $\|$ symbol denotes the periodic cell boundary.

Sorting starts from the first element in the cell, i.e. right of $\|$, and inverts elements as in the aperiodic variant. The rules are, however, more complicated due to the presence of the boundary $\|$:

| $(\leq)$ | stop inverting if neighbors are ordered; |
|---|---|
| $(\|\bullet)$ | current element left of $\|$ is below 0 (lower period boundary); in this case, decrement element's period, decrease its coordinate by $s_x$ and move $\|$ right; |
| $(\bullet\|)$ | current element right of $\|$ is above $s_x$ (upper period boundary); increment element's period, increase its coordinate by $s_x$ and move $\|$ left; |
| $(\|\!\|)$ | inversion across $\|$ must subtract $s_x$ from the left coordinate during comparison. If the elements are not in order, they are swapped, but they must have their periods changed as they traverse $\|$. Apply $(\|\bullet)$ if necessary; |
| $(\|\circ)$ | if after $(\|\!\|)$ the element that is now right of $\|$ has $x_i < s_x$, decrease its coordinate by $s_x$ and decrement its period. Do not move $\|$. |

In the first step, $(\|\bullet)$ is applied, and inversion with $12_2$ happens; then we stop because of $(\leq)$:

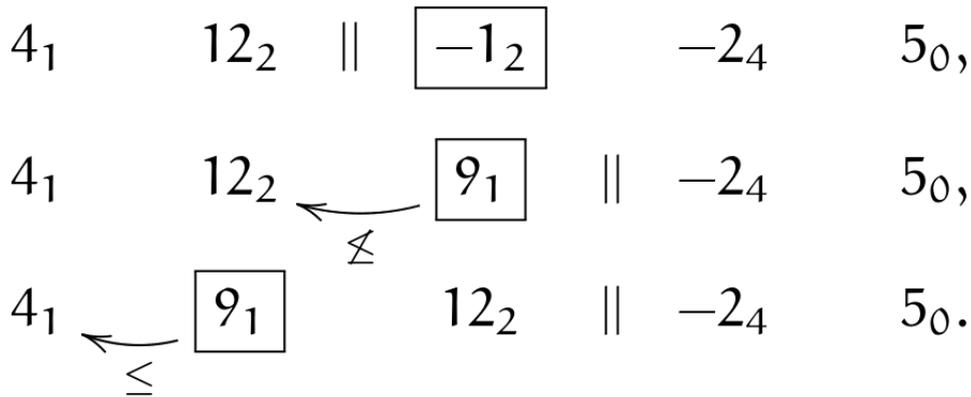

We move to next element $\boxed{-2_4}$; first, we apply $(\|\bullet)$, then invert until $(\leq)$:

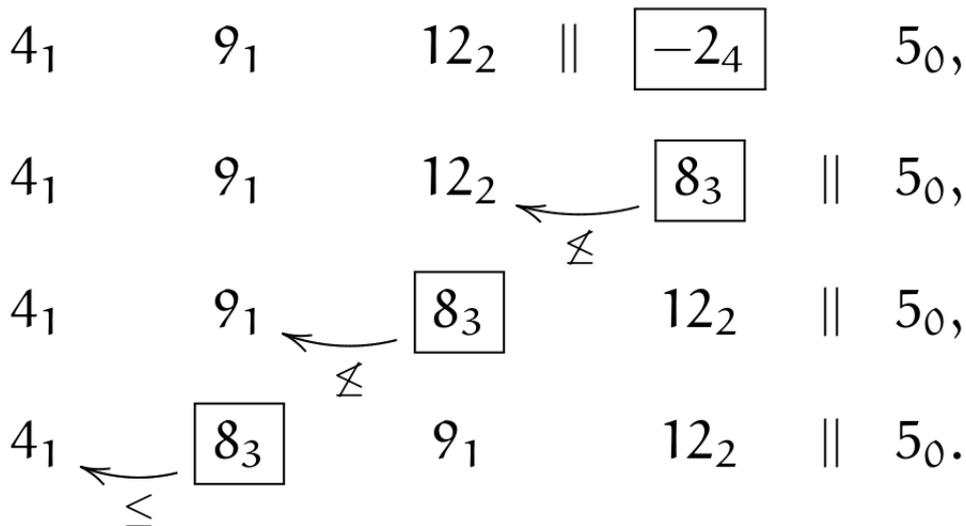

The next element is $\boxed{5_0}$; we satisfy $(\|\!\|)$, therefore instead of comparing $12_2 > 5_0$, we must do $(12_2 - s_x) = 2_3 \leq 5$; we adjust periods when swapping over $\|$ and apply $(\|\circ)$, turning $12_2$ into $2_3$; then we keep inverting, until $(\leq)$:

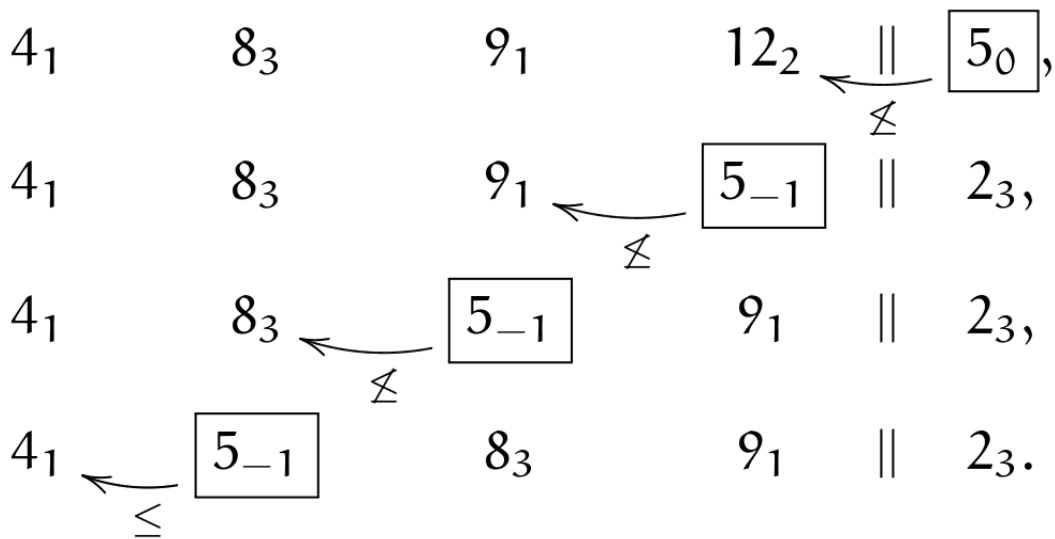

We move (wrapping around) to $\boxed{4_1}$, which is ordered:

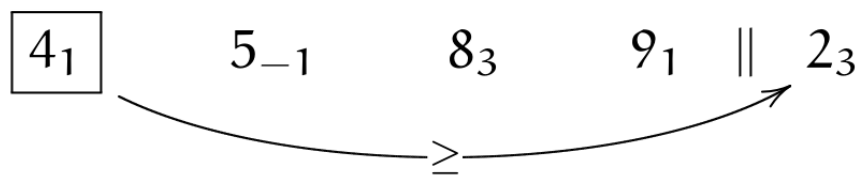

and so is the last element

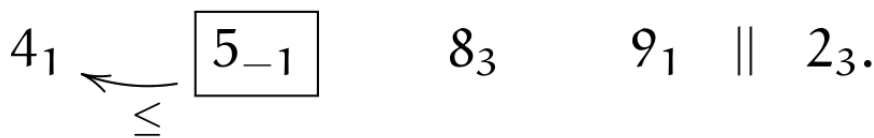

### 2.1.7 Computational aspects

**Cost**

The DEM computation using an explicit integration scheme demands a relatively high number of steps during simulation, compared to implicit scehemes. The total computation time $Z$ of simulation spanning $T$ seconds (of simulated time), containing $N$ particles in volume $V$ depends on:

- linearly, the number of steps $i = T/(s_t \Delta t_{cr})$, where $s_t$ is timestep safety factor; $\Delta t_{cr}$ can be estimated by p-wave velocity using $E$ and $\rho$ (sect. *Estimation of \Dtcr by wave propagation speed*) as $\Delta t_{cr}^{(p)} = r\sqrt{\frac{\rho}{E}}$. Therefore

$$i = \frac{T}{s_t r}\sqrt{\frac{E}{\rho}}.$$

- the number of particles $N$; for fixed value of simulated domain volume $V$ and particle radius $r$

$$N = p\frac{V}{\frac{4}{3}\pi r^3},$$

where $p$ is packing porosity, roughly $\frac{1}{2}$ for dense irregular packings of spheres of similar radius.

The dependency is not strictly linear (which would be the best case), as some algorithms do not scale linearly; a case in point is the sweep and prune collision detection algorithm introduced in sect. *Sweep and prune*, with scaling roughly $\mathcal{O}(N \log N)$.

The number of interactions scales with $N$, as long as packing characteristics are the same.

- the number of computational cores $n_{cpu}$; in the ideal case, the dependency would be inverse-linear were all algorithms parallelized (in Yade, collision detection is not).

Let us suppose linear scaling. Additionally, let us suppose that the material to be simulated $(E, \rho)$ and the simulation setup $(V, T)$ are given in advance. Finally, dimensionless constants $s_t$, $p$ and $n_{cpu}$ will have a fixed value. This leaves us with one last degree of freedom, $r$. We may write

$$Z \propto iN \frac{1}{n_{cpu}} = \frac{T}{s_t \, r} \sqrt{\frac{E}{\rho}} p \frac{V}{\frac{4}{3}\pi r^3} \frac{1}{n_{cpu}} \propto \frac{1}{r} \frac{1}{r^3} = \frac{1}{r^4}.$$

This (rather trivial) result is essential to realize DEM scaling; if we want to have finer results, refining the "mesh" by halving $r$, the computation time will grow $2^4 = 16$ times.

For very crude estimates, one can use a known simulation to obtain a machine "constant"

$$\mu = \frac{Z}{Ni}$$

with the meaning of time per particle and per timestep (in the order of $10^{-6}$ s for current machines). $\mu$ will be only useful if simulation characteristics are similar and non-linearities in scaling do not have major influence, i.e. $N$ should be in the same order of magnitude as in the reference case.

**Result indeterminism**

It is naturally expected that running the same simulation several times will give exactly the same results: although the computation is done with finite precision, round-off errors would be deterministically the same at every run. While this is true for *single-threaded* computation where exact order of all operations is given by the simulation itself, it is not true anymore in *multi-threaded* computation which is described in detail in later sections.

The straight-forward manner of parallel processing in explicit DEM is given by the possibility of treating interactions in arbitrary order. Strain and stress is evaluated for each interaction independently, but forces from interactions have to be summed up. If summation order is also arbitrary (in Yade, forces are accumulated for each thread in the order interactions are processed, then summed together), then the results can be slightly different. For instance

```
(1/10.)+(1/13.)+(1/17.)=0.23574660633484162
(1/17.)+(1/13.)+(1/10.)=0.23574660633484165
```

As forces generated by interactions are assigned to bodies in quasi-random order, summary force $F_i$ on the body can be different between single-threaded and multi-threaded computations, but also between different runs of multi-threaded computation with exactly the same parameters. Exact thread scheduling by the kernel is not predictable since it depends on asynchronous events (hardware interrupts) and other unrelated tasks running on the system; and it is thread scheduling that ultimately determines summation order of force contributions from interactions.

## 2.2 User's manual

### 2.2.1 Scene construction

**Adding particles**

The *BodyContainer* holds *Body* objects in the simulation; it is accessible as `O.bodies`.

### Creating Body objects

*Body* objects are only rarely constructed by hand by their components (*Shape*, *Bound*, *State*, *Material*); instead, convenience functions *sphere*, *facet* and *wall* are used to create them. Using these functions also ensures better future compatibility, if internals of *Body* change in some way. These functions receive geometry of the particle and several other characteristics. See their documentation for details. If the same *Material* is used for several (or many) bodies, it can be shared by adding it in `O.materials`, as explained below.

### Defining materials

The `O.materials` object (instance of *Omega.materials*) holds defined shared materials for bodies. It only supports addition, and will typically hold only a few instances (though there is no limit).

`label` given to each material is optional, but can be passed to *sphere* and other functions for constructing body. The value returned by `O.materials.append` is an `id` of the material, which can be also passed to *sphere* – it is a little bit faster than using label, though not noticeable for small number of particles and perhaps less convenient.

If no *Material* is specified when calling *sphere*, the *last* defined material is used; that is a convenient default. If no material is defined yet (hence there is no last material), a default material will be created: FrictMat(density=1e3,young=1e7,poisson=.3,frictionAngle=.5). This should not happen for serious simulations, but is handy in simple scripts, where exact material properties are more or less irrelevant.

```
Yade [1]: len(O.materials)
Out[1]: 0

Yade [2]: idConcrete=O.materials.append(FrictMat(young=30e9,poisson=.2,frictionAngle=.6,label=
 ↪"concrete"))

Yade [3]: O.materials[idConcrete]
Out[3]: <FrictMat instance at 0x3f01be0>

# uses the last defined material
Yade [4]: O.bodies.append(sphere(center=(0,0,0),radius=1))
Out[4]: 0

# material given by id
Yade [5]: O.bodies.append(sphere((0,0,2),1,material=idConcrete))
Out[5]: 1

# material given by label
Yade [6]: O.bodies.append(sphere((0,2,0),1,material="concrete"))
Out[6]: 2

Yade [7]: idSteel=O.materials.append(FrictMat(young=210e9,poisson=.25,frictionAngle=.8,label=
 ↪"steel"))

Yade [8]: len(O.materials)
Out[8]: 2

# implicitly uses "steel" material, as it is the last one now
Yade [9]: O.bodies.append(facet([(1,0,0),(0,1,0),(-1,-1,0)]))
Out[9]: 3
```

### Adding multiple particles

As shown above, bodies are added one by one or several at the same time using the `append` method:

```
Yade [10]: O.bodies.append(sphere((0,10,0),1))
Out[10]: 0

Yade [11]: O.bodies.append(sphere((0,0,2),1))
Out[11]: 1

# this is the same, but in one function call
Yade [12]: O.bodies.append([
   ....:     sphere((0,0,0),1),
   ....:     sphere((1,1,3),1)
   ....: ])
   ....:
Out[12]: [2, 3]
```

Many functions introduced in next sections return list of bodies which can be readily added to the simulation, including

- packing generators, such as *pack.randomDensePack*, *pack.regularHexa*
- surface function *pack.gtsSurface2Facets*
- import functions *ymport.gmsh*, *ymport.stl*, ...

As those functions use *sphere* and *facet* internally, they accept additional arguments passed to those functions. In particular, material for each body is selected following the rules above (last one if not specified, by label, by index, etc.).

### Clumping particles together

In some cases, you might want to create rigid aggregate of individual particles (i.e. particles will retain their mutual position during simulation). This we call a *clump*. A clump is internally represented by a special *body*, referenced by *clumpId* of its members (see also *isClump*, *isClumpMember* and *isStandalone*). Like every body a clump has a *position*, which is the (mass) balance point between all members. A clump body itself has no *interactions* with other bodies. Interactions between clumps is represented by interactions between clump members. There are no interactions between clump members of the same clump.

YADE supports different ways of creating clumps:

- Create clumps and spheres (clump members) directly with one command:

The function *appendClumped()* is designed for this task. For instance, we might add 2 spheres tied together:

```
Yade [13]: O.bodies.appendClumped([
   ....:     sphere([0,0,0],1),
   ....:     sphere([0,0,2],1)
   ....: ])
   ....:
Out[13]: (2, [0, 1])

Yade [14]: len(O.bodies)
Out[14]: 3

Yade [15]: O.bodies[1].isClumpMember, O.bodies[2].clumpId
Out[15]: (True, 2)

Yade [16]: O.bodies[2].isClump, O.bodies[2].clumpId
Out[16]: (True, 2)
```

-> *appendClumped()* returns a tuple of ids (clumpId,[memberId1,memberId2,...])

- Use existing spheres and clump them together:

For this case the function *clump()* can be applied on a list of existing bodies:

```
Yade [17]: bodyList = []

Yade [18]: for ii in range(0,5):
    ....:        bodyList.append(O.bodies.append(sphere([ii,0,1],.5)))#create a "chain" of 5 spheres
    ....:

Yade [19]: print(bodyList)
[0, 1, 2, 3, 4]

Yade [20]: idClump=O.bodies.clump(bodyList)
```

-> *clump()* returns clumpId

- Another option is to replace *standalone* spheres from a given packing (see *SpherePack* and *make-Cloud*) by clumps using clump templates.

This is done by a function called *replaceByClumps()*. This function takes a list of *clumpTemplates()* and a list of amounts and replaces spheres by clumps. The volume of a new clump will be the same as the volume of the sphere, that was replaced (clump volume/mass/inertia is accounting for overlaps assuming that there are only pair overlaps).

-> *replaceByClumps()* returns a list of tuples: [(clumpId1,[memberId1,memberId2,...]),(clumpId2, [memberId1,memberId2,...]),...]

It is also possible to *add* bodies to a clump and *release* bodies from a clump. Also you can *erase* the clump (clump members will become standalone).

Additionally YADE allows to achieve the *roundness* of a clump or roundness coefficient of a packing. Parts of the packing can be excluded from roundness measurement via exclude list.

```
Yade [21]: bodyList = []

Yade [22]: for ii in range(1,5):
    ....:        bodyList.append(O.bodies.append(sphere([ii,ii,ii],.5)))
    ....:

Yade [23]: O.bodies.clump(bodyList)
Out[23]: 4

Yade [24]: RC=O.bodies.getRoundness()

Yade [25]: print(RC)
0.25619141423166986
```

-> *getRoundness()* returns roundness coefficient RC of a packing or a part of the packing

**Note:** Have a look at examples/clumps/ folder. There you will find some examples, that show usage of different functions for clumps.

### Sphere packings

Representing a solid of an arbitrary shape by arrangement of spheres presents the problem of sphere packing, i.e. spatial arrangement of spheres such that a given solid is approximately filled with them. For the purposes of DEM simulation, there can be several requirements.

1. Distribution of spheres' radii. Arbitrary volume can be filled completely with spheres provided there are no restrictions on their radius; in such case, number of spheres can be infinite and their radii approach zero. Since both number of particles and minimum sphere radius (via critical timestep) determine computation cost, radius distribution has to be given mandatorily. The most typical distribution is uniform: mean±dispersion; if dispersion is zero, all spheres will have the same radius.

2. Smooth boundary. Some algorithms treat boundaries in such way that spheres are aligned on them, making them smoother as surface.

3. Packing density, or the ratio of spheres volume and solid size. It is closely related to radius distribution.

4. Coordination number, (average) number of contacts per sphere.

5. Isotropy (related to regularity/irregularity); packings with preferred directions are usually not desirable, unless the modeled solid also has such preference.

6. Permissible Spheres' overlap; some algorithms might create packing where spheres slightly overlap; since overlap usually causes forces in DEM, overlap-free packings are sometimes called "stress-free .

### Volume representation

There are 2 methods for representing exact volume of the solid in question in Yade: boundary representation and constructive solid geometry. Despite their fundamental differences, they are abstracted in Yade in the *Predicate* class. Predicate provides the following functionality:

1. defines axis-aligned bounding box for the associated solid (optionally defines oriented bounding box);

2. can decide whether given point is inside or outside the solid; most predicates can also (exactly or approximately) tell whether the point is inside *and* satisfies some given padding distance from the represented solid boundary (so that sphere of that volume doesn't stick out of the solid).

### Constructive Solid Geometry (CSG)

CSG approach describes volume by geometric *primitives* or primitive solids (sphere, cylinder, box, cone, ...) and boolean operations on them. Primitives defined in Yade include *inCylinder*, *inSphere*, *inEllipsoid*, *inHyperboloid*, *notInNotch*.

For instance, *hyperboloid* (dogbone) specimen for tension-compression test can be constructed in this way (shown at img. *img-hyperboloid*):

```python
from yade import pack

## construct the predicate first
pred=pack.inHyperboloid(centerBottom=(0,0,-.1),centerTop=(0,0,.1),radius=.05,skirt=.03)
## alternatively: pack.inHyperboloid((0,0,-.1),(0,0,.1),.05,.03)

## pack the predicate with spheres (will be explained later)
spheres=pack.randomDensePack(pred,spheresInCell=2000,radius=3.5e-3)

## add spheres to simulation
O.bodies.append(spheres)
```

### Boundary representation (BREP)

Representing a solid by its boundary is much more flexible than CSG volumes, but is mostly only approximate. Yade interfaces to GNU Triangulated Surface Library (GTS) to import surfaces readable by GTS, but also to construct them explicitly from within simulation scripts. This makes possible parametric construction of rather complicated shapes; there are functions to create set of 3d polylines from 2d polyline (*pack.revolutionSurfaceMeridians*), to triangulate surface between such set of 3d polylines (*pack.sweptPolylines2gtsSurface*).

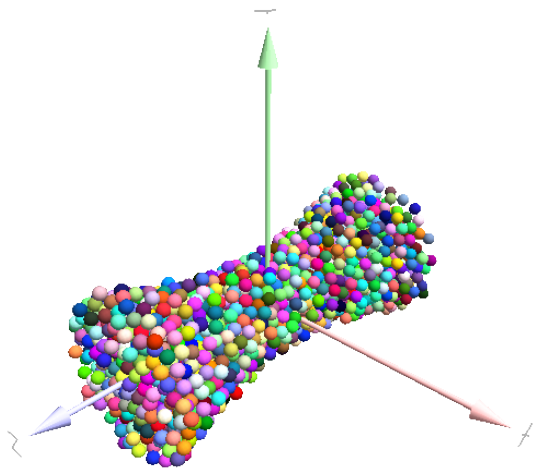

Fig. 8: Specimen constructed with the *pack.inHyperboloid* predicate, packed with *pack.randomDensePack*.

For example, we can construct a simple funnel (examples/funnel.py, shown at *img-funnel*):

```python
from numpy import linspace
from yade import pack

# angles for points on circles
thetas=linspace(0,2*pi,num=16,endpoint=True)

# creates list of polylines in 3d from list of 2d projections
# turned from 0 to π
meridians=pack.revolutionSurfaceMeridians(
        [[(3+rad*sin(th),10*rad+rad*cos(th)) for th in thetas] for rad in linspace(1,2,
        num=10)],
        linspace(0,pi,num=10)
)

# create surface
surf=pack.sweptPolylines2gtsSurface(
        meridians
        +[[Vector3(5*sin(-th),-10+5*cos(-th),30) for th in thetas]]  # add funnel top
)

# add to simulation
O.bodies.append(pack.gtsSurface2Facets(surf))
```

GTS surface objects can be used for 2 things:

1. *pack.gtsSurface2Facets* function can create the triangulated surface (from *Facet* particles) in the simulation itself, as shown in the funnel example. (Triangulated surface can also be imported directly from a STL file using *ymport.stl*.)

2. *pack.inGtsSurface* predicate can be created, using the surface as boundary representation of the enclosed volume.

The examples/gts-horse/gts-horse.py (img. *img-horse*) shows both possibilities; first, a GTS surface is imported:

```python
import gts
surf=gts.read(open('horse.coarse.gts'))
```

That surface object is used as predicate for packing:

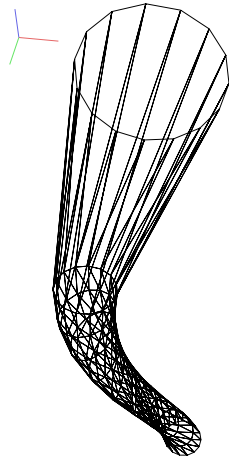

Fig. 9: Triangulated funnel, constructed with the examples/funnel.py script.

```
pred=pack.inGtsSurface(surf)
aabb=pred.aabb()
radius=(aabb[1][0]-aabb[0][0])/40
O.bodies.append(pack.regularHexa(pred,radius=radius,gap=radius/4.))
```

and then, after being translated, as base for triangulated surface in the simulation itself:

```
surf.translate(0,0,-(aabb[1][2]-aabb[0][2]))
O.bodies.append(pack.gtsSurface2Facets(surf,wire=True))
```

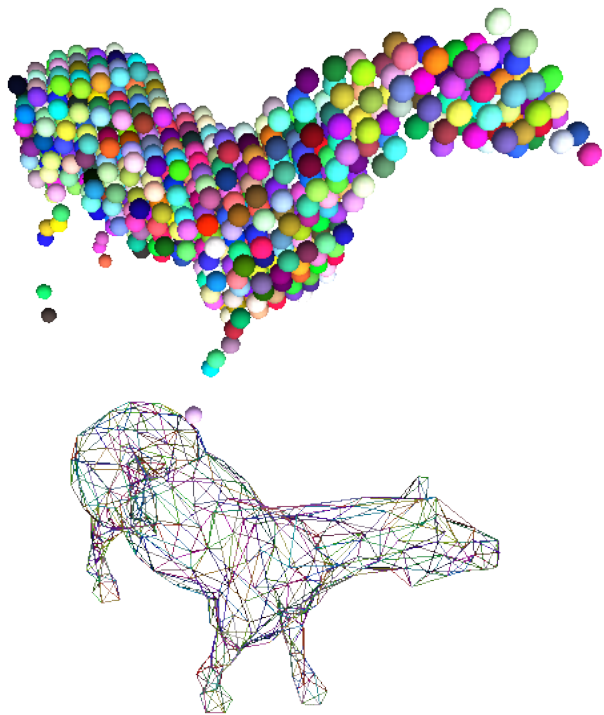

Fig. 10: Imported GTS surface (horse) used as packing predicate (top) and surface constructed from *facets* (bottom). See http://www.youtube.com/watch?v=PZVruIlUX1A for movie of this simulation.

**Boolean operations on predicates**

Boolean operations on pair of predicates (noted `A` and `B`) are defined:

- *intersection* `A & B` (conjunction): point must be in both predicates involved.

- *union* `A | B` (disjunction): point must be in the first or in the second predicate.

- *difference* `A - B` (conjunction with second predicate negated): the point must be in the first predicate and not in the second one.

- *symmetric difference* `A ^ B` (exclusive disjunction): point must be in exactly one of the two predicates.

Composed predicates also properly define their bounding box. For example, we can take box and remove cylinder from inside, using the `A - B` operation (img. *img-predicate-difference*):

```
pred=pack.inAlignedBox((-2,-2,-2),(2,2,2))-pack.inCylinder((0,-2,0),(0,2,0),1)
spheres=pack.randomDensePack(pred,spheresInCell=2000,radius=.1,rRelFuzz=.4,
↪returnSpherePack=True)
spheres.toSimulation()
```

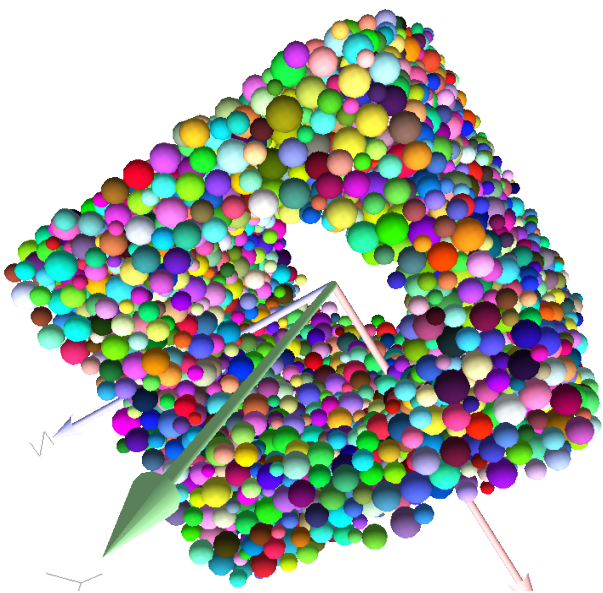

Fig. 11: Box with cylinder removed from inside, using difference of these two predicates.

**Packing algorithms**

Algorithms presented below operate on geometric spheres, defined by their center and radius. With a few exception documented below, the procedure is as follows:

1. Sphere positions and radii are computed (some functions use volume predicate for this, some do not)

2. *sphere* is called for each position and radius computed; it receives extra keyword arguments of the packing function (i.e. arguments that the packing function doesn't specify in its definition; they are noted `**kw`). Each *sphere* call creates actual *Body* objects with *Sphere shape*. List of *Body* objects is returned.

3. List returned from the packing function can be added to simulation using *toSimulation()*. Legacy code used a call to *O.bodies.append*.

Taking the example of pierced box:

```
pred=pack.inAlignedBox((-2,-2,-2),(2,2,2))-pack.inCylinder((0,-2,0),(0,2,0),1)
spheres=pack.randomDensePack(pred,spheresInCell=2000,radius=.1,rRelFuzz=.4,wire=True,color=(0,
↪0,1),material=1,returnSpherePack=True)
```

Keyword arguments `wire`, `color` and `material` are not declared in *pack.randomDensePack*, therefore will be passed to *sphere*, where they are also documented. `spheres` is now a *SpherePack* object.:

```
spheres.toSimulation()
```

Packing algorithms described below produce dense packings. If one needs loose packing, *SpherePack* class provides functions for generating loose packing, via its *makeCloud()* method. It is used internally for generating initial configuration in dynamic algorithms. For instance:

```
from yade import pack
sp=pack.SpherePack()
sp.makeCloud(minCorner=(0,0,0),maxCorner=(3,3,3),rMean=.2,rRelFuzz=.5)
```

will fill given box with spheres, until no more spheres can be placed. The object can be used to add spheres to simulation:

```
sp.toSimulation()
```

### Geometric

Geometric algorithms compute packing without performing dynamic simulation; among their advantages are

- speed;
- spheres touch exactly, there are no overlaps (what some people call "stress-free" packing);

their chief disadvantage is that radius distribution cannot be prescribed exactly, save in specific cases (regular packings); sphere radii are given by the algorithm, which already makes the system determined. If exact radius distribution is important for your problem, consider dynamic algorithms instead.

### Regular

Yade defines packing generators for spheres with constant radii, which can be used with volume predicates as described above. They are dense orthogonal packing (*pack.regularOrtho*) and dense hexagonal packing (*pack.regularHexa*). The latter creates so-called "hexagonal close packing", which achieves maximum density (http://en.wikipedia.org/wiki/Close-packing_of_spheres).

Clear disadvantage of regular packings is that they have very strong directional preferences, which might not be an issue in some cases.

### Irregular

Random geometric algorithms do not integrate at all with volume predicates described above; rather, they take their own boundary/volume definition, which is used during sphere positioning. On the other hand, this makes it possible for them to respect boundary in the sense of making spheres touch it at appropriate places, rather than leaving empty space in-between.

**GenGeo** is library (python module) for packing generation developed with ESyS-Particle. It creates packing by random insertion of spheres with given radius range. Inserted spheres touch each other exactly and, more importantly, they also touch the boundary, if in its neighbourhood. Boundary is represented as special object of the GenGeo library (Sphere, cylinder, box, convex polyhedron, ...). Therefore, GenGeo cannot be used with volume represented by yade predicates as explained above.

Packings generated by this module can be imported directly via *ymport.gengeo*, or from saved file via *ymport.gengeoFile*. There is an example script examples/test/genCylLSM.py. Full documentation for GenGeo can be found at ESyS documentation website.

There are debian packages esys-particle and python-demgengeo.

### Dynamic

The most versatile algorithm for random dense packing is provided by *pack.randomDensePack*. Initial loose packing of non-overlapping spheres is generated by randomly placing them in cuboid volume, with radii given by requested (currently only uniform) radius distribution. When no more spheres can be inserted, the packing is compressed and then uncompressed (see py/pack/pack.py for exact values of these "stresses") by running a DEM simulation; *Omega.switchScene* is used to not affect existing simulation). Finally, resulting packing is clipped using provided predicate, as explained above.

By its nature, this method might take relatively long; and there are 2 provisions to make the computation time shorter:

- If number of spheres using the `spheresInCell` parameter is specified, only smaller specimen with *periodic* boundary is created and then repeated as to fill the predicate. This can provide high-quality packing with low regularity, depending on the `spheresInCell` parameter (value of several thousands is recommended).

- Providing `memoizeDb` parameter will make *pack.randomDensePack* first look into provided file (SQLite database) for packings with similar parameters. On success, the packing is simply read from database and returned. If there is no similar pre-existent packing, normal procedure is run, and the result is saved in the database before being returned, so that subsequent calls with same parameters will return quickly.

If you need to obtain full periodic packing (rather than packing clipped by predicate), you can use *pack.randomPeriPack*.

In case of specific needs, you can create packing yourself, "by hand". For instance, packing boundary can be constructed from *facets*, letting randomly positioned spheres in space fall down under gravity.

### Triangulated surfaces

Yade integrates with the the GNU Triangulated Surface library, exposed in python via GTS module. GTS provides variety of functions for surface manipulation (coarsening, tesselation, simplification, import), to be found in its documentation.

GTS surfaces are geometrical objects, which can be inserted into simulation as set of particles whose *Body.shape* is of type *Facet* – single triangulation elements. *pack.gtsSurface2Facets* can be used to convert GTS surface triangulation into list of *bodies* ready to be inserted into simulation via `O.bodies.append`.

Facet particles are created by default as non-*Body.dynamic* (they have zero inertial mass). That means that they are fixed in space and will not move if subject to forces. You can however

- prescribe arbitrary movement to facets using a *PartialEngine* (such as *TranslationEngine* or *RotationEngine*);

- assign explicitly *mass* and *inertia* to that particle;

- make that particle part of a clump and assign *mass* and *inertia* of the clump itself (described below).

**Note:** Facets can only (currently) interact with *spheres*, not with other facets, even if they are *dynamic*. Collision of 2 *facets* will not create interaction, therefore no forces on facets.

**Import**

Yade currently offers 3 formats for importing triangulated surfaces from external files, in the *ymport* module:

*ymport.gts*  text file in native GTS format.

*ymport.stl*  STereoLitography format, in either text or binary form; exported from Blender, but from many CAD systems as well.

*ymport.gmsh.*  text file in native format for GMSH, popular open-source meshing program.

If you need to manipulate surfaces before creating list of facets, you can study the py/ymport.py file where the import functions are defined. They are rather simple in most cases.

**Parametric construction**

The GTS module provides convenient way of creating surface by vertices, edges and triangles.

Frequently, though, the surface can be conveniently described as surface between polylines in space. For instance, cylinder is surface between two polygons (closed polylines). The *pack.sweptPolylines2gtsSurface* offers the functionality of connecting several polylines with triangulation.

---

**Note:**  The implementation of *pack.sweptPolylines2gtsSurface* is rather simplistic: all polylines must be of the same length, and they are connected with triangles between points following their indices within each polyline (not by distance). On the other hand, points can be co-incident, if the `threshold` parameter is positive: degenerate triangles with vertices closer that `threshold` are automatically eliminated.

---

Manipulating lists efficiently (in terms of code length) requires being familiar with list comprehensions in python.

Another examples can be found in examples/mill.py (fully parametrized) or examples/funnel.py (with hardcoded numbers).

**Creating interactions**

In typical cases, interactions are created during simulations as particles collide. This is done by a *Collider* detecting approximate contact between particles and then an *IGeomFunctor* detecting exact collision.

Some material models (such as the *concrete model*) rely on initial interaction network which is denser than geometrical contact of spheres: sphere's radii as "enlarged" by a dimensionless factor called *inter-action radius* (or *interaction ratio*) to create this initial network. This is done typically in this way (see examples/concrete/uniax.py for an example):

1. Approximate collision detection is adjusted so that approximate contacts are detected also between particles within the interaction radius. This consists in setting value of *Bo1_Sphere_-Aabb.aabbEnlargeFactor* to the interaction radius value.

2. The geometry functor (`Ig2`) would normally say that "there is no contact" if given 2 spheres that are not in contact. Therefore, the same value as for *Bo1_Sphere_Aabb.aabbEnlargeFactor* must be given to it (*Ig2_Sphere_Sphere_ScGeom.interactionDetectionFactor* ).

   Note that only *Sphere + Sphere* interactions are supported; there is no parameter analogous to *distFactor* in *Ig2_Facet_Sphere_ScGeom*. This is on purpose, since the interaction radius is meaningful in bulk material represented by sphere packing, whereas facets usually represent boundary conditions which should be exempt from this dense interaction network.

3. Run one single step of the simulation so that the initial network is created.

4. Reset interaction radius in both `Bo1` and `Ig2` functors to their default value again.

---

5. Continue the simulation; interactions that are already established will not be deleted (the `Law2` functor in use permitting).

In code, such scenario might look similar to this one (labeling is explained in *Labeling things*):

```
intRadius=1.5
damping=0.05

O.engines=[
   ForceResetter(),
   InsertionSortCollider([
       # enlarge here
       Bo1_Sphere_Aabb(aabbEnlargeFactor=intRadius,label='bo1s'),
       Bo1_Facet_Aabb(),
          ]),
   InteractionLoop(
       [
           # enlarge here
           Ig2_Sphere_Sphere_ScGeom(interactionDetectionFactor=intRadius,label='ig2ss'),
           Ig2_Facet_Sphere_ScGeom(),
       ],
       [Ip2_CpmMat_CpmMat_CpmPhys()],
       [Law2_ScGeom_CpmPhys_Cpm(epsSoft=0)], # deactivated
   ),
   NewtonIntegrator(damping=damping,label='damper'),
]

# run one single step
O.step()

# reset interaction radius to the default value
bo1s.aabbEnlargeFactor=1.0
ig2ss.interactionDetectionFactor=1.0

# now continue simulation
O.run()
```

### Individual interactions on demand

It is possible to create an interaction between a pair of particles independently of collision detection using *createInteraction*. This function looks for and uses matching `Ig2` and `Ip2` functors. Interaction will be created regardless of distance between given particles (by passing a special parameter to the `Ig2` functor to force creation of the interaction even without any geometrical contact). Appropriate constitutive law should be used to avoid deletion of the interaction at the next simulation step.

```
Yade [26]: O.materials.append(FrictMat(young=3e10,poisson=.2,density=1000))
Out[26]: 0

Yade [27]: O.bodies.append([
   ....:      sphere([0,0,0],1),
   ....:      sphere([0,0,1000],1)
   ....: ])
   ....:
Out[27]: [0, 1]

# only add InteractionLoop, no other engines are needed now
Yade [28]: O.engines=[
   ....:      InteractionLoop(
   ....:          [Ig2_Sphere_Sphere_ScGeom(),],
   ....:          [Ip2_FrictMat_FrictMat_FrictPhys()],
   ....:          [] # not needed now
```

```
   ....:     )
   ....: ]
   ....:

Yade [29]: i=createInteraction(0,1)

# created by functors in InteractionLoop
Yade [30]: i.geom, i.phys
Out[30]: (<ScGeom instance at 0x37acbe0>, <FrictPhys instance at 0x3ed45b0>)
```

This method will be rather slow if many interactions are to be created (the functor lookup will be repeated for each of them). In such case, ask on yade-dev@lists.launchpad.net to have the *createInteraction* function accept list of pairs id's as well.

### Base engines

A typical DEM simulation in Yade does at least the following at each step (see *Function components* for details):

1. Reset forces from previous step

2. Detect new collisions

3. Handle interactions

4. Apply forces and update positions of particles

Each of these points corresponds to one or several engines:

```
O.engines=[
    ForceResetter(),              # reset forces
    InsertionSortCollider([...]),  # approximate collision detection
    InteractionLoop([...],[...],[...]) # handle interactions
    NewtonIntegrator()            # apply forces and update positions
]
```

The order of engines is important. In majority of cases, you will put any additional engine after *InteractionLoop*:

- if it applies force, it should come before *NewtonIntegrator*, otherwise the force will never be effective.

- if it makes use of bodies' positions, it should also come before *NewtonIntegrator*, otherwise, positions at the next step will be used (this might not be critical in many cases, such as output for visualization with *VTKRecorder*).

The *O.engines* sequence must be always assigned at once (the reason is in the fact that although engines themselves are passed by reference, the sequence is *copied* from c++ to Python or from Python to c++). This includes modifying an existing `O.engines`; therefore

```
O.engines.append(SomeEngine()) # wrong
```

will not work;

```
O.engines=O.engines+[SomeEngine()] # ok
```

must be used instead. For inserting an engine after position #2 (for example), use python slice notation:

```
O.engines=O.engines[:2]+[SomeEngine()]+O.engines[2:]
```

---

**Note:** When Yade starts, O.engines is filled with a reasonable default list, so that it is not strictly necessary to redefine it when trying simple things. The default scene will handle spheres, boxes, and

---

facets with *frictional* properties correctly, and adjusts the timestep dynamically. You can find an example in *examples/simple-scene/simple-scene-default-engines.py*.

---

**Functors choice**

In the above example, we omited functors, only writing ellipses `...` instead. As explained in *Dispatchers and functors*, there are 4 kinds of functors and associated dispatchers. User can choose which ones to use, though the choice must be consistent.

**Bo1 functors**

`Bo1` functors must be chosen depending on the collider in use; they are given directly to the collider (which internally uses *BoundDispatcher*).

At this moment (January 2019), the most common choice is *InsertionSortCollider*, which uses *Aabb*; functors creating *Aabb* must be used in that case. Depending on particle *shapes* in your simulation, choose appropriate functors:

```
O.engines=[...,
    InsertionSortCollider([Bo1_Sphere_Aabb(),Bo1_Facet_Aabb()]),
    ...
]
```

Using more functors than necessary (such as *Bo1_Facet_Aabb* if there are no *facets* in the simulation) has no performance penalty. On the other hand, missing functors for existing *shapes* will cause those bodies to not collide with other bodies (they will freely interpenetrate).

There are other *colliders* as well, though their usage is only experimental:

- *SpatialQuickSortCollider* is correctness-reference collider operating on *Aabb*; it is significantly slower than *InsertionSortCollider*.

- *PersistentTriangulationCollider* only works on spheres; it does not use a *BoundDispatcher*, as it operates on spheres directly.

- *FlatGridCollider* is proof-of-concept grid-based collider, which computes grid positions internally (no *BoundDispatcher* either)

**Ig2 functors**

`Ig2` functor choice (all of them derive from *IGeomFunctor*) depends on

1. shape combinations that should collide; for instance:

   ```
   InteractionLoop([Ig2_Sphere_Sphere_ScGeom()],[],[])
   ```

   will handle collisions for *Sphere* + *Sphere*, but not for *Facet* + *Sphere* – if that is desired, an additional functor must be used:

   ```
   InteractionLoop([
       Ig2_Sphere_Sphere_ScGeom(),
       Ig2_Facet_Sphere_ScGeom()
   ],[],[])
   ```

   Again, missing combination will cause given shape combinations to freely interpenetrate one another. There are several possible choices of a functor for each pair, hence they cannot be put into *InsertionSortCollider* by default. A common mistake for bodies going through each other is that the necessary functor was not added.

2. *IGeom* type accepted by the `Law2` functor (below); it is the first part of functor's name after `Law2` (for instance, *Law2_ScGeom_CpmPhys_Cpm* accepts *ScGeom*).

**Ip2 functors**

Ip2 functors (deriving from *IPhysFunctor*) must be chosen depending on

1. *Material* combinations within the simulation. In most cases, `Ip2` functors handle 2 instances of the same *Material* class (such as *Ip2_FrictMat_FrictMat_FrictPhys* for 2 bodies with *FrictMat*)

2. *IPhys* accepted by the constitutive law (`Law2` functor), which is the second part of the `Law2` functor's name (e.g. *Law2_ScGeom_FrictPhys_CundallStack* accepts *FrictPhys*)

---

**Note:** Unlike with `Bo1` and `Ig2` functors, unhandled combination of *Materials* is an error condition signaled by an exception.

---

**Law2 functor(s)**

`Law2` functor was the ultimate criterion for the choice of `Ig2` and `Ip2` functors; there are no restrictions on its choice in itself, as it only applies forces without creating new objects.

In most simulations, only one `Law2` functor will be in use; it is possible, though, to have several of them, dispatched based on combination of *IGeom* and *IPhys* produced previously by `Ig2` and `Ip2` functors respectively (in turn based on combination of *Shapes* and *Materials*).

---

**Note:** As in the case of `Ip2` functors, receiving a combination of *IGeom* and *IPhys* which is not handled by any `Law2` functor is an error.

---

**Warning:** Many `Law2` exist in Yade, and new ones can appear at any time. In some cases different functors are only different implementations of the same contact law (e.g. *Law2_ScGeom_FrictPhys_-CundallStack* and *Law2_L3Geom_FrictPhys_ElPerfPl*). Also, sometimes, the peculiarity of one functor may be reproduced as a special case of a more general one. Therefore, for a given constitutive behavior, the user may have the choice between different functors. It is strongly recommended to favor the most used and most validated implementation when facing such choice. A list of available functors classified from mature to unmaintained is updated here to guide this choice.

**Examples**

Let us give several examples of the chain of created and accepted types.

**Basic DEM model**

Suppose we want to use the *Law2_ScGeom_FrictPhys_CundallStack* constitutive law. We see that

1. the `Ig2` functors must create *ScGeom*. If we have for instance *spheres* and *boxes* in the simulation, we will need functors accepting *Sphere + Sphere* and *Box + Sphere* combinations. We don't want interactions between boxes themselves (as a matter of fact, there is no such functor anyway). That gives us *Ig2_Sphere_Sphere_ScGeom* and *Ig2_Box_Sphere_ScGeom*.

2. the `Ip2` functors should create *FrictPhys*. Looking at *InteractionPhysicsFunctors*, there is only *Ip2_FrictMat_FrictMat_FrictPhys*. That obliges us to use *FrictMat* for particles.

The result will be therefore:

```
InteractionLoop(
    [Ig2_Sphere_Sphere_ScGeom(),Ig2_Box_Sphere_ScGeom()],
    [Ip2_FrictMat_FrictMat_FrictPhys()],
    [Law2_ScGeom_FrictPhys_CundallStrack()]
)
```

### Concrete model

In this case, our goal is to use the *Law2_ScGeom_CpmPhys_Cpm* constitutive law.

- We use *spheres* and *facets* in the simulation, which selects `Ig2` functors accepting those types and producing *ScGeom*: *Ig2_Sphere_Sphere_ScGeom* and *Ig2_Facet_Sphere_ScGeom*.

- We have to use *Material* which can be used for creating *CpmPhys*. We find that *CpmPhys* is only created by *Ip2_CpmMat_CpmMat_CpmPhys*, which determines the choice of *CpmMat* for all particles.

Therefore, we will use:

```
InteractionLoop(
    [Ig2_Sphere_Sphere_ScGeom(),Ig2_Facet_Sphere_ScGeom()],
    [Ip2_CpmMat_CpmMat_CpmPhys()],
    [Law2_ScGeom_CpmPhys_Cpm()]
)
```

### Imposing conditions

In most simulations, it is not desired that all particles float freely in space. There are several ways of imposing boundary conditions that block movement of all or some particles with regard to global space.

### Motion constraints

- *Body.dynamic* determines whether a body will be accelerated by *NewtonIntegrator*; it is mandatory to make it false for bodies with zero mass, where applying non-zero force would result in infinite displacement.

  *Facets* are case in the point: *facet* makes them non-dynamic by default, as they have zero volume and zero mass (this can be changed, by passing `dynamic=True` to *facet* or setting *Body.dynamic*; setting *State.mass* to a non-zero value must be done as well). The same is true for *wall*.

  Making sphere non-dynamic is achieved simply by:

  ```
  b = sphere([x,y,z],radius,dynamic=False)
  b.dynamic=True #revert the previous
  ```

- *State.blockedDOFs* permits selective blocking of any of 6 degrees of freedom in global space. For instance, a sphere can be made to move only in the xy plane by saying:

  ```
  Yade [31]: O.bodies.append(sphere((0,0,0),1))
  Out[31]: 0

  Yade [32]: O.bodies[0].state.blockedDOFs='zXY'
  ```

  In contrast to *Body.dynamic*, *blockedDOFs* will only block forces (and acceleration) in selected directions. Actually, `b.dynamic=False` is nearly only a shorthand for `b.state.blockedDOFs='xyzXYZ'`. A subtle difference is that the former does reset the velocity components automaticaly, while the latest does not. If you prescribed linear or angular velocity, they will be applied regardless of *blockedDOFs*. It also implies that if the velocity is not zero when degrees of freedom are blocked via blockedDOFs assignments, the body will keep moving at the velocity it has at the time of blocking. The differences are shown below:

```
Yade [33]: b1 = sphere([0,0,0],1,dynamic=True)

Yade [34]: b1.state.blockedDOFs
Out[34]: ''

Yade [35]: b1.state.vel = Vector3(1,0,0) #we want it to move...

Yade [36]: b1.dynamic = False #... at a constant velocity

Yade [37]: print(b1.state.blockedDOFs, b1.state.vel)
xyzXYZ Vector3(0,0,0)

Yade [38]: # oops, velocity has been reset when setting dynamic=False

Yade [39]: b1.state.vel = (1,0,0) # we can still assign it now

Yade [40]: print(b1.state.blockedDOFs, b1.state.vel)
xyzXYZ Vector3(1,0,0)

Yade [41]: b2 = sphere([0,0,0],1,dynamic=True) #another try

Yade [42]: b2.state.vel = (1,0,0)

Yade [43]: b2.state.blockedDOFs = "xyzXYZ" #this time we assign blockedDOFs directly,
↪velocity is unchanged

Yade [44]: print(b2.state.blockedDOFs, b2.state.vel)
xyzXYZ Vector3(1,0,0)
```

It might be desirable to constrain motion of some particles constructed from a generated sphere packing, following some condition, such as being at the bottom of a specimen; this can be done by looping over all bodies with a conditional:

```
for b in O.bodies:
    # block all particles with z coord below .5:
    if b.state.pos[2]<.5: b.dynamic=False
```

Arbitrary spatial predicates introduced above can be expoited here as well:

```
from yade import pack
pred=pack.inAlignedBox(lowerCorner,upperCorner)
for b in O.bodies:
    if not isinstance(b.shape,Sphere): continue # skip non-spheres
    # ask the predicate if we are inside
    if pred(b.state.pos,b.shape.radius): b.dynamic=False
```

### Imposing motion and forces

### Imposed velocity

If a degree of freedom is blocked and a velocity is assigned along that direction (translational or rotational velocity), then the body will move at constant velocity. This is the simpler and recommended method to impose the motion of a body. This, for instance, will result in a constant velocity along x (it can still be freely accelerated along y and z):

```
O.bodies.append(sphere((0,0,0),1))
O.bodies[0].state.blockedDOFs='x'
O.bodies[0].state.vel=(10,0,0)
```

Conversely, modifying the position directly is likely to break Yade's algorithms, especially those related to collision detection and contact laws, as they are based on bodies velocities. Therefore, unless you really know what you are doing, don't do that for imposing a motion:

```
O.bodies.append(sphere((0,0,0),1))
O.bodies[0].state.blockedDOFs='x'
O.bodies[0].state.pos=10*O.dt #REALLY BAD! Don't assign position
```

### Imposed force

Applying a force or a torque on a body is done via functions of the *ForceContainer*. It is as simple as this:

```
O.forces.addF(0,(1,0,0)) #applies for one step
```

This way, the force applies for one time step only, and is resetted at the beginning of each step. For this reason, imposing a force at the begining of one step will have no effect at all, since it will be immediately resetted. The only way is to place a *PyRunner* inside the simulation loop.

Applying the force permanently is possible with another function (in this case it does not matter if the command comes at the begining of the time step):

```
O.forces.setPermF(0,(1,0,0)) #applies permanently
```

The force will persist across iterations, until it is overwritten by another call to `O.forces.setPermF(id, f)` or erased by `O.forces.reset(resetAll=True)`. The permanent force on a body can be checked with `O.forces.permF(id)`.

### Boundary controllers

Engines deriving from *BoundaryController* impose boundary conditions during simulation, either directly, or by influencing several bodies. You are referred to their individual documentation for details, though you might find interesting in particular

- *UniaxialStrainer* for applying strain along one axis at constant rate; useful for plotting strain-stress diagrams for uniaxial loading case. See examples/concrete/uniax.py for an example.

- *TriaxialStressController* which applies prescribed stress/strain along 3 perpendicular axes on cuboid-shaped packing using 6 walls (*Box* objects)

- *PeriTriaxController* for applying stress/strain along 3 axes independently, for simulations using periodic boundary conditions (*Cell*)

### Field appliers

Engines deriving from *FieldApplier* are acting on all particles. The one most used is *GravityEngine* applying uniform acceleration field (*GravityEngine* is deprecated, use *NewtonIntegrator.gravity* instead).

### Partial engines

Engines deriving from *PartialEngine* define the *ids* attribute determining bodies which will be affected. Several of them warrant explicit mention here:

- *TranslationEngine* and *RotationEngine* for applying constant speed linear and rotational motion on subscribers.

- *ForceEngine* and *TorqueEngine* applying given values of force/torque on subscribed bodies at every step.

- *StepDisplacer* for applying generalized displacement delta at every timestep; designed for precise control of motion when testing constitutive laws on 2 particles.

The real value of partial engines is when you need to prescribe a complex type of force or displacement field. For moving a body at constant velocity or for imposing a single force, the methods explained in *Imposing motion and forces* are much simpler. There are several interpolating engines (*InterpolatingDirectedForceEngine* for applying force with varying magnitude, *InterpolatingHelixEngine* for applying spiral displacement with varying angular velocity; see examples/test/helix.py and possibly others); writing a new interpolating engine is rather simple using examples of those that already exist.

### Convenience features

#### Labeling things

Engines and functors can define a `label` attribute. Whenever the `O.engines` sequence is modified, python variables of those names are created/updated; since it happens in the `__builtins__` namespaces, these names are immediately accessible from anywhere. This was used in *Creating interactions* to change interaction radius in multiple functors at once.

---
> **Warning:** Make sure you do not use label that will overwrite (or shadow) an object that you already use under that variable name. Take care not to use syntactically wrong names, such as "er*452" or "my engine"; only variable names permissible in Python can be used.
---

#### Simulation tags

*Omega.tags* is a dictionary (it behaves like a dictionary, although the implementation in C++ is different) mapping keys to labels. Contrary to regular python dictionaries that you could create,

- `O.tags` is *saved and loaded with simulation*;

- `O.tags` has some values pre-initialized.

After Yade startup, `O.tags` contains the following:

```
Yade [45]: dict(O.tags) # convert to real dictionary
Out[45]:
{'author': 'bchareyre-(bchareyre@HP-ZBook-15-G3)',
 'isoTime': '20220726T141512',
 'id': '20220726T141512p61547',
 'd.id': '20220726T141512p61547',
 'id.d': '20220726T141512p61547'}
```

**author** Real name, username and machine as obtained from your system at simulation creation

**id** Unique identifier of this Yade instance (or of the instance which created a loaded simulation). It is composed of date, time and process number. Useful if you run simulations in parallel and want to avoid overwriting each other's outputs; embed `O.tags['id']` in output filenames (either as directory name, or as part of the file's name itself) to avoid it. This is explained in *Separating output files from jobs* in detail.

**isoTime** Time when simulation was created (with second resolution).

**d.id, id.d** Simulation description and id joined by period (and vice-versa). Description is used in batch jobs; in non-batch jobs, these tags are identical to id.

---

You can add your own tags by simply assigning value, with the restriction that the left-hand side object must be a string and must not contain =.

```
Yade [46]: O.tags['anythingThat I lik3']='whatever'

Yade [47]: O.tags['anythingThat I lik3']
Out[47]: 'whatever'
```

### Saving python variables

Python variable lifetime is limited; in particular, if you save simulation, variables will be lost after reloading. Yade provides limited support for data persistence for this reason (internally, it uses special values of `O.tags`). The functions in question are *saveVars* and *loadVars*.

*saveVars* takes dictionary (variable names and their values) and a *mark* (identification string for the variable set); it saves the dictionary inside the simulation. These variables can be re-created (after the simulation was loaded from a XML file, for instance) in the `yade.params.`*mark* namespace by calling *loadVars* with the same identification *mark*:

```
Yade [48]: a=45; b=pi/3

Yade [49]: saveVars('ab',a=a,b=b)

# save simulation (we could save to disk just as well)
Yade [49]: O.saveTmp()

Yade [51]: O.loadTmp()

Yade [52]: loadVars('ab')

Yade [53]: yade.params.ab.a
Out[53]: 45

# import like this
Yade [54]: from yade.params import ab

Yade [55]: ab.a, ab.b
Out[55]: (45, 1.0471975511965976)

# also possible
Yade [56]: from yade.params import *

Yade [57]: ab.a, ab.b
Out[57]: (45, 1.0471975511965976)
```

Enumeration of variables can be tedious if they are many; creating local scope (which is a function definition in Python, for instance) can help:

```
def setGeomVars():
        radius=4
        thickness=22
        p_t=4/3*pi
        dim=Vector3(1.23,2.2,3)
        #
        # define as much as you want here
        # it all appears in locals() (and nothing else does)
        #
        saveVars('geom',loadNow=True,**locals())

setGeomVars()
```

```
from yade.params.geom import *
# use the variables now
```

---

**Note:** Only types that can be *pickled* can be passed to *saveVars*.

---

## 2.2.2 Controlling simulation

### Tracking variables

### Running python code

A special engine *PyRunner* can be used to periodically call python code, specified via the `command` parameter. Periodicity can be controlled by specifying computation time (`realPeriod`), virtual time (`virtPeriod`) or iteration number (`iterPeriod`).

For instance, to print kinetic energy (using *kineticEnergy*) every 5 seconds, the following engine will be put to `O.engines`:

```
PyRunner(command="print('kinetic energy',kineticEnergy())",realPeriod=5)
```

For running more complex commands, it is convenient to define an external function and only call it from within the engine. Since the `command` is run in the script's namespace, functions defined within scripts can be called. Let us print information on interaction between bodies 0 and 1 periodically:

```
def intrInfo(id1,id2):
        try:
                i=O.interactions[id1,id2]
                # assuming it is a CpmPhys instance
                print (d1,id2,i.phys.sigmaN)
        except:
                # in case the interaction doesn't exist (yet?)
                print("No interaction between",id1,id2)
O.engines=[...,
        PyRunner(command="intrInfo(0,1)",realPeriod=5)
]
```

---

**Warning:** If a function was declared inside a *live* yade session (ipython) then an error `NameError: name 'intrInfo' is not defined` will occur unless python globals() are updated with command

```
globals().update(locals())
```

---

More useful examples will be given below.

The *plot* module provides simple interface and storage for tracking various data. Although originally conceived for plotting only, it is widely used for tracking variables in general.

The data are in *plot.data* dictionary, which maps variable names to list of their values; the *plot.addData* function is used to add them.

```
Yade [58]: from yade import plot

Yade [59]: plot.data
Out[59]: {}

Yade [60]: plot.addData(sigma=12,eps=1e-4)
```

```
# not adding sigma will add a NaN automatically
# this assures all variables have the same number of records
Yade [61]: plot.addData(eps=1e-3)

# adds NaNs to already existing sigma and eps columns
Yade [62]: plot.addData(force=1e3)

Yade [63]: plot.data
Out[63]:
{'sigma': [12, nan, nan],
 'eps': [0.0001, 0.001, nan],
 'force': [nan, nan, 1000.0]}

# retrieve only one column
Yade [64]: plot.data['eps']
Out[64]: [0.0001, 0.001, nan]

# get maximum eps
Yade [65]: max(plot.data['eps'])
Out[65]: 0.001
```

New record is added to all columns at every time *plot.addData* is called; this assures that lines in different columns always match. The special value **nan** or **NaN** (Not a Number) is inserted to mark the record invalid.

---

**Note:** It is not possible to have two columns with the same name, since data are stored as a dictionary.

---

To record data periodically, use *PyRunner*. This will record the $z$ coordinate and velocity of body #1, iteration number and simulation time (every 20 iterations):

```
O.engines=O.engines+[PyRunner(command='myAddData()', iterPeriod=20)]

from yade import plot
def myAddData():
        b=O.bodies[1]
        plot.addData(z1=b.state.pos[2], v1=b.state.vel.norm(), i=O.iter, t=O.time)
```

---

**Note:** Arbitrary string can be used as a column label for *plot.data*. However if the name has spaces inside (e.g. my funny column) or is a reserved python keyword (e.g. for) the only way to pass it to *plot.addData* is to use a dictionary:

```
plot.addData(**{'my funny column':1e3, 'for':0.3})
```

An exception are columns having leading of trailing whitespaces. They are handled specially in *plot.plots* and should not be used (see below).

---

Labels can be conveniently used to access engines in the **myAddData** function:

```
O.engines=[...,
        UniaxialStrainer(...,label='strainer')
]
def myAddData():
        plot.addData(sigma=strainer.avgStress,eps=strainer.strain)
```

In that case, naturally, the labeled object must define attributes which are used (*UniaxialStrainer.strain* and *UniaxialStrainer.avgStress* in this case).

---

**Plotting variables**

Above, we explained how to track variables by storing them using *plot.addData*. These data can be readily used for plotting. Yade provides a simple, quick to use, plotting in the *plot* module. Naturally, since direct access to underlying data is possible via *plot.data*, these data can be processed in any other way.

The *plot.plots* dictionary is a simple specification of plots. Keys are x-axis variable, and values are tuple of y-axis variables, given as strings that were used for *plot.addData*; each entry in the dictionary represents a separate figure:

```
plot.plots={
        'i':('t',),      # plot t(i)
        't':('z1','v1') # z1(t) and v1(t)
}
```

Actual plot using data in *plot.data* and plot specification of *plot.plots* can be triggered by invoking the *plot.plot* function.

**Live updates of plots**

Yade features live-updates of figures during calculations. It is controlled by following settings:

- *plot.live* - By setting `yade.plot.live=True` you can watch the plot being updated while the calculations run. Set to `False` otherwise.
- *plot.liveInterval* - This is the interval in seconds between the plot updates.
- *plot.autozoom* - When set to `True` the plot will be automatically rezoomed.

**Controlling line properties**

In this subsection let us use a *basic complete script* like examples/simple-scene/simple-scene-plot.py, which we will later modify to make the plots prettier. Line of interest from that file is, and generates a picture presented below:

```
plot.plots={'i':('t'),'t':('z_sph',None,('v_sph','go-'),'z_sph_half')}
```

The line plots take an optional second string argument composed of a line color (eg. `'r'`, `'g'` or `'b'`), a line style (eg. `'-'`, `'--'` or `':'`) and a line marker (`'o'`, `'s'` or `'d'`). A red dotted line with circle markers is created with 'ro:' argument. For a listing of all options please have a look at http://matplotlib.sourceforge.net/api/pyplot_api.html#matplotlib.pyplot.plot

For example using following plot.plots() command, will produce a following graph:

```
plot.plots={'i':(('t','xr:'),),'t':(('z_sph','r:'),None,('v_sph','g--'),('z_sph_half','b-.'))}
```

And this one will produce a following graph:

```
plot.plots={'i':(('t','xr:'),),'t':(('z_sph','Hr:'),None,('v_sph','+g--'),('z_sph_half','*b-.
↪'))}
```

---

**Note:** You can learn more in matplotlib tutorial http://matplotlib.sourceforge.net/users/pyplot_tutorial.html and documentation http://matplotlib.sourceforge.net/users/pyplot_tutorial.html#controlling-line-properties

---

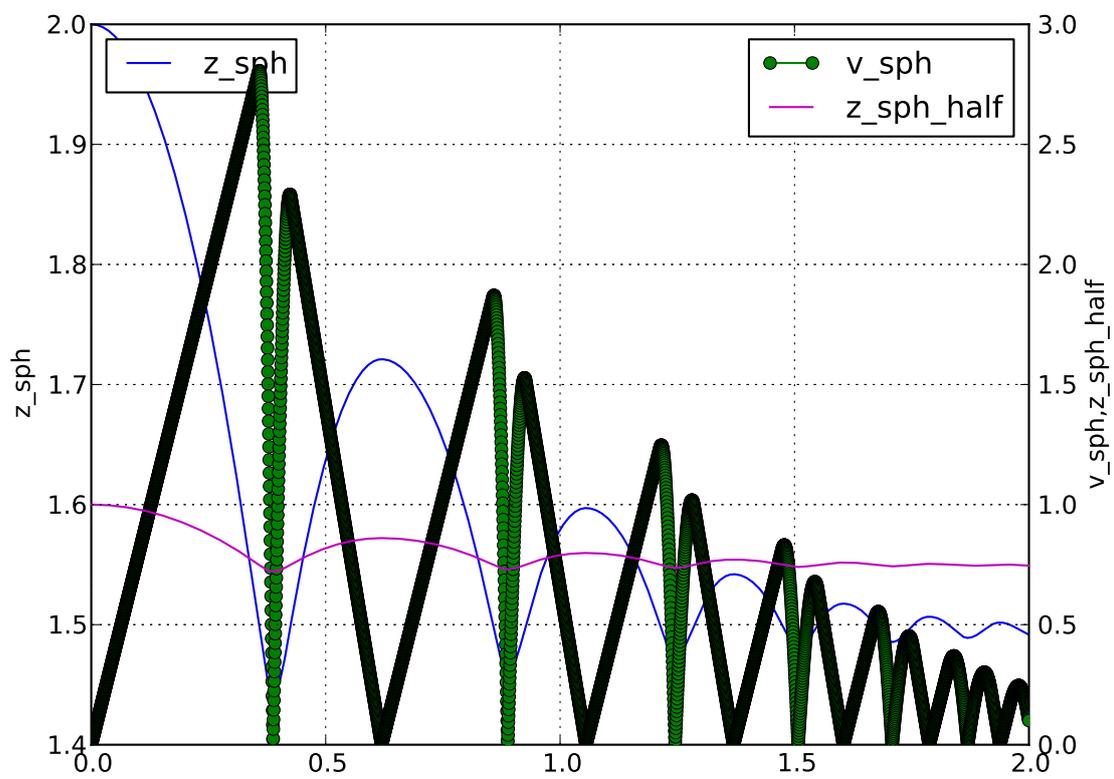

Fig. 12: Figure generated by examples/simple-scene/simple-scene-plot.py.

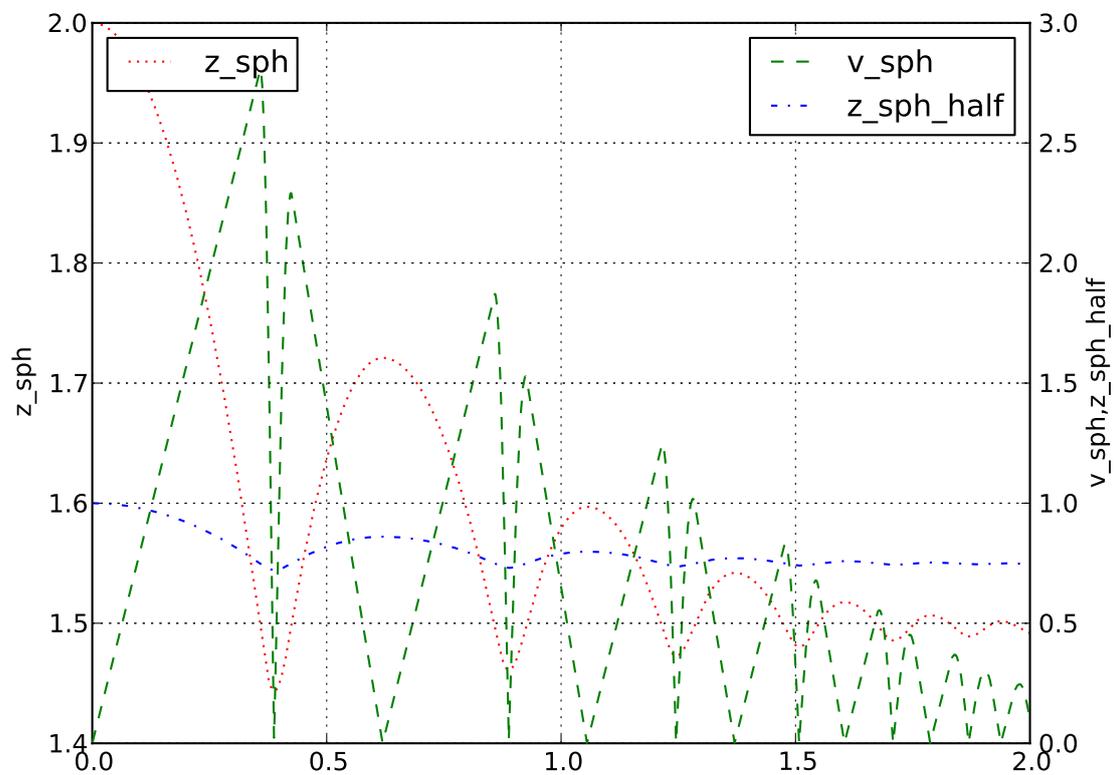

Fig. 13: Figure generated by changing parameters to plot.plots as above.

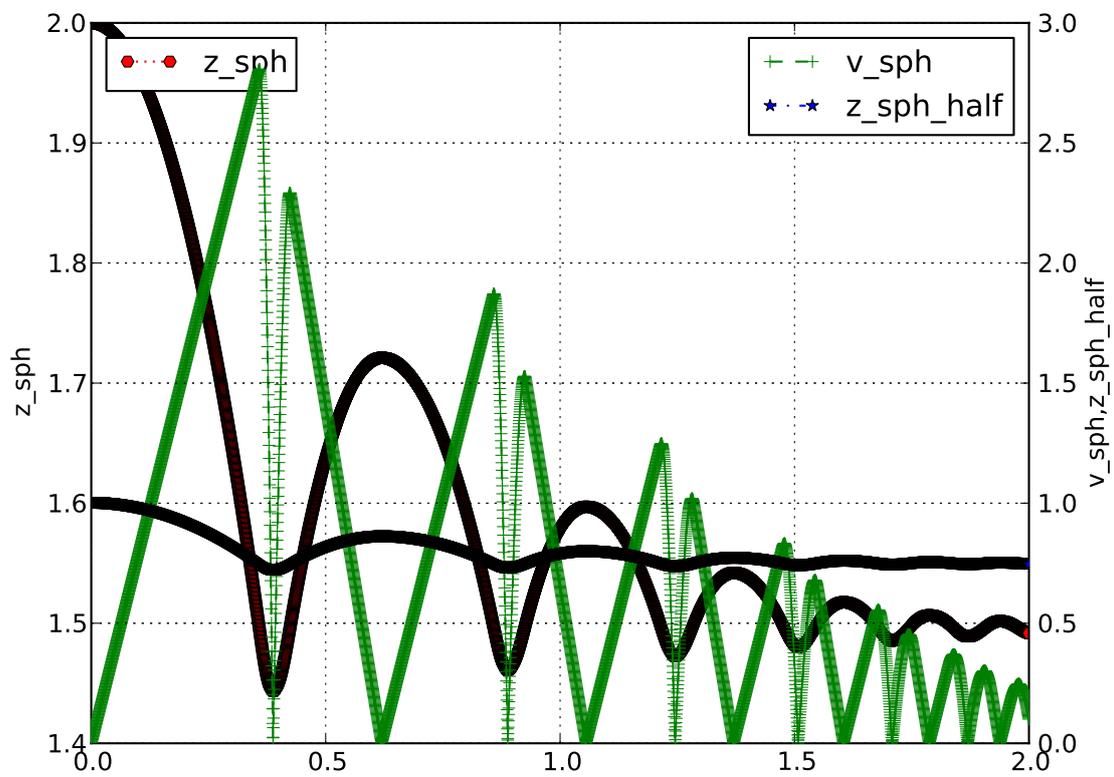

Fig. 14: Figure generated by changing parameters to plot.plots as above.

**Note:** Please note that there is an extra `,` in `'i':(('t','xr':),)`, otherwise the `'xr:'` wouldn't be recognized as a line style parameter, but would be treated as an extra data to plot.

### Controlling text labels

It is possible to use TeX syntax in plot labels. For example using following two lines in examples/simple-scene/simple-scene-plot.py, will produce a following picture:

```
plot.plots={'i':(('t','xr:'),),'t':(('z_sph','r:'),None,('v_sph','g--'),('z_sph_half','b-.'))}
plot.labels={'z_sph':'$z_{sph}$' , 'v_sph':'$v_{sph}$' , 'z_sph_half':'$z_{sph}/2$'}
```

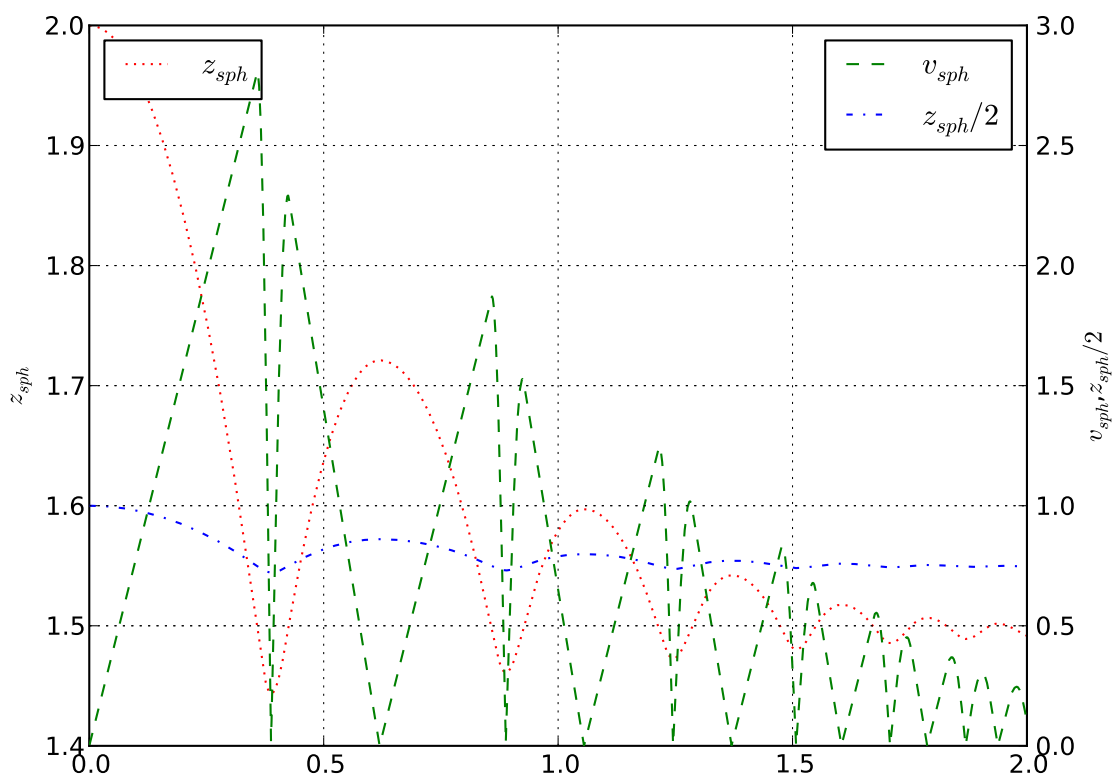

Fig. 15: Figure generated by examples/simple-scene/simple-scene-plot.py, with TeX labels.

Greek letters are simply a `'$\alpha$'`, `'$\beta$'` etc. in those labels. To change the font style a following command could be used:

```
yade.plot.matplotlib.rc('mathtext', fontset='stixsans')
```

But this is not part of yade, but a part of matplotlib, and if you want something more complex you really should have a look at matplotlib users manual http://matplotlib.sourceforge.net/users/index.html

### Multiple figures

Since *plot.plots* is a dictionary, multiple entries with the same key (x-axis variable) would not be possible, since they overwrite each other:

```
Yade [66]: plot.plots={
   ....:     'i':('t',),
   ....:     'i':('z1','v1')
   ....: }
   ....:

Yade [67]: plot.plots
Out[67]: {'i': ('z1', 'v1')}
```

You can, however, distinguish them by prepending/appending space to the x-axis variable, which will be removed automatically when looking for the variable in *plot.data* – both x-axes will use the **i** column:

```
Yade [68]: plot.plots={
   ....:     'i':('t',),
   ....:     'i ':('z1','v1') # note the space in 'i '
   ....: }
   ....:

Yade [69]: plot.plots
Out[69]: {'i': ('t',), 'i ': ('z1', 'v1')}
```

### Split y1 y2 axes

To avoid big range differences on the y axis, it is possible to have left and right **y** axes separate (like **axes x1y2** in gnuplot). This is achieved by inserting **None** to the plot specifier; variables coming before will be plot normally (on the left *y*-axis), while those after will appear on the right:

```
plot.plots={'i':('z1',None,'v1')}
```

### Exporting

Plots and data can be exported to external files for later post-processing in *Gnuplot <http://www.gnuplot.info/>* via that *plot.saveGnuplot* function. Note that all data you added via plot.addData is saved - even data that you don't plot live during simulation. By editing the generated .gnuplot file you can plot any of the added Data afterwards.

- Data file is saved (compressed using bzip2) separately from the gnuplot file, so any other programs can be used to process them. In particular, the **numpy.genfromtxt** (documented here) can be useful to import those data back to python; the decompression happens automatically.

- The gnuplot file can be run through gnuplot to produce the figure; see *plot.saveGnuplot* documentation for details.

For post-processing with other tools than gnuplot, saved data can also be exported in another kind of text file with *plot.saveDataTxt*.

### Stop conditions

For simulations with a pre-determined number of steps, it can be prescribed:

```
# absolute iteration number
O.stopAtIter=35466
O.run()
O.wait()
```

or

```
# number of iterations to run from now
O.run(35466,True) # wait=True
```

causes the simulation to run 35466 iterations, then stopping.

Frequently, decisions have to be made based on evolution of the simulation itself, which is not yet known. In such case, a function checking some specific condition is called periodically; if the condition is satisfied, `O.pause` or other functions can be called to stop the stimulation. See documentation for *Omega.run*, *Omega.pause*, *Omega.step*, *Omega.stopAtIter* for details.

For simulations that seek static equilibrium, the *unbalancedForce* can provide a useful metrics (see its documentation for details); for a desired value of `1e-2` or less, for instance, we can use:

```
def checkUnbalanced():
        if unbalancedForce<1e-2: O.pause()

O.engines=O.engines+[PyRunner(command="checkUnbalanced()",iterPeriod=100)]

# this would work as well, without the function defined apart:
#   PyRunner(command="if unbalancedForce<1e-2: O.pause()",iterPeriod=100)

O.run(); O.wait()
# will continue after O.pause() will have been called
```

Arbitrary functions can be periodically checked, and they can also use history of variables tracked via *plot.addData*. For example, this is a simplified version of damage control in examples/concrete/uniax.py; it stops when current stress is lower than half of the peak stress:

```
O.engines=[...,
        UniaxialStrainer=(...,label='strainer'),
        PyRunner(command='myAddData()',iterPeriod=100),
        PyRunner(command='stopIfDamaged()',iterPeriod=100)
]

def myAddData():
        plot.addData(t=O.time,eps=strainer.strain,sigma=strainer.stress)

def stopIfDamaged():
        currSig=plot.data['sigma'][-1] # last sigma value
        maxSig=max(plot.data['sigma']) # maximum sigma value
        # print something in any case, so that we know what is happening
        print(plot.data['eps'][-1],currSig)
        if currSig<.5*maxSig:
                print("Damaged, stopping")
                print('gnuplot',plot.saveGnuplot(O.tags['id']))
                import sys
                sys.exit(0)

O.run(); O.wait()
# this place is never reached, since we call sys.exit(0) directly
```

**Checkpoints**

Occasionally, it is useful to revert to simulation at some past point and continue from it with different parameters. For instance, tension/compression test will use the same initial state but load it in 2 different directions. Two functions, *Omega.saveTmp* and *Omega.loadTmp* are provided for this purpose; *memory* is used as storage medium, which means that saving is faster, and also that the simulation will disappear when Yade finishes.

```
O.saveTmp()
# do something
O.saveTmp('foo')
O.loadTmp()      # loads the first state
O.loadTmp('foo') # loads the second state
```

> **Warning:** `O.loadTmp` cannot be called from inside an engine, since *before* loading a simulation, the old one must finish the current iteration; it would lead to deadlock, since `O.loadTmp` would wait for the current iteration to finish, while the current iteration would be blocked on `O.loadTmp`.
>
> A special trick must be used: a separate function to be run after the current iteration is defined and is invoked from an independent thread launched only for that purpose:
>
> ```
> O.engines=[...,PyRunner('myFunc()',iterPeriod=345)]
>
> def myFunc():
>         if someCondition:
>                 import thread
>                 # the () are arguments passed to the function
>                 thread.start_new_thread(afterIterFunc,())
> def afterIterFunc():
>         O.pause(); O.wait() # wait till the iteration really finishes
>         O.loadTmp()
>
> O.saveTmp()
> O.run()
> ```

## Remote control

Yade can be controlled remotely over network. At yade startup, the following lines appear, among other messages:

```
TCP python prompt on localhost:9000, auth cookie `dcekyu'
TCP info provider on localhost:21000
```

They inform about 2 ports on which connection of 2 different kind is accepted.

## Python prompt

`TCP python prompt` is telnet server with authenticated connection, providing full python command-line. It listens on port 9000, or higher if already occupied (by another yade instance, for example).

Using the authentication cookie, connection can be made using telnet:

```
$ telnet localhost 9000
Trying 127.0.0.1...
Connected to localhost.
Escape character is '^]'.
Enter auth cookie: dcekyu
__   __   ____              __  _____ ____ ____
\ \ / / /_  |  _ \   ___     ___   / / |_   _/ ___| _  \
 \ V / / _` | | | |/ _ \   / _ \ / /    | || |  | |_) |
  | | | (_| | | | |  __/  | (_) / /     | || |___|  __/
  |_|\__,_|____/ \___|  \___/_/      |_| \____|_|

(connected from 127.0.0.1:40372)
>>>
```

The python pseudo-prompt `>>>` lets you write commands to manipulate simulation in variety of ways as usual. Two things to notice:

1. The new python interpreter (`>>>`) lives in a namespace separate from `Yade [1]:` command-line. For your convenience, `from yade import *` is run in the new python instance first, but local and global variables are not accessible (only builtins are).

2. The (fake) `>>>` interpreter does not have rich interactive feature of IPython, which handles the usual command-line `Yade [1]:`; therefore, you will have no command history, `?` help and so on.

---

**Note:** By giving access to python interpreter, full control of the system (including reading user's files) is possible. For this reason, **connection is only allowed from localhost**, not over network remotely. Of course you can log into the system via SSH over network to get remote access.

---

> **Warning:** Authentication cookie is trivial to crack via bruteforce attack. Although the listener stalls for 5 seconds after every failed login attempt (and disconnects), the cookie could be guessed by trial-and-error during very long simulations on a shared computer.

### Info provider

`TCP Info provider` listens at port 21000 (or higher) and returns some basic information about current simulation upon connection; the connection terminates immediately afterwards. The information is python dictionary represented as string (serialized) using standard pickle module.

This functionality is used by the batch system (described below) to be informed about individual simulation progress and estimated times. If you want to access this information yourself, you can study core/main/yade-batch.in for details.

### Batch queuing and execution (yade-batch)

Yade features light-weight system for running one simulation with different parameters; it handles assignment of parameter values to python variables in simulation script, scheduling jobs based on number of available and required cores and more. The whole batch consists of 2 files:

**simulation script** regular Yade script, which calls *readParamsFromTable* to obtain parameters from parameter table. In order to make the script runnable outside the batch, *readParamsFromTable* takes default values of parameters, which might be overridden from the parameter table.

> *readParamsFromTable* knows which parameter file and which line to read by inspecting the `PARAM_-TABLE` environment variable, set by the batch system.

**parameter table** simple text file, each line representing one parameter set. This file is read by *readParamsFromTable* (using *TableParamReader* class), called from simulation script, as explained above. For better reading of the text file you can make use of tabulators, these will be ignored by *readParamsFromTable*. Parameters are not restricted to numerical values. You can also make use of strings by `"quoting"` them (`' '` may also be used instead of `" "`). This can be useful for nominal parameters.

The batch can be run as

```
yade-batch parameters.table simulation.py
```

and it will intelligently run one simulation for each parameter table line. A minimal example is found in examples/test/batch/params.table and examples/test/batch/sim.py, another example follows.

---

**Example**

Suppose we want to study influence of parameters *density* and *initialVelocity* on position of a sphere falling on fixed box. We create parameter table like this:

```
description density initialVelocity # first non-empty line are column headings
reference    2400    10
hi_v         =       20                  # = to use value from previous line
lo_v         =        5
# comments are allowed
hi_rho       5000    10
# blank lines as well:

hi_rho_v     =       20
hi_rh0_lo_v  =        5
```

Each line give one combination of these 2 parameters and assigns (which is optional) a *description* of this simulation.

In the simulation file, we read parameters from table, at the beginning of the script; each parameter has default value, which is used if not specified in the parameters file:

```
readParamsFromTable(
        gravity=-9.81,
        density=2400,
        initialVelocity=20,
        noTableOk=True     # use default values if not run in batch
)
from yade.params.table import *
print(gravity, density, initialVelocity)
```

after the call to *readParamsFromTable*, corresponding python variables are created in the `yade.params.table` module and can be readily used in the script, e.g.

```
GravityEngine(gravity=(0,0,gravity))
```

Let us see what happens when running the batch:

```
$ yade-batch batch.table batch.py
Will run `/usr/local/bin/yade-trunk' on `batch.py' with nice value 10, output redirected to␣
↪`batch.@.log', 4 jobs at a time.
Will use table `batch.table', with available lines 2, 3, 4, 5, 6, 7.
Will use lines  2 (reference), 3 (hi_v), 4 (lo_v), 5 (hi_rho), 6 (hi_rho_v), 7 (hi_rh0_lo_v).
Master process pid 7030
```

These lines inform us about general batch information: nice level, log file names, how many cores will be used (4); table name, and line numbers that contain parameters; finally, which lines will be used; master PID is useful for killing (stopping) the whole batch with the `kill` command.

```
Job summary:
    #0 (reference/4): PARAM_TABLE=batch.table:2 DISPLAY=  /usr/local/bin/yade-trunk --threads=4␣
↪--nice=10 -x batch.py > batch.reference.log 2>&1
    #1 (hi_v/4): PARAM_TABLE=batch.table:3 DISPLAY=  /usr/local/bin/yade-trunk --threads=4 --
↪nice=10 -x batch.py > batch.hi_v.log 2>&1
    #2 (lo_v/4): PARAM_TABLE=batch.table:4 DISPLAY=  /usr/local/bin/yade-trunk --threads=4 --
↪nice=10 -x batch.py > batch.lo_v.log 2>&1
    #3 (hi_rho/4): PARAM_TABLE=batch.table:5 DISPLAY=  /usr/local/bin/yade-trunk --threads=4 --
↪nice=10 -x batch.py > batch.hi_rho.log 2>&1
    #4 (hi_rho_v/4): PARAM_TABLE=batch.table:6 DISPLAY=  /usr/local/bin/yade-trunk --threads=4 -
↪-nice=10 -x batch.py > batch.hi_rho_v.log 2>&1
    #5 (hi_rh0_lo_v/4): PARAM_TABLE=batch.table:7 DISPLAY=  /usr/local/bin/yade-trunk --
↪threads=4 --nice=10 -x batch.py > batch.hi_rh0_lo_v.log 2>&1
```

displays all jobs with command-lines that will be run for each of them. At this moment, the batch starts to be run.

```
#0 (reference/4) started on Tue Apr 13 13:59:32 2010
#0 (reference/4) done    (exit status 0), duration 00:00:01, log batch.reference.log
#1 (hi_v/4) started on Tue Apr 13 13:59:34 2010
#1 (hi_v/4) done    (exit status 0), duration 00:00:01, log batch.hi_v.log
#2 (lo_v/4) started on Tue Apr 13 13:59:35 2010
#2 (lo_v/4) done    (exit status 0), duration 00:00:01, log batch.lo_v.log
#3 (hi_rho/4) started on Tue Apr 13 13:59:37 2010
#3 (hi_rho/4) done    (exit status 0), duration 00:00:01, log batch.hi_rho.log
#4 (hi_rho_v/4) started on Tue Apr 13 13:59:38 2010
#4 (hi_rho_v/4) done    (exit status 0), duration 00:00:01, log batch.hi_rho_v.log
#5 (hi_rh0_lo_v/4) started on Tue Apr 13 13:59:40 2010
#5 (hi_rh0_lo_v/4) done    (exit status 0), duration 00:00:01, log batch.hi_rh0_lo_v.log
```

information about job status changes is being printed, until:

```
All jobs finished, total time  00:00:08
Log files:
batch.reference.log batch.hi_v.log batch.lo_v.log batch.hi_rho.log batch.hi_rho_v.log batch.hi_
↪rh0_lo_v.log
Bye.
```

### Separating output files from jobs

As one might output data to external files during simulation (using classes such as *VTKRecorder*), it is important to name files in such way that they are not overwritten by next (or concurrent) job in the same batch. A special tag `O.tags['id']` is provided for such purposes: it is comprised of date, time and PID, which makes it always unique (e.g. `20100413T144723p7625`); additional advantage is that alphabetical order of the `id` tag is also chronological. To add the used parameter set or the description of the job, if set, you could add O.tags['params'] to the filename.

For smaller simulations, prepending all output file names with `O.tags['id']` can be sufficient:

```
saveGnuplot(O.tags['id'])
```

For larger simulations, it is advisable to create separate directory of that name first, putting all files inside afterwards:

```
os.mkdir(O.tags['id'])
O.engines=[
        # …
        VTKRecorder(fileName=O.tags['id']+'/'+'vtk'),
        # …
]
# …
O.saveGnuplot(O.tags['id']+'/'+'graph1')
```

### Controlling parallel computation

Default total number of available cores is determined from `/proc/cpuinfo` (provided by Linux kernel); in addition, if `OMP_NUM_THREADS` environment variable is set, minimum of these two is taken. The `-j`/`--jobs` option can be used to override this number.

By default, each job uses all available cores for itself, which causes jobs to be effectively run in parallel. Number of cores per job can be globally changed via the `--job-threads` option.

Table column named `!OMP_NUM_THREADS` (! prepended to column generally means to assign *environment variable*, rather than python variable) controls number of threads for each job separately, if it exists.

If number of cores for a job exceeds total number of cores, warning is issued and only the total number of cores is used instead.

### Merging gnuplot from individual jobs

Frequently, it is desirable to obtain single figure for all jobs in the batch, for comparison purposes. Somewhat heuristic way for this functionality is provided by the batch system. `yade-batch` must be run with the `--gnuplot` option, specifying some file name that will be used for the merged figure:

```
yade-trunk --gnuplot merged.gnuplot batch.table batch.py
```

Data are collected in usual way during the simulation (using *plot.addData*) and saved to gnuplot file via *plot.saveGnuplot* (it creates 2 files: gnuplot command file and compressed data file). The batch system *scans*, once the job is finished, log file for line of the form `gnuplot [something]`. Therefore, in order to print this *magic line* we put:

```
print('gnuplot',plot.saveGnuplot(O.tags['id']))
```

and the end of the script (even after waitIfBatch()) , which prints:

```
gnuplot 20100413T144723p7625.gnuplot
```

to the output (redirected to log file).

This file itself contains single graph:

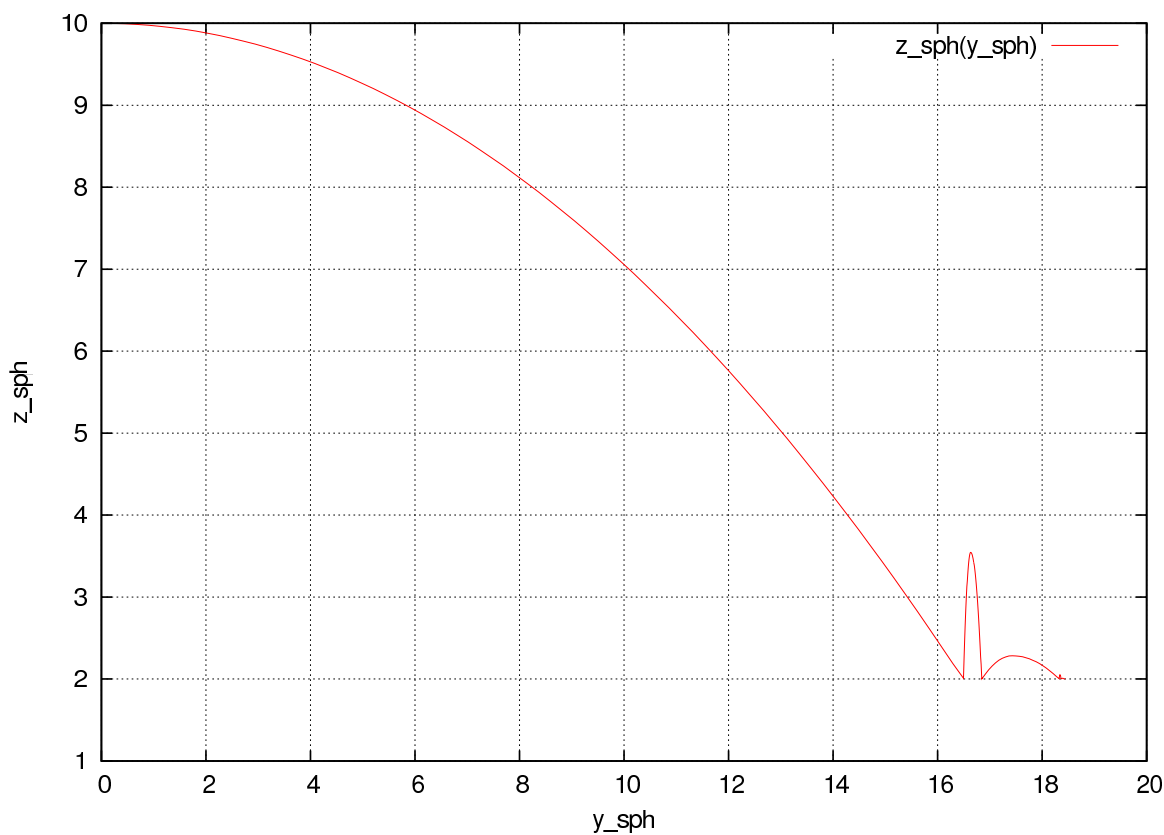

Fig. 16: Figure from single job in the batch.

At the end, the batch system knows about all gnuplot files and tries to merge them together, by assembling the `merged.gnuplot` file.

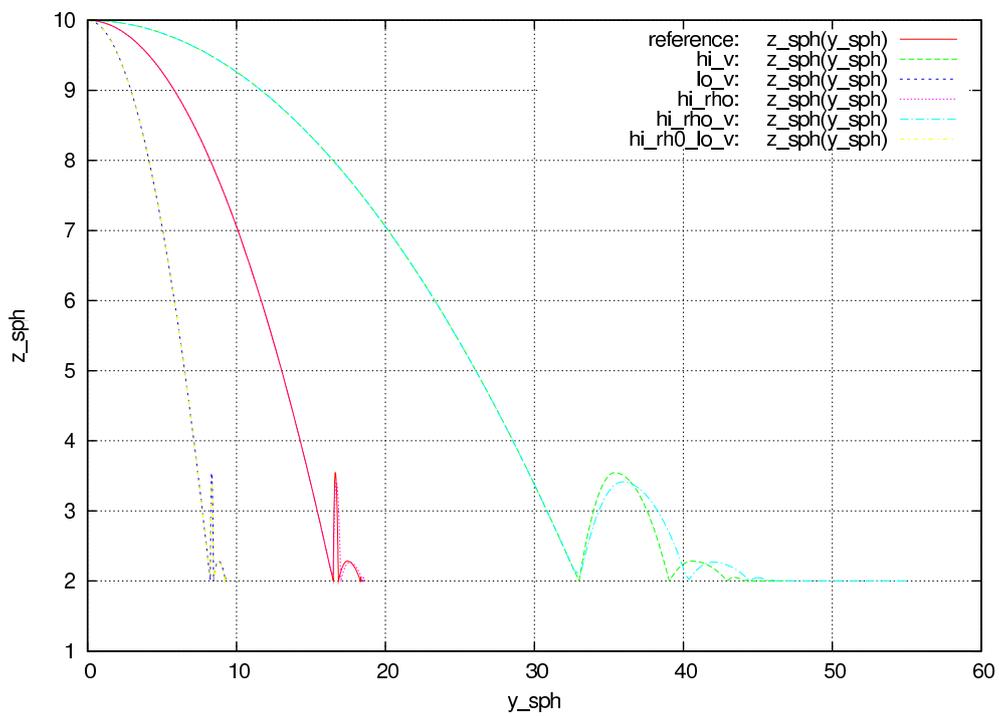

Fig. 17: Merged figure from all jobs in the batch. Note that labels are prepended by job description to make lines distinguishable.

### HTTP overview

While job is running, the batch system presents progress via simple HTTP server running at port 9080, which can be acessed from a regular web browser (or e.g. `lynx` for a terminal usage) by requesting the `http://localhost:9080` URL. This page can be accessed remotely over network as well.

Running for 00:10:19, since Tue Apr 13 16:17:11 2010.

Pid 9873

4 slots available, 4 used, 0 free.

## Jobs

**4** total, **2** running, **1** done

| id | status | info | slots | command |
|---|---|---|---|---|
| _geomType=B | 00:10:19 | 96.33% done step 9180/9530 avg 14.9596/sec 10267 bodies 65506 intrs | 2 | PARAM_TABLE=iParams.table:2 DISPLAY= /usr/local/bin/yade-trunk --threads=2 --nice=10 -x indent.py > indent._geomType=B.log 2>&1 |
| _geomType=smallA | 00:09:53 | (no info) | 2 | PARAM_TABLE=iParams.table:3 DISPLAY= /usr/local/bin/yade-trunk --threads=2 --nice=10 -x indent.py > indent._geomType=smallA.log 2>&1 |
| _geomType=smallB | 00:00:24 | 6.95% done step 694/9985 avg 35.8212/sec 9021 bodies 58352 intrs | 2 | PARAM_TABLE=iParams.table:4 DISPLAY= /usr/local/bin/yade-trunk --threads=2 --nice=10 -x indent.py > indent._geomType=smallB.log 2>&1 |
| _geomType=smallC | (pending) | (no info) | 2 | PARAM_TABLE=iParams.table:5 DISPLAY= /usr/local/bin/yade-trunk --threads=2 --nice=10 -x indent.py > indent._geomType=smallC.log 2>&1 |

Fig. 18: Summary page available at port 9080 as batch is processed (updates every 5 seconds automatically). Possible job statuses are pending, running, done, failed.

### Batch execution on Job-based clusters (OAR)

On High Performance Computation clusters with a scheduling system, the following script might be useful. Exactly like yade-batch, it handles assignemnt of parameters value to python variables in simulation script from a parameter table, and job submission. This script is written for oar-based system , and may be extended to others ones. On those system, usually, a job can't run forever and has a specific duration allocation. The whole job submission consists of 3 files:

**Simulation script:** Regular Yade script, which calls *readParamsFromTable* to obtain parameters from parameter table. In order to make the script runnable outside the batch, *readParamsFromTable* takes default values of parameters, which might be overridden from the parameter table.

*readParamsFromTable* knows which parameter file and which line to read by inspecting the `PARAM_-TABLE` environment variable, set by the batch system.

**Parameter table:** Simple text file, each line representing one parameter set. This file is read by *readParamsFromTable* (using *TableParamReader* class), called from simulation script, as explained above. For better reading of the text file you can make use of tabulators, these will be ignored by *readParamsFromTable*. Parameters are not restricted to numerical values. You can also make use of strings by `"quoting"` them (` ' ` may also be used instead of `" "`). This can be useful for nominal parameters.

**Job script:** Bash script, which calls yade on computing nodes. This script eventually creates temp folders, save data to storage server etc. The script must be formatted as a template where some tags will be replaced by specific values at the execution time:

- `__YADE_COMMAND__` will be replaced by the actual yade run command

- `__YADE_LOGFILE__` will be replaced by the log file path (output to stdout)

- `__YADE_ERRFILE__` will be replaced by the error file path (output to stderr)

- `__YADE_JOBNO__` will be replaced by an identifier composed as (launch script pid)-(job order)

- `__YADE_JOBID__` will be replaced by an identifier composed of all parameters values

The batch can be run as

```
yade-oar --oar-project=<your project name> --oar-script=job.sh --oar-walltime=hh:mm:ss␣
↪parameters.table simulation.py
```

and it will generate one launch script and submit one job for each parameter table line. A minimal example is found in examples/oar/params.table examples/oar/job.sh and examples/oar/sim.py.

---

**Note:** You have to specify either –oar-walltime or a !WALLTIME column in params.table. !WALLTIME will override –oar-walltime

---

┌─────────────────────────────────────────────────────────────────────────────┐
**Warning:** yade-oar is not compiled by default. Use -DENABLE_OAR=1 option to cmake to enable it.
└─────────────────────────────────────────────────────────────────────────────┘

## 2.2.3 Postprocessing

### 3d rendering & videos

There are multiple ways to produce a video of simulation:

1. Capture screen output (the 3d rendering window) during the simulation — there are tools available for that (such as Istanbul or RecordMyDesktop, which are also packaged for most Linux distributions). The output is "what you see is what you get", with all the advantages and disadvantages.

2. Periodic frame snapshot using *SnapshotEngine* (see examples/test/force-network-video.py, examples/bulldozer/bulldozer.py or examples/test/beam-l6geom.py for a complete example):

```
O.engines=[
    #...
    SnapshotEngine(iterPeriod=100,fileBase='/tmp/bulldozer-',viewNo=0,label='snapshooter')
]
```

which will save numbered files like `/tmp/bulldozer-0000.png`. These files can be processed externally (with mencoder and similar tools) or directly with the *makeVideo*:

```
makeVideo(frameSpec,out,renameNotOverwrite=True,fps=24,kbps=6000,bps=None)
```

The video is encoded using the default mencoder codec (mpeg4).

3. Specialized post-processing tools, notably Paraview. This is described in more detail in the following section.

### Paraview

#### Saving data during the simulation

Paraview is based on the Visualization Toolkit, which defines formats for saving various types of data. One of them (with the `.vtu` extension) can be written by a special engine *VTKRecorder*. It is added to the simulation loop:

```
O.engines=[
        # ...
        VTKRecorder(iterPeriod=100,recorders=['spheres','facets','colors'],fileName='/tmp/p1-')
]
```

- *iterPeriod* determines how often to save simulation data (besides *iterPeriod*, you can also use *virtPeriod* or *realPeriod*). If the period is too high (and data are saved only few times), the video will have few frames.
- *fileName* is the prefix for files being saved. In this case, output files will be named `/tmp/p1-spheres.0.vtu` and `/tmp/p1-facets.0.vtu`, where the number is the number of iteration; many files are created, putting them in a separate directory is advisable.
- *recorders* determines what data to save

*export.VTKExporter* plays a similar role, with the difference that it is more flexible. It will save any user defined variable associated to the bodies.

#### Loading data into Paraview

All sets of files (`spheres`, `facets`, …) must be opened one-by-one in Paraview. The open dialogue automatically collapses numbered files in one, making it easy to select all of them:

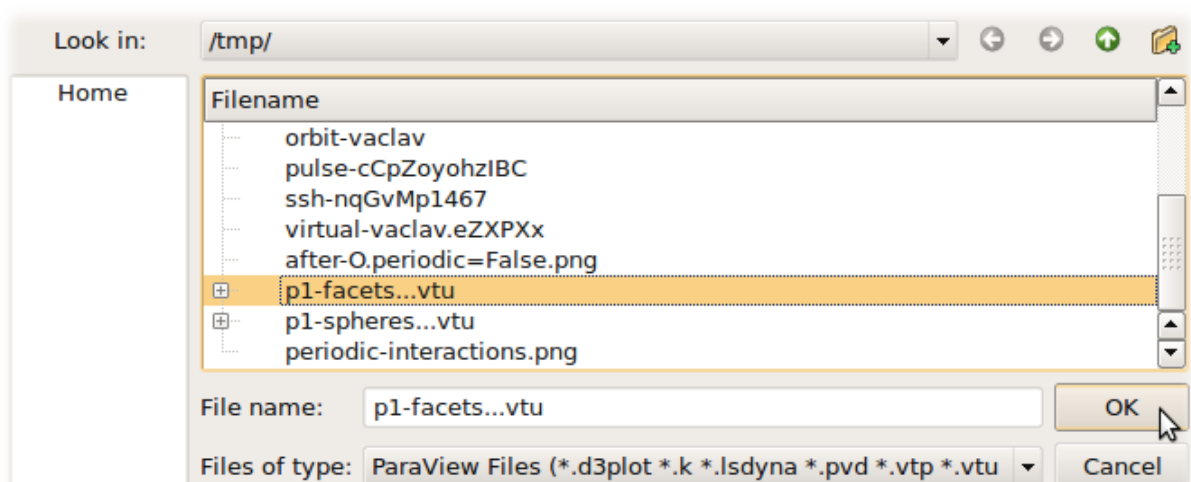

Click on the "Apply" button in the "Object inspector" sub-window to make loaded objects visible. You can see tree of displayed objects in the "Pipeline browser":

#### Rendering spherical particles. Glyphs

Spheres will only appear as points. To make them look as spheres, you have to add "glyph" to the 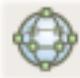 p1-spheres.* item in the pipeline using the icon. Then set (in the Object inspector)

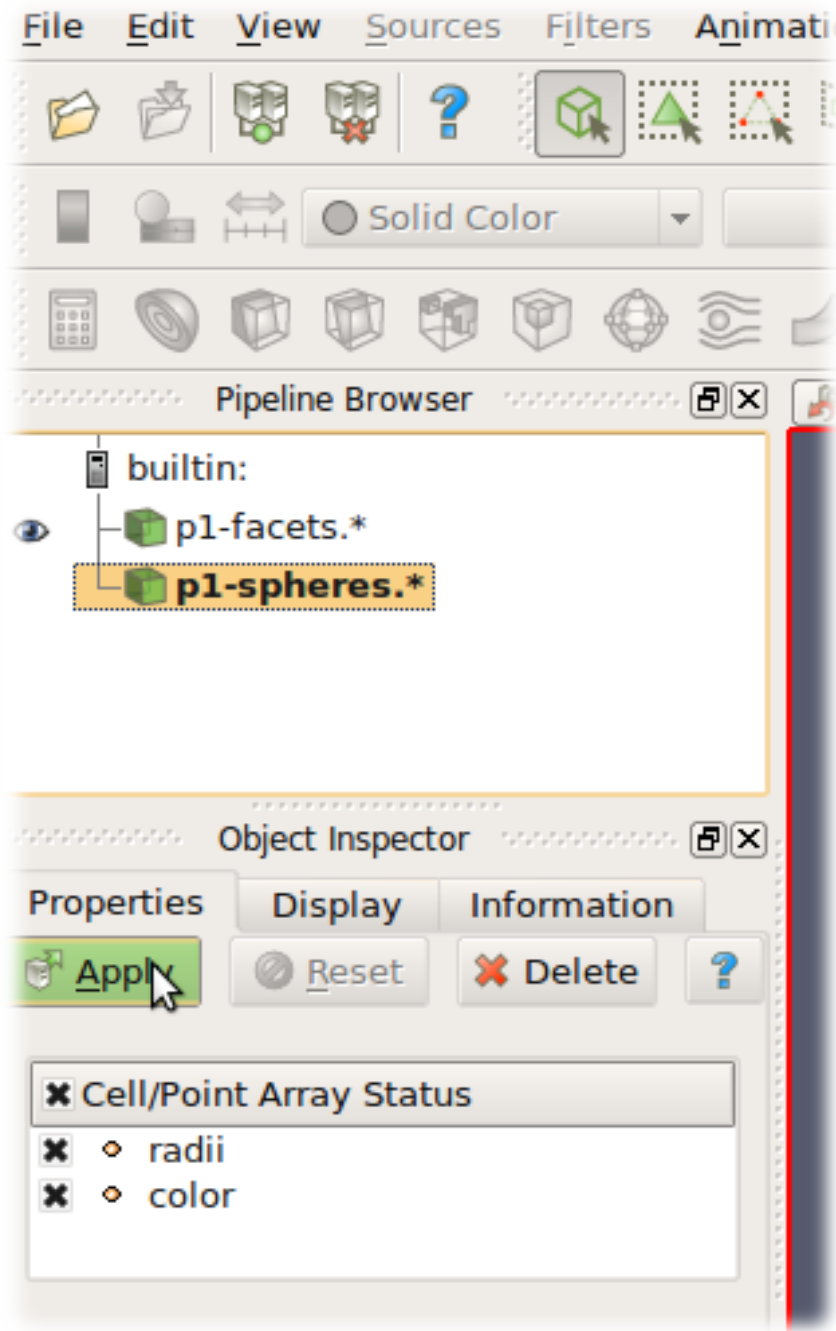

- "Glyph type" to *Sphere*

- "Radius" to *1*

- "Scale mode" to *Scalar* (*Scalar* is set above to be the *radii* value saved in the file, therefore spheres with radius *1* will be scaled by their true radius)

- "Set scale factor" to *1*

- optionally uncheck "Mask points" and "Random mode" (they make some particles not to be rendered for performance reasons, controlled by the "Maximum Number of Points")

After clicking "Apply", spheres will appear. They will be rendered over the original white points, which you can disable by clicking on the eye icon next to `p1-spheres.*` in the Pipeline browser.

### Rendering spherical particles. PointSprite

Another opportunity to display spheres is by using *PointSprite* plugin. This technique requires much less RAM in comparison to Glyphs.

- "Tools -> Manage Plugins"

- "PointSprite_Plugin -> Load selected -> Close"

- Load VTU-files

- "Representation -> Point Sprite"

- "Point Sprite -> Scale By -> radii"

- "Edit Radius Transfer Function -> Proportional -> Multiplier = 1.0 -> Close"

### Rendering interactions as force chain

Data saved by `VTKRecorder` (the steps below generates cones rather than tubes) or `export.VTKExporter(...).exportInteractions(what=dict(forceN='i.phys.normalForce.norm()'))` (the steps below generates per interaction tubes with constant radius):

- Load interactions VTP or VTK files

- Filters -> Cell Data To Point Data

- Filters -> Tube

- Set color by "forceN"

- Set "Vary Radius" to "By Scalar"

- Set "Radius" and "Radius Factor" such that the result looks OK (in 3D postprocessing tutorial script, Radius=0.0005 and Radius Factor=100 looks reasonably)

### Facet transparency

If you want to make facet objects transparent, select `p1-facets.*` in the Pipeline browser, then go to the Object inspector on the Display tab. Under "Style", you can set the "Opacity" value to something smaller than 1.

### Animation

You can move between frames (snapshots that were saved) via the "Animation" menu. After setting the view angle, zoom etc to your satisfaction, the animation can be saved with *File/Save animation*.

**Micro-stress and micro-strain**

It is sometimes useful to visualize a DEM simulation through equivalent strain fields or stress fields. This is possible with *TesselationWrapper*. This class handles the triangulation of spheres in a scene, build tesselation on request, and give access to computed quantities: volume, porosity and local deformation for each sphere. The definition of microstrain and microstress is at the scale of particle-centered subdomains shown below, as explained in [Catalano2014a] .

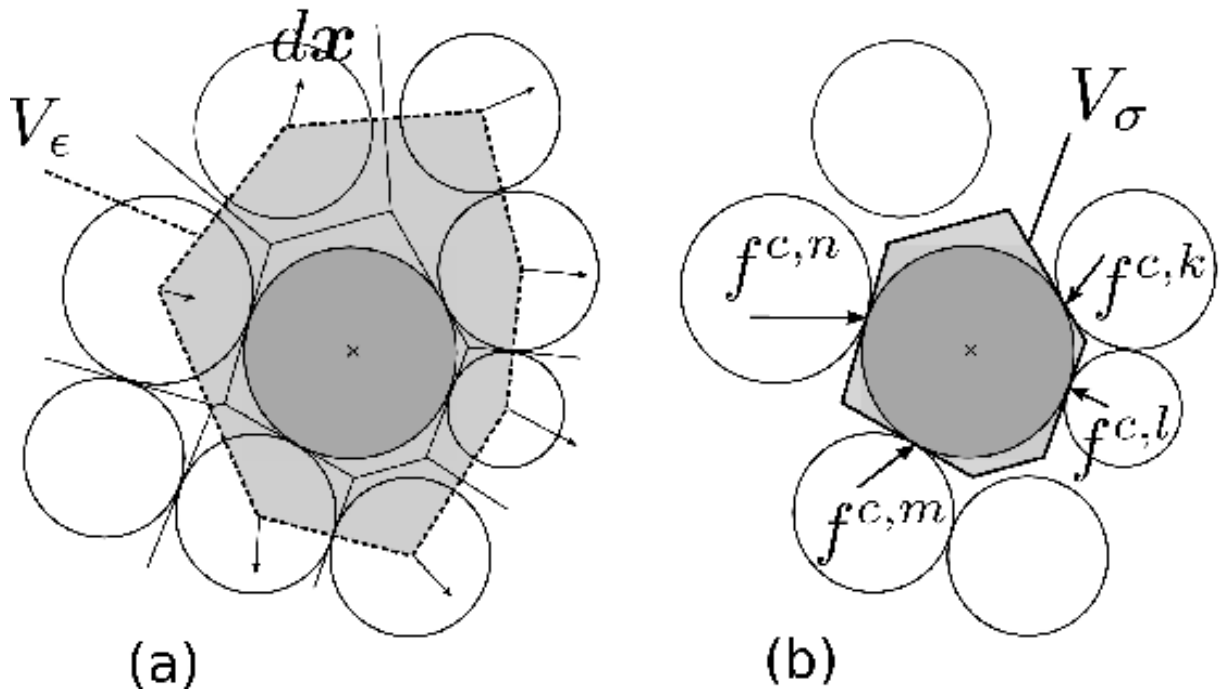

**Micro-strain**

Below is an output of the *defToVtk* function visualized with paraview (in this case Yade's TesselationWrapper was used to process experimental data obtained on sand by Edward Ando at Grenoble University, 3SR lab.). The output is visualized with paraview, as explained in the previous section. Similar results can be generated from simulations:

```
tt=TriaxialTest()
tt.generate("test.yade")
O.load("test.yade")
O.run(100,True)
TW=TesselationWrapper()
TW.triangulate()          #compute regular Delaunay triangulation, don't construct tesselation
TW.computeVolumes()       #will silently tesselate the packing, then compute volume of each
↪Voronoi cell
TW.volume(10)             #get volume associated to sphere of id 10
TW.setState(0)            #store current positions internaly for later use as the "0" state
O.run(100,True)           #make particles move a little (let's hope they will!)
TW.setState(1)            #store current positions internaly in the "1" (deformed) state
#Now we can define strain by comparing states 0 and 1, and average them at the particles scale
TW.defToVtk("strain.vtk")
```

**Micro-stress**

Stress fields can be generated by combining the volume returned by TesselationWrapper to per-particle stress given by *bodyStressTensors*. Since the stress σ from bodyStressTensor implies a division by the

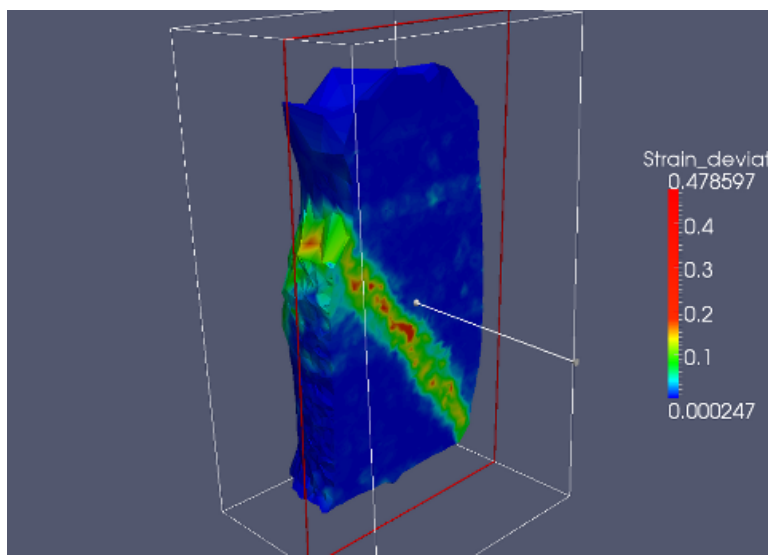

volume $V_b$ of the solid particle, one has to re-normalize it in order to obtain the micro-stress as defined in [Catalano2014a] (equation 39 therein), i.e. $\overline{\sigma}^k = \sigma^k \times V_b^k / V_\sigma^k$ where $V_\sigma^k$ is the volume assigned to particle k in the tesselation. For instance:

```
#"b" being a body
TW=TesselationWrapper()
TW.setState()
TW.computeVolumes()
s=bodyStressTensors()
stress = s[b.id]*4.*pi/3.*b.shape.radius**3/TW.volume(b.id)
```

As any other value, the stress can be exported to a vtk file for display in Paraview using *export.VTKExporter*.

### 2.2.4 Python specialties and tricks

**Importing Yade in other Python applications**

Yade can be imported in other Python applications. To do so, you need somehow to make yade executable .py extended. The easiest way is to create a symbolic link, i.e. (suppose your Yade executable file is called "yade-trunk" and you want make it "yadeimport.py"):

```
$ cd /path/where/you/want/yadeimport
$ ln -s /path/to/yade/executable/yade-trunk yadeimport.py
```

Then you need to make your yadeimport.py findable by Python. You can export PYTHONPATH environment variable, or simply use sys.path directly in Python script:

```
import sys
sys.path.append('/path/where/you/want/yadeimport')
from yadeimport import *

print(Matrix3(1,2,3, 4,5,6, 7,8,9))
print(O.bodies)
# any other Yade code
```

### 2.2.5 Extending Yade

- new particle shape

- new constitutive law

### 2.2.6 Troubleshooting

**Crashes**

It is possible that you encounter crash of Yade, i.e. Yade terminates with error message such as

```
Segmentation fault (core dumped)
```

without further explanation. Frequent causes of such conditions are

- program error in Yade itself;

- fatal condition in your particular simulation (such as impossible dispatch);

- problem with graphics card driver.

Try to reproduce the error (run the same script) with debug-enabled version of Yade. Debugger will be automatically launched at crash, showing backtrace of the code (in this case, we triggered crash by hand):

```
Yade [1]: import os,signal
Yade [2]: os.kill(os.getpid(),signal.SIGSEGV)
SIGSEGV/SIGABRT handler called; gdb batch file is `/tmp/yade-YwtfRY/tmp-0'
GNU gdb (GDB) 7.1-ubuntu
Copyright (C) 2010 Free Software Foundation, Inc.
License GPLv3+: GNU GPL version 3 or later <http://gnu.org/licenses/gpl.html>
This is free software: you are free to change and redistribute it.
There is NO WARRANTY, to the extent permitted by law.  Type "show copying"
and "show warranty" for details.
This GDB was configured as "x86_64-linux-gnu".
For bug reporting instructions, please see:
<http://www.gnu.org/software/gdb/bugs/>.
[Thread debugging using libthread_db enabled]
[New Thread 0x7f0fb1268710 (LWP 16471)]
[New Thread 0x7f0fb29f2710 (LWP 16470)]
[New Thread 0x7f0fb31f3710 (LWP 16469)]

…
```

What looks as cryptic message is valuable information for developers to locate source of the bug. In particular, there is (usually) line `<signal handler called>`; lines below it are source of the bug (at least very likely so):

```
Thread 1 (Thread 0x7f0fcee53700 (LWP 16465)):
#0  0x00007f0fcd8f4f7d in __libc_waitpid (pid=16497, stat_loc=<value optimized out>,
↪options=0) at ../sysdeps/unix/sysv/linux/waitpid.c:41
#1  0x00007f0fcd88c7e9 in do_system (line=<value optimized out>) at ../sysdeps/posix/system.
↪c:149
#2  0x00007f0fcd88cb20 in __libc_system (line=<value optimized out>) at ../sysdeps/posix/
↪system.c:190
#3  0x00007f0fcd0b4b23 in crashHandler (sig=11) at core/main/pyboot.cpp:45
#4  <signal handler called>
#5  0x00007f0fcd87ed57 in kill () at ../sysdeps/unix/syscall-template.S:82
#6  0x000000000051336d in posix_kill (self=<value optimized out>, args=<value optimized out>)
↪at ../Modules/posixmodule.c:4046
#7  0x00000000004a7c5e in call_function (f=Frame 0x1c54620, for file <ipython console>, line 1,
↪ in <module> (), throwflag=<value optimized out>) at ../Python/ceval.c:3750
#8  PyEval_EvalFrameEx (f=Frame 0x1c54620, for file <ipython console>, line 1, in <module> (),
↪throwflag=<value optimized out>) at ../Python/ceval.c:2412
```

If you think this might be error in Yade, file a bug report as explained below. Do not forget to attach *full* yade output from terminal, including startup messages and debugger output – select with right mouse button, with middle button paste the bugreport to a file and attach it. Attach your simulation script as well.

### Reporting bugs

Bugs are general name for defects (functionality shortcomings, misdocumentation, crashes) or feature requests. They are tracked at https://gitlab.com/yade-dev/trunk/issues.

When reporting a new bug, be as specific as possible; state version of yade you use, system version and the output of *printAllVersions()*, as explained in the above section on crashes.

### Getting help

### Questions and answers

---

**Hint:** Please use Launchpad interface at https://answers.launchpad.net/yade/ for asking questions about Yade.

---

In case you're not familiar with computer oriented discussion lists, please read this wiki page (a Yade-oriented and shortened version of How To Ask Questions The Smart Way) before posting, in order to increase your chances getting help. Do not forget to state what *version* of Yade you use (shown when you start Yade, or even better as printed by function *libVersions.printAllVersions*), whether you installed it from source code or a package, what operating system (such as Ubuntu 18.04), and if you have done any local modifications to source code in case of compiled version.

### Mailing lists

In addition to the Q&A Launchpad interface, Yade has two mailing-lists. Both are hosted at http://www.launchpad.net and before posting, you must register to Launchpad and subscribe to the list by adding yourself to "team" of the same name running the list.

**yade-users@lists.launchpad.net** is a general discussion list for all Yade users. Add yourself to yade-users team so that you can post messages. List archives:

- https://lists.launchpad.net/yade-users/

- http://www.mail-archive.com/yade-users@lists.launchpad.net/

**yade-dev@lists.launchpad.net** is for discussions about Yade development; you must be member of yade-dev team to post. This list is archived in two places:

- https://lists.launchpad.net/yade-dev/

- http://www.mail-archive.com/yade-dev@lists.launchpad.net/

### Wiki

http://www.yade-dem.org/wiki/

### Private and/or paid support

You might contact developers by their private mail (rather than by mailing list) if you do not want to disclose details on the mailing list. This is also a suitable method for proposing financial reward for

---

implementation of a substantial feature that is not yet in Yade – typically, though, we will request this feature to be part of the public codebase once completed, so that the rest of the community can benefit from it as well.

## 2.3 Yade wrapper class reference

### 2.3.1 Bodies

**Body**

**class yade.wrapper.Body**(*inherits Serializable*)
A particle, basic element of simulation; interacts with other bodies.

**aspherical**(*=false*)
Whether this body has different inertia along principal axes; *NewtonIntegrator* makes use of this flag to call rotation integration routine for aspherical bodies, which is more expensive.

**bound**(*=uninitalized*)
*Bound*, approximating volume for the purposes of collision detection.

**bounded**(*=true*)
Whether this body should have *Body.bound* created. Note that bodies without a *bound* do not participate in collision detection. (In c++, use `Body::isBounded`/`Body::setBounded`)

**chain**
Returns Id of chain to which the body belongs.

**clumpId**
Id of clump this body makes part of; invalid number if not part of clump; see *Body::isStandalone*, *Body::isClump*, *Body::isClumpMember* properties.

Not meant to be modified directly from Python, use *O.bodies.appendClumped* instead.

**dict**(*(Serializable)arg1*) → dict :
Return dictionary of attributes.

**dynamic**(*=true*)
Whether this body will be moved by forces. (In c++, use `Body::isDynamic`/`Body::setDynamic`)

**flags**(*=FLAG_BOUNDED*)
Bits of various body-related flags. *Do not access directly.* In c++, use isDynamic/setDynamic, isBounded/setBounded, isAspherical/setAspherical. In python, use *Body.dynamic*, *Body.bounded*, *Body.aspherical*.

**groupMask**(*=1*)
Bitmask for interaction detection purposes: it is required that two bodies have at least one bit in common in their groupMask for their interaction to be possible from the *Collider* point of view.

**id**(*=Body::ID_NONE*)
Unique id of this body.

**intrs**(*(Body)arg1*) → list :
Return list of all real interactions in which this body participates.

**isClump**
True if this body is clump itself, false otherwise.

**isClumpMember**
True if this body is clump member, false otherwise.

**isStandalone**
:   True if this body is neither clump, nor clump member; false otherwise.

**iterBorn**(*=-1*)
:   Step number at which the body was added to simulation.

**mask**
:   Shorthand for *Body::groupMask*

**mat**
:   Shorthand for *Body::material*

**material**(*=uninitalized*)
:   *Material* instance associated with this body.

**shape**(*=uninitalized*)
:   Geometrical *Shape*.

**state**(*=new State*)
:   Physical *state*.

**timeBorn**(*=-1*)
:   Time at which the body was added to simulation.

**updateAttrs**(*(Serializable)arg1, (dict)arg2*) → None :
:   Update object attributes from given dictionary

## Shape

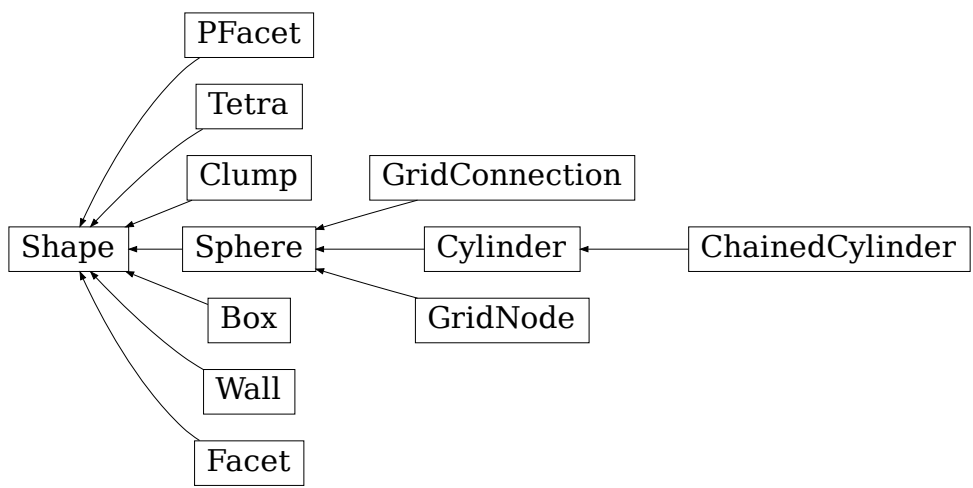

Fig. 19: Inheritance graph of Shape. See also: *Box*, *ChainedCylinder*, *Clump*, *Cylinder*, *Facet*, *GridConnection*, *GridNode*, *PFacet*, *Sphere*, *Tetra*, *Wall*.

**class yade.wrapper.Shape**(*inherits Serializable*)
:   Geometry of a body

    **color**(*=Vector3r(1, 1, 1)*)
    :   Color for rendering (normalized RGB).

    **dict**(*(Serializable)arg1*) → dict :
    :   Return dictionary of attributes.

    **dispHierarchy**(*(Shape)arg1*[, *(bool)names=True*]) → list :
    :   Return list of dispatch classes (from down upwards), starting with the class instance itself, top-level indexable at last. If names is true (default), return class names rather than numerical indices.

**dispIndex**
Return class index of this instance.

**highlight**(*=false*)
Whether this Shape will be highlighted when rendered.

**updateAttrs**(*(Serializable)arg1, (dict)arg2*) → None :
Update object attributes from given dictionary

**wire**(*=false*)
Whether this Shape is rendered using color surfaces, or only wireframe (can still be overridden by global config of the renderer).

**class yade.wrapper.Box**(*inherits* *Shape* → *Serializable*)

**color**(*=Vector3r(1, 1, 1)*)
Color for rendering (normalized RGB).

**dict**(*(Serializable)arg1*) → dict :
Return dictionary of attributes.

**dispHierarchy**(*(Shape)arg1*[, *(bool)names=True*]) → list :
Return list of dispatch classes (from down upwards), starting with the class instance itself, top-level indexable at last. If names is true (default), return class names rather than numerical indices.

**dispIndex**
Return class index of this instance.

**extents**(*=uninitalized*)
Half-size of the cuboid

**highlight**(*=false*)
Whether this Shape will be highlighted when rendered.

**updateAttrs**(*(Serializable)arg1, (dict)arg2*) → None :
Update object attributes from given dictionary

**wire**(*=false*)
Whether this Shape is rendered using color surfaces, or only wireframe (can still be overridden by global config of the renderer).

**class yade.wrapper.ChainedCylinder**(*inherits* *Cylinder* → *Sphere* → *Shape* → *Serializable*)
Geometry of a deformable chained cylinder, using geometry *Cylinder*.

**chainedOrientation**(*=Quaternionr::Identity()*)
Deviation of node1 orientation from node-to-node vector

**color**(*=Vector3r(1, 1, 1)*)
Color for rendering (normalized RGB).

**dict**(*(Serializable)arg1*) → dict :
Return dictionary of attributes.

**dispHierarchy**(*(Shape)arg1*[, *(bool)names=True*]) → list :
Return list of dispatch classes (from down upwards), starting with the class instance itself, top-level indexable at last. If names is true (default), return class names rather than numerical indices.

**dispIndex**
Return class index of this instance.

**highlight**(*=false*)
Whether this Shape will be highlighted when rendered.

**initLength**(*=0*)
tensile-free length, used as reference for tensile strain

**length**(*=NaN*)
> Length [m]

**radius**(*=NaN*)
> Radius [m]

**segment**(*=Vector3r::Zero()*)
> Length vector

**updateAttrs**(*(Serializable)arg1, (dict)arg2*) → None :
> Update object attributes from given dictionary

**wire**(*=false*)
> Whether this Shape is rendered using color surfaces, or only wireframe (can still be overridden by global config of the renderer).

**class yade.wrapper.Clump**(*inherits Shape → Serializable*)
> Rigid aggregate of bodies

**color**(*=Vector3r(1, 1, 1)*)
> Color for rendering (normalized RGB).

**dict**(*(Serializable)arg1*) → dict :
> Return dictionary of attributes.

**dispHierarchy**(*(Shape)arg1*[, *(bool)names=True*]) → list :
> Return list of dispatch classes (from down upwards), starting with the class instance itself, top-level indexable at last. If names is true (default), return class names rather than numerical indices.

**dispIndex**
> Return class index of this instance.

**highlight**(*=false*)
> Whether this Shape will be highlighted when rendered.

**ids**(*=uninitalized*)
> Ids of constituent particles (only informative; direct modifications will have no effect).

**members**
> Return clump members as {'id1':(relPos,relOri),...}

**updateAttrs**(*(Serializable)arg1, (dict)arg2*) → None :
> Update object attributes from given dictionary

**wire**(*=false*)
> Whether this Shape is rendered using color surfaces, or only wireframe (can still be overridden by global config of the renderer).

**class yade.wrapper.Cylinder**(*inherits Sphere → Shape → Serializable*)
> Geometry of a cylinder, as Minkowski sum of line and sphere.

**color**(*=Vector3r(1, 1, 1)*)
> Color for rendering (normalized RGB).

**dict**(*(Serializable)arg1*) → dict :
> Return dictionary of attributes.

**dispHierarchy**(*(Shape)arg1*[, *(bool)names=True*]) → list :
> Return list of dispatch classes (from down upwards), starting with the class instance itself, top-level indexable at last. If names is true (default), return class names rather than numerical indices.

**dispIndex**
> Return class index of this instance.

**highlight**(*=false*)
> Whether this Shape will be highlighted when rendered.

**length**(*=NaN*)
>   Length [m]

**radius**(*=NaN*)
>   Radius [m]

**segment**(*=Vector3r::Zero()*)
>   Length vector

**updateAttrs**(*(Serializable)arg1, (dict)arg2*) → None :
>   Update object attributes from given dictionary

**wire**(*=false*)
>   Whether this Shape is rendered using color surfaces, or only wireframe (can still be overridden by global config of the renderer.)

**class yade.wrapper.Facet**(*inherits* *Shape* → *Serializable*)
>   Facet (triangular particle) geometry.

>   **area**(*=NaN*)
>   >   Facet's area

>   **color**(*=Vector3r(1, 1, 1)*)
>   >   Color for rendering (normalized RGB).

>   **dict**(*(Serializable)arg1*) → dict :
>   >   Return dictionary of attributes.

>   **dispHierarchy**(*(Shape)arg1*[, *(bool)names=True*]) → list :
>   >   Return list of dispatch classes (from down upwards), starting with the class instance itself, top-level indexable at last. If names is true (default), return class names rather than numerical indices.

>   **dispIndex**
>   >   Return class index of this instance.

>   **highlight**(*=false*)
>   >   Whether this Shape will be highlighted when rendered.

>   **normal**(*=Vector3r(NaN, NaN, NaN)*)
>   >   Facet's normal (in local coordinate system)

>   **setVertices**(*(Facet)arg1, (Vector3)arg2, (Vector3)arg3, (Vector3)arg4*) → None :
>   >   TODO

>   **updateAttrs**(*(Serializable)arg1, (dict)arg2*) → None :
>   >   Update object attributes from given dictionary

>   **vertices**(*=vector<Vector3r>(3, Vector3r(NaN, NaN, NaN))*)
>   >   Vertex positions in local coordinates.

>   **wire**(*=false*)
>   >   Whether this Shape is rendered using color surfaces, or only wireframe (can still be overridden by global config of the renderer.)

**class yade.wrapper.GridConnection**(*inherits* *Sphere* → *Shape* → *Serializable*)
>   GridConnection shape (see [Effeindzourou2016], [Bourrier2013]). Component of a grid designed to link two *GridNodes*. It is highly recommended to use *gridpfacet.gridConnection* to generate correct *GridConnections*.

>   **addPFacet**(*(GridConnection)arg1, (Body)Body*) → None :
>   >   Add a PFacet to the GridConnection.

>   **cellDist**(*=Vector3i(0, 0, 0)*)
>   >   Distance of bodies in cell size units, if using periodic boundary conditions. Note that periodic boundary conditions for GridConnections have not yet been fully implemented.

**color**(*=Vector3r(1, 1, 1)*)
   Color for rendering (normalized RGB).

**dict**(*(Serializable)arg1*) → dict :
   Return dictionary of attributes.

**dispHierarchy**(*(Shape)arg1*[, *(bool)names=True*]) → list :
   Return list of dispatch classes (from down upwards), starting with the class instance itself, top-level indexable at last. If names is true (default), return class names rather than numerical indices.

**dispIndex**
   Return class index of this instance.

**getPFacets**(*(GridConnection)arg1*) → object :
   get list of linked PFacets.

**highlight**(*=false*)
   Whether this Shape will be highlighted when rendered.

**node1**(*=uninitalized*)
   First *Body* the GridConnection is connected to.

**node2**(*=uninitalized*)
   Second *Body* the GridConnection is connected to.

**periodic**(*=false*)
   true if two nodes from different periods are connected.

**radius**(*=NaN*)
   Radius [m]

**updateAttrs**(*(Serializable)arg1, (dict)arg2*) → None :
   Update object attributes from given dictionary

**wire**(*=false*)
   Whether this Shape is rendered using color surfaces, or only wireframe (can still be overridden by global config of the renderer).

**class yade.wrapper.GridNode**(*inherits Sphere → Shape → Serializable*)
   GridNode shape, component of a grid. To create a Grid, place the nodes first, they will define the spacial discretisation of it. It is highly recommended to use *gridpfacet.gridNode* to generate correct *GridNodes*. Note that the GridNodes should only be in an Interaction with other GridNodes. The Sphere-Grid contact is only handled by the *GridConnections*.

**addConnection**(*(GridNode)arg1, (Body)Body*) → None :
   Add a GridConnection to the GridNode.

**addPFacet**(*(GridNode)arg1, (Body)Body*) → None :
   Add a PFacet to the GridNode.

**color**(*=Vector3r(1, 1, 1)*)
   Color for rendering (normalized RGB).

**dict**(*(Serializable)arg1*) → dict :
   Return dictionary of attributes.

**dispHierarchy**(*(Shape)arg1*[, *(bool)names=True*]) → list :
   Return list of dispatch classes (from down upwards), starting with the class instance itself, top-level indexable at last. If names is true (default), return class names rather than numerical indices.

**dispIndex**
   Return class index of this instance.

**getConnections**(*(GridNode)arg1*) → object :
   get list of linked *GridConnection*'s.

**getPFacets**(*(GridNode)arg1*) → object :
>   get list of linked *PFacet*'s.

**highlight**(*=false*)
>   Whether this Shape will be highlighted when rendered.

**radius**(*=NaN*)
>   Radius [m]

**updateAttrs**(*(Serializable)arg1, (dict)arg2*) → None :
>   Update object attributes from given dictionary

**wire**(*=false*)
>   Whether this Shape is rendered using color surfaces, or only wireframe (can still be overridden by global config of the renderer).

**class yade.wrapper.PFacet**(*inherits Shape → Serializable*)
>   PFacet (particle facet) geometry (see [Effeindzourou2016], [Effeindzourou2015a]). It is highly recommended to use the helper functions in *gridpfacet* (e.g., gridpfacet.pfacetCreator1-4) to generate correct *PFacet* elements.

**area**(*=NaN*)
>   PFacet's area

**cellDist**(*=Vector3i(0, 0, 0)*)
>   Distance of bodies in cell size units, if using periodic boundary conditions. Note that periodic boundary conditions for PFacets have not yet been fully implemented.

**color**(*=Vector3r(1, 1, 1)*)
>   Color for rendering (normalized RGB).

**conn1**(*=uninitalized*)
>   First *Body* the Pfacet is connected to.

**conn2**(*=uninitalized*)
>   Second *Body* the Pfacet is connected to.

**conn3**(*=uninitalized*)
>   third *Body* the Pfacet is connected to.

**dict**(*(Serializable)arg1*) → dict :
>   Return dictionary of attributes.

**dispHierarchy**(*(Shape)arg1*[, *(bool)names=True*]) → list :
>   Return list of dispatch classes (from down upwards), starting with the class instance itself, top-level indexable at last. If names is true (default), return class names rather than numerical indices.

**dispIndex**
>   Return class index of this instance.

**highlight**(*=false*)
>   Whether this Shape will be highlighted when rendered.

**node1**(*=uninitalized*)
>   First *Body* the Pfacet is connected to.

**node2**(*=uninitalized*)
>   Second *Body* the Pfacet is connected to.

**node3**(*=uninitalized*)
>   third *Body* the Pfacet is connected to.

**normal**(*=Vector3r(NaN, NaN, NaN)*)
>   PFacet's normal (in local coordinate system)

**radius**(*=-1*)
>   PFacet's radius

**updateAttrs**(*(Serializable)arg1, (dict)arg2*) → None :
> Update object attributes from given dictionary

**wire**(*=false*)
> Whether this Shape is rendered using color surfaces, or only wireframe (can still be overridden by global config of the renderer).

**class yade.wrapper.Sphere**(*inherits Shape → Serializable*)
> Geometry of spherical particle.

**color**(*=Vector3r(1, 1, 1)*)
> Color for rendering (normalized RGB).

**dict**(*(Serializable)arg1*) → dict :
> Return dictionary of attributes.

**dispHierarchy**(*(Shape)arg1[, (bool)names=True]*) → list :
> Return list of dispatch classes (from down upwards), starting with the class instance itself, top-level indexable at last. If names is true (default), return class names rather than numerical indices.

**dispIndex**
> Return class index of this instance.

**highlight**(*=false*)
> Whether this Shape will be highlighted when rendered.

**radius**(*=NaN*)
> Radius [m]

**updateAttrs**(*(Serializable)arg1, (dict)arg2*) → None :
> Update object attributes from given dictionary

**wire**(*=false*)
> Whether this Shape is rendered using color surfaces, or only wireframe (can still be overridden by global config of the renderer).

**class yade.wrapper.Tetra**(*inherits Shape → Serializable*)
> Tetrahedron geometry.

**color**(*=Vector3r(1, 1, 1)*)
> Color for rendering (normalized RGB).

**dict**(*(Serializable)arg1*) → dict :
> Return dictionary of attributes.

**dispHierarchy**(*(Shape)arg1[, (bool)names=True]*) → list :
> Return list of dispatch classes (from down upwards), starting with the class instance itself, top-level indexable at last. If names is true (default), return class names rather than numerical indices.

**dispIndex**
> Return class index of this instance.

**highlight**(*=false*)
> Whether this Shape will be highlighted when rendered.

**updateAttrs**(*(Serializable)arg1, (dict)arg2*) → None :
> Update object attributes from given dictionary

**v**(*=std::vector<Vector3r>(4)*)
> Tetrahedron vertices (in local coordinate system).

**wire**(*=false*)
> Whether this Shape is rendered using color surfaces, or only wireframe (can still be overridden by global config of the renderer).

**class** `yade.wrapper.`**Wall**(*inherits Shape → Serializable*)

> Object representing infinite plane aligned with the coordinate system (axis-aligned wall).

> **axis**(*=0*)
> > Axis of the normal; can be 0,1,2 for +x, +y, +z respectively (Body's orientation is disregarded for walls)

> **color**(*=Vector3r(1, 1, 1)*)
> > Color for rendering (normalized RGB).

> **dict**(*(Serializable)arg1*) → dict :
> > Return dictionary of attributes.

> **dispHierarchy**(*(Shape)arg1*[, *(bool)names=True*]) → list :
> > Return list of dispatch classes (from down upwards), starting with the class instance itself, top-level indexable at last. If names is true (default), return class names rather than numerical indices.

> **dispIndex**
> > Return class index of this instance.

> **highlight**(*=false*)
> > Whether this Shape will be highlighted when rendered.

> **sense**(*=0*)
> > Which side of the wall interacts: -1 for negative only, 0 for both, +1 for positive only

> **updateAttrs**(*(Serializable)arg1, (dict)arg2*) → None :
> > Update object attributes from given dictionary

> **wire**(*=false*)
> > Whether this Shape is rendered using color surfaces, or only wireframe (can still be overridden by global config of the renderer).

### State

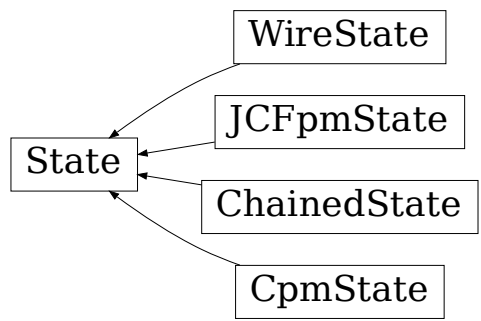

Fig. 20: Inheritance graph of State. See also: *ChainedState*, *CpmState*, *JCFpmState*, *WireState*.

**class** `yade.wrapper.`**State**(*inherits Serializable*)

> State of a body (spatial configuration, internal variables).

> **angMom**(*=Vector3r::Zero()*)
> > Current angular momentum

> **angVel**(*=Vector3r::Zero()*)
> > Current angular velocity

> **blockedDOFs**
> > Degress of freedom where linear/angular velocity will be always constant (equal to zero, or to an user-defined value), regardless of applied force/torque. String that may contain 'xyzXYZ' (translations and rotations).

**densityScaling**(*=-1*)
> *(auto-updated)* see *GlobalStiffnessTimeStepper::targetDt*.

**dict**(*(Serializable)arg1*) → dict :
> Return dictionary of attributes.

**dispHierarchy**(*(State)arg1*[, *(bool)names=True*]) → list :
> Return list of dispatch classes (from down upwards), starting with the class instance itself, top-level indexable at last. If names is true (default), return class names rather than numerical indices.

**dispIndex**
> Return class index of this instance.

**displ**(*(State)arg1*) → Vector3 :
> Displacement from *reference position* (*pos* - *refPos*)

**inertia**(*=Vector3r::Zero()*)
> Inertia of associated body, in local coordinate system.

**isDamped**(*=true*)
> Damping in *NewtonIntegrator* can be deactivated for individual particles by setting this variable to FALSE. E.g. damping is inappropriate for particles in free flight under gravity but it might still be applicable to other particles in the same simulation.

**mass**(*=0*)
> Mass of this body

**ori**
> Current orientation.

**pos**
> Current position.

**refOri**(*=Quaternionr::Identity()*)
> Reference orientation

**refPos**(*=Vector3r::Zero()*)
> Reference position

**rot**(*(State)arg1*) → Vector3 :
> Rotation from *reference orientation* (as rotation vector)

**se3**(*=Se3r(Vector3r::Zero(), Quaternionr::Identity())*)
> Position and orientation as one object.

**updateAttrs**(*(Serializable)arg1, (dict)arg2*) → None :
> Update object attributes from given dictionary

**vel**(*=Vector3r::Zero()*)
> Current linear velocity.

**class yade.wrapper.ChainedState**(*inherits State → Serializable*)
> State of a chained bodies, containing information on connectivity in order to track contacts jumping over contiguous elements. Chains are 1D lists from which id of chained bodies are retrieved via *rank* and *chainNumber*.

**addToChain**(*(ChainedState)arg1, (int)bodyId*) → None :
> Add body to current active chain

**angMom**(*=Vector3r::Zero()*)
> Current angular momentum

**angVel**(*=Vector3r::Zero()*)
> Current angular velocity

**bId**(*=-1*)
> id of the body containing - for postLoad operations only.

**blockedDOFs**
> Degress of freedom where linear/angular velocity will be always constant (equal to zero, or to an user-defined value), regardless of applied force/torque. String that may contain 'xyzXYZ' (translations and rotations).

**chainNumber**(*=0*)
> chain id.

**currentChain = 0**

**densityScaling**(*=-1*)
> *(auto-updated)* see *GlobalStiffnessTimeStepper::targetDt*.

**dict**(*(Serializable)arg1*) → dict :
> Return dictionary of attributes.

**dispHierarchy**(*(State)arg1*[, *(bool)names=True*]) → list :
> Return list of dispatch classes (from down upwards), starting with the class instance itself, top-level indexable at last. If names is true (default), return class names rather than numerical indices.

**dispIndex**
> Return class index of this instance.

**displ**(*(State)arg1*) → Vector3 :
> Displacement from *reference position* (*pos* - *refPos*)

**inertia**(*=Vector3r::Zero()*)
> Inertia of associated body, in local coordinate system.

**isDamped**(*=true*)
> Damping in *NewtonIntegrator* can be deactivated for individual particles by setting this variable to FALSE. E.g. damping is inappropriate for particles in free flight under gravity but it might still be applicable to other particles in the same simulation.

**mass**(*=0*)
> Mass of this body

**ori**
> Current orientation.

**pos**
> Current position.

**rank**(*=0*)
> rank in the chain.

**refOri**(*=Quaternionr::Identity()*)
> Reference orientation

**refPos**(*=Vector3r::Zero()*)
> Reference position

**rot**(*(State)arg1*) → Vector3 :
> Rotation from *reference orientation* (as rotation vector)

**se3**(*=Se3r(Vector3r::Zero(), Quaternionr::Identity())*)
> Position and orientation as one object.

**updateAttrs**(*(Serializable)arg1, (dict)arg2*) → None :
> Update object attributes from given dictionary

**vel**(*=Vector3r::Zero()*)
> Current linear velocity.

**class yade.wrapper.CpmState**(*inherits State → Serializable*)
> State information about body use by *cpm-model*.

None of that is used for computation (at least not now), only for post-processing.

**angMom**(*=Vector3r::Zero()*)
    Current angular momentum

**angVel**(*=Vector3r::Zero()*)
    Current angular velocity

**blockedDOFs**
    Degress of freedom where linear/angular velocity will be always constant (equal to zero, or to an user-defined value), regardless of applied force/torque. String that may contain 'xyzXYZ' (translations and rotations).

**damageTensor**(*=Matrix3r::Zero()*)
    Damage tensor computed with microplane theory averaging. state.damageTensor.trace() = state.normDmg

**densityScaling**(*=-1*)
    *(auto-updated)* see *GlobalStiffnessTimeStepper::targetDt*.

**dict**(*(Serializable)arg1*) → dict :
    Return dictionary of attributes.

**dispHierarchy**(*(State)arg1*[, *(bool)names=True*]) → list :
    Return list of dispatch classes (from down upwards), starting with the class instance itself, top-level indexable at last. If names is true (default), return class names rather than numerical indices.

**dispIndex**
    Return class index of this instance.

**displ**(*(State)arg1*) → Vector3 :
    Displacement from *reference position* (*pos* - *refPos*)

**epsVolumetric**(*=0*)
    Volumetric strain around this body (unused for now)

**inertia**(*=Vector3r::Zero()*)
    Inertia of associated body, in local coordinate system.

**isDamped**(*=true*)
    Damping in *NewtonIntegrator* can be deactivated for individual particles by setting this variable to FALSE. E.g. damping is inappropriate for particles in free flight under gravity but it might still be applicable to other particles in the same simulation.

**mass**(*=0*)
    Mass of this body

**normDmg**(*=0*)
    Average damage including already deleted contacts (it is really not damage, but 1-relResidualStrength now)

**numBrokenCohesive**(*=0*)
    Number of (cohesive) contacts that damaged completely

**numContacts**(*=0*)
    Number of contacts with this body

**ori**
    Current orientation.

**pos**
    Current position.

**refOri**(*=Quaternionr::Identity()*)
    Reference orientation

**refPos**(*=Vector3r::Zero()*)
  Reference position

**rot**(*(State)arg1*) → Vector3 :
  Rotation from *reference orientation* (as rotation vector)

**se3**(*=Se3r(Vector3r::Zero(), Quaternionr::Identity())*)
  Position and orientation as one object.

**stress**(*=Matrix3r::Zero()*)
  Stress tensor of the spherical particle (under assumption that particle volume = pi*r*r*r*4/3.)
  for packing fraction 0.62

**updateAttrs**(*(Serializable)arg1, (dict)arg2*) → None :
  Update object attributes from given dictionary

**vel**(*=Vector3r::Zero()*)
  Current linear velocity.

**class yade.wrapper.JCFpmState**(*inherits State → Serializable*)
  JCFpm state information about each body.

  **angMom**(*=Vector3r::Zero()*)
    Current angular momentum

  **angVel**(*=Vector3r::Zero()*)
    Current angular velocity

  **blockedDOFs**
    Degress of freedom where linear/angular velocity will be always constant (equal to zero, or to
    an user-defined value), regardless of applied force/torque. String that may contain 'xyzXYZ'
    (translations and rotations).

  **damageIndex**(*=0*)
    Ratio of broken bonds over initial bonds. [-]

  **densityScaling**(*=-1*)
    *(auto-updated)* see *GlobalStiffnessTimeStepper::targetDt*.

  **dict**(*(Serializable)arg1*) → dict :
    Return dictionary of attributes.

  **dispHierarchy**(*(State)arg1*[, *(bool)names=True*]) → list :
    Return list of dispatch classes (from down upwards), starting with the class instance itself,
    top-level indexable at last. If names is true (default), return class names rather than numerical
    indices.

  **dispIndex**
    Return class index of this instance.

  **displ**(*(State)arg1*) → Vector3 :
    Displacement from *reference position* (*pos* - *refPos*)

  **inertia**(*=Vector3r::Zero()*)
    Inertia of associated body, in local coordinate system.

  **isDamped**(*=true*)
    Damping in *NewtonIntegrator* can be deactivated for individual particles by setting this vari-
    able to FALSE. E.g. damping is inappropriate for particles in free flight under gravity but it
    might still be applicable to other particles in the same simulation.

  **joint**(*=0*)
    Indicates the number of joint surfaces to which the particle belongs (0-> no joint, 1->1 joint,
    etc..). [-]

  **jointNormal1**(*=Vector3r::Zero()*)
    Specifies the normal direction to the joint plane 1. Rk: the ideal here would be to create a

vector of vector wich size is defined by the joint integer (as much joint normals as joints). However, it needs to make the pushback function works with python since joint detection is done through a python script. lines 272 to 312 of cpp file should therefore be adapted. [-]

**jointNormal2**(*=Vector3r::Zero()*)
Specifies the normal direction to the joint plane 2. [-]

**jointNormal3**(*=Vector3r::Zero()*)
Specifies the normal direction to the joint plane 3. [-]

**mass**(*=0*)
Mass of this body

**nbBrokenBonds**(*=0*)
Number of broken bonds. [-]

**nbInitBonds**(*=0*)
Number of initial bonds. [-]

**onJoint**(*=false*)
Identifies if the particle is on a joint surface.

**ori**
Current orientation.

**pos**
Current position.

**refOri**(*=Quaternionr::Identity()*)
Reference orientation

**refPos**(*=Vector3r::Zero()*)
Reference position

**rot**(*(State)arg1*) → Vector3 :
Rotation from *reference orientation* (as rotation vector)

**se3**(*=Se3r(Vector3r::Zero(), Quaternionr::Identity())*)
Position and orientation as one object.

**updateAttrs**(*(Serializable)arg1, (dict)arg2*) → None :
Update object attributes from given dictionary

**vel**(*=Vector3r::Zero()*)
Current linear velocity.

**class yade.wrapper.WireState**(*inherits State → Serializable*)
Wire state information of each body.

None of that is used for computation (at least not now), only for post-processing.

**angMom**(*=Vector3r::Zero()*)
Current angular momentum

**angVel**(*=Vector3r::Zero()*)
Current angular velocity

**blockedDOFs**
Degress of freedom where linear/angular velocity will be always constant (equal to zero, or to an user-defined value), regardless of applied force/torque. String that may contain 'xyzXYZ' (translations and rotations).

**densityScaling**(*=-1*)
*(auto-updated)* see *GlobalStiffnessTimeStepper::targetDt*.

**dict**(*(Serializable)arg1*) → dict :
Return dictionary of attributes.

**dispHierarchy**(*(State)arg1*[, *(bool)names=True*]) → list :
> Return list of dispatch classes (from down upwards), starting with the class instance itself, top-level indexable at last. If names is true (default), return class names rather than numerical indices.

**dispIndex**
> Return class index of this instance.

**displ**(*(State)arg1*) → Vector3 :
> Displacement from *reference position* (*pos* - *refPos*)

**inertia**(*=Vector3r::Zero()*)
> Inertia of associated body, in local coordinate system.

**isDamped**(*=true*)
> Damping in *NewtonIntegrator* can be deactivated for individual particles by setting this variable to FALSE. E.g. damping is inappropriate for particles in free flight under gravity but it might still be applicable to other particles in the same simulation.

**mass**(*=0*)
> Mass of this body

**numBrokenLinks**(*=0*)
> Number of broken links (e.g. number of wires connected to the body which are broken). [-]

**ori**
> Current orientation.

**pos**
> Current position.

**refOri**(*=Quaternionr::Identity()*)
> Reference orientation

**refPos**(*=Vector3r::Zero()*)
> Reference position

**rot**(*(State)arg1*) → Vector3 :
> Rotation from *reference orientation* (as rotation vector)

**se3**(*=Se3r(Vector3r::Zero(), Quaternionr::Identity())*)
> Position and orientation as one object.

**updateAttrs**(*(Serializable)arg1, (dict)arg2*) → None :
> Update object attributes from given dictionary

**vel**(*=Vector3r::Zero()*)
> Current linear velocity.

## Material

**class yade.wrapper.Material**(*inherits Serializable*)
> Material properties of a *body*.

**density**(*=1000*)
> Density of the material [kg/m$^3$]

**dict**(*(Serializable)arg1*) → dict :
> Return dictionary of attributes.

**dispHierarchy**(*(Material)arg1*[, *(bool)names=True*]) → list :
> Return list of dispatch classes (from down upwards), starting with the class instance itself, top-level indexable at last. If names is true (default), return class names rather than numerical indices.

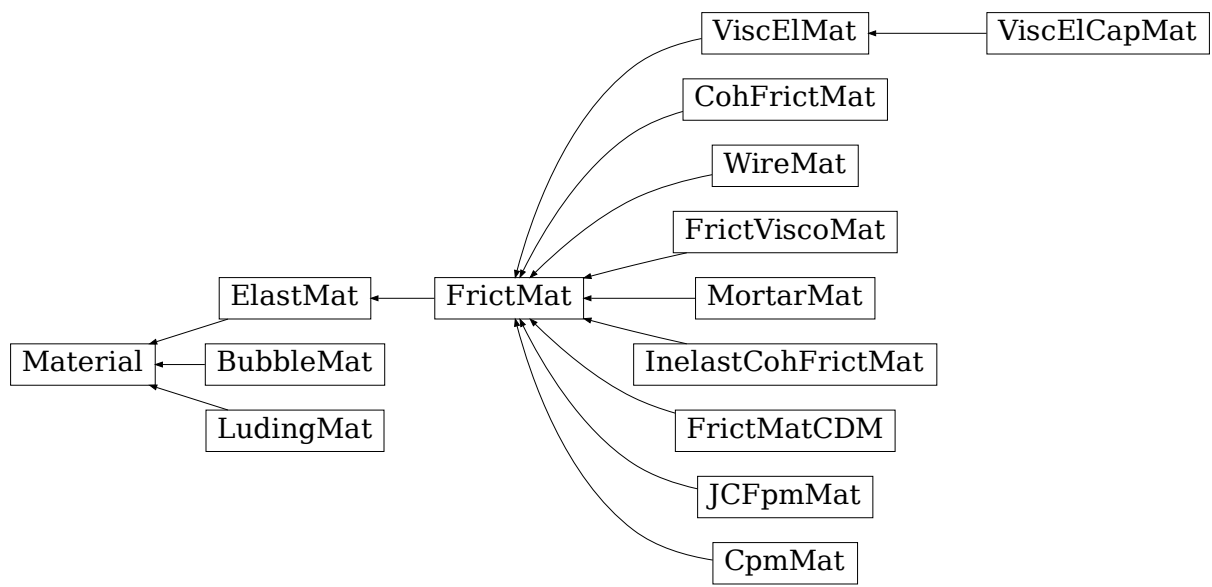

Fig. 21: Inheritance graph of Material. See also: *BubbleMat*, *CohFrictMat*, *CpmMat*, *ElastMat*, *Frict-Mat*, *FrictMatCDM*, *FrictViscoMat*, *InelastCohFrictMat*, *JCFpmMat*, *LudingMat*, *MortarMat*, *ViscEl-CapMat*, *ViscElMat*, *WireMat*.

**dispIndex**
>   Return class index of this instance.

**id**(*=-1, not shared*)
>   Numeric id of this material; is non-negative only if this Material is shared (i.e. in O.materials),
>   -1 otherwise. This value is set automatically when the material is inserted to the simulation
>   via *O.materials.append*. (This id was necessary since before boost::serialization was used,
>   shared pointers were not tracked properly; it might disappear in the future)

**label**(*=uninitalized*)
>   Textual identifier for this material; can be used for shared materials lookup in *MaterialCon-tainer*.

**newAssocState**(*(Material)arg1*) → State :
>   Return new *State* instance, which is associated with this *Material*. Some materials have
>   special requirement on *Body::state* type and calling this function when the body is created
>   will ensure that they match. (This is done automatically if you use utils.sphere, … functions
>   from python).

**updateAttrs**(*(Serializable)arg1, (dict)arg2*) → None :
>   Update object attributes from given dictionary

**class yade.wrapper.BubbleMat**(*inherits Material → Serializable*)
>   material for bubble interactions, for use with other Bubble classes

**density**(*=1000*)
>   Density of the material [kg/m³]

**dict**(*(Serializable)arg1*) → dict :
>   Return dictionary of attributes.

**dispHierarchy**(*(Material)arg1[, (bool)names=True]*) → list :
>   Return list of dispatch classes (from down upwards), starting with the class instance itself,
>   top-level indexable at last. If names is true (default), return class names rather than numerical
>   indices.

**dispIndex**
>   Return class index of this instance.

**id**(*=-1, not shared*)
    Numeric id of this material; is non-negative only if this Material is shared (i.e. in O.materials), -1 otherwise. This value is set automatically when the material is inserted to the simulation via *O.materials.append*. (This id was necessary since before boost::serialization was used, shared pointers were not tracked properly; it might disappear in the future)

**label**(*=uninitalized*)
    Textual identifier for this material; can be used for shared materials lookup in *MaterialContainer*.

**newAssocState**(*(Material)arg1*) → State :
    Return new *State* instance, which is associated with this *Material*. Some materials have special requirement on *Body::state* type and calling this function when the body is created will ensure that they match. (This is done automatically if you use utils.sphere, … functions from python).

**surfaceTension**(*=0.07197*)
    The surface tension in the fluid surrounding the bubbles. The default value is that of water at 25 degrees Celcius.

**updateAttrs**(*(Serializable)arg1, (dict)arg2*) → None :
    Update object attributes from given dictionary

**class yade.wrapper.CohFrictMat**(*inherits FrictMat → ElastMat → Material → Serializable*)
    Material description extending *FrictMat* with cohesive properties and rotational stiffness. For use e.g. with *Law2_ScGeom6D_CohFrictPhys_CohesionMoment*.

**alphaKr**(*=2.0*)
    Dimensionless rolling stiffness.

**alphaKtw**(*=2.0*)
    Dimensionless twist stiffness.

**density**(*=1000*)
    Density of the material [kg/m$^3$]

**dict**(*(Serializable)arg1*) → dict :
    Return dictionary of attributes.

**dispHierarchy**(*(Material)arg1[, (bool)names=True]*) → list :
    Return list of dispatch classes (from down upwards), starting with the class instance itself, top-level indexable at last. If names is true (default), return class names rather than numerical indices.

**dispIndex**
    Return class index of this instance.

**etaRoll**(*=-1.*)
    Dimensionless rolling (aka 'bending') strength. If negative, rolling moment will be elastic.

**etaTwist**(*=-1.*)
    Dimensionless twisting strength. If negative, twist moment will be elastic.

**fragile**(*=true*)
    do cohesion disappear when contact strength is exceeded

**frictionAngle**(*=.5*)
    Contact friction angle (in radians). Hint : use 'radians(degreesValue)' in python scripts.

**id**(*=-1, not shared*)
    Numeric id of this material; is non-negative only if this Material is shared (i.e. in O.materials), -1 otherwise. This value is set automatically when the material is inserted to the simulation via *O.materials.append*. (This id was necessary since before boost::serialization was used, shared pointers were not tracked properly; it might disappear in the future)

**isCohesive**(*=true*)

> Whether this body can form possibly cohesive interactions (if true and depending on other parameters such as *Ip2_CohFrictMat_CohFrictMat_CohFrictPhys.setCohesionNow*).

**label**(*=uninitialized*)

> Textual identifier for this material; can be used for shared materials lookup in *MaterialContainer*.

**momentRotationLaw**(*=false*)

> Use bending/twisting moment at contact. The contact may have moments only if both bodies have this flag true. See *Law2_ScGeom6D_CohFrictPhys_CohesionMoment.always_use_-moment_law* for details.

**newAssocState**(*(Material)arg1*) → State :

> Return new *State* instance, which is associated with this *Material*. Some materials have special requirement on *Body::state* type and calling this function when the body is created will ensure that they match. (This is done automatically if you use utils.sphere, ... functions from python).

**normalCohesion**(*=-1*)

> Tensile strength, homogeneous to a pressure. If negative the normal force is purely elastic.

**poisson**(*=.25*)

> Poisson's ratio or the ratio between shear and normal stiffness [-]. It has different meanings depending on the Ip functor.

**shearCohesion**(*=-1*)

> Shear strength, homogeneous to a pressure. If negative the shear force is purely elastic.

**updateAttrs**(*(Serializable)arg1, (dict)arg2*) → None :

> Update object attributes from given dictionary

**young**(*=1e9*)

> elastic modulus [Pa]. It has different meanings depending on the Ip functor.

**class yade.wrapper.CpmMat**(*inherits FrictMat → ElastMat → Material → Serializable*)

> Concrete material, for use with other Cpm classes.

---

**Note:** *Density* is initialized to 4800 kgm $^3$ automatically, which gives approximate 2800 kgm $^3$ on 0.5 density packing.

---

Concrete Particle Model (CPM)

*CpmMat* is particle material, *Ip2_CpmMat_CpmMat_CpmPhys* averages two particles' materials, creating *CpmPhys*, which is then used in interaction resultion by *Law2_ScGeom_CpmPhys_Cpm*. *CpmState* is associated to *CpmMat* and keeps state defined on particles rather than interactions (such as number of completely damaged interactions).

The model is contained in externally defined macro CPM_MATERIAL_MODEL, which features damage in tension, plasticity in shear and compression and rate-dependence. For commercial reasons, rate-dependence and compression-plasticity is not present in reduced version of the model, used when CPM_MATERIAL_MODEL is not defined. The full model will be described in detail in my (Václav Šmilauer) thesis along with calibration procedures (rigidity, poisson's ratio, compressive/tensile strength ratio, fracture energy, behavior under confinement, rate-dependent behavior).

Even the public model is useful enough to run simulation on concrete samples, such as uniaxial tension-compression test.

**damLaw**(*=1*)

> Law for damage evolution in uniaxial tension. 0 for linear stress-strain softening branch, 1 (default) for exponential damage evolution law

---

**density**(*=1000*)
:   Density of the material [kg/m³]

**dict**(*(Serializable)arg1*) → dict :
:   Return dictionary of attributes.

**dispHierarchy**(*(Material)arg1*[, *(bool)names=True*]) → list :
:   Return list of dispatch classes (from down upwards), starting with the class instance itself, top-level indexable at last. If names is true (default), return class names rather than numerical indices.

**dispIndex**
:   Return class index of this instance.

**dmgRateExp**(*=0*)
:   Exponent for normal viscosity function. [-]

**dmgTau**(*=-1, deactivated if negative*)
:   Characteristic time for normal viscosity. [s]

**epsCrackOnset**(*=NaN*)
:   Limit elastic strain [-]

**equivStrainShearContrib**(*=0*)
:   Coefficient of shear contribution to equivalent strain

**frictionAngle**(*=.5*)
:   Contact friction angle (in radians). Hint : use 'radians(degreesValue)' in python scripts.

**id**(*=-1, not shared*)
:   Numeric id of this material; is non-negative only if this Material is shared (i.e. in O.materials), -1 otherwise. This value is set automatically when the material is inserted to the simulation via *O.materials.append*. (This id was necessary since before boost::serialization was used, shared pointers were not tracked properly; it might disappear in the future)

**isoPrestress**(*=0*)
:   Isotropic prestress of the whole specimen. [Pa]

**label**(*=uninitalized*)
:   Textual identifier for this material; can be used for shared materials lookup in *MaterialContainer*.

**neverDamage**(*=false*)
:   If true, no damage will occur (for testing only).

**newAssocState**(*(Material)arg1*) → State :
:   Return new *State* instance, which is associated with this *Material*. Some materials have special requirement on *Body::state* type and calling this function when the body is created will ensure that they match. (This is done automatically if you use utils.sphere, ... functions from python).

**plRateExp**(*=0*)
:   Exponent for visco-plasticity function. [-]

**plTau**(*=-1, deactivated if negative*)
:   Characteristic time for visco-plasticity. [s]

**poisson**(*=.25*)
:   Poisson's ratio or the ratio between shear and normal stiffness [-]. It has different meanings depending on the Ip functor.

**relDuctility**(*=NaN*)
:   relative ductility of bonds in normal direction

**sigmaT**(*=NaN*)
:   Initial cohesion [Pa]

**updateAttrs**(*(Serializable)arg1, (dict)arg2*) → None :
> Update object attributes from given dictionary

**young**(*=1e9*)
> elastic modulus [Pa]. It has different meanings depending on the Ip functor.

**class yade.wrapper.ElastMat**(*inherits Material → Serializable*)
> Purely elastic material. The material parameters may have different meanings depending on the *IPhysFunctor* used : true Young and Poisson in *Ip2_FrictMat_FrictMat_MindlinPhys*, or contact stiffnesses in *Ip2_FrictMat_FrictMat_FrictPhys*.

**density**(*=1000*)
> Density of the material [kg/m³]

**dict**(*(Serializable)arg1*) → dict :
> Return dictionary of attributes.

**dispHierarchy**(*(Material)arg1*[, *(bool)names=True*]) → list :
> Return list of dispatch classes (from down upwards), starting with the class instance itself, top-level indexable at last. If names is true (default), return class names rather than numerical indices.

**dispIndex**
> Return class index of this instance.

**id**(*=-1, not shared*)
> Numeric id of this material; is non-negative only if this Material is shared (i.e. in O.materials), -1 otherwise. This value is set automatically when the material is inserted to the simulation via *O.materials.append*. (This id was necessary since before boost::serialization was used, shared pointers were not tracked properly; it might disappear in the future)

**label**(*=uninitialized*)
> Textual identifier for this material; can be used for shared materials lookup in *MaterialContainer*.

**newAssocState**(*(Material)arg1*) → State :
> Return new *State* instance, which is associated with this *Material*. Some materials have special requirement on *Body::state* type and calling this function when the body is created will ensure that they match. (This is done automatically if you use utils.sphere, ... functions from python).

**poisson**(*=.25*)
> Poisson's ratio or the ratio between shear and normal stiffness [-]. It has different meanings depending on the Ip functor.

**updateAttrs**(*(Serializable)arg1, (dict)arg2*) → None :
> Update object attributes from given dictionary

**young**(*=1e9*)
> elastic modulus [Pa]. It has different meanings depending on the Ip functor.

**class yade.wrapper.FrictMat**(*inherits ElastMat → Material → Serializable*)
> Elastic material with contact friction. See also *ElastMat*.

**density**(*=1000*)
> Density of the material [kg/m³]

**dict**(*(Serializable)arg1*) → dict :
> Return dictionary of attributes.

**dispHierarchy**(*(Material)arg1*[, *(bool)names=True*]) → list :
> Return list of dispatch classes (from down upwards), starting with the class instance itself, top-level indexable at last. If names is true (default), return class names rather than numerical indices.

**dispIndex**
    Return class index of this instance.

**frictionAngle**(*=.5*)
    Contact friction angle (in radians). Hint : use 'radians(degreesValue)' in python scripts.

**id**(*=-1, not shared*)
    Numeric id of this material; is non-negative only if this Material is shared (i.e. in O.materials), -1 otherwise. This value is set automatically when the material is inserted to the simulation via *O.materials.append*. (This id was necessary since before boost::serialization was used, shared pointers were not tracked properly; it might disappear in the future)

**label**(*=uninitialized*)
    Textual identifier for this material; can be used for shared materials lookup in *MaterialContainer*.

**newAssocState**(*(Material)arg1*) → State :
    Return new *State* instance, which is associated with this *Material*. Some materials have special requirement on *Body::state* type and calling this function when the body is created will ensure that they match. (This is done automatically if you use utils.sphere, … functions from python).

**poisson**(*=.25*)
    Poisson's ratio or the ratio between shear and normal stiffness [-]. It has different meanings depending on the Ip functor.

**updateAttrs**(*(Serializable)arg1, (dict)arg2*) → None :
    Update object attributes from given dictionary

**young**(*=1e9*)
    elastic modulus [Pa]. It has different meanings depending on the Ip functor.

**class yade.wrapper.FrictMatCDM**(*inherits FrictMat → ElastMat → Material → Serializable*)
Material to be used for extended Hertz-Mindlin contact law. Normal direction: parameters for Conical Damage Model (Harkness et al. 2016, Suhr & Six 2017). Tangential direction: parameters for stress dependent interparticle friction coefficient (Suhr & Six 2016). Both models can be switched on/off separately.

**alpha**(*=1e-6*)
    [rad] angle of conical asperities, alpha in (0, pi/2)

**c1**(*=0.0*)
    [-] parameter of pressure dependent friction model c1, choose 0 for constant interparticle friction coefficient

**c2**(*=0.0*)
    [-] parameter of pressure dependent friction model c2, choose 0 for constant interparticle friction coefficient

**density**(*=1000*)
    Density of the material [kg/m$^3$]

**dict**(*(Serializable)arg1*) → dict :
    Return dictionary of attributes.

**dispHierarchy**(*(Material)arg1*[, *(bool)names=True*]) → list :
    Return list of dispatch classes (from down upwards), starting with the class instance itself, top-level indexable at last. If names is true (default), return class names rather than numerical indices.

**dispIndex**
    Return class index of this instance.

**frictionAngle**(*=.5*)
    Contact friction angle (in radians). Hint : use 'radians(degreesValue)' in python scripts.

**id**(*=-1, not shared*)

   Numeric id of this material; is non-negative only if this Material is shared (i.e. in O.materials), -1 otherwise. This value is set automatically when the material is inserted to the simulation via *O.materials.append*. (This id was necessary since before boost::serialization was used, shared pointers were not tracked properly; it might disappear in the future)

**label**(*=uninitalized*)

   Textual identifier for this material; can be used for shared materials lookup in *MaterialContainer*.

**newAssocState**(*(Material)arg1*) → State :

   Return new *State* instance, which is associated with this *Material*. Some materials have special requirement on *Body::state* type and calling this function when the body is created will ensure that they match. (This is done automatically if you use utils.sphere, … functions from python).

**poisson**(*=.25*)

   Poisson's ratio or the ratio between shear and normal stiffness [-]. It has different meanings depending on the Ip functor.

**sigmaMax**(*=1e99*)

   >0 [Pa] max compressive strength of material, choose 1e99 to switch off conical damage model

**updateAttrs**(*(Serializable)arg1, (dict)arg2*) → None :

   Update object attributes from given dictionary

**young**(*=1e9*)

   elastic modulus [Pa]. It has different meanings depending on the Ip functor.

**class yade.wrapper.FrictViscoMat**(*inherits FrictMat → ElastMat → Material → Serializable*)

   Material for use with the FrictViscoPM classes

**betan**(*=0.*)

   Fraction of the viscous damping coefficient in normal direction equal to $\frac{c_n}{C_{n,crit}}$.

**density**(*=1000*)

   Density of the material [kg/m$^3$]

**dict**(*(Serializable)arg1*) → dict :

   Return dictionary of attributes.

**dispHierarchy**(*(Material)arg1[, (bool)names=True]*) → list :

   Return list of dispatch classes (from down upwards), starting with the class instance itself, top-level indexable at last. If names is true (default), return class names rather than numerical indices.

**dispIndex**

   Return class index of this instance.

**frictionAngle**(*=.5*)

   Contact friction angle (in radians). Hint : use 'radians(degreesValue)' in python scripts.

**id**(*=-1, not shared*)

   Numeric id of this material; is non-negative only if this Material is shared (i.e. in O.materials), -1 otherwise. This value is set automatically when the material is inserted to the simulation via *O.materials.append*. (This id was necessary since before boost::serialization was used, shared pointers were not tracked properly; it might disappear in the future)

**label**(*=uninitalized*)

   Textual identifier for this material; can be used for shared materials lookup in *MaterialContainer*.

**newAssocState**(*(Material)arg1*) → State :

   Return new *State* instance, which is associated with this *Material*. Some materials have special requirement on *Body::state* type and calling this function when the body is created will ensure that they match. (This is done automatically if you use utils.sphere, … functions from python).

**poisson**(*=.25*)

Poisson's ratio or the ratio between shear and normal stiffness [-]. It has different meanings depending on the Ip functor.

**updateAttrs**(*(Serializable)arg1, (dict)arg2*) → None :

Update object attributes from given dictionary

**young**(*=1e9*)

elastic modulus [Pa]. It has different meanings depending on the Ip functor.

**class yade.wrapper.InelastCohFrictMat**(*inherits* *FrictMat → ElastMat → Material → Serializable*)

**alphaKr**(*=2.0*)

Dimensionless coefficient used for the rolling stiffness.

**alphaKtw**(*=2.0*)

Dimensionless coefficient used for the twist stiffness.

**compressionModulus**(*=0.0*)

Compresion elasticity modulus

**creepBending**(*=0.0*)

Bending creeping coefficient. Usual values between 0 and 1.

**creepTension**(*=0.0*)

Tension/compression creeping coefficient. Usual values between 0 and 1.

**creepTwist**(*=0.0*)

Twist creeping coefficient. Usual values between 0 and 1.

**density**(*=1000*)

Density of the material [kg/m$^3$]

**dict**(*(Serializable)arg1*) → dict :

Return dictionary of attributes.

**dispHierarchy**(*(Material)arg1* [*, (bool)names=True* ]) → list :

Return list of dispatch classes (from down upwards), starting with the class instance itself, top-level indexable at last. If names is true (default), return class names rather than numerical indices.

**dispIndex**

Return class index of this instance.

**epsilonMaxCompression**(*=0.0*)

Maximal plastic strain compression

**epsilonMaxTension**(*=0.0*)

Maximal plastic strain tension

**etaMaxBending**(*=0.0*)

Maximal plastic bending strain

**etaMaxTwist**(*=0.0*)

Maximal plastic twist strain

**frictionAngle**(*=.5*)

Contact friction angle (in radians). Hint : use 'radians(degreesValue)' in python scripts.

**id**(*=-1, not shared*)

Numeric id of this material; is non-negative only if this Material is shared (i.e. in O.materials), -1 otherwise. This value is set automatically when the material is inserted to the simulation via *O.materials.append*. (This id was necessary since before boost::serialization was used, shared pointers were not tracked properly; it might disappear in the future)

**label**(*=uninitalized*)
  Textual identifier for this material; can be used for shared materials lookup in *MaterialContainer*.

**newAssocState**(*(Material)arg1*) → State :
  Return new *State* instance, which is associated with this *Material*. Some materials have special requirement on *Body::state* type and calling this function when the body is created will ensure that they match. (This is done automatically if you use utils.sphere, … functions from python).

**nuBending**(*=0.0*)
  Bending elastic stress limit

**nuTwist**(*=0.0*)
  Twist elastic stress limit

**poisson**(*=.25*)
  Poisson's ratio or the ratio between shear and normal stiffness [-]. It has different meanings depending on the Ip functor.

**shearCohesion**(*=0.0*)
  Shear elastic stress limit

**shearModulus**(*=0.0*)
  shear elasticity modulus

**sigmaCompression**(*=0.0*)
  Compression elastic stress limit

**sigmaTension**(*=0.0*)
  Tension elastic stress limit

**tensionModulus**(*=0.0*)
  Tension elasticity modulus

**unloadBending**(*=0.0*)
  Bending plastic unload coefficient. Usual values between 0 and +infinity.

**unloadTension**(*=0.0*)
  Tension/compression plastic unload coefficient. Usual values between 0 and +infinity.

**unloadTwist**(*=0.0*)
  Twist plastic unload coefficient. Usual values between 0 and +infinity.

**updateAttrs**(*(Serializable)arg1, (dict)arg2*) → None :
  Update object attributes from given dictionary

**young**(*=1e9*)
  elastic modulus [Pa]. It has different meanings depending on the Ip functor.

**class yade.wrapper.JCFpmMat**(*inherits FrictMat → ElastMat → Material → Serializable*)
  Possibly jointed, cohesive frictional material, for use with other JCFpm classes

**cohesion**(*=0.*)
  Defines the maximum admissible tangential force in shear, for Fn=0, in the matrix (*FsMax* = cohesion * *crossSection*). [Pa]

**density**(*=1000*)
  Density of the material [kg/m³]

**dict**(*(Serializable)arg1*) → dict :
  Return dictionary of attributes.

**dispHierarchy**(*(Material)arg1*[, *(bool)names=True*]) → list :
  Return list of dispatch classes (from down upwards), starting with the class instance itself, top-level indexable at last. If names is true (default), return class names rather than numerical indices.

**dispIndex**
    Return class index of this instance.

**frictionAngle**(*=.5*)
    Contact friction angle (in radians). Hint : use 'radians(degreesValue)' in python scripts.

**id**(*=-1, not shared*)
    Numeric id of this material; is non-negative only if this Material is shared (i.e. in O.materials), -1 otherwise. This value is set automatically when the material is inserted to the simulation via *O.materials.append*. (This id was necessary since before boost::serialization was used, shared pointers were not tracked properly; it might disappear in the future)

**jointCohesion**(*=0.*)
    Defines the *maximum admissible tangential force in shear*, for Fn=0, on the joint surface. [Pa]

**jointDilationAngle**(*=0*)
    Defines the dilatancy of the joint surface (only valid for *smooth contact logic*). [rad]

**jointFrictionAngle**(*=-1*)
    Defines Coulomb friction on the joint surface. [rad]

**jointNormalStiffness**(*=0.*)
    Defines the normal stiffness on the joint surface. [Pa/m]

**jointShearStiffness**(*=0.*)
    Defines the shear stiffness on the joint surface. [Pa/m]

**jointTensileStrength**(*=0.*)
    Defines the *maximum admissible normal force in traction* on the joint surface. [Pa]

**label**(*=uninitalized*)
    Textual identifier for this material; can be used for shared materials lookup in *MaterialContainer*.

**newAssocState**(*(Material)arg1*) → State :
    Return new *State* instance, which is associated with this *Material*. Some materials have special requirement on *Body::state* type and calling this function when the body is created will ensure that they match. (This is done automatically if you use utils.sphere, … functions from python).

**poisson**(*=.25*)
    Poisson's ratio or the ratio between shear and normal stiffness [-]. It has different meanings depending on the Ip functor.

**residualFrictionAngle**(*=-1.*)
    Defines the residual friction angle (when contacts are not cohesive). residualFrictionAngle=frictionAngle if not specified. [rad]

**tensileStrength**(*=0.*)
    Defines the maximum admissible normal force in traction in the matrix (*FnMax = tensileStrength * crossSection*). [Pa]

**type**(*=0*)
    If particles of two different types interact, it will be with friction only (no cohesion).[-]

**updateAttrs**(*(Serializable)arg1, (dict)arg2*) → None :
    Update object attributes from given dictionary

**young**(*=1e9*)
    elastic modulus [Pa]. It has different meanings depending on the Ip functor.

**class yade.wrapper.LudingMat**(*inherits Material → Serializable*)
    Material for simple Luding's model of contact [Luding2008] ,[Singh2013]__ .

    **G0**(*=NaN*)
        Viscous damping

**PhiF**(*=NaN*)
> Dimensionless plasticity depth

**density**(*=1000*)
> Density of the material [kg/m³]

**dict**(*(Serializable)arg1*) → dict :
> Return dictionary of attributes.

**dispHierarchy**(*(Material)arg1*[, *(bool)names=True*]) → list :
> Return list of dispatch classes (from down upwards), starting with the class instance itself, top-level indexable at last. If names is true (default), return class names rather than numerical indices.

**dispIndex**
> Return class index of this instance.

**frictionAngle**(*=NaN*)
> Friction angle [rad]

**id**(*=-1, not shared*)
> Numeric id of this material; is non-negative only if this Material is shared (i.e. in O.materials), -1 otherwise. This value is set automatically when the material is inserted to the simulation via *O.materials.append*. (This id was necessary since before boost::serialization was used, shared pointers were not tracked properly; it might disappear in the future)

**k1**(*=NaN*)
> Slope of loading plastic branch

**kc**(*=NaN*)
> Slope of irreversible, tensile adhesive branch

**kp**(*=NaN*)
> Slope of unloading and reloading limit elastic branch

**ks**(*=NaN*)
> Shear stiffness

**label**(*=uninitalized*)
> Textual identifier for this material; can be used for shared materials lookup in *MaterialContainer*.

**newAssocState**(*(Material)arg1*) → State :
> Return new *State* instance, which is associated with this *Material*. Some materials have special requirement on *Body::state* type and calling this function when the body is created will ensure that they match. (This is done automatically if you use utils.sphere, … functions from python).

**updateAttrs**(*(Serializable)arg1, (dict)arg2*) → None :
> Update object attributes from given dictionary

**class yade.wrapper.MortarMat**(*inherits FrictMat → ElastMat → Material → Serializable*)
Material for mortar interface, used in Ip2_MortarMat_MortarMat_MortarPhys and Law2_ScGeom_MortarPhys_Lourenco. Default values according to

**cohesion**(*=1e6*)
> cohesion [Pa]

**compressiveStrength**(*=10e6*)
> compressiveStrength [Pa]

**density**(*=1000*)
> Density of the material [kg/m³]

**dict**(*(Serializable)arg1*) → dict :
> Return dictionary of attributes.

**dispHierarchy**(*(Material)arg1*[, *(bool)names=True*]) → list :
    Return list of dispatch classes (from down upwards), starting with the class instance itself, top-level indexable at last. If names is true (default), return class names rather than numerical indices.

**dispIndex**
    Return class index of this instance.

**ellAspect**(*=3*)
    aspect ratio of elliptical 'cap'. Value >1 means the ellipse is longer along normal stress axis.

**frictionAngle**(*=.25*)
    Friction angle

**id**(*=-1, not shared*)
    Numeric id of this material; is non-negative only if this Material is shared (i.e. in O.materials), -1 otherwise. This value is set automatically when the material is inserted to the simulation via *O.materials.append*. (This id was necessary since before boost::serialization was used, shared pointers were not tracked properly; it might disappear in the future)

**label**(*=uninitalized*)
    Textual identifier for this material; can be used for shared materials lookup in *MaterialContainer*.

**neverDamage**(*=false*)
    If true, interactions remain elastic regardless stresses

**newAssocState**(*(Material)arg1*) → State :
    Return new *State* instance, which is associated with this *Material*. Some materials have special requirement on *Body::state* type and calling this function when the body is created will ensure that they match. (This is done automatically if you use utils.sphere, … functions from python).

**poisson**(*=1*)
    Shear to normal modulus ratio

**tensileStrength**(*=1e6*)
    tensileStrength [Pa]

**updateAttrs**(*(Serializable)arg1, (dict)arg2*) → None :
    Update object attributes from given dictionary

**young**(*=1e9*)
    Normal elastic modulus [Pa]

**class yade.wrapper.ViscElCapMat**(*inherits ViscElMat → FrictMat → ElastMat → Material → Serializable*)
    Material for extended viscoelastic model of contact with capillary parameters.

**Capillar**(*=false*)
    True, if capillar forces need to be added.

**CapillarType**(*=""*)
    Different types of capillar interaction: Willett_numeric, Willett_analytic [Willett2000] , Weigert [Weigert1999] , Rabinovich [Rabinov2005] , Lambert (simplified, corrected Rabinovich model) [Lambert2008]

**Vb**(*=0.0*)
    Liquid bridge volume [m^3]

**cn**(*=NaN*)
    Normal viscous constant. Attention, this parameter cannot be set if tc, en or es is defined!

**cs**(*=NaN*)
    Shear viscous constant. Attention, this parameter cannot be set if tc, en or es is defined!

**dcap**(*=0.0*)

    Damping coefficient for the capillary phase [-]

**density**(*=1000*)

    Density of the material [kg/m$^3$]

**dict**(*(Serializable)arg1*) → dict :

    Return dictionary of attributes.

**dispHierarchy**(*(Material)arg1*[, *(bool)names=True*]) → list :

    Return list of dispatch classes (from down upwards), starting with the class instance itself, top-level indexable at last. If names is true (default), return class names rather than numerical indices.

**dispIndex**

    Return class index of this instance.

**en**(*=NaN*)

    Restitution coefficient in normal direction

**et**(*=NaN*)

    Restitution coefficient in tangential direction

**frictionAngle**(*=.5*)

    Contact friction angle (in radians). Hint : use 'radians(degreesValue)' in python scripts.

**gamma**(*=0.0*)

    Surface tension [N/m]

**id**(*=-1, not shared*)

    Numeric id of this material; is non-negative only if this Material is shared (i.e. in O.materials), -1 otherwise. This value is set automatically when the material is inserted to the simulation via *O.materials.append*. (This id was necessary since before boost::serialization was used, shared pointers were not tracked properly; it might disappear in the future)

**kn**(*=NaN*)

    Normal elastic stiffness. Attention, this parameter cannot be set if tc, en or es is defined!

**ks**(*=NaN*)

    Shear elastic stiffness. Attention, this parameter cannot be set if tc, en or es is defined!

**label**(*=uninitalized*)

    Textual identifier for this material; can be used for shared materials lookup in *MaterialContainer*.

**lubrication**(*=false*)

    option to apply lubrication forces when material is defined from young, poisson and en (restitution coefficient).

**mR**(*=0.0*)

    Rolling resistance, see [Zhou1999536].

**mRtype**(*=1*)

    Rolling resistance type, see [Zhou1999536]. mRtype=1 - equation (3) in [Zhou1999536]; mRtype=2 - equation (4) in [Zhou1999536].

**newAssocState**(*(Material)arg1*) → State :

    Return new *State* instance, which is associated with this *Material*. Some materials have special requirement on *Body::state* type and calling this function when the body is created will ensure that they match. (This is done automatically if you use utils.sphere, ... functions from python).

**poisson**(*=.25*)

    Poisson's ratio or the ratio between shear and normal stiffness [-]. It has different meanings depending on the Ip functor.

**roughnessScale**(*=1e-3*)
   if lubrication is activated, roughness scale considered for the particles to evaluate the effective restitution coefficient.

**tc**(*=NaN*)
   Contact time

**theta**(*=0.0*)
   Contact angle [°]

**updateAttrs**(*(Serializable)arg1, (dict)arg2*) → None :
   Update object attributes from given dictionary

**viscoDyn**(*=1e-3*)
   if lubrication is activated, surrounding fluid dynamic viscosity considered to evaluate the effective restitution coefficient as a function of the local Stokes number of the collision.

**young**(*=1e9*)
   elastic modulus [Pa]. It has different meanings depending on the Ip functor.

**class yade.wrapper.ViscElMat**(*inherits FrictMat → ElastMat → Material → Serializable*)
   Material for simple viscoelastic model of contact from analytical solution of a pair spheres interaction problem [Pournin2001] .

**cn**(*=NaN*)
   Normal viscous constant. Attention, this parameter cannot be set if tc, en or es is defined!

**cs**(*=NaN*)
   Shear viscous constant. Attention, this parameter cannot be set if tc, en or es is defined!

**density**(*=1000*)
   Density of the material [kg/m³]

**dict**(*(Serializable)arg1*) → dict :
   Return dictionary of attributes.

**dispHierarchy**(*(Material)arg1[, (bool)names=True]*) → list :
   Return list of dispatch classes (from down upwards), starting with the class instance itself, top-level indexable at last. If names is true (default), return class names rather than numerical indices.

**dispIndex**
   Return class index of this instance.

**en**(*=NaN*)
   Restitution coefficient in normal direction

**et**(*=NaN*)
   Restitution coefficient in tangential direction

**frictionAngle**(*=.5*)
   Contact friction angle (in radians). Hint : use 'radians(degreesValue)' in python scripts.

**id**(*=-1, not shared*)
   Numeric id of this material; is non-negative only if this Material is shared (i.e. in O.materials), -1 otherwise. This value is set automatically when the material is inserted to the simulation via *O.materials.append*. (This id was necessary since before boost::serialization was used, shared pointers were not tracked properly; it might disappear in the future)

**kn**(*=NaN*)
   Normal elastic stiffness. Attention, this parameter cannot be set if tc, en or es is defined!

**ks**(*=NaN*)
   Shear elastic stiffness. Attention, this parameter cannot be set if tc, en or es is defined!

**label**(*=uninitalized*)
 Textual identifier for this material; can be used for shared materials lookup in *MaterialContainer*.

**lubrication**(*=false*)
 option to apply lubrication forces when material is defined from young, poisson and en (restitution coefficient).

**mR**(*=0.0*)
 Rolling resistance, see [Zhou1999536].

**mRtype**(*=1*)
 Rolling resistance type, see [Zhou1999536]. mRtype=1 - equation (3) in [Zhou1999536]; mRtype=2 - equation (4) in [Zhou1999536].

**newAssocState**(*(Material)arg1*) → State :
 Return new *State* instance, which is associated with this *Material*. Some materials have special requirement on *Body::state* type and calling this function when the body is created will ensure that they match. (This is done automatically if you use utils.sphere, … functions from python).

**poisson**(*=.25*)
 Poisson's ratio or the ratio between shear and normal stiffness [-]. It has different meanings depending on the Ip functor.

**roughnessScale**(*=1e-3*)
 if lubrication is activated, roughness scale considered for the particles to evaluate the effective restitution coefficient.

**tc**(*=NaN*)
 Contact time

**updateAttrs**(*(Serializable)arg1, (dict)arg2*) → None :
 Update object attributes from given dictionary

**viscoDyn**(*=1e-3*)
 if lubrication is activated, surrounding fluid dynamic viscosity considered to evaluate the effective restitution coefficient as a function of the local Stokes number of the collision.

**young**(*=1e9*)
 elastic modulus [Pa]. It has different meanings depending on the Ip functor.

**class yade.wrapper.WireMat**(*inherits FrictMat → ElastMat → Material → Serializable*)
Material for use with the Wire classes. In conjunction with the corresponding functors it can be used to model steel wire meshes [Thoeni2014], geotextiles [Cheng2016] and more.

**as**(*=0.*)
 Cross-section area of a single wire used to transform stress into force. [$m^2$]

**density**(*=1000*)
 Density of the material [kg/$m^3$]

**diameter**(*=0.0027*)
 Diameter of the single wire in [m] (the diameter is used to compute the cross-section area of the wire).

**dict**(*(Serializable)arg1*) → dict :
 Return dictionary of attributes.

**dispHierarchy**(*(Material)arg1*[, *(bool)names=True*]) → list :
 Return list of dispatch classes (from down upwards), starting with the class instance itself, top-level indexable at last. If names is true (default), return class names rather than numerical indices.

**dispIndex**
 Return class index of this instance.

**frictionAngle**(*=.5*)

    Contact friction angle (in radians). Hint : use 'radians(degreesValue)' in python scripts.

**id**(*=-1, not shared*)

    Numeric id of this material; is non-negative only if this Material is shared (i.e. in O.materials), -1 otherwise. This value is set automatically when the material is inserted to the simulation via *O.materials.append*. (This id was necessary since before boost::serialization was used, shared pointers were not tracked properly; it might disappear in the future)

**isDoubleTwist**(*=false*)

    Type of the mesh. If true two particles of the same material which body ids differ by one will be considered as double-twisted interaction.

**label**(*=uninitalized*)

    Textual identifier for this material; can be used for shared materials lookup in *MaterialContainer*.

**lambdaEps**(*=0.47*)

    Parameter between 0 and 1 to reduce strain at failure of a double-twisted wire (as used by [Bertrand2008]). [-]

**lambdaF**(*=1.0*)

    Parameter between 0 and 1 introduced by [Thoeni2013] which defines where the shifted force-displacement curve intersects with the new initial stiffness: $F^* = \lambda_F F_{elastic}$. [-]

**lambdak**(*=0.73*)

    Parameter between 0 and 1 to compute the elastic stiffness of a double-twisted wire (as used by [Bertrand2008]): $k^D = 2(\lambda_k k_h + (1 - \lambda_k)k^S)$. [-]

**lambdau**(*=0.2*)

    Parameter between 0 and 1 introduced by [Thoeni2013] which defines the maximum shift of the force-displacement curve in order to take an additional initial elongation (e.g. wire distortion/imperfections, slipping, system flexibility) into account: $\Delta l^* = \lambda_u l_0 \text{rnd(seed)}$. [-]

**newAssocState**(*(Material)arg1*) → State :

    Return new *State* instance, which is associated with this *Material*. Some materials have special requirement on *Body::state* type and calling this function when the body is created will ensure that they match. (This is done automatically if you use utils.sphere, … functions from python).

**poisson**(*=.25*)

    Poisson's ratio or the ratio between shear and normal stiffness [-]. It has different meanings depending on the Ip functor.

**seed**(*=12345*)

    Integer used to initialize the random number generator for the calculation of the distortion. If the integer is equal to 0 a internal seed number based on the time is computed. [-]

**strainStressValues**(*=uninitalized*)

    Piecewise linear definition of the stress-strain curve by set of points (strain[-]>0,stress[Pa]>0) for one single wire. Tension only is considered and the point (0,0) is not needed! NOTE: Vector needs to be initialized!

**strainStressValuesDT**(*=uninitalized*)

    Piecewise linear definition of the stress-strain curve by set of points (strain[-]>0,stress[Pa]>0) for the double twist. Tension only is considered and the point (0,0) is not needed! If this value is given the calculation will be based on two different stress-strain curves without considering the parameter introduced by [Bertrand2008] (see [Thoeni2013]).

**type**

    Three different types are considered:

| 0 | Corresponds to Bertrand's approach (see [Bertrand2008]): only one stress-strain curve is used |
|---|---|
| 1 | New approach: two separate stress-strain curves can be used (see [Thoeni2013]) |
| 2 | New approach with stochastically distorted contact model: two separate stress-strain curves with changed initial stiffness and horizontal shift (shift is random if seed $\geq$ 0, for more details see [Thoeni2013]) |

By default the type is 0.

**updateAttrs**(*(Serializable)arg1, (dict)arg2*) → None :
    Update object attributes from given dictionary

**young**(*=1e9*)
    elastic modulus [Pa]. It has different meanings depending on the Ip functor.

## Bound

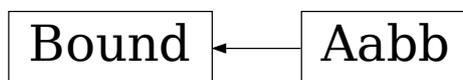

Fig. 22: Inheritance graph of Bound. See also: *Aabb*.

**class yade.wrapper.Bound**(*inherits Serializable*)
    Object bounding part of space taken by associated body; might be larger, used to optimalize collision detection

**color**(*=Vector3r(1, 1, 1)*)
    Color for rendering this object

**dict**(*(Serializable)arg1*) → dict :
    Return dictionary of attributes.

**dispHierarchy**(*(Bound)arg1*[, *(bool)names=True*]) → list :
    Return list of dispatch classes (from down upwards), starting with the class instance itself, top-level indexable at last. If names is true (default), return class names rather than numerical indices.

**dispIndex**
    Return class index of this instance.

**lastUpdateIter**(*=0*)
    record iteration of last reference position update *(auto-updated)*

**max**(*=Vector3r(NaN, NaN, NaN)*)
    Upper corner of box containing this bound (and the *Body* as well)

**min**(*=Vector3r(NaN, NaN, NaN)*)
    Lower corner of box containing this bound (and the *Body* as well)

**refPos**(*=Vector3r(NaN, NaN, NaN)*)
    Reference position, updated at current body position each time the bound dispatcher update bounds *(auto-updated)*

**sweepLength**(*=0*)
    The length used to increase the bounding boxe size, can be adjusted on the basis of previous displacement if *BoundDispatcher::targetInterv*>0. *(auto-updated)*

**updateAttrs**(*(Serializable)arg1, (dict)arg2*) → None :
    Update object attributes from given dictionary

**class** yade.wrapper.**Aabb**(*inherits Bound → Serializable*)

Axis-aligned bounding box, for use with *InsertionSortCollider*. (This class is quasi-redundant since min,max are already contained in *Bound* itself. That might change at some point, though.)

**color**(*=Vector3r(1, 1, 1)*)
Color for rendering this object

**dict**(*(Serializable)arg1*) → dict :
Return dictionary of attributes.

**dispHierarchy**(*(Bound)arg1*[, *(bool)names=True*]) → list :
Return list of dispatch classes (from down upwards), starting with the class instance itself, top-level indexable at last. If names is true (default), return class names rather than numerical indices.

**dispIndex**
Return class index of this instance.

**lastUpdateIter**(*=0*)
record iteration of last reference position update *(auto-updated)*

**max**(*=Vector3r(NaN, NaN, NaN)*)
Upper corner of box containing this bound (and the *Body* as well)

**min**(*=Vector3r(NaN, NaN, NaN)*)
Lower corner of box containing this bound (and the *Body* as well)

**refPos**(*=Vector3r(NaN, NaN, NaN)*)
Reference position, updated at current body position each time the bound dispatcher update bounds *(auto-updated)*

**sweepLength**(*=0*)
The length used to increase the bounding boxe size, can be adjusted on the basis of previous displacement if *BoundDispatcher::targetInterv*>0. *(auto-updated)*

**updateAttrs**(*(Serializable)arg1, (dict)arg2*) → None :
Update object attributes from given dictionary

### 2.3.2 Interactions

**Interaction**

**class** yade.wrapper.**Interaction**(*inherits Serializable*)

Interaction between pair of bodies.

**cellDist**
Distance of bodies in cell size units, if using periodic boundary conditions; id2 is shifted by this number of cells from its *State::pos* coordinates for this interaction to exist. Assigned by the collider.

> **Warning:** (internal) cellDist must survive Interaction::reset(), it is only initialized in ctor. Interaction that was cancelled by the constitutive law, was reset() and became only potential must have the period information if the geometric functor again makes it real. Good to know after few days of debugging that :-)

**dict**(*(Serializable)arg1*) → dict :
Return dictionary of attributes.

**geom**(*=uninitialized*)
Geometry part of the interaction.

**id1**(*=0*)
> *Id* of the first body in this interaction.

**id2**(*=0*)
> *Id* of the second body in this interaction.

**isActive**
> True if this interaction is active. Otherwise the forces from this interaction will not be taken into account. True by default.

**isReal**
> True if this interaction has both *geom* and *phys*; False otherwise.

**iterBorn**(*=-1*)
> Step number at which the interaction was added to simulation.

**iterMadeReal**(*=-1*)
> Step number at which the interaction was fully (in the sense of geom and phys) created. (Should be touched only by *IPhysDispatcher* and *InteractionLoop*, therefore they are made friends of Interaction

**phys**(*=uninitalized*)
> Physical (material) part of the interaction.

**updateAttrs**(*(Serializable)arg1, (dict)arg2*) → None :
> Update object attributes from given dictionary

### IGeom

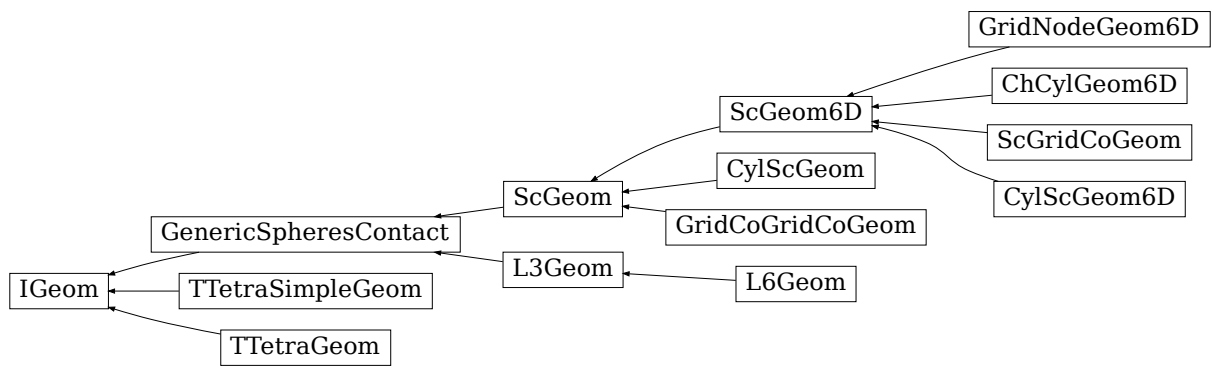

Fig. 23: Inheritance graph of IGeom. See also: *ChCylGeom6D*, *CylScGeom*, *CylScGeom6D*, *GenericSpheresContact*, *GridCoGridCoGeom*, *GridNodeGeom6D*, *L3Geom*, *L6Geom*, *ScGeom*, *ScGeom6D*, *ScGridCoGeom*, *TTetraGeom*, *TTetraSimpleGeom*.

**class yade.wrapper.IGeom**(*inherits Serializable*)
> Geometrical configuration of interaction

**dict**(*(Serializable)arg1*) → dict :
> Return dictionary of attributes.

**dispHierarchy**(*(IGeom)arg1*[, *(bool)names=True*]) → list :
> Return list of dispatch classes (from down upwards), starting with the class instance itself, top-level indexable at last. If names is true (default), return class names rather than numerical indices.

**dispIndex**
> Return class index of this instance.

**updateAttrs**(*(Serializable)arg1, (dict)arg2*) → None :
> Update object attributes from given dictionary **class yade.wrapper.ChCylGeom6D**(*inherits ScGeom6D → ScGeom → GenericSpheresContact → IGeom → Serializable*)

    Test

    **bending**(*=Vector3r::Zero()*)

        Bending at contact as a vector defining axis of rotation and angle (angle=norm).

    **contactPoint**(*=uninitialized*)

        some reference point for the interaction (usually in the middle). *(auto-computed)*

    **dict**(*(Serializable)arg1*) → dict :

        Return dictionary of attributes.

    **dispHierarchy**(*(IGeom)arg1*[, *(bool)names=True*]) → list :

        Return list of dispatch classes (from down upwards), starting with the class instance itself, top-level indexable at last. If names is true (default), return class names rather than numerical indices.

    **dispIndex**

        Return class index of this instance.

    **incidentVel**(*(ScGeom)arg1, (Interaction)i*[, *(bool)avoidGranularRatcheting=True*]) → Vector3 :

        Return incident velocity of the interaction (see also *Ig2_Sphere_Sphere_ScGeom.avoidGranularRatcheting* for explanation of the ratcheting argument).

    **initialOrientation1**(*=Quaternionr(1.0, 0.0, 0.0, 0.0)*)

        Orientation of body 1 one at initialisation time *(auto-updated)*

    **initialOrientation2**(*=Quaternionr(1.0, 0.0, 0.0, 0.0)*)

        Orientation of body 2 one at initialisation time *(auto-updated)*

    **normal**(*=uninitialized*)

        Unit vector oriented along the interaction, from particle #1, towards particle #2. *(auto-updated)*

    **penetrationDepth**(*=NaN*)

        Penetration distance of spheres (positive if overlapping)

    **refR1**(*=uninitialized*)

        Reference radius of particle #1. *(auto-computed)*

    **refR2**(*=uninitialized*)

        Reference radius of particle #2. *(auto-computed)*

    **relAngVel**(*(ScGeom)arg1, (Interaction)i*) → Vector3 :

        Return relative angular velocity of the interaction.

    **shearInc**(*=Vector3r::Zero()*)

        Shear displacement increment in the last step

    **twist**(*=0*)

        Elastic twist angle (around *normal axis*) of the contact.

    **twistCreep**(*=Quaternionr(1.0, 0.0, 0.0, 0.0)*)

        Stored creep, substracted from total relative rotation for computation of elastic moment *(auto-updated)*

    **updateAttrs**(*(Serializable)arg1, (dict)arg2*) → None :

        Update object attributes from given dictionary

**class yade.wrapper.CylScGeom**(*inherits ScGeom → GenericSpheresContact → IGeom → Serializable*)

    Geometry of a cylinder-sphere contact.

    **contactPoint**(*=uninitialized*)

        some reference point for the interaction (usually in the middle). *(auto-computed)*

**dict**(*(Serializable)arg1*) → dict :
    Return dictionary of attributes.

**dispHierarchy**(*(IGeom)arg1*[, *(bool)names=True*]) → list :
    Return list of dispatch classes (from down upwards), starting with the class instance itself,
    top-level indexable at last. If names is true (default), return class names rather than numerical
    indices.

**dispIndex**
    Return class index of this instance.

**end**(*=Vector3r::Zero()*)
    position of 2nd node *(auto-updated)*

**id3**(*=0*)
    id of next chained cylinder *(auto-updated)*

**incidentVel**(*(ScGeom)arg1, (Interaction)i*[, *(bool)avoidGranularRatcheting=True*]) → Vector3 :
    Return incident velocity of the interaction (see also *Ig2_Sphere_Sphere_ScGeom.avoidGranularRatcheting* for explanation of the ratcheting argument).

**isDuplicate**(*=0*)
    this flag is turned true (1) automatically if the contact is shared between two chained cylinders.
    A duplicated interaction will be skipped once by the constitutive law, so that only one contact
    at a time is effective. If isDuplicate=2, it means one of the two duplicates has no longer
    geometric interaction, and should be erased by the constitutive laws.

**normal**(*=uninitalized*)
    Unit vector oriented along the interaction, from particle #1, towards particle #2. *(auto-updated)*

**onNode**(*=false*)
    contact on node?

**penetrationDepth**(*=NaN*)
    Penetration distance of spheres (positive if overlapping)

**refR1**(*=uninitalized*)
    Reference radius of particle #1. *(auto-computed)*

**refR2**(*=uninitalized*)
    Reference radius of particle #2. *(auto-computed)*

**relAngVel**(*(ScGeom)arg1, (Interaction)i*) → Vector3 :
    Return relative angular velocity of the interaction.

**relPos**(*=0*)
    position of the contact on the cylinder (0: node-, 1:node+) *(auto-updated)*

**shearInc**(*=Vector3r::Zero()*)
    Shear displacement increment in the last step

**start**(*=Vector3r::Zero()*)
    position of 1st node *(auto-updated)*

**trueInt**(*=-1*)
    Defines the body id of the cylinder where the contact is real, when *CylScGeom::isDuplicate*>0.

**updateAttrs**(*(Serializable)arg1, (dict)arg2*) → None :
    Update object attributes from given dictionary

**class yade.wrapper.CylScGeom6D**(*inherits ScGeom6D → ScGeom → GenericSpheresContact → IGeom → Serializable*)
    Class representing *geometry* of two *bodies* in contact. The contact has 6 DOFs (normal, 2×shear,
    twist, 2xbending) and uses *ScGeom* incremental algorithm for updating shear.

**bending**(*=Vector3r::Zero()*)
> Bending at contact as a vector defining axis of rotation and angle (angle=norm).

**contactPoint**(*=uninitalized*)
> some reference point for the interaction (usually in the middle). *(auto-computed)*

**dict**(*(Serializable)arg1*) → dict :
> Return dictionary of attributes.

**dispHierarchy**(*(IGeom)arg1*[*, (bool)names=True*]) → list :
> Return list of dispatch classes (from down upwards), starting with the class instance itself, top-level indexable at last. If names is true (default), return class names rather than numerical indices.

**dispIndex**
> Return class index of this instance.

**end**(*=Vector3r::Zero()*)
> position of 2nd node *(auto-updated)*

**id3**(*=0*)
> id of next chained cylinder *(auto-updated)*

**incidentVel**(*(ScGeom)arg1, (Interaction)i*[*, (bool)avoidGranularRatcheting=True*]) → Vector3 :
> Return incident velocity of the interaction (see also *Ig2_Sphere_Sphere_ScGeom.avoidGranularRatcheting* for explanation of the ratcheting argument).

**initialOrientation1**(*=Quaternionr(1.0, 0.0, 0.0, 0.0)*)
> Orientation of body 1 one at initialisation time *(auto-updated)*

**initialOrientation2**(*=Quaternionr(1.0, 0.0, 0.0, 0.0)*)
> Orientation of body 2 one at initialisation time *(auto-updated)*

**isDuplicate**(*=0*)
> this flag is turned true (1) automatically if the contact is shared between two chained cylinders. A duplicated interaction will be skipped once by the constitutive law, so that only one contact at a time is effective. If isDuplicate=2, it means one of the two duplicates has no longer geometric interaction, and should be erased by the constitutive laws.

**normal**(*=uninitalized*)
> Unit vector oriented along the interaction, from particle #1, towards particle #2. *(auto-updated)*

**onNode**(*=false*)
> contact on node?

**penetrationDepth**(*=NaN*)
> Penetration distance of spheres (positive if overlapping)

**refR1**(*=uninitalized*)
> Reference radius of particle #1. *(auto-computed)*

**refR2**(*=uninitalized*)
> Reference radius of particle #2. *(auto-computed)*

**relAngVel**(*(ScGeom)arg1, (Interaction)i*) → Vector3 :
> Return relative angular velocity of the interaction.

**relPos**(*=0*)
> position of the contact on the cylinder (0: node-, 1:node+) *(auto-updated)*

**shearInc**(*=Vector3r::Zero()*)
> Shear displacement increment in the last step

**start**(*=Vector3r::Zero()*)
> position of 1st node *(auto-updated)*

**trueInt**(*=-1*)
 Defines the body id of the cylinder where the contact is real, when *CylScGeom::isDuplicate*>0.

**twist**(*=0*)
 Elastic twist angle (around *normal axis*) of the contact.

**twistCreep**(*=Quaternionr(1.0, 0.0, 0.0, 0.0)*)
 Stored creep, substracted from total relative rotation for computation of elastic moment *(auto-updated)*

**updateAttrs**(*(Serializable)arg1, (dict)arg2*) → None :
 Update object attributes from given dictionary

**class yade.wrapper.GenericSpheresContact**(*inherits IGeom → Serializable*)
 Class uniting *ScGeom* and *L3Geom*, for the purposes of *GlobalStiffnessTimeStepper*. (It might be removed in the future). Do not use this class directly.

**contactPoint**(*=uninitalized*)
 some reference point for the interaction (usually in the middle). *(auto-computed)*

**dict**(*(Serializable)arg1*) → dict :
 Return dictionary of attributes.

**dispHierarchy**(*(IGeom)arg1*[, *(bool)names=True*]) → list :
 Return list of dispatch classes (from down upwards), starting with the class instance itself, top-level indexable at last. If names is true (default), return class names rather than numerical indices.

**dispIndex**
 Return class index of this instance.

**normal**(*=uninitalized*)
 Unit vector oriented along the interaction, from particle #1, towards particle #2. *(auto-updated)*

**refR1**(*=uninitalized*)
 Reference radius of particle #1. *(auto-computed)*

**refR2**(*=uninitalized*)
 Reference radius of particle #2. *(auto-computed)*

**updateAttrs**(*(Serializable)arg1, (dict)arg2*) → None :
 Update object attributes from given dictionary

**class yade.wrapper.GridCoGridCoGeom**(*inherits ScGeom → GenericSpheresContact → IGeom → Serializable*)
 Geometry of a *GridConnection*-*GridConnection* contact.

**contactPoint**(*=uninitalized*)
 some reference point for the interaction (usually in the middle). *(auto-computed)*

**dict**(*(Serializable)arg1*) → dict :
 Return dictionary of attributes.

**dispHierarchy**(*(IGeom)arg1*[, *(bool)names=True*]) → list :
 Return list of dispatch classes (from down upwards), starting with the class instance itself, top-level indexable at last. If names is true (default), return class names rather than numerical indices.

**dispIndex**
 Return class index of this instance.

**incidentVel**(*(ScGeom)arg1, (Interaction)i*[, *(bool)avoidGranularRatcheting=True*]) → Vector3 :
 Return incident velocity of the interaction (see also *Ig2_Sphere_Sphere_ScGeom.avoidGranularRatcheting* for explanation of the ratcheting argument).

**normal**(*=uninitalized*)

    Unit vector oriented along the interaction, from particle #1, towards particle #2. *(auto-updated)*

**penetrationDepth**(*=NaN*)

    Penetration distance of spheres (positive if overlapping)

**refR1**(*=uninitalized*)

    Reference radius of particle #1. *(auto-computed)*

**refR2**(*=uninitalized*)

    Reference radius of particle #2. *(auto-computed)*

**relAngVel**(*(ScGeom)arg1, (Interaction)i*) → Vector3 :

    Return relative angular velocity of the interaction.

**relPos1**(*=0*)

    position of the contact on the first connection (0: node-, 1:node+) *(auto-updated)*

**relPos2**(*=0*)

    position of the contact on the first connection (0: node-, 1:node+) *(auto-updated)*

**shearInc**(*=Vector3r::Zero()*)

    Shear displacement increment in the last step

**updateAttrs**(*(Serializable)arg1, (dict)arg2*) → None :

    Update object attributes from given dictionary

**class yade.wrapper.GridNodeGeom6D**(*inherits ScGeom6D → ScGeom → GenericSpheresContact → IGeom → Serializable*)

    Geometry of a *GridNode-GridNode* contact. Inherits almost everything from *ScGeom6D*.

**bending**(*=Vector3r::Zero()*)

    Bending at contact as a vector defining axis of rotation and angle (angle=norm).

**connectionBody**(*=uninitalized*)

    Reference to the *GridNode Body* who is linking the two *GridNodes*.

**contactPoint**(*=uninitalized*)

    some reference point for the interaction (usually in the middle). *(auto-computed)*

**dict**(*(Serializable)arg1*) → dict :

    Return dictionary of attributes.

**dispHierarchy**(*(IGeom)arg1*[, *(bool)names=True*]) → list :

    Return list of dispatch classes (from down upwards), starting with the class instance itself, top-level indexable at last. If names is true (default), return class names rather than numerical indices.

**dispIndex**

    Return class index of this instance.

**incidentVel**(*(ScGeom)arg1, (Interaction)i*[, *(bool)avoidGranularRatcheting=True*]) → Vector3 :

    Return incident velocity of the interaction (see also *Ig2_Sphere_Sphere_ScGeom.avoidGranularRatcheting* for explanation of the ratcheting argument).

**initialOrientation1**(*=Quaternionr(1.0, 0.0, 0.0, 0.0)*)

    Orientation of body 1 one at initialisation time *(auto-updated)*

**initialOrientation2**(*=Quaternionr(1.0, 0.0, 0.0, 0.0)*)

    Orientation of body 2 one at initialisation time *(auto-updated)*

**normal**(*=uninitalized*)

    Unit vector oriented along the interaction, from particle #1, towards particle #2. *(auto-updated)*

**penetrationDepth**(*=NaN*)
 Penetration distance of spheres (positive if overlapping)

**refR1**(*=uninitalized*)
 Reference radius of particle #1. *(auto-computed)*

**refR2**(*=uninitalized*)
 Reference radius of particle #2. *(auto-computed)*

**relAngVel**((*ScGeom)arg1, (Interaction)i*) → Vector3 :
 Return relative angular velocity of the interaction.

**shearInc**(*=Vector3r::Zero()*)
 Shear displacement increment in the last step

**twist**(*=0*)
 Elastic twist angle (around *normal axis*) of the contact.

**twistCreep**(*=Quaternionr(1.0, 0.0, 0.0, 0.0)*)
 Stored creep, substracted from total relative rotation for computation of elastic moment *(auto-updated)*

**updateAttrs**((*Serializable)arg1, (dict)arg2*) → None :
 Update object attributes from given dictionary

**class yade.wrapper.L3Geom**(*inherits GenericSpheresContact → IGeom → Serializable*)
 Geometry of contact given in local coordinates with 3 degress of freedom: normal and two in shear plane. [experimental]

 **F**(*=Vector3r::Zero()*)
  Applied force in local coordinates [debugging only, will be removed]

 **contactPoint**(*=uninitalized*)
  some reference point for the interaction (usually in the middle). *(auto-computed)*

 **dict**(*(Serializable)arg1*) → dict :
  Return dictionary of attributes.

 **dispHierarchy**((*IGeom)arg1* [, *(bool)names=True*]) → list :
  Return list of dispatch classes (from down upwards), starting with the class instance itself, top-level indexable at last. If names is true (default), return class names rather than numerical indices.

 **dispIndex**
  Return class index of this instance.

 **normal**(*=uninitalized*)
  Unit vector oriented along the interaction, from particle #1, towards particle #2. *(auto-updated)*

 **refR1**(*=uninitalized*)
  Reference radius of particle #1. *(auto-computed)*

 **refR2**(*=uninitalized*)
  Reference radius of particle #2. *(auto-computed)*

 **trsf**(*=Matrix3r::Identity()*)
  Transformation (rotation) from global to local coordinates. (the translation part is in *GenericSpheresContact.contactPoint*)

 **u**(*=Vector3r::Zero()*)
  Displacement components, in local coordinates. *(auto-updated)*

 **u0**
  Zero displacement value; u0 should be always subtracted from the *geometrical* displacement *u* computed by appropriate *IGeomFunctor*, resulting in *u*. This value can be changed for instance

1. by *IGeomFunctor*, e.g. to take in account large shear displacement value unrepresentable by underlying geomeric algorithm based on quaternions)

2. by *LawFunctor*, to account for normal equilibrium position different from zero geometric overlap (set once, just after the interaction is created)

3. by *LawFunctor* to account for plastic slip.

---

**Note:** Never set an absolute value of *u0*, only increment, since both *IGeomFunctor* and *LawFunctor* use it. If you need to keep track of plastic deformation, store it in *IPhys* isntead (this might be changed: have *u0* for *LawFunctor* exclusively, and a separate value stored (when that is needed) inside classes deriving from *L3Geom*.

---

**updateAttrs**(*(Serializable)arg1, (dict)arg2*) → None :
    Update object attributes from given dictionary

**class yade.wrapper.L6Geom**(*inherits L3Geom → GenericSpheresContact → IGeom → Serializable*)
    Geometric of contact in local coordinates with 6 degrees of freedom. [experimental]

**F**(*=Vector3r::Zero()*)
    Applied force in local coordinates [debugging only, will be removed]

**contactPoint**(*=uninitalized*)
    some reference point for the interaction (usually in the middle). *(auto-computed)*

**dict**(*(Serializable)arg1*) → dict :
    Return dictionary of attributes.

**dispHierarchy**(*(IGeom)arg1[, (bool)names=True]*) → list :
    Return list of dispatch classes (from down upwards), starting with the class instance itself, top-level indexable at last. If names is true (default), return class names rather than numerical indices.

**dispIndex**
    Return class index of this instance.

**normal**(*=uninitalized*)
    Unit vector oriented along the interaction, from particle #1, towards particle #2. *(auto-updated)*

**phi**(*=Vector3r::Zero()*)
    Rotation components, in local coordinates. *(auto-updated)*

**phi0**(*=Vector3r::Zero()*)
    Zero rotation, should be always subtracted from *phi* to get the value. See *L3Geom.u0*.

**refR1**(*=uninitalized*)
    Reference radius of particle #1. *(auto-computed)*

**refR2**(*=uninitalized*)
    Reference radius of particle #2. *(auto-computed)*

**trsf**(*=Matrix3r::Identity()*)
    Transformation (rotation) from global to local coordinates. (the translation part is in *GenericSpheresContact.contactPoint*)

**u**(*=Vector3r::Zero()*)
    Displacement components, in local coordinates. *(auto-updated)*

**u0**
    Zero displacement value; u0 should be always subtracted from the *geometrical* displacement *u* computed by appropriate *IGeomFunctor*, resulting in *u*. This value can be changed for instance

---

1. by *IGeomFunctor*, e.g. to take in account large shear displacement value unrepresentable by underlying geomeric algorithm based on quaternions)

2. by *LawFunctor*, to account for normal equilibrium position different from zero geometric overlap (set once, just after the interaction is created)

3. by *LawFunctor* to account for plastic slip.

---

**Note:** Never set an absolute value of *u0*, only increment, since both *IGeomFunctor* and *LawFunctor* use it. If you need to keep track of plastic deformation, store it in *IPhys* isntead (this might be changed: have *u0* for *LawFunctor* exclusively, and a separate value stored (when that is needed) inside classes deriving from *L3Geom*.

---

**updateAttrs**(*(Serializable)arg1, (dict)arg2*) → None :
  Update object attributes from given dictionary

**class yade.wrapper.ScGeom**(*inherits GenericSpheresContact → IGeom → Serializable*)
  Class representing *geometry* of a contact point between two *bodies*. It is more general than sphere-sphere contact even though it is primarily focused on spheres contact interactions (reason for the 'Sc' naming); it is also used for representing contacts of a *Sphere* with non-spherical bodies (*Facet*, **Plane**, *Box*, *ChainedCylinder*), or between two non-spherical bodies (*ChainedCylinder*). The contact has 3 DOFs (normal and 2×shear) and uses incremental algorithm for updating shear.

  We use symbols $\mathbf{x}$, $\mathbf{v}$, $\boldsymbol{\omega}$ respectively for position, linear and angular velocities (all in global coordinates) and $\mathbf{r}$ for particles radii; subscripted with 1 or 2 to distinguish 2 spheres in contact. Then we define branch length and unit contact normal

  $$l = \|\mathbf{x}_2 - \mathbf{x}_1\|, \mathbf{n} = \frac{\mathbf{x}_2 - \mathbf{x}_1}{\|\mathbf{x}_2 - \mathbf{x}_1\|}$$

  The relative velocity of the spheres is then

  $$\mathbf{v}_{12} = \frac{\mathbf{r}_1 + \mathbf{r}_2}{l}(\mathbf{v}_2 - \mathbf{v}_1) - (\mathbf{r}_2\boldsymbol{\omega}_2 + \mathbf{r}_1\boldsymbol{\omega}_1) \times \mathbf{n}$$

  where the fraction multiplying translational velocities is to make the definition objective and avoid ratcheting effects (see *Ig2_Sphere_Sphere_ScGeom.avoidGranularRatcheting*). The shear component is

  $$\mathbf{v}_{12}^s = \mathbf{v}_{12} - (\mathbf{n} \cdot \mathbf{v}_{12})\mathbf{n}.$$

  Tangential displacement increment over last step then reads

  $$\Delta\mathbf{x}_{12}^s = \Delta t\mathbf{v}_{12}^s.$$

**contactPoint**(*=uninitalized*)
  some reference point for the interaction (usually in the middle). *(auto-computed)*

**dict**(*(Serializable)arg1*) → dict :
  Return dictionary of attributes.

**dispHierarchy**(*(IGeom)arg1[, (bool)names=True]*) → list :
  Return list of dispatch classes (from down upwards), starting with the class instance itself, top-level indexable at last. If names is true (default), return class names rather than numerical indices.

**dispIndex**
  Return class index of this instance.

**incidentVel**(*(ScGeom)arg1, (Interaction)i[, (bool)avoidGranularRatcheting=True]*) → Vector3 :
  Return incident velocity of the interaction (see also *Ig2_Sphere_Sphere_ScGeom.avoidGranularRatcheting* for explanation of the ratcheting argument).

**normal**(=*uninitalized*)
> Unit vector oriented along the interaction, from particle #1, towards particle #2. *(auto-updated)*

**penetrationDepth**(=*NaN*)
> Penetration distance of spheres (positive if overlapping)

**refR1**(=*uninitalized*)
> Reference radius of particle #1. *(auto-computed)*

**refR2**(=*uninitalized*)
> Reference radius of particle #2. *(auto-computed)*

**relAngVel**(*(ScGeom)arg1, (Interaction)i*) → Vector3 :
> Return relative angular velocity of the interaction.

**shearInc**(=*Vector3r::Zero()*)
> Shear displacement increment in the last step

**updateAttrs**(*(Serializable)arg1, (dict)arg2*) → None :
> Update object attributes from given dictionary

**class yade.wrapper.ScGeom6D**(*inherits ScGeom → GenericSpheresContact → IGeom → Serializable*)
Class representing *geometry* of two *bodies* in contact. The contact has 6 DOFs (normal, 2×shear, twist, 2xbending) and uses *ScGeom* incremental algorithm for updating shear.

**bending**(=*Vector3r::Zero()*)
> Bending at contact as a vector defining axis of rotation and angle (angle=norm).

**contactPoint**(=*uninitalized*)
> some reference point for the interaction (usually in the middle). *(auto-computed)*

**dict**(*(Serializable)arg1*) → dict :
> Return dictionary of attributes.

**dispHierarchy**(*(IGeom)arg1*[, *(bool)names=True*]) → list :
> Return list of dispatch classes (from down upwards), starting with the class instance itself, top-level indexable at last. If names is true (default), return class names rather than numerical indices.

**dispIndex**
> Return class index of this instance.

**incidentVel**(*(ScGeom)arg1, (Interaction)i*[, *(bool)avoidGranularRatcheting=True*]) → Vector3 :
> Return incident velocity of the interaction (see also *Ig2_Sphere_Sphere_ScGeom.avoidGranularRatcheting* for explanation of the ratcheting argument).

**initialOrientation1**(=*Quaternionr(1.0, 0.0, 0.0, 0.0)*)
> Orientation of body 1 one at initialisation time *(auto-updated)*

**initialOrientation2**(=*Quaternionr(1.0, 0.0, 0.0, 0.0)*)
> Orientation of body 2 one at initialisation time *(auto-updated)*

**normal**(=*uninitalized*)
> Unit vector oriented along the interaction, from particle #1, towards particle #2. *(auto-updated)*

**penetrationDepth**(=*NaN*)
> Penetration distance of spheres (positive if overlapping)

**refR1**(=*uninitalized*)
> Reference radius of particle #1. *(auto-computed)*

**refR2**(=*uninitalized*)
> Reference radius of particle #2. *(auto-computed)*

**relAngVel**(*(ScGeom)arg1, (Interaction)i*) → Vector3 :
   Return relative angular velocity of the interaction.

**shearInc**(*=Vector3r::Zero()*)
   Shear displacement increment in the last step

**twist**(*=0*)
   Elastic twist angle (around *normal axis*) of the contact.

**twistCreep**(*=Quaternionr(1.0, 0.0, 0.0, 0.0)*)
   Stored creep, substracted from total relative rotation for computation of elastic moment *(auto-updated)*

**updateAttrs**(*(Serializable)arg1, (dict)arg2*) → None :
   Update object attributes from given dictionary

**class yade.wrapper.ScGridCoGeom**(*inherits ScGeom6D → ScGeom → GenericSpheresContact → IGeom → Serializable*)
   Geometry of a *GridConnection-Sphere* contact.

**bending**(*=Vector3r::Zero()*)
   Bending at contact as a vector defining axis of rotation and angle (angle=norm).

**contactPoint**(*=uninitialized*)
   some reference point for the interaction (usually in the middle). *(auto-computed)*

**dict**(*(Serializable)arg1*) → dict :
   Return dictionary of attributes.

**dispHierarchy**(*(IGeom)arg1*[, *(bool)names=True*]) → list :
   Return list of dispatch classes (from down upwards), starting with the class instance itself, top-level indexable at last. If names is true (default), return class names rather than numerical indices.

**dispIndex**
   Return class index of this instance.

**id3**(*=0*)
   id of the first *GridNode*. *(auto-updated)*

**id4**(*=0*)
   id of the second *GridNode*. *(auto-updated)*

**id5**(*=-1*)
   id of the third *GridNode*. *(auto-updated)*

**incidentVel**(*(ScGeom)arg1, (Interaction)i*[, *(bool)avoidGranularRatcheting=True*]) → Vector3 :
   Return incident velocity of the interaction (see also *Ig2_Sphere_Sphere_ScGeom.avoidGranularRatcheting* for explanation of the ratcheting argument).

**initialOrientation1**(*=Quaternionr(1.0, 0.0, 0.0, 0.0)*)
   Orientation of body 1 one at initialisation time *(auto-updated)*

**initialOrientation2**(*=Quaternionr(1.0, 0.0, 0.0, 0.0)*)
   Orientation of body 2 one at initialisation time *(auto-updated)*

**isDuplicate**(*=0*)
   this flag is turned true (1) automatically if the contact is shared between two Connections. A duplicated interaction will be skipped once by the constitutive law, so that only one contact at a time is effective. If isDuplicate=2, it means one of the two duplicates has no longer geometric interaction, and should be erased by the constitutive laws.

**normal**(*=uninitialized*)
   Unit vector oriented along the interaction, from particle #1, towards particle #2. *(auto-updated)*

**penetrationDepth**(*=NaN*)
    Penetration distance of spheres (positive if overlapping)

**refR1**(*=uninitalized*)
    Reference radius of particle #1. *(auto-computed)*

**refR2**(*=uninitalized*)
    Reference radius of particle #2. *(auto-computed)*

**relAngVel**(*(ScGeom)arg1, (Interaction)i*) → Vector3 :
    Return relative angular velocity of the interaction.

**relPos**(*=0*)
    position of the contact on the connection (0: node-, 1:node+) *(auto-updated)*

**shearInc**(*=Vector3r::Zero()*)
    Shear displacement increment in the last step

**trueInt**(*=-1*)
    Defines the body id of the *GridConnection* where the contact is real, when *ScGridCo-Geom::isDuplicate*>0.

**twist**(*=0*)
    Elastic twist angle (around *normal axis*) of the contact.

**twistCreep**(*=Quaternionr(1.0, 0.0, 0.0, 0.0)*)
    Stored creep, substracted from total relative rotation for computation of elastic moment *(auto-updated)*

**updateAttrs**(*(Serializable)arg1, (dict)arg2*) → None :
    Update object attributes from given dictionary

**weight**(*=Vector3r(0, 0, 0)*)
    barycentric coordinates of the projection point *(auto-updated)*

**class yade.wrapper.TTetraGeom**(*inherits IGeom → Serializable*)
    Geometry of interaction between 2 *tetrahedra*, including volumetric characteristics

**contactPoint**(*=uninitalized*)
    Contact point (global coords)

**dict**(*(Serializable)arg1*) → dict :
    Return dictionary of attributes.

**dispHierarchy**(*(IGeom)arg1*[, *(bool)names=True*]) → list :
    Return list of dispatch classes (from down upwards), starting with the class instance itself, top-level indexable at last. If names is true (default), return class names rather than numerical indices.

**dispIndex**
    Return class index of this instance.

**equivalentCrossSection**(*=NaN*)
    Cross-section of the overlap (perpendicular to the axis of least inertia

**equivalentPenetrationDepth**(*=NaN*)
    ??

**maxPenetrationDepthA**(*=NaN*)
    ??

**maxPenetrationDepthB**(*=NaN*)
    ??

**normal**(*=uninitalized*)
    Normal of the interaction, directed in the sense of least inertia of the overlap volume

**penetrationVolume**(*=NaN*)
> Volume of overlap [m³]

**updateAttrs**(*(Serializable)arg1, (dict)arg2*) → None :
> Update object attributes from given dictionary

**class yade.wrapper.TTetraSimpleGeom**(*inherits IGeom → Serializable*)
> EXPERIMENTAL. Geometry of interaction between 2 *tetrahedra*

**contactPoint**(*=uninitalized*)
> Contact point (global coords)

**dict**(*(Serializable)arg1*) → dict :
> Return dictionary of attributes.

**dispHierarchy**(*(IGeom)arg1*[, *(bool)names=True*]) → list :
> Return list of dispatch classes (from down upwards), starting with the class instance itself, top-level indexable at last. If names is true (default), return class names rather than numerical indices.

**dispIndex**
> Return class index of this instance.

**flag**(*=0*)
> TODO

**normal**(*=uninitalized*)
> Normal of the interaction TODO

**penetrationVolume**(*=NaN*)
> Volume of overlap [m³]

**updateAttrs**(*(Serializable)arg1, (dict)arg2*) → None :
> Update object attributes from given dictionary

## IPhys

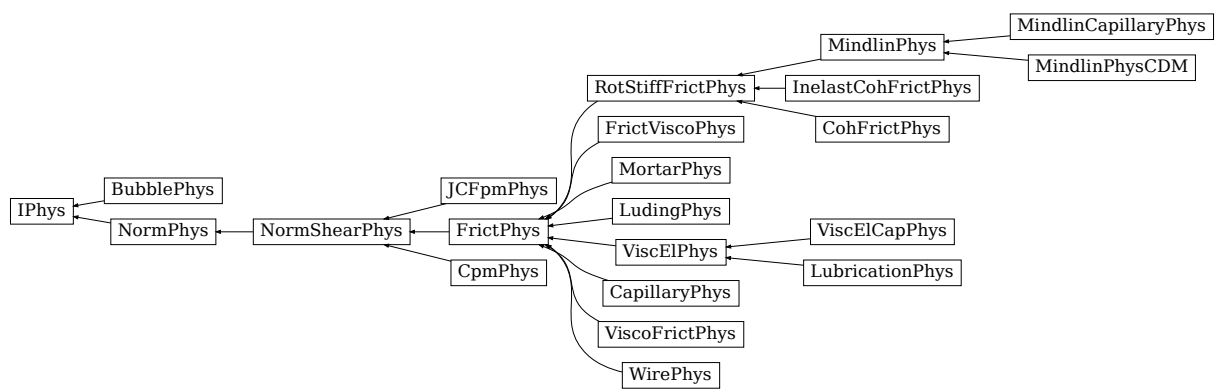

Fig. 24: Inheritance graph of IPhys. See also: *BubblePhys*, *CapillaryPhys*, *CohFrictPhys*, *CpmPhys*, *FrictPhys*, *FrictViscoPhys*, *InelastCohFrictPhys*, *JCFpmPhys*, *LubricationPhys*, *LudingPhys*, *Mindlin-CapillaryPhys*, *MindlinPhys*, *MindlinPhysCDM*, *MortarPhys*, *NormPhys*, *NormShearPhys*, *RotStiff-FrictPhys*, *ViscElCapPhys*, *ViscElPhys*, *ViscoFrictPhys*, *WirePhys*.

**class yade.wrapper.IPhys**(*inherits Serializable*)
> Physical (material) properties of *interaction*.

**dict**(*(Serializable)arg1*) → dict :
> Return dictionary of attributes.

**dispHierarchy**(*(IPhys)arg1*[, *(bool)names=True*]) → list :
Return list of dispatch classes (from down upwards), starting with the class instance itself, top-level indexable at last. If names is true (default), return class names rather than numerical indices.

**dispIndex**
Return class index of this instance.

**updateAttrs**(*(Serializable)arg1*, *(dict)arg2*) → None :
Update object attributes from given dictionary

**class yade.wrapper.BubblePhys**(*inherits IPhys → Serializable*)
Physics of bubble-bubble interactions, for use with BubbleMat

**Dmax**(*=NaN*)
Maximum penetrationDepth of the bubbles before the force displacement curve changes to an artificial exponential curve. Setting this value will have no effect. See Law2_ScGeom_-BubblePhys_Bubble::pctMaxForce for more information

**static computeForce**(*(float)arg1*, *(float)arg2*, *(float)arg3*, *(int)arg4*, *(float)arg5*, *(float)arg6*, *(float)arg7*, *(BubblePhys)arg8*) → float :
Computes the normal force acting between the two interacting bubbles using the Newton-Rhapson method

**dict**(*(Serializable)arg1*) → dict :
Return dictionary of attributes.

**dispHierarchy**(*(IPhys)arg1*[, *(bool)names=True*]) → list :
Return list of dispatch classes (from down upwards), starting with the class instance itself, top-level indexable at last. If names is true (default), return class names rather than numerical indices.

**dispIndex**
Return class index of this instance.

**fN**(*=NaN*)
Contact normal force

**newtonIter**(*=50*)
Maximum number of force iterations allowed

**newtonTol**(*=1e-6*)
Convergence criteria for force iterations

**normalForce**(*=Vector3r::Zero()*)
Normal force

**rAvg**(*=NaN*)
Average radius of the two interacting bubbles

**surfaceTension**(*=NaN*)
Surface tension of the surrounding liquid

**updateAttrs**(*(Serializable)arg1*, *(dict)arg2*) → None :
Update object attributes from given dictionary

**class yade.wrapper.CapillaryPhys**(*inherits FrictPhys → NormShearPhys → NormPhys → IPhys → Serializable*)
Physics (of interaction) for *Law2_ScGeom_CapillaryPhys_Capillarity*.

**Delta1**(*=0.*)
Defines the surface area wetted by the meniscus on the smallest grains of radius R1 (R1<R2)

**Delta2**(*=0.*)
Defines the surface area wetted by the meniscus on the biggest grains of radius R2 (R1<R2)

**capillaryPressure**(*=0.*)
> Value of the capillary pressure Uc. Defined as Ugas-Uliquid, obtained from *corresponding Law2 parameter*

**dict**(*(Serializable)arg1*) → dict :
> Return dictionary of attributes.

**dispHierarchy**(*(IPhys)arg1*[, *(bool)names=True*]) → list :
> Return list of dispatch classes (from down upwards), starting with the class instance itself, top-level indexable at last. If names is true (default), return class names rather than numerical indices.

**dispIndex**
> Return class index of this instance.

**fCap**(*=Vector3r::Zero()*)
> Capillary force produced by the presence of the meniscus. This is the force acting on particle #2

**fusionNumber**(*=0.*)
> Indicates the number of meniscii that overlap with this one

**isBroken**(*=false*)
> Might be set to true by the user to make liquid bridge inactive (capillary force is zero)

**kn**(*=0*)
> Normal stiffness

**ks**(*=0*)
> Shear stiffness

**meniscus**(*=false*)
> True when a meniscus with a non-zero liquid volume (*vMeniscus*) has been computed for this interaction

**nn11**(*=0.*)
> $\iint_A \mathfrak{n}_1\mathfrak{n}_1 \, \mathrm{d}S = \iint_A \mathfrak{n}_2\mathfrak{n}_2 \, \mathrm{d}S$, A being the liquid-gas surface of the meniscus, $\mathfrak{n}$ the associated normal, and $(1,2,3)$ a local basis with 3 the meniscus orientation (*ScGeom.normal*). NB: A = 2 *nn11* + *nn33*.

**nn33**(*=0.*)
> $\iint_A \mathfrak{n}_3\mathfrak{n}_3 \, \mathrm{d}S$, A being the liquid-gas surface of the meniscus, $\mathfrak{n}$ the associated normal, and $(1,2,3)$ a local basis with 3 the meniscus orientation (*ScGeom.normal*). NB: A = 2 *nn11* + *nn33*.

**normalForce**(*=Vector3r::Zero()*)
> Normal force after previous step (in global coordinates), as sustained by particle #2 (from particle #1).

**shearForce**(*=Vector3r::Zero()*)
> Shear force after previous step (in global coordinates), as sustained by particle #2 (from particle #1).

**tangensOfFrictionAngle**(*=NaN*)
> tan of angle of friction

**updateAttrs**(*(Serializable)arg1, (dict)arg2*) → None :
> Update object attributes from given dictionary

**vMeniscus**(*=0.*)
> Volume of the meniscus

**class yade.wrapper.CohFrictPhys**(*inherits RotStiffFrictPhys → FrictPhys → NormShearPhys → NormPhys → IPhys → Serializable*)
> An *interaction physics* that extends *RotStiffFrictPhys* adding a breakable cohesive nature. Used e.g. by *Law2__ScGeom6D__CohFrictPhys__CohesionMoment*.

**cohesionBroken**(*=true*)
> is cohesion active? Set to false at the creation of a cohesive contact, and set to true when a fragile contact is broken

**cohesionDisablesFriction**(*=false*)
> is shear strength the sum of friction and adhesion or only adhesion?

**creep_viscosity**(*=-1*)
> creep viscosity [Pa.s/m].

**dict**(*(Serializable)arg1*) → dict :
> Return dictionary of attributes.

**dispHierarchy**(*(IPhys)arg1*[, *(bool)names=True*]) → list :
> Return list of dispatch classes (from down upwards), starting with the class instance itself, top-level indexable at last. If names is true (default), return class names rather than numerical indices.

**dispIndex**
> Return class index of this instance.

**fragile**(*=true*)
> do cohesion disappear when contact strength is exceeded?

**initCohesion**(*=false*)
> Initialize the cohesive behaviour with current state as equilibrium state (same as *Ip2_CohFrictMat_CohFrictMat_CohFrictPhys::setCohesionNow* but acting on only one interaction)

**kn**(*=0*)
> Normal stiffness

**kr**(*=0*)
> rotational stiffness [N.m/rad]

**ks**(*=0*)
> Shear stiffness

**ktw**(*=0*)
> twist stiffness [N.m/rad]

**maxRollPl**(*=0.0*)
> Coefficient of rolling friction (negative means elastic).

**maxTwistPl**(*=0.0*)
> Coefficient of twisting friction (negative means elastic).

**momentRotationLaw**(*=false*)
> set from *CohFrictMat::momentRotationLaw* in order to possibly use bending/twisting moment at contacts (if true). See *Law2_ScGeom6D_CohFrictPhys_CohesionMoment::always_use_moment_law* for details.

**moment_bending**(*=Vector3r(0, 0, 0)*)
> Bending moment

**moment_twist**(*=Vector3r(0, 0, 0)*)
> Twist moment

**normalAdhesion**(*=0*)
> tensile strength

**normalForce**(*=Vector3r::Zero()*)
> Normal force after previous step (in global coordinates), as sustained by particle #2 (from particle #1).

**shearAdhesion**(*=0*)
> cohesive part of the shear strength (a frictional term might be added depending on *CohFrictPhys::cohesionDisablesFriction*)

---

**shearForce**(*=Vector3r::Zero()*)
    Shear force after previous step (in global coordinates), as sustained by particle #2 (from particle #1).

**tangensOfFrictionAngle**(*=NaN*)
    tan of angle of friction

**unp**(*=0*)
    plastic normal displacement, only used for tensile behaviour and if *CohFrictPhys::fragile* =false.

**unpMax**(*=0*)
    maximum value of plastic normal displacement (counted positively), after that the interaction breaks even if *CohFrictPhys::fragile* =false. A negative value (i.e. -1) means no maximum.

**updateAttrs**(*(Serializable)arg1, (dict)arg2*) → None :
    Update object attributes from given dictionary

**class yade.wrapper.CpmPhys**(*inherits NormShearPhys → NormPhys → IPhys → Serializable*)
    Representation of a single interaction of the Cpm type: storage for relevant parameters.

    Evolution of the contact is governed by *Law2_ScGeom_CpmPhys_Cpm*, that includes damage effects and chages of parameters inside CpmPhys. See *cpm-model* for details.

**E**(*=NaN*)
    normal modulus (stiffness / crossSection) [Pa]

**Fn**
    Magnitude of normal force *(auto-updated)*

**Fs**
    Magnitude of shear force *(auto-updated)*

**G**(*=NaN*)
    shear modulus [Pa]

**crossSection**(*=NaN*)
    equivalent cross-section associated with this contact [m$^2$]

**cummBetaCount = 0**

**cummBetaIter = 0**

**damLaw**(*=1*)
    Law for softening part of uniaxial tension. 0 for linear, 1 for exponential (default)

**dict**(*(Serializable)arg1*) → dict :
    Return dictionary of attributes.

**dispHierarchy**(*(IPhys)arg1*[, *(bool)names=True*]) → list :
    Return list of dispatch classes (from down upwards), starting with the class instance itself, top-level indexable at last. If names is true (default), return class names rather than numerical indices.

**dispIndex**
    Return class index of this instance.

**dmgOverstress**(*=0*)
    damage viscous overstress (at previous step or at current step)

**dmgRateExp**(*=0*)
    exponent in the rate-dependent damage evolution

**dmgStrain**(*=0*)
    damage strain (at previous or current step)

**dmgTau**(*=-1*)
    characteristic time for damage (if non-positive, the law without rate-dependence is used)

**epsCrackOnset**(*=NaN*)
> strain at which the material starts to behave non-linearly

**epsFracture**(*=NaN*)
> strain at which the bond is fully broken [-]

**epsN**
> Current normal strain *(auto-updated)*

**epsNPl**
> normal plastic strain (initially zero) *(auto-updated)*

**epsT**
> Current shear strain *(auto-updated)*

**epsTPl**
> shear plastic strain (initially zero) *(auto-updated)*

**equivStrainShearContrib**(*=NaN*)
> Coefficient of shear contribution to equivalent strain

**static funcG**(*(float)kappaD*,          *(float)epsCrackOnset*,          *(float)epsFracture* [, *(bool)neverDamage=False* [, *(int)damLaw=1* ] ] ) → float :
> Damage evolution law, evaluating the $\omega$ parameter. $\kappa_D$ is historically maximum strain, *epsCrackOnset* ($\varepsilon_0$) = *CpmPhys.epsCrackOnset*, *epsFracture* = *CpmPhys.epsFracture*; if *neverDamage* is True, the value returned will always be 0 (no damage). TODO

**static funcGInv**(*(float)omega*,          *(float)epsCrackOnset*,          *(float)epsFracture* [, *(bool)neverDamage=False* [, *(int)damLaw=1* ] ] ) → float :
> Inversion of damage evolution law, evaluating the $\kappa_D$ parameter. $\omega$ is damage, for other parameters see funcG function

**isCohesive**(*=false*)
> if not cohesive, interaction is deleted when distance is greater than zero.

**isoPrestress**(*=0*)
> "prestress" of this link (used to simulate isotropic stress)

**kappaD**
> Up to now maximum normal strain (semi-norm), non-decreasing in time *(auto-updated)*

**kn**(*=0*)
> Normal stiffness

**ks**(*=0*)
> Shear stiffness

**neverDamage**(*=false*)
> the damage evolution function will always return virgin state

**normalForce**(*=Vector3r::Zero()*)
> Normal force after previous step (in global coordinates), as sustained by particle #2 (from particle #1).

**omega**
> Damage internal variable *(auto-updated)*

**plRateExp**(*=0*)
> exponent in the rate-dependent viscoplasticity

**plTau**(*=-1*)
> characteristic time for viscoplasticity (if non-positive, no rate-dependence for shear)

**refLength**(*=NaN*)
> initial length of interaction [m]

**refPD**(*=NaN*)
> initial penetration depth of interaction [m] (used with ScGeom)

**relDuctility**(*=NaN*)
> Relative ductility of bonds in normal direction

**relResidualStrength**
> Relative residual strength *(auto-updated)*

**setDamage**(*(CpmPhys)arg1, (float)arg2*) → None :
> TODO

**setRelResidualStrength**(*(CpmPhys)arg1, (float)arg2*) → None :
> TODO

**shearForce**(*=Vector3r::Zero()*)
> Shear force after previous step (in global coordinates), as sustained by particle #2 (from particle #1).

**sigmaN**
> Current normal stress *(auto-updated)*

**sigmaT**
> Current shear stress *(auto-updated)*

**tanFrictionAngle**(*=NaN*)
> tangens of internal friction angle [-]

**undamagedCohesion**(*=NaN*)
> virgin material cohesion [Pa]

**updateAttrs**(*(Serializable)arg1, (dict)arg2*) → None :
> Update object attributes from given dictionary

**class yade.wrapper.FrictPhys**(*inherits NormShearPhys → NormPhys → IPhys → Serializable*)
> The simple linear elastic-plastic interaction with friction angle, like in the traditional [CundallStrack1979]

**dict**(*(Serializable)arg1*) → dict :
> Return dictionary of attributes.

**dispHierarchy**(*(IPhys)arg1*[*, (bool)names=True*]) → list :
> Return list of dispatch classes (from down upwards), starting with the class instance itself, top-level indexable at last. If names is true (default), return class names rather than numerical indices.

**dispIndex**
> Return class index of this instance.

**kn**(*=0*)
> Normal stiffness

**ks**(*=0*)
> Shear stiffness

**normalForce**(*=Vector3r::Zero()*)
> Normal force after previous step (in global coordinates), as sustained by particle #2 (from particle #1).

**shearForce**(*=Vector3r::Zero()*)
> Shear force after previous step (in global coordinates), as sustained by particle #2 (from particle #1).

**tangensOfFrictionAngle**(*=NaN*)
> tan of angle of friction

**updateAttrs**(*(Serializable)arg1, (dict)arg2*) → None :
> Update object attributes from given dictionary **class yade.wrapper.FrictViscoPhys**(*inherits* *FrictPhys → NormShearPhys → NormPhys →* *IPhys → Serializable*)

Representation of a single interaction of the FrictViscoPM type, storage for relevant parameters

**cn**(*=NaN*)

Normal viscous constant defined as $\mathfrak{n} = c_{n,crit}\beta_n$.

**cn_crit**(*=NaN*)

Normal viscous constant for ctitical damping defined as $\mathfrak{n} = C_{n,crit}\beta_n$.

**dict**(*(Serializable)arg1*) → dict :

Return dictionary of attributes.

**dispHierarchy**(*(IPhys)arg1*[, *(bool)names=True*]) → list :

Return list of dispatch classes (from down upwards), starting with the class instance itself, top-level indexable at last. If names is true (default), return class names rather than numerical indices.

**dispIndex**

Return class index of this instance.

**kn**(*=0*)

Normal stiffness

**ks**(*=0*)

Shear stiffness

**normalForce**(*=Vector3r::Zero()*)

Normal force after previous step (in global coordinates), as sustained by particle #2 (from particle #1).

**normalViscous**(*=Vector3r::Zero()*)

Normal viscous component

**shearForce**(*=Vector3r::Zero()*)

Shear force after previous step (in global coordinates), as sustained by particle #2 (from particle #1).

**tangensOfFrictionAngle**(*=NaN*)

tan of angle of friction

**updateAttrs**(*(Serializable)arg1, (dict)arg2*) → None :

Update object attributes from given dictionary

**class yade.wrapper.InelastCohFrictPhys**(*inherits* *RotStiffFrictPhys → FrictPhys → NormShearPhys → NormPhys → IPhys → Serializable*)

**cohesionBroken**(*=false*)

is cohesion active? will be set false when a fragile contact is broken

**dict**(*(Serializable)arg1*) → dict :

Return dictionary of attributes.

**dispHierarchy**(*(IPhys)arg1*[, *(bool)names=True*]) → list :

Return list of dispatch classes (from down upwards), starting with the class instance itself, top-level indexable at last. If names is true (default), return class names rather than numerical indices.

**dispIndex**

Return class index of this instance.

**isBroken**(*=false*)

true if compression plastic fracture achieved

**kDam**(*=0*)

Damage coefficient on bending, computed from maximum bending moment reached and pure creep behaviour. Its values will vary between *InelastCohFrictPhys::kr* and *InelastCohFrict-Phys::kRCrp* .

**kRCrp**(*=0.0*)
   Bending creep stiffness

**kRUnld**(*=0.0*)
   Bending plastic unload stiffness

**kTCrp**(*=0.0*)
   Tension/compression creep stiffness

**kTUnld**(*=0.0*)
   Tension/compression plastic unload stiffness

**kTwCrp**(*=0.0*)
   Twist creep stiffness

**kTwUnld**(*=0.0*)
   Twist plastic unload stiffness

**kn**(*=0*)
   Normal stiffness

**knC**(*=0*)
   compression stiffness

**knT**(*=0*)
   tension stiffness

**kr**(*=0*)
   rotational stiffness [N.m/rad]

**ks**(*=0*)
   shear stiffness

**ktw**(*=0*)
   twist stiffness [N.m/rad]

**maxBendMom**(*=0.0*)
   Plastic failure bending moment.

**maxContract**(*=0.0*)
   Plastic failure contraction (shrinkage).

**maxCrpRchdB**(*=Vector3r(0, 0, 0)*)
   maximal bending moment reached on plastic deformation.

**maxCrpRchdC**(*=Vector2r(0, 0)*)
   maximal compression reached on plastic deformation. maxCrpRchdC[0] stores un and max-CrpRchdC[1] stores Fn.

**maxCrpRchdT**(*=Vector2r(0, 0)*)
   maximal extension reached on plastic deformation. maxCrpRchdT[0] stores un and maxCr-pRchdT[1] stores Fn.

**maxCrpRchdTw**(*=Vector2r(0, 0)*)
   maximal twist reached on plastic deformation. maxCrpRchdTw[0] stores twist angle and maxCrpRchdTw[1] stores twist moment.

**maxElB**(*=0.0*)
   Maximum bending elastic moment.

**maxElC**(*=0.0*)
   Maximum compression elastic force.

**maxElT**(*=0.0*)
   Maximum tension elastic force.

**maxElTw**(*=0.0*)
    Maximum twist elastic moment.

**maxExten**(*=0.0*)
    Plastic failure extension (stretching).

**maxTwist**(*=0.0*)
    Plastic failure twist angle

**moment_bending**(*=Vector3r(0, 0, 0)*)
    Bending moment

**moment_twist**(*=Vector3r(0, 0, 0)*)
    Twist moment

**normalForce**(*=Vector3r::Zero()*)
    Normal force after previous step (in global coordinates), as sustained by particle #2 (from particle #1).

**onPlastB**(*=false*)
    true if plasticity achieved on bending

**onPlastC**(*=false*)
    true if plasticity achieved on compression

**onPlastT**(*=false*)
    true if plasticity achieved on traction

**onPlastTw**(*=false*)
    true if plasticity achieved on twisting

**pureCreep**(*=Vector3r(0, 0, 0)*)
    Pure creep curve, used for comparison in calculation.

**shearAdhesion**(*=0*)
    Maximum elastic shear force (cohesion).

**shearForce**(*=Vector3r::Zero()*)
    Shear force after previous step (in global coordinates), as sustained by particle #2 (from particle #1).

**tangensOfFrictionAngle**(*=NaN*)
    tan of angle of friction

**twp**(*=0*)
    plastic twist penetration depth describing the equilibrium state.

**unp**(*=0*)
    plastic normal penetration depth describing the equilibrium state.

**updateAttrs**(*(Serializable)arg1, (dict)arg2*) → None :
    Update object attributes from given dictionary

**class yade.wrapper.JCFpmPhys**(*inherits NormShearPhys → NormPhys → IPhys → Serializable*)
    Representation of a single interaction of the JCFpm type, storage for relevant parameters

    **FnMax**(*=0.*)
        positiv value computed from *tensile strength* (or joint variant) to define the maximum admissible normal force in traction: Fn >= -FnMax. [N]

    **FsMax**(*=0.*)
        computed from *cohesion* (or jointCohesion) to define the maximum admissible tangential force in shear, for Fn=0. [N]

    **checkedForCluster**(*=false*)
        Have we checked if this int belongs in cluster?

    **clusterInts**(*=uninitalized*)
        vector of pointers to the broken interactions nearby constituting a cluster

**clusteredEvent**(*=false*)
   is this interaction part of a cluster?

**computedCentroid**(*=false*)
   Flag for moment calculation

**crackJointAperture**(*=0.*)
   Relative displacement between 2 spheres (in case of a crack it is equivalent of the crack aperture)

**crossSection**(*=0.*)
   crossSection=pi*Rmin^2. [m2]

**dict**(*(Serializable)arg1*) → dict :
   Return dictionary of attributes.

**dilation**(*=0.*)
   defines the normal displacement in the joint after sliding treshold. [m]

**dispHierarchy**(*(IPhys)arg1*[, *(bool)names=True*]) → list :
   Return list of dispatch classes (from down upwards), starting with the class instance itself, top-level indexable at last. If names is true (default), return class names rather than numerical indices.

**dispIndex**
   Return class index of this instance.

**elapsedIter**(*=0*)
   number of elapsed iterations for moment calculation

**eventBeginTime**(*=0*)
   The time at which event initiated

**eventNumber**(*=0*)
   cluster event number

**firstMomentCalc**(*=true*)
   Flag for moment calculation *(auto-updated)*

**initD**(*=0.*)
   equilibrium distance for interacting particles. Computed as the interparticular distance at first contact detection.

**interactionsAdded**(*=false*)
   have we added the ints associated with this event?

**isBroken**(*=false*)
   flag for broken interactions

**isCohesive**(*=false*)
   If false, particles interact in a frictional way. If true, particles are bonded regarding the given *cohesion* and *tensile strength* (or their jointed variants).

**isOnJoint**(*=false*)
   defined as true when both interacting particles are *on joint* and are in opposite sides of the joint surface. In this case, mechanical parameters of the interaction are derived from the ''joint...'' material properties of the particles. Furthermore, the normal of the interaction may be re-oriented (see *Law2_ScGeom_JCFpmPhys_JointedCohesiveFrictionalPM.smoothJoint*).

**isOnSlot**(*=false*)
   defined as true when interaction is located in the perforation slot (surface).

**jointCumulativeSliding**(*=0.*)
   sliding distance for particles interacting on a joint. Used, when  is true, to take into account dilatancy due to shearing. [-]

**jointNormal**(*=Vector3r::Zero()*)
    normal direction to the joint, deduced from e.g. .

**kineticEnergy**(*=0*)
    kinetic energy of the two spheres participating in the interaction (easiest to store this value with interaction instead of spheres since we are using this information for moment magnitude estimations and associated interaction searches)

**kn**(*=0*)
    Normal stiffness

**ks**(*=0*)
    Shear stiffness

**momentBroken**(*=false*)
    Flag for moment calculation

**momentCalculated**(*=false*)
    Flag for moment calculation to avoid repeating twice the operations *(auto-updated)*

**momentCentroid**(*=Vector3r::Zero()*)
    centroid of the AE event (avg location of clustered breaks)

**momentEnergy**(*=0*)
    reference strain (or kinetic) energy of surrounding interactions (particles)

**momentEnergyChange**(*=0*)
    storage of the maximum strain (or kinetic) energy change for surrounding interactions (particles)

**momentMagnitude**(*=0*)
    Moment magnitude of a failed interaction

**more**(*=false*)
    specifies if the interaction is crossed by more than 3 joints. If true, interaction is deleted (temporary solution).

**nearbyFound**(*=0*)
    Count used to debug moment calc

**nearbyInts**(*=uninitalized*)
    vector of pointers to the nearby ints used for moment calc

**normalForce**(*=Vector3r::Zero()*)
    Normal force after previous step (in global coordinates), as sustained by particle #2 (from particle #1).

**originalClusterEvent**(*=false*)
    the original AE event for a cluster

**originalEvent**(*=uninitalized*)
    pointer to the original interaction of a cluster

**shearForce**(*=Vector3r::Zero()*)
    Shear force after previous step (in global coordinates), as sustained by particle #2 (from particle #1).

**strainEnergy**(*=0*)
    strain energy of interaction

**tanDilationAngle**(*=0.*)
    tangent of the angle defining the dilatancy of the joint surface (auto. computed from *JCFp-mMat.jointDilationAngle*). [-]

**tanFrictionAngle**(*=0.*)
    tangent of Coulomb friction angle for this interaction (auto. computed). [-]

**temporalWindow**(*=0*)
    temporal window for the clustering algorithm

**updateAttrs**(*(Serializable)arg1, (dict)arg2*) → None :
    Update object attributes from given dictionary

**class yade.wrapper.LubricationPhys**(*inherits* *ViscElPhys* → *FrictPhys* → *NormShearPhys* → *NormPhys* → *IPhys* → *Serializable*)
    IPhys class for Lubrication w/o FlowEngine. Used by Law2_ScGeom_ImplicitLubricationPhys.

**Fn**(*=0.0*)
    Normal force of the contact

**Fv**(*=0.0*)
    Viscous force of the contact

**a**(*=0.*)
    Mean radius [m]

**cn**(*=NaN*)
    Normal viscous constant

**contact**(*=false*)
    The spheres are in contact

**cs**(*=NaN*)
    Shear viscous constant

**delta**(*=0*)
    log($\mathfrak{u}$) - used for scheme with $\delta = \log(\mathfrak{u})$ variable change

**dict**(*(Serializable)arg1*) → dict :
    Return dictionary of attributes.

**dispHierarchy**(*(IPhys)arg1*[, *(bool)names=True*]) → list :
    Return list of dispatch classes (from down upwards), starting with the class instance itself, top-level indexable at last. If names is true (default), return class names rather than numerical indices.

**dispIndex**
    Return class index of this instance.

**eps**(*=0.001*)
    Roughness: fraction of radius used as roughness [-]

**eta**(*=1*)
    Fluid viscosity [Pa.s]

**keps**(*=1*)
    stiffness coefficient of the asperities [N/m]. Only used with resolution method=0, with resolution>0 it is always equal to kn.

**kn**(*=0*)
    Normal stiffness

**kno**(*=0.0*)
    Coefficient for normal stiffness (Hertzian-like contact) [N/m^(3/2)]

**ks**(*=0*)
    Shear stiffness

**mR**(*=0.0*)
    Rolling resistance, see [Zhou1999536].

**mRtype**(*=1*)
    Rolling resistance type, see [Zhou1999536]. mRtype=1 - equation (3) in [Zhou1999536]; mRtype=2 - equation (4) in [Zhou1999536]

**mum**(*=0.3*)
>    Friction coefficient [-]

**normalContactForce**(*=Vector3r::Zero()*)
>    Normal contact force [N]

**normalForce**(*=Vector3r::Zero()*)
>    Normal force after previous step (in global coordinates), as sustained by particle #2 (from particle #1).

**normalLubricationForce**(*=Vector3r::Zero()*)
>    Normal lubrication force [N]

**normalPotentialForce**(*=Vector3r::Zero()*)
>    Normal force from potential other than contact [N]

**nun**(*=0.0*)
>    Coefficient for normal lubrication [N.s]

**prevDotU**(*=0*)
>    du/dt from previous integration - used for trapezoidal scheme (see *Law2_ScGeom_ImplicitLubricationPhys::resolution* for choosing resolution scheme)

**prev_un**(*=0*)
>    Nondeformed distance (un) at t-dt [m]

**shearContactForce**(*=Vector3r::Zero()*)
>    Frictional contact force [N]

**shearForce**(*=Vector3r::Zero()*)
>    Shear force after previous step (in global coordinates), as sustained by particle #2 (from particle #1).

**shearLubricationForce**(*=Vector3r::Zero()*)
>    Shear lubrication force [N]

**slip**(*=false*)
>    The contact is slipping

**tangensOfFrictionAngle**(*=NaN*)
>    tan of angle of friction

**u**(*=-1*)
>    Interfacial distance (u) at t-dt [m]

**ue**(*=0.*)
>    Surface deflection (ue) at t-dt [m]

**updateAttrs**(*(Serializable)arg1, (dict)arg2*) → None :
>    Update object attributes from given dictionary

**class yade.wrapper.LudingPhys**(*inherits FrictPhys → NormShearPhys → NormPhys → IPhys → Serializable*)
>    IPhys created from *LudingMat*, for use with *Law2_ScGeom_LudingPhys_Basic*.

**DeltMax**(*=NaN*)
>    Maximum overlap between particles for a collision

**DeltMin**(*=NaN*)
>    MinimalDelta value of delta

**DeltNull**(*=NaN*)
>    Force free overlap, plastic contact deformation

**DeltPMax**(*=NaN*)
>    Maximum overlap between particles for the limit case

**DeltPNull**(*=NaN*)
>    Max force free overlap, plastic contact deformation

**DeltPrev**(*=NaN*)
> Previous value of delta

**G0**(*=NaN*)
> Viscous damping

**PhiF**(*=NaN*)
> Dimensionless plasticity depth

**dict**(*(Serializable)arg1*) → dict :
> Return dictionary of attributes.

**dispHierarchy**(*(IPhys)arg1*[, *(bool)names=True*]) → list :
> Return list of dispatch classes (from down upwards), starting with the class instance itself, top-level indexable at last. If names is true (default), return class names rather than numerical indices.

**dispIndex**
> Return class index of this instance.

**k1**(*=NaN*)
> Slope of loading plastic branch

**k2**(*=NaN*)
> Slope of unloading and reloading elastic branch

**kc**(*=NaN*)
> Slope of irreversible, tensile adhesive branch

**kn**(*=0*)
> Normal stiffness

**kp**(*=NaN*)
> Slope of unloading and reloading limit elastic branch

**ks**(*=0*)
> Shear stiffness

**normalForce**(*=Vector3r::Zero()*)
> Normal force after previous step (in global coordinates), as sustained by particle #2 (from particle #1).

**shearForce**(*=Vector3r::Zero()*)
> Shear force after previous step (in global coordinates), as sustained by particle #2 (from particle #1).

**tangensOfFrictionAngle**(*=NaN*)
> tan of angle of friction

**updateAttrs**(*(Serializable)arg1, (dict)arg2*) → None :
> Update object attributes from given dictionary

**class yade.wrapper.MindlinCapillaryPhys**(*inherits* *MindlinPhys* → *RotStiffFrictPhys* → *FrictPhys* → *NormShearPhys* → *NormPhys* → *IPhys* → *Serializable*)

Adds capillary physics to Mindlin's interaction physics.

**Delta1**(*=0.*)
> Defines the surface area wetted by the meniscus on the smallest grains of radius R1 (R1<R2)

**Delta2**(*=0.*)
> Defines the surface area wetted by the meniscus on the biggest grains of radius R2 (R1<R2)

**Fs**(*=Vector2r::Zero()*)
> Shear force in local axes (computed incrementally)

**adhesionForce**(*=0.0*)
> Force of adhesion as predicted by DMT

**alpha**(*=0.0*)

Constant coefficient to define contact viscous damping for non-linear elastic force-displacement relationship.

**betan**(*=0.0*)

Normal Damping Ratio. Fraction of the viscous damping coefficient (normal direction) equal to $\frac{c_n}{C_{n,crit}}$.

**betas**(*=0.0*)

Shear Damping Ratio. Fraction of the viscous damping coefficient (shear direction) equal to $\frac{c_s}{C_{s,crit}}$.

**capillaryPressure**(*=0.*)

Value of the capillary pressure Uc. Defined as Ugas-Uliquid, obtained from *corresponding Law2 parameter*

**dict**(*(Serializable)arg1*) → dict :

Return dictionary of attributes.

**dispHierarchy**(*(IPhys)arg1*[, *(bool)names=True*]) → list :

Return list of dispatch classes (from down upwards), starting with the class instance itself, top-level indexable at last. If names is true (default), return class names rather than numerical indices.

**dispIndex**

Return class index of this instance.

**fCap**(*=Vector3r::Zero()*)

Capillary Force produces by the presence of the meniscus. This is the force acting on particle #2

**fusionNumber**(*=0.*)

Indicates the number of meniscii that overlap with this one

**isAdhesive**(*=false*)

bool to identify if the contact is adhesive, that is to say if the contact force is attractive

**isBroken**(*=false*)

Might be set to true by the user to make liquid bridge inactive (capillary force is zero)

**isSliding**(*=false*)

check if the contact is sliding (useful to calculate the ratio of sliding contacts)

**kn**(*=0*)

Normal stiffness

**kno**(*=0.0*)

Constant value in the formulation of the normal stiffness

**kr**(*=0*)

rotational stiffness [N.m/rad]

**ks**(*=0*)

Shear stiffness

**kso**(*=0.0*)

Constant value in the formulation of the tangential stiffness

**ktw**(*=0*)

twist stiffness [N.m/rad]

**maxBendPl**(*=0.0*)

Coefficient to determine the maximum plastic moment to apply at the contact

**meniscus**(*=false*)

True when a meniscus with a non-zero liquid volume (*vMeniscus*) has been computed for this interaction

**momentBend**(=*Vector3r::Zero()*)
  Artificial bending moment to provide rolling resistance in order to account for some degree of interlocking between particles

**momentTwist**(=*Vector3r::Zero()*)
  Artificial twisting moment (no plastic condition can be applied at the moment)

**normalForce**(=*Vector3r::Zero()*)
  Normal force after previous step (in global coordinates), as sustained by particle #2 (from particle #1).

**normalViscous**(=*Vector3r::Zero()*)
  Normal viscous component

**prevU**(=*Vector3r::Zero()*)
  Previous local displacement; only used with *Law2_L3Geom_FrictPhys_HertzMindlin.*

**radius**(=*NaN*)
  Contact radius (only computed with *Law2_ScGeom_MindlinPhys_Mindlin::calcEnergy*)

**shearElastic**(=*Vector3r::Zero()*)
  Total elastic shear force

**shearForce**(=*Vector3r::Zero()*)
  Shear force after previous step (in global coordinates), as sustained by particle #2 (from particle #1).

**shearViscous**(=*Vector3r::Zero()*)
  Shear viscous component

**tangensOfFrictionAngle**(=*NaN*)
  tan of angle of friction

**updateAttrs**(*(Serializable)arg1, (dict)arg2*) → None :
  Update object attributes from given dictionary

**usElastic**(=*Vector3r::Zero()*)
  Total elastic shear displacement (only elastic part)

**usTotal**(=*Vector3r::Zero()*)
  Total elastic shear displacement (elastic+plastic part)

**vMeniscus**(=*0.*)
  Volume of the meniscus

**class yade.wrapper.MindlinPhys**(*inherits RotStiffFrictPhys → FrictPhys → NormShearPhys → NormPhys → IPhys → Serializable*)
  Representation of an interaction of the Hertz-Mindlin type.

**Fs**(=*Vector2r::Zero()*)
  Shear force in local axes (computed incrementally)

**adhesionForce**(=*0.0*)
  Force of adhesion as predicted by DMT

**alpha**(=*0.0*)
  Constant coefficient to define contact viscous damping for non-linear elastic force-displacement relationship.

**betan**(=*0.0*)
  Normal Damping Ratio. Fraction of the viscous damping coefficient (normal direction) equal to $\frac{c_n}{C_{n,crit}}$.

**betas**(=*0.0*)
  Shear Damping Ratio. Fraction of the viscous damping coefficient (shear direction) equal to $\frac{c_s}{C_{s,crit}}$.

**dict**(*(Serializable)arg1*) → dict :
    Return dictionary of attributes.

**dispHierarchy**(*(IPhys)arg1*[, *(bool)names=True*]) → list :
    Return list of dispatch classes (from down upwards), starting with the class instance itself, top-level indexable at last. If names is true (default), return class names rather than numerical indices.

**dispIndex**
    Return class index of this instance.

**isAdhesive**(*=false*)
    bool to identify if the contact is adhesive, that is to say if the contact force is attractive

**isSliding**(*=false*)
    check if the contact is sliding (useful to calculate the ratio of sliding contacts)

**kn**(*=0*)
    Normal stiffness

**kno**(*=0.0*)
    Constant value in the formulation of the normal stiffness

**kr**(*=0*)
    rotational stiffness [N.m/rad]

**ks**(*=0*)
    Shear stiffness

**kso**(*=0.0*)
    Constant value in the formulation of the tangential stiffness

**ktw**(*=0*)
    twist stiffness [N.m/rad]

**maxBendPl**(*=0.0*)
    Coefficient to determine the maximum plastic moment to apply at the contact

**momentBend**(*=Vector3r::Zero()*)
    Artificial bending moment to provide rolling resistance in order to account for some degree of interlocking between particles

**momentTwist**(*=Vector3r::Zero()*)
    Artificial twisting moment (no plastic condition can be applied at the moment)

**normalForce**(*=Vector3r::Zero()*)
    Normal force after previous step (in global coordinates), as sustained by particle #2 (from particle #1).

**normalViscous**(*=Vector3r::Zero()*)
    Normal viscous component

**prevU**(*=Vector3r::Zero()*)
    Previous local displacement; only used with *Law2_L3Geom_FrictPhys_HertzMindlin*.

**radius**(*=NaN*)
    Contact radius (only computed with *Law2_ScGeom_MindlinPhys_Mindlin::calcEnergy*)

**shearElastic**(*=Vector3r::Zero()*)
    Total elastic shear force

**shearForce**(*=Vector3r::Zero()*)
    Shear force after previous step (in global coordinates), as sustained by particle #2 (from particle #1).

**shearViscous**(*=Vector3r::Zero()*)
    Shear viscous component

**tangensOfFrictionAngle**(*=NaN*)
> tan of angle of friction

**updateAttrs**(*(Serializable)arg1, (dict)arg2*) → None :
> Update object attributes from given dictionary

**usElastic**(*=Vector3r::Zero()*)
> Total elastic shear displacement (only elastic part)

**usTotal**(*=Vector3r::Zero()*)
> Total elastic shear displacement (elastic+plastic part)

**class yade.wrapper.MindlinPhysCDM**(*inherits MindlinPhys → RotStiffFrictPhys → FrictPhys*
> *→ NormShearPhys → NormPhys → IPhys → Serializable*)

Representation of an interaction of an extended Hertz-Mindlin type. Normal direction: parameters for Conical Damage Model (Harkness et al. 2016, Suhr & Six 2017). Tangential direction: parameters for stress dependent interparticle friction coefficient (Suhr & Six 2016). Both models can be switched on/off separately, see FrictMatCDM.

**E**(*=0.0*)
> [Pa] equiv. Young's modulus

**Fs**(*=Vector2r::Zero()*)
> Shear force in local axes (computed incrementally)

**G**(*=0.0*)
> [Pa] equiv. shear modulus

**R**(*=0.0*)
> [m] contact radius in conical damage model

**adhesionForce**(*=0.0*)
> Force of adhesion as predicted by DMT

**alpha**(*=0.0*)
> Constant coefficient to define contact viscous damping for non-linear elastic force-displacement relationship.

**alphaFac**(*=0.0*)
> factor considering angle of conical asperities

**betan**(*=0.0*)
> Normal Damping Ratio. Fraction of the viscous damping coefficient (normal direction) equal to $\frac{c_n}{C_{n,crit}}$.

**betas**(*=0.0*)
> Shear Damping Ratio. Fraction of the viscous damping coefficient (shear direction) equal to $\frac{c_s}{C_{s,crit}}$.

**c1**(*=0.0*)
> [-] parameter of pressure dependent friction model c1

**c2**(*=0.0*)
> [-] parameter of pressure dependent friction model c2

**dict**(*(Serializable)arg1*) → dict :
> Return dictionary of attributes.

**dispHierarchy**(*(IPhys)arg1*[, *(bool)names=True*]) → list :
> Return list of dispatch classes (from down upwards), starting with the class instance itself, top-level indexable at last. If names is true (default), return class names rather than numerical indices.

**dispIndex**
> Return class index of this instance.

**isAdhesive**(=*false*)
    bool to identify if the contact is adhesive, that is to say if the contact force is attractive

**isSliding**(=*false*)
    check if the contact is sliding (useful to calculate the ratio of sliding contacts)

**isYielding**(=*false*)
    bool: is contact currently yielding?

**kn**(=*0*)
    Normal stiffness

**kno**(=*0.0*)
    Constant value in the formulation of the normal stiffness

**kr**(=*0*)
    rotational stiffness [N.m/rad]

**ks**(=*0*)
    Shear stiffness

**kso**(=*0.0*)
    Constant value in the formulation of the tangential stiffness

**ktw**(=*0*)
    twist stiffness [N.m/rad]

**maxBendPl**(=*0.0*)
    Coefficient to determine the maximum plastic moment to apply at the contact

**momentBend**(=*Vector3r::Zero()*)
    Artificial bending moment to provide rolling resistance in order to account for some degree of interlocking between particles

**momentTwist**(=*Vector3r::Zero()*)
    Artificial twisting moment (no plastic condition can be applied at the moment)

**mu0**(=*0.0*)
    [-] parameter of pressure dependent friction model mu0

**normalForce**(=*Vector3r::Zero()*)
    Normal force after previous step (in global coordinates), as sustained by particle #2 (from particle #1).

**normalViscous**(=*Vector3r::Zero()*)
    Normal viscous component

**prevU**(=*Vector3r::Zero()*)
    Previous local displacement; only used with *Law2_L3Geom_FrictPhys_HertzMindlin*.

**radius**(=*NaN*)
    Contact radius (only computed with *Law2_ScGeom_MindlinPhys_Mindlin::calcEnergy*)

**shearElastic**(=*Vector3r::Zero()*)
    Total elastic shear force

**shearForce**(=*Vector3r::Zero()*)
    Shear force after previous step (in global coordinates), as sustained by particle #2 (from particle #1).

**shearViscous**(=*Vector3r::Zero()*)
    Shear viscous component

**sigmaMax**(=*0.0*)
    [Pa] max compressive strength of material

**tangensOfFrictionAngle**(=*NaN*)
    tan of angle of friction

**updateAttrs**(*(Serializable)arg1, (dict)arg2*) → None :
    Update object attributes from given dictionary

**usElastic**(*=Vector3r::Zero()*)
    Total elastic shear displacement (only elastic part)

**usTotal**(*=Vector3r::Zero()*)
    Total elastic shear displacement (elastic+plastic part)

**class yade.wrapper.MortarPhys**(*inherits FrictPhys → NormShearPhys → NormPhys → IPhys → Serializable*)
    IPhys class containing parameters of MortarMat. Used by Law2_ScGeom_MortarPhys_Lourenco.

**cohesion**(*=NaN*)
    cohesion [Pa]

**compressiveStrength**(*=NaN*)
    compressiveStrength [Pa]

**crossSection**(*=NaN*)
    Crosssection of interaction

**dict**(*(Serializable)arg1*) → dict :
    Return dictionary of attributes.

**dispHierarchy**(*(IPhys)arg1*[, *(bool)names=True*]) → list :
    Return list of dispatch classes (from down upwards), starting with the class instance itself, top-level indexable at last. If names is true (default), return class names rather than numerical indices.

**dispIndex**
    Return class index of this instance.

**ellAspect**(*=NaN*)
    aspect ratio of elliptical 'cap'. Value >1 means the ellipse is longer along normal stress axis.

**failureCondition**(*(MortarPhys)arg1, (float)arg2, (float)arg3*) → bool :
    Failure condition from normal stress and norm of shear stress (false=elastic, true=damaged)

**kn**(*=0*)
    Normal stiffness

**ks**(*=0*)
    Shear stiffness

**neverDamage**(*=false*)
    If true, interactions remain elastic regardless stresses

**normalForce**(*=Vector3r::Zero()*)
    Normal force after previous step (in global coordinates), as sustained by particle #2 (from particle #1).

**shearForce**(*=Vector3r::Zero()*)
    Shear force after previous step (in global coordinates), as sustained by particle #2 (from particle #1).

**sigmaN**
    Current normal stress *(auto-updated)*

**sigmaT**
    Current shear stress *(auto-updated)*

**tangensOfFrictionAngle**(*=NaN*)
    tan of angle of friction

**tensileStrength**(*=NaN*)
    tensileStrength [Pa]

> **updateAttrs**(*(Serializable)arg1, (dict)arg2*) → None :
> > Update object attributes from given dictionary

**class yade.wrapper.NormPhys**(*inherits IPhys → Serializable*)
> Abstract class for interactions that have normal stiffness.

> **dict**(*(Serializable)arg1*) → dict :
> > Return dictionary of attributes.

> **dispHierarchy**(*(IPhys)arg1*[, *(bool)names=True*]) → list :
> > Return list of dispatch classes (from down upwards), starting with the class instance itself,
> > top-level indexable at last. If names is true (default), return class names rather than numerical
> > indices.

> **dispIndex**
> > Return class index of this instance.

> **kn**(*=0*)
> > Normal stiffness

> **normalForce**(*=Vector3r::Zero()*)
> > Normal force after previous step (in global coordinates), as sustained by particle #2 (from
> > particle #1).

> **updateAttrs**(*(Serializable)arg1, (dict)arg2*) → None :
> > Update object attributes from given dictionary

**class yade.wrapper.NormShearPhys**(*inherits NormPhys → IPhys → Serializable*)
> Abstract class for interactions that have shear stiffnesses, in addition to normal stiffness. This class
> is used in the PFC3d-style *stiffness timestepper*.

> **dict**(*(Serializable)arg1*) → dict :
> > Return dictionary of attributes.

> **dispHierarchy**(*(IPhys)arg1*[, *(bool)names=True*]) → list :
> > Return list of dispatch classes (from down upwards), starting with the class instance itself,
> > top-level indexable at last. If names is true (default), return class names rather than numerical
> > indices.

> **dispIndex**
> > Return class index of this instance.

> **kn**(*=0*)
> > Normal stiffness

> **ks**(*=0*)
> > Shear stiffness

> **normalForce**(*=Vector3r::Zero()*)
> > Normal force after previous step (in global coordinates), as sustained by particle #2 (from
> > particle #1).

> **shearForce**(*=Vector3r::Zero()*)
> > Shear force after previous step (in global coordinates), as sustained by particle #2 (from
> > particle #1).

> **updateAttrs**(*(Serializable)arg1, (dict)arg2*) → None :
> > Update object attributes from given dictionary

**class yade.wrapper.RotStiffFrictPhys**(*inherits FrictPhys → NormShearPhys → NormPhys*
> > *→ IPhys → Serializable*)
> Version of *FrictPhys* with a rotational stiffness

> **dict**(*(Serializable)arg1*) → dict :
> > Return dictionary of attributes.

**dispHierarchy**(*(IPhys)arg1*[, *(bool)names=True*]) → list :
> Return list of dispatch classes (from down upwards), starting with the class instance itself, top-level indexable at last. If names is true (default), return class names rather than numerical indices.

**dispIndex**
> Return class index of this instance.

**kn**(*=0*)
> Normal stiffness

**kr**(*=0*)
> rotational stiffness [N.m/rad]

**ks**(*=0*)
> Shear stiffness

**ktw**(*=0*)
> twist stiffness [N.m/rad]

**normalForce**(*=Vector3r::Zero()*)
> Normal force after previous step (in global coordinates), as sustained by particle #2 (from particle #1).

**shearForce**(*=Vector3r::Zero()*)
> Shear force after previous step (in global coordinates), as sustained by particle #2 (from particle #1).

**tangensOfFrictionAngle**(*=NaN*)
> tan of angle of friction

**updateAttrs**(*(Serializable)arg1, (dict)arg2*) → None :
> Update object attributes from given dictionary

**class yade.wrapper.ViscElCapPhys**(*inherits* *ViscElPhys → FrictPhys → NormShearPhys →*
*NormPhys → IPhys → Serializable*)
> IPhys created from *ViscElCapMat*, for use with *Law2_ScGeom_ViscElCapPhys_Basic*.

**Capillar**(*=false*)
> True, if capillar forces need to be added.

**CapillarType**(*=None_Capillar*)
> Different types of capillar interaction: Willett_numeric, Willett_analytic, Weigert, Rabinovich, Lambert, Soulie

**Fn**(*=0.0*)
> Normal force of the contact

**Fv**(*=0.0*)
> Viscous force of the contact

**Vb**(*=0.0*)
> Liquid bridge volume [m^3]

**cn**(*=NaN*)
> Normal viscous constant

**cs**(*=NaN*)
> Shear viscous constant

**dcap**(*=0.0*)
> Damping coefficient for the capillary phase [-]

**dict**(*(Serializable)arg1*) → dict :
> Return dictionary of attributes.

**dispHierarchy**(*(IPhys)arg1*[, *(bool)names=True*]) → list :
> Return list of dispatch classes (from down upwards), starting with the class instance itself,

top-level indexable at last. If names is true (default), return class names rather than numerical indices.

**dispIndex**
    Return class index of this instance.

**gamma**(*=0.0*)
    Surface tension [N/m]

**kn**(*=0*)
    Normal stiffness

**ks**(*=0*)
    Shear stiffness

**liqBridgeActive**(*=false*)
    Whether liquid bridge is active at the moment

**liqBridgeCreated**(*=false*)
    Whether liquid bridge was created, only after a normal contact of spheres

**mR**(*=0.0*)
    Rolling resistance, see [Zhou1999536].

**mRtype**(*=1*)
    Rolling resistance type, see [Zhou1999536]. mRtype=1 - equation (3) in [Zhou1999536]; mRtype=2 - equation (4) in [Zhou1999536]

**normalForce**(*=Vector3r::Zero()*)
    Normal force after previous step (in global coordinates), as sustained by particle #2 (from particle #1).

**sCrit**(*=false*)
    Critical bridge length [m]

**shearForce**(*=Vector3r::Zero()*)
    Shear force after previous step (in global coordinates), as sustained by particle #2 (from particle #1).

**tangensOfFrictionAngle**(*=NaN*)
    tan of angle of friction

**theta**(*=0.0*)
    Contact angle [rad]

**updateAttrs**(*(Serializable)arg1, (dict)arg2*) → None :
    Update object attributes from given dictionary

**class yade.wrapper.ViscElPhys**(*inherits FrictPhys → NormShearPhys → NormPhys → IPhys → Serializable*)
    IPhys created from *ViscElMat*, for use with *Law2_ScGeom_ViscElPhys_Basic*.

**Fn**(*=0.0*)
    Normal force of the contact

**Fv**(*=0.0*)
    Viscous force of the contact

**cn**(*=NaN*)
    Normal viscous constant

**cs**(*=NaN*)
    Shear viscous constant

**dict**(*(Serializable)arg1*) → dict :
    Return dictionary of attributes.

**dispHierarchy**(*(IPhys)arg1*[, *(bool)names=True*]) → list :
Return list of dispatch classes (from down upwards), starting with the class instance itself, top-level indexable at last. If names is true (default), return class names rather than numerical indices.

**dispIndex**
Return class index of this instance.

**kn**(*=0*)
Normal stiffness

**ks**(*=0*)
Shear stiffness

**mR**(*=0.0*)
Rolling resistance, see [Zhou1999536].

**mRtype**(*=1*)
Rolling resistance type, see [Zhou1999536]. mRtype=1 - equation (3) in [Zhou1999536]; mRtype=2 - equation (4) in [Zhou1999536]

**normalForce**(*=Vector3r::Zero()*)
Normal force after previous step (in global coordinates), as sustained by particle #2 (from particle #1).

**shearForce**(*=Vector3r::Zero()*)
Shear force after previous step (in global coordinates), as sustained by particle #2 (from particle #1).

**tangensOfFrictionAngle**(*=NaN*)
tan of angle of friction

**updateAttrs**(*(Serializable)arg1*, *(dict)arg2*) → None :
Update object attributes from given dictionary

**class yade.wrapper.ViscoFrictPhys**(*inherits FrictPhys → NormShearPhys → NormPhys →
IPhys → Serializable*)
Temporary version of *FrictPhys* for compatibility reasons

**creepedShear**(*=Vector3r(0, 0, 0)*)
Creeped force (parallel)

**dict**(*(Serializable)arg1*) → dict :
Return dictionary of attributes.

**dispHierarchy**(*(IPhys)arg1*[, *(bool)names=True*]) → list :
Return list of dispatch classes (from down upwards), starting with the class instance itself, top-level indexable at last. If names is true (default), return class names rather than numerical indices.

**dispIndex**
Return class index of this instance.

**kn**(*=0*)
Normal stiffness

**ks**(*=0*)
Shear stiffness

**normalForce**(*=Vector3r::Zero()*)
Normal force after previous step (in global coordinates), as sustained by particle #2 (from particle #1).

**shearForce**(*=Vector3r::Zero()*)
Shear force after previous step (in global coordinates), as sustained by particle #2 (from particle #1).

**tangensOfFrictionAngle**(*=NaN*)
  tan of angle of friction

**updateAttrs**(*(Serializable)arg1*, *(dict)arg2*) → None :
  Update object attributes from given dictionary

**class yade.wrapper.WirePhys**(*inherits FrictPhys → NormShearPhys → NormPhys → IPhys → Serializable*)
  Representation of a single interaction of the WirePM type, storage for relevant parameters

**dL**(*=0.*)
  Additional wire length for considering the distortion for *WireMat* type=2 (see [Thoeni2013]).

**dict**(*(Serializable)arg1*) → dict :
  Return dictionary of attributes.

**dispHierarchy**(*(IPhys)arg1*[, *(bool)names=True*]) → list :
  Return list of dispatch classes (from down upwards), starting with the class instance itself, top-level indexable at last. If names is true (default), return class names rather than numerical indices.

**dispIndex**
  Return class index of this instance.

**displForceValues**(*=uninitalized*)
  Defines the values for force-displacement curve.

**initD**(*=0.*)
  Equilibrium distance for particles. Computed as the initial inter-particular distance when particle are linked.

**isDoubleTwist**(*=false*)
  If true the properties of the interaction will be defined as a double-twisted wire.

**isLinked**(*=false*)
  If true particles are linked and will interact. Interactions are linked automatically by the definition of the corresponding interaction radius. The value is false if the wire breaks (no more interaction).

**isShifted**(*=false*)
  If true *WireMat* type=2 and the force-displacement curve will be shifted.

**kn**(*=0*)
  Normal stiffness

**ks**(*=0*)
  Shear stiffness

**limitFactor**(*=0.*)
  This value indicates on how far from failing the wire is, e.g. actual normal displacement divided by admissible normal displacement.

**normalForce**(*=Vector3r::Zero()*)
  Normal force after previous step (in global coordinates), as sustained by particle #2 (from particle #1).

**plastD**
  Plastic part of the inter-particular distance of the previous step.

---

**Note:** Only elastic displacements are reversible (the elastic stiffness is used for unloading) and compressive forces are inadmissible. The compressive stiffness is assumed to be equal to zero.

---

**shearForce**(*=Vector3r::Zero()*)

    Shear force after previous step (in global coordinates), as sustained by particle #2 (from particle #1).

**stiffnessValues**(*=uninitalized*)

    Defines the values for the various stiffnesses (the elastic stiffness is stored as kn).

**tangensOfFrictionAngle**(*=NaN*)

    tan of angle of friction

**updateAttrs**(*(Serializable)arg1, (dict)arg2*) → None :

    Update object attributes from given dictionary

### 2.3.3 Global engines

**GlobalEngine**

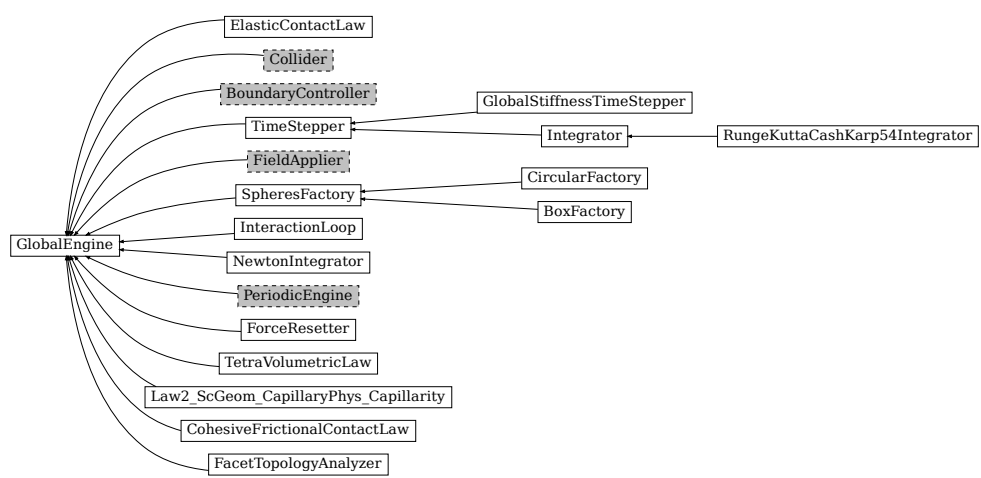

Fig. 25: Inheritance graph of GlobalEngine, gray dashed classes are discussed in their own sections: *Collider*, *BoundaryController*, *FieldApplier*, *PeriodicEngine*. See also: *BoxFactory*, *CircularFactory*, *CohesiveFrictionalContactLaw*, *ElasticContactLaw*, *FacetTopologyAnalyzer*, *ForceResetter*, *GlobalStiffnessTimeStepper*, *Integrator*, *InteractionLoop*, *Law2_ScGeom_CapillaryPhys_Capillarity*, *NewtonIntegrator*, *RungeKuttaCashKarp54Integrator*, *SpheresFactory*, *TetraVolumetricLaw*, *TimeStepper*.

**class yade.wrapper.GlobalEngine**(*inherits Engine → Serializable*)

    Engine that will generally affect the whole simulation (contrary to *PartialEngine*).

**dead**(*=false*)

    If true, this engine will not run at all; can be used for making an engine temporarily deactivated and only resurrect it at a later point.

**dict**(*(Serializable)arg1*) → dict :

    Return dictionary of attributes.

**execCount**

    Cumulative count this engine was run (only used if *O.timingEnabled*==True).

**execTime**

    Cumulative time in nanoseconds this Engine took to run (only used if *O.timingEnabled*==True).

**label**(*=uninitalized*)

    Textual label for this object; must be valid python identifier, you can refer to it directly from python.

**ompThreads**(*=-1*)

> Number of threads to be used in the engine. If ompThreads<0 (default), the number will be typically OMP_NUM_THREADS or the number N defined by 'yade -jN' (this behavior can depend on the engine though). This attribute will only affect engines whose code includes openMP parallel regions (e.g. *InteractionLoop*). This attribute is mostly useful for experiments or when combining *ParallelEngine* with engines that run parallel regions, resulting in nested OMP loops with different number of threads at each level.

**timingDeltas**

> Detailed information about timing inside the Engine itself. Empty unless enabled in the source code and *O.timingEnabled*==**True**.

**updateAttrs**(*(Serializable)arg1, (dict)arg2*) → None :

> Update object attributes from given dictionary

**class yade.wrapper.BoxFactory**(*inherits SpheresFactory → GlobalEngine → Engine → Serializable*)

> Box geometry of the SpheresFactory region, given by extents and center

**PSDcalculateMass**(*=true*)

> PSD-Input is in mass (true), otherwise the number of particles will be considered.

**PSDcum**(*=uninitalized*)

> PSD-dispersion, cumulative procent meanings [-]

**PSDsizes**(*=uninitalized*)

> PSD-dispersion, sizes of cells, Diameter [m]

**blockedDOFs**(*=""*)

> Blocked degress of freedom

**center**(*=Vector3r(NaN, NaN, NaN)*)

> Center of the region

**color**(*=Vector3r(-1, -1, -1)*)

> Use the color for newly created particles, if specified

**dead**(*=false*)

> If true, this engine will not run at all; can be used for making an engine temporarily deactivated and only resurrect it at a later point.

**dict**(*(Serializable)arg1*) → dict :

> Return dictionary of attributes.

**exactDiam**(*=true*)

> If true, the particles only with the defined in PSDsizes diameters will be created. Otherwise the diameter will be randomly chosen in the range [PSDsizes[i-1]:PSDsizes[i]], in this case the length of PSDsizes should be more on 1, than the length of PSDcum.

**execCount**

> Cumulative count this engine was run (only used if *O.timingEnabled*==**True**).

**execTime**

> Cumulative time in nanoseconds this Engine took to run (only used if *O.timingEnabled*==**True**).

**extents**(*=Vector3r(NaN, NaN, NaN)*)

> Extents of the region

**goalMass**(*=0*)

> Total mass that should be attained at the end of the current step. *(auto-updated)*

**ids**(*=uninitalized*)

> ids of created bodies

**label**(*=uninitalized*)
> Textual label for this object; must be valid python identifier, you can refer to it directly from python.

**mask**(*=-1*)
> groupMask to apply for newly created spheres

**massFlowRate**(*=NaN*)
> Mass flow rate [kg/s]

**materialId**(*=-1*)
> Shared material id to use for newly created spheres (can be negative to count from the end)

**maxAttempt**(*=5000*)
> Maximum number of attempts to position a new sphere randomly.

**maxMass**(*=-1*)
> Maximal mass at which to stop generating new particles regardless of massFlowRate. if maxMass=-1 - this parameter is ignored.

**maxParticles**(*=100*)
> The number of particles at which to stop generating new ones regardless of massFlowRate. if maxParticles=-1 - this parameter is ignored .

**normal**(*=Vector3r(NaN, NaN, NaN)*)
> Orientation of the region's geometry, direction of particle's velocites if normalVel is not set.

**normalVel**(*=Vector3r(NaN, NaN, NaN)*)
> Direction of particle's velocites.

**numParticles**(*=0*)
> Cummulative number of particles produces so far *(auto-updated)*

**ompThreads**(*=-1*)
> Number of threads to be used in the engine. If ompThreads<0 (default), the number will be typically OMP_NUM_THREADS or the number N defined by 'yade -jN' (this behavior can depend on the engine though). This attribute will only affect engines whose code includes openMP parallel regions (e.g. *InteractionLoop*). This attribute is mostly useful for experiments or when combining *ParallelEngine* with engines that run parallel regions, resulting in nested OMP loops with different number of threads at each level.

**rMax**(*=NaN*)
> Maximum radius of generated spheres (uniform distribution)

**rMin**(*=NaN*)
> Minimum radius of generated spheres (uniform distribution)

**silent**(*=false*)
> If true no complain about excessing maxAttempt but disable the factory (by set massFlowRate=0).

**stopIfFailed**(*=true*)
> If true, the SpheresFactory stops (sets massFlowRate=0), when maximal number of attempts to insert particle exceed.

**timingDeltas**
> Detailed information about timing inside the Engine itself. Empty unless enabled in the source code and *O.timingEnabled*==`True`.

**totalMass**(*=0*)
> Mass of spheres that was produced so far. *(auto-updated)*

**totalVolume**(*=0*)
> Volume of spheres that was produced so far. *(auto-updated)*

**updateAttrs**(*(Serializable)arg1, (dict)arg2*) → None :
> Update object attributes from given dictionary

**vAngle**(*=NaN*)
    Maximum angle by which the initial sphere velocity deviates from the normal.

**vMax**(*=NaN*)
    Maximum velocity norm of generated spheres (uniform distribution)

**vMin**(*=NaN*)
    Minimum velocity norm of generated spheres (uniform distribution)

**class yade.wrapper.CircularFactory**(*inherits SpheresFactory → GlobalEngine → Engine → Serializable*)
    Circular geometry of the SpheresFactory region. It can be disk (given by radius and center), or cylinder (given by radius, length and center).

**PSDcalculateMass**(*=true*)
    PSD-Input is in mass (true), otherwise the number of particles will be considered.

**PSDcum**(*=uninitalized*)
    PSD-dispersion, cumulative procent meanings [-]

**PSDsizes**(*=uninitalized*)
    PSD-dispersion, sizes of cells, Diameter [m]

**blockedDOFs**(*=""*)
    Blocked degress of freedom

**center**(*=Vector3r(NaN, NaN, NaN)*)
    Center of the region

**color**(*=Vector3r(-1, -1, -1)*)
    Use the color for newly created particles, if specified

**dead**(*=false*)
    If true, this engine will not run at all; can be used for making an engine temporarily deactivated and only resurrect it at a later point.

**dict**(*(Serializable)arg1*) → dict :
    Return dictionary of attributes.

**exactDiam**(*=true*)
    If true, the particles only with the defined in PSDsizes diameters will be created. Otherwise the diameter will be randomly chosen in the range [PSDsizes[i-1]:PSDsizes[i]], in this case the length of PSDsizes should be more on 1, than the length of PSDcum.

**execCount**
    Cumulative count this engine was run (only used if *O.timingEnabled*==True).

**execTime**
    Cumulative time in nanoseconds this Engine took to run (only used if *O.timingEnabled*==True).

**goalMass**(*=0*)
    Total mass that should be attained at the end of the current step. *(auto-updated)*

**ids**(*=uninitalized*)
    ids of created bodies

**label**(*=uninitalized*)
    Textual label for this object; must be valid python identifier, you can refer to it directly from python.

**length**(*=0*)
    Length of the cylindrical region (0 by default)

**mask**(*=-1*)
    groupMask to apply for newly created spheres

**massFlowRate** (*=NaN*)

    Mass flow rate [kg/s]

**materialId** (*=-1*)

    Shared material id to use for newly created spheres (can be negative to count from the end)

**maxAttempt** (*=5000*)

    Maximum number of attempts to position a new sphere randomly.

**maxMass** (*=-1*)

    Maximal mass at which to stop generating new particles regardless of massFlowRate. if maxMass=-1 - this parameter is ignored.

**maxParticles** (*=100*)

    The number of particles at which to stop generating new ones regardless of massFlowRate. if maxParticles=-1 - this parameter is ignored .

**normal** (*=Vector3r(NaN, NaN, NaN)*)

    Orientation of the region's geometry, direction of particle's velocites if normalVel is not set.

**normalVel** (*=Vector3r(NaN, NaN, NaN)*)

    Direction of particle's velocites.

**numParticles** (*=0*)

    Cummulative number of particles produces so far *(auto-updated)*

**ompThreads** (*=-1*)

    Number of threads to be used in the engine. If ompThreads<0 (default), the number will be typically OMP_NUM_THREADS or the number N defined by 'yade -jN' (this behavior can depend on the engine though). This attribute will only affect engines whose code includes openMP parallel regions (e.g. *InteractionLoop*). This attribute is mostly useful for experiments or when combining *ParallelEngine* with engines that run parallel regions, resulting in nested OMP loops with different number of threads at each level.

**rMax** (*=NaN*)

    Maximum radius of generated spheres (uniform distribution)

**rMin** (*=NaN*)

    Minimum radius of generated spheres (uniform distribution)

**radius** (*=NaN*)

    Radius of the region

**silent** (*=false*)

    If true no complain about excessing maxAttempt but disable the factory (by set massFlowRate=0).

**stopIfFailed** (*=true*)

    If true, the SpheresFactory stops (sets massFlowRate=0), when maximal number of attempts to insert particle exceed.

**timingDeltas**

    Detailed information about timing inside the Engine itself. Empty unless enabled in the source code and *O.timingEnabled*==True.

**totalMass** (*=0*)

    Mass of spheres that was produced so far. *(auto-updated)*

**totalVolume** (*=0*)

    Volume of spheres that was produced so far. *(auto-updated)*

**updateAttrs** (*(Serializable)arg1, (dict)arg2*) → None :

    Update object attributes from given dictionary

**vAngle** (*=NaN*)

    Maximum angle by which the initial sphere velocity deviates from the normal.

**vMax**(*=NaN*)

Maximum velocity norm of generated spheres (uniform distribution)

**vMin**(*=NaN*)

Minimum velocity norm of generated spheres (uniform distribution)

**class yade.wrapper.CohesiveFrictionalContactLaw**(*inherits GlobalEngine → Engine → Serializable*)

[DEPRECATED] Loop over interactions applying *Law2_ScGeom6D_CohFrictPhys_CohesionMoment* on all interactions.

---

**Note:** Use *InteractionLoop* and *Law2_ScGeom6D_CohFrictPhys_CohesionMoment* instead of this class for performance reasons.

---

**always_use_moment_law**(*=false*)

If true, use bending/twisting moments at all contacts. If false, compute moments only for cohesive contacts.

**creep_viscosity**(*=false*)

creep viscosity [Pa.s/m]. probably should be moved to Ip2_CohFrictMat_CohFrictMat_-CohFrictPhys...

**dead**(*=false*)

If true, this engine will not run at all; can be used for making an engine temporarily deactivated and only resurrect it at a later point.

**dict**(*(Serializable)arg1*) → dict :

Return dictionary of attributes.

**execCount**

Cumulative count this engine was run (only used if *O.timingEnabled*==True).

**execTime**

Cumulative time in nanoseconds this Engine took to run (only used if *O.timingEnabled*==True).

**label**(*=uninitalized*)

Textual label for this object; must be valid python identifier, you can refer to it directly from python.

**neverErase**(*=false*)

Keep interactions even if particles go away from each other (only in case another constitutive law is in the scene, e.g. *Law2_ScGeom_CapillaryPhys_Capillarity*)

**ompThreads**(*=-1*)

Number of threads to be used in the engine. If ompThreads<0 (default), the number will be typically OMP_NUM_THREADS or the number N defined by 'yade -jN' (this behavior can depend on the engine though). This attribute will only affect engines whose code includes openMP parallel regions (e.g. *InteractionLoop*). This attribute is mostly useful for experiments or when combining *ParallelEngine* with engines that run parallel regions, resulting in nested OMP loops with different number of threads at each level.

**shear_creep**(*=false*)

activate creep on the shear force, using *CohesiveFrictionalContactLaw::creep_viscosity*.

**timingDeltas**

Detailed information about timing inside the Engine itself. Empty unless enabled in the source code and *O.timingEnabled*==True.

**twist_creep**(*=false*)

activate creep on the twisting moment, using *CohesiveFrictionalContactLaw::creep_viscosity*.

**updateAttrs**(*(Serializable)arg1, (dict)arg2*) → None :

Update object attributes from given dictionary

---

**class** `yade.wrapper.ElasticContactLaw`(*inherits GlobalEngine → Engine → Serializable*)
[DEPRECATED] Loop over interactions applying *Law2_ScGeom_FrictPhys_CundallStrack* on all interactions.

---

**Note:** Use *InteractionLoop* and *Law2_ScGeom_FrictPhys_CundallStrack* instead of this class for performance reasons.

---

**dead**(*=false*)
    If true, this engine will not run at all; can be used for making an engine temporarily deactivated and only resurrect it at a later point.

**dict**(*(Serializable)arg1*) → dict :
    Return dictionary of attributes.

**execCount**
    Cumulative count this engine was run (only used if *O.timingEnabled*==`True`).

**execTime**
    Cumulative time in nanoseconds this Engine took to run (only used if *O.timingEnabled*==`True`).

**label**(*=uninitialized*)
    Textual label for this object; must be valid python identifier, you can refer to it directly from python.

**neverErase**(*=false*)
    Keep interactions even if particles go away from each other (only in case another constitutive law is in the scene, e.g. *Law2_ScGeom_CapillaryPhys_Capillarity*)

**ompThreads**(*=-1*)
    Number of threads to be used in the engine. If ompThreads<0 (default), the number will be typically OMP_NUM_THREADS or the number N defined by 'yade -jN' (this behavior can depend on the engine though). This attribute will only affect engines whose code includes openMP parallel regions (e.g. *InteractionLoop*). This attribute is mostly useful for experiments or when combining *ParallelEngine* with engines that run parallel regions, resulting in nested OMP loops with different number of threads at each level.

**timingDeltas**
    Detailed information about timing inside the Engine itself. Empty unless enabled in the source code and *O.timingEnabled*==`True`.

**updateAttrs**(*(Serializable)arg1, (dict)arg2*) → None :
    Update object attributes from given dictionary

**class** `yade.wrapper.FacetTopologyAnalyzer`(*inherits GlobalEngine → Engine → Serializable*)
Initializer for filling adjacency geometry data for facets.

Common vertices and common edges are identified and mutual angle between facet faces is written to Facet instances. If facets don't move with respect to each other, this must be done only at the beginng.

**commonEdgesFound**(*=0*)
    how many common edges were identified during last run. *(auto-updated)*

**commonVerticesFound**(*=0*)
    how many common vertices were identified during last run. *(auto-updated)*

**dead**(*=false*)
    If true, this engine will not run at all; can be used for making an engine temporarily deactivated and only resurrect it at a later point.

**dict**(*(Serializable)arg1*) → dict :
    Return dictionary of attributes.

**execCount**
    Cumulative count this engine was run (only used if *O.timingEnabled*==`True`).

**execTime**
    Cumulative time in nanoseconds this Engine took to run (only used if *O.timingEnabled*==`True`).

**label**(*=uninitalized*)
    Textual label for this object; must be valid python identifier, you can refer to it directly from python.

**ompThreads**(*=-1*)
    Number of threads to be used in the engine. If ompThreads<0 (default), the number will be typically OMP_NUM_THREADS or the number N defined by 'yade -jN' (this behavior can depend on the engine though). This attribute will only affect engines whose code includes openMP parallel regions (e.g. *InteractionLoop*). This attribute is mostly useful for experiments or when combining *ParallelEngine* with engines that run parallel regions, resulting in nested OMP loops with different number of threads at each level.

**projectionAxis**(*=Vector3r::UnitX()*)
    Axis along which to do the initial vertex sort

**relTolerance**(*=1e-4*)
    maximum distance of 'identical' vertices, relative to minimum facet size

**timingDeltas**
    Detailed information about timing inside the Engine itself. Empty unless enabled in the source code and *O.timingEnabled*==`True`.

**updateAttrs**(*(Serializable)arg1, (dict)arg2*) → None :
    Update object attributes from given dictionary

**class yade.wrapper.ForceResetter**(*inherits GlobalEngine → Engine → Serializable*)
Reset all forces stored in Scene::forces (`O.forces` in python). Typically, this is the first engine to be run at every step. In addition, reset those energies that should be reset, if energy tracing is enabled.

**dead**(*=false*)
    If true, this engine will not run at all; can be used for making an engine temporarily deactivated and only resurrect it at a later point.

**dict**(*(Serializable)arg1*) → dict :
    Return dictionary of attributes.

**execCount**
    Cumulative count this engine was run (only used if *O.timingEnabled*==`True`).

**execTime**
    Cumulative time in nanoseconds this Engine took to run (only used if *O.timingEnabled*==`True`).

**label**(*=uninitalized*)
    Textual label for this object; must be valid python identifier, you can refer to it directly from python.

**ompThreads**(*=-1*)
    Number of threads to be used in the engine. If ompThreads<0 (default), the number will be typically OMP_NUM_THREADS or the number N defined by 'yade -jN' (this behavior can depend on the engine though). This attribute will only affect engines whose code includes openMP parallel regions (e.g. *InteractionLoop*). This attribute is mostly useful for experiments or when combining *ParallelEngine* with engines that run parallel regions, resulting in nested OMP loops with different number of threads at each level.

**timingDeltas**
    Detailed information about timing inside the Engine itself. Empty unless enabled in the source code and *O.timingEnabled*==`True`.

**updateAttrs**(*(Serializable)arg1, (dict)arg2*) → None :
    Update object attributes from given dictionary

**class** yade.wrapper.**GlobalStiffnessTimeStepper**(*inherits TimeStepper → GlobalEngine → Engine → Serializable*)
    An engine assigning the time-step as a fraction of the minimum eigen-period in the problem. The derivation is detailed in the chapter on *DEM formulation*. The viscEl option enables to evaluate the timestep in a similar way for the visco-elastic contact law *Law2_ScGeom_ViscElPhys_Basic*, more detail in *GlobalStiffnessTimestepper::viscEl*.

**active**(*=true*)
    is the engine active?

**dead**(*=false*)
    If true, this engine will not run at all; can be used for making an engine temporarily deactivated and only resurrect it at a later point.

**defaultDt**(*=-1*)
    used as the initial value of the timestep (especially useful in the first steps when no contact exist). If negative, it will be defined by *utils.PWaveTimeStep * GlobalStiffnessTimeStepper::timestepSafetyCoefficient*

**densityScaling**(*=false*)
    *(auto-updated)* don't modify this value if you don't plan to modify the scaling factor manually for some bodies. In most cases, it is enough to set *NewtonIntegrator::densityScaling* and let this one be adjusted automatically.

**dict**(*(Serializable)arg1*) → dict :
    Return dictionary of attributes.

**execCount**
    Cumulative count this engine was run (only used if *O.timingEnabled*==True).

**execTime**
    Cumulative time in nanoseconds this Engine took to run (only used if *O.timingEnabled*==True).

**label**(*=uninitalized*)
    Textual label for this object; must be valid python identifier, you can refer to it directly from python.

**maxDt**(*=Mathr::MAX_REAL*)
    if positive, used as max value of the timestep whatever the computed value

**ompThreads**(*=-1*)
    Number of threads to be used in the engine. If ompThreads<0 (default), the number will be typically OMP_NUM_THREADS or the number N defined by 'yade -jN' (this behavior can depend on the engine though). This attribute will only affect engines whose code includes openMP parallel regions (e.g. *InteractionLoop*). This attribute is mostly useful for experiments or when combining *ParallelEngine* with engines that run parallel regions, resulting in nested OMP loops with different number of threads at each level.

**previousDt**(*=Mathr::MAX_REAL*)
    last computed dt *(auto-updated)*

**targetDt**(*=1*)
    if *NewtonIntegrator::densityScaling* is active, this value will be used as the simulation timestep and the scaling will use this value of dt as the target value. The value of targetDt is arbitrary and should have no effect in the result in general. However if some bodies have imposed velocities, for instance, they will move more or less per each step depending on this value.

**timeStepUpdateInterval**(*=1*)
    dt update interval

**timestepSafetyCoefficient**(*=0.8*)
    safety factor between the minimum eigen-period and the final assigned dt (less than 1)

**timingDeltas**
> Detailed information about timing inside the Engine itself. Empty unless enabled in the source code and *O.timingEnabled*==`True`.

**updateAttrs**(*(Serializable)arg1, (dict)arg2*) → None :
> Update object attributes from given dictionary

**viscEl**(*=false*)
> To use with *ViscElPhys*. if True, evaluate separetly the minimum eigen-period in the problem considering only the elastic contribution on one hand (spring only), and only the viscous contribution on the other hand (dashpot only). Take then the minimum of the two and use the safety coefficient *GlobalStiffnessTimestepper::timestepSafetyCoefficient* to take into account the possible coupling between the two contribution.

**class yade.wrapper.Integrator**(*inherits TimeStepper → GlobalEngine → Engine → Serializable*)
> Integration Engine Interface.

**active**(*=true*)
> is the engine active?

**dead**(*=false*)
> If true, this engine will not run at all; can be used for making an engine temporarily deactivated and only resurrect it at a later point.

**dict**(*(Serializable)arg1*) → dict :
> Return dictionary of attributes.

**execCount**
> Cumulative count this engine was run (only used if *O.timingEnabled*==`True`).

**execTime**
> Cumulative time in nanoseconds this Engine took to run (only used if *O.timingEnabled*==`True`).

**integrationsteps**(*=uninitalized*)
> all integrationsteps count as all succesfull substeps

**label**(*=uninitalized*)
> Textual label for this object; must be valid python identifier, you can refer to it directly from python.

**maxVelocitySq**(*=NaN*)
> store square of max. velocity, for informative purposes; computed again at every step. *(auto-updated)*

**ompThreads**(*=-1*)
> Number of threads to be used in the engine. If ompThreads<0 (default), the number will be typically OMP_NUM_THREADS or the number N defined by 'yade -jN' (this behavior can depend on the engine though). This attribute will only affect engines whose code includes openMP parallel regions (e.g. *InteractionLoop*). This attribute is mostly useful for experiments or when combining *ParallelEngine* with engines that run parallel regions, resulting in nested OMP loops with different number of threads at each level.

**slaves**
> List of lists of Engines to calculate the force acting on the particles; to obtain the derivatives of the states, engines inside will be run sequentially.

**timeStepUpdateInterval**(*=1*)
> dt update interval

**timingDeltas**
> Detailed information about timing inside the Engine itself. Empty unless enabled in the source code and *O.timingEnabled*==`True`.

> **updateAttrs**(*(Serializable)arg1, (dict)arg2*) → None :
> Update object attributes from given dictionary

**class** yade.wrapper.**InteractionLoop**(*inherits GlobalEngine → Engine → Serializable*)
Unified dispatcher for handling interaction loop at every step, for parallel performance reasons.

---

**Special constructor**

Constructs from 3 lists of *Ig2*, *Ip2*, *Law2* functors respectively; they will be passed to internal dispatchers, which you might retrieve as *geomDispatcher*, *physDispatcher*, *lawDispatcher* respectively.

---

> **callbacks**(*=uninitalized*)
> *Callbacks* which will be called for every *Interaction*, if activated.

> **dead**(*=false*)
> If true, this engine will not run at all; can be used for making an engine temporarily deactivated and only resurrect it at a later point.

> **dict**(*(Serializable)arg1*) → dict :
> Return dictionary of attributes.

> **execCount**
> Cumulative count this engine was run (only used if *O.timingEnabled*==**True**).

> **execTime**
> Cumulative time in nanoseconds this Engine took to run (only used if *O.timingEnabled*==**True**).

> **geomDispatcher**(*=new IGeomDispatcher*)
> *IGeomDispatcher* object that is used for dispatch.

> **label**(*=uninitalized*)
> Textual label for this object; must be valid python identifier, you can refer to it directly from python.

> **lawDispatcher**(*=new LawDispatcher*)
> *LawDispatcher* object used for dispatch.

> **loopOnSortedInteractions**(*=false*)
> If true, the main interaction loop will occur on a sorted list of interactions. This is SLOW but useful to workaround floating point force addition non reproducibility when debugging parallel implementations of yade.

> **ompThreads**(*=-1*)
> Number of threads to be used in the engine. If ompThreads<0 (default), the number will be typically OMP_NUM_THREADS or the number N defined by 'yade -jN' (this behavior can depend on the engine though). This attribute will only affect engines whose code includes openMP parallel regions (e.g. *InteractionLoop*). This attribute is mostly useful for experiments or when combining *ParallelEngine* with engines that run parallel regions, resulting in nested OMP loops with different number of threads at each level.

> **physDispatcher**(*=new IPhysDispatcher*)
> *IPhysDispatcher* object used for dispatch.

> **timingDeltas**
> Detailed information about timing inside the Engine itself. Empty unless enabled in the source code and *O.timingEnabled*==**True**.

> **updateAttrs**(*(Serializable)arg1, (dict)arg2*) → None :
> Update object attributes from given dictionary

**class** yade.wrapper.**Law2_ScGeom_CapillaryPhys_Capillarity**(*inherits GlobalEngine → Engine → Serializable*)
This law allows one to take into account capillary forces/effects between spheres coming from the presence of interparticular liquid bridges (menisci).

---

The control parameter is the *capillary pressure* (or suction) Uc = Ugas - Uliquid. Liquid bridges properties (volume V, extent over interacting grains delta1 and delta2) are computed as a result of the defined capillary pressure and of the interacting geometry (spheres radii and interparticular distance).

References: in english [Scholtes2009b]; more detailed, but in french [Scholtes2009d].

The law needs ascii files M(r=i) with i=R1/R2 to work (see https://yade-dem.org/wiki/ CapillaryTriaxialTest). These ASCII files contain a set of results from the resolution of the Laplace-Young equation for different configurations of the interacting geometry, assuming a null wetting angle.

In order to allow capillary forces between distant spheres, it is necessary to enlarge the bounding boxes using *Bo1_Sphere_Aabb::aabbEnlargeFactor* and make the Ig2 define define distant interactions via *interactionDetectionFactor*. It is also necessary to disable interactions removal by the constitutive law (*Law2*). The only combinations of laws supported are currently capillary law + *Law2_ScGeom_FrictPhys_CundallStrack* and capillary law + *Law2_ScGeom_MindlinPhys_-Mindlin* (and the other variants of Hertz-Mindlin).

See CapillaryPhys-example.py for an example script.

**binaryFusion**(*=true*)
    If true, capillary forces are set to zero as soon as, at least, 1 overlap (menisci fusion) is detected. Otherwise *fCap = fCap* / (*fusionNumber* + 1 )

**capillaryPressure**(*=0.*)
    Value of the capillary pressure Uc defined as Uc=Ugas-Uliquid

**createDistantMeniscii**(*=false*)
    Generate meniscii between distant spheres? Else only maintain the existing ones. For modeling a wetting path this flag should always be false. For a drying path it should be true for one step (initialization) then false, as in the logic of [Scholtes2009c]

**dead**(*=false*)
    If true, this engine will not run at all; can be used for making an engine temporarily deactivated and only resurrect it at a later point.

**dict**(*(Serializable)arg1*) → dict :
    Return dictionary of attributes.

**execCount**
    Cumulative count this engine was run (only used if *O.timingEnabled*==True).

**execTime**
    Cumulative time in nanoseconds this Engine took to run (only used if *O.timingEnabled*==True).

**fusionDetection**(*=false*)
    If true potential menisci overlaps are checked, computing *fusionNumber* for each capillary interaction, and reducing *fCap* according to *binaryFusion*

**label**(*=uninitalized*)
    Textual label for this object; must be valid python identifier, you can refer to it directly from python.

**ompThreads**(*=-1*)
    Number of threads to be used in the engine. If ompThreads<0 (default), the number will be typically OMP_NUM_THREADS or the number N defined by 'yade -jN' (this behavior can depend on the engine though). This attribute will only affect engines whose code includes openMP parallel regions (e.g. *InteractionLoop*). This attribute is mostly useful for experiments or when combining *ParallelEngine* with engines that run parallel regions, resulting in nested OMP loops with different number of threads at each level.

**suffCapFiles**(*=""*)
    Capillary files suffix: M(r=X)suffCapFiles

**surfaceTension**(*=0.073*)
> Value of considered surface tension

**timingDeltas**
> Detailed information about timing inside the Engine itself. Empty unless enabled in the source code and *O.timingEnabled*==`True`.

**updateAttrs**(*(Serializable)arg1, (dict)arg2*) → None :
> Update object attributes from given dictionary

**class yade.wrapper.NewtonIntegrator**(*inherits GlobalEngine → Engine → Serializable*)
> Engine integrating newtonian motion equations.

**dampGravity**(*=true*)
> By default, numerical damping applies to ALL forces, even gravity. If this option is set to false, then the gravity forces calculated based on NewtonIntegrator.gravity are excluded from the damping calculation. This option has no effect on gravity forces added by GravityEngine.

**damping**(*=0.2*)
> damping coefficient for Cundall's non viscous damping (see *Numerical damping* and [Chareyre2005])

**dead**(*=false*)
> If true, this engine will not run at all; can be used for making an engine temporarily deactivated and only resurrect it at a later point.

**densityScaling**
> if True, then density scaling [Pfc3dManual30] will be applied in order to have a critical timestep equal to *GlobalStiffnessTimeStepper::targetDt* for all bodies. This option makes the simulation unrealistic from a dynamic point of view, but may speedup quasistatic simulations. In rare situations, it could be useful to not set the scalling factor automatically for each body (which the time-stepper does). In such case revert *GlobalStiffnessTimeStepper.densityScaling* to False.

**dict**(*(Serializable)arg1*) → dict :
> Return dictionary of attributes.

**exactAsphericalRot**(*=true*)
> Enable more exact body rotation integrator for *aspherical bodies only*, using formulation from [Allen1989], pg. 89.

**execCount**
> Cumulative count this engine was run (only used if *O.timingEnabled*==`True`).

**execTime**
> Cumulative time in nanoseconds this Engine took to run (only used if *O.timingEnabled*==`True`).

**gravity**(*=Vector3r::Zero()*)
> Gravitational acceleration (effectively replaces GravityEngine).

**kinSplit**(*=false*)
> Whether to separately track translational and rotational kinetic energy.

**label**(*=uninitalized*)
> Textual label for this object; must be valid python identifier, you can refer to it directly from python.

**mask**(*=-1*)
> If mask defined and the bitwise AND between mask and body's groupMask gives 0, the body will not move/rotate. Velocities and accelerations will be calculated not paying attention to this parameter.

**maxVelocitySq**(*=0*)
> stores max. displacement, based on which we trigger collision detection. *(auto-updated)*

**ompThreads**(*=-1*)
> Number of threads to be used in the engine. If ompThreads<0 (default), the number will be typically OMP_NUM_THREADS or the number N defined by 'yade -jN' (this behavior can depend on the engine though). This attribute will only affect engines whose code includes openMP parallel regions (e.g. *InteractionLoop*). This attribute is mostly useful for experiments or when combining *ParallelEngine* with engines that run parallel regions, resulting in nested OMP loops with different number of threads at each level.

**prevVelGrad**(=*Matrix3r::Zero()*)
Store previous velocity gradient (*Cell::velGrad*) to track acceleration. *(auto-updated)*

**timingDeltas**
Detailed information about timing inside the Engine itself. Empty unless enabled in the source code and *O.timingEnabled*==`True`.

**updateAttrs**(*(Serializable)arg1, (dict)arg2*) → None :
Update object attributes from given dictionary

**warnNoForceReset**(=*true*)
Warn when forces were not resetted in this step by *ForceResetter*; this mostly points to *ForceResetter* being forgotten incidentally and should be disabled only with a good reason.

**class** yade.wrapper.**RungeKuttaCashKarp54Integrator**(*inherits Integrator* → *TimeStepper* → *GlobalEngine* → *Engine* → *Serializable*)
RungeKuttaCashKarp54Integrator engine.

**__init__**(*(object)arg1*) → None
object ___init___(tuple args, dict kwds)

___init___( (object)arg1, (list)arg2) -> object : Construct from (possibly nested) list of slaves.

**a_dxdt**(=*1.0*)

**a_x**(=*1.0*)

**abs_err**(=*1e-6*)
Relative integration tolerance

**active**(=*true*)
is the engine active?

**dead**(=*false*)
If true, this engine will not run at all; can be used for making an engine temporarily deactivated and only resurrect it at a later point.

**dict**(*(Serializable)arg1*) → dict :
Return dictionary of attributes.

**execCount**
Cumulative count this engine was run (only used if *O.timingEnabled*==`True`).

**execTime**
Cumulative time in nanoseconds this Engine took to run (only used if *O.timingEnabled*==`True`).

**integrationsteps**(=*uninitalized*)
all integrationsteps count as all succesfull substeps

**label**(=*uninitalized*)
Textual label for this object; must be valid python identifier, you can refer to it directly from python.

**maxVelocitySq**(=*NaN*)
store square of max. velocity, for informative purposes; computed again at every step. *(auto-updated)*

**ompThreads**(=*-1*)
Number of threads to be used in the engine. If ompThreads<0 (default), the number will be typically OMP_NUM_THREADS or the number N defined by 'yade -jN' (this behavior can depend on the engine though). This attribute will only affect engines whose code includes openMP parallel regions (e.g. *InteractionLoop*). This attribute is mostly useful for experiments or when combining *ParallelEngine* with engines that run parallel regions, resulting in nested OMP loops with different number of threads at each level.

**rel_err**(*=1e-6*)
> Absolute integration tolerance

**slaves**
> List of lists of Engines to calculate the force acting on the particles; to obtain the derivatives of the states, engines inside will be run sequentially.

**stepsize**(*=1e-6*)
> It is not important for an adaptive integration but important for the observer for setting the found states after integration

**timeStepUpdateInterval**(*=1*)
> dt update interval

**timingDeltas**
> Detailed information about timing inside the Engine itself. Empty unless enabled in the source code and *O.timingEnabled*==**True**.

**updateAttrs**(*(Serializable)arg1, (dict)arg2*) → None :
> Update object attributes from given dictionary

**class yade.wrapper.SpheresFactory**(*inherits GlobalEngine → Engine → Serializable*)
> Engine for spitting spheres based on mass flow rate, particle size distribution etc. Initial velocity of particles is given by *vMin*, *vMax*, the *massFlowRate* determines how many particles to generate at each step. When *goalMass* is attained or positive *maxParticles* is reached, the engine does not produce particles anymore. Geometry of the region should be defined in a derived engine by overridden SpheresFactory::pickRandomPosition().

> A sample script for this engine is in scripts/spheresFactory.py.

**PSDcalculateMass**(*=true*)
> PSD-Input is in mass (true), otherwise the number of particles will be considered.

**PSDcum**(*=uninitialized*)
> PSD-dispersion, cumulative procent meanings [-]

**PSDsizes**(*=uninitialized*)
> PSD-dispersion, sizes of cells, Diameter [m]

**blockedDOFs**(*=""*)
> Blocked degress of freedom

**color**(*=Vector3r(-1, -1, -1)*)
> Use the color for newly created particles, if specified

**dead**(*=false*)
> If true, this engine will not run at all; can be used for making an engine temporarily deactivated and only resurrect it at a later point.

**dict**(*(Serializable)arg1*) → dict :
> Return dictionary of attributes.

**exactDiam**(*=true*)
> If true, the particles only with the defined in PSDsizes diameters will be created. Otherwise the diameter will be randomly chosen in the range [PSDsizes[i-1]:PSDsizes[i]], in this case the length of PSDsizes should be more on 1, than the length of PSDcum.

**execCount**
> Cumulative count this engine was run (only used if *O.timingEnabled*==**True**).

**execTime**
> Cumulative time in nanoseconds this Engine took to run (only used if *O.timingEnabled*==True).

**goalMass**(*=0*)
> Total mass that should be attained at the end of the current step. *(auto-updated)*

**ids**(*=uninitalized*)
> ids of created bodies

**label**(*=uninitalized*)
> Textual label for this object; must be valid python identifier, you can refer to it directly from python.

**mask**(*=-1*)
> groupMask to apply for newly created spheres

**massFlowRate**(*=NaN*)
> Mass flow rate [kg/s]

**materialId**(*=-1*)
> Shared material id to use for newly created spheres (can be negative to count from the end)

**maxAttempt**(*=5000*)
> Maximum number of attempts to position a new sphere randomly.

**maxMass**(*=-1*)
> Maximal mass at which to stop generating new particles regardless of massFlowRate. if maxMass=-1 - this parameter is ignored.

**maxParticles**(*=100*)
> The number of particles at which to stop generating new ones regardless of massFlowRate. if maxParticles=-1 - this parameter is ignored .

**normal**(*=Vector3r(NaN, NaN, NaN)*)
> Orientation of the region's geometry, direction of particle's velocites if normalVel is not set.

**normalVel**(*=Vector3r(NaN, NaN, NaN)*)
> Direction of particle's velocites.

**numParticles**(*=0*)
> Cummulative number of particles produces so far *(auto-updated)*

**ompThreads**(*=-1*)
> Number of threads to be used in the engine. If ompThreads<0 (default), the number will be typically OMP_NUM_THREADS or the number N defined by 'yade -jN' (this behavior can depend on the engine though). This attribute will only affect engines whose code includes openMP parallel regions (e.g. *InteractionLoop*). This attribute is mostly useful for experiments or when combining *ParallelEngine* with engines that run parallel regions, resulting in nested OMP loops with different number of threads at each level.

**rMax**(*=NaN*)
> Maximum radius of generated spheres (uniform distribution)

**rMin**(*=NaN*)
> Minimum radius of generated spheres (uniform distribution)

**silent**(*=false*)
> If true no complain about excessing maxAttempt but disable the factory (by set massFlowRate=0).

**stopIfFailed**(*=true*)
> If true, the SpheresFactory stops (sets massFlowRate=0), when maximal number of attempts to insert particle exceed.

**timingDeltas**
>   Detailed information about timing inside the Engine itself. Empty unless enabled in the source code and *O.timingEnabled*==`True`.

**totalMass**(*=0*)
>   Mass of spheres that was produced so far. *(auto-updated)*

**totalVolume**(*=0*)
>   Volume of spheres that was produced so far. *(auto-updated)*

**updateAttrs**(*(Serializable)arg1, (dict)arg2*) → None :
>   Update object attributes from given dictionary

**vAngle**(*=NaN*)
>   Maximum angle by which the initial sphere velocity deviates from the normal.

**vMax**(*=NaN*)
>   Maximum velocity norm of generated spheres (uniform distribution)

**vMin**(*=NaN*)
>   Minimum velocity norm of generated spheres (uniform distribution)

**class yade.wrapper.TetraVolumetricLaw**(*inherits GlobalEngine → Engine → Serializable*)
>   Calculate physical response of 2 *tetrahedra* in interaction, based on penetration configuration given by *TTetraGeom*.

**dead**(*=false*)
>   If true, this engine will not run at all; can be used for making an engine temporarily deactivated and only resurrect it at a later point.

**dict**(*(Serializable)arg1*) → dict :
>   Return dictionary of attributes.

**execCount**
>   Cumulative count this engine was run (only used if *O.timingEnabled*==`True`).

**execTime**
>   Cumulative time in nanoseconds this Engine took to run (only used if *O.timingEnabled*==`True`).

**label**(*=uninitalized*)
>   Textual label for this object; must be valid python identifier, you can refer to it directly from python.

**ompThreads**(*=-1*)
>   Number of threads to be used in the engine. If ompThreads<0 (default), the number will be typically OMP_NUM_THREADS or the number N defined by 'yade -jN' (this behavior can depend on the engine though). This attribute will only affect engines whose code includes openMP parallel regions (e.g. *InteractionLoop*). This attribute is mostly useful for experiments or when combining *ParallelEngine* with engines that run parallel regions, resulting in nested OMP loops with different number of threads at each level.

**timingDeltas**
>   Detailed information about timing inside the Engine itself. Empty unless enabled in the source code and *O.timingEnabled*==`True`.

**updateAttrs**(*(Serializable)arg1, (dict)arg2*) → None :
>   Update object attributes from given dictionary

**class yade.wrapper.TimeStepper**(*inherits GlobalEngine → Engine → Serializable*)
>   Engine defining time-step (fundamental class)

**active**(*=true*)
>   is the engine active?

**dead**(*=false*)
> If true, this engine will not run at all; can be used for making an engine temporarily deactivated and only resurrect it at a later point.

**dict**(*(Serializable)arg1*) → dict :
> Return dictionary of attributes.

**execCount**
> Cumulative count this engine was run (only used if *O.timingEnabled*==`True`).

**execTime**
> Cumulative time in nanoseconds this Engine took to run (only used if *O.timingEnabled*==`True`).

**label**(*=uninitalized*)
> Textual label for this object; must be valid python identifier, you can refer to it directly from python.

**ompThreads**(*=-1*)
> Number of threads to be used in the engine. If ompThreads<0 (default), the number will be typically OMP_NUM_THREADS or the number N defined by 'yade -jN' (this behavior can depend on the engine though). This attribute will only affect engines whose code includes openMP parallel regions (e.g. *InteractionLoop*). This attribute is mostly useful for experiments or when combining *ParallelEngine* with engines that run parallel regions, resulting in nested OMP loops with different number of threads at each level.

**timeStepUpdateInterval**(*=1*)
> dt update interval

**timingDeltas**
> Detailed information about timing inside the Engine itself. Empty unless enabled in the source code and *O.timingEnabled*==`True`.

**updateAttrs**(*(Serializable)arg1, (dict)arg2*) → None :
> Update object attributes from given dictionary

**PeriodicEngine**

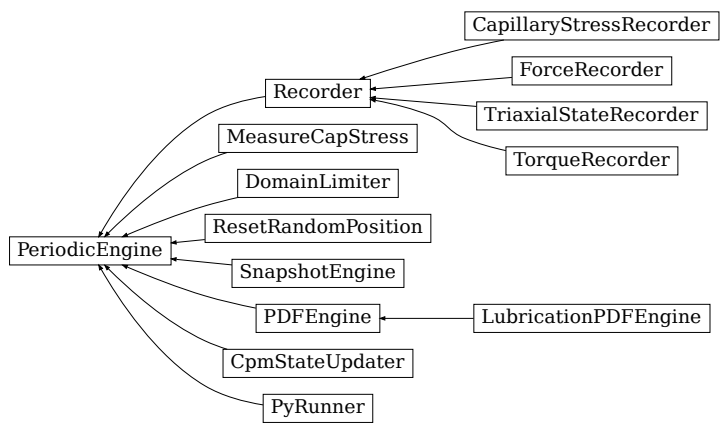

Fig. 26: Inheritance graph of PeriodicEngine. See also: *CapillaryStressRecorder*, *CpmStateUpdater*, *DomainLimiter*, *ForceRecorder*, *LubricationPDFEngine*, *MeasureCapStress*, *PDFEngine*, *PyRunner*, *Recorder*, *ResetRandomPosition*, *SnapshotEngine*, *TorqueRecorder*, *TriaxialStateRecorder*.

**class yade.wrapper.PeriodicEngine**(*inherits GlobalEngine → Engine → Serializable*)
> Run Engine::action with given fixed periodicity real time (=wall clock time, computation time), virtual time (simulation time), iteration number), by setting any of those criteria (virtPeriod, realPeriod, iterPeriod) to a positive value. They are all negative (inactive) by default.

The number of times this engine is activated can be limited by setting nDo>0. If the number of activations will have been already reached, no action will be called even if an active period has elapsed.

If initRun is set (false by default), the engine will run when called for the first time; otherwise it will only start counting period (realLast, etc, interval variables) from that point, but without actually running, and will run only once a period has elapsed since the initial run.

This class should not be used directly; rather, derive your own engine which you want to be run periodically.

Derived engines should override Engine::action(), which will be called periodically. If the derived Engine overrides also Engine::isActivated, it should also take in account return value from PeriodicEngine::isActivated, since otherwise the periodicity will not be functional.

Example with *PyRunner*, which derives from PeriodicEngine; likely to be encountered in python scripts:

```
PyRunner(realPeriod=5,iterPeriod=10000,command='print O.iter')
```

will print iteration number every 10000 iterations or every 5 seconds of wall clock time, whichever comes first since it was last run.

**dead**(*=false*)
    If true, this engine will not run at all; can be used for making an engine temporarily deactivated and only resurrect it at a later point.

**dict**(*(Serializable)arg1*) → dict :
    Return dictionary of attributes.

**execCount**
    Cumulative count this engine was run (only used if *O.timingEnabled*==`True`).

**execTime**
    Cumulative time in nanoseconds this Engine took to run (only used if *O.timingEnabled*==`True`).

**firstIterRun**(*=0*)
    Sets the step number, at each an engine should be executed for the first time (disabled by default).

**initRun**(*=false*)
    Run the first time we are called as well.

**iterLast**(*=0*)
    Tracks step number of last run *(auto-updated)*.

**iterPeriod**(*=0, deactivated*)
    Periodicity criterion using step number (deactivated if <= 0)

**label**(*=uninitialized*)
    Textual label for this object; must be valid python identifier, you can refer to it directly from python.

**nDo**(*=-1, deactivated*)
    Limit number of executions by this number (deactivated if negative)

**nDone**(*=0*)
    Track number of executions (cummulative) *(auto-updated)*.

**ompThreads**(*=-1*)
    Number of threads to be used in the engine. If ompThreads<0 (default), the number will be typically OMP_NUM_THREADS or the number N defined by 'yade -jN' (this behavior can depend on the engine though). This attribute will only affect engines whose code includes

openMP parallel regions (e.g. *InteractionLoop*). This attribute is mostly useful for experiments or when combining *ParallelEngine* with engines that run parallel regions, resulting in nested OMP loops with different number of threads at each level.

**realLast**(*=0*)
Tracks real time of last run *(auto-updated)*.

**realPeriod**(*=0, deactivated*)
Periodicity criterion using real (wall clock, computation, human) time in seconds (deactivated if <=0)

**timingDeltas**
Detailed information about timing inside the Engine itself. Empty unless enabled in the source code and *O.timingEnabled*==`True`.

**updateAttrs**(*(Serializable)arg1, (dict)arg2*) → None :
Update object attributes from given dictionary

**virtLast**(*=0*)
Tracks virtual time of last run *(auto-updated)*.

**virtPeriod**(*=0, deactivated*)
Periodicity criterion using virtual (simulation) time (deactivated if <= 0)

**class yade.wrapper.CapillaryStressRecorder**(*inherits Recorder → PeriodicEngine → GlobalEngine → Engine → Serializable*)
Records information from capillary meniscii on samples submitted to triaxial compressions. Classical sign convention (tension positiv) is used for capillary stresses. -> New formalism needs to be tested!!!

**addIterNum**(*=false*)
Adds an iteration number to the file name, when the file was created. Useful for creating new files at each call (false by default)

**dead**(*=false*)
If true, this engine will not run at all; can be used for making an engine temporarily deactivated and only resurrect it at a later point.

**dict**(*(Serializable)arg1*) → dict :
Return dictionary of attributes.

**execCount**
Cumulative count this engine was run (only used if *O.timingEnabled*==`True`).

**execTime**
Cumulative time in nanoseconds this Engine took to run (only used if *O.timingEnabled*==`True`).

**file**(*=uninitialized*)
Name of file to save to; must not be empty.

**firstIterRun**(*=0*)
Sets the step number, at each an engine should be executed for the first time (disabled by default).

**initRun**(*=false*)
Run the first time we are called as well.

**iterLast**(*=0*)
Tracks step number of last run *(auto-updated)*.

**iterPeriod**(*=0, deactivated*)
Periodicity criterion using step number (deactivated if <= 0)

**label**(*=uninitialized*)
Textual label for this object; must be valid python identifier, you can refer to it directly from python.

**nDo**(*=-1, deactivated*)
  Limit number of executions by this number (deactivated if negative)

**nDone**(*=0*)
  Track number of executions (cummulative) *(auto-updated)*.

**ompThreads**(*=-1*)
  Number of threads to be used in the engine. If ompThreads<0 (default), the number will be typically OMP_NUM_THREADS or the number N defined by 'yade -jN' (this behavior can depend on the engine though). This attribute will only affect engines whose code includes openMP parallel regions (e.g. *InteractionLoop*). This attribute is mostly useful for experiments or when combining *ParallelEngine* with engines that run parallel regions, resulting in nested OMP loops with different number of threads at each level.

**realLast**(*=0*)
  Tracks real time of last run *(auto-updated)*.

**realPeriod**(*=0, deactivated*)
  Periodicity criterion using real (wall clock, computation, human) time in seconds (deactivated if <=0)

**timingDeltas**
  Detailed information about timing inside the Engine itself. Empty unless enabled in the source code and *O.timingEnabled*==**True**.

**truncate**(*=false*)
  Whether to delete current file contents, if any, when opening (false by default)

**updateAttrs**(*(Serializable)arg1, (dict)arg2*) → None :
  Update object attributes from given dictionary

**virtLast**(*=0*)
  Tracks virtual time of last run *(auto-updated)*.

**virtPeriod**(*=0, deactivated*)
  Periodicity criterion using virtual (simulation) time (deactivated if <= 0)

**class yade.wrapper.CpmStateUpdater**(*inherits PeriodicEngine → GlobalEngine → Engine → Serializable*)
  Update *CpmState* of bodies based on state variables in *CpmPhys* of interactions with this bod. In particular, bodies' colors and *CpmState::normDmg* depending on average *damage* of their interactions and number of interactions that were already fully broken and have disappeared is updated. This engine contains its own loop (2 loops, more precisely) over all bodies and should be run periodically to update colors during the simulation, if desired.

**avgRelResidual**(*=NaN*)
  Average residual strength at last run.

**dead**(*=false*)
  If true, this engine will not run at all; can be used for making an engine temporarily deactivated and only resurrect it at a later point.

**dict**(*(Serializable)arg1*) → dict :
  Return dictionary of attributes.

**execCount**
  Cumulative count this engine was run (only used if *O.timingEnabled*==**True**).

**execTime**
  Cumulative time in nanoseconds this Engine took to run (only used if *O.timingEnabled*==**True**).

**firstIterRun**(*=0*)
  Sets the step number, at each an engine should be executed for the first time (disabled by default).

**initRun**(*=false*)
    Run the first time we are called as well.

**iterLast**(*=0*)
    Tracks step number of last run *(auto-updated)*.

**iterPeriod**(*=0, deactivated*)
    Periodicity criterion using step number (deactivated if <= 0)

**label**(*=uninitalized*)
    Textual label for this object; must be valid python identifier, you can refer to it directly from python.

**maxOmega**(*=NaN*)
    Globally maximum damage parameter at last run.

**nDo**(*=-1, deactivated*)
    Limit number of executions by this number (deactivated if negative)

**nDone**(*=0*)
    Track number of executions (cummulative) *(auto-updated)*.

**ompThreads**(*=-1*)
    Number of threads to be used in the engine. If ompThreads<0 (default), the number will be typically OMP_NUM_THREADS or the number N defined by 'yade -jN' (this behavior can depend on the engine though). This attribute will only affect engines whose code includes openMP parallel regions (e.g. *InteractionLoop*). This attribute is mostly useful for experiments or when combining *ParallelEngine* with engines that run parallel regions, resulting in nested OMP loops with different number of threads at each level.

**realLast**(*=0*)
    Tracks real time of last run *(auto-updated)*.

**realPeriod**(*=0, deactivated*)
    Periodicity criterion using real (wall clock, computation, human) time in seconds (deactivated if <=0)

**timingDeltas**
    Detailed information about timing inside the Engine itself. Empty unless enabled in the source code and *O.timingEnabled*==`True`.

**updateAttrs**(*(Serializable)arg1, (dict)arg2*) → None :
    Update object attributes from given dictionary

**virtLast**(*=0*)
    Tracks virtual time of last run *(auto-updated)*.

**virtPeriod**(*=0, deactivated*)
    Periodicity criterion using virtual (simulation) time (deactivated if <= 0)

**class yade.wrapper.DomainLimiter**(*inherits PeriodicEngine → GlobalEngine → Engine → Serializable*)
    Delete particles that are out of axis-aligned box given by *lo* and *hi*.

**dead**(*=false*)
    If true, this engine will not run at all; can be used for making an engine temporarily deactivated and only resurrect it at a later point.

**dict**(*(Serializable)arg1*) → dict :
    Return dictionary of attributes.

**execCount**
    Cumulative count this engine was run (only used if *O.timingEnabled*==`True`).

**execTime**
    Cumulative time in nanoseconds this Engine took to run (only used if *O.timingEnabled*==`True`).

**firstIterRun**(*=0*)
    Sets the step number, at each an engine should be executed for the first time (disabled by default).

**hi**(*=Vector3r(0, 0, 0)*)
    Upper corner of the domain.

**initRun**(*=false*)
    Run the first time we are called as well.

**iterLast**(*=0*)
    Tracks step number of last run *(auto-updated)*.

**iterPeriod**(*=0, deactivated*)
    Periodicity criterion using step number (deactivated if <= 0)

**label**(*=uninitalized*)
    Textual label for this object; must be valid python identifier, you can refer to it directly from python.

**lo**(*=Vector3r(0, 0, 0)*)
    Lower corner of the domain.

**mDeleted**(*=0*)
    Mass of deleted particles.

**mask**(*=-1*)
    If mask is defined, only particles with corresponding groupMask will be deleted.

**nDeleted**(*=0*)
    Cummulative number of particles deleted.

**nDo**(*=-1, deactivated*)
    Limit number of executions by this number (deactivated if negative)

**nDone**(*=0*)
    Track number of executions (cummulative) *(auto-updated)*.

**ompThreads**(*=-1*)
    Number of threads to be used in the engine. If ompThreads<0 (default), the number will be typically OMP_NUM_THREADS or the number N defined by 'yade -jN' (this behavior can depend on the engine though). This attribute will only affect engines whose code includes openMP parallel regions (e.g. *InteractionLoop*). This attribute is mostly useful for experiments or when combining *ParallelEngine* with engines that run parallel regions, resulting in nested OMP loops with different number of threads at each level.

**realLast**(*=0*)
    Tracks real time of last run *(auto-updated)*.

**realPeriod**(*=0, deactivated*)
    Periodicity criterion using real (wall clock, computation, human) time in seconds (deactivated if <=0)

**timingDeltas**
    Detailed information about timing inside the Engine itself. Empty unless enabled in the source code and *O.timingEnabled*==**True**.

**updateAttrs**(*(Serializable)arg1, (dict)arg2*) → None :
    Update object attributes from given dictionary

**vDeleted**(*=0*)
    Volume of deleted spheres (clumps not counted, in that case check *mDeleted*)

**virtLast**(*=0*)
    Tracks virtual time of last run *(auto-updated)*.

**virtPeriod**(*=0, deactivated*)
    Periodicity criterion using virtual (simulation) time (deactivated if <= 0)

**class yade.wrapper.ForceRecorder**(*inherits Recorder → PeriodicEngine → GlobalEngine →
Engine → Serializable*)

Engine saves the resultant force affecting to bodies, listed in *ids*. For instance, can be useful for
defining the forces, which affects to __buldozer__ during its work.

**addIterNum**(*=false*)
    Adds an iteration number to the file name, when the file was created. Useful for creating new
    files at each call (false by default)

**dead**(*=false*)
    If true, this engine will not run at all; can be used for making an engine temporarily deactivated
    and only resurrect it at a later point.

**dict**(*(Serializable)arg1*) → dict :
    Return dictionary of attributes.

**execCount**
    Cumulative count this engine was run (only used if *O.timingEnabled*==True).

**execTime**
    Cumulative time in nanoseconds this Engine took to run (only used if
    *O.timingEnabled*==True).

**file**(*=uninitalized*)
    Name of file to save to; must not be empty.

**firstIterRun**(*=0*)
    Sets the step number, at each an engine should be executed for the first time (disabled by
    default).

**ids**(*=uninitalized*)
    List of bodies whose state will be measured

**initRun**(*=false*)
    Run the first time we are called as well.

**iterLast**(*=0*)
    Tracks step number of last run *(auto-updated)*.

**iterPeriod**(*=0, deactivated*)
    Periodicity criterion using step number (deactivated if <= 0)

**label**(*=uninitalized*)
    Textual label for this object; must be valid python identifier, you can refer to it directly from
    python.

**nDo**(*=-1, deactivated*)
    Limit number of executions by this number (deactivated if negative)

**nDone**(*=0*)
    Track number of executions (cummulative) *(auto-updated)*.

**ompThreads**(*=-1*)
    Number of threads to be used in the engine. If ompThreads<0 (default), the number will be
    typically OMP_NUM_THREADS or the number N defined by 'yade -jN' (this behavior can
    depend on the engine though). This attribute will only affect engines whose code includes
    openMP parallel regions (e.g. *InteractionLoop*). This attribute is mostly useful for experi-
    ments or when combining *ParallelEngine* with engines that run parallel regions, resulting in
    nested OMP loops with different number of threads at each level.

**realLast**(*=0*)
    Tracks real time of last run *(auto-updated)*.

**realPeriod**(*=0, deactivated*)
    Periodicity criterion using real (wall clock, computation, human) time in seconds (deactivated
    if <=0)

**timingDeltas**
    Detailed information about timing inside the Engine itself. Empty unless enabled in the source code and *O.timingEnabled*==`True`.

**totalForce**(*=Vector3r::Zero()*)
    Resultant force, returning by the function.

**truncate**(*=false*)
    Whether to delete current file contents, if any, when opening (false by default)

**updateAttrs**(*(Serializable)arg1, (dict)arg2*) → None :
    Update object attributes from given dictionary

**virtLast**(*=0*)
    Tracks virtual time of last run *(auto-updated)*.

**virtPeriod**(*=0, deactivated*)
    Periodicity criterion using virtual (simulation) time (deactivated if <= 0)

**class yade.wrapper.LubricationPDFEngine**(*inherits PDFEngine → PeriodicEngine → GlobalEngine → Engine → Serializable*)
    Implementation of *PDFEngine* for Lubrication law

**dead**(*=false*)
    If true, this engine will not run at all; can be used for making an engine temporarily deactivated and only resurrect it at a later point.

**dict**(*(Serializable)arg1*) → dict :
    Return dictionary of attributes.

**execCount**
    Cumulative count this engine was run (only used if *O.timingEnabled*==`True`).

**execTime**
    Cumulative time in nanoseconds this Engine took to run (only used if *O.timingEnabled*==`True`).

**filename**(*="PDF.txt"*)
    Filename

**firstIterRun**(*=0*)
    Sets the step number, at each an engine should be executed for the first time (disabled by default).

**initRun**(*=false*)
    Run the first time we are called as well.

**iterLast**(*=0*)
    Tracks step number of last run *(auto-updated)*.

**iterPeriod**(*=0, deactivated*)
    Periodicity criterion using step number (deactivated if <= 0)

**label**(*=uninitalized*)
    Textual label for this object; must be valid python identifier, you can refer to it directly from python.

**nDo**(*=-1, deactivated*)
    Limit number of executions by this number (deactivated if negative)

**nDone**(*=0*)
    Track number of executions (cummulative) *(auto-updated)*.

**numDiscretizeAnglePhi**(*=20*)
    Number of sector for phi-angle

**numDiscretizeAngleTheta**(*=20*)
    Number of sector for theta-angle

**ompThreads**(*=-1*)

    Number of threads to be used in the engine. If ompThreads<0 (default), the number will be typically OMP_NUM_THREADS or the number N defined by 'yade -jN' (this behavior can depend on the engine though). This attribute will only affect engines whose code includes openMP parallel regions (e.g. *InteractionLoop*). This attribute is mostly useful for experiments or when combining *ParallelEngine* with engines that run parallel regions, resulting in nested OMP loops with different number of threads at each level.

**realLast**(*=0*)

    Tracks real time of last run *(auto-updated)*.

**realPeriod**(*=0, deactivated*)

    Periodicity criterion using real (wall clock, computation, human) time in seconds (deactivated if <=0)

**timingDeltas**

    Detailed information about timing inside the Engine itself. Empty unless enabled in the source code and *O.timingEnabled*==`True`.

**updateAttrs**(*(Serializable)arg1, (dict)arg2*) → None :

    Update object attributes from given dictionary

**virtLast**(*=0*)

    Tracks virtual time of last run *(auto-updated)*.

**virtPeriod**(*=0, deactivated*)

    Periodicity criterion using virtual (simulation) time (deactivated if <= 0)

**warnedOnce**(*=false*)

    For one-time warning. May trigger usefull warnings

**class yade.wrapper.MeasureCapStress**(*inherits PeriodicEngine → GlobalEngine → Engine → Serializable*)

Post-processing engine giving *the capillary stress tensor* (the fluids mixture contribution to the total stress in unsaturated, i.e. triphasic, conditions) according to the μUNSAT expression detailed in [Duriez2017c]. Although this expression differs in nature from the one of utils.getCapillaryStress (consideration of distributed integrals herein, vs resultant capillary force therein), both are equivalent [Duriez2016b], [Duriez2017], [Duriez2017c]. The REV volume V entering the expression is automatically measured, from the *Cell* for periodic conditions, or from utils.aabbExtrema function otherwise.

**capillaryPressure**(*=0*)

    Capillary pressure $u_c$, to be defined equal to *Law2_ScGeom_CapillaryPhys_CapillarityPressure.capillaryPressure*.

**dead**(*=false*)

    If true, this engine will not run at all; can be used for making an engine temporarily deactivated and only resurrect it at a later point.

**debug**(*=0*)

    To output some debugging messages.

**dict**(*(Serializable)arg1*) → dict :

    Return dictionary of attributes.

**execCount**

    Cumulative count this engine was run (only used if *O.timingEnabled*==`True`).

**execTime**

    Cumulative time in nanoseconds this Engine took to run (only used if *O.timingEnabled*==`True`).

**firstIterRun**(*=0*)

    Sets the step number, at each an engine should be executed for the first time (disabled by default).

**initRun**(*=false*)

> Run the first time we are called as well.

**iterLast**(*=0*)

> Tracks step number of last run *(auto-updated)*.

**iterPeriod**(*=0, deactivated*)

> Periodicity criterion using step number (deactivated if <= 0)

**label**(*=uninitalized*)

> Textual label for this object; must be valid python identifier, you can refer to it directly from python.

**muGamma**(*=Matrix3r::Zero()*)

> Tensorial contribution to *sigmaCap* from the contact lines $\Gamma$: $\boldsymbol{\mu_\Gamma} = \int_\Gamma \boldsymbol{\nu_{nw}} \otimes \mathbf{x} \, dl$ with $\boldsymbol{\nu_{nw}}$ the fluid-fluid interface conormal [Duriez2017c], and $\mathbf{x}$ the position. *(auto-updated)*

**muSnw**(*=Matrix3r::Zero()*)

> Tensorial contribution to *sigmaCap* from the wetting/non-wetting (e.g. liquid/gas) interface Snw: $\boldsymbol{\mu_{Snw}} = \int_{Snw} (\boldsymbol{\delta} - \mathbf{n} \otimes \mathbf{n}) dS$ with $\mathbf{n}$ the outward normal and $\boldsymbol{\delta}$ the identity tensor. *(auto-updated)*

**muSsw**(*=Matrix3r::Zero()*)

> Tensorial contribution to *sigmaCap* from the wetted solid surfaces Ssw: $\boldsymbol{\mu_{Ssw}} = \int_{Ssw} \mathbf{n} \otimes \mathbf{x} dS$ with $\mathbf{n}$ the outward normal and $\mathbf{x}$ the position. *(auto-updated)*

**muVw**(*=Matrix3r::Zero()*)

> Tensorial contribution (spherical i.e. isotropic) to *sigmaCap* from the wetting fluid volume: $\boldsymbol{\mu_{Vw}} = V_w \boldsymbol{\delta}$ with $V_w = vW$ and $\boldsymbol{\delta}$ the identity tensor. *(auto-updated)*

**nDo**(*=-1, deactivated*)

> Limit number of executions by this number (deactivated if negative)

**nDone**(*=0*)

> Track number of executions (cummulative) *(auto-updated)*.

**ompThreads**(*=-1*)

> Number of threads to be used in the engine. If ompThreads<0 (default), the number will be typically OMP_NUM_THREADS or the number N defined by 'yade -jN' (this behavior can depend on the engine though). This attribute will only affect engines whose code includes openMP parallel regions (e.g. *InteractionLoop*). This attribute is mostly useful for experiments or when combining *ParallelEngine* with engines that run parallel regions, resulting in nested OMP loops with different number of threads at each level.

**realLast**(*=0*)

> Tracks real time of last run *(auto-updated)*.

**realPeriod**(*=0, deactivated*)

> Periodicity criterion using real (wall clock, computation, human) time in seconds (deactivated if <=0)

**sigmaCap**(*=Matrix3r::Zero()*)

> The capillary stress tensor $\boldsymbol{\sigma^{cap}}$ itself, expressed as $\boldsymbol{\sigma^{cap}} = 1/V [u_c(\boldsymbol{\mu_{Vw}} + \boldsymbol{\mu_{Ssw}}) + \gamma_{nw}(\boldsymbol{\mu_{Snw}} + \boldsymbol{\mu_\Gamma})]$ where the four microstructure tensors $\boldsymbol{\mu_{Vw}}, \boldsymbol{\mu_{Ssw}}, \boldsymbol{\mu_{Snw}}, \boldsymbol{\mu_\Gamma}$ correspond to *muVw*, *muSsw*, *muSnw* and *muGamma* attributes. *(auto-updated)*

**surfaceTension**(*=0.073*)

> Fluid-fluid surface tension $\gamma_{nw}$, to be defined equal to *Law2_ScGeom_CapillaryPhys_Capillarity.surfaceTension*.

**timingDeltas**

> Detailed information about timing inside the Engine itself. Empty unless enabled in the source code and *O.timingEnabled*==**True**.

**updateAttrs**(*(Serializable)arg1, (dict)arg2*) → None :

> Update object attributes from given dictionary

**vW**(*=0*)
Wetting fluid volume, summing *menisci volumes* (faster here than through python loops). *(auto-updated)*

**virtLast**(*=0*)
Tracks virtual time of last run *(auto-updated)*.

**virtPeriod**(*=0, deactivated*)
Periodicity criterion using virtual (simulation) time (deactivated if <= 0)

**wettAngle**(*=0*)
Wetting, i.e. contact, angle value (radians). To be defined consistently with the value upon which the capillary files (used by *Law2_ScGeom_CapillaryPhys_Capillarity*) rely.

**class yade.wrapper.PDFEngine**(*inherits PeriodicEngine → GlobalEngine → Engine → Serializable*)
Base class for spectrums calculations. Compute Probability Density Functions of normalStress, shearStress, distance, velocity and interactions in spherical coordinates and write result to a file. Column name format is: Data(theta, phi). Convention used: x: phi = 0, y: theta = 0, z: phi = pi/2

**dead**(*=false*)
If true, this engine will not run at all; can be used for making an engine temporarily deactivated and only resurrect it at a later point.

**dict**(*(Serializable)arg1*) → dict :
Return dictionary of attributes.

**execCount**
Cumulative count this engine was run (only used if *O.timingEnabled*==**True**).

**execTime**
Cumulative time in nanoseconds this Engine took to run (only used if *O.timingEnabled*==**True**).

**filename**(*="PDF.txt"*)
Filename

**firstIterRun**(*=0*)
Sets the step number, at each an engine should be executed for the first time (disabled by default).

**initRun**(*=false*)
Run the first time we are called as well.

**iterLast**(*=0*)
Tracks step number of last run *(auto-updated)*.

**iterPeriod**(*=0, deactivated*)
Periodicity criterion using step number (deactivated if <= 0)

**label**(*=uninitialized*)
Textual label for this object; must be valid python identifier, you can refer to it directly from python.

**nDo**(*=-1, deactivated*)
Limit number of executions by this number (deactivated if negative)

**nDone**(*=0*)
Track number of executions (cummulative) *(auto-updated)*.

**numDiscretizeAnglePhi**(*=20*)
Number of sector for phi-angle

**numDiscretizeAngleTheta**(*=20*)
Number of sector for theta-angle

**ompThreads**(*=-1*)
> Number of threads to be used in the engine. If ompThreads<0 (default), the number will be typically OMP_NUM_THREADS or the number N defined by 'yade -jN' (this behavior can depend on the engine though). This attribute will only affect engines whose code includes openMP parallel regions (e.g. *InteractionLoop*). This attribute is mostly useful for experiments or when combining *ParallelEngine* with engines that run parallel regions, resulting in nested OMP loops with different number of threads at each level.

**realLast**(*=0*)
> Tracks real time of last run *(auto-updated)*.

**realPeriod**(*=0, deactivated*)
> Periodicity criterion using real (wall clock, computation, human) time in seconds (deactivated if <=0)

**timingDeltas**
> Detailed information about timing inside the Engine itself. Empty unless enabled in the source code and *O.timingEnabled*==`True`.

**updateAttrs**(*(Serializable)arg1, (dict)arg2*) → None :
> Update object attributes from given dictionary

**virtLast**(*=0*)
> Tracks virtual time of last run *(auto-updated)*.

**virtPeriod**(*=0, deactivated*)
> Periodicity criterion using virtual (simulation) time (deactivated if <= 0)

**warnedOnce**(*=false*)
> For one-time warning. May trigger usefull warnings

**class yade.wrapper.PyRunner**(*inherits PeriodicEngine → GlobalEngine → Engine → Serializable*)
> Execute a python command periodically, with defined (and adjustable) periodicity. See *PeriodicEngine* documentation for details.

**command**(*=""*)
> Command to be run by python interpreter. Not run if empty.

**dead**(*=false*)
> If true, this engine will not run at all; can be used for making an engine temporarily deactivated and only resurrect it at a later point.

**dict**(*(Serializable)arg1*) → dict :
> Return dictionary of attributes.

**execCount**
> Cumulative count this engine was run (only used if *O.timingEnabled*==`True`).

**execTime**
> Cumulative time in nanoseconds this Engine took to run (only used if *O.timingEnabled*==`True`).

**firstIterRun**(*=0*)
> Sets the step number, at each an engine should be executed for the first time (disabled by default).

**ignoreErrors**(*=false*)
> Debug only: set this value to true to tell PyRunner to ignore any errors encountered during command execution.

**initRun**(*=false*)
> Run the first time we are called as well.

**iterLast**(*=0*)
> Tracks step number of last run *(auto-updated)*.

**iterPeriod**(*=0, deactivated*)
> Periodicity criterion using step number (deactivated if <= 0)

**label**(*=uninitalized*)
> Textual label for this object; must be valid python identifier, you can refer to it directly from python.

**nDo**(*=-1, deactivated*)
> Limit number of executions by this number (deactivated if negative)

**nDone**(*=0*)
> Track number of executions (cummulative) *(auto-updated)*.

**ompThreads**(*=-1*)
> Number of threads to be used in the engine. If ompThreads<0 (default), the number will be typically OMP_NUM_THREADS or the number N defined by 'yade -jN' (this behavior can depend on the engine though). This attribute will only affect engines whose code includes openMP parallel regions (e.g. *InteractionLoop*). This attribute is mostly useful for experiments or when combining *ParallelEngine* with engines that run parallel regions, resulting in nested OMP loops with different number of threads at each level.

**realLast**(*=0*)
> Tracks real time of last run *(auto-updated)*.

**realPeriod**(*=0, deactivated*)
> Periodicity criterion using real (wall clock, computation, human) time in seconds (deactivated if <=0)

**timingDeltas**
> Detailed information about timing inside the Engine itself. Empty unless enabled in the source code and *O.timingEnabled*==`True`.

**updateAttrs**(*(Serializable)arg1, (dict)arg2*) → None :
> Update object attributes from given dictionary

**updateGlobals**
> Whether to workaround ipython not recognizing local variables by calling `globals().update(locals())`. If `true` then PyRunner is able to call functions declared later locally in a running **live** yade session. The `PyRunner` call is a bit slower because it updates `globals()` with recently declared python functions.

> > **Warning:**
> > When `updateGlobals==False` and a function was declared inside a *live* yade session (ipython) then an error `NameError: name 'command' is not defined` will occur unless python `globals()` are updated with command
> > ```
> > globals().update(locals())
> > ```

**virtLast**(*=0*)
> Tracks virtual time of last run *(auto-updated)*.

**virtPeriod**(*=0, deactivated*)
> Periodicity criterion using virtual (simulation) time (deactivated if <= 0)

**class yade.wrapper.Recorder**(*inherits PeriodicEngine → GlobalEngine → Engine → Serializable*)
Engine periodically storing some data to (one) external file. In addition PeriodicEngine, it handles opening the file as needed. See *PeriodicEngine* for controlling periodicity.

**addIterNum**(*=false*)
> Adds an iteration number to the file name, when the file was created. Useful for creating new files at each call (false by default)

**dead**(*=false*)
    If true, this engine will not run at all; can be used for making an engine temporarily deactivated and only resurrect it at a later point.

**dict**(*(Serializable)arg1*) → dict :
    Return dictionary of attributes.

**execCount**
    Cumulative count this engine was run (only used if *O.timingEnabled*==`True`).

**execTime**
    Cumulative time in nanoseconds this Engine took to run (only used if *O.timingEnabled*==`True`).

**file**(*=uninitalized*)
    Name of file to save to; must not be empty.

**firstIterRun**(*=0*)
    Sets the step number, at each an engine should be executed for the first time (disabled by default).

**initRun**(*=false*)
    Run the first time we are called as well.

**iterLast**(*=0*)
    Tracks step number of last run *(auto-updated)*.

**iterPeriod**(*=0, deactivated*)
    Periodicity criterion using step number (deactivated if <= 0)

**label**(*=uninitalized*)
    Textual label for this object; must be valid python identifier, you can refer to it directly from python.

**nDo**(*=-1, deactivated*)
    Limit number of executions by this number (deactivated if negative)

**nDone**(*=0*)
    Track number of executions (cummulative) *(auto-updated)*.

**ompThreads**(*=-1*)
    Number of threads to be used in the engine. If ompThreads<0 (default), the number will be typically OMP_NUM_THREADS or the number N defined by 'yade -jN' (this behavior can depend on the engine though). This attribute will only affect engines whose code includes openMP parallel regions (e.g. *InteractionLoop*). This attribute is mostly useful for experiments or when combining *ParallelEngine* with engines that run parallel regions, resulting in nested OMP loops with different number of threads at each level.

**realLast**(*=0*)
    Tracks real time of last run *(auto-updated)*.

**realPeriod**(*=0, deactivated*)
    Periodicity criterion using real (wall clock, computation, human) time in seconds (deactivated if <=0)

**timingDeltas**
    Detailed information about timing inside the Engine itself. Empty unless enabled in the source code and *O.timingEnabled*==`True`.

**truncate**(*=false*)
    Whether to delete current file contents, if any, when opening (false by default)

**updateAttrs**(*(Serializable)arg1, (dict)arg2*) → None :
    Update object attributes from given dictionary

**virtLast**(*=0*)
    Tracks virtual time of last run *(auto-updated)*.

**virtPeriod**(*=0, deactivated*)
    Periodicity criterion using virtual (simulation) time (deactivated if <= 0)

**class yade.wrapper.ResetRandomPosition**(*inherits* *PeriodicEngine* → *GlobalEngine* → *Engine*
                                        → *Serializable*)
Creates spheres during simulation, placing them at random positions. Every time called, one new
sphere will be created and inserted in the simulation.

**angularVelocity**(*=Vector3r::Zero()*)
    Mean angularVelocity of spheres.

**angularVelocityRange**(*=Vector3r::Zero()*)
    Half size of a angularVelocity distribution interval. New sphere will have random angularVe-
    locity within the range angularVelocity±angularVelocityRange.

**dead**(*=false*)
    If true, this engine will not run at all; can be used for making an engine temporarily deactivated
    and only resurrect it at a later point.

**dict**(*(Serializable)arg1*) → dict :
    Return dictionary of attributes.

**execCount**
    Cumulative count this engine was run (only used if *O.timingEnabled*==`True`).

**execTime**
    Cumulative time in nanoseconds this Engine took to run (only used if
    *O.timingEnabled*==`True`).

**factoryFacets**(*=uninitalized*)
    The geometry of the section where spheres will be placed; they will be placed on facets or in
    volume between them depending on *volumeSection* flag.

**firstIterRun**(*=0*)
    Sets the step number, at each an engine should be executed for the first time (disabled by
    default).

**initRun**(*=false*)
    Run the first time we are called as well.

**iterLast**(*=0*)
    Tracks step number of last run *(auto-updated)*.

**iterPeriod**(*=0, deactivated*)
    Periodicity criterion using step number (deactivated if <= 0)

**label**(*=uninitalized*)
    Textual label for this object; must be valid python identifier, you can refer to it directly from
    python.

**maxAttempts**(*=20*)
    Max attempts to place sphere. If placing the sphere in certain random position would cause
    an overlap with any other physical body in the model, SpheresFactory will try to find another
    position.

**nDo**(*=-1, deactivated*)
    Limit number of executions by this number (deactivated if negative)

**nDone**(*=0*)
    Track number of executions (cummulative) *(auto-updated)*.

**normal**(*=Vector3r(0, 1, 0)*)
    ??

**ompThreads**(*=-1*)
    Number of threads to be used in the engine. If ompThreads<0 (default), the number will be
    typically OMP_NUM_THREADS or the number N defined by 'yade -jN' (this behavior can

depend on the engine though). This attribute will only affect engines whose code includes openMP parallel regions (e.g. *InteractionLoop*). This attribute is mostly useful for experiments or when combining *ParallelEngine* with engines that run parallel regions, resulting in nested OMP loops with different number of threads at each level.

**point**(=*Vector3r::Zero()*)
    ??

**realLast**(=*0*)
    Tracks real time of last run *(auto-updated)*.

**realPeriod**(=*0, deactivated*)
    Periodicity criterion using real (wall clock, computation, human) time in seconds (deactivated if <=0)

**subscribedBodies**(=*uninitalized*)
    Affected bodies.

**timingDeltas**
    Detailed information about timing inside the Engine itself. Empty unless enabled in the source code and *O.timingEnabled*==`True`.

**updateAttrs**(*(Serializable)arg1, (dict)arg2*) → None :
    Update object attributes from given dictionary

**velocity**(=*Vector3r::Zero()*)
    Mean velocity of spheres.

**velocityRange**(=*Vector3r::Zero()*)
    Half size of a velocities distribution interval. New sphere will have random velocity within the range velocity±velocityRange.

**virtLast**(=*0*)
    Tracks virtual time of last run *(auto-updated)*.

**virtPeriod**(=*0, deactivated*)
    Periodicity criterion using virtual (simulation) time (deactivated if <= 0)

**volumeSection**(=*false, define factory by facets.*)
    Create new spheres inside factory volume rather than on its surface.

**class yade.wrapper.SnapshotEngine**(*inherits PeriodicEngine → GlobalEngine → Engine → Serializable*)
Periodically save snapshots of GLView(s) as .png files. Files are named *fileBase + counter +* '.png' (counter is left-padded by 0s, i.e. snap00004.png).

**counter**(=*0*)
    Number that will be appended to fileBase when the next snapshot is saved (incremented at every save). *(auto-updated)*

**dead**(=*false*)
    If true, this engine will not run at all; can be used for making an engine temporarily deactivated and only resurrect it at a later point.

**deadTimeout**(=*3*)
    Timeout for 3d operations (opening new view, saving snapshot); after timing out, throw exception (or only report error if *ignoreErrors*) and make myself *dead*. [s]

**dict**(*(Serializable)arg1*) → dict :
    Return dictionary of attributes.

**execCount**
    Cumulative count this engine was run (only used if *O.timingEnabled*==`True`).

**execTime**
    Cumulative time in nanoseconds this Engine took to run (only used if *O.timingEnabled*==`True`).

**fileBase**(*=""*)
> Basename for snapshots

**firstIterRun**(*=0*)
> Sets the step number, at each an engine should be executed for the first time (disabled by default).

**format**(*="PNG"*)
> Format of snapshots (one of JPEG, PNG, EPS, PS, PPM, BMP) QGLViewer documentation. File extension will be lowercased *format*. Validity of format is not checked.

**ignoreErrors**(*=true*)
> Only report errors instead of throwing exceptions, in case of timeouts.

**initRun**(*=false*)
> Run the first time we are called as well.

**iterLast**(*=0*)
> Tracks step number of last run *(auto-updated)*.

**iterPeriod**(*=0, deactivated*)
> Periodicity criterion using step number (deactivated if <= 0)

**label**(*=uninitalized*)
> Textual label for this object; must be valid python identifier, you can refer to it directly from python.

**msecSleep**(*=0*)
> number of msec to sleep after snapshot (to prevent 3d hw problems) [ms]

**nDo**(*=-1, deactivated*)
> Limit number of executions by this number (deactivated if negative)

**nDone**(*=0*)
> Track number of executions (cummulative) *(auto-updated)*.

**ompThreads**(*=-1*)
> Number of threads to be used in the engine. If ompThreads<0 (default), the number will be typically OMP_NUM_THREADS or the number N defined by 'yade -jN' (this behavior can depend on the engine though). This attribute will only affect engines whose code includes openMP parallel regions (e.g. *InteractionLoop*). This attribute is mostly useful for experiments or when combining *ParallelEngine* with engines that run parallel regions, resulting in nested OMP loops with different number of threads at each level.

**plot**(*=uninitalized*)
> Name of field in *plot.imgData* to which taken snapshots will be appended automatically.

**realLast**(*=0*)
> Tracks real time of last run *(auto-updated)*.

**realPeriod**(*=0, deactivated*)
> Periodicity criterion using real (wall clock, computation, human) time in seconds (deactivated if <=0)

**snapshots**(*=uninitalized*)
> Files that have been created so far

**timingDeltas**
> Detailed information about timing inside the Engine itself. Empty unless enabled in the source code and *O.timingEnabled*==**True**.

**updateAttrs**(*(Serializable)arg1, (dict)arg2*) → None :
> Update object attributes from given dictionary

**virtLast**(*=0*)
> Tracks virtual time of last run *(auto-updated)*.

**virtPeriod**(*=0, deactivated*)
    Periodicity criterion using virtual (simulation) time (deactivated if <= 0)

**class** yade.wrapper.**TorqueRecorder**(*inherits Recorder → PeriodicEngine → GlobalEngine →*
                                    *Engine → Serializable*)
Engine saves the total torque according to the given axis and ZeroPoint, the force is taken from
bodies, listed in *ids* For instance, can be useful for defining the torque, which affects on ball mill
during its work.

**addIterNum**(*=false*)
    Adds an iteration number to the file name, when the file was created. Useful for creating new
    files at each call (false by default)

**dead**(*=false*)
    If true, this engine will not run at all; can be used for making an engine temporarily deactivated
    and only resurrect it at a later point.

**dict**(*(Serializable)arg1*) → dict :
    Return dictionary of attributes.

**execCount**
    Cumulative count this engine was run (only used if *O.timingEnabled*==True).

**execTime**
    Cumulative time in nanoseconds this Engine took to run (only used if
    *O.timingEnabled*==True).

**file**(*=uninitalized*)
    Name of file to save to; must not be empty.

**firstIterRun**(*=0*)
    Sets the step number, at each an engine should be executed for the first time (disabled by
    default).

**ids**(*=uninitalized*)
    List of bodies whose state will be measured

**initRun**(*=false*)
    Run the first time we are called as well.

**iterLast**(*=0*)
    Tracks step number of last run *(auto-updated)*.

**iterPeriod**(*=0, deactivated*)
    Periodicity criterion using step number (deactivated if <= 0)

**label**(*=uninitalized*)
    Textual label for this object; must be valid python identifier, you can refer to it directly from
    python.

**nDo**(*=-1, deactivated*)
    Limit number of executions by this number (deactivated if negative)

**nDone**(*=0*)
    Track number of executions (cummulative) *(auto-updated)*.

**ompThreads**(*=-1*)
    Number of threads to be used in the engine. If ompThreads<0 (default), the number will be
    typically OMP_NUM_THREADS or the number N defined by 'yade -jN' (this behavior can
    depend on the engine though). This attribute will only affect engines whose code includes
    openMP parallel regions (e.g. *InteractionLoop*). This attribute is mostly useful for experi-
    ments or when combining *ParallelEngine* with engines that run parallel regions, resulting in
    nested OMP loops with different number of threads at each level.

**realLast**(*=0*)
    Tracks real time of last run *(auto-updated)*.

**realPeriod**(*=0, deactivated*)
> Periodicity criterion using real (wall clock, computation, human) time in seconds (deactivated if <=0)

**rotationAxis**(*=Vector3r::UnitX()*)
> Rotation axis

**timingDeltas**
> Detailed information about timing inside the Engine itself. Empty unless enabled in the source code and *O.timingEnabled*==`True`.

**totalTorque**(*=0*)
> Resultant torque, returning by the function.

**truncate**(*=false*)
> Whether to delete current file contents, if any, when opening (false by default)

**updateAttrs**(*(Serializable)arg1, (dict)arg2*) → None :
> Update object attributes from given dictionary

**virtLast**(*=0*)
> Tracks virtual time of last run *(auto-updated)*.

**virtPeriod**(*=0, deactivated*)
> Periodicity criterion using virtual (simulation) time (deactivated if <= 0)

**zeroPoint**(*=Vector3r::Zero()*)
> Point of rotation center

**class** `yade.wrapper.`**TriaxialStateRecorder**(*inherits* *Recorder* → *PeriodicEngine* → *GlobalEngine* → *Engine* → *Serializable*)
Engine recording triaxial variables (see the variables list in the first line of the output file). This recorder needs *TriaxialCompressionEngine* or *ThreeDTriaxialEngine* present in the simulation.

**addIterNum**(*=false*)
> Adds an iteration number to the file name, when the file was created. Useful for creating new files at each call (false by default)

**dead**(*=false*)
> If true, this engine will not run at all; can be used for making an engine temporarily deactivated and only resurrect it at a later point.

**dict**(*(Serializable)arg1*) → dict :
> Return dictionary of attributes.

**execCount**
> Cumulative count this engine was run (only used if *O.timingEnabled*==`True`).

**execTime**
> Cumulative time in nanoseconds this Engine took to run (only used if *O.timingEnabled*==`True`).

**file**(*=uninitialized*)
> Name of file to save to; must not be empty.

**firstIterRun**(*=0*)
> Sets the step number, at each an engine should be executed for the first time (disabled by default).

**initRun**(*=false*)
> Run the first time we are called as well.

**iterLast**(*=0*)
> Tracks step number of last run *(auto-updated)*.

**iterPeriod**(*=0, deactivated*)
> Periodicity criterion using step number (deactivated if <= 0)

**label** *(=uninitalized)*
　　Textual label for this object; must be valid python identifier, you can refer to it directly from python.

**nDo** *(=-1, deactivated)*
　　Limit number of executions by this number (deactivated if negative)

**nDone** *(=0)*
　　Track number of executions (cummulative) *(auto-updated)*.

**ompThreads** *(=-1)*
　　Number of threads to be used in the engine. If ompThreads<0 (default), the number will be typically OMP_NUM_THREADS or the number N defined by 'yade -jN' (this behavior can depend on the engine though). This attribute will only affect engines whose code includes openMP parallel regions (e.g. *InteractionLoop*). This attribute is mostly useful for experiments or when combining *ParallelEngine* with engines that run parallel regions, resulting in nested OMP loops with different number of threads at each level.

**porosity** *(=1)*
　　porosity of the packing [-]

**realLast** *(=0)*
　　Tracks real time of last run *(auto-updated)*.

**realPeriod** *(=0, deactivated)*
　　Periodicity criterion using real (wall clock, computation, human) time in seconds (deactivated if <=0)

**timingDeltas**
　　Detailed information about timing inside the Engine itself. Empty unless enabled in the source code and *O.timingEnabled*==`True`.

**truncate** *(=false)*
　　Whether to delete current file contents, if any, when opening (false by default)

**updateAttrs** *((Serializable)arg1, (dict)arg2)* → None :
　　Update object attributes from given dictionary

**virtLast** *(=0)*
　　Tracks virtual time of last run *(auto-updated)*.

**virtPeriod** *(=0, deactivated)*
　　Periodicity criterion using virtual (simulation) time (deactivated if <= 0)

## BoundaryController

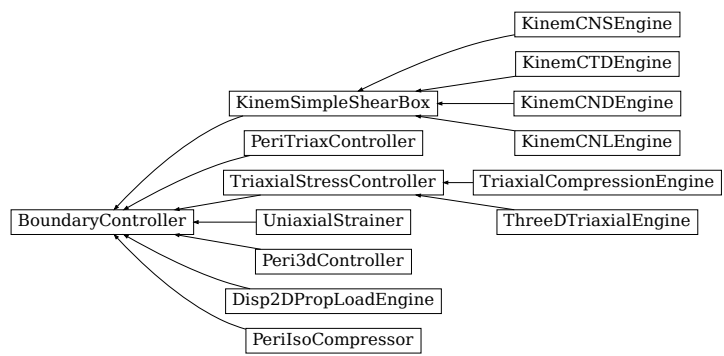

Fig. 27: Inheritance graph of BoundaryController. See also: *Disp2DPropLoadEngine*, *KinemCNDEngine*, *KinemCNLEngine*, *KinemCNSEngine*, *KinemCTDEngine*, *KinemSimpleShearBox*, *Peri3dController*, *PeriIsoCompressor*, *PeriTriaxController*, *ThreeDTriaxialEngine*, *TriaxialCompressionEngine*, *TriaxialStressController*, *UniaxialStrainer*.

`class yade.wrapper.`**`BoundaryController`**`(`*inherits* *GlobalEngine* → *Engine* → *Serializable*`)`
  Base for engines controlling boundary conditions of simulations. Not to be used directly.

  **`dead`**`(`*=false*`)`
      If true, this engine will not run at all; can be used for making an engine temporarily deactivated and only resurrect it at a later point.

  **`dict`**`(`*(Serializable)arg1*`)` → dict :
      Return dictionary of attributes.

  **`execCount`**
      Cumulative count this engine was run (only used if *O.timingEnabled*==`True`).

  **`execTime`**
      Cumulative time in nanoseconds this Engine took to run (only used if *O.timingEnabled*==`True`).

  **`label`**`(`*=uninitalized*`)`
      Textual label for this object; must be valid python identifier, you can refer to it directly from python.

  **`ompThreads`**`(`*=-1*`)`
      Number of threads to be used in the engine. If ompThreads<0 (default), the number will be typically OMP_NUM_THREADS or the number N defined by 'yade -jN' (this behavior can depend on the engine though). This attribute will only affect engines whose code includes openMP parallel regions (e.g. *InteractionLoop*). This attribute is mostly useful for experiments or when combining *ParallelEngine* with engines that run parallel regions, resulting in nested OMP loops with different number of threads at each level.

  **`timingDeltas`**
      Detailed information about timing inside the Engine itself. Empty unless enabled in the source code and *O.timingEnabled*==`True`.

  **`updateAttrs`**`(`*(Serializable)arg1, (dict)arg2*`)` → None :
      Update object attributes from given dictionary

`class yade.wrapper.`**`Disp2DPropLoadEngine`**`(`*inherits* *BoundaryController* → *GlobalEngine* → *Engine* → *Serializable*`)`
  Disturbs a simple shear sample in a given displacement direction

  This engine allows one to apply, on a simple shear sample, a loading controlled by du/dgamma = cste, which is equivalent to du + cste' * dgamma = 0 (proportionnal path loadings). To do so, the upper plate of the simple shear box is moved in a given direction (corresponding to a given du/dgamma), whereas lateral plates are moved so that the box remains closed. This engine can easily be used to perform directionnal probes, with a python script launching successivly the same .xml which contains this engine, after having modified the direction of loading (see *theta* attribute). That's why this Engine contains a *saveData* procedure which can save data on the state of the sample at the end of the loading (in case of successive loadings - for successive directions - through a python script, each line would correspond to one direction of loading).

  **`Key`**`(`*=""*`)`
      string to add at the names of the saved files, and of the output file filled by *saveData*

  **`LOG`**`(`*=false*`)`
      boolean controling the output of messages on the screen

  **`dead`**`(`*=false*`)`
      If true, this engine will not run at all; can be used for making an engine temporarily deactivated and only resurrect it at a later point.

  **`dict`**`(`*(Serializable)arg1*`)` → dict :
      Return dictionary of attributes.

  **`execCount`**
      Cumulative count this engine was run (only used if *O.timingEnabled*==`True`).

**execTime**
> Cumulative time in nanoseconds this Engine took to run (only used if *O.timingEnabled*==**True**).

**id_boxback**(*=4*)
> the id of the wall at the back of the sample

**id_boxbas**(*=1*)
> the id of the lower wall

**id_boxfront**(*=5*)
> the id of the wall in front of the sample

**id_boxleft**(*=0*)
> the id of the left wall

**id_boxright**(*=2*)
> the id of the right wall

**id_topbox**(*=3*)
> the id of the upper wall

**label**(*=uninitalized*)
> Textual label for this object; must be valid python identifier, you can refer to it directly from python.

**nbre_iter**(*=0*)
> the number of iterations of loading to perform

**ompThreads**(*=-1*)
> Number of threads to be used in the engine. If ompThreads<0 (default), the number will be typically OMP_NUM_THREADS or the number N defined by 'yade -jN' (this behavior can depend on the engine though). This attribute will only affect engines whose code includes openMP parallel regions (e.g. *InteractionLoop*). This attribute is mostly useful for experiments or when combining *ParallelEngine* with engines that run parallel regions, resulting in nested OMP loops with different number of threads at each level.

**theta**(*=0.0*)
> the angle, in a (gamma,h=-u) plane from the gamma - axis to the perturbation vector (trigo wise) [degrees]

**timingDeltas**
> Detailed information about timing inside the Engine itself. Empty unless enabled in the source code and *O.timingEnabled*==**True**.

**updateAttrs**(*(Serializable)arg1, (dict)arg2*) → None :
> Update object attributes from given dictionary

**v**(*=0.0*)
> the speed at which the perturbation is imposed. In case of samples which are more sensitive to normal loadings than tangential ones, one possibility is to take v = V_shear - | (V_shear-V_comp)*sin(theta) | => v=V_shear in shear; V_comp in compression [m/s]

**class yade.wrapper.KinemCNDEngine**(*inherits KinemSimpleShearBox → BoundaryController → GlobalEngine → Engine → Serializable*)

To apply a Constant Normal Displacement (CND) shear for a parallelogram box

This engine, designed for simulations implying a simple shear box (*SimpleShear* Preprocessor or scripts/simpleShear.py), allows one to perform a constant normal displacement shear, by translating horizontally the upper plate, while the lateral ones rotate so that they always keep contact with the lower and upper walls.

**Key**(*=""*)
> string to add at the names of the saved files

**LOG**(*=false*)
> boolean controling the output of messages on the screen

**alpha**(*=Mathr::PI/2.0*)
> the angle from the lower box to the left box (trigo wise). Measured by this Engine. Has to be saved, but not to be changed by the user.

**dead**(*=false*)
> If true, this engine will not run at all; can be used for making an engine temporarily deactivated and only resurrect it at a later point.

**dict**(*(Serializable)arg1*) → dict :
> Return dictionary of attributes.

**execCount**
> Cumulative count this engine was run (only used if *O.timingEnabled*==True).

**execTime**
> Cumulative time in nanoseconds this Engine took to run (only used if *O.timingEnabled*==True).

**f0**(*=0.0*)
> the (vertical) force acting on the upper plate on the very first time step (determined by the Engine). Controls of the loadings in case of *KinemCNSEngine* or *KinemCNLEngine* will be done according to this initial value [N]. Has to be saved, but not to be changed by the user.

**firstRun**(*=true*)
> boolean set to false as soon as the engine has done its job one time : useful to know if initial height of, and normal force sustained by, the upper box are known or not (and thus if they have to be initialized). Has to be saved, but not to be changed by the user.

**gamma**(*=0.0*)
> the current value of the tangential displacement

**gamma_save**(*=uninitalized*)
> vector with the values of gamma at which a save of the simulation is performed [m]

**gammalim**(*=0.0*)
> the value of the tangential displacement at wich the displacement is stopped [m]

**id_boxback**(*=4*)
> the id of the wall at the back of the sample

**id_boxbas**(*=1*)
> the id of the lower wall

**id_boxfront**(*=5*)
> the id of the wall in front of the sample

**id_boxleft**(*=0*)
> the id of the left wall

**id_boxright**(*=2*)
> the id of the right wall

**id_topbox**(*=3*)
> the id of the upper wall

**label**(*=uninitalized*)
> Textual label for this object; must be valid python identifier, you can refer to it directly from python.

**max_vel**(*=1.0*)
> to limit the speed of the vertical displacements done to control σ (CNL or CNS cases) [m/s]

**ompThreads**(*=-1*)
> Number of threads to be used in the engine. If ompThreads<0 (default), the number will be typically OMP_NUM_THREADS or the number N defined by 'yade -jN' (this behavior can depend on the engine though). This attribute will only affect engines whose code includes openMP parallel regions (e.g. *InteractionLoop*). This attribute is mostly useful for experiments or when combining *ParallelEngine* with engines that run parallel regions, resulting in nested OMP loops with different number of threads at each level.

**shearSpeed**(*=0.0*)
    the speed at which the shear is performed : speed of the upper plate [m/s]

**temoin_save**(*=uninitialized*)
    vector (same length as 'gamma_save' for ex), with 0 or 1 depending whether the save for the corresponding value of gamma has been done (1) or not (0). Has to be saved, but not to be changed by the user.

**timingDeltas**
    Detailed information about timing inside the Engine itself. Empty unless enabled in the source code and *O.timingEnabled*==`True`.

**updateAttrs**(*(Serializable)arg1, (dict)arg2*) → None :
    Update object attributes from given dictionary

**wallDamping**(*=0.2*)
    the vertical displacements done to to control σ (CNL or CNS cases) are in fact damped, through this wallDamping

**y0**(*=0.0*)
    the height of the upper plate at the very first time step : the engine finds its value [m]. Has to be saved, but not to be changed by the user.

**class yade.wrapper.KinemCNLEngine**(*inherits KinemSimpleShearBox → BoundaryController → GlobalEngine → Engine → Serializable*)

To apply a constant normal stress shear (i.e. Constant Normal Load : CNL) for a parallelogram box (simple shear box : *SimpleShear* Preprocessor or scripts/simpleShear.py)

This engine allows one to translate horizontally the upper plate while the lateral ones rotate so that they always keep contact with the lower and upper walls.

In fact the upper plate can move not only horizontally but also vertically, so that the normal stress acting on it remains constant (this constant value is not chosen by the user but is the one that exists at the beginning of the simulation)

The right vertical displacements which will be allowed are computed from the rigidity Kn of the sample over the wall (so to cancel a deltaSigma, a normal dplt deltaSigma*S/(Kn) is set)

The movement is moreover controlled by the user via a *shearSpeed* which will be the speed of the upper wall, and by a maximum value of horizontal displacement *gammalim*, after which the shear stops.

---

**Note:** Not only the positions of walls are updated but also their speeds, which is all but useless considering the fact that in the contact laws these velocities of bodies are used to compute values of tangential relative displacements.

---

> **Warning:** Because of this last point, if you want to use later saves of simulations executed with this Engine, but without that stopMovement was executed, your boxes will keep their speeds => you will have to cancel them 'by hand' in the .xml.

**Key**(*=""*)
    string to add at the names of the saved files

**LOG**(*=false*)
    boolean controling the output of messages on the screen

---

**alpha**(*=Mathr::PI/2.0*)
    the angle from the lower box to the left box (trigo wise). Measured by this Engine. Has to be saved, but not to be changed by the user.

**dead**(*=false*)
    If true, this engine will not run at all; can be used for making an engine temporarily deactivated and only resurrect it at a later point.

**dict**(*(Serializable)arg1*) → dict :
    Return dictionary of attributes.

**execCount**
    Cumulative count this engine was run (only used if *O.timingEnabled*==True).

**execTime**
    Cumulative time in nanoseconds this Engine took to run (only used if *O.timingEnabled*==True).

**f0**(*=0.0*)
    the (vertical) force acting on the upper plate on the very first time step (determined by the Engine). Controls of the loadings in case of *KinemCNSEngine* or *KinemCNLEngine* will be done according to this initial value [N]. Has to be saved, but not to be changed by the user.

**firstRun**(*=true*)
    boolean set to false as soon as the engine has done its job one time : useful to know if initial height of, and normal force sustained by, the upper box are known or not (and thus if they have to be initialized). Has to be saved, but not to be changed by the user.

**gamma**(*=0.0*)
    current value of tangential displacement [m]

**gamma_save**(*=uninitalized*)
    vector with the values of gamma at which a save of the simulation is performed [m]

**gammalim**(*=0.0*)
    the value of tangential displacement (of upper plate) at wich the shearing is stopped [m]

**id_boxback**(*=4*)
    the id of the wall at the back of the sample

**id_boxbas**(*=1*)
    the id of the lower wall

**id_boxfront**(*=5*)
    the id of the wall in front of the sample

**id_boxleft**(*=0*)
    the id of the left wall

**id_boxright**(*=2*)
    the id of the right wall

**id_topbox**(*=3*)
    the id of the upper wall

**label**(*=uninitalized*)
    Textual label for this object; must be valid python identifier, you can refer to it directly from python.

**max_vel**(*=1.0*)
    to limit the speed of the vertical displacements done to control σ (CNL or CNS cases) [m/s]

**ompThreads**(*=-1*)
    Number of threads to be used in the engine. If ompThreads<0 (default), the number will be typically OMP_NUM_THREADS or the number N defined by 'yade -jN' (this behavior can depend on the engine though). This attribute will only affect engines whose code includes

openMP parallel regions (e.g. *InteractionLoop*). This attribute is mostly useful for experiments or when combining *ParallelEngine* with engines that run parallel regions, resulting in nested OMP loops with different number of threads at each level.

**shearSpeed**(*=0.0*)
 the speed at wich the shearing is performed : speed of the upper plate [m/s]

**temoin_save**(*=uninitialized*)
 vector (same length as 'gamma_save' for ex), with 0 or 1 depending whether the save for the corresponding value of gamma has been done (1) or not (0). Has to be saved, but not to be changed by the user.

**timingDeltas**
 Detailed information about timing inside the Engine itself. Empty unless enabled in the source code and *O.timingEnabled*==`True`.

**updateAttrs**(*(Serializable)arg1, (dict)arg2*) → None :
 Update object attributes from given dictionary

**wallDamping**(*=0.2*)
 the vertical displacements done to to control σ (CNL or CNS cases) are in fact damped, through this wallDamping

**y0**(*=0.0*)
 the height of the upper plate at the very first time step : the engine finds its value [m]. Has to be saved, but not to be changed by the user.

**class yade.wrapper.KinemCNSEngine**(*inherits KinemSimpleShearBox → BoundaryController → GlobalEngine → Engine → Serializable*)

To apply a Constant Normal Stifness (CNS) shear for a parallelogram box (simple shear)

This engine, useable in simulations implying one deformable parallelepipedic box, allows one to translate horizontally the upper plate while the lateral ones rotate so that they always keep contact with the lower and upper walls. The upper plate can move not only horizontally but also vertically, so that the normal rigidity defined by DeltaF(upper plate)/DeltaU(upper plate) = constant (= *KnC* defined by the user).

The movement is moreover controlled by the user via a *shearSpeed* which is the horizontal speed of the upper wall, and by a maximum value of horizontal displacement *gammalim* (of the upper plate), after which the shear stops.

---

**Note:** not only the positions of walls are updated but also their speeds, which is all but useless considering the fact that in the contact laws these velocities of bodies are used to compute values of tangential relative displacements.

---

> **Warning:** But, because of this last point, if you want to use later saves of simulations executed with this Engine, but without that stopMovement was executed, your boxes will keep their speeds => you will have to cancel them by hand in the .xml

**Key**(*=""*)
 string to add at the names of the saved files

**KnC**(*=10.0e6*)
 the normal rigidity chosen by the user [MPa/mm] - the conversion in Pa/m will be made

**LOG**(*=false*)
 boolean controling the output of messages on the screen

**alpha**(*=Mathr::PI/2.0*)
 the angle from the lower box to the left box (trigo wise). Measured by this Engine. Has to be saved, but not to be changed by the user.

**dead**(*=false*)
    If true, this engine will not run at all; can be used for making an engine temporarily deactivated and only resurrect it at a later point.

**dict**(*(Serializable)arg1*) → dict :
    Return dictionary of attributes.

**execCount**
    Cumulative count this engine was run (only used if *O.timingEnabled*==`True`).

**execTime**
    Cumulative time in nanoseconds this Engine took to run (only used if *O.timingEnabled*==`True`).

**f0**(*=0.0*)
    the (vertical) force acting on the upper plate on the very first time step (determined by the Engine). Controls of the loadings in case of *KinemCNSEngine* or *KinemCNLEngine* will be done according to this initial value [N]. Has to be saved, but not to be changed by the user.

**firstRun**(*=true*)
    boolean set to false as soon as the engine has done its job one time : useful to know if initial height of, and normal force sustained by, the upper box are known or not (and thus if they have to be initialized). Has to be saved, but not to be changed by the user.

**gamma**(*=0.0*)
    current value of tangential displacement [m]

**gammalim**(*=0.0*)
    the value of tangential displacement (of upper plate) at wich the shearing is stopped [m]

**id_boxback**(*=4*)
    the id of the wall at the back of the sample

**id_boxbas**(*=1*)
    the id of the lower wall

**id_boxfront**(*=5*)
    the id of the wall in front of the sample

**id_boxleft**(*=0*)
    the id of the left wall

**id_boxright**(*=2*)
    the id of the right wall

**id_topbox**(*=3*)
    the id of the upper wall

**label**(*=uninitialized*)
    Textual label for this object; must be valid python identifier, you can refer to it directly from python.

**max_vel**(*=1.0*)
    to limit the speed of the vertical displacements done to control σ (CNL or CNS cases) [m/s]

**ompThreads**(*=-1*)
    Number of threads to be used in the engine. If ompThreads<0 (default), the number will be typically OMP_NUM_THREADS or the number N defined by 'yade -jN' (this behavior can depend on the engine though). This attribute will only affect engines whose code includes openMP parallel regions (e.g. *InteractionLoop*). This attribute is mostly useful for experiments or when combining *ParallelEngine* with engines that run parallel regions, resulting in nested OMP loops with different number of threads at each level.

**shearSpeed**(*=0.0*)
    the speed at wich the shearing is performed : speed of the upper plate [m/s]

**temoin_save**(*=uninitialized*)
> vector (same length as 'gamma_save' for ex), with 0 or 1 depending whether the save for the corresponding value of gamma has been done (1) or not (0). Has to be saved, but not to be changed by the user.

**timingDeltas**
> Detailed information about timing inside the Engine itself. Empty unless enabled in the source code and *O.timingEnabled*==`True`.

**updateAttrs**(*(Serializable)arg1, (dict)arg2*) → None :
> Update object attributes from given dictionary

**wallDamping**(*=0.2*)
> the vertical displacements done to to control σ (CNL or CNS cases) are in fact damped, through this wallDamping

**y0**(*=0.0*)
> the height of the upper plate at the very first time step : the engine finds its value [m]. Has to be saved, but not to be changed by the user.

**class yade.wrapper.KinemCTDEngine**(*inherits KinemSimpleShearBox → BoundaryController → GlobalEngine → Engine → Serializable*)

To compress a simple shear sample by moving the upper box in a vertical way only, so that the tangential displacement (defined by the horizontal gap between the upper and lower boxes) remains constant (thus, the CTD = Constant Tangential Displacement). The lateral boxes move also to keep always contact. All that until this box is submitted to a given stress (*targetSigma*). Moreover saves are executed at each value of stresses stored in the vector *sigma_save*, and at *targetSigma*

**Key**(*=""*)
> string to add at the names of the saved files

**LOG**(*=false*)
> boolean controling the output of messages on the screen

**alpha**(*=Mathr::PI/2.0*)
> the angle from the lower box to the left box (trigo wise). Measured by this Engine. Has to be saved, but not to be changed by the user.

**compSpeed**(*=0.0*)
> (vertical) speed of the upper box : >0 for real compression, <0 for unloading [m/s]

**dead**(*=false*)
> If true, this engine will not run at all; can be used for making an engine temporarily deactivated and only resurrect it at a later point.

**dict**(*(Serializable)arg1*) → dict :
> Return dictionary of attributes.

**execCount**
> Cumulative count this engine was run (only used if *O.timingEnabled*==`True`).

**execTime**
> Cumulative time in nanoseconds this Engine took to run (only used if *O.timingEnabled*==`True`).

**f0**(*=0.0*)
> the (vertical) force acting on the upper plate on the very first time step (determined by the Engine). Controls of the loadings in case of *KinemCNSEngine* or *KinemCNLEngine* will be done according to this initial value [N]. Has to be saved, but not to be changed by the user.

**firstRun**(*=true*)
> boolean set to false as soon as the engine has done its job one time : useful to know if initial height of, and normal force sustained by, the upper box are known or not (and thus if they have to be initialized). Has to be saved, but not to be changed by the user.

**id_boxback**(=*4*)
    the id of the wall at the back of the sample

**id_boxbas**(=*1*)
    the id of the lower wall

**id_boxfront**(=*5*)
    the id of the wall in front of the sample

**id_boxleft**(=*0*)
    the id of the left wall

**id_boxright**(=*2*)
    the id of the right wall

**id_topbox**(=*3*)
    the id of the upper wall

**label**(=*uninitalized*)
    Textual label for this object; must be valid python identifier, you can refer to it directly from python.

**max_vel**(=*1.0*)
    to limit the speed of the vertical displacements done to control $\sigma$ (CNL or CNS cases) [m/s]

**ompThreads**(=*-1*)
    Number of threads to be used in the engine. If ompThreads<0 (default), the number will be typically OMP_NUM_THREADS or the number N defined by 'yade -jN' (this behavior can depend on the engine though). This attribute will only affect engines whose code includes openMP parallel regions (e.g. *InteractionLoop*). This attribute is mostly useful for experiments or when combining *ParallelEngine* with engines that run parallel regions, resulting in nested OMP loops with different number of threads at each level.

**sigma_save**(=*uninitalized*)
    vector with the values of sigma at which a save of the simulation should be performed [kPa]

**targetSigma**(=*0.0*)
    the value of sigma at which the compression should stop [kPa]

**temoin_save**(=*uninitalized*)
    vector (same length as 'gamma_save' for ex), with 0 or 1 depending whether the save for the corresponding value of gamma has been done (1) or not (0). Has to be saved, but not to be changed by the user.

**timingDeltas**
    Detailed information about timing inside the Engine itself. Empty unless enabled in the source code and *O.timingEnabled*==**True**.

**updateAttrs**(*(Serializable)arg1, (dict)arg2*) → None :
    Update object attributes from given dictionary

**wallDamping**(=*0.2*)
    the vertical displacements done to to control $\sigma$ (CNL or CNS cases) are in fact damped, through this wallDamping

**y0**(=*0.0*)
    the height of the upper plate at the very first time step : the engine finds its value [m]. Has to be saved, but not to be changed by the user.

**class yade.wrapper.KinemSimpleShearBox**(*inherits BoundaryController → GlobalEngine → Engine → Serializable*)
    This class is supposed to be a mother class for all Engines performing loadings on the simple shear box of *SimpleShear*. It is not intended to be used by itself, but its declaration and implentation will thus contain all what is useful for all these Engines. The script simpleShear.py illustrates the use of the various corresponding Engines.

**Key**(=*""*)
  string to add at the names of the saved files

**LOG**(=*false*)
  boolean controling the output of messages on the screen

**alpha**(=*Mathr::PI/2.0*)
  the angle from the lower box to the left box (trigo wise). Measured by this Engine. Has to be saved, but not to be changed by the user.

**dead**(=*false*)
  If true, this engine will not run at all; can be used for making an engine temporarily deactivated and only resurrect it at a later point.

**dict**(*(Serializable)arg1*) → dict :
  Return dictionary of attributes.

**execCount**
  Cumulative count this engine was run (only used if *O.timingEnabled*==True).

**execTime**
  Cumulative time in nanoseconds this Engine took to run (only used if *O.timingEnabled*==True).

**f0**(=*0.0*)
  the (vertical) force acting on the upper plate on the very first time step (determined by the Engine). Controls of the loadings in case of *KinemCNSEngine* or *KinemCNLEngine* will be done according to this initial value [N]. Has to be saved, but not to be changed by the user.

**firstRun**(=*true*)
  boolean set to false as soon as the engine has done its job one time : useful to know if initial height of, and normal force sustained by, the upper box are known or not (and thus if they have to be initialized). Has to be saved, but not to be changed by the user.

**id_boxback**(=*4*)
  the id of the wall at the back of the sample

**id_boxbas**(=*1*)
  the id of the lower wall

**id_boxfront**(=*5*)
  the id of the wall in front of the sample

**id_boxleft**(=*0*)
  the id of the left wall

**id_boxright**(=*2*)
  the id of the right wall

**id_topbox**(=*3*)
  the id of the upper wall

**label**(=*uninitialized*)
  Textual label for this object; must be valid python identifier, you can refer to it directly from python.

**max_vel**(=*1.0*)
  to limit the speed of the vertical displacements done to control σ (CNL or CNS cases) [m/s]

**ompThreads**(=*-1*)
  Number of threads to be used in the engine. If ompThreads<0 (default), the number will be typically OMP_NUM_THREADS or the number N defined by 'yade -jN' (this behavior can depend on the engine though). This attribute will only affect engines whose code includes openMP parallel regions (e.g. *InteractionLoop*). This attribute is mostly useful for experiments or when combining *ParallelEngine* with engines that run parallel regions, resulting in nested OMP loops with different number of threads at each level.

**temoin_save**(*=uninitalized*)
> vector (same length as 'gamma_save' for ex), with 0 or 1 depending whether the save for the corresponding value of gamma has been done (1) or not (0). Has to be saved, but not to be changed by the user.

**timingDeltas**
> Detailed information about timing inside the Engine itself. Empty unless enabled in the source code and *O.timingEnabled*==`True`.

**updateAttrs**(*(Serializable)arg1, (dict)arg2*) → None :
> Update object attributes from given dictionary

**wallDamping**(*=0.2*)
> the vertical displacements done to to control σ (CNL or CNS cases) are in fact damped, through this wallDamping

**y0**(*=0.0*)
> the height of the upper plate at the very first time step : the engine finds its value [m]. Has to be saved, but not to be changed by the user.

**class yade.wrapper.Peri3dController**(*inherits* *BoundaryController* → *GlobalEngine* → *Engine* → *Serializable*)

Class for controlling independently all 6 components of "engineering" *stress* and *strain* of periodic *Cell*. *goal* are the goal values, while *stressMask* determines which components prescribe stress and which prescribe strain.

If the strain is prescribed, appropriate strain rate is directly applied. If the stress is prescribed, the strain predictor is used: from stress values in two previous steps the value of strain rate is prescribed so as the value of stress in the next step is as close as possible to the ideal one. Current algorithm is extremly simple and probably will be changed in future, but is robust enough and mostly works fine.

Stress error (difference between actual and ideal stress) is evaluated in current and previous steps ($d\sigma_i, d\sigma_{i-1}$). Linear extrapolation is used to estimate error in the next step

$$d\sigma_{i+1} = 2d\sigma_i - d\sigma_{i-1}$$

According to this error, the strain rate is modified by *mod* parameter

$$d\sigma_{i+1} \begin{cases} > 0 \to \dot\varepsilon_{i+1} = \dot\varepsilon_i - \max(\mathrm{abs}(\dot\varepsilon_i)) \cdot \mathrm{mod} \\ < 0 \to \dot\varepsilon_{i+1} = \dot\varepsilon_i + \max(\mathrm{abs}(\dot\varepsilon_i)) \cdot \mathrm{mod} \end{cases}$$

According to this fact, the prescribed stress will (almost) never have exact prescribed value, but the difference would be very small (and decreasing for increasing *nSteps*. This approach works good if one of the dominant strain rates is prescribed. If all stresses are prescribed or if all goal strains is prescribed as zero, a good estimation is needed for the first step, therefore the compliance matrix is estimated (from user defined estimations of macroscopic material parameters *youngEstimation* and *poissonEstimation*) and respective strain rates is computed form prescribed stress rates and compliance matrix (the estimation of compliance matrix could be computed automaticaly avoiding user inputs of this kind).

The simulation on rotated periodic cell is also supported. Firstly, the polar decomposition is performed on cell's transformation matrix *trsf* $\mathcal{T} = \mathbf{U}\mathbf{P}$, where $\mathbf{U}$ is orthogonal (unitary) matrix representing rotation and $\mathbf{P}$ is a positive semi-definite Hermitian matrix representing strain. A logarithm of $\mathbf{P}$ should be used to obtain realistic values at higher strain values (not implemented yet). A prescribed strain increment in global coordinates $dt \cdot \dot\varepsilon$ is properly rotated to cell's local coordinates and added to $\mathbf{P}$

$$\mathbf{P}_{i+1} = \mathbf{P} + \mathbf{U}^T dt \cdot \dot\varepsilon \mathbf{U}$$

The new value of *trsf* is computed at $\mathbf{T}_{i+1} = \mathbf{U}\mathbf{P}_{i+1}$. From current and next *trsf* the cell's velocity gradient *velGrad* is computed (according to its definition) as

$$\mathbf{V} = (\mathbf{T}_{i+1}\mathbf{T}^{-1} - \mathbf{I})/dt$$

Current implementation allow user to define independent loading "path" for each prescribed component. i.e. define the prescribed value as a function of time (or *progress* or steps). See *Paths*.

Examples *examples/test/peri3dController_example1.py* and *examples/test/peri3dController_-triaxialCompression.py* explain usage and inputs of Peri3dController, *examples/test/peri3dController_shear.py* is an example of using shear components and also simulation on rotated cell.

**dead**(*=false*)
> If true, this engine will not run at all; can be used for making an engine temporarily deactivated and only resurrect it at a later point.

**dict**(*(Serializable)arg1*) → dict :
> Return dictionary of attributes.

**doneHook**(*=uninitalized*)
> Python command (as string) to run when *nSteps* is achieved. If empty, the engine will be set *dead*.

**execCount**
> Cumulative count this engine was run (only used if *O.timingEnabled*==`True`).

**execTime**
> Cumulative time in nanoseconds this Engine took to run (only used if *O.timingEnabled*==`True`).

**goal**(*=Vector6r::Zero()*)
> Goal state; only the upper triangular matrix is considered; each component is either prescribed stress or strain, depending on *stressMask*.

**label**(*=uninitalized*)
> Textual label for this object; must be valid python identifier, you can refer to it directly from python.

**lenPe**(*=0*)
> Peri3dController internal variable

**lenPs**(*=0*)
> Peri3dController internal variable

**maxStrain**(*=1e6*)
> Maximal asolute value of *strain* allowed in the simulation. If reached, the simulation is considered as finished

**maxStrainRate**(*=1e3*)
> Maximal absolute value of strain rate (both normal and shear components of *strain*)

**mod**(*=.1*)
> Predictor modificator, by trail-and-error analysis the value 0.1 was found as the best.

**nSteps**(*=1000*)
> Number of steps of the simulation.

**ompThreads**(*=-1*)
> Number of threads to be used in the engine. If ompThreads<0 (default), the number will be typically OMP_NUM_THREADS or the number N defined by 'yade -jN' (this behavior can depend on the engine though). This attribute will only affect engines whose code includes openMP parallel regions (e.g. *InteractionLoop*). This attribute is mostly useful for experiments or when combining *ParallelEngine* with engines that run parallel regions, resulting in nested OMP loops with different number of threads at each level.

**pathSizes**(*=Vector6i::Zero()*)
> Peri3dController internal variable

**pathsCounter**(*=Vector6i::Zero()*)
> Peri3dController internal variable

**pe**(*=Vector6i::Zero()*)
    Peri3dController internal variable

**poissonEstimation**(*=.25*)
    Estimation of macroscopic Poisson's ratio, used used for the first simulation step

**progress**(*=0.*)
    Actual progress of the simulation with Controller.

**ps**(*=Vector6i::Zero()*)
    Peri3dController internal variable

**strain**(*=Vector6r::Zero()*)
    Current strain (deformation) vector ($\varepsilon_x,\varepsilon_y,\varepsilon_z,\gamma_{yz},\gamma_{zx},\gamma_{xy}$) *(auto-updated)*.

**strainGoal**(*=Vector6r::Zero()*)
    Peri3dController internal variable

**strainRate**(*=Vector6r::Zero()*)
    Current strain rate vector.

**stress**(*=Vector6r::Zero()*)
    Current stress vector ($\sigma_x,\sigma_y,\sigma_z,\tau_{yz},\tau_{zx},\tau_{xy}$)|yupdate|.

**stressGoal**(*=Vector6r::Zero()*)
    Peri3dController internal variable

**stressIdeal**(*=Vector6r::Zero()*)
    Ideal stress vector at current time step.

**stressMask**(*=0, all strains*)
    mask determining whether components of *goal* are strain (0) or stress (1). The order is 00,11,22,12,02,01 from the least significant bit. (e.g. 0b000011 is stress 00 and stress 11).

**stressRate**(*=Vector6r::Zero()*)
    Current stress rate vector (that is prescribed, the actual one slightly differ).

**timingDeltas**
    Detailed information about timing inside the Engine itself. Empty unless enabled in the source code and *O.timingEnabled*==**True**.

**updateAttrs**(*(Serializable)arg1, (dict)arg2*) → None :
    Update object attributes from given dictionary

**xxPath**
    "Time function" (piecewise linear) for xx direction. Sequence of couples of numbers. First number is time, second number desired value of respective quantity (stress or strain). The last couple is considered as final state (equal to (*nSteps*, *goal*)), other values are relative to this state.

    Example: nSteps=1000, goal[0]=300, xxPath=((2,3),(4,1),(5,2))

    at step 400 (=5*1000/2) the value is 450 (=3*300/2),

    at step 800 (=4*1000/5) the value is 150 (=1*300/2),

    at step 1000 (=5*1000/5=nSteps) the value is 300 (=2*300/2=goal[0]).

    See example scripts/test/peri3dController_example1 for illusration.

**xyPath**(*=vector<Vector2r>(1, Vector2r::Ones())*)
    Time function for xy direction, see *xxPath*

**youngEstimation**(*=1e20*)
    Estimation of macroscopic Young's modulus, used for the first simulation step

**yyPath**(*=vector<Vector2r>(1, Vector2r::Ones())*)
    Time function for yy direction, see *xxPath*

**yzPath**(*=vector<Vector2r>(1, Vector2r::Ones())*)
   Time function for yz direction, see *xxPath*

**zxPath**(*=vector<Vector2r>(1, Vector2r::Ones())*)
   Time function for zx direction, see *xxPath*

**zzPath**(*=vector<Vector2r>(1, Vector2r::Ones())*)
   Time function for zz direction, see *xxPath*

**class yade.wrapper.PeriIsoCompressor**(*inherits BoundaryController → GlobalEngine → Engine → Serializable*)

Compress/decompress cloud of spheres by controlling periodic cell size until it reaches prescribed average stress, then moving to next stress value in given stress series.

**charLen**(*=-1.*)
   Characteristic length, should be something like mean particle diameter (default -1=invalid value))

**currUnbalanced**
   Current value of unbalanced force

**dead**(*=false*)
   If true, this engine will not run at all; can be used for making an engine temporarily deactivated and only resurrect it at a later point.

**dict**(*(Serializable)arg1*) → dict :
   Return dictionary of attributes.

**doneHook**(*=""*)
   Python command to be run when reaching the last specified stress

**execCount**
   Cumulative count this engine was run (only used if *O.timingEnabled*==True).

**execTime**
   Cumulative time in nanoseconds this Engine took to run (only used if *O.timingEnabled*==True).

**globalUpdateInt**(*=20*)
   how often to recompute average stress, stiffness and unbalanced force

**keepProportions**(*=true*)
   Exactly keep proportions of the cell (stress is controlled based on average, not its components

**label**(*=uninitalized*)
   Textual label for this object; must be valid python identifier, you can refer to it directly from python.

**maxSpan**(*=-1.*)
   Maximum body span in terms of bbox, to prevent periodic cell getting too small. *(auto-computed)*

**maxUnbalanced**(*=1e-4*)
   if actual unbalanced force is smaller than this number, the packing is considered stable,

**ompThreads**(*=-1*)
   Number of threads to be used in the engine. If ompThreads<0 (default), the number will be typically OMP_NUM_THREADS or the number N defined by 'yade -jN' (this behavior can depend on the engine though). This attribute will only affect engines whose code includes openMP parallel regions (e.g. *InteractionLoop*). This attribute is mostly useful for experiments or when combining *ParallelEngine* with engines that run parallel regions, resulting in nested OMP loops with different number of threads at each level.

**sigma**
   Current stress value

**state**(*=0*)
> Where are we at in the stress series

**stresses**(*=uninitialized*)
> Stresses that should be reached, one after another

**timingDeltas**
> Detailed information about timing inside the Engine itself. Empty unless enabled in the source code and *O.timingEnabled*==**True**.

**updateAttrs**(*(Serializable)arg1, (dict)arg2*) → None :
> Update object attributes from given dictionary

**class yade.wrapper.PeriTriaxController**(*inherits* *BoundaryController* → *GlobalEngine* → *Engine* → *Serializable*)
Engine for independently controlling stress or strain in periodic simulations.

> *PeriTriaxController.goal* contains absolute values for the controlled quantity, and *PeriTriaxController.stressMask* determines meaning of those values (0 for strain, 1 for stress): e.g. ( 1<<0 | 1<<2 ) = 1 | 4 = 5 means that goal[0] and goal[2] are stress values, and goal[1] is strain.

See scripts/test/periodic-triax.py for a simple example.

**absStressTol**(*=1e3*)
> Absolute stress tolerance

**currUnbalanced**(*=NaN*)
> current unbalanced force (updated every globUpdate) *(auto-updated)*

**dead**(*=false*)
> If true, this engine will not run at all; can be used for making an engine temporarily deactivated and only resurrect it at a later point.

**dict**(*(Serializable)arg1*) → dict :
> Return dictionary of attributes.

**doneHook**(*=uninitialized*)
> python command to be run when the desired state is reached

**dynCell**(*=false*)
> Imposed stress can be controlled using the packing stiffness or by applying the laws of dynamic (dynCell=true). Don't forget to assign a *mass* to the cell.

**execCount**
> Cumulative count this engine was run (only used if *O.timingEnabled*==**True**).

**execTime**
> Cumulative time in nanoseconds this Engine took to run (only used if *O.timingEnabled*==**True**).

**externalWork**(*=0*)
> Work input from boundary controller.

**globUpdate**(*=5*)
> How often to recompute average stress, stiffness and unbalaced force.

**goal**
> Desired stress or strain values (depending on stressMask), strains defined as strain(i)=log(Fii).

> > **Warning:** Strains are relative to the *O.cell.refSize* (reference cell size), not the current one (e.g. at the moment when the new strain value is set).

**growDamping**(*=.25*)
    Damping of cell resizing (0=perfect control, 1=no control at all); see also `wallDamping` in
    *TriaxialStressController*.

**label**(*=uninitalized*)
    Textual label for this object; must be valid python identifier, you can refer to it directly from
    python.

**mass**(*=NaN*)
    mass of the cell (user set); if not set and *dynCell* is used, it will be computed as sum of masses
    of all particles.

**maxBodySpan**(*=Vector3r::Zero()*)
    maximum body dimension *(auto-computed)*

**maxStrainRate**(*=Vector3r(1, 1, 1)*)
    Maximum strain rate of the periodic cell.

**maxUnbalanced**(*=1e-4*)
    maximum unbalanced force.

**ompThreads**(*=-1*)
    Number of threads to be used in the engine. If ompThreads<0 (default), the number will be
    typically OMP_NUM_THREADS or the number N defined by 'yade -jN' (this behavior can
    depend on the engine though). This attribute will only affect engines whose code includes
    openMP parallel regions (e.g. *InteractionLoop*). This attribute is mostly useful for experi-
    ments or when combining *ParallelEngine* with engines that run parallel regions, resulting in
    nested OMP loops with different number of threads at each level.

**prevGrow**(*=Vector3r::Zero()*)
    previous cell grow

**relStressTol**(*=3e-5*)
    Relative stress tolerance

**stiff**(*=Vector3r::Zero()*)
    average stiffness (only every globUpdate steps recomputed from interactions) *(auto-updated)*

**strain**(*=Vector3r::Zero()*)
    cell strain *(auto-updated)*

**strainRate**(*=Vector3r::Zero()*)
    cell strain rate *(auto-updated)*

**stress**(*=Vector3r::Zero()*)
    diagonal terms of the stress tensor

**stressMask**(*=0, all strains*)
    mask determining strain/stress (0/1) meaning for goal components

**stressTensor**(*=Matrix3r::Zero()*)
    average stresses, updated at every step (only every globUpdate steps recomputed from inter-
    actions if !dynCell)

**timingDeltas**
    Detailed information about timing inside the Engine itself. Empty unless enabled in the source
    code and *O.timingEnabled*==`True`.

**updateAttrs**(*(Serializable)arg1, (dict)arg2*) → None :
    Update object attributes from given dictionary

**class yade.wrapper.ThreeDTriaxialEngine**(*inherits    TriaxialStressController  →  Bound-
                                            aryController  →  GlobalEngine  →  Engine  →
                                            Serializable*)
The engine perform a triaxial compression with a control in direction 'i' in stress (if stressControl_i)
else in strain.

For a stress control the imposed stress is specified by 'sigma_i' with a 'max_veli' depending on 'strainRatei'. To obtain the same strain rate in stress control than in strain control you need to set 'wallDamping = 0.8'. For a strain control the imposed strain is specified by 'strainRatei'. With this engine you can also perform internal compaction by growing the size of particles by using `TriaxialStressController::controlInternalStress`. For that, just switch on 'internalCompaction=1' and fix sigma_iso=value of mean pressure that you want at the end of the internal compaction.

> **Warning:** This engine is deprecated, please switch to TriaxialStressController if you expect long term support.

**Key**(*=""*)
    A string appended at the end of all files, use it to name simulations.

**UnbalancedForce**(*=1*)
    mean resultant forces divided by mean contact force

**boxVolume**
    Total packing volume.

**computeStressStrainInterval**(*=10*)

**currentStrainRate1**(*=0*)
    current strain rate in direction 1 - converging to *ThreeDTriaxialEngine::strainRate1* (./s)

**currentStrainRate2**(*=0*)
    current strain rate in direction 2 - converging to *ThreeDTriaxialEngine::strainRate2* (./s)

**currentStrainRate3**(*=0*)
    current strain rate in direction 3 - converging to *ThreeDTriaxialEngine::strainRate3* (./s)

**dead**(*=false*)
    If true, this engine will not run at all; can be used for making an engine temporarily deactivated and only resurrect it at a later point.

**depth**(*=0*)
    size of the box (2-axis) *(auto-updated)*

**depth0**(*=0*)
    Reference size for strain definition. See *TriaxialStressController::depth*

**dict**(*(Serializable)arg1*) → dict :
    Return dictionary of attributes.

**execCount**
    Cumulative count this engine was run (only used if *O.timingEnabled*==`True`).

**execTime**
    Cumulative time in nanoseconds this Engine took to run (only used if *O.timingEnabled*==`True`).

**externalWork**(*=0*)
    Mechanical work associated to the boundary conditions, i.e. $\int_{\partial\Omega} \mathbf{T} \cdot \mathbf{u} ds$ with $\mathbf{T}$ the surface traction and $\mathbf{u}$ the displacement at the boundary. *(auto-updated)*

**finalMaxMultiplier**(*=1.00001*)
    max multiplier of diameters during internal compaction (secondary precise adjustment - *TriaxialStressController::maxMultiplier* is used in the initial stage)

**frictionAngleDegree**(*=-1*)
    Value of friction used in the simulation if (updateFrictionAngle)

**goal1**(*=0*)
    prescribed stress/strain rate on axis 1, as defined by *TriaxialStressController::stressMask*

**goal2**(*=0*)
  prescribed stress/strain rate on axis 2, as defined by *TriaxialStressController::stressMask*

**goal3**(*=0*)
  prescribed stress/strain rate on axis 3, as defined by *TriaxialStressController::stressMask*

**height**(*=0*)
  size of the box (1-axis) *(auto-updated)*

**height0**(*=0*)
  Reference size for strain definition. See *TriaxialStressController::height*

**internalCompaction**(*=true*)
  Switch between 'external' (walls) and 'internal' (growth of particles) compaction.

**label**(*=uninitalized*)
  Textual label for this object; must be valid python identifier, you can refer to it directly from python.

**maxMultiplier**(*=1.001*)
  max multiplier of diameters during internal compaction (initial fast increase - *TriaxialStressController::finalMaxMultiplier* is used in a second stage)

**max_vel**(*=1*)
  Maximum allowed walls velocity [m/s]. This value superseeds the one assigned by the stress controller if the later is higher. max_vel can be set to infinity in many cases, but sometimes helps stabilizing packings. Based on this value, different maxima are computed for each axis based on the dimensions of the sample, so that if each boundary moves at its maximum velocity, the strain rate will be isotropic (see e.g. *TriaxialStressController::max_vel1*).

**max_vel1**
  see *TriaxialStressController::max_vel* *(auto-computed)*

**max_vel2**
  see *TriaxialStressController::max_vel* *(auto-computed)*

**max_vel3**
  see *TriaxialStressController::max_vel* *(auto-computed)*

**meanStress**(*=0*)
  Mean stress in the packing. *(auto-updated)*

**ompThreads**(*=-1*)
  Number of threads to be used in the engine. If ompThreads<0 (default), the number will be typically OMP_NUM_THREADS or the number N defined by 'yade -jN' (this behavior can depend on the engine though). This attribute will only affect engines whose code includes openMP parallel regions (e.g. *InteractionLoop*). This attribute is mostly useful for experiments or when combining *ParallelEngine* with engines that run parallel regions, resulting in nested OMP loops with different number of threads at each level.

**particlesVolume**
  Total volume of particles (clumps and *dynamic* spheres). *(auto-computed)*

**porosity**
  Porosity of the packing, computed from *particlesVolume* and *boxVolume*. *(auto-updated)*

**previousMultiplier**(*=1*)
  *(auto-updated)*

**previousStress**(*=0*)
  *(auto-updated)*

**radiusControlInterval**(*=10*)

**setContactProperties**(*(ThreeDTriaxialEngine)arg1, (float)arg2*) → None :
  Assign a new friction angle (degrees) to dynamic bodies and relative interactions

**spheresVolume**
Shorthand for *TriaxialStressController::particlesVolume*

**stiffnessUpdateInterval**(=*10*)
iteration period for measuring the resultant packing-boundaries stiffnesses, for stress servo-control

**strain**
Current strain in a vector (exx,eyy,ezz). The values reflect true (logarithmic) strain.

**strainDamping**(=*0.9997*)
factor used for smoothing changes in effective strain rate. If target rate is TR, then (1-damping)*(TR-currentRate) will be added at each iteration. With damping=0, rate=target all the time. With damping=1, it doesn't change.

**strainRate**
Current strain rate in a vector d/dt(exx,eyy,ezz).

**strainRate1**(=*0*)
target strain rate in direction 1 (./s, >0 for compression)

**strainRate2**(=*0*)
target strain rate in direction 2 (./s, >0 for compression)

**strainRate3**(=*0*)
target strain rate in direction 3 (./s, >0 for compression)

**stress**(*(TriaxialStressController)arg1, (int)id*) → Vector3 :
Returns the average stress on boundary 'id'. Here, 'id' refers to the internal numbering of boundaries, between 0 and 5.

**stressControl_1**(=*true*)
Switch to choose a stress or a strain control in directions 1

**stressControl_2**(=*true*)
Switch to choose a stress or a strain control in directions 2

**stressControl_3**(=*true*)
Switch to choose a stress or a strain control in directions 3

**stressDamping**(=*0.25*)
wall damping coefficient for the stress control - wallDamping=0 implies a (theoretical) perfect control, wallDamping=1 means no movement

**stressMask**(=*7*)
Bitmask determining wether the imposed *goal* values are stresses (0 for none, 7 for all, 1 for direction 1, 5 for directions 1 and 3, etc.) or strain rates

**thickness**(=*-1*)
thickness of boxes (needed by some functions)

**timingDeltas**
Detailed information about timing inside the Engine itself. Empty unless enabled in the source code and *O.timingEnabled*==`True`.

**updateAttrs**(*(Serializable)arg1, (dict)arg2*) → None :
Update object attributes from given dictionary

**updateFrictionAngle**(=*false*)
Switch to activate the update of the intergranular frictionto the value *ThreeDTriaxialEngine::frictionAngleDegree*.

**updatePorosity**(=*false*)
If true, *solid volume* will be updated once (will automatically reset to false after one calculation step) e.g. for porosity calculation purpose. Can be used when volume of particles changes during the simulation (e.g. when particles are erased or when clumps are created).

**volumetricStrain**(=*0*)
> Volumetric strain (see *TriaxialStressController::strain*). *(auto-updated)*

**wall_back_activated**(=*true*)
> if true, this wall moves according to the target value (stress or strain rate).

**wall_back_id**(=*4*)
> id of boundary ; coordinate 2- (default value is ok if aabbWalls are appended BEFORE spheres.)

**wall_bottom_activated**(=*true*)
> if true, this wall moves according to the target value (stress or strain rate).

**wall_bottom_id**(=*2*)
> id of boundary ; coordinate 1- (default value is ok if aabbWalls are appended BEFORE spheres.)

**wall_front_activated**(=*true*)
> if true, this wall moves according to the target value (stress or strain rate).

**wall_front_id**(=*5*)
> id of boundary ; coordinate 2+ (default value is ok if aabbWalls are appended BEFORE spheres.)

**wall_left_activated**(=*true*)
> if true, this wall moves according to the target value (stress or strain rate).

**wall_left_id**(=*0*)
> id of boundary ; coordinate 0- (default value is ok if aabbWalls are appended BEFORE spheres.)

**wall_right_activated**(=*true*)
> if true, this wall moves according to the target value (stress or strain rate).

**wall_right_id**(=*1*)
> id of boundary ; coordinate 0+ (default value is ok if aabbWalls are appended BEFORE spheres.)

**wall_top_activated**(=*true*)
> if true, this wall moves according to the target value (stress or strain rate).

**wall_top_id**(=*3*)
> id of boundary ; coordinate 1+ (default value is ok if aabbWalls are appended BEFORE spheres.)

**width**(=*0*)
> size of the box (0-axis) *(auto-updated)*

**width0**(=*0*)
> Reference size for strain definition. See *TriaxialStressController::width*

**class yade.wrapper.TriaxialCompressionEngine**(*inherits TriaxialStressController → BoundaryController → GlobalEngine → Engine → Serializable*)

The engine is a state machine with the following states; transitions my be automatic, see below.

1. STATE_ISO_COMPACTION: isotropic compaction (compression) until the prescribed mean pressue sigmaIsoCompaction is reached and the packing is stable. The compaction happens either by straining the walls (!internalCompaction) or by growing size of grains (internalCompaction).

2. STATE_ISO_UNLOADING: isotropic unloading from the previously reached state, until the mean pressure sigmaLateralConfinement is reached (and stabilizes).

> **Note:** this state will be skipped if sigmaLateralConfinement == sigmaIsoCompaction.

3. STATE_TRIAX_LOADING: confined uniaxial compression: constant sigmaLateralConfinement is kept at lateral walls (left, right, front, back), while top and bottom walls load the packing in their axis (by straining), until the value of epsilonMax (deformation along the loading axis) is reached. At this point, the simulation is stopped.

4. STATE_FIXED_POROSITY_COMPACTION: isotropic compaction (compression) until a chosen porosity value (parameter:fixedPorosity). The six walls move with a chosen translation speed (parameter StrainRate).

5. STATE_TRIAX_LIMBO: currently unused, since simulation is hard-stopped in the previous state.

Transition from COMPACTION to UNLOADING is done automatically if autoUnload==true;

> Transition from (UNLOADING to LOADING) or from (COMPACTION to LOADING: if UNLOADING is skipped) is done automatically if autoCompressionActivation=true; Both autoUnload and autoCompressionActivation are true by default.

> **Note:** Most of the algorithms used have been developed initialy for simulations reported in [Chareyre2002a] and [Chareyre2005]. They have been ported to Yade in a second step and used in e.g. [Kozicki2008],[Scholtes2009b]_,[Jerier2010b].

> **Warning:** This engine is deprecated, please switch to TriaxialStressController if you expect long term support.

**Key**(=*""*)
    A string appended at the end of all files, use it to name simulations.

**StabilityCriterion**(=*0.001*)
    tolerance in terms of *TriaxialCompressionEngine::UnbalancedForce* to consider the packing is stable

**UnbalancedForce**(=*1*)
    mean resultant forces divided by mean contact force

**autoCompressionActivation**(=*true*)
    Auto-switch from isotropic compaction (or unloading state if sigmaLateralConfinement<sigmaIsoCompaction) to deviatoric loading

**autoStopSimulation**(=*false*)
    Stop the simulation when the sample reach STATE_LIMBO, or keep running

**autoUnload**(=*true*)
    Auto-switch from isotropic compaction to unloading

**boxVolume**
    Total packing volume.

**computeStressStrainInterval**(=*10*)

**currentState**(=*1*)
    There are 5 possible states in which TriaxialCompressionEngine can be. See above *wrapper.TriaxialCompressionEngine*

**currentStrainRate**(=*0*)
    current strain rate - converging to *TriaxialCompressionEngine::strainRate* (./s)

**dead**(*=false*)
    If true, this engine will not run at all; can be used for making an engine temporarily deactivated and only resurrect it at a later point.

**depth**(*=0*)
    size of the box (2-axis) *(auto-updated)*

**depth0**(*=0*)
    Reference size for strain definition. See *TriaxialStressController::depth*

**dict**(*(Serializable)arg1*) → dict :
    Return dictionary of attributes.

**epsilonMax**(*=0.5*)
    Value of axial deformation for which the loading must stop

**execCount**
    Cumulative count this engine was run (only used if *O.timingEnabled*==True).

**execTime**
    Cumulative time in nanoseconds this Engine took to run (only used if *O.timingEnabled*==True).

**externalWork**(*=0*)
    Mechanical work associated to the boundary conditions, i.e. $\int_{\partial\Omega} \mathbf{T} \cdot \mathbf{u} \, ds$ with $\mathbf{T}$ the surface traction and $\mathbf{u}$ the displacement at the boundary. *(auto-updated)*

**finalMaxMultiplier**(*=1.00001*)
    max multiplier of diameters during internal compaction (secondary precise adjustment - *TriaxialStressController::maxMultiplier* is used in the initial stage)

**fixedPoroCompaction**(*=false*)
    A special type of compaction with imposed final porosity *TriaxialCompressionEngine::fixedPorosity* (WARNING : can give unrealistic results!)

**fixedPorosity**(*=0*)
    Value of porosity chosen by the user

**frictionAngleDegree**(*=-1*)
    Value of friction assigned just before the deviatoric loading

**goal1**(*=0*)
    prescribed stress/strain rate on axis 1, as defined by *TriaxialStressController::stressMask*

**goal2**(*=0*)
    prescribed stress/strain rate on axis 2, as defined by *TriaxialStressController::stressMask*

**goal3**(*=0*)
    prescribed stress/strain rate on axis 3, as defined by *TriaxialStressController::stressMask*

**height**(*=0*)
    size of the box (1-axis) *(auto-updated)*

**height0**(*=0*)
    Reference size for strain definition. See *TriaxialStressController::height*

**internalCompaction**(*=true*)
    Switch between 'external' (walls) and 'internal' (growth of particles) compaction.

**isAxisymmetric**(*=false*)
    if true, sigma_iso is assigned to sigma1, 2 and 3 (applies at each iteration and overrides user-set values of s1,2,3)

**label**(*=uninitalized*)
    Textual label for this object; must be valid python identifier, you can refer to it directly from python.

**maxMultiplier**(*=1.001*)
  max multiplier of diameters during internal compaction (initial fast increase - *TriaxialStress-Controller::finalMaxMultiplier* is used in a second stage)

**maxStress**(*=0*)
  Max absolute value of axial stress during the simulation (for post-processing)

**max_vel**(*=1*)
  Maximum allowed walls velocity [m/s]. This value superseeds the one assigned by the stress controller if the later is higher. max_vel can be set to infinity in many cases, but sometimes helps stabilizing packings. Based on this value, different maxima are computed for each axis based on the dimensions of the sample, so that if each boundary moves at its maximum velocity, the strain rate will be isotropic (see e.g. *TriaxialStressController::max_vel1*).

**max_vel1**
  see *TriaxialStressController::max_vel* *(auto-computed)*

**max_vel2**
  see *TriaxialStressController::max_vel* *(auto-computed)*

**max_vel3**
  see *TriaxialStressController::max_vel* *(auto-computed)*

**meanStress**(*=0*)
  Mean stress in the packing. *(auto-updated)*

**noFiles**(*=false*)
  If true, no files will be generated (*.xml, *.spheres,...)

**ompThreads**(*=-1*)
  Number of threads to be used in the engine. If ompThreads<0 (default), the number will be typically OMP_NUM_THREADS or the number N defined by 'yade -jN' (this behavior can depend on the engine though). This attribute will only affect engines whose code includes openMP parallel regions (e.g. *InteractionLoop*). This attribute is mostly useful for experiments or when combining *ParallelEngine* with engines that run parallel regions, resulting in nested OMP loops with different number of threads at each level.

**particlesVolume**
  Total volume of particles (clumps and *dynamic* spheres). *(auto-computed)*

**porosity**
  Porosity of the packing, computed from *particlesVolume* and *boxVolume*. *(auto-updated)*

**previousMultiplier**(*=1*)
  *(auto-updated)*

**previousSigmaIso**(*=1*)
  Previous value of inherited sigma_iso (used to detect manual changes of the confining pressure)

**previousState**(*=1*)
  Previous state (used to detect manual changes of the state in .xml)

**previousStress**(*=0*)
  *(auto-updated)*

**radiusControlInterval**(*=10*)

**setContactProperties**(*(TriaxialCompressionEngine)arg1, (float)arg2*) → None :
  Assign a new friction angle (degrees) to dynamic bodies and relative interactions

**sigmaIsoCompaction**(*=1*)
  Prescribed isotropic pressure during the compaction phase (< 0 for real - compressive - compaction)

**sigmaLateralConfinement**(*=1*)
  Prescribed confining pressure in the deviatoric loading (< 0 for classical compressive cases); might be different from *TriaxialCompressionEngine::sigmaIsoCompaction*

**sigma_iso**(*=0*)
    prescribed confining stress (see :yref:'TriaxialCompressionEngine::isAxisymetric')

**spheresVolume**
    Shorthand for *TriaxialStressController::particlesVolume*

**stiffnessUpdateInterval**(*=10*)
    iteration period for measuring the resultant packing-boundaries stiffness, for stress servo-control

**strain**
    Current strain in a vector (exx,eyy,ezz). The values reflect true (logarithmic) strain.

**strainDamping**(*=0.99*)
    coefficient used for smoother transitions in the strain rate. The rate reaches the target value like $d^n$ reaches 0, where $d$ is the damping coefficient and $n$ is the number of steps

**strainRate**(*=0*)
    target strain rate (./s, >0 for compression)

**stress**(*(TriaxialStressController)arg1, (int)id*) → Vector3 :
    Returns the average stress on boundary 'id'. Here, 'id' refers to the internal numbering of boundaries, between 0 and 5.

**stressDamping**(*=0.25*)
    wall damping coefficient for the stress control - wallDamping=0 implies a (theoretical) perfect control, wallDamping=1 means no movement

**stressMask**(*=7*)
    Bitmask determining wether the imposed *goal* values are stresses (0 for none, 7 for all, 1 for direction 1, 5 for directions 1 and 3, etc.) or strain rates

**testEquilibriumInterval**(*=20*)
    interval of checks for transition between phases, higher than 1 saves computation time.

**thickness**(*=-1*)
    thickness of boxes (needed by some functions)

**timingDeltas**
    Detailed information about timing inside the Engine itself. Empty unless enabled in the source code and *O.timingEnabled*==**True**.

**translationAxis**(*=TriaxialStressController::normal[wall_bottom]*)
    compression axis

**uniaxialEpsilonCurr**(*=1*)
    Current value of axial deformation during confined loading (is reference to strain[1])

**updateAttrs**(*(Serializable)arg1, (dict)arg2*) → None :
    Update object attributes from given dictionary

**updatePorosity**(*=false*)
    If true, *solid volume* will be updated once (will automatically reset to false after one calculation step) e.g. for porosity calculation purpose. Can be used when volume of particles changes during the simulation (e.g. when particles are erased or when clumps are created).

**volumetricStrain**(*=0*)
    Volumetric strain (see *TriaxialStressController::strain*). *(auto-updated)*

**wall_back_activated**(*=true*)
    if true, this wall moves according to the target value (stress or strain rate).

**wall_back_id**(*=4*)
    id of boundary ; coordinate 2- (default value is ok if aabbWalls are appended BEFORE spheres.)

**wall_bottom_activated**(*=true*)
    if true, this wall moves according to the target value (stress or strain rate).

**wall_bottom_id**(*=2*)
    id of boundary ; coordinate 1- (default value is ok if aabbWalls are appended BEFORE spheres.)

**wall_front_activated**(*=true*)
    if true, this wall moves according to the target value (stress or strain rate).

**wall_front_id**(*=5*)
    id of boundary ; coordinate 2+ (default value is ok if aabbWalls are appended BEFORE spheres.)

**wall_left_activated**(*=true*)
    if true, this wall moves according to the target value (stress or strain rate).

**wall_left_id**(*=0*)
    id of boundary ; coordinate 0- (default value is ok if aabbWalls are appended BEFORE spheres.)

**wall_right_activated**(*=true*)
    if true, this wall moves according to the target value (stress or strain rate).

**wall_right_id**(*=1*)
    id of boundary ; coordinate 0+ (default value is ok if aabbWalls are appended BEFORE spheres.)

**wall_top_activated**(*=true*)
    if true, this wall moves according to the target value (stress or strain rate).

**wall_top_id**(*=3*)
    id of boundary ; coordinate 1+ (default value is ok if aabbWalls are appended BEFORE spheres.)

**warn**(*=0*)
    counter used for sending a deprecation warning once

**width**(*=0*)
    size of the box (0-axis) *(auto-updated)*

**width0**(*=0*)
    Reference size for strain definition. See *TriaxialStressController::width*

**class yade.wrapper.TriaxialStressController**(*inherits BoundaryController → GlobalEngine → Engine → Serializable*)
An engine maintaining constant stresses or constant strain rates on some boundaries of a parallelipipedic packing. The stress/strain control is defined for each axis using *TriaxialStressController::stressMask* (a bitMask) and target values are defined by goal1,goal2, and goal3. The sign conventions of continuum mechanics are used for strains and stresses (positive traction).

---

**Note:** The algorithms used have been developed initialy for simulations reported in [Chareyre2002a] and [Chareyre2005]. They have been ported to Yade in a second step and used in e.g. [Kozicki2008],[Scholtes2009b]\_,[Jerier2010b].

---

**boxVolume**
    Total packing volume.

**computeStressStrainInterval**(*=10*)

**dead**(*=false*)
    If true, this engine will not run at all; can be used for making an engine temporarily deactivated and only resurrect it at a later point.

**depth**(*=0*)
    size of the box (2-axis) *(auto-updated)*

**depth0**(*=0*)
    Reference size for strain definition. See *TriaxialStressController::depth*

**dict**(*(Serializable)arg1*) → dict :
    Return dictionary of attributes.

**execCount**
    Cumulative count this engine was run (only used if *O.timingEnabled*==`True`).

**execTime**
    Cumulative time in nanoseconds this Engine took to run (only used if *O.timingEnabled*==`True`).

**externalWork**(*=0*)
    Mechanical work associated to the boundary conditions, i.e. $\int_{\partial\Omega} \mathbf{T} \cdot \mathbf{u}\,ds$ with $\mathbf{T}$ the surface traction and $\mathbf{u}$ the displacement at the boundary. *(auto-updated)*

**finalMaxMultiplier**(*=1.00001*)
    max multiplier of diameters during internal compaction (secondary precise adjustment - *TriaxialStressController::maxMultiplier* is used in the initial stage)

**goal1**(*=0*)
    prescribed stress/strain rate on axis 1, as defined by *TriaxialStressController::stressMask*

**goal2**(*=0*)
    prescribed stress/strain rate on axis 2, as defined by *TriaxialStressController::stressMask*

**goal3**(*=0*)
    prescribed stress/strain rate on axis 3, as defined by *TriaxialStressController::stressMask*

**height**(*=0*)
    size of the box (1-axis) *(auto-updated)*

**height0**(*=0*)
    Reference size for strain definition. See *TriaxialStressController::height*

**internalCompaction**(*=true*)
    Switch between 'external' (walls) and 'internal' (growth of particles) compaction.

**label**(*=uninitialized*)
    Textual label for this object; must be valid python identifier, you can refer to it directly from python.

**maxMultiplier**(*=1.001*)
    max multiplier of diameters during internal compaction (initial fast increase - *TriaxialStressController::finalMaxMultiplier* is used in a second stage)

**max_vel**(*=1*)
    Maximum allowed walls velocity [m/s]. This value superseeds the one assigned by the stress controller if the later is higher. max_vel can be set to infinity in many cases, but sometimes helps stabilizing packings. Based on this value, different maxima are computed for each axis based on the dimensions of the sample, so that if each boundary moves at its maximum velocity, the strain rate will be isotropic (see e.g. *TriaxialStressController::max_vel1*).

**max_vel1**
    see *TriaxialStressController::max_vel* *(auto-computed)*

**max_vel2**
    see *TriaxialStressController::max_vel* *(auto-computed)*

**max_vel3**
    see *TriaxialStressController::max_vel* *(auto-computed)*

**meanStress**(*=0*)
    Mean stress in the packing. *(auto-updated)*

**ompThreads**(=-1)

Number of threads to be used in the engine. If ompThreads<0 (default), the number will be typically OMP_NUM_THREADS or the number N defined by 'yade -jN' (this behavior can depend on the engine though). This attribute will only affect engines whose code includes openMP parallel regions (e.g. *InteractionLoop*). This attribute is mostly useful for experiments or when combining *ParallelEngine* with engines that run parallel regions, resulting in nested OMP loops with different number of threads at each level.

**particlesVolume**

Total volume of particles (clumps and *dynamic* spheres). *(auto-computed)*

**porosity**

Porosity of the packing, computed from *particlesVolume* and *boxVolume*. *(auto-updated)*

**previousMultiplier**(=1)

*(auto-updated)*

**previousStress**(=0)

*(auto-updated)*

**radiusControlInterval**(=10)

**spheresVolume**

Shorthand for *TriaxialStressController::particlesVolume*

**stiffnessUpdateInterval**(=10)

iteration period for measuring the resultant packing-boundaries stiffnesses, for stress servo-control

**strain**

Current strain in a vector (exx,eyy,ezz). The values reflect true (logarithmic) strain.

**strainDamping**(=0.99)

coefficient used for smoother transitions in the strain rate. The rate reaches the target value like $d^n$ reaches 0, where `d` is the damping coefficient and `n` is the number of steps

**strainRate**

Current strain rate in a vector d/dt(exx,eyy,ezz).

**stress**(*(TriaxialStressController)arg1*, *(int)id*) → Vector3 :

Returns the average stress on boundary 'id'. Here, 'id' refers to the internal numbering of boundaries, between 0 and 5.

**stressDamping**(=0.25)

wall damping coefficient for the stress control - wallDamping=0 implies a (theoretical) perfect control, wallDamping=1 means no movement

**stressMask**(=7)

Bitmask determining wether the imposed *goal* values are stresses (0 for none, 7 for all, 1 for direction 1, 5 for directions 1 and 3, etc.) or strain rates

**thickness**(=-1)

thickness of boxes (needed by some functions)

**timingDeltas**

Detailed information about timing inside the Engine itself. Empty unless enabled in the source code and *O.timingEnabled*==`True`.

**updateAttrs**(*(Serializable)arg1*, *(dict)arg2*) → None :

Update object attributes from given dictionary

**updatePorosity**(=false)

If true, *solid volume* will be updated once (will automatically reset to false after one calculation step) e.g. for porosity calculation purpose. Can be used when volume of particles changes during the simulation (e.g. when particles are erased or when clumps are created).

**volumetricStrain**(*=0*)
    Volumetric strain (see *TriaxialStressController::strain*). *(auto-updated)*

**wall_back_activated**(*=true*)
    if true, this wall moves according to the target value (stress or strain rate).

**wall_back_id**(*=4*)
    id of boundary ; coordinate 2- (default value is ok if aabbWalls are appended BEFORE spheres.)

**wall_bottom_activated**(*=true*)
    if true, this wall moves according to the target value (stress or strain rate).

**wall_bottom_id**(*=2*)
    id of boundary ; coordinate 1- (default value is ok if aabbWalls are appended BEFORE spheres.)

**wall_front_activated**(*=true*)
    if true, this wall moves according to the target value (stress or strain rate).

**wall_front_id**(*=5*)
    id of boundary ; coordinate 2+ (default value is ok if aabbWalls are appended BEFORE spheres.)

**wall_left_activated**(*=true*)
    if true, this wall moves according to the target value (stress or strain rate).

**wall_left_id**(*=0*)
    id of boundary ; coordinate 0- (default value is ok if aabbWalls are appended BEFORE spheres.)

**wall_right_activated**(*=true*)
    if true, this wall moves according to the target value (stress or strain rate).

**wall_right_id**(*=1*)
    id of boundary ; coordinate 0+ (default value is ok if aabbWalls are appended BEFORE spheres.)

**wall_top_activated**(*=true*)
    if true, this wall moves according to the target value (stress or strain rate).

**wall_top_id**(*=3*)
    id of boundary ; coordinate 1+ (default value is ok if aabbWalls are appended BEFORE spheres.)

**width**(*=0*)
    size of the box (0-axis) *(auto-updated)*

**width0**(*=0*)
    Reference size for strain definition. See *TriaxialStressController::width*

**class yade.wrapper.UniaxialStrainer**(*inherits BoundaryController → GlobalEngine → Engine → Serializable*)
Axial displacing two groups of bodies in the opposite direction with given strain rate.

**absSpeed**(*=NaN*)
    alternatively, absolute speed of boundary motion can be specified; this is effective only at the beginning and if strainRate is not set; changing absSpeed directly during simulation wil have no effect. [ms $^{1}$]

**active**(*=true*)
    Whether this engine is activated

**asymmetry**(*=0, symmetric*)
    If 0, straining is symmetric for negIds and posIds; for 1 (or -1), only posIds are strained and negIds don't move (or vice versa)

**avgStress**(*=0*)
:   Current average stress *(auto-updated)* [Pa]

**axis**(*=2*)
:   The axis which is strained (0,1,2 for x,y,z)

**blockDisplacements**(*=false*)
:   Whether displacement of boundary bodies perpendicular to the strained axis are blocked or are free

**blockRotations**(*=false*)
:   Whether rotations of boundary bodies are blocked.

**crossSectionArea**(*=NaN*)
:   crossSection perpendicular to he strained axis; must be given explicitly [m$^2$]

**currentStrainRate**(*=NaN*)
:   Current strain rate (update automatically). *(auto-updated)*

**dead**(*=false*)
:   If true, this engine will not run at all; can be used for making an engine temporarily deactivated and only resurrect it at a later point.

**dict**(*(Serializable)arg1*) → dict :
:   Return dictionary of attributes.

**execCount**
:   Cumulative count this engine was run (only used if *O.timingEnabled*==`True`).

**execTime**
:   Cumulative time in nanoseconds this Engine took to run (only used if *O.timingEnabled*==`True`).

**idleIterations**(*=0*)
:   Number of iterations that will pass without straining activity after stopStrain has been reached

**initAccelTime**(*=-200*)
:   Time for strain reaching the requested value (linear interpolation). If negative, the time is dt*(-initAccelTime), where dt is the timestep at the first iteration. [s]

**label**(*=uninitalized*)
:   Textual label for this object; must be valid python identifier, you can refer to it directly from python.

**limitStrain**(*=0, disabled*)
:   Invert the sense of straining (sharply, without transition) one this value of strain is reached. Not effective if 0.

**negIds**(*=uninitalized*)
:   Bodies on which strain will be applied (on the negative end along the axis)

**notYetReversed**(*=true*)
:   Flag whether the sense of straining has already been reversed (only used internally).

**ompThreads**(*=-1*)
:   Number of threads to be used in the engine. If ompThreads<0 (default), the number will be typically OMP_NUM_THREADS or the number N defined by 'yade -jN' (this behavior can depend on the engine though). This attribute will only affect engines whose code includes openMP parallel regions (e.g. *InteractionLoop*). This attribute is mostly useful for experiments or when combining *ParallelEngine* with engines that run parallel regions, resulting in nested OMP loops with different number of threads at each level.

**originalLength**(*=NaN*)
:   Distance of reference bodies in the direction of axis before straining started (computed automatically) [m]

**posIds**(*=uninitalized*)
> Bodies on which strain will be applied (on the positive end along the axis)

**setSpeeds**(*=false*)
> should we set speeds at the beginning directly, instead of increasing strain rate progressively?

**stopStrain**(*=NaN*)
> Strain at which we will pause simulation; inactive (nan) by default; must be reached from below (in absolute value)

**strain**(*=0*)
> Current strain value, elongation/originalLength *(auto-updated)* [-]

**strainRate**(*=NaN*)
> Rate of strain, starting at 0, linearly raising to strainRate. [-]

**stressUpdateInterval**(*=10*)
> How often to recompute stress on supports.

**timingDeltas**
> Detailed information about timing inside the Engine itself. Empty unless enabled in the source code and *O.timingEnabled*==True.

**updateAttrs**(*(Serializable)arg1, (dict)arg2*) → None :
> Update object attributes from given dictionary

## Collider

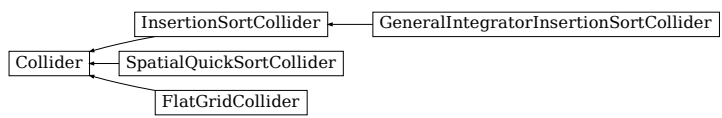

Fig. 28: Inheritance graph of Collider. See also: *FlatGridCollider*, *GeneralIntegratorInsertionSortCollider*, *InsertionSortCollider*, *SpatialQuickSortCollider*.

**class yade.wrapper.Collider**(*inherits GlobalEngine → Engine → Serializable*)
> Abstract class for finding spatial collisions between bodies.

---

### Special constructor

Derived colliders (unless they override `pyHandleCustomCtorArgs`) can be given list of *BoundFunctors* which is used to initialize the internal *boundDispatcher* instance.

---

**avoidSelfInteractionMask**(*=0*)
> This mask is used to avoid the interactions inside a group of particles. To do so, the particles must have the exact same mask and that mask should have one bit in common with this *avoidSelfInteractionMask* as for their binary representations.

**boundDispatcher**(*=new BoundDispatcher*)
> *BoundDispatcher* object that is used for creating *bounds* on collider's request as necessary.

**dead**(*=false*)
> If true, this engine will not run at all; can be used for making an engine temporarily deactivated and only resurrect it at a later point.

**dict**(*(Serializable)arg1*) → dict :
> Return dictionary of attributes.

**execCount**
> Cumulative count this engine was run (only used if *O.timingEnabled*==True).

**execTime**
> Cumulative time in nanoseconds this Engine took to run (only used if *O.timingEnabled*==**True**).

**label**(*=uninitalized*)
> Textual label for this object; must be valid python identifier, you can refer to it directly from python.

**ompThreads**(*=-1*)
> Number of threads to be used in the engine. If ompThreads<0 (default), the number will be typically OMP_NUM_THREADS or the number N defined by 'yade -jN' (this behavior can depend on the engine though). This attribute will only affect engines whose code includes openMP parallel regions (e.g. *InteractionLoop*). This attribute is mostly useful for experiments or when combining *ParallelEngine* with engines that run parallel regions, resulting in nested OMP loops with different number of threads at each level.

**timingDeltas**
> Detailed information about timing inside the Engine itself. Empty unless enabled in the source code and *O.timingEnabled*==**True**.

**updateAttrs**(*(Serializable)arg1, (dict)arg2*) → None :
> Update object attributes from given dictionary

**class yade.wrapper.FlatGridCollider**(*inherits Collider → GlobalEngine → Engine → Serializable*)
> Non-optimized grid collider, storing grid as dense flat array. Each body is assigned to (possibly multiple) cells, which are arranged in regular grid between *aabbMin* and *aabbMax*, with cell size *step* (same in all directions). Bodies outsize (*aabbMin, aabbMax*) are handled gracefully, assigned to closest cells (this will create spurious potential interactions). *verletDist* determines how much is each body enlarged to avoid collision detection at every step.

---
**Note:** This collider keeps all cells in linear memory array, therefore will be memory-inefficient for sparse simulations.

---

> **Warning:** objects *Body::bound* are not used, *BoundFunctors* are not used either: assigning cells to bodies is hard-coded internally. Currently handles *Shapes* are: *Sphere*.

---
**Note:** Periodic boundary is not handled (yet).

---

**aabbMax**(*=Vector3r::Zero()*)
> Upper corner of grid (approximate, might be rouded up to *minStep*.

**aabbMin**(*=Vector3r::Zero()*)
> Lower corner of grid.

**avoidSelfInteractionMask**(*=0*)
> This mask is used to avoid the interactions inside a group of particles. To do so, the particles must have the exact same mask and that mask should have one bit in common with this *avoidSelfInteractionMask* as for their binary representations.

**boundDispatcher**(*=new BoundDispatcher*)
> *BoundDispatcher* object that is used for creating *bounds* on collider's request as necessary.

**dead**(*=false*)
> If true, this engine will not run at all; can be used for making an engine temporarily deactivated and only resurrect it at a later point.

**dict**(*(Serializable)arg1*) → dict :
> Return dictionary of attributes.

**execCount**

 Cumulative count this engine was run (only used if *O.timingEnabled*==`True`).

**execTime**

 Cumulative time in nanoseconds this Engine took to run (only used if *O.timingEnabled*==`True`).

**label**(*=uninitalized*)

 Textual label for this object; must be valid python identifier, you can refer to it directly from python.

**ompThreads**(*=-1*)

 Number of threads to be used in the engine. If ompThreads<0 (default), the number will be typically OMP_NUM_THREADS or the number N defined by 'yade -jN' (this behavior can depend on the engine though). This attribute will only affect engines whose code includes openMP parallel regions (e.g. *InteractionLoop*). This attribute is mostly useful for experiments or when combining *ParallelEngine* with engines that run parallel regions, resulting in nested OMP loops with different number of threads at each level.

**step**(*=0*)

 Step in the grid (cell size)

**timingDeltas**

 Detailed information about timing inside the Engine itself. Empty unless enabled in the source code and *O.timingEnabled*==`True`.

**updateAttrs**(*(Serializable)arg1, (dict)arg2*) → None :

 Update object attributes from given dictionary

**verletDist**(*=0*)

 Length by which enlarge space occupied by each particle; avoids running collision detection at every step.

**class yade.wrapper.GeneralIntegratorInsertionSortCollider**(*inherits* *InsertionSort-Collider* → *Collider* → *GlobalEngine* → *Engine* → *Serializable*)

This class is the adaptive version of the InsertionSortCollider and changes the NewtonIntegrator dependency of the collider algorithms to the Integrator interface which is more general.

**allowBiggerThanPeriod**

 If true, tests on bodies sizes will be disabled, and the simulation will run normaly even if bodies larger than period are found. It can be useful when the periodic problem include e.g. a floor modelized with wall/box/facet. Be sure you know what you are doing if you touch this flag. The result is undefined if one large body moves out of the (0,0,0) period.

**avoidSelfInteractionMask**(*=0*)

 This mask is used to avoid the interactions inside a group of particles. To do so, the particles must have the exact same mask and that mask should have one bit in common with this *avoidSelfInteractionMask* as for their binary representations.

**boundDispatcher**(*=new BoundDispatcher*)

 *BoundDispatcher* object that is used for creating *bounds* on collider's request as necessary.

**dead**(*=false*)

 If true, this engine will not run at all; can be used for making an engine temporarily deactivated and only resurrect it at a later point.

**dict**(*(Serializable)arg1*) → dict :

 Return dictionary of attributes.

**doSort**(*=false*)

 Do forced resorting of interactions.

**dumpBounds**(*(InsertionSortCollider)arg1*) → tuple :

Return representation of the internal sort data. The format is ([...],[...],[...]) for 3 axes, where each ... is a list of entries (bounds). The entry is a tuple with the flowing items:

- coordinate (float)

- body id (int), but negated for negative bounds

- period numer (int), if the collider is in the periodic regime.

**execCount**

Cumulative count this engine was run (only used if *O.timingEnabled*==True).

**execTime**

Cumulative time in nanoseconds this Engine took to run (only used if *O.timingEnabled*==True).

**fastestBodyMaxDist**(*=0*)

Normalized maximum displacement of the fastest body since last run; if >= 1, we could get out of bboxes and will trigger full run. *(auto-updated)*

**isActivated**(*(InsertionSortCollider)arg1*) → bool :

Return true if collider needs execution at next iteration.

**keepListsShort**(*=false*)

if true remove bounds of non-existent or unbounded bodies from the lists *(auto-updated)*; turned true automatically in MPI mode and if bodies are erased with *BodyContainer.enableRedirection'=True. :ydefault:'false*

**label**(*=uninitalized*)

Textual label for this object; must be valid python identifier, you can refer to it directly from python.

**minSweepDistFactor**(*=0.1*)

Minimal distance by which enlarge all bounding boxes; superseeds computed value of verlet-Dist when lower that (minSweepDistFactor x verletDist).

**newton**(*=shared_ptr<NewtonIntegrator>()*)

reference to active *Newton integrator*. *(auto-updated)*

**numAction**(*=0*)

Cummulative number of collision detection.

**numReinit**(*=0*)

Cummulative number of bound array re-initialization.

**ompThreads**(*=-1*)

Number of threads to be used in the engine. If ompThreads<0 (default), the number will be typically OMP_NUM_THREADS or the number N defined by 'yade -jN' (this behavior can depend on the engine though). This attribute will only affect engines whose code includes openMP parallel regions (e.g. *InteractionLoop*). This attribute is mostly useful for experiments or when combining *ParallelEngine* with engines that run parallel regions, resulting in nested OMP loops with different number of threads at each level.

**overlapTolerance**(*=1e-7*)

Tolerance on determining overlap. In rare cases different parts of the code can inconsistently lead to different results in terms of overlap, with false negative by spatialOverlapPeri possibly leading to nasty bugs in contact detection (false positive are harmless). This tolerance is to avoid false negative, the value can be understood as relative to 1 (i.e. independent of particle size or any other reference length). The default should be ok.

**periodic**

Whether the collider is in periodic mode (read-only; for debugging) *(auto-updated)*

**smartInsertErase**(*=false*)

Use an algorithm optimized for heavy insert/delete (avoid initSort) - experimental.

**sortAxis**(*=0*)
   Axis for the initial contact detection.

**sortThenCollide**(*=false*)
   Separate sorting and colliding phase; it is MUCH slower, but all interactions are processed at every step; this effectively makes the collider non-persistent, not remembering last state. (The default behavior relies on the fact that inversions during insertion sort are overlaps of bounding boxes that just started/ceased to exist, and only processes those; this makes the collider much more efficient.)

**strideActive**
   Whether striding is active (read-only; for debugging). *(auto-updated)*

**targetInterv**(*=100*)
   (experimental) Target number of iterations between bound update, used to define a smaller sweep distance for slower grains if >0, else always use 1*verletDist. Useful in simulations with strong velocity contrasts between slow bodies and fast bodies.

**timingDeltas**
   Detailed information about timing inside the Engine itself. Empty unless enabled in the source code and *O.timingEnabled*==**True**.

**updateAttrs**(*(Serializable)arg1, (dict)arg2*) → None :
   Update object attributes from given dictionary

**updatingDispFactor**(*=-1*)
   (experimental) Displacement factor used to trigger bound update: the bound is updated only if updatingDispFactor*disp>sweepDist when >0, else all bounds are updated.

**verletDist**(*=-.5, Automatically initialized*)
   Length by which to enlarge particle bounds, to avoid running collider at every step. Stride disabled if zero. Negative value will trigger automatic computation, so that the real value will be *verletDist* × minimum spherical particle radius; if there are no spherical particles, it will be disabled. The actual length added to one bound can be only a fraction of verletDist when *InsertionSortCollider::targetInterv* is > 0.

**class yade.wrapper.InsertionSortCollider**(*inherits Collider → GlobalEngine → Engine → Serializable*)
   Collider with O(n log(n)) complexity, using *Aabb* for bounds.

   At the initial step, Bodies' bounds (along *sortAxis*) are first std::sort'ed along this (sortAxis) axis, then collided. The initial sort has $O(n^2)$ complexity, see Colliders' performance for some information (There are scripts in examples/collider-perf for measurements).

   Insertion sort is used for sorting the bound list that is already pre-sorted from last iteration, where each inversion calls checkOverlap which then handles either overlap (by creating interaction if necessary) or its absence (by deleting interaction if it is only potential).

   Bodies without bounding volume (such as clumps) are handled gracefully and never collide. Deleted bodies are handled gracefully as well.

   This collider handles periodic boundary conditions. There are some limitations, notably:

   1. No body can have Aabb larger than cell's half size in that respective dimension. You get exception if it does and gets in interaction. One way to explicitly by-pass this restriction is offered by **allowBiggerThanPeriod**, which can be turned on to insert a floor in the form of a very large box for instance (see examples/periodicSandPile.py).

   2. No body can travel more than cell's distance in one step; this would mean that the simulation is numerically exploding, and it is only detected in some cases.

   **Stride** can be used to avoid running collider at every step by enlarging the particle's bounds, tracking their displacements and only re-run if they might have gone out of that bounds (see Verlet list for brief description and background) . This requires cooperation from *NewtonIntegrator* as well as *BoundDispatcher*, which will be found among engines automatically (exception is thrown if they are not found).

If you wish to use strides, set `verletDist` (length by which bounds will be enlarged in all directions) to some value, e.g. 0.05 × typical particle radius. This parameter expresses the tradeoff between many potential interactions (running collider rarely, but with longer exact interaction resolution phase) and few potential interactions (running collider more frequently, but with less exact resolutions of interactions); it depends mainly on packing density and particle radius distribution.

If `targetInterv` is >1, not all particles will have their bound enlarged by `verletDist`; instead, they will have bounds increased by a length in order to trigger a new colliding after `targetInterv` iteration, assuming they move at almost constant velocity. Ideally in this method, all particles would reach their bounds at the sime iteration. This is of course not the case as soon as velocities fluctuate in time. *Bound::sweepLength* is tuned on the basis of the displacement recorded between the last two runs of the collider. In this situation, `verletDist` defines the maximum sweep length.

**allowBiggerThanPeriod**
> If true, tests on bodies sizes will be disabled, and the simulation will run normaly even if bodies larger than period are found. It can be useful when the periodic problem include e.g. a floor modelized with wall/box/facet. Be sure you know what you are doing if you touch this flag. The result is undefined if one large body moves out of the (0,0,0) period.

**avoidSelfInteractionMask**(*=0*)
> This mask is used to avoid the interactions inside a group of particles. To do so, the particles must have the exact same mask and that mask should have one bit in common with this *avoidSelfInteractionMask* as for their binary representations.

**boundDispatcher**(*=new BoundDispatcher*)
> *BoundDispatcher* object that is used for creating *bounds* on collider's request as necessary.

**dead**(*=false*)
> If true, this engine will not run at all; can be used for making an engine temporarily deactivated and only resurrect it at a later point.

**dict**(*(Serializable)arg1*) → dict :
> Return dictionary of attributes.

**doSort**(*=false*)
> Do forced resorting of interactions.

**dumpBounds**(*(InsertionSortCollider)arg1*) → tuple :
> Return representation of the internal sort data. The format is (`[...]`,`[...]`,`[...]`) for 3 axes, where each `...` is a list of entries (bounds). The entry is a tuple with the fllowing items:
>
> - coordinate (float)
> - body id (int), but negated for negative bounds
> - period numer (int), if the collider is in the periodic regime.

**execCount**
> Cumulative count this engine was run (only used if *O.timingEnabled*==`True`).

**execTime**
> Cumulative time in nanoseconds this Engine took to run (only used if *O.timingEnabled*==`True`).

**fastestBodyMaxDist**(*=0*)
> Normalized maximum displacement of the fastest body since last run; if >= 1, we could get out of bboxes and will trigger full run. *(auto-updated)*

**isActivated**(*(InsertionSortCollider)arg1*) → bool :
> Return true if collider needs execution at next iteration.

**keepListsShort**(*=false*)
> if true remove bounds of non-existent or unbounded bodies from the lists *(auto-updated)*; turned true automatically in MPI mode and if bodies are erased with *BodyContainer.enableRedirection‘=True. :ydefault:‘false*

**label**(*=uninitalized*)
> Textual label for this object; must be valid python identifier, you can refer to it directly from python.

**minSweepDistFactor**(*=0.1*)
> Minimal distance by which enlarge all bounding boxes; superseeds computed value of verlet-Dist when lower that (minSweepDistFactor x verletDist).

**newton**(*=shared_ptr<NewtonIntegrator>()*)
> reference to active *Newton integrator*. *(auto-updated)*

**numAction**(*=0*)
> Cummulative number of collision detection.

**numReinit**(*=0*)
> Cummulative number of bound array re-initialization.

**ompThreads**(*=-1*)
> Number of threads to be used in the engine. If ompThreads<0 (default), the number will be typically OMP_NUM_THREADS or the number N defined by 'yade -jN' (this behavior can depend on the engine though). This attribute will only affect engines whose code includes openMP parallel regions (e.g. *InteractionLoop*). This attribute is mostly useful for experiments or when combining *ParallelEngine* with engines that run parallel regions, resulting in nested OMP loops with different number of threads at each level.

**overlapTolerance**(*=1e-7*)
> Tolerance on determining overlap. In rare cases different parts of the code can inconsistently lead to different results in terms of overlap, with false negative by spatialOverlapPeri possibly leading to nasty bugs in contact detection (false positive are harmless). This tolerance is to avoid false negative, the value can be understood as relative to 1 (i.e. independent of particle size or any other reference length). The default should be ok.

**periodic**
> Whether the collider is in periodic mode (read-only; for debugging) *(auto-updated)*

**smartInsertErase**(*=false*)
> Use an algorithm optimized for heavy insert/delete (avoid initSort) - experimental.

**sortAxis**(*=0*)
> Axis for the initial contact detection.

**sortThenCollide**(*=false*)
> Separate sorting and colliding phase; it is MUCH slower, but all interactions are processed at every step; this effectively makes the collider non-persistent, not remembering last state. (The default behavior relies on the fact that inversions during insertion sort are overlaps of bounding boxes that just started/ceased to exist, and only processes those; this makes the collider much more efficient.)

**strideActive**
> Whether striding is active (read-only; for debugging). *(auto-updated)*

**targetInterv**(*=100*)
> (experimental) Target number of iterations between bound update, used to define a smaller sweep distance for slower grains if >0, else always use 1*verletDist. Useful in simulations with strong velocity contrasts between slow bodies and fast bodies.

**timingDeltas**
> Detailed information about timing inside the Engine itself. Empty unless enabled in the source code and *O.timingEnabled*==`True`.

**updateAttrs**(*(Serializable)arg1, (dict)arg2*) → None :
> Update object attributes from given dictionary

**updatingDispFactor**(*=-1*)
> (experimental) Displacement factor used to trigger bound update: the bound is updated only if updatingDispFactor*disp>sweepDist when >0, else all bounds are updated.

**verletDist**(*=-.5, Automatically initialized*)
> Length by which to enlarge particle bounds, to avoid running collider at every step. Stride disabled if zero. Negative value will trigger automatic computation, so that the real value will be *verletDist* × minimum spherical particle radius; if there are no spherical particles, it will be disabled. The actual length added to one bound can be only a fraction of verletDist when *InsertionSortCollider::targetInterv* is > 0.

**class yade.wrapper.SpatialQuickSortCollider**(*inherits Collider → GlobalEngine → Engine → Serializable*)
> Collider using quicksort along axes at each step, using *Aabb* bounds.

> Its performance is lower than that of *InsertionSortCollider* (see Colliders' performance), but the algorithm is simple enought to make it good for checking other collider's correctness.

**avoidSelfInteractionMask**(*=0*)
> This mask is used to avoid the interactions inside a group of particles. To do so, the particles must have the exact same mask and that mask should have one bit in common with this *avoidSelfInteractionMask* as for their binary representations.

**boundDispatcher**(*=new BoundDispatcher*)
> *BoundDispatcher* object that is used for creating *bounds* on collider's request as necessary.

**dead**(*=false*)
> If true, this engine will not run at all; can be used for making an engine temporarily deactivated and only resurrect it at a later point.

**dict**(*(Serializable)arg1*) → dict :
> Return dictionary of attributes.

**execCount**
> Cumulative count this engine was run (only used if *O.timingEnabled*==True).

**execTime**
> Cumulative time in nanoseconds this Engine took to run (only used if *O.timingEnabled*==True).

**label**(*=uninitalized*)
> Textual label for this object; must be valid python identifier, you can refer to it directly from python.

**ompThreads**(*=-1*)
> Number of threads to be used in the engine. If ompThreads<0 (default), the number will be typically OMP_NUM_THREADS or the number N defined by 'yade -jN' (this behavior can depend on the engine though). This attribute will only affect engines whose code includes openMP parallel regions (e.g. *InteractionLoop*). This attribute is mostly useful for experiments or when combining *ParallelEngine* with engines that run parallel regions, resulting in nested OMP loops with different number of threads at each level.

**timingDeltas**
> Detailed information about timing inside the Engine itself. Empty unless enabled in the source code and *O.timingEnabled*==True.

**updateAttrs**(*(Serializable)arg1, (dict)arg2*) → None :
> Update object attributes from given dictionary

### FieldApplier

**class yade.wrapper.FieldApplier**(*inherits GlobalEngine → Engine → Serializable*)
> Base for engines applying force files on particles. Not to be used directly.

**dead**(*=false*)
> If true, this engine will not run at all; can be used for making an engine temporarily deactivated and only resurrect it at a later point.

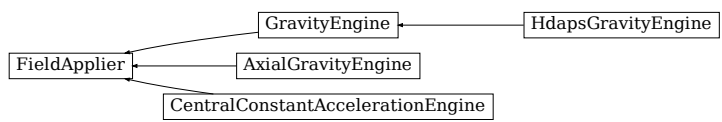

Fig. 29: Inheritance graph of FieldApplier. See also: *AxialGravityEngine*, *CentralConstantAcceleration-Engine*, *GravityEngine*, *HdapsGravityEngine*.

**dict**(*(Serializable)arg1*) → dict :
 Return dictionary of attributes.

**execCount**
 Cumulative count this engine was run (only used if *O.timingEnabled*==True).

**execTime**
 Cumulative time in nanoseconds this Engine took to run (only used if *O.timingEnabled*==True).

**label**(*=uninitalized*)
 Textual label for this object; must be valid python identifier, you can refer to it directly from python.

**ompThreads**(*=-1*)
 Number of threads to be used in the engine. If ompThreads<0 (default), the number will be typically OMP_NUM_THREADS or the number N defined by 'yade -jN' (this behavior can depend on the engine though). This attribute will only affect engines whose code includes openMP parallel regions (e.g. *InteractionLoop*). This attribute is mostly useful for experiments or when combining *ParallelEngine* with engines that run parallel regions, resulting in nested OMP loops with different number of threads at each level.

**timingDeltas**
 Detailed information about timing inside the Engine itself. Empty unless enabled in the source code and *O.timingEnabled*==True.

**updateAttrs**(*(Serializable)arg1, (dict)arg2*) → None :
 Update object attributes from given dictionary

**class yade.wrapper.AxialGravityEngine**(*inherits FieldApplier → GlobalEngine → Engine → Serializable*)
 Apply acceleration (independent of distance) directed towards an axis.

**acceleration**(*=0*)
 Acceleration magnitude [kgms $^2$]

**axisDirection**(*=Vector3r::UnitX()*)
 direction of the gravity axis (will be normalized automatically)

**axisPoint**(*=Vector3r::Zero()*)
 Point through which the axis is passing.

**dead**(*=false*)
 If true, this engine will not run at all; can be used for making an engine temporarily deactivated and only resurrect it at a later point.

**dict**(*(Serializable)arg1*) → dict :
 Return dictionary of attributes.

**execCount**
 Cumulative count this engine was run (only used if *O.timingEnabled*==True).

**execTime**
 Cumulative time in nanoseconds this Engine took to run (only used if *O.timingEnabled*==True).

**label**(*=uninitalized*)
> Textual label for this object; must be valid python identifier, you can refer to it directly from python.

**mask**(*=0*)
> If mask defined, only bodies with corresponding groupMask will be affected by this engine. If 0, all bodies will be affected.

**ompThreads**(*=-1*)
> Number of threads to be used in the engine. If ompThreads<0 (default), the number will be typically OMP_NUM_THREADS or the number N defined by 'yade -jN' (this behavior can depend on the engine though). This attribute will only affect engines whose code includes openMP parallel regions (e.g. *InteractionLoop*). This attribute is mostly useful for experiments or when combining *ParallelEngine* with engines that run parallel regions, resulting in nested OMP loops with different number of threads at each level.

**timingDeltas**
> Detailed information about timing inside the Engine itself. Empty unless enabled in the source code and *O.timingEnabled*==**True**.

**updateAttrs**(*(Serializable)arg1, (dict)arg2*) → None :
> Update object attributes from given dictionary

**class** yade.wrapper.**CentralConstantAccelerationEngine**(*inherits   FieldApplier  →  GlobalEngine  →  Engine  →  Serializable*)
Engine applying constant acceleration to all bodies, towards a central body. Ignoring the distance between them.

**accel**(*=0*)
> Acceleration magnitude [kgms $^2$]

**centralBody**(*=Body::ID_NONE*)
> The *body* towards which all other bodies are attracted.

**dead**(*=false*)
> If true, this engine will not run at all; can be used for making an engine temporarily deactivated and only resurrect it at a later point.

**dict**(*(Serializable)arg1*) → dict :
> Return dictionary of attributes.

**execCount**
> Cumulative count this engine was run (only used if *O.timingEnabled*==**True**).

**execTime**
> Cumulative time in nanoseconds this Engine took to run (only used if *O.timingEnabled*==**True**).

**label**(*=uninitalized*)
> Textual label for this object; must be valid python identifier, you can refer to it directly from python.

**mask**(*=0*)
> If mask defined, only bodies with corresponding groupMask will be affected by this engine. If 0, all bodies will be affected.

**ompThreads**(*=-1*)
> Number of threads to be used in the engine. If ompThreads<0 (default), the number will be typically OMP_NUM_THREADS or the number N defined by 'yade -jN' (this behavior can depend on the engine though). This attribute will only affect engines whose code includes openMP parallel regions (e.g. *InteractionLoop*). This attribute is mostly useful for experiments or when combining *ParallelEngine* with engines that run parallel regions, resulting in nested OMP loops with different number of threads at each level.

**reciprocal**(*=false*)
    If true, acceleration will be applied on the central body as well.

**timingDeltas**
    Detailed information about timing inside the Engine itself. Empty unless enabled in the source code and *O.timingEnabled*==`True`.

**updateAttrs**(*(Serializable)arg1, (dict)arg2*) → None :
    Update object attributes from given dictionary

**class yade.wrapper.GravityEngine**(*inherits* *FieldApplier* → *GlobalEngine* → *Engine* → *Serializable*)
    Engine applying constant acceleration to all bodies. DEPRECATED, use *Newton::gravity* unless you need energy tracking or selective gravity application using groupMask).

**dead**(*=false*)
    If true, this engine will not run at all; can be used for making an engine temporarily deactivated and only resurrect it at a later point.

**dict**(*(Serializable)arg1*) → dict :
    Return dictionary of attributes.

**execCount**
    Cumulative count this engine was run (only used if *O.timingEnabled*==`True`).

**execTime**
    Cumulative time in nanoseconds this Engine took to run (only used if *O.timingEnabled*==`True`).

**gravity**(*=Vector3r::Zero()*)
    Acceleration [kgms $^2$]

**label**(*=uninitalized*)
    Textual label for this object; must be valid python identifier, you can refer to it directly from python.

**mask**(*=0*)
    If mask defined, only bodies with corresponding groupMask will be affected by this engine. If 0, all bodies will be affected.

**ompThreads**(*=-1*)
    Number of threads to be used in the engine. If ompThreads<0 (default), the number will be typically OMP_NUM_THREADS or the number N defined by 'yade -jN' (this behavior can depend on the engine though). This attribute will only affect engines whose code includes openMP parallel regions (e.g. *InteractionLoop*). This attribute is mostly useful for experiments or when combining *ParallelEngine* with engines that run parallel regions, resulting in nested OMP loops with different number of threads at each level.

**timingDeltas**
    Detailed information about timing inside the Engine itself. Empty unless enabled in the source code and *O.timingEnabled*==`True`.

**updateAttrs**(*(Serializable)arg1, (dict)arg2*) → None :
    Update object attributes from given dictionary

**warnOnce**(*=true*)
    For deprecation warning once.

**class yade.wrapper.HdapsGravityEngine**(*inherits* *GravityEngine* → *FieldApplier* → *GlobalEngine* → *Engine* → *Serializable*)
    Read accelerometer in Thinkpad laptops (HDAPS and accordingly set gravity within the simulation. This code draws from hdaps-gl . See scripts/test/hdaps.py for an example.

**accel**(*=Vector2i::Zero()*)
    reading from the sysfs file

**calibrate**(*=Vector2i::Zero()*)
    Zero position; if NaN, will be read from the *hdapsDir* / calibrate.

**calibrated**(*=false*)
    Whether *calibrate* was already updated. Do not set to `True` by hand unless you also give a meaningful value for *calibrate*.

**dead**(*=false*)
    If true, this engine will not run at all; can be used for making an engine temporarily deactivated and only resurrect it at a later point.

**dict**(*(Serializable)arg1*) → dict :
    Return dictionary of attributes.

**execCount**
    Cumulative count this engine was run (only used if *O.timingEnabled*==`True`).

**execTime**
    Cumulative time in nanoseconds this Engine took to run (only used if *O.timingEnabled*==`True`).

**gravity**(*=Vector3r::Zero()*)
    Acceleration [kgms $^2$]

**hdapsDir**(*="/sys/devices/platform/hdaps"*)
    Hdaps directory; contains `position` (with accelerometer readings) and `calibration` (zero acceleration).

**label**(*=uninitalized*)
    Textual label for this object; must be valid python identifier, you can refer to it directly from python.

**mask**(*=0*)
    If mask defined, only bodies with corresponding groupMask will be affected by this engine. If 0, all bodies will be affected.

**msecUpdate**(*=50*)
    How often to update the reading.

**ompThreads**(*=-1*)
    Number of threads to be used in the engine. If ompThreads<0 (default), the number will be typically OMP_NUM_THREADS or the number N defined by 'yade -jN' (this behavior can depend on the engine though). This attribute will only affect engines whose code includes openMP parallel regions (e.g. *InteractionLoop*). This attribute is mostly useful for experiments or when combining *ParallelEngine* with engines that run parallel regions, resulting in nested OMP loops with different number of threads at each level.

**timingDeltas**
    Detailed information about timing inside the Engine itself. Empty unless enabled in the source code and *O.timingEnabled*==`True`.

**updateAttrs**(*(Serializable)arg1, (dict)arg2*) → None :
    Update object attributes from given dictionary

**updateThreshold**(*=4*)
    Minimum difference of reading from the file before updating gravity, to avoid jitter.

**warnOnce**(*=true*)
    For deprecation warning once.

**zeroGravity**(*=Vector3r(0, 0, -1)*)
    Gravity if the accelerometer is in flat (zero) position.

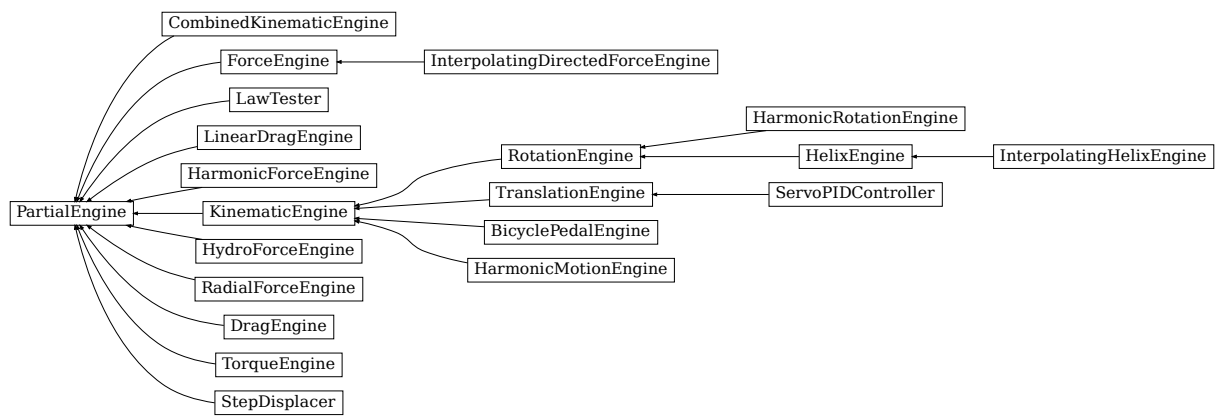

Fig. 30: Inheritance graph of PartialEngine. See also: *BicyclePedalEngine*, *CombinedKinematicEngine*, *DragEngine*, *ForceEngine*, *HarmonicForceEngine*, *HarmonicMotionEngine*, *HarmonicRotationEngine*, *HelixEngine*, *HydroForceEngine*, *InterpolatingDirectedForceEngine*, *InterpolatingHelixEngine*, *KinematicEngine*, *LawTester*, *LinearDragEngine*, *RadialForceEngine*, *RotationEngine*, *ServoPIDController*, *StepDisplacer*, *TorqueEngine*, *TranslationEngine*.

### 2.3.4 Partial engines

**class** `yade.wrapper.`**PartialEngine**(*inherits Engine → Serializable*)

Engine affecting only particular bodies in the simulation, namely those defined in *ids attribute*. See also *GlobalEngine*.

**dead**(*=false*)

If true, this engine will not run at all; can be used for making an engine temporarily deactivated and only resurrect it at a later point.

**dict**(*(Serializable)arg1*) → dict :

Return dictionary of attributes.

**execCount**

Cumulative count this engine was run (only used if *O.timingEnabled*==`True`).

**execTime**

Cumulative time in nanoseconds this Engine took to run (only used if *O.timingEnabled*==`True`).

**ids**(*=uninitialized*)

*Ids* list of bodies affected by this PartialEngine.

**label**(*=uninitialized*)

Textual label for this object; must be valid python identifier, you can refer to it directly from python.

**ompThreads**(*=-1*)

Number of threads to be used in the engine. If ompThreads<0 (default), the number will be typically OMP_NUM_THREADS or the number N defined by 'yade -jN' (this behavior can depend on the engine though). This attribute will only affect engines whose code includes openMP parallel regions (e.g. *InteractionLoop*). This attribute is mostly useful for experiments or when combining *ParallelEngine* with engines that run parallel regions, resulting in nested OMP loops with different number of threads at each level.

**timingDeltas**

Detailed information about timing inside the Engine itself. Empty unless enabled in the source code and *O.timingEnabled*==`True`.

**updateAttrs**(*(Serializable)arg1, (dict)arg2*) → None :

Update object attributes from given dictionary **class yade.wrapper.BicyclePedalEngine**(*inherits* *KinematicEngine* → *PartialEngine* → *Engine* → *Serializable*)

Engine applying the linear motion of `bicycle pedal` e.g. moving points around the axis without rotation

**angularVelocity**(*=0*)

Angular velocity. [rad/s]

**dead**(*=false*)

If true, this engine will not run at all; can be used for making an engine temporarily deactivated and only resurrect it at a later point.

**dict**(*(Serializable)arg1*) → dict :

Return dictionary of attributes.

**execCount**

Cumulative count this engine was run (only used if *O.timingEnabled*==`True`).

**execTime**

Cumulative time in nanoseconds this Engine took to run (only used if *O.timingEnabled*==`True`).

**fi**(*=Mathr::PI/2.0*)

Initial phase [radians]

**ids**(*=uninitalized*)

*Ids* list of bodies affected by this PartialEngine.

**label**(*=uninitalized*)

Textual label for this object; must be valid python identifier, you can refer to it directly from python.

**ompThreads**(*=-1*)

Number of threads to be used in the engine. If ompThreads<0 (default), the number will be typically OMP_NUM_THREADS or the number N defined by 'yade -jN' (this behavior can depend on the engine though). This attribute will only affect engines whose code includes openMP parallel regions (e.g. *InteractionLoop*). This attribute is mostly useful for experiments or when combining *ParallelEngine* with engines that run parallel regions, resulting in nested OMP loops with different number of threads at each level.

**radius**(*=-1.0*)

Rotation radius. [m]

**rotationAxis**(*=Vector3r::UnitX()*)

Axis of rotation (direction); will be normalized automatically.

**timingDeltas**

Detailed information about timing inside the Engine itself. Empty unless enabled in the source code and *O.timingEnabled*==`True`.

**updateAttrs**(*(Serializable)arg1, (dict)arg2*) → None :

Update object attributes from given dictionary

**class yade.wrapper.CombinedKinematicEngine**(*inherits* *PartialEngine* → *Engine* → *Serializable*)

Engine for applying combined displacements on pre-defined bodies. Constructed using + operator on regular *KinematicEngines*. The `ids` operated on are those of the first engine in the combination (assigned automatically).

**comb**(*=uninitalized*)

Kinematic engines that will be combined by this one, run in the order given.

**dead**(*=false*)

If true, this engine will not run at all; can be used for making an engine temporarily deactivated and only resurrect it at a later point.

**dict**(*(Serializable)arg1*) → dict :
    Return dictionary of attributes.

**execCount**
    Cumulative count this engine was run (only used if *O.timingEnabled*==`True`).

**execTime**
    Cumulative time in nanoseconds this Engine took to run (only used if *O.timingEnabled*==`True`).

**ids**(*=uninitalized*)
    *Ids* list of bodies affected by this PartialEngine.

**label**(*=uninitalized*)
    Textual label for this object; must be valid python identifier, you can refer to it directly from python.

**ompThreads**(*=-1*)
    Number of threads to be used in the engine. If ompThreads<0 (default), the number will be typically OMP_NUM_THREADS or the number N defined by 'yade -jN' (this behavior can depend on the engine though). This attribute will only affect engines whose code includes openMP parallel regions (e.g. *InteractionLoop*). This attribute is mostly useful for experiments or when combining *ParallelEngine* with engines that run parallel regions, resulting in nested OMP loops with different number of threads at each level.

**timingDeltas**
    Detailed information about timing inside the Engine itself. Empty unless enabled in the source code and *O.timingEnabled*==`True`.

**updateAttrs**(*(Serializable)arg1, (dict)arg2*) → None :
    Update object attributes from given dictionary

**class yade.wrapper.DragEngine**(*inherits PartialEngine → Engine → Serializable*)
    Apply drag force on some particles at each step, decelerating them proportionally to their linear velocities. The applied force reads

$$F_d = -\frac{\boldsymbol{v}}{|\boldsymbol{v}|}\frac{1}{2}\rho|\boldsymbol{v}|^2 C_d A$$

where $\rho$ is the medium density (*density*), $v$ is particle's velocity, $A$ is particle projected area (disc), $C_d$ is the drag coefficient (0.47 for *Sphere*),

---

**Note:**  Drag force is only applied to spherical particles, listed in ids.

---

**Cd**(*=0.47*)
    Drag coefficient <http://en.wikipedia.org/wiki/Drag_coefficient>'_.

**Rho**(*=1.225*)
    Density of the medium (fluid or air), by default - the density of the air.

**dead**(*=false*)
    If true, this engine will not run at all; can be used for making an engine temporarily deactivated and only resurrect it at a later point.

**dict**(*(Serializable)arg1*) → dict :
    Return dictionary of attributes.

**execCount**
    Cumulative count this engine was run (only used if *O.timingEnabled*==`True`).

**execTime**
    Cumulative time in nanoseconds this Engine took to run (only used if *O.timingEnabled*==`True`).

**ids**(*=uninitalized*)
> *Ids* list of bodies affected by this PartialEngine.

**label**(*=uninitalized*)
> Textual label for this object; must be valid python identifier, you can refer to it directly from python.

**ompThreads**(*=-1*)
> Number of threads to be used in the engine. If ompThreads<0 (default), the number will be typically OMP_NUM_THREADS or the number N defined by 'yade -jN' (this behavior can depend on the engine though). This attribute will only affect engines whose code includes openMP parallel regions (e.g. *InteractionLoop*). This attribute is mostly useful for experiments or when combining *ParallelEngine* with engines that run parallel regions, resulting in nested OMP loops with different number of threads at each level.

**timingDeltas**
> Detailed information about timing inside the Engine itself. Empty unless enabled in the source code and *O.timingEnabled*==`True`.

**updateAttrs**(*(Serializable)arg1, (dict)arg2*) → None :
> Update object attributes from given dictionary

**class yade.wrapper.ForceEngine**(*inherits* *PartialEngine* → *Engine* → *Serializable*)
> Apply contact force on some particles at each step.

**dead**(*=false*)
> If true, this engine will not run at all; can be used for making an engine temporarily deactivated and only resurrect it at a later point.

**dict**(*(Serializable)arg1*) → dict :
> Return dictionary of attributes.

**execCount**
> Cumulative count this engine was run (only used if *O.timingEnabled*==`True`).

**execTime**
> Cumulative time in nanoseconds this Engine took to run (only used if *O.timingEnabled*==`True`).

**force**(*=Vector3r::Zero()*)
> Force to apply.

**ids**(*=uninitalized*)
> *Ids* list of bodies affected by this PartialEngine.

**label**(*=uninitalized*)
> Textual label for this object; must be valid python identifier, you can refer to it directly from python.

**ompThreads**(*=-1*)
> Number of threads to be used in the engine. If ompThreads<0 (default), the number will be typically OMP_NUM_THREADS or the number N defined by 'yade -jN' (this behavior can depend on the engine though). This attribute will only affect engines whose code includes openMP parallel regions (e.g. *InteractionLoop*). This attribute is mostly useful for experiments or when combining *ParallelEngine* with engines that run parallel regions, resulting in nested OMP loops with different number of threads at each level.

**timingDeltas**
> Detailed information about timing inside the Engine itself. Empty unless enabled in the source code and *O.timingEnabled*==`True`.

**updateAttrs**(*(Serializable)arg1, (dict)arg2*) → None :
> Update object attributes from given dictionary **class** yade.wrapper.**HarmonicForceEngine**(*inherits PartialEngine → Engine → Serializable*)

This engine adds a harmonic (sinusoidal) force to a set of bodies. It is identical to *Harmonic-MotionEngine* except a force amplitude is prescribed instead of motion, see also the dynamics of harmonic motion

**A**(*=Vector3r::Zero()*)
Amplitude [N]

**dead**(*=false*)
If true, this engine will not run at all; can be used for making an engine temporarily deactivated and only resurrect it at a later point.

**dict**(*(Serializable)arg1*) → dict :
Return dictionary of attributes.

**execCount**
Cumulative count this engine was run (only used if *O.timingEnabled*==True).

**execTime**
Cumulative time in nanoseconds this Engine took to run (only used if *O.timingEnabled*==True).

**f**(*=Vector3r::Zero()*)
Frequency [hertz]

**fi**(*=Vector3r::Zero()*)
Initial phase [radians]. By default, the phase is zero such that the force starts at zero.

**ids**(*=uninitalized*)
*Ids* list of bodies affected by this PartialEngine.

**label**(*=uninitalized*)
Textual label for this object; must be valid python identifier, you can refer to it directly from python.

**ompThreads**(*=-1*)
Number of threads to be used in the engine. If ompThreads<0 (default), the number will be typically OMP_NUM_THREADS or the number N defined by 'yade -jN' (this behavior can depend on the engine though). This attribute will only affect engines whose code includes openMP parallel regions (e.g. *InteractionLoop*). This attribute is mostly useful for experiments or when combining *ParallelEngine* with engines that run parallel regions, resulting in nested OMP loops with different number of threads at each level.

**timingDeltas**
Detailed information about timing inside the Engine itself. Empty unless enabled in the source code and *O.timingEnabled*==True.

**updateAttrs**(*(Serializable)arg1, (dict)arg2*) → None :
Update object attributes from given dictionary

**class** yade.wrapper.**HarmonicMotionEngine**(*inherits KinematicEngine → PartialEngine → Engine → Serializable*)

This engine implements the harmonic oscillation of bodies. See also *HarmonicForceEngine* that applies a harmonic force, see also the dynamics of harmonic motion

**A**(*=Vector3r::Zero()*)
Amplitude [m]

**dead**(*=false*)
If true, this engine will not run at all; can be used for making an engine temporarily deactivated and only resurrect it at a later point.

**dict**(*(Serializable)arg1*) → dict :
Return dictionary of attributes.

**execCount**
Cumulative count this engine was run (only used if *O.timingEnabled*==True).

**execTime**
  Cumulative time in nanoseconds this Engine took to run (only used if *O.timingEnabled*==`True`).

**f**(=*Vector3r::Zero()*)
  Frequency [hertz]

**fi**(=*Vector3r(Mathr::PI/2.0, Mathr::PI/2.0, Mathr::PI/2.0)*)
  Initial phase [radians]. By default, the body oscillates around initial position.

**ids**(=*uninitalized*)
  *Ids* list of bodies affected by this PartialEngine.

**label**(=*uninitalized*)
  Textual label for this object; must be valid python identifier, you can refer to it directly from python.

**ompThreads**(=*-1*)
  Number of threads to be used in the engine. If ompThreads<0 (default), the number will be typically OMP_NUM_THREADS or the number N defined by 'yade -jN' (this behavior can depend on the engine though). This attribute will only affect engines whose code includes openMP parallel regions (e.g. *InteractionLoop*). This attribute is mostly useful for experiments or when combining *ParallelEngine* with engines that run parallel regions, resulting in nested OMP loops with different number of threads at each level.

**timingDeltas**
  Detailed information about timing inside the Engine itself. Empty unless enabled in the source code and *O.timingEnabled*==`True`.

**updateAttrs**(*(Serializable)arg1, (dict)arg2*) → None :
  Update object attributes from given dictionary

**class yade.wrapper.HarmonicRotationEngine**(*inherits RotationEngine → KinematicEngine → PartialEngine → Engine → Serializable*)
This engine implements the harmonic-rotation oscillation of bodies, see also the dynamics of harmonic motion ; please, set dynamic=False for bodies, droven by this engine, otherwise amplitude will be 2x more, than awaited.

**A**(=*0*)
  Amplitude [rad]

**angularVelocity**(=*0*)
  Angular velocity. [rad/s]

**dead**(=*false*)
  If true, this engine will not run at all; can be used for making an engine temporarily deactivated and only resurrect it at a later point.

**dict**(*(Serializable)arg1*) → dict :
  Return dictionary of attributes.

**execCount**
  Cumulative count this engine was run (only used if *O.timingEnabled*==`True`).

**execTime**
  Cumulative time in nanoseconds this Engine took to run (only used if *O.timingEnabled*==`True`).

**f**(=*0*)
  Frequency [hertz]

**fi**(=*Mathr::PI/2.0*)
  Initial phase [radians]. By default, the body oscillates around initial position.

**ids**(=*uninitalized*)
  *Ids* list of bodies affected by this PartialEngine.

**label**(*=uninitalized*)
　　Textual label for this object; must be valid python identifier, you can refer to it directly from python.

**ompThreads**(*=-1*)
　　Number of threads to be used in the engine. If ompThreads<0 (default), the number will be typically OMP_NUM_THREADS or the number N defined by 'yade -jN' (this behavior can depend on the engine though). This attribute will only affect engines whose code includes openMP parallel regions (e.g. *InteractionLoop*). This attribute is mostly useful for experiments or when combining *ParallelEngine* with engines that run parallel regions, resulting in nested OMP loops with different number of threads at each level.

**rotateAroundZero**(*=false*)
　　If True, bodies will not rotate around their centroids, but rather around `zeroPoint`.

**rotationAxis**(*=Vector3r::UnitX()*)
　　Axis of rotation (direction); will be normalized automatically.

**timingDeltas**
　　Detailed information about timing inside the Engine itself. Empty unless enabled in the source code and *O.timingEnabled*==`True`.

**updateAttrs**(*(Serializable)arg1, (dict)arg2*) → None :
　　Update object attributes from given dictionary

**zeroPoint**(*=Vector3r::Zero()*)
　　Point around which bodies will rotate if `rotateAroundZero` is True

**class yade.wrapper.HelixEngine**(*inherits* *RotationEngine* → *KinematicEngine* → *PartialEngine* → *Engine* → *Serializable*)
Engine applying both rotation and translation, along the same axis, whence the name HelixEngine

**angleTurned**(*=0*)
　　How much have we turned so far. *(auto-updated)* [rad]

**angularVelocity**(*=0*)
　　Angular velocity. [rad/s]

**dead**(*=false*)
　　If true, this engine will not run at all; can be used for making an engine temporarily deactivated and only resurrect it at a later point.

**dict**(*(Serializable)arg1*) → dict :
　　Return dictionary of attributes.

**execCount**
　　Cumulative count this engine was run (only used if *O.timingEnabled*==`True`).

**execTime**
　　Cumulative time in nanoseconds this Engine took to run (only used if *O.timingEnabled*==`True`).

**ids**(*=uninitalized*)
　　*Ids* list of bodies affected by this PartialEngine.

**label**(*=uninitalized*)
　　Textual label for this object; must be valid python identifier, you can refer to it directly from python.

**linearVelocity**(*=0*)
　　Linear velocity [m/s]

**ompThreads**(*=-1*)
　　Number of threads to be used in the engine. If ompThreads<0 (default), the number will be typically OMP_NUM_THREADS or the number N defined by 'yade -jN' (this behavior can depend on the engine though). This attribute will only affect engines whose code includes openMP parallel regions (e.g. *InteractionLoop*). This attribute is mostly useful for experiments or when combining *ParallelEngine* with engines that run parallel regions, resulting in nested OMP loops with different number of threads at each level.

**rotateAroundZero**(*=false*)
  If True, bodies will not rotate around their centroids, but rather around `zeroPoint`.

**rotationAxis**(*=Vector3r::UnitX()*)
  Axis of rotation (direction); will be normalized automatically.

**timingDeltas**
  Detailed information about timing inside the Engine itself. Empty unless enabled in the source code and *O.timingEnabled*==`True`.

**updateAttrs**(*(Serializable)arg1, (dict)arg2*) → None :
  Update object attributes from given dictionary

**zeroPoint**(*=Vector3r::Zero()*)
  Point around which bodies will rotate if `rotateAroundZero` is True

**class yade.wrapper.HydroForceEngine**(*inherits PartialEngine → Engine → Serializable*)

**Engine performing a coupling of the DEM with a volume-averaged 1D fluid resolution to simulate s**
  The engine can be decomposed in three different parts: (i) It applies the fluid force on the particles imposed by the fluid velocity profiles and fluid properties, (ii) It evaluates averaged solid depth profiles necessary for the fluid force application and for the fluid resolution, (iii) It solve the volume-averaged 1D fluid momentum balance.

The three different functions are detailed below:

(i) Fluid force on particles Apply to each particles, buoyancy, drag and lift force due to a 1D fluid flow and can apply lubrication force between two particles. The applied drag force reads

$$F_d = \frac{1}{2} C_d A \rho^f |\mathbf{v_f} - \mathbf{v}| \mathbf{v_f} - \mathbf{v}$$

where $\rho$ is the fluid density (*densFluid*), $\mathbf{v}$ is particle's velocity, $\mathbf{v_f}$ is the velocity of the fluid at the particle center (taken from the fluid velocity profile *vxFluid*), $A = \pi d^2/4$ is particle projected area (disc), $C_d$ is the drag coefficient. The formulation of the drag coefficient depends on the local particle reynolds number and the solid volume fraction. The formulation of the drag is [Dallavalle1948] [RevilBaudard2013] with a correction of Richardson-Zaki [Richardson1954] to take into account the hindrance effect. This law is classical in sediment transport. The possibly activated lubrication force (with parameter:yref:*lubrication<HydroForceEngine.lubrication>* put to True) reads: $F_{lubrication} = \frac{6\pi \eta^f v_r \epsilon l^n}{\delta^n + \epsilon_r}$, with $\eta^f$ the fluid dynamic viscosity *viscoDyn*, $v_r \epsilon l^n$ the normal relative velocity of the two particles, $\delta^n$ the distance between the two particles surface, and $\epsilon_r$ the roughness scale of the particle (*roughnessPartScale*).

**It is possible to activate a fluctuation of the drag force for each particle which account for the turb**
  The formulation of the lift is taken from [Wiberg1985] and is such that :

$$F_L = \frac{1}{2} C_L A \rho^f ((v_f - v)^2_{top} - (v_f - v)^2_{bottom})$$

Where the subscript top and bottom means evaluated at the top (respectively the bottom) of the sphere considered. This formulation of the lift account for the difference of pressure at the top and the bottom of the particle inside a turbulent shear flow. As this formulation is controversial when approaching the threshold of motion [Schmeeckle2007] it is possible to desactivate it with the variable *lift*. The buoyancy is taken into account through the buoyant weight :

$$F_{buoyancy} = -\rho^f V^p g$$

, where g is the gravity vector along the vertical, and $V^p$ is the volume of the particle. In the case where the fluid flow is steady and uniform, the buoyancy reduces to its wall-normal component (see [Maurin2018] for a full explanation), and one should put *steadyFlow* to true in order to kill the streamwise component.

(ii) Averaged solid depth profiles The function averageProfile evaluates the volume averaged depth profiles (1D) of particle velocity, particle solid volume fraction and particle drag force. It uses a volume-weighting average following [Maurin2015PhD]__[Maurin2015b]__, i.e. the average of a variable $A^p$ associated to particles at a given discretized wall-normal position $z$ is given by:

$$\langle A \rangle^s (z) = \frac{\sum\limits_{p|z^p \in [z-dz/2,\, z+dz/2]} A^p(t)V_z^p}{\sum\limits_{p|z^p \in [z-dz/2,\, z+dz/2]} V_z^p}$$

Where the sums are over the particles contained inside the slice between the wall-normal position $z - dz/2$ and $z + dz/2$, and $V^p$ represents the part of the volume of the given particle effectively contained inside the slice. For more details, see [Maurin2015PhD].

(iii) 1D volume-average fluid resolution The fluid resolution is based on the resolution of the 1D volume-averaged fluid momentum balance. It assumes by definition (unidirectional) that the fluid flow is steady and uniform. It is the same fluid resolution as [RevilBaudard2013]. Details can be found in this paper and in [Maurin2015PhD] [Maurin2015b].

The three different component can be used independently, e.g. applying a fluid force due to an imposed fluid profile or solving the fluid momentum balance for a given concentration of particles.

**Cl**(=*0.2*)
Value of the lift coefficient taken from [Wiberg1985]

**ReynoldStresses**(=*uninitialized*)
Vector of size equal to *nCell* containing the Reynolds stresses as a function of the depth. ReynoldStresses(z) = $\rho^f < u_z' u_z' > (z)^2$

**averageDrag**(=*uninitialized*)
Discretized average drag depth profile. No role in the engine, output parameter. For practical reason, it can be evaluated directly inside the engine, calling from python the averageProfile() method of the engine

**averageDrag1**(=*uninitialized*)
Discretized average drag depth profile of particles of type 1. Evaluated when *twoSize* is set to True.

**averageDrag2**(=*uninitialized*)
Discretized average drag depth profile of particles of type 2. Evaluated when *twoSize* is set to True.

**averageProfile**(*(HydroForceEngine)arg1*) → None :
Compute and store the particle velocity (*vxPart*, *vyPart*, *vzPart*) and solid volume fraction (*phiPart*) depth profile. For each defined cell z, the k component of the average particle velocity reads:

$< v_k >^z = \sum_p V^p v_k^p / \sum_p V^p$,

where the sum is made over the particles contained in the cell, $v_k^p$ is the k component of the velocity associated to particle p, and $V^p$ is the part of the volume of the particle p contained inside the cell. This definition allows to smooth the averaging, and is equivalent to taking into account the center of the particles only when there is a lot of particles in each cell. As for the solid volume fraction, it is evaluated in the same way: for each defined cell z, it reads:

$< \varphi >^z = \frac{1}{V_{cell}} \sum_p V^p$, where $V_{cell}$ is the volume of the cell considered, and $V^p$ is the volume of particle p contained in cell z. This function gives depth profiles of average velocity and solid volume fraction, returning the average quantities in each cell of height dz, from the reference horizontal plane at elevation *zRef* (input parameter) until the plane of elevation *zRef* plus

*nCell* times *deltaZ* (input parameters). When the option *twoSize* is set to True, evaluate in addition the average drag (*averageDrag1* and *averageDrag2*) and solid volume fraction (*phiPart1* and *phiPart2*) depth profiles considering only the particles of radius respectively *radiusPart1* and *radiusPart2* in the averaging.

`bedElevation`(*=0.*)
    Elevation of the bed above which the fluid flow is turbulent and the particles undergo turbulent velocity fluctuation.

`channelWidth`(*=1.*)
    Fluid resolution: Channel width for the evaluation of the fluid wall friction inside the fluid resolution.

`compatibilityOldVersion`(*=false*)
    Option to make HydroForceEngine compatible with former scripts. Slow down slightly the calculation and will eventually be removed.

`computeRadiusParts`(*(HydroForceEngine)arg1*) → None :
    compute the different class of radius present in the simulation.

`convAcc`(*=uninitalized*)
    Convective acceleration, depth dependent

`convAccOption`(*=false*)
    To activate the convective acceleration option in order to account for a convective acceleration term inside the momentum balance.

`dead`(*=false*)
    If true, this engine will not run at all; can be used for making an engine temporarily deactivated and only resurrect it at a later point.

`deltaZ`(*=uninitalized*)
    Height of the discretization cell.

`densFluid`(*=1000*)
    Density of the fluid, by default - density of water

`dict`(*(Serializable)arg1*) → dict :
    Return dictionary of attributes.

`dpdx`(*=0.*)
    pressure gradient along streamwise direction

`dtFluct`(*=uninitalized*)
    Execution time step of the turbulent fluctuation model.

`enableMultiClassAverage`(*=false*)
    Enables specific averaging for all the different particle size. Uses a lot of memory if using a lots of different particle size

`execCount`
    Cumulative count this engine was run (only used if *O.timingEnabled*==`True`).

`execTime`
    Cumulative time in nanoseconds this Engine took to run (only used if *O.timingEnabled*==`True`).

`expoRZ`(*=3.1*)
    Value of the Richardson-Zaki exponent, for the drag correction due to hindrance

`fluctTime`(*=uninitalized*)
    Vector containing the time of life of the fluctuations associated to each particles.

`fluidFrictionCoef`(*=1.*)
    Fluid resolution: fitting coefficient for the fluid wall friction

**fluidResolution**(*(HydroForceEngine)arg1, (float)arg2, (float)arg3*) → None :
    Solve the 1D volume-averaged fluid momentum balance on the defined mesh (*nCell*, *deltaZ*) from the volume-averaged solid profiles (*phiPart*,;yref:*vxPart<HydroForceEngine.vxPart>*,;yref:*averageDrag<HydroForceEngine.averageDrag>*), which can be evaluated with the averageProfile function.

**fluidWallFriction**(*=false*)
    Fluid resolution: if set to true, introduce a sink term to account for the fluid friction at the wall, see [Maurin2015] for details. Requires to set the width of the channel. It might slow down significantly the calculation.

**gravity**(*=Vector3r(0, 0, -9.81)*)
    Gravity vector

**ids**(*=uninitalized*)
    *Ids* list of bodies affected by this PartialEngine.

**ilm**(*=2*)
    Fluid resolution: type of mixing length resolution applied: 0: classical Prandtl mixing length, 1: Prandtl mixing length with free-surface effects, 2: Damp turbulence accounting for the presence of particles [Li1995], see [RevilBaudard2013] for more details.

**initialization**(*(HydroForceEngine)arg1*) → None :
    Initialize the necessary parameters to make HydroForceEngine run. Necessary to execute before any simulation run, otherwise it crashes

**irheolf**(*=0*)
    Fluid resolution: effective fluid viscosity option: 0: pure fluid viscosity, 1: Einstein viscosity.

**iturbu**(*=1*)
    Fluid resolution: activate the turbulence resolution, 1, or not, 0

**iusl**(*=1*)
    Fluid resolution: option to set the boundary condition at the top of the fluid, 0: Dirichlet, fixed ($\mathtt{u} = \mathtt{uTop}$ en $z = \mathtt{h}$), 1:Neumann, free-surface ($\mathtt{du/dz} = \mathtt{0}$ en $z = \mathtt{h}$).

**kappa**(*=0.41*)
    Fluid resolution: Von Karman constant. Can be tuned to account for the effect of particles on the fluid turbulence, see e.g. [RevilBaudard2015]

**label**(*=uninitalized*)
    Textual label for this object; must be valid python identifier, you can refer to it directly from python.

**lift**(*=false*)
    Option to activate or not the evaluation of the lift

**lubrication**(*=false*)
    Condition to activate the calculation of the lubrication force.

**multiDragPart**(*=uninitalized*)
    Spatial-averaged mean drag force for each class of particle. Un-used ? Or just for debug.

**multiPhiPart**(*=uninitalized*)
    Spatial-averaged solid volume fraction for each class of particle.

**multiVxPart**(*=uninitalized*)
    Spatial-averaged velocity in x direction for each class of particle.

**multiVyPart**(*=uninitalized*)
    Spatial-averaged velocity in y direction for each class of particle.

**multiVzPart**(*=uninitalized*)
    Spatial-averaged velocity in z direction for each class of particle.

**nCell**(*=1*)
    Number of cell in the depth

**nbAverageT**(*=0*)
　　If >0, perform a time-averaging (in addition to the spatial averaging) over nbAverage steps.

**ompThreads**(*=-1*)
　　Number of threads to be used in the engine. If ompThreads<0 (default), the number will be typically OMP_NUM_THREADS or the number N defined by 'yade -jN' (this behavior can depend on the engine though). This attribute will only affect engines whose code includes openMP parallel regions (e.g. *InteractionLoop*). This attribute is mostly useful for experiments or when combining *ParallelEngine* with engines that run parallel regions, resulting in nested OMP loops with different number of threads at each level.

**phiBed**(*=0.08*)
　　Turbulence modelling parameter. Associated with mixing length modelling ilm = 5.

**phiMax**(*=0.64*)
　　Fluid resolution: maximum solid volume fraction.

**phiPart**(*=uninitalized*)
　　Discretized solid volume fraction depth profile. Can be taken as input parameter or evaluated directly inside the engine, calling from python the averageProfile() function

**phiPart1**(*=uninitalized*)
　　Discretized solid volume fraction depth profile of particles of type 1. Evaluated when *twoSize* is set to True.

**phiPart2**(*=uninitalized*)
　　Discretized solid volume fraction depth profile of particles of type 2. Evaluated when *twoSize* is set to True.

**pointParticleAverage**(*=false*)
　　Evaluate the averaged with a point particle method. If False, consider the particle extent and weigth the averaged by the volume contained in each averaging cell.

**radiusPart**(*=0.*)
　　Reference particle radius

**radiusPart1**(*=0.*)
　　Radius of the particles of type 1. Useful only when *twoSize* is set to True.

**radiusPart2**(*=0.*)
　　Radius of the particles of type 2. Useful only when *twoSize* is set to True.

**radiusParts**(*=uninitalized*)
　　Variables containing the number of different radius of particles in the simulation. Allow to perform class averaging by particle size.

**roughnessPartScale**(*=1e-3*)
　　Roughness length scale of the particle. In practice, the lubrication force is cut off when the two particles are at a distance roughnessPartScale.

**steadyFlow**(*=true*)
　　Condition to modify the buoyancy force according to the physical difference between a fluid at rest and a steady fluid flow. For more details see [Maurin2018]

**taufsi**(*=uninitalized*)
　　Fluid Resolution: Create Taufsi/rhof = dragTerm/(rhof(vf-vxp)) to transmit to the fluid code

**timingDeltas**
　　Detailed information about timing inside the Engine itself. Empty unless enabled in the source code and *O.timingEnabled*==`True`.

**turbulentFluctuation**(*(HydroForceEngine)arg1*) → None :
　　Apply a discrete random walk model to the evaluation of the drag force to account for the fluid velocity turbulent fluctuations. Very simple model applying fluctuations from the values of the Reynolds stresses in order to recover the property $< u'_x u'_z > (z) = < R^f_{xz} > (z)/\rho^f$. The random fluctuations are modified over a time scale given by the eddy turn over time.

**turbulentFluctuationZDep**(*(HydroForceEngine)arg1*) → None :
  Apply turbulent fluctuation to the problem similarly to turbulentFluctuation but with an update of the fluctuation depending on the particle position.

**turbulentViscosity**(*=uninitalized*)
  Fluid Resolution: turbulent viscocity as a function of the depth

**twoSize**(*=false*)
  Not maintained anymore. Option to activate when considering two particle size in the simulation. When activated evaluate the average solid volume fraction and drag force for the two type of particles of diameter diameterPart1 and diameterPart2 independently.

**uTop**(*=1.*)
  Fluid resolution: fluid velocity at the top boundary when iusl = 0

**unCorrelatedFluctuations**(*=false*)
  Condition to generate uncorrelated fluid fluctuations. Default case represent in free-surface flows, for which the vertical and streamwise fluid velocity fluctuations are correlated (see e.g. reference book of Nezu & Nakagawa 1992, turbulence in open channel flows).

**updateAttrs**(*(Serializable)arg1, (dict)arg2*) → None :
  Update object attributes from given dictionary

**vCell**(*=uninitalized*)
  Volume of averaging cell

**vFluctX**(*=uninitalized*)
  Vector associating a streamwise fluid velocity fluctuation to each particle. Fluctuation calculated in the C++ code from the discrete random walk model

**vFluctY**(*=uninitalized*)
  Vector associating a spanwise fluid velocity fluctuation to each particle. Fluctuation calculated in the C++ code from the discrete random walk model

**vFluctZ**(*=uninitalized*)
  Vector associating a normal fluid velocity fluctuation to each particle. Fluctuation calculated in the C++ code from the discrete random walk model

**vPart**(*=uninitalized*)
  Discretized streamwise solid velocity depth profile, in x, y and z direction. Only the x direction measurement is taken into account in the 1D fluid coupling resolution. The two other can be used as output parameters. The x component can be taken as input parameter, or evaluated directly inside the engine, calling from python the averageProfile() function

**velFluct**(*=false*)
  If true, activate the determination of turbulent fluid velocity fluctuation for the next time step only at the position of each particle, using a simple discrete random walk (DRW) model based on the Reynolds stresses profile (*ReynoldStresses*)

**viscoDyn**(*=1e-3*)
  Dynamic viscosity of the fluid, by default - viscosity of water

**viscousSubLayer**(*=0*)
  Fluid resolution: solve the viscous sublayer close to the bottom boundary if set to 1

**vxFluid**(*=uninitalized*)
  Discretized streamwise fluid velocity depth profile at t

**vxPart**(*=uninitalized*)
  Discretized streamwise solid velocity depth profile. Can be taken as input parameter, or evaluated directly inside the engine, calling from python the averageProfile() function

**vxPart1**(*=uninitalized*)
  Discretized solid streamwise velocity depth profile of particles of type 1. Evaluated when *twoSize* is set to True.

**vxPart2**(*=uninitalized*)
>    Discretized solid streamwise velocity depth profile of particles of type 2. Evaluated when *twoSize* is set to True.

**vyPart**(*=uninitalized*)
>    Discretized spanwise solid velocity depth profile. Can be taken as input parameter, or evaluated directly inside the engine, calling from python the averageProfile() function

**vyPart1**(*=uninitalized*)
>    Discretized solid spanwise velocity depth profile of particles of type 1. Evaluated when *twoSize* is set to True.

**vyPart2**(*=uninitalized*)
>    Discretized solid spanwise velocity depth profile of particles of type 2. Evaluated when *twoSize* is set to True.

**vzPart**(*=uninitalized*)
>    Discretized wall-normal solid velocity depth profile. Can be taken as input parameter, or evaluated directly inside the engine, calling from python the averageProfile() function

**vzPart1**(*=uninitalized*)
>    Discretized solid wall-normal velocity depth profile of particles of type 1. Evaluated when *twoSize* is set to True.

**vzPart2**(*=uninitalized*)
>    Discretized solid wall-normal velocity depth profile of particles of type 2. Evaluated when *twoSize* is set to True.

**zRef**(*=0.*)
>    Position of the reference point which correspond to the first value of the fluid velocity, i.e. to the ground.

**class yade.wrapper.InterpolatingDirectedForceEngine**(*inherits* *ForceEngine* → *PartialEngine* → *Engine* → *Serializable*)

Engine for applying force of varying magnitude but constant direction on subscribed bodies. times and magnitudes must have the same length, direction (normalized automatically) gives the orientation.

As usual with interpolating engines: the first magnitude is used before the first time point, last magnitude is used after the last time point. Wrap specifies whether time wraps around the last time point to the first time point.

**dead**(*=false*)
>    If true, this engine will not run at all; can be used for making an engine temporarily deactivated and only resurrect it at a later point.

**dict**(*(Serializable)arg1*) → dict :
>    Return dictionary of attributes.

**direction**(*=Vector3r::UnitX()*)
>    Contact force direction (normalized automatically)

**execCount**
>    Cumulative count this engine was run (only used if *O.timingEnabled*==True).

**execTime**
>    Cumulative time in nanoseconds this Engine took to run (only used if *O.timingEnabled*==True).

**force**(*=Vector3r::Zero()*)
>    Force to apply.

**ids**(*=uninitalized*)
>    *Ids* list of bodies affected by this PartialEngine.

**label**(*=uninitalized*)
> Textual label for this object; must be valid python identifier, you can refer to it directly from python.

**magnitudes**(*=uninitalized*)
> Force magnitudes readings [N]

**ompThreads**(*=-1*)
> Number of threads to be used in the engine. If ompThreads<0 (default), the number will be typically OMP_NUM_THREADS or the number N defined by 'yade -jN' (this behavior can depend on the engine though). This attribute will only affect engines whose code includes openMP parallel regions (e.g. *InteractionLoop*). This attribute is mostly useful for experiments or when combining *ParallelEngine* with engines that run parallel regions, resulting in nested OMP loops with different number of threads at each level.

**times**(*=uninitalized*)
> Time readings [s]

**timingDeltas**
> Detailed information about timing inside the Engine itself. Empty unless enabled in the source code and *O.timingEnabled*==`True`.

**updateAttrs**(*(Serializable)arg1, (dict)arg2*) → None :
> Update object attributes from given dictionary

**wrap**(*=false*)
> wrap to the beginning of the sequence if beyond the last time point

**class yade.wrapper.InterpolatingHelixEngine**(*inherits HelixEngine → RotationEngine → KinematicEngine → PartialEngine → Engine → Serializable*)
Engine applying spiral motion, finding current angular velocity by linearly interpolating in times and velocities and translation by using slope parameter.

The interpolation assumes the margin value before the first time point and last value after the last time point. If wrap is specified, time will wrap around the last times value to the first one (note that no interpolation between last and first values is done).

**angleTurned**(*=0*)
> How much have we turned so far. *(auto-updated)* [rad]

**angularVelocities**(*=uninitalized*)
> List of angular velocities; manadatorily of same length as times. [rad/s]

**angularVelocity**(*=0*)
> Angular velocity. [rad/s]

**dead**(*=false*)
> If true, this engine will not run at all; can be used for making an engine temporarily deactivated and only resurrect it at a later point.

**dict**(*(Serializable)arg1*) → dict :
> Return dictionary of attributes.

**execCount**
> Cumulative count this engine was run (only used if *O.timingEnabled*==`True`).

**execTime**
> Cumulative time in nanoseconds this Engine took to run (only used if *O.timingEnabled*==`True`).

**ids**(*=uninitalized*)
> *Ids* list of bodies affected by this PartialEngine.

**label**(*=uninitalized*)
> Textual label for this object; must be valid python identifier, you can refer to it directly from python.

**linearVelocity**(*=0*)
> Linear velocity [m/s]

**ompThreads**(*=-1*)
> Number of threads to be used in the engine. If ompThreads<0 (default), the number will be typically OMP_NUM_THREADS or the number N defined by 'yade -jN' (this behavior can depend on the engine though). This attribute will only affect engines whose code includes openMP parallel regions (e.g. *InteractionLoop*). This attribute is mostly useful for experiments or when combining *ParallelEngine* with engines that run parallel regions, resulting in nested OMP loops with different number of threads at each level.

**rotateAroundZero**(*=false*)
> If True, bodies will not rotate around their centroids, but rather around **zeroPoint**.

**rotationAxis**(*=Vector3r::UnitX()*)
> Axis of rotation (direction); will be normalized automatically.

**slope**(*=0*)
> Axial translation per radian turn (can be negative) [m/rad]

**times**(*=uninitalized*)
> List of time points at which velocities are given; must be increasing [s]

**timingDeltas**
> Detailed information about timing inside the Engine itself. Empty unless enabled in the source code and *O.timingEnabled*==True.

**updateAttrs**(*(Serializable)arg1, (dict)arg2*) → None :
> Update object attributes from given dictionary

**wrap**(*=false*)
> Wrap t if t>times_n, i.e. t_wrapped=t-N*(times_n-times_0)

**zeroPoint**(*=Vector3r::Zero()*)
> Point around which bodies will rotate if **rotateAroundZero** is True

**class yade.wrapper.KinematicEngine**(*inherits PartialEngine → Engine → Serializable*)
Abstract engine for applying prescribed displacement.

---

**Note:** Derived classes should override the **apply** with given list of **ids** (not **action** with *PartialEngine.ids*), so that they work when combined together; *velocity* and *angular velocity* of all subscribed bodies is reset before the **apply** method is called, it should therefore only increment those quantities.

---

**dead**(*=false*)
> If true, this engine will not run at all; can be used for making an engine temporarily deactivated and only resurrect it at a later point.

**dict**(*(Serializable)arg1*) → dict :
> Return dictionary of attributes.

**execCount**
> Cumulative count this engine was run (only used if *O.timingEnabled*==True).

**execTime**
> Cumulative time in nanoseconds this Engine took to run (only used if *O.timingEnabled*==True).

**ids**(*=uninitalized*)
> *Ids* list of bodies affected by this PartialEngine.

**label**(*=uninitalized*)
> Textual label for this object; must be valid python identifier, you can refer to it directly from python.

**ompThreads**(*=-1*)

> Number of threads to be used in the engine. If ompThreads<0 (default), the number will be typically OMP_NUM_THREADS or the number N defined by 'yade -jN' (this behavior can depend on the engine though). This attribute will only affect engines whose code includes openMP parallel regions (e.g. *InteractionLoop*). This attribute is mostly useful for experiments or when combining *ParallelEngine* with engines that run parallel regions, resulting in nested OMP loops with different number of threads at each level.

**timingDeltas**

> Detailed information about timing inside the Engine itself. Empty unless enabled in the source code and *O.timingEnabled*==**True**.

**updateAttrs**(*(Serializable)arg1, (dict)arg2*) → None :

> Update object attributes from given dictionary

**class yade.wrapper.LawTester**(*inherits PartialEngine → Engine → Serializable*)

> Prescribe and apply deformations of an interaction in terms of local mutual displacements and rotations. The loading path is specified either using *path* (as sequence of 6-vectors containing generalized displacements $u_x$, $u_y$, $u_z$, $\varphi_x$, $\varphi_y$, $\varphi_z$) or *disPath* ($u_x$, $u_y$, $u_z$) and *rotPath* ($\varphi_x$, $\varphi_y$, $\varphi_z$). Time function with time values (step numbers) corresponding to points on loading path is given by *pathSteps*. Loading values are linearly interpolated between given loading path points, and starting zero-value (the initial configuration) is assumed for both *path* and *pathSteps*. *hooks* can specify python code to run when respective point on the path is reached; when the path is finished, *doneHook* will be run.

> LawTester should be placed between *InteractionLoop* and *NewtonIntegrator* in the simulation loop, since it controls motion via setting linear/angular velocities on particles; those velocities are integrated by *NewtonIntegrator* to yield an actual position change, which in turn causes *IGeom* to be updated (and *contact law* applied) when *InteractionLoop* is executed. Constitutive law generating forces on particles will not affect prescribed particle motion, since both particles have all *DoFs blocked* when first used with LawTester.

> LawTester uses, as much as possible, *IGeom* to provide useful data (such as local coordinate system), but is able to compute those independently if absent in the respective *IGeom*:

| *IGeom* | #DoFs | LawTester support level |
|---------|-------|-------------------------|
| *L3Geom* | 3 | full |
| *L6Geom* | 6 | full |
| *ScGeom* | 3 | emulate local coordinate system |
| *ScGeom6D* | 6 | emulate local coordinate system |

> Depending on *IGeom*, 3 ($u_x$, $u_y$, $u_z$) or 6 ($u_x$, $u_y$, $u_z$, $\varphi_x$, $\varphi_y$, $\varphi_z$) degrees of freedom (DoFs) are controlled with LawTester, by prescribing linear and angular velocities of both particles in contact. All DoFs controlled with LawTester are orthogonal (fully decoupled) and are controlled independently.

> When 3 DoFs are controlled, *rotWeight* controls whether local shear is applied by moving particle on arc around the other one, or by rotating without changing position; although such rotation induces mutual rotation on the interaction, it is ignored with *IGeom* with only 3 DoFs. When 6 DoFs are controlled, only arc-displacement is applied for shear, since otherwise mutual rotation would occur.

> *idWeight* distributes prescribed motion between both particles (resulting local deformation is the same if **id1** is moved towards **id2** or **id2** towards **id1**). This is true only for $u_x$, $u_y$, $u_z$, $\varphi_x$ however ; bending rotations $\varphi_y$, $\varphi_z$ are nevertheless always distributed regardless of **idWeight** to both spheres in inverse proportion to their radii, so that there is no shear induced.

> LawTester knows current contact deformation from 2 sources: from its own internal data (which are used for prescribing the displacement at every step), which can be accessed in *uTest*, and from *IGeom* itself (depending on which data it provides), which is stored in *uGeom*. These two values should be identical (disregarding numerical precision), and it is a way to test whether *IGeom* and related functors compute what they are supposed to compute.

LawTester-operated interactions can be rendered with *GlExtra_LawTester* renderer.

See scripts/test/law-test.py for an example.

**dead**(*=false*)
    If true, this engine will not run at all; can be used for making an engine temporarily deactivated and only resurrect it at a later point.

**dict**(*(Serializable)arg1*) → dict :
    Return dictionary of attributes.

**disPath**(*=uninitalized*)
    Loading path, where each Vector3 contains desired normal displacement and two components of the shear displacement (in local coordinate system, which is being tracked automatically. If shorter than *rotPath*, the last value is repeated.

**displIsRel**(*=true*)
    Whether displacement values in *disPath* are normalized by reference contact length (r1+r2 for 2 spheres).

**doneHook**(*=uninitalized*)
    Python command (as string) to run when end of the path is achieved. If empty, the engine will be set *dead*.

**execCount**
    Cumulative count this engine was run (only used if *O.timingEnabled*==True).

**execTime**
    Cumulative time in nanoseconds this Engine took to run (only used if *O.timingEnabled*==True).

**hooks**(*=uninitalized*)
    Python commands to be run when the corresponding point in path is reached, before doing other things in that particular step. See also *doneHook*.

**idWeight**(*=1*)
    Float, usually ⟨0,1⟩, determining on how are displacements distributed between particles (0 for id1, 1 for id2); intermediate values will apply respective part to each of them. This parameter is ignored with 6-DoFs *IGeom*.

**ids**(*=uninitalized*)
    *Ids* list of bodies affected by this PartialEngine.

**label**(*=uninitalized*)
    Textual label for this object; must be valid python identifier, you can refer to it directly from python.

**ompThreads**(*=-1*)
    Number of threads to be used in the engine. If ompThreads<0 (default), the number will be typically OMP_NUM_THREADS or the number N defined by 'yade -jN' (this behavior can depend on the engine though). This attribute will only affect engines whose code includes openMP parallel regions (e.g. *InteractionLoop*). This attribute is mostly useful for experiments or when combining *ParallelEngine* with engines that run parallel regions, resulting in nested OMP loops with different number of threads at each level.

**pathSteps**(*=vector<int>(1, 1), (constant step)*)
    Step number for corresponding values in *path*; if shorter than path, distance between last 2 values is used for the rest.

**refLength**(*=0*)
    Reference contact length, for rendering only.

**renderLength**(*=0*)
    Characteristic length for the purposes of rendering, set equal to the smaller radius.

**rotPath**(*=uninitalized*)
> Rotational components of the loading path, where each item contains torsion and two bending rotations in local coordinates. If shorter than *path*, the last value is repeated.

**rotWeight**(*=1*)
> Float ⟨0,1⟩ determining whether shear displacement is applied as rotation or displacement on arc (0 is displacement-only, 1 is rotation-only). Not effective when mutual rotation is specified.

**step**(*=1*)
> Step number in which this engine is active; determines position in path, using pathSteps.

**timingDeltas**
> Detailed information about timing inside the Engine itself. Empty unless enabled in the source code and *O.timingEnabled*==True.

**trsf**(*=uninitalized*)
> Transformation matrix for the local coordinate system. *(auto-updated)*

**uGeom**(*=Vector6r::Zero()*)
> Current generalized displacements (3 displacements, 3 rotations), as stored in the interation itself. They should corredpond to *uTest*, otherwise a bug is indicated.

**uTest**(*=Vector6r::Zero()*)
> Current generalized displacements (3 displacements, 3 rotations), as they should be according to this *LawTester*. Should correspond to *uGeom*.

**updateAttrs**(*(Serializable)arg1, (dict)arg2*) → None :
> Update object attributes from given dictionary

**uuPrev**(*=Vector6r::Zero()*)
> Generalized displacement values reached in the previous step, for knowing which increment to apply in the current step.

**class yade.wrapper.LinearDragEngine**(*inherits PartialEngine → Engine → Serializable*)
> Apply viscous resistance or linear drag on some particles at each step, decelerating them proportionally to their linear velocities. The applied force reads

$$\mathbf{F_d} = -b\boldsymbol{\nu}$$

where $b$ is the linear drag, $\boldsymbol{\nu}$ is particle's velocity.

$$b = 6\pi\nu r$$

where $\nu$ is the medium viscosity, $r$ is the Stokes radius of the particle (but in this case we accept it equal to sphere radius for simplification),

---

**Note:** linear drag is only applied to spherical particles, listed in ids.

---

**dead**(*=false*)
> If true, this engine will not run at all; can be used for making an engine temporarily deactivated and only resurrect it at a later point.

**dict**(*(Serializable)arg1*) → dict :
> Return dictionary of attributes.

**execCount**
> Cumulative count this engine was run (only used if *O.timingEnabled*==True).

**execTime**
> Cumulative time in nanoseconds this Engine took to run (only used if *O.timingEnabled*==True).

**ids**(*=uninitalized*)
> *Ids* list of bodies affected by this PartialEngine.

**label**(*=uninitalized*)

> Textual label for this object; must be valid python identifier, you can refer to it directly from python.

**nu**(*=0.001*)

> Viscosity of the medium.

**ompThreads**(*=-1*)

> Number of threads to be used in the engine. If ompThreads<0 (default), the number will be typically OMP_NUM_THREADS or the number N defined by 'yade -jN' (this behavior can depend on the engine though). This attribute will only affect engines whose code includes openMP parallel regions (e.g. *InteractionLoop*). This attribute is mostly useful for experiments or when combining *ParallelEngine* with engines that run parallel regions, resulting in nested OMP loops with different number of threads at each level.

**timingDeltas**

> Detailed information about timing inside the Engine itself. Empty unless enabled in the source code and *O.timingEnabled*==**True**.

**updateAttrs**(*(Serializable)arg1, (dict)arg2*) → None :

> Update object attributes from given dictionary

**class yade.wrapper.RadialForceEngine**(*inherits PartialEngine → Engine → Serializable*)

> Apply force of given magnitude directed away from spatial axis.

**axisDir**(*=Vector3r::UnitX()*)

> Axis direction (normalized automatically)

**axisPt**(*=Vector3r::Zero()*)

> Point on axis

**dead**(*=false*)

> If true, this engine will not run at all; can be used for making an engine temporarily deactivated and only resurrect it at a later point.

**dict**(*(Serializable)arg1*) → dict :

> Return dictionary of attributes.

**execCount**

> Cumulative count this engine was run (only used if *O.timingEnabled*==**True**).

**execTime**

> Cumulative time in nanoseconds this Engine took to run (only used if *O.timingEnabled*==**True**).

**fNorm**(*=0*)

> Applied force magnitude

**ids**(*=uninitalized*)

> *Ids* list of bodies affected by this PartialEngine.

**label**(*=uninitalized*)

> Textual label for this object; must be valid python identifier, you can refer to it directly from python.

**ompThreads**(*=-1*)

> Number of threads to be used in the engine. If ompThreads<0 (default), the number will be typically OMP_NUM_THREADS or the number N defined by 'yade -jN' (this behavior can depend on the engine though). This attribute will only affect engines whose code includes openMP parallel regions (e.g. *InteractionLoop*). This attribute is mostly useful for experiments or when combining *ParallelEngine* with engines that run parallel regions, resulting in nested OMP loops with different number of threads at each level.

**timingDeltas**

> Detailed information about timing inside the Engine itself. Empty unless enabled in the source code and *O.timingEnabled*==**True**.

**updateAttrs**(*(Serializable)arg1, (dict)arg2*) → None :
   Update object attributes from given dictionary

**class yade.wrapper.RotationEngine**(*inherits KinematicEngine → PartialEngine → Engine → Serializable*)
   Engine applying rotation (by setting angular velocity) to subscribed bodies. If *rotateAroundZero* is set, then each body is also displaced around *zeroPoint*.

**angularVelocity**(*=0*)
   Angular velocity. [rad/s]

**dead**(*=false*)
   If true, this engine will not run at all; can be used for making an engine temporarily deactivated and only resurrect it at a later point.

**dict**(*(Serializable)arg1*) → dict :
   Return dictionary of attributes.

**execCount**
   Cumulative count this engine was run (only used if *O.timingEnabled*==True).

**execTime**
   Cumulative time in nanoseconds this Engine took to run (only used if *O.timingEnabled*==True).

**ids**(*=uninitalized*)
   *Ids* list of bodies affected by this PartialEngine.

**label**(*=uninitalized*)
   Textual label for this object; must be valid python identifier, you can refer to it directly from python.

**ompThreads**(*=-1*)
   Number of threads to be used in the engine. If ompThreads<0 (default), the number will be typically OMP_NUM_THREADS or the number N defined by 'yade -jN' (this behavior can depend on the engine though). This attribute will only affect engines whose code includes openMP parallel regions (e.g. *InteractionLoop*). This attribute is mostly useful for experiments or when combining *ParallelEngine* with engines that run parallel regions, resulting in nested OMP loops with different number of threads at each level.

**rotateAroundZero**(*=false*)
   If True, bodies will not rotate around their centroids, but rather around **zeroPoint**.

**rotationAxis**(*=Vector3r::UnitX()*)
   Axis of rotation (direction); will be normalized automatically.

**timingDeltas**
   Detailed information about timing inside the Engine itself. Empty unless enabled in the source code and *O.timingEnabled*==True.

**updateAttrs**(*(Serializable)arg1, (dict)arg2*) → None :
   Update object attributes from given dictionary

**zeroPoint**(*=Vector3r::Zero()*)
   Point around which bodies will rotate if **rotateAroundZero** is True

**class yade.wrapper.ServoPIDController**(*inherits TranslationEngine → KinematicEngine → PartialEngine → Engine → Serializable*)
   PIDController servo-engine for applying prescribed force on bodies. http://en.wikipedia.org/wiki/PID_controller

**axis**(*=Vector3r::Zero()*)
   Unit vector along which apply the velocity [-]

**curVel**(*=0.0*)
   Current applied velocity [m/s]

**current**(*=Vector3r::Zero()*)
    Current value for the controller [N]

**dead**(*=false*)
    If true, this engine will not run at all; can be used for making an engine temporarily deactivated and only resurrect it at a later point.

**dict**(*(Serializable)arg1*) → dict :
    Return dictionary of attributes.

**errorCur**(*=0.0*)
    Current error [N]

**errorPrev**(*=0.0*)
    Previous error [N]

**execCount**
    Cumulative count this engine was run (only used if *O.timingEnabled*==**True**).

**execTime**
    Cumulative time in nanoseconds this Engine took to run (only used if *O.timingEnabled*==**True**).

**iTerm**(*=0.0*)
    Integral term [N]

**ids**(*=uninitalized*)
    *Ids* list of bodies affected by this PartialEngine.

**iterPeriod**(*=100.0*)
    Periodicity criterion of velocity correlation [-]

**iterPrevStart**(*=-1.0*)
    Previous iteration of velocity correlation [-]

**kD**(*=0.0*)
    Derivative gain/coefficient for the PID-controller [-]

**kI**(*=0.0*)
    Integral gain/coefficient for the PID-controller [-]

**kP**(*=0.0*)
    Proportional gain/coefficient for the PID-controller [-]

**label**(*=uninitalized*)
    Textual label for this object; must be valid python identifier, you can refer to it directly from python.

**maxVelocity**(*=0.0*)
    Velocity [m/s]

**ompThreads**(*=-1*)
    Number of threads to be used in the engine. If ompThreads<0 (default), the number will be typically OMP_NUM_THREADS or the number N defined by 'yade -jN' (this behavior can depend on the engine though). This attribute will only affect engines whose code includes openMP parallel regions (e.g. *InteractionLoop*). This attribute is mostly useful for experiments or when combining *ParallelEngine* with engines that run parallel regions, resulting in nested OMP loops with different number of threads at each level.

**target**(*=0.0*)
    Target value for the controller [N]

**timingDeltas**
    Detailed information about timing inside the Engine itself. Empty unless enabled in the source code and *O.timingEnabled*==**True**.

**translationAxis**(*=uninitalized*)
    Direction of imposed translation [Vector3]

> **updateAttrs**(*(Serializable)arg1, (dict)arg2*) → None :
>> Update object attributes from given dictionary

> **velocity**(*=uninitalized*)
>> Scalar value of the imposed velocity [m/s]. Imposed vector velocity is *velocity * axis*

**class yade.wrapper.StepDisplacer**(*inherits PartialEngine → Engine → Serializable*)

Apply generalized displacement (displacement or rotation) stepwise on subscribed bodies. Could be used for purposes of contact law tests (by moving one sphere compared to another), but in this case, see rather *LawTester*

> **dead**(*=false*)
>> If true, this engine will not run at all; can be used for making an engine temporarily deactivated and only resurrect it at a later point.

> **dict**(*(Serializable)arg1*) → dict :
>> Return dictionary of attributes.

> **execCount**
>> Cumulative count this engine was run (only used if *O.timingEnabled*==True).

> **execTime**
>> Cumulative time in nanoseconds this Engine took to run (only used if *O.timingEnabled*==True).

> **ids**(*=uninitalized*)
>> *Ids* list of bodies affected by this PartialEngine.

> **label**(*=uninitalized*)
>> Textual label for this object; must be valid python identifier, you can refer to it directly from python.

> **mov**(*=Vector3r::Zero()*)
>> Linear displacement step to be applied per iteration, by addition to *State.pos*.

> **ompThreads**(*=-1*)
>> Number of threads to be used in the engine. If ompThreads<0 (default), the number will be typically OMP_NUM_THREADS or the number N defined by 'yade -jN' (this behavior can depend on the engine though). This attribute will only affect engines whose code includes openMP parallel regions (e.g. *InteractionLoop*). This attribute is mostly useful for experiments or when combining *ParallelEngine* with engines that run parallel regions, resulting in nested OMP loops with different number of threads at each level.

> **rot**(*=Quaternionr::Identity()*)
>> Rotation step to be applied per iteration (via rotation composition with *State.ori*).

> **setVelocities**(*=false*)
>> If false, positions and orientations are directly updated, without changing the speeds of concerned bodies. If true, only velocity and angularVelocity are modified. In this second case *integrator* is supposed to be used, so that, thanks to this Engine, the bodies will have the prescribed jump over one iteration (dt).

> **timingDeltas**
>> Detailed information about timing inside the Engine itself. Empty unless enabled in the source code and *O.timingEnabled*==True.

> **updateAttrs**(*(Serializable)arg1, (dict)arg2*) → None :
>> Update object attributes from given dictionary

**class yade.wrapper.TorqueEngine**(*inherits PartialEngine → Engine → Serializable*)

Apply given torque (momentum) value at every subscribed particle, at every step.

> **dead**(*=false*)
>> If true, this engine will not run at all; can be used for making an engine temporarily deactivated and only resurrect it at a later point.

**dict**(*(Serializable)arg1*) → dict :
  Return dictionary of attributes.

**execCount**
  Cumulative count this engine was run (only used if *O.timingEnabled*==`True`).

**execTime**
  Cumulative time in nanoseconds this Engine took to run (only used if *O.timingEnabled*==`True`).

**ids**(*=uninitalized*)
  *Ids* list of bodies affected by this PartialEngine.

**label**(*=uninitalized*)
  Textual label for this object; must be valid python identifier, you can refer to it directly from python.

**moment**(*=Vector3r::Zero()*)
  Torque value to be applied.

**ompThreads**(*=-1*)
  Number of threads to be used in the engine. If ompThreads<0 (default), the number will be typically OMP_NUM_THREADS or the number N defined by 'yade -jN' (this behavior can depend on the engine though). This attribute will only affect engines whose code includes openMP parallel regions (e.g. *InteractionLoop*). This attribute is mostly useful for experiments or when combining *ParallelEngine* with engines that run parallel regions, resulting in nested OMP loops with different number of threads at each level.

**timingDeltas**
  Detailed information about timing inside the Engine itself. Empty unless enabled in the source code and *O.timingEnabled*==`True`.

**updateAttrs**(*(Serializable)arg1, (dict)arg2*) → None :
  Update object attributes from given dictionary

**class yade.wrapper.TranslationEngine**(*inherits KinematicEngine → PartialEngine → Engine → Serializable*)
  Engine applying translation motion (by setting linear velocity) to subscribed bodies.

**dead**(*=false*)
  If true, this engine will not run at all; can be used for making an engine temporarily deactivated and only resurrect it at a later point.

**dict**(*(Serializable)arg1*) → dict :
  Return dictionary of attributes.

**execCount**
  Cumulative count this engine was run (only used if *O.timingEnabled*==`True`).

**execTime**
  Cumulative time in nanoseconds this Engine took to run (only used if *O.timingEnabled*==`True`).

**ids**(*=uninitalized*)
  *Ids* list of bodies affected by this PartialEngine.

**label**(*=uninitalized*)
  Textual label for this object; must be valid python identifier, you can refer to it directly from python.

**ompThreads**(*=-1*)
  Number of threads to be used in the engine. If ompThreads<0 (default), the number will be typically OMP_NUM_THREADS or the number N defined by 'yade -jN' (this behavior can depend on the engine though). This attribute will only affect engines whose code includes openMP parallel regions (e.g. *InteractionLoop*). This attribute is mostly useful for experiments or when combining *ParallelEngine* with engines that run parallel regions, resulting in nested OMP loops with different number of threads at each level.

**timingDeltas**
Detailed information about timing inside the Engine itself. Empty unless enabled in the source code and *O.timingEnabled*==`True`.

**translationAxis**(*=uninitalized*)
Direction of imposed translation [Vector3]

**updateAttrs**(*(Serializable)arg1, (dict)arg2*) → None :
Update object attributes from given dictionary

**velocity**(*=uninitalized*)
Scalar value of the imposed velocity [m/s]. Imposed vector velocity is *velocity * axis*

## 2.3.5 Dispatchers

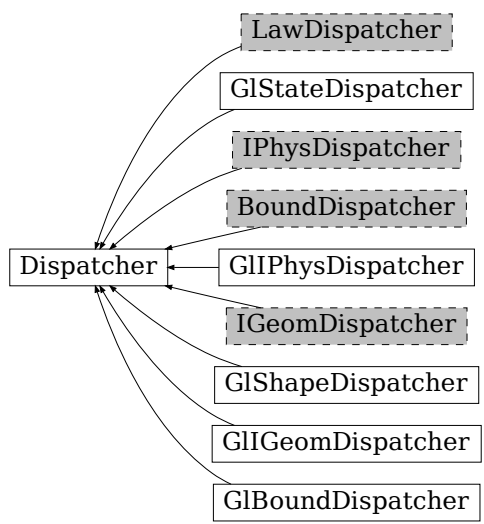

Fig. 31: Inheritance graph of Dispatcher, gray dashed classes are discussed in their own sections: *LawDispatcher*, *IPhysDispatcher*, *BoundDispatcher*, *IGeomDispatcher*. See also: *GlBoundDispatcher*, *GlIGeomDispatcher*, *GlIPhysDispatcher*, *GlShapeDispatcher*, *GlStateDispatcher*.

**class yade.wrapper.Dispatcher**(*inherits Engine → Serializable*)
Engine dispatching control to its associated functors, based on types of argument it receives. This abstract base class provides no functionality in itself.

**dead**(*=false*)
If true, this engine will not run at all; can be used for making an engine temporarily deactivated and only resurrect it at a later point.

**dict**(*(Serializable)arg1*) → dict :
Return dictionary of attributes.

**execCount**
Cumulative count this engine was run (only used if *O.timingEnabled*==`True`).

**execTime**
Cumulative time in nanoseconds this Engine took to run (only used if *O.timingEnabled*==`True`).

**label**(*=uninitalized*)
Textual label for this object; must be valid python identifier, you can refer to it directly from python.

**ompThreads**(*=-1*)
> Number of threads to be used in the engine. If ompThreads<0 (default), the number will be typically OMP_NUM_THREADS or the number N defined by 'yade -jN' (this behavior can depend on the engine though). This attribute will only affect engines whose code includes openMP parallel regions (e.g. *InteractionLoop*). This attribute is mostly useful for experiments or when combining *ParallelEngine* with engines that run parallel regions, resulting in nested OMP loops with different number of threads at each level.

**timingDeltas**
> Detailed information about timing inside the Engine itself. Empty unless enabled in the source code and *O.timingEnabled*==`True`.

**updateAttrs**(*(Serializable)arg1, (dict)arg2*) → None :
> Update object attributes from given dictionary

**class yade.wrapper.GlBoundDispatcher**(*inherits Dispatcher → Engine → Serializable*)
> Dispatcher calling *functors* based on received argument type(s).

**dead**(*=false*)
> If true, this engine will not run at all; can be used for making an engine temporarily deactivated and only resurrect it at a later point.

**dict**(*(Serializable)arg1*) → dict :
> Return dictionary of attributes.

**dispFunctor**(*(GlBoundDispatcher)arg1, (Bound)arg2*) → GlBoundFunctor :
> Return functor that would be dispatched for given argument(s); None if no dispatch; ambiguous dispatch throws.

**dispMatrix**(*(GlBoundDispatcher)arg1*[, *(bool)names=True*]) → dict :
> Return dictionary with contents of the dispatch matrix.

**execCount**
> Cumulative count this engine was run (only used if *O.timingEnabled*==`True`).

**execTime**
> Cumulative time in nanoseconds this Engine took to run (only used if *O.timingEnabled*==`True`).

**functors**
> Functors associated with this dispatcher.

**label**(*=uninitalized*)
> Textual label for this object; must be valid python identifier, you can refer to it directly from python.

**ompThreads**(*=-1*)
> Number of threads to be used in the engine. If ompThreads<0 (default), the number will be typically OMP_NUM_THREADS or the number N defined by 'yade -jN' (this behavior can depend on the engine though). This attribute will only affect engines whose code includes openMP parallel regions (e.g. *InteractionLoop*). This attribute is mostly useful for experiments or when combining *ParallelEngine* with engines that run parallel regions, resulting in nested OMP loops with different number of threads at each level.

**timingDeltas**
> Detailed information about timing inside the Engine itself. Empty unless enabled in the source code and *O.timingEnabled*==`True`.

**updateAttrs**(*(Serializable)arg1, (dict)arg2*) → None :
> Update object attributes from given dictionary

**class yade.wrapper.GlIGeomDispatcher**(*inherits Dispatcher → Engine → Serializable*)
> Dispatcher calling *functors* based on received argument type(s).

**dead**(*=false*)
　　If true, this engine will not run at all; can be used for making an engine temporarily deactivated and only resurrect it at a later point.

**dict**(*(Serializable)arg1*) → dict :
　　Return dictionary of attributes.

**dispFunctor**(*(GlIGeomDispatcher)arg1, (IGeom)arg2*) → GlIGeomFunctor :
　　Return functor that would be dispatched for given argument(s); None if no dispatch; ambiguous dispatch throws.

**dispMatrix**(*(GlIGeomDispatcher)arg1*[, *(bool)names=True*]) → dict :
　　Return dictionary with contents of the dispatch matrix.

**execCount**
　　Cumulative count this engine was run (only used if *O.timingEnabled*==True).

**execTime**
　　Cumulative time in nanoseconds this Engine took to run (only used if *O.timingEnabled*==True).

**functors**
　　Functors associated with this dispatcher.

**label**(*=uninitalized*)
　　Textual label for this object; must be valid python identifier, you can refer to it directly from python.

**ompThreads**(*=-1*)
　　Number of threads to be used in the engine. If ompThreads<0 (default), the number will be typically OMP_NUM_THREADS or the number N defined by 'yade -jN' (this behavior can depend on the engine though). This attribute will only affect engines whose code includes openMP parallel regions (e.g. *InteractionLoop*). This attribute is mostly useful for experiments or when combining *ParallelEngine* with engines that run parallel regions, resulting in nested OMP loops with different number of threads at each level.

**timingDeltas**
　　Detailed information about timing inside the Engine itself. Empty unless enabled in the source code and *O.timingEnabled*==True.

**updateAttrs**(*(Serializable)arg1, (dict)arg2*) → None :
　　Update object attributes from given dictionary

**class yade.wrapper.GlIPhysDispatcher**(*inherits Dispatcher → Engine → Serializable*)
　　Dispatcher calling *functors* based on received argument type(s).

**dead**(*=false*)
　　If true, this engine will not run at all; can be used for making an engine temporarily deactivated and only resurrect it at a later point.

**dict**(*(Serializable)arg1*) → dict :
　　Return dictionary of attributes.

**dispFunctor**(*(GlIPhysDispatcher)arg1, (IPhys)arg2*) → GlIPhysFunctor :
　　Return functor that would be dispatched for given argument(s); None if no dispatch; ambiguous dispatch throws.

**dispMatrix**(*(GlIPhysDispatcher)arg1*[, *(bool)names=True*]) → dict :
　　Return dictionary with contents of the dispatch matrix.

**execCount**
　　Cumulative count this engine was run (only used if *O.timingEnabled*==True).

**execTime**
　　Cumulative time in nanoseconds this Engine took to run (only used if *O.timingEnabled*==True).

**functors**
Functors associated with this dispatcher.

**label**(*=uninitalized*)
Textual label for this object; must be valid python identifier, you can refer to it directly from python.

**ompThreads**(*=-1*)
Number of threads to be used in the engine. If ompThreads<0 (default), the number will be typically OMP_NUM_THREADS or the number N defined by 'yade -jN' (this behavior can depend on the engine though). This attribute will only affect engines whose code includes openMP parallel regions (e.g. *InteractionLoop*). This attribute is mostly useful for experiments or when combining *ParallelEngine* with engines that run parallel regions, resulting in nested OMP loops with different number of threads at each level.

**timingDeltas**
Detailed information about timing inside the Engine itself. Empty unless enabled in the source code and *O.timingEnabled*==True.

**updateAttrs**(*(Serializable)arg1, (dict)arg2*) → None :
Update object attributes from given dictionary

**class yade.wrapper.GlShapeDispatcher**(*inherits Dispatcher → Engine → Serializable*)
Dispatcher calling *functors* based on received argument type(s).

**dead**(*=false*)
If true, this engine will not run at all; can be used for making an engine temporarily deactivated and only resurrect it at a later point.

**dict**(*(Serializable)arg1*) → dict :
Return dictionary of attributes.

**dispFunctor**(*(GlShapeDispatcher)arg1, (Shape)arg2*) → GlShapeFunctor :
Return functor that would be dispatched for given argument(s); None if no dispatch; ambiguous dispatch throws.

**dispMatrix**(*(GlShapeDispatcher)arg1*[, *(bool)names=True*]) → dict :
Return dictionary with contents of the dispatch matrix.

**execCount**
Cumulative count this engine was run (only used if *O.timingEnabled*==True).

**execTime**
Cumulative time in nanoseconds this Engine took to run (only used if *O.timingEnabled*==True).

**functors**
Functors associated with this dispatcher.

**label**(*=uninitalized*)
Textual label for this object; must be valid python identifier, you can refer to it directly from python.

**ompThreads**(*=-1*)
Number of threads to be used in the engine. If ompThreads<0 (default), the number will be typically OMP_NUM_THREADS or the number N defined by 'yade -jN' (this behavior can depend on the engine though). This attribute will only affect engines whose code includes openMP parallel regions (e.g. *InteractionLoop*). This attribute is mostly useful for experiments or when combining *ParallelEngine* with engines that run parallel regions, resulting in nested OMP loops with different number of threads at each level.

**timingDeltas**
Detailed information about timing inside the Engine itself. Empty unless enabled in the source code and *O.timingEnabled*==True.

**updateAttrs**(*(Serializable)arg1, (dict)arg2*) → None :
    Update object attributes from given dictionary

**class yade.wrapper.GlStateDispatcher**(*inherits Dispatcher → Engine → Serializable*)
    Dispatcher calling *functors* based on received argument type(s).

    **dead**(*=false*)
        If true, this engine will not run at all; can be used for making an engine temporarily deactivated and only resurrect it at a later point.

    **dict**(*(Serializable)arg1*) → dict :
        Return dictionary of attributes.

    **dispFunctor**(*(GlStateDispatcher)arg1, (State)arg2*) → GlStateFunctor :
        Return functor that would be dispatched for given argument(s); None if no dispatch; ambiguous dispatch throws.

    **dispMatrix**(*(GlStateDispatcher)arg1*[, *(bool)names=True*]) → dict :
        Return dictionary with contents of the dispatch matrix.

    **execCount**
        Cumulative count this engine was run (only used if *O.timingEnabled*==True).

    **execTime**
        Cumulative time in nanoseconds this Engine took to run (only used if *O.timingEnabled*==True).

    **functors**
        Functors associated with this dispatcher.

    **label**(*=uninitalized*)
        Textual label for this object; must be valid python identifier, you can refer to it directly from python.

    **ompThreads**(*=-1*)
        Number of threads to be used in the engine. If ompThreads<0 (default), the number will be typically OMP_NUM_THREADS or the number N defined by 'yade -jN' (this behavior can depend on the engine though). This attribute will only affect engines whose code includes openMP parallel regions (e.g. *InteractionLoop*). This attribute is mostly useful for experiments or when combining *ParallelEngine* with engines that run parallel regions, resulting in nested OMP loops with different number of threads at each level.

    **timingDeltas**
        Detailed information about timing inside the Engine itself. Empty unless enabled in the source code and *O.timingEnabled*==True.

    **updateAttrs**(*(Serializable)arg1, (dict)arg2*) → None :
        Update object attributes from given dictionary

## 2.3.6 Functors

**class yade.wrapper.Functor**(*inherits Serializable*)
    Function-like object that is called by Dispatcher, if types of arguments match those the Functor declares to accept.

    **bases**
        Ordered list of types (as strings) this functor accepts.

    **dict**(*(Serializable)arg1*) → dict :
        Return dictionary of attributes.

    **label**(*=uninitalized*)
        Textual label for this object; must be a valid python identifier, you can refer to it directly from python.

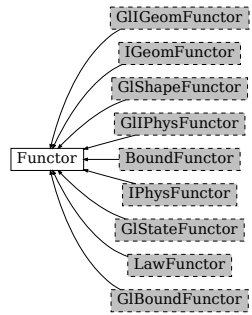

Fig. 32: Inheritance graph of Functor, gray dashed classes are discussed in their own sections: *GlIGeom-Functor*, *IGeomFunctor*, *GlShapeFunctor*, *GlIPhysFunctor*, *BoundFunctor*, *IPhysFunctor*, *GlStateFunc-tor*, *LawFunctor*, *GlBoundFunctor*.

**timingDeltas**
Detailed information about timing inside the Dispatcher itself. Empty unless enabled in the source code and O.timingEnabled==True.

**updateAttrs**(*(Serializable)arg1, (dict)arg2*) → None :
Update object attributes from given dictionary

### 2.3.7 Bounding volume creation

**BoundFunctor**

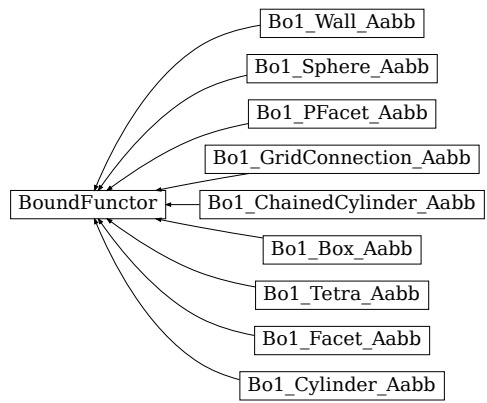

Fig. 33: Inheritance graph of BoundFunctor. See also: *Bo1_Box_Aabb*, *Bo1_ChainedCylinder_-Aabb*, *Bo1_Cylinder_Aabb*, *Bo1_Facet_Aabb*, *Bo1_GridConnection_Aabb*, *Bo1_PFacet_Aabb*, *Bo1_-Sphere_Aabb*, *Bo1_Tetra_Aabb*, *Bo1_Wall_Aabb*.

**class yade.wrapper.BoundFunctor**(*inherits Functor → Serializable*)
Functor for creating/updating *Body::bound*.

**bases**
Ordered list of types (as strings) this functor accepts.

**dict**(*(Serializable)arg1*) → dict :
Return dictionary of attributes.

**label**(*=uninitalized*)
Textual label for this object; must be a valid python identifier, you can refer to it directly from python.

**timingDeltas**
> Detailed information about timing inside the Dispatcher itself. Empty unless enabled in the source code and O.timingEnabled==True.

**updateAttrs**(*(Serializable)arg1*, *(dict)arg2*) → None :
> Update object attributes from given dictionary

**class yade.wrapper.Bo1_Box_Aabb**(*inherits BoundFunctor → Functor → Serializable*)
> Create/update an *Aabb* of a *Box*.

**bases**
> Ordered list of types (as strings) this functor accepts.

**dict**(*(Serializable)arg1*) → dict :
> Return dictionary of attributes.

**label**(*=uninitalized*)
> Textual label for this object; must be a valid python identifier, you can refer to it directly from python.

**timingDeltas**
> Detailed information about timing inside the Dispatcher itself. Empty unless enabled in the source code and O.timingEnabled==True.

**updateAttrs**(*(Serializable)arg1*, *(dict)arg2*) → None :
> Update object attributes from given dictionary

**class yade.wrapper.Bo1_ChainedCylinder_Aabb**(*inherits BoundFunctor → Functor → Serializable*)
> Functor creating *Aabb* from *ChainedCylinder*.

**aabbEnlargeFactor**
> Relative enlargement of the bounding box; deactivated if negative.

> ---
> **Note:** This attribute is used to create distant interaction, but is only meaningful with an *IGeomFunctor* which will not simply discard such interactions: *Ig2_Cylinder_Cylinder_ScGeom::interactionDetectionFactor* should have the same value as *aabbEnlargeFactor*.
> ---

**bases**
> Ordered list of types (as strings) this functor accepts.

**dict**(*(Serializable)arg1*) → dict :
> Return dictionary of attributes.

**label**(*=uninitalized*)
> Textual label for this object; must be a valid python identifier, you can refer to it directly from python.

**timingDeltas**
> Detailed information about timing inside the Dispatcher itself. Empty unless enabled in the source code and O.timingEnabled==True.

**updateAttrs**(*(Serializable)arg1*, *(dict)arg2*) → None :
> Update object attributes from given dictionary

**class yade.wrapper.Bo1_Cylinder_Aabb**(*inherits BoundFunctor → Functor → Serializable*)
> Functor creating *Aabb* from *Cylinder*.

**aabbEnlargeFactor**
> Relative enlargement of the bounding box; deactivated if negative.

> ---
> **Note:** This attribute is used to create distant interaction, but is only meaningful with an *IGeomFunctor* which will not simply discard such interactions: *Ig2__Cylinder__Cylinder__-ScGeom::interactionDetectionFactor* should have the same value as *aabbEnlargeFactor*.

**bases**
Ordered list of types (as strings) this functor accepts.

**dict**(*(Serializable)arg1*) → dict :
Return dictionary of attributes.

**label**(*=uninitalized*)
Textual label for this object; must be a valid python identifier, you can refer to it directly from python.

**timingDeltas**
Detailed information about timing inside the Dispatcher itself. Empty unless enabled in the source code and O.timingEnabled==True.

**updateAttrs**(*(Serializable)arg1, (dict)arg2*) → None :
Update object attributes from given dictionary

**class yade.wrapper.Bo1_Facet_Aabb**(*inherits BoundFunctor → Functor → Serializable*)
Creates/updates an *Aabb* of a *Facet*.

**bases**
Ordered list of types (as strings) this functor accepts.

**dict**(*(Serializable)arg1*) → dict :
Return dictionary of attributes.

**label**(*=uninitalized*)
Textual label for this object; must be a valid python identifier, you can refer to it directly from python.

**timingDeltas**
Detailed information about timing inside the Dispatcher itself. Empty unless enabled in the source code and O.timingEnabled==True.

**updateAttrs**(*(Serializable)arg1, (dict)arg2*) → None :
Update object attributes from given dictionary

**class yade.wrapper.Bo1_GridConnection_Aabb**(*inherits BoundFunctor → Functor → Serializable*)
Functor creating *Aabb* from a *GridConnection*.

**aabbEnlargeFactor**(*=-1, deactivated*)
Relative enlargement of the bounding box; deactivated if negative.

**bases**
Ordered list of types (as strings) this functor accepts.

**dict**(*(Serializable)arg1*) → dict :
Return dictionary of attributes.

**label**(*=uninitalized*)
Textual label for this object; must be a valid python identifier, you can refer to it directly from python.

**timingDeltas**
Detailed information about timing inside the Dispatcher itself. Empty unless enabled in the source code and O.timingEnabled==True.

**updateAttrs**(*(Serializable)arg1, (dict)arg2*) → None :
Update object attributes from given dictionary

**class yade.wrapper.Bo1_PFacet_Aabb**(*inherits BoundFunctor → Functor → Serializable*)
Functor creating *Aabb* from a *PFacet*.

**aabbEnlargeFactor**(*=-1, deactivated*)
    Relative enlargement of the bounding box; deactivated if negative.

**bases**
    Ordered list of types (as strings) this functor accepts.

**dict**(*(Serializable)arg1*) → dict :
    Return dictionary of attributes.

**label**(*=uninitalized*)
    Textual label for this object; must be a valid python identifier, you can refer to it directly from python.

**timingDeltas**
    Detailed information about timing inside the Dispatcher itself. Empty unless enabled in the source code and O.timingEnabled==True.

**updateAttrs**(*(Serializable)arg1, (dict)arg2*) → None :
    Update object attributes from given dictionary

**class yade.wrapper.Bo1_Sphere_Aabb**(*inherits BoundFunctor → Functor → Serializable*)
    Functor creating *Aabb* from *Sphere*.

    **aabbEnlargeFactor**
        Relative enlargement of the bounding box; deactivated if negative.

> **Note:** This attribute is used to create distant interaction, but is only meaningful with an *IGeomFunctor* which will not simply discard such interactions: *Ig2_Sphere_Sphere_-ScGeom::interactionDetectionFactor* should have the same value as *aabbEnlargeFactor*.

    **bases**
        Ordered list of types (as strings) this functor accepts.

    **dict**(*(Serializable)arg1*) → dict :
        Return dictionary of attributes.

    **label**(*=uninitalized*)
        Textual label for this object; must be a valid python identifier, you can refer to it directly from python.

    **timingDeltas**
        Detailed information about timing inside the Dispatcher itself. Empty unless enabled in the source code and O.timingEnabled==True.

    **updateAttrs**(*(Serializable)arg1, (dict)arg2*) → None :
        Update object attributes from given dictionary

**class yade.wrapper.Bo1_Tetra_Aabb**(*inherits BoundFunctor → Functor → Serializable*)
    Create/update *Aabb* of a *Tetra*

    **bases**
        Ordered list of types (as strings) this functor accepts.

    **dict**(*(Serializable)arg1*) → dict :
        Return dictionary of attributes.

    **label**(*=uninitalized*)
        Textual label for this object; must be a valid python identifier, you can refer to it directly from python.

    **timingDeltas**
        Detailed information about timing inside the Dispatcher itself. Empty unless enabled in the source code and O.timingEnabled==True.

    **updateAttrs**(*(Serializable)arg1, (dict)arg2*) → None :
        Update object attributes from given dictionary

**class** **yade.wrapper.Bo1_Wall_Aabb**(*inherits BoundFunctor → Functor → Serializable*)
>    Creates/updates an *Aabb* of a *Wall*

>    **bases**
>    >    Ordered list of types (as strings) this functor accepts.

>    **dict**(*(Serializable)arg1*) → dict :
>    >    Return dictionary of attributes.

>    **label**(*=uninitalized*)
>    >    Textual label for this object; must be a valid python identifier, you can refer to it directly
>    >    from python.

>    **timingDeltas**
>    >    Detailed information about timing inside the Dispatcher itself. Empty unless enabled in the
>    >    source code and O.timingEnabled==True.

>    **updateAttrs**(*(Serializable)arg1, (dict)arg2*) → None :
>    >    Update object attributes from given dictionary

## BoundDispatcher

**class** **yade.wrapper.BoundDispatcher**(*inherits Dispatcher → Engine → Serializable*)
>    Dispatcher calling *functors* based on received argument type(s).

>    **activated**(*=true*)
>    >    Whether the engine is activated (only should be changed by the collider)

>    **dead**(*=false*)
>    >    If true, this engine will not run at all; can be used for making an engine temporarily deactivated
>    >    and only resurrect it at a later point.

>    **dict**(*(Serializable)arg1*) → dict :
>    >    Return dictionary of attributes.

>    **dispFunctor**(*(BoundDispatcher)arg1, (Shape)arg2*) → BoundFunctor :
>    >    Return functor that would be dispatched for given argument(s); None if no dispatch; ambiguous dispatch throws.

>    **dispMatrix**(*(BoundDispatcher)arg1*[, *(bool)names=True*]) → dict :
>    >    Return dictionary with contents of the dispatch matrix.

>    **execCount**
>    >    Cumulative count this engine was run (only used if *O.timingEnabled*==True).

>    **execTime**
>    >    Cumulative time in nanoseconds this Engine took to run (only used if
>    >    *O.timingEnabled*==True).

>    **functors**
>    >    Functors associated with this dispatcher.

>    **label**(*=uninitalized*)
>    >    Textual label for this object; must be valid python identifier, you can refer to it directly from
>    >    python.

>    **minSweepDistFactor**(*=0.2*)
>    >    Minimal distance by which enlarge all bounding boxes; superseeds computed value of sweep-
>    >    Dist when lower that (minSweepDistFactor x sweepDist). Updated by the collider. *(auto-
>    >    updated)*.

>    **ompThreads**(*=-1*)
>    >    Number of threads to be used in the engine. If ompThreads<0 (default), the number will be
>    >    typically OMP_NUM_THREADS or the number N defined by 'yade -jN' (this behavior can
>    >    depend on the engine though). This attribute will only affect engines whose code includes openMP parallel regions (e.g. *InteractionLoop*). This attribute is mostly useful for experiments or when combining *ParallelEngine* with engines that run parallel regions, resulting in nested OMP loops with different number of threads at each level.

**sweepDist** (*=0*)
    Distance by which enlarge all bounding boxes, to prevent collider from being run at every step (only should be changed by the collider).

**targetInterv** (*=-1*)
    see *InsertionSortCollider::targetInterv (auto-updated)*

**timingDeltas**
    Detailed information about timing inside the Engine itself. Empty unless enabled in the source code and *O.timingEnabled*==**True**.

**updateAttrs** (*(Serializable)arg1, (dict)arg2*) → None :
    Update object attributes from given dictionary

**updatingDispFactor** (*=-1*)
    see *InsertionSortCollider::updatingDispFactor (auto-updated)*

### 2.3.8 Interaction Geometry creation

**IGeomFunctor**

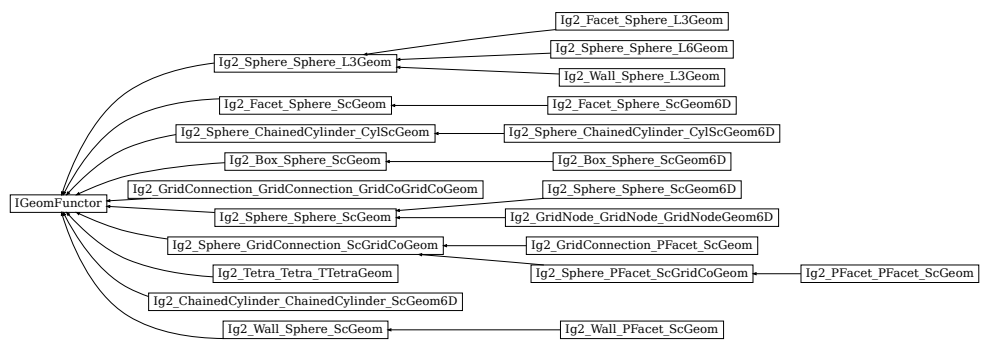

Fig. 34: Inheritance graph of IGeomFunctor. See also: *Ig2_Box_Sphere_ScGeom*, *Ig2_Box_Sphere_ScGeom6D*, *Ig2_ChainedCylinder_ChainedCylinder_ScGeom6D*, *Ig2_Facet_Sphere_L3Geom*, *Ig2_Facet_Sphere_ScGeom*, *Ig2_Facet_Sphere_ScGeom6D*, *Ig2_GridConnection_GridConnection_GridCoGridCoGeom*, *Ig2_GridConnection_PFacet_ScGeom*, *Ig2_GridNode_GridNode_GridNodeGeom6D*, *Ig2_PFacet_PFacet_ScGeom*, *Ig2_Sphere_ChainedCylinder_CylScGeom*, *Ig2_Sphere_ChainedCylinder_CylScGeom6D*, *Ig2_Sphere_GridConnection_ScGridCoGeom*, *Ig2_Sphere_PFacet_ScGridCoGeom*, *Ig2_Sphere_Sphere_L3Geom*, *Ig2_Sphere_Sphere_L6Geom*, *Ig2_Sphere_Sphere_ScGeom*, *Ig2_Sphere_Sphere_ScGeom6D*, *Ig2_Tetra_Tetra_TTetraGeom*, *Ig2_Wall_PFacet_ScGeom*, *Ig2_Wall_Sphere_L3Geom*, *Ig2_Wall_Sphere_ScGeom*.

**class** yade.wrapper.**IGeomFunctor** (*inherits Functor → Serializable*)
    Functor for creating/updating *Interaction::geom* objects.

**bases**
    Ordered list of types (as strings) this functor accepts.

**dict** (*(Serializable)arg1*) → dict :
    Return dictionary of attributes.

**label** (*=uninitialized*)
    Textual label for this object; must be a valid python identifier, you can refer to it directly from python.

**timingDeltas**
> Detailed information about timing inside the Dispatcher itself. Empty unless enabled in the source code and O.timingEnabled==True.

**updateAttrs**(*(Serializable)arg1, (dict)arg2*) → None :
> Update object attributes from given dictionary

**class** yade.wrapper.**Ig2_Box_Sphere_ScGeom**(*inherits IGeomFunctor → Functor → Serializable*)
> Create an interaction geometry *ScGeom* from *Box* and *Sphere*, representing the box with a projected virtual sphere of same radius.

**bases**
> Ordered list of types (as strings) this functor accepts.

**dict**(*(Serializable)arg1*) → dict :
> Return dictionary of attributes.

**interactionDetectionFactor**
> Enlarge sphere radii by this factor (if >1), to permit creation of distant interactions.
>
> InteractionGeometry will be computed when interactionDetectionFactor*(rad) > distance.
>
> ---
> **Note:** This parameter is functionally coupled with *Bo1_Sphere_Aabb::aabbEnlargeFactor*, which will create larger bounding boxes and should be of the same value.
> ---

**label**(*=uninitalized*)
> Textual label for this object; must be a valid python identifier, you can refer to it directly from python.

**timingDeltas**
> Detailed information about timing inside the Dispatcher itself. Empty unless enabled in the source code and O.timingEnabled==True.

**updateAttrs**(*(Serializable)arg1, (dict)arg2*) → None :
> Update object attributes from given dictionary

**class** yade.wrapper.**Ig2_Box_Sphere_ScGeom6D**(*inherits Ig2_Box_Sphere_ScGeom → IGeomFunctor → Functor → Serializable*)
> Create an interaction geometry *ScGeom6D* from *Box* and *Sphere*, representing the box with a projected virtual sphere of same radius.

**bases**
> Ordered list of types (as strings) this functor accepts.

**dict**(*(Serializable)arg1*) → dict :
> Return dictionary of attributes.

**interactionDetectionFactor**
> Enlarge sphere radii by this factor (if >1), to permit creation of distant interactions.
>
> InteractionGeometry will be computed when interactionDetectionFactor*(rad) > distance.
>
> ---
> **Note:** This parameter is functionally coupled with *Bo1_Sphere_Aabb::aabbEnlargeFactor*, which will create larger bounding boxes and should be of the same value.
> ---

**label**(*=uninitalized*)
> Textual label for this object; must be a valid python identifier, you can refer to it directly from python.

**timingDeltas**
> Detailed information about timing inside the Dispatcher itself. Empty unless enabled in the source code and O.timingEnabled==True.

> **updateAttrs**(*(Serializable)arg1, (dict)arg2*) → None :
>> Update object attributes from given dictionary

**class** yade.wrapper.**Ig2_ChainedCylinder_ChainedCylinder_ScGeom6D**(*inherits IGeomFunctor → Functor → Serializable*)

> Create/update a *ScGeom* instance representing connexion between *chained cylinders*.

> **bases**
>> Ordered list of types (as strings) this functor accepts.

> **dict**(*(Serializable)arg1*) → dict :
>> Return dictionary of attributes.

> **halfLengthContacts**(*=true*)
>> If True, Cylinders nodes interact like spheres of radius 0.5*length, else one node has size length while the other has size 0. The difference is mainly the locus of rotation definition.

> **interactionDetectionFactor**(*=1*)
>> Enlarge both radii by this factor (if >1), to permit creation of distant interactions.

> **label**(*=uninitalized*)
>> Textual label for this object; must be a valid python identifier, you can refer to it directly from python.

> **timingDeltas**
>> Detailed information about timing inside the Dispatcher itself. Empty unless enabled in the source code and O.timingEnabled==True.

> **updateAttrs**(*(Serializable)arg1, (dict)arg2*) → None :
>> Update object attributes from given dictionary

**class** yade.wrapper.**Ig2_Facet_Sphere_L3Geom**(*inherits Ig2_Sphere_Sphere_L3Geom → IGeomFunctor → Functor → Serializable*)

> Incrementally compute *L3Geom* for contact between *Facet* and *Sphere*. Uses attributes of *Ig2_Sphere_Sphere_L3Geom*.

> **approxMask**
>> Selectively enable geometrical approximations (bitmask); add the values for approximations to be enabled.

| | |
|---|---|
| 1 | use previous transformation to transform velocities (which are known at mid-steps), instead of mid-step transformation computed as quaternion slerp at t=0.5. |
| 2 | do not take average (mid-step) normal when computing relative shear displacement, use previous value instead |
| 4 | do not re-normalize average (mid-step) normal, if used.... |

> **By default, the mask is zero, wherefore none of these approximations is used.**

> **bases**
>> Ordered list of types (as strings) this functor accepts.

> **dict**(*(Serializable)arg1*) → dict :
>> Return dictionary of attributes.

> **distFactor**(*=1*)
>> Create interaction if spheres are not futher than *distFactor* *(r1+r2). If negative, zero normal deformation will be set to be the initial value (otherwise, the geometrical distance is the ''zero'' one).

> **label**(*=uninitalized*)
>> Textual label for this object; must be a valid python identifier, you can refer to it directly from python.

**noRatch**(*=true*)
> See *Ig2_Sphere_Sphere_ScGeom.avoidGranularRatcheting*.

**timingDeltas**
> Detailed information about timing inside the Dispatcher itself. Empty unless enabled in the source code and O.timingEnabled==True.

**trsfRenorm**(*=100*)
> How often to renormalize *trsf*; if non-positive, never renormalized (simulation might be unstable)

**updateAttrs**(*(Serializable)arg1, (dict)arg2*) → None :
> Update object attributes from given dictionary

**class yade.wrapper.Ig2_Facet_Sphere_ScGeom**(*inherits IGeomFunctor → Functor → Serializable*)

Create/update a *ScGeom* instance representing intersection of *Facet* and *Sphere*. The equivalent radius for the Facet (*ScGeom.refR1*) is chosen as twice the Sphere's one.

**bases**
> Ordered list of types (as strings) this functor accepts.

**dict**(*(Serializable)arg1*) → dict :
> Return dictionary of attributes.

**hertzian**(*=false*)
> The equivalent radius for the Facet (*ScGeom.refR1*) is chosen as 1e8 times the Sphere's radius (closer to Hertzian theory, where it is infinite).

**label**(*=uninitalized*)
> Textual label for this object; must be a valid python identifier, you can refer to it directly from python.

**shrinkFactor**(*=0, no shrinking*)
> The radius of the inscribed circle of the facet is decreased by the value of the sphere's radius multiplied by *shrinkFactor*. From the definition of contact point on the surface made of facets, the given surface is not continuous and becomes in effect surface covered with triangular tiles, with gap between the separate tiles equal to the sphere's radius multiplied by 2×*shrinkFactor*. If zero, no shrinking is done.

**timingDeltas**
> Detailed information about timing inside the Dispatcher itself. Empty unless enabled in the source code and O.timingEnabled==True.

**updateAttrs**(*(Serializable)arg1, (dict)arg2*) → None :
> Update object attributes from given dictionary

**class yade.wrapper.Ig2_Facet_Sphere_ScGeom6D**(*inherits Ig2_Facet_Sphere_ScGeom → IGeomFunctor → Functor → Serializable*)

Create an interaction geometry *ScGeom6D* from *Facet* and *Sphere*, representing the Facet with a projected virtual sphere of same radius.

**bases**
> Ordered list of types (as strings) this functor accepts.

**dict**(*(Serializable)arg1*) → dict :
> Return dictionary of attributes.

**hertzian**(*=false*)
> The equivalent radius for the Facet (*ScGeom.refR1*) is chosen as 1e8 times the Sphere's radius (closer to Hertzian theory, where it is infinite).

**label**(*=uninitalized*)
> Textual label for this object; must be a valid python identifier, you can refer to it directly from python.

**shrinkFactor**(*=0, no shrinking*)
> The radius of the inscribed circle of the facet is decreased by the value of the sphere's radius multiplied by *shrinkFactor*. From the definition of contact point on the surface made of facets, the given surface is not continuous and becomes in effect surface covered with triangular tiles, with gap between the separate tiles equal to the sphere's radius multiplied by 2×*shrinkFactor*. If zero, no shrinking is done.

**timingDeltas**
> Detailed information about timing inside the Dispatcher itself. Empty unless enabled in the source code and O.timingEnabled==True.

**updateAttrs**(*(Serializable)arg1, (dict)arg2*) → None :
> Update object attributes from given dictionary

**class yade.wrapper.Ig2_GridConnection_GridConnection_GridCoGridCoGeom**(*inherits IGeomFunctor → Functor → Serializable*)

Create/update a *GridCoGridCoGeom* instance representing the geometry of a contact point between two *GridConnection* , including relative rotations.

**bases**
> Ordered list of types (as strings) this functor accepts.

**dict**(*(Serializable)arg1*) → dict :
> Return dictionary of attributes.

**label**(*=uninitalized*)
> Textual label for this object; must be a valid python identifier, you can refer to it directly from python.

**timingDeltas**
> Detailed information about timing inside the Dispatcher itself. Empty unless enabled in the source code and O.timingEnabled==True.

**updateAttrs**(*(Serializable)arg1, (dict)arg2*) → None :
> Update object attributes from given dictionary

**class yade.wrapper.Ig2_GridConnection_PFacet_ScGeom**(*inherits Ig2_Sphere_GridConnection_ScGridCoGeom → IGeomFunctor → Functor → Serializable*)

Create/update a *ScGeom* instance representing intersection of *Facet* and *GridConnection*.

**bases**
> Ordered list of types (as strings) this functor accepts.

**dict**(*(Serializable)arg1*) → dict :
> Return dictionary of attributes.

**interactionDetectionFactor**(*=1*)
> Enlarge both radii by this factor (if >1), to permit creation of distant interactions.

**label**(*=uninitalized*)
> Textual label for this object; must be a valid python identifier, you can refer to it directly from python.

**shrinkFactor**(*=0, no shrinking*)
> The radius of the inscribed circle of the facet is decreased by the value of the sphere's radius multipled by *shrinkFactor*. From the definition of contact point on the surface made of facets, the given surface is not continuous and becomes in effect surface covered with triangular tiles, with gap between the separate tiles equal to the sphere's radius multiplied by 2×*shrinkFactor*. If zero, no shrinking is done.

**timingDeltas**
> Detailed information about timing inside the Dispatcher itself. Empty unless enabled in the source code and O.timingEnabled==True.

**updateAttrs**(*(Serializable)arg1, (dict)arg2*) → None :
    Update object attributes from given dictionary

**class** yade.wrapper.**Ig2_GridNode_GridNode_GridNodeGeom6D**(*inherits Ig2_Sphere_Sphere_-ScGeom → IGeomFunctor → Functor → Serializable*)

Create/update a *GridNodeGeom6D* instance representing the geometry of a contact point between two *GridNode*, including relative rotations.

**avoidGranularRatcheting**
    Define relative velocity so that ratcheting is avoided. It applies for sphere-sphere contacts. It eventualy also apply for sphere-emulating interactions (i.e. convertible into the ScGeom type), if the virtual sphere's motion is defined correctly (see e.g. *Ig2_Sphere_ChainedCylinder_-CylScGeom*).

    Short explanation of what we want to avoid :

    Numerical ratcheting is best understood considering a small elastic cycle at a contact between two grains : assuming b1 is fixed, impose this displacement to b2 :

      1. translation $dx$ in the normal direction

      2. rotation $a$

      3. translation $-dx$ (back to the initial position)

      4. rotation $-a$ (back to the initial orientation)

    If the branch vector used to define the relative shear in rotation×branch is not constant (typically if it is defined from the vector center→contactPoint), then the shear displacement at the end of this cycle is not zero: rotations $a$ and $-a$ are multiplied by branches of different lengths.

    It results in a finite contact force at the end of the cycle even though the positions and orientations are unchanged, in total contradiction with the elastic nature of the problem. It could also be seen as an *inconsistent energy creation or loss.* Given that DEM simulations tend to generate oscillations around equilibrium (damped mass-spring), it can have a significant impact on the evolution of the packings, resulting for instance in slow creep in iterations under constant load.

    The solution adopted here to avoid ratcheting is as proposed by McNamara and co-workers. They analyzed the ratcheting problem in detail - even though they comment on the basis of a cycle that differs from the one shown above. One will find interesting discussions in e.g. [McNamara2008], even though solution it suggests is not fully applied here (equations of motion are not incorporating alpha, in contradiction with what is suggested by McNamara et al.).

**bases**
    Ordered list of types (as strings) this functor accepts.

**creep**(*=false*)
    Substract rotational creep from relative rotation. The rotational creep *ScGeom6D::twistCreep* is a quaternion and has to be updated inside a constitutive law, see for instance *Law2_-ScGeom6D_CohFrictPhys_CohesionMoment*.

**dict**(*(Serializable)arg1*) → dict :
    Return dictionary of attributes.

**interactionDetectionFactor**
    Enlarge both radii by this factor (if >1), to permit creation of distant interactions.

    InteractionGeometry will be computed when interactionDetectionFactor*(rad1+rad2) > distance.

---

**Note:** This parameter is functionally coupled with *Bo1_Sphere_Aabb::aabbEnlargeFactor*, which will create larger bounding boxes and should be of the same value.

---

**label**(*=uninitalized*)
　Textual label for this object; must be a valid python identifier, you can refer to it directly from python.

**timingDeltas**
　Detailed information about timing inside the Dispatcher itself. Empty unless enabled in the source code and O.timingEnabled==True.

**updateAttrs**(*(Serializable)arg1, (dict)arg2*) → None :
　Update object attributes from given dictionary

**updateRotations**(*=true*)
　Precompute relative rotations. Turning this false can speed up simulations when rotations are not needed in constitutive laws (e.g. when spheres are compressed without cohesion and moment in early stage of a triaxial test), but is not foolproof. Change this value only if you know what you are doing.

**class yade.wrapper.Ig2_PFacet_PFacet_ScGeom**(*inherits Ig2_Sphere_PFacet_ScGridCoGeom → Ig2_Sphere_GridConnection_ScGridCoGeom → IGeomFunctor → Functor → Serializable*)
　Create/update a *ScGridCoGeom* instance representing intersection of *Facet* and *Sphere*.

**bases**
　Ordered list of types (as strings) this functor accepts.

**dict**(*(Serializable)arg1*) → dict :
　Return dictionary of attributes.

**interactionDetectionFactor**(*=1*)
　Enlarge both radii by this factor (if >1), to permit creation of distant interactions.

**label**(*=uninitalized*)
　Textual label for this object; must be a valid python identifier, you can refer to it directly from python.

**shrinkFactor**(*=0, no shrinking*)
　The radius of the inscribed circle of the facet is decreased by the value of the sphere's radius multiplied by *shrinkFactor*. From the definition of contact point on the surface made of facets, the given surface is not continuous and becomes in effect surface covered with triangular tiles, with gap between the separate tiles equal to the sphere's radius multiplied by 2×*shrinkFactor*. If zero, no shrinking is done.

**timingDeltas**
　Detailed information about timing inside the Dispatcher itself. Empty unless enabled in the source code and O.timingEnabled==True.

**updateAttrs**(*(Serializable)arg1, (dict)arg2*) → None :
　Update object attributes from given dictionary

**class yade.wrapper.Ig2_Sphere_ChainedCylinder_CylScGeom**(*inherits IGeomFunctor → Functor → Serializable*)
　Create/update a *ScGeom* instance representing intersection of two *Spheres*.

**bases**
　Ordered list of types (as strings) this functor accepts.

**dict**(*(Serializable)arg1*) → dict :
　Return dictionary of attributes.

---

**interactionDetectionFactor**(*=1*)
> Enlarge both radii by this factor (if >1), to permit creation of distant interactions.

**label**(*=uninitialized*)
> Textual label for this object; must be a valid python identifier, you can refer to it directly from python.

**timingDeltas**
> Detailed information about timing inside the Dispatcher itself. Empty unless enabled in the source code and O.timingEnabled==True.

**updateAttrs**(*(Serializable)arg1, (dict)arg2*) → None :
> Update object attributes from given dictionary

**class yade.wrapper.Ig2_Sphere_ChainedCylinder_CylScGeom6D**(*inherits Ig2_Sphere_-
ChainedCylinder_CylSc-
Geom → IGeomFunctor →
Functor → Serializable*)

Create/update a *ScGeom6D* instance representing the geometry of a contact point between two *Spheres*, including relative rotations.

**bases**
> Ordered list of types (as strings) this functor accepts.

**creep**(*=false*)
> Substract rotational creep from relative rotation. The rotational creep *ScGeom6D::twistCreep* is a quaternion and has to be updated inside a constitutive law, see for instance *Law2_-
ScGeom6D_CohFrictPhys_CohesionMoment*.

**dict**(*(Serializable)arg1*) → dict :
> Return dictionary of attributes.

**interactionDetectionFactor**(*=1*)
> Enlarge both radii by this factor (if >1), to permit creation of distant interactions.

**label**(*=uninitialized*)
> Textual label for this object; must be a valid python identifier, you can refer to it directly from python.

**timingDeltas**
> Detailed information about timing inside the Dispatcher itself. Empty unless enabled in the source code and O.timingEnabled==True.

**updateAttrs**(*(Serializable)arg1, (dict)arg2*) → None :
> Update object attributes from given dictionary

**updateRotations**(*=false*)
> Precompute relative rotations. Turning this false can speed up simulations when rotations are not needed in constitutive laws (e.g. when spheres are compressed without cohesion and moment in early stage of a triaxial test), but is not foolproof. Change this value only if you know what you are doing.

**class yade.wrapper.Ig2_Sphere_GridConnection_ScGridCoGeom**(*inherits IGeomFunctor →
Functor → Serializable*)

Create/update a *ScGridCoGeom6D* instance representing the geometry of a contact point between a *GricConnection* and a *Sphere* including relative rotations.

**bases**
> Ordered list of types (as strings) this functor accepts.

**dict**(*(Serializable)arg1*) → dict :
> Return dictionary of attributes.

**interactionDetectionFactor**(*=1*)
> Enlarge both radii by this factor (if >1), to permit creation of distant interactions.

**label**(*=uninitalized*)
Textual label for this object; must be a valid python identifier, you can refer to it directly from python.

**timingDeltas**
Detailed information about timing inside the Dispatcher itself. Empty unless enabled in the source code and O.timingEnabled==True.

**updateAttrs**(*(Serializable)arg1, (dict)arg2*) → None :
Update object attributes from given dictionary

**class yade.wrapper.Ig2_Sphere_PFacet_ScGridCoGeom**(*inherits    Ig2_Sphere_GridConnection_ScGridCoGeom → IGeomFunctor → Functor → Serializable*)

Create/update a *ScGridCoGeom* instance representing intersection of *PFacet* and *Sphere*.

**bases**
Ordered list of types (as strings) this functor accepts.

**dict**(*(Serializable)arg1*) → dict :
Return dictionary of attributes.

**interactionDetectionFactor**(*=1*)
Enlarge both radii by this factor (if >1), to permit creation of distant interactions.

**label**(*=uninitalized*)
Textual label for this object; must be a valid python identifier, you can refer to it directly from python.

**shrinkFactor**(*=0, no shrinking*)
The radius of the inscribed circle of the facet is decreased by the value of the sphere's radius multiplied by *shrinkFactor*. From the definition of contact point on the surface made of facets, the given surface is not continuous and becomes in effect surface covered with triangular tiles, with gap between the separate tiles equal to the sphere's radius multiplied by 2×*shrinkFactor*. If zero, no shrinking is done.

**timingDeltas**
Detailed information about timing inside the Dispatcher itself. Empty unless enabled in the source code and O.timingEnabled==True.

**updateAttrs**(*(Serializable)arg1, (dict)arg2*) → None :
Update object attributes from given dictionary

**class yade.wrapper.Ig2_Sphere_Sphere_L3Geom**(*inherits IGeomFunctor → Functor → Serializable*)

Functor for computing incrementally configuration of 2 *Spheres* stored in *L3Geom*; the configuration is positioned in global space by local origin **c** (contact point) and rotation matrix **T** (orthonormal transformation matrix), and its degrees of freedom are local displacement **u** (in one normal and two shear directions); with *Ig2_Sphere_Sphere_L6Geom* and *L6Geom*, there is additionally **φ**. The first row of **T**, i.e. local x-axis, is the contact normal noted **n** for brevity. Additionally, quasi-constant values of $\mathbf{u}_0$ (and $\mathbf{φ}_0$) are stored as shifted origins of **u** (and **φ**); therefore, current value of displacement is always $\mathbf{u}^\circ - \mathbf{u}_0$.

Suppose two spheres with radii $r_i$, positions $\mathbf{x}_i$, velocities $\mathbf{v}_i$, angular velocities $\mathbf{\omega}_i$.

When there is not yet contact, it will be created if $\mathbf{u}_N = |\mathbf{x}_2^\circ - \mathbf{x}_1^\circ| - |f_d|(r_1 + r2) < 0$, where $f_d$ is *distFactor* (sometimes also called "interaction radius"). If $f_d > 0$, then $\mathbf{u}_{0x}$ will be initalized to $\mathbf{u}_N$, otherwise to 0. In another words, contact will be created if spheres enlarged by $|f_d|$ touch, and the "equilibrium distance" (where $\mathbf{u}_x - \mathbf{u} - 0x$ is zero) will be set to the current distance if $f_d$ is positive, and to the geometrically-touching distance if negative.

Local axes (rows of **T**) are initially defined as follows:

- local x-axis is $\mathbf{n} = \mathbf{x}_1 = \widehat{\mathbf{x}_2 - \mathbf{x}_1}$;

- local y-axis positioned arbitrarily, but in a deterministic manner: aligned with the xz plane (if $\mathbf{n}_y < \mathbf{n}_z$) or xy plane (otherwise);

- local $z$-axis $\mathbf{z}_l = \mathbf{x}_l \times \mathbf{y}_l$.

If there has already been contact between the two spheres, it is updated to keep track of rigid motion of the contact (one that does not change mutual configuration of spheres) and mutual configuration changes. Rigid motion transforms local coordinate system and can be decomposed in rigid translation (affecting $\mathbf{c}$), and rigid rotation (affecting $\mathbf{T}$), which can be split in rotation $\mathbf{o_r}$ perpendicular to the normal and rotation $\mathbf{o_t}$ ("twist") parallel with the normal:

$$\mathbf{o_r^{\ominus}} = \mathbf{n^-} \times \mathbf{n^{\circ}}.$$

Since velocities are known at previous midstep $(\mathrm{t} - \Delta \mathrm{t}/2)$, we consider mid-step normal

$$\mathbf{n^{\ominus}} = \frac{\mathbf{n^-} + \mathbf{n^{\circ}}}{2}.$$

For the sake of numerical stability, $\mathbf{n^{\ominus}}$ is re-normalized after being computed, unless prohibited by *approxMask*. If *approxMask* has the appropriate bit set, the mid-normal is not compute, and we simply use $\mathbf{n^{\ominus}} \approx \mathbf{n^-}$.

Rigid rotation parallel with the normal is

$$\mathbf{o_t^{\ominus}} = \mathbf{n^{\ominus}} \left( \mathbf{n^{\ominus}} \cdot \frac{\boldsymbol{\omega}_1^{\ominus} + \boldsymbol{\omega}_2^{\ominus}}{2} \right) \Delta \mathrm{t}.$$

*Branch vectors* $\mathbf{b}_1$, $\mathbf{b}_2$ (connecting $\mathbf{x}_1^{\circ}$, $\mathbf{x}_2^{\circ}$ with $\mathbf{c}^{\circ}$ are computed depending on *noRatch* (see *here*).

$$\mathbf{b}_1 = \begin{cases} r_1 \mathbf{n^{\circ}} & \text{with } \texttt{noRatch} \\ \mathbf{c^{\circ}} - \mathbf{x}_1^{\circ} & \text{otherwise} \end{cases}$$

$$\mathbf{b}_2 = \begin{cases} -r_2 \mathbf{n^{\circ}} & \text{with } \texttt{noRatch} \\ \mathbf{c^{\circ}} - \mathbf{x}_2^{\circ} & \text{otherwise} \end{cases}$$

Relative velocity at $\mathbf{c}^{\circ}$ can be computed as

$$\boldsymbol{\nu}_\mathrm{r}^{\ominus} = (\tilde{\boldsymbol{\nu}}_2^{\ominus} + \boldsymbol{\omega}_2 \times \mathbf{b}_2) - (\boldsymbol{\nu}_1 + \boldsymbol{\omega}_1 \times \mathbf{b}_1)$$

where $\tilde{\boldsymbol{\nu}}_2$ is $\boldsymbol{\nu}_2$ without mean-field velocity gradient in periodic boundary conditions (see *Cell.homoDeform*). In the numerial implementation, the normal part of incident velocity is removed (since it is computed directly) with $\boldsymbol{\nu}_{\mathrm{r}2}^{\ominus} = \boldsymbol{\nu}_\mathrm{r}^{\ominus} - (\mathbf{n^{\ominus}} \cdot \boldsymbol{\nu}_\mathrm{r}^{\ominus}) \mathbf{n^{\ominus}}$.

Any vector $\mathbf{a}$ expressed in global coordinates transforms during one timestep as

$$\mathbf{a^{\circ}} = \mathbf{a^-} + \boldsymbol{\nu}_\mathrm{r}^{\ominus} \Delta \mathrm{t} - \mathbf{a^-} \times \mathbf{o_r^{\ominus}} - \mathbf{a^-} \times \mathbf{t_t^{\ominus}}$$

where the increments have the meaning of relative shear, rigid rotation normal to $\mathbf{n}$ and rigid rotation parallel with $\mathbf{n}$. Local coordinate system orientation, rotation matrix $\mathbf{T}$, is updated by rows, i.e.

$$\mathbf{T^{\circ}} = \begin{pmatrix} \mathbf{n}_\mathrm{x}^{\circ} & \mathbf{n}_\mathrm{y}^{\circ} & \mathbf{n}_\mathrm{z}^{\circ} \\ \mathbf{T}_{1,\bullet}^- - \mathbf{T}_{1,\bullet}^- \times \mathbf{o_r^{\ominus}} - \mathbf{T}_{1,\bullet}^- \times \mathbf{o_t^{\ominus}} \\ \mathbf{T}_{2,\bullet}^- - \mathbf{T}_{2,\bullet}^- \times \mathbf{o_r^{\ominus}} - \mathbf{T}_{\bullet,\bullet}^- \times \mathbf{o_t^{\ominus}} \end{pmatrix}$$

This matrix is re-normalized (unless prevented by *approxMask*) and mid-step transformation is computed using quaternion spherical interpolation as

$$\mathbf{T^{\ominus}} = \mathrm{Slerp} \left( \mathbf{T^-}; \mathbf{T^{\circ}}; \mathrm{t} = 1/2 \right).$$

Depending on *approxMask*, this computation can be avoided by approximating $\mathbf{T^{\ominus}} = \mathbf{T^-}$.

Finally, current displacement is evaluated as

$$\mathbf{u^{\circ}} = \mathbf{u^-} + \mathbf{T^{\ominus}} \boldsymbol{\nu}_\mathrm{r}^{\ominus} \Delta \mathrm{t}.$$

For the normal component, non-incremental evaluation is preferred, giving

$$\mathbf{u}_x^\circ = |\mathbf{x}_2^\circ - \mathbf{x}_1^\circ| - (r_1 + r_2)$$

If this functor is called for *L6Geom*, local rotation is updated as

$$\boldsymbol{\varphi}^\circ = \boldsymbol{\varphi}^- + \mathbf{T}^\ominus \Delta t(\boldsymbol{\omega}_2 - \boldsymbol{\omega}_1)$$

**approxMask**
> Selectively enable geometrical approximations (bitmask); add the values for approximations to be enabled.

| 1 | use previous transformation to transform velocities (which are known at mid-steps), instead of mid-step transformation computed as quaternion slerp at t=0.5. |
|---|---|
| 2 | do not take average (mid-step) normal when computing relative shear displacement, use previous value instead |
| 4 | do not re-normalize average (mid-step) normal, if used.... |

**By default, the mask is zero, wherefore none of these approximations is used.**

**bases**
> Ordered list of types (as strings) this functor accepts.

**dict**(*(Serializable)arg1*) → dict :
> Return dictionary of attributes.

**distFactor**(*=1*)
> Create interaction if spheres are not futher than *distFactor* \*(r1+r2). If negative, zero normal deformation will be set to be the initial value (otherwise, the geometrical distance is the ''zero'' one).

**label**(*=uninitalized*)
> Textual label for this object; must be a valid python identifier, you can refer to it directly from python.

**noRatch**(*=true*)
> See *Ig2_Sphere_Sphere_ScGeom.avoidGranularRatcheting*.

**timingDeltas**
> Detailed information about timing inside the Dispatcher itself. Empty unless enabled in the source code and O.timingEnabled==True.

**trsfRenorm**(*=100*)
> How often to renormalize *trsf*; if non-positive, never renormalized (simulation might be unstable)

**updateAttrs**(*(Serializable)arg1, (dict)arg2*) → None :
> Update object attributes from given dictionary

**class yade.wrapper.Ig2_Sphere_Sphere_L6Geom**(*inherits Ig2_Sphere_Sphere_L3Geom → IGeomFunctor → Functor → Serializable*)
Incrementally compute *L6Geom* for contact of 2 spheres.

**approxMask**
> Selectively enable geometrical approximations (bitmask); add the values for approximations to be enabled.

| 1 | use previous transformation to transform velocities (which are known at mid-steps), instead of mid-step transformation computed as quaternion slerp at t=0.5. |
|---|---|
| 2 | do not take average (mid-step) normal when computing relative shear displacement, use previous value instead |
| 4 | do not re-normalize average (mid-step) normal, if used.... |

**By default, the mask is zero, wherefore none of these approximations is used.**

**bases**
Ordered list of types (as strings) this functor accepts.

**dict**(*(Serializable)arg1*) → dict :
Return dictionary of attributes.

**distFactor**(*=1*)
Create interaction if spheres are not futher than *distFactor* \*(r1+r2). If negative, zero normal deformation will be set to be the initial value (otherwise, the geometrical distance is the ''zero'' one).

**label**(*=uninitalized*)
Textual label for this object; must be a valid python identifier, you can refer to it directly from python.

**noRatch**(*=true*)
See *Ig2_Sphere_Sphere_ScGeom.avoidGranularRatcheting*.

**timingDeltas**
Detailed information about timing inside the Dispatcher itself. Empty unless enabled in the source code and O.timingEnabled==True.

**trsfRenorm**(*=100*)
How often to renormalize *trsf*; if non-positive, never renormalized (simulation might be unstable)

**updateAttrs**(*(Serializable)arg1, (dict)arg2*) → None :
Update object attributes from given dictionary

**class yade.wrapper.Ig2_Sphere_Sphere_ScGeom**(*inherits IGeomFunctor → Functor → Serializable*)
Create/update a *ScGeom* instance representing the geometry of a contact point between two *Spheres* s.

**avoidGranularRatcheting**
Define relative velocity so that ratcheting is avoided. It applies for sphere-sphere contacts. It eventualy also apply for sphere-emulating interactions (i.e. convertible into the ScGeom type), if the virtual sphere's motion is defined correctly (see e.g. *Ig2_Sphere_ChainedCylinder_-CylScGeom*).

Short explanation of what we want to avoid :

Numerical ratcheting is best understood considering a small elastic cycle at a contact between two grains : assuming b1 is fixed, impose this displacement to b2 :

1. translation *dx* in the normal direction

2. rotation *a*

3. translation -*dx* (back to the initial position)

4. rotation -*a* (back to the initial orientation)

If the branch vector used to define the relative shear in rotation×branch is not constant (typically if it is defined from the vector center→contactPoint), then the shear displacement at the end of this cycle is not zero: rotations *a* and -*a* are multiplied by branches of different lengths.

It results in a finite contact force at the end of the cycle even though the positions and orientations are unchanged, in total contradiction with the elastic nature of the problem. It could also be seen as an *inconsistent energy creation or loss.* Given that DEM simulations tend to generate oscillations around equilibrium (damped mass-spring), it can have a significant impact on the evolution of the packings, resulting for instance in slow creep in iterations under constant load.

The solution adopted here to avoid ratcheting is as proposed by McNamara and co-workers. They analyzed the ratcheting problem in detail - even though they comment on the basis of a cycle that differs from the one shown above. One will find interesting discussions in e.g. [McNamara2008], even though solution it suggests is not fully applied here (equations of motion are not incorporating alpha, in contradiction with what is suggested by McNamara et al.).

**bases**
 Ordered list of types (as strings) this functor accepts.

**dict**(*(Serializable)arg1*) → dict :
 Return dictionary of attributes.

**interactionDetectionFactor**
 Enlarge both radii by this factor (if >1), to permit creation of distant interactions.

 InteractionGeometry will be computed when interactionDetectionFactor*(rad1+rad2) > distance.

---

**Note:** This parameter is functionally coupled with *Bo1_Sphere_Aabb::aabbEnlargeFactor*, which will create larger bounding boxes and should be of the same value.

---

**label**(*=uninitalized*)
 Textual label for this object; must be a valid python identifier, you can refer to it directly from python.

**timingDeltas**
 Detailed information about timing inside the Dispatcher itself. Empty unless enabled in the source code and O.timingEnabled==True.

**updateAttrs**(*(Serializable)arg1, (dict)arg2*) → None :
 Update object attributes from given dictionary

**class yade.wrapper.Ig2_Sphere_Sphere_ScGeom6D**(*inherits Ig2_Sphere_Sphere_ScGeom → IGeomFunctor → Functor → Serializable*)
 Create/update a *ScGeom6D* instance representing the geometry of a contact point between two *Spheres*, including relative rotations.

**avoidGranularRatcheting**
 Define relative velocity so that ratcheting is avoided. It applies for sphere-sphere contacts. It eventualy also apply for sphere-emulating interactions (i.e. convertible into the ScGeom type), if the virtual sphere's motion is defined correctly (see e.g. *Ig2_Sphere_ChainedCylinder_-CylScGeom*).

 Short explanation of what we want to avoid :

 Numerical ratcheting is best understood considering a small elastic cycle at a contact between two grains : assuming b1 is fixed, impose this displacement to b2 :

 1. translation $dx$ in the normal direction

 2. rotation $a$

 3. translation $-dx$ (back to the initial position)

 4. rotation $-a$ (back to the initial orientation)

 If the branch vector used to define the relative shear in rotation×branch is not constant (typically if it is defined from the vector center→contactPoint), then the shear displacement at the end of this cycle is not zero: rotations $a$ and $-a$ are multiplied by branches of different lengths.

 It results in a finite contact force at the end of the cycle even though the positions and orientations are unchanged, in total contradiction with the elastic nature of the problem. It could also be seen as an *inconsistent energy creation or loss*. Given that DEM simulations tend to generate oscillations around equilibrium (damped mass-spring), it can have a significant impact on the evolution of the packings, resulting for instance in slow creep in iterations under constant load.

The solution adopted here to avoid ratcheting is as proposed by McNamara and co-workers. They analyzed the ratcheting problem in detail - even though they comment on the basis of a cycle that differs from the one shown above. One will find interesting discussions in e.g. [McNamara2008], even though solution it suggests is not fully applied here (equations of motion are not incorporating alpha, in contradiction with what is suggested by McNamara et al.).

**bases**
    Ordered list of types (as strings) this functor accepts.

**creep**(*=false*)
    Substract rotational creep from relative rotation. The rotational creep *ScGeom6D::twistCreep* is a quaternion and has to be updated inside a constitutive law, see for instance *Law2_-ScGeom6D_CohFrictPhys_CohesionMoment*.

**dict**(*(Serializable)arg1*) → dict :
    Return dictionary of attributes.

**interactionDetectionFactor**
    Enlarge both radii by this factor (if >1), to permit creation of distant interactions.

    InteractionGeometry will be computed when interactionDetectionFactor*(rad1+rad2) > distance.

---

**Note:** This parameter is functionally coupled with *Bo1_Sphere_Aabb::aabbEnlargeFactor*, which will create larger bounding boxes and should be of the same value.

---

**label**(*=uninitalized*)
    Textual label for this object; must be a valid python identifier, you can refer to it directly from python.

**timingDeltas**
    Detailed information about timing inside the Dispatcher itself. Empty unless enabled in the source code and O.timingEnabled==True.

**updateAttrs**(*(Serializable)arg1, (dict)arg2*) → None :
    Update object attributes from given dictionary

**updateRotations**(*=true*)
    Precompute relative rotations. Turning this false can speed up simulations when rotations are not needed in constitutive laws (e.g. when spheres are compressed without cohesion and moment in early stage of a triaxial test), but is not foolproof. Change this value only if you know what you are doing.

**class yade.wrapper.Ig2_Tetra_Tetra_TTetraGeom**(*inherits IGeomFunctor → Functor → Serializable*)
    Create/update geometry of collision between 2 *tetrahedra* (*TTetraGeom* instance)

**bases**
    Ordered list of types (as strings) this functor accepts.

**dict**(*(Serializable)arg1*) → dict :
    Return dictionary of attributes.

**label**(*=uninitalized*)
    Textual label for this object; must be a valid python identifier, you can refer to it directly from python.

**timingDeltas**
    Detailed information about timing inside the Dispatcher itself. Empty unless enabled in the source code and O.timingEnabled==True.

**updateAttrs**(*(Serializable)arg1, (dict)arg2*) → None :
    Update object attributes from given dictionary

**class yade.wrapper.Ig2_Wall_PFacet_ScGeom**(*inherits Ig2_Wall_Sphere_ScGeom → IGeom-Functor → Functor → Serializable*)
    Create/update a *ScGeom* instance representing intersection of *Wall* and *PFacet*.

**bases**
    Ordered list of types (as strings) this functor accepts.

**dict**(*(Serializable)arg1*) → dict :
    Return dictionary of attributes.

**hertzian**(*=false*)
    The equivalent radius for the Wall (*ScGeom.refR1*) is chosen as 1e8 times the Sphere's radius (closer to Hertzian theory, where it is infinite).

**label**(*=uninitialized*)
    Textual label for this object; must be a valid python identifier, you can refer to it directly from python.

**noRatch**(*=true*)
    Avoid granular ratcheting

**timingDeltas**
    Detailed information about timing inside the Dispatcher itself. Empty unless enabled in the source code and O.timingEnabled==True.

**updateAttrs**(*(Serializable)arg1, (dict)arg2*) → None :
    Update object attributes from given dictionary

**class yade.wrapper.Ig2_Wall_Sphere_L3Geom**(*inherits Ig2_Sphere_Sphere_L3Geom → IGeomFunctor → Functor → Serializable*)
    Incrementally compute *L3Geom* for contact between *Wall* and *Sphere*. Uses attributes of *Ig2_Sphere_Sphere_L3Geom*.

**approxMask**
    Selectively enable geometrical approximations (bitmask); add the values for approximations to be enabled.

| | |
|---|---|
| 1 | use previous transformation to transform velocities (which are known at mid-steps), instead of mid-step transformation computed as quaternion slerp at t=0.5. |
| 2 | do not take average (mid-step) normal when computing relative shear displacement, use previous value instead |
| 4 | do not re-normalize average (mid-step) normal, if used.... |

**By default, the mask is zero, wherefore none of these approximations is used.**

**bases**
    Ordered list of types (as strings) this functor accepts.

**dict**(*(Serializable)arg1*) → dict :
    Return dictionary of attributes.

**distFactor**(*=1*)
    Create interaction if spheres are not further than *distFactor* *(r1+r2). If negative, zero normal deformation will be set to be the initial value (otherwise, the geometrical distance is the ''zero'' one).

**label**(*=uninitalized*)
> Textual label for this object; must be a valid python identifier, you can refer to it directly from python.

**noRatch**(*=true*)
> See *Ig2_Sphere_Sphere_ScGeom.avoidGranularRatcheting*.

**timingDeltas**
> Detailed information about timing inside the Dispatcher itself. Empty unless enabled in the source code and O.timingEnabled==True.

**trsfRenorm**(*=100*)
> How often to renormalize *trsf*; if non-positive, never renormalized (simulation might be unstable)

**updateAttrs**(*(Serializable)arg1, (dict)arg2*) → None :
> Update object attributes from given dictionary

**class yade.wrapper.Ig2_Wall_Sphere_ScGeom**(*inherits IGeomFunctor → Functor → Serializable*)
> Create/update a *ScGeom* instance representing intersection of *Wall* and *Sphere*. The equivalent radius for the Wall (*ScGeom.refR1*) is chosen equal to the Sphere's radius.

**bases**
> Ordered list of types (as strings) this functor accepts.

**dict**(*(Serializable)arg1*) → dict :
> Return dictionary of attributes.

**hertzian**(*=false*)
> The equivalent radius for the Wall (*ScGeom.refR1*) is chosen as 1e8 times the Sphere's radius (closer to Hertzian theory, where it is infinite).

**label**(*=uninitalized*)
> Textual label for this object; must be a valid python identifier, you can refer to it directly from python.

**noRatch**(*=true*)
> Avoid granular ratcheting

**timingDeltas**
> Detailed information about timing inside the Dispatcher itself. Empty unless enabled in the source code and O.timingEnabled==True.

**updateAttrs**(*(Serializable)arg1, (dict)arg2*) → None :
> Update object attributes from given dictionary

### IGeomDispatcher

**class yade.wrapper.IGeomDispatcher**(*inherits Dispatcher → Engine → Serializable*)
> Dispatcher calling *functors* based on received argument type(s).

**dead**(*=false*)
> If true, this engine will not run at all; can be used for making an engine temporarily deactivated and only resurrect it at a later point.

**dict**(*(Serializable)arg1*) → dict :
> Return dictionary of attributes.

**dispFunctor**(*(IGeomDispatcher)arg1, (Shape)arg2, (Shape)arg3*) → IGeomFunctor :
> Return functor that would be dispatched for given argument(s); None if no dispatch; ambiguous dispatch throws.

**dispMatrix**(*(IGeomDispatcher)arg1[, (bool)names=True]*) → dict :
> Return dictionary with contents of the dispatch matrix.

**execCount**
> Cumulative count this engine was run (only used if *O.timingEnabled*==`True`).

**execTime**
> Cumulative time in nanoseconds this Engine took to run (only used if *O.timingEnabled*==`True`).

**functors**
> Functors associated with this dispatcher.

**label**(*=uninitalized*)
> Textual label for this object; must be valid python identifier, you can refer to it directly from python.

**ompThreads**(*=-1*)
> Number of threads to be used in the engine. If ompThreads<0 (default), the number will be typically OMP_NUM_THREADS or the number N defined by 'yade -jN' (this behavior can depend on the engine though). This attribute will only affect engines whose code includes openMP parallel regions (e.g. *InteractionLoop*). This attribute is mostly useful for experiments or when combining *ParallelEngine* with engines that run parallel regions, resulting in nested OMP loops with different number of threads at each level.

**timingDeltas**
> Detailed information about timing inside the Engine itself. Empty unless enabled in the source code and *O.timingEnabled*==`True`.

**updateAttrs**(*(Serializable)arg1, (dict)arg2*) → None :
> Update object attributes from given dictionary

## 2.3.9 Interaction Physics creation

**IPhysFunctor**

**class** yade.wrapper.**IPhysFunctor**(*inherits Functor → Serializable*)
> Functor for creating/updating *Interaction::phys* objects from *bodies' material* properties.

> **bases**
> > Ordered list of types (as strings) this functor accepts.

> **dict**(*(Serializable)arg1*) → dict :
> > Return dictionary of attributes.

> **label**(*=uninitalized*)
> > Textual label for this object; must be a valid python identifier, you can refer to it directly from python.

> **timingDeltas**
> > Detailed information about timing inside the Dispatcher itself. Empty unless enabled in the source code and O.timingEnabled==True.

> **updateAttrs**(*(Serializable)arg1, (dict)arg2*) → None :
> > Update object attributes from given dictionary

**class** yade.wrapper.**Ip2_2xInelastCohFrictMat_InelastCohFrictPhys**(*inherits IPhysFunctor → Functor → Serializable*)
> Generates cohesive-frictional interactions with moments. Used in the contact law *Law2_ScGeom6D_InelastCohFrictPhys_CohesionMoment*.

> **bases**
> > Ordered list of types (as strings) this functor accepts.

> **dict**(*(Serializable)arg1*) → dict :
> > Return dictionary of attributes.

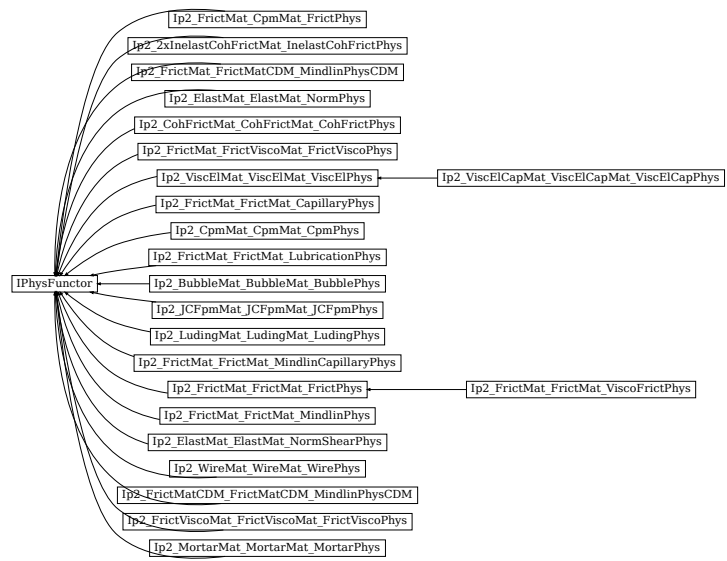

Fig. 35: Inheritance graph of IPhysFunctor. See also: *Ip2_2xInelastCohFrictMat_Inelast-CohFrictPhys*, *Ip2_BubbleMat_BubbleMat_BubblePhys*, *Ip2_CohFrictMat_CohFrictMat_CohFrict-Phys*, *Ip2_CpmMat_CpmMat_CpmPhys*, *Ip2_ElastMat_ElastMat_NormPhys*, *Ip2_ElastMat_Elast-Mat_NormShearPhys*, *Ip2_FrictMatCDM_FrictMatCDM_MindlinPhysCDM*, *Ip2_FrictMat_Cpm-Mat_FrictPhys*, *Ip2_FrictMat_FrictMatCDM_MindlinPhysCDM*, *Ip2_FrictMat_FrictMat_Capillary-Phys*, *Ip2_FrictMat_FrictMat_FrictPhys*, *Ip2_FrictMat_FrictMat_LubricationPhys*, *Ip2_FrictMat_-FrictMat_MindlinCapillaryPhys*, *Ip2_FrictMat_FrictMat_MindlinPhys*, *Ip2_FrictMat_FrictMat_-ViscoFrictPhys*, *Ip2_FrictMat_FrictViscoMat_FrictViscoPhys*, *Ip2_FrictViscoMat_FrictViscoMat_-FrictViscoPhys*, *Ip2_JCFpmMat_JCFpmMat_JCFpmPhys*, *Ip2_LudingMat_LudingMat_LudingPhys*, *Ip2_MortarMat_MortarMat_MortarPhys*, *Ip2_ViscElCapMat_ViscElCapMat_ViscElCapPhys*, *Ip2_-ViscElMat_ViscElMat_ViscElPhys*, *Ip2_WireMat_WireMat_WirePhys*.

**label**(*=uninitalized*)
  Textual label for this object; must be a valid python identifier, you can refer to it directly from python.

**timingDeltas**
  Detailed information about timing inside the Dispatcher itself. Empty unless enabled in the source code and O.timingEnabled==True.

**updateAttrs**(*(Serializable)arg1, (dict)arg2*) → None :
  Update object attributes from given dictionary

**class yade.wrapper.Ip2_BubbleMat_BubbleMat_BubblePhys**(*inherits IPhysFunctor → Functor → Serializable*)
Generates bubble interactions.Used in the contact law Law2_ScGeom__BubblePhys_Bubble.

**bases**
  Ordered list of types (as strings) this functor accepts.

**dict**(*(Serializable)arg1*) → dict :
  Return dictionary of attributes.

**label**(*=uninitalized*)
  Textual label for this object; must be a valid python identifier, you can refer to it directly from python.

**timingDeltas**
  Detailed information about timing inside the Dispatcher itself. Empty unless enabled in the source code and O.timingEnabled==True.

**updateAttrs**(*(Serializable)arg1, (dict)arg2*) → None :
  Update object attributes from given dictionary

**class yade.wrapper.Ip2_CohFrictMat_CohFrictMat_CohFrictPhys**(*inherits IPhysFunctor → Functor → Serializable*)
Generates cohesive-frictional interactions with moments, used in the contact law *Law2_ScGeom6D_CohFrictPhys_CohesionMoment*. The normal/shear stiffness and friction definitions are the same as in *Ip2_FrictMat_FrictMat_FrictPhys*, check the documentation there for details.

Adhesions related to the normal and the shear components are calculated from *CohFrictMat::normalCohesion* ($C_n$) and *CohFrictMat::shearCohesion* ($C_s$). For particles of size $R_1, R_2$ the adhesion will be $a_i = C_i \min(R_1, R_2)^2, i = n, s$.

Twist and rolling stiffnesses are proportional to the shear stiffness through dimensionless factors alphaKtw and alphaKr, such that the rotational stiffnesses are defined by $k_s \alpha_i R_1 R_2, i = tw\, r$

**bases**
  Ordered list of types (as strings) this functor accepts.

**dict**(*(Serializable)arg1*) → dict :
  Return dictionary of attributes.

**frictAngle**(*=uninitalized*)
  Instance of *MatchMaker* determining how to compute interaction's friction angle. If None, minimum value is used.

**label**(*=uninitalized*)
  Textual label for this object; must be a valid python identifier, you can refer to it directly from python.

**normalCohesion**(*=uninitalized*)
  Instance of *MatchMaker* determining tensile strength

**setCohesionNow**(*=false*)
  If true, assign cohesion to all existing contacts in current time-step. The flag is turned false automatically, so that assignment is done in the current timestep only.

**setCohesionOnNewContacts**(*=false*)
  If true, assign cohesion at all new contacts. If false, only existing contacts can be cohesive (also see *Ip2_CohFrictMat_CohFrictMat_CohFrictPhys::setCohesionNow*), and new contacts are only frictional.

**shearCohesion**(*=uninitalized*)
> Instance of *MatchMaker* determining cohesive part of the shear strength (a frictional term might be added depending on *CohFrictPhys::cohesionDisablesFriction*)

**timingDeltas**
> Detailed information about timing inside the Dispatcher itself. Empty unless enabled in the source code and O.timingEnabled==True.

**updateAttrs**(*(Serializable)arg1*, *(dict)arg2*) → None :
> Update object attributes from given dictionary

**class yade.wrapper.Ip2_CpmMat_CpmMat_CpmPhys**(*inherits IPhysFunctor → Functor → Serializable*)
> Convert 2 *CpmMat* instances to *CpmPhys* with corresponding parameters. Uses simple (arithmetic) averages if material are different. Simple copy of parameters is performed if the *material* is shared between both particles. See *cpm-model* for detals.

**E**(*=uninitalized*)
> Instance of *MatchMaker* determining how to compute interaction's normal modulus. If `None`, average value is used.

**bases**
> Ordered list of types (as strings) this functor accepts.

**cohesiveThresholdIter**(*=10*)
> Should new contacts be cohesive? They will before this iter#, they will not be afterwards. If 0, they will never be. If negative, they will always be created as cohesive (10 by default).

**dict**(*(Serializable)arg1*) → dict :
> Return dictionary of attributes.

**label**(*=uninitalized*)
> Textual label for this object; must be a valid python identifier, you can refer to it directly from python.

**timingDeltas**
> Detailed information about timing inside the Dispatcher itself. Empty unless enabled in the source code and O.timingEnabled==True.

**updateAttrs**(*(Serializable)arg1*, *(dict)arg2*) → None :
> Update object attributes from given dictionary

**class yade.wrapper.Ip2_ElastMat_ElastMat_NormPhys**(*inherits IPhysFunctor → Functor → Serializable*)
> Create a *NormPhys* from two *ElastMats*. TODO. EXPERIMENTAL

**bases**
> Ordered list of types (as strings) this functor accepts.

**dict**(*(Serializable)arg1*) → dict :
> Return dictionary of attributes.

**label**(*=uninitalized*)
> Textual label for this object; must be a valid python identifier, you can refer to it directly from python.

**timingDeltas**
> Detailed information about timing inside the Dispatcher itself. Empty unless enabled in the source code and O.timingEnabled==True.

**updateAttrs**(*(Serializable)arg1*, *(dict)arg2*) → None :
> Update object attributes from given dictionary **class yade.wrapper.Ip2_ElastMat_ElastMat_NormShearPhys**(*inherits IPhysFunctor → Functor → Serializable*)

    Create a *NormShearPhys* from two *ElastMats*. TODO. EXPERIMENTAL

    **bases**
        Ordered list of types (as strings) this functor accepts.

    **dict**(*(Serializable)arg1*) → dict :
        Return dictionary of attributes.

    **label**(*=uninitalized*)
        Textual label for this object; must be a valid python identifier, you can refer to it directly
        from python.

    **timingDeltas**
        Detailed information about timing inside the Dispatcher itself. Empty unless enabled in the
        source code and O.timingEnabled==True.

    **updateAttrs**(*(Serializable)arg1, (dict)arg2*) → None :
        Update object attributes from given dictionary

**class yade.wrapper.Ip2_FrictMatCDM_FrictMatCDM_MindlinPhysCDM**(*inherits IPhysFunctor → Functor → Serializable*)

    Create a *MindlinPhysCDM* from two *FrictMatCDMsExts*.

    **bases**
        Ordered list of types (as strings) this functor accepts.

    **dict**(*(Serializable)arg1*) → dict :
        Return dictionary of attributes.

    **frictAngle**(*=uninitalized*)
        Instance of *MatchMaker* determining how to compute interaction's friction angle. If `None`,
        minimum value is used.

    **label**(*=uninitalized*)
        Textual label for this object; must be a valid python identifier, you can refer to it directly
        from python.

    **timingDeltas**
        Detailed information about timing inside the Dispatcher itself. Empty unless enabled in the
        source code and O.timingEnabled==True.

    **updateAttrs**(*(Serializable)arg1, (dict)arg2*) → None :
        Update object attributes from given dictionary

**class yade.wrapper.Ip2_FrictMat_CpmMat_FrictPhys**(*inherits IPhysFunctor → Functor → Serializable*)

    Convert *CpmMat* instance and *FrictMat* instance to *FrictPhys* with corresponding parameters
    (young, poisson, frictionAngle). Uses simple (arithmetic) averages if material parameters are different.

    **bases**
        Ordered list of types (as strings) this functor accepts.

    **dict**(*(Serializable)arg1*) → dict :
        Return dictionary of attributes.

    **frictAngle**(*=uninitalized*)
        See *Ip2_FrictMat_FrictMat_FrictPhys*.

    **label**(*=uninitalized*)
        Textual label for this object; must be a valid python identifier, you can refer to it directly
        from python.

**timingDeltas**
> Detailed information about timing inside the Dispatcher itself. Empty unless enabled in the source code and O.timingEnabled==True.

**updateAttrs**(*(Serializable)arg1, (dict)arg2*) → None :
> Update object attributes from given dictionary

**class yade.wrapper.Ip2_FrictMat_FrictMatCDM_MindlinPhysCDM**(*inherits IPhysFunctor → Functor → Serializable*)

Create a *MindlinPhysCDM* from one FrictMat and one FrictMatCDM instance.

**bases**
> Ordered list of types (as strings) this functor accepts.

**dict**(*(Serializable)arg1*) → dict :
> Return dictionary of attributes.

**frictAngle**(*=uninitalized*)
> Instance of *MatchMaker* determining how to compute interaction's friction angle. If None, minimum value is used.

**label**(*=uninitalized*)
> Textual label for this object; must be a valid python identifier, you can refer to it directly from python.

**timingDeltas**
> Detailed information about timing inside the Dispatcher itself. Empty unless enabled in the source code and O.timingEnabled==True.

**updateAttrs**(*(Serializable)arg1, (dict)arg2*) → None :
> Update object attributes from given dictionary

**class yade.wrapper.Ip2_FrictMat_FrictMat_CapillaryPhys**(*inherits IPhysFunctor → Functor → Serializable*)

RelationShips to use with *Law2_ScGeom_CapillaryPhys_Capillarity*.

> In these RelationShips all the interaction attributes are computed.

> **Warning:** as in the others *Ip2 functors*, most of the attributes are computed only once, when the interaction is new.

**bases**
> Ordered list of types (as strings) this functor accepts.

**dict**(*(Serializable)arg1*) → dict :
> Return dictionary of attributes.

**label**(*=uninitalized*)
> Textual label for this object; must be a valid python identifier, you can refer to it directly from python.

**timingDeltas**
> Detailed information about timing inside the Dispatcher itself. Empty unless enabled in the source code and O.timingEnabled==True.

**updateAttrs**(*(Serializable)arg1, (dict)arg2*) → None :
> Update object attributes from given dictionary

**class yade.wrapper.Ip2_FrictMat_FrictMat_FrictPhys**(*inherits IPhysFunctor → Functor → Serializable*)

Create a *FrictPhys* from two *FrictMats*. The compliance of one sphere under point load is defined here as $1/(E.D)$, with $E$ the stiffness of the sphere and $D$ its diameter. The compliance of the contact itself is taken as the sum of compliances from each sphere, i.e. $1/(E_1.D_1) + 1/(E_2.D_2)$ in the general case, or $2/(E.D)$ in the special case of equal sizes and equal stiffness. Note that summing compliances is equivalent to summing the harmonic average of stiffnesses. This reasoning is applied in both the normal and the tangential directions (as in e.g. [Scholtes2009a]), hence the general form of the contact stiffness:

$k = \frac{E_1 D_1 * E_2 D_2}{E_1 D_1 + E_2 D_2} = \frac{k_1 * k_2}{k_1 + k_2}$, with $k_i = E_i D_i$.

In the above equation $E_i$ is taken equal to *FrictMat::young* of sphere $i$ for the normal stiffness, and *FrictMat::young* × *ElastMat::poisson* for the shear stiffness. In the case of a contact between a *ViscElMat* and a *FrictMat*, be sure to set *FrictMat::young* and *FrictMat::poisson*, otherwise the default value will be used.

The contact friction is defined according to *Ip2_FrictMat_FrictMat_FrictPhys::frictAngle* (minimum of the two materials by default).

**bases**
> Ordered list of types (as strings) this functor accepts.

**dict**(*(Serializable)arg1*) → dict :
> Return dictionary of attributes.

**frictAngle**(*=uninitalized*)
> Instance of *MatchMaker* determining how to compute interaction's friction angle. If `None`, minimum value is used.

**kn**(*=uninitalized*)
> Instance of *MatchMaker* determining how to compute interaction's normal stiffness. If `None`, harmonic average is used.

**ks**(*=uninitalized*)
> Instance of *MatchMaker* determining how to compute interaction's shear stiffness. If `None`, harmonic average is used.

**label**(*=uninitalized*)
> Textual label for this object; must be a valid python identifier, you can refer to it directly from python.

**timingDeltas**
> Detailed information about timing inside the Dispatcher itself. Empty unless enabled in the source code and O.timingEnabled==True.

**updateAttrs**(*(Serializable)arg1*, *(dict)arg2*) → None :
> Update object attributes from given dictionary

**class yade.wrapper.Ip2_FrictMat_FrictMat_LubricationPhys**(*inherits IPhysFunctor → Functor → Serializable*)
Ip2 creating LubricationPhys from two Material instances.

**bases**
> Ordered list of types (as strings) this functor accepts.

**dict**(*(Serializable)arg1*) → dict :
> Return dictionary of attributes.

**eps**(*=0.001*)
> Roughness: fraction of radius enlargement for contact asperities

**eta**(*=1*)
> Fluid viscosity [Pa.s]

**keps**(*=1*)
> Dimensionless stiffness coefficient of the asperities, relative to the stiffness of the surface (the final stiffness will be keps*kn). Only used with resolution method=0, with resolution>0 it is always equal to 1. [-]

**label**(*=uninitalized*)
> Textual label for this object; must be a valid python identifier, you can refer to it directly from python.

**timingDeltas**
    Detailed information about timing inside the Dispatcher itself. Empty unless enabled in the source code and O.timingEnabled==True.

**updateAttrs**(*(Serializable)arg1, (dict)arg2*) → None :
    Update object attributes from given dictionary

**class yade.wrapper.Ip2_FrictMat_FrictMat_MindlinCapillaryPhys**(*inherits  IPhysFunctor → Functor → Serializable*)

RelationShips to use with Law2_ScGeom_CapillaryPhys_Capillarity

    In these RelationShips all the interaction attributes are computed.

> **Warning:** as in the others *Ip2 functors*, most of the attributes are computed only once, when the interaction is new.

**bases**
    Ordered list of types (as strings) this functor accepts.

**betan**(*=uninitalized*)
    Normal viscous damping ratio $\beta_n$.

**betas**(*=uninitalized*)
    Shear viscous damping ratio $\beta_s$.

**dict**(*(Serializable)arg1*) → dict :
    Return dictionary of attributes.

**en**(*=uninitalized*)
    Normal coefficient of restitution $e_n$.

**es**(*=uninitalized*)
    Shear coefficient of restitution $e_s$.

**eta**(*=0.0*)
    Coefficient to determine the plastic bending moment

**gamma**(*=0.0*)
    Surface energy parameter [J/m^2] per each unit contact surface, to derive DMT formulation from HM

**krot**(*=0.0*)
    Rotational stiffness for moment contact law

**ktwist**(*=0.0*)
    Torsional stiffness for moment contact law

**label**(*=uninitalized*)
    Textual label for this object; must be a valid python identifier, you can refer to it directly from python.

**timingDeltas**
    Detailed information about timing inside the Dispatcher itself. Empty unless enabled in the source code and O.timingEnabled==True.

**updateAttrs**(*(Serializable)arg1, (dict)arg2*) → None :
    Update object attributes from given dictionary

**class yade.wrapper.Ip2_FrictMat_FrictMat_MindlinPhys**(*inherits IPhysFunctor → Functor → Serializable*)

Calculate some physical parameters needed to obtain the normal and shear stiffnesses according to the Hertz-Mindlin formulation (as implemented in PFC). The viscous damping coefficients $c_n$, $c_s$ can be specified either using viscous damping ratios ($\beta_n$, $\beta_s$) or coefficients of restitution ($e_n$, $e_s$).

\# If the viscous damping ratio $\beta_n$ ($\beta_s$) is given, it is assigned directly to *MindlinPhys.betan* (*MindlinPhys.betas*) and the viscous damping coefficient is calculated as $c_n = 2 \cdot \beta_n \cdot \sqrt{m_{bar} \cdot k_n}$ ($c_s = 2 \cdot \beta_s \cdot \sqrt{m_{bar} \cdot k_s}$), where $k_n$ ($k_s$) the tangential normal (shear) stiffness. Replacing $k_n = 3/2 \cdot k_{no} \cdot u_N^{0.5}$ ($k_s = k_{so} \cdot u_N^{0.5}$) and $k_{no} = 4/3 \cdot E \cdot \sqrt{R}$ ($k_{so} = 2 \cdot \sqrt{4 \cdot R} \cdot G/(2-\nu)$), we get $c_n = 2 \cdot \beta_n \cdot \sqrt{m_{bar}} \cdot \sqrt{2 \cdot E \cdot \sqrt{R}} \cdot u_N^{0.25}$ ($c_s = 2 \cdot \beta_s \cdot \sqrt{m_{bar}} \cdot \sqrt{4 \cdot \sqrt{R} \cdot G/(2-\nu)} \cdot u_N^{0.25}$), where $m_{bar}$, R, E, G the effective mass and mean radius, elastic and shear moduli of the interacting particles.

\# If the coefficient of restitution $\underline{e_n}$ is given instead, the normal viscous damping ratio is computed using $\beta_n = -(\log e_n)/\sqrt{\pi^2 + (\log e_n)^2}$. The shear coefficient of restitution is considered as $e_s = e_n$ and the viscous damping coefficient is calculated as $c_n = c_s = \alpha \cdot \sqrt{m_{bar}} \cdot u_N^{0.25}$, where $\alpha = 2 \cdot \sqrt{5/6} \cdot \beta_n \cdot \sqrt{2 \cdot E \cdot \sqrt{R}}$, i.e. $c_n = c_s = 2 \cdot \sqrt{5/6} \cdot \beta_n \cdot \sqrt{m_{bar}} \cdot \sqrt{2 \cdot E \cdot \sqrt{R}} \cdot u_N^{0.25}$.

In both cases, the viscous forces are calculated as $F_{n,viscous} = c_n \cdot v_n$ ($F_{s,viscous} = c_s \cdot v_s$), where $v_n$ ($v_s$) the normal (shear) component of the relative velocity. The following rules apply:

\# If $\beta_n$ and $\beta_s$ are used, then *MindlinPhys.alpha* =0; if $e_n$ is defined instead, then *Mindlin-Phys.betan = MindlinPhys.betan* =0.0.

\# It is an error (exception) to specify both $e_n$ and $\beta_n$ ($e_s$ and $\beta_s$).

\# If neither $e_n$ nor $\beta_n$ is given, zero value for *MindlinPhys.betan* is used; there will be no viscous effects.

\# If neither $e_s$ nor $\beta_s$ is given, the value of *MindlinPhys.betan* is used for *MindlinPhys.betas* as well.

\# To consider different viscous coefficients in the normal and shear contact directions, use $\beta_n$, $\beta_s$, instead of $e_n$.

The $e_n$, $\beta_n$, $e_s$, $\beta_s$ are *MatchMaker* objects; they can be constructed from float values to always return constant values. See scripts/examples/spheresFactory.py for an example of specifying $e_n$ based on combination of parameters, for different materials in contact.

**bases**
> Ordered list of types (as strings) this functor accepts.

**betan**(*=uninitalized*)
> Normal viscous damping ratio $\beta_n$.

**betas**(*=uninitalized*)
> Shear viscous damping ratio $\beta_s$.

**dict**(*(Serializable)arg1*) → dict :
> Return dictionary of attributes.

**en**(*=uninitalized*)
> Normal coefficient of restitution $e_n$.

**es**(*=uninitalized*)
> Shear coefficient of restitution $e_s$.

**eta**(*=0.0*)
> Coefficient to determine the plastic bending moment

**frictAngle**(*=uninitalized*)
> Instance of *MatchMaker* determining how to compute the friction angle of an interaction. If `None`, minimum value is used.

**gamma**(*=0.0*)
> Surface energy parameter [J/m^2] per each unit contact surface, to derive DMT formulation from HM

**krot**(*=0.0*)
> Rotational stiffness for moment contact law

**ktwist**(*=0.0*)
> Torsional stiffness for moment contact law

**label**(*=uninitalized*)
> Textual label for this object; must be a valid python identifier, you can refer to it directly from python.

**timingDeltas**
> Detailed information about timing inside the Dispatcher itself. Empty unless enabled in the source code and O.timingEnabled==True.

**updateAttrs**(*(Serializable)arg1, (dict)arg2*) → None :
> Update object attributes from given dictionary

**class yade.wrapper.Ip2_FrictMat_FrictMat_ViscoFrictPhys**(*inherits  Ip2_FrictMat_Frict-Mat_FrictPhys → IPhysFunc-tor → Functor → Serializable*)

Create a *FrictPhys* from two *FrictMats*. The compliance of one sphere under symetric point loads is defined here as 1/(E.r), with E the stiffness of the sphere and r its radius, and corresponds to a compliance 1/(2.E.r)=1/(E.D) from each contact point. The compliance of the contact itself will be the sum of compliances from each sphere, i.e. 1/(E.D1)+1/(E.D2) in the general case, or 1/(E.r) in the special case of equal sizes. Note that summing compliances corresponds to an harmonic average of stiffnesss, which is how kn is actually computed in the *Ip2_FrictMat_-FrictMat_FrictPhys* functor.

The shear stiffness ks of one sphere is defined via the material parameter *ElastMat::poisson*, as ks=poisson*kn, and the resulting shear stiffness of the interaction will be also an harmonic average.

**bases**
> Ordered list of types (as strings) this functor accepts.

**dict**(*(Serializable)arg1*) → dict :
> Return dictionary of attributes.

**frictAngle**(*=uninitalized*)
> Instance of *MatchMaker* determining how to compute interaction's friction angle. If `None`, minimum value is used.

**kn**(*=uninitalized*)
> Instance of *MatchMaker* determining how to compute interaction's normal stiffness. If `None`, harmonic average is used.

**ks**(*=uninitalized*)
> Instance of *MatchMaker* determining how to compute interaction's shear stiffness. If `None`, harmonic average is used.

**label**(*=uninitalized*)
> Textual label for this object; must be a valid python identifier, you can refer to it directly from python.

**timingDeltas**
> Detailed information about timing inside the Dispatcher itself. Empty unless enabled in the source code and O.timingEnabled==True.

**updateAttrs**(*(Serializable)arg1, (dict)arg2*) → None :
> Update object attributes from given dictionary

**class yade.wrapper.Ip2_FrictMat_FrictViscoMat_FrictViscoPhys**(*inherits IPhysFunctor → Functor → Serializable*)

Converts a *FrictMat* and *FrictViscoMat* instance to *FrictViscoPhys* with corresponding parameters. Basically this functor corresponds to *Ip2_FrictMat_FrictMat_FrictPhys* with the only difference that damping in normal direction can be considered.

**bases**
> Ordered list of types (as strings) this functor accepts.

**dict**(*(Serializable)arg1*) → dict :
  Return dictionary of attributes.

**frictAngle**(*=uninitalized*)
  Instance of *MatchMaker* determining how to compute interaction's friction angle. If `None`, minimum value is used.

**kRatio**(*=uninitalized*)
  Instance of *MatchMaker* determining how to compute interaction's shear contact stiffnesses. If this value is not given the elastic properties (i.e. poisson) of the two colliding materials are used to calculate the stiffness.

**kn**(*=uninitalized*)
  Instance of *MatchMaker* determining how to compute interaction's normal contact stiffnesses. If this value is not given the elastic properties (i.e. young) of the two colliding materials are used to calculate the stiffness.

**label**(*=uninitalized*)
  Textual label for this object; must be a valid python identifier, you can refer to it directly from python.

**timingDeltas**
  Detailed information about timing inside the Dispatcher itself. Empty unless enabled in the source code and O.timingEnabled==True.

**updateAttrs**(*(Serializable)arg1, (dict)arg2*) → None :
  Update object attributes from given dictionary

**class yade.wrapper.Ip2_FrictViscoMat_FrictViscoMat_FrictViscoPhys**(*inherits IPhys-Functor → Func-tor → Serializ-able*)

Converts 2 *FrictViscoMat* instances to *FrictViscoPhys* with corresponding parameters. Basically this functor corresponds to *Ip2_FrictMat_FrictMat_FrictPhys* with the only difference that damping in normal direction can be considered.

**bases**
  Ordered list of types (as strings) this functor accepts.

**dict**(*(Serializable)arg1*) → dict :
  Return dictionary of attributes.

**frictAngle**(*=uninitalized*)
  Instance of *MatchMaker* determining how to compute interaction's friction angle. If `None`, minimum value is used.

**kRatio**(*=uninitalized*)
  Instance of *MatchMaker* determining how to compute interaction's shear contact stiffnesses. If this value is not given the elastic properties (i.e. poisson) of the two colliding materials are used to calculate the stiffness.

**kn**(*=uninitalized*)
  Instance of *MatchMaker* determining how to compute interaction's normal contact stiffnesses. If this value is not given the elastic properties (i.e. young) of the two colliding materials are used to calculate the stiffness.

**label**(*=uninitalized*)
  Textual label for this object; must be a valid python identifier, you can refer to it directly from python.

**timingDeltas**
  Detailed information about timing inside the Dispatcher itself. Empty unless enabled in the source code and O.timingEnabled==True.

**updateAttrs**(*(Serializable)arg1, (dict)arg2*) → None :
  Update object attributes from given dictionary **class yade.wrapper.Ip2_JCFpmMat_JCFpmMat_JCFpmPhys**(*inherits IPhysFunctor → Functor → Serializable*)

Converts 2 *JCFpmMat* instances to one *JCFpmPhys* instance, with corresponding parameters. See *JCFpmMat* and [Duriez2016] for details

**bases**
    Ordered list of types (as strings) this functor accepts.

**cohesiveTresholdIteration**(*=1*)
    should new contacts be cohesive? If strictly negativ, they will in any case. If positiv, they will before this iter, they won't afterward.

**dict**(*(Serializable)arg1*) → dict :
    Return dictionary of attributes.

**label**(*=uninitalized*)
    Textual label for this object; must be a valid python identifier, you can refer to it directly from python.

**timingDeltas**
    Detailed information about timing inside the Dispatcher itself. Empty unless enabled in the source code and O.timingEnabled==True.

**updateAttrs**(*(Serializable)arg1, (dict)arg2*) → None :
    Update object attributes from given dictionary

**weibullCutOffMax**(*=10*)
    Factor that cuts off the largest values of the weibull distributed interaction areas.

**weibullCutOffMin**(*=0.*)
    Factor that cuts off the smallest values of the weibull distributed interaction areas.

**xSectionWeibullScaleParameter**(*=1*)
    Scale parameter used to generate interaction radii for the crosssectional areas (changing strength criteria only) according to Weibull distribution. Activated for any value other than 0. Needs to be combined with a *shape parameter*

**xSectionWeibullShapeParameter**(*=0*)
    Shape parameter used to generate interaction radii for the crossSectional areas (changing strength criteria only) according to Weibull distribution. Activated for any value other than 0. Needs to be combined with a *scale parameter*)

**class yade.wrapper.Ip2_LudingMat_LudingMat_LudingPhys**(*inherits IPhysFunctor → Functor → Serializable*)

Convert 2 instances of *LudingMat* to *LudingPhys* using the rule of consecutive connection.

**bases**
    Ordered list of types (as strings) this functor accepts.

**dict**(*(Serializable)arg1*) → dict :
    Return dictionary of attributes.

**label**(*=uninitalized*)
    Textual label for this object; must be a valid python identifier, you can refer to it directly from python.

**timingDeltas**
    Detailed information about timing inside the Dispatcher itself. Empty unless enabled in the source code and O.timingEnabled==True.

**updateAttrs**(*(Serializable)arg1, (dict)arg2*) → None :
    Update object attributes from given dictionary

**class yade.wrapper.Ip2_MortarMat_MortarMat_MortarPhys**(*inherits IPhysFunctor → Functor → Serializable*)

Ip2 creating MortarPhys from two MortarMat instances.

**bases**
> Ordered list of types (as strings) this functor accepts.

**cohesiveThresholdIter**(*=2*)
> Should new contacts be cohesive? They will before this iter#, they will not be afterwards. If <=0, they will never be.

**dict**(*(Serializable)arg1*) → dict :
> Return dictionary of attributes.

**label**(*=uninitalized*)
> Textual label for this object; must be a valid python identifier, you can refer to it directly from python.

**timingDeltas**
> Detailed information about timing inside the Dispatcher itself. Empty unless enabled in the source code and O.timingEnabled==True.

**updateAttrs**(*(Serializable)arg1*, *(dict)arg2*) → None :
> Update object attributes from given dictionary

**class yade.wrapper.Ip2_ViscElCapMat_ViscElCapMat_ViscElCapPhys**(*inherits   Ip2_ViscEl-Mat_ViscElMat_Vis-cElPhys  →  IPhys-Functor  →  Functor → Serializable*)

Convert 2 instances of *ViscElCapMat* to *ViscElCapPhys* using the rule of consecutive connection.

**bases**
> Ordered list of types (as strings) this functor accepts.

**dict**(*(Serializable)arg1*) → dict :
> Return dictionary of attributes.

**en**(*=uninitalized*)
> Instance of *MatchMaker* determining restitution coefficient in normal direction

**et**(*=uninitalized*)
> Instance of *MatchMaker* determining restitution coefficient in tangential direction

**frictAngle**(*=uninitalized*)
> Instance of *MatchMaker* determining how to compute interaction's friction angle. If `None`, minimum value is used.

**label**(*=uninitalized*)
> Textual label for this object; must be a valid python identifier, you can refer to it directly from python.

**tc**(*=uninitalized*)
> Instance of *MatchMaker* determining contact time

**timingDeltas**
> Detailed information about timing inside the Dispatcher itself. Empty unless enabled in the source code and O.timingEnabled==True.

**updateAttrs**(*(Serializable)arg1*, *(dict)arg2*) → None :
> Update object attributes from given dictionary

**class yade.wrapper.Ip2_ViscElMat_ViscElMat_ViscElPhys**(*inherits IPhysFunctor → Functor → Serializable*)

Convert 2 instances of *ViscElMat* to *ViscElPhys* using the rule of consecutive connection.

**bases**
> Ordered list of types (as strings) this functor accepts.

**dict**(*(Serializable)arg1*) → dict :
> Return dictionary of attributes.

**en**(*=uninitalized*)
Instance of *MatchMaker* determining restitution coefficient in normal direction

**et**(*=uninitalized*)
Instance of *MatchMaker* determining restitution coefficient in tangential direction

**frictAngle**(*=uninitalized*)
Instance of *MatchMaker* determining how to compute interaction's friction angle. If `None`, minimum value is used.

**label**(*=uninitalized*)
Textual label for this object; must be a valid python identifier, you can refer to it directly from python.

**tc**(*=uninitalized*)
Instance of *MatchMaker* determining contact time

**timingDeltas**
Detailed information about timing inside the Dispatcher itself. Empty unless enabled in the source code and O.timingEnabled==True.

**updateAttrs**(*(Serializable)arg1, (dict)arg2*) → None :
Update object attributes from given dictionary

**class yade.wrapper.Ip2_WireMat_WireMat_WirePhys**(*inherits IPhysFunctor → Functor → Serializable*)
Converts 2 *WireMat* instances to *WirePhys* with corresponding parameters.

**bases**
Ordered list of types (as strings) this functor accepts.

**dict**(*(Serializable)arg1*) → dict :
Return dictionary of attributes.

**label**(*=uninitalized*)
Textual label for this object; must be a valid python identifier, you can refer to it directly from python.

**linkThresholdIteration**(*=1*)
Iteration to create the link.

**timingDeltas**
Detailed information about timing inside the Dispatcher itself. Empty unless enabled in the source code and O.timingEnabled==True.

**updateAttrs**(*(Serializable)arg1, (dict)arg2*) → None :
Update object attributes from given dictionary

### IPhysDispatcher

**class yade.wrapper.IPhysDispatcher**(*inherits Dispatcher → Engine → Serializable*)
Dispatcher calling *functors* based on received argument type(s).

**dead**(*=false*)
If true, this engine will not run at all; can be used for making an engine temporarily deactivated and only resurrect it at a later point.

**dict**(*(Serializable)arg1*) → dict :
Return dictionary of attributes.

**dispFunctor**(*(IPhysDispatcher)arg1, (Material)arg2, (Material)arg3*) → IPhysFunctor :
Return functor that would be dispatched for given argument(s); None if no dispatch; ambiguous dispatch throws.

**dispMatrix**(*(IPhysDispatcher)arg1*[, *(bool)names=True*]) → dict :
Return dictionary with contents of the dispatch matrix.

**execCount**
Cumulative count this engine was run (only used if *O.timingEnabled*==`True`).

**execTime**
Cumulative time in nanoseconds this Engine took to run (only used if *O.timingEnabled*==`True`).

**functors**
Functors associated with this dispatcher.

**label**(*=uninitialized*)
Textual label for this object; must be valid python identifier, you can refer to it directly from python.

**ompThreads**(*=-1*)
Number of threads to be used in the engine. If ompThreads<0 (default), the number will be typically OMP_NUM_THREADS or the number N defined by 'yade -jN' (this behavior can depend on the engine though). This attribute will only affect engines whose code includes openMP parallel regions (e.g. *InteractionLoop*). This attribute is mostly useful for experiments or when combining *ParallelEngine* with engines that run parallel regions, resulting in nested OMP loops with different number of threads at each level.

**timingDeltas**
Detailed information about timing inside the Engine itself. Empty unless enabled in the source code and *O.timingEnabled*==`True`.

**updateAttrs**(*(Serializable)arg1, (dict)arg2*) → None :
Update object attributes from given dictionary

## 2.3.10 Constitutive laws

**LawFunctor**

**class** yade.wrapper.**LawFunctor**(*inherits Functor → Serializable*)
Functor for applying constitutive laws on *interactions*.

**bases**
Ordered list of types (as strings) this functor accepts.

**dict**(*(Serializable)arg1*) → dict :
Return dictionary of attributes.

**label**(*=uninitialized*)
Textual label for this object; must be a valid python identifier, you can refer to it directly from python.

**timingDeltas**
Detailed information about timing inside the Dispatcher itself. Empty unless enabled in the source code and O.timingEnabled==True.

**updateAttrs**(*(Serializable)arg1, (dict)arg2*) → None :
Update object attributes from given dictionary

**class** yade.wrapper.**Law2_ChCylGeom6D_CohFrictPhys_CohesionMoment**(*inherits LawFunctor → Functor → Serializable*)
Law for linear compression, and Mohr-Coulomb plasticity surface without cohesion. This law implements the classical linear elastic-plastic law from [CundallStrack1979] (see also [Pfc3dManual30]). The normal force is (with the convention of positive tensile forces) $F_n = \min(k_n u_n, 0)$. The shear force is $F_s = k_s u_s$, the plasticity condition defines the maximum value of the shear force : $F_s^{max} = F_n \tan(\varphi)$, with $\varphi$ the friction angle.

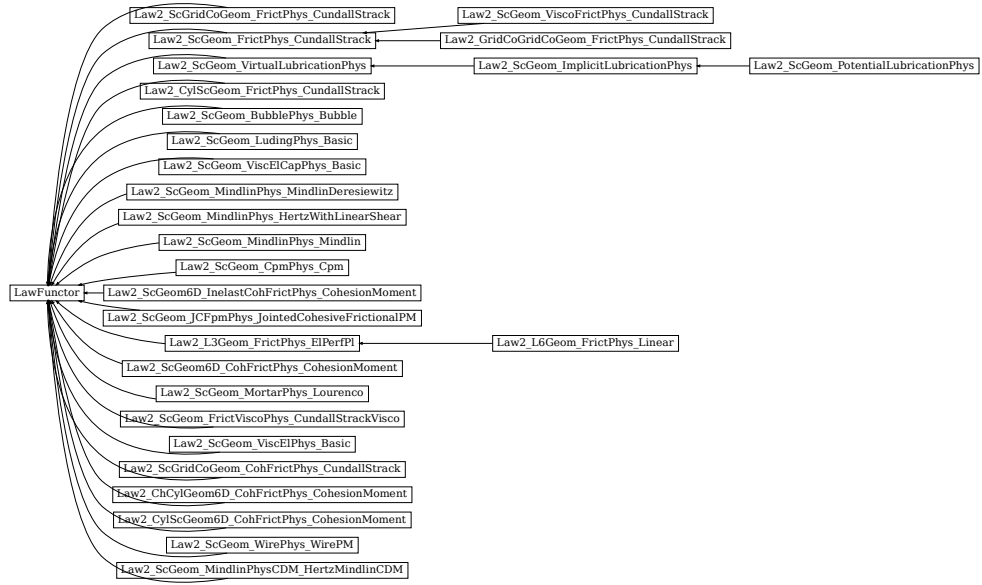

Fig. 36: Inheritance graph of LawFunctor. See also: *Law2__ChCylGeom6D__CohFrictPhys__CohesionMoment*, *Law2__CylScGeom6D__CohFrictPhys__CohesionMoment*, *Law2__CylScGeom__FrictPhys__CundallStrack*, *Law2__GridCoGridCoGeom__FrictPhys__CundallStrack*, *Law2__L3Geom__FrictPhys__ElPerfPl*, *Law2__L6Geom__FrictPhys__Linear*, *Law2__ScGeom6D__CohFrictPhys__CohesionMoment*, *Law2__ScGeom6D__InelastCohFrictPhys__CohesionMoment*, *Law2__ScGeom__BubblePhys__Bubble*, *Law2__ScGeom__CpmPhys__Cpm*, *Law2__ScGeom__FrictPhys__CundallStrack*, *Law2__ScGeom__FrictViscoPhys__CundallStrackVisco*, *Law2__ScGeom__ImplicitLubricationPhys*, *Law2__ScGeom__JCFpmPhys__JointedCohesiveFrictionalPM*, *Law2__ScGeom__LudingPhys__Basic*, *Law2__ScGeom__MindlinPhysCDM__HertzMindlinCDM*, *Law2__ScGeom__MindlinPhys__HertzWithLinearShear*, *Law2__ScGeom__MindlinPhys__Mindlin*, *Law2__ScGeom__MindlinPhys__MindlinDeresiewitz*, *Law2__ScGeom__MortarPhys__Lourenco*, *Law2__ScGeom__PotentialLubricationPhys*, *Law2__ScGeom__VirtualLubricationPhys*, *Law2__ScGeom__ViscElCapPhys__Basic*, *Law2__ScGeom__ViscElPhys__Basic*, *Law2__ScGeom__ViscoFrictPhys__CundallStrack*, *Law2__ScGeom__WirePhys__WirePM*, *Law2__ScGridCoGeom__CohFrictPhys__CundallStrack*, *Law2__ScGridCoGeom__FrictPhys__CundallStrack*.

---

**Note:** This law is well tested in the context of triaxial simulation, and has been used for a number of published results (see e.g. [Scholtes2009b] and other papers from the same authors). It is generalised by *Law2_ScGeom6D_CohFrictPhys_CohesionMoment*, which adds cohesion and moments at contact.

---

**always_use_moment_law**(*=false*)
    If true, use bending/twisting moments at all contacts. If false, compute moments only for cohesive contacts.

**bases**
    Ordered list of types (as strings) this functor accepts.

**creep_viscosity**(*=1*)
    creep viscosity [Pa.s/m]. probably should be moved to Ip2_CohFrictMat_CohFrictMat_-CohFrictPhys…

**dict**(*(Serializable)arg1*) → dict :
    Return dictionary of attributes.

**label**(*=uninitalized*)
    Textual label for this object; must be a valid python identifier, you can refer to it directly from python.

**neverErase**(*=false*)
    Keep interactions even if particles go away from each other (only in case another constitutive law is in the scene, e.g. *Law2_ScGeom_CapillaryPhys_Capillarity*)

**shear_creep**(*=false*)
    activate creep on the shear force, using *CohesiveFrictionalContactLaw::creep_viscosity*.

**timingDeltas**
    Detailed information about timing inside the Dispatcher itself. Empty unless enabled in the source code and O.timingEnabled==True.

**twist_creep**(*=false*)
    activate creep on the twisting moment, using *CohesiveFrictionalContactLaw::creep_viscosity*.

**updateAttrs**(*(Serializable)arg1, (dict)arg2*) → None :
    Update object attributes from given dictionary

**useIncrementalForm**(*=false*)
    use the incremental formulation to compute bending and twisting moments. Creep on the twisting moment is not included in such a case.

**class yade.wrapper.Law2_CylScGeom6D_CohFrictPhys_CohesionMoment**(*inherits LawFunctor → Functor → Serializable*)
    This law generalises *Law2_CylScGeom_FrictPhys_CundallStrack* by adding cohesion and moments at contact.

**always_use_moment_law**(*=false*)
    If true, use bending/twisting moments at all contacts. If false, compute moments only for cohesive contacts.

**bases**
    Ordered list of types (as strings) this functor accepts.

**creep_viscosity**(*=1*)
    creep viscosity [Pa.s/m]. probably should be moved to Ip2_CohFrictMat_CohFrictMat_-CohFrictPhys…

**dict**(*(Serializable)arg1*) → dict :
    Return dictionary of attributes.

**label**(*=uninitalized*)
>   Textual label for this object; must be a valid python identifier, you can refer to it directly from python.

**neverErase**(*=false*)
>   Keep interactions even if particles go away from each other (only in case another constitutive law is in the scene, e.g. *Law2_ScGeom_CapillaryPhys_Capillarity*)

**shear_creep**(*=false*)
>   activate creep on the shear force, using *CohesiveFrictionalContactLaw::creep_viscosity*.

**timingDeltas**
>   Detailed information about timing inside the Dispatcher itself. Empty unless enabled in the source code and O.timingEnabled==True.

**twist_creep**(*=false*)
>   activate creep on the twisting moment, using *CohesiveFrictionalContactLaw::creep_viscosity*.

**updateAttrs**(*(Serializable)arg1, (dict)arg2*) → None :
>   Update object attributes from given dictionary

**useIncrementalForm**(*=false*)
>   use the incremental formulation to compute bending and twisting moments. Creep on the twisting moment is not included in such a case.

**class yade.wrapper.Law2_CylScGeom_FrictPhys_CundallStrack**(*inherits* *LawFunctor* → *Functor* → *Serializable*)

Law for linear compression, and Mohr-Coulomb plasticity surface without cohesion. This law implements the classical linear elastic-plastic law from [CundallStrack1979] (see also [Pfc3dManual30]). The normal force is (with the convention of positive tensile forces) $F_n = \min(k_n u_n, 0)$. The shear force is $F_s = k_s u_s$, the plasticity condition defines the maximum value of the shear force : $F_s^{\max} = F_n \tan(\varphi)$, with $\varphi$ the friction angle.

---

**Note:** This law uses *ScGeom*.

---

---

**Note:** This law is well tested in the context of triaxial simulation, and has been used for a number of published results (see e.g. [Scholtes2009b] and other papers from the same authors). It is generalised by *Law2_ScGeom6D_CohFrictPhys_CohesionMoment*, which adds cohesion and moments at contact.

---

**bases**
>   Ordered list of types (as strings) this functor accepts.

**dict**(*(Serializable)arg1*) → dict :
>   Return dictionary of attributes.

**label**(*=uninitalized*)
>   Textual label for this object; must be a valid python identifier, you can refer to it directly from python.

**neverErase**(*=false*)
>   Keep interactions even if particles go away from each other (only in case another constitutive law is in the scene, e.g. *Law2_ScGeom_CapillaryPhys_Capillarity*)

**timingDeltas**
>   Detailed information about timing inside the Dispatcher itself. Empty unless enabled in the source code and O.timingEnabled==True.

**updateAttrs**(*(Serializable)arg1, (dict)arg2*) → None :
>   Update object attributes from given dictionary **class** yade.wrapper.**Law2_GridCoGridCoGeom_FrictPhys_CundallStrack**(*inherits Law2_Sc-GeoM__FrictPhys_-CundallStrack → LawFunctor → Functor → Serializ-able*)

Frictional elastic contact law between two *gridConnection* . See *Law2__ScGeom__FrictPhys__Cun-dallStrack* for more details.

**bases**
Ordered list of types (as strings) this functor accepts.

**dict**(*(Serializable)arg1*) → dict :
Return dictionary of attributes.

**elasticEnergy**(*(Law2_ScGeom__FrictPhys_CundallStrack)arg1*) → float :
Compute and return the total elastic energy in all "FrictPhys" contacts

**initPlasticDissipation**(*(Law2_ScGeom__FrictPhys_CundallStrack)arg1, (float)arg2*) → None :
Initialize cummulated plastic dissipation to a value (0 by default).

**label**(*=uninitalized*)
Textual label for this object; must be a valid python identifier, you can refer to it directly from python.

**neverErase**(*=false*)
Keep interactions even if particles go away from each other (only in case another constitutive law is in the scene, e.g. *Law2__ScGeom__CapillaryPhys__Capillarity*)

**plasticDissipation**(*(Law2_ScGeom__FrictPhys_CundallStrack)arg1*) → float :
Total energy dissipated in plastic slips at all FrictPhys contacts. Computed only if *Law2_-ScGeom__FrictPhys__CundallStrack::traceEnergy* is true.

**sphericalBodies**(*=true*)
If true, compute branch vectors from radii (faster), else use contactPoint-position. Turning this flag true is safe for sphere-sphere contacts and a few other specific cases. It will give wrong values of torques on facets or boxes.

**timingDeltas**
Detailed information about timing inside the Dispatcher itself. Empty unless enabled in the source code and O.timingEnabled==True.

**traceEnergy**(*=false*)
Define the total energy dissipated in plastic slips at all contacts. This will trace only plastic energy in this law, see O.trackEnergy for a more complete energies tracing

**updateAttrs**(*(Serializable)arg1, (dict)arg2*) → None :
Update object attributes from given dictionary

**class** yade.wrapper.**Law2_L3Geom_FrictPhys_ElPerfPl**(*inherits LawFunctor → Functor → Se-rializable*)

Basic law for testing *L3Geom*; it bears no cohesion (unless *noBreak* is True), and plastic slip obeys the Mohr-Coulomb criterion (unless *noSlip* is True).

**bases**
Ordered list of types (as strings) this functor accepts.

**dict**(*(Serializable)arg1*) → dict :
Return dictionary of attributes.

**label**(*=uninitalized*)
Textual label for this object; must be a valid python identifier, you can refer to it directly from python.

**noBreak**(*=false*)
Do not break contacts when particles separate.

**noSlip**(*=false*)
No plastic slipping.

**timingDeltas**
Detailed information about timing inside the Dispatcher itself. Empty unless enabled in the source code and O.timingEnabled==True.

**updateAttrs**(*(Serializable)arg1, (dict)arg2*) → None :
Update object attributes from given dictionary

**class** yade.wrapper.**Law2_L6Geom_FrictPhys_Linear**(*inherits* *Law2_L3Geom_FrictPhys_-ElPerfPl → LawFunctor → Functor → Serializable*)
Basic law for testing *L6Geom* – linear in both normal and shear sense, without slip or breakage.

**bases**
Ordered list of types (as strings) this functor accepts.

**charLen**(*=1*)
Characteristic length with the meaning of the stiffness ratios bending/shear and torsion/normal.

**dict**(*(Serializable)arg1*) → dict :
Return dictionary of attributes.

**label**(*=uninitalized*)
Textual label for this object; must be a valid python identifier, you can refer to it directly from python.

**noBreak**(*=false*)
Do not break contacts when particles separate.

**noSlip**(*=false*)
No plastic slipping.

**timingDeltas**
Detailed information about timing inside the Dispatcher itself. Empty unless enabled in the source code and O.timingEnabled==True.

**updateAttrs**(*(Serializable)arg1, (dict)arg2*) → None :
Update object attributes from given dictionary

**class** yade.wrapper.**Law2_ScGeom6D_CohFrictPhys_CohesionMoment**(*inherits* *LawFunctor → Functor → Serializable*)
Law for linear traction-compression-bending-twisting, with cohesion+friction and Mohr-Coulomb plasticity surface. This law adds adhesion and moments to *Law2_ScGeom_FrictPhys_Cundall-Strack*.

The normal force is (with the convention of positive tensile forces) $F_n = \min(k_n * (u_n - u_n^p), a_n)$, with $a_n$ the normal adhesion and $u_n^p$ the plastic part of normal displacement. The shear force is $F_s = k_s * u_s$, the plasticity condition defines the maximum value of the shear force, by default $F_s^{max} = F_n * \tan(\varphi) + a_s$, with $\varphi$ the friction angle and $a_s$ the shear adhesion. If *CohFrict-Phys::cohesionDisablesFriction* is True, friction is ignored as long as adhesion is active, and the maximum shear force is only $F_s^{max} = a_s$.

If the maximum tensile or maximum shear force is reached and *CohFrictPhys::fragile* =True (default), the cohesive link is broken, and $a_n, a_s$ are set back to zero. If a tensile force is present, the contact is lost, else the shear strength is $F_s^{max} = F_n * \tan(\varphi)$. If *CohFrictPhys::fragile* =False, the behaviour is perfectly plastic, and the shear strength is kept constant.

If *Law2_ScGeom6D_CohFrictPhys_CohesionMoment::momentRotationLaw* =True, bending and twisting moments are computed using a linear law with moduli respectively $k_t$ and $k_r$, so that the moments are : $M_b = k_b * \Theta_b$ and $M_t = k_t * \Theta_t$, with $\Theta_{b,t}$ the relative rotations between interacting bodies (details can be found in [Bourrier2013]). The maximum value of moments can be defined and takes the form of rolling friction. Cohesive -type moment may also be included in the future.

Creep at contact is implemented in this law, as defined in [Hassan2010]. If activated, there is a viscous behaviour of the shear and twisting components, and the evolution of the elastic parts of shear displacement and relative twist is given by $du_{s,e}/dt = -F_s/\nu_s$ and $d\Theta_{t,e}/dt = -M_t/\nu_t$.

**always_use_moment_law**(=*false*)
> If true, use bending/twisting moments at all contacts. If false, compute moments only for cohesive contacts. Both cases also require *CohFrictPhys::momentRotationLaw* to be true.

**bases**
> Ordered list of types (as strings) this functor accepts.

**bendingElastEnergy**(*(Law2_ScGeom6D_CohFrictPhys_CohesionMoment)arg1*) → float :
> Compute bending elastic energy.

**creep_viscosity**(=*1*)
> creep viscosity [Pa.s/m]. probably should be moved to Ip2_CohFrictMat_CohFrictMat_-CohFrictPhys.

**dict**(*(Serializable)arg1*) → dict :
> Return dictionary of attributes.

**elasticEnergy**(*(Law2_ScGeom6D_CohFrictPhys_CohesionMoment)arg1*) → float :
> Compute total elastic energy.

**initPlasticDissipation**(*(Law2_ScGeom6D_CohFrictPhys_CohesionMoment)arg1*,
>                          *(float)arg2*) → None :
> Initialize cummulated plastic dissipation to a value (0 by default).

**label**(=*uninitialized*)
> Textual label for this object; must be a valid python identifier, you can refer to it directly from python.

**neverErase**(=*false*)
> Keep interactions even if particles go away from each other (only in case another constitutive law is in the scene, e.g. *Law2_ScGeom_CapillaryPhys_Capillarity*)

**normElastEnergy**(*(Law2_ScGeom6D_CohFrictPhys_CohesionMoment)arg1*) → float :
> Compute normal elastic energy.

**plasticDissipation**(*(Law2_ScGeom6D_CohFrictPhys_CohesionMoment)arg1*) → float :
> Total energy dissipated in plastic slips at all CohFrictPhys contacts. Computed only if *Law2_-ScGeom_FrictPhys_CundallStrack::traceEnergy* is true.

**shearElastEnergy**(*(Law2_ScGeom6D_CohFrictPhys_CohesionMoment)arg1*) → float :
> Compute shear elastic energy.

**shear_creep**(=*false*)
> activate creep on the shear force, using *CohesiveFrictionalContactLaw::creep_viscosity*.

**timingDeltas**
> Detailed information about timing inside the Dispatcher itself. Empty unless enabled in the source code and O.timingEnabled==True.

**traceEnergy**(=*false*)
> Define the total energy dissipated in plastic slips at contacts. Note that it will not reflect any energy associated to de-bonding, as it may occur for fragile contacts, nor does it include plastic dissipation in traction.

**twistElastEnergy**(*(Law2_ScGeom6D_CohFrictPhys_CohesionMoment)arg1*) → float :
> Compute twist elastic energy.

**twist_creep**(=*false*)
> activate creep on the twisting moment, using *CohesiveFrictionalContactLaw::creep_viscosity*.

**updateAttrs**(*(Serializable)arg1, (dict)arg2*) → None :
> Update object attributes from given dictionary

**useIncrementalForm**(*=false*)
> use the incremental formulation to compute bending and twisting moments. Creep on the twisting moment is not included in such a case.

**class yade.wrapper.Law2_ScGeom6D_InelastCohFrictPhys_CohesionMoment**(*inherits Law-Functor → Functor → Serializable*)

This law is currently under developpement. Final version and documentation will come before the end of 2014.

**bases**
> Ordered list of types (as strings) this functor accepts.

**dict**(*(Serializable)arg1*) → dict :
> Return dictionary of attributes.

**label**(*=uninitalized*)
> Textual label for this object; must be a valid python identifier, you can refer to it directly from python.

**normElastEnergy**(*(Law2_ScGeom6D_InelastCohFrictPhys_CohesionMoment)arg1*) → float :
> Compute normal elastic energy.

**shearElastEnergy**(*(Law2_ScGeom6D_InelastCohFrictPhys_CohesionMoment)arg1*) → float :
> Compute shear elastic energy.

**timingDeltas**
> Detailed information about timing inside the Dispatcher itself. Empty unless enabled in the source code and O.timingEnabled==True.

**updateAttrs**(*(Serializable)arg1, (dict)arg2*) → None :
> Update object attributes from given dictionary

**class yade.wrapper.Law2_ScGeom_BubblePhys_Bubble**(*inherits LawFunctor → Functor → Serializable*)

Constitutive law for Bubble model.

**bases**
> Ordered list of types (as strings) this functor accepts.

**dict**(*(Serializable)arg1*) → dict :
> Return dictionary of attributes.

**label**(*=uninitalized*)
> Textual label for this object; must be a valid python identifier, you can refer to it directly from python.

**pctMaxForce**(*=0.1*)
> Chan[2011] states the contact law is valid only for small interferences; therefore an exponential force-displacement curve models the contact stiffness outside that regime (large penetration). This artificial stiffening ensures that bubbles will not pass through eachother or completely overlap during the simulation. The maximum force is Fmax = (2*pi*surfaceTension*rAvg). pctMaxForce is the percentage of the maximum force dictates the separation threshold, Dmax, for each contact. Penetrations less than Dmax calculate the reaction force from the derived contact law, while penetrations equal to or greater than Dmax calculate the reaction force from the artificial exponential curve.

**surfaceTension**(*=0.07197*)
> The surface tension in the liquid surrounding the bubbles. The default value is that of water at 25 degrees Celcius.

**timingDeltas**
> Detailed information about timing inside the Dispatcher itself. Empty unless enabled in the source code and O.timingEnabled==True.

> **updateAttrs**(*(Serializable)arg1, (dict)arg2*) → None :
>> Update object attributes from given dictionary

**class yade.wrapper.Law2_ScGeom_CpmPhys_Cpm**(*inherits LawFunctor → Functor → Serializable*)

Constitutive law for the *cpm-model*.

> **bases**
>> Ordered list of types (as strings) this functor accepts.

> **dict**(*(Serializable)arg1*) → dict :
>> Return dictionary of attributes.

> **elasticEnergy**(*(Law2_ScGeom_CpmPhys_Cpm)arg1*) → float :
>> Compute and return the total elastic energy in all "CpmPhys" contacts

> **epsSoft**(*=1., approximates confinement (for -3e-3) -20MPa precisely, -100MPa a little over, -200 and -400 are OK (secant)*)
>> Strain at which softening in compression starts (non-negative to deactivate). The default value is such that plasticity does not occur

> **label**(*=uninitalized*)
>> Textual label for this object; must be a valid python identifier, you can refer to it directly from python.

> **omegaThreshold**(*=1., >=1. to deactivate, i.e. never delete any contacts*)
>> damage after which the contact disappears (<1), since omega reaches 1 only for strain $\to +\infty$

> **relKnSoft**(*=.3*)
>> Relative rigidity of the softening branch in compression (0=perfect elastic-plastic, <0 softening, >0 hardening)

> **timingDeltas**
>> Detailed information about timing inside the Dispatcher itself. Empty unless enabled in the source code and O.timingEnabled==True.

> **updateAttrs**(*(Serializable)arg1, (dict)arg2*) → None :
>> Update object attributes from given dictionary

> **yieldEllipseShift**(*=NaN*)
>> horizontal scaling of the ellipse (shifts on the +x axis as interactions with +y are given)

> **yieldLogSpeed**(*=.1*)
>> scaling in the logarithmic yield surface (should be <1 for realistic results; >=0 for meaningful results)

> **yieldSigmaTMagnitude**(*(Law2_ScGeom_CpmPhys_Cpm)arg1, (float)sigmaN, (float)omega, (float)undamagedCohesion, (float)tanFrictionAngle*) → float :
>> Return radius of yield surface for given material and state parameters; uses attributes of the current instance (*yieldSurfType* etc), change them before calling if you need that.

> **yieldSurfType**(*=2*)
>> yield function: 0: mohr-coulomb (original); 1: parabolic; 2: logarithmic, 3: log+lin_tension, 4: elliptic, 5: elliptic+log

**class yade.wrapper.Law2_ScGeom_FrictPhys_CundallStrack**(*inherits LawFunctor → Functor → Serializable*)

Law for linear compression, and Mohr-Coulomb plasticity surface without cohesion. This law implements the classical linear elastic-plastic law from [CundallStrack1979] (see also [Pfc3dManual30]). The normal force is (with the convention of positive tensile forces) $F_n = \min(k_n u_n, 0)$. The shear force is $F_s = k_s u_s$, the plasticity condition defines the maximum value of the shear force : $F_s^{max} = F_n \tan(\varphi)$, with $\varphi$ the friction angle.

This law is well tested in the context of triaxial simulation, and has been used for a number of published results (see e.g. [Scholtes2009b] and other papers from the same authors). It is generalised by *Law2_ScGeom6D_CohFrictPhys_CohesionMoment*, which adds cohesion and moments at contact.

**bases**
> Ordered list of types (as strings) this functor accepts.

**dict**(*(Serializable)arg1*) → dict :
> Return dictionary of attributes.

**elasticEnergy**(*(Law2_ScGeom_FrictPhys_CundallStrack)arg1*) → float :
> Compute and return the total elastic energy in all "FrictPhys" contacts

**initPlasticDissipation**(*(Law2_ScGeom_FrictPhys_CundallStrack)arg1, (float)arg2*) →
> None :
> Initialize cummulated plastic dissipation to a value (0 by default).

**label**(*=uninitalized*)
> Textual label for this object; must be a valid python identifier, you can refer to it directly
> from python.

**neverErase**(*=false*)
> Keep interactions even if particles go away from each other (only in case another constitutive
> law is in the scene, e.g. *Law2_ScGeom_CapillaryPhys_Capillarity*)

**plasticDissipation**(*(Law2_ScGeom_FrictPhys_CundallStrack)arg1*) → float :
> Total energy dissipated in plastic slips at all FrictPhys contacts. Computed only if *Law2_-
> ScGeom_FrictPhys_CundallStrack::traceEnergy* is true.

**sphericalBodies**(*=true*)
> If true, compute branch vectors from radii (faster), else use contactPoint-position. Turning
> this flag true is safe for sphere-sphere contacts and a few other specific cases. It will give
> wrong values of torques on facets or boxes.

**timingDeltas**
> Detailed information about timing inside the Dispatcher itself. Empty unless enabled in the
> source code and O.timingEnabled==True.

**traceEnergy**(*=false*)
> Define the total energy dissipated in plastic slips at all contacts. This will trace only plastic
> energy in this law, see O.trackEnergy for a more complete energies tracing

**updateAttrs**(*(Serializable)arg1, (dict)arg2*) → None :
> Update object attributes from given dictionary

**class** yade.wrapper.**Law2_ScGeom_FrictViscoPhys_CundallStrackVisco**(*inherits LawFunctor → Functor → Serializable*)
> Constitutive law for the FrictViscoPM. Corresponds to *Law2_ScGeom_FrictPhys_CundallStrack*
> with the only difference that viscous damping in normal direction can be considered.

**bases**
> Ordered list of types (as strings) this functor accepts.

**dict**(*(Serializable)arg1*) → dict :
> Return dictionary of attributes.

**elasticEnergy**(*(Law2_ScGeom_FrictViscoPhys_CundallStrackVisco)arg1*) → float :
> Compute and return the total elastic energy in all "FrictViscoPhys" contacts

**initPlasticDissipation**(*(Law2_ScGeom_FrictViscoPhys_CundallStrackVisco)arg1,*
> *(float)arg2*) → None :
> Initialize cummulated plastic dissipation to a value (0 by default).

**label**(*=uninitalized*)
> Textual label for this object; must be a valid python identifier, you can refer to it directly
> from python.

**neverErase**(*=false*)
> Keep interactions even if particles go away from each other (only in case another constitutive
> law is in the scene, e.g. *Law2_ScGeom_CapillaryPhys_Capillarity*)

**plasticDissipation**(*(Law2_ScGeom_FrictViscoPhys_CundallStrackVisco)arg1*) → float :
    Total energy dissipated in plastic slips at all FrictPhys contacts. Computed only if
    :yref:Law2_ScGeom_FrictViscoPhys_CundallStrackVisco::traceEnergy' is true.

**sphericalBodies**(*=true*)
    If true, compute branch vectors from radii (faster), else use contactPoint-position. Turning
    this flag true is safe for sphere-sphere contacts and a few other specific cases. It will give
    wrong values of torques on facets or boxes.

**timingDeltas**
    Detailed information about timing inside the Dispatcher itself. Empty unless enabled in the
    source code and O.timingEnabled==True.

**traceEnergy**(*=false*)
    Define the total energy dissipated in plastic slips at all contacts. This will trace only plastic
    energy in this law, see O.trackEnergy for a more complete energies tracing

**updateAttrs**(*(Serializable)arg1, (dict)arg2*) → None :
    Update object attributes from given dictionary

**class yade.wrapper.Law2_ScGeom_ImplicitLubricationPhys**(*inherits Law2_ScGeom__Virtu-
                                                      alLubricationPhys → LawFunc-
                                                      tor → Functor → Serializable*)
Material law for lubrication and contact between two spheres, solved using implicit method. The
full description of this contact law is available in [Chevremont2020] . Several resolution methods are
available. Iterative exact, solving the 2nd order polynomia. Other resolutions methods are nu-
merical (Newton-Rafson and Dichotomy) with a variable change $\delta = \log(\mathfrak{u})$, solved in dimentionless
coordinates.

**MaxDist**(*=2.*)
    Maximum distance (d/a) for the interaction

**MaxIter**(*=30*)
    Maximum iterations for numerical resolution (Dichotomy and Newton-Rafson)

**SolutionTol**(*=1.e-8*)
    Tolerance for numerical resolution (Dichotomy and Newton-Rafson)

**activateRollLubrication**(*=true*)
    Activate roll lubrication (default: true)

**activateTangencialLubrication**(*=true*)
    Activate tangencial lubrication (default: true)

**activateTwistLubrication**(*=true*)
    Activate twist lubrication (default: true)

**bases**
    Ordered list of types (as strings) this functor accepts.

**dict**(*(Serializable)arg1*) → dict :
    Return dictionary of attributes.

**static getStressForEachBody**() → tuple :
    Get stresses tensors for each bodies: normal contact stress, shear contact stress, normal
    lubrication stress, shear lubrication stress, stress from additionnal potential forces.

**static getTotalStresses**() → tuple :
    Get total stresses tensors: normal contact stress, shear contact stress, normal lubrication
    stress, shear lubrication stress, stress from additionnal potential forces.

**label**(*=uninitalized*)
    Textual label for this object; must be a valid python identifier, you can refer to it directly
    from python.

**maxSubSteps**(*=4*)
max recursion depth of adaptative timestepping in the theta-method, the minimal time interval is thus $Omega::dt/2^{\mathtt{depth}}$. If still not converged the integrator will switch to backward Euler.

**resolution**(*=0*)
Change normal component resolution method, 0: Iterative exact resolution with substepping (theta method, linear contact), 1: Newton-Rafson dimensionless resolution (theta method, linear contact), 2: (default) Dichotomy dimensionless resolution (theta method, linear contact), 3: Exact dimensionless solution with contact prediction (theta method, linear contact). Method 3 is better if the volumic fraction is not too high. Use 2 otherwise.

**theta**(*=0.55*)
parameter of the 'theta'-method, 1: backward Euler, 0.5: trapezoidal rule, 0: not used, 0.55: suggested optimum)

**timingDeltas**
Detailed information about timing inside the Dispatcher itself. Empty unless enabled in the source code and O.timingEnabled==True.

**updateAttrs**(*(Serializable)arg1, (dict)arg2*) → None :
Update object attributes from given dictionary

**class yade.wrapper.Law2_ScGeom_JCFpmPhys_JointedCohesiveFrictionalPM**(*inherits Law-Functor → Functor → Serializable*)

Interaction law for cohesive frictional material, e.g. rock, possibly presenting joint surfaces, that can be mechanically described with a smooth contact logic [Ivars2011] (implemented in Yade in [Scholtes2012]). See examples/jointedCohesiveFrictionalPM for script examples. Joint surface definitions (through stl meshes or direct definition with gts module) are illustrated there.

**Key**(*=""*)
string specifying the name of saved file 'cracks____.txt', when *recordCracks* is true.

**bases**
Ordered list of types (as strings) this functor accepts.

**clusterMoments**(*=true*)
computer clustered moments? (on by default

**computedCentroid**(*=false*)
computer clustered moments?

**cracksFileExist**(*=false*)
if true (and if *recordCracks*), data are appended to an existing 'cracksKey' text file; otherwise its content is reset.

**dict**(*(Serializable)arg1*) → dict :
Return dictionary of attributes.

**eventNumber**(*=0*)
cluster event number (used for clustering and paraview visualization of groups).

**label**(*=uninitalized*)
Textual label for this object; must be a valid python identifier, you can refer to it directly from python.

**momentFudgeFactor**(*=1.*)
Fudge factor used by Hazzard and Damjanac 2013 to improve moment size accuracy (set to 1 for no impact by default)

**momentRadiusFactor**(*=5.*)
Average particle diameter multiplier for moment magnitude calculation

**momentsFileExist**(*=false*)

    if true (and if *recordCracks*), data are appended to an existing 'momentsKey' text file; otherwise its content is reset.

**nbShearCracks**(*=0*)

    number of shear microcracks.

**nbTensCracks**(*=0*)

    number of tensile microcracks.

**neverErase**(*=false*)

    Keep interactions even if particles go away from each other (only in case another constitutive law is in the scene

**recordCracks**(*=false*)

    if true, data about interactions that lose their cohesive feature are stored in the text file cracksKey.txt (see *Key* and *cracksFileExist*). It contains 9 columns: the break iteration, the 3 coordinates of the contact point, the type (1 means shear break, while 0 corresponds to tensile break), the ''cross section'' (mean radius of the 2 spheres) and the 3 coordinates of the contact normal.

**recordMoments**(*=false*)

    Combines with :yref: *Key<Law2ScGeom__JCFpmPhys__JointedCohesiveFrictionalPM.Key>* to compute acoustic emissions according to clustered broken bond method? (off by default)

**smoothJoint**(*=false*)

    if true, interactions of particles belonging to joint surface (*JCFpmPhys.isOnJoint*) are handled according to a smooth contact logic [Ivars2011], [Scholtes2012].

**timingDeltas**

    Detailed information about timing inside the Dispatcher itself. Empty unless enabled in the source code and O.timingEnabled==True.

**totalCracksSurface**(*=0.*)

    calculate the total cracked surface.

**totalShearCracksE**(*=0.*)

    calculate the overall energy dissipated by interparticle microcracking in shear.

**totalTensCracksE**(*=0.*)

    calculate the overall energy dissipated by interparticle microcracking in tension.

**updateAttrs**(*(Serializable)arg1, (dict)arg2*) → None :

    Update object attributes from given dictionary

**useStrainEnergy**(*=true*)

    use strain energy for moment magnitude estimation (if false, use kinetic energy)

**class yade.wrapper.Law2_ScGeom_LudingPhys_Basic**(*inherits LawFunctor → Functor → Serializable*)

    Linear viscoelastic model operating on *ScGeom* and *LudingPhys*. See [Luding2008] ,[Singh2013]_- for more details.

**bases**

    Ordered list of types (as strings) this functor accepts.

**dict**(*(Serializable)arg1*) → dict :

    Return dictionary of attributes.

**label**(*=uninitalized*)

    Textual label for this object; must be a valid python identifier, you can refer to it directly from python.

**timingDeltas**

    Detailed information about timing inside the Dispatcher itself. Empty unless enabled in the source code and O.timingEnabled==True.

**updateAttrs**(*(Serializable)arg1, (dict)arg2*) → None :
Update object attributes from given dictionary

**class** yade.wrapper.**Law2_ScGeom_MindlinPhysCDM_HertzMindlinCDM**(*inherits LawFunctor →*
*Functor → Serializ-*
*able*)

Hertz-Mindlin model extended: Normal direction: conical damage model from Harkness et al.
2016./ Suhr & Six 2017. Tangential direction: stress dependent interparticle friction coefficient,
Suhr & Six 2016. Both models can be switched on/off separately. In this version there is NO
damping (neither viscous nor linear), NO adhesion and NO calc_energy, NO includeMoment, NO
preventGranularRatcheting. NOT tested for periodic simulations.

**bases**
Ordered list of types (as strings) this functor accepts.

**dict**(*(Serializable)arg1*) → dict :
Return dictionary of attributes.

**label**(*=uninitalized*)
Textual label for this object; must be a valid python identifier, you can refer to it directly
from python.

**neverErase**(*=false*)
Keep interactions even if particles go away from each other (only in case another constitutive
law is in the scene, e.g. *Law2_ScGeom_CapillaryPhys_Capillarity*)

**ratioSlidingContacts**(*(Law2_ScGeom_MindlinPhysCDM_HertzMindlinCDM)arg1*) →
float :
Return the ratio between the number of contacts sliding to the total number at a given time.

**ratioYieldingContacts**(*(Law2_ScGeom_MindlinPhysCDM_HertzMindlinCDM)arg1*) →
float :
Return the ratio between the number of contacts yielding to the total number at a given time.

**timingDeltas**
Detailed information about timing inside the Dispatcher itself. Empty unless enabled in the
source code and O.timingEnabled==True.

**updateAttrs**(*(Serializable)arg1, (dict)arg2*) → None :
Update object attributes from given dictionary

**class** yade.wrapper.**Law2_ScGeom_MindlinPhys_HertzWithLinearShear**(*inherits LawFunctor*
*→ Functor → Serial-*
*izable*)

Constitutive law for the Hertz formulation (using *MindlinPhys.kno*) and linear behavior in shear
(using *MindlinPhys.kso* for stiffness and *FrictPhys.tangensOfFrictionAngle*).

---

**Note:** No viscosity or damping. If you need those, look at *Law2_ScGeom_MindlinPhys_Mindlin*,
which also includes non-linear Mindlin shear.

---

**bases**
Ordered list of types (as strings) this functor accepts.

**dict**(*(Serializable)arg1*) → dict :
Return dictionary of attributes.

**label**(*=uninitalized*)
Textual label for this object; must be a valid python identifier, you can refer to it directly
from python.

**neverErase**(*=false*)
Keep interactions even if particles go away from each other (only in case another constitutive
law is in the scene, e.g. *Law2_ScGeom_CapillaryPhys_Capillarity*)

**nonLin**(*=0*)
> Shear force nonlinearity (the value determines how many features of the non-linearity are taken in account). 1: ks as in HM 2: shearElastic increment computed as in HM 3. granular ratcheting disabled.

**timingDeltas**
> Detailed information about timing inside the Dispatcher itself. Empty unless enabled in the source code and O.timingEnabled==True.

**updateAttrs**(*(Serializable)arg1, (dict)arg2*) → None :
> Update object attributes from given dictionary

**class yade.wrapper.Law2_ScGeom_MindlinPhys_Mindlin**(*inherits LawFunctor → Functor → Serializable*)

Constitutive law for the Hertz-Mindlin formulation. It includes non linear elasticity in the normal direction as predicted by Hertz for two non-conforming elastic contact bodies. In the shear direction, instead, it reseambles the simplified case without slip discussed in Mindlin's paper, where a linear relationship between shear force and tangential displacement is provided. Finally, the Mohr-Coulomb criterion is employed to established the maximum friction force which can be developed at the contact. Moreover, it is also possible to include the effect of linear viscous damping through the definition of the parameters $\beta_n$ and $\beta_s$.

**bases**
> Ordered list of types (as strings) this functor accepts.

**calcEnergy**(*=false*)
> bool to calculate energy terms (shear potential energy, dissipation of energy due to friction and dissipation of energy due to normal and tangential damping)

**contactsAdhesive**(*(Law2_ScGeom_MindlinPhys_Mindlin)arg1*) → float :
> Compute total number of adhesive contacts.

**dict**(*(Serializable)arg1*) → dict :
> Return dictionary of attributes.

**frictionDissipation**(*=uninitalized*)
> Energy dissipation due to sliding

**includeAdhesion**(*=false*)
> bool to include the adhesion force following the DMT formulation. If true, also the normal elastic energy takes into account the adhesion effect.

**includeMoment**(*=false*)
> bool to consider rolling resistance (if *Ip2_FrictMat_FrictMat_MindlinPhys::eta* is 0.0, no plastic condition is applied.)

**label**(*=uninitalized*)
> Textual label for this object; must be a valid python identifier, you can refer to it directly from python.

**neverErase**(*=false*)
> Keep interactions even if particles go away from each other (only in case another constitutive law is in the scene, e.g. *Law2_ScGeom_CapillaryPhys_Capillarity*)

**normDampDissip**(*=uninitalized*)
> Energy dissipated by normal damping

**normElastEnergy**(*(Law2_ScGeom_MindlinPhys_Mindlin)arg1*) → float :
> Compute normal elastic potential energy. It handles the DMT formulation if *Law2_ScGeom_MindlinPhys_Mindlin::includeAdhesion* is set to true.

**preventGranularRatcheting**(*=true*)
> bool to avoid granular ratcheting

**ratioSlidingContacts**(*(Law2_ScGeom_MindlinPhys_Mindlin)arg1*) → float :
> Return the ratio between the number of contacts sliding to the total number at a given time.

> **shearDampDissip**(*=uninitalized*)
>> Energy dissipated by tangential damping

> **shearEnergy**(*=uninitalized*)
>> Shear elastic potential energy

> **timingDeltas**
>> Detailed information about timing inside the Dispatcher itself. Empty unless enabled in the source code and O.timingEnabled==True.

> **updateAttrs**(*(Serializable)arg1, (dict)arg2*) → None :
>> Update object attributes from given dictionary

**class** yade.wrapper.**Law2_ScGeom_MindlinPhys_MindlinDeresiewitz**(*inherits LawFunctor →*
*Functor → Serializable*)

Hertz-Mindlin contact law with partial slip solution, as described in [Thornton1991].

> **bases**
>> Ordered list of types (as strings) this functor accepts.

> **dict**(*(Serializable)arg1*) → dict :
>> Return dictionary of attributes.

> **label**(*=uninitalized*)
>> Textual label for this object; must be a valid python identifier, you can refer to it directly from python.

> **neverErase**(*=false*)
>> Keep interactions even if particles go away from each other (only in case another constitutive law is in the scene, e.g. *Law2_ScGeom_CapillaryPhys_Capillarity*)

> **timingDeltas**
>> Detailed information about timing inside the Dispatcher itself. Empty unless enabled in the source code and O.timingEnabled==True.

> **updateAttrs**(*(Serializable)arg1, (dict)arg2*) → None :
>> Update object attributes from given dictionary

**class** yade.wrapper.**Law2_ScGeom_MortarPhys_Lourenco**(*inherits LawFunctor → Functor →*
*Serializable*)

Material law for mortar layer according to [Lourenco1994]. The contact behaves elastic until brittle failure when reaching strength envelope. The envelope has three parts.

Tensile with condition $\sigma_N - f_t$.

Shear part with Mohr-Coulomb condition $|\sigma_T| + \sigma_N \tan\varphi - c$.

Compressive part with condition $\sigma_N^2 + A^2\sigma_T^2 - f_c^2$

The main idea is to begin simulation with this model and when the contact is broken, to use standard non-cohesive Law2_PolyhedraGeom_PolyhedraPhys_Volumetric.

> **bases**
>> Ordered list of types (as strings) this functor accepts.

> **dict**(*(Serializable)arg1*) → dict :
>> Return dictionary of attributes.

> **label**(*=uninitalized*)
>> Textual label for this object; must be a valid python identifier, you can refer to it directly from python.

> **timingDeltas**
>> Detailed information about timing inside the Dispatcher itself. Empty unless enabled in the source code and O.timingEnabled==True.

**updateAttrs**(*(Serializable)arg1, (dict)arg2*) → None :
    Update object attributes from given dictionary

**class** yade.wrapper.**Law2_ScGeom_PotentialLubricationPhys**(*inherits Law2_ScGeom_-ImplicitLubricationPhys → Law2_ScGeom_VirtualLubricationPhys → LawFunctor → Functor → Serializable*)

Material law for lubrication + potential between two spheres. The potential model include contact. This material law will solve the system with lubrication and the provided potential.

**MaxDist**(*=2.*)
    Maximum distance (d/a) for the interaction

**MaxIter**(*=30*)
    Maximum iterations for numerical resolution (Dichotomy and Newton-Rafson)

**SolutionTol**(*=1.e-8*)
    Tolerance for numerical resolution (Dichotomy and Newton-Rafson)

**activateRollLubrication**(*=true*)
    Activate roll lubrication (default: true)

**activateTangencialLubrication**(*=true*)
    Activate tangencial lubrication (default: true)

**activateTwistLubrication**(*=true*)
    Activate twist lubrication (default: true)

**bases**
    Ordered list of types (as strings) this functor accepts.

**dict**(*(Serializable)arg1*) → dict :
    Return dictionary of attributes.

**static getStressForEachBody**() → tuple :
    Get stresses tensors for each bodies: normal contact stress, shear contact stress, normal lubrication stress, shear lubrication stress, stress from additionnal potential forces.

**static getTotalStresses**() → tuple :
    Get total stresses tensors: normal contact stress, shear contact stress, normal lubrication stress, shear lubrication stress, stress from additionnal potential forces.

**label**(*=uninitalized*)
    Textual label for this object; must be a valid python identifier, you can refer to it directly from python.

**maxSubSteps**(*=4*)
    max recursion depth of adaptative timestepping in the theta-method, the minimal time interval is thus $Omega::dt/2^{\text{depth}}$. If still not converged the integrator will switch to backward Euler.

**potential**(*=new GenericPotential()*)
    Physical potential force between spheres.

**resolution**(*=0*)
    Change normal component resolution method, 0: Iterative exact resolution with substepping (theta method, linear contact), 1: Newton-Rafson dimensionless resolution (theta method, linear contact), 2: (default) Dichotomy dimensionless resolution (theta method, linear contact), 3: Exact dimensionless solution with contact prediction (theta method, linear contact). Method 3 is better if the volumic fraction is not too high. Use 2 otherwise.

**theta**(*=0.55*)
    parameter of the 'theta'-method, 1: backward Euler, 0.5: trapezoidal rule, 0: not used, 0.55: suggested optimum)

**timingDeltas**
> Detailed information about timing inside the Dispatcher itself. Empty unless enabled in the source code and O.timingEnabled==True.

**updateAttrs**(*(Serializable)arg1, (dict)arg2*) → None :
> Update object attributes from given dictionary

**class yade.wrapper.Law2_ScGeom_VirtualLubricationPhys**(*inherits LawFunctor → Functor → Serializable*)
Virtual class for sheared lubrication functions. This don't do any computation and shouldn't be used directly!

**MaxDist**(*=2.*)
> Maximum distance (d/a) for the interaction

**activateRollLubrication**(*=true*)
> Activate roll lubrication (default: true)

**activateTangencialLubrication**(*=true*)
> Activate tangencial lubrication (default: true)

**activateTwistLubrication**(*=true*)
> Activate twist lubrication (default: true)

**bases**
> Ordered list of types (as strings) this functor accepts.

**dict**(*(Serializable)arg1*) → dict :
> Return dictionary of attributes.

**static getStressForEachBody**() → tuple :
> Get stresses tensors for each bodies: normal contact stress, shear contact stress, normal lubrication stress, shear lubrication stress, stress from additionnal potential forces.

**static getTotalStresses**() → tuple :
> Get total stresses tensors: normal contact stress, shear contact stress, normal lubrication stress, shear lubrication stress, stress from additionnal potential forces.

**label**(*=uninitalized*)
> Textual label for this object; must be a valid python identifier, you can refer to it directly from python.

**timingDeltas**
> Detailed information about timing inside the Dispatcher itself. Empty unless enabled in the source code and O.timingEnabled==True.

**updateAttrs**(*(Serializable)arg1, (dict)arg2*) → None :
> Update object attributes from given dictionary

**class yade.wrapper.Law2_ScGeom_ViscElCapPhys_Basic**(*inherits LawFunctor → Functor → Serializable*)
Extended version of Linear viscoelastic model with capillary parameters.

**NLiqBridg**(*=uninitalized*)
> The total number of liquid bridges

**VLiqBridg**(*=uninitalized*)
> The total volume of liquid bridges

**bases**
> Ordered list of types (as strings) this functor accepts.

**dict**(*(Serializable)arg1*) → dict :
> Return dictionary of attributes.

**label**(*=uninitalized*)
> Textual label for this object; must be a valid python identifier, you can refer to it directly from python.

**timingDeltas**
> Detailed information about timing inside the Dispatcher itself. Empty unless enabled in the source code and O.timingEnabled==True.

**updateAttrs**(*(Serializable)arg1, (dict)arg2*) → None :
> Update object attributes from given dictionary

**class yade.wrapper.Law2_ScGeom_ViscElPhys_Basic**(*inherits LawFunctor → Functor → Serializable*)

Linear viscoelastic model operating on ScGeom and ViscElPhys. The contact law is visco-elastic in the normal direction, and visco-elastic frictional in the tangential direction. The normal contact is modelled as a spring of equivalent stiffness $k_n$, placed in parallel with a viscous damper of equivalent viscosity $c_n$. As for the tangential contact, it is made of a spring-dashpot system (in parallel with equivalent stiffness $k_s$ and viscosity $c_s$) in serie with a slider of friction coefficient $\mu = \tan \varphi$.

The friction coefficient $\mu = \tan \varphi$ is always evaluated as $\tan(\min(\varphi_1, \varphi_2))$, where $\varphi_1$ and $\varphi_2$ are respectively the friction angle of particle 1 and 2. For the other parameters, depending on the material input, the equivalent parameters of the contact $(K_n, C_n, K_s, C_s, \varphi)$ are evaluated differently. In the following, the quantities in parenthesis are the material constant which are precised for each particle. They are then associated to particle 1 and 2 (e.g. $kn_1, kn_2, cn_1...$), and should not be confused with the equivalent parameters of the contact $(K_n, C_n, K_s, C_s, \varphi)$.

- If contact time (tc), normal and tangential restitution coefficient (en,et) are precised, the equivalent parameters are evaluated following the formulation of Pournin [Pournin2001].

- If normal and tangential stiffnesses (kn, ks) and damping constant (cn,cs) of each particle are precised, the equivalent stiffnesses and damping constants of each contact made of two particles 1 and 2 is made $A = 2\frac{a_1 a_2}{a_1 + a_2}$, where A is $K_n$, $K_s$, $C_n$ and $C_s$, and 1 and 2 refer to the value associated to particle 1 and 2.

- Alternatively it is possible to precise the Young's modulus (young) and Poisson's ratio (poisson) instead of the normal and spring constant (kn and ks). In this case, the equivalent parameters are evaluated the same way as the previous case with $kn_x = E_x d_x$, $ks_x = \nu_x kn_x$, where $E_x$, $\nu_x$ and $d_x$ are Young's modulus, Poisson's ratio and diameter of particle x.

- If Young's modulus (young), Poisson's ratio (poisson), normal and tangential restitution coefficient (en,et)are precised, the equivalent stiffnesses are evaluated as previously: $K_n = 2\frac{kn_1 kn_2}{kn_1 + kn_2}$, $kn_x = E_x d_x$, $K_s = 2(ks_1 ks_2)/(ks_1 + ks_2)$, $ks_x = \nu kn_x$. The damping constant is computed at each contact in order to fulfill the normal restitution coefficient $e_n = (en_1 + en_2)/2$. This is achieved resolving numerically equation 21 of [Schwager2007] (There is in fact a mistake in the article from equation 18 to 19, so that there is a change in sign). Be careful in this configuration the tangential restitution coefficient is set to 1 (no tangential damping). This formulation imposes directly the normal restitution coefficient of the collisions instead of the damping constant.

**bases**
> Ordered list of types (as strings) this functor accepts.

**dict**(*(Serializable)arg1*) → dict :
> Return dictionary of attributes.

**label**(*=uninitalized*)
> Textual label for this object; must be a valid python identifier, you can refer to it directly from python.

**timingDeltas**
> Detailed information about timing inside the Dispatcher itself. Empty unless enabled in the source code and O.timingEnabled==True.

**updateAttrs**(*(Serializable)arg1, (dict)arg2*) → None :
> Update object attributes from given dictionary **class yade.wrapper.Law2_ScGeom_ViscoFrictPhys_CundallStrack**(*inherits Law2_ScGeom_-FrictPhys_CundallStrack → LawFunctor → Functor → Serializable*)

Law similar to *Law2_ScGeom_FrictPhys_CundallStrack* with the addition of shear creep at contacts.

**bases**
> Ordered list of types (as strings) this functor accepts.

**creepStiffness**(*=1*)

**dict**(*(Serializable)arg1*) → dict :
> Return dictionary of attributes.

**elasticEnergy**(*(Law2_ScGeom_FrictPhys_CundallStrack)arg1*) → float :
> Compute and return the total elastic energy in all "FrictPhys" contacts

**initPlasticDissipation**(*(Law2_ScGeom_FrictPhys_CundallStrack)arg1, (float)arg2*) → None :
> Initialize cummulated plastic dissipation to a value (0 by default).

**label**(*=uninitalized*)
> Textual label for this object; must be a valid python identifier, you can refer to it directly from python.

**neverErase**(*=false*)
> Keep interactions even if particles go away from each other (only in case another constitutive law is in the scene, e.g. *Law2_ScGeom_CapillaryPhys_Capillarity*)

**plasticDissipation**(*(Law2_ScGeom_FrictPhys_CundallStrack)arg1*) → float :
> Total energy dissipated in plastic slips at all FrictPhys contacts. Computed only if *Law2_-ScGeom_FrictPhys_CundallStrack::traceEnergy* is true.

**shearCreep**(*=false*)

**sphericalBodies**(*=true*)
> If true, compute branch vectors from radii (faster), else use contactPoint-position. Turning this flag true is safe for sphere-sphere contacts and a few other specific cases. It will give wrong values of torques on facets or boxes.

**timingDeltas**
> Detailed information about timing inside the Dispatcher itself. Empty unless enabled in the source code and O.timingEnabled==True.

**traceEnergy**(*=false*)
> Define the total energy dissipated in plastic slips at all contacts. This will trace only plastic energy in this law, see O.trackEnergy for a more complete energies tracing

**updateAttrs**(*(Serializable)arg1, (dict)arg2*) → None :
> Update object attributes from given dictionary

**viscosity**(*=1*)

**class yade.wrapper.Law2_ScGeom_WirePhys_WirePM**(*inherits LawFunctor → Functor → Serializable*)

Constitutive law for the wire model.

**bases**
> Ordered list of types (as strings) this functor accepts.

**dict**(*(Serializable)arg1*) → dict :
> Return dictionary of attributes.

**label**(*=uninitalized*)
> Textual label for this object; must be a valid python identifier, you can refer to it directly from python.

**linkThresholdIteration**(*=1*)
> Iteration to create the link.

**timingDeltas**
> Detailed information about timing inside the Dispatcher itself. Empty unless enabled in the source code and O.timingEnabled==True.

**updateAttrs**(*(Serializable)arg1, (dict)arg2*) → None :
> Update object attributes from given dictionary

**class yade.wrapper.Law2_ScGridCoGeom_CohFrictPhys_CundallStrack**(*inherits LawFunctor → Functor → Serializable*)

Law between a cohesive frictional *GridConnection* and a cohesive frictional *Sphere*. Almost the same than *Law2_ScGeom6D_CohFrictPhys_CohesionMoment*, but THE ROTATIONAL MOMENTS ARE NOT COMPUTED.

**bases**
> Ordered list of types (as strings) this functor accepts.

**dict**(*(Serializable)arg1*) → dict :
> Return dictionary of attributes.

**label**(*=uninitalized*)
> Textual label for this object; must be a valid python identifier, you can refer to it directly from python.

**neverErase**(*=false*)
> Keep interactions even if particles go away from each other (only in case another constitutive law is in the scene, e.g. *Law2_ScGeom_CapillaryPhys_Capillarity*)

**timingDeltas**
> Detailed information about timing inside the Dispatcher itself. Empty unless enabled in the source code and O.timingEnabled==True.

**updateAttrs**(*(Serializable)arg1, (dict)arg2*) → None :
> Update object attributes from given dictionary

**class yade.wrapper.Law2_ScGridCoGeom_FrictPhys_CundallStrack**(*inherits LawFunctor → Functor → Serializable*)

Law between a frictional *GridConnection* and a frictional *Sphere*. Almost the same than *Law2_ScGeom_FrictPhys_CundallStrack*, but the force is divided and applied on the two *GridNodes* only.

**bases**
> Ordered list of types (as strings) this functor accepts.

**dict**(*(Serializable)arg1*) → dict :
> Return dictionary of attributes.

**label**(*=uninitalized*)
> Textual label for this object; must be a valid python identifier, you can refer to it directly from python.

**neverErase**(*=false*)
> Keep interactions even if particles go away from each other (only in case another constitutive law is in the scene, e.g. *Law2_ScGeom_CapillaryPhys_Capillarity*)

**timingDeltas**
> Detailed information about timing inside the Dispatcher itself. Empty unless enabled in the source code and O.timingEnabled==True.

**updateAttrs**(*(Serializable)arg1, (dict)arg2*) → None :
> Update object attributes from given dictionary

**LawDispatcher**

**class** yade.wrapper.**LawDispatcher**(*inherits Dispatcher → Engine → Serializable*)
Dispatcher calling *functors* based on received argument type(s).

**dead**(*=false*)
If true, this engine will not run at all; can be used for making an engine temporarily deactivated and only resurrect it at a later point.

**dict**(*(Serializable)arg1*) → dict :
Return dictionary of attributes.

**dispFunctor**(*(LawDispatcher)arg1, (IGeom)arg2, (IPhys)arg3*) → LawFunctor :
Return functor that would be dispatched for given argument(s); None if no dispatch; ambiguous dispatch throws.

**dispMatrix**(*(LawDispatcher)arg1*[, *(bool)names=True*]) → dict :
Return dictionary with contents of the dispatch matrix.

**execCount**
Cumulative count this engine was run (only used if *O.timingEnabled*==True).

**execTime**
Cumulative time in nanoseconds this Engine took to run (only used if *O.timingEnabled*==True).

**functors**
Functors associated with this dispatcher.

**label**(*=uninitalized*)
Textual label for this object; must be valid python identifier, you can refer to it directly from python.

**ompThreads**(*=-1*)
Number of threads to be used in the engine. If ompThreads<0 (default), the number will be typically OMP_NUM_THREADS or the number N defined by 'yade -jN' (this behavior can depend on the engine though). This attribute will only affect engines whose code includes openMP parallel regions (e.g. *InteractionLoop*). This attribute is mostly useful for experiments or when combining *ParallelEngine* with engines that run parallel regions, resulting in nested OMP loops with different number of threads at each level.

**timingDeltas**
Detailed information about timing inside the Engine itself. Empty unless enabled in the source code and *O.timingEnabled*==True.

**updateAttrs**(*(Serializable)arg1, (dict)arg2*) → None :
Update object attributes from given dictionary

## 2.3.11 Internal forces

**InternalForceFunctor**

**InternalForceDispatcher**

## 2.3.12 Callbacks

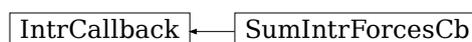

Fig. 37: Inheritance graph of IntrCallback. See also: *SumIntrForcesCb*.

**class** `yade.wrapper.`**IntrCallback**(*inherits Serializable*)

Abstract callback object which will be called for every (real) *Interaction* after the interaction has been processed by *InteractionLoop*.

At the beginning of the interaction loop, `stepInit` is called, initializing the object; it returns either `NULL` (to deactivate the callback during this time step) or pointer to function, which will then be passed (1) pointer to the callback object itself and (2) pointer to *Interaction*.

> **Note:** (NOT YET DONE) This functionality is accessible from python by passing 4th argument to *InteractionLoop* constructor, or by appending the callback object to *InteractionLoop::callbacks*.

**dict**(*(Serializable)arg1*) → dict :
    Return dictionary of attributes.

**updateAttrs**(*(Serializable)arg1, (dict)arg2*) → None :
    Update object attributes from given dictionary

**class** `yade.wrapper.`**SumIntrForcesCb**(*inherits IntrCallback → Serializable*)

Callback summing magnitudes of forces over all interactions. *IPhys* of interactions must derive from *NormShearPhys* (responsability fo the user).

**dict**(*(Serializable)arg1*) → dict :
    Return dictionary of attributes.

**updateAttrs**(*(Serializable)arg1, (dict)arg2*) → None :
    Update object attributes from given dictionary

## 2.3.13 Preprocessors

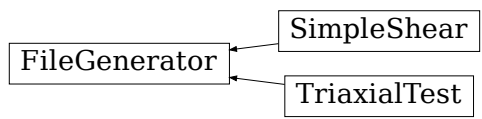

Fig. 38: Inheritance graph of FileGenerator. See also: *SimpleShear*, *TriaxialTest*.

**class** `yade.wrapper.`**FileGenerator**(*inherits Serializable*)

Base class for scene generators, preprocessors.

**dict**(*(Serializable)arg1*) → dict :
    Return dictionary of attributes.

**generate**(*(FileGenerator)arg1, (str)out*) → None :
    Generate scene, save to given file

**load**(*(FileGenerator)arg1*) → None :
    Generate scene, save to temporary file and load immediately

**updateAttrs**(*(Serializable)arg1, (dict)arg2*) → None :
    Update object attributes from given dictionary

**class** `yade.wrapper.`**SimpleShear**(*inherits FileGenerator → Serializable*)

Preprocessor for a simple shear box model. The packing initially conforms a gas-like, very loose, state (see utils.makeCloud function), but importing some existing packing from a text file can be also performed after little change in the source code. In its current state, the preprocessor carries out an oedometric compression, until a value of normal stress equal to 2 MPa (and a stable mechanical state). Others Engines such as *KinemCNDEngine*, *KinemCNSEngine* and *KinemCNLEngine*, could be used to apply resp. constant normal displacement, constant normal rigidity and constant normal stress paths using such a simple shear box.

**density**(*=2600*)
    density of the spheres [kg/m$^3$]

**dict**(*(Serializable)arg1*) → dict :
    Return dictionary of attributes.

**generate**(*(FileGenerator)arg1, (str)out*) → None :
    Generate scene, save to given file

**gravApplied**(*=false*)
    depending on this, *GravityEngine* is added or not to the scene to take into account the weight of particles

**gravity**(*=Vector3r(0, -9.81, 0)*)
    vector corresponding to used gravity (if *gravApplied*) [m/s$^2$]

**height**(*=0.02*)
    initial height (along y-axis) of the shear box [m]

**length**(*=0.1*)
    initial length (along x-axis) of the shear box [m]

**load**(*(FileGenerator)arg1*) → None :
    Generate scene, save to temporary file and load immediately

**matFrictionDeg**(*=37*)
    value of *FrictMat.frictionAngle* within the packing and for the two horizontal boundaries (friction is zero along other boundaries) [°] (the necessary conversion in [rad] is done automatically)

**matPoissonRatio**(*=0.04*)
    value of *FrictMat.poisson* for the bodies [-]

**matYoungModulus**(*=4.0e9*)
    value of *FrictMat.young* for the bodies [Pa]

**thickness**(*=0.001*)
    thickness of the boxes constituting the shear box [m]

**timeStepUpdateInterval**(*=50*)
    value of *TimeStepper::timeStepUpdateInterval* for the *TimeStepper* used here

**updateAttrs**(*(Serializable)arg1, (dict)arg2*) → None :
    Update object attributes from given dictionary

**width**(*=0.04*)
    initial width (along z-axis) of the shear box [m]

**class yade.wrapper.TriaxialTest**(*inherits FileGenerator → Serializable*)
    Create a scene for triaxal test.

    **Introduction** Yade includes tools to simulate triaxial tests on particles assemblies. This preprocessor (and variants like e.g. *CapillaryTriaxialTest*) illustrate how to use them. It generates a scene which will - by default - go through the following steps :

- generate random loose packings in a parallelepiped.

- compress the packing isotropicaly, either squeezing the packing between moving rigid boxes or expanding the particles while boxes are fixed (depending on flag *internalCompaction*). The confining pressure in this stage is defined via *sigmaIsoCompaction*.

- when the packing is dense and stable, simulate a loading path and get the mechanical response as a result.

    The default loading path corresponds to a constant lateral stress (*sigmaLateralConfinement*) in 2 directions and constant strain rate on the third direction. This default loading path is performed when the flag *autoCompressionActivation* it `True`, otherwise the simulation stops after isotropic compression.

Different loading paths might be performed. In order to define them, the user can modify the flags found in engine *TriaxialStressController* at any point in the simulation (in c++). If `TriaxialStressController.wall_X_activated` is `true` boundary X is moved automatically to maintain the defined stress level *sigmaN* (see axis conventions below). If `false` the boundary is not controlled by the engine at all. In that case the user is free to prescribe fixed position, constant velocity, or more complex conditions.

---

**Note:** *Axis conventions.* Boundaries perpendicular to the *x* axis are called "left" and "right", *y* corresponds to "top" and "bottom", and axis *z* to "front" and "back". In the default loading path, strain rate is assigned along *y*, and constant stresses are assigned on *x* and *z*.

---

**Essential engines**

1. The *TriaxialCompressionEngine* is used for controlling the state of the sample and simulating loading paths. *TriaxialCompressionEngine* inherits from *TriaxialStressController*, which computes stress- and strain-like quantities in the packing and maintain a constant level of stress at each boundary. *TriaxialCompressionEngine* has few more members in order to impose constant strain rate and control the transition between isotropic compression and triaxial test. Transitions are defined by changing some flags of the *TriaxialStressController*, switching from/to imposed strain rate to/from imposed stress.

2. The class *TriaxialStateRecorder* is used to write to a file the history of stresses and strains.

3. *TriaxialTest* is using *GlobalStiffnessTimeStepper* to compute an appropriate Δt for the numerical scheme.

---

**Note:** `TriaxialStressController::ComputeUnbalancedForce` returns a value that can be useful for evaluating the stability of the packing. It is defined as (mean force on particles)/(mean contact force), so that it tends to 0 in a stable packing. This parameter is checked by *TriaxialCompressionEngine* to switch from one stage of the simulation to the next one (e.g. stop isotropic confinement and start axial loading)

---

**Frequently Asked Questions**

1. **How is generated the packing? How to change particles sizes distribution? Why do I have a m**
   The initial positioning of spheres is done by generating random (x,y,z) in a box and checking if a sphere of radius R (R also randomly generated with respect to a uniform distribution between mean*(1-std_dev) and mean*(1+std_dev) can be inserted at this location without overlapping with others.

   If the sphere overlaps, new (x,y,z)'s are generated until a free position for the new sphere is found. This explains the message you have: after 3000 trial-and-error, the sphere couldn't be placed, and the algorithm stops.

   You get the message above if you try to generate an initialy dense packing, which is not possible with this algorithm. It can only generate clouds. You should keep the default value of porosity (n~0.7), or even increase if it is still to low in some cases. The dense state will be obtained in the second step (compaction, see below).

2. **How is the compaction done, what are the parameters *maxWallVelocity* and *finalMaxMultiplie***

   **Compaction is done**

   1. by moving rigid boxes or

   2. by increasing the sizes of the particles (decided using the option *internalCompaction* size increase).

Both algorithm needs numerical parameters to prevent instabilities. For instance, with the method (1) *maxWallVelocity* is the maximum wall velocity, with method (2) *finalMaxMultiplier* is the max value of the multiplier applied on sizes at each iteration (always something like 1.00001).

3. **During the simulation of triaxial compression test, the wall in one direction moves with an incr**
   The control of stress on a boundary is based on the total stiffness $K$ of all contacts between the packing and this boundary. In short, at each step, displacement=stress_-error/K. This algorithm is implemented in *TriaxialStressController*, and the control itself is in `TriaxialStressController::ControlExternalStress`. The control can be turned off independently for each boundary, using the flags `wall_XXX_activated`, with *XXX* {*top, bottom, left, right, back, front*}. The imposed sress is a unique value (*sigma_iso*) for all directions if *TriaxialStressController.isAxisymetric*, or 3 independent values *sigma1, sigma2, sigma3*.

4. **Which value of friction angle do you use during the compaction phase of the Triaxial Test?**
   The friction during the compaction (whether you are using the expansion method or the compression one for the specimen generation) can be anything between 0 and the final value used during the Triaxial phase. Note that higher friction than the final one would result in volumetric collapse at the beginning of the test. The purpose of using a different value of friction during this phase is related to the fact that the final porosity you get at the end of the sample generation essentially depends on it as well as on the assumed Particle Size Distribution. Changing the initial value of friction will get to a different value of the final porosity.

5. **Which is the aim of the bool isRadiusControlIteration?** This internal variable (updated automatically) is true each $N$ timesteps (with $N=$*radiusControlInterval*). For other timesteps, there is no expansion. Cycling without expanding is just a way to speed up the simulation, based on the idea that 1% increase each 10 iterations needs less operations than 0.1% at each iteration, but will give similar results.

6. **How comes the unbalanced force reaches a low value only after many timesteps in the compact**
   The value of unbalanced force (dimensionless) is expected to reach low value (i.e. identifying a static-equilibrium condition for the specimen) only at the end of the compaction phase. The code is not aiming at simulating a quasistatic isotropic compaction process, it is only giving a stable packing at the end of it.

---

`Key`(=*""*)
   A code that is added to output filenames.

`StabilityCriterion`(=*0.01*)
   Value of unbalanced force for which the system is considered stable. Used in conditionals to switch between loading stages.

`WallStressRecordFile`(=*"./WallStresses"+Key*)

`autoCompressionActivation`(=*true*)
   Do we just want to generate a stable packing under isotropic pressure (false) or do we want the triaxial loading to start automatically right after compaction stage (true)?

`autoStopSimulation`(=*false*)
   freeze the simulation when conditions are reached (don't activate this if you want to be able to run/stop from Qt GUI)

`autoUnload`(=*true*)
   auto adjust the isotropic stress state from *TriaxialTest::sigmaIsoCompaction* to *TriaxialTest::sigmaLateralConfinement* if they have different values. See docs for *TriaxialCompressionEngine::autoUnload*

`boxFrictionDeg`(=*0.0*)
   Friction angle [°] of boundaries contacts.

**boxKsDivKn**(*=0.5*)
    Ratio of shear vs. normal contact stiffness for boxes.

**boxYoungModulus**(*=15000000.0*)
    Stiffness of boxes.

**compactionFrictionDeg**(*=sphereFrictionDeg*)
    Friction angle [°] of spheres during compaction (different values result in different porosities)]. This value is overridden by *TriaxialTest::sphereFrictionDeg* before triaxial testing.

**dampingForce**(*=0.2*)
    Coefficient of Cundal-Non-Viscous damping (applied on on the 3 components of forces)

**dampingMomentum**(*=0.2*)
    Coefficient of Cundal-Non-Viscous damping (applied on on the 3 components of torques)

**defaultDt**(*=-1*)
    Max time-step. Used as initial value if defined. Latter adjusted by the time stepper.

**density**(*=2600*)
    density of spheres

**dict**(*(Serializable)arg1*) → dict :
    Return dictionary of attributes.

**facetWalls**(*=false*)
    Use facets for boundaries (not tested)

**finalMaxMultiplier**(*=1.001*)
    max multiplier of diameters during internal compaction (secondary precise adjustment)

**fixedBoxDims**(*=""*)
    string that contains some subset (max. 2) of {'x','y','z'} ; contains axes will have box dimension hardcoded, even if box is scaled as mean_radius is prescribed: scaling will be applied on the rest.

**generate**(*(FileGenerator)arg1, (str)out*) → None :
    Generate scene, save to given file

**importFilename**(*=""*)
    File with positions and sizes of spheres.

**internalCompaction**(*=false*)
    flag for choosing between moving boundaries or increasing particles sizes during the compaction stage.

**load**(*(FileGenerator)arg1*) → None :
    Generate scene, save to temporary file and load immediately

**lowerCorner**(*=Vector3r(0, 0, 0)*)
    Lower corner of the box.

**maxMultiplier**(*=1.01*)
    max multiplier of diameters during internal compaction (initial fast increase)

**maxWallVelocity**(*=10*)
    max velocity of boundaries. Usually useless, but can help stabilizing the system in some cases.

**noFiles**(*=false*)
    Do not create any files during run (.xml, .spheres, wall stress records)

**numberOfGrains**(*=400*)
    Number of generated spheres.

**radiusControlInterval**(*=10*)
    interval between size changes when growing spheres.

**radiusMean**(*=-1*)
  Mean radius. If negative (default), autocomputed to as a function of box size and *Triaxial-Test::numberOfGrains*

**radiusStdDev**(*=0.3*)
  Normalized standard deviation of generated sizes.

**recordIntervalIter**(*=20*)
  interval between file outputs

**seed**(*=0*)
  Seed used for the call to makeCloud

**sigmaIsoCompaction**(*=-50000*)
  Confining stress during isotropic compaction ($< 0$ for real - compressive - compaction).

**sigmaLateralConfinement**(*=-50000*)
  Lateral stress during triaxial loading ($< 0$ for classical compressive cases). An isotropic unloading is performed if the value is not equal to *TriaxialTest::sigmaIsoCompaction*.

**sphereFrictionDeg**(*=18.0*)
  Friction angle [°] of spheres assigned just before triaxial testing.

**sphereKsDivKn**(*=0.5*)
  Ratio of shear vs. normal contact stiffness for spheres.

**sphereYoungModulus**(*=15000000.0*)
  Stiffness of spheres.

**strainRate**(*=0.1*)
  Strain rate in triaxial loading.

**thickness**(*=0.001*)
  thickness of boundaries. It is arbitrary and should have no effect

**timeStepUpdateInterval**(*=50*)
  interval for *GlobalStiffnessTimeStepper*

**updateAttrs**(*(Serializable)arg1, (dict)arg2*) → None :
  Update object attributes from given dictionary

**upperCorner**(*=Vector3r(1, 1, 1)*)
  Upper corner of the box.

**wallOversizeFactor**(*=1.3*)
  Make boundaries larger than the packing to make sure spheres don't go out during deformation.

**wallStiffnessUpdateInterval**(*=10*)
  interval for updating the stiffness of sample/boundaries contacts

**wallWalls**(*=false*)
  Use walls for boundaries (not tested)

## 2.3.14 Rendering

### OpenGLRenderer

**class** yade.wrapper.**OpenGLRenderer**(*inherits Serializable*)
  Class responsible for rendering scene on OpenGL devices.

**bgColor**(*=Vector3r(.2, .2, .2)*)
  Color of the background canvas (RGB)

**blinkHighlight**(*=BlinkHighlight::NORMAL*)
  Adjust blinking of the body selected in the 'Simulation Inspection' window.

**bound**(*=false*)
>   Render body *Bound*

**cellColor**(*=Vector3r(1, 1, 0)*)
>   Color of the periodic cell (RGB).

**clipPlaneActive**(*=vector<bool>(numClipPlanes, false)*)
>   Activate/deactivate respective clipping planes

**clipPlaneSe3**(*=vector<Se3r>(numClipPlanes,     Se3r(Vector3r::Zero(),     Quaternionr::Identity()))*)
>   Position and orientation of clipping planes

**dict**(*(Serializable)arg1*) → dict :
>   Return dictionary of attributes.

**dispScale**(*=Vector3r::Ones(), disable scaling*)
>   Artificially enlarge (scale) dispalcements from bodies' *reference positions* by this relative amount, so that they become better visible (independently in 3 dimensions). Disbled if (1,1,1).

**dof**(*=false*)
>   Show which degrees of freedom are blocked for each body

**extraDrawers**(*=uninitalized*)
>   Additional rendering components (*GlExtraDrawer*).

**ghosts**(*=true*)
>   Render objects crossing periodic cell edges by cloning them in multiple places (periodic simulations only).

**hideBody**(*(OpenGLRenderer)arg1, (int)id*) → None :
>   Hide body from id (see *OpenGLRenderer::showBody*)

**id**(*=false*)
>   Show body id's

**intrAllWire**(*=false*)
>   Draw wire for all interactions, blue for potential and green for real ones (mostly for debugging)

**intrGeom**(*=false*)
>   Render *Interaction::geom* objects.

**intrPhys**(*=false*)
>   Render *Interaction::phys* objects

**intrWire**(*=false*)
>   If rendering interactions, use only wires to represent them.

**light1**(*=true*)
>   Turn light 1 on.

**light2**(*=true*)
>   Turn light 2 on.

**light2Color**(*=Vector3r(0.5, 0.5, 0.1)*)
>   Per-color intensity of secondary light (RGB).

**light2Pos**(*=Vector3r(-130, 75, 30)*)
>   Position of secondary OpenGL light source in the scene.

**lightColor**(*=Vector3r(0.6, 0.6, 0.6)*)
>   Per-color intensity of primary light (RGB).

**lightPos**(*=Vector3r(75, 130, 0)*)
>   Position of OpenGL light source in the scene.

**mask**(*=~0, draw everything*)
>   Bitmask for showing only bodies where ((mask & *Body::mask*)!=0)

**render**(*(OpenGLRenderer)arg1*) → None :
    Render the scene in the current OpenGL context.

**rotScale**(*=1., disable scaling*)
    Artificially enlarge (scale) rotations of bodies relative to their *reference orientation*, so the
    they are better visible.

**selId**(*=Body::ID_NONE*)
    Id of particle that was selected by the user.

**setRefSe3**(*(OpenGLRenderer)arg1*) → None :
    Make current positions and orientation reference for scaleDisplacements and scaleRotations.

**shape**(*=true*)
    Render body *Shape*

**showBody**(*(OpenGLRenderer)arg1, (int)id*) → None :
    Make body visible (see *OpenGLRenderer::hideBody*)

**updateAttrs**(*(Serializable)arg1, (dict)arg2*) → None :
    Update object attributes from given dictionary

**wire**(*=false*)
    Render all bodies with wire only (faster)

### GlShapeFunctor

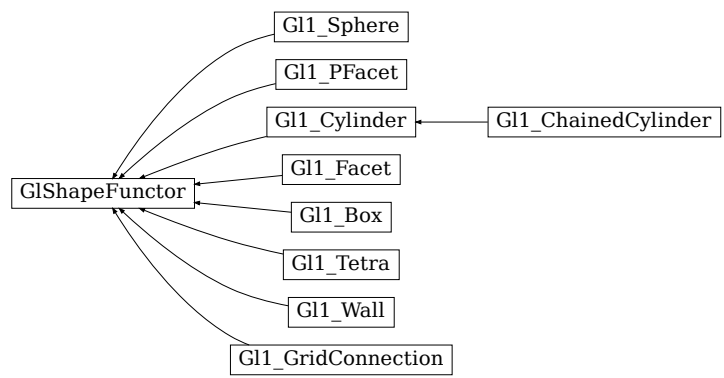

Fig. 39: Inheritance graph of GlShapeFunctor. See also: *Gl1_Box*, *Gl1_ChainedCylinder*, *Gl1_Cylinder*, *Gl1_Facet*, *Gl1_GridConnection*, *Gl1_PFacet*, *Gl1_Sphere*, *Gl1_Tetra*, *Gl1_Wall*.

**class yade.wrapper.GlShapeFunctor**(*inherits Functor → Serializable*)
    Abstract functor for rendering *Shape* objects.

**bases**
    Ordered list of types (as strings) this functor accepts.

**dict**(*(Serializable)arg1*) → dict :
    Return dictionary of attributes.

**label**(*=uninitalized*)
    Textual label for this object; must be a valid python identifier, you can refer to it directly
    from python.

**timingDeltas**
    Detailed information about timing inside the Dispatcher itself. Empty unless enabled in the
    source code and O.timingEnabled==True.

**updateAttrs**(*(Serializable)arg1, (dict)arg2*) → None :
    Update object attributes from given dictionary

**class** yade.wrapper.**Gl1_Box**(*inherits GlShapeFunctor → Functor → Serializable*)
Renders *Box* object

> **bases**
> > Ordered list of types (as strings) this functor accepts.
>
> **dict**(*(Serializable)arg1*) → dict :
> > Return dictionary of attributes.
>
> **label**(*=uninitalized*)
> > Textual label for this object; must be a valid python identifier, you can refer to it directly from python.
>
> **timingDeltas**
> > Detailed information about timing inside the Dispatcher itself. Empty unless enabled in the source code and O.timingEnabled==True.
>
> **updateAttrs**(*(Serializable)arg1, (dict)arg2*) → None :
> > Update object attributes from given dictionary

**class** yade.wrapper.**Gl1_ChainedCylinder**(*inherits Gl1_Cylinder → GlShapeFunctor → Functor → Serializable*)
Renders *ChainedCylinder* object including a shift for compensating flexion.

> **bases**
> > Ordered list of types (as strings) this functor accepts.
>
> **dict**(*(Serializable)arg1*) → dict :
> > Return dictionary of attributes.
>
> **glutNormalize = True**
>
> **glutSlices = 8**
>
> **glutStacks = 4**
>
> **label**(*=uninitalized*)
> > Textual label for this object; must be a valid python identifier, you can refer to it directly from python.
>
> **timingDeltas**
> > Detailed information about timing inside the Dispatcher itself. Empty unless enabled in the source code and O.timingEnabled==True.
>
> **updateAttrs**(*(Serializable)arg1, (dict)arg2*) → None :
> > Update object attributes from given dictionary
>
> **wire = False**

**class** yade.wrapper.**Gl1_Cylinder**(*inherits GlShapeFunctor → Functor → Serializable*)
Renders *Cylinder* object

> **wire**(*=false*) **[static]**
> > Only show wireframe (controlled by glutSlices and glutStacks.
>
> **glutNormalize**(*=true*) **[static]**
> > Fix normals for non-wire rendering
>
> **glutSlices**(*=8*) **[static]**
> > Number of sphere slices.
>
> **glutStacks**(*=4*) **[static]**
> > Number of sphere stacks.
>
> **bases**
> > Ordered list of types (as strings) this functor accepts.
>
> **dict**(*(Serializable)arg1*) → dict :
> > Return dictionary of attributes.

**glutNormalize = True**

**glutSlices = 8**

**glutStacks = 4**

**label**(*=uninitalized*)
> Textual label for this object; must be a valid python identifier, you can refer to it directly from python.

**timingDeltas**
> Detailed information about timing inside the Dispatcher itself. Empty unless enabled in the source code and O.timingEnabled==True.

**updateAttrs**(*(Serializable)arg1, (dict)arg2*) → None :
> Update object attributes from given dictionary

**wire = False**

**class yade.wrapper.Gl1_Facet**(*inherits GlShapeFunctor → Functor → Serializable*)
> Renders *Facet* object

**normals**(*=false*) **[static]**
> In wire mode, render normals of facets and edges; facet's *colors* are disregarded in that case.

**bases**
> Ordered list of types (as strings) this functor accepts.

**dict**(*(Serializable)arg1*) → dict :
> Return dictionary of attributes.

**label**(*=uninitalized*)
> Textual label for this object; must be a valid python identifier, you can refer to it directly from python.

**normals = False**

**timingDeltas**
> Detailed information about timing inside the Dispatcher itself. Empty unless enabled in the source code and O.timingEnabled==True.

**updateAttrs**(*(Serializable)arg1, (dict)arg2*) → None :
> Update object attributes from given dictionary

**class yade.wrapper.Gl1_GridConnection**(*inherits GlShapeFunctor → Functor → Serializable*)
> Renders *Cylinder* object

**wire**(*=false*) **[static]**
> Only show wireframe (controlled by **glutSlices** and **glutStacks**.

**glutNormalize**(*=true*) **[static]**
> Fix normals for non-wire rendering

**glutSlices**(*=8*) **[static]**
> Number of cylinder slices.

**glutStacks**(*=4*) **[static]**
> Number of cylinder stacks.

**bases**
> Ordered list of types (as strings) this functor accepts.

**dict**(*(Serializable)arg1*) → dict :
> Return dictionary of attributes.

**glutNormalize = True**

**glutSlices = 8**

**glutStacks = 4**

**label**(*=uninitalized*)
> Textual label for this object; must be a valid python identifier, you can refer to it directly from python.

**timingDeltas**
> Detailed information about timing inside the Dispatcher itself. Empty unless enabled in the source code and O.timingEnabled==True.

**updateAttrs**(*(Serializable)arg1, (dict)arg2*) → None :
> Update object attributes from given dictionary

**wire = False**

**class yade.wrapper.Gl1_PFacet**(*inherits GlShapeFunctor → Functor → Serializable*)
> Renders *Facet* object

**wire**(*=false*) **[static]**
> Only show wireframe (controlled by `glutSlices` and `glutStacks`.

**bases**
> Ordered list of types (as strings) this functor accepts.

**dict**(*(Serializable)arg1*) → dict :
> Return dictionary of attributes.

**label**(*=uninitalized*)
> Textual label for this object; must be a valid python identifier, you can refer to it directly from python.

**timingDeltas**
> Detailed information about timing inside the Dispatcher itself. Empty unless enabled in the source code and O.timingEnabled==True.

**updateAttrs**(*(Serializable)arg1, (dict)arg2*) → None :
> Update object attributes from given dictionary

**wire = False**

**class yade.wrapper.Gl1_Sphere**(*inherits GlShapeFunctor → Functor → Serializable*)
> Renders *Sphere* object

**quality**(*=1.0*) **[static]**
> Change discretization level of spheres. quality>1 for better image quality, at the price of more cpu/gpu usage, 0<quality<1 for faster rendering. If mono-color spheres are displayed (*Gl1_-Sphere::stripes* = False), quality mutiplies *Gl1_Sphere::glutSlices* and *Gl1_Sphere::glutStacks*. If striped spheres are displayed (*Gl1_Sphere::stripes* = True), only integer increments are meaningfull : quality=1 and quality=1.9 will give the same result, quality=2 will give finer result.

**wire**(*=false*) **[static]**
> Only show wireframe (controlled by `glutSlices` and `glutStacks`.

**stripes**(*=false*) **[static]**
> In non-wire rendering, show stripes clearly showing particle rotation.

**localSpecView**(*=true*) **[static]**
> Compute specular light in local eye coordinate system.

**glutSlices**(*=12*) **[static]**
> Base number of sphere slices, multiplied by *Gl1_Sphere::quality* before use); not used with `stripes` (see glut{Solid,Wire}Sphere reference)

**glutStacks**(*=6*) **[static]**
> Base number of sphere stacks, multiplied by *Gl1_Sphere::quality* before use; not used with `stripes` (see glut{Solid,Wire}Sphere reference)

**circleView**(*=false*) **[static]**
> For 2D simulations : display tori instead of spheres, so they will appear like circles if the viewer is looking in the right direction. In this case, remember to disable perspective by pressing "t"-key in the viewer.

**circleRelThickness**(*=0.2*) **[static]**
> If *Gl1_Sphere::circleView* is enabled, this is the torus diameter relative to the sphere radius (i.e. the circle relative thickness).

**circleAllowedRotationAxis**(*='z'*) **[static]**
> If *Gl1_Sphere::circleView* is enabled, this is the only axis ('x', 'y' or 'z') along which rotation is allowed for the 2D simulation. It allows right orientation of the tori to appear like circles in the viewer. For example, if circleAllowedRotationAxis='x' is set, blockedDOFs="YZ" should also be set for all your particles.

**bases**
> Ordered list of types (as strings) this functor accepts.

**circleAllowedRotationAxis = 'z'**

**circleRelThickness = 0.2**

**circleView = False**

**dict**(*(Serializable)arg1*) → dict :
> Return dictionary of attributes.

**glutSlices = 12**

**glutStacks = 6**

**label**(*=uninitalized*)
> Textual label for this object; must be a valid python identifier, you can refer to it directly from python.

**localSpecView = True**

**quality = 1.0**

**stripes = False**

**timingDeltas**
> Detailed information about timing inside the Dispatcher itself. Empty unless enabled in the source code and O.timingEnabled==True.

**updateAttrs**(*(Serializable)arg1, (dict)arg2*) → None :
> Update object attributes from given dictionary

**wire = False**

**class yade.wrapper.Gl1_Tetra**(*inherits GlShapeFunctor → Functor → Serializable*)
> Renders *Tetra* object

**wire**(*=true*) **[static]**
> TODO

**bases**
> Ordered list of types (as strings) this functor accepts.

**dict**(*(Serializable)arg1*) → dict :
> Return dictionary of attributes.

**label**(*=uninitalized*)
> Textual label for this object; must be a valid python identifier, you can refer to it directly from python.

**timingDeltas**
> Detailed information about timing inside the Dispatcher itself. Empty unless enabled in the source code and O.timingEnabled==True.

> **updateAttrs**(*(Serializable)arg1, (dict)arg2*) → None :
> > Update object attributes from given dictionary

> **wire = True**

**class** yade.wrapper.**Gl1_Wall**(*inherits GlShapeFunctor → Functor → Serializable*)
> Renders *Wall* object

> **div**(*=20*) **[static]**
> > Number of divisions of the wall inside visible scene part.

> **bases**
> > Ordered list of types (as strings) this functor accepts.

> **dict**(*(Serializable)arg1*) → dict :
> > Return dictionary of attributes.

> **div = 20**

> **label**(*=uninitalized*)
> > Textual label for this object; must be a valid python identifier, you can refer to it directly
> > from python.

> **timingDeltas**
> > Detailed information about timing inside the Dispatcher itself. Empty unless enabled in the
> > source code and O.timingEnabled==True.

> **updateAttrs**(*(Serializable)arg1, (dict)arg2*) → None :
> > Update object attributes from given dictionary

## GlStateFunctor

**class** yade.wrapper.**GlStateFunctor**(*inherits Functor → Serializable*)
> Abstract functor for rendering *State* objects.

> **bases**
> > Ordered list of types (as strings) this functor accepts.

> **dict**(*(Serializable)arg1*) → dict :
> > Return dictionary of attributes.

> **label**(*=uninitalized*)
> > Textual label for this object; must be a valid python identifier, you can refer to it directly
> > from python.

> **timingDeltas**
> > Detailed information about timing inside the Dispatcher itself. Empty unless enabled in the
> > source code and O.timingEnabled==True.

> **updateAttrs**(*(Serializable)arg1, (dict)arg2*) → None :
> > Update object attributes from given dictionary

## GlBoundFunctor

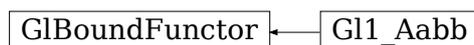

Fig. 40: Inheritance graph of GlBoundFunctor. See also: *Gl1_Aabb*.

**class** yade.wrapper.**GlBoundFunctor**(*inherits Functor → Serializable*)
> Abstract functor for rendering *Bound* objects.

> **bases**
> > Ordered list of types (as strings) this functor accepts.

**dict**(*(Serializable)arg1*) → dict :
> Return dictionary of attributes.

**label**(*=uninitalized*)
> Textual label for this object; must be a valid python identifier, you can refer to it directly from python.

**timingDeltas**
> Detailed information about timing inside the Dispatcher itself. Empty unless enabled in the source code and O.timingEnabled==True.

**updateAttrs**(*(Serializable)arg1*, *(dict)arg2*) → None :
> Update object attributes from given dictionary

**class yade.wrapper.Gl1_Aabb**(*inherits GlBoundFunctor → Functor → Serializable*)
> Render Axis-aligned bounding box (*Aabb*).

**bases**
> Ordered list of types (as strings) this functor accepts.

**dict**(*(Serializable)arg1*) → dict :
> Return dictionary of attributes.

**label**(*=uninitalized*)
> Textual label for this object; must be a valid python identifier, you can refer to it directly from python.

**timingDeltas**
> Detailed information about timing inside the Dispatcher itself. Empty unless enabled in the source code and O.timingEnabled==True.

**updateAttrs**(*(Serializable)arg1*, *(dict)arg2*) → None :
> Update object attributes from given dictionary

### GlIGeomFunctor

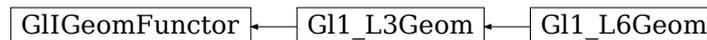

Fig. 41: Inheritance graph of GlIGeomFunctor. See also: *Gl1_L3Geom*, *Gl1_L6Geom*.

**class yade.wrapper.GlIGeomFunctor**(*inherits Functor → Serializable*)
> Abstract functor for rendering *IGeom* objects.

**bases**
> Ordered list of types (as strings) this functor accepts.

**dict**(*(Serializable)arg1*) → dict :
> Return dictionary of attributes.

**label**(*=uninitalized*)
> Textual label for this object; must be a valid python identifier, you can refer to it directly from python.

**timingDeltas**
> Detailed information about timing inside the Dispatcher itself. Empty unless enabled in the source code and O.timingEnabled==True.

**updateAttrs**(*(Serializable)arg1*, *(dict)arg2*) → None :
> Update object attributes from given dictionary

**class yade.wrapper.Gl1_L3Geom**(*inherits GlIGeomFunctor → Functor → Serializable*)
> Render *L3Geom* geometry.

**axesLabels**(*=false*) **[static]**
> Whether to display labels for local axes (x,y,z)

**axesScale**(*=1.*) **[static]**
> Scale local axes, their reference length being half of the minimum radius.

**axesWd**(*=1.*) **[static]**
> Width of axes lines, in pixels; not drawn if non-positive

**uPhiWd**(*=2.*) **[static]**
> Width of lines for drawing displacements (and rotations for *L6Geom*); not drawn if non-positive.

**uScale**(*=1.*) **[static]**
> Scale local displacements ($u$ - $u0$); 1 means the true scale, 0 disables drawing local displacements; negative values are permissible.

**axesLabels = False**

**axesScale = 1.0**

**axesWd = 1.0**

**bases**
> Ordered list of types (as strings) this functor accepts.

**dict**(*(Serializable)arg1*) → dict :
> Return dictionary of attributes.

**label**(*=uninitalized*)
> Textual label for this object; must be a valid python identifier, you can refer to it directly from python.

**timingDeltas**
> Detailed information about timing inside the Dispatcher itself. Empty unless enabled in the source code and O.timingEnabled==True.

**uPhiWd = 2.0**

**uScale = 1.0**

**updateAttrs**(*(Serializable)arg1, (dict)arg2*) → None :
> Update object attributes from given dictionary

**class yade.wrapper.Gl1_L6Geom**(*inherits Gl1_L3Geom → GlIGeomFunctor → Functor → Serializable*)
Render *L6Geom* geometry.

**phiScale**(*=1.*) **[static]**
> Scale local rotations ($phi$ - $phi0$). The default scale is to draw $\pi$ rotation with length equal to minimum radius.

**axesLabels = False**

**axesScale = 1.0**

**axesWd = 1.0**

**bases**
> Ordered list of types (as strings) this functor accepts.

**dict**(*(Serializable)arg1*) → dict :
> Return dictionary of attributes.

**label**(*=uninitalized*)
> Textual label for this object; must be a valid python identifier, you can refer to it directly from python.

**phiScale = 1.0**

**timingDeltas**
    Detailed information about timing inside the Dispatcher itself. Empty unless enabled in the source code and O.timingEnabled==True.

**uPhiWd = 2.0**

**uScale = 1.0**

**updateAttrs**(*(Serializable)arg1, (dict)arg2*) → None :
    Update object attributes from given dictionary

### GlIPhysFunctor

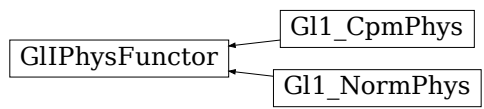

Fig. 42: Inheritance graph of GlIPhysFunctor. See also: *Gl1__CpmPhys*, *Gl1__NormPhys*.

**class yade.wrapper.GlIPhysFunctor**(*inherits Functor → Serializable*)
    Abstract functor for rendering *IPhys* objects.

**bases**
    Ordered list of types (as strings) this functor accepts.

**dict**(*(Serializable)arg1*) → dict :
    Return dictionary of attributes.

**label**(*=uninitalized*)
    Textual label for this object; must be a valid python identifier, you can refer to it directly from python.

**timingDeltas**
    Detailed information about timing inside the Dispatcher itself. Empty unless enabled in the source code and O.timingEnabled==True.

**updateAttrs**(*(Serializable)arg1, (dict)arg2*) → None :
    Update object attributes from given dictionary

**class yade.wrapper.Gl1_CpmPhys**(*inherits GlIPhysFunctor → Functor → Serializable*)
    Render *CpmPhys* objects of interactions.

**contactLine**(*=true*) **[static]**
    Show contact line

**dmgLabel**(*=true*) **[static]**
    Numerically show contact damage parameter

**dmgPlane**(*=false*) **[static]**
    [what is this?]

**epsT**(*=false*) **[static]**
    Show shear strain

**epsTAxes**(*=false*) **[static]**
    Show axes of shear plane

**normal**(*=false*) **[static]**
    Show contact normal

**colorStrainRatio**(*=-1*) **[static]**
    If positive, set the interaction (wire) color based on $\varepsilon_N$ normalized by $\varepsilon_0$ x *colorStrainRatio* ($\varepsilon_0$ = *CpmPhys.epsCrackOnset* ). Otherwise, color based on the residual strength.

---

**epsNLabel**(*=false*) **[static]**
>   Numerically show normal strain

**bases**
>   Ordered list of types (as strings) this functor accepts.

**colorStrainRatio = -1.0**

**contactLine = True**

**dict**(*(Serializable)arg1*) → dict :
>   Return dictionary of attributes.

**dmgLabel = True**

**dmgPlane = False**

**epsNLabel = False**

**epsT = False**

**epsTAxes = False**

**label**(*=uninitalized*)
>   Textual label for this object; must be a valid python identifier, you can refer to it directly
>   from python.

**normal = False**

**timingDeltas**
>   Detailed information about timing inside the Dispatcher itself. Empty unless enabled in the
>   source code and O.timingEnabled==True.

**updateAttrs**(*(Serializable)arg1, (dict)arg2*) → None :
>   Update object attributes from given dictionary

**class yade.wrapper.Gl1_NormPhys**(*inherits GlIPhysFunctor → Functor → Serializable*)
>   Renders *NormPhys* objects as cylinders of which diameter and color depends on *Norm-
>   Phys.normalForce* magnitude.

**maxFn**(*=0*) **[static]**
>   Value of *NormPhys.normalForce* corresponding to *maxRadius*. This value will be increased
>   (but *not decreased* ) automatically.

**signFilter**(*=0*) **[static]**
>   If non-zero, only display contacts with negative (-1) or positive (+1) normal forces; if zero,
>   all contacts will be displayed.

**refRadius**(*=std::numeric_limits<Real>::infinity()*) **[static]**
>   Reference (minimum) particle radius; used only if *maxRadius* is negative. This value will be
>   decreased (but *not increased* ) automatically. *(auto-updated)*

**maxRadius**(*=-1*) **[static]**
>   Cylinder radius corresponding to the maximum normal force. If negative, auto-updated *re-
>   fRadius* will be used instead.

**slices**(*=6*) **[static]**
>   Number of sphere slices; (see glutCylinder reference)

**stacks**(*=1*) **[static]**
>   Number of sphere stacks; (see glutCylinder reference)

**maxWeakFn**(*=NaN*) **[static]**
>   Value that divides contacts by their normal force into the 'weak fabric' and 'strong fabric'.
>   This value is set as side-effect by *utils.fabricTensor*.

**weakFilter**(*=0*) **[static]**
>   If non-zero, only display contacts belonging to the 'weak' (-1) or 'strong' (+1) fabric.

**weakScale**(*=1.*) **[static]**
>   If *maxWeakFn* is set, scale radius of the weak fabric by this amount (usually smaller than 1). If zero, 1 pixel line is displayed. Colors are not affected by this value.

**bases**
>   Ordered list of types (as strings) this functor accepts.

**dict**(*(Serializable)arg1*) → dict :
>   Return dictionary of attributes.

**label**(*=uninitalized*)
>   Textual label for this object; must be a valid python identifier, you can refer to it directly from python.

**maxFn = 0.0**

**maxRadius = -1.0**

**maxWeakFn = nan**

**refRadius = inf**

**signFilter = 0**

**slices = 6**

**stacks = 1**

**timingDeltas**
>   Detailed information about timing inside the Dispatcher itself. Empty unless enabled in the source code and O.timingEnabled==True.

**updateAttrs**(*(Serializable)arg1, (dict)arg2*) → None :
>   Update object attributes from given dictionary

**weakFilter = 0**

**weakScale = 1.0**

## 2.3.15 Simulation data

### Omega

**class** yade.wrapper.**Omega**

>   **addScene**(*(Omega)arg1*) → int :
>   >   Add new scene to Omega, returns its number
>
>   **bodies**
>   >   Bodies in the current simulation (container supporting index access by id and iteration)
>
>   **cell**
>   >   Periodic cell of the current scene (None if the scene is aperiodic).
>
>   **childClassesNonrecursive**(*(Omega)arg1, (str)arg2*) → list :
>   >   Return list of all classes deriving from given class, as registered in the class factory
>
>   **disableGdb**(*(Omega)arg1*) → None :
>   >   Revert SEGV and ABRT handlers to system defaults.
>
>   **dt**
>   >   Current timestep (Δt) value. See *dynDt* for enabling/disabling automatic Δt updates through a *TimeStepper*.
>
>   **dynDt**
>   >   Whether a *TimeStepper* (when present in *O.engines*) is used for dynamic Δt control.

**dynDtAvailable**

    Whether a *TimeStepper* is amongst *O.engines*, activated or not.

**energy**

    *EnergyTracker* of the current simulation. (meaningful only with *O.trackEnergy*)

**engines**

    List of engines in the simulation (corresponds to Scene::engines in C++ source code).

**exitNoBacktrace**(*(Omega)arg1*[, *(int)status=0*]) → None :

    Disable SEGV handler and exit, optionally with given status number.

**filename**

    Filename under which the current simulation was saved (None if never saved).

**forceSyncCount**

    Counter for number of syncs in ForceContainer, for profiling purposes.

**forces**

    *ForceContainer* (forces, torques, displacements) in the current simulation.

**interactions**

    Access to *interactions* of simulation, by using

    1. id's of both *Bodies* of the interactions, e.g. `O.interactions[23,65]`

    2. iteration over the whole container:

```
for i in O.interactions: print i.id1,i.id2
```

---

**Note:** Iteration silently skips interactions that are not *real*.

---

**isChildClassOf**(*(Omega)arg1*, *(str)arg2*, *(str)arg3*) → bool :

    Tells whether the first class derives from the second one (both given as strings).

**iter**

    Get current step number

**labeledEngine**(*(Omega)arg1*, *(str)arg2*) → object :

    Return instance of engine/functor with the given label. This function shouldn't be called by the user directly; every change in O.engines will assign respective global python variables according to labels.

    For example:

        *O.engines=[InsertionSortCollider(label='collider')]*

        *collider.nBins=5 # collider has become a variable after assignment to O.engines automatically*

**load**(*(Omega)arg1*, *(str)file*[, *(bool)quiet=False*]) → None :

    Load simulation from file. The file should have been *saved* in the same version of Yade built or compiled with the same features, otherwise compatibility is not guaranteed. Compatibility may also be affected by different versions of external libraries such as Boost

**loadTmp**(*(Omega)arg1*[, *(str)mark=''*[, *(bool)quiet=False*]]) → None :

    Load simulation previously stored in memory by saveTmp. *mark* optionally distinguishes multiple saved simulations

**lsTmp**(*(Omega)arg1*) → list :

    Return list of all memory-saved simulations.

**materials**

    Shared materials; they can be accessed by id or by label

**miscParams**
  MiscParams in the simulation (Scene::mistParams), usually used to save serializables that don't fit anywhere else, like GL functors

**numThreads**
  Get maximum number of threads openMP can use.

**pause**(*(Omega)arg1*) → None :
  Stop simulation execution. (May be called from within the loop, and it will stop after the current step).

**periodic**
  Get/set whether the scene is periodic or not (True/False).

**plugins**(*(Omega)arg1*) → list :
  Return list of all plugins registered in the class factory.

**realtime**
  Return clock (human world) time the simulation has been running.

**reload**(*(Omega)arg1*[, *(bool)quiet=False*]) → None :
  Reload current simulation

**reset**(*(Omega)arg1*) → None :
  Reset simulations completely (including another scenes!).

**resetAllScenes**(*(Omega)arg1*) → None :
  Reset all scenes.

**resetCurrentScene**(*(Omega)arg1*) → None :
  Reset current scene.

**resetThisScene**(*(Omega)arg1*) → None :
  Reset current scene.

**resetTime**(*(Omega)arg1*) → None :
  Reset simulation time: step number, virtual and real time. (Doesn't touch anything else, including timings).

**run**(*(Omega)arg1*[, *(int)nSteps=-1*[, *(bool)wait=False*]]) → None :
  Run the simulation. *nSteps* how many steps to run, then stop (if positive); *wait* will cause not returning to python until simulation will have stopped.

**runEngine**(*(Omega)arg1, (Engine)arg2*) → None :
  Run given engine exactly once; simulation time, step number etc. will not be incremented (use only if you know what you do).

**running**
  Whether background thread is currently running a simulation.

**save**(*(Omega)arg1, (str)file*[, *(bool)quiet=False*]) → None :
  Save current simulation to file (should be .xml or .xml.bz2 or .yade or .yade.gz). .xml files are bigger than .yade, but can be more or less easily (due to their size) opened and edited, e.g. with text editors. .bz2 and .gz correspond both to compressed versions. There are software requirements for successful reloads, see *O.load*.

**saveTmp**(*(Omega)arg1*[, *(str)mark=""*[, *(bool)quiet=False*]]) → None :
  Save simulation to memory (disappears at shutdown), can be loaded later with loadTmp. *mark* optionally distinguishes different memory-saved simulations.

**sceneToString**(*(Omega)arg1*) → object :
  Return the entire scene as a string. Equivalent to using O.save(…) except that the scene goes to a string instead of a file. (see also stringToScene())

**speed**
  Return current calculation speed [iter/sec].

**step**(*(Omega)arg1*) → None :
  Advance the simulation by one step. Returns after the step will have finished.

**stopAtIter**
  Get/set number of iteration after which the simulation will stop.

**stopAtTime**
  Get/set time after which the simulation will stop.

**stringToScene**(*(Omega)arg1, (str)arg2*[, *(str)mark=''*]) → None :
  Load simulation from a string passed as argument (see also sceneToString).

**subStep**
  Get the current subStep number (only meaningful if O.subStepping==True); -1 when outside the loop, otherwise either 0 (O.subStepping==False) or number of engine to be run (O.subStepping==True)

**subStepping**
  Get/set whether subStepping is active.

**switchScene**(*(Omega)arg1*) → None :
  Switch to alternative simulation (while keeping the old one). Calling the function again switches back to the first one. Note that most variables from the first simulation will still refer to the first simulation even after the switch (e.g. b=O.bodies[4]; O.switchScene(); [b still refers to the body in the first simulation here])

**switchToScene**(*(Omega)arg1, (int)arg2*) → None :
  Switch to defined scene. Default scene has number 0, other scenes have to be created by addScene method.

**tags**
  Tags (string=string dictionary) of the current simulation (container supporting string-index access/assignment)

**thisScene**
  Return current scene's id.

**time**
  Return virtual (model world) time of the simulation.

**timingEnabled**
  Globally enable/disable timing services (see documentation of the *timing module*).

**tmpFilename**(*(Omega)arg1*) → str :
  Return unique name of file in temporary directory which will be deleted when yade exits.

**tmpToFile**(*(Omega)arg1, (str)fileName*[, *(str)mark=''*]) → None :
  Save XML of *saveTmp*'d simulation into *fileName*.

**tmpToString**(*(Omega)arg1*[, *(str)mark=''*]) → str :
  Return XML of *saveTmp*'d simulation as string.

**trackEnergy**
  When energy tracking is enabled or disabled in this simulation.

**wait**(*(Omega)arg1*) → None :
  Don't return until the simulation will have been paused. (Returns immediately if not running).

### BodyContainer

**class yade.wrapper.BodyContainer**

  **__init__**(*(object)arg1, (BodyContainer)arg2*) → None

**addToClump**(*(BodyContainer)arg1, (object)arg2, (int)arg3*$\Big[$, *(int)discretization=0*$\Big]$) → None

Add body b (or a list of bodies) to an existing clump c. c must be clump and b may not be a clump member of c. Clump masses and inertia are adapted automatically (for details see *clump()*).

See examples/clumps/addToClump-example.py for an example script.

---

**Note:** If b is a clump itself, then all members will be added to c and b will be deleted. If b is a clump member of clump d, then all members from d will be added to c and d will be deleted. If you need to add just clump member b, *release* this member from d first.

---

**append**(*(BodyContainer)arg1, (Body)arg2*) → int :
Append one Body instance, return its id.

> **append( (BodyContainer)arg1, (object)arg2) -> object :** Append list of Body instance, return list of ids

**appendClumped**(*(BodyContainer)arg1, (object)arg2*$\Big[$, *(int)discretization=0*$\Big]$) → tuple :
Append given list of bodies as a clump (rigid aggregate); returns a tuple of (`clumpId`, [`memberId1,memberId2,...`]). Clump masses and inertia are computed automatically depending upon *discretization* (for details see *clump()*).

**clear**(*(BodyContainer)arg1*) → None :
Remove all bodies (interactions not checked)

**clump**(*(BodyContainer)arg1, (object)arg2*$\Big[$, *(int)discretization=0*$\Big]$) → int :
Clump given bodies together (creating a rigid aggregate); returns `clumpId`. A precise definition of clump masses and inertia when clump members overlap requires *discretization*>0 and is achieved in this case by integration/summation over mass points using a regular grid of cells (grid cells length is defined as $L_{min}$/`discretization`, where $L_{min}$ is the minimum length of an Axis-Aligned Bounding Box. If *discretization*<=0 sum of inertias from members is simply used, which is faster but accurate only for non-overlapping members).

**deleteClumpBody**(*(BodyContainer)arg1, (Body)arg2*) → None :
Erase clump member.

**deleteClumpMember**(*(BodyContainer)arg1, (Body)arg2, (Body)arg3*) → None :
Erase clump member.

**enableRedirection**
let collider switch to optimized algorithm with body redirection when bodies are erased - true by default

**erase**(*(BodyContainer)arg1, (int)arg2*$\Big[$, *(bool)eraseClumpMembers=0*$\Big]$) → bool :
Erase body with the given id; all interaction will be deleted by InteractionLoop in the next step. If a clump is erased use *O.bodies.erase(clumpId,True)* to erase the clump AND its members.

**getRoundness**(*(BodyContainer)arg1*$\Big[$, *(list)excludeList=[]*$\Big]$) → float :
Returns roundness coefficient RC = R2/R1. R1 is the equivalent sphere radius of a clump. R2 is the minimum radius of a sphere, that imbeds the clump. If just spheres are present RC = 1. If clumps are present 0 < RC < 1. Bodies can be excluded from the calculation by giving a list of ids: *O.bodies.getRoundness([ids])*.

See examples/clumps/replaceByClumps-example.py for an example script.

**insertAtId**(*(BodyContainer)arg1, (Body)arg2, (int)insertatid*) → int :
Insert a body at theid, (no body should exist in this id)

**releaseFromClump**(*(BodyContainer)arg1, (int)arg2, (int)arg3*$\Big[$, *(int)discretization=0*$\Big]$) → None :
Release body b from clump c. b must be a clump member of c. Clump masses and inertia are adapted automatically (for details see *clump()*).

See examples/clumps/releaseFromClump-example.py for an example script.

---

**Note:** If c contains only 2 members b will not be released and a warning will appear. In this case clump c should be *erased*.

---

**replace**(*(BodyContainer)arg1, (object)arg2*) → object

**replaceByClumps**(*(BodyContainer)arg1, (list)arg2, (object)arg3*[, *(int)discretization=0*]) → list :
    Replace spheres by clumps using a list of clump templates and a list of amounts; returns a list of tuples: `[(clumpId1,[memberId1,memberId2,...]),(clumpId2,[memberId1,memberId2, ...]),...]`. A new clump will have the same volume as the sphere, that was replaced. Clump masses and inertia are adapted automatically (for details see *clump()*).

    *O.bodies.replaceByClumps( [utils.clumpTemplate([1,1],[.5,.5])] , [.9] ) #will replace 90 % of all standalone spheres by 'dyads'*

See examples/clumps/replaceByClumps-example.py for an example script.

**updateClumpProperties**(*(BodyContainer)arg1*[, *(list)excludeList=[]*[, *(int)discretization=5*]]) → None :
    Manually force Yade to update clump properties mass, volume and inertia (for details of 'discretization' value see *clump()*). Can be used, when clumps are modified or erased during a simulation. Clumps can be excluded from the calculation by giving a list of ids: *O.bodies.updateProperties([ids])*.

**useRedirection**
    true if the scene uses up-to-date lists for boundedBodies and realBodies; turned true automatically 1/ after removal of bodies if *enableRedirection=True* , and 2/ in MPI execution. *(auto-updated)*

## InteractionContainer

**class yade.wrapper.InteractionContainer**
    Access to *interactions* of simulation, by using

    1. id's of both *Bodies* of the interactions, e.g. `O.interactions[23,65]`

    2. iteration over the whole container:

```
for i in O.interactions: print i.id1,i.id2
```

---

**Note:** Iteration silently skips interactions that are virtual i.e. not *real*.

---

**__init__**(*(object)arg1, (InteractionContainer)arg2*) → None

**all**(*(InteractionContainer)arg1*[, *(bool)onlyReal=False*]) → list :
    Return list of all interactions. Virtual interaction are filtered out if onlyReal=True, else (default) it dumps the full content.

**clear**(*(InteractionContainer)arg1*) → None :
    Remove all interactions, and invalidate persistent collider data (if the collider supports it).

**countReal**(*(InteractionContainer)arg1*) → int :
    Return number of interactions that are *real*.

**erase**(*(InteractionContainer)arg1, (int)arg2, (int)arg3*) → None :
    Erase one interaction, given by id1, id2 (internally, `requestErase` is called – the interaction might still exist as potential, if the *Collider* decides so).

**eraseNonReal**(*(InteractionContainer)arg1*) → None :
    Erase all interactions that are not *real* .

**has**(*(InteractionContainer)arg1, (int)id1, (int)id2*[, *(bool)onlyReal=False*]) → bool :
    Tell if a pair of ids *id1, id2* corresponds to an existing interaction (*real* or not depending on *onlyReal*)

**nth**(*(InteractionContainer)arg1, (int)arg2*) → Interaction :
    Return n-th interaction from the container (usable for picking random interaction). The virtual interactions are not reached.

**serializeSorted**

**withBody**(*(InteractionContainer)arg1, (int)arg2*) → list :
    Return list of *real* interactions of given body.

**withBodyAll**(*(InteractionContainer)arg1, (int)arg2*) → list :
    Return list of all (*real* as well as non-real) interactions of given body.

## ForceContainer

`class yade.wrapper.ForceContainer`

**__init__**(*(object)arg1, (ForceContainer)arg2*) → None

**addF**(*(ForceContainer)arg1, (int)id, (Vector3)f*[, *(bool)permanent=False*]) → None :
    Apply force on body (accumulates). The force applies for one iteration, then it is reset by ForceResetter. # permanent parameter is deprecated, instead of addF(…,permanent=True) use setPermF(…).

**addT**(*(ForceContainer)arg1, (int)id, (Vector3)t*[, *(bool)permanent=False*]) → None :
    Apply torque on body (accumulates). The torque applies for one iteration, then it is reset by ForceResetter. # permanent parameter is deprecated, instead of addT(…,permanent=True) use setPermT(…).

**f**(*(ForceContainer)arg1, (int)id*[, *(bool)sync=False*]) → Vector3 :
    Resultant force on body, excluding *gravity*. For clumps in openMP, synchronize the force container with sync=True, else the value will be wrong.

**getPermForceUsed**(*(ForceContainer)arg1*) → bool :
    Check wether permanent forces are present.

**m**(*(ForceContainer)arg1, (int)id*[, *(bool)sync=False*]) → Vector3 :
    Deprecated alias for t (torque).

**permF**(*(ForceContainer)arg1, (int)id*) → Vector3 :
    read the value of permanent force on body (set with setPermF()).

**permT**(*(ForceContainer)arg1, (int)id*) → Vector3 :
    read the value of permanent torque on body (set with setPermT()).

**reset**(*(ForceContainer)arg1*[, *(bool)resetAll=True*]) → None :
    Reset the force container, including user defined permanent forces/torques. resetAll=False will keep permanent forces/torques unchanged.

**setPermF**(*(ForceContainer)arg1, (int)arg2, (Vector3)arg3*) → None :
    set the value of permanent force on body.

**setPermT**(*(ForceContainer)arg1, (int)arg2, (Vector3)arg3*) → None :
    set the value of permanent torque on body.

**syncCount**
    Number of synchronizations of ForceContainer (cummulative); if significantly higher than number of steps, there might be unnecessary syncs hurting performance.

**t**(*(ForceContainer)arg1, (int)id*$\big[$*, (bool)sync=False*$\big]$) → Vector3 :
> Torque applied on body. For clumps in openMP, synchronize the force container with sync=True, else the value will be wrong.

## MaterialContainer

**class yade.wrapper.MaterialContainer**
> Container for *Materials*. A material can be accessed using

> 1. numerical index in range(0,len(cont)), like cont[2];

> 2. textual label that was given to the material, like cont['steel']. This entails traversing all materials and should not be used frequently.

> **__init__**(*(object)arg1, (MaterialContainer)arg2*) → None

> **append**(*(MaterialContainer)arg1, (Material)arg2*) → int :
> > Add new shared *Material*; changes its id and return it.

> > **append( (MaterialContainer)arg1, (object)arg2) -> object :** Append list of *Material* instances, return list of ids.

> **index**(*(MaterialContainer)arg1, (str)arg2*) → int :
> > Return id of material, given its label.

## Scene

**class yade.wrapper.Scene**(*inherits Serializable*)
> Object comprising the whole simulation.

> **dict**(*(Serializable)arg1*) → dict :
> > Return dictionary of attributes.

> **doSort**(*=false*)
> > Used, when new body is added to the scene.

> **dt**(*=1e-8*)
> > Current timestep for integration.

> **isPeriodic**(*=false*)
> > Whether periodic boundary conditions are active.

> **iter**(*=0*)
> > Current iteration (computational step) number

> **selectedBody**(*=-1*)
> > Id of body that is selected by the user

> **speed**(*=0*)
> > Current calculation speed [iter/s]

> **stopAtIter**(*=0*)
> > Iteration after which to stop the simulation.

> **stopAtTime**(*=0*)
> > Time after which to stop the simulation

> **subStep**(*=-1*)
> > Number of sub-step; not to be changed directly. -1 means to run loop prologue (cell integration), 0...n-1 runs respective engines (n is number of engines), n runs epilogue (increment step number and time.

> **subStepping**(*=false*)
> > Whether we currently advance by one engine in every step (rather than by single run through all engines).

**tags**(*=uninitialized*)
> Arbitrary key=value associations (tags like mp3 tags: author, date, version, description etc.)

**time**(*=0*)
> Simulation time (virtual time) [s]

**trackEnergy**(*=false*)
> Whether energies are being traced.

**updateAttrs**(*(Serializable)arg1, (dict)arg2*) → None :
> Update object attributes from given dictionary

## Cell

**class** yade.wrapper.**Cell**(*inherits Serializable*)
> Parameters of periodic boundary conditions. Only applies if O.isPeriodic==True.

**dict**(*(Serializable)arg1*) → dict :
> Return dictionary of attributes.

**getDefGrad**(*(Cell)arg1*) → Matrix3 :
> Returns deformation gradient tensor $\mathbf{F}$ of the cell deformation (http://en.wikipedia.org/wiki/Finite_strain_theory)

**getEulerianAlmansiStrain**(*(Cell)arg1*) → Matrix3 :
> Returns Eulerian-Almansi strain tensor $\mathbf{e} = \frac{1}{2}(\mathbf{I} - \mathbf{b}^{-1}) = \frac{1}{2}(\mathbf{I} - (\mathbf{F}\mathbf{F}^\mathsf{T})^{-1})$ of the cell (http://en.wikipedia.org/wiki/Finite_strain_theory)

**getLCauchyGreenDef**(*(Cell)arg1*) → Matrix3 :
> Returns left Cauchy-Green deformation tensor $\mathbf{b} = \mathbf{F}\mathbf{F}^\mathsf{T}$ of the cell (http://en.wikipedia.org/wiki/Finite_strain_theory)

**getLagrangianStrain**(*(Cell)arg1*) → Matrix3 :
> Returns Lagrangian strain tensor $\mathbf{E} = \frac{1}{2}(\mathbf{C} - \mathbf{I}) = \frac{1}{2}(\mathbf{F}^\mathsf{T}\mathbf{F} - \mathbf{I}) = \frac{1}{2}(\mathbf{U}^2 - \mathbf{I})$ of the cell (http://en.wikipedia.org/wiki/Finite_strain_theory)

**getLeftStretch**(*(Cell)arg1*) → Matrix3 :
> Returns left (spatial) stretch tensor of the cell (matrix $\mathbf{U}$ from polar decomposition $\mathbf{F} = \mathbf{R}\mathbf{U}$ )

**getPolarDecOfDefGrad**(*(Cell)arg1*) → tuple :
> Returns orthogonal matrix $\mathbf{R}$ and symmetric positive semi-definite matrix $\mathbf{U}$ as polar decomposition of deformation gradient $\mathbf{F}$ of the cell ( $\mathbf{F} = \mathbf{R}\mathbf{U}$ )

**getRCauchyGreenDef**(*(Cell)arg1*) → Matrix3 :
> Returns right Cauchy-Green deformation tensor $\mathbf{C} = \mathbf{F}^\mathsf{T}\mathbf{F}$ of the cell (http://en.wikipedia.org/wiki/Finite_strain_theory)

**getRightStretch**(*(Cell)arg1*) → Matrix3 :
> Returns right (material) stretch tensor of the cell (matrix $\mathbf{V}$ from polar decomposition $\mathbf{F} = \mathbf{R}\mathbf{U} = \mathbf{V}\mathbf{R} \rightarrow \mathbf{V} = \mathbf{F}\mathbf{R}^\mathsf{T}$ )

**getRotation**(*(Cell)arg1*) → Matrix3 :
> Returns rotation of the cell (orthogonal matrix $\mathbf{R}$ from polar decomposition $\mathbf{F} = \mathbf{R}\mathbf{U}$ )

**getSmallStrain**(*(Cell)arg1*) → Matrix3 :
> Returns small strain tensor $\boldsymbol{\varepsilon} = \frac{1}{2}(\mathbf{F} + \mathbf{F}^\mathsf{T}) - \mathbf{I}$ of the cell (http://en.wikipedia.org/wiki/Finite_strain_theory)

**getSpin**(*(Cell)arg1*) → Vector3 :
> Returns the spin defined by the skew symmetric part of *velGrad*

**hSize**
> Base cell vectors (columns of the matrix), updated at every step from *velGrad* (*trsf* accumulates applied *velGrad* transformations). Setting *hSize* during a simulation is not supported by most contact laws, it is only meant to be used at iteration 0 before any interactions have been created.

**hSize0**

Value of untransformed hSize, with respect to current *trsf* (computed as *trsf* $^{-1}$ × *hSize*.

**homoDeform**(*=2*)

If >0, deform (*velGrad*) the cell homothetically by adjusting positions and velocities of bodies. The velocity change is obtained by deriving the expression v= v.x, where v is the macroscopic velocity gradient, giving in an incremental form: Δv=Δ v x + v Δx. As a result, velocities are modified as soon as `velGrad` changes, according to the first term: Δv(t)=Δ v x(t), while the 2nd term reflects a convective term: Δv'= v v(t-dt/2). The second term is neglected if homoDeform=1. All terms are included if homoDeform=2 (default)

**nextVelGrad**(*=Matrix3r::Zero()*)

see *Cell.velGrad*.

**prevHSize**(*=Matrix3r::Identity()*)

*hSize* from the previous step, used in the definition of relative velocity across periods.

**prevVelGrad**(*=Matrix3r::Zero()*)

Velocity gradient in the previous step.

**refHSize**(*=Matrix3r::Identity()*)

Reference cell configuration, only used with *OpenGLRenderer.dispScale*. Updated automatically when *hSize* or *trsf* is assigned directly; also modified by *utils.setRefSe3* (called e.g. by the `Reference` button in the UI).

**refSize**

Reference size of the cell (lengths of initial cell vectors, i.e. column norms of *hSize*).

---

**Note:** Modifying this value is deprecated, use *setBox* instead.

---

**setBox**(*(Cell)arg1, (Vector3)arg2*) → None :

Set *Cell* shape to be rectangular, with dimensions along axes specified by given argument. Shorthand for assigning diagonal matrix with respective entries to *hSize*.

**setBox( (Cell)arg1, (float)arg2, (float)arg3, (float)arg4) -> None :** Set *Cell* shape to be rectangular, with dimensions along x, y, z specified by arguments. Shorthand for assigning diagonal matrix with the respective entries to *hSize*.

**shearPt**(*(Cell)arg1, (Vector3)arg2*) → Vector3 :

Apply shear (cell skew+rot) on the point

**shearTrsf**

Current skew+rot transformation (no resize)

**size**

Current size of the cell, i.e. lengths of the 3 cell lateral vectors contained in *Cell.hSize* columns. Updated automatically at every step.

**trsf**

Current transformation matrix of the cell, obtained from time integration of *Cell.velGrad*.

**unshearPt**(*(Cell)arg1, (Vector3)arg2*) → Vector3 :

Apply inverse shear on the point (removes skew+rot of the cell)

**unshearTrsf**

Inverse of the current skew+rot transformation (no resize)

**updateAttrs**(*(Serializable)arg1, (dict)arg2*) → None :

Update object attributes from given dictionary

---

**velGrad**

Velocity gradient of the transformation; used in *NewtonIntegrator*. Values of *velGrad* accumulate in *trsf* at every step.

note: changing velGrad at the beginning of a timestep would lead to inaccurate integration for that step, as it should normally be changed after the contact laws (but before Newton). To avoid this problem, assignment is deferred automatically. The assigned value is internaly stored in *Cell.nextVelGrad* and will be applied right in time by Newton integrator.

> **Warning:** Assigning individual components as in *O.cell.velGrad[0,0]=1* is not possible (it will not return any error but it will have no effect). Instead, the whole matrix should be assigned, as in *O.cell.velGrad=Matrix3(...)*. Alternatively *nextVelGrad* can be assigned directly (both per-component or as a whole) and the effect should be the same.

**velGradChanged**(*=false*)

true when velGrad has been changed manually (see also *Cell.nextVelGrad*)

**volume**

Current volume of the cell.

**wrap**(*(Cell)arg1, (Vector3)arg2*) → Vector3 :

Transform an arbitrary point into a point in the reference cell

**wrapPt**(*(Cell)arg1, (Vector3)arg2*) → Vector3 :

Wrap point inside the reference cell, assuming the cell has no skew+rot.

## 2.3.16 Other classes

**class** yade.wrapper.**TimingDeltas**

**data**

Get timing data as list of tuples (label, execTime[nsec], execCount) (one tuple per checkpoint)

**reset**(*(TimingDeltas)arg1*) → None :

Reset timing information

**class** yade.wrapper.**Serializable**

**dict**(*(Serializable)arg1*) → dict :

Return dictionary of attributes.

**updateAttrs**(*(Serializable)arg1, (dict)arg2*) → None :

Update object attributes from given dictionary

**class** yade.wrapper.**GlExtra_LawTester**(*inherits GlExtraDrawer → Serializable*)

Find an instance of *LawTester* and show visually its data.

**dead**(*=false*)

Deactivate the object (on error/exception).

**dict**(*(Serializable)arg1*) → dict :

Return dictionary of attributes.

**tester**(*=uninitalized*)

Associated *LawTester* object.

**updateAttrs**(*(Serializable)arg1, (dict)arg2*) → None :

Update object attributes from given dictionary **class** `yade.wrapper.MatchMaker`(*inherits Serializable*)

Class matching pair of ids to return pre-defined (for a pair of ids defined in *matches*) or derived value (computed using *algo*) of a scalar parameter. It can be called (`id1, id2, val1=NaN, val2=NaN`) in both python and c++.

---

**Note:** There is a *converter* from python number defined for this class, which creates a new *MatchMaker* returning the value of that number; instead of giving the object instance therefore, you can only pass the number value and it will be converted automatically.

---

**algo**

Algorithm used to compute value when no match for ids is found. Possible values are

- 'avg' (arithmetic average)

- 'min' (minimum value)

- 'max' (maximum value)

- 'harmAvg' (harmonic average)

The following algo algorithms do *not* require meaningful input values in order to work:

- 'val' (return value specified by *val*)

- 'zero' (always return 0.)

**computeFallback**(*(MatchMaker)arg1, (float)val1, (float)val2*) → float :
Compute algo value for *val1* and *val2*, using algorithm specified by *algo*.

**dict**(*(Serializable)arg1*) → dict :
Return dictionary of attributes.

**matches**(*=uninitalized*)
Array of (`id1,id2,value`) items; queries matching `id1` + `id2` or `id2` + `id1` will return `value`

**updateAttrs**(*(Serializable)arg1, (dict)arg2*) → None :
Update object attributes from given dictionary

**val**(*=NaN*)
Constant value returned if there is no match and *algo* is `val`

**class** `yade.wrapper.Engine`(*inherits Serializable*)

Basic execution unit of simulation, called from the simulation loop (O.engines)

**dead**(*=false*)
If true, this engine will not run at all; can be used for making an engine temporarily deactivated and only resurrect it at a later point.

**dict**(*(Serializable)arg1*) → dict :
Return dictionary of attributes.

**execCount**
Cumulative count this engine was run (only used if *O.timingEnabled*==`True`).

**execTime**
Cumulative time in nanoseconds this Engine took to run (only used if *O.timingEnabled*==`True`).

**label**(*=uninitalized*)
Textual label for this object; must be valid python identifier, you can refer to it directly from python.

**ompThreads**(*=-1*)

    Number of threads to be used in the engine. If ompThreads<0 (default), the number will be typically OMP_NUM_THREADS or the number N defined by 'yade -jN' (this behavior can depend on the engine though). This attribute will only affect engines whose code includes openMP parallel regions (e.g. *InteractionLoop*). This attribute is mostly useful for experiments or when combining *ParallelEngine* with engines that run parallel regions, resulting in nested OMP loops with different number of threads at each level.

**timingDeltas**

    Detailed information about timing inside the Engine itself. Empty unless enabled in the source code and *O.timingEnabled*==`True`.

**updateAttrs**(*(Serializable)arg1, (dict)arg2*) → None :

    Update object attributes from given dictionary

**class yade.wrapper.EnergyTracker**(*inherits Serializable*)

    Storage for tracing energies. Only to be used if *O.trackEnergy* is True.

    **clear**(*(EnergyTracker)arg1*) → None :

        Clear all stored values.

    **dict**(*(Serializable)arg1*) → dict :

        Return dictionary of attributes.

    **energies**(*=uninitalized*)

        Energy values, in linear array

    **items**(*(EnergyTracker)arg1*) → list :

        Return contents as list of (name,value) tuples.

    **keys**(*(EnergyTracker)arg1*) → list :

        Return defined energies.

    **total**(*(EnergyTracker)arg1*) → float :

        Return sum of all energies.

    **updateAttrs**(*(Serializable)arg1, (dict)arg2*) → None :

        Update object attributes from given dictionary

**class yade.wrapper.LinExponentialPotential**(*inherits CundallStrackPotential → GenericPotential → Serializable*)

    LinExponential Potential with only Cundall-and-Strack-like contact. The LinExponential potential formula is $F(u) = \frac{k*(x_e - x_0)}{x_e}(u/a - x_0)\exp\left(\frac{-(u/a)}{x_e - x_0}\right)$. Where k is the slope at the origin, $x_0$ is the position where the potential cross 0 and $x_e$ is the position of the extremum.

    **F0**(*=1*)

        Force at contact. Force when $F_0 = F(u = 0)$ (LinExponential)

    **Fe**(*=1*)

        Extremum force. Value of force at extremum. (LinExponential)

    **alpha**(*=1*)

        Bulk-to-roughness stiffness ratio

    **computeParametersFromF0**(*(LinExponentialPotential)arg1, (float)F0, (float)xe, (float)k*) → None :

        Set parameters of the potential, with k computed from $F_0$

    **computeParametersFromF0Fe**(*(LinExponentialPotential)arg1, (float)xe, (float)Fe, (float)F0*) → None :

        Set parameters of the potential, with k and $x_0$ computed from $F_0$ and $F_e$

    **dict**(*(Serializable)arg1*) → dict :

        Return dictionary of attributes.

    **k**(*=1*)

        Slope at the origin (stiffness). (LinExponential)

**potential**(*(LinExponentialPotential)arg1, (float)u*) → float :
  Get potential value at any point.

**setParameters**(*(LinExponentialPotential)arg1, (float)x0, (float)xe, (float)k*) → None :
  Set parameters of the potential

**updateAttrs**(*(Serializable)arg1, (dict)arg2*) → None :
  Update object attributes from given dictionary

**x0**(*=0*)
  Equilibrium distance. Potential force is 0 at $x_0$ (LinExponential)

**xe**(*=1*)
  Extremum position. Position of local max/min of force. (LinExponential)

**class yade.wrapper.CundallStrackPotential**(*inherits* *GenericPotential* → *Serializable*)
  Potential with only Cundall-and-Strack-like contact.

**alpha**(*=1*)
  Bulk-to-roughness stiffness ratio

**dict**(*(Serializable)arg1*) → dict :
  Return dictionary of attributes.

**updateAttrs**(*(Serializable)arg1, (dict)arg2*) → None :
  Update object attributes from given dictionary

**class yade.wrapper.GlExtra_OctreeCubes**(*inherits* *GlExtraDrawer* → *Serializable*)
  Render boxed read from file

**boxesFile**(*=uninitalized*)
  File to read boxes from; ascii files with `x0 y0 z0 x1 y1 z1 c` records, where `c` is an integer specifying fill (0 for wire, 1 for filled).

**dead**(*=false*)
  Deactivate the object (on error/exception).

**dict**(*(Serializable)arg1*) → dict :
  Return dictionary of attributes.

**fillRangeDraw**(*=Vector2i(-2, 2)*)
  Range of fill indices that will be rendered.

**fillRangeFill**(*=Vector2i(2, 2)*)
  Range of fill indices that will be filled.

**levelRangeDraw**(*=Vector2i(-2, 2)*)
  Range of levels that will be rendered.

**noFillZero**(*=true*)
  Do not fill 0-fill boxed (those that are further subdivided)

**updateAttrs**(*(Serializable)arg1, (dict)arg2*) → None :
  Update object attributes from given dictionary

**class yade.wrapper.ParallelEngine**(*inherits* *Engine* → *Serializable*)
  Engine for running other Engine in parallel.

**__init__**(*(object)arg1*) → None
  object ___init___(tuple args, dict kwds)

  **___init___( (object)arg1, (list)arg2) -> object :** Construct from (possibly nested) list of slaves.

**dead**(*=false*)
  If true, this engine will not run at all; can be used for making an engine temporarily deactivated and only resurrect it at a later point.

**dict**(*(Serializable)arg1*) → dict :
   Return dictionary of attributes.

**execCount**
   Cumulative count this engine was run (only used if *O.timingEnabled*==`True`).

**execTime**
   Cumulative time in nanoseconds this Engine took to run (only used if *O.timingEnabled*==`True`).

**label**(*=uninitalized*)
   Textual label for this object; must be valid python identifier, you can refer to it directly from python.

**ompThreads**(*=-1*)
   Number of threads to be used in the engine. If ompThreads<0 (default), the number will be typically OMP_NUM_THREADS or the number N defined by 'yade -jN' (this behavior can depend on the engine though). This attribute will only affect engines whose code includes openMP parallel regions (e.g. *InteractionLoop*). This attribute is mostly useful for experiments or when combining *ParallelEngine* with engines that run parallel regions, resulting in nested OMP loops with different number of threads at each level.

**slaves**
   List of lists of Engines; each top-level group will be run in parallel with other groups, while Engines inside each group will be run sequentially, in given order.

**timingDeltas**
   Detailed information about timing inside the Engine itself. Empty unless enabled in the source code and *O.timingEnabled*==`True`.

**updateAttrs**(*(Serializable)arg1, (dict)arg2*) → None :
   Update object attributes from given dictionary

**class yade.wrapper.Cell**(*inherits Serializable*)
   Parameters of periodic boundary conditions. Only applies if O.isPeriodic==True.

**dict**(*(Serializable)arg1*) → dict :
   Return dictionary of attributes.

**getDefGrad**(*(Cell)arg1*) → Matrix3 :
   Returns deformation gradient tensor $\mathbf{F}$ of the cell deformation (http://en.wikipedia.org/wiki/Finite_strain_theory)

**getEulerianAlmansiStrain**(*(Cell)arg1*) → Matrix3 :
   Returns Eulerian-Almansi strain tensor $\mathbf{e} = \frac{1}{2}(\mathbf{I} - \mathbf{b}^{-1}) = \frac{1}{2}(\mathbf{I} - (\mathbf{FF}^\mathsf{T})^{-1})$ of the cell (http://en.wikipedia.org/wiki/Finite_strain_theory)

**getLCauchyGreenDef**(*(Cell)arg1*) → Matrix3 :
   Returns left Cauchy-Green deformation tensor $\mathbf{b} = \mathbf{FF}^\mathsf{T}$ of the cell (http://en.wikipedia.org/wiki/Finite_strain_theory)

**getLagrangianStrain**(*(Cell)arg1*) → Matrix3 :
   Returns Lagrangian strain tensor $\mathbf{E} = \frac{1}{2}(\mathbf{C} - \mathbf{I}) = \frac{1}{2}(\mathbf{F}^\mathsf{T}\mathbf{F} - \mathbf{I}) = \frac{1}{2}(\mathbf{U}^2 - \mathbf{I})$ of the cell (http://en.wikipedia.org/wiki/Finite_strain_theory)

**getLeftStretch**(*(Cell)arg1*) → Matrix3 :
   Returns left (spatial) stretch tensor of the cell (matrix $\mathbf{U}$ from polar decomposition $\mathbf{F} = \mathbf{RU}$ )

**getPolarDecOfDefGrad**(*(Cell)arg1*) → tuple :
   Returns orthogonal matrix $\mathbf{R}$ and symmetric positive semi-definite matrix $\mathbf{U}$ as polar decomposition of deformation gradient $\mathbf{F}$ of the cell ( $\mathbf{F} = \mathbf{RU}$ )

**getRCauchyGreenDef**(*(Cell)arg1*) → Matrix3 :
   Returns right Cauchy-Green deformation tensor $\mathbf{C} = \mathbf{F}^\mathsf{T}\mathbf{F}$ of the cell (http://en.wikipedia.org/wiki/Finite_strain_theory)

**getRightStretch**(*(Cell)arg1*) → Matrix3 :

Returns right (material) stretch tensor of the cell (matrix $\mathbf{V}$ from polar decomposition $\mathbf{F} = \mathbf{RU} = \mathbf{VR} \rightarrow \mathbf{V} = \mathbf{FR}^\mathsf{T}$ )

**getRotation**(*(Cell)arg1*) → Matrix3 :

Returns rotation of the cell (orthogonal matrix $\mathbf{R}$ from polar decomposition $\mathbf{F} = \mathbf{RU}$ )

**getSmallStrain**(*(Cell)arg1*) → Matrix3 :

Returns small strain tensor $\boldsymbol{\varepsilon} = \frac{1}{2}(\mathbf{F}+\mathbf{F}^\mathsf{T})-\mathbf{I}$ of the cell (http://en.wikipedia.org/wiki/Finite_strain_theory)

**getSpin**(*(Cell)arg1*) → Vector3 :

Returns the spin defined by the skew symmetric part of *velGrad*

**hSize**

Base cell vectors (columns of the matrix), updated at every step from *velGrad* (*trsf* accumulates applied *velGrad* transformations). Setting *hSize* during a simulation is not supported by most contact laws, it is only meant to be used at iteration 0 before any interactions have been created.

**hSize0**

Value of untransformed hSize, with respect to current *trsf* (computed as $trsf^{-1} \times hSize$.

**homoDeform**(*=2*)

If >0, deform (*velGrad*) the cell homothetically by adjusting positions and velocities of bodies. The velocity change is obtained by deriving the expression v= v.x, where v is the macroscopic velocity gradient, giving in an incremental form: Δv=Δ v x + v Δx. As a result, velocities are modified as soon as **velGrad** changes, according to the first term: Δv(t)=Δ v x(t), while the 2nd term reflects a convective term: Δv'= v v(t-dt/2). The second term is neglected if homoDeform=1. All terms are included if homoDeform=2 (default)

**nextVelGrad**(*=Matrix3r::Zero()*)

see *Cell.velGrad*.

**prevHSize**(*=Matrix3r::Identity()*)

*hSize* from the previous step, used in the definition of relative velocity across periods.

**prevVelGrad**(*=Matrix3r::Zero()*)

Velocity gradient in the previous step.

**refHSize**(*=Matrix3r::Identity()*)

Reference cell configuration, only used with *OpenGLRenderer.dispScale*. Updated automatically when *hSize* or *trsf* is assigned directly; also modified by *utils.setRefSe3* (called e.g. by the **Reference** button in the UI).

**refSize**

Reference size of the cell (lengths of initial cell vectors, i.e. column norms of *hSize*).

---

**Note:** Modifying this value is deprecated, use *setBox* instead.

---

**setBox**(*(Cell)arg1, (Vector3)arg2*) → None :

Set *Cell* shape to be rectangular, with dimensions along axes specified by given argument. Shorthand for assigning diagonal matrix with respective entries to *hSize*.

**setBox( (Cell)arg1, (float)arg2, (float)arg3, (float)arg4) -> None :** Set *Cell* shape to be rectangular, with dimensions along x, y, z specified by arguments. Shorthand for assigning diagonal matrix with the respective entries to *hSize*.

**shearPt**(*(Cell)arg1, (Vector3)arg2*) → Vector3 :

Apply shear (cell skew+rot) on the point

**shearTrsf**

Current skew+rot transformation (no resize)

**size**
> Current size of the cell, i.e. lengths of the 3 cell lateral vectors contained in *Cell.hSize* columns. Updated automatically at every step.

**trsf**
> Current transformation matrix of the cell, obtained from time integration of *Cell.velGrad*.

**unshearPt**(*(Cell)arg1, (Vector3)arg2*) → Vector3 :
> Apply inverse shear on the point (removes skew+rot of the cell)

**unshearTrsf**
> Inverse of the current skew+rot transformation (no resize)

**updateAttrs**(*(Serializable)arg1, (dict)arg2*) → None :
> Update object attributes from given dictionary

**velGrad**
> Velocity gradient of the transformation; used in *NewtonIntegrator*. Values of *velGrad* accumulate in *trsf* at every step.
>
> > note: changing velGrad at the beginning of a timestep would lead to inaccurate integration for that step, as it should normally be changed after the contact laws (but before Newton). To avoid this problem, assignment is deferred automatically. The assigned value is internaly stored in *Cell.nextVelGrad* and will be applied right in time by Newton integrator.
>
> > **Warning:** Assigning individual components as in *O.cell.velGrad[0,0]=1* is not possible (it will not return any error but it will have no effect). Instead, the whole matrix should be assigned, as in *O.cell.velGrad=Matrix3(...)*. Alternatively *nextVelGrad* can be assigned directly (both per-component or as a whole) and the effect should be the same.

**velGradChanged**(*=false*)
> true when velGrad has been changed manually (see also *Cell.nextVelGrad*)

**volume**
> Current volume of the cell.

**wrap**(*(Cell)arg1, (Vector3)arg2*) → Vector3 :
> Transform an arbitrary point into a point in the reference cell

**wrapPt**(*(Cell)arg1, (Vector3)arg2*) → Vector3 :
> Wrap point inside the reference cell, assuming the cell has no skew+rot.

**class yade.wrapper.CundallStrackAdhesivePotential**(*inherits* *CundallStrackPotential* → *GenericPotential* → *Serializable*)
> CundallStrack model with adhesive part. Contact is created when $u/a - \varepsilon < 0$ and released when $u/a - \varepsilon > l_{adh}$, where $l_{adh} = f_{adh}/k_n$. This lead to an hysteretic attractive part.

**alpha**(*=1*)
> Bulk-to-roughness stiffness ratio

**dict**(*(Serializable)arg1*) → dict :
> Return dictionary of attributes.

**fadh**(*=0*)
> Adhesion force.

**updateAttrs**(*(Serializable)arg1, (dict)arg2*) → None :
> Update object attributes from given dictionary

**class yade.wrapper.GenericPotential**(*inherits* *Serializable*)
> Generic class for potential representation in PotentialLubrication law. Don't do anything. If set as potential, the result will be a lubrication-only simulation.

> **dict**(*(Serializable)arg1*) → dict :
>> Return dictionary of attributes.

> **updateAttrs**(*(Serializable)arg1, (dict)arg2*) → None :
>> Update object attributes from given dictionary

**class** `yade.wrapper.GlExtraDrawer`(*inherits Serializable*)
> Performing arbitrary OpenGL drawing commands; called from *OpenGLRenderer* (see *OpenGLRenderer.extraDrawers*) once regular rendering routines will have finished.

> This class itself does not render anything, derived classes should override the *render* method.

> **dead**(*=false*)
>> Deactivate the object (on error/exception).

> **dict**(*(Serializable)arg1*) → dict :
>> Return dictionary of attributes.

> **updateAttrs**(*(Serializable)arg1, (dict)arg2*) → None :
>> Update object attributes from given dictionary

## 2.4 Yade modules reference

### 2.4.1 yade.bodiesHandling module

Miscellaneous functions, which are useful for handling bodies.

`yade.bodiesHandling.`**facetsDimensions**(*idFacets=[], mask=-1*)
> The function accepts the list of facet id's or list of facets and calculates max and min dimensions, geometrical center.

>> **Parameters**
>>
>> • **idFacets** (`list`) – list of spheres
>>
>> • **mask** (`int`) – *Body.mask* for the checked bodies
>
>> **Returns** dictionary with keys `min` (minimal dimension, Vector3), `max` (maximal dimension, Vector3), `minId` (minimal dimension facet Id, Vector3), `maxId` (maximal dimension facet Id, Vector3), `center` (central point of bounding box, Vector3), `extends` (sizes of bounding box, Vector3), `number` (number of facets, int),

`yade.bodiesHandling.`**sphereDuplicate**(*idSphere*)
> The functions makes a copy of sphere

`yade.bodiesHandling.`**spheresModify**(*idSpheres=[], mask=-1, shift=Vector3(0, 0, 0), scale=1.0, orientation=Quaternion((1, 0, 0), 0), copy=False*)
> The function accepts the list of spheres id's or list of bodies and modifies them: rotating, scaling, shifting. if copy=True copies bodies and modifies them. Also the mask can be given. If idSpheres not empty, the function affects only bodies, where the mask passes. If idSpheres is empty, the function search for bodies, where the mask passes.

>> **Parameters**
>>
>> • **shift** (`Vector3`) – Vector3(X,Y,Z) parameter moves spheres.
>>
>> • **scale** (`float`) – factor scales given spheres.
>>
>> • **orientation** (`Quaternion`) – orientation of spheres
>>
>> • **mask** (`int`) – *Body.mask* for the checked bodies
>
>> **Returns** list of bodies if copy=True, and Boolean value if copy=False

`yade.bodiesHandling.`**`spheresPackDimensions`**(*idSpheres=[], mask=-1*)

    The function accepts the list of spheres id's or list of bodies and calculates max and min dimensions, geometrical center.

        **Parameters**

- **idSpheres** (*list*) – list of spheres
- **mask** (*int*) – *Body.mask* for the checked bodies

        **Returns** dictionary with keys **min** (minimal dimension, Vector3), **max** (maximal dimension, Vector3), **minId** (minimal dimension sphere Id, Vector3), **maxId** (maximal dimension sphere Id, Vector3), **center** (central point of bounding box, Vector3), **extends** (sizes of bounding box, Vector3), **volume** (volume of spheres, Real), **mass** (mass of spheres, Real), **number** (number of spheres, int),

## 2.4.2 yade.export module

Export (not only) geometry to various formats.

**class** `yade.export.`**`VTKExporter`**(*inherits object*)

    Class for exporting data to VTK Simple Legacy File (for example if, for some reason, you are not able to use *VTKRecorder*). Supported export of:

- spheres
- facets
- polyhedra
- PotentialBlocks
- interactions
- contact points
- periodic cell

Usage:

- create object **vtkExporter** = VTKExporter('baseFileName'),
- add to **O.engines** a PyRunner with **command='vtkExporter.exportSomething(...)'**
- alternatively, just use **vtkExporter.exportSomething(...)** at the end of the script for instance

Example: examples/test/vtk-exporter/vtkExporter.py, examples/test/unv-read/unvReadVTKExport.py.

        **Parameters**

- **baseName** (*string*) – name of the exported files. The files would be named, e.g., baseName-spheres-snapNb.vtk or baseName-facets-snapNb.vtk
- **startSnap** (*int*) – the numbering of files will start form **startSnap**

**exportContactPoints**(*ids='all', what={}, useRef={}, comment='comment', numLabel=None*)

    exports contact points (CPs) and defined properties.

        **Parameters**

- **ids** (*[(int,int)]*) – see *exportInteractions()*
- **what** (*dictionary*) – see *exportInteractions()*
- **useRef** (*bool*) – see *exportInteractions()*
- **comment** (*string*) – comment to add to vtk file

- **numLabel** (*int*) – number of file (e.g. time step), if unspecified, the last used value + 1 will be used

**exportFacets**(*ids='all'*, *what={}*, *comment='comment'*, *numLabel=None*)
    exports facets (positions) and defined properties. Facets are exported with multiplicated nodes

      **Parameters**

- **ids** (*[int]/"all"*) – if "all", then export all facets, otherwise only facets from integer list
- **what** (*dictionary*) – see *exportSpheres()*
- **comment** (*string*) – comment to add to vtk file
- **numLabel** (*int*) – number of file (e.g. time step), if unspecified, the last used value + 1 will be used

**exportFacetsAsMesh**(*ids='all'*, *connectivityTable=None*, *what={}*, *comment='comment'*, *numLabel=None*)
    exports facets (positions) and defined properties. Facets are exported as mesh (not with multiplicated nodes). Therefore additional parameters connectivityTable is needed

      **Parameters**

- **ids** (*[int]/"all"*) – if "all", then export all facets, otherwise only facets from integer list
- **what** (*dictionary*) – see *exportSpheres()*
- **comment** (*string*) – comment to add to vtk file
- **numLabel** (*int*) – number of file (e.g. time step), if unspecified, the last used value + 1 will be used
- **nodes** (*[(float,float,float)/Vector3]*) – list of coordinates of nodes
- **connectivityTable** (*[(int,int,int)]*) – list of node ids of individual elements (facets)

**exportInteractions**(*ids='all'*, *what={}*, *verticesWhat={}*, *comment='comment'*, *numLabel=None*, *useRef=False*)
    exports interactions and defined properties.

      **Parameters**

- **ids** (*[(int,int)]/"all"*) – if "all", then export all interactions, otherwise only interactions from (int,int) list
- **what** (*dictionary*) – what to export. parameter is a name->command dictionary. Name is string under which it is saved to vtk, command is string to evaluate. Note that the interactions are labeled as i in this function. Scalar, vector and tensor variables are supported. For example, to export the stiffness difference (named as **dStiff**) from a certain value (1e9) you should write: **what=dict(dStiff='i.phys.kn-1e9', ... )**
- **verticesWhat** (*dictionary*) – what to export on connected bodies. Bodies are labeled as **b** (or **b1** and **b2** if you need to treat both bodies differently)
- **comment** (*string*) – comment to add to vtk file
- **numLabel** (*int*) – number of file (e.g. time step), if unspecified, the last used value + 1 will be used
- **useRef** (*bool*) – if False (default), use current position of the bodies for export, use reference position otherwise

**exportPeriodicCell**(*comment='comment'*, *numLabel=None*)
    exports the *Cell* geometry for periodic simulations.

**Parameters**

- **comment** (*string*) – comment to add to vtk file

- **numLabel** (*int*) – number of file (e.g. time step), if unspecified, the last used value + 1 will be used

**exportPolyhedra**(*ids='all', what={}, comment='comment', numLabel=None, useRef=False*)

Exports polyhedrons and defined properties.

**Parameters**

- **ids** (*[int] / "all"*) – if "all", then export all polyhedrons, otherwise only polyhedrons from integer list

- **what** (*dictionary*) – which additional quantities (in addition to the positions) to export. parameter is name->command dictionary. Name is string under which it is saved to vtk, command is string to evaluate. Note that the bodies are labeled as b in this function. Scalar, vector and tensor variables are supported. For example, to export velocity (named as particleVelocity) and the distance from point (0,0,0) (named as dist) you should write: `what=dict(particleVelocity='b.state.vel',dist='b.state.pos.norm()', ... )`

- **comment** (*string*) – comment to add to vtk file

- **numLabel** (*int*) – number of file (e.g. time step), if unspecified, the last used value + 1 will be used

**exportPotentialBlocks**(*ids='all', what={}, comment='comment', numLabel=None, useRef=False*)

Exports Potential Blocks and defined properties.

**Parameters**

- **ids** (*[int] / "all"*) – if "all", then export all Potential Blocks, otherwise only Potential Blocks from integer list

- **what** (*dictionary*) – which additional quantities (in addition to the positions) to export. parameter is name->command dictionary. Name is string under which it is saved to vtk, command is string to evaluate. Note that the bodies are labeled as b in this function. Scalar, vector and tensor variables are supported. For example, to export velocity (named as particleVelocity) and the distance from point (0,0,0) (named as dist) you should write: `what=dict(particleVelocity='b.state.vel',dist='b.state.pos.norm()', ... )`

- **comment** (*string*) – comment to add to vtk file

- **numLabel** (*int*) – number of file (e.g. time step), if unspecified, the last used value + 1 will be used

**exportSpheres**(*ids='all', what={}, comment='comment', numLabel=None, useRef=False*)

exports spheres (positions and radius) and defined properties.

**Parameters**

- **ids** (*[int]/"all"*) – if "all", then export all spheres, otherwise only spheres from integer list

- **what** (*dictionary*) – which additional quantities (other than the position and the radius) to export. parameter is name->command dictionary. Name is string under which it is save to vtk, command is string to evaluate. Note that the bodies are labeled as b in this function. Scalar, vector and tensor variables are supported. For example, to export velocity (with name particleVelocity) and the distance form point (0,0,0) (named as dist) you should write: `what=dict(particleVelocity='b.state.vel',dist='b.state.pos.norm()', ... )`

- **comment** (*string*) – comment to add to vtk file

- **numLabel** (*int*) – number of file (e.g. time step), if unspecified, the last used value + 1 will be used

- **useRef** (*bool*) – if False (default), use current position of the spheres for export, use reference position otherwise

**class** yade.export.**VTKWriter**(*inherits object*)

    USAGE: create object vtk_writer = VTKWriter('base_file_name'), add to engines PyRunner with command='vtk_writer.snapshot()'

    **snapshot()**

yade.export.**gmshGeo**(*filename*, *comment=''*, *mask=-1*, *accuracy=-1*)

    Save spheres in geo-file for the following using in GMSH (http://www.geuz.org/gmsh/doc/texinfo/) program. The spheres can be there meshed.

    **Parameters**

- **filename** (*string*) – the name of the file, where sphere coordinates will be exported.

- **mask** (*int*) – export only spheres with the corresponding mask export only spheres with the corresponding mask

- **accuracy** (*float*) – the accuracy parameter, which will be set for the poinst in geo-file. By default: 1./10. of the minimal sphere diameter.

    **Returns** number of spheres which were exported.

    **Return type** int

yade.export.**text**(*filename*, *mask=-1*)

    Save sphere coordinates into a text file; the format of the line is: x y z r. Non-spherical bodies are silently skipped. Example added to examples/regular-sphere-pack/regular-sphere-pack.py

    **Parameters**

- **filename** (*string*) – the name of the file, where sphere coordinates will be exported.

- **mask** (*int*) – export only spheres with the corresponding mask

    **Returns** number of spheres which were written.

    **Return type** int

yade.export.**text2vtk**(*inFileName*, *outFileName*, *comment='comment'*)

    Converts text file (created by *export.textExt* function) into vtk file. See examples/test/paraview-spheres-solid-section/export_text.py example

    **Parameters**

- **inFileName** (*str*) – name of input text file

- **outFileName** (*str*) – name of output vtk file

- **comment** (*str*) – optional comment in vtk file

yade.export.**text2vtkSection**(*inFileName*, *outFileName*, *point*, *normal=(1, 0, 0)*)

    Converts section through spheres from text file (created by *export.textExt* function) into vtk file. See examples/test/paraview-spheres-solid-section/export_text.py example

    **Parameters**

- **inFileName** (*str*) – name of input text file

- **outFileName** (*str*) – name of output vtk file

- **point** (*Vector3/(float,float,float)*) – coordinates of a point lying on the section plane

- **normal** (*Vector3/(float,float,float)*) – normal vector of the section plane

yade.export.**textClumps**(*filename, format='x_y_z_r_clumpId', comment='', mask=-1*)

　　Save clumps-members into a text file. Non-clumps members are bodies are silently skipped.

　　**Parameters**

- **filename** (*string*) – the name of the file, where sphere coordinates will be exported.

- **comment** (*string*) – the text, which will be added as a comment at the top of file. If you want to create several lines of text, please use '\n#' for next lines.

- **mask** (*int*) – export only spheres with the corresponding mask export only spheres with the corresponding mask

　　**Returns** number of clumps, number of spheres which were written.

　　**Return type** int

yade.export.**textExt**(*filename, format='x_y_z_r', comment='', mask=-1, attrs=[]*)

　　Save sphere coordinates and other parameters into a text file in specific format. Non-spherical bodies are silently skipped. Users can add here their own specific format, giving meaningful names. The first file row will contain the format name. Be sure to add the same format specification in ymport.textExt.

　　**Parameters**

- **filename** (*string*) – the name of the file, where sphere coordinates will be exported.

- **format** (*string*) – the name of output format. Supported 'x_y_z_r'(default), 'x_y_z_r_matId', 'x_y_z_r_attrs' (use proper comment)

- **comment** (*string*) – the text, which will be added as a comment at the top of file. If you want to create several lines of text, please use '\n#' for next lines. With 'x_y_z_r_attrs' format, the last (or only) line should consist of column headers of quantities passed as attrs (1 comment word for scalars, 3 comment words for vectors and 9 comment words for matrices)

- **mask** (*int*) – export only spheres with the corresponding mask export only spheres with the corresponding mask

- **attrs** (*[str]*) – attributes to be exported with 'x_y_z_r_attrs' format. Each str in the list is evaluated for every body exported with body=b (i.e. 'b.state.pos.norm()' would stand for distance of body from coordinate system origin)

　　**Returns** number of spheres which were written.

　　**Return type** int

yade.export.**textPolyhedra**(*fileName, comment='', mask=-1, explanationComment=True, attrs=[]*)

　　Save polyhedra into a text file. Non-polyhedra bodies are silently skipped.

　　**Parameters**

- **filename** (*string*) – the name of the output file

- **comment** (*string*) – the text, which will be added as a comment at the top of file. If you want to create several lines of text, please use '\n#' for next lines.

- **mask** (*int*) – export only polyhedra with the corresponding mask

- **explanationComment** (*str*) – inclde explanation of format to the beginning of file

> **Returns** number of polyhedra which were written.
>
> **Return type** int

### 2.4.3 yade.geom module

Creates geometry objects from facets.

`yade.geom.facetBox`(*center, extents, orientation=Quaternion((1, 0, 0), 0), wallMask=63, **kw*)
> Create arbitrarily-aligned box composed of facets, with given center, extents and orientation. If any of the box dimensions is zero, corresponding facets will not be created. The facets are oriented outwards from the box.
>
> > **Parameters**
> >
> > - **center** (`Vector3`) – center of the box
> >
> > - **extents** (`Vector3`) – half lengths of the box sides
> >
> > - **orientation** (`Quaternion`) – orientation of the box
> >
> > - **wallMask** (*bitmask*) – determines which walls will be created, in the order -x (1), +x (2), -y (4), +y (8), -z (16), +z (32). The numbers are ANDed; the default 63 means to create all walls
> >
> > - **\*\*kw** – (unused keyword arguments) passed to *utils.facet*
> >
> > **Returns** list of facets forming the box

`yade.geom.facetBunker`(*center, dBunker, dOutput, hBunker, hOutput, hPipe=0.0, orientation=Quaternion((1, 0, 0), 0), segmentsNumber=10, wallMask=4, angleRange=None, closeGap=False, **kw*)
> Create arbitrarily-aligned bunker, composed of facets, with given center, radii, heights and orientation. Return List of facets forming the bunker;

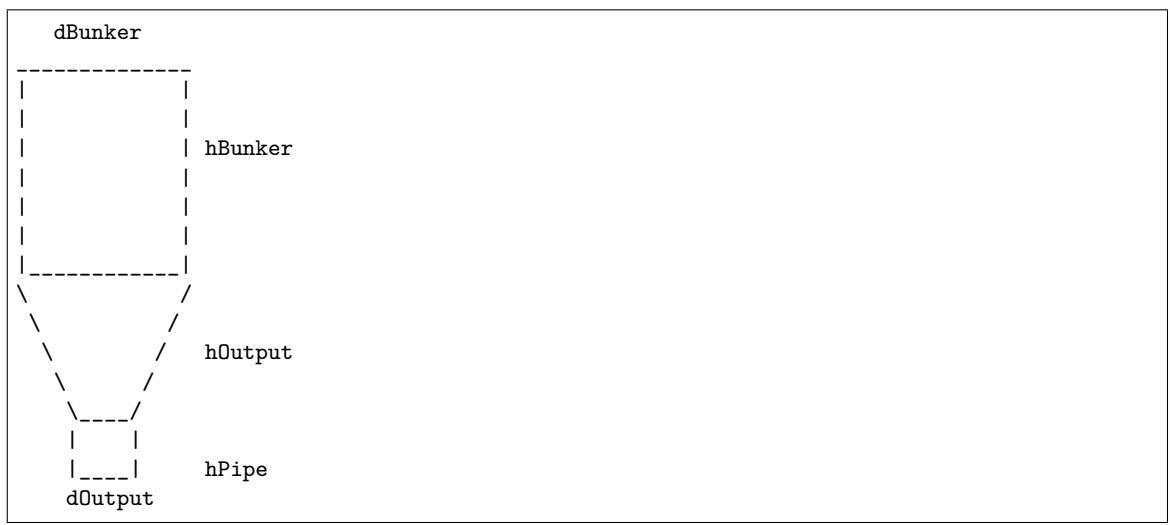

> > **Parameters**
> >
> > - **center** (`Vector3`) – center of the created bunker
> >
> > - **dBunker** (*float*) – bunker diameter, top
> >
> > - **dOutput** (*float*) – bunker output diameter
> >
> > - **hBunker** (*float*) – bunker height
> >
> > - **hOutput** (*float*) – bunker output height
> >
> > - **hPipe** (*float*) – bunker pipe height

- **orientation** (`Quaternion`) – orientation of the bunker; the reference orientation has axis along the +x axis.

- **segmentsNumber** (`int`) – number of edges on the bunker surface (>=5)

- **wallMask** (`bitmask`) – determines which walls will be created, in the order up (1), down (2), side (4). The numbers are ANDed; the default 7 means to create all walls

- **angleRange** (*(ϑmin,Θmax)*) – allows one to create only part of bunker by specifying range of angles; if `None`, (0,2*pi) is assumed.

- **closeGap** (`bool`) – close range skipped in angleRange with triangular facets at cylinder bases.

- **\*\*kw** – (unused keyword arguments) passed to utils.facet;

yade.geom.**facetCone**(*center, radiusTop, radiusBottom, height, orientation=Quaternion((1, 0, 0), 0), segmentsNumber=10, wallMask=7, angleRange=None, closeGap=False, radiusTopInner=-1, radiusBottomInner=-1, \*\*kw)*)
Create arbitrarily-aligned cone composed of facets, with given center, radius, height and orientation. Return List of facets forming the cone;

**Parameters**

- **center** (`Vector3`) – center of the created cylinder

- **radiusTop** (`float`) – cone top radius

- **radiusBottom** (`float`) – cone bottom radius

- **radiusTopInner** (`float`) – inner radius of cones top, -1 by default

- **radiusBottomInner** (`float`) – inner radius of cones bottom, -1 by default

- **height** (`float`) – cone height

- **orientation** (`Quaternion`) – orientation of the cone; the reference orientation has axis along the +x axis.

- **segmentsNumber** (`int`) – number of edges on the cone surface (>=5)

- **wallMask** (`bitmask`) – determines which walls will be created, in the order up (1), down (2), side (4). The numbers are ANDed; the default 7 means to create all walls

- **angleRange** (*(ϑmin,Θmax)*) – allows one to create only part of cone by specifying range of angles; if `None`, (0,2*pi) is assumed.

- **closeGap** (`bool`) – close range skipped in angleRange with triangular facets at cylinder bases.

- **\*\*kw** – (unused keyword arguments) passed to utils.facet;

yade.geom.**facetCylinder**(*center, radius, height, orientation=Quaternion((1, 0, 0), 0), segmentsNumber=10, wallMask=7, angleRange=None, closeGap=False, radiusTopInner=-1, radiusBottomInner=-1, \*\*kw)*)
Create arbitrarily-aligned cylinder composed of facets, with given center, radius, height and orientation. Return List of facets forming the cylinder;

**Parameters**

- **center** (`Vector3`) – center of the created cylinder

- **radius** (`float`) – cylinder radius

- **height** (`float`) – cylinder height

- **radiusTopInner** (`float`) – inner radius of cylinders top, -1 by default

- **radiusBottomInner** (`float`) – inner radius of cylinders bottom, -1 by default

- **orientation** (*Quaternion*) – orientation of the cylinder; the reference orientation has axis along the +x axis.

- **segmentsNumber** (*int*) – number of edges on the cylinder surface (>=5)

- **wallMask** (*bitmask*) – determines which walls will be created, in the order up (1), down (2), side (4). The numbers are ANDed; the default 7 means to create all walls

- **angleRange** (*(ϑmin,Θmax)*) – allows one to create only part of bunker by specifying range of angles; if None, (0,2*pi) is assumed.

- **closeGap** (*bool*) – close range skipped in angleRange with triangular facets at cylinder bases.

- ****kw** – (unused keyword arguments) passed to utils.facet;

yade.geom.**facetCylinderConeGenerator**(*center, radiusTop, height, orientation=Quaternion((1, 0, 0), 0), segmentsNumber=10, wallMask=7, angleRange=None, closeGap=False, radiusBottom=-1, radiusTopInner=-1, radiusBottomInner=-1, **kw*)

Please, do not use this function directly! Use geom.facetCylinder and geom.facetCone instead. This is the base function for generating cylinders and cones from facets.

**Parameters**

- **radiusTop** (*float*) – top radius

- **radiusBottom** (*float*) – bottom radius

- ****kw** – (unused keyword arguments) passed to utils.facet;

yade.geom.**facetHelix**(*center, radiusOuter, pitch, orientation=Quaternion((1, 0, 0), 0), segmentsNumber=10, angleRange=None, radiusInner=0, **kw*)

Create arbitrarily-aligned helix composed of facets, with given center, radius (outer and inner), pitch and orientation. Return List of facets forming the helix;

**Parameters**

- **center** (*Vector3*) – center of the created cylinder

- **radiusOuter** (*float*) – outer radius

- **radiusInner** (*float*) – inner height (can be 0)

- **orientation** (*Quaternion*) – orientation of the helix; the reference orientation has axis along the +x axis.

- **segmentsNumber** (*int*) – number of edges on the helix surface (>=3)

- **angleRange** (*(ϑmin,Θmax)*) – range of angles; if None, (0,2*pi) is assumed.

- ****kw** – (unused keyword arguments) passed to utils.facet;

yade.geom.**facetParallelepiped**(*center, extents, height, orientation=Quaternion((1, 0, 0), 0), wallMask=63, **kw*)

Create arbitrarily-aligned Parallelepiped composed of facets, with given center, extents, height and orientation. If any of the parallelepiped dimensions is zero, corresponding facets will not be created. The facets are oriented outwards from the parallelepiped.

**Parameters**

- **center** (*Vector3*) – center of the parallelepiped

- **extents** (*Vector3*) – half lengths of the parallelepiped sides

- **height** (*Real*) – height of the parallelepiped (along axis z)

- **orientation** (*Quaternion*) – orientation of the parallelepiped

- **wallMask** (`bitmask`) – determines which walls will be created, in the order -x (1), +x (2), -y (4), +y (8), -z (16), +z (32). The numbers are ANDed; the default 63 means to create all walls

- **\*\*kw** – (unused keyword arguments) passed to *utils.facet*

  **Returns** list of facets forming the parallelepiped

yade.geom.**facetPolygon**(*center, radiusOuter, orientation=Quaternion((1, 0, 0), 0), seg­mentsNumber=10, angleRange=None, radiusInner=0, \*\*kw*)

  Create arbitrarily-aligned polygon composed of facets, with given center, radius (outer and inner) and orientation. Return List of facets forming the polygon;

  **Parameters**

- **center** (`Vector3`) – center of the created cylinder

- **radiusOuter** (`float`) – outer radius

- **radiusInner** (`float`) – inner height (can be 0)

- **orientation** (`Quaternion`) – orientation of the polygon; the reference orienta­tion has axis along the +x axis.

- **segmentsNumber** (`int`) – number of edges on the polygon surface (>=3)

- **angleRange** (`(ϑmin,Θmax)`) – allows one to create only part of polygon by specifying range of angles; if `None`, (0,2*pi) is assumed.

- **\*\*kw** – (unused keyword arguments) passed to utils.facet;

yade.geom.**facetPolygonHelixGenerator**(*center, radiusOuter, pitch=0, orienta­tion=Quaternion((1, 0, 0), 0), segmentsNumber=10, angleRange=None, radiusInner=0, \*\*kw*)

  Please, do not use this function directly! Use geom.facetPloygon and geom.facetHelix instead. This is the base function for generating polygons and helixes from facets.

yade.geom.**facetSphere**(*center, radius, thetaResolution=8, phiResolution=8, returnEle­mentMap=False, \*\*kw*)

  Create arbitrarily-aligned sphere composed of facets, with given center, radius and orientation. Return List of facets forming the sphere. Parameters inspired by ParaView sphere glyph

  **Parameters**

- **center** (`Vector3`) – center of the created sphere

- **radius** (`float`) – sphere radius

- **thetaResolution** (`int`) – number of facets around "equator"

- **phiResolution** (`int`) – number of facets between "poles" + 1

- **returnElementMap** (`bool`) – returns also tuple of nodes ((x1,y1,z1),(x2,y2,z2),...) and elements ((id01,id02,id03),(id11,id12,id13),...) if true, only facets otherwise

- **\*\*kw** – (unused keyword arguments) passed to utils.facet;

### 2.4.4  yade.gridpfacet module

Helper functions for creating cylinders, grids and membranes. For more details on this type of elements see [Effeindzourou2016], [Effeindzourou2015a], [Bourrier2013],.

For examples using *GridConnections*, see

- examples/grids/CohesiveGridConnectionSphere.py

- examples/grids/GridConnection__Spring.py

- examples/grids/Simple__Grid__Falling.py

- examples/grids/Simple__GridConnection__Falling.py

For examples using *PFacets*, see

- examples/pfacet/gts-pfacet.py

- examples/pfacet/mesh-pfacet.py

- examples/pfacet/pfacetcreators.py

yade.gridpfacet.**chainedCylinder**(*begin=Vector3(0, 0, 0), end=Vector3(1, 0, 0), radius=0.2, dynamic=None, fixed=False, wire=False, color=None, high-light=False, material=-1, mask=1*)

Create and connect a chainedCylinder with given parameters. The shape generated by repeted calls of this function is the Minkowski sum of polyline and sphere.

**Parameters**

- **radius** (`Real`) – radius of sphere in the Minkowski sum.

- **begin** (`Vector3`) – first point positioning the line in the Minkowski sum

- **last** (`Vector3`) – last point positioning the line in the Minkowski sum

In order to build a correct chain, last point of element of rank N must correspond to first point of element of rank N+1 in the same chain (with some tolerance, since bounding boxes will be used to create connections).

**Returns** Body object with the *ChainedCylinder shape*.

---

**Note:** *ChainedCylinder* is deprecated and will be removed in the future, use *GridConnection* instead. See *gridpfacet.cylinder* and *gridpfacet.cylinderConnection*.

---

yade.gridpfacet.**cylinder**(*begin=Vector3(0, 0, 0), end=Vector3(1, 0, 0), radius=0.2, nodesIds=[], cylIds=[], dynamic=None, fixed=False, wire=False, color=None, highlight=False, intMaterial=-1, extMaterial=-1, mask=1*)

Create a cylinder with given parameters. The shape corresponds to the Minkowski sum of line-segment and sphere, hence, the cylinder has rounded vertices. The cylinder (*GridConnection*) and its corresponding nodes (yref:*GridNodes<GridNode>*) are automatically added to the simulation. The lists with nodes and cylinder ids will be updated automatically.

**Parameters**

- **begin** (`Vector3`) – first point of the Minkowski sum in the global coordinate system.

- **end** (`Vector3`) – last point of the Minkowski sum in the global coordinate system.

- **radius** (`Real`) – radius of sphere in the Minkowski sum.

- **nodesIds** (`list`) – list with ids of already existing *GridNodes*. New ids will be added.

- **cylIds** (`list`) – list with ids of already existing *GridConnections*. New id will be added.

- **intMaterial** – *Body.material* used to create the interaction physics between the two GridNodes

- **extMaterial** – *Body.material* used to create the interaction physics between the Cylinder (GridConnection) and other bodies (e.g., spheres interaction with the cylinder)

See *utils.sphere*'s documentation for meaning of other parameters.

`yade.gridpfacet.`**`cylinderConnection`**(*vertices, radius=0.2, nodesIds=[], cylIds=[], dynamic=None, fixed=False, wire=False, color=None, highlight=False, intMaterial=-1, extMaterial=-1, mask=1*)

Create a chain of cylinders with given parameters. The cylinders (*GridConnection*) and its corresponding nodes (yref:*GridNodes<GridNode>*) are automatically added to the simulation. The lists with nodes and cylinder ids will be updated automatically.

> **Parameters vertices** (*[Vector3]*) – coordinates of vertices to connect in the global coordinate system.

See *gridpfacet.cylinder* documentation for meaning of other parameters.

`yade.gridpfacet.`**`gmshPFacet`**(*meshfile='file.mesh', shift=Vector3(0, 0, 0), scale=1.0, orientation=Quaternion((1, 0, 0), 0), radius=1.0, wire=True, fixed=True, materialNodes=-1, material=-1, color=None*)

Imports mesh geometry from .mesh file and automatically creates connected PFacet elements. For an example see examples/pfacet/mesh-pfacet.py.

> **Parameters**
>
> - **filename** (*string*) – .gts file to read.
>
> - **shift** (*[float,float,float]*) – [X,Y,Z] parameter shifts the mesh.
>
> - **scale** (*float*) – factor scales the mesh.
>
> - **orientation** (*quaternion*) – orientation of the imported geometry.
>
> - **radius** (*float*) – radius used to create the *PFacets*.
>
> - **materialNodes** – specify *Body.material* of *GridNodes*. This material is used to make the internal connections.
>
> - **material** – specify *Body.material* of *PFacets*. This material is used for interactions with external bodies.

See documentation of *utils.sphere* for meaning of other parameters.

> **Returns** lists of *GridNode* ids *nodesIds*, *GridConnection* ids *cylIds*, and *PFacet* ids *pfIds*

mesh files can easily be created with GMSH.

Additional examples of mesh-files can be downloaded from http://www-roc.inria.fr/gamma/download/download.php

`yade.gridpfacet.`**`gridConnection`**(*id1, id2, radius, wire=False, color=None, highlight=False, material=-1, mask=1, cellDist=None*)

Create a *GridConnection* by connecting two *GridNodes*.

> **Parameters**
>
> - **id1,id2** – the two *GridNodes* forming the cylinder.
>
> - **radius** (*float*) – radius of the cylinder. Note that the radius needs to be the same as the one for the *GridNodes*.
>
> - **cellDist** (*Vector3*) – for periodic boundary conditions, see *Interaction.cellDist*. Note: periodic boundary conditions for gridConnections are not yet implemented!

See documentation of *utils.sphere* for meaning of other parameters.

> **Returns** Body object with the *GridConnection shape*.

---

**Note:** The material of the *GridNodes* will be used to set the constitutive behaviour of the internal connection, i.e., the constitutive behaviour of the cylinder. The material of the *GridConnection* is used for interactions with other (external) bodies.

---

`yade.gridpfacet.``gridNode`(*center,    radius,    dynamic=None,    fixed=False,    wire=False,
color=None, highlight=False, material=-1*)

Create a *GridNode* which is needed to set up *GridConnections*.

See documentation of *utils.sphere* for meaning of parameters.

> **Returns** Body object with the *gridNode shape*.

`yade.gridpfacet.``gtsPFacet`(*meshfile, shift=Vector3(0, 0, 0), scale=1.0, radius=1, wire=True,
fixed=True, materialNodes=-1, material=-1, color=None*)

Imports mesh geometry from .gts file and automatically creates connected *PFacet3* elements. For
an example see examples/pfacet/gts-pfacet.py.

> **Parameters**
>
> - `filename` (*string*) – .gts file to read.
>
> - `shift` (*[float,float,float]*) – [X,Y,Z] parameter shifts the mesh.
>
> - `scale` (*float*) – factor scales the mesh.
>
> - `radius` (*float*) – radius used to create the *PFacets*.
>
> - `materialNodes` – specify *Body.material* of *GridNodes*. This material is used to
>   make the internal connections.
>
> - `material` – specify *Body.material* of *PFacets*. This material is used for interac-
>   tions with external bodies.

See documentation of *utils.sphere* for meaning of other parameters.

> **Returns** lists of *GridNode* ids *nodesIds*, *GridConnection* ids *cylIds*, and *PFacet* ids *pfIds*

`yade.gridpfacet.``pfacet`(*id1, id2, id3, wire=True, color=None, highlight=False, material=-1,
mask=1, cellDist=None*)

Create a *PFacet* element from 3 *GridNodes* which are already connected via 3 *GridConnections*:

> **Parameters**
>
> - `id1,id2,id3` – already with *GridConnections* connected *GridNodes*
>
> - `wire` (*bool*) – if `True`, top and bottom facet are shown as skeleton; otherwise
>   facets are filled.
>
> - `color` (*Vector3-or-None*) – color of the PFacet; random color will be assigned
>   if `None`.
>
> - `cellDist` (*Vector3*) – for periodic boundary conditions, see *Interaction.cellDist*.
>   Note: periodic boundary conditions are not yet implemented for PFacets!

See documentation of *utils.sphere* for meaning of other parameters.

> **Returns** Body object with the *PFacet shape*.

---

**Note:**  *GridNodes* and *GridConnections* need to have the same radius. This is also the radius
used to create the *PFacet*

---

`yade.gridpfacet.``pfacetCreator1`(*vertices, radius, nodesIds=[], cylIds=[], pfIds=[], wire=False,
fixed=True, materialNodes=-1, material=-1, color=None*)

Create a *PFacet* element from 3 vertices and automatically append to simulation. The function
uses the vertices to create *GridNodes* and automatically checks for existing nodes.

> **Parameters**
>
> - `vertices` (*[Vector3,Vector3,Vector3]*) – coordinates of vertices in the global
>   coordinate system.
>
> - `radius` (*float*) – radius used to create the *PFacets*.

- **nodesIds** (*list*) – list with ids of already existing *GridNodes*. New ids will be added.

- **cylIds** (*list*) – list with ids of already existing *GridConnections*. New ids will be added.

- **pfIds** (*list*) – list with ids of already existing *PFacets*. New ids will be added.

- **materialNodes** – specify *Body.material* of *GridNodes*. This material is used to make the internal connections.

- **material** – specify *Body.material* of *PFacets*. This material is used for interactions with external bodies.

See documentation of *utils.sphere* for meaning of other parameters.

yade.gridpfacet.**pfacetCreator2**(*id1,  id2,  vertex,  radius,  nodesIds=[],  wire=True, materialNodes=-1, material=-1, color=None, fixed=True*)

Create a *PFacet* element from 2 already existing and connected *GridNodes* and one vertex. The element is automatically appended to the simulation.

> **Parameters**
>
> - **id1,id2** (*int*) – ids of already with *GridConnection* connected *GridNodes*.
>
> - **vertex** (Vector3) – coordinates of the vertex in the global coordinate system.

See documentation of *gridpfacet.pfacetCreator1* for meaning of other parameters.

yade.gridpfacet.**pfacetCreator3**(*id1,  id2,  id3,  cylIds=[],  pfIds=[],  wire=True,  material=-1, color=None, fixed=True, mask=-1*)

Create a *PFacet* element from 3 already existing *GridNodes* which are not yet connected. The element is automatically appended to the simulation.

> **Parameters id1,id2,id3** (*int*) – id of the 3 *GridNodes* forming the *PFacet*.

See documentation of *gridpfacet.pfacetCreator1* for meaning of other parameters.

yade.gridpfacet.**pfacetCreator4**(*id1, id2, id3, pfIds=[], wire=True, material=-1, color=None, fixed=True, mask=-1*)

Create a *PFacet* element from 3 already existing *GridConnections*. The element is automatically appended to the simulation.

> **Parameters id1,id2,id3** (*int*) – id of the 3 *GridConnections* forming the *PFacet*.

See documentation of *gridpfacet.pfacetCreator1* for meaning of other parameters.

## 2.4.5 yade.libVersions module

The **yade.libVersions** module tracks versions of all libraries it was compiled with. Example usage is as follows:

```
from yade.libVersions import *
if(getVersion('cgal') > (4,9,0)):
        …
else:
        …
```

To obtain a list of all libraries use the function *libVersions.printAllVersions*.

All libraries listed in *prerequisites* are detected by this module.

---

**Note:** If we need a version of some library not listed in *prerequisites*, then it must also be added to that list.

---

When adding a new version please have a look at these three files:

1. py/_libVersions.cpp: detection of versions from `#include` files by C++.

2. py/libVersions.py.in: python module which is constructed by cmake during compilation. All `*.in` files are processed by cmake.

3. cMake/FindMissingVersions.cmake: forced detection of library with undetectable version.

---

**Hint:** The safest way to compare versions is to use builtin python tuple comparison e.g. `if(cgalVer > (4,9,0) and cgalVer < (5,1,1)):`.

---

yade.libVersions.**getAllVersions**(*rstFormat=False*)

> **Returns** `str` - this function returns the result of *printAllVersions(rstFormat)* call inside a string variable.

yade.libVersions.**getAllVersionsCmake**()

> This function returns library versions as provided by cmake during compilation.

> **Returns** dictionary in following format: `{ "libName" : [ (major, minor, patchlevel) , "versionString" ] }`

As an example the dict below reflects what libraries this documentation was compiled with (here are only those detected by CMAKE):

```
Yade [1]: from yade.libVersions import *

Yade [2]: getAllVersionsCmake()
Out[2]:
{'cmake': [(3, 16, 3), '3.16.3'],
 'compiler': [(9, 4, 0), '/usr/bin/c++ 9.4.0'],
 'boost': [(1, 71, 0), '107100'],
 'freeglut': [(2, 8, 1), '2.8.1'],
 'python': [(3, 8, 10), '3.8.10'],
 'eigen': [(3, 3, 7), '3.3.7'],
 'ipython': [(7, 13, 0), '7.13.0'],
 'sphinx': [(1, 8, 5), '1.8.5-final-0'],
 'mpi4py': [(3, 0, 3), '3.0.3'],
 'mpmath': [(1, 1, 0), '1.1.0']}
```

---

**Note:** Please add here detection of other libraries when yade starts using them or if you discover how to extract from cmake a version which I didn't add here.

---

yade.libVersions.**getArchitecture**()

> **Returns** string containing processor architecture name, as reported by `uname -m` call or from `CMAKE_HOST_SYSTEM_PROCESSOR` cmake variable.

yade.libVersions.**getLinuxVersion**()

> **Returns** string containing linux release and version, preferably the value of `PRETTY_-NAME` from file `/etc/os-release`.

yade.libVersions.**getVersion**(*libName*)

> This function returns the tuple (`major, minor, patchlevel`) with library version number. The `yade --test` in file py/tests/libVersions.py tests that this version is the same as detected by cmake and C++. If only one of those could detect the library version, then this number is used.

> **Parameters** `libName` (*string*) – the name of the library

> **Returns** tuple in format (`major, minor, patchlevel`) if `libName` exists. Otherwise it returns `None`.

---

**Note:** library openblas has no properly defined version in header files, this function will return (0,0,0) for openblas. Parsing the version string would be unreliable. The mpi version detected by cmake sometimes is different than version detected by C++, this needs further investigation.

---

`yade.libVersions.`**`printAllVersions`**`(`*rstFormat=False*`)`
    This function prints a nicely formatted table with library versions.

        **Parameters rstFormat** (`bool`) – whether to print table using the reStructuredText formatting. Defaults to `False` and prints using Gitlab markdown rules so that it is easy to paste into gitlab discussions.

    As an example the table below actually reflects with what libraries this documentation was compiled:

```
Yade [1]: printAllVersions()

```
Yade version   :  2021-11-19.git-639e121
Yade features  :  QT5
Yade config dir:  ~/.yadeflip
Yade precision :  53 bits, 15 decimal places, without mpmath, PrecisionDouble
```

Libraries used :

| library     | cmake               | C++        |
| ----------- | ------------------- | ---------- |
| boost       | 107100              | 1.71.0     |
| cmake       | 3.16.3              |            |
| compiler    | /usr/bin/c++ 9.4.0  | gcc 9.4.0  |
| eigen       | 3.3.7               | 3.3.7      |
| freeglut    | 2.8.1               |            |
| gl          |                     | 20190805   |
| ipython     | 7.13.0              |            |
| mpi4py      | 3.0.3               |            |
| mpmath      | 1.1.0               |            |
| python      | 3.8.10              | 3.8.10     |
| qglviewer   |                     | 2.6.3      |
| qt          |                     | 5.12.8     |
| sphinx      | 1.8.5-final-0       |            |
| sqlite      |                     | 3.31.1     |

```
Linux version  :  Ubuntu 20.04.4 LTS
Architecture   :  amd64
Little endian  :  True
```
```

---

**Note:** For convenience at startup `from yade.libVersions import printAllVersions` is executed, so that this function is readily accessible.

---

`yade._libVersions.`**`getAllVersionsCpp`**`()` → dict
    This function returns library versions as discovered by C++ during compilation from all the `#include` headers. This can be useful in debugging to detect some library `.so` conflicts.

        **Returns** dictionary in folowing format: `{ "libName" : [ (major, minor, patch) , "versionString" ] }`

    As an example the dict below reflects what libraries this documentation was compiled with (here are only those detected by C++):

```
Yade [1]: from yade.libVersions import *

Yade [2]: getAllVersionsCpp()
Out[2]:
{'compiler': [(9, 4, 0), 'gcc 9.4.0'],
 'boost': [(1, 71, 0), '1.71.0'],
 'qt': [(5, 12, 8), '5.12.8'],
 'gl': [(2019, 8, 5), '20190805'],
 'qglviewer': [(2, 6, 3), '2.6.3'],
 'python': [(3, 8, 10), '3.8.10'],
 'eigen': [(3, 3, 7), '3.3.7'],
 'sqlite': [(3, 31, 1), '3.31.1'],
 'vtk': [],
 'cgal': [],
 'suitesparse': [],
 'openblas': [],
 'metis': [],
 'mpi': [],
 'clp': [],
 'coinutils': [],
 'mpfr': [],
 'mpc': []}
```

**Note:** Please add here C++ detection of other libraries when yade starts using them.

### 2.4.6 yade.linterpolation module

Module for rudimentary support of manipulation with piecewise-linear functions (which are usually interpolations of higher-order functions, whence the module name). Interpolation is always given as two lists of the same length, where the x-list must be increasing.

Periodicity is supported by supposing that the interpolation can wrap from the last x-value to the first x-value (which should be 0 for meaningful results).

Non-periodic interpolation can be converted to periodic one by padding the interpolation with constant head and tail using the sanitizeInterpolation function.

There is a c++ template function for interpolating on such sequences in pkg/common/Engine/PartialEngine/LinearInterpolate.hpp (stateful, therefore fast for sequential reads).

TODO: Interpolating from within python is not (yet) supported.

yade.linterpolation.**integral**(*x, y*)
    Return integral of piecewise-linear function given by points x0,x1,… and y0,y1,…

yade.linterpolation.**revIntegrateLinear**(*I, x0, y0, x1, y1*)
    Helper function, returns value of integral variable x for linear function f passing through (x0,y0),(x1,y1) such that 1. x [x0,x1] 2. __x0^x f dx=I and raise exception if such number doesn't exist or the solution is not unique (possible?)

yade.linterpolation.**sanitizeInterpolation**(*x, y, x0, x1*)
    Extends piecewise-linear function in such way that it spans at least the x0…x1 interval, by adding constant padding at the beginning (using y0) and/or at the end (using y1) or not at all.

yade.linterpolation.**xFractionalFromIntegral**(*integral, x, y*)
    Return x within range x0…xn such that __x0^x f dx==integral. Raises error if the integral value is not reached within the x-range.

`yade.linterpolation.xFromIntegral`(*integralValue, x, y*)

> Return x such that __x0^x f dx==integral. x wraps around at xn. For meaningful results, therefore, x0 should == 0

### 2.4.7 yade.log module

The `yade.log` module serves as an interface to yade logging framework implemented on top of boost::log. For full documentation see *debugging section*. Example usage in python is as follows:

```
import yade.log
yade.log.setLevel('PeriTriaxController',yade.log.TRACE)
```

Example usage in C++ is as follows:

```
LOG_WARN("Something: "<<something)
```

`yade._log.defaultConfigFileName`() → str

> **Returns** the default log config file, which is loaded at startup, if it exists.

`yade._log.getAllLevels`() → dict

> **Returns** A python dictionary with all known loggers in yade. Those without a debug level set will have value -1 to indicate that `Default` filter log level is to be used for them.

`yade._log.getDefaultLogLevel`() → int

> **Returns** The current `Default` filter log level.

`yade._log.getMaxLevel`() → int

> **Returns** the MAX_LOG_LEVEL of the current yade build.

`yade._log.getUsedLevels`() → dict

> **Returns** A python dictionary with all used log levels in yade. Those without a debug level (value -1) are omitted.

`yade._log.readConfigFile`(*(str)arg1*) → None

> Loads the given configuration file.
>
> > **Parameters fname** (*str*) – the config file to be loaded.

`yade._log.resetOutputStream`() → None

> Resets log output stream to default state: all logs are printed on `std::clog` channel, which usually redirects to `std::cerr`.

`yade._log.saveConfigFile`(*(str)arg1*) → None

> Saves log config to specified file.
>
> > **Parameters fname** (*str*) – the config file to be saved.

`yade._log.setDefaultLogLevel`(*(int)arg1*) → None

> > **Parameters level** (*int*) – Sets the `Default` filter log level, same as calling `log.setLevel("Default",level)`.

`yade._log.setLevel`(*(str)arg1, (int)arg2*) → None

> Set filter level (constants `TRACE` (6), `DEBUG` (5), `INFO` (4), `WARN` (3), `ERROR` (2), `FATAL` (1), `NOFILTER` (0)) for given logger.
>
> > **Parameters**
> >
> > - **className** (*str*) – The logger name for which the filter level is to be set. Use name `Default` to change the default filter level.
> >
> > - **level** (*int*) – The filter level to be set.

---

> **Warning:** setting `Default` log level higher than `MAX_LOG_LEVEL` provided during compilation
> will have no effect. Logs will not be printed because they are removed during compilation.

---

yade._log.**setOutputStream**(*(str)arg1, (bool)arg2*) → None

**Parameters**

- **streamName** (*str*) – sets the output stream, special names `cout`, `cerr`, `clog`
  use the `std::cout`, `std::cerr`, `std::clog` counterpart (`std::clog` the is the
  default output stream). Every other name means that log will be written to a
  file with name provided in the argument.

- **reset** (*bool*) – dictates whether all previously set output streams are to be
  removed. When set to false: the new output stream is set additionally to the
  current one.

yade._log.**setUseColors**(*(bool)arg1*) → None
   Turn on/off colors in log messages. By default is on. If logging to a file then it is better to be
   turned off.

yade._log.**testAllLevels**() → None
   This function prints test messages on all log levels. Can be used to see how filtering works and to
   what streams the logs are written.

yade._log.**testOnceLevels**() → None
   This function prints test messages on all log levels using LOG__ONCE__* macro family.

yade._log.**testTimedLevels**() → None
   This function prints timed test messages on all log levels. In this test the log levels   [0...2] are
   timed to print every 2 seconds, levels   [3,4] every 1 second and levels   [5,6] every 0.5 seconds. The
   loop lasts for 2.1 seconds. Can be used to see how timed filtering works and to what streams the
   logs are written.

yade._log.**unsetLevel**(*(str)arg1*) → None

**Parameters className** (*str*) – The logger name for which the filter level is to be unset,
   so that a `Default` will be used instead. Unsetting the `Default` level will change it
   to max level and print everything.

## 2.4.8 yade.math module

This python module exposes all C++ math functions for Real and Complex types to python. In fact it
sort of duplicates `import math`, `import cmath` or `import mpmath`. Also it facilitates migration of old
python scripts to high precision calculations.

This module has following purposes:

1. To reliably *test* all C++ math functions of arbitrary precision Real and Complex types against
   mpmath.

2. To act as a "migration helper" for python scripts which call python mathematical functions that
   do not work well with mpmath. As an example see *math.linspace* below and this merge request

3. To allow writing python math code in a way that mirrors C++ math code in Yade. As a bonus
   it will be faster than mpmath because mpmath is a purely python library (which was one of the
   main difficulties when writing lib/high-precision/ToFromPythonConverter.hpp)

4. To test Eigen NumTraits

5. To test CGAL NumTraits

If another `C++` *math function* is needed it should be added to following files:

1. lib/high-precision/MathFunctions.hpp

---

2. py/high-precision/_math.cpp

3. py/tests/testMath.py

4. py/tests/testMathHelper.py

If another **python** math function does not work well with **mpmath** it should be added below, and original calls to this function should call this function instead, e.g. **numpy.linspace(…)** is replaced with **yade.math.linspace(…)**.

The **RealHP<n>** *higher precision* math functions can be accessed in python by using the **.HPn** module scope. For example:

```python
import yade.math as mth
mth.HP2.sqrt(2)    # produces square root of 2 using RealHP<2> precision
mth.sqrt(2)        # without using HPn module scope it defaults to RealHP<1>
```

**yade.math.Real**(*arg*)

> This function is for compatibility of calls like: **g = yade.math.toHP1("-9.81")**. If yade is compiled with default **Real** precision set as **double**, then python won't accept string arguments as numbers. However when using higher precisions only calls **yade.math.toHP1("1.23456789012345678901234567901234567890")** do not cut to the first 15 decimal places. The calls such as **yade.math.toHP1(1.23456789012345678901234567901234567890)** will use default **python double** conversion and will cut the number to its first 15 digits.
>
> If you are debugging a high precision python script, and have difficulty finding places where such cuts have happened you should use **yade.math.toHP1(string)** for declaring all python floating point numbers which are physically important in the simulation. This function will throw exception if bad conversion is about to take place.
>
> Also see example high precision check checkGravityRungeKuttaCashKarp54.py.

**yade.math.Real1**(*arg*)

> This function is for compatibility of calls like: **g = yade.math.toHP1("-9.81")**. If yade is compiled with default **Real** precision set as **double**, then python won't accept string arguments as numbers. However when using higher precisions only calls **yade.math.toHP1("1.23456789012345678901234567901234567890")** do not cut to the first 15 decimal places. The calls such as **yade.math.toHP1(1.23456789012345678901234567901234567890)** will use default **python double** conversion and will cut the number to its first 15 digits.
>
> If you are debugging a high precision python script, and have difficulty finding places where such cuts have happened you should use **yade.math.toHP1(string)** for declaring all python floating point numbers which are physically important in the simulation. This function will throw exception if bad conversion is about to take place.
>
> Also see example high precision check checkGravityRungeKuttaCashKarp54.py.

**yade.math.degrees**(*arg*)

> **Returns** arg in radians converted to degrees, using yade.math.Real precision.

**yade.math.degreesHP1**(*arg*)

> **Returns** arg in radians converted to degrees, using yade.math.Real precision.

**yade.math.getRealHPCppDigits10**()

> **Returns** tuple containing amount of decimal digits supported on C++ side by Eigen and CGAL.

**yade.math.getRealHPPythonDigits10**()

> **Returns** tuple containing amount of decimal digits supported on python side by yade.minieigenHP.

**yade.math.linspace**(*a*, *b*, *num*)

> This function calls **numpy.linspace(…)** or **mpmath.linspace(…)**, because **numpy.linspace** function does not work with mpmath.

`yade.math.`**`needsMpmathAtN`**`(N)`

>    **Parameters** **`N`** – The **`int`** N value of `RealHP<N>` in question. Must be **`N >= 1`**.

>    **Returns** **`True`** or **`False`** with information if using **`mpmath`** is necessary to avoid losing precision when working with `RealHP<N>`.

`yade.math.`**`radians`**`(arg)`

>   The default python function **`import math ; math.radians(arg)`** only works on 15 digit **`double`** precision. If you want to carry on calculations in higher precision it is advisable to use this function **`yade.math.radiansHP1(arg)`** instead. It uses full yade **`Real`** precision numbers.

>   NOTE: in the future this function may replace **`radians(…)`** function which is called in yade in many scripts, and which in fact is a call to native python **`math.radians`**. We only need to find the best backward compatible approach for this. The function **`yade.math.radiansHP1(arg)`** will remain as the function which uses native yade **`Real`** precision.

`yade.math.`**`radiansHP1`**`(arg)`

>   The default python function **`import math ; math.radians(arg)`** only works on 15 digit **`double`** precision. If you want to carry on calculations in higher precision it is advisable to use this function **`yade.math.radiansHP1(arg)`** instead. It uses full yade **`Real`** precision numbers.

>   NOTE: in the future this function may replace **`radians(…)`** function which is called in yade in many scripts, and which in fact is a call to native python **`math.radians`**. We only need to find the best backward compatible approach for this. The function **`yade.math.radiansHP1(arg)`** will remain as the function which uses native yade **`Real`** precision.

`yade.math.`**`toHP1`**`(arg)`

>   This function is for compatibility of calls like: **`g = yade.math.toHP1("-9.81")`**. If yade is compiled with default **`Real`** precision set as **`double`**, then python won't accept string arguments as numbers. However when using higher precisions only calls **`yade.math.toHP1("1.23456789012345678901234567890123456789")`** do not cut to the first 15 decimal places. The calls such as **`yade.math.toHP1(1.23456789012345678901234567890123456789)`** will use default **`python double`** conversion and will cut the number to its first 15 digits.

>   If you are debugging a high precision python script, and have difficulty finding places where such cuts have happened you should use **`yade.math.toHP1(string)`** for declaring all python floating point numbers which are physically important in the simulation. This function will throw exception if bad conversion is about to take place.

>   Also see example high precision check checkGravityRungeKuttaCashKarp54.py.

`yade.math.`**`usesHP`**`()`

>    **Returns** **`True`** if yade is using default **`Real`** precision higher than 15 digit (53 bits) **`double`** type.

`yade._math.`**`Catalan`**`([(int)Precision=53]) → float`

>    **Returns** **`Real`** the catalan constant, exposed to python for testing of eigen numerical traits.

`yade._math.`**`Euler`**`([(int)Precision=53]) → float`

>    **Returns** **`Real`** The Euler–Mascheroni constant, exposed to python for testing of eigen numerical traits.

**class** `yade._math.`**`HP1`**

>   `AddCost = 1`

>   **`Catalan`**`([(int)Precision=53]) → float` :

>>    **Returns** **`Real`** The catalan constant, exposed to python for testing of eigen numerical traits.

>   `ComplexAddCost = 2`

`ComplexMulCost = 6`

`ComplexReadCost = 2`

`Euler(`$\big[(int)Precision=53\big]$`)` → float :

> **Returns** `Real` The Euler–Mascheroni constant, exposed to python for testing of eigen numerical traits.

`IsComplex = 0`

`IsInteger = 0`

`IsSigned = 1`

`Log2(`$\big[(int)Precision=53\big]$`)` → float :

> **Returns** `Real` natural logarithm of 2, exposed to python for testing of eigen numerical traits.

`MulCost = 1`

`Pi(`$\big[(int)Precision=53\big]$`)` → float :

> **Returns** `Real` The π constant, exposed to python for testing of eigen numerical traits.

`ReadCost = 1`

`RequireInitialization = 0`

`class Var`
> The `Var` class is used to test to/from python converters for arbitrary precision `Real`
>
> **cpl**
> > one `Complex` variable to test reading from and writing to it.
>
> **val**
> > one `Real` variable for testing.

`abs(`*(complex)x*`)` → float :

> **Returns** the `Real` absolute value of the `Complex` argument. Depending on compilation options wraps `::boost::multiprecision::abs(…)` or std::abs(…) function.
>
> > **abs( (float)x) -> float :**
> >
> > > **return** the `Real` absolute value of the `Real` argument. Depending on compilation options wraps `::boost::multiprecision::abs(…)` or std::abs(…) function.

`acos(`*(complex)x*`)` → complex :

> **Returns** `Complex` the arc-cosine of the `Complex` argument in radians. Depending on compilation options wraps `::boost::multiprecision::acos(…)` or std::acos(…) function.
>
> > **acos( (float)x) -> float :**
> >
> > > **return** `Real` the arcus cosine of the argument. Depending on compilation options wraps `::boost::multiprecision::acos(…)` or std::acos(…) function.

`acosh(`*(complex)x*`)` → complex :

> **Returns** `Complex` the arc-hyperbolic cosine of the `Complex` argument in radians. Depending on compilation options wraps `::boost::multiprecision::acosh(…)` or std::acosh(…) function.
>
> > **acosh( (float)x) -> float :**

> > **return** `Real` the hyperbolic arcus cosine of the argument. Depending on compilation options wraps `::boost::multiprecision::acosh(…)` or std::acosh(…) function.

**arg**(*(complex)x*) → float :

> > **Returns** `Real` the arg (Phase angle of complex in radians) of the `Complex` argument in radians. Depending on compilation options wraps `::boost::multiprecision::arg(…)` or std::arg(…) function.

**asin**(*(complex)x*) → complex :

> > **Returns** `Complex` the arc-sine of the `Complex` argument in radians. Depending on compilation options wraps `::boost::multiprecision::asin(…)` or std::asin(…) function.

> **asin( (float)x) -> float :**

> > **return** `Real` the arcus sine of the argument. Depending on compilation options wraps `::boost::multiprecision::asin(…)` or std::asin(…) function.

**asinh**(*(complex)x*) → complex :

> > **Returns** `Complex` the arc-hyperbolic sine of the `Complex` argument in radians. Depending on compilation options wraps `::boost::multiprecision::asinh(…)` or std::asinh(…) function.

> **asinh( (float)x) -> float :**

> > **return** `Real` the hyperbolic arcus sine of the argument. Depending on compilation options wraps `::boost::multiprecision::asinh(…)` or std::asinh(…) function.

**atan**(*(complex)x*) → complex :

> > **Returns** `Complex` the arc-tangent of the `Complex` argument in radians. Depending on compilation options wraps `::boost::multiprecision::atan(…)` or std::atan(…) function.

> **atan( (float)x) -> float :**

> > **return** `Real` the arcus tangent of the argument. Depending on compilation options wraps `::boost::multiprecision::atan(…)` or std::atan(…) function.

**atan2**(*(float)x, (float)y*) → float :

> > **Returns** `Real` the arc tangent of y/x using the signs of the arguments x and y to determine the correct quadrant. Depending on compilation options wraps `::boost::multiprecision::atan2(…)` or std::atan2(…) function.

**atanh**(*(complex)x*) → complex :

> > **Returns** `Complex` the arc-hyperbolic tangent of the `Complex` argument in radians. Depending on compilation options wraps `::boost::multiprecision::atanh(…)` or std::atanh(…) function.

> **atanh( (float)x) -> float :**

> > **return** `Real` the hyperbolic arcus tangent of the argument. Depending on compilation options wraps `::boost::multiprecision::atanh(…)` or std::atanh(…) function.

**cbrt**(*(float)x*) → float :

> **Returns** `Real` cubic root of the argument. Depending on compilation options wraps
> `::boost::multiprecision::cbrt(…)` or std::cbrt(…) function.

**ceil**(*(float)x*) → float :

> **Returns** `Real` Computes the smallest integer value not less than arg. Depending on
> compilation options wraps `::boost::multiprecision::ceil(…)` or std::ceil(…)
> function.

**conj**(*(complex)x*) → complex :

> **Returns** the complex conjugation a `Complex` argument. Depending on compilation
> options wraps `::boost::multiprecision::conj(…)` or std::conj(…) function.

**cos**(*(complex)x*) → complex :

> **Returns** `Complex` the cosine of the `Complex` argument in radians. Depending on
> compilation options wraps `::boost::multiprecision::cos(…)` or std::cos(…)
> function.

> **cos( (float)x) -> float :**
>
> > **return** `Real` the cosine of the `Real` argument in radians. Depending on compil-
> > ation options wraps `::boost::multiprecision::cos(…)` or std::cos(…) func-
> > tion.

**cosh**(*(complex)x*) → complex :

> **Returns** `Complex` the hyperbolic cosine of the `Complex` argument in radians. De-
> pending on compilation options wraps `::boost::multiprecision::cosh(…)` or
> std::cosh(…) function.

> **cosh( (float)x) -> float :**
>
> > **return** `Real` the hyperbolic cosine of the `Real` argument in radians. Depend-
> > ing on compilation options wraps `::boost::multiprecision::cosh(…)` or
> > std::cosh(…) function.

**cylBesselJ**(*(int)k, (float)x*) → float :

> **Returns** `Real` the Bessel Functions of the First Kind of the order k and the `Real` ar-
> gument. See: <https://www.boost.org/doc/libs/1_77_0/libs/math/doc/html/
> math_toolkit/bessel/bessel_first.html>'___

**defprec = 53**

**dummy_precision**() → float :

> **Returns** similar to the function `epsilon`, but assumes that last 10% of bits con-
> tain the numerical error only. This is sometimes used by Eigen when calling
> `isEqualFuzzy` to determine if values differ a lot or if they are vaguely close to
> each other.

**epsilon**($\big[$*(int)Precision=53*$\big]$) → float :

> **Returns** `Real` returns the difference between `1.0` and the next representable value
> of the `Real` type. Wraps std::numeric_limits<Real>::epsilon() function.

> **epsilon( (float)x) -> float :**
>
> > **return** `Real` returns the difference between `1.0` and the next representable value
> > of the `Real` type. Wraps std::numeric_limits<Real>::epsilon() function.

**erf**(*(float)x*) → float :

> **Returns** `Real` Computes the error function of argument. Depending on compilation
> options wraps `::boost::multiprecision::erf(…)` or std::erf(…) function.

**erfc(**(float)x**)** → float :

> **Returns** Real Computes the complementary error function of argument, that is 1.0-erf(arg). Depending on compilation options wraps ::boost::multiprecision::erfc(…) or std::erfc(…) function.

**exp(**(complex)x**)** → complex :

> **Returns** the base $e$ exponential of a Complex argument. Depending on compilation options wraps ::boost::multiprecision::exp(…) or std::exp(…) function.

> **exp( (float)x) -> float :**

>> **return** the base $e$ exponential of a Real argument. Depending on compilation options wraps ::boost::multiprecision::exp(…) or std::exp(…) function.

**exp2(**(float)x**)** → float :

> **Returns** the base $2$ exponential of a Real argument. Depending on compilation options wraps ::boost::multiprecision::exp2(…) or std::exp2(…) function.

**expm1(**(float)x**)** → float :

> **Returns** the base $e$ exponential of a Real argument minus 1.0. Depending on compilation options wraps ::boost::multiprecision::expm1(…) or std::expm1(…) function.

**fabs(**(float)x**)** → float :

> **Returns** the Real absolute value of the argument. Depending on compilation options wraps ::boost::multiprecision::abs(…) or std::abs(…) function.

**factorial(**(int)x**)** → float :

> **Returns** Real the factorial of the Real argument. See: <https://www.boost.org/doc/libs/1_77_0/libs/math/doc/html/math_toolkit/factorials/sf_factorial.html>‘___

**floor(**(float)x**)** → float :

> **Returns** Real Computes the largest integer value not greater than arg. Depending on compilation options wraps ::boost::multiprecision::floor(…) or std::floor(…) function.

**fma(**(float)x, (float)y, (float)z**)** → float :

> **Returns** Real - computes (x*y) + z as if to infinite precision and rounded only once to fit the result type. Depending on compilation options wraps ::boost::multiprecision::fma(…) or std::fma(…) function.

**fmod(**(float)x, (float)y**)** → float :

> **Returns** Real the floating-point remainder of the division operation x/y of the arguments x and y. Depending on compilation options wraps ::boost::multiprecision::fmod(…) or std::fmod(…) function.

**frexp(**(float)x**)** → tuple :

> **Returns** tuple of (Real,int), decomposes given floating point Real argument into a normalized fraction and an integral power of two. Depending on compilation options wraps ::boost::multiprecision::frexp(…) or std::frexp(…) function.

**fromBits(**(str)bits[, (int)exp=0[, (int)sign=1]]**)** → float :

> **Parameters**

> - **bits** – str - a string containing '0', '1' characters.
> - **exp** – int - the binary exponent which shifts the bits.

- **sign** – **int** - the sign, should be -1 or +1, but it is not checked. It multiplies the result when construction from bits is finished.

  **Returns** **RealHP<N>** constructed from string containing '0', '1' bits. This is for debugging purposes, rather slow.

**getDecomposedReal**(*(float)x*) → dict :

  **Returns** **dict** - the dictionary with the debug information how the DecomposedReal class sees this type. This is for debugging purposes, rather slow. Includes result from *fpclassify* function call, a binary representation and other useful info. See also *fromBits*.

**getDemangledName**() → str :

  **Returns** **string** - the demangled C++ typename of **RealHP<N>**.

**getDemangledNameComplex**() → str :

  **Returns** **string** - the demangled C++ typename of **ComplexHP<N>**.

**getFloatDistanceULP**(*(float)arg1, (float)arg2*) → float :

  **Returns** an integer value stored in **RealHP<N>**, the ULP distance calculated by boost::math::float_distance, also see Floating-point Comparison and Prof. Kahan paper about this topic.

> **Warning:** The returned value is the **directed distance** between two arguments, this means that it can be negative.

**getRawBits**(*(float)x*) → str :

  **Returns** **string** - the raw bits in memory representing this type. Be careful: it only checks the system endianness and either prints bytes in reverse order or not. Does not make any attempts to further interpret the bits of: sign, exponent or significand (on a typical x86 processor they are printed in that order), and different processors might store them differently. It is not useful for types which internally use a pointer because for them this function prints not the floating point number but a pointer. This is for debugging purposes.

**hasInfinityNan = True**

**highest**([*(int)Precision=53*]) → float :

  **Returns** **Real** returns the largest finite value of the **Real** type. Wraps std::numeric_limits<Real>::max() function.

**hypot**(*(float)x, (float)y*) → float :

  **Returns** **Real** the square root of the sum of the squares of **x** and **y**, without undue overflow or underflow at intermediate stages of the computation. Depending on compilation options wraps ::boost::multiprecision::hypot(…) or std::hypot(…) function.

**ilogb**(*(float)x*) → float :

  **Returns** **Real** extracts the value of the unbiased exponent from the floating-point argument arg, and returns it as a signed integer value. Depending on compilation options wraps ::boost::multiprecision::ilogb(…) or std::ilogb(…) function.

**imag**(*(complex)x*) → float :

  **Returns** the imag part of a **Complex** argument. Depending on compilation options wraps ::boost::multiprecision::imag(…) or std::imag(…) function.

**isApprox**(*(float)a, (float)b, (float)eps*) → bool :

>   **Returns** `bool`, True if `a` is approximately equal `b` with provided `eps`, see also here

**isApproxOrLessThan**(*(float)a, (float)b, (float)eps*) → bool :

>   **Returns** `bool`, True if `a` is approximately less than or equal `b` with provided `eps`, see also here

**isEqualFuzzy**(*(float)arg1, (float)arg2, (float)arg3*) → bool :

>   **Returns** `bool`, True if the absolute difference between two numbers is smaller than std::numeric_limits<Real>::epsilon()

**isMuchSmallerThan**(*(float)a, (float)b, (float)eps*) → bool :

>   **Returns** `bool`, True if `a` is less than `b` with provided `eps`, see also here

**isfinite**(*(float)x*) → bool :

>   **Returns** `bool` indicating if the `Real` argument is Inf. Depending on compilation options wraps `::boost::multiprecision::isfinite(…)` or std::isfinite(…) function.

**isinf**(*(float)x*) → bool :

>   **Returns** `bool` indicating if the `Real` argument is Inf. Depending on compilation options wraps `::boost::multiprecision::isinf(…)` or std::isinf(…) function.

**isnan**(*(float)x*) → bool :

>   **Returns** `bool` indicating if the `Real` argument is NaN. Depending on compilation options wraps `::boost::multiprecision::isnan(…)` or std::isnan(…) function.

**laguerre**(*(int)n, (int)m, (float)x*) → float :

>   **Returns** `Real` the Laguerre polynomial of the orders `n`, `m` and the `Real` argument. See: <https://www.boost.org/doc/libs/1_77_0/libs/math/doc/html/math_toolkit/sf_poly/laguerre.html>'___

**ldexp**(*(float)x, (int)y*) → float :

>   **Returns** Multiplies a floating point value `x` by the number 2 raised to the `exp` power. Depending on compilation options wraps `::boost::multiprecision::ldexp(…)` or std::ldexp(…) function.

**lgamma**(*(float)x*) → float :

>   **Returns** `Real` Computes the natural logarithm of the absolute value of the gamma function of arg. Depending on compilation options wraps `::boost::multiprecision::lgamma(…)` or std::lgamma(…) function.

**log**(*(complex)x*) → complex :

>   **Returns** the `Complex` natural (base *e*) logarithm of a complex value z with a branch cut along the negative real axis. Depending on compilation options wraps `::boost::multiprecision::log(…)` or std::log(…) function.

>   **log( (float)x) -> float :**
>
>>   **return** the `Real` natural (base *e*) logarithm of a real value. Depending on compilation options wraps `::boost::multiprecision::log(…)` or std::log(…) function.

**log10**(*(complex)x*) → complex :

>   **Returns** the `Complex` (base *10*) logarithm of a complex value z with a branch cut along the negative real axis. Depending on compilation options wraps `::boost::multiprecision::log10(…)` or std::log10(…) function.

>   **log10( (float)x) -> float :**

**return** the `Real` decimal (base 10) logarithm of a real value. Depending on compilation options wraps `::boost::multiprecision::log10(…)` or std::log10(…) function.

`log1p(`*(float)x*`)` → float :

**Returns** the `Real` natural (base $e$) logarithm of `1+argument`. Depending on compilation options wraps `::boost::multiprecision::log1p(…)` or std::log1p(…) function.

`log2(`*(float)x*`)` → float :

**Returns** the `Real` binary (base 2) logarithm of a real value. Depending on compilation options wraps `::boost::multiprecision::log2(…)` or std::log2(…) function.

`logb(`*(float)x*`)` → float :

**Returns** Extracts the value of the unbiased radix-independent exponent from the floating-point argument arg, and returns it as a floating-point value. Depending on compilation options wraps `::boost::multiprecision::logb(…)` or std::logb(…) function.

`lowest(`[*(int)Precision=53*]`)` → float :

**Returns** `Real` returns the lowest (negative) finite value of the `Real` type. Wraps std::numeric_limits<Real>::lowest() function.

`max(`*(float)x*`, `*(float)y*`)` → float :

**Returns** `Real` larger of the two arguments. Depending on compilation options wraps `::boost::multiprecision::max(…)` or std::max(…) function.

**max_exp2 = 1024**

`min(`*(float)x*`, `*(float)y*`)` → float :

**Returns** `Real` smaller of the two arguments. Depending on compilation options wraps `::boost::multiprecision::min(…)` or std::min(…) function.

`modf(`*(float)x*`)` → tuple :

**Returns** tuple of (`Real,Real`), decomposes given floating point `Real` into integral and fractional parts, each having the same type and sign as x. Depending on compilation options wraps `::boost::multiprecision::modf(…)` or std::modf(…) function.

`polar(`*(float)x*`, `*(float)y*`)` → complex :

**Returns** `Complex` the polar (Complex from polar components) of the `Real` rho (length), `Real` theta (angle) arguments in radians. Depending on compilation options wraps `::boost::multiprecision::polar(…)` or std::polar(…) function.

`pow(`*(complex)x*`, `*(complex)pow*`)` → complex :

**Returns** the `Complex` complex arg1 raised to the `Complex` power arg2. Depending on compilation options wraps `::boost::multiprecision::pow(…)` or std::pow(…) function.

**pow( (float)x, (float)y) -> float :**

**return** `Real` the value of `base` raised to the power exp. Depending on compilation options wraps `::boost::multiprecision::pow(…)` or std::pow(…) function.

`proj(`*(complex)x*`)` → complex :

**Returns** `Complex` the proj (projection of the complex number onto the Riemann sphere) of the `Complex` argument in radians. Depending on compilation options wraps `::boost::multiprecision::proj(…)` or std::proj(…) function.

**random()** → float :

**Returns** `Real` a symmetric random number in interval `(-1,1)`. Used by Eigen.

**random( (float)a, (float)b) -> float :**

**return** `Real` a random number in interval `(a,b)`. Used by Eigen.

**real(**_(complex)x_**)** → float :

**Returns** the real part of a `Complex` argument. Depending on compilation options wraps `::boost::multiprecision::real(…)` or std::real(…) function.

**remainder(**_(float)x, (float)y_**)** → float :

**Returns** `Real` the IEEE remainder of the floating point division operation `x/y`. Depending on compilation options wraps `::boost::multiprecision::remainder(…)` or std::remainder(…) function.

**remquo(**_(float)x, (float)y_**)** → tuple :

**Returns** tuple of `(Real,long)`, the floating-point remainder of the division operation `x/y` as the std::remainder() function does. Additionally, the sign and at least the three of the last bits of `x/y` are returned, sufficient to determine the octant of the result within a period. Depending on compilation options wraps `::boost::multiprecision::remquo(…)` or std::remquo(…) function.

**rint(**_(float)x_**)** → float :

**Returns** Rounds the floating-point argument arg to an integer value (in floating-point format), using the current rounding mode. Depending on compilation options wraps `::boost::multiprecision::rint(…)` or std::rint(…) function.

**round(**_(float)x_**)** → float :

**Returns** `Real` the nearest integer value to arg (in floating-point format), rounding halfway cases away from zero, regardless of the current rounding mode.. Depending on compilation options wraps `::boost::multiprecision::round(…)` or std::round(…) function.

**roundTrip(**_(float)x_**)** → float :

**Returns** `Real` returns the argument x. Can be used to convert type to native RealHP<N> accuracy.

**sgn(**_(float)x_**)** → int :

**Returns** `int` the sign of the argument: `-1`, `0` or `1`.

**sign(**_(float)x_**)** → int :

**Returns** `int` the sign of the argument: `-1`, `0` or `1`.

**sin(**_(complex)x_**)** → complex :

**Returns** `Complex` the sine of the `Complex` argument in radians. Depending on compilation options wraps `::boost::multiprecision::sin(…)` or std::sin(…) function.

**sin( (float)x -> float :**

**return** `Real` the sine of the `Real` argument in radians. Depending on compilation options wraps `::boost::multiprecision::sin(…)` or std::sin(…) function.

**sinh(**_(complex)x_**)** → complex :

> > **Returns** `Complex` the hyperbolic sine of the `Complex` argument in radians. Depending on compilation options wraps `::boost::multiprecision::sinh(…)` or std::sinh(…) function.

> **sinh( (float)x) -> float :**

> > **return** `Real` the hyperbolic sine of the `Real` argument in radians. Depending on compilation options wraps `::boost::multiprecision::sinh(…)` or std::sinh(…) function.

**smallest_positive()** → float :

> **Returns** `Real` the smallest number greater than zero. Wraps std::numeric_limits<Real>::min()

**sphericalHarmonic(**(int)l, (int)m, (float)theta, (float)phi**)** → complex :

> **Returns** `Real` the spherical harmonic polynomial of the orders `l` (unsigned int), `m` (signed int) and the `Real` arguments `theta` and `phi`. See: <https://www.boost.org/doc/libs/1_77_0/libs/math/doc/html/math_toolkit/sf_poly/sph_harm.html>'___

**sqrt(**(complex)x**)** → complex :

> **Returns** the `Complex` square root of `Complex` argument. Depending on compilation options wraps `::boost::multiprecision::sqrt(…)` or std::sqrt(…) function.

> **sqrt( (float)x) -> float :**

> > **return** `Real` square root of the argument. Depending on compilation options wraps `::boost::multiprecision::sqrt(…)` or std::sqrt(…) function.

**squaredNorm(**(complex)x**)** → float :

> **Returns** `Real` the norm (squared magnitude) of the `Complex` argument in radians. Depending on compilation options wraps `::boost::multiprecision::norm(…)` or std::norm(…) function.

**tan(**(complex)x**)** → complex :

> **Returns** `Complex` the tangent of the `Complex` argument in radians. Depending on compilation options wraps `::boost::multiprecision::tan(…)` or std::tan(…) function.

> **tan( (float)x) -> float :**

> > **return** `Real` the tangent of the `Real` argument in radians. Depending on compilation options wraps `::boost::multiprecision::tan(…)` or std::tan(…) function.

**tanh(**(complex)x**)** → complex :

> **Returns** `Complex` the hyperbolic tangent of the `Complex` argument in radians. Depending on compilation options wraps `::boost::multiprecision::tanh(…)` or std::tanh(…) function.

> **tanh( (float)x) -> float :**

> > **return** `Real` the hyperbolic tangent of the `Real` argument in radians. Depending on compilation options wraps `::boost::multiprecision::tanh(…)` or std::tanh(…) function.

**testArray()** → None :
This function tests call to `std::vector::data(…)` function in order to extract the array.

`testCgalNumTraits = False`

`testConstants()` → None :
This function tests lib/high-precision/Constants.hpp, the yade::math::ConstantsHP<N>, former yade::Mathr constants.

`tgamma(`*(float)x*`)` → float :

**Returns** `Real` Computes the gamma function of arg. Depending on compilation options wraps `::boost::multiprecision::tgamma(…)` or `std::tgamma(…)` function.

`toDouble(`*(float)x*`)` → float :

**Returns** `float` converts `Real` type to `double` and returns a native python `float`.

`toHP1(`*(float)x*`)` → float :

**Returns** `RealHP<1>` converted from argument `RealHP<1>` as a result of `static_cast<RealHP<1>>(arg)`.

`toInt(`*(float)x*`)` → int :

**Returns** `int` converts `Real` type to `int` and returns a native python `int`.

`toLong(`*(float)x*`)` → int :

**Returns** `int` converts `Real` type to `long int` and returns a native python `int`.

`toLongDouble(`*(float)x*`)` → float :

**Returns** `float` converts `Real` type to `long double` and returns a native python `float`.

`trunc(`*(float)x*`)` → float :

**Returns** `Real` the nearest integer not greater in magnitude than arg. Depending on compilation options wraps `::boost::multiprecision::trunc(…)` or `std::trunc(…)` function.

`yade._math.Log2(`$\big[$*(int)Precision=53*$\big]$`)` → float

**Returns** `Real` natural logarithm of 2, exposed to python for testing of eigen numerical traits.

`yade._math.Pi(`$\big[$*(int)Precision=53*$\big]$`)` → float

**Returns** `Real` The π constant, exposed to python for testing of eigen numerical traits.

`class yade._math.RealHPConfig`
RealHPConfig class provides information about RealHP<N> type.

**Variables**

- *extraStringDigits10* – this static variable allows to control how many extra digits to use when converting to decimal strings. Assign a different value to it to affect the string conversion done in C++ python conversions as well as in all other conversions. Be careful, because values smaller than 3 can fail the round trip conversion test.

- *isFloat128Broken* – provides runtime information if Yade was compiled with g++ version < 9.2.1 and thus `boost::multiprecision::float128` cannot work.

- *isEnabledRealHP* – provides runtime information `RealHP<N>` is available for N higher than 1.

- *workaroundSlowBoostBinFloat* – boost::multiprecision::cpp_bin_float has some problem that importing it in python is very slow when these functions are exported: erf, erfc, lgamma, tgamma. In such case the python `import yade.math` can take more than minute. The workaround is to make them unavailable in python for higher N values. See invocation of IfConstexprForSlowFunctions in _math.cpp. This variable contains the highest N in which these functions are available. It equals to highest N when `boost::multiprecision::cpp_bin_-float` is not used.

**extraStringDigits10 = 4**

**getDigits10**(*(int)N*) → int :
>    This is a yade.math.RealHPConfig diagnostic function.

>    **Parameters**    `N` – `int` - the value of `N` in `RealHP<N>`.

>    **Returns**    the `int` representing `std::numeric_limits<RealHP<N>>::digits10`

**getDigits2**(*(int)N*) → int :
>    This is a yade.math.RealHPConfig diagnostic function.

>    **Parameters**    `N` – `int` - the value of `N` in `RealHP<N>`.

>    **Returns**    the `int` representing `std::numeric_limits<RealHP<N>>::digits`, which corresponds to the number of significand bits used by this type.

**getSupportedByEigenCgal**() → tuple :

>    **Returns**    the `tuple` containing N from `RealHP<N>` precisions supported by Eigen and CGAL

**getSupportedByMinieigen**() → tuple :

>    **Returns**    the `tuple` containing N from `RealHP<N>` precisions supported by minieigenHP

**isEnabledRealHP = False**

**isFloat128Broken = False**

**isFloat128Present = False**

**workaroundSlowBoostBinFloat = 1**

**class yade._math.Var**
>    The `Var` class is used to test to/from python converters for arbitrary precision `Real`

>    **cpl**
>    >    one `Complex` variable to test reading from and writing to it.

>    **val**
>    >    one `Real` variable for testing.

**yade._math.abs**(*(complex)x*) → float :

>    >    **return** the `Real` absolute value of the `Complex` argument. Depending on compilation options wraps `::boost::multiprecision::abs(…)` or std::abs(…) function.

>    **abs( (float)x)** → **float** :

>    >    **return** the `Real` absolute value of the `Real` argument. Depending on compilation options wraps `::boost::multiprecision::abs(…)` or std::abs(…) function.

**yade._math.acos**(*(complex)x*) → complex :

>    >    **return** `Complex` the arc-cosine of the `Complex` argument in radians. Depending on compilation options wraps `::boost::multiprecision::acos(…)` or std::acos(…) function.

>    **acos( (float)x)** → **float** :

>    >    **return** `Real` the arcus cosine of the argument. Depending on compilation options wraps `::boost::multiprecision::acos(…)` or std::acos(…) function.

`yade._math.acosh(`*(complex)x*`)` → complex

> **return** `Complex` the arc-hyperbolic cosine of the `Complex` argument in radians. Depending on compilation options wraps `::boost::multiprecision::acosh(…)` or std::acosh(…) function.

**acosh( (float)x)** → **float :**

> **return** `Real` the hyperbolic arcus cosine of the argument. Depending on compilation options wraps `::boost::multiprecision::acosh(…)` or std::acosh(…) function.

`yade._math.`**arg**`(`*(complex)x*`)` → float

> **Returns** `Real` the arg (Phase angle of complex in radians) of the `Complex` argument in radians. Depending on compilation options wraps `::boost::multiprecision::arg(…)` or std::arg(…) function.

`yade._math.`**asin**`(`*(complex)x*`)` → complex

> **return** `Complex` the arc-sine of the `Complex` argument in radians. Depending on compilation options wraps `::boost::multiprecision::asin(…)` or std::asin(…) function.

**asin( (float)x)** → **float :**

> **return** `Real` the arcus sine of the argument. Depending on compilation options wraps `::boost::multiprecision::asin(…)` or std::asin(…) function.

`yade._math.`**asinh**`(`*(complex)x*`)` → complex

> **return** `Complex` the arc-hyperbolic sine of the `Complex` argument in radians. Depending on compilation options wraps `::boost::multiprecision::asinh(…)` or std::asinh(…) function.

**asinh( (float)x)** → **float :**

> **return** `Real` the hyperbolic arcus sine of the argument. Depending on compilation options wraps `::boost::multiprecision::asinh(…)` or std::asinh(…) function.

`yade._math.`**atan**`(`*(complex)x*`)` → complex

> **return** `Complex` the arc-tangent of the `Complex` argument in radians. Depending on compilation options wraps `::boost::multiprecision::atan(…)` or std::atan(…) function.

**atan( (float)x)** → **float :**

> **return** `Real` the arcus tangent of the argument. Depending on compilation options wraps `::boost::multiprecision::atan(…)` or std::atan(…) function.

`yade._math.`**atan2**`(`*(float)x, (float)y*`)` → float

> **Returns** `Real` the arc tangent of y/x using the signs of the arguments `x` and `y` to determine the correct quadrant. Depending on compilation options wraps `::boost::multiprecision::atan2(…)` or std::atan2(…) function.

`yade._math.`**atanh**`(`*(complex)x*`)` → complex

> **return** `Complex` the arc-hyperbolic tangent of the `Complex` argument in radians. Depending on compilation options wraps `::boost::multiprecision::atanh(…)` or std::atanh(…) function.

**atanh( (float)x)** → **float :**

> > **return** `Real` the hyperbolic arcus tangent of the argument. Depending on compilation options wraps `::boost::multiprecision::atanh(…)` or std::atanh(…) function.

`yade._math.cbrt`(*(float)x*) → float

> **Returns** `Real` cubic root of the argument. Depending on compilation options wraps `::boost::multiprecision::cbrt(…)` or std::cbrt(…) function.

`yade._math.ceil`(*(float)x*) → float

> **Returns** `Real` Computes the smallest integer value not less than arg. Depending on compilation options wraps `::boost::multiprecision::ceil(…)` or std::ceil(…) function.

`yade._math.conj`(*(complex)x*) → complex

> **Returns** the complex conjugation a `Complex` argument. Depending on compilation options wraps `::boost::multiprecision::conj(…)` or std::conj(…) function.

`yade._math.cos`(*(complex)x*) → complex

> > **return** `Complex` the cosine of the `Complex` argument in radians. Depending on compilation options wraps `::boost::multiprecision::cos(…)` or std::cos(…) function.

> **cos( (float)x) → float :**

> > **return** `Real` the cosine of the `Real` argument in radians. Depending on compilation options wraps `::boost::multiprecision::cos(…)` or std::cos(…) function.

`yade._math.cosh`(*(complex)x*) → complex

> > **return** `Complex` the hyperbolic cosine of the `Complex` argument in radians. Depending on compilation options wraps `::boost::multiprecision::cosh(…)` or std::cosh(…) function.

> **cosh( (float)x) → float :**

> > **return** `Real` the hyperbolic cosine of the `Real` argument in radians. Depending on compilation options wraps `::boost::multiprecision::cosh(…)` or std::cosh(…) function.

`yade._math.cylBesselJ`(*(int)k*, *(float)x*) → float

> **Returns** `Real` the Bessel Functions of the First Kind of the order `k` and the `Real` argument. See: <https://www.boost.org/doc/libs/1_77_0/libs/math/doc/html/math_toolkit/bessel/bessel_first.html>`___

`yade._math.dummy_precision`() → float

> **Returns** similar to the function `epsilon`, but assumes that last 10% of bits contain the numerical error only. This is sometimes used by Eigen when calling `isEqualFuzzy` to determine if values differ a lot or if they are vaguely close to each other.

`yade._math.epsilon`([*(int)Precision=53*]) → float

> > **return** `Real` returns the difference between `1.0` and the next representable value of the `Real` type. Wraps std::numeric_limits<Real>::epsilon() function.

> **epsilon( (float)x) → float :**

> > **return** `Real` returns the difference between `1.0` and the next representable value of the `Real` type. Wraps std::numeric_limits<Real>::epsilon() function.

`yade._math.erf`(*(float)x*) → float

>   **Returns** Real Computes the error function of argument. Depending on compilation options wraps `::boost::multiprecision::erf(…)` or std::erf(…) function.

`yade._math.erfc`(*(float)x*) → float

>   **Returns** Real Computes the complementary error function of argument, that is `1.0-erf(arg)`. Depending on compilation options wraps `::boost::multiprecision::erfc(…)` or std::erfc(…) function.

`yade._math.exp`(*(complex)x*) → complex

>   **return** the base *e* exponential of a `Complex` argument. Depending on compilation options wraps `::boost::multiprecision::exp(…)` or std::exp(…) function.

>   **exp( (float)x) → float :**

>   >   **return** the base *e* exponential of a `Real` argument. Depending on compilation options wraps `::boost::multiprecision::exp(…)` or std::exp(…) function.

`yade._math.exp2`(*(float)x*) → float

>   **Returns** the base *2* exponential of a `Real` argument. Depending on compilation options wraps `::boost::multiprecision::exp2(…)` or std::exp2(…) function.

`yade._math.expm1`(*(float)x*) → float

>   **Returns** the base *e* exponential of a `Real` argument minus `1.0`. Depending on compilation options wraps `::boost::multiprecision::expm1(…)` or std::expm1(…) function.

`yade._math.fabs`(*(float)x*) → float

>   **Returns** the `Real` absolute value of the argument. Depending on compilation options wraps `::boost::multiprecision::abs(…)` or std::abs(…) function.

`yade._math.factorial`(*(int)x*) → float

>   **Returns** Real the factorial of the Real argument. See: <https://www.boost.org/doc/libs/1_77_0/libs/math/doc/html/math_toolkit/factorials/sf_factorial.html>'___

`yade._math.floor`(*(float)x*) → float

>   **Returns** Real Computes the largest integer value not greater than arg. Depending on compilation options wraps `::boost::multiprecision::floor(…)` or std::floor(…) function.

`yade._math.fma`(*(float)x, (float)y, (float)z*) → float

>   **Returns** Real - computes `(x*y) + z` as if to infinite precision and rounded only once to fit the result type. Depending on compilation options wraps `::boost::multiprecision::fma(…)` or std::fma(…) function.

`yade._math.fmod`(*(float)x, (float)y*) → float

>   **Returns** Real the floating-point remainder of the division operation `x/y` of the arguments x and y. Depending on compilation options wraps `::boost::multiprecision::fmod(…)` or std::fmod(…) function.

`yade._math.frexp`(*(float)x*) → tuple

>   **Returns** tuple of `(Real,int)`, decomposes given floating point `Real` argument into a normalized fraction and an integral power of two. Depending on compilation options wraps `::boost::multiprecision::frexp(…)` or std::frexp(…) function.

`yade._math.fromBits`(*(str)bits*[, *(int)exp=0*[, *(int)sign=1*]]) → float

**Parameters**

- **bits** – `str` - a string containing '0', '1' characters.

- **exp** – `int` - the binary exponent which shifts the bits.

- **sign** – `int` - the sign, should be -1 or +1, but it is not checked. It multiplies the result when construction from bits is finished.

**Returns** `RealHP<N>` constructed from string containing '0', '1' bits. This is for debugging purposes, rather slow.

yade._math.**getDecomposedReal**(*(float)x*) → dict

**Returns** `dict` - the dictionary with the debug information how the DecomposedReal class sees this type. This is for debugging purposes, rather slow. Includes result from fpclassify function call, a binary representation and other useful info. See also *fromBits*.

yade._math.**getDemangledName**() → str

**Returns** `string` - the demangled C++ typename of `RealHP<N>`.

yade._math.**getDemangledNameComplex**() → str

**Returns** `string` - the demangled C++ typename of `ComplexHP<N>`.

yade._math.**getEigenFlags**() → dict

**Returns** A python dictionary listing flags for all types, see: https://eigen.tuxfamily. org/dox/group___flags.html

yade._math.**getEigenStorageOrders**() → dict

**Returns** A python dictionary listing options for all types, see: https://eigen.tuxfamily. org/dox/group___TopicStorageOrders.html

yade._math.**getFloatDistanceULP**(*(float)arg1, (float)arg2*) → float

**Returns** an integer value stored in `RealHP<N>`, the ULP distance calculated by boost::math::float_distance, also see Floating-point Comparison and Prof. Kahan paper about this topic.

The returned value is the **directed distance** between two arguments, this means that it can be negative.

yade._math.**getRawBits**(*(float)x*) → str

**Returns** `string` - the raw bits in memory representing this type. Be careful: it only checks the system endianness and either prints bytes in reverse order or not. Does not make any attempts to further interpret the bits of: sign, exponent or significand (on a typical x86 processor they are printed in that order), and different processors might store them differently. It is not useful for types which internally use a pointer because for them this function prints not the floating point number but a pointer. This is for debugging purposes.

yade._math.**getRealHPErrors**(*(list)testLevelsHP[, (int)testCount=10[, (float)minX=-10.0[, (float)maxX=10.0[, (bool)useRandomArgs=False[, (int)printEveryNth=1000[, (bool)collectArgs=False[, (bool)extraChecks=False]]]]]]]*) → dict

Tests mathematical functions against the highest precision in argument `testLevelsHP` and returns the largest ULP distance found with *getFloatDistanceULP*. A `testCount` randomized tries with function arguments in range `minX ... maxX` are performed on the `RealHP<N>` types where `N` is from the list provided in `testLevelsHP` argument.

**Parameters**

- **testLevelsHP** – a list of **int** values consisting of high precision levels **N** (in **RealHP<N>**) for which the tests should be done. Must consist at least of two elements so that there is a higher precision type available against which to perform the tests.

- **testCount** – **int** - specifies how many randomized tests of each function to perform.

- **minX** – **Real** - start of the range in which the random arguments are generated.

- **maxX** – **Real** - end of that range.

- **useRandomArgs** – If **False** (default) then **minX ... maxX** is divided into **testCount** equidistant points. If **True** then each call is a random number. This applies only to the first argument of a function, if a function takes more than one argument, then remaining arguments are random - 2D scans are not performed.

- **printEveryNth** – will *print using* **LOG_INFO** the progress information every Nth step in the **testCount** loop. To see it e.g. start using **yade -f6**, also see *logger documentation*.

- **collectArgs** – if **True** then in returned results will be a longer list of arguments that produce incorrect results.

- **extraChecks** – will perform extra checks while executing this funcion. Useful only for debugging of *getRealHPErrors*.

**Returns** A python dictionary with the largest ULP distance to the correct function value. For each function name there is a dictionary consisting of: how many binary digits (bits) are in the tested **RealHP<N>** type, the worst arguments for this function, and the ULP distance to the reference value.

The returned ULP error is an absolute value, as opposed to *getFloatDistanceULP* which is signed.

**yade._math.highest**($\big[$*(int)Precision=53*$\big]$) → float

**Returns** **Real** returns the largest finite value of the **Real** type. Wraps std::numeric_-limits<Real>::max() function.

**yade._math.hypot**(*(float)x, (float)y*) → float

**Returns** **Real** the square root of the sum of the squares of **x** and **y**, without undue overflow or underflow at intermediate stages of the computation. Depending on compilation options wraps ::boost::multiprecision::hypot(...) or std::hypot(...) function.

**yade._math.ilogb**(*(float)x*) → float

**Returns** **Real** extracts the value of the unbiased exponent from the floating-point argument arg, and returns it as a signed integer value. Depending on compilation options wraps ::boost::multiprecision::ilogb(...) or std::ilogb(...) function.

**yade._math.imag**(*(complex)x*) → float

**Returns** the imag part of a **Complex** argument. Depending on compilation options wraps ::boost::multiprecision::imag(...) or std::imag(...) function.

**yade._math.isApprox**(*(float)a, (float)b, (float)eps*) → bool

**Returns** **bool**, True if **a** is approximately equal **b** with provided **eps**, see also here

**yade._math.isApproxOrLessThan**(*(float)a, (float)b, (float)eps*) → bool

**Returns** **bool**, True if **a** is approximately less than or equal **b** with provided **eps**, see also here

**yade._math.isEqualFuzzy**(*(float)arg1, (float)arg2, (float)arg3*) → bool

> **Returns** `bool`, `True` if the absolute difference between two numbers is smaller than std::numeric_limits<Real>::epsilon()

yade._math.**isMuchSmallerThan**(*(float)a, (float)b, (float)eps*) → bool

> **Returns** `bool`, True if `a` is less than b with provided `eps`, see also here

yade._math.**isThisSystemLittleEndian**() → bool

> **Returns** `True` if this system uses little endian architecture, `False` otherwise.

yade._math.**isfinite**(*(float)x*) → bool

> **Returns** `bool` indicating if the `Real` argument is Inf. Depending on compilation options wraps ::boost::multiprecision::isfinite(…) or std::isfinite(…) function.

yade._math.**isinf**(*(float)x*) → bool

> **Returns** `bool` indicating if the `Real` argument is Inf. Depending on compilation options wraps ::boost::multiprecision::isinf(…) or std::isinf(…) function.

yade._math.**isnan**(*(float)x*) → bool

> **Returns** `bool` indicating if the `Real` argument is NaN. Depending on compilation options wraps ::boost::multiprecision::isnan(…) or std::isnan(…) function.

yade._math.**laguerre**(*(int)n, (int)m, (float)x*) → float

> **Returns** `Real` the Laguerre polynomial of the orders `n`, `m` and the `Real` argument. See: <https://www.boost.org/doc/libs/1_77_0/libs/math/doc/html/math_toolkit/sf_poly/laguerre.html>'___

yade._math.**ldexp**(*(float)x, (int)y*) → float

> **Returns** Multiplies a floating point value `x` by the number 2 raised to the `exp` power. Depending on compilation options wraps ::boost::multiprecision::ldexp(…) or std::ldexp(…) function.

yade._math.**lgamma**(*(float)x*) → float

> **Returns** `Real` Computes the natural logarithm of the absolute value of the gamma function of arg. Depending on compilation options wraps ::boost::multiprecision::lgamma(…) or std::lgamma(…) function.

yade._math.**log**(*(complex)x*) → complex

> **return** the `Complex` natural (base *e*) logarithm of a complex value z with a branch cut along the negative real axis. Depending on compilation options wraps ::boost::multiprecision::log(…) or std::log(…) function.

**log( (float)x)** → **float :**

> **return** the `Real` natural (base *e*) logarithm of a real value. Depending on compilation options wraps ::boost::multiprecision::log(…) or std::log(…) function.

yade._math.**log10**(*(complex)x*) → complex

> **return** the `Complex` (base *10*) logarithm of a complex value z with a branch cut along the negative real axis. Depending on compilation options wraps ::boost::multiprecision::log10(…) or std::log10(…) function.

**log10( (float)x)** → **float :**

> **return** the `Real` decimal (base 10) logarithm of a real value. Depending on compilation options wraps ::boost::multiprecision::log10(…) or std::log10(…) function.

yade._math.**log1p**(*(float)x*) → float

> **Returns** the `Real` natural (base *e*) logarithm of `1+argument`. Depending on compilation options wraps `::boost::multiprecision::log1p(…)` or std::log1p(…) function.

yade._math.**log2**(*(float)x*) → float

> **Returns** the `Real` binary (base 2) logarithm of a real value. Depending on compilation options wraps `::boost::multiprecision::log2(…)` or std::log2(…) function.

yade._math.**logb**(*(float)x*) → float

> **Returns** Extracts the value of the unbiased radix-independent exponent from the floating-point argument arg, and returns it as a floating-point value. Depending on compilation options wraps `::boost::multiprecision::logb(…)` or std::logb(…) function.

yade._math.**lowest**([*(int)Precision=53*]) → float

> **Returns** Real returns the lowest (negative) finite value of the Real type. Wraps std::numeric_limits<Real>::lowest() function.

yade._math.**max**(*(float)x, (float)y*) → float

> **Returns** Real larger of the two arguments. Depending on compilation options wraps `::boost::multiprecision::max(…)` or std::max(…) function.

yade._math.**min**(*(float)x, (float)y*) → float

> **Returns** Real smaller of the two arguments. Depending on compilation options wraps `::boost::multiprecision::min(…)` or std::min(…) function.

yade._math.**modf**(*(float)x*) → tuple

> **Returns** tuple of (`Real,Real`), decomposes given floating point `Real` into integral and fractional parts, each having the same type and sign as x. Depending on compilation options wraps `::boost::multiprecision::modf(…)` or std::modf(…) function.

yade._math.**polar**(*(float)x, (float)y*) → complex

> **Returns** `Complex` the polar (Complex from polar components) of the `Real` rho (length), `Real` theta (angle) arguments in radians. Depending on compilation options wraps `::boost::multiprecision::polar(…)` or std::polar(…) function.

yade._math.**pow**(*(complex)x, (complex)pow*) → complex

> **return** the `Complex` complex arg1 raised to the `Complex` power arg2. Depending on compilation options wraps `::boost::multiprecision::pow(…)` or std::pow(…) function.

> **pow( (float)x, (float)y) → float :**

> > **return** Real the value of `base` raised to the power `exp`. Depending on compilation options wraps `::boost::multiprecision::pow(…)` or std::pow(…) function.

yade._math.**proj**(*(complex)x*) → complex

> **Returns** `Complex` the proj (projection of the complex number onto the Riemann sphere) of the `Complex` argument in radians. Depending on compilation options wraps `::boost::multiprecision::proj(…)` or std::proj(…) function.

yade._math.**random**() → float

> **return** Real a symmetric random number in interval (`-1,1`). Used by Eigen.

> **random( (float)a, (float)b) → float :**

> > **return** Real a random number in interval (`a,b`). Used by Eigen.

yade._math.**real**(*(complex)x*) → float

> **Returns** the real part of a `Complex` argument. Depending on compilation options wraps
> `::boost::multiprecision::real(…)` or std::real(…) function.

yade._math.**remainder**(*(float)x, (float)y*) → float

> **Returns** `Real` the IEEE remainder of the floating point division operation x/y. Depending on compilation options wraps `::boost::multiprecision::remainder(…)` or std::remainder(…) function.

yade._math.**remquo**(*(Real)x, (float)y*) → tuple

> **Returns** tuple of (`Real`,`long`), the floating-point remainder of the division operation x/y as the std::remainder() function does. Additionally, the sign and at least the three of the last bits of x/y are returned, sufficient to determine the octant of the result within a period. Depending on compilation options wraps `::boost::multiprecision::remquo(…)` or std::remquo(…) function.

yade._math.**rint**(*(float)x*) → float

> **Returns** Rounds the floating-point argument arg to an integer value (in floating-point format), using the current rounding mode. Depending on compilation options wraps `::boost::multiprecision::rint(…)` or std::rint(…) function.

yade._math.**round**(*(float)x*) → float

> **Returns** `Real` the nearest integer value to arg (in floating-point format), rounding halfway cases away from zero, regardless of the current rounding mode.. Depending on compilation options wraps `::boost::multiprecision::round(…)` or std::round(…) function.

yade._math.**roundTrip**(*(float)x*) → float

> **Returns** `Real` returns the argument x. Can be used to convert type to native RealHP<N> accuracy.

yade._math.**sgn**(*(float)x*) → int

> **Returns** `int` the sign of the argument: −1, 0 or 1.

yade._math.**sign**(*(float)x*) → int

> **Returns** `int` the sign of the argument: −1, 0 or 1.

yade._math.**sin**(*(complex)x*) → complex

> > **return** `Complex` the sine of the `Complex` argument in radians. Depending on compilation options wraps `::boost::multiprecision::sin(…)` or std::sin(…) function.

> **sin( (float)x) → float :**

> > **return** `Real` the sine of the `Real` argument in radians. Depending on compilation options wraps `::boost::multiprecision::sin(…)` or std::sin(…) function.

yade._math.**sinh**(*(complex)x*) → complex

> > **return** `Complex` the hyperbolic sine of the `Complex` argument in radians. Depending on compilation options wraps `::boost::multiprecision::sinh(…)` or std::sinh(…) function.

> **sinh( (float)x) → float :**

> > **return** `Real` the hyperbolic sine of the `Real` argument in radians. Depending on compilation options wraps `::boost::multiprecision::sinh(…)` or std::sinh(…) function.

yade._math.**smallest_positive**() → float

> > **Returns** `Real` the smallest number greater than zero. Wraps std::numeric_limits<Real>::min()

`yade._math.`**`sphericalHarmonic`**`((int)l, (int)m, (float)theta, (float)phi)` → complex

> > **Returns** `Real` the spherical harmonic polynomial of the orders l (unsigned int), m (signed int) and the `Real` arguments `theta` and `phi`. See: <https://www.boost.org/doc/libs/1_77_0/libs/math/doc/html/math_toolkit/sf_poly/sph_harm.html>`___

`yade._math.`**`sqrt`**`((complex)x)` → complex

> > return the `Complex` square root of `Complex` argument. Depending on compilation options wraps `::boost::multiprecision::sqrt(…)` or std::sqrt(…) function.

> **sqrt( (float)x) → float :**

> > return `Real` square root of the argument. Depending on compilation options wraps `::boost::multiprecision::sqrt(…)` or std::sqrt(…) function.

`yade._math.`**`squaredNorm`**`((complex)x)` → float

> > **Returns** `Real` the norm (squared magnitude) of the `Complex` argument in radians. Depending on compilation options wraps `::boost::multiprecision::norm(…)` or std::norm(…) function.

`yade._math.`**`tan`**`((complex)x)` → complex

> > return `Complex` the tangent of the `Complex` argument in radians. Depending on compilation options wraps `::boost::multiprecision::tan(…)` or std::tan(…) function.

> **tan( (float)x) → float :**

> > return `Real` the tangent of the `Real` argument in radians. Depending on compilation options wraps `::boost::multiprecision::tan(…)` or std::tan(…) function.

`yade._math.`**`tanh`**`((complex)x)` → complex

> > return `Complex` the hyperbolic tangent of the `Complex` argument in radians. Depending on compilation options wraps `::boost::multiprecision::tanh(…)` or std::tanh(…) function.

> **tanh( (float)x) → float :**

> > return `Real` the hyperbolic tangent of the `Real` argument in radians. Depending on compilation options wraps `::boost::multiprecision::tanh(…)` or std::tanh(…) function.

`yade._math.`**`testArray`**`()` → None
  This function tests call to `std::vector::data(…)` function in order to extract the array.

`yade._math.`**`testConstants`**`()` → None
  This function tests lib/high-precision/Constants.hpp, the yade::math::ConstantsHP<N>, former yade::Mathr constants.

`yade._math.`**`testLoopRealHP`**`()` → None
  This function tests lib/high-precision/Constants.hpp, but the C++ side: all precisions, even those inaccessible from python

`yade._math.`**`tgamma`**`((float)x)` → float

> > **Returns** `Real` Computes the gamma function of arg. Depending on compilation options wraps `::boost::multiprecision::tgamma(…)` or std::tgamma(…) function.

`yade._math.toDouble(`*(float)x*`)` → float

> **Returns** float converts Real type to `double` and returns a native python `float`.

`yade._math.toHP1(`*(float)x*`)` → float

> **Returns** RealHP<1> converted from argument `RealHP<1>` as a result of `static_-cast<RealHP<1>>(arg)`.

`yade._math.toInt(`*(float)x*`)` → int

> **Returns** int converts Real type to `int` and returns a native python `int`.

`yade._math.toLong(`*(float)x*`)` → int

> **Returns** int converts Real type to `long int` and returns a native python `int`.

`yade._math.toLongDouble(`*(float)x*`)` → float

> **Returns** float converts Real type to `long double` and returns a native python `float`.

`yade._math.trunc(`*(float)x*`)` → float

> **Returns** Real the nearest integer not greater in magnitude than arg. Depending on compilation options wraps `::boost::multiprecision::trunc(…)` or std::trunc(…) function.

### 2.4.9 yade.minieigenHP module

When yade uses high-precision number as `Real` type the usual (old):

```python
from minieigen import *
```

has to be replaced with:

```python
from yade.minieigenHP import *
```

This command ensures backward compatibility between both. It is then guaranteed that python uses the same number of decimal places as yade is using everywhere else.

Please note that used precision can be very arbitrary, because `cpp_bin_float` or `mpfr` take it as a *compile-time argument*. Hence such `yade.minieigenHP` cannot be separately precompiled as a package. Though it could be precompiled for some special types such as `boost::multiprecision::float128`.

The RealHP<n> *higher precision* vectors and matrices can be accessed in python by using the `.HPn` module scope. For example:

```python
import yade.minieigenHP as mne
mne.HP2.Vector3(1,2,3) # produces Vector3 using RealHP<2> precision
mne.Vector3(1,2,3)     # without using HPn module scope it defaults to RealHP<1>
```

miniEigen is wrapper for a small part of the Eigen library. Refer to its documentation for details. All classes in this module support pickling.

**class** `yade._minieigenHP.AlignedBox2`

> Axis-aligned box object in 2d, defined by its minimum and maximum corners

> `__init__(`*(object)arg1*`)` → None
>> \_\_init\_\_( (object)arg1, (AlignedBox2)other) -> None
>>
>> \_\_init\_\_( (object)arg1, (Vector2)min, (Vector2)max) -> None

> `center(`*(AlignedBox2)arg1*`)` → Vector2

> `clamp(`*(AlignedBox2)arg1, (AlignedBox2)arg2*`)` → None

> `contains(`*(AlignedBox2)arg1, (Vector2)arg2*`)` → bool
>> contains( (AlignedBox2)arg1, (AlignedBox2)arg2) -> bool

**empty**(*(AlignedBox2)arg1*) → bool

**extend**(*(AlignedBox2)arg1, (Vector2)arg2*) → None
     extend( (AlignedBox2)arg1, (AlignedBox2)arg2) -> None

**intersection**(*(AlignedBox2)arg1, (AlignedBox2)arg2*) → AlignedBox2

**max**

**merged**(*(AlignedBox2)arg1, (AlignedBox2)arg2*) → AlignedBox2

**min**

**sizes**(*(AlignedBox2)arg1*) → Vector2

**volume**(*(AlignedBox2)arg1*) → float

**class yade._minieigenHP.AlignedBox3**
    Axis-aligned box object, defined by its minimum and maximum corners

    **__init__**(*(object)arg1*) → None
        ___init___( (object)arg1, (AlignedBox3)other) -> None

        ___init___( (object)arg1, (Vector3)min, (Vector3)max) -> None

    **center**(*(AlignedBox3)arg1*) → Vector3

    **clamp**(*(AlignedBox3)arg1, (AlignedBox3)arg2*) → None

    **contains**(*(AlignedBox3)arg1, (Vector3)arg2*) → bool
        contains( (AlignedBox3)arg1, (AlignedBox3)arg2) -> bool

    **empty**(*(AlignedBox3)arg1*) → bool

    **extend**(*(AlignedBox3)arg1, (Vector3)arg2*) → None
        extend( (AlignedBox3)arg1, (AlignedBox3)arg2) -> None

    **intersection**(*(AlignedBox3)arg1, (AlignedBox3)arg2*) → AlignedBox3

    **max**

    **merged**(*(AlignedBox3)arg1, (AlignedBox3)arg2*) → AlignedBox3

    **min**

    **sizes**(*(AlignedBox3)arg1*) → Vector3

    **volume**(*(AlignedBox3)arg1*) → float

**class yade._minieigenHP.HP1**

    **class AlignedBox2**
        Axis-aligned box object in 2d, defined by its minimum and maximum corners

        **__init__**(*(object)arg1*) → None
            ___init___( (object)arg1, (AlignedBox2)other) -> None

            ___init___( (object)arg1, (Vector2)min, (Vector2)max) -> None

        **center**(*(AlignedBox2)arg1*) → Vector2

        **clamp**(*(AlignedBox2)arg1, (AlignedBox2)arg2*) → None

        **contains**(*(AlignedBox2)arg1, (Vector2)arg2*) → bool
            contains( (AlignedBox2)arg1, (AlignedBox2)arg2) -> bool

        **empty**(*(AlignedBox2)arg1*) → bool

        **extend**(*(AlignedBox2)arg1, (Vector2)arg2*) → None
            extend( (AlignedBox2)arg1, (AlignedBox2)arg2) -> None

        **intersection**(*(AlignedBox2)arg1, (AlignedBox2)arg2*) → AlignedBox2

**max**

**merged**(*(AlignedBox2)arg1, (AlignedBox2)arg2*) → AlignedBox2

**min**

**sizes**(*(AlignedBox2)arg1*) → Vector2

**volume**(*(AlignedBox2)arg1*) → float

**class AlignedBox3**

Axis-aligned box object, defined by its minimum and maximum corners

**__init__**(*(object)arg1*) → None
   ___init___( (object)arg1, (AlignedBox3)other) -> None

   ___init___( (object)arg1, (Vector3)min, (Vector3)max) -> None

**center**(*(AlignedBox3)arg1*) → Vector3

**clamp**(*(AlignedBox3)arg1, (AlignedBox3)arg2*) → None

**contains**(*(AlignedBox3)arg1, (Vector3)arg2*) → bool
   contains( (AlignedBox3)arg1, (AlignedBox3)arg2) -> bool

**empty**(*(AlignedBox3)arg1*) → bool

**extend**(*(AlignedBox3)arg1, (Vector3)arg2*) → None
   extend( (AlignedBox3)arg1, (AlignedBox3)arg2) -> None

**intersection**(*(AlignedBox3)arg1, (AlignedBox3)arg2*) → AlignedBox3

**max**

**merged**(*(AlignedBox3)arg1, (AlignedBox3)arg2*) → AlignedBox3

**min**

**sizes**(*(AlignedBox3)arg1*) → Vector3

**volume**(*(AlignedBox3)arg1*) → float

**class Matrix3**

3x3 float matrix.

Supported operations (`m` is a Matrix3, `f` if a float/int, `v` is a Vector3): `-m, m+m, m+=m, m-m, m-=m, m*f, f*m, m*=f, m/f, m/=f, m*m, m*=m, m*v, v*m, m==m, m!=m`.

Static attributes: `Zero`, `Ones`, `Identity`.

**Identity = Matrix3(1,0,0, 0,1,0, 0,0,1)**

**Ones = Matrix3(1,1,1, 1,1,1, 1,1,1)**

**static Random()** → Matrix3 :
   Return an object where all elements are randomly set to values between 0 and 1.

**Zero = Matrix3(0,0,0, 0,0,0, 0,0,0)**

**__init__**(*(object)arg1*) → None
   ___init___( (object)arg1, (Quaternion)q) -> None

   ___init___( (object)arg1, (Matrix3)other) -> None

   ___init___( (object)arg1, (Vector3)diag) -> object

   ___init___( (object)arg1, (float)m00, (float)m01, (float)m02, (float)m10, (float)m11, (float)m12, (float)m20, (float)m21, (float)m22) -> object

   ___init___( (object)arg1, (Vector3)r0, (Vector3)r1, (Vector3)r2 [, (bool)cols=False]) -> object

**col**(*(Matrix3)arg1, (int)col*) → Vector3 :
   Return column as vector.

**cols**(*(Matrix3)arg1*) → int :
    Number of columns.

**computeUnitaryPositive**(*(Matrix3)arg1*) → tuple :
    Compute polar decomposition (unitary matrix U and positive semi-definite symmetric matrix P such that self=U*P).

**determinant**(*(Matrix3)arg1*) → float :
    Return matrix determinant.

**diagonal**(*(Matrix3)arg1*) → Vector3 :
    Return diagonal as vector.

**inverse**(*(Matrix3)arg1*) → Matrix3 :
    Return inverted matrix.

**isApprox**(*(Matrix3)arg1, (Matrix3)other*[, *(float)prec=1e-12*]) → bool :
    Approximate comparison with precision *prec*.

**jacobiSVD**(*(Matrix3)arg1*) → tuple :
    Compute SVD decomposition of square matrix, retuns (U,S,V) such that self=U*S*V.transpose()

**maxAbsCoeff**(*(Matrix3)arg1*) → float :
    Maximum absolute value over all elements.

**maxCoeff**(*(Matrix3)arg1*) → float :
    Maximum value over all elements.

**mean**(*(Matrix3)arg1*) → float :
    Mean value over all elements.

**minCoeff**(*(Matrix3)arg1*) → float :
    Minimum value over all elements.

**norm**(*(Matrix3)arg1*) → float :
    Euclidean norm.

**normalize**(*(Matrix3)arg1*) → None :
    Normalize this object in-place.

**normalized**(*(Matrix3)arg1*) → Matrix3 :
    Return normalized copy of this object

**polarDecomposition**(*(Matrix3)arg1*) → tuple :
    Alias for *computeUnitaryPositive*.

**prod**(*(Matrix3)arg1*) → float :
    Product of all elements.

**pruned**(*(Matrix3)arg1*[, *(float)absTol=1e-06*]) → Matrix3 :
    Zero all elements which are greater than *absTol*. Negative zeros are not pruned.

**row**(*(Matrix3)arg1, (int)row*) → Vector3 :
    Return row as vector.

**rows**(*(Matrix3)arg1*) → int :
    Number of rows.

**selfAdjointEigenDecomposition**(*(Matrix3)arg1*) → tuple :
    Compute eigen (spectral) decomposition of symmetric matrix, returns (eigVecs,eigVals). eigVecs is orthogonal Matrix3 with columns ar normalized eigenvectors, eigVals is Vector3 with corresponding eigenvalues. self=eigVecs*diag(eigVals)*eigVecs.transpose().

**spectralDecomposition**(*(Matrix3)arg1*) → tuple :
    Alias for *selfAdjointEigenDecomposition*.

**squaredNorm**(*(Matrix3)arg1*) → float :
    Square of the Euclidean norm.

**sum**(*(Matrix3)arg1*) → float :
    Sum of all elements.

**svd**(*(Matrix3)arg1*) → tuple :
    Alias for `jacobiSVD`.

**trace**(*(Matrix3)arg1*) → float :
    Return sum of diagonal elements.

**transpose**(*(Matrix3)arg1*) → Matrix3 :
    Return transposed matrix.

**class Matrix3c**
    */TODO/*

    **Identity = Matrix3c(1,0,0, 0,1,0, 0,0,1)**

    **Ones = Matrix3c(1,1,1, 1,1,1, 1,1,1)**

    **static Random()** → Matrix3c :
        Return an object where all elements are randomly set to values between 0 and 1.

    **Zero = Matrix3c(0,0,0, 0,0,0, 0,0,0)**

    **__init__**(*(object)arg1*) → None
        ___init___( (object)arg1, (Matrix3c)other) -> None

        ___init___( (object)arg1, (Vector3c)diag) -> object

        ___init___( (object)arg1, (complex)m00, (complex)m01, (complex)m02, (complex)m10, (complex)m11, (complex)m12, (complex)m20, (complex)m21, (complex)m22) -> object

        ___init___( (object)arg1, (Vector3c)r0, (Vector3c)r1, (Vector3c)r2 [, (bool)cols=False]) -> object

    **col**(*(Matrix3c)arg1, (int)col*) → Vector3c :
        Return column as vector.

    **cols**(*(Matrix3c)arg1*) → int :
        Number of columns.

    **determinant**(*(Matrix3c)arg1*) → complex :
        Return matrix determinant.

    **diagonal**(*(Matrix3c)arg1*) → Vector3c :
        Return diagonal as vector.

    **inverse**(*(Matrix3c)arg1*) → Matrix3c :
        Return inverted matrix.

    **isApprox**(*(Matrix3c)arg1, (Matrix3c)other*[, *(float)prec=1e-12*]) → bool :
        Approximate comparison with precision *prec*.

    **maxAbsCoeff**(*(Matrix3c)arg1*) → float :
        Maximum absolute value over all elements.

    **mean**(*(Matrix3c)arg1*) → complex :
        Mean value over all elements.

    **norm**(*(Matrix3c)arg1*) → float :
        Euclidean norm.

    **normalize**(*(Matrix3c)arg1*) → None :
        Normalize this object in-place.

    **normalized**(*(Matrix3c)arg1*) → Matrix3c :
        Return normalized copy of this object

    **prod**(*(Matrix3c)arg1*) → complex :
        Product of all elements.

**pruned**(*(Matrix3c)arg1*[, *(float)absTol=1e-06*]) → Matrix3c :
    Zero all elements which are greater than *absTol*. Negative zeros are not pruned.

**row**(*(Matrix3c)arg1, (int)row*) → Vector3c :
    Return row as vector.

**rows**(*(Matrix3c)arg1*) → int :
    Number of rows.

**squaredNorm**(*(Matrix3c)arg1*) → float :
    Square of the Euclidean norm.

**sum**(*(Matrix3c)arg1*) → complex :
    Sum of all elements.

**trace**(*(Matrix3c)arg1*) → complex :
    Return sum of diagonal elements.

**transpose**(*(Matrix3c)arg1*) → Matrix3c :
    Return transposed matrix.

## class Matrix6

6x6 float matrix. Constructed from 4 3x3 sub-matrices, from 6xVector6 (rows).

Supported operations (m is a Matrix6, f if a float/int, v is a Vector6): -m, m+m, m+=m, m-m, m-=m, m*f, f*m, m*=f, m/f, m/=f, m*m, m*=m, m*v, v*m, m==m, m!=m.

Static attributes: Zero, Ones, Identity.

**Identity = Matrix6(** (1,0,0,0,0,0), (0,1,0,0,0,0), (0,0,1,0,0,0), (0,0,0,1,0,0), (0,0,0,0,1,

**Ones = Matrix6(** (1,1,1,1,1,1), (1,1,1,1,1,1), (1,1,1,1,1,1), (1,1,1,1,1,1), (1,1,1,1,1,1),

**static Random()** → Matrix6 :
    Return an object where all elements are randomly set to values between 0 and 1.

**Zero = Matrix6(** (0,0,0,0,0,0), (0,0,0,0,0,0), (0,0,0,0,0,0), (0,0,0,0,0,0), (0,0,0,0,0,0),

**__init__**(*(object)arg1*) → None
    ___init___( (object)arg1, (Matrix6)other) -> None

    ___init___( (object)arg1, (Vector6)diag) -> object

    ___init___( (object)arg1, (Matrix3)ul, (Matrix3)ur, (Matrix3)ll, (Matrix3)lr) -> object

    ___init___( (object)arg1, (Vector6)l0, (Vector6)l1, (Vector6)l2, (Vector6)l3, (Vector6)l4, (Vector6)l5 [, (bool)cols=False]) -> object

**col**(*(Matrix6)arg1, (int)col*) → Vector6 :
    Return column as vector.

**cols**(*(Matrix6)arg1*) → int :
    Number of columns.

**computeUnitaryPositive**(*(Matrix6)arg1*) → tuple :
    Compute polar decomposition (unitary matrix U and positive semi-definite symmetric matrix P such that self=U*P).

**determinant**(*(Matrix6)arg1*) → float :
    Return matrix determinant.

**diagonal**(*(Matrix6)arg1*) → Vector6 :
    Return diagonal as vector.

**inverse**(*(Matrix6)arg1*) → Matrix6 :
    Return inverted matrix.

**isApprox**(*(Matrix6)arg1, (Matrix6)other*[, *(float)prec=1e-12*]) → bool :
    Approximate comparison with precision *prec*.

**jacobiSVD**(*(Matrix6)arg1*) → tuple :
>   Compute SVD decomposition of square matrix, retuns (U,S,V) such that self=U*S*V.transpose()

**ll**(*(Matrix6)arg1*) → Matrix3 :
>   Return lower-left 3x3 block

**lr**(*(Matrix6)arg1*) → Matrix3 :
>   Return lower-right 3x3 block

**maxAbsCoeff**(*(Matrix6)arg1*) → float :
>   Maximum absolute value over all elements.

**maxCoeff**(*(Matrix6)arg1*) → float :
>   Maximum value over all elements.

**mean**(*(Matrix6)arg1*) → float :
>   Mean value over all elements.

**minCoeff**(*(Matrix6)arg1*) → float :
>   Minimum value over all elements.

**norm**(*(Matrix6)arg1*) → float :
>   Euclidean norm.

**normalize**(*(Matrix6)arg1*) → None :
>   Normalize this object in-place.

**normalized**(*(Matrix6)arg1*) → Matrix6 :
>   Return normalized copy of this object

**polarDecomposition**(*(Matrix6)arg1*) → tuple :
>   Alias for *computeUnitaryPositive*.

**prod**(*(Matrix6)arg1*) → float :
>   Product of all elements.

**pruned**(*(Matrix6)arg1*[, *(float)absTol=1e-06*]) → Matrix6 :
>   Zero all elements which are greater than *absTol*. Negative zeros are not pruned.

**row**(*(Matrix6)arg1, (int)row*) → Vector6 :
>   Return row as vector.

**rows**(*(Matrix6)arg1*) → int :
>   Number of rows.

**selfAdjointEigenDecomposition**(*(Matrix6)arg1*) → tuple :
>   Compute eigen (spectral) decomposition of symmetric matrix, returns (eigVecs,eigVals). eigVecs is orthogonal Matrix3 with columns ar normalized eigenvectors, eigVals is Vector3 with corresponding eigenvalues. self=eigVecs*diag(eigVals)*eigVecs.transpose().

**spectralDecomposition**(*(Matrix6)arg1*) → tuple :
>   Alias for *selfAdjointEigenDecomposition*.

**squaredNorm**(*(Matrix6)arg1*) → float :
>   Square of the Euclidean norm.

**sum**(*(Matrix6)arg1*) → float :
>   Sum of all elements.

**svd**(*(Matrix6)arg1*) → tuple :
>   Alias for *jacobiSVD*.

**trace**(*(Matrix6)arg1*) → float :
>   Return sum of diagonal elements.

**transpose**(*(Matrix6)arg1*) → Matrix6 :
>   Return transposed matrix.

> **ul**(*(Matrix6)arg1*) → Matrix3 :
>> Return upper-left 3x3 block
>
> **ur**(*(Matrix6)arg1*) → Matrix3 :
>> Return upper-right 3x3 block

**class Matrix6c**
> */TODO/*
>
> **Identity** = Matrix6c( (1,0,0,0,0,0), (0,1,0,0,0,0), (0,0,1,0,0,0), (0,0,0,1,0,0), (0,0,0,0,1
>
> **Ones** = Matrix6c( (1,1,1,1,1,1), (1,1,1,1,1,1), (1,1,1,1,1,1), (1,1,1,1,1,1), (1,1,1,1,1,1),
>
> **static Random**() → Matrix6c :
>> Return an object where all elements are randomly set to values between 0 and 1.
>
> **Zero** = Matrix6c( (0,0,0,0,0,0), (0,0,0,0,0,0), (0,0,0,0,0,0), (0,0,0,0,0,0), (0,0,0,0,0,0),
>
> **__init__**(*(object)arg1*) → None
>> ___init___( (object)arg1, (Matrix6c)other) -> None
>>
>> ___init___( (object)arg1, (Vector6c)diag) -> object
>>
>> ___init___( (object)arg1, (Matrix3c)ul, (Matrix3c)ur, (Matrix3c)ll, (Matrix3c)lr) -> object
>>
>> ___init___( (object)arg1, (Vector6c)l0, (Vector6c)l1, (Vector6c)l2, (Vector6c)l3, (Vector6c)l4, (Vector6c)l5 [, (bool)cols=False]) -> object
>
> **col**(*(Matrix6c)arg1, (int)col*) → Vector6c :
>> Return column as vector.
>
> **cols**(*(Matrix6c)arg1*) → int :
>> Number of columns.
>
> **determinant**(*(Matrix6c)arg1*) → complex :
>> Return matrix determinant.
>
> **diagonal**(*(Matrix6c)arg1*) → Vector6c :
>> Return diagonal as vector.
>
> **inverse**(*(Matrix6c)arg1*) → Matrix6c :
>> Return inverted matrix.
>
> **isApprox**(*(Matrix6c)arg1, (Matrix6c)other*[, *(float)prec=1e-12*]) → bool :
>> Approximate comparison with precision *prec*.
>
> **ll**(*(Matrix6c)arg1*) → Matrix3c :
>> Return lower-left 3x3 block
>
> **lr**(*(Matrix6c)arg1*) → Matrix3c :
>> Return lower-right 3x3 block
>
> **maxAbsCoeff**(*(Matrix6c)arg1*) → float :
>> Maximum absolute value over all elements.
>
> **mean**(*(Matrix6c)arg1*) → complex :
>> Mean value over all elements.
>
> **norm**(*(Matrix6c)arg1*) → float :
>> Euclidean norm.
>
> **normalize**(*(Matrix6c)arg1*) → None :
>> Normalize this object in-place.
>
> **normalized**(*(Matrix6c)arg1*) → Matrix6c :
>> Return normalized copy of this object
>
> **prod**(*(Matrix6c)arg1*) → complex :
>> Product of all elements.

**pruned**(*(Matrix6c)arg1*[, *(float)absTol=1e-06*]) → Matrix6c :
    Zero all elements which are greater than *absTol*. Negative zeros are not pruned.

**row**(*(Matrix6c)arg1, (int)row*) → Vector6c :
    Return row as vector.

**rows**(*(Matrix6c)arg1*) → int :
    Number of rows.

**squaredNorm**(*(Matrix6c)arg1*) → float :
    Square of the Euclidean norm.

**sum**(*(Matrix6c)arg1*) → complex :
    Sum of all elements.

**trace**(*(Matrix6c)arg1*) → complex :
    Return sum of diagonal elements.

**transpose**(*(Matrix6c)arg1*) → Matrix6c :
    Return transposed matrix.

**ul**(*(Matrix6c)arg1*) → Matrix3c :
    Return upper-left 3x3 block

**ur**(*(Matrix6c)arg1*) → Matrix3c :
    Return upper-right 3x3 block

**class MatrixX**
    XxX (dynamic-sized) float matrix. Constructed from list of rows (as VectorX).

    Supported operations (m is a MatrixX, f if a float/int, v is a VectorX): -m, m+m, m+=m, m-m, m-=m, m*f, f*m, m*=f, m/f, m/=f, m*m, m*=m, m*v, v*m, m==m, m!=m.

    **static Identity**(*(int)arg1, (int)rank*) → MatrixX :
        Create identity matrix with given rank (square).

    **static Ones**(*(int)rows, (int)cols*) → MatrixX :
        Create matrix of given dimensions where all elements are set to 1.

    **static Random**(*(int)rows, (int)cols*) → MatrixX :
        Create matrix with given dimensions where all elements are set to number between 0 and 1 (uniformly-distributed).

    **static Zero**(*(int)rows, (int)cols*) → MatrixX :
        Create zero matrix of given dimensions

    **__init__**(*(object)arg1*) → None :
        ___init___( (object)arg1, (MatrixX)other) -> None

        ___init___( (object)arg1, (VectorX)diag) -> object

        ___init___( (object)arg1 [, (VectorX)r0=VectorX() [, (VectorX)r1=VectorX() [, (VectorX)r2=VectorX() [, (VectorX)r3=VectorX() [, (VectorX)r4=VectorX() [, (VectorX)r5=VectorX() [, (VectorX)r6=VectorX() [, (VectorX)r7=VectorX() [, (VectorX)r8=VectorX() [, (VectorX)r9=VectorX() [, (bool)cols=False]]]]]]]]]]]) -> object

        ___init___( (object)arg1, (object)rows [, (bool)cols=False]) -> object

    **col**(*(MatrixX)arg1, (int)col*) → VectorX :
        Return column as vector.

    **cols**(*(MatrixX)arg1*) → int :
        Number of columns.

    **computeUnitaryPositive**(*(MatrixX)arg1*) → tuple :
        Compute polar decomposition (unitary matrix U and positive semi-definite symmetric matrix P such that self=U*P).

**determinant**(*(MatrixX)arg1*) → float :
    Return matrix determinant.

**diagonal**(*(MatrixX)arg1*) → VectorX :
    Return diagonal as vector.

**inverse**(*(MatrixX)arg1*) → MatrixX :
    Return inverted matrix.

**isApprox**(*(MatrixX)arg1, (MatrixX)other*[*, (float)prec=1e-12*]) → bool :
    Approximate comparison with precision *prec*.

**jacobiSVD**(*(MatrixX)arg1*) → tuple :
    Compute SVD decomposition of square matrix, retuns (U,S,V) such that
    self=U*S*V.transpose()

**maxAbsCoeff**(*(MatrixX)arg1*) → float :
    Maximum absolute value over all elements.

**maxCoeff**(*(MatrixX)arg1*) → float :
    Maximum value over all elements.

**mean**(*(MatrixX)arg1*) → float :
    Mean value over all elements.

**minCoeff**(*(MatrixX)arg1*) → float :
    Minimum value over all elements.

**norm**(*(MatrixX)arg1*) → float :
    Euclidean norm.

**normalize**(*(MatrixX)arg1*) → None :
    Normalize this object in-place.

**normalized**(*(MatrixX)arg1*) → MatrixX :
    Return normalized copy of this object

**polarDecomposition**(*(MatrixX)arg1*) → tuple :
    Alias for *computeUnitaryPositive*.

**prod**(*(MatrixX)arg1*) → float :
    Product of all elements.

**pruned**(*(MatrixX)arg1*[*, (float)absTol=1e-06*]) → MatrixX :
    Zero all elements which are greater than *absTol*. Negative zeros are not pruned.

**resize**(*(MatrixX)arg1, (int)rows, (int)cols*) → None :
    Change size of the matrix, keep values of elements which exist in the new matrix

**row**(*(MatrixX)arg1, (int)row*) → VectorX :
    Return row as vector.

**rows**(*(MatrixX)arg1*) → int :
    Number of rows.

**selfAdjointEigenDecomposition**(*(MatrixX)arg1*) → tuple :
    Compute eigen (spectral) decomposition of symmetric matrix, returns (eigVecs,eigVals).
    eigVecs is orthogonal Matrix3 with columns ar normalized eigenvectors, eigVals is Vector3
    with corresponding eigenvalues. self=eigVecs*diag(eigVals)*eigVecs.transpose().

**spectralDecomposition**(*(MatrixX)arg1*) → tuple :
    Alias for *selfAdjointEigenDecomposition*.

**squaredNorm**(*(MatrixX)arg1*) → float :
    Square of the Euclidean norm.

**sum**(*(MatrixX)arg1*) → float :
    Sum of all elements.

**svd**(*(MatrixX)arg1*) → tuple :
    Alias for `jacobiSVD`.

**trace**(*(MatrixX)arg1*) → float :
    Return sum of diagonal elements.

**transpose**(*(MatrixX)arg1*) → MatrixX :
    Return transposed matrix.

**class MatrixXc**
    */TODO/*

    **static Identity**(*(int)arg1, (int)rank*) → MatrixXc :
        Create identity matrix with given rank (square).

    **static Ones**(*(int)rows, (int)cols*) → MatrixXc :
        Create matrix of given dimensions where all elements are set to 1.

    **static Random**(*(int)rows, (int)cols*) → MatrixXc :
        Create matrix with given dimensions where all elements are set to number between 0 and 1 (uniformly-distributed).

    **static Zero**(*(int)rows, (int)cols*) → MatrixXc :
        Create zero matrix of given dimensions

    **__init__**(*(object)arg1*) → None
        ___init___( (object)arg1, (MatrixXc)other) -> None

        ___init___( (object)arg1, (VectorXc)diag) -> object

        ___init___( (object)arg1 [, (VectorXc)r0=VectorXc() [, (VectorXc)r1=VectorXc() [, (VectorXc)r2=VectorXc() [, (VectorXc)r3=VectorXc() [, (VectorXc)r4=VectorXc() [, (VectorXc)r5=VectorXc() [, (VectorXc)r6=VectorXc() [, (VectorXc)r7=VectorXc() [, (VectorXc)r8=VectorXc() [, (VectorXc)r9=VectorXc() [, (bool)cols=False]]]]]]]]]]]) -> object

        ___init___( (object)arg1, (object)rows [, (bool)cols=False]) -> object

    **col**(*(MatrixXc)arg1, (int)col*) → VectorXc :
        Return column as vector.

    **cols**(*(MatrixXc)arg1*) → int :
        Number of columns.

    **determinant**(*(MatrixXc)arg1*) → complex :
        Return matrix determinant.

    **diagonal**(*(MatrixXc)arg1*) → VectorXc :
        Return diagonal as vector.

    **inverse**(*(MatrixXc)arg1*) → MatrixXc :
        Return inverted matrix.

    **isApprox**(*(MatrixXc)arg1, (MatrixXc)other*[, *(float)prec=1e-12*]) → bool :
        Approximate comparison with precision *prec*.

    **maxAbsCoeff**(*(MatrixXc)arg1*) → float :
        Maximum absolute value over all elements.

    **mean**(*(MatrixXc)arg1*) → complex :
        Mean value over all elements.

    **norm**(*(MatrixXc)arg1*) → float :
        Euclidean norm.

    **normalize**(*(MatrixXc)arg1*) → None :
        Normalize this object in-place.

    **normalized**(*(MatrixXc)arg1*) → MatrixXc :
        Return normalized copy of this object

**prod**(*(MatrixXc)arg1*) → complex :
  Product of all elements.

**pruned**(*(MatrixXc)arg1*[, *(float)absTol=1e-06*]) → MatrixXc :
  Zero all elements which are greater than *absTol*. Negative zeros are not pruned.

**resize**(*(MatrixXc)arg1, (int)rows, (int)cols*) → None :
  Change size of the matrix, keep values of elements which exist in the new matrix

**row**(*(MatrixXc)arg1, (int)row*) → VectorXc :
  Return row as vector.

**rows**(*(MatrixXc)arg1*) → int :
  Number of rows.

**squaredNorm**(*(MatrixXc)arg1*) → float :
  Square of the Euclidean norm.

**sum**(*(MatrixXc)arg1*) → complex :
  Sum of all elements.

**trace**(*(MatrixXc)arg1*) → complex :
  Return sum of diagonal elements.

**transpose**(*(MatrixXc)arg1*) → MatrixXc :
  Return transposed matrix.

**class Quaternion**
  Quaternion representing rotation.

  Supported operations (q is a Quaternion, v is a Vector3): `q*q` (rotation composition), `q*=q`, `q*v` (rotating v by q), `q==q`, `q!=q`.

  Static attributes: `Identity`.

---

**Note:** Quaternion is represented as axis-angle when printed (e.g. `Identity` is `Quaternion((1,0,0),0)`, and can also be constructed from the axis-angle representation. This is however different from the data stored inside, which can be accessed by indices `[0]` (x), `[1]` (y), `[2]` (z), `[3]` (w). To obtain axis-angle programatically, use *Quaternion.toAxisAngle* which returns the tuple.

---

`Identity = Quaternion((1,0,0),0)`

**Rotate**(*(Quaternion)arg1, (Vector3)v*) → Vector3

**__init__**(*(object)arg1*) → None
  ___init___( (object)arg1, (Vector3)axis, (float)angle) -> object

  ___init___( (object)arg1, (float)angle, (Vector3)axis) -> object

  ___init___( (object)arg1, (Vector3)u, (Vector3)v) -> object
  **___init___( (object)arg1, (float)w, (float)x, (float)y, (float)z) -> None :**
  Initialize from coefficients.

---

  **Note:** The order of coefficients is *w, x, y, z*. The [] operator numbers them differently, 0...4 for *x y z w*!

---

  ___init___( (object)arg1, (Matrix3)rotMatrix) -> None

  ___init___( (object)arg1, (Quaternion)other) -> None

**angularDistance**(*(Quaternion)arg1, (Quaternion)arg2*) → float

**conjugate**(*(Quaternion)arg1*) → Quaternion

**inverse**(*(Quaternion)arg1*) → Quaternion

**norm**(*(Quaternion)arg1*) → float

**normalize**(*(Quaternion)arg1*) → None

**normalized**(*(Quaternion)arg1*) → Quaternion

**setFromTwoVectors**(*(Quaternion)arg1, (Vector3)u, (Vector3)v*) → None

**slerp**(*(Quaternion)arg1, (float)t, (Quaternion)other*) → Quaternion

**toAngleAxis**(*(Quaternion)arg1*) → tuple

**toAxisAngle**(*(Quaternion)arg1*) → tuple

**toRotationMatrix**(*(Quaternion)arg1*) → Matrix3

**toRotationVector**(*(Quaternion)arg1*) → Vector3

**class Vector2**
    3-dimensional float vector.

    Supported operations (`f` if a float/int, `v` is a Vector3): `-v`, `v+v`, `v+=v`, `v-v`, `v-=v`, `v*f`, `f*v`, `v*=f`, `v/f`, `v/=f`, `v==v`, `v!=v`.

    Implicit conversion from sequence (list, tuple, …) of 2 floats.

    Static attributes: `Zero`, `Ones`, `UnitX`, `UnitY`.

    **Identity = Vector2(1,0)**

    **Ones = Vector2(1,1)**

    **static Random()** → Vector2 :
        Return an object where all elements are randomly set to values between 0 and 1.

    **static Unit**(*(int)arg1*) → Vector2

    **UnitX = Vector2(1,0)**

    **UnitY = Vector2(0,1)**

    **Zero = Vector2(0,0)**

    **__init__**(*(object)arg1*) → None
        ___init___( (object)arg1, (Vector2)other) -> None

        ___init___( (object)arg1, (float)x, (float)y) -> None

    **asDiagonal**(*(Vector2)arg1*) → object :
        Return diagonal matrix with this vector on the diagonal.

    **cols**(*(Vector2)arg1*) → int :
        Number of columns.

    **dot**(*(Vector2)arg1, (Vector2)other*) → float :
        Dot product with *other*.

    **isApprox**(*(Vector2)arg1, (Vector2)other*[, *(float)prec=1e-12*]) → bool :
        Approximate comparison with precision *prec*.

    **maxAbsCoeff**(*(Vector2)arg1*) → float :
        Maximum absolute value over all elements.

    **maxCoeff**(*(Vector2)arg1*) → float :
        Maximum value over all elements.

    **mean**(*(Vector2)arg1*) → float :
        Mean value over all elements.

    **minCoeff**(*(Vector2)arg1*) → float :
        Minimum value over all elements.

**norm**(*(Vector2)arg1*) → float :
  Euclidean norm.

**normalize**(*(Vector2)arg1*) → None :
  Normalize this object in-place.

**normalized**(*(Vector2)arg1*) → Vector2 :
  Return normalized copy of this object

**outer**(*(Vector2)arg1, (Vector2)other*) → object :
  Outer product with *other*.

**prod**(*(Vector2)arg1*) → float :
  Product of all elements.

**pruned**(*(Vector2)arg1*[, *(float)absTol=1e-06*]) → Vector2 :
  Zero all elements which are greater than *absTol*. Negative zeros are not pruned.

**rows**(*(Vector2)arg1*) → int :
  Number of rows.

**squaredNorm**(*(Vector2)arg1*) → float :
  Square of the Euclidean norm.

**sum**(*(Vector2)arg1*) → float :
  Sum of all elements.

**class Vector2c**
  */TODO/*

  **Identity = Vector2c(1,0)**

  **Ones = Vector2c(1,1)**

  **static Random**() → Vector2c :
    Return an object where all elements are randomly set to values between 0 and 1.

  **static Unit**(*(int)arg1*) → Vector2c

  **UnitX = Vector2c(1,0)**

  **UnitY = Vector2c(0,1)**

  **Zero = Vector2c(0,0)**

  **__init__**(*(object)arg1*) → None :
    __init__( (object)arg1, (Vector2c)other) -> None

    __init__( (object)arg1, (complex)x, (complex)y) -> None

  **asDiagonal**(*(Vector2c)arg1*) → object :
    Return diagonal matrix with this vector on the diagonal.

  **cols**(*(Vector2c)arg1*) → int :
    Number of columns.

  **dot**(*(Vector2c)arg1, (Vector2c)other*) → complex :
    Dot product with *other*.

  **isApprox**(*(Vector2c)arg1, (Vector2c)other*[, *(float)prec=1e-12*]) → bool :
    Approximate comparison with precision *prec*.

  **maxAbsCoeff**(*(Vector2c)arg1*) → float :
    Maximum absolute value over all elements.

  **mean**(*(Vector2c)arg1*) → complex :
    Mean value over all elements.

  **norm**(*(Vector2c)arg1*) → float :
    Euclidean norm.

**normalize**(*(Vector2c)arg1*) → None :
    Normalize this object in-place.

**normalized**(*(Vector2c)arg1*) → Vector2c :
    Return normalized copy of this object

**outer**(*(Vector2c)arg1, (Vector2c)other*) → object :
    Outer product with *other*.

**prod**(*(Vector2c)arg1*) → complex :
    Product of all elements.

**pruned**(*(Vector2c)arg1*[, *(float)absTol=1e-06*]) → Vector2c :
    Zero all elements which are greater than *absTol*. Negative zeros are not pruned.

**rows**(*(Vector2c)arg1*) → int :
    Number of rows.

**squaredNorm**(*(Vector2c)arg1*) → float :
    Square of the Euclidean norm.

**sum**(*(Vector2c)arg1*) → complex :
    Sum of all elements.

**class Vector2i**
    2-dimensional integer vector.

    Supported operations (i if an int, v is a Vector2i): `-v, v+v, v+=v, v-v, v-=v, v*i, i*v, v*=i, v==v, v!=v`.

    Implicit conversion from sequence (list, tuple, ...) of 2 integers.

    Static attributes: `Zero, Ones, UnitX, UnitY`.

    **Identity = Vector2i(1,0)**

    **Ones = Vector2i(1,1)**

    **static Random**() → Vector2i :
        Return an object where all elements are randomly set to values between 0 and 1.

    **static Unit**(*(int)arg1*) → Vector2i

    **UnitX = Vector2i(1,0)**

    **UnitY = Vector2i(0,1)**

    **Zero = Vector2i(0,0)**

    **__init__**(*(object)arg1*) → None
        ___init___( (object)arg1, (Vector2i)other) -> None

        ___init___( (object)arg1, (int)x, (int)y) -> None

    **asDiagonal**(*(Vector2i)arg1*) → object :
        Return diagonal matrix with this vector on the diagonal.

    **cols**(*(Vector2i)arg1*) → int :
        Number of columns.

    **dot**(*(Vector2i)arg1, (Vector2i)other*) → int :
        Dot product with *other*.

    **isApprox**(*(Vector2i)arg1, (Vector2i)other*[, *(int)prec=0*]) → bool :
        Approximate comparison with precision *prec*.

    **maxAbsCoeff**(*(Vector2i)arg1*) → int :
        Maximum absolute value over all elements.

    **maxCoeff**(*(Vector2i)arg1*) → int :
        Maximum value over all elements.

**mean**(*(Vector2i)arg1*) → int :
> Mean value over all elements.

**minCoeff**(*(Vector2i)arg1*) → int :
> Minimum value over all elements.

**outer**(*(Vector2i)arg1, (Vector2i)other*) → object :
> Outer product with *other*.

**prod**(*(Vector2i)arg1*) → int :
> Product of all elements.

**rows**(*(Vector2i)arg1*) → int :
> Number of rows.

**sum**(*(Vector2i)arg1*) → int :
> Sum of all elements.

**class Vector3**
> 3-dimensional float vector.
>
> Supported operations (`f` if a float/int, `v` is a Vector3): `-v, v+v, v+=v, v-v, v-=v, v*f, f*v, v*=f, v/f, v/=f, v==v, v!=v`, plus operations with `Matrix3` and `Quaternion`.
>
> Implicit conversion from sequence (list, tuple, …) of 3 floats.
>
> Static attributes: `Zero, Ones, UnitX, UnitY, UnitZ`.
>
> **Identity = Vector3(1,0,0)**
>
> **Ones = Vector3(1,1,1)**
>
> **static Random**() → Vector3 :
> > Return an object where all elements are randomly set to values between 0 and 1.
>
> **static Unit**(*(int)arg1*) → Vector3
>
> **UnitX = Vector3(1,0,0)**
>
> **UnitY = Vector3(0,1,0)**
>
> **UnitZ = Vector3(0,0,1)**
>
> **Zero = Vector3(0,0,0)**
>
> **__init__**(*(object)arg1*) → None
> > ___init___( (object)arg1, (Vector3)other) -> None
> >
> > ___init___( (object)arg1 [, (float)x=0.0 [, (float)y=0.0 [, (float)z=0.0]]]) -> None
>
> **asDiagonal**(*(Vector3)arg1*) → Matrix3 :
> > Return diagonal matrix with this vector on the diagonal.
>
> **cols**(*(Vector3)arg1*) → int :
> > Number of columns.
>
> **cross**(*(Vector3)arg1, (Vector3)arg2*) → Vector3
>
> **dot**(*(Vector3)arg1, (Vector3)other*) → float :
> > Dot product with *other*.
>
> **isApprox**(*(Vector3)arg1, (Vector3)other*[, *(float)prec=1e-12*]) → bool :
> > Approximate comparison with precision *prec*.
>
> **maxAbsCoeff**(*(Vector3)arg1*) → float :
> > Maximum absolute value over all elements.
>
> **maxCoeff**(*(Vector3)arg1*) → float :
> > Maximum value over all elements.

**mean**(*(Vector3)arg1*) → float :
    Mean value over all elements.

**minCoeff**(*(Vector3)arg1*) → float :
    Minimum value over all elements.

**norm**(*(Vector3)arg1*) → float :
    Euclidean norm.

**normalize**(*(Vector3)arg1*) → None :
    Normalize this object in-place.

**normalized**(*(Vector3)arg1*) → Vector3 :
    Return normalized copy of this object

**outer**(*(Vector3)arg1, (Vector3)other*) → Matrix3 :
    Outer product with *other*.

**prod**(*(Vector3)arg1*) → float :
    Product of all elements.

**pruned**(*(Vector3)arg1*[, *(float)absTol=1e-06*]) → Vector3 :
    Zero all elements which are greater than *absTol*. Negative zeros are not pruned.

**rows**(*(Vector3)arg1*) → int :
    Number of rows.

**squaredNorm**(*(Vector3)arg1*) → float :
    Square of the Euclidean norm.

**sum**(*(Vector3)arg1*) → float :
    Sum of all elements.

**xy**(*(Vector3)arg1*) → Vector2

**xz**(*(Vector3)arg1*) → Vector2

**yx**(*(Vector3)arg1*) → Vector2

**yz**(*(Vector3)arg1*) → Vector2

**zx**(*(Vector3)arg1*) → Vector2

**zy**(*(Vector3)arg1*) → Vector2

**class Vector3c**
    /*TODO*/

**Identity = Vector3c(1,0,0)**

**Ones = Vector3c(1,1,1)**

**static Random**() → Vector3c :
    Return an object where all elements are randomly set to values between 0 and 1.

**static Unit**(*(int)arg1*) → Vector3c

**UnitX = Vector3c(1,0,0)**

**UnitY = Vector3c(0,1,0)**

**UnitZ = Vector3c(0,0,1)**

**Zero = Vector3c(0,0,0)**

**__init__**(*(object)arg1*) → None
    \_\_init\_\_( (object)arg1, (Vector3c)other) -> None

    \_\_init\_\_( (object)arg1 [, (complex)x=0j [, (complex)y=0j [, (complex)z=0j]]]) -> None

**asDiagonal**(*(Vector3c)arg1*) → Matrix3c :
    Return diagonal matrix with this vector on the diagonal.

> **cols**(*(Vector3c)arg1*) → int :
>> Number of columns.

> **cross**(*(Vector3c)arg1*, *(Vector3c)arg2*) → Vector3c

> **dot**(*(Vector3c)arg1*, *(Vector3c)other*) → complex :
>> Dot product with *other*.

> **isApprox**(*(Vector3c)arg1*, *(Vector3c)other*$\left[$, *(float)prec=1e-12*$\right]$) → bool :
>> Approximate comparison with precision *prec*.

> **maxAbsCoeff**(*(Vector3c)arg1*) → float :
>> Maximum absolute value over all elements.

> **mean**(*(Vector3c)arg1*) → complex :
>> Mean value over all elements.

> **norm**(*(Vector3c)arg1*) → float :
>> Euclidean norm.

> **normalize**(*(Vector3c)arg1*) → None :
>> Normalize this object in-place.

> **normalized**(*(Vector3c)arg1*) → Vector3c :
>> Return normalized copy of this object

> **outer**(*(Vector3c)arg1*, *(Vector3c)other*) → Matrix3c :
>> Outer product with *other*.

> **prod**(*(Vector3c)arg1*) → complex :
>> Product of all elements.

> **pruned**(*(Vector3c)arg1*$\left[$, *(float)absTol=1e-06*$\right]$) → Vector3c :
>> Zero all elements which are greater than *absTol*. Negative zeros are not pruned.

> **rows**(*(Vector3c)arg1*) → int :
>> Number of rows.

> **squaredNorm**(*(Vector3c)arg1*) → float :
>> Square of the Euclidean norm.

> **sum**(*(Vector3c)arg1*) → complex :
>> Sum of all elements.

> **xy**(*(Vector3c)arg1*) → Vector2c

> **xz**(*(Vector3c)arg1*) → Vector2c

> **yx**(*(Vector3c)arg1*) → Vector2c

> **yz**(*(Vector3c)arg1*) → Vector2c

> **zx**(*(Vector3c)arg1*) → Vector2c

> **zy**(*(Vector3c)arg1*) → Vector2c

**class Vector3i**
> 3-dimensional integer vector.

> Supported operations (i if an int, v is a Vector3i): `-v`, `v+v`, `v+=v`, `v-v`, `v-=v`, `v*i`, `i*v`, `v*=i`, `v==v`, `v!=v`.

> Implicit conversion from sequence (list, tuple, …) of 3 integers.

> Static attributes: `Zero`, `Ones`, `UnitX`, `UnitY`, `UnitZ`.

> **Identity = Vector3i(1,0,0)**

> **Ones = Vector3i(1,1,1)**

**static Random()** → Vector3i :
   Return an object where all elements are randomly set to values between 0 and 1.

**static Unit(***(int)arg1***)** → Vector3i

**UnitX = Vector3i(1,0,0)**

**UnitY = Vector3i(0,1,0)**

**UnitZ = Vector3i(0,0,1)**

**Zero = Vector3i(0,0,0)**

**__init__(***(object)arg1***)** → None
   ___init___( (object)arg1, (Vector3i)other) -> None

   ___init___( (object)arg1 [, (int)x=0 [, (int)y=0 [, (int)z=0]]]) -> None

**asDiagonal(***(Vector3i)arg1***)** → object :
   Return diagonal matrix with this vector on the diagonal.

**cols(***(Vector3i)arg1***)** → int :
   Number of columns.

**cross(***(Vector3i)arg1, (Vector3i)arg2***)** → Vector3i

**dot(***(Vector3i)arg1, (Vector3i)other***)** → int :
   Dot product with *other*.

**isApprox(***(Vector3i)arg1, (Vector3i)other*$\left[\right.$*, (int)prec=0*$\left.\right]$***)** → bool :
   Approximate comparison with precision *prec*.

**maxAbsCoeff(***(Vector3i)arg1***)** → int :
   Maximum absolute value over all elements.

**maxCoeff(***(Vector3i)arg1***)** → int :
   Maximum value over all elements.

**mean(***(Vector3i)arg1***)** → int :
   Mean value over all elements.

**minCoeff(***(Vector3i)arg1***)** → int :
   Minimum value over all elements.

**outer(***(Vector3i)arg1, (Vector3i)other***)** → object :
   Outer product with *other*.

**prod(***(Vector3i)arg1***)** → int :
   Product of all elements.

**rows(***(Vector3i)arg1***)** → int :
   Number of rows.

**sum(***(Vector3i)arg1***)** → int :
   Sum of all elements.

**xy(***(Vector3i)arg1***)** → Vector2i

**xz(***(Vector3i)arg1***)** → Vector2i

**yx(***(Vector3i)arg1***)** → Vector2i

**yz(***(Vector3i)arg1***)** → Vector2i

**zx(***(Vector3i)arg1***)** → Vector2i

**zy(***(Vector3i)arg1***)** → Vector2i

**class Vector4**
   4-dimensional float vector.

Supported operations (`f` if a float/int, `v` is a Vector3): `-v, v+v, v+=v, v-v, v-=v, v*f, f*v, v*=f, v/f, v/=f, v==v, v!=v`.

Implicit conversion from sequence (list, tuple, …) of 4 floats.

Static attributes: `Zero`, `Ones`.

**`Identity = Vector4(1,0,0, 0)`**

**`Ones = Vector4(1,1,1, 1)`**

**`static Random()`** → Vector4 :
    Return an object where all elements are randomly set to values between 0 and 1.

**`static Unit`**(*(int)arg1*) → Vector4

**`Zero = Vector4(0,0,0, 0)`**

**`__init__`**(*(object)arg1*) → None
    ___init___( (object)arg1, (Vector4)other) -> None

    ___init___( (object)arg1, (float)v0, (float)v1, (float)v2, (float)v3) -> None

**`asDiagonal`**(*(Vector4)arg1*) → object :
    Return diagonal matrix with this vector on the diagonal.

**`cols`**(*(Vector4)arg1*) → int :
    Number of columns.

**`dot`**(*(Vector4)arg1, (Vector4)other*) → float :
    Dot product with *other*.

**`isApprox`**(*(Vector4)arg1, (Vector4)other*[, *(float)prec=1e-12*]) → bool :
    Approximate comparison with precision *prec*.

**`maxAbsCoeff`**(*(Vector4)arg1*) → float :
    Maximum absolute value over all elements.

**`maxCoeff`**(*(Vector4)arg1*) → float :
    Maximum value over all elements.

**`mean`**(*(Vector4)arg1*) → float :
    Mean value over all elements.

**`minCoeff`**(*(Vector4)arg1*) → float :
    Minimum value over all elements.

**`norm`**(*(Vector4)arg1*) → float :
    Euclidean norm.

**`normalize`**(*(Vector4)arg1*) → None :
    Normalize this object in-place.

**`normalized`**(*(Vector4)arg1*) → Vector4 :
    Return normalized copy of this object

**`outer`**(*(Vector4)arg1, (Vector4)other*) → object :
    Outer product with *other*.

**`prod`**(*(Vector4)arg1*) → float :
    Product of all elements.

**`pruned`**(*(Vector4)arg1*[, *(float)absTol=1e-06*]) → Vector4 :
    Zero all elements which are greater than *absTol*. Negative zeros are not pruned.

**`rows`**(*(Vector4)arg1*) → int :
    Number of rows.

**`squaredNorm`**(*(Vector4)arg1*) → float :
    Square of the Euclidean norm.

**sum**(*(Vector4)arg1*) → float :
    Sum of all elements.

**class Vector6**
    6-dimensional float vector.

    Supported operations (`f` if a float/int, `v` is a Vector6): `-v, v+v, v+=v, v-v, v-=v, v*f, f*v, v*=f, v/f, v/=f, v==v, v!=v`.

    Implicit conversion from sequence (list, tuple, …) of 6 floats.

    Static attributes: `Zero`, `Ones`.

    **Identity = Vector6(1,0,0, 0,0,0)**

    **Ones = Vector6(1,1,1, 1,1,1)**

    **static Random**() → Vector6 :
        Return an object where all elements are randomly set to values between 0 and 1.

    **static Unit**(*(int)arg1*) → Vector6

    **Zero = Vector6(0,0,0, 0,0,0)**

    **__init__**(*(object)arg1*) → None
        ___init___( (object)arg1, (Vector6)other) -> None

        ___init___( (object)arg1, (float)v0, (float)v1, (float)v2, (float)v3, (float)v4, (float)v5) -> object

        ___init___( (object)arg1, (Vector3)head, (Vector3)tail) -> object

    **asDiagonal**(*(Vector6)arg1*) → Matrix6 :
        Return diagonal matrix with this vector on the diagonal.

    **cols**(*(Vector6)arg1*) → int :
        Number of columns.

    **dot**(*(Vector6)arg1, (Vector6)other*) → float :
        Dot product with *other*.

    **head**(*(Vector6)arg1*) → Vector3

    **isApprox**(*(Vector6)arg1, (Vector6)other*[, *(float)prec=1e-12*]) → bool :
        Approximate comparison with precision *prec*.

    **maxAbsCoeff**(*(Vector6)arg1*) → float :
        Maximum absolute value over all elements.

    **maxCoeff**(*(Vector6)arg1*) → float :
        Maximum value over all elements.

    **mean**(*(Vector6)arg1*) → float :
        Mean value over all elements.

    **minCoeff**(*(Vector6)arg1*) → float :
        Minimum value over all elements.

    **norm**(*(Vector6)arg1*) → float :
        Euclidean norm.

    **normalize**(*(Vector6)arg1*) → None :
        Normalize this object in-place.

    **normalized**(*(Vector6)arg1*) → Vector6 :
        Return normalized copy of this object

    **outer**(*(Vector6)arg1, (Vector6)other*) → Matrix6 :
        Outer product with *other*.

**prod**(*(Vector6)arg1*) → float :
    Product of all elements.

**pruned**(*(Vector6)arg1*[, *(float)absTol=1e-06*]) → Vector6 :
    Zero all elements which are greater than *absTol*. Negative zeros are not pruned.

**rows**(*(Vector6)arg1*) → int :
    Number of rows.

**squaredNorm**(*(Vector6)arg1*) → float :
    Square of the Euclidean norm.

**sum**(*(Vector6)arg1*) → float :
    Sum of all elements.

**tail**(*(Vector6)arg1*) → Vector3

**class Vector6c**
    */TODO/*

**Identity = Vector6c(1,0,0, 0,0,0)**

**Ones = Vector6c(1,1,1, 1,1,1)**

**static Random**() → Vector6c :
    Return an object where all elements are randomly set to values between 0 and 1.

**static Unit**(*(int)arg1*) → Vector6c

**Zero = Vector6c(0,0,0, 0,0,0)**

**__init__**(*(object)arg1*) → None
    ___init___( (object)arg1, (Vector6c)other) -> None

    ___init___( (object)arg1, (complex)v0, (complex)v1, (complex)v2, (complex)v3, (complex)v4, (complex)v5) -> object

    ___init___( (object)arg1, (Vector3c)head, (Vector3c)tail) -> object

**asDiagonal**(*(Vector6c)arg1*) → Matrix6c :
    Return diagonal matrix with this vector on the diagonal.

**cols**(*(Vector6c)arg1*) → int :
    Number of columns.

**dot**(*(Vector6c)arg1, (Vector6c)other*) → complex :
    Dot product with *other*.

**head**(*(Vector6c)arg1*) → Vector3c

**isApprox**(*(Vector6c)arg1, (Vector6c)other*[, *(float)prec=1e-12*]) → bool :
    Approximate comparison with precision *prec*.

**maxAbsCoeff**(*(Vector6c)arg1*) → float :
    Maximum absolute value over all elements.

**mean**(*(Vector6c)arg1*) → complex :
    Mean value over all elements.

**norm**(*(Vector6c)arg1*) → float :
    Euclidean norm.

**normalize**(*(Vector6c)arg1*) → None :
    Normalize this object in-place.

**normalized**(*(Vector6c)arg1*) → Vector6c :
    Return normalized copy of this object

**outer**(*(Vector6c)arg1, (Vector6c)other*) → Matrix6c :
    Outer product with *other*.

**prod**(*(Vector6c)arg1*) → complex :
> Product of all elements.

**pruned**(*(Vector6c)arg1*[, *(float)absTol=1e-06*]) → Vector6c :
> Zero all elements which are greater than *absTol*. Negative zeros are not pruned.

**rows**(*(Vector6c)arg1*) → int :
> Number of rows.

**squaredNorm**(*(Vector6c)arg1*) → float :
> Square of the Euclidean norm.

**sum**(*(Vector6c)arg1*) → complex :
> Sum of all elements.

**tail**(*(Vector6c)arg1*) → Vector3c

**class Vector6i**
> 6-dimensional float vector.

> Supported operations (f if a float/int, v is a Vector6): -v, v+v, v+=v, v-v, v-=v, v*f, f*v, v*=f, v/f, v/=f, v==v, v!=v.

> Implicit conversion from sequence (list, tuple, ...) of 6 ints.

> Static attributes: Zero, Ones.

> **Identity = Vector6i(1,0,0, 0,0,0)**

> **Ones = Vector6i(1,1,1, 1,1,1)**

> **static Random**() → Vector6i :
> > Return an object where all elements are randomly set to values between 0 and 1.

> **static Unit**(*(int)arg1*) → Vector6i

> **Zero = Vector6i(0,0,0, 0,0,0)**

> **__init__**(*(object)arg1*) → None
> > ___init___( (object)arg1, (Vector6i)other) -> None

> > ___init___( (object)arg1, (int)v0, (int)v1, (int)v2, (int)v3, (int)v4, (int)v5) -> object

> > ___init___( (object)arg1, (Vector3i)head, (Vector3i)tail) -> object

> **asDiagonal**(*(Vector6i)arg1*) → object :
> > Return diagonal matrix with this vector on the diagonal.

> **cols**(*(Vector6i)arg1*) → int :
> > Number of columns.

> **dot**(*(Vector6i)arg1, (Vector6i)other*) → int :
> > Dot product with *other*.

> **head**(*(Vector6i)arg1*) → Vector3i

> **isApprox**(*(Vector6i)arg1, (Vector6i)other*[, *(int)prec=0*]) → bool :
> > Approximate comparison with precision *prec*.

> **maxAbsCoeff**(*(Vector6i)arg1*) → int :
> > Maximum absolute value over all elements.

> **maxCoeff**(*(Vector6i)arg1*) → int :
> > Maximum value over all elements.

> **mean**(*(Vector6i)arg1*) → int :
> > Mean value over all elements.

> **minCoeff**(*(Vector6i)arg1*) → int :
> > Minimum value over all elements.

**outer**(*(Vector6i)arg1, (Vector6i)other*) → object :
    Outer product with *other*.

**prod**(*(Vector6i)arg1*) → int :
    Product of all elements.

**rows**(*(Vector6i)arg1*) → int :
    Number of rows.

**sum**(*(Vector6i)arg1*) → int :
    Sum of all elements.

**tail**(*(Vector6i)arg1*) → Vector3i

**class VectorX**
    Dynamic-sized float vector.

    Supported operations (**f** if a float/int, **v** is a VectorX): `-v, v+v, v+=v, v-v, v-=v, v*f, f*v, v*=f, v/f, v/=f, v==v, v!=v.`

    Implicit conversion from sequence (list, tuple, …) of X floats.

**static Ones**(*(int)arg1*) → VectorX

**static Random**(*(int)len*) → VectorX :
    Return vector of given length with all elements set to values between 0 and 1 randomly.

**static Unit**(*(int)arg1, (int)arg2*) → VectorX

**static Zero**(*(int)arg1*) → VectorX

**__init__**(*(object)arg1*) → None
    ___init___( (object)arg1, (VectorX)other) -> None

    ___init___( (object)arg1, (object)vv) -> object

**asDiagonal**(*(VectorX)arg1*) → MatrixX :
    Return diagonal matrix with this vector on the diagonal.

**cols**(*(VectorX)arg1*) → int :
    Number of columns.

**dot**(*(VectorX)arg1, (VectorX)other*) → float :
    Dot product with *other*.

**isApprox**(*(VectorX)arg1, (VectorX)other*[, *(float)prec=1e-12*]) → bool :
    Approximate comparison with precision *prec*.

**maxAbsCoeff**(*(VectorX)arg1*) → float :
    Maximum absolute value over all elements.

**maxCoeff**(*(VectorX)arg1*) → float :
    Maximum value over all elements.

**mean**(*(VectorX)arg1*) → float :
    Mean value over all elements.

**minCoeff**(*(VectorX)arg1*) → float :
    Minimum value over all elements.

**norm**(*(VectorX)arg1*) → float :
    Euclidean norm.

**normalize**(*(VectorX)arg1*) → None :
    Normalize this object in-place.

**normalized**(*(VectorX)arg1*) → VectorX :
    Return normalized copy of this object

**outer**(*(VectorX)arg1, (VectorX)other*) → MatrixX :
  Outer product with *other*.

**prod**(*(VectorX)arg1*) → float :
  Product of all elements.

**pruned**(*(VectorX)arg1*[, *(float)absTol=1e-06*]) → VectorX :
  Zero all elements which are greater than *absTol*. Negative zeros are not pruned.

**resize**(*(VectorX)arg1, (int)arg2*) → None

**rows**(*(VectorX)arg1*) → int :
  Number of rows.

**squaredNorm**(*(VectorX)arg1*) → float :
  Square of the Euclidean norm.

**sum**(*(VectorX)arg1*) → float :
  Sum of all elements.

**class VectorXc**
  */TODO/*

  **static Ones**(*(int)arg1*) → VectorXc

  **static Random**(*(int)len*) → VectorXc :
    Return vector of given length with all elements set to values between 0 and 1 randomly.

  **static Unit**(*(int)arg1, (int)arg2*) → VectorXc

  **static Zero**(*(int)arg1*) → VectorXc

  **__init__**(*(object)arg1*) → None
    ___init___( (object)arg1, (VectorXc)other) -> None

    ___init___( (object)arg1, (object)vv) -> object

  **asDiagonal**(*(VectorXc)arg1*) → MatrixXc :
    Return diagonal matrix with this vector on the diagonal.

  **cols**(*(VectorXc)arg1*) → int :
    Number of columns.

  **dot**(*(VectorXc)arg1, (VectorXc)other*) → complex :
    Dot product with *other*.

  **isApprox**(*(VectorXc)arg1, (VectorXc)other*[, *(float)prec=1e-12*]) → bool :
    Approximate comparison with precision *prec*.

  **maxAbsCoeff**(*(VectorXc)arg1*) → float :
    Maximum absolute value over all elements.

  **mean**(*(VectorXc)arg1*) → complex :
    Mean value over all elements.

  **norm**(*(VectorXc)arg1*) → float :
    Euclidean norm.

  **normalize**(*(VectorXc)arg1*) → None :
    Normalize this object in-place.

  **normalized**(*(VectorXc)arg1*) → VectorXc :
    Return normalized copy of this object

  **outer**(*(VectorXc)arg1, (VectorXc)other*) → MatrixXc :
    Outer product with *other*.

  **prod**(*(VectorXc)arg1*) → complex :
    Product of all elements.

> **pruned**(*(VectorXc)arg1*[, *(float)absTol=1e-06*]) → VectorXc :
>> Zero all elements which are greater than *absTol*. Negative zeros are not pruned.

> **resize**(*(VectorXc)arg1, (int)arg2*) → None

> **rows**(*(VectorXc)arg1*) → int :
>> Number of rows.

> **squaredNorm**(*(VectorXc)arg1*) → float :
>> Square of the Euclidean norm.

> **sum**(*(VectorXc)arg1*) → complex :
>> Sum of all elements.

**vectorize = False**

**class yade._minieigenHP.Matrix3**
> 3x3 float matrix.

> Supported operations (m is a Matrix3, f if a float/int, v is a Vector3): `-m, m+m, m+=m, m-m, m-=m, m*f, f*m, m*=f, m/f, m/=f, m*m, m*=m, m*v, v*m, m==m, m!=m`.

> Static attributes: `Zero, Ones, Identity`.

> **Identity = Matrix3(1,0,0, 0,1,0, 0,0,1)**

> **Ones = Matrix3(1,1,1, 1,1,1, 1,1,1)**

> **static Random()** → Matrix3 :
>> Return an object where all elements are randomly set to values between 0 and 1.

> **Zero = Matrix3(0,0,0, 0,0,0, 0,0,0)**

> **__init__**(*(object)arg1*) → None
>> ___init___( (object)arg1, (Quaternion)q) -> None

>> ___init___( (object)arg1, (Matrix3)other) -> None

>> ___init___( (object)arg1, (Vector3)diag) -> object

>> ___init___( (object)arg1, (float)m00, (float)m01, (float)m02, (float)m10, (float)m11, (float)m12, (float)m20, (float)m21, (float)m22) -> object

>> ___init___( (object)arg1, (Vector3)r0, (Vector3)r1, (Vector3)r2 [, (bool)cols=False]) -> object

> **col**(*(Matrix3)arg1, (int)col*) → Vector3 :
>> Return column as vector.

> **cols**(*(Matrix3)arg1*) → int :
>> Number of columns.

> **computeUnitaryPositive**(*(Matrix3)arg1*) → tuple :
>> Compute polar decomposition (unitary matrix U and positive semi-definite symmetric matrix P such that self=U*P).

> **determinant**(*(Matrix3)arg1*) → float :
>> Return matrix determinant.

> **diagonal**(*(Matrix3)arg1*) → Vector3 :
>> Return diagonal as vector.

> **inverse**(*(Matrix3)arg1*) → Matrix3 :
>> Return inverted matrix.

> **isApprox**(*(Matrix3)arg1, (Matrix3)other*[, *(float)prec=1e-12*]) → bool :
>> Approximate comparison with precision *prec*.

> **jacobiSVD**(*(Matrix3)arg1*) → tuple :
>> Compute SVD decomposition of square matrix, retuns (U,S,V) such that self=U*S*V.transpose()

**maxAbsCoeff**(*(Matrix3)arg1*) → float :
    Maximum absolute value over all elements.

**maxCoeff**(*(Matrix3)arg1*) → float :
    Maximum value over all elements.

**mean**(*(Matrix3)arg1*) → float :
    Mean value over all elements.

**minCoeff**(*(Matrix3)arg1*) → float :
    Minimum value over all elements.

**norm**(*(Matrix3)arg1*) → float :
    Euclidean norm.

**normalize**(*(Matrix3)arg1*) → None :
    Normalize this object in-place.

**normalized**(*(Matrix3)arg1*) → Matrix3 :
    Return normalized copy of this object

**polarDecomposition**(*(Matrix3)arg1*) → tuple :
    Alias for *computeUnitaryPositive*.

**prod**(*(Matrix3)arg1*) → float :
    Product of all elements.

**pruned**(*(Matrix3)arg1*[, *(float)absTol=1e-06*]) → Matrix3 :
    Zero all elements which are greater than *absTol*. Negative zeros are not pruned.

**row**(*(Matrix3)arg1, (int)row*) → Vector3 :
    Return row as vector.

**rows**(*(Matrix3)arg1*) → int :
    Number of rows.

**selfAdjointEigenDecomposition**(*(Matrix3)arg1*) → tuple :
    Compute eigen (spectral) decomposition of symmetric matrix, returns (eigVecs,eigVals). eigVecs is orthogonal Matrix3 with columns ar normalized eigenvectors, eigVals is Vector3 with corresponding eigenvalues. self=eigVecs*diag(eigVals)*eigVecs.transpose().

**spectralDecomposition**(*(Matrix3)arg1*) → tuple :
    Alias for *selfAdjointEigenDecomposition*.

**squaredNorm**(*(Matrix3)arg1*) → float :
    Square of the Euclidean norm.

**sum**(*(Matrix3)arg1*) → float :
    Sum of all elements.

**svd**(*(Matrix3)arg1*) → tuple :
    Alias for *jacobiSVD*.

**trace**(*(Matrix3)arg1*) → float :
    Return sum of diagonal elements.

**transpose**(*(Matrix3)arg1*) → Matrix3 :
    Return transposed matrix.

**class yade._minieigenHP.Matrix3c**
    */TODO/*

    **Identity = Matrix3c(1,0,0, 0,1,0, 0,0,1)**

    **Ones = Matrix3c(1,1,1, 1,1,1, 1,1,1)**

    **static Random**() → Matrix3c :
        Return an object where all elements are randomly set to values between 0 and 1.

    **Zero = Matrix3c(0,0,0, 0,0,0, 0,0,0)**

**__init__**(*(object)arg1*) → None
    ___init___( (object)arg1, (Matrix3c)other) -> None

    ___init___( (object)arg1, (Vector3c)diag) -> object

    ___init___( (object)arg1, (complex)m00, (complex)m01, (complex)m02, (complex)m10, (complex)m11, (complex)m12, (complex)m20, (complex)m21, (complex)m22) -> object

    ___init___( (object)arg1, (Vector3c)r0, (Vector3c)r1, (Vector3c)r2 [, (bool)cols=False]) -> object

**col**(*(Matrix3c)arg1, (int)col*) → Vector3c :
    Return column as vector.

**cols**(*(Matrix3c)arg1*) → int :
    Number of columns.

**determinant**(*(Matrix3c)arg1*) → complex :
    Return matrix determinant.

**diagonal**(*(Matrix3c)arg1*) → Vector3c :
    Return diagonal as vector.

**inverse**(*(Matrix3c)arg1*) → Matrix3c :
    Return inverted matrix.

**isApprox**(*(Matrix3c)arg1, (Matrix3c)other*[, *(float)prec=1e-12*]) → bool :
    Approximate comparison with precision *prec*.

**maxAbsCoeff**(*(Matrix3c)arg1*) → float :
    Maximum absolute value over all elements.

**mean**(*(Matrix3c)arg1*) → complex :
    Mean value over all elements.

**norm**(*(Matrix3c)arg1*) → float :
    Euclidean norm.

**normalize**(*(Matrix3c)arg1*) → None :
    Normalize this object in-place.

**normalized**(*(Matrix3c)arg1*) → Matrix3c :
    Return normalized copy of this object

**prod**(*(Matrix3c)arg1*) → complex :
    Product of all elements.

**pruned**(*(Matrix3c)arg1*[, *(float)absTol=1e-06*]) → Matrix3c :
    Zero all elements which are greater than *absTol*. Negative zeros are not pruned.

**row**(*(Matrix3c)arg1, (int)row*) → Vector3c :
    Return row as vector.

**rows**(*(Matrix3c)arg1*) → int :
    Number of rows.

**squaredNorm**(*(Matrix3c)arg1*) → float :
    Square of the Euclidean norm.

**sum**(*(Matrix3c)arg1*) → complex :
    Sum of all elements.

**trace**(*(Matrix3c)arg1*) → complex :
    Return sum of diagonal elements.

**transpose**(*(Matrix3c)arg1*) → Matrix3c :
    Return transposed matrix.

```
class yade._minieigenHP.Matrix6
```
   6x6 float matrix. Constructed from 4 3x3 sub-matrices, from 6xVector6 (rows).

   Supported operations (`m` is a Matrix6, `f` if a float/int, `v` is a Vector6): `-m, m+m, m+=m, m-m, m-=m, m*f, f*m, m*=f, m/f, m/=f, m*m, m*=m, m*v, v*m, m==m, m!=m`.

   Static attributes: `Zero, Ones, Identity`.

   **Identity = Matrix6( (1,0,0,0,0,0), (0,1,0,0,0,0), (0,0,1,0,0,0), (0,0,0,1,0,0), (0,0,0,0,1,0),**

   **Ones = Matrix6( (1,1,1,1,1,1), (1,1,1,1,1,1), (1,1,1,1,1,1), (1,1,1,1,1,1), (1,1**

   **static Random()** → Matrix6 :
       Return an object where all elements are randomly set to values between 0 and 1.

   **Zero = Matrix6( (0,0,0,0,0,0), (0,0,0,0,0,0), (0,0,0,0,0,0), (0,0,0,0,0,0), (0,0**

   **__init__**(*(object)arg1*) → None
       ___init___( (object)arg1, (Matrix6)other) -> None

       ___init___( (object)arg1, (Vector6)diag) -> object

       ___init___( (object)arg1, (Matrix3)ul, (Matrix3)ur, (Matrix3)ll, (Matrix3)lr) -> object

       ___init___( (object)arg1, (Vector6)l0, (Vector6)l1, (Vector6)l2, (Vector6)l3, (Vector6)l4, (Vector6)l5 [, (bool)cols=False]) -> object

   **col**(*(Matrix6)arg1, (int)col*) → Vector6 :
       Return column as vector.

   **cols**(*(Matrix6)arg1*) → int :
       Number of columns.

   **computeUnitaryPositive**(*(Matrix6)arg1*) → tuple :
       Compute polar decomposition (unitary matrix U and positive semi-definite symmetric matrix P such that self=U*P).

   **determinant**(*(Matrix6)arg1*) → float :
       Return matrix determinant.

   **diagonal**(*(Matrix6)arg1*) → Vector6 :
       Return diagonal as vector.

   **inverse**(*(Matrix6)arg1*) → Matrix6 :
       Return inverted matrix.

   **isApprox**(*(Matrix6)arg1, (Matrix6)other*[, *(float)prec=1e-12*]) → bool :
       Approximate comparison with precision *prec*.

   **jacobiSVD**(*(Matrix6)arg1*) → tuple :
       Compute SVD decomposition of square matrix, retuns (U,S,V) such that self=U*S*V.transpose()

   **ll**(*(Matrix6)arg1*) → Matrix3 :
       Return lower-left 3x3 block

   **lr**(*(Matrix6)arg1*) → Matrix3 :
       Return lower-right 3x3 block

   **maxAbsCoeff**(*(Matrix6)arg1*) → float :
       Maximum absolute value over all elements.

   **maxCoeff**(*(Matrix6)arg1*) → float :
       Maximum value over all elements.

   **mean**(*(Matrix6)arg1*) → float :
       Mean value over all elements.

   **minCoeff**(*(Matrix6)arg1*) → float :
       Minimum value over all elements.

**norm**(*(Matrix6)arg1*) → float :
    Euclidean norm.

**normalize**(*(Matrix6)arg1*) → None :
    Normalize this object in-place.

**normalized**(*(Matrix6)arg1*) → Matrix6 :
    Return normalized copy of this object

**polarDecomposition**(*(Matrix6)arg1*) → tuple :
    Alias for *computeUnitaryPositive*.

**prod**(*(Matrix6)arg1*) → float :
    Product of all elements.

**pruned**(*(Matrix6)arg1*[, *(float)absTol=1e-06*]) → Matrix6 :
    Zero all elements which are greater than *absTol*. Negative zeros are not pruned.

**row**(*(Matrix6)arg1, (int)row*) → Vector6 :
    Return row as vector.

**rows**(*(Matrix6)arg1*) → int :
    Number of rows.

**selfAdjointEigenDecomposition**(*(Matrix6)arg1*) → tuple :
    Compute eigen (spectral) decomposition of symmetric matrix, returns (eigVecs,eigVals).
    eigVecs is orthogonal Matrix3 with columns ar normalized eigenvectors, eigVals is Vector3
    with corresponding eigenvalues. self=eigVecs*diag(eigVals)*eigVecs.transpose().

**spectralDecomposition**(*(Matrix6)arg1*) → tuple :
    Alias for *selfAdjointEigenDecomposition*.

**squaredNorm**(*(Matrix6)arg1*) → float :
    Square of the Euclidean norm.

**sum**(*(Matrix6)arg1*) → float :
    Sum of all elements.

**svd**(*(Matrix6)arg1*) → tuple :
    Alias for *jacobiSVD*.

**trace**(*(Matrix6)arg1*) → float :
    Return sum of diagonal elements.

**transpose**(*(Matrix6)arg1*) → Matrix6 :
    Return transposed matrix.

**ul**(*(Matrix6)arg1*) → Matrix3 :
    Return upper-left 3x3 block

**ur**(*(Matrix6)arg1*) → Matrix3 :
    Return upper-right 3x3 block

**class yade._minieigenHP.Matrix6c**
    */TODO/*

    **Identity** = Matrix6c( (1,0,0,0,0,0), (0,1,0,0,0,0), (0,0,1,0,0,0), (0,0,0,1,0,0), (0,0,0,0,1,0),

    **Ones** = Matrix6c( (1,1,1,1,1,1), (1,1,1,1,1,1), (1,1,1,1,1,1), (1,1,1,1,1,1), (1,

    **static Random**() → Matrix6c :
        Return an object where all elements are randomly set to values between 0 and 1.

    **Zero** = Matrix6c( (0,0,0,0,0,0), (0,0,0,0,0,0), (0,0,0,0,0,0), (0,0,0,0,0,0), (0,0

    **__init__**(*(object)arg1*) → None
        ___init___( (object)arg1, (Matrix6c)other) -> None

        ___init___( (object)arg1, (Vector6c)diag) -> object

___init___( (object)arg1, (Matrix3c)ul, (Matrix3c)ur, (Matrix3c)ll, (Matrix3c)lr) -> object

___init___( (object)arg1, (Vector6c)l0, (Vector6c)l1, (Vector6c)l2, (Vector6c)l3, (Vector6c)l4, (Vector6c)l5 [, (bool)cols=False]) -> object

**col**(*(Matrix6c)arg1, (int)col*) → Vector6c :
    Return column as vector.

**cols**(*(Matrix6c)arg1*) → int :
    Number of columns.

**determinant**(*(Matrix6c)arg1*) → complex :
    Return matrix determinant.

**diagonal**(*(Matrix6c)arg1*) → Vector6c :
    Return diagonal as vector.

**inverse**(*(Matrix6c)arg1*) → Matrix6c :
    Return inverted matrix.

**isApprox**(*(Matrix6c)arg1, (Matrix6c)other*[, *(float)prec=1e-12*]) → bool :
    Approximate comparison with precision *prec*.

**ll**(*(Matrix6c)arg1*) → Matrix3c :
    Return lower-left 3x3 block

**lr**(*(Matrix6c)arg1*) → Matrix3c :
    Return lower-right 3x3 block

**maxAbsCoeff**(*(Matrix6c)arg1*) → float :
    Maximum absolute value over all elements.

**mean**(*(Matrix6c)arg1*) → complex :
    Mean value over all elements.

**norm**(*(Matrix6c)arg1*) → float :
    Euclidean norm.

**normalize**(*(Matrix6c)arg1*) → None :
    Normalize this object in-place.

**normalized**(*(Matrix6c)arg1*) → Matrix6c :
    Return normalized copy of this object

**prod**(*(Matrix6c)arg1*) → complex :
    Product of all elements.

**pruned**(*(Matrix6c)arg1*[, *(float)absTol=1e-06*]) → Matrix6c :
    Zero all elements which are greater than *absTol*. Negative zeros are not pruned.

**row**(*(Matrix6c)arg1, (int)row*) → Vector6c :
    Return row as vector.

**rows**(*(Matrix6c)arg1*) → int :
    Number of rows.

**squaredNorm**(*(Matrix6c)arg1*) → float :
    Square of the Euclidean norm.

**sum**(*(Matrix6c)arg1*) → complex :
    Sum of all elements.

**trace**(*(Matrix6c)arg1*) → complex :
    Return sum of diagonal elements.

**transpose**(*(Matrix6c)arg1*) → Matrix6c :
    Return transposed matrix.

**ul**(*(Matrix6c)arg1*) → Matrix3c :
    Return upper-left 3x3 block

**ur**(*(Matrix6c)arg1*) → Matrix3c :
    Return upper-right 3x3 block

**class yade._minieigenHP.MatrixX**

    XxX (dynamic-sized) float matrix. Constructed from list of rows (as VectorX).

    Supported operations (m is a MatrixX, f if a float/int, v is a VectorX): -m, m+m, m+=m, m-m, m-=m, m*f, f*m, m*=f, m/f, m/=f, m*m, m*=m, m*v, v*m, m==m, m!=m.

**static Identity**(*(int)arg1, (int)rank*) → MatrixX :
    Create identity matrix with given rank (square).

**static Ones**(*(int)rows, (int)cols*) → MatrixX :
    Create matrix of given dimensions where all elements are set to 1.

**static Random**(*(int)rows, (int)cols*) → MatrixX :
    Create matrix with given dimensions where all elements are set to number between 0 and 1 (uniformly-distributed).

**static Zero**(*(int)rows, (int)cols*) → MatrixX :
    Create zero matrix of given dimensions

**__init__**(*(object)arg1*) → None

    ___init___( (object)arg1, (MatrixX)other) -> None

    ___init___( (object)arg1, (VectorX)diag) -> object

    ___init___( (object)arg1 [, (VectorX)r0=VectorX() [, (VectorX)r1=VectorX() [, (VectorX)r2=VectorX() [, (VectorX)r3=VectorX() [, (VectorX)r4=VectorX() [, (VectorX)r5=VectorX() [, (VectorX)r6=VectorX() [, (VectorX)r7=VectorX() [, (VectorX)r8=VectorX() [, (VectorX)r9=VectorX() [, (bool)cols=False]]]]]]]]]]]) -> object

    ___init___( (object)arg1, (object)rows [, (bool)cols=False]) -> object

**col**(*(MatrixX)arg1, (int)col*) → VectorX :
    Return column as vector.

**cols**(*(MatrixX)arg1*) → int :
    Number of columns.

**computeUnitaryPositive**(*(MatrixX)arg1*) → tuple :
    Compute polar decomposition (unitary matrix U and positive semi-definite symmetric matrix P such that self=U*P).

**determinant**(*(MatrixX)arg1*) → float :
    Return matrix determinant.

**diagonal**(*(MatrixX)arg1*) → VectorX :
    Return diagonal as vector.

**inverse**(*(MatrixX)arg1*) → MatrixX :
    Return inverted matrix.

**isApprox**(*(MatrixX)arg1, (MatrixX)other*[*, (float)prec=1e-12*]) → bool :
    Approximate comparison with precision *prec*.

**jacobiSVD**(*(MatrixX)arg1*) → tuple :
    Compute SVD decomposition of square matrix, retuns (U,S,V) such that self=U*S*V.transpose()

**maxAbsCoeff**(*(MatrixX)arg1*) → float :
    Maximum absolute value over all elements.

**maxCoeff**(*(MatrixX)arg1*) → float :
    Maximum value over all elements.

**mean**(*(MatrixX)arg1*) → float :
    Mean value over all elements.

**minCoeff**(*(MatrixX)arg1*) → float :
    Minimum value over all elements.

**norm**(*(MatrixX)arg1*) → float :
    Euclidean norm.

**normalize**(*(MatrixX)arg1*) → None :
    Normalize this object in-place.

**normalized**(*(MatrixX)arg1*) → MatrixX :
    Return normalized copy of this object

**polarDecomposition**(*(MatrixX)arg1*) → tuple :
    Alias for *computeUnitaryPositive*.

**prod**(*(MatrixX)arg1*) → float :
    Product of all elements.

**pruned**(*(MatrixX)arg1*[, *(float)absTol=1e-06*]) → MatrixX :
    Zero all elements which are greater than *absTol*. Negative zeros are not pruned.

**resize**(*(MatrixX)arg1*, *(int)rows*, *(int)cols*) → None :
    Change size of the matrix, keep values of elements which exist in the new matrix

**row**(*(MatrixX)arg1*, *(int)row*) → VectorX :
    Return row as vector.

**rows**(*(MatrixX)arg1*) → int :
    Number of rows.

**selfAdjointEigenDecomposition**(*(MatrixX)arg1*) → tuple :
    Compute eigen (spectral) decomposition of symmetric matrix, returns (eigVecs,eigVals).
    eigVecs is orthogonal Matrix3 with columns ar normalized eigenvectors, eigVals is Vector3
    with corresponding eigenvalues. self=eigVecs*diag(eigVals)*eigVecs.transpose().

**spectralDecomposition**(*(MatrixX)arg1*) → tuple :
    Alias for *selfAdjointEigenDecomposition*.

**squaredNorm**(*(MatrixX)arg1*) → float :
    Square of the Euclidean norm.

**sum**(*(MatrixX)arg1*) → float :
    Sum of all elements.

**svd**(*(MatrixX)arg1*) → tuple :
    Alias for *jacobiSVD*.

**trace**(*(MatrixX)arg1*) → float :
    Return sum of diagonal elements.

**transpose**(*(MatrixX)arg1*) → MatrixX :
    Return transposed matrix.

**class yade._minieigenHP.MatrixXc**
    */TODO/*

    **static Identity**(*(int)arg1*, *(int)rank*) → MatrixXc :
        Create identity matrix with given rank (square).

    **static Ones**(*(int)rows*, *(int)cols*) → MatrixXc :
        Create matrix of given dimensions where all elements are set to 1.

    **static Random**(*(int)rows*, *(int)cols*) → MatrixXc :
        Create matrix with given dimensions where all elements are set to number between 0 and 1
        (uniformly-distributed).

    **static Zero**(*(int)rows*, *(int)cols*) → MatrixXc :
        Create zero matrix of given dimensions

**__init__**(*(object)arg1*) → None
    ___init___( (object)arg1, (MatrixXc)other) -> None

    ___init___( (object)arg1, (VectorXc)diag) -> object

    ___init___( (object)arg1 [, (VectorXc)r0=VectorXc() [, (VectorXc)r1=VectorXc() [, (VectorXc)r2=VectorXc() [, (VectorXc)r3=VectorXc() [, (VectorXc)r4=VectorXc() [, (VectorXc)r5=VectorXc() [, (VectorXc)r6=VectorXc() [, (VectorXc)r7=VectorXc() [, (VectorXc)r8=VectorXc() [, (VectorXc)r9=VectorXc() [, (bool)cols=False]]]]]]]]]]]) -> object

    ___init___( (object)arg1, (object)rows [, (bool)cols=False]) -> object

**col**(*(MatrixXc)arg1, (int)col*) → VectorXc :
    Return column as vector.

**cols**(*(MatrixXc)arg1*) → int :
    Number of columns.

**determinant**(*(MatrixXc)arg1*) → complex :
    Return matrix determinant.

**diagonal**(*(MatrixXc)arg1*) → VectorXc :
    Return diagonal as vector.

**inverse**(*(MatrixXc)arg1*) → MatrixXc :
    Return inverted matrix.

**isApprox**(*(MatrixXc)arg1, (MatrixXc)other*[, *(float)prec=1e-12*]) → bool :
    Approximate comparison with precision *prec*.

**maxAbsCoeff**(*(MatrixXc)arg1*) → float :
    Maximum absolute value over all elements.

**mean**(*(MatrixXc)arg1*) → complex :
    Mean value over all elements.

**norm**(*(MatrixXc)arg1*) → float :
    Euclidean norm.

**normalize**(*(MatrixXc)arg1*) → None :
    Normalize this object in-place.

**normalized**(*(MatrixXc)arg1*) → MatrixXc :
    Return normalized copy of this object

**prod**(*(MatrixXc)arg1*) → complex :
    Product of all elements.

**pruned**(*(MatrixXc)arg1*[, *(float)absTol=1e-06*]) → MatrixXc :
    Zero all elements which are greater than *absTol*. Negative zeros are not pruned.

**resize**(*(MatrixXc)arg1, (int)rows, (int)cols*) → None :
    Change size of the matrix, keep values of elements which exist in the new matrix

**row**(*(MatrixXc)arg1, (int)row*) → VectorXc :
    Return row as vector.

**rows**(*(MatrixXc)arg1*) → int :
    Number of rows.

**squaredNorm**(*(MatrixXc)arg1*) → float :
    Square of the Euclidean norm.

**sum**(*(MatrixXc)arg1*) → complex :
    Sum of all elements.

**trace**(*(MatrixXc)arg1*) → complex :
    Return sum of diagonal elements.

**transpose**(*(MatrixXc)arg1*) → MatrixXc :
    Return transposed matrix.

**class** yade._minieigenHP.**Quaternion**
    Quaternion representing rotation.

    Supported operations (q is a Quaternion, v is a Vector3): `q*q` (rotation composition), `q*=q`, `q*v` (rotating v by q), `q==q`, `q!=q`.

    Static attributes: `Identity`.

---

**Note:** Quaternion is represented as axis-angle when printed (e.g. `Identity` is `Quaternion((1, 0,0),0)`, and can also be constructed from the axis-angle representation. This is however different from the data stored inside, which can be accessed by indices `[0]` (x), `[1]` (y), `[2]` (z), `[3]` (w). To obtain axis-angle programatically, use *Quaternion.toAxisAngle* which returns the tuple.

---

    `Identity = Quaternion((1,0,0),0)`

    **Rotate**(*(Quaternion)arg1, (Vector3)v*) → Vector3

    **__init__**(*(object)arg1*) → None
        ___init___( (object)arg1, (Vector3)axis, (float)angle) -> object

        ___init___( (object)arg1, (float)angle, (Vector3)axis) -> object

        ___init___( (object)arg1, (Vector3)u, (Vector3)v) -> object

        **___init___( (object)arg1, (float)w, (float)x, (float)y, (float)z) -> None :** Initialize from coefficients.

---

        **Note:** The order of coefficients is $w$, $x$, $y$, $z$. The [] operator numbers them differently, 0...4 for $x$ $y$ $z$ $w$!

---

        ___init___( (object)arg1, (Matrix3)rotMatrix) -> None

        ___init___( (object)arg1, (Quaternion)other) -> None

    **angularDistance**(*(Quaternion)arg1, (Quaternion)arg2*) → float

    **conjugate**(*(Quaternion)arg1*) → Quaternion

    **inverse**(*(Quaternion)arg1*) → Quaternion

    **norm**(*(Quaternion)arg1*) → float

    **normalize**(*(Quaternion)arg1*) → None

    **normalized**(*(Quaternion)arg1*) → Quaternion

    **setFromTwoVectors**(*(Quaternion)arg1, (Vector3)u, (Vector3)v*) → None

    **slerp**(*(Quaternion)arg1, (float)t, (Quaternion)other*) → Quaternion

    **toAngleAxis**(*(Quaternion)arg1*) → tuple

    **toAxisAngle**(*(Quaternion)arg1*) → tuple

    **toRotationMatrix**(*(Quaternion)arg1*) → Matrix3

    **toRotationVector**(*(Quaternion)arg1*) → Vector3

**class** yade._minieigenHP.**Vector2**
    3-dimensional float vector.

    Supported operations (`f` if a float/int, `v` is a Vector3): `-v`, `v+v`, `v+=v`, `v-v`, `v-=v`, `v*f`, `f*v`, `v*=f`, `v/f`, `v/=f`, `v==v`, `v!=v`.

    Implicit conversion from sequence (list, tuple, ...) of 2 floats.

Static attributes: `Zero`, `Ones`, `UnitX`, `UnitY`.

`Identity = Vector2(1,0)`

`Ones = Vector2(1,1)`

`static Random()` → Vector2 :
    Return an object where all elements are randomly set to values between 0 and 1.

`static Unit(`*(int)arg1*`)` → Vector2

`UnitX = Vector2(1,0)`

`UnitY = Vector2(0,1)`

`Zero = Vector2(0,0)`

`__init__(`*(object)arg1*`)` → None
    ___init___( (object)arg1, (Vector2)other) -> None

    ___init___( (object)arg1, (float)x, (float)y) -> None

`asDiagonal(`*(Vector2)arg1*`)` → object :
    Return diagonal matrix with this vector on the diagonal.

`cols(`*(Vector2)arg1*`)` → int :
    Number of columns.

`dot(`*(Vector2)arg1*, *(Vector2)other*`)` → float :
    Dot product with *other*.

`isApprox(`*(Vector2)arg1*, *(Vector2)other*$\left[\right.$, *(float)prec=1e-12*$\left.\right]$`)` → bool :
    Approximate comparison with precision *prec*.

`maxAbsCoeff(`*(Vector2)arg1*`)` → float :
    Maximum absolute value over all elements.

`maxCoeff(`*(Vector2)arg1*`)` → float :
    Maximum value over all elements.

`mean(`*(Vector2)arg1*`)` → float :
    Mean value over all elements.

`minCoeff(`*(Vector2)arg1*`)` → float :
    Minimum value over all elements.

`norm(`*(Vector2)arg1*`)` → float :
    Euclidean norm.

`normalize(`*(Vector2)arg1*`)` → None :
    Normalize this object in-place.

`normalized(`*(Vector2)arg1*`)` → Vector2 :
    Return normalized copy of this object

`outer(`*(Vector2)arg1*, *(Vector2)other*`)` → object :
    Outer product with *other*.

`prod(`*(Vector2)arg1*`)` → float :
    Product of all elements.

`pruned(`*(Vector2)arg1*$\left[\right.$, *(float)absTol=1e-06*$\left.\right]$`)` → Vector2 :
    Zero all elements which are greater than *absTol*. Negative zeros are not pruned.

`rows(`*(Vector2)arg1*`)` → int :
    Number of rows.

`squaredNorm(`*(Vector2)arg1*`)` → float :
    Square of the Euclidean norm.

**sum**(*(Vector2)arg1*) → float :
 Sum of all elements.

**class yade._minieigenHP.Vector2c**
 */TODO/*

 **Identity = Vector2c(1,0)**

 **Ones = Vector2c(1,1)**

 **static Random**() → Vector2c :
  Return an object where all elements are randomly set to values between 0 and 1.

 **static Unit**(*(int)arg1*) → Vector2c

 **UnitX = Vector2c(1,0)**

 **UnitY = Vector2c(0,1)**

 **Zero = Vector2c(0,0)**

 **__init__**(*(object)arg1*) → None
  ___init___( (object)arg1, (Vector2c)other) -> None

  ___init___( (object)arg1, (complex)x, (complex)y) -> None

 **asDiagonal**(*(Vector2c)arg1*) → object :
  Return diagonal matrix with this vector on the diagonal.

 **cols**(*(Vector2c)arg1*) → int :
  Number of columns.

 **dot**(*(Vector2c)arg1, (Vector2c)other*) → complex :
  Dot product with *other*.

 **isApprox**(*(Vector2c)arg1, (Vector2c)other*[, *(float)prec=1e-12*]) → bool :
  Approximate comparison with precision *prec*.

 **maxAbsCoeff**(*(Vector2c)arg1*) → float :
  Maximum absolute value over all elements.

 **mean**(*(Vector2c)arg1*) → complex :
  Mean value over all elements.

 **norm**(*(Vector2c)arg1*) → float :
  Euclidean norm.

 **normalize**(*(Vector2c)arg1*) → None :
  Normalize this object in-place.

 **normalized**(*(Vector2c)arg1*) → Vector2c :
  Return normalized copy of this object

 **outer**(*(Vector2c)arg1, (Vector2c)other*) → object :
  Outer product with *other*.

 **prod**(*(Vector2c)arg1*) → complex :
  Product of all elements.

 **pruned**(*(Vector2c)arg1*[, *(float)absTol=1e-06*]) → Vector2c :
  Zero all elements which are greater than *absTol*. Negative zeros are not pruned.

 **rows**(*(Vector2c)arg1*) → int :
  Number of rows.

 **squaredNorm**(*(Vector2c)arg1*) → float :
  Square of the Euclidean norm.

 **sum**(*(Vector2c)arg1*) → complex :
  Sum of all elements.

**class** yade._minieigenHP.**Vector2i**

    2-dimensional integer vector.

    Supported operations (`i` if an int, `v` is a Vector2i): `-v, v+v, v+=v, v-v, v-=v, v*i, i*v, v*=i, v==v, v!=v`.

    Implicit conversion from sequence (list, tuple, …) of 2 integers.

    Static attributes: `Zero, Ones, UnitX, UnitY`.

    **Identity = Vector2i(1,0)**

    **Ones = Vector2i(1,1)**

    **static Random**() → Vector2i :

        Return an object where all elements are randomly set to values between 0 and 1.

    **static Unit**(*(int)arg1*) → Vector2i

    **UnitX = Vector2i(1,0)**

    **UnitY = Vector2i(0,1)**

    **Zero = Vector2i(0,0)**

    **__init__**(*(object)arg1*) → None

        ___init___( (object)arg1, (Vector2i)other) -> None

        ___init___( (object)arg1, (int)x, (int)y) -> None

    **asDiagonal**(*(Vector2i)arg1*) → object :

        Return diagonal matrix with this vector on the diagonal.

    **cols**(*(Vector2i)arg1*) → int :

        Number of columns.

    **dot**(*(Vector2i)arg1, (Vector2i)other*) → int :

        Dot product with *other*.

    **isApprox**(*(Vector2i)arg1, (Vector2i)other*[, *(int)prec=0*]) → bool :

        Approximate comparison with precision *prec*.

    **maxAbsCoeff**(*(Vector2i)arg1*) → int :

        Maximum absolute value over all elements.

    **maxCoeff**(*(Vector2i)arg1*) → int :

        Maximum value over all elements.

    **mean**(*(Vector2i)arg1*) → int :

        Mean value over all elements.

    **minCoeff**(*(Vector2i)arg1*) → int :

        Minimum value over all elements.

    **outer**(*(Vector2i)arg1, (Vector2i)other*) → object :

        Outer product with *other*.

    **prod**(*(Vector2i)arg1*) → int :

        Product of all elements.

    **rows**(*(Vector2i)arg1*) → int :

        Number of rows.

    **sum**(*(Vector2i)arg1*) → int :

        Sum of all elements.

**class** yade._minieigenHP.**Vector3**

    3-dimensional float vector.

    Supported operations (`f` if a float/int, `v` is a Vector3): `-v, v+v, v+=v, v-v, v-=v, v*f, f*v, v*=f, v/f, v/=f, v==v, v!=v`, plus operations with `Matrix3` and `Quaternion`.

Implicit conversion from sequence (list, tuple, ...) of 3 floats.

Static attributes: `Zero, Ones, UnitX, UnitY, UnitZ`.

**Identity = Vector3(1,0,0)**

**Ones = Vector3(1,1,1)**

**static Random()** → Vector3 :
    Return an object where all elements are randomly set to values between 0 and 1.

**static Unit(***(int)arg1***)** → Vector3

**UnitX = Vector3(1,0,0)**

**UnitY = Vector3(0,1,0)**

**UnitZ = Vector3(0,0,1)**

**Zero = Vector3(0,0,0)**

**__init__(***(object)arg1***)** → None
    ___init___( (object)arg1, (Vector3)other) -> None

    ___init___( (object)arg1 [, (float)x=0.0 [, (float)y=0.0 [, (float)z=0.0]]]) -> None

**asDiagonal(***(Vector3)arg1***)** → Matrix3 :
    Return diagonal matrix with this vector on the diagonal.

**cols(***(Vector3)arg1***)** → int :
    Number of columns.

**cross(***(Vector3)arg1, (Vector3)arg2***)** → Vector3

**dot(***(Vector3)arg1, (Vector3)other***)** → float :
    Dot product with *other*.

**isApprox(***(Vector3)arg1, (Vector3)other***[**, *(float)prec=1e-12***]**) → bool :
    Approximate comparison with precision *prec*.

**maxAbsCoeff(***(Vector3)arg1***)** → float :
    Maximum absolute value over all elements.

**maxCoeff(***(Vector3)arg1***)** → float :
    Maximum value over all elements.

**mean(***(Vector3)arg1***)** → float :
    Mean value over all elements.

**minCoeff(***(Vector3)arg1***)** → float :
    Minimum value over all elements.

**norm(***(Vector3)arg1***)** → float :
    Euclidean norm.

**normalize(***(Vector3)arg1***)** → None :
    Normalize this object in-place.

**normalized(***(Vector3)arg1***)** → Vector3 :
    Return normalized copy of this object

**outer(***(Vector3)arg1, (Vector3)other***)** → Matrix3 :
    Outer product with *other*.

**prod(***(Vector3)arg1***)** → float :
    Product of all elements.

**pruned(***(Vector3)arg1***[**, *(float)absTol=1e-06***]**) → Vector3 :
    Zero all elements which are greater than *absTol*. Negative zeros are not pruned.

**rows**(*(Vector3)arg1*) → int :
> Number of rows.

**squaredNorm**(*(Vector3)arg1*) → float :
> Square of the Euclidean norm.

**sum**(*(Vector3)arg1*) → float :
> Sum of all elements.

**xy**(*(Vector3)arg1*) → Vector2

**xz**(*(Vector3)arg1*) → Vector2

**yx**(*(Vector3)arg1*) → Vector2

**yz**(*(Vector3)arg1*) → Vector2

**zx**(*(Vector3)arg1*) → Vector2

**zy**(*(Vector3)arg1*) → Vector2

**class yade._minieigenHP.Vector3c**
> */TODO/*

> **Identity = Vector3c(1,0,0)**

> **Ones = Vector3c(1,1,1)**

> **static Random**() → Vector3c :
>> Return an object where all elements are randomly set to values between 0 and 1.

> **static Unit**(*(int)arg1*) → Vector3c

> **UnitX = Vector3c(1,0,0)**

> **UnitY = Vector3c(0,1,0)**

> **UnitZ = Vector3c(0,0,1)**

> **Zero = Vector3c(0,0,0)**

> **__init__**(*(object)arg1*) → None
>> ___init___( (object)arg1, (Vector3c)other) -> None

>> ___init___( (object)arg1 [, (complex)x=0j [, (complex)y=0j [, (complex)z=0j]]]) -> None

> **asDiagonal**(*(Vector3c)arg1*) → Matrix3c :
>> Return diagonal matrix with this vector on the diagonal.

> **cols**(*(Vector3c)arg1*) → int :
>> Number of columns.

> **cross**(*(Vector3c)arg1, (Vector3c)arg2*) → Vector3c

> **dot**(*(Vector3c)arg1, (Vector3c)other*) → complex :
>> Dot product with *other*.

> **isApprox**(*(Vector3c)arg1, (Vector3c)other*[, *(float)prec=1e-12*]) → bool :
>> Approximate comparison with precision *prec*.

> **maxAbsCoeff**(*(Vector3c)arg1*) → float :
>> Maximum absolute value over all elements.

> **mean**(*(Vector3c)arg1*) → complex :
>> Mean value over all elements.

> **norm**(*(Vector3c)arg1*) → float :
>> Euclidean norm.

> **normalize**(*(Vector3c)arg1*) → None :
>> Normalize this object in-place.

**normalized**(*(Vector3c)arg1*) → Vector3c :
    Return normalized copy of this object

**outer**(*(Vector3c)arg1, (Vector3c)other*) → Matrix3c :
    Outer product with *other*.

**prod**(*(Vector3c)arg1*) → complex :
    Product of all elements.

**pruned**(*(Vector3c)arg1*[, *(float)absTol=1e-06*]) → Vector3c :
    Zero all elements which are greater than *absTol*. Negative zeros are not pruned.

**rows**(*(Vector3c)arg1*) → int :
    Number of rows.

**squaredNorm**(*(Vector3c)arg1*) → float :
    Square of the Euclidean norm.

**sum**(*(Vector3c)arg1*) → complex :
    Sum of all elements.

**xy**(*(Vector3c)arg1*) → Vector2c

**xz**(*(Vector3c)arg1*) → Vector2c

**yx**(*(Vector3c)arg1*) → Vector2c

**yz**(*(Vector3c)arg1*) → Vector2c

**zx**(*(Vector3c)arg1*) → Vector2c

**zy**(*(Vector3c)arg1*) → Vector2c

**class yade._minieigenHP.Vector3i**
    3-dimensional integer vector.

    Supported operations (`i` if an int, `v` is a Vector3i): `-v`, `v+v`, `v+=v`, `v-v`, `v-=v`, `v*i`, `i*v`, `v*=i`, `v==v`, `v!=v`.

    Implicit conversion from sequence (list, tuple, …) of 3 integers.

    Static attributes: `Zero`, `Ones`, `UnitX`, `UnitY`, `UnitZ`.

    **Identity = Vector3i(1,0,0)**

    **Ones = Vector3i(1,1,1)**

    **static Random**() → Vector3i :
        Return an object where all elements are randomly set to values between 0 and 1.

    **static Unit**(*(int)arg1*) → Vector3i

    **UnitX = Vector3i(1,0,0)**

    **UnitY = Vector3i(0,1,0)**

    **UnitZ = Vector3i(0,0,1)**

    **Zero = Vector3i(0,0,0)**

    **__init__**(*(object)arg1*) → None
        ___init___( (object)arg1, (Vector3i)other) -> None

        ___init___( (object)arg1 [, (int)x=0 [, (int)y=0 [, (int)z=0]]]) -> None

    **asDiagonal**(*(Vector3i)arg1*) → object :
        Return diagonal matrix with this vector on the diagonal.

    **cols**(*(Vector3i)arg1*) → int :
        Number of columns.

    **cross**(*(Vector3i)arg1, (Vector3i)arg2*) → Vector3i

**dot**(*(Vector3i)arg1, (Vector3i)other*) → int :
    Dot product with *other*.

**isApprox**(*(Vector3i)arg1, (Vector3i)other*[, *(int)prec=0*]) → bool :
    Approximate comparison with precision *prec*.

**maxAbsCoeff**(*(Vector3i)arg1*) → int :
    Maximum absolute value over all elements.

**maxCoeff**(*(Vector3i)arg1*) → int :
    Maximum value over all elements.

**mean**(*(Vector3i)arg1*) → int :
    Mean value over all elements.

**minCoeff**(*(Vector3i)arg1*) → int :
    Minimum value over all elements.

**outer**(*(Vector3i)arg1, (Vector3i)other*) → object :
    Outer product with *other*.

**prod**(*(Vector3i)arg1*) → int :
    Product of all elements.

**rows**(*(Vector3i)arg1*) → int :
    Number of rows.

**sum**(*(Vector3i)arg1*) → int :
    Sum of all elements.

**xy**(*(Vector3i)arg1*) → Vector2i

**xz**(*(Vector3i)arg1*) → Vector2i

**yx**(*(Vector3i)arg1*) → Vector2i

**yz**(*(Vector3i)arg1*) → Vector2i

**zx**(*(Vector3i)arg1*) → Vector2i

**zy**(*(Vector3i)arg1*) → Vector2i

**class yade._minieigenHP.Vector4**
    4-dimensional float vector.

    Supported operations (`f` if a float/int, `v` is a Vector3): `-v, v+v, v+=v, v-v, v-=v, v*f, f*v, v*=f, v/f, v/=f, v==v, v!=v`.

    Implicit conversion from sequence (list, tuple, …) of 4 floats.

    Static attributes: `Zero, Ones`.

    **Identity = Vector4(1,0,0, 0)**

    **Ones = Vector4(1,1,1, 1)**

    **static Random**() → Vector4 :
        Return an object where all elements are randomly set to values between 0 and 1.

    **static Unit**(*(int)arg1*) → Vector4

    **Zero = Vector4(0,0,0, 0)**

    **__init__**(*(object)arg1*) → None
        ___init___( (object)arg1, (Vector4)other) -> None

        ___init___( (object)arg1, (float)v0, (float)v1, (float)v2, (float)v3) -> None

    **asDiagonal**(*(Vector4)arg1*) → object :
        Return diagonal matrix with this vector on the diagonal.

**cols**(*(Vector4)arg1*) → int :
    Number of columns.

**dot**(*(Vector4)arg1, (Vector4)other*) → float :
    Dot product with *other*.

**isApprox**(*(Vector4)arg1, (Vector4)other*[, *(float)prec=1e-12*]) → bool :
    Approximate comparison with precision *prec*.

**maxAbsCoeff**(*(Vector4)arg1*) → float :
    Maximum absolute value over all elements.

**maxCoeff**(*(Vector4)arg1*) → float :
    Maximum value over all elements.

**mean**(*(Vector4)arg1*) → float :
    Mean value over all elements.

**minCoeff**(*(Vector4)arg1*) → float :
    Minimum value over all elements.

**norm**(*(Vector4)arg1*) → float :
    Euclidean norm.

**normalize**(*(Vector4)arg1*) → None :
    Normalize this object in-place.

**normalized**(*(Vector4)arg1*) → Vector4 :
    Return normalized copy of this object

**outer**(*(Vector4)arg1, (Vector4)other*) → object :
    Outer product with *other*.

**prod**(*(Vector4)arg1*) → float :
    Product of all elements.

**pruned**(*(Vector4)arg1*[, *(float)absTol=1e-06*]) → Vector4 :
    Zero all elements which are greater than *absTol*. Negative zeros are not pruned.

**rows**(*(Vector4)arg1*) → int :
    Number of rows.

**squaredNorm**(*(Vector4)arg1*) → float :
    Square of the Euclidean norm.

**sum**(*(Vector4)arg1*) → float :
    Sum of all elements.

**class yade._minieigenHP.Vector6**
    6-dimensional float vector.

    Supported operations (`f` if a float/int, `v` is a Vector6): `-v, v+v, v+=v, v-v, v-=v, v*f, f*v, v*=f, v/f, v/=f, v==v, v!=v`.

    Implicit conversion from sequence (list, tuple, …) of 6 floats.

    Static attributes: `Zero, Ones`.

    **Identity = Vector6(1,0,0, 0,0,0)**

    **Ones = Vector6(1,1,1, 1,1,1)**

    **static Random**() → Vector6 :
        Return an object where all elements are randomly set to values between 0 and 1.

    **static Unit**(*(int)arg1*) → Vector6

    **Zero = Vector6(0,0,0, 0,0,0)**

**\_\_init\_\_**(*(object)arg1*) → None
 \_\_init\_\_( (object)arg1, (Vector6)other) -> None

 \_\_init\_\_( (object)arg1, (float)v0, (float)v1, (float)v2, (float)v3, (float)v4, (float)v5) -> object

 \_\_init\_\_( (object)arg1, (Vector3)head, (Vector3)tail) -> object

**asDiagonal**(*(Vector6)arg1*) → Matrix6 :
 Return diagonal matrix with this vector on the diagonal.

**cols**(*(Vector6)arg1*) → int :
 Number of columns.

**dot**(*(Vector6)arg1, (Vector6)other*) → float :
 Dot product with *other*.

**head**(*(Vector6)arg1*) → Vector3

**isApprox**(*(Vector6)arg1, (Vector6)other*[, *(float)prec=1e-12*]) → bool :
 Approximate comparison with precision *prec*.

**maxAbsCoeff**(*(Vector6)arg1*) → float :
 Maximum absolute value over all elements.

**maxCoeff**(*(Vector6)arg1*) → float :
 Maximum value over all elements.

**mean**(*(Vector6)arg1*) → float :
 Mean value over all elements.

**minCoeff**(*(Vector6)arg1*) → float :
 Minimum value over all elements.

**norm**(*(Vector6)arg1*) → float :
 Euclidean norm.

**normalize**(*(Vector6)arg1*) → None :
 Normalize this object in-place.

**normalized**(*(Vector6)arg1*) → Vector6 :
 Return normalized copy of this object

**outer**(*(Vector6)arg1, (Vector6)other*) → Matrix6 :
 Outer product with *other*.

**prod**(*(Vector6)arg1*) → float :
 Product of all elements.

**pruned**(*(Vector6)arg1*[, *(float)absTol=1e-06*]) → Vector6 :
 Zero all elements which are greater than *absTol*. Negative zeros are not pruned.

**rows**(*(Vector6)arg1*) → int :
 Number of rows.

**squaredNorm**(*(Vector6)arg1*) → float :
 Square of the Euclidean norm.

**sum**(*(Vector6)arg1*) → float :
 Sum of all elements.

**tail**(*(Vector6)arg1*) → Vector3

**class** yade.\_minieigenHP.**Vector6c**
 */TODO/*

 **Identity = Vector6c(1,0,0, 0,0,0)**

 **Ones = Vector6c(1,1,1, 1,1,1)**

**static Random()** → Vector6c :
    Return an object where all elements are randomly set to values between 0 and 1.

**static Unit**(*(int)arg1*) → Vector6c

**Zero = Vector6c(0,0,0, 0,0,0)**

**__init__**(*(object)arg1*) → None
    ___init___( (object)arg1, (Vector6c)other) -> None

    ___init___( (object)arg1, (complex)v0, (complex)v1, (complex)v2, (complex)v3, (complex)v4, (complex)v5) -> object

    ___init___( (object)arg1, (Vector3c)head, (Vector3c)tail) -> object

**asDiagonal**(*(Vector6c)arg1*) → Matrix6c :
    Return diagonal matrix with this vector on the diagonal.

**cols**(*(Vector6c)arg1*) → int :
    Number of columns.

**dot**(*(Vector6c)arg1, (Vector6c)other*) → complex :
    Dot product with *other*.

**head**(*(Vector6c)arg1*) → Vector3c

**isApprox**(*(Vector6c)arg1, (Vector6c)other*[, *(float)prec=1e-12*]) → bool :
    Approximate comparison with precision *prec*.

**maxAbsCoeff**(*(Vector6c)arg1*) → float :
    Maximum absolute value over all elements.

**mean**(*(Vector6c)arg1*) → complex :
    Mean value over all elements.

**norm**(*(Vector6c)arg1*) → float :
    Euclidean norm.

**normalize**(*(Vector6c)arg1*) → None :
    Normalize this object in-place.

**normalized**(*(Vector6c)arg1*) → Vector6c :
    Return normalized copy of this object

**outer**(*(Vector6c)arg1, (Vector6c)other*) → Matrix6c :
    Outer product with *other*.

**prod**(*(Vector6c)arg1*) → complex :
    Product of all elements.

**pruned**(*(Vector6c)arg1*[, *(float)absTol=1e-06*]) → Vector6c :
    Zero all elements which are greater than *absTol*. Negative zeros are not pruned.

**rows**(*(Vector6c)arg1*) → int :
    Number of rows.

**squaredNorm**(*(Vector6c)arg1*) → float :
    Square of the Euclidean norm.

**sum**(*(Vector6c)arg1*) → complex :
    Sum of all elements.

**tail**(*(Vector6c)arg1*) → Vector3c

**class yade._minieigenHP.Vector6i**
    6-dimensional float vector.

Supported operations (`f` if a float/int, `v` is a Vector6): `-v, v+v, v+=v, v-v, v-=v, v*f, f*v, v*=f, v/f, v/=f, v==v, v!=v`.

Implicit conversion from sequence (list, tuple, ...) of 6 ints.

Static attributes: `Zero`, `Ones`.

**`Identity = Vector6i(1,0,0, 0,0,0)`**

**`Ones = Vector6i(1,1,1, 1,1,1)`**

**`static Random()`** → Vector6i :
    Return an object where all elements are randomly set to values between 0 and 1.

**`static Unit`**(*(int)arg1*) → Vector6i

**`Zero = Vector6i(0,0,0, 0,0,0)`**

**`__init__`**(*(object)arg1*) → None
    ___init___( (object)arg1, (Vector6i)other) -> None

    ___init___( (object)arg1, (int)v0, (int)v1, (int)v2, (int)v3, (int)v4, (int)v5) -> object

    ___init___( (object)arg1, (Vector3i)head, (Vector3i)tail) -> object

**`asDiagonal`**(*(Vector6i)arg1*) → object :
    Return diagonal matrix with this vector on the diagonal.

**`cols`**(*(Vector6i)arg1*) → int :
    Number of columns.

**`dot`**(*(Vector6i)arg1, (Vector6i)other*) → int :
    Dot product with *other*.

**`head`**(*(Vector6i)arg1*) → Vector3i

**`isApprox`**(*(Vector6i)arg1, (Vector6i)other*[, *(int)prec=0*]) → bool :
    Approximate comparison with precision *prec*.

**`maxAbsCoeff`**(*(Vector6i)arg1*) → int :
    Maximum absolute value over all elements.

**`maxCoeff`**(*(Vector6i)arg1*) → int :
    Maximum value over all elements.

**`mean`**(*(Vector6i)arg1*) → int :
    Mean value over all elements.

**`minCoeff`**(*(Vector6i)arg1*) → int :
    Minimum value over all elements.

**`outer`**(*(Vector6i)arg1, (Vector6i)other*) → object :
    Outer product with *other*.

**`prod`**(*(Vector6i)arg1*) → int :
    Product of all elements.

**`rows`**(*(Vector6i)arg1*) → int :
    Number of rows.

**`sum`**(*(Vector6i)arg1*) → int :
    Sum of all elements.

**`tail`**(*(Vector6i)arg1*) → Vector3i

**class `yade._minieigenHP.VectorX`**
    Dynamic-sized float vector.

    Supported operations (`f` if a float/int, `v` is a VectorX): `-v`, `v+v`, `v+=v`, `v-v`, `v-=v`, `v*f`, `f*v`, `v*=f`, `v/f`, `v/=f`, `v==v`, `v!=v`.

    Implicit conversion from sequence (list, tuple, ...) of X floats.

    **`static Ones`**(*(int)arg1*) → VectorX

**static Random**(*(int)len*) → VectorX :
> Return vector of given length with all elements set to values between 0 and 1 randomly.

**static Unit**(*(int)arg1, (int)arg2*) → VectorX

**static Zero**(*(int)arg1*) → VectorX

**__init__**(*(object)arg1*) → None
> \_\_init\_\_( (object)arg1, (VectorX)other) -> None

> \_\_init\_\_( (object)arg1, (object)vv) -> object

**asDiagonal**(*(VectorX)arg1*) → MatrixX :
> Return diagonal matrix with this vector on the diagonal.

**cols**(*(VectorX)arg1*) → int :
> Number of columns.

**dot**(*(VectorX)arg1, (VectorX)other*) → float :
> Dot product with *other*.

**isApprox**(*(VectorX)arg1, (VectorX)other*[, *(float)prec=1e-12*]) → bool :
> Approximate comparison with precision *prec*.

**maxAbsCoeff**(*(VectorX)arg1*) → float :
> Maximum absolute value over all elements.

**maxCoeff**(*(VectorX)arg1*) → float :
> Maximum value over all elements.

**mean**(*(VectorX)arg1*) → float :
> Mean value over all elements.

**minCoeff**(*(VectorX)arg1*) → float :
> Minimum value over all elements.

**norm**(*(VectorX)arg1*) → float :
> Euclidean norm.

**normalize**(*(VectorX)arg1*) → None :
> Normalize this object in-place.

**normalized**(*(VectorX)arg1*) → VectorX :
> Return normalized copy of this object

**outer**(*(VectorX)arg1, (VectorX)other*) → MatrixX :
> Outer product with *other*.

**prod**(*(VectorX)arg1*) → float :
> Product of all elements.

**pruned**(*(VectorX)arg1*[, *(float)absTol=1e-06*]) → VectorX :
> Zero all elements which are greater than *absTol*. Negative zeros are not pruned.

**resize**(*(VectorX)arg1, (int)arg2*) → None

**rows**(*(VectorX)arg1*) → int :
> Number of rows.

**squaredNorm**(*(VectorX)arg1*) → float :
> Square of the Euclidean norm.

**sum**(*(VectorX)arg1*) → float :
> Sum of all elements.

**class yade._minieigenHP.VectorXc**
> /TODO/

**static Ones**(*(int)arg1*) → VectorXc

**static Random**(*(int)len*) → VectorXc :
    Return vector of given length with all elements set to values between 0 and 1 randomly.

**static Unit**(*(int)arg1, (int)arg2*) → VectorXc

**static Zero**(*(int)arg1*) → VectorXc

**__init__**(*(object)arg1*) → None
    ___init___( (object)arg1, (VectorXc)other) -> None

    ___init___( (object)arg1, (object)vv) -> object

**asDiagonal**(*(VectorXc)arg1*) → MatrixXc :
    Return diagonal matrix with this vector on the diagonal.

**cols**(*(VectorXc)arg1*) → int :
    Number of columns.

**dot**(*(VectorXc)arg1, (VectorXc)other*) → complex :
    Dot product with *other*.

**isApprox**(*(VectorXc)arg1, (VectorXc)other*$\Big[$, *(float)prec=1e-12*$\Big]$) → bool :
    Approximate comparison with precision *prec*.

**maxAbsCoeff**(*(VectorXc)arg1*) → float :
    Maximum absolute value over all elements.

**mean**(*(VectorXc)arg1*) → complex :
    Mean value over all elements.

**norm**(*(VectorXc)arg1*) → float :
    Euclidean norm.

**normalize**(*(VectorXc)arg1*) → None :
    Normalize this object in-place.

**normalized**(*(VectorXc)arg1*) → VectorXc :
    Return normalized copy of this object

**outer**(*(VectorXc)arg1, (VectorXc)other*) → MatrixXc :
    Outer product with *other*.

**prod**(*(VectorXc)arg1*) → complex :
    Product of all elements.

**pruned**(*(VectorXc)arg1*$\Big[$, *(float)absTol=1e-06*$\Big]$) → VectorXc :
    Zero all elements which are greater than *absTol*. Negative zeros are not pruned.

**resize**(*(VectorXc)arg1, (int)arg2*) → None

**rows**(*(VectorXc)arg1*) → int :
    Number of rows.

**squaredNorm**(*(VectorXc)arg1*) → float :
    Square of the Euclidean norm.

**sum**(*(VectorXc)arg1*) → complex :
    Sum of all elements.

## 2.4.10 yade.mpy module

This module defines mpirun(), a parallel implementation of run() using a distributed memory approach. Message passing is done with mpi4py mainly, however some messages are also handled in c++ (with openmpi).

**Note:** Many internals of the mpy module listed on this page are not helpful to the user. Instead, please find *introductory material on mpy module* in user manual.

**Logic:**

The logic for an initially centralized scene is as follows:

1. Instanciate a complete, ordinary, yade scene

2. Insert subdomains as special yade bodies. This is somehow similar to adding a clump body on the top of clump members

3. Broadcast this scene to all workers. In the initialization phase the workers will:

   - define the bounding box of their assigned bodies and return it to other workers

   - detect which assigned bodies are virtually in interaction with other domains (based on their bounding boxes) and communicate the lists to the relevant workers

   - erase the bodies which are neither assigned nor virtually interacting with the subdomain

4. Run a number of 'regular' iterations without re-running collision detection (verlet dist mechanism). In each regular iteration the workers will:

   - calculate internal and cross-domains interactions

   - execute Newton on assigned bodies (modified Newton skips other domains)

   - send updated positions to other workers and partial force on floor to master

5. When one worker triggers collision detection all workers will follow. It will result in updating the intersections between subdomains.

6. If enabled, bodies may be re-allocated to different domains just after a collision detection, based on a filter. Custom filters are possible. One is predidefined here (medianFilter)

**Rules:**

   #- intersections[0] has 0-bodies (to which we need to send force) #- intersections[thisDomain] has ids of the other domains overlapping the current ones #- intersections[otherDomain] has ids of bodies in _current_ domain which are overlapping with other domain (for which we need to send updated pos/vel)

**Hints:**

   #- handle subD.intersections with care (same for mirrorIntersections). subD.intersections.append() will not reach the c++ object. subD.intersections can only be assigned (a list of list of int)

yade.mpy.**MAX_RANK_OUTPUT = 5**
   larger ranks will be skipped in mprint

yade.mpy.**REALLOCATE_FILTER**(*i, j, giveAway*)
   Returns bodies in "i" to be assigned to "j" based on median split between the center points of subdomain's AABBs If giveAway!=0, positive or negative, "i" will give/acquire this number to "j" with nothing in return (for load balancing purposes)

class yade.mpy.**Timing_comm**(*inherits object*)

   **Allgather**(*timing_name, *args, **kwargs*)

**Gather**(*timing_name, \*args, \*\*kwargs*)

**Gatherv**(*timing_name, \*args, \*\*kwargs*)

**allreduce**(*timing_name, \*args, \*\*kwargs*)

**bcast**(*timing_name, \*args, \*\*kwargs*)

**clear**()

**enable_timing**()

**mpiSendStates**(*timing_name, \*args, \*\*kwargs*)

**mpiWait**(*timing_name, \*args, \*\*kwargs*)

**mpiWaitReceived**(*timing_name, \*args, \*\*kwargs*)

**print_all**()

**recv**(*timing_name, \*args, \*\*kwargs*)

**send**(*timing_name, \*args, \*\*kwargs*)

yade.mpy.**bodyErase**(*ids*)

> The parallel version of O.bodies.erase(id), should be called collectively else the distributed scenes become inconsistent with each other (even the subdomains which don't have 'id' can call safely). For performance, better call on a list: bodyErase([i,j,k]).

yade.mpy.**checkAndCollide**()

> return true if collision detection needs activation in at least one SD, else false. If COPY_MIRROR_BODIES_WHEN_COLLIDE run collider when needed, and in that case return False.

yade.mpy.**colorDomains**()

> Apply color to body to reflect their subdomain idx

yade.mpy.**configure**()

> Import MPI and define context, configure will no spawn workers by itself, that is done by initialize() openmpi environment variables needs to be set before calling configure()

yade.mpy.**declareMasterInteractive**()

> This is to signal that we are in interactive session, so TIMEOUT will be reset to 0 (ignored)

yade.mpy.**disconnect**()

> Kill all mpi processes, leaving python interpreter to rank 0 as in single-threaded execution. The scenes in workers are lost since further reconnexion to mpi will just spawn new processes. The scene in master thread is left unchanged.

yade.mpy.**eraseRemote**()

yade.mpy.**genLocalIntersections**(*subdomains*)

> Defines sets of bodies within current domain overlapping with other domains.
>
> The structure of the data for domain 'k' is: [[id1, id2, …], <——— intersections[0] = ids of bodies in domain k interacting with master domain (subdomain k itself excluded) [id3, id4, …], <——— intersections[1] = ids of bodies in domain k interacting with domain rank=1 (subdomain k itself excluded) … [domain1, domain2, domain3, …], <———- intersections[k] = ranks (not ids!) of external domains interacting with domain k … ]

yade.mpy.**genUpdatedStates**(*b_ids*)

> return list of [id,state] (or [id,state,shape] conditionnaly) to be sent to other workers

yade.mpy.**initialize**(*np*)

yade.mpy.**isendRecvForces**()

> Communicate forces from subdomain to master Warning: the sending sides (everyone but master) must wait() the returned list of requests

yade.mpy.**makeColorScale**(*n=None*)

yade.mpy.**makeMpiArgv**()

`yade.mpy.maskedConnection`(*b*, *boolArray*)

>   List bodies within a facet selectively, the ones marked 'True' in boolArray (i.e. already selected from another facet) are discarded

`yade.mpy.maskedPFacet`(*b*, *boolArray*)

>   List bodies within a facet selectively, the ones marked 'True' in boolArray (i.e. already selected from another facet) are discarded

`yade.mpy.medianFilter`(*i*, *j*, *giveAway*)

>   Returns bodies in "i" to be assigned to "j" based on median split between the center points of subdomain's AABBs If giveAway!=0, positive or negative, "i" will give/acquire this number to "j" with nothing in return (for load balancing purposes)

`yade.mpy.mergeScene`()

`yade.mpy.migrateBodies`(*ids*, *origin*, *destination*)

>   Reassign bodies from origin to destination. The function has to be called by both origin (send) and destination (recv). Note: subD.completeSendBodies() will have to be called after a series of reassignement since subD.sendBodies() is non-blocking

`yade.mpy.mpiStats`()

`yade.mpy.mpirun`(*nSteps*, *np=None*, *withMerge=False*)

>   Parallel version of O.run() using MPI domain decomposition.
>
>   Parameters
>
>   nSteps : The numer of steps to compute np : number of mpi workers (master+subdomains), if=1 the function fallback to O.run() withMerge : wether subdomains should be merged into master at the end of the run (default False). If True the scene in the master process is exactly in the same state as after O.run(nSteps,True). The merge can be time consumming, it is recommended to activate only if post-processing or other similar tasks require it.

`yade.mpy.mprint`(*\*args*, *force=False*)

>   Print with rank-reflecting color regardless of mpy.VERBOSE_OUTPUT, still limited to rank<=mpy.MAX_RANK_OUTPUT

`yade.mpy.pairOp`(*talkTo*)

`yade.mpy.parallelCollide`()

`yade.mpy.probeRecvMessage`(*source*, *tag*)

`yade.mpy.projectedBounds`(*i*, *j*)

>   Returns sorted list of projections of bounds on a given axis, with bounds taken in i->j and j->i intersections

`yade.mpy.reallocateBodiesPairWiseBlocking`(*_filter*, *otherDomain*)

>   Re-assign bodies from/to otherDomain based on '_filter' argument. Requirement: '_filter' is a function taking ranks of origin and destination and returning the list of bodies (by index) to be moved. That's where the decomposition strategy is defined. See example medianFilter (used by default).

`yade.mpy.reallocateBodiesToSubdomains`(*_filter=<function medianFilter>*, *blocking=True*)

>   Re-assign bodies to subdomains based on '_filter' argument. Requirement: '_filter' is a function taking ranks of origin and destination and returning the list of bodies (by index) to be moved. That's where the decomposition strategy is defined. See example medianFilter (used by default). This function must be called in parallel, hence if ran interactively the command needs to be sent explicitely: mp.sendCommand("all","reallocateBodiesToSubdomains(medianFilter)",True)

`yade.mpy.reboundRemoteBodies`(*ids*)

>   update states of bodies handled by other workers, argument 'states' is a list of [id,state] (or [id,state,shape] conditionnaly)

`yade.mpy.receiveForces`(*subdomains*)

>   Accumulate forces from subdomains (only executed by master process), should happen after

ForceResetter but before Newton and before any other force-dependent engine (e.g. StressController), could be inserted via yade's pyRunner.

**yade.mpy.recordMpiTiming**(*name*, *val*)
    append val to a list of values defined by 'name' in the dictionnary timing.mpi

**yade.mpy.runOnSynchronouslPairs**(*workers*, *command*)
    Locally (from one worker POV), this function runs interactive mpi tasks defined by 'command' on a list of other workers (typically the list of interacting subdomains). Overall, peer-to-peer connexions are established so so that 'command' is executed symmetrically and simultaneously on both sides of each worker pair. I.e. if worker "i" executes "command" with argument "j" (index of another worker), then by design "j" will execute the same thing with argument "i" *simultaneously*.

    In many cases a similar series of data exchanges can be obtained more simply (and fastly) with asynchronous irecv+send like below.

    **for w in workers:** m=comm.irecv(w) comm.send(data,dest=w)

    The above only works if the messages are all known in advance locally, before any communication. If the interaction with workers[1] depends on the result of a previous interaction with workers[0] OTOH, it needs synchronous execution, hence this function. Synchronicity is also required if more than one blocking call is present in 'command', else an obvious deadlock as if 'irecv' was replaced by 'recv' in that naive loop. Both cases occur with the 'medianFilter' algorithm, hence why we need this synchronous method.

    In this function pair connexions are established by the workers in a non-supervized and non-deterministic manner. Each time an interactive communication (i,j) is established 'command' is executed simultaneously by i and j. It is guaranted that all possible pairs are visited.

    The function can be used for all-to-all operations (N^2 pairs), but more interestingly it works with workers=intersections[rank] (O(N) pairs). It can be tested with the dummy funtion 'pairOp': runOnSynchronouslPairs(range(numThreads),pairOp)

    **command:** a function taking index of another worker as argument, can include blocking communications with the other worker since runOnSynchronouslPairs guarantee that the other worker will be running the command symmetrically.

**yade.mpy.sendCommand**(*executors*, *command*, *wait=True*, *workerToWorker=False*)
    Send a command to a worker (or list of) from master or from another worker. Accepted executors are "i", "[i,j,k]", "slaves", "all" (then even master will execute the command).

**yade.mpy.sendRecvStates()**

**yade.mpy.shrinkIntersections()**
    Reduce intersections and mirrorIntersections to bodies effectively interacting with another statefull body form current subdomain This will reduce the number of updates in sendRecvStates Initial lists are backed-up and need to be restored (and all states updated) before collision detection (see checkAndCollide())

**yade.mpy.spawnedProcessWaitCommand()**

**yade.mpy.splitScene()**
    Split a monolithic scene into distributed scenes on threads.

    Precondition: the bodies have subdomain no. set in user script

**yade.mpy.unboundRemoteBodies()**
    Turn bounding boxes on/off depending on rank

**yade.mpy.updateAllIntersections()**

**yade.mpy.updateDomainBounds**(*subdomains*)
    Update bounds of current subdomain, broadcast, and receive updated bounds from other subdomains Precondition: collider.boundDispatcher.___call___()

**yade.mpy.updateMirrorOwners()**

`yade.mpy.`**updateRemoteStates**(*states, setBounded=False*)

> update states of bodies handled by other workers, argument 'states' is a list of [id,state] (or [id,state,shape] conditionnaly)

`yade.mpy.`**waitForces**()

> wait until all forces are sent to master. O.freqs is empty for master, and for all threads if not ACCUMULATE_FORCES

`yade.mpy.`**wprint**(*\*args*)

> Print with rank-reflecting color, *only if* mpy.VERBOSE_OUTPUT=True (else see *mpy.mprint*), limited to rank<=mpy.MAX_RANK_OUTPUT

## 2.4.11 yade.pack module

Creating packings and filling volumes defined by boundary representation or constructive solid geometry.

For examples, see

- examples/gts-horse/gts-operators.py
- examples/gts-horse/gts-random-pack-obb.py
- examples/gts-horse/gts-random-pack.py
- examples/test/pack-cloud.py
- examples/test/pack-predicates.py
- examples/packs/packs.py
- examples/gts-horse/gts-horse.py
- examples/WireMatPM/wirepackings.py

`yade.pack.`**SpherePack_toSimulation**(*self, rot=Matrix3(1, 0, 0, 0, 1, 0, 0, 0, 1), \*\*kw*)

> Append spheres directly to the simulation. In addition calling *O.bodies.append*, this method also appropriately sets periodic cell information of the simulation.

```
>>> from yade import pack; from math import *
>>> sp=pack.SpherePack()
```

Create random periodic packing with 20 spheres:

```
>>> sp.makeCloud((0,0,0),(5,5,5),rMean=.5,rRelFuzz=.5,periodic=True,num=20)
20
```

Virgin simulation is aperiodic:

```
>>> O.reset()
>>> O.periodic
False
```

Add generated packing to the simulation, rotated by 45° along +z

```
>>> sp.toSimulation(rot=Quaternion((0,0,1),pi/4),color=(0,0,1))
[0, 1, 2, 3, 4, 5, 6, 7, 8, 9, 10, 11, 12, 13, 14, 15, 16, 17, 18, 19]
```

Periodic properties are transferred to the simulation correctly, including rotation (this could be avoided by explicitly passing "hSize=O.cell.hSize" as an argument):

```
>>> O.periodic
True
>>> O.cell.refSize
Vector3(5,5,5)
```

(continues on next page)

```
>>> O.cell.hSize # doctest: +SKIP
Matrix3(3.53553,-3.53553,0, 3.53553,3.53553,0, 0,0,5)
```

The current state (even if rotated) is taken as mechanically undeformed, i.e. with identity transformation:

```
>>> O.cell.trsf
Matrix3(1,0,0, 0,1,0, 0,0,1)
```

> **Parameters**
>
> - **rot** (*Quaternion/Matrix3*) – rotation of the packing, which will be applied on spheres and will be used to set *Cell.trsf* as well.
>
> - **\*\*kw** – passed to *utils.sphere*
>
> **Returns** list of body ids added (like *O.bodies.append*)

yade.pack.**filterSpherePack**(*predicate, spherePack, returnSpherePack=None, \*\*kw*)
> Using given SpherePack instance, return spheres that satisfy predicate. It returns either a *pack.SpherePack* (if returnSpherePack) or a list. The packing will be recentered to match the predicate and warning is given if the predicate is larger than the packing.

yade.pack.**gtsSurface2Facets**(*surf, \*\*kw*)
> Construct facets from given GTS surface. \*\*kw is passed to utils.facet.

yade.pack.**gtsSurfaceBestFitOBB**(*surf*)
> Return (Vector3 center, Vector3 halfSize, Quaternion orientation) describing best-fit oriented bounding box (OBB) for the given surface. See cloudBestFitOBB for details.

yade.pack.**hexaNet**(*radius, cornerCoord=[0, 0, 0], xLength=1.0, yLength=0.5, mos=0.08, a=0.04, b=0.04, startAtCorner=True, isSymmetric=False, \*\*kw*)
> Definition of the particles for a hexagonal wire net in the x-y-plane for the WireMatPM.
>
> > **Parameters**
> >
> > - **radius** – radius of the particle
> >
> > - **cornerCoord** – coordinates of the lower left corner of the net
> >
> > - **xLenght** – net length in x-direction
> >
> > - **yLenght** – net length in y-direction
> >
> > - **mos** – mesh opening size (horizontal distance between the double twists)
> >
> > - **a** – length of double-twist
> >
> > - **b** – height of single wire section
> >
> > - **startAtCorner** – if true the generation starts with a double-twist at the lower left corner
> >
> > - **isSymmetric** – defines if the net is symmetric with respect to the y-axis
> >
> > **Returns** set of spheres which defines the net (net) and exact dimensions of the net (lx,ly).

---

**Note:** This packing works for the WireMatPM only. The particles at the corner are always generated first. For examples on how to use this packing see examples/WireMatPM. In order to create the proper interactions for the net the interaction radius has to be adapted in the simulation.

---

*class* yade.pack.**inConvexPolyhedron**(*inherits Predicate*)

> **aabb**(*(Predicate)arg1*) → tuple
> > aabb( (Predicate)arg1) -> None
>
> **center**(*(Predicate)arg1*) → Vector3
>
> **dim**(*(Predicate)arg1*) → Vector3

**class yade.pack.inGtsSurface_py**(*inherits Predicate*)

> This class was re-implemented in c++, but should stay here to serve as reference for implementing Predicates in pure python code. C++ allows us to play dirty tricks in GTS which are not accessible through pygts itself; the performance penalty of pygts comes from fact that if constructs and destructs bb tree for the surface at every invocation of gts.Point().is_inside(). That is cached in the c++ code, provided that the surface is not manipulated with during lifetime of the object (user's responsibility).
>
> —
>
> Predicate for GTS surfaces. Constructed using an already existing surfaces, which must be closed.
>
> > import gts surf=gts.read(open('horse.gts')) inGtsSurface(surf)
>
> ---
>
> **Note:** Padding is optionally supported by testing 6 points along the axes in the pad distance. This must be enabled in the ctor by saying doSlowPad=True. If it is not enabled and pad is not zero, warning is issued.
>
> ---
>
> **aabb**(*(Predicate)arg1*) → tuple
> > aabb( (Predicate)arg1) -> None
>
> **center**(*(Predicate)arg1*) → Vector3
>
> **dim**(*(Predicate)arg1*) → Vector3

**class yade.pack.inHalfSpace**(*inherits Predicate*)

> Predicate returning True any points, with infinite bounding box.
>
> **aabb**(*(Predicate)arg1*) → tuple
> > aabb( (Predicate)arg1) -> None
>
> **center**(*(Predicate)arg1*) → Vector3
>
> **dim**(*(Predicate)arg1*) → Vector3

**class yade.pack.inSpace**(*inherits Predicate*)

> Predicate returning True for any points, with infinite bounding box.
>
> **aabb**(*(Predicate)arg1*) → tuple
> > aabb( (Predicate)arg1) -> None
>
> **center**(*(Predicate)arg1*) → Vector3
>
> **dim**(*(Predicate)arg1*) → Vector3

**yade.pack.randomDensePack**(*predicate, radius, material=-1, dim=None, cropLayers=0, rRelFuzz=0.0, spheresInCell=0, memoizeDb=None, useOBB=False, memoDbg=False, color=None, returnSpherePack=None, seed=-1*)

> Generator of random dense packing with given geometry properties, using TriaxialTest (aperiodic) or PeriIsoCompressor (periodic). The periodicity depens on whether the spheresInCell parameter is given.
>
> *O.switchScene()* magic is used to have clean simulation for TriaxialTest without deleting the original simulation. This function therefore should never run in parallel with some code accessing your simulation.
>
> > **Parameters**
> >
> > • **predicate** – solid-defining predicate for which we generate packing

- **spheresInCell** – if given, the packing will be periodic, with given number of spheres in the periodic cell.

- **radius** – mean radius of spheres

- **rRelFuzz** – relative fuzz of the radius – e.g. radius=10, rRelFuzz=.2, then spheres will have radii $10 \pm (10*.2)$), with an uniform distribution. 0 by default, meaning all spheres will have exactly the same radius.

- **cropLayers** – (aperiodic only) how many layers of spheres will be added to the computed dimension of the box so that there no (or not so much, at least) boundary effects at the boundaries of the predicate.

- **dim** – dimension of the packing, to override dimensions of the predicate (if it is infinite, for instance)

- **memoizeDb** – name of sqlite database (existent or nonexistent) to find an already generated packing or to store the packing that will be generated, if not found (the technique of caching results of expensive computations is known as memoization). Fuzzy matching is used to select suitable candidate – packing will be scaled, rRelFuzz and dimensions compared. Packing that are too small are dictarded. From the remaining candidate, the one with the least number spheres will be loaded and returned.

- **useOBB** – effective only if a inGtsSurface predicate is given. If true (not default), oriented bounding box will be computed first; it can reduce substantially number of spheres for the triaxial compression (like $10\times$ depending on how much asymmetric the body is), see examples/gts-horse/gts-random-pack-obb.py

- **memoDbg** – show packings that are considered and reasons why they are rejected/accepted

- **returnSpherePack** – see the corresponding argument in *pack.filterSpherePack*

   **Returns** SpherePack object with spheres, filtered by the predicate.

yade.pack.**randomPeriPack**(*radius, initSize, rRelFuzz=0.0, memoizeDb=None, noPrint=False, seed=-1*)
   Generate periodic dense packing.

   A cell of initSize is stuffed with as many spheres as possible, then we run periodic compression with PeriIsoCompressor, just like with randomDensePack.

      **Parameters**

- **radius** – mean sphere radius

- **rRelFuzz** – relative fuzz of sphere radius (equal distribution); see the same param for randomDensePack.

- **initSize** – initial size of the periodic cell.

   **Returns** SpherePack object, which also contains periodicity information.

yade.pack.**regularHexa**(*predicate, radius, gap, **kw*)
   Return set of spheres in regular hexagonal grid, clipped inside solid given by predicate. Created spheres will have given radius and will be separated by gap space.

yade.pack.**regularOrtho**(*predicate, radius, gap, **kw*)
   Return set of spheres in regular orthogonal grid, clipped inside solid given by predicate. Created spheres will have given radius and will be separated by gap space.

yade.pack.**revolutionSurfaceMeridians**(*sects, angles, origin=Vector3(0, 0, 0), orientation=Quaternion((1, 0, 0), 0)*)
   Revolution surface given sequences of 2d points and sequence of corresponding angles, returning sequences of 3d points representing meridian sections of the revolution surface. The 2d sections are turned around z-axis, but they can be transformed using the origin and orientation arguments to give arbitrary orientation.

`yade.pack.`**`sweptPolylines2gtsSurface`**`(pts, threshold=0, capStart=False, capEnd=False)`
  Create swept suface (as GTS triangulation) given same-length sequences of points (as 3-tuples).

  If threshold is given (>0), then

  - degenerate faces (with edges shorter than threshold) will not be created

  - gts.Surface().cleanup(threshold) will be called before returning, which merges vertices mutually closer than threshold. In case your pts are closed (last point concident with the first one) this will be surface strip of triangles. If you additionally have capStart==True and capEnd==True, the surface will be closed.

---

**Note:** capStart and capEnd make the most naive polygon triangulation (diagonals) and will perhaps fail for non-convex sections.

---

> **Warning:** the algorithm connects points sequentially; if two polylines are mutually rotated or have inverse sense, the algorithm will not detect it and connect them regardless in their given order.

Creation, manipulation, IO for generic sphere packings.

**class** `yade._packSpheres.`**`SpherePack`**
  Set of spheres represented as centers and radii. This class is returned by *pack.randomDensePack*, *pack.randomPeriPack* and others. The object supports iteration over spheres, as in

```
>>> sp=SpherePack()
>>> for center,radius in sp: print center,radius
```

```
>>> for sphere in sp: print sphere[0],sphere[1]   ## same, but without unpacking the
↪tuple automatically
```

```
>>> for i in range(0,len(sp)): print sp[i][0], sp[i][1]   ## same, but accessing spheres
↪by index
```

---

**Special constructors**

Construct from list of `[(c1,r1),(c2,r2),…]`. To convert two same-length lists of `centers` and `radii`, construct with `zip(centers,radii)`.

---

**`__init__`**`((object)arg1[, (list)list])` → None :
  Empty constructor, optionally taking list [ ((cx,cy,cz),r), … ] for initial data.

**`aabb`**`((SpherePack)arg1)` → tuple :
  Get axis-aligned bounding box coordinates, as 2 3-tuples.

**`add`**`((SpherePack)arg1, (Vector3)arg2, (float)arg3)` → None :
  Add single sphere to packing, given center as 3-tuple and radius

**`appliedPsdScaling`**
  A factor between 0 and 1, uniformly applied on all sizes of the PSD.

**`cellFill`**`((SpherePack)arg1, (Vector3)arg2)` → None :
  Repeat the packing (if periodic) so that the results has dim() >= given size. The packing retains periodicity, but changes cellSize. Raises exception for non-periodic packing.

**`cellRepeat`**`((SpherePack)arg1, (Vector3i)arg2)` → None :
  Repeat the packing given number of times in each dimension. Periodicity is retained, cellSize changes. Raises exception for non-periodic packing.

**cellSize**

Size of periodic cell; is Vector3(0,0,0) if not periodic. (Change this property only if you know what you're doing).

**center**(*(SpherePack)arg1*) → Vector3 :

Return coordinates of the bounding box center.

**dim**(*(SpherePack)arg1*) → Vector3 :

Return dimensions of the packing in terms of aabb(), as a 3-tuple.

**fromList**(*(SpherePack)arg1, (list)arg2*) → None :

Make packing from given list, same format as for constructor. Discards current data.

**fromList( (SpherePack)arg1, (object)centers, (object)radii) -> None :** Make packing from given list, same format as for constructor. Discards current data.

**fromSimulation**(*(SpherePack)arg1*) → None :

Make packing corresponding to the current simulation. Discards current data.

**getClumps**(*(SpherePack)arg1*) → tuple :

Return lists of sphere ids sorted by clumps they belong to. The return value is (standalones,[clump1,clump2,...]), where each item is list of id's of spheres.

**hasClumps**(*(SpherePack)arg1*) → bool :

Whether this object contains clumps.

**isPeriodic**

was the packing generated in periodic boundaries?

**load**(*(SpherePack)arg1, (str)fileName*) → None :

Load packing from external text file (current data will be discarded).

**makeCloud**(*(SpherePack)arg1*$\big[$, *(Vector3)minCorner=Vector3(0, 0, 0)*$\big[$, *(Vector3)maxCorner=Vector3(0, 0, 0)*$\big[$, *(float)rMean=-1*$\big[$, *(float)rRelFuzz=0*$\big[$, *(int)num=-1*$\big[$, *(bool)periodic=False*$\big[$, *(float)porosity=0.65*$\big[$, *(object)psdSizes=[]*$\big[$, *(object)psdCumm=[]*$\big[$, *(bool)distributeMass=False*$\big[$, *(int)seed=-1*$\big[$, *(Matrix3)hSize=Matrix3(0, 0, 0, 0, 0, 0, 0, 0, 0)*$\big]\big]\big]\big]\big]\big]\big]\big]\big]\big]\big]\big]$) → int :

Create a random cloud of particles enclosed in a parallelepiped. The resulting packing is a gas-like state with no contacts between particles initially. Usually used as a first step before reaching a dense packing.

**Parameters**

- **minCorner** (`Vector3`) – lower corner of an axis-aligned box

- **maxCorner** (`Vector3`) – upper corner of an axis-aligned box

- **hSize** (`Matrix3`) – base vectors of a generalized box (arbitrary parallelepiped, typically *Cell::hSize*), superseeds minCorner and maxCorner if defined. For periodic boundaries only.

- **rMean** (`float`) – mean radius or spheres

- **rRelFuzz** (`float`) – dispersion of radius relative to rMean

- **num** (`int`) – number of spheres to be generated. If negative (default), generate as many as possible with stochastic sizes, ending after a fixed number of tries to place the sphere in space, else generate exactly **num** spheres with deterministic size distribution.

- **periodic** (`bool`) – whether the packing to be generated should be periodic

- **porosity** (`float`) – initial guess for the iterative generation procedure (if **num**>1). The algorithm will be retrying until the number of generated spheres is **num**. The first iteration tries with the provided porosity, but next iterations increase it if necessary (hence an initialy high porosity can speed-up the algorithm). If `psdSizes` is not defined, `rRelFuzz` (z) and `num` (N) are used so that the porosity given ($\rho$) is approximately achieved at the end of generation, $r_m = \sqrt[3]{\frac{V(1-\rho)}{\frac{4}{3}\pi(1+z^2)N}}$. The default is $\rho$=0.5. The optimal value depends on `rRelFuzz` or `psdSizes`.

- **psdSizes** – sieve sizes (particle diameters) when particle size distribution (PSD) is specified.

- **psdCumm** – cummulative fractions of particle sizes given by `psdSizes`; must be the same length as *psdSizes* and should be non-decreasing.

- **distributeMass** (*bool*) – if `True`, given distribution will be used to distribute sphere's mass rather than radius of them.

- **seed** – number used to initialize the random number generator.

**Returns** number of created spheres, which can be lower than `num` depending on the method used.

---

**Note:**

- Works in 2D if `minCorner[k]=maxCorner[k]` for one coordinate.

- If `num` is defined, then sizes generation is deterministic, giving the best fit of target distribution. It enables spheres placement in descending size order, thus giving lower porosity than the random generation.

- By default (with `distributeMass==False`), the distribution is applied to particle count (i.e. particle count percent passing). The typical geomechanics sense of "particle size distribution" is the distribution of *mass fraction* (i.e. mass percent passing); this can be achieved with `distributeMass=True`.

- Sphere radius distribution can be specified using one of the following ways:

  1. `rMean`, `rRelFuzz` and `num` gives uniform radius distribution in `rMean`×(1±`rRelFuzz`). Less than `num` spheres can be generated if it is too high.

  2. `rRelFuzz`, `num` and (optional) `porosity`, which estimates mean radius so that `porosity` is attained at the end. `rMean` must be less than 0 (default). `porosity` is only an initial guess for the generation algorithm, which will retry with higher porosity until the prescibed *num* is obtained.

  3. `psdSizes` and `psdCumm`, two arrays specifying points of the particle size distribution function. As many spheres as possible are generated.

  4. `psdSizes`, `psdCumm`, `num`, and (optional) `porosity`, like above but if `num` is not obtained, `psdSizes` will be scaled down uniformly, until `num` is obtained (see *appliedPsdScaling*).

---

**makeClumpCloud**(*(SpherePack)arg1*, *(Vector3)minCorner*, *(Vector3)maxCorner*, *(object)clumps*[, *(bool)periodic=False*[, *(int)num=-1*[, *(int)seed=-1*]]]) → int

Create a random loose packing of clumps the same way `makeCloud` does with spheres. The parameters `minCorner`, `maxCorner`, `periodic`, `num` and `seed` are the same as in `makeCloud`. The parameter `clumps` is a list containing all the different clumps to be appended as `SpherePack` objects. Here is an exemple that shows how to create a cloud made of 10 identical clumps :

```
clp = SpherePack([((0,0,0), 1e-2), ((1e-2,0,0), 1e-2)]) # The clump we want a cloud
↪of
sp = SpherePack()
sp.makeClumpCloud((0,0,0), (1,1,1), [clp], num=10, seed=42)
sp.toSimulation() # All the particles in the cloud are now appended to O.bodies
```

**psd**(*(SpherePack)arg1*[, *(int)bins=50*[, *(bool)mass=True*]]) → tuple :
    Return particle size distribution of the packing.

> **Parameters**
>> • **bins** (`int`) – number of bins between minimum and maximum diameter
>>
>> • **mass** – Compute relative mass rather than relative particle count for each bin. Corresponds to *distributeMass parameter for makeCloud.*
>
> **Returns** tuple of (`cumm,edges`), where `cumm` are cummulative fractions for respective diameters and `edges` are those diameter values. Dimension of both arrays is equal to `bins+1`.

**relDensity**(*(SpherePack)arg1*) → float :
    Relative packing density, measured as sum of spheres' volumes / aabb volume. (Sphere overlaps are ignored.)

**rotate**(*(SpherePack)arg1*, *(Vector3)axis*, *(float)angle*) → None :
    Rotate all spheres around packing center (in terms of aabb()), given axis and angle of the rotation.

**save**(*(SpherePack)arg1*, *(str)fileName*) → None :
    Save packing to external text file (will be overwritten).

**scale**(*(SpherePack)arg1*, *(float)arg2*) → None :
    Scale the packing around its center (in terms of aabb()) by given factor (may be negative).

**toList**(*(SpherePack)arg1*) → list :
    Return packing data as python list.

**toSimulation**(*rot=Matrix3(1, 0, 0, 0, 1, 0, 0, 0, 1)*, *\*\*kw*)
    Append spheres directly to the simulation. In addition calling *O.bodies.append*, this method also appropriately sets periodic cell information of the simulation.

```
>>> from yade import pack; from math import *
>>> sp=pack.SpherePack()
```

Create random periodic packing with 20 spheres:

```
>>> sp.makeCloud((0,0,0),(5,5,5),rMean=.5,rRelFuzz=.5,periodic=True,num=20)
20
```

Virgin simulation is aperiodic:

```
>>> O.reset()
>>> O.periodic
False
```

Add generated packing to the simulation, rotated by 45° along +z

```
>>> sp.toSimulation(rot=Quaternion((0,0,1),pi/4),color=(0,0,1))
[0, 1, 2, 3, 4, 5, 6, 7, 8, 9, 10, 11, 12, 13, 14, 15, 16, 17, 18, 19]
```

Periodic properties are transferred to the simulation correctly, including rotation (this could be avoided by explicitly passing "hSize=O.cell.hSize" as an argument):

```
>>> O.periodic
True
>>> O.cell.refSize
Vector3(5,5,5)
>>> O.cell.hSize # doctest: +SKIP
Matrix3(3.53553,-3.53553,0, 3.53553,3.53553,0, 0,0,5)
```

The current state (even if rotated) is taken as mechanically undeformed, i.e. with identity transformation:

```
>>> O.cell.trsf
Matrix3(1,0,0, 0,1,0, 0,0,1)
```

**Parameters**

- **rot** (*Quaternion/Matrix3*) – rotation of the packing, which will be applied on spheres and will be used to set *Cell.trsf* as well.

- **\*\*kw** – passed to *utils.sphere*

**Returns** list of body ids added (like *O.bodies.append*)

**translate**(*(SpherePack)arg1, (Vector3)arg2*) → None :
    Translate all spheres by given vector.

**class yade._packSpheres.SpherePackIterator**

**__init__**(*(object)arg1, (SpherePackIterator)arg2*) → None

**next**()
    ___next___( (SpherePackIterator)arg1) -> tuple

Spatial predicates for volumes (defined analytically or by triangulation).

**class yade._packPredicates.Predicate**

**aabb**(*(Predicate)arg1*) → tuple
    aabb( (Predicate)arg1) -> None

**center**(*(Predicate)arg1*) → Vector3

**dim**(*(Predicate)arg1*) → Vector3

**class yade._packPredicates.PredicateBoolean**(*inherits Predicate*)
    Boolean operation on 2 predicates (abstract class)

**A**

**B**

**__init__**()
    Raises an exception This class cannot be instantiated from Python

**aabb**(*(Predicate)arg1*) → tuple
    aabb( (Predicate)arg1) -> None

**center**(*(Predicate)arg1*) → Vector3

**dim**(*(Predicate)arg1*) → Vector3

**class yade._packPredicates.PredicateDifference**(*inherits PredicateBoolean → Predicate*)
    Difference (conjunction with negative predicate) of 2 predicates. A point has to be inside the first and outside the second predicate. Can be constructed using the - operator on predicates: **pred1 - pred2**.

**A**

**B**

**__init__**(*(object)arg1, (object)arg2, (object)arg3*) → None

**aabb**(*(Predicate)arg1*) → tuple
    aabb( (Predicate)arg1) -> None

**center**(*(Predicate)arg1*) → Vector3

**dim**(*(Predicate)arg1*) → Vector3

**class yade._packPredicates.PredicateIntersection**(*inherits PredicateBoolean → Predicate*)
Intersection (conjunction) of 2 predicates. A point has to be inside both predicates. Can be constructed using the `&` operator on predicates: `pred1 & pred2`.

**A**

**B**

**__init__**(*(object)arg1, (object)arg2, (object)arg3*) → None

**aabb**(*(Predicate)arg1*) → tuple
aabb( (Predicate)arg1) -> None

**center**(*(Predicate)arg1*) → Vector3

**dim**(*(Predicate)arg1*) → Vector3

**class yade._packPredicates.PredicateSymmetricDifference**(*inherits PredicateBoolean →*
*Predicate*)
SymmetricDifference (exclusive disjunction) of 2 predicates. A point has to be in exactly one predicate of the two. Can be constructed using the `^` operator on predicates: `pred1 ^ pred2`.

**A**

**B**

**__init__**(*(object)arg1, (object)arg2, (object)arg3*) → None

**aabb**(*(Predicate)arg1*) → tuple
aabb( (Predicate)arg1) -> None

**center**(*(Predicate)arg1*) → Vector3

**dim**(*(Predicate)arg1*) → Vector3

**class yade._packPredicates.PredicateUnion**(*inherits PredicateBoolean → Predicate*)
Union (non-exclusive disjunction) of 2 predicates. A point has to be inside any of the two predicates to be inside. Can be constructed using the `|` operator on predicates: `pred1 | pred2`.

**A**

**B**

**__init__**(*(object)arg1, (object)arg2, (object)arg3*) → None

**aabb**(*(Predicate)arg1*) → tuple
aabb( (Predicate)arg1) -> None

**center**(*(Predicate)arg1*) → Vector3

**dim**(*(Predicate)arg1*) → Vector3

**class yade._packPredicates.inAlignedBox**(*inherits Predicate*)
Axis-aligned box predicate

**__init__**(*(object)arg1, (Vector3)minAABB, (Vector3)maxAABB*) → None :
Ctor taking minumum and maximum points of the box (as 3-tuples).

**aabb**(*(Predicate)arg1*) → tuple
aabb( (Predicate)arg1) -> None

**center**(*(Predicate)arg1*) → Vector3

**dim**(*(Predicate)arg1*) → Vector3

**class yade._packPredicates.inCylinder**(*inherits Predicate*)
Cylinder predicate

**__init__**(*(object)arg1, (Vector3)centerBottom, (Vector3)centerTop, (float)radius*) → None :
Ctor taking centers of the lateral walls (as 3-tuples) and radius.

      **aabb**(*(Predicate)arg1*) → tuple
          aabb( (Predicate)arg1) -> None

      **center**(*(Predicate)arg1*) → Vector3

      **dim**(*(Predicate)arg1*) → Vector3

**class yade._packPredicates.inEllipsoid**(*inherits Predicate*)
      Ellipsoid predicate

      **__init__**(*(object)arg1, (Vector3)centerPoint, (Vector3)abc*) → None :
          Ctor taking center of the ellipsoid (3-tuple) and its 3 radii (3-tuple).

      **aabb**(*(Predicate)arg1*) → tuple
          aabb( (Predicate)arg1) -> None

      **center**(*(Predicate)arg1*) → Vector3

      **dim**(*(Predicate)arg1*) → Vector3

**class yade._packPredicates.inHyperboloid**(*inherits Predicate*)
      Hyperboloid predicate

      **__init__**(*(object)arg1, (Vector3)centerBottom, (Vector3)centerTop, (float)radius, (float)skirt*) → None :
          Ctor taking centers of the lateral walls (as 3-tuples), radius at bases and skirt (middle radius).

      **aabb**(*(Predicate)arg1*) → tuple
          aabb( (Predicate)arg1) -> None

      **center**(*(Predicate)arg1*) → Vector3

      **dim**(*(Predicate)arg1*) → Vector3

**class yade._packPredicates.inParallelepiped**(*inherits Predicate*)
      Parallelepiped predicate

      **__init__**(*(object)arg1, (Vector3)o, (Vector3)a, (Vector3)b, (Vector3)c*) → None :
          Ctor taking four points: `o` (for origin) and then `a`, `b`, `c` which define endpoints of 3 respective edges from `o`.

      **aabb**(*(Predicate)arg1*) → tuple
          aabb( (Predicate)arg1) -> None

      **center**(*(Predicate)arg1*) → Vector3

      **dim**(*(Predicate)arg1*) → Vector3

**class yade._packPredicates.inSphere**(*inherits Predicate*)
      Sphere predicate.

      **__init__**(*(object)arg1, (Vector3)center, (float)radius*) → None :
          Ctor taking center (as a 3-tuple) and radius

      **aabb**(*(Predicate)arg1*) → tuple
          aabb( (Predicate)arg1) -> None

      **center**(*(Predicate)arg1*) → Vector3

      **dim**(*(Predicate)arg1*) → Vector3

**class yade._packPredicates.notInNotch**(*inherits Predicate*)
      Outside of infinite, rectangle-shaped notch predicate

      **__init__**(*(object)arg1, (Vector3)centerPoint, (Vector3)edge, (Vector3)normal, (float)aperture*) → None :
          Ctor taking point in the symmetry plane, vector pointing along the edge, plane normal and aperture size. The side inside the notch is edge×normal. Normal is made perpendicular to the edge. All vectors are normalized at construction time.

    **aabb**(*(Predicate)arg1*) → tuple

        aabb( (Predicate)arg1) -> None

    **center**(*(Predicate)arg1*) → Vector3

    **dim**(*(Predicate)arg1*) → Vector3

Computation of oriented bounding box for cloud of points.

**yade._packObb.cloudBestFitOBB**(*(tuple)arg1*) → tuple

    Return (Vector3 center, Vector3 halfSize, Quaternion orientation) of best-fit oriented bounding-box for given tuple of points (uses brute-force velome minimization, do not use for very large clouds).

## 2.4.12 yade.plot module

Module containing utility functions for plotting inside yade. See examples/simple-scene/simple-scene-plot.py or examples/concrete/uniax.py for example of usage.

**yade.plot.data = {'eps': [0.0001, 0.001, nan], 'force': [nan, nan, 1000.0], 'sigma': [12, nan, nan]**

    Global dictionary containing all data values, common for all plots, in the form {'name':[value,…],…}. Data should be added using plot.addData function. All [value,…] columns have the same length, they are padded with NaN if unspecified.

**yade.plot.plots = {'i': ('t',), 'i ': ('z1', 'v1')}**

    dictionary x-name -> (yspec,…), where yspec is either y-name or (y-name,'line-specification'). If (**yspec,...**) is **None**, then the plot has meaning of image, which will be taken from respective field of *plot.imgData*.

**yade.plot.labels = {}**

    Dictionary converting names in data to human-readable names (TeX names, for instance); if a variable is not specified, it is left untranslated.

**yade.plot.live = True**

    Enable/disable live plot updating.

**yade.plot.liveInterval = 1**

    Interval for the live plot updating, in seconds.

**yade.plot.setLiveForceAlwaysUpdate**(*forceLiveUpdate*)

    The *plot.liveInterval* and *plot.live* control live refreshing of the plot during calculations. The refreshing is done in a separate thread, so that it does not interfere with calculations. Drawing the data will not work when at exactly the same time it is being updated in other thread. Use **yade.plot.setLiveForceAlwaysUpdate(True)** if you want calculations to **PAUSE** during the plot updates. This function returns current **bool** value of forced updates if the call was a success, otherwise it returns a **str** with explanation why it failed. It is guaranteed to work if simulation was paused with *O.pause()* call.

**yade.plot.autozoom = True**

    Enable/disable automatic plot rezooming after data update. Sometimes rezooming must be skipped unless a call to *plot.setLiveForceAlwaysUpdate* forces it to work.

**yade.plot.plot**(*noShow=False, subPlots=True*)

    Do the actual plot, which is either shown on screen (and nothing is returned: if *noShow* is **False** - note that your yade compilation should present qt4 feature so that figures can be displayed) or, if *noShow* is **True**, returned as matplotlib's Figure object or list of them.

    You can use

```
>>> from yade import plot
>>> plot.resetData()
>>> plot.plots={'foo':('bar',)}
>>> plot.plot(noShow=True).savefig('someFile.pdf')
>>> import os
```

```
>>> os.path.exists('someFile.pdf')
True
>>> os.remove('someFile.pdf')
```

to save the figure to file automatically.

---

**Note:** For backwards compatibility reasons, *noShow* option will return list of figures for multiple figures but a single figure (rather than list with 1 element) if there is only 1 figure.

---

yade.plot.**reset**()
> Reset all plot-related variables (data, plots, labels)

yade.plot.**resetData**()
> Reset all plot data; keep plots and labels intact.

yade.plot.**splitData**()
> Make all plots discontinuous at this point (adds nan's to all data fields)

yade.plot.**reverseData**()
> Reverse yade.plot.data order.
>
> Useful for tension-compression test, where the initial (zero) state is loaded and, to make data continuous, last part must *end* in the zero state.

yade.plot.**addData**(*d_in, **kw*)
> Add data from arguments name1=value1,name2=value2 to yade.plot.data. (the old {'name1':value1,'name2':value2} is deprecated, but still supported)
>
> New data will be padded with nan's, unspecified data will be nan (nan's don't appear in graphs). This way, equal length of all data is assured so that they can be plotted one against any other.

```
>>> from yade import plot
>>> from pprint import pprint
>>> plot.resetData()
>>> plot.addData(a=1)
>>> plot.addData(b=2)
>>> plot.addData(a=3,b=4)
>>> pprint(plot.data)
{'a': [1, nan, 3], 'b': [nan, 2, 4]}
```

> Some sequence types can be given to addData; they will be saved in synthesized columns for individual components.

```
>>> plot.resetData()
>>> plot.addData(c=Vector3(5,6,7),d=Matrix3(8,9,10, 11,12,13, 14,15,16))
>>> pprint(plot.data)
{'c_x': [5.0],
 'c_y': [6.0],
 'c_z': [7.0],
 'd_xx': [8.0],
 'd_xy': [9.0],
 'd_xz': [10.0],
 'd_yx': [11.0],
 'd_yy': [12.0],
 'd_yz': [13.0],
 'd_zx': [14.0],
 'd_zy': [15.0],
 'd_zz': [16.0]}
```

yade.plot.**addAutoData**()

Add data by evaluating contents of *plot.plots*. Expressions rasing exceptions will be handled gracefully, but warning is printed for each.

```
>>> from yade import plot
>>> from pprint import pprint
>>> O.reset()
>>> plot.resetData()
>>> plot.plots={'O.iter':('O.time',None,'numParticles=len(O.bodies)')}
>>> plot.addAutoData()
>>> pprint(plot.data)
{'O.iter': [0], 'O.time': [0.0], 'numParticles': [0]}
```

Note that each item in *plot.plots* can be

- an expression to be evaluated (using the `eval` builtin);

- `name=expression` string, where `name` will appear as label in plots, and expression will be evaluated each time;

- a dictionary-like object – current keys are labels of plots and current values are added to *plot.data*. The contents of the dictionary can change over time, in which case new lines will be created as necessary.

A simple simulation with plot can be written in the following way; note how the energy plot is specified.

```
>>> from yade import plot, utils
>>> plot.plots={'i=O.iter':(O.energy,None,'total energy=O.energy.total()')}
>>> # we create a simple simulation with one ball falling down
>>> plot.resetData()
>>> O.bodies.append(utils.sphere((0,0,0),1))
0
>>> O.dt=utils.PWaveTimeStep()
>>> O.engines=[
...     ForceResetter(),
...     GravityEngine(gravity=(0,0,-10),warnOnce=False),
...     NewtonIntegrator(damping=.4,kinSplit=True),
...     # get data required by plots at every step
...     PyRunner(command='yade.plot.addAutoData()',iterPeriod=1,initRun=True)
... ]
>>> O.trackEnergy=True
>>> O.run(2,True)
>>> pprint(plot.data)    #doctest: +ELLIPSIS
{'gravWork': [0.0, -25.13274...],
 'i': [0, 1],
 'kinRot': [0.0, 0.0],
 'kinTrans': [0.0, 7.5398...],
 'nonviscDamp': [0.0, 10.0530...],
 'total energy': [0.0, -7.5398...]}
```

yade.plot.**saveGnuplot**(*baseName*, *term='wxt'*, *extension=None*, *timestamp=False*, *comment=None, title=None, varData=False*)

Save data added with *plot.addData* into (compressed) file and create .gnuplot file that attempts to mimick plots specified with *plot.plots*.

**Parameters**

- `baseName` – used for creating baseName.gnuplot (command file for gnuplot), associated `baseName.data.bz2` (data) and output files (if applicable) in the form `baseName.[plot number].extension`

- `term` – specify the gnuplot terminal; defaults to `x11`, in which case gnuplot will draw persistent windows to screen and terminate; other useful terminals are `png`, `cairopdf` and so on

- **extension** – extension for `baseName` defaults to terminal name; fine for png for example; if you use `cairopdf`, you should also say `extension='pdf'` however

- **timestamp** (`bool`) – append numeric time to the basename

- **varData** (`bool`) – whether file to plot will be declared as variable or be in-place in the plot expression

- **comment** – a user comment (may be multiline) that will be embedded in the control file

**Returns** name of the gnuplot file created.

`yade.plot.`**`saveDataTxt`**(*fileName*, *vars=None*, *headers=None*)

Save plot data into a (optionally compressed) text file. The first line contains a comment (starting with **#**) giving variable name for each of the columns. This format is suitable for being loaded for further processing (outside yade) with **numpy.genfromtxt** function, which recognizes those variable names (creating numpy array with named entries) and handles decompression transparently.

```
>>> from yade import plot
>>> from pprint import pprint
>>> plot.reset()
>>> plot.addData(a=1,b=11,c=21,d=31)    # add some data here
>>> plot.addData(a=2,b=12,c=22,d=32)
>>> pprint(plot.data)
{'a': [1, 2], 'b': [11, 12], 'c': [21, 22], 'd': [31, 32]}
>>> plot.saveDataTxt('/tmp/dataFile.txt.tar.gz',vars=('a','b','c'))
>>> import numpy
>>> d=numpy.genfromtxt('/tmp/dataFile.txt.tar.gz',dtype=None,names=True)
>>> d['a']
array([1, 2])
>>> d['b']
array([11, 12])
>>> import os # cleanup
>>> os.remove('/tmp/dataFile.txt.tar.gz')
```

**Parameters**

- **fileName** – file to save data to; if it ends with `.bz2` / `.gz`, the file will be compressed using bzip2 / gzip.

- **vars** – Sequence (tuple/list/set) of variable names to be saved. If `None` (default), all variables in *plot.plot* are saved.

- **headers** – Set of parameters to write on header

`yade.plot.`**`savePlotSequence`**(*fileBase*, *stride=1*, *imgRatio=(5, 7)*, *title=None*, *titleFrames=20*, *lastFrames=30*)

Save sequence of plots, each plot corresponding to one line in history. It is especially meant to be used for *utils.makeVideo*.

**Parameters**

- **stride** – only consider every stride-th line of history (default creates one frame per each line)

- **title** – Create title frame, where lines of title are separated with newlines (`\n`) and optional subtitle is separated from title by double newline.

- **titleFrames** (`int`) – Create this number of frames with title (by repeating its filename), determines how long the title will stand in the movie.

- **lastFrames** (`int`) – Repeat the last frame this number of times, so that the movie does not end abruptly.

**Returns** List of filenames with consecutive frames.

## 2.4.13 yade.polyhedra_utils module

## 2.4.14 yade.post2d module

Module for 2d postprocessing, containing classes to project points from 3d to 2d in various ways, providing basic but flexible framework for extracting arbitrary scalar values from bodies/interactions and plotting the results. There are 2 basic components: flatteners and extractors.

The algorithms operate on bodies (default) or interactions, depending on the `intr` parameter of post2d.data.

**Flatteners**

Instance of classes that convert 3d (model) coordinates to 2d (plot) coordinates. Their interface is defined by the *post2d.Flatten* class (`__call__`, `planar`, `normal`).

**Extractors**

Callable objects returning scalar or vector value, given a body/interaction object. If a 3d vector is returned, Flattener.planar is called, which should return only in-plane components of the vector.

**Example**

This example can be found in examples/concrete/uniax-post.py

```python
from yade import post2d
import pylab # the matlab-like interface of matplotlib

O.load('/tmp/uniax-tension.xml.bz2')

# flattener that project to the xz plane
flattener=post2d.AxisFlatten(useRef=False,axis=1)
# return scalar given a Body instance
extractDmg=lambda b: b.state.normDmg
# will call flattener.planar implicitly
# the same as: extractVelocity=lambda b: flattener.planar(b,b.state.vel)
extractVelocity=lambda b: b.state.vel

# create new figure
pylab.figure()
# plot raw damage
post2d.plot(post2d.data(extractDmg,flattener))

# plot smooth damage into new figure
pylab.figure(); ax,map=post2d.plot(post2d.data(extractDmg,flattener,stDev=2e-3))
# show color scale
pylab.colorbar(map,orientation='horizontal')

# raw velocity (vector field) plot
pylab.figure(); post2d.plot(post2d.data(extractVelocity,flattener))

# smooth velocity plot; data are sampled at regular grid
pylab.figure(); ax,map=post2d.plot(post2d.data(extractVelocity,flattener,stDev=1e-3))
# save last (current) figure to file
pylab.gcf().savefig('/tmp/foo.png')

# show the figures
pylab.show()
```

**class** `yade.post2d.AxisFlatten`(*inherits Flatten → object*)

> **__init__**(*useRef=False, axis=2*)
>
> > **Parameters**
> >
> > - **useRef** (`bool`) – use reference positions rather than actual positions (only meaningful when operating on Bodies)
> > - **axis** (`{0,1,2}`) – axis normal to the plane; the return value will be simply position with this component dropped.
>
> **normal**(*pos, vec*)
> > Given position and vector value, return lenght of the vector normal to the flat plane.
>
> **planar**(*pos, vec*)
> > Given position and vector value, project the vector value to the flat plane and return its 2 in-plane components.

**class** `yade.post2d.CylinderFlatten`(*inherits Flatten → object*)
> Class for converting 3d point to 2d based on projection onto plane from circle. The y-axis in the projection corresponds to the rotation axis; the x-axis is distance form the axis.
>
> > **__init__**(*useRef, axis=2*)
> >
> > > **Parameters**
> > >
> > > - **useRef** – (bool) use reference positions rather than actual positions
> > > - **axis** – axis of the cylinder, {0,1,2}
> >
> > **normal**(*b, vec*)
> > > Given position and vector value, return lenght of the vector normal to the flat plane.
> >
> > **planar**(*b, vec*)
> > > Given position and vector value, project the vector value to the flat plane and return its 2 in-plane components.

**class** `yade.post2d.Flatten`(*inherits object*)
> Abstract class for converting 3d point into 2d. Used by post2d.data2d.
>
> **normal**(*pos, vec*)
> > Given position and vector value, return lenght of the vector normal to the flat plane.
>
> **planar**(*pos, vec*)
> > Given position and vector value, project the vector value to the flat plane and return its 2 in-plane components.

**class** `yade.post2d.HelixFlatten`(*inherits Flatten → object*)
> Class converting 3d point to 2d based on projection from helix. The y-axis in the projection corresponds to the rotation axis
>
> > **__init__**(*useRef, thetaRange, dH_dTheta, axis=2, periodStart=0*)
> >
> > > **Parameters**
> > >
> > > - **useRef** (`bool`) – use reference positions rather than actual positions
> > > - **thetaRange** (`(ϑmin,ϑmax)`) – bodies outside this range will be discarded
> > > - **dH_dTheta** (`float`) – inclination of the spiral (per radian)
> > > - **axis** (`{0,1,2}`) – axis of rotation of the spiral
> > > - **periodStart** (`float`) – height of the spiral for zero angle
>
> **normal**(*pos, vec*)
> > Given position and vector value, return lenght of the vector normal to the flat plane.

**planar**(*b, vec*)

Given position and vector value, project the vector value to the flat plane and return its 2 in-plane components.

**yade.post2d.data**(*extractor, flattener, intr=False, onlyDynamic=True, stDev=None, relThreshold=3.0, perArea=0, div=(50, 50), margin=(0, 0), radius=1*)

Filter all bodies/interactions, project them to 2d and extract required scalar value; return either discrete array of positions and values, or smoothed data, depending on whether the stDev value is specified.

The **intr** parameter determines whether we operate on bodies or interactions; the extractor provided should expect to receive body/interaction.

> **Parameters**
>
> - **extractor** (*callable*) – receives *Body* (or *Interaction*, if **intr** is **True**) instance, should return scalar, a 2-tuple (vector fields) or None (to skip that body/interaction)
>
> - **flattener** (*callable*) – *post2d.Flatten* instance, receiving body/interaction, returns its 2d coordinates or **None** (to skip that body/interaction)
>
> - **intr** (*bool*) – operate on interactions rather than bodies
>
> - **onlyDynamic** (*bool*) – skip all non-dynamic bodies
>
> - **stDev** (*float/None*) – standard deviation for averaging, enables smoothing; **None** (default) means raw mode, where discrete points are returned
>
> - **relThreshold** (*float*) – threshold for the gaussian weight function relative to stDev (smooth mode only)
>
> - **perArea** (*int*) – if 1, compute weightedSum/weightedArea rather than weighted average (weightedSum/sumWeights); the first is useful to compute average stress; if 2, compute averages on subdivision elements, not using weight function
>
> - **div** (*(int,int)*) – number of cells for the gaussian grid (smooth mode only)
>
> - **margin** (*(float,float)*) – x,y margins around bounding box for data (smooth mode only)
>
> - **radius** (*float/callable*) – Fallback value for radius (for raw plotting) for non-spherical bodies or interactions; if a callable, receives body/interaction and returns radius

> **Returns** dictionary

Returned dictionary always containing keys 'type' (one of 'rawScalar','rawVector','smoothScalar','smoothVector', depending on value of smooth and on return value from extractor), 'x', 'y', 'bbox'.

Raw data further contains 'radii'.

Scalar fields contain 'val' (value from *extractor*), vector fields have 'valX' and 'valY' (2 components returned by the *extractor*).

**yade.post2d.plot**(*data, axes=None, alpha=0.5, clabel=True, cbar=False, aspect='equal', **kw*)

Given output from post2d.data, plot the scalar as discrete or smooth plot.

For raw discrete data, plot filled circles with radii of particles, colored by the scalar value.

For smooth discrete data, plot image with optional contours and contour labels.

For vector data (raw or smooth), plot quiver (vector field), with arrows colored by the magnitude.

> **Parameters**
>
> - **axes** – matplotlib.axesinstance where the figure will be plotted; if None, will be created from scratch.

- **data** – value returned by *post2d.data*
- **clabel** (*bool*) – show contour labels (smooth mode only), or annotate cells with numbers inside (with perArea==2)
- **cbar** (*bool*) – show colorbar (equivalent to calling pylab.colorbar(mappable) on the returned mappable)

**Returns** tuple of (`axes,mappable`); mappable can be used in further calls to pylab.colorbar.

## 2.4.15 yade.qt module

Common initialization core for yade.

This file is executed when anything is imported from yade for the first time. It loads yade plugins and injects c++ class constructors to the ___builtins___ (that might change in the future, though) namespace, making them available everywhere.

**class** yade.qt._GLViewer.GLViewer

**__init__()**
Raises an exception This class cannot be instantiated from Python

**axes**
Show arrows for axes.

**center**(*(GLViewer)arg1*[, *(bool)median=True*[, *(float)suggestedRadius=-1.0*]]) → None :
Center view. View is centered either so that all bodies fit inside (*median* = False), or so that 75% of bodies fit inside (*median* = True). If radius cannot be determined automatically then suggestedRadius is used.

**close**(*(GLViewer)arg1*) → None

**eyePosition**
Camera position.

**fitAABB**(*(GLViewer)arg1, (Vector3)mn, (Vector3)mx*) → None :
Adjust scene bounds so that Axis-aligned bounding box given by its lower and upper corners *mn, mx* fits in.

**fitSphere**(*(GLViewer)arg1, (Vector3)center, (float)radius*) → None :
Adjust scene bounds so that sphere given by *center* and *radius* fits in.

**fps**
Show frames per second indicator.

**grid**
Display square grid in zero planes, as 3-tuple of bools for yz, xz, xy planes.

**loadState**(*(GLViewer)arg1*[, *(str)stateFilename='.qglviewer.xml'*]) → None :
Load display parameters from file saved previously into.

**lookAt**
Point at which camera is directed.

**ortho**
Whether orthographic projection is used; if false, use perspective projection.

**saveSnapshot**(*(GLViewer)arg1, (str)filename*) → None :
Save the current view to image file

**saveState**(*(GLViewer)arg1*[, *(str)stateFilename='.qglviewer.xml'*]) → None :
Save display parameters into a file. Saves state for both *GLViewer* and associated *OpenGLRenderer*.

**scale**
> Scale of the view (?)

**sceneRadius**
> Visible scene radius.

**screenSize**
> Size of the viewer's window, in screen pixels

**selection**

**showEntireScene**(*(GLViewer)arg1*) → None

**timeDisp**
> Time displayed on in the window; is a string composed of characters $r$, $v$, $i$ standing respectively for real time, virtual time, iteration number.

**upVector**
> Vector that will be shown oriented up on the screen.

**viewDir**
> Camera orientation (as vector).

**yade.qt._GLViewer.Renderer**() → OpenGLRenderer
> Return the active *OpenGLRenderer* object.

**yade.qt._GLViewer.View**( $\big[$*(float)timeout=5.0*$\big]$ ) → GLViewer
> Create a new 3d view.

**yade.qt._GLViewer.center**( $\big[$*(float)suggestedRadius=-1.0*$\big[$, *(Vector3)gridOrigin=Vector3(0, 0, 0)*$\big[$, *(Vector3)suggestedCenter=Vector3(0, 0, 0)*$\big[$, *(int)gridDecimalPlaces=4*$\big]\big]\big]\big]$ ) → None
> Center all views.

> **Parameters**
>
> - **suggestedRadius** – optional parameter, if provided and positive then it will be used instead of automatic radius detection. This parameter affects the (1) size of grid being drawn (2) the Z-clipping distance in OpenGL, it means that if clipping is too large and some of your scene is not being drawn but is "cut" or "sliced" then this parameter needs to be bigger.
>
> - **gridOrigin** – optional parameter, if provided it will be used as the origin for drawing the grid. Meaning the intersection of all three grids will not be at 0,0,0; but at the provided coordinate rounded to the nearest gridStep.
>
> - **suggestedCenter** – optional parameter, if provided other than (0,0,0) then it will be used instead of automatic calculation of scene center using bounding boxes. This parameter affects the drawn rotation-center. If you try to rotate the view, and the rotation is around some strange point, then this parameter needs to be changed.
>
> - **gridDecimalPlaces** – default value=4, determines the number of decimal places to be shown on grid labels using stringstream (extra zeros are not shown).

> **Note:** You can get the current values of all these four arguments by invoking command: *qt.centerValues()*

**yade.qt._GLViewer.centerValues**() → dict

> **Returns** a dictionary with all parameters currently used by **yade.qt.center**(...), see *qt.center* or type **yade.qt.center?** for details. Returns zeros if view is closed.

**yade.qt._GLViewer.views**() → list

> **Returns** a list of all open *qt.GLViewer* objects

If one needs to exactly copy camera position and settings between two different yade sessions, the following commands can be used:

```
v=yade.qt.views()[0]                          ## to obtain a handle of currently opened
↪view.
v.lookAt, v.viewDir, v.eyePosition, v.upVector ## to print the current camera parameters
↪of the view.

## Then copy the output of this command into the second yade session to reposition the
↪camera.
v.lookAt, v.viewDir, v.eyePosition, v.upVector = (Vector3(-0.5,1.6,0.47),Vector3(-0.5,0.6,
↪0.4),Vector3(0.015,0.98,-0.012),Vector3(0.84,0.46,0.27))
## Since these parameters depend on each other it might be necessary to execute this
↪command twice.
```

Also one can call *qt.centerValues()* to obtain current settings of axis and scene radius (if defaults are not used) and apply them via call to *qt.center* in the second yade session.

This cumbersome method above may be improved in the future.

## 2.4.16 yade.timing module

Functions for accessing timing information stored in engines and functors.

See *Timing* section of the programmer's manual, wiki page for some examples.

yade.timing.**reset**()
 Zero all timing data.

yade.timing.**runtime**()
 Return total running time (same as last line in the output of stats()) in nanoseconds

yade.timing.**stats**()
 Print summary table of timing information from engines and functors. Absolute times as well as percentages are given. Sample output:

```
Name                                                Count              Time
↪    Rel. time
-------------------------------------------------------------------------------
↪-------------
ForceResetter                                       102                2150us
↪  0.02%
"collider"                                          5                  64200us
↪  0.60%
InteractionLoop                                     102                10571887us
↪ 98.49%
"combEngine"                                        102                8362us
↪  0.08%
"newton"                                            102                73166us
↪  0.68%
"cpmStateUpdater"                                   1                  9605us
↪  0.09%
PyRunner                                            1                  136us
↪  0.00%
"plotDataCollector"                                 1                  291us
↪  0.00%
TOTAL                                                                  10733564us
↪100.00%
```

sample output (compiled with -DENABLE_PROFILING=1 option):

```
Name                                            Count              Time
↪    Rel. time
--------------------------------------------------------------------------------
↪-------------
ForceResetter                                   102                 2150us
↪  0.02%
"collider"                                      5                  64200us
↪  0.60%
InteractionLoop                                 102              10571887us
↪ 98.49%
  Ig2_Sphere_Sphere_ScGeom                      1222186           1723168us
↪   16.30%
    Ig2_Sphere_Sphere_ScGeom                    1222186           1723168us
↪    100.00%
  Ig2_Facet_Sphere_ScGeom                       753                 1157us
↪   0.01%
    Ig2_Facet_Sphere_ScGeom                     753                 1157us
↪    100.00%
  Ip2_CpmMat_CpmMat_CpmPhys                     11712              26015us
↪   0.25%
    end of Ip2_CpmPhys                          11712              26015us
↪    100.00%
  Ip2_FrictMat_CpmMat_FrictPhys                 0                     0us
↪   0.00%
  Law2_ScGeom_CpmPhys_Cpm                       3583872          4819289us
↪   45.59%
    GO A                                        1194624          1423738us
↪    29.54%
    GO B                                        1194624          1801250us
↪    37.38%
    rest                                        1194624          1594300us
↪    33.08%
    TOTAL                                       3583872          4819289us
↪    100.00%
  Law2_ScGeom_FrictPhys_CundallStrack           0                     0us
↪   0.00%
"combEngine"                                    102                 8362us
↪  0.08%
"newton"                                        102                73166us
↪  0.68%
"cpmStateUpdater"                               1                   9605us
↪  0.09%
PyRunner                                        1                    136us
↪  0.00%
"plotDataCollector"                             1                    291us
↪  0.00%
TOTAL                                                           10733564us
↪100.00%
```

## 2.4.17 yade.utils module

Heap of functions that don't (yet) fit anywhere else.

Devs: please DO NOT ADD more functions here, it is getting too crowded!

yade.utils.**NormalRestitution2DampingRate**($en$)

Compute the normal damping rate as a function of the normal coefficient of restitution $e_n$. For $e_n \in \langle 0, 1 \rangle$ damping rate equals

$$-\frac{\log e_n}{\sqrt{e_n^2 + \pi^2}}$$

**yade.utils.SpherePWaveTimeStep**(*radius*, *density*, *young*)

Compute P-wave critical timestep for a single (presumably representative) sphere, using formula for P-Wave propagation speed $\Delta t_c = \frac{r}{\sqrt{E/\rho}}$. If you want to compute minimum critical timestep for all spheres in the simulation, use *utils.PWaveTimeStep* instead.

```
>>> SpherePWaveTimeStep(1e-3,2400,30e9)
2.8284271247461903e-07
```

**class yade.utils.TableParamReader**(*inherits object*)

Class for reading simulation parameters from text file.

Each parameter is represented by one column, each parameter set by one line. Colums are separated by blanks (no quoting).

First non-empty line contains column titles (without quotes). You may use special column named 'description' to describe this parameter set; if such colum is absent, description will be built by concatenating column names and corresponding values (param1=34,param2=12.22,param4=foo)

- from columns ending in **!** (the **!** is not included in the column name)

- from all columns, if no columns end in **!**.

Empty lines within the file are ignored (although counted); **#** starts comment till the end of line. Number of blank-separated columns must be the same for all non-empty lines.

A special value **=** can be used instead of parameter value; value from the previous non-empty line will be used instead (works recursively).

This class is used by *utils.readParamsFromTable*.

**__init__**(*file*)

Setup the reader class, read data into memory.

**paramDict**()

Return dictionary containing data from file given to constructor. Keys are line numbers (which might be non-contiguous and refer to real line numbers that one can see in text editors), values are dictionaries mapping parameter names to their values given in the file. The special value '=' has already been interpreted, **!** (bangs) (if any) were already removed from column titles, **description** column has already been added (if absent).

**yade.utils.aabbDim**(*cutoff=0.0*, *centers=False*)

Return dimensions of the axis-aligned bounding box, optionally with relative part *cutoff* cut away.

**yade.utils.aabbExtrema2d**(*pts*)

Return 2d bounding box for a sequence of 2-tuples.

**yade.utils.aabbWalls**(*extrema=None*, *thickness=0*, *oversizeFactor=1.5*, ***kw*)

Return 6 boxes that will wrap existing packing as walls from all sides.

**Parameters**

- **extrema** – extremal points of the Aabb of the packing, as a list of two Vector3, or any equivalent type (will be calculated if not specified)

- **thickness** (*float*) – is wall thickness (will be 1/10 of the X-dimension if not specified)

- **oversizeFactor** (*float*) – factor to enlarge walls in their plane.

**Returns** a list of 6 wall Bodies enclosing the packing, in the order minX,maxX,minY,maxY,minZ,maxZ.

**yade.utils.avgNumInteractions**(*cutoff=0.0*, *skipFree=False*, *considerClumps=False*)

Return average numer of interactions per particle, also known as *coordination number* Z. This number is defined as

$$Z = 2C/N$$

where C is number of contacts and N is number of particles. When clumps are present, number of particles is the sum of standalone spheres plus the sum of clumps. Clumps are considered in the calculation if cutoff != 0 or skipFree = True. If cutoff=0 (default) and skipFree=False (default) one needs to set considerClumps=True to consider clumps in the calculation.

With *skipFree*, particles not contributing to stable state of the packing are skipped, following equation (8) given in [Thornton2000]:

$$Z_m = \frac{2C - N_1}{N - N_0 - N_1}$$

> **Parameters**
>
> - **cutoff** – cut some relative part of the sample's bounding box away.
>
> - **skipFree** – see above.
>
> - **considerClumps** – also consider clumps if cutoff=0 and skipFree=False; for further explanation see above.

yade.utils.**box**(*center, extents, orientation=Quaternion((1, 0, 0), 0), dynamic=None, fixed=False, wire=False, color=None, highlight=False, material=-1, mask=1*)
  Create box (cuboid) with given parameters.

> **Parameters**
>
> - **extents** (Vector3) – half-sizes along x,y,z axes. Use can be made of *orientation* parameter in case those box-related axes do not conform the simulation axes
>
> - **orientation** (Quaternion) – assigned to the *body's orientation*, which corresponds to rotating the *extents* axes

See *utils.sphere*'s documentation for meaning of other parameters.

**class** yade.utils.**clumpTemplate**(*inherits object*)
  Create a clump template by a list of relative radii and a list of relative positions. Both lists must have the same length.

> **Parameters**
>
> - **relRadii** (*[float,float,..]*) – list of relative radii (minimum length = 2)
>
> - **relPositions** (*[Vector3,Vector3,..]*) – list of relative positions (minimum length = 2)

yade.utils.**defaultMaterial**()
  Return default material, when creating bodies with *utils.sphere* and friends, material is unspecified and there is no shared material defined yet. By default, this function returns

```
FrictMat(density=1e3,young=1e7,poisson=.3,frictionAngle=.5,label='defaultMat')
```

yade.utils.**facet**(*vertices, dynamic=None, fixed=True, wire=True, color=None, highlight=False, noBound=False, material=-1, mask=1, chain=-1*)
  Create facet with given parameters.

> **Parameters**
>
> - **vertices** (*[Vector3,Vector3,Vector3]*) – coordinates of vertices in the global coordinate system.
>
> - **wire** (*bool*) – if True, facets are shown as skeleton; otherwise facets are filled
>
> - **noBound** (*bool*) – set *Body.bounded*
>
> - **color** (*Vector3-or-None*) – color of the facet; random color will be assigned if None.

See *utils.sphere*'s documentation for meaning of other parameters.

`yade.utils.`**`fractionalBox`**(*fraction=1.0, minMax=None*)

 Return (min,max) that is the original minMax box (or aabb of the whole simulation if not specified) linearly scaled around its center to the fraction factor

`yade.utils.`**`levelSetBody`**(*shape='', center=Vector3(0, 0, 0), radius=0, extents=Vector3(0, 0, 0), epsilons=Vector2(0, 0), clump=None, spacing=0.1, grid=None, distField=[], nSurfNodes=27, nodesPath=2, nodesTol=50, orientation=Quaternion((1, 0, 0), 0), dynamic=True, material=-1)*)

 Creates a *LevelSet* shaped body through various workflows: one can choose among pre-defined shapes (through *shape* and related attributes), or to mimick a *Clump* instance (*clump* attribute, for comparison purposes), or directly assign the discrete distance field on some grid (*distField* and *grid* attributes) :param string shape: use this argument to enjoy predefined shapes among 'sphere', 'box' (for a rectangular parallelepiped), 'disk' (for a 2D analysis in (x,y) plane), or 'superellipsoid'; in conjunction with *extents* or *radius* attributes. Superellipsoid surfaces are defined in local axes (inertial frame) by the following equation: $f(x, y, z) = (|x/r_x|^{2/\varepsilon_e} + |y/r_y|^{2/\varepsilon_e})^{\varepsilon_e/\varepsilon_n} + |z/r_z|^{2/\varepsilon_n} = 1$ and their distance field is obtained thanks to a *Fast Marching Method*. :param Vector3 center: (initial) position of that body :param Clump clump: pass here a multi-sphere instance to mimick, if desired :param Real radius: imposed radius in case *shape* = 'sphere' or 'disk' :param Vector3 extents: half extents along the local axes in case *shape* = 'box' or 'superellipsoid' ($r_x, r_y, r_z$ for the latter) :param Vector2 epsilons: in case *shape* = 'superellipsoid', the ($\varepsilon_e, \varepsilon_n$) exponents :param Real spacing: spatial increment of the *level set grid*, if you picked a pre-defined *shape* or a *clump* :param list distField: the *discrete distance field* on *grid* (if given) as a list (of list of list; use .tolist() if working initially with 3D numpy arrays), where distField[i][j][k] is the distance value at grid.gridPoint(i,j,k) :param RegularGrid grid: the *grid carrying the distance field*, when the latter is directly assigned through *distField* :param int nSurfNodes: number of boundary nodes, passed to *LevelSet.nSurfNodes* :param int nodesPath: path for the boundary nodes, passed to *LevelSet.nodesPath* :param Real nodesTol: tolerance while ray tracing boundary nodes, passed to *LevelSet.nodesTol* :param Quaternion orientation: the initial orientation of the body :param bool dynamic: passed to *Body.dynamic* :param Material material: passed to *Body.material* :return: a corresponding body instance

`yade.utils.`**`loadVars`**(*mark=None*)

 Load variables from *utils.saveVars*, which are saved inside the simulation. If `mark==None`, all save variables are loaded. Otherwise only those with the mark passed.

`yade.utils.`**`makeVideo`**(*frameSpec, out, renameNotOverwrite=True, fps=24, kbps=6000, bps=None*)

 Create a video from external image files using mencoder. Two-pass encoding using the default mencoder codec (mpeg4) is performed, running multi-threaded with number of threads equal to number of OpenMP threads allocated for Yade.

 **Parameters**

- **frameSpec** – wildcard | sequence of filenames. If list or tuple, filenames to be encoded in given order; otherwise wildcard understood by mencoder's mf:// URI option (shell wildcards such as `/tmp/snap-*.png` or and printf-style pattern like `/tmp/snap-%05d.png`)

- **out** (*str*) – file to save video into

- **renameNotOverwrite** (*bool*) – if True, existing same-named video file will have -*number* appended; will be overwritten otherwise.

- **fps** (*int*) – Frames per second (`-mf fps=…`)

- **kbps** (*int*) – Bitrate (`-lavcopts vbitrate=…`) in kb/s

`yade.utils.`**`perpendicularArea`**(*axis*)

 Return area perpendicular to given axis (0=x,1=y,2=z) generated by bodies for which the function consider returns True (defaults to returning True always) and which is of the type *Sphere*.

`yade.utils.`**`phiIniPy`**(*ioPyFn, grid*)

 Returns a 3D discrete field appropriate to serve as *FastMarchingMethod.phiIni* (LS_DEM feature required), applying a user-made Python function *ioPyFn*

**Parameters**

- **ioPyFn** – an existing inside-outside Python function that takes three numbers (cartesian coordinates) as arguments

- **grid** (*RegularGrid*) – the *RegularGrid* instance to operate on

**Return list** an appropriate 3D discrete field to pass at *FastMarchingMethod.phiIni*

yade.utils.**plotDirections**(*aabb=(), mask=0, bins=20, numHist=True, noShow=False, sph-Sph=False*)
    Plot 3 histograms for distribution of interaction directions, in yz,xz and xy planes and (optional but default) histogram of number of interactions per body. If sphSph only sphere-sphere interactions are considered for the 3 directions histograms.

        **Returns** If *noShow* is False, displays the figure and returns nothing. If *noShow*, the figure object is returned without being displayed (works the same way as *plot.plot*).

yade.utils.**plotNumInteractionsHistogram**(*cutoff=0.0*)
    Plot histogram with number of interactions per body, optionally cutting away *cutoff* relative axis-aligned box from specimen margin.

yade.utils.**polyhedron**(*vertices, fixed=False, wire=True, color=None, highlight=False, noBound=False, material=-1, mask=1, chain=-1*)
    Create polyhedron with given parameters.

        **Parameters vertices** (*[Vector3]*) – coordinates of vertices in the global coordinate system.

    See *utils.sphere*'s documentation for meaning of other parameters.

yade.utils.**psd**(*bins=5, mass=True, mask=-1*)
    Calculates particle size distribution.

    **Parameters**

- **bins** (*int*) – number of bins

- **mass** (*bool*) – if true, the mass-PSD will be calculated

- **mask** (*int*) – *Body.mask* for the body

    **Returns**

- binsSizes: list of bin's sizes

- binsProc: how much material (in percents) are in the bin, cumulative

- binsSumCum: how much material (in units) are in the bin, cumulative

    binsSizes, binsProc, binsSumCum

yade.utils.**randomColor**(*seed=None*)
    Return random Vector3 with each component in interval 0…1 (uniform distribution)

yade.utils.**randomOrientation**()
    Returns (uniformly distributed) random orientation. Taken from Eigen::Quaternion::UnitRandom() source code. Uses standard Python random.random() function(s), you can random.seed() it

yade.utils.**randomizeColors**(*onlyDynamic=False*)
    Assign random colors to *Shape::color*.

    If onlyDynamic is true, only dynamic bodies will have the color changed.

yade.utils.**readParamsFromTable**(*tableFileLine=None, noTableOk=True, unknownOk=False, **kw*)
    Read parameters from a file and assign them to ___builtin___ variables.

    The format of the file is as follows (commens starting with # and empty lines allowed):

```
# commented lines allowed anywhere
name1 name2 … # first non-blank line are column headings
                      # empty line is OK, with or without comment
val1    val2    … # 1st parameter set
val2    val2    … # 2nd
…
```

Assigned tags (the `description` column is synthesized if absent, see *utils.TableParamReader*);

> O.tags['description']=… # assigns the description column; might be synthesized O.tags['params']="name1=val1,name2=val2,…" # all explicitly assigned parameters O.tags['defaultParams']="unassignedName1=defaultValue1,…" # parameters that were left at their defaults O.tags['d.id']=O.tags['id']+'.'+O.tags['description'] O.tags['id.d']=O.tags['description']+'.'+O.tags['id']

All parameters (default as well as settable) are saved using *utils.saveVars*(`'table'`).

> **Parameters**
>
> - **tableFileLine** – string attribute to define which line number (as seen in a text editor) from wich text file (with one value per blank-separated columns) to get the values from. A ':' should appear between the two informations, e.g. 'file.table:4' to read the 4th line from file.table file
>
> - **noTableOk** (`bool`) – if False, raise exception if the file cannot be open; use default values otherwise
>
> - **unknownOk** (`bool`) – do not raise exception if unknown column name is found in the file, and assign it as well
>
> **Returns** number of assigned parameters

**yade.utils.replaceCollider**(*colliderEngine*)

> Replaces collider (Collider) engine with the engine supplied. Raises error if no collider is in engines.

**yade.utils.runningInBatch**()

> Tell whether we are running inside the batch or separately.

**yade.utils.saveVars**(*mark=''*, *loadNow=True*, *\*\*kw*)

> Save passed variables into the simulation so that it can be recovered when the simulation is loaded again.
>
> For example, variables *a*, *b* and *c* are defined. To save them, use:

```
>>> saveVars('something',a=1,b=2,c=3)
>>> from yade.params.something import *
>>> a,b,c
(1, 2, 3)
```

those variables will be save in the .xml file, when the simulation itself is saved. To recover those variables once the .xml is loaded again, use `loadVars('something')` and they will be defined in the yade.params.*mark* module. The *loadNow* parameter calls *utils.loadVars* after saving automatically. If 'something' already exists, given variables will be inserted.

**yade.utils.sphere**(*center*, *radius*, *dynamic=None*, *fixed=False*, *wire=False*, *color=None*, *highlight=False*, *material=-1*, *mask=1*)

> Create sphere with given parameters; mass and inertia computed automatically.
>
> Last assigned material is used by default (*material* = -1), and utils.defaultMaterial() will be used if no material is defined at all.
>
> **Parameters**
>
> - **center** (`Vector3`) – center
>
> - **radius** (`float`) – radius

- **dynamic** (*float*) – deprecated, see "fixed"

- **fixed** (*float*) – generate the body with all DOFs blocked?

- **material** –

  **specify** *Body.material*; **different types are accepted:**

  – int: O.materials[material] will be used; as a special case, if material==-1 and there is no shared materials defined, utils.defaultMaterial() will be assigned to O.materials[0]

  – string: label of an existing material that will be used

  – *Material* instance: this instance will be used

  – callable: will be called without arguments; returned Material value will be used (Material factory object, if you like)

- **mask** (*int*) – *Body.mask* for the body

- **wire** – display as wire sphere?

- **highlight** – highlight this body in the viewer?

- **Vector3-or-None** – body's color, as normalized RGB; random color will be assigned if **None**.

**Returns** A Body instance with desired characteristics.

Creating default shared material if none exists neither is given:

```
>>> O.reset()
>>> from yade import utils
>>> len(O.materials)
0
>>> s0=utils.sphere([2,0,0],1)
>>> len(O.materials)
1
```

Instance of material can be given:

```
>>> s1=utils.sphere([0,0,0],1,wire=False,color=(0,1,0),material=ElastMat(young=30e9,
↪density=2e3))
>>> s1.shape.wire
False
>>> s1.shape.color
Vector3(0,1,0)
>>> s1.mat.density
2000.0
```

Material can be given by label:

```
>>> O.materials.append(FrictMat(young=10e9,poisson=.11,label='myMaterial'))
1
>>> s2=utils.sphere([0,0,2],1,material='myMaterial')
>>> s2.mat.label
'myMaterial'
>>> s2.mat.poisson
0.11
```

Finally, material can be a callable object (taking no arguments), which returns a Material instance. Use this if you don't call this function directly (for instance, through yade.pack.randomDensePack), passing only 1 *material* parameter, but you don't want material to be shared.

For instance, randomized material properties can be created like this:

```
>>> import random
>>> def matFactory(): return ElastMat(young=1e10*random.random(),density=1e3+1e3*random.
↪random())
...
>>> s3=utils.sphere([0,2,0],1,material=matFactory)
>>> s4=utils.sphere([1,2,0],1,material=matFactory)
```

**yade.utils.tetra**(*vertices, strictCheck=True, fixed=False, wire=True, color=None, highlight=False, noBound=False, material=-1, mask=1, chain=-1*)

> Create tetrahedron with given parameters.
>
> > **Parameters**
> >
> > - **vertices** (*[Vector3,Vector3,Vector3,Vector3]*) – coordinates of vertices in the global coordinate system.
> >
> > - **strictCheck** (*bool*) – checks vertices order, raise RuntimeError for negative volume
>
> See *utils.sphere*'s documentation for meaning of other parameters.

**yade.utils.tetraPoly**(*vertices, fixed=False, wire=True, color=None, highlight=False, noBound=False, material=-1, mask=1, chain=-1*)

> Create tetrahedron (actually simple Polyhedra) with given parameters.
>
> > **Parameters vertices** (*[Vector3,Vector3,Vector3,Vector3]*) – coordinates of vertices in the global coordinate system.
>
> See *utils.sphere*'s documentation for meaning of other parameters.

**yade.utils.trackPerfomance**(*updateTime=5*)

> Track perfomance of a simulation. (Experimental) Will create new thread to produce some plots. Useful for track perfomance of long run simulations (in bath mode for example).

**yade.utils.typedEngine**(*name*)

> Return first engine from current O.engines, identified by its type (as string). For example:

```
>>> from yade import utils
>>> O.engines=[InsertionSortCollider(),NewtonIntegrator(),GravityEngine()]
>>> utils.typedEngine("NewtonIntegrator") == O.engines[1]
True
```

**yade.utils.uniaxialTestFeatures**(*filename=None, areaSections=10, axis=-1, distFactor=2.2, **kw*)

> Get some data about the current packing useful for uniaxial test:
>
> 1. Find the dimensions that is the longest (uniaxial loading axis)
>
> 2. Find the minimum cross-section area of the specimen by examining several (areaSections) sections perpendicular to axis, computing area of the convex hull for each one. This will work also for non-prismatic specimen.
>
> 3. Find the bodies that are on the negative/positive boundary, to which the straining condition should be applied.
>
> > **Parameters**
> >
> > - **filename** – if given, spheres will be loaded from this file (ASCII format); if not, current simulation will be used.
> >
> > - **areaSection** (*float*) – number of section that will be used to estimate cross-section
> >
> > - **axis** (*{0,1,2}*) – if given, force strained axis, rather than computing it from predominant length
> >
> > **Returns** dictionary with keys **negIds**, **posIds**, **axis**, **area**.

> **Warning:** The function *utils.approxSectionArea* uses convex hull algorithm to find the area, but the implementation is reported to be *buggy* (bot works in some cases). Always check this number, or fix the convex hull algorithm (it is documented in the source, see py/__utils.cpp).

**yade.utils.vmData()**
    Return memory usage data from Linux's /proc/[pid]/status, line VmData.

**yade.utils.voxelPorosityTriaxial**(*triax, resolution=200, offset=0*)
    Calculate the porosity of a sample, given the TriaxialCompressionEngine.

    A function *utils.voxelPorosity* is invoked, with the volume of a box enclosed by TriaxialCompressionEngine walls. The additional parameter offset allows using a smaller volume inside the box, where each side of the volume is at offset distance from the walls. By this way it is possible to find a more precise porosity of the sample, since at walls' contact the porosity is usually reduced.

    A recommended value of offset is bigger or equal to the average radius of spheres inside.

    The value of resolution depends on size of spheres used. It can be calibrated by invoking voxelPorosityTriaxial with offset=0 and comparing the result with TriaxialCompressionEngine.porosity. After calibration, the offset can be set to radius, or a bigger value, to get the result.

    **Parameters**

    - **triax** – the TriaxialCompressionEngine handle
    - **resolution** – voxel grid resolution
    - **offset** – offset distance

    **Returns** the porosity of the sample inside given volume

    Example invocation:

```
from yade import utils
rAvg=0.03
TriaxialTest(numberOfGrains=200,radiusMean=rAvg).load()
O.dt=-1
O.run(1000)
O.engines[4].porosity
0.44007807740143889
utils.voxelPorosityTriaxial(O.engines[4],200,0)
0.44055412500000002
utils.voxelPorosityTriaxial(O.engines[4],200,rAvg)
0.36798199999999998
```

**yade.utils.waitIfBatch()**
    Block the simulation if running inside a batch. Typically used at the end of script so that it does not finish prematurely in batch mode (the execution would be ended in such a case).

**yade.utils.wall**(*position, axis, sense=0, color=None, material=-1, mask=1*)
    Return ready-made wall body.

    **Parameters**

    - **position** (*float-or-Vector3*) – center of the wall. If float, it is the position along given axis, the other 2 components being zero
    - **axis** ( *{0,1,2}*) – orientation of the wall normal (0,1,2) for x,y,z (sc. planes yz, xz, xy)
    - **sense** ( *{-1,0,1}*) – sense in which to interact (0: both, -1: negative, +1: positive; see *Wall*)

    See *utils.sphere*'s documentation for meaning of other parameters.

`yade.utils.xMirror`(*half*)

    Mirror a sequence of 2d points around the x axis (changing sign on the y coord). The sequence should start up and then it will wrap from y downwards (or vice versa). If the last point's x coord is zero, it will not be duplicated.

`yade._utils.PWaveTimeStep`() $\rightarrow$ float

    Get timestep accoring to the velocity of P-Wave propagation; computed from sphere radii, rigidities and masses.

`yade._utils.RayleighWaveTimeStep`() $\rightarrow$ float

    Determination of time step according to Rayleigh wave speed of force propagation.

`yade._utils.TetrahedronCentralInertiaTensor`(*(object)arg1*) $\rightarrow$ Matrix3

    TODO

`yade._utils.TetrahedronInertiaTensor`(*(object)arg1*) $\rightarrow$ Matrix3

    TODO

`yade._utils.TetrahedronSignedVolume`(*(object)arg1*) $\rightarrow$ float

    TODO

`yade._utils.TetrahedronVolume`(*(object)arg1*) $\rightarrow$ float

    TODO

`yade._utils.TetrahedronWithLocalAxesPrincipal`(*(Body)arg1*) $\rightarrow$ Quaternion

    TODO

`yade._utils.aabbExtrema`([*(float)cutoff=0.0*[, *(bool)centers=False*]]) $\rightarrow$ tuple

    Return coordinates of box enclosing all spherical bodies

        **Parameters**

            • **centers** (`bool`) – do not take sphere radii in account, only their centroids

            • **cutoff** (`float (0...1)`) – relative dimension by which the box will be cut away at its boundaries.

        **Returns** [lower corner, upper corner] as [Vector3,Vector3]

`yade._utils.angularMomentum`([*(Vector3)origin=Vector3(0, 0, 0)*]) $\rightarrow$ Vector3

    TODO

`yade._utils.approxSectionArea`(*(float)arg1, (int)arg2*) $\rightarrow$ float

    Compute area of convex hull when when taking (swept) spheres crossing the plane at coord, perpendicular to axis.

`yade._utils.bodyNumInteractionsHistogram`(*(tuple)aabb*) $\rightarrow$ tuple

`yade._utils.bodyStressTensors`() $\rightarrow$ list

    Compute and return a table with per-particle stress tensors. Each tensor represents the average stress in one particle, obtained from the contour integral of applied load as detailed below. This definition is considering each sphere as a continuum. It can be considered exact in the context of spheres at static equilibrium, interacting at contact points with negligible volume changes of the solid phase (this last assumption is not restricting possible deformations and volume changes at the packing scale).

    Proof:

    First, we remark the identity: $\sigma_{ij} = \delta_{ik}\sigma_{kj} = x_{i,k}\sigma_{kj} = (x_i\sigma_{kj})_{,k} - x_i\sigma_{kj,k}$.

    At equilibrium, the divergence of stress is null: $\sigma_{kj,k} = \mathbf{0}$. Consequently, after divergence theorem: $\frac{1}{V}\int_V \sigma_{ij}dV = \frac{1}{V}\int_V (x_i\sigma_{kj})_{,k}dV = \frac{1}{V}\int_{\partial V} x_i\sigma_{kj}n_k dS = \frac{1}{V}\sum_b x_i^b f_j^b$.

    The last equality is implicitly based on the representation of external loads as Dirac distributions whose zeros are the so-called *contact points*: 0-sized surfaces on which the *contact forces* are applied, located at $x_i$ in the deformed configuration.

A weighted average of per-body stresses will give the average stress inside the solid phase. There is a simple relation between the stress inside the solid phase and the stress in an equivalent continuum in the absence of fluid pressure. For porosity $n$, the relation reads: $\sigma_{ij}^{equ.} = (1-n)\sigma_{ij}^{solid}$.

This last relation may not be very useful if porosity is not homogeneous. If it happens, one can define the equivalent bulk stress a the particles scale by assigning a volume to each particle. This volume can be obtained from *Tesselation Wrapper* (see e.g. [Catalano2014a])

`yade._utils.calm(`*[(int)mask=-1 ]*`)` → None
> Set translational and rotational velocities of bodies to zero. Applied to all bodies by default. To calm only some bodies, use mask parameter, it will calm only bodies with groupMask compatible to given value

`yade._utils.coordsAndDisplacements(`*(int)axis*`[`*, (tuple)Aabb=() ]*`)` → tuple
> Return tuple of 2 same-length lists for coordinates and displacements (coordinate minus reference coordinate) along given axis (1st arg); if the Aabb=((x_min,y_min,z_min),(x_max,y_max,z_-max)) box is given, only bodies within this box will be considered.

`yade._utils.createInteraction(`*(int)id1, (int)id2*`[`*, (bool)virtualI=False ]*`)` → Interaction
> Create interaction between given bodies by hand.

> If *virtualI=False*, current engines are searched for *IGeomDispatcher* and *IPhysDispatcher* (might be both hidden in *InteractionLoop*). Geometry is created using `force` parameter of the *geometry dispatcher*, wherefore the interaction will exist even if bodies do not spatially overlap and the functor would return `false` under normal circumstances.

> If *virtualI=True* the interaction is left in a virtual state.

---

**Warning:** This function will very likely behave incorrectly for periodic simulations (though it could be extended it to handle it fairily easily).

---

`yade._utils.fabricTensor(`*[(float)cutoff=0.0*`[`                    *(bool)splitTensor=False*`[`*, (float)thresholdForce=nan ] ] ]*`)` → tuple
> Computes the fabric tensor $F_{ij} = \frac{1}{n_c}\sum_c n_i n_j$ [Satake1982], for all interactions `c`.

> **Parameters**

> - **cutoff** (*Real*) – intended to disregard boundary effects: to define in [0;1] to focus on the interactions located in the centered inner (1-cutoff)^3*V part of the spherical packing $V$.

> - **splitTensor** (*bool*) – split the fabric tensor into two parts related to the strong (greatest compressive normal forces) and weak contact forces respectively.

> - **thresholdForce** (*Real*) – if the fabric tensor is split into two parts, a threshold value can be specified otherwise the mean contact force is considered by default. Use negative signed values for compressive states. To note that this value could be set to zero if one wanted to make distinction between compressive and tensile forces.

`yade._utils.flipCell(`*[(Matrix3)flip=Matrix3(0, 0, 0, 0, 0, 0, 0, 0, 0) ]*`)` → Matrix3
> Flip periodic cell so that angles between $R^3$ axes and transformed axes are as small as possible, using the two following facts:1. repeating in $R^3$ space the corners of a periodic cell defines a regular grid; 2. two cells leading through this process to a unique grid are equivalent and can be flipped one over another. Flipping necessitates adjustment of *Interaction.cellDist* for interactions that cross the boundary and didn't before (or vice versa), and re-initialization of collider. The *flip* argument can be used to specify desired flip: integers, each column for one axis; if zero matrix, best fit (minimizing the angles) is computed automatically.

> In c++, this function is accessible as `Shop::flipCell`.

`yade._utils.forcesOnCoordPlane(`*(float)arg1, (int)arg2*`)` → Vector3

---

`yade._utils.forcesOnPlane`(*(Vector3)planePt*, *(Vector3)normal*) → Vector3

Find all interactions deriving from *NormShearPhys* that cross given plane and sum forces (both normal and shear) on them.

**Parameters**

- **planePt** (`Vector3`) – a point on the plane

- **normal** (`Vector3`) – plane normal (will be normalized).

`yade._utils.getBodyIdsContacts`([*(int)bodyID=0*]) → list

Get a list of body-ids, which contacts the given body.

`yade._utils.getCapillaryStress`([*(float)volume=0*[, *(bool)mindlin=False*]]) → Matrix3

Compute and return Love-Weber capillary stress tensor:

$\sigma_{ij}^{cap} = \frac{1}{V} \sum_b l_i^b f_j^{cap,b}$, where the sum is over all interactions, with $l$ the branch vector (joining centers of the bodies) and $f^{cap}$ is the capillary force. $V$ can be passed to the function. If it is not, it will be equal to one in non-periodic cases, or equal to the volume of the cell in periodic cases. Only the CapillaryPhys interaction type is supported presently. Using this function with physics MindlinCapillaryPhys needs to pass True as second argument.

`yade._utils.getDepthProfiles`(*(float)volume*, *(int)nCell*, *(float)dz*, *(float)zRef*, *(bool)activateCond*, *(float)radiusPy*, *(int)direction*) → tuple

Compute and return the particle velocity and solid volume fraction (porosity) depth profile along the direction specified (default is z; 0=>x,1=>y,2=>z). For each defined cell z, the k component of the average particle velocity reads:

$<v_k>^z = \sum_p V^p v_k^p / \sum_p V^p$,

where the sum is made over the particles contained in the cell, $v_k^p$ is the k component of the velocity associated to particle p, and $V^p$ is the part of the volume of the particle p contained inside the cell. This definition allows to smooth the averaging, and is equivalent to taking into account the center of the particles only when there is a lot of particles in each cell. As for the solid volume fraction, it is evaluated in the same way: for each defined cell z, it reads:

$<\varphi>^z = \frac{1}{V_{cell}} \sum_p V^p$, where $V_{cell}$ is the volume of the cell considered, and $V^p$ is the volume of partic

This function gives depth profiles of average velocity and solid volume fraction, returning the average quantities in each cell of height dz, from the reference horizontal plane at elevation zRef (input parameter) until the plane of elevation zRef+nCell*dz (input parameters). If the argument activateCond is set to true, do the average only on particles of radius equal to radiusPy (input parameter)

`yade._utils.getDepthProfiles_center`(*(float)volume*, *(int)nCell*, *(float)dz*, *(float)zRef*, *(bool)activateCond*, *(float)radiusPy*) → tuple

Same as getDepthProfiles but taking into account particles as points located at the particle center.

`yade._utils.getDynamicStress`() → list

Compute the dynamic stress tensor for each body: $\sigma_D^p = -\frac{1}{V^p} m^p u'^p \otimes u'^p$

`yade._utils.getSpheresMass`([*(int)mask=-1*]) → float

Compute the total mass of spheres in the simulation, mask parameter is considered

`yade._utils.getSpheresVolume`([*(int)mask=-1*]) → float

Compute the total volume of spheres in the simulation, mask parameter is considered

`yade._utils.getSpheresVolume2D`([*(int)mask=-1*]) → float

Compute the total volume of discs in the simulation, mask parameter is considered

`yade._utils.getStress`([*(float)volume=0*]) → Matrix3

Compute and return Love-Weber stress tensor:

$\sigma_{ij} = \frac{1}{V} \sum_b f_i^b l_j^b$, where the sum is over all interactions, with $f$ the contact force and $l$ the branch vector (joining centers of the bodies). Stress is negativ for repulsive contact forces, i.e. compression. V can be passed to the function. If it is not, it will be equal to the volume of the cell in periodic cases, or to the one deduced from utils.aabbDim() in non-periodic cases.

`yade._utils.getStressAndTangent(`*([float]volume=0[, (bool)symmetry=True]]*`)` → tuple

Compute overall stress of periodic cell using the same equation as function getStress. In addition, the tangent operator is calculated using the equation published in [Kruyt and Rothenburg1998]_:

$$S_{ijkl} = \frac{1}{V} \sum_c (k_n n_i l_j n_k l_l + k_t t_i l_j t_k l_l)$$

**Parameters**

- **volume** (*float*) – same as in function getStress

- **symmetry** (*bool*) – make the tensors symmetric.

**Returns** macroscopic stress tensor and tangent operator as py::tuple

`yade._utils.getStressProfile(`*([float]volume, (int)nCell, (float)dz, (float)zRef, (object)vPartAverageX, (object)vPartAverageY, (object)vPartAverageZ)*`)` → tuple

Compute and return the stress tensor depth profile, including the contribution from Love-Weber stress tensor and the dynamic stress tensor taking into account the effect of particles inertia. For each defined cell z, the stress tensor reads:

$$\sigma_{ij}^z = \frac{1}{V} \sum_c f_i^c l_j^{c,z} - \frac{1}{V} \sum_p m^p u_i'^p u_j'^p,$$

where the first sum is made over the contacts which are contained or cross the cell z, f^c is the contact force from particle 1 to particle 2, and l^{c,z} is the part of the branch vector from particle 2 to particle 1, contained in the cell. The second sum is made over the particles, and u'^p is the velocity fluctuations of the particle p with respect to the spatial averaged particle velocity at this point (given as input parameters). The expression of the stress tensor is the same as the one given in getStress plus the inertial contribution. Apart from that, the main difference with getStress stands in the fact that it gives a depth profile of stress tensor, i.e. from the reference horizontal plane at elevation zRef (input parameter) until the plane of elevation zRef+nCell*dz (input parameters), it is computing the stress tensor for each cell of height dz. For the love-Weber stress contribution, the branch vector taken into account in the calculations is only the part of the branch vector contained in the cell considered. To validate the formulation, it has been checked that activating only the Love-Weber stress tensor, and suming all the contributions at the different altitude, we recover the same stress tensor as when using getStress. For my own use, I have troubles with strong overlap between fixed object, so that I made a condition to exclude the contribution to the stress tensor of the fixed objects, this can be desactivated easily if needed (and should be desactivated for the comparison with getStress).

`yade._utils.getStressProfile_contact(`*([float]volume, (int)nCell, (float)dz, (float)zRef)*`)` → tuple same as getStressProfile, only contact contribution.

`yade._utils.getTotalDynamicStress(`*([float]volume=0]*`)` → Matrix3

Compute the total dynamic stress tensor : $\sigma_D = -\frac{1}{V} \sum_p m^p u'^p \otimes u'^p$. The volume have to be provided for non-periodic simulations. It is computed from cell volume for periodic simulations.

`yade._utils.getViscoelasticFromSpheresInteraction(`*([float]tc, (float)en, (float)es)*`)` → dict

Attention! The function is deprecated! Compute viscoelastic interaction parameters from analytical solution of a pair spheres collision problem:

$$k_n = \frac{m}{t_c^2} \left( \pi^2 + (\ln e_n)^2 \right)$$

$$c_n = -\frac{2m}{t_c} \ln e_n$$

$$k_t = \frac{2}{7} \frac{m}{t_c^2} \left( \pi^2 + (\ln e_t)^2 \right)$$

$$c_t = -\frac{2}{7} \frac{m}{t_c} \ln e_t$$

where $k_n$, $c_n$ are normal elastic and viscous coefficients and $k_t$, $c_t$ shear elastic and viscous coefficients. For details see [Pournin2001].

> **Parameters**
> - **m** (*float*) – sphere mass $m$
> - **tc** (*float*) – collision time $t_c$
> - **en** (*float*) – normal restitution coefficient $e_n$
> - **es** (*float*) – tangential restitution coefficient $e_s$
>
> **Returns** dictionary with keys $kn$ (the value of $k_n$), $cn$ ($c_n$), $kt$ ($k_t$), $ct$ ($c_t$).

`yade._utils.growParticle`(*(int)bodyID, (float)multiplier*[, *(bool)updateMass=True*]) → None
> Change the size of a single sphere (to be implemented: single clump). If updateMass=True, then the mass is updated.

`yade._utils.growParticles`(*(float)multiplier*[, *(bool)updateMass=True*[, *(bool)dynamicOnly=True*]]) → None
> Change the size of spheres and clumps of spheres by the multiplier. If updateMass=True, then the mass and inertia are updated. dynamicOnly=True will select dynamic bodies.

`yade._utils.highlightNone`() → None
> Reset *highlight* on all bodies.

`yade._utils.initMPI`() → None
> Initialize MPI communicator, for Foam Coupling

`yade._utils.inscribedCircleCenter`(*(Vector3)v1, (Vector3)v2, (Vector3)v3*) → Vector3
> Return center of inscribed circle for triangle given by its vertices *v1, v2, v3*.

`yade._utils.interactionAnglesHistogram`(*(int)axis*[, *(int)mask=0*[, *(int)bins=20*[, *(tuple)aabb=()*[, *(bool)sphSph=0*[, *(float)minProjLen=1e-06*]]]]]]) → tuple

`yade._utils.intrsOfEachBody`() → list
> returns list of lists of interactions of each body

`yade._utils.kineticEnergy`([*(bool)findMaxId=False*]) → object
> Compute overall kinetic energy of the simulation as

$$\sum \frac{1}{2} \left( m_i v_i^2 + \omega(I\omega^T) \right).$$

For *aspherical* bodies, the inertia tensor $I$ is transformed to global frame, before multiplied by $\omega$, therefore the value should be accurate.

`yade._utils.maxOverlapRatio`() → float
> Return maximum overlap ration in interactions (with *ScGeom*) of two *spheres*. The ratio is computed as $\frac{u_N}{2(r_1 r_2)/r_1+r_2}$, where $u_N$ is the current overlap distance and $r_1$, $r_2$ are radii of the two spheres in contact.

`yade._utils.`**`momentum`**`()` → Vector3
　　TODO

`yade._utils.`**`negPosExtremeIds`**`((int)axis, (float)distFactor)` → tuple
　　Return list of ids for spheres (only) that are on extremal ends of the specimen along given axis; distFactor multiplies their radius so that sphere that do not touch the boundary coordinate can also be returned.

`yade._utils.`**`normalShearStressTensors`**`([(bool)compressionPositive=False[,`
　　　　　　　　　　　　　　　　　　　　　　`(bool)splitNormalTensor=False[,`
　　　　　　　　　　　　　　　　　　　　　　`(float)thresholdForce=nan]]])` → tuple
Compute overall stress tensor of the periodic cell decomposed in 2 parts, one contributed by normal forces, the other by shear forces. The formulation can be found in [Thornton2000], eq. (3):

$$\sigma_{ij} = \frac{2}{V} \sum RN\mathbf{n}_i\mathbf{n}_j + \frac{2}{V} \sum RT\mathbf{n}_i\mathbf{t}_j$$

where $V$ is the cell volume, $R$ is "contact radius" (in our implementation, current distance between particle centroids), $\mathbf{n}$ is the normal vector, $\mathbf{t}$ is a vector perpendicular to $\mathbf{n}$, $N$ and $T$ are norms of normal and shear forces.

　　**Parameters**

　　　　• **splitNormalTensor** (`bool`) – if true the function returns normal stress tensor split into two parts according to the two subnetworks of strong an weak forces.

　　　　• **thresholdForce** (`Real`) – threshold value according to which the normal stress tensor can be split (e.g. a zero value would make distinction between tensile and compressive forces).

`yade._utils.`**`numIntrsOfEachBody`**`()` → list
　　returns list of number of interactions of each body

`yade._utils.`**`pointInsidePolygon`**`((tuple)arg1, (object)arg2)` → bool

`yade._utils.`**`porosity`**`([(float)volume=-1])` → float
　　Compute packing porosity $\frac{V-V_s}{V}$ where $V$ is overall volume and $V_s$ is volume of spheres.

　　**Parameters volume** (`float`) – overall volume $V$. For periodic simulations, current volume of the *Cell* is used. For aperiodic simulations, the value deduced from utils.aabbDim() is used. For compatibility reasons, positive values passed by the user are also accepted in this case.

`yade._utils.`**`ptInAABB`**`((Vector3)arg1, (Vector3)arg2, (Vector3)arg3)` → bool
　　Return True/False whether the point p is within box given by its min and max corners

`yade._utils.`**`scalarOnColorScale`**`((float)x[, (float)xmin=0[, (float)xmax=1]])` → Vector3
Map scalar variable to color scale.

　　**Parameters**

　　　　• **x** (`float`) – scalar value which the function applies to.

　　　　• **xmin** (`float`) – minimum value for the color scale, with a return value of (0,0,1) for $x \le xmin$, i.e. blue color in RGB.

　　　　• **xmax** (`float`) – maximum value, with a return value of (1,0,0) for $x \ge xmax$, i.e. red color in RGB.

　　**Returns** a Vector3 depending on the relative position of $x$ on a $[xmin;*xmax*]$ scale.

`yade._utils.`**`setBodyAngularVelocity`**`((int)id, (Vector3)angVel)` → None
　　Set a body angular velocity from its id and a new Vector3r.

　　**Parameters**

　　　　• **id** (`int`) – the body id.

- **angVel** (`Vector3`) – the desired updated angular velocity.

yade._utils.**setBodyColor**(*(int)id, (Vector3)color*) → None
 Set a body color from its id and a new Vector3r.

  **Parameters**

  - **id** (`int`) – the body id.

  - **color** (`Vector3`) – the desired updated color.

yade._utils.**setBodyOrientation**(*(int)id, (Quaternion)ori*) → None
 Set a body orientation from its id and a new Quaternionr.

  **Parameters**

  - **id** (`int`) – the body id.

  - **ori** (`Quaternion`) – the desired updated orientation.

yade._utils.**setBodyPosition**(*(int)id, (Vector3)pos*[, *(str)axis='xyz'*]) → None
 Set a body position from its id and a new vector3r.

  **Parameters**

  - **id** (`int`) – the body id.

  - **pos** (`Vector3`) – the desired updated position.

  - **axis** (`str`) – the axis along which the position has to be updated (ex: if axis=="xy" and pos==Vector3r(r0,r1,r2), r2 will be ignored and the position along z will not be updated).

yade._utils.**setBodyVelocity**(*(int)id, (Vector3)vel*[, *(str)axis='xyz'*]) → None
 Set a body velocity from its id and a new vector3r.

  **Parameters**

  - **id** (`int`) – the body id.

  - **vel** (`Vector3`) – the desired updated velocity.

  - **axis** (`str`) – the axis along which the velocity has to be updated (ex: if axis=="xy" and vel==Vector3r(r0,r1,r2), r2 will be ignored and the velocity along z will not be updated).

yade._utils.**setContactFriction**(*(float)angleRad*) → None
 Modify the friction angle (in radians) inside the material classes and existing contacts. The friction for non-dynamic bodies is not modified.

yade._utils.**setNewVerticesOfFacet**(*(Body)b, (Vector3)v1, (Vector3)v2, (Vector3)v3*) → None
 Sets new vertices (in global coordinates) to given facet.

yade._utils.**setRefSe3**() → None
 Set reference *positions* and *orientations* of all *bodies* equal to their current *positions* and *orientations*.

yade._utils.**shiftBodies**(*(list)ids, (Vector3)shift*) → float
 Shifts bodies listed in ids without updating their velocities.

yade._utils.**spiralProject**(*(Vector3)pt, (float)dH_dTheta*[, *(int)axis=2*[, *(float)periodStart=nan*[, *(float)theta0=0*]]]) → tuple

yade._utils.**sumFacetNormalForces**(*(object)ids*[, *(int)axis=-1*]) → float
 Sum force magnitudes on given bodies (must have *shape* of the *Facet* type), considering only part of forces perpendicular to each *facet's* face; if *axis* has positive value, then the specified axis (0=x, 1=y, 2=z) will be used instead of facet's normals.

`yade._utils.`**`sumForces`**`((list)ids, (Vector3)direction)` → float
    Return summary force on bodies with given *ids*, projected on the *direction* vector.

`yade._utils.`**`sumTorques`**`((list)ids, (Vector3)axis, (Vector3)axisPt)` → float
    Sum forces and torques on bodies given in *ids* with respect to axis specified by a point *axisPt* and
    its direction *axis*.

`yade._utils.`**`totalForceInVolume`**`()` → tuple
    Return summed forces on all interactions and average isotropic stiffness, as tuple (Vector3,float)

`yade._utils.`**`unbalancedForce`**`([(bool)useMaxForce=False])` → float
    Compute the ratio of mean (or maximum, if *useMaxForce*) summary force on bodies and mean force
    magnitude on interactions. For perfectly static equilibrium, summary force on all bodies is zero
    (since forces from interactions cancel out and induce no acceleration of particles); this ratio will tend
    to zero as simulation stabilizes, though zero is never reached because of finite precision computation.
    Sufficiently small value can be e.g. 1e-2 or smaller, depending on how much equilibrium it should
    be.

`yade._utils.`**`voidratio2D`**`([(float)zlen=1])` → float
    Compute 2D packing void ratio $\frac{V-V_s}{V_s}$ where V is overall volume and $V_s$ is volume of disks.

    **Parameters zlen** (*float*) – length in the third direction.

`yade._utils.`**`voxelPorosity`**`([(int)resolution=200[, (Vector3)start=Vector3(0, 0, 0)[, (Vector3)end=Vector3(0, 0, 0)]]])` → float
    Compute packing porosity $\frac{V-V_v}{V}$ where V is a specified volume (from start to end) and $V_v$ is volume
    of voxels that fall inside any sphere. The calculation method is to divide whole volume into a dense
    grid of voxels (at given resolution), and count the voxels that fall inside any of the spheres. This
    method allows one to calculate porosity in any given sub-volume of a whole sample. It is properly
    excluding part of a sphere that does not fall inside a specified volume.

    **Parameters**
    - **resolution** (*int*) – voxel grid resolution, values bigger than resolution=1600
      require a 64 bit operating system, because more than 4GB of RAM is used, a
      resolution=800 will use 500MB of RAM.
    - **start** (*Vector3*) – start corner of the volume.
    - **end** (*Vector3*) – end corner of the volume.

`yade._utils.`**`wireAll`**`()` → None
    Set *Shape::wire* on all bodies to True, rendering them with wireframe only.

`yade._utils.`**`wireNoSpheres`**`()` → None
    Set *Shape::wire* to True on non-spherical bodies (*Facets*, *Walls*).

`yade._utils.`**`wireNone`**`()` → None
    Set *Shape::wire* on all bodies to False, rendering them as solids.

## 2.4.18 yade.ymport module

Import geometry from various formats ('import' is python keyword, hence the name 'ymport').

`yade.ymport.`**`ele`**`(nodeFileName, eleFileName, shift=(0, 0, 0), scale=1.0, **kw)`
    Import tetrahedral mesh from .ele file, return list of created tetrahedrons.

    **Parameters**
    - **nodeFileName** (*string*) – name of .node file
    - **eleFileName** (*string*) – name of .ele file
    - **shift** (*(float,float,float)|Vector3*) – (X,Y,Z) parameter moves the speci-
      men.

- **scale** (`float`) – factor scales the given data.

- **\*\*kw** – (unused keyword arguments) is passed to *utils.polyhedron*

`yade.ymport.gengeo`(*mntable, shift=Vector3(0, 0, 0), scale=1.0, \*\*kw*)

Imports geometry from LSMGenGeo library and creates spheres. Since 2012 the package is available in Debian/Ubuntu and known as python-demgengeo http://packages.qa.debian.org/p/python-demgengeo.html

> **Parameters**
>
> > **mntable**: **mntable** object, which creates by LSMGenGeo library, see example
> >
> > **shift**: [**float,float,float**] [X,Y,Z] parameter moves the specimen.
> >
> > **scale**: **float** factor scales the given data.
> >
> > **\*\*kw**: **(unused keyword arguments)** is passed to *utils.sphere*

LSMGenGeo library allows one to create pack of spheres with given [Rmin:Rmax] with null stress inside the specimen. Can be useful for Mining Rock simulation.

Example: examples/packs/packs.py, usage of LSMGenGeo library in examples/test/genCylLSM.py.

- https://answers.launchpad.net/esys-particle/+faq/877

- http://www.access.edu.au/lsmgengeo__python_doc/current/pythonapi/html/GenGeo-module.html

- https://svn.esscc.uq.edu.au/svn/esys3/lsm/contrib/LSMGenGeo/

`yade.ymport.gengeoFile`(*fileName='file.geo', shift=Vector3(0, 0, 0), scale=1.0, orientation=Quaternion((1, 0, 0), 0), \*\*kw*)

Imports geometry from LSMGenGeo .geo file and creates spheres. Since 2012 the package is available in Debian/Ubuntu and known as python-demgengeo http://packages.qa.debian.org/p/python-demgengeo.html

> **Parameters**
>
> > **filename**: **string** file which has 4 colums [x, y, z, radius].
> >
> > **shift**: **Vector3** Vector3(X,Y,Z) parameter moves the specimen.
> >
> > **scale**: **float** factor scales the given data.
> >
> > **orientation**: **quaternion** orientation of the imported geometry
> >
> > **\*\*kw**: **(unused keyword arguments)** is passed to *utils.sphere*
>
> **Returns** list of spheres.

LSMGenGeo library allows one to create pack of spheres with given [Rmin:Rmax] with null stress inside the specimen. Can be useful for Mining Rock simulation.

Example: examples/packs/packs.py, usage of LSMGenGeo library in examples/test/genCylLSM.py.

- https://answers.launchpad.net/esys-particle/+faq/877

- http://www.access.edu.au/lsmgengeo__python_doc/current/pythonapi/html/GenGeo-module.html

- https://svn.esscc.uq.edu.au/svn/esys3/lsm/contrib/LSMGenGeo/

`yade.ymport.gmsh`(*meshfile='file.mesh', shift=Vector3(0, 0, 0), scale=1.0, orientation=Quaternion((1, 0, 0), 0), \*\*kw*)

Imports geometry from .mesh file and creates facets.

> **Parameters**
>
> > **shift**: [**float,float,float**] [X,Y,Z] parameter moves the specimen.

> ***scale*: float** factor scales the given data.
>
> ***orientation*: quaternion** orientation of the imported mesh
>
> ***\*\*kw*: (unused keyword arguments)** is passed to *utils.facet*

> **Returns** list of facets forming the specimen.

mesh files can easily be created with GMSH. Example added to examples/packs/packs.py

Additional examples of mesh-files can be downloaded from http://www-roc.inria.fr/gamma/download/download.php

yade.ymport.**gts**(*meshfile, shift=Vector3(0, 0, 0), scale=1.0, \*\*kw*)
    Read given meshfile in gts format.

> **Parameters**

> ***meshfile*: string** name of the input file.
>
> ***shift*: [float,float,float]** [X,Y,Z] parameter moves the specimen.
>
> ***scale*: float** factor scales the given data.
>
> ***\*\*kw*: (unused keyword arguments)** is passed to *utils.facet*

> **Returns** list of facets.

yade.ymport.**iges**(*fileName, shift=(0, 0, 0), scale=1.0, returnConnectivityTable=False, \*\*kw*)
    Import triangular mesh from .igs file, return list of created facets.

> **Parameters**

> - **fileName** (*string*) – name of iges file
> - **shift** (*(float,float,float)|Vector3*) – (X,Y,Z) parameter moves the specimen.
> - **scale** (*float*) – factor scales the given data.
> - **\*\*kw** – (unused keyword arguments) is passed to *utils.facet*
> - **returnConnectivityTable** (*bool*) – if True, apart from facets returns also nodes (list of (x,y,z) nodes coordinates) and elements (list of (id1,id2,id3) element nodes ids). If False (default), returns only facets

yade.ymport.**stl**(*file, dynamic=None, fixed=True, wire=True, color=None, highlight=False, noBound=False, material=-1, scale=1.0, shift=Vector3(0, 0, 0)*)
    Import a .stl geometry in the form of a set of *Facet*-shaped bodies.

> **Parameters**

> - **file** (*string*) – the .stl file serving as geometry input
> - **dynamic** (*bool*) – controls *Body.dynamic*
> - **fixed** (*bool*) – controls *Body.dynamic* (with fixed = True imposing *Body.dynamic* = False) if *dynamic* attribute is not given
> - **wire** (*bool*) – rendering option, passed to *Facet.wire*
> - **color** – rendering option, passed to *Facet.color*
> - **highlight** (*bool*) – rendering option, passed to *Facet.highlight*
> - **noBound** (*bool*) – sets *Body.bounded* to False if True, preventing collision detection (and vice-versa)
> - **material** – defines *material* properties, see *Defining materials* for usage
> - **scale** (*float*) – scaling factor to e.g. dilate the geometry if $> 1$
> - **shift** (*Vector3*) – for translating the geometry

> **Returns** a corresponding list of *Facet*-shaped bodies

`yade.ymport.text`(*fileName, shift=Vector3(0, 0, 0), scale=1.0, \*\*kw*)

> Load sphere coordinates from file, returns a list of corresponding bodies; that may be inserted to the simulation with O.bodies.append().
>
> > **Parameters**
> >
> > - **filename** (`string`) – file which has 4 colums [x, y, z, radius].
> >
> > - **shift** (`[float,float,float]`) – [X,Y,Z] parameter moves the specimen.
> >
> > - **scale** (`float`) – factor scales the given data.
> >
> > - **\*\*kw** – (unused keyword arguments) is passed to *utils.sphere*
> >
> > **Returns** list of spheres.
>
> Lines starting with # are skipped

`yade.ymport.textClumps`(*fileName, shift=Vector3(0, 0, 0), discretization=0, orientation=Quaternion((1, 0, 0), 0), scale=1.0, \*\*kw*)

> Load clumps-members from file in a format selected by the `format` argument, insert them to the simulation.
>
> > **Parameters**
> >
> > - **filename** (`str`) – file name
> >
> > - **format** (`str`) – selected input format. Supported `'x_y_z_r'``(default),
> > ``'x_y_z_r_clumpId'`
> >
> > - **shift** (`[float,float,float]`) – [X,Y,Z] parameter moves the specimen.
> >
> > - **scale** (`float`) – factor scales the given data.
> >
> > - **\*\*kw** – (unused keyword arguments) is passed to *utils.sphere*
> >
> > **Returns** list of spheres.
>
> Lines starting with # are skipped

`yade.ymport.textExt`(*fileName, format='x_y_z_r', shift=Vector3(0, 0, 0), scale=1.0, attrs=[], \*\*kw*)

> Load sphere coordinates from file in a format selected by the `format` argument, returns a list of corresponding bodies; that may be inserted to the simulation with O.bodies.append().
>
> > **Parameters**
> >
> > - **filename** (`str`) – file name
> >
> > - **format** (`str`) – selected input format. Supported `'x_y_z_r'``(default),
> > ``'x_y_z_r_matId', 'x_y_z_r_attrs'`
> >
> > - **shift** (`[float,float,float]`) – [X,Y,Z] parameter moves the specimen.
> >
> > - **scale** (`float`) – factor scales the given data.
> >
> > - **attrs** (`list`) – attrs read from file if export.textExt(format='x_y_z_r_attrs')
> > were used ('passed by reference' style)
> >
> > - **\*\*kw** – (unused keyword arguments) is passed to *utils.sphere*
> >
> > **Returns** list of spheres.
>
> Lines starting with # are skipped

`yade.ymport.textFacets`(*fileName, format='x1_y1_z1_x2_y2_z2_x3_y3_z3', shift=Vector3(0, 0, 0), scale=1.0, attrs=[], \*\*kw*)

> Load facet coordinates from file in a format selected by the `format` argument, returns a list of corresponding bodies; that may be inserted to the simulation with O.bodies.append().
>
> > **Parameters**

- **filename** (*str*) – file name

- **format** (*str*) – selected input format. Supported `'x1_y1_z1_x2_y2_-z2_x3_y3_z3'``(default)`, `` ``'x1_y1_z1_x2_y2_z2_x3_y3_z3_matId'`, `'id_x1_y1_z1_x2_y2_z2_x3_y3_z3_matId'` or `'x1_y1_z1_x2_y2_z2_x3_y3_-z3_attrs'`

- **shift** (*[float,float,float]*) – [X,Y,Z] parameter moves the specimen.

- **scale** (*float*) – factor scales the given data.

- **attrs** (*list*) – attrs read from file ('passed by reference' style)

- **\*\*kw** – (unused keyword arguments) is passed to *utils.facet*

    **Returns** list of facets.

Lines starting with # are skipped

`yade.ymport.`**`textPolyhedra`**(*fileName, material, shift=Vector3(0, 0, 0), scale=1.0, orientation=Quaternion((1, 0, 0), 0), \*\*kw*)
  Load polyhedra from a text file.

    **Parameters**

    - **filename** (*str*) – file name. Expected file format is the one output by export.textPolyhedra.

    - **shift** (*[float,float,float]*) – [X,Y,Z] parameter moves the specimen.

    - **scale** (*float*) – factor scales the given data.

    - **orientation** (*quaternion*) – orientation of the imported polyhedra

    - **\*\*kw** – (unused keyword arguments) is passed to *polyhedra__utils.polyhedra*

    **Returns** list of polyhedras.

Lines starting with # are skipped

`yade.ymport.`**`unv`**(*fileName, shift=(0, 0, 0), scale=1.0, returnConnectivityTable=False, \*\*kw*)
  Import geometry from unv file, return list of created facets.

    **param string fileName** name of unv file

    **param (float,float,float)|Vector3 shift** (X,Y,Z) parameter moves the specimen.

    **param float scale** factor scales the given data.

    **param \*\*kw** (unused keyword arguments) is passed to *utils.facet*

    **param bool returnConnectivityTable** if True, apart from facets returns also list of (x,y,z) nodes coordinates) and elements (list of (id1,id2,id3) element nodes ids). If False (default), returns only facets

unv files are mainly used for FEM analyses (are used by OOFEM and Abaqus), but triangular elements can be imported as facets. These files cen be created e.g. with open-source free software Salome.

Example: examples/test/unv-read/unvRead.py.

## 2.5 Installation

- Linux systems: Yade can be installed from packages (pre-compiled binaries) or source code. The choice depends on what you need: if you don't plan to modify Yade itself, package installation is easier. In the contrary case, you must download and install the source code.

- Other Operating Systems: Jump to the last section of this page.

- 64 bit Operating Systems required; no support for 32 bit (i386).

## 2.5.1 Packages

**Stable packages**

Since 2011, all Ubuntu (starting from 11.10, Oneiric) and Debian (starting from Wheezy) versions have Yade in their main repositories. There are only stable releases in place. To install Yade, run the following:

```
sudo apt-get install yade
```

After that you can normally start Yade using the command `yade` or `yade-batch`.

This image shows versions and up to date status of Yade in some repositories.

**Daily packages**

Pre-built packages updated more frequently than the stable versions are provided for all currently supported Debian and Ubuntu versions and available on yade-dem.org/packages .

These are "daily" versions of the packages which are being updated regularly and, hence, include all the newly added features.

To install the daily-version you need to add the repository to your /etc/apt/sources.list.

- Debian 9 **stretch**:

```
sudo bash -c 'echo "deb http://www.yade-dem.org/packages/ stretch main" >> /etc/apt/
↪sources.list'
```

- Debian 10 **buster**:

```
sudo bash -c 'echo "deb http://www.yade-dem.org/packages/ buster main" >> /etc/apt/
↪sources.list'
```

- Debian 11 **bullseye**:

```
sudo bash -c 'echo "deb http://www.yade-dem.org/packages/ bullseye main" >> /etc/apt/
↪sources.list'
```

- Debian 12 **bookworm**:

```
sudo bash -c 'echo "deb http://www.yade-dem.org/packages/ bookworm main" >> /etc/apt/
↪sources.list'
```

- Ubuntu 16.04 **xenial**:

```
sudo bash -c 'echo "deb http://www.yade-dem.org/packages/ xenial main" >> /etc/apt/
↪sources.list'
```

- Ubuntu 18.04 **bionic**:

```
sudo bash -c 'echo "deb http://www.yade-dem.org/packages/ bionic main" >> /etc/apt/
↪sources.list'
```

- Ubuntu 20.04 **focal**:

```
sudo bash -c 'echo "deb http://www.yade-dem.org/packages/ focal main" >> /etc/apt/sources.
↪list'
```

- Ubuntu 22.04 **jammy**:

```
sudo bash -c 'echo "deb http://www.yade-dem.org/packages/ jammy main" >> /etc/apt/sources.
↪list'
```

Add the PGP-key AA915EEB as trusted and install `yadedaily`:

```
wget -O - http://www.yade-dem.org/packages/yadedev_pub.gpg | sudo apt-key add -
sudo apt-get update
sudo apt-get install yadedaily
```

After that you can normally start Yade using the command `yadedaily` or `yadedaily-batch`. `yadedaily` on older distributions can have some disabled features due to older library versions, shipped with particular distribution.

The Git-repository for packaging stuff is available on GitLab.

If you do not need `yadedaily`-package anymore, just remove the corresponding line in /etc/apt/sources.list and the package itself:

```
sudo apt-get remove yadedaily
```

To remove our key from keyring, execute the following command:

```
sudo apt-key remove AA915EEB
```

Daily and stable Yade versions can coexist without any conflicts, i.e., you can use `yade` and `yadedaily` at the same time.

### 2.5.2 Docker

Yade can be installed using docker images, which are daily built. Images contain both stable and dialy versions of packages. Docker images are based on supported distributions:

- Debian 9 **stretch**:

```
docker run -it registry.gitlab.com/yade-dev/docker-prod:debian-stretch
```

- Debian 10 **buster**:

```
docker run -it registry.gitlab.com/yade-dev/docker-prod:debian-buster
```

- Debian 11 **bullseye**:

```
docker run -it registry.gitlab.com/yade-dev/docker-prod:debian-bullseye
```

- Debian 12 **bookworm**:

```
docker run -it registry.gitlab.com/yade-dev/docker-prod:debian-bookworm
```

- Ubuntu 16.04 **xenial**:

```
docker run -it registry.gitlab.com/yade-dev/docker-prod:ubuntu16.04
```

- Ubuntu 18.04 **bionic**:

```
docker run -it registry.gitlab.com/yade-dev/docker-prod:ubuntu18.04
```

- Ubuntu 20.04 **focal**:

```
docker run -it registry.gitlab.com/yade-dev/docker-prod:ubuntu20.04
```

- Ubuntu 22.04 **jammy**:

```
docker run -it registry.gitlab.com/yade-dev/docker-prod:ubuntu22.04
```

After the container is pulled and is running, Yade functionality can be checked:

```
yade --test
yade --check
yadedaily --test
yadedaily --check
```

### 2.5.3 Source code

Installation from source code is reasonable, when you want to add or modify constitutive laws, engines, functions etc. Installing the latest trunk version allows one to use newly added features, which are not yet available in packaged versions.

#### Download

If you want to install from source, you can install either a release (numbered version, which is frozen) or the current development version (updated by the developers frequently). You should download the development version (called `trunk`) if you want to modify the source code, as you might encounter problems that will be fixed by the developers. Release versions will not be updated (except for updates due to critical and easy-to-fix bugs), but generally they are more stable than the trunk.

1. Releases can be downloaded from the download page, as compressed archive. Uncompressing the archive gives you a directory with the sources.

2. The development version (`trunk`) can be obtained from the code repository at GitLab.

We use GIT (the `git` command) for code management (install the `git` package on your system and create a GitLab account):

```
git clone git@gitlab.com:yade-dev/trunk.git
```

will download the whole code repository of the `trunk`. Check out *Yade on GitLab* for more details on how to collaborate using `git`.

Alternatively, a read-only checkout is possible via https without a GitLab account (easier if you don't want to modify the trunk version):

```
git clone https://gitlab.com/yade-dev/trunk.git
```

For those behind a firewall, you can download the sources from our GitLab repository as compressed archive.

Release and trunk sources are compiled in exactly the same way.

#### Prerequisites

Yade relies on a number of external software to run; they are checked before the compilation starts. Some of them are only optional.

- cmake build system
- gcc compiler (g++); other compilers will not work; you need g++>=4.2 for openMP support
- boost 1.47 or later
- Qt library
- freeglut3
- libQGLViewer
- python, numpy, ipython, sphinx, mpi4py
- matplotlib

- eigen algebra library (minimal required version 3.2.1)

- gdb debugger

- sqlite3 database engine

- VTK library (optional but recommended)

- CGAL library (optional)

- SuiteSparse sparse algebra library (fluid coupling, optional, requires eigen>=3.1)

- OpenBLAS optimized and parallelized alternative to the standard blas+lapack (fluid coupling *FlowEngine*, optional)

- Metis matrix preconditioning (fluid coupling, optional)

- OpenMPI library for parallel distributed computing (For MPI and OpenFOAM coupling, optional)

- python3-mpi4py MPI for Python (For MPI, optional)

- coin-or COIN-OR Linear Programming Solver (For *PotentialBlock*, optional)

- mpfr in C++ and mpmath in `python` for high precision `Real` or for CGAL exact predicates (optional)

- mpc is an MPFR extension to complex numbers. It is used explicitly together with MPFR.

Most of the list above is very likely already packaged for your distribution. In case you are confronted with some errors concerning not available packages (e.g., package libmetis-dev is not available) it may be necessary to add yade external ppa from https://launchpad.net/~yade-users/+archive/external (see below) as well as http://www.yade-dem.org/packages (see the top of this page):

```
sudo add-apt-repository ppa:yade-users/external
sudo apt-get update
```

The following commands have to be executed in the command line of your corresponding distribution. Just copy&paste to the terminal. Note, to execute these commands you need root privileges.

- **Ubuntu 20.04, 18.04, Debian 9, 10, 11** and their derivatives:

```
sudo apt install cmake git freeglut3-dev libloki-dev libboost-all-dev fakeroot \
dpkg-dev build-essential g++ python3-dev python3-ipython python3-matplotlib \
libsqlite3-dev python3-numpy python3-tk gnuplot libgts-dev python3-pygraphviz \
libvtk6-dev libeigen3-dev python3-xlib python3-pyqt5 pyqt5-dev-tools python3-mpi4py \
python3-pyqt5.qtwebkit gtk2-engines-pixbuf python3-pyqt5.qtsvg libqglviewer-dev-qt5 \
python3-pil libjs-jquery python3-sphinx python3-git libxmu-dev libxi-dev libcgal-dev \
help2man libbz2-dev zlib1g-dev libopenblas-dev libsuitesparse-dev \
libmetis-dev python3-bibtexparser python3-future coinor-clp coinor-libclp-dev \
python3-mpmath libmpfr-dev libmpfrc++-dev libmpc-dev
```

- For **Ubuntu 16.04** `libqglviewer-dev-qt5` is to be replaced by `libqglviewer-dev` and `python3-ipython` by `ipython3`.

- The packages `python3-mpmath libmpfr-dev libmpfrc++-dev` in above list are required only if one wants to use high precision calculations. The latter two only if mpfr will be used. See *high precision documentation* for more details.

- For building documentation (the `make doc` invocation explained below) additional package `texlive-xetex` is required. On some multi-language systems an error `Building format(s) --all. This may take some time... fmtutil failed.` may occur, in that case a package `locales-all` is required.

Some of the packages (for example, cmake, eigen3) are mandatory, some of them are optional. Watch for notes and warnings/errors, which are shown by `cmake` during the configuration step. If the missing package is optional, some of Yade features will be disabled (see the messages at the end of the configuration).

Some packages listed here are relatively new and they can be absent in your distribution (for example, libmetis-dev). They can be installed from yade-dem.org/packages or from our external PPA. If not installed the related features will be disabled automatically.

If you are using other distributions than Debian or its derivatives you should install the software packages listed above. Their names in other distributions can differ from the names of the Debian-packages.

> **Warning:** If you have Ubuntu 14.04 Trusty, you need to add -DCMAKE_CXX_FLAGS=-frounding-math during the configuration step of compilation (see below) or to install libcgal-dev from our external PPA. Otherwise the following error occurs on AMD64 architectures:
>
> ```
> terminate called after throwing an instance of 'CGAL::Assertion_exception'
> what(): CGAL ERROR: assertion violation!
> Expr: -CGAL_IA_MUL(-1.1, 10.1) != CGAL_IA_MUL(1.1, 10.1)
> File: /usr/include/CGAL/Interval_nt.h
> Line: 209
> Explanation: Wrong rounding: did you forget the -frounding-math option if you use GCC (or
> ↪ -fp-model strict for Intel)?
> Aborted
> ```

## Compilation

You should create a separate build-place-folder, where Yade will be configured and where the source code will be compiled. Here is an example for a folder structure:

```
myYade/         ## base directory
    trunk/      ## folder for source code in which you use git
    build/      ## folder in which the sources will be compiled; build-directory; use
↪cmake here
    install/    ## install folder; contains the executables
```

Then, inside this build-directory you should call **cmake** to configure the compilation process:

```
cmake -DCMAKE_INSTALL_PREFIX=/path/to/installfolder /path/to/sources
```

For the folder structure given above call the following command in the folder "build":

```
cmake -DCMAKE_INSTALL_PREFIX=../install ../trunk
```

Additional options can be configured in the same line with the following syntax:

```
cmake -DOPTION1=VALUE1 -DOPTION2=VALUE2
```

For example:

```
cmake -DENABLE_POTENTIAL_BLOCKS=ON
```

The following cmake options are available: (see the source code for a most up-to-date list)

- CMAKE_INSTALL_PREFIX: path where Yade should be installed (/usr/local by default)

- LIBRARY_OUTPUT_PATH: path to install libraries (lib by default)

- DEBUG: compile in debug-mode (OFF by default)

- MAX_LOG_LEVEL: *set maximum level* for LOG_* macros compiled with ENABLE_LOGGER, (default is 5)

- CMAKE_VERBOSE_MAKEFILE: output additional information during compiling (OFF by default)

- SUFFIX: suffix, added after binary-names (version number by default)

- NOSUFFIX: do not add a suffix after binary-name (OFF by default)

- YADE_VERSION: explicitly set version number (is defined from git-directory by default)

- ENABLE_ASAN: AddressSanitizer allows detection of memory errors, memory leaks, heap corruption errors and out-of-bounds accesses (but it is slow)

- ENABLE_CGAL: enable CGAL option (ON by default)

- ENABLE_COMPLEX_MP: use boost multiprecision complex for `ComplexHP<N>`, otherwise use `std::complex<RealHP<N>>`. See *high precision documentation* for additional details. (ON by default if possible: requires boost >= 1.71)

- ENABLE_DEFORM: enable constant volume deformation engine (OFF by default)

- ENABLE_FAST_NATIVE: use maximum optimization compiler flags including `-Ofast` and `-mtune=native`. Note: `native` means that code will **only** run on the same processor type on which it was compiled. Observed speedup was 2% (below standard deviation measurement error) and above 5% if clang compiler was used. (OFF by default)

- ENABLE_FEMLIKE: enable meshed solids, FEM-like (ON by default)

- ENABLE_GL2PS: enable GL2PS-option (ON by default)

- ENABLE_GTS: enable GTS-option (ON by default)

- ENABLE_GUI: enable GUI option (ON by default)

- ENABLE_LBMFLOW: enable LBMFLOW-option, LBM_ENGINE (ON by default)

- ENABLE_LS_DEM: enable a *LevelSet* shape description (ON by default)

- ENABLE_LINSOLV: enable LINSOLV-option (ON by default)

- ENABLE_LIQMIGRATION: enable LIQMIGRATION-option, see [Mani2013] for details (OFF by default)

- ENABLE_LOGGER: use boost::log library for *logging* separately for each class (ON by default)

- ENABLE_MASK_ARBITRARY: enable MASK_ARBITRARY option (OFF by default)

- ENABLE_MPFR: use mpfr in `C++` and mpmath in `python`. It can be used for higher precision `Real` or for CGAL exact predicates (OFF by default)

- ENABLE_MPI: Enable MPI enviroment and communication, required distributed memory and for Yade-OpenFOAM coupling (ON by default)

- ENABLE_OAR: generate a script for oar-based task scheduler (OFF by default)

- ENABLE_OPENMP: enable OpenMP-parallelizing option (ON by default)

- ENABLE_PARTIALSAT : enable the partially saturated clay engine, under construction (ON by default)

- ENABLE_PFVFLOW: enable PFVFLOW-option, FlowEngine (ON by default)

- ENABLE_POTENTIAL_BLOCKS: enable potential blocks option (ON by default)

- ENABLE_POTENTIAL_PARTICLES: enable potential particles option (ON by default)

- ENABLE_PROFILING: enable profiling, e.g., shows some more metrics, which can define bottlenecks of the code (OFF by default)

- ENABLE_REAL_HP: allow using twice, quadruple or higher precisions of `Real` as `RealHP<2>`, `RealHP<4>` or `RealHP<N>` in computationally demanding sections of `C++` code. See *high precision documentation* for additional details (ON by default).

- ENABLE_SPH: enable SPH-option, Smoothed Particle Hydrodynamics (OFF by default)

- ENABLE_THERMAL : enable thermal engine (ON by default, experimental)"

- ENABLE_TWOPHASEFLOW: enable TWOPHASEFLOW-option, TwoPhaseFlowEngine (ON by default)

- ENABLE_USEFUL_ERRORS: enable useful compiler errors which help a lot in error-free development (ON by default)

- ENABLE_VTK: enable VTK-export option (ON by default)

- REAL_PRECISION_BITS, REAL_DECIMAL_PLACES: specify either of them to use a custom calculation precision of `Real` type. By default `double` (64 bits, 15 decimal places) precision is used as the `Real` type. See *high precision documentation* for additional details.

- runtimePREFIX: used for packaging, when install directory is not the same as runtime directory (/usr/local by default)

- VECTORIZE: enables vectorization and alignment in Eigen3 library, experimental (OFF by default)

- USE_QT5: use QT5 for GUI (ON by default)

- CHOLMOD_GPU link Yade to custom SuiteSparse installation and activate GPU accelerated PFV (OFF by default)

- SUITESPARSEPATH: define this variable with the path to a custom suitesparse install

- PYTHON_VERSION: force Python version to the given one, e.g. `-DPYTHON_VERSION=3.5`. Set to -1 to automatically use the last version on the system (-1 by default)

It is possible to disable all options to create the slim build:

```
cmake -DDISABLE_ALL=ON
```

In this case all available options will be switched off. In this case some required options can be enabled explicitely:

```
cmake -DDISABLE_ALL=ON -DENABLE_VTK=ON
```

For using more extended parameters of cmake, please follow the corresponding documentation on https://cmake.org/documentation.

> **Warning:** Only Qt5 is supported. On Debian/Ubuntu operating systems libQGLViewer of version 2.6.3 and higher are compiled against Qt5 (for other operating systems refer to the package archive of your distribution). If you mix Qt-versions a `Segmentation fault` will appear just after Yade is started. To provide necessary build dependencies for Qt5, install `python-pyqt5 pyqt5-dev-tools`.

If cmake finishes without errors, you will see all enabled and disabled options at the end. Then start the actual compilation process with:

```
make
```

The compilation process can take a considerable amount of time, be patient. If you are using a multi-core systems you can use the parameter `-j` to speed-up the compilation and split the compilation onto many cores. For example, on 4-core machines it would be reasonable to set the parameter `-j4`. Note, Yade requires approximately 3GB RAM per core for compilation, otherwise the swap-file will be used and compilation time dramatically increases.

The installation is performed with the following command:

```
make install
```

The `install` command will in fact also recompile if source files have been modified. Hence there is no absolute need to type the two commands separately. You may receive make errors if you don't have permission to write into the target folder. These errors are not critical but without writing permissions Yade won't be installed in /usr/local/bin/.

After the compilation finished successfully, the new built can be started by navigating to /path/to/installfolder/bin and calling yade via (based on version yade-2014-02-20.git-a7048f4):

```
cd /path/to/installfolder/bin
./yade-2014-02-20.git-a7048f4
```

For building the documentation you should at first execute the command `make install` and then `make doc` to build it. The generated files will be stored in your current install directory /path/to/installfolder/share/doc/yade-your-version. Once again writing permissions are necessary for installing into /usr/local/share/doc/. To open your local documentation go into the folder html and open the file index.html with a browser.

`make manpage` command generates and moves manpages in a standard place. `make check` command executes standard test to check the functionality of the compiled program.

Yade can be compiled not only by GCC-compiler, but also by CLANG front-end for the LLVM compiler. For that you set the environment variables CC and CXX upon detecting the C and C++ compiler to use:

```
export CC=/usr/bin/clang
export CXX=/usr/bin/clang++
cmake -DOPTION1=VALUE1 -DOPTION2=VALUE2
```

Clang does not support OpenMP-parallelizing for the moment, that is why the feature will be disabled.

### Supported linux releases

Currently supported[1] linux releases and their respective docker files are:

- Ubuntu 16.04 xenial
- Ubuntu 18.04 bionic
- Debian 9 stretch
- Debian 10 buster
- openSUSE 15

These are the bash commands used to prepare the linux distribution and environment for installing and testing yade. These instructions are automatically performed using the gitlab continuous integration service after each merge to master. This makes sure that yade always works correctly on these linux distributions. In fact yade can be installed manually by following step by step these instructions in following order:

1. Bash commands in the respective Dockerfile to install necessary packages,
2. do `git clone https://gitlab.com/yade-dev/trunk.git`,
3. then the `cmake_*` commands in the .gitlab-ci.yml file for respective distribution,
4. then the `make_*` commands to compile yade,
5. and finally the `--check` and `--test` commands.
6. Optionally documentation can be built with `make doc` command, however currently it is not guaranteed to work on all linux distributions due to frequent interface changes in sphinx.

These instructions use `ccache` and `ld.gold` to *speed-up compilation* as described below.

### Python 2 backward compatibility

Following the end of Python 2 support (beginning of 2020), Yade compilation on a Python 2 ecosystem is no longer garanteed since the 6e097e95 trunk version. Python 2-compilation of the latter is still possible using the above `PYTHON_VERSION` cmake option, requiring Python 2 version of prerequisites packages whose list can be found in the corresponding paragraph (Python 2 backward compatibility) of the historical doc.

---

[1] To see details of the latest build log click on the *master* branch.

Ongoing development of Yade now assumes a Python 3 environment, and you may refer to some notes about *converting Python 2 scripts into Python 3* if needed.

### 2.5.4 Speed-up compilation

**Compile with ccache**

Caching previous compilations with ccache can significantly speed up re-compilation:

```
cmake -DCMAKE_CXX_COMPILER_LAUNCHER=ccache [options as usual]
```

Additionally one can check current ccache status with command `ccache --show-stats` (`ccache -s` for short) or change the default cache size stored in file `~/.ccache/ccache.conf`.

**Compile with distcc**

When spliting the compilation on many cores (`make -jN`), N is limited by the available cores and memory. It is possible to use more cores if remote computers are available, distributing the compilation with distcc (see distcc documentation for configuring slaves and master):

```
export CC="distcc gcc"
export CXX="distcc g++"
cmake [options as usual]
make -jN
```

The two tools can be combined, adding to the above exports:

```
export CCACHE_PREFIX="distcc"
```

**Compile with cmake UNITY_BUILD**

This option concatenates source files in batches containing several `*.cpp` each, in order to share the overhead of include directives (since most source files include the same boost headers, typically). It accelerates full compilation from scratch (quite significantly). It is activated by adding the following to cmake command, `CMAKE_UNITY_BUILD_BATCH_SIZE` defines the maximum number of files to be concatenated together (the higher the better, main limitation might be available RAM):

```
-DCMAKE_UNITY_BUILD=ON -DCMAKE_UNITY_BUILD_BATCH_SIZE=18
```

This method is helpless for incremental re-compilation and might even be detrimental since a full batch has to be recompiled each time a single file is modified. If it is anticipated that specific files will need incremental compilation they can be excluded from the unity build by assigning their full path to cmake flag `NO_UNITY` (a single file or a comma-separated list):

```
-DCMAKE_UNITY_BUILD=ON -DCMAKE_UNITY_BUILD_BATCH_SIZE=18 -DNO_UNITY=../trunk/pkg/dem/
↪CohesiveFrictionalContactLaw.cpp
```

**Link time**

The link time can be reduced by changing the default linker from `ld` to `ld.gold`. They are both in the same package `binutils` (on opensuse15 it is package `binutils-gold`). To perform the switch execute these commands as root:

```
ld --version
update-alternatives --install "/usr/bin/ld" "ld" "/usr/bin/ld.gold" 20
update-alternatives --install "/usr/bin/ld" "ld" "/usr/bin/ld.bfd" 10
ld --version
```

To switch back run the commands above with reversed priorities 10 20. Alternatively a manual selection can be performed by command: `update-alternatives --config ld`.

Note: `ld.gold` is incompatible with the compiler wrapper `mpicxx` in some distributions, which is manifested as an error in the `cmake` stage. We do not use `mpicxx` for our gitlab builds currently. If you want to use it then disable `ld.gold`. Cmake MPI-related failures have also been reported without the `mpicxx` compiler, if it happens then the only solution is to disable either `ld.gold` or the MPI feature.

### 2.5.5 Cloud Computing

It is possible to exploit cloud computing services to run Yade. The combo Yade/Amazon Web Service has been found to work well, namely. Detailed instructions for migrating to amazon can be found in the section *Using YADE with cloud computing on Amazon EC2*.

### 2.5.6 GPU Acceleration

The FlowEngine can be accelerated with CHOLMOD's GPU accelerated solver. The specific hardware and software requirements are outlined in the section *Accelerating Yade's FlowEngine with GPU*.

### 2.5.7 Special builds

The software can be compiled by a special way to find some specific bugs and problems in it: memory corruptions, data races, undefined behaviour etc.

The listed sanitizers are runtime-detectors. They can only find the problems in the code, if the particular part of the code is executed. If you have written a new C++ class (constitutive law, engine etc.) try to run your Python script with the sanitized software to check, whether the problem in your code exist.

**AddressSanitizer**

AddressSanitizer is a memory error detector, which helps to find heap corruptions, out-of-bounds errors and many other memory errors, leading to crashes and even wrong results.

To compile Yade with this type of sanitizer, use ENABLE_ASAN option:

```
cmake -DENABLE_ASAN=1
```

The compilation time, memory consumption during build and the size of build-files are much higher than during the normall build. Monitor RAM and disk usage during compilation to prevent out-of-RAM problems.

To find the proper libasan library in your particular distribution, use `locate` or `find /usr -iname "libasan*so"` command. Then, launch your yade executable in connection with that libasan library, e.g.:

```
LD_PRELOAD=/some/path/to/libasan.so yade
```

By default the leak detector is enabled in the asan build. Yade is producing a lot of leak warnings at the moment. To mute those warnings and concentrate on other memory errors, one can use detect_leaks=0 option. Accounting for the latter, the full command to run tests with the AddressSanitized-Yade on Debian 10 Buster is:

```
ASAN_OPTIONS=detect_leaks=0:verify_asan_link_order=false yade --test
```

If you add a new check script, it is being run automatically through the AddressSanitizer in the CI-pipeline.

### 2.5.8 Yubuntu

If you are not running a Linux system there is a way to create an Ubuntu live-usb on any usb mass-storage device (minimum size 10GB). It is a way to boot the computer on a linux system with Yadedaily pre-installed without affecting the original system. More informations about this alternative are available here (see the README file first).

Alternatively, images of a linux virtual machine can be downloaded, here again, and they should run on any system with a virtualization software (tested with VirtualBox and VMWare).

## 2.6 Acknowledging Yade

We kindly ask Yade users to cite this documentation as a whole in scientific publications as a way to assess Yade's contribution to their field. It can be done using the following reference:

- V. Šmilauer et al. (2021), Yade Documentation 3rd ed. The Yade Project. DOI:10.5281/zenodo.5705394 (http://yade-dem.org/doc/)

Beyond acknowledging the work of the developpers, it helps finding new use cases or new users by tracking the citations on Yade's Scholar profile.

# Chapter 3

# Yade for programmers

## 3.1 Programmer's manual

### 3.1.1 Build system

Yade uses cmake the cross-platform, open-source build system for managing the build process. It takes care of configuration, compilation and installation. CMake is used to control the software compilation process using simple platform and compiler independent configuration files. CMake generates native makefiles and workspaces that can be used in the compiler environment of your choice.

#### Building

The structure of Yade source tree is presented below. We shall call each top-level component *module* (excluding, **doc**, **examples** and **scripts** which don't participate in the build process). Some subdirectories of *modules* are skipped for brevity, see `README.rst` files therein for more information:

```
cMake/              ## cmake files used to detect compilation requirements
core/               ## core simulation building blocks
data/               ## data files used by yade, packaged separately
doc/                ## this documentation
examples/           ## examples directory
gui/                ## user interfaces
    qt5/                ## same, but for qt5
lib/                ## support libraries, not specific to simulations
preprocessing/      ## files associated with creation or generation of the simulation
    dem/                ## creating a DEM simulation
    potential/          ## creating a PotentialBlocks or PotentialParticles simulation
    README.rst          ## more information about this directory
pkg/                ## simulation-specific files
    common/             ## generally useful classes
    dem/                ## classes for Discrete Element Method
    README.rst          ## more information about this directory
postprocessing/     ## files associated with extracting results for postprocessing
    dem/                ## general data extraction from DEM, no particular data target
    image/              ## creating images from simulation
    vtk/                ## extracting data for VTK
    README.rst          ## more information about this directory
py/                 ## python modules
scripts/            ## helper scripts including packaging and checks-and-tests
```

**Header installation**

CMAKE uses the original source layout and it is advised to use `#include <module/Class.hpp>` style of inclusion rather than `#include "Class.hpp"` even if you are in the same directory. The following table gives a few examples:

| Original header location | Included as |
|---|---|
| `core/Scene.hpp` | `#include <core/Scene.hpp>` |
| `lib/base/Logging.hpp` | `#include <lib/base/Logging.hpp>` |
| `lib/serialization/Serializable.hpp` | `#include <lib/serialization/Serializable.hpp>` |
| `pkg/dem/SpherePack.hpp` | `#include <pkg/dem/SpherePack.hpp>` |

**Automatic compilation**

In the `pkg/` directory, situation is different. In order to maximally ease addition of modules to yade, all `*.cpp` files are *automatically scanned recursively* by CMAKE and considered for compilation.

To enable/disable some component use the cmake flags `ENABLE_FEATURE`, which are listed in:

1. *compilation instructions*.
2. CMakeLists.txt.

When some component is enabled an extra `#define` flag `YADE_FEATURE` is passed from cmake to the compiler. Then inside the code both the `.cpp` and `.hpp` files which contain the `FEATURE` feature should have an `#ifdef YADE_FEATURE` guard at the beginning.

**Linking**

The order in which modules might depend on each other is given as follows:

| module | resulting shared library | dependencies |
|---|---|---|
| lib | `libyade-support.so` | can depend on external libraries, may **not** depend on any other part of Yade. |
| core | `libcore.so` | yade-support; *may* depend on external libraries. |
| pkg | `libplugins.so` | core, yade-support |
| gui | `libQtGUI.so`, `libPythonUI.so` | lib, core, pkg |
| py | (many files) | lib, core, pkg, external |

## 3.1.2 Development tools

**Integrated Development Environment and other tools**

A frequently used IDE is Kdevelop. We recommend using this software for navigating in the sources, compiling and debugging. Other useful tools for debugging and profiling are Valgrind and KCachegrind. A series of wiki pages is dedicated to these tools in the development section of the wiki.

**Hosting and versioning**

The Yade project is kindly hosted at Launchpad and GitLab:

- source code on gitlab
- issue and bug tracking on gitlab

- release downloads on launchpad

- yade-dev mailing list on launchpad: yade-dev@lists.launchpad.net

- yade-users mailing list on launchpad: yade-users@lists.launchpad.net

- questions and answers on launchpad

The versioning software used is GIT, for which a short tutorial can be found in *Yade on GitLab*. GIT is a distributed revision control system. It is available packaged for all major linux distributions.

The source code is periodically imported to Launchpad for building PPA-packages. The repository can be http-browsed.

**Build robot**

A build robot hosted at UMS Gricad is tracking source code changes via gitlab pipeline mechanism. Each time a change in the source code is committed to the main development branch via GIT, or a Merge Request (MR) is submitted the "buildbot" downloads and compiles the new version, and then starts a series of tests.

If a compilation error has been introduced, it will be notified to the yade-dev mailing list and to the committer, thus helping to fix problems quickly. If the compilation is successful, the buildbot starts unit regression tests and "check tests" (see below) and report the results. If all tests are passed, a new version of the documentation is generated and uploaded to the website in html and pdf formats. As a consequence, those two links always point to the documentation (the one you are reading now) of the last successful build, and the delay between commits and documentation updates are very short (minutes). The buildbot activity and logs can be browsed online.

The output of each particular build is directly accessible by clicking the green "Passed" button, and then clicking "Browse" in the "Job Artifacts" on the right.

### 3.1.3 Debugging

For yade debugging two tools are available:

1. Use the debug build so that the stack trace provides complete information about potential crash. This can be achieved in two ways:

   a) Compiling yade with cmake option `-DDEBUG=ON`,

   b) Installing `yade-dbgsym` debian/ubuntu package (this option will be available after this task is completed).

2. Use *Logging* framework described below.

These tools can be used in conjunction with other software. A detailed discussion of these is on yade wiki. These tools include: kdevelop, valgrind, alleyoop, kcachegrind, ddd, gdb, kompare, kdiff3, meld.

---

**Note:** On some linux systems stack trace will not be shown and a message `ptrace: Operation not permitted` will appear instead. To enable stack trace issue command: `sudo echo 0 > /proc/sys/kernel/yama/ptrace_scope`. To disable stack trace issue command `sudo echo 1 > /proc/sys/kernel/yama/ptrace_scope`.

---

**Hint:** When debugging make sure there is enough free space in /tmp.

---

### Logging

Yade uses boost::log library for flexible logging levels and per-class debugging. See also description of *log module*. A cmake compilation option `-DENABLE_LOGGER=ON` must be supplied during compilation[1].

Figure *imgLogging* shows example use of logging framework. Usually a `ClassName` appears in place of `_log.cpp` shown on the screenshot. It is there because the `yade.log` module uses `CREATE_CPP_LOCAL_LOGGER` macro instead of the regular `DECLARE_LOGGER` and `CREATE_LOGGER`, which are *discussed below*.

---

**Note:** Default format of log message is:

```
<severity level> ClassName:LineNumber FunctionName: Log Message
```

special macro `LOG_NOFILTER` is printed without `ClassName` because it lacks one.

---

Config files can be saved and loaded via *readConfigFile* and *saveConfigFile*. The *defaultConfigFileName* is read upon startup if it exists. The filter level setting `-f` supplied from command line will override the setting in config file.

### Log levels

Following debug levels are supported:

---

[1] Without `-DENABLE_LOGGER=ON` cmake option the debug macros in /lib/base/Logging.hpp use regular `std::cerr` for output, per-class logging and log levels do not work.

Table 1: Yade logging verbosity levels.

| macro name | filter name | option | explanation |
|---|---|---|---|
| `LOG_NOFILTER` | `log.NOFILTER` | `-f0` | Will print only the unfiltered messages. The LOG_-NOFILTER macro is for developer use only, so basically `-f0` means that nothing will be printed. This log level is not useful unless a very silent mode is necessary. |
| `LOG_FATAL` | `log.FATAL` | `-f1` | Will print only critical errors. Even a throw to yade python interface will not recover from this situation. This is usually followed by yade exiting to shell. |
| `LOG_ERROR` | `log.ERROR` | `-f2` | Will also print errors which do not require to throw to yade python interface. Calculations will continue, but very likely the results will be all wrong. |
| `LOG_WARN` | `log.WARN` | `-f3` | Will also print warnings about recoverable problems that you should be notified about (e.g., invalid value in a configuration file, so yade fell back to the default value). |
| `LOG_INFO` | `log.INFO` | `-f4` | Will also print all informational messages (e.g. something was loaded, something was called, etc.). |
| `LOG_DEBUG` | `log.DEBUG` | `-f5` | Will also print debug messages. A yade developer puts them everywhere, and yade user enables them on *per-class basis* to provide some extra debug info. |
| `LOG_TRACE` | `log.TRACE` | `-f6` | Trace messages, they capture every possible detail about yade behavior. |

Yade default log level is `yade.log.WARN` which is the same as invoking `yade -f3`.

### Setting a filter level

> **Warning:** The messages (such as `a << b << " message."`) given as arguments to `LOG_*` macros are used only if the message passes the filter level. **Do not use such messages to perform mission critical calculations**.

There are two settings for the filter level, the `Default` level used when no `ClassName` (or `"filename.cpp"`) specific filter is set and a filter level set for specific `ClassName` (or `"filename.cpp"`). They can be set with following means:

1. When starting yade with `yade -fN` command, where `N` sets the `Default` filter level. The default value is `yade.log.WARN` (3).

2. To change `Default` filter level during runtime invoke command `log.setLevel("Default",value)` or `log.setDefaultLogLevel(value)`:

```
Yade [1]: import log

Yade [2]: log.setLevel("Default",log.WARN)

Yade [3]: log.setLevel("Default",3)

Yade [4]: log.setDefaultLogLevel(log.WARN)

Yade [5]: log.setDefaultLogLevel(3)
```

3. To change filter level for `SomeClass` invoke command:

```
Yade [6]: import log
```

```
Yade [7]: log.setLevel("NewtonIntegrator",log.TRACE)

Yade [8]: log.setLevel("NewtonIntegrator",6)
```

4. To change the filter level for **"filename.cpp"** use the name specified when creating it. For example manipulating filter log level of **"_log.cpp"** might look like following:

```
Yade [9]: import log

Yade [10]: log.getUsedLevels()
Out[10]: {}

Yade [11]: log.setLevel("_log.cpp",log.WARN)

Yade [12]: log.getUsedLevels()
Out[12]: {}

Yade [13]: log.getAllLevels()["_log.cpp"]
---------------------------------------------------------------------
KeyError                                Traceback (most recent call last)
~/yade/lib/x86_64-linux-gnu/yadeflip/py/yade/__init__.py in <module>
----> 1 log.getAllLevels()["_log.cpp"]

KeyError: '_log.cpp'
```

### Debug macros

To enable debugging for particular class the **DECLARE_LOGGER;** macro should be put in class definition inside header to create a separate named logger for that class. Then the **CREATE_LOGGER(ClassName);** macro must be used in the class implementation **.cpp** file to create the static variable. Sometimes a logger is necessary outside the class, such named logger can be created inside a **.cpp** file and by convention its name should correspond to the name of the file, use the macro **CREATE_CPP_LOCAL_LOGGER("filename.cpp");** for this. On rare occasions logging is necessary inside **.hpp** file outside of a class (where the local class named logger is unavailable), then the solution is to use **LOG_NOFILTER(…)** macro, because it is the only one that can work without a named logger. If the need arises this solution can be improved, see Logging.cpp for details.

All debug macros (**LOG_TRACE**, **LOG_DEBUG**, **LOG_INFO**, **LOG_WARN**, **LOG_ERROR**, **LOG_FATAL**, **LOG_NOFILTER**) listed in section above accept the **std::ostream** syntax inside the brackets, such as **LOG_TRACE( a << b << " text" )**. The **LOG_NOFILTER** is special because it is always printed regardless of debug level, hence it should be used only in development branches.

Additionally seven macros for printing variables at **LOG_TRACE** level are available: **TRVAR1**, **TRVAR2**, **TRVAR3**, **TRVAR4**, **TRVAR5**, **TRVAR6** and **TRVARn**. They print the variables, e.g.: **TRVAR3(testInt,testStr, testReal);** or **TRVARn((testInt)(testStr)(testReal))**. See function testAllLevels for example use.

The macro **TRACE;** prints a **"Been here"** message at **TRACE** log filter level, and can be used for quick debugging.

### Utility debug macros

The **LOG_TIMED_*** family of macros:

In some situations it is useful to debug variables inside a **very fast**, or maybe a **multithreaded**, loop. In such situations it would be useful to:

1. Avoid spamming console with very fast printed messages and add some print timeout to them, preferably specified with units of seconds or milliseconds.

2. Make sure that each separate thread has opportunity to print message, without interleaving such messages with other threads.

To use above functionality one must `#include <lib/base/LoggingUtils.hpp>` in the `.cpp` file which provides the `LOG_TIMED_*` and `TIMED_TRVAR*` macro family. Example usage can be found in function testTimedLevels.

To satisfy the first requirement all `LOG_TIMED_*` macros accept **two arguments**, where the first argument is the wait timeout, using standard C++14 / C++20 time units, example use is `LOG_TIMED_INFO( 2s , "test int: " << testInt++);` to print every 2 seconds. But only seconds and milliseconds are accepted (this can be changed if necessary).

To satisfy the second requirement a thread_local static Timer variable is used. This way each thread in a parallel loop can print a message every `500ms` or `10s` e.g. in this parallel loop. The time of last print to console is stored independently for each thread and an extra code block which checks time is added. It means that a bit more checks are done than typical `LOG_*` which only perform an integer comparison to check filter level. Therefore suggested use is only during heavy debugging. When debugging is finished then better to remove them.

---

**Note:** The `*_TRACE` family of macros are removed by compiler during the release builds, because the default `-DMAX_LOG_LEVEL` is 5. So those are very safe to use, but to have them working locally make sure to compile yade with `cmake -DMAX_LOG_LEVEL=6` option.

---

The `LOG_ONCE_*` family of macros:

In a similar manner a `LOG_ONCE_*` and `ONCE_TRVAR*` family of macros is provided inside file LoggingUtils.hpp. Then the message is printed only once.

All debug macros are summarized in the table below:

Table 2: Yade debug macros.

| macro name | explanation |
|---|---|
| `DECLARE_LOGGER;` | Declares logger variable inside class definition in `.hpp` file. |
| `CREATE_LOGGER(ClassName);` | Creates logger static variable (with name `"ClassName"`) inside class implementation in `.cpp` file. |
| `TEMPLATE_CREATE_-LOGGER(ClassName<OtherClass>);` | Creates logger static variable (with name `"ClassName<OtherClass>"`) inside class implementation in a `.cpp` file. Use this for templated classes. |
| `CREATE_CPP_LOCAL_LOGGER("filename.cpp");` | Creates logger static variable outside of any class (with name `"filename.cpp"`) inside the `filename.cpp` file. |
| `LOG_TRACE, LOG_TIMED_TRACE, LOG_ONCE_TRACE, LOG_DEBUG, LOG_TIMED_DEBUG, LOG_ONCE_DEBUG, LOG_INFO, LOG_TIMED_INFO, LOG_ONCE_INFO, LOG_WARN, LOG_TIMED_WARN, LOG_ONCE_WARN, LOG_ERROR, LOG_TIMED_ERROR, LOG_ONCE_ERROR, LOG_FATAL, LOG_TIMED_FATAL, LOG_ONCE_FATAL, LOG_NOFILTER, LOG_TIMED_NOFILTER, LOG_ONCE_NOFILTER` | Prints message using $\mathtt{std::ostream}$ syntax, like: `LOG_TRACE( a << b << " text" )` `LOG_TIMED_TRACE( 5s , a << b << " text" );`, prints every 5 seconds `LOG_TIMED_DEBUG( 500ms , a );`, prints every 500 milliseconds `LOG_ONCE_TRACE( a << b << " text" );`, prints just once `LOG_ONCE_DEBUG( a );`, prints only once |
| `TRVAR1, TIMED_TRVAR1, ONCE_TRVAR1, TRVAR2, TIMED_TRVAR2, ONCE_TRVAR2, TRVAR3, TIMED_TRVAR3, ONCE_TRVAR3, TRVAR4, TIMED_TRVAR4, ONCE_TRVAR4, TRVAR5, TIMED_TRVAR5, ONCE_TRVAR5, TRVAR6, TIMED_TRVAR6, ONCE_TRVAR6, TRVARn, TIMED_TRVARn, ONCE_TRVARn` | Prints provided variables like: `TRVAR3(testInt,testStr,testReal);` `TRVARn((testInt)(testStr)(testReal));` `TIMED_TRVAR3( 10s , testInt , testStr , testReal);` `ONCE_TRVARn( (testInt)(testStr)(testReal));` See file py/__log.cpp for example use. |
| `TRACE;` | Prints a `"Been here"` message at `TRACE` log filter level. |
| `LOG_TIMED_6, LOG_6_TRACE, LOG_ONCE_6, LOG_TIMED_5, LOG_5_DEBUG, LOG_ONCE_5, LOG_TIMED_4, LOG_4_INFO, LOG_ONCE_4, LOG_TIMED_3, LOG_3_WARN, LOG_ONCE_3, LOG_TIMED_2, LOG_2_ERROR, LOG_ONCE_2, LOG_TIMED_1, LOG_1_FATAL, LOG_ONCE_1, LOG_TIMED_0, LOG_0_NOFILTER, LOG_ONCE_0, LOG_TIMED_6_TRACE, LOG_6, LOG_ONCE_6_TRACE, LOG_TIMED_5_DEBUG, LOG_5, LOG_ONCE_5_DEBUG, LOG_TIMED_4_INFO, LOG_4, LOG_ONCE_4_INFO, LOG_TIMED_3_WARN, LOG_3, LOG_ONCE_3_WARN, LOG_TIMED_2_ERROR, LOG_2, LOG_ONCE_2_ERROR, LOG_TIMED_1_FATAL, LOG_1, LOG_ONCE_1_FATAL, LOG_TIMED_0_NOFILTER, LOG_0 LOG_ONCE_0_NOFILTER,` | Additional macro aliases for easier use in editors with tab completion. They have have a filter level number in their name. |

**Maximum log level**

Using boost::log for log filtering means that each call to `LOG_*` macro must perform a single integer comparison to determine if the message passes current filter level. For production use calculations should be as fast as possible and this filtering is not optimal, because the macros are *not optimized out*, as they can be re-enabled with a simple call to `log.setLevel("Default",log.TRACE)` or `log.setLevel("Default",6)`. The remedy is to use the cmake compilation option `MAX_LOG_LEVEL=4` (or 3) which will remove macros higher than the specified level during compilation. The code will run slightly faster and the command `log.setLevel("Default",6)` will only print a warning that such high log level (which can be checked with *log.getMaxLevel()* call) is impossible to obtain with current build.

**Note:** At the time when logging was introduced into yade the speed-up gain was so small, that it turned out to be impossible to measure with `yade -f0 --stdperformance` command. Hence this option `MAX_LOG_LEVEL` was introduced only on principle.

The upside of this approach is that yade can be compiled in a non-debug build, and the log filtering framework can be still used.

## 3.1.4 Regression tests

Yade contains two types of regression tests, some are unit tests while others are testing more complex simulations. Although both types can be considered regression tests, the usage is that we name the first simply "regression tests", while the latest are called "check tests". Both series of tests can be ran at yade startup by passing the options "test" or "checkall"

```
yade --test
yade --checkall
```

The `yade --checkall` is a complete check. To skip checks lasting more than 30 seconds one can use this command

```
yade --check
```

**Unit regression tests**

Unit regression tests are testing the output of individual functors and engines in well defined conditions. They are defined in the folder py/tests/. The purpose of unit testing is to make sure that the behaviour of the most important classes remains correct during code development. Since they test classes one by one, unit tests can't detect problems coming from the interaction between different engines in a typical simulation. That is why check tests have been introduced.

To add a new test, the following steps must be performed:

1. Place a new file such as py/tests/dummyTest.py.

2. Add the file name such as `dummyTest` to the py/tests/___init___.py file.

3. If necessary modify the `import` and `allModules` lines in py/tests/___init___.py.

4. According to instructions in python unittest documentation use commands such as `self.assertTrue(…)`, `self.assertFalse(…)` or `self.assertRaises(…,…)` to report possible errors.

**Note:** It is important that all variables used in the test are stored inside the class (using the `self.` accessor), and that all preparations are done inside the function `setUp()`.

**Check tests**

Check tests (also see README) perform comparisons of simulation results between different versions of yade, as discussed here. They differ with regression tests in the sense that they simulate more complex situations and combinations of different engines, and usually don't have a mathematical proof (though there is no restriction on the latest). They compare the values obtained in version N with values obtained in a previous version or any other "expected" results. The reference values must be hardcoded in the script itself or in data files provided with the script. Check tests are based on regular yade scripts, so that users can easily commit their own scripts to trunk in order to get some automatized testing after commits from other developers.

When check fails the script should return an error message via python command `raise YadeCheckError(messageString)` telling what went wrong. If the script itself fails for some reason and can't generate an output, the log will contain only "scriptName failure". If the script defines differences on obtained and awaited data, it should print some useful information about the problem. After this occurs, the automatic test will stop the execution with error message.

An example dummy check test scripts/checks-and-tests/checks/checkTestDummy.py demonstrates a minimal empty test. A little more functional example check test can be found in scripts/checks-and-tests/checks/checkTestTriax.py. It shows results comparison, output, and how to define the path to data files using `checksPath`. Users are encouraged to add their own scripts into the scripts/checks-and-tests/checks/ folder. Discussion of some specific checktests design in questions and answers is welcome. Note that re-compiling is required before the newly added scripts can be launched by `yade --check` (or direct changes have to be performed in "lib" subfolders). A check test should never need more than a few seconds to run. If your typical script needs more, try to reduce the number of elements or the number of steps.

To add a new check, the following steps must be performed:

1. Place a new file such as scripts/checks-and-tests/checks/checkTestDummy.py,

2. Inside the new script use `checksPath` when it is necessary to load some data file, like scripts/checks-and-tests/checks/data/100spheres

3. When error occurs raise exception with command `raise YadeCheckError(messageString)`

**GUI Tests**

In order to add a new GUI test one needs to add a file to scripts/checks-and-tests/gui directory. File must be named according to the following convention: `testGuiName.py` with an appropriate test `Name` in place (the `testGui.sh` script is searching for files matching this pattern). The scripts/checks-and-tests/gui/testGuiBilliard.py may serve as a boilerplate example. The important "extra" parts of the code (taken from e.g. example directory) are:

1. `from testGuiHelper import TestGUIHelper`

2. `scr = TestGUIHelper("Billiard")`, make sure to put the chosen test `Name` in place of `Billiard`.

3. Establish a reasonable value of `guiIterPeriod` which makes the test finish in less than 30 seconds.

4. Inside `O.engines` there has to be a call at the end of the loop to `PyRunner(iterPeriod=guiIterPeriod, command='scr.screenshotEngine()')`.

5. The last command in the script should be `O.run(guiIterPeriod * scr.getTestNum() + 1)` to start the test process.

6. Make sure to push to yade-data repository the reference screenshots (for dealing with `./data` dir see Yade on GitLab). These screenshots can be also obtained from artifacts by clicking "Download" button in the gitlab pipeline, next to the "Browse" button in the right pane.

These tests can be run locally, after adjusting the paths at the start of testGui.sh script. Two modes of operation are possible:

1. Launch on the local desktop via command: `scripts/checks-and-tests/gui/testGui.sh`, in this case the screenshots will be different from those used during the test.

2. Or launch inside a virtual xserver via command: `xvfb-run -a -s "-screen 0 1600x1200x24" scripts/checks-and-tests/gui/testGui.sh`, then the screenshots will be similar to those used in the test, but still there may be some differences in the font size. In such case it is recommended to use the reference screenshots downloaded from the artifacts in the gitlab pipeline (see point 6. above).

Care should be taken to not use random colors of bodies used in the test. Also no windows such as 3d View or Inspector view should be opened in the script `testGuiName.py`, because they are opened during the test by the `TestGUIHelper` class.

**Note:** It is not possible to call GUI tests from a call such as `yade --test` because of the necessity to launch YADE inside a virtual xserver.

### 3.1.5 Conventions

The following coding rules should be respected; documentation is treated separately.

- general

  - C++ source files have `.hpp` and `.cpp` extensions (for headers and implementation, respectively). In rare cases `.ipp` is used for pure template code.

  - All header files should have the `#pragma once` multiple-inclusion guard.

  - Do not type `using namespace …` in header files, this can lead to obscure bugs due to namespace pollution.

  - Avoid `using std::something` in `.hpp` files. Feel free to use them as much as you like inside `.cpp` files. But remember that the usual problems with this practice still apply: wrong type or function might be used instead of the one that you would expect. But since it's limited to a single `.cpp` file, it will be easier to debug and the convenience might outweight the associated dangers.

  - Use tabs for indentation. While this is merely visual in `C++`, it has semantic meaning in python; inadvertently mixing tabs and spaces can result in syntax errors.

- capitalization style

  - Types should be always capitalized. Use CamelCase for composed class and typenames (`GlobalEngine`). Underscores should be used only in special cases, such as functor names.

  - Class data members and methods must not be capitalized, composed names should use lowercase camelCase (`glutSlices`). The same applies for functions in python modules.

  - Preprocessor macros are uppercase, separated by underscores; those that are used outside the core take (with exceptions) the form `YADE_*`, such as *YADE_CLASS_BASE_DOC_* macro family*.

- programming style

  - Be defensive, if it has no significant performance impact. Use assertions abundantly: they don't affect performance (in the optimized build) and make spotting error conditions much easier.

  - Use `YADE_CAST` and `YADE_PTR_CAST` where you want type-check during debug builds, but fast casting in optimized build.

  - Initialize all class variables in the default constructor. This avoids bugs that may manifest randomly and are difficult to fix. Initializing with NaN's will help you find otherwise unitialized variable. (This is taken care of by *YADE_CLASS_BASE_DOC_* macro family* macros for user classes)

### Using clang-format

The file .clang-format contains the config which should produce always the same results. It works with `clang-format --version >= 10`. The aim is to eliminate commits that change formatting. The script scripts/clang-formatter.sh can be invoked on either file or a directory and will do the reformatting. Usually this can be integrated with the editor, see clang-format documentation (except that for vim `py3f` command has to be used), and in kdevelop it is added as a custom formatter.

The script scripts/python-formatter.sh applies our coding conventions to formatting of python scripts. It should be used before committing changes to python scripts.

For more help see:

1. clang-format documentation

2. yapf3 documentation

Sometimes it is useful to disable formatting in a small section of the file. In order to do so, put the guards around this section:

1. In `C++` use:

```
// clang-format off
……
// clang-format on
```

2. In `Python` use:

```
# yapf: disable
……
# yapf: enable
```

### Class naming

Although for historical reasons the naming scheme is not completely consistent, these rules should be obeyed especially when adding a new class.

***GlobalEngines*** **and** ***PartialEngines*** GlobalEngines should be named in a way suggesting that it is a performer of certain action (like *ForceResetter*, *InsertionSortCollider*, *Recorder*); if this is not appropriate, append the `Engine` to the characteristics name (e.g. *GravityEngine*). *PartialEngines* have no special naming convention different from *GlobalEngines*.

***Dispatchers*** Names of all dispatchers end in `Dispatcher`. The name is composed of type it creates or, in case it doesn't create any objects, its main characteristics. Currently, the following dispatchers[2] are defined:

| dispatcher | arity | dispatch types | created type | functor type | functor prefix |
|---|---|---|---|---|---|
| *BoundDispatcher* | 1 | *Shape* | *Bound* | *BoundFunctor* | Bo1 |
| *IGeomDispatcher* | 2 (symetric) | 2 × *Shape* | *IGeom* | *IGeomFunctor* | Ig2 |
| *IPhysDispatcher* | 2 (symetric) | 2 × *Material* | *IPhys* | *IPhysFunctor* | Ip2 |
| *LawDispatcher* | 2 (asymetric) | *IGeom* *IPhys* | *(none)* | *LawFunctor* | Law2 |

Respective abstract functors for each dispatchers are *BoundFunctor*, *IGeomFunctor*, *IPhysFunctor* and *LawFunctor*.

---

[2] Not considering OpenGL dispatchers, which might be replaced by regular virtual functions in the future.

*Functors* Functor name is composed of 3 parts, separated by underscore.

1. prefix, composed of abbreviated functor type and arity (see table above)

2. Types entering the dispatcher logic (1 for unary and 2 for binary functors)

3. Return type for functors that create instances, simple characteristics for functors that don't create instances.

To give a few examples:

- *Bo1_Sphere_Aabb* is a *BoundFunctor* which is called for *Sphere*, creating an instance of *Aabb*.

- *Ig2_Facet_Sphere_ScGeom* is binary functor called for *Facet* and *Sphere*, creating and instace of *ScGeom*.

- *Law2_ScGeom_CpmPhys_Cpm* is binary functor (*LawFunctor*) called for types *ScGeom (Geom)* and *CpmPhys*.

### Documentation

**Documenting code properly is one of the most important aspects of sustained development.**

Read it again.

Most code in research software like Yade is not only used, but also read, by developers or even by regular users. Therefore, when adding new class, always mention the following in the documentation:

- purpose

- details of the functionality, unless obvious (algorithms, internal logic)

- limitations (by design, by implementation), bugs

- bibliographical reference, if using non-trivial published algorithms (see below)

- references to other related classes

- hyperlinks to bugs, blueprints, wiki or mailing list about this particular feature.

As much as it is meaningful, you should also

- update any other documentation affected

- provide a simple python script demonstrating the new functionality in `scripts/test`.

### Sphinx documentation

Most c++ classes are wrapped in Python, which provides good introspection and interactive documentation (try writing `Material?` in the ipython prompt; or `help(CpmState)`).

Syntax of documentation is ReST (reStructuredText, see reStructuredText Primer). It is the same for c++ and python code.

- Documentation of c++ classes exposed to python is given as 3rd argument to *YADE_CLASS_-BASE_DOC_*_* macro family* introduced below.

- Python classes/functions are documented using regular python docstrings. Besides explaining functionality, meaning and types of all arguments should also be documented. Short pieces of code might be very helpful. See the *utils* module for an example.

---

**Note:** Use C++ string literal when writing docstrings in C++. By convention the `R"""(raw text)"""` is used. For example see *here* and here.

---

---

**Note:** Remember that inside C++ docstrings it is possible to invoke python commands which are executed by yade when documentation is being compiled. For example compare this source docstring with the final effect.

---

In addition to standard ReST syntax, yade provides several shorthand macros:

**:yref:** creates hyperlink to referenced term, for instance:

> ```
> :yref:`CpmMat`
> ```

> becomes *CpmMat*; link name and target can be different:

> ```
> :yref:`Material used in the CPM model<CpmMat>`
> ```

> yielding *Material used in the CPM model*.

**:ysrc:** creates hyperlink to file within the source tree (to its latest version in the repository), for instance core/Cell.hpp. Just like with **:yref:**, alternate text can be used with

> ```
> :ysrc:`Link text<target/file>`
> ```

> like this. This cannot be used to link to a specified line number, since changing the file will cause the line numbers to become outdated. To link to a line number use **:ysrccommit:** described below.

**:ysrccommit:** creates hyperlink to file within the source tree at the specified commit hash. This allows to link to the line numbers using for example **#L121** at the end of the link. Use it just like the **:ysrc:** except that commit hash must be provided at the beginning:

> ```
> :ysrccommit:`Link text<commithash/target/file#Lnumber>`
> ```

> ```
> :ysrccommit:`default engines<775ae7436/py/__init__.py.in#L112>`
> ```

> becomes default engines.

**Linking to inheritanceGraph\*** To link to an inheritance graph of some base class a *global anchor* is created with name **inheritanceGraph\*** added in front of the class name, for example **:ref:`Shape<inheritanceGraphShape>`** yields link to *inheritance graph of Shape*.

**|ycomp|** is used in attribute description for those that should not be provided by the user, but are auto-computed instead; **|ycomp|** expands to *(auto-computed)*.

**|yupdate|** marks attributes that are periodically updated, being subset of the previous. **|yupdate|** expands to *(auto-updated)*.

**$...$** delimits inline math expressions; they will be replaced by:

> ```
> :math:`...`
> ```

> and rendered via LaTeX. To write a single dollar sign, escape it with backslash \\$.

> Displayed mathematics (standalone equations) can be inserted as explained in Math support for HTML outputs in Sphinx.

As a reminder in the standard ReST syntax the references are:

**:ref:** is the the standard restructured text reference to an anchor placed elsewere in the text. For instance an anchor **.. _NumericalDamping:** is placed in formulation.rst then it is linked to with **:ref:`NumericalDamping`** inside the source code.

**.. _anchor-name:** is used to place anchors in the text, to be referenced from elsewhere in the text. Symbol _ is forbidden in the anchor name, because it has a special meaning: **_anchor** specifies anchor, while **anchor_** links to it, see below.

---

**anchor-name_** is used to place a link to anchor within the same file. It is a shorter form compared to the one which works between different files: `:ref:`. For example usage on anchor `imgQtGui` see here and here.

---

**Note:** The command `:scale: NN %` (with percent) does not work well with `.html` + `.pdf` output, better to specify `:width: NN cm`. Then it is the same size in `.html` and `.pdf.`. For example see here which becomes *this picture*. But bear in mind that maximum picture width in `.pdf` is `16.2 cm`.

---

### Bibliographical references

As in any scientific documentation, references to publications are very important. To cite an article, first add it in BibTeX format to files doc/references.bib or doc/yade-*.bib depending whether that reference used Yade (the latter cases) or not (the former). Please adhere to the following conventions:

1. Keep entries in the form `Author2008` (`Author` is the first author), `Author2008b` etc if multiple articles from one author;

2. Try to fill mandatory fields for given type of citation;

3. Do not use `\'{i}` funny escapes for accents, since they will not work with the HTML output; put everything in straight utf-8.

In your docstring, the `Author2008` article can be then cited by `[Author2008]_`; for example:

```
According to [Allen1989]_, the integration scheme …
```

will be rendered as

> According to [Allen1989], the integration scheme …

### Separate class/function documentation

Some c++ might have long or content-rich documentation, which is rather inconvenient to type in the c++ source itself as string literals. Yade provides a way to write documentation separately in py/_-extraDocs.py file: it is executed after loading c++ plugins and can set `__doc__` attribute of any object directly, overwriting docstring from c++. In such (exceptional) cases:

1. Provide at least a brief description of the class in the c++ code nevertheless, for people only reading the code.

2. Add notice saying "This class is documented in detail in the py/_extraDocs.py file".

3. Add documentation to py/_extraDocs.py in this way:

```
module.YourClass.__doc__='''
        This is the docstring for YourClass.

        Class, methods and functions can be documented this way.

        .. note:: It can use any syntax features you like.

'''
```

---

**Note:** Boost::python embeds function signatures in the docstring (before the one provided by the user). Therefore, before creating separate documentation of your function, have a look at its `__doc__` attribute and copy the first line (and the blank line afterwards) in the separate docstring. The first line is then used to create the function signature (arguments and return value).

---

**Internal c++ documentation**

doxygen was used for automatic generation of c++ code. Since user-visible classes are defined with sphinx now, it is not meaningful to use doxygen to generate overall documentation. However, take care to document well internal parts of code using regular comments, including public and private data members.

## 3.1.6 Support framework

Besides the framework provided by the c++ standard library (including STL), boost and other dependencies, Yade provides its own specific services.

### Pointers

#### Shared pointers

Yade makes extensive use of shared pointers `shared_ptr`.[3] Although it probably has some performance impacts, it greatly simplifies memory management, ownership management of c++ objects in python and so forth. To obtain raw pointer from a `shared_ptr`, use its `get()` method; raw pointers should be used in case the object will be used only for short time (during a function call, for instance) and not stored anywhere.

Python defines thin wrappers for most c++ Yade classes (for all those registered with *YADE_CLASS_-BASE_DOC_\* macro family* and several others), which can be constructed from `shared_ptr`; in this way, Python reference counting blends with the `shared_ptr` reference counting model, preventing crashes due to python objects pointing to c++ objects that were destructed in the meantime.

#### Typecasting

Frequently, pointers have to be typecast; there is choice between static and dynamic casting.

- `dynamic_cast` (`dynamic_pointer_cast` for a `shared_ptr`) assures cast admissibility by checking runtime type of its argument and returns NULL if the cast is invalid; such check obviously costs time. Invalid cast is easily caught by checking whether the pointer is NULL or not; even if such check (e.g. `assert`) is absent, dereferencing NULL pointer is easily spotted from the stacktrace (debugger output) after crash. Moreover, `shared_ptr` checks that the pointer is non-NULL before dereferencing in debug build and aborts with "Assertion 'px!=0' failed." if the check fails.

- `static_cast` is fast but potentially dangerous (`static_pointer_cast` for `shared_ptr`). Static cast will return non-NULL pointer even if types don't allow the cast (such as casting from `State*` to `Material*`); the consequence of such cast is interpreting garbage data as instance of the class cast to, leading very likely to invalid memory access (segmentation fault, "crash" for short).

To have both speed and safety, Yade provides 2 macros:

`YADE_CAST` expands to `static_cast` in optimized builds and to `dynamic_cast` in debug builds.

`YADE_PTR_CAST` expands to `static_pointer_cast` in optimized builds and to `dynamic_pointer_cast` in debug builds.

### Basic numerics

The floating point type to use in Yade is `Real`, which is by default typedef for `double` (64 bits, 15 decimal places).[4]

---

[3] Either `boost::shared_ptr` or `tr1::shared_ptr` is used, but it is always imported with the `using` statement so that unqualified `shared_ptr` can be used.

[4] See *high precision documentation* for additional details.

Yade uses the Eigen library for computations. It provides classes for 2d and 3d vectors, quaternions and 3x3 matrices templated by number type; their specialization for the `Real` type are typedef'ed with the "r" suffix, and occasionally useful integer types with the "i" suffix:

- `Vector2r`, `Vector2i`

- `Vector3r`, `Vector3i`

- `Quaternionr`

- `Matrix3r`

Yade additionally defines a class named *Se3r*, which contains spatial position (`Vector3r Se3r::position`) and orientation (`Quaternionr Se3r::orientation`), since they are frequently used one with another, and it is convenient to pass them as single parameter to functions.

Eigen provides full rich linear algebra functionality. Some code further uses the [cgal] library for computational geometry.

In Python, basic numeric types are wrapped and imported from the `yade.minieigenHP` module; the types drop the `r` type qualifier at the end, the syntax is otherwise similar. `Se3r` is not wrapped at all, only converted automatically, rarely as it is needed, from/to a (`Vector3`,`Quaternion`) tuple/list. See *high precision section* for more details.

```
# cross product
Yade [14]: Vector3(1,2,3).cross(Vector3(0,0,1))
Out[14]: Vector3(2,-1,0)

# construct quaternion from axis and angle
Yade [15]: Quaternion(Vector3(0,0,1),pi/2)
Out[15]: Quaternion((0,0,1),1.570796326794896558)
```

---

**Note:** Quaternions are internally stored as 4 numbers. Their usual human-readable representation is, however, (normalized) axis and angle of rotation around that axis, and it is also how they are input/output in Python. Raw internal values can be accessed using the `[0]` … `[3]` element access (or `.W()`, `.X()`, `.Y()` and `.Z()` methods), in both c++ and Python.

---

### Run-time type identification (RTTI)

Since serialization and dispatchers need extended type and inheritance information, which is not sufficiently provided by standard RTTI. Each yade class is therefore derived from `Factorable` and it must use macro to override its virtual functions providing this extended RTTI:

`YADE_CLASS_BASE_DOC(Foo,Bar Baz,"Docstring")` creates the following virtual methods (mediated via the `REGISTER_CLASS_AND_BASE` macro, which is not user-visible and should not be used directly):

- `std::string getClassName()` returning class name (`Foo`) as string. (There is the `typeid(instanceOrType).name()` standard c++ construct, but the name returned is compiler-dependent.)

- `unsigned getBaseClassNumber()` returning number of base classes (in this case, 2).

- `std::string getBaseClassName(unsigned i=0)` returning name of *i*-th base class (here, `Bar` for i=0 and `Baz` for i=1).

---

**Warning:** RTTI relies on virtual functions; in order for virtual functions to work, at least one virtual method must be present in the implementation (`.cpp`) file. Otherwise, virtual method table (vtable) will not be generated for this class by the compiler, preventing virtual methods from functioning properly.

---

Some RTTI information can be accessed from python:

```
Yade [16]: yade.system.childClasses('Shape')
Out[16]:
{'Box',
 'ChainedCylinder',
 'Clump',
 'Cylinder',
 'Facet',
 'GridConnection',
 'GridNode',
 'PFacet',
 'Sphere',
 'Tetra',
 'Wall'}

Yade [17]: Sphere().__class__.__name__              ## getClassName()
Out[17]: 'Sphere'
```

## Serialization

Serialization serves to save simulation to file and restore it later. This process has several necessary conditions:

- classes know which attributes (data members) they have and what are their names (as strings);
- creating class instances based solely on its name;
- knowing what classes are defined inside a particular shared library (plugin).

This functionality is provided by 3 macros and 4 optional methods; details are provided below.

**Serializable::preLoad, Serializable::preSave, Serializable::postLoad, Serializable::postSave**
   Prepare attributes before serialization (saving) or deserialization (loading) or process them after serialization or deserialization.

   See *Attribute registration*.

**YADE_CLASS_BASE_DOC_\*** Inside the class declaration (i.e. in the `.hpp` file within the `class Foo { /* … */};` block). See *Attribute registration*.

   Enumerate class attributes that should be saved and loaded; associate each attribute with its literal name, which can be used to retrieve it. See *YADE_CLASS_BASE_DOC_\* macro family*.

   Additionally documents the class in python, adds methods for attribute access from python, and documents each attribute.

**REGISTER_SERIALIZABLE** In header file, but *after* the class declaration block. See *Class factory*.

   Associate literal name of the class with functions that will create its new instance (`ClassFactory`).

   Must be declared inside `namespace yade`.

**YADE_PLUGIN** In the implementation `.cpp` file. See *Plugin registration*.

   Declare what classes are declared inside a particular plugin at time the plugin is being loaded (yade startup).

   Must be declared inside `namespace yade`.

## Attribute registration

All (serializable) types in Yade are one of the following:

- Type deriving from *Serializable*, which provide information on how to serialize themselves via overriding the `Serializable::registerAttributes` method; it declares data members that should be serialized along with their literal names, by which they are identified. This method then invokes `registerAttributes` of its base class (until `Serializable` itself is reached); in this way, derived classes properly serialize data of their base classes.

  This funcionality is hidden behind the macro *YADE_CLASS_BASE_DOC_* macro family* used in class declaration body (header file), which takes base class and list of attributes:

  ```
  YADE_CLASS_BASE_DOC_ATTRS(ThisClass,BaseClass,"class documentation",((type1,attribute1,
  ↪initValue1,,"Documentation for attribute 1"))((type2,attribute2,initValue2,,
  ↪"Documentation for attribute 2")));
  ```

  Note that attributes are encoded in double parentheses, not separated by commas. Empty attribute list can be given simply by `YADE_CLASS_BASE_DOC_ATTRS(ThisClass,BaseClass, "documentation",)` (the last comma is mandatory), or by omiting `ATTRS` from macro name and last parameter altogether.

- Fundamental type: strings, various number types, booleans, `Vector3r` and others. Their "handlers" (serializers and deserializers) are defined in `lib/serialization`.

- Standard container of any serializable objects.

- Shared pointer to serializable object.

Yade uses the excellent boost::serialization library internally for serialization of data.

---

**Note:** `YADE_CLASS_BASE_DOC_ATTRS` also generates code for attribute access from python; this will be *discussed later*. Since this macro serves both purposes, the consequence is that attributes that are serialized can always be accessed from python.

---

Yade also provides callback for before/after (de) serialization, virtual functions *Serializable::preProcessAttributes* and *Serializable::postProcessAttributes*, which receive one `bool deserializing` argument (`true` when deserializing, `false` when serializing). Their default implementation in *Serializable* doesn't do anything, but their typical use is:

- converting some non-serializable internal data structure of the class (such as multi-dimensional array, hash table, array of pointers) into a serializable one (pre-processing) and fill this non-serializable structure back after deserialization (post-processing); for instance, InteractionContainer uses these hooks to ask its concrete implementation to store its contents to a unified storage (`vector<shared_ptr<Interaction> >`) before serialization and to restore from it after deserialization.

- precomputing non-serialized attributes from the serialized values; e.g. *Facet* computes its (local) edge normals and edge lengths from vertices' coordinates.

### Class factory

Each serializable class must use `REGISTER_SERIALIZABLE`, which defines function to create that class by `ClassFactory`. `ClassFactory` is able to instantiate a class given its name (as string), which is necessary for deserialization.

Although mostly used internally by the serialization framework, programmer can ask for a class instantiation using `shared_ptr<Factorable> f=ClassFactory::instance().createShared("ClassName");`, casting the returned `shared_ptr<Factorable>` to desired type afterwards. *Serializable* itself derives from `Factorable`, i.e. all serializable types are also factorable.

---

**Note:** Both macros `REGISTER_SERIALIZABLE` and `YADE_PLUGIN` have to be declared inside yade namespace.

---

## Plugin registration

Yade loads dynamic libraries containing all its functionality at startup. ClassFactory must be taught about classes each particular file provides. `YADE_PLUGIN` serves this purpose and, contrary to *YADE_-CLASS_BASE_DOC_* macro family*, must be placed in the implementation (.cpp) file, inside yade namespace. It simply enumerates classes that are provided by this file:

```
YADE_PLUGIN((ClassFoo)(ClassBar));
```

**Note:** You must use parentheses around the class name even if there is only one class (preprocessor limitation): `YADE_PLUGIN((classFoo));`. If there is no class in this file, do not use this macro at all.

Internally, this macro creates function `registerThisPluginClasses_` declared specially as `__-attribute__((constructor))` (see GCC Function Attributes); this attributes makes the function being executed when the plugin is loaded via `dlopen` from `ClassFactory::load(...)`. It registers all factorable classes from that file in the *Class factory*.

**Note:** Classes that do not derive from `Factorable`, such as `Shop` or `SpherePack`, are not declared with `YADE_PLUGIN`.

This is an example of a serializable class header:

```cpp
namespace yade {
/*! Homogeneous gravity field; applies gravity×mass force on all bodies. */
class GravityEngine: public GlobalEngine{
        public:
                virtual void action();
                // registering class and its base for the RTTI system
                YADE_CLASS_BASE_DOC_ATTRS(GravityEngine,GlobalEngine,
                        // documentation visible from python and generated reference documentation
                        "Homogeneous gravity field; applies gravity×mass force on all bodies.",
                        // enumerating attributes here, include documentation
                        ((Vector3r,gravity,Vector3r::ZERO,"acceleration, zero by default [kgms$^2$]"))
                );
};
// registration function for ClassFactory
REGISTER_SERIALIZABLE(GravityEngine);
} // namespace yade
```

and this is the implementation:

```cpp
#include <pkg/common/GravityEngine.hpp>
#include <core/Scene.hpp>

namespace yade {
// registering the plugin
YADE_PLUGIN((GravityEngine));

void GravityEngine::action(){
        /* do the work here */
}
} // namespace yade
```

We can create a mini-simulation (with only one GravityEngine):

```
Yade [18]: O.engines=[GravityEngine(gravity=Vector3(0,0,-9.81))]

Yade [19]: O.save('abc.xml')
```

and the XML save looks like this:

```xml
<?xml version="1.0" encoding="UTF-8" standalone="yes" ?>
<!DOCTYPE boost_serialization>
<boost_serialization signature="serialization::archive" version="17">
<scene class_id="0" tracking_level="0" version="1">
        <px class_id="1" tracking_level="1" version="0" object_id="_0">
                <Serializable class_id="2" tracking_level="1" version="0" object_id="_1"></
↪Serializable>
                <dt>1.0000000000000002e-08</dt>
                <iter>0</iter>
                <subStepping>0</subStepping>
                <subStep>-1</subStep>
                <time>0.0000000000000000e+00</time>
                <speed>0.0000000000000000e+00</speed>
                <stopAtIter>0</stopAtIter>
                <stopAtTime>0.0000000000000000e+00</stopAtTime>
                <isPeriodic>0</isPeriodic>
                <trackEnergy>0</trackEnergy>
                <doSort>0</doSort>
                <runInternalConsistencyChecks>1</runInternalConsistencyChecks>
                <selectedBody>-1</selectedBody>
                <tags class_id="3" tracking_level="0" version="0">
                        <count>5</count>
                        <item_version>0</item_version>
                        <item>author=bchareyre~(bchareyre@HP-ZBook-15-G3)</item>
                        <item>isoTime=20220726T141500</item>
                        <item>id=20220726T141500p61547</item>
                        <item>d.id=20220726T141500p61547</item>
                        <item>id.d=20220726T141500p61547</item>
                </tags>
                <engines class_id="4" tracking_level="0" version="0">
                        <count>1</count>
                        <item_version>1</item_version>
                        <item class_id="5" tracking_level="0" version="1">
                                <px class_id="7" class_name="yade::GravityEngine" tracking_
↪level="1" version="0" object_id="_2">
                                        <FieldApplier class_id="8" tracking_level="1" version=
↪"0" object_id="_3">
                                                <GlobalEngine class_id="9" tracking_level="1"␣
↪version="0" object_id="_4">
                                                        <Engine class_id="6" tracking_level="1
↪" version="0" object_id="_5">
                                                                <Serializable object_id="_6"></
↪Serializable>
                                                                <dead>0</dead>
                                                                <ompThreads>-1</ompThreads>
                                                                <label></label>
                                                        </Engine>
                                                </GlobalEngine>
                                        </FieldApplier>
                                        <gravity class_id="10" tracking_level="0" version="0">
                                                <x>0.0000000000000000e+00</x>
                                                <y>0.0000000000000000e+00</y>
                                                <z>-9.8100000000000050e+00</z>
                                        </gravity>
                                        <mask>0</mask>
                                        <warnOnce>1</warnOnce>
```

```xml
                            </px>
                    </item>
            </engines>
            <_nextEngines>
                    <count>0</count>
                    <item_version>1</item_version>
            </_nextEngines>
            <bodies class_id="11" tracking_level="0" version="1">
                    <px class_id="12" tracking_level="1" version="0" object_id="_7">
                            <Serializable object_id="_8"></Serializable>
                            <body class_id="13" tracking_level="0" version="0">
                                    <count>0</count>
                                    <item_version>1</item_version>
                            </body>
                            <insertedBodies>
                                    <count>0</count>
                                    <item_version>0</item_version>
                            </insertedBodies>
                            <erasedBodies>
                                    <count>0</count>
                                    <item_version>0</item_version>
                            </erasedBodies>
                            <realBodies>
                                    <count>0</count>
                                    <item_version>0</item_version>
                            </realBodies>
                            <useRedirection>0</useRedirection>
                            <enableRedirection>1</enableRedirection>
                    </px>
            </bodies>
            <interactions class_id="15" tracking_level="0" version="1">
                    <px class_id="16" tracking_level="1" version="0" object_id="_9">
                            <Serializable object_id="_10"></Serializable>
                            <interaction class_id="17" tracking_level="0" version="0">
                                    <count>0</count>
                                    <item_version>1</item_version>
                            </interaction>
                            <serializeSorted>0</serializeSorted>
                            <dirty>1</dirty>
                    </px>
            </interactions>
            <energy class_id="18" tracking_level="0" version="1">
                    <px class_id="19" tracking_level="1" version="0" object_id="_11">
                            <Serializable object_id="_12"></Serializable>
                            <energies class_id="20" tracking_level="0" version="0">
                                    <size>0</size>
                            </energies>
                            <names class_id="21" tracking_level="0" version="0">
                                    <count>0</count>
                                    <item_version>0</item_version>
                            </names>
                            <resetStep>
                                    <count>0</count>
                            </resetStep>
                    </px>
            </energy>
            <materials class_id="23" tracking_level="0" version="0">
                    <count>0</count>
                    <item_version>1</item_version>
            </materials>
            <bound class_id="24" tracking_level="0" version="1">
```

```xml
                    <px class_id="-1"></px>
            </bound>
    <cell class_id="26" tracking_level="0" version="1">
            <px class_id="27" tracking_level="1" version="0" object_id="_13">
                    <Serializable object_id="_14"></Serializable>
                    <trsf class_id="28" tracking_level="0" version="0">
                            <m00>1.00000000000000000e+00</m00>
                            <m01>0.00000000000000000e+00</m01>
                            <m02>0.00000000000000000e+00</m02>
                            <m10>0.00000000000000000e+00</m10>
                            <m11>1.00000000000000000e+00</m11>
                            <m12>0.00000000000000000e+00</m12>
                            <m20>0.00000000000000000e+00</m20>
                            <m21>0.00000000000000000e+00</m21>
                            <m22>1.00000000000000000e+00</m22>
                    </trsf>
                    <refHSize>
                            <m00>1.00000000000000000e+00</m00>
                            <m01>0.00000000000000000e+00</m01>
                            <m02>0.00000000000000000e+00</m02>
                            <m10>0.00000000000000000e+00</m10>
                            <m11>1.00000000000000000e+00</m11>
                            <m12>0.00000000000000000e+00</m12>
                            <m20>0.00000000000000000e+00</m20>
                            <m21>0.00000000000000000e+00</m21>
                            <m22>1.00000000000000000e+00</m22>
                    </refHSize>
                    <hSize>
                            <m00>1.00000000000000000e+00</m00>
                            <m01>0.00000000000000000e+00</m01>
                            <m02>0.00000000000000000e+00</m02>
                            <m10>0.00000000000000000e+00</m10>
                            <m11>1.00000000000000000e+00</m11>
                            <m12>0.00000000000000000e+00</m12>
                            <m20>0.00000000000000000e+00</m20>
                            <m21>0.00000000000000000e+00</m21>
                            <m22>1.00000000000000000e+00</m22>
                    </hSize>
                    <prevHSize>
                            <m00>1.00000000000000000e+00</m00>
                            <m01>0.00000000000000000e+00</m01>
                            <m02>0.00000000000000000e+00</m02>
                            <m10>0.00000000000000000e+00</m10>
                            <m11>1.00000000000000000e+00</m11>
                            <m12>0.00000000000000000e+00</m12>
                            <m20>0.00000000000000000e+00</m20>
                            <m21>0.00000000000000000e+00</m21>
                            <m22>1.00000000000000000e+00</m22>
                    </prevHSize>
                    <velGrad>
                            <m00>0.00000000000000000e+00</m00>
                            <m01>0.00000000000000000e+00</m01>
                            <m02>0.00000000000000000e+00</m02>
                            <m10>0.00000000000000000e+00</m10>
                            <m11>0.00000000000000000e+00</m11>
                            <m12>0.00000000000000000e+00</m12>
                            <m20>0.00000000000000000e+00</m20>
                            <m21>0.00000000000000000e+00</m21>
                            <m22>0.00000000000000000e+00</m22>
                    </velGrad>
                    <nextVelGrad>
```

```
                                <m00>0.0000000000000000e+00</m00>
                                <m01>0.0000000000000000e+00</m01>
                                <m02>0.0000000000000000e+00</m02>
                                <m10>0.0000000000000000e+00</m10>
                                <m11>0.0000000000000000e+00</m11>
                                <m12>0.0000000000000000e+00</m12>
                                <m20>0.0000000000000000e+00</m20>
                                <m21>0.0000000000000000e+00</m21>
                                <m22>0.0000000000000000e+00</m22>
                        </nextVelGrad>
                        <prevVelGrad>
                                <m00>0.0000000000000000e+00</m00>
                                <m01>0.0000000000000000e+00</m01>
                                <m02>0.0000000000000000e+00</m02>
                                <m10>0.0000000000000000e+00</m10>
                                <m11>0.0000000000000000e+00</m11>
                                <m12>0.0000000000000000e+00</m12>
                                <m20>0.0000000000000000e+00</m20>
                                <m21>0.0000000000000000e+00</m21>
                                <m22>0.0000000000000000e+00</m22>
                        </prevVelGrad>
                        <homoDeform>2</homoDeform>
                        <velGradChanged>0</velGradChanged>
                </px>
        </cell>
        <miscParams class_id="29" tracking_level="0" version="0">
                <count>0</count>
                <item_version>1</item_version>
        </miscParams>
        <dispParams class_id="30" tracking_level="0" version="0">
                <count>0</count>
                <item_version>1</item_version>
        </dispParams>
    </px>
</px>
</scene>
</boost_serialization>
```

> **Warning:** Since XML files closely reflect implementation details of Yade, they will not be compatible between different versions. Use them only for short-term saving of scenes. Python is *the* high-level description Yade uses.

### Python attribute access

The macro *YADE_CLASS_BASE_DOC_* macro family* introduced above is (behind the scenes) also used to create functions for accessing attributes from Python. As already noted, set of serialized attributes and set of attributes accessible from Python are identical. Besides attribute access, these wrapper classes imitate also some functionality of regular python dictionaries:

```
Yade [20]: s=Sphere()

Yade [21]: s.radius            ## read-access
Out[21]: nan

Yade [22]: s.radius=4.         ## write access

Yade [23]: s.dict().keys()              ## show all available keys
Out[23]: dict_keys(['radius', 'color', 'wire', 'highlight'])
```

```
Yade [24]: for k in s.dict().keys(): print(s.dict()[k])    ## iterate over keys, print their␣
↪values
   ....:
4.0
Vector3(1,1,1)
False
False

Yade [25]: s.dict()['radius']                ## same as: 'radius' in s.keys()
Out[25]: 4.0

Yade [26]: s.dict()                           ## show dictionary of both attributes and values
Out[26]: {'radius': 4.0, 'color': Vector3(1,1,1), 'wire': False, 'highlight': False}
```

### YADE_CLASS_BASE_DOC_* macro family

There is several macros that hide behind them the functionality of *Sphinx documentation*, *Run-time type identification (RTTI)*, *Attribute registration*, *Python attribute access*, plus automatic attribute initialization and documentation. They are all defined as shorthands for base macro `YADE_CLASS_BASE_DOC_-ATTRS_INIT_CTOR_PY` with some arguments left out. They must be placed in class declaration's body (`.hpp` file):

```
#define YADE_CLASS_BASE_DOC(klass,base,doc) \
        YADE_CLASS_BASE_DOC_ATTRS(klass,base,doc,)
#define YADE_CLASS_BASE_DOC_ATTRS(klass,base,doc,attrs) \
        YADE_CLASS_BASE_DOC_ATTRS_CTOR(klass,base,doc,attrs,)
#define YADE_CLASS_BASE_DOC_ATTRS_CTOR(klass,base,doc,attrs,ctor) \
        YADE_CLASS_BASE_DOC_ATTRS_CTOR_PY(klass,base,doc,attrs,ctor,)
#define YADE_CLASS_BASE_DOC_ATTRS_CTOR_PY(klass,base,doc,attrs,ctor,py) \
        YADE_CLASS_BASE_DOC_ATTRS_INIT_CTOR_PY(klass,base,doc,attrs,,ctor,py)
#define YADE_CLASS_BASE_DOC_ATTRS_INIT_CTOR_PY(klass,base,doc,attrs,init,ctor,py) \
        YADE_CLASS_BASE_DOC_ATTRS_INIT_CTOR_PY(klass,base,doc,attrs,inits,ctor,py)
```

Expected parameters are indicated by macro name components separated with underscores. Their meaning is as follows:

**klass** (unquoted) name of this class (used for RTTI and python)

**base** (unquoted) name of the base class (used for RTTI and python)

**doc** docstring of this class, written in the ReST syntax. This docstring will appear in generated documentation (such as *CpmMat*). It can be as long as necessary, use string literal to avoid sequences interpreted by c++ compiler (so that some backslashes don't have to be doubled, like in $\sigma = \varepsilon E$) instead of writing this:

```
":math:`\\sigma=\\epsilon E`"
```

Write following: R"""(:math:`\sigma=\epsilon E`)""". When the R"""(raw text)""" is used the escaped characters \n and \t do not have to be written. Newlines and tabs can be used instead. For example see *here* and here. Hyperlink the documentation abundantly with **yref** (all references to other classes should be hyperlinks). See *previous section* about syntax on using references and anchors.

**attrs** Attribute must be written in the form of parethesized list:

```
((type1,attr1,initValue1,attrFlags,"Attribute 1 documentation"))
((type2,attr2,,,"Attribute 2 documentation"))  // initValue and attrFlags unspecified
```

This will expand to

1. data members declaration in c++ (note that all attributes are *public*):

```
public: type1 attr1;
        type2 attr2;
```

2. Initializers of the default (argument-less) constructor, for attributes that have non-empty initValue:

```
Klass(): attr1(initValue1), attr2() { /* constructor body */ }
```

No initial value will be assigned for attribute of which initial value is left empty (as is for attr2 in the above example). Note that you still have to write the commas.

3. Registration of the attribute in the serialization system (unless disabled by attrFlags *– see below*)

4. **Registration of the attribute in python (unless disabled by attrFlags), so that it can be accessed**
The attribute is read-write by default, see attrFlags to change that.

This attribute will carry the docstring provided, along with knowledge of the initial value. You can add text description to the default value using the comma operator of c++ and casting the char* to (void):

```
((Real,dmgTau,((void)"deactivated if negative",-1),,"Characteristic time for␣
↪normal viscosity. [s]"))
```

leading to *CpmMat::dmgTau*.

The attribute is registered via `boost::python::add_property` specifying `return_by_-value` policy rather than `return_internal_reference`, which is the default when using `def_readwrite`. The reason is that we need to honor custom converters for those values; see note in *Custom converters* for details.

---

**Attribute flags**

By default, an attribute will be serialized and will be read-write from python. There is a number of flags that can be passed as the 4th argument (empty by default) to change that:

- `Attr::noSave` avoids serialization of the attribute (while still keeping its accessibility from Python)

- `Attr::readonly` makes the attribute read-only from Python

- `Attr::triggerPostLoad` will trigger call to `postLoad` function to handle attribute change after its value is set from Python; this is to ensure consistency of other pre-computed data which depend on this value (such as `Cell.trsf` and such)

- `Attr::hidden` will not expose the attribute to Python at all

- `Attr::noResize` will not permit changing size of the array from Python [not yet used]

Flags can be combined as usual using bitwise disjunction | (such as `Attr::noSave` | `Attr::readonly`), though in such case the value should be parenthesized to avoid a warning with some compilers (g++ specifically), i.e. (`Attr::noSave` | `Attr::readonly`).

Currently, the flags logic handled at runtime; that means that even for attributes with `Attr::noSave`, their serialization template must be defined (although it will never be used). In the future, the implementation might be template-based, avoiding this necessity.

---

**deprec** List of deprecated attribute names. The syntax is

```
((oldName1,newName1,"Explanation why renamed etc."))
((oldName2,newName2,"! Explanation why removed and what to do instead."))
```

---

This will make accessing **oldName1** attribute *from Python* return value of **newName**, but displaying warning message about the attribute name change, displaying provided explanation. This happens whether the access is read or write.

If the explanation's first character is ! (*bang*), the message will be displayed upon attribute access, but exception will be thrown immediately. Use this in cases where attribute is no longer meaningful or was not straightfordsly replaced by another, but more complex adaptation of user's script is needed. You still have to give **newName2**, although its value will never be used – you can use any variable you like, but something must be given for syntax reasons).

> **Warning:** Due to compiler limitations, this feature only works if Yade is compiled with gcc >= 4.4. In the contrary case, deprecated attribute functionality is disabled, even if such attributes are declared.

**init** Parethesized list of the form:

```
((attr3,value3)) ((attr4,value4))
```

which will be expanded to initializers in the default ctor:

```
Klass(): /* attributes declared with the attrs argument */ attr4(value4), attr5(value5) {␣
↪/* constructor body */ }
```

The purpose of this argument is to make it possible to initialize constants and references (which are not declared as attributes using this macro themselves, but separately), as that cannot be done in constructor body. This argument is rarely used, though.

**ctor** will be put directly into the generated constructor's body. Mostly used for calling createIndex(); in the constructor.

> **Note:** The code must not contain commas outside parentheses (since preprocessor uses commas to separate macro arguments). If you need complex things at construction time, create a separate init() function and call it from the constructor instead.

**py** will be appended directly after generated python code that registers the class and all its attributes. You can use it to access class methods from python, for instance, to override an existing attribute with the same name etc:

```
.def_readonly("omega",&CpmPhys::omega,"Damage internal variable")
.def_readonly("Fn",&CpmPhys::Fn,"Magnitude of normal force.")
```

**def_readonly** will not work for custom types (such as std::vector), as it bypasses conversion registry; see *Custom converters* for details.

### Exposing function-attributes to GUI

Usually to expose a more complex data a getter and setter functions are used, for example *Body::mask*. They are accessible from python. To make them visible in GUI without a corresponding variable at all a function **virtual ::boost::python::dict pyDictCustom() const { …… };** must be overridden. For example see Interaction.hpp where a special attribute **isReal** is exposed to GUI. To mark such attribute as readonly an extra information has to be added to its docstring: **:yattrflags:`2`**. Normally it is put there by the *class attribute registration macros*. But since it is not a variable, such attribute has to be added manually.

**Special python constructors**

The Python wrapper automatically creates constructor that takes keyword (named) arguments corresponding to instance attributes; those attributes are set to values provided in the constructor. In some cases, more flexibility is desired (such as *InteractionLoop*, which takes 3 lists of functors). For such cases, you can override the function `Serializable::pyHandleCustomCtorArgs`, which can arbitrarily modify the new (already existing) instance. It should modify in-place arguments given to it, as they will be passed further down to the routine which sets attribute values. In such cases, you should document the constructor:

```
.. admonition:: Special constructor

        Constructs from lists of …
```

which then appears in the documentation similar to *InteractionLoop*.

**Enums**

It is possible to expose `enum` and `enum class` (the `enum class` is the preferred one because it has stronger type safety to protect programmer from mistakes) in GUI in a dropdown menu. This approach is backward compatible, an assignment of `int` value in an old python script will work the same as before. Additionally it will be possible to assign the `string` type values to an enum. To enable the dropdown menu one must put a macro `YADE_ENUM( Scope , EnumName , (ValueName1)(ValueName2)(ValueName3)(ValueName4) )` in a `.cpp` file. Where each macro argument means:

1. `Scope` is the full scope name in which the enum resides. For example the scope of `yade::OpenGLRenderer::BlinkHighlight` is `yade::OpenGLRenderer`.

2. `EnumName` is the name of the enum to be registered

3. `ValueName` are all enum values that are to be exposed to python. They have to be updated if the C++ enum declaration in `.hpp` file changes.

After it is registered, like for example in OpenGLRenderer.cpp it is available for use. Additionally the registered enum class type definitions are exposed in `yade.EnumClass_*` scope, for example one can check the `names` and `values` dictionaries:

```
Yade [27]: yade.EnumClass_BlinkHighlight.names
Out[27]:
{'NEVER': yade.EnumClass_BlinkHighlight.NEVER,
 'NORMAL': yade.EnumClass_BlinkHighlight.NORMAL,
 'WEAK': yade.EnumClass_BlinkHighlight.WEAK}

Yade [28]: yade.EnumClass_BlinkHighlight.values
Out[28]:
{0: yade.EnumClass_BlinkHighlight.NEVER,
 1: yade.EnumClass_BlinkHighlight.NORMAL,
 2: yade.EnumClass_BlinkHighlight.WEAK}
```

Keep in mind that these are **not the variable instances** hence trying to assign something to them will not change the blinkHighlight setting in GUI. To change enum value from python the respective variable must be assigned to, such as *yade.qt.Renderer().blinkHighlight*. Trying to assign an incorrect value will throw an exception. For example:

```
Yade [29]: r = yade.FlowEngine() # this is only a test of enum, not of FlowEngine
---------------------------------------------------------------------------
AttributeError                            Traceback (most recent call last)
~/yade/lib/x86_64-linux-gnu/yadeflip/py/yade/__init__.py in <module>
----> 1 r = yade.FlowEngine() # this is only a test of enum, not of FlowEngine
```

(continues on next page)

```
AttributeError: module 'yade' has no attribute 'FlowEngine'

Yade [30]: r.useSolver
---------------------------------------------------------------------------
NameError                                 Traceback (most recent call last)
~/yade/lib/x86_64-linux-gnu/yadeflip/py/yade/__init__.py in <module>
----> 1 r.useSolver

NameError: name 'r' is not defined

Yade [31]: r.useSolver = 'GaussSeidel'
---------------------------------------------------------------------------
NameError                                 Traceback (most recent call last)
~/yade/lib/x86_64-linux-gnu/yadeflip/py/yade/__init__.py in <module>
----> 1 r.useSolver = 'GaussSeidel'

NameError: name 'r' is not defined

Yade [32]: try:
   ....:     r.useSolver = 20    # assigning incorrect value has no effect
   ....: except:
   ....:     print("Error, value is still equal to:",r.useSolver)
   ....:
---------------------------------------------------------------------------
NameError                                 Traceback (most recent call last)
~/yade/lib/x86_64-linux-gnu/yadeflip/py/yade/__init__.py in <module>
      1 try:
----> 2     r.useSolver = 20    # assigning incorrect value has no effect
      3 except:

NameError: name 'r' is not defined

During handling of the above exception, another exception occurred:

NameError                                 Traceback (most recent call last)
~/yade/lib/x86_64-linux-gnu/yadeflip/py/yade/__init__.py in <module>
      2     r.useSolver = 20    # assigning incorrect value has no effect
      3 except:
----> 4     print("Error, value is still equal to:",r.useSolver)

NameError: name 'r' is not defined

Yade [33]: r.useSolver
---------------------------------------------------------------------------
NameError                                 Traceback (most recent call last)
~/yade/lib/x86_64-linux-gnu/yadeflip/py/yade/__init__.py in <module>
----> 1 r.useSolver

NameError: name 'r' is not defined
```

Alternatively the dropdown menu in GUI can be used for the same effect.

### Static attributes

Some classes (such as OpenGL functors) are instantiated automatically; since we want their attributes to be persistent throughout the session, they are static. To expose class with static attributes, use the `YADE_CLASS_BASE_DOC_STATICATTRS` macro. Attribute syntax is the same as for `YADE_CLASS_BASE_DOC_ATTRS`:

```
class SomeClass: public BaseClass{
        YADE_CLASS_BASE_DOC_STATICATTRS(SomeClass,BaseClass,"Documentation of SomeClass",
                ((Type1,attr1,default1,"doc for attr1"))
                ((Type2,attr2,default2,"doc for attr2"))
        );
};
```

additionally, you *have* to allocate memory for static data members in the `.cpp` file (otherwise, error about undefined symbol will appear when the plugin is loaded):

There is no way to expose class that has both static and non-static attributes using `YADE_CLASS_BASE_*` macros. You have to expose non-static attributes normally and wrap static attributes separately in the `py` parameter.

### Returning attribute by value or by reference

When attribute is passed from c++ to python, it can be passed either as

- value: new python object representing the original c++ object is constructed, but not bound to it; changing the python object doesn't modify the c++ object, unless explicitly assigned back to it, where inverse conversion takes place and the c++ object is replaced.

- reference: only reference to the underlying c++ object is given back to python; modifying python object will make the c++ object modified automatically.

The way of passing attributes given to `YADE_CLASS_BASE_DOC_ATTRS` in the `attrs` parameter is determined automatically in the following manner:

- **Vector3, Vector3i, Vector2, Vector2i, Matrix3 and Quaternion objects are passed by *reference*. For in**
        O.bodies[0].state.pos[0]=1.33

    will assign correct value to x component of position, without changing the other ones.

- **Yade classes (all that use shared_ptr when declared in python: all classes deriving from *Serializable*
        O.engines[4].damping=.3

    will change *damping* parameter on the original engine object, not on its copy.

- **All other types are passed by *value*. This includes, most importantly, sequence types declared in *C*
        O.engines[4]=NewtonIntegrator()

    will *not* work as expected; it will replace 5th element of a *copy* of the sequence, and this change will not propagate back to c++.

### Multiple dispatch

Multiple dispatch is generalization of virtual methods: a *Dispatcher* decides based on type(s) of its argument(s) which of its *Functors* to call. Number of arguments (currently 1 or 2) determines *arity* of the dispatcher (and of the functor): unary or binary. For example:

```
InsertionSortCollider([Bo1_Sphere_Aabb(),Bo1_Facet_Aabb()])
```

creates *InsertionSortCollider*, which internally contains *Collider.boundDispatcher*, a *BoundDispatcher* (a *Dispatcher*), with 2 functors; they receive `Sphere` or `Facet` instances and create `Aabb`. This code would look like this in c++:

```
shared_ptr<InsertionSortCollider> collider=(new InsertionSortCollider);
collider->boundDispatcher->add(new Bo1_Sphere_Aabb());
collider->boundDispatcher->add(new Bo1_Facet_Aabb());
```

There are currently 4 predefined dispatchers (see *dispatcher-names*) and corresponding functor types. They are inherited from template instantiations of `Dispatcher1D` or `Dispatcher2D` (for functors,

`Functor1D` or `Functor2D`). These templates themselves derive from `DynlibDispatcher` (for dispatchers) and `FunctorWrapper` (for functors).

### Example: IGeomDispatcher

Let's take (the most complicated perhaps) *IGeomDispatcher*. *IGeomFunctor*, which is dispatched based on types of 2 *Shape* instances (a *Functor*), takes a number of arguments and returns bool. The functor "call" is always provided by its overridden `Functor::go` method; it always receives the dispatched instances as first argument(s) (2 × `const shared_ptr<Shape>&`) and a number of other arguments it needs:

```
class IGeomFunctor: public Functor2D<
    bool,                             //return type
    TYPELIST_7(const shared_ptr<Shape>&, // 1st class for dispatch
        const shared_ptr<Shape>&,     // 2nd class for dispatch
        const State&,                 // other arguments passed to ::go
        const State&,                 // …
        const Vector3r&,              // …
        const bool&,                  // …
        const shared_ptr<Interaction>&  // …
    )
>
```

The dispatcher is declared as follows:

```
class IGeomDispatcher: public Dispatcher2D<
    Shape,                  // 1st class for dispatch
    Shape,                  // 2nd class for dispatch
    IGeomFunctor,  // functor type
    bool,                   // return type of the functor

    // follow argument types for functor call
    // they must be exactly the same as types
    // given to the IGeomFunctor above.
    TYPELIST_7(const shared_ptr<Shape>&,
        const shared_ptr<Shape>&,
        const State&,
        const State&,
        const Vector3r&,
        const bool &,
        const shared_ptr<Interaction>&
    ),

    // handle symetry automatically
    // (if the dispatcher receives Sphere+Facet,
    // the dispatcher might call functor for Facet+Sphere,
    // reversing the arguments)
    false
>
{ /* … */ }
```

Functor derived from IGeomFunctor must then

- override the ::go method with appropriate arguments (they must match exactly types given to `TYPELIST_*` macro);

- declare what types they should be dispatched for, and in what order if they are not the same.

```
class Ig2_Facet_Sphere_ScGeom: public IGeomFunctor{
    public:
```

```cpp
    // override the IGeomFunctor::go
    //   (it is really inherited from FunctorWrapper template,
    //    therefore not declare explicitly in the
    //    IGeomFunctor declaration as such)
    // since dispatcher dispatches only for declared types
    //   (or types derived from them), we can do
    //   static_cast<Facet>(shape1) and static_cast<Sphere>(shape2)
    //   in the ::go body, without worrying about types being wrong.
    virtual bool go(
        // objects for dispatch
        const shared_ptr<Shape>& shape1, const shared_ptr<Shape>& shape2,
        // other arguments
        const State& state1, const State& state2, const Vector3r& shift2,
        const bool& force, const shared_ptr<Interaction>& c
    );
    /* … */

    // this declares the type we want to be dispatched for, matching
    //   first 2 arguments to ::go and first 2 classes in TYPELIST_7 above
    //   shape1 is a Facet and shape2 is a Sphere
    //   (or vice versa, see lines below)
    FUNCTOR2D(Facet,Sphere);

    // declare how to swap the arguments
    //   so that we can receive those as well
    DEFINE_FUNCTOR_ORDER_2D(Facet,Sphere);
    /* … */
};
```

**Dispatch resolution**

The dispatcher doesn't always have functors that exactly match the actual types it receives. In the same way as virtual methods, it tries to find the closest match in such way that:

1. the actual instances are derived types of those the functor accepts, or exactly the accepted types;

2. sum of distances from actual to accepted types is sharp-minimized (each step up in the class hierarchy counts as 1)

If no functor is able to accept given types (first condition violated) or multiple functors have the same distance (in condition 2), an exception is thrown.

This resolution mechanism makes it possible, for instance, to have a hierarchy of *ScGeom* classes (for different combination of shapes), but only provide a *LawFunctor* accepting `ScGeom`, rather than having different laws for each shape combination.

**Note:** Performance implications of dispatch resolution are relatively low. The dispatcher lookup is only done once, and uses fast lookup matrix (1D or 2D); then, the functor found for this type(s) is cached within the `Interaction` (or `Body`) instance. Thus, regular functor call costs the same as dereferencing pointer and calling virtual method. There is blueprint to avoid virtual function call as well.

**Note:** At the beginning, the dispatch matrix contains just entries exactly matching given functors. Only when necessary (by passing other types), appropriate entries are filled in as well.

**Indexing dispatch types**

Classes entering the dispatch mechanism must provide for fast identification of themselves and of their parent class.[5] This is called class indexing and all such classes derive from *Indexable*. There are `top-level` Indexables (types that the dispatchers accept) and each derived class registers its index related to this top-level Indexable. Currently, there are:

| Top-level Indexable | used by |
|---|---|
| *Shape* | *BoundFunctor*, *IGeomDispatcher* |
| *Material* | *IPhysDispatcher* |
| *IPhys* | *LawDispatcher* |
| *IGeom* | *LawDispatcher* |

The top-level Indexable must use the `REGISTER_INDEX_COUNTER` macro, which sets up the machinery for identifying types of derived classes; they must then use the `REGISTER_CLASS_INDEX` macro *and* call `createIndex()` in their constructor. For instance, taking the *Shape* class (which is a top-level Indexable):

```
// derive from Indexable
class Shape: public Serializable, public Indexable {
    // never call createIndex() in the top-level Indexable ctor!
    /* … */

    // allow index registration for classes deriving from ``Shape``
    REGISTER_INDEX_COUNTER(Shape);
};
```

Now, all derived classes (such as *Sphere* or *Facet*) use this:

```
class Sphere: public Shape{
    /* … */
    YADE_CLASS_BASE_DOC_ATTRS_CTOR(Sphere,Shape,"docstring",
        ((Type1,attr1,default1,"docstring1"))
        /* … */,
        // this is the CTOR argument
            // important; assigns index to the class at runtime
            createIndex();
    );
    // register index for this class, and give name of the immediate parent class
    //   (i.e. if there were a class deriving from Sphere, it would use
    //     REGISTER_CLASS_INDEX(SpecialSphere,Sphere),
    //     not REGISTER_CLASS_INDEX(SpecialSphere,Shape)!)
    REGISTER_CLASS_INDEX(Sphere,Shape);
};
```

At runtime, each class within the top-level Indexable hierarchy has its own unique numerical index. These indices serve to build the dispatch matrix for each dispatcher.

**Inspecting dispatch in python**

If there is a need to debug/study multiple dispatch, python provides convenient interface for this low-level functionality.

We can inspect indices with the `dispIndex` property (note that the top-level indexable `Shape` has negative (invalid) class index; we purposively didn't call `createIndex` in its constructor):

---

[5] The functionality described in *Run-time type identification (RTTI)* serves a different purpose (serialization) and would hurt the performance here. For this reason, classes provide numbers (indices) in addition to strings.

```
Yade [34]: Sphere().dispIndex, Facet().dispIndex, Wall().dispIndex
Out[34]: (1, 5, 10)

Yade [35]: Shape().dispIndex                        # top-level indexable
Out[35]: -1
```

Dispatch hierarchy for a particular class can be shown with the `dispHierarchy()` function, returning list of class names; 0th element is the instance itself, last element is the top-level indexable (again, with invalid index); for instance:

```
Yade [36]: ScGeom().dispHierarchy()          # parent class of all other ScGeom_ classes
Out[36]: ['ScGeom', 'GenericSpheresContact', 'IGeom']

Yade [37]: ScGridCoGeom().dispHierarchy(), ScGeom6D().dispHierarchy(), CylScGeom().
           ↪dispHierarchy()
Out[37]:
(['ScGridCoGeom', 'ScGeom6D', 'ScGeom', 'GenericSpheresContact', 'IGeom'],
 ['ScGeom6D', 'ScGeom', 'GenericSpheresContact', 'IGeom'],
 ['CylScGeom', 'ScGeom', 'GenericSpheresContact', 'IGeom'])

Yade [38]: CylScGeom().dispHierarchy(names=False)   # show numeric indices instead
Out[38]: [4, 1, 0, -1]
```

Dispatchers can also be inspected, using the .dispMatrix() method:

```
Yade [39]: ig=IGeomDispatcher([
      ....:     Ig2_Sphere_Sphere_ScGeom(),
      ....:     Ig2_Facet_Sphere_ScGeom(),
      ....:     Ig2_Wall_Sphere_ScGeom()
      ....: ])
      ....:

Yade [40]: ig.dispMatrix()
Out[40]:
{('Sphere', 'Sphere'): 'Ig2_Sphere_Sphere_ScGeom',
 ('Sphere', 'Facet'): 'Ig2_Facet_Sphere_ScGeom',
 ('Sphere', 'Wall'): 'Ig2_Wall_Sphere_ScGeom',
 ('Facet', 'Sphere'): 'Ig2_Facet_Sphere_ScGeom',
 ('Wall', 'Sphere'): 'Ig2_Wall_Sphere_ScGeom'}

Yade [41]: ig.dispMatrix(False)                # don't convert to class names
Out[41]:
{(1, 1): 'Ig2_Sphere_Sphere_ScGeom',
 (1, 5): 'Ig2_Facet_Sphere_ScGeom',
 (1, 10): 'Ig2_Wall_Sphere_ScGeom',
 (5, 1): 'Ig2_Facet_Sphere_ScGeom',
 (10, 1): 'Ig2_Wall_Sphere_ScGeom'}
```

We can see that functors make use of symmetry (i.e. that Sphere+Wall are dispatched to the same functor as Wall+Sphere).

Finally, dispatcher can be asked to return functor suitable for given argument(s):

```
Yade [42]: ld=LawDispatcher([Law2_ScGeom_CpmPhys_Cpm()])

Yade [43]: ld.dispMatrix()
Out[43]: {('GenericSpheresContact', 'CpmPhys'): 'Law2_ScGeom_CpmPhys_Cpm'}

# see how the entry for ScGridCoGeom will be filled after this request
Yade [44]: ld.dispFunctor(ScGridCoGeom(),CpmPhys())
Out[44]: <Law2_ScGeom_CpmPhys_Cpm instance at 0x397baf0>
```

(continues on next page)

```
Yade [45]: ld.dispMatrix()
Out[45]:
{('GenericSpheresContact', 'CpmPhys'): 'Law2_ScGeom_CpmPhys_Cpm',
 ('ScGridCoGeom', 'CpmPhys'): 'Law2_ScGeom_CpmPhys_Cpm'}
```

### OpenGL functors

OpenGL rendering is being done also by 1D functors (dispatched for the type to be rendered). Since it is sufficient to have exactly one class for each rendered type, the functors are found automatically. Their base functor types are `GlShapeFunctor`, `GlBoundFunctor`, `GlIGeomFunctor` and so on. These classes register the type they render using the `RENDERS` macro:

```
namespace yade { // Cannot have #include directive inside.
class Gl1_Sphere: public GlShapeFunctor {
    public :
        virtual void go(const shared_ptr<Shape>&,
            const shared_ptr<State>&,
            bool wire,
            const GLViewInfo&
        );
        RENDERS(Sphere);
        YADE_CLASS_BASE_DOC_STATICATTRS(Gl1_Sphere,GlShapeFunctor,"docstring",
            ((Type1,staticAttr1,informativeDefault,"docstring"))
            /* … */
        );
};
REGISTER_SERIALIZABLE(Gl1_Sphere);
} // namespace yade
```

You can list available functors of a particular type by querying child classes of the base functor:

```
Yade [46]: yade.system.childClasses('GlShapeFunctor')
Out[46]:
{'Gl1_Box',
 'Gl1_ChainedCylinder',
 'Gl1_Cylinder',
 'Gl1_Facet',
 'Gl1_GridConnection',
 'Gl1_PFacet',
 'Gl1_Sphere',
 'Gl1_Tetra',
 'Gl1_Wall'}
```

**Note:** OpenGL functors may disappear in the future, being replaced by virtual functions of each class that can be rendered.

### Parallel execution

Yade was originally not designed with parallel computation in mind, but rather with maximum flexibility (for good or for bad). Parallel execution was added later; in order to not have to rewrite whole Yade from scratch, relatively non-instrusive way of parallelizing was used: OpenMP. OpenMP is standardized shared-memory parallel execution environment, where parallel sections are marked by special `#pragma` in the code (which means that they can compile with compiler that doesn't support OpenMP) and a few functions to query/manipulate OpenMP runtime if necessary.

There is parallelism at 3 levels:

- Computation, interaction (python, GUI) and rendering threads are separate. This is done via regular threads (boost::threads) and is not related to OpenMP.

- *ParallelEngine* can run multiple engine groups (which are themselves run serially) in parallel; it rarely finds use in regular simulations, but it could be used for example when coupling with an independent expensive computation:

```
ParallelEngine([
        [Engine1(),Engine2()],      # Engine1 will run before Engine2
        [Engine3()]                 # Engine3() will run in parallel with the group␣
↪[Engine1(),Engine2()]

                                    # arbitrary number of groups can be used
])
```

  `Engine2` will be run after `Engine1`, but in parallel with `Engine3`.

  > **Warning:** It is your reponsibility to avoid concurrent access to data when using ParallelEngine. Make sure you understand *very well* what the engines run in parallel do.

- Parallelism inside Engines. Some loops over bodies or interactions are parallelized (notably *InteractionLoop* and *NewtonIntegrator*, which are treated in detail later (FIXME: link)):

```
#pragma omp parallel for
for(long id=0; id<size; id++){
   const shared_ptr<Body>& b(scene->bodies[id]);
   /* … */
}
```

  > **Note:** OpenMP requires loops over contiguous range of integers (OpenMP 3 also accepts containers with random-access iterators).

  If you consider running parallelized loop in your engine, always evalue its benefits. OpenMP has some overhead fo creating threads and distributing workload, which is proportionally more expensive if the loop body execution is fast. The results are highly hardware-dependent (CPU caches, RAM controller).

Maximum number of OpenMP threads is determined by the `OMP_NUM_THREADS` environment variable and is constant throughout the program run. Yade main program also sets this variable (before loading OpenMP libraries) if you use the `-j`/`--threads` option. It can be queried at runtime with the `omp_-get_max_threads` function.

At places which are susceptible of being accessed concurrently from multiple threads, Yade provides some mutual exclusion mechanisms, discussed elsewhere (FIXME):

- simultaneously writeable container for *ForceContainer*,

- mutex for *Body::state*.

### Timing

Yade provides 2 services for measuring time spent in different parts of the code. One has the granularity of engine and can be enabled at runtime. The other one is finer, but requires adjusting and recompiling the code being measured.

### Per-engine timing

The coarser timing works by merely accumulating number of invocations and time (with the precision of the `clock_gettime` function) spent in each engine, which can be then post-processed by associated Python module `yade.timing`. There is a static bool variable controlling whether such measurements take place (disabled by default), which you can change

```
TimingInfo::enabled=True;          // in c++
```

```
O.timingEnabled=True               ## in python
```

After running the simulation, `yade.timing.stats()` function will show table with the results and percentages:

```
Yade [47]: TriaxialTest(numberOfGrains=100).load()

Yade [48]: O.engines[0].label='firstEngine'   ## labeled engines will show by labels in the
  ↪stats table

Yade [49]: import yade.timing;

Yade [50]: O.timingEnabled=True

Yade [51]: yade.timing.reset()                 ## not necessary if used for the first time

Yade [52]: O.run(50); O.wait()

Yade [53]: yade.timing.stats()
Name                                           Count              Time               ↪
  ↪Rel. time
-------------------------------------------------------------------------------------------
  ↪--------
"firstEngine"                                  50                 15.278us           0.
  ↪26%
InsertionSortCollider                          25                 1866.469us         32.
  ↪16%
InteractionLoop                                50                 2849.897us         49.
  ↪11%
GlobalStiffnessTimeStepper                     2                  15.417us           0.
  ↪27%
TriaxialCompressionEngine                      50                 262.545us          4.
  ↪52%
TriaxialStateRecorder                          3                  145.883us          2.
  ↪51%
NewtonIntegrator                               50                 647.859us          11.
  ↪16%
  forces sync                                  50                 7.259us                 ↪
  ↪1.12%
  motion integration                           50                 624.789us               ↪
  ↪96.44%
  sync max vel                                 50                 3.264us                 ↪
  ↪0.50%
  terminate                                    50                 2.337us                 ↪
  ↪0.36%
  TOTAL                                        200                637.649us               ↪
  ↪98.42%
TOTAL                                                             5803.348us         100.
  ↪00%
```

Exec count and time can be accessed and manipulated through `Engine::timingInfo` from c++ or `Engine().execCount` and `Engine().execTime` properties in Python.

**In-engine and in-functor timing**

Timing within engines (and functors) is based on *TimingDeltas* class which is by default instantiated in engines and functors as Engine::timingDeltas and Functor::timingDeltas (*Engine.timingDeltas* and *Functor.timingDeltas* in Python). It is made for timing loops (functors' loop is in their respective dispatcher) and stores cummulatively time differences between *checkpoints*.

---

**Note:** Fine timing with `TimingDeltas` will only work if timing is enabled globally (see previous section). The code would still run, but giving zero times and exec counts.

---

1. Preferably define the timingDeltas attributes in the constructor:

```
// header file
class Law2_ScGeom_CpmPhys_Cpm: public LawFunctor {
    /* … */
    YADE_CLASS_BASE_DOC_ATTRS_CTOR(Law2_ScGeom_CpmPhys_Cpm,LawFunctor,"docstring",
        /* attrs */,
        /* constructor */
        timingDeltas=shared_ptr<TimingDeltas>(new TimingDeltas); // timingDeltas␣
↪object is automatically initialized when using -DENABLE_PROFILING=1 cmake␣
↪option
    );
    // ...
};
```

2. Inside the loop, start the timing by calling `timingDeltas->start();`

3. At places of interest, call `timingDeltas->checkpoint("label")`. The label is used only for post-processing, data are stored based on the checkpoint position, not the label.

> **Warning:** Checkpoints must be always reached in the same order, otherwise the timing data will be garbage. Your code can still branch, but you have to put checkpoints to places which are in common.

```
void Law2_ScGeom_CpmPhys_Cpm::go(shared_ptr<IGeom>& _geom,
                                 shared_ptr<IPhys>& _phys,
                                 Interaction* I,
                                 Scene* scene)
{
    timingDeltas->start();                          // the point at which the first␣
↪timing starts
    // prepare some variables etc here
    timingDeltas->checkpoint("setup");
    // find geometrical data (deformations) here
    timingDeltas->checkpoint("geom");
    // compute forces here
    timingDeltas->checkpoint("material");
    // apply forces, cleanup here
    timingDeltas->checkpoint("rest");
}
```

4. **Alternatively, you can compile Yade using -DENABLE_PROFILING=1 cmake option and use pre**

```
void Law2_ScGeom_CpmPhys_Cpm::go(shared_ptr<IGeom>& _geom,
                                 shared_ptr<IPhys>& _phys,
                                 Interaction* I,
                                 Scene* scene)
```

```
{
    TIMING_DELTAS_START();
    // prepare some variables etc here
    TIMING_DELTAS_CHECKPOINT("setup")
    // find geometrical data (deformations) here
    TIMING_DELTAS_CHECKPOINT("geom")
    // compute forces here
    TIMING_DELTAS_CHECKPOINT("material")
    // apply forces, cleanup here
    TIMING_DELTAS_CHECKPOINT("rest")
}
```

The output might look like this (note that functors are nested inside dispatchers and `TimingDeltas` inside their engine/functor):

| Name | Count | Time | Rel. time |
|------|-------|------|-----------|
| ForceReseter | 400 | 9449μs | 0.01% |
| BoundDispatcher | 400 | 1171770μs | 1.15% |
| InsertionSortCollider | 400 | 9433093μs | 9.24% |
| IGeomDispatcher | 400 | 15177607μs | 14.87% |
| IPhysDispatcher | 400 | 9518738μs | 9.33% |
| LawDispatcher | 400 | 64810867μs | 63.49% |
|   Law2_ScGeom_CpmPhys_Cpm | | | |
|     setup | 4926145 | 7649131μs | 15.25% |
|     geom | 4926145 | 23216292μs | 46.28% |
|     material | 4926145 | 8595686μs | 17.14% |
|     rest | 4926145 | 10700007μs | 21.33% |
|     TOTAL | | 50161117μs | 100.00% |
| NewtonIntegrator | 400 | 1866816μs | 1.83% |
| "strainer" | 400 | 21589μs | 0.02% |
| "plotDataCollector" | 160 | 64284μs | 0.06% |
| "damageChecker" | 9 | 3272μs | 0.00% |
| TOTAL | | 102077490μs | 100.00% |

> **Warning:** Do not use *TimingDeltas* in parallel sections, results might not be meaningful. In particular, avoid timing functors inside *InteractionLoop* when running with multiple OpenMP threads.

`TimingDeltas` data are accessible from Python as list of (*label*,*time*,*count*) tuples, one tuple representing each checkpoint:

```
deltas=someEngineOrFunctor.timingDeltas.data()
deltas[0][0]  # 0th checkpoint label
deltas[0][1]  # 0th checkpoint time in nanoseconds
deltas[0][2]  # 0th checkpoint execution count
deltas[1][0]  # 1st checkpoint label
              # ...
deltas.reset()
```

### Timing overhead

The overhead of the coarser, per-engine timing, is very small. For simulations with at least several hundreds of elements, they are below the usual time variance (a few percent).

The finer *TimingDeltas* timing can have major performance impact and should be only used during debugging and performance-tuning phase. The parts that are file-timed will take disproportionally longer time that the rest of engine; in the output presented above, *LawDispatcher* takes almost of total simulation time in average, but the number would be twice of thrice lower typically (note that each checkpoint was timed almost 5 million times in this particular case).

### OpenGL Rendering

Yade provides 3d rendering based on QGLViewer. It is not meant to be full-featured rendering and post-processing, but rather a way to quickly check that scene is as intended or that simulation behaves sanely.

---

**Note:** Although 3d rendering runs in a separate thread, it has performance impact on the computation itself, since interaction container requires mutual exclusion for interaction creation/deletion. The `InteractionContainer::drawloopmutex` is either held by the renderer (*OpenGLRenderingEngine*) or by the insertion/deletion routine.

---

---

**Warning:** There are 2 possible causes of crash, which are not prevented because of serious performance penalty that would result:

1. access to *BodyContainer*, in particular deleting bodies from simulation; this is a rare operation, though.

2. deleting Interaction::phys or Interaction::geom.

---

Renderable entities (*Shape*, *State*, *Bound*, *IGeom*, *IPhys*) have their associated *OpenGL functors*. An entity is rendered if

1. Rendering such entities is enabled by appropriate attribute in *OpenGLRenderingEngine*

2. Functor for that particular entity type is found via the *dispatch mechanism*.

`Gl1_*` functors operating on Body's attributes (*Shape*, *State*, *Bound*) are called with the OpenGL context translated and rotated according to *State::pos* and *State::ori*. Interaction functors work in global coordinates.

## 3.1.7 Simulation framework

Besides the support framework mentioned in the previous section, some functionality pertaining to simulation itself is also provided.

There are special containers for storing bodies, interactions and (generalized) forces. Their internal functioning is normally opaque to the programmer, but should be understood as it can influence performance.

### Scene

`Scene` is the object containing the whole simulation. Although multiple scenes can be present in the memory, only one of them is active. Saving and loading (serializing and deserializing) the `Scene` object should make the simulation run from the point where it left off.

---

**Note:** All *Engines* and functors have interally a `Scene* scene` pointer which is updated regularly by engine/functor callers; this ensures that the current scene can be accessed from within user code.

For outside functions (such as those called from python, or static functions in `Shop`), you can use `Omega::instance().getScene` to retrieve a `shared_ptr<Scene>` of the current scene.

---

## Body container

Body container is linear storage of bodies. Each body in the simulation has its unique *id*, under which it must be found in the *BodyContainer*. Body that is not yet part of the simulation typically has id equal to invalid value `Body::ID_NONE`, and will have its `id` assigned upon insertion into the container. The requirements on *BodyContainer* are

- O(1) access to elements,

- linear-addressability (0...n indexability),

- store `shared_ptr`, not objects themselves,

- *no* mutual exclusion for insertion/removal (this must be assured by the called, if desired),

- intelligent allocation of `id` for new bodies (tracking removed bodies),

- easy iteration over all bodies.

**Note:** Currently, there is "abstract" class `BodyContainer`, from which derive concrete implementations; the initial idea was the ability to select at runtime which implementation to use (to find one that performs the best for given simulation). This incurs the penalty of many virtual function calls, and will probably change in the future. All implementations of BodyContainer were removed in the meantime, except `BodyVector` (internally a `vector<shared_ptr<Body> >` plus a few methods around), which is the fastest.

### Insertion/deletion

Body insertion is typically used in *FileGenerator*'s:

```
shared_ptr<Body> body(new Body);
// ... (body setup)
scene->bodies->insert(body); // assigns the id
```

Bodies are deleted only rarely:

```
scene->bodies->erase(id);
```

**Warning:** Since mutual exclusion is not assured, never insert/erase bodies from parallel sections, unless you explicitly assure there will be no concurrent access.

### Iteration

The container can be iterated over using `for(const auto& ... : ... )` C++ syntax:

```
for(const auto& b : *scene->bodies){
    if(!b) continue;                      // skip deleted bodies, nullptr-check
    /* do something here */
}
```

The same loop can be also written by using the type `const shared_ptr<Body>&` explicitly:

```
for(const shared_ptr<Body>& b : *scene->bodies){
    if(!b) continue;                      // skip deleted bodies, nullptr-check
    /* do something here */
}
```

---

> **Warning:** The previously used macro `FOREACH` is now deprecated.

---

Note a few important things:

1. Always use `const shared_ptr<Body>&` (const reference); that avoids incrementing and decrementing the reference count on each `shared_ptr`.

2. Take care to skip NULL bodies (`if(!b) continue`): deleted bodies are deallocated from the container, but since body id's must be persistent, their place is simply held by an empty `shared_ptr<Body>()` object, which is implicitly convertible to `false`.

In python, the BodyContainer wrapper also has iteration capabilities; for convenience (which is different from the c++ iterator), NULL bodies as silently skipped:

```
Yade [54]: O.bodies.append([Body(),Body(),Body()])
Out[54]: [0, 1, 2]

Yade [55]: O.bodies.erase(1)
Out[55]: True

Yade [56]: [b.id for b in O.bodies]
Out[56]: [0, 2]
```

In loops parallelized using OpenMP, the loop must traverse integer interval (rather than using iterators):

```
const long size=(long)bodies.size();        // store this value, since it doesn't change during
↪the loop
#pragma omp parallel for
for(long _id=0; _id<size; _id++){
    const shared_ptr<Body>& b(bodies[_id]);
    if(!b) continue;
    /* … */
}
```

### InteractionContainer

Interactions are stored in special container, and each interaction must be uniquely identified by pair of ids (id1,id2).

- O(1) access to elements,
- linear-addressability (0…n indexability),
- store `shared_ptr`, not objects themselves,
- mutual exclusion for insertion/removal,
- easy iteration over all interactions,
- addressing symmetry, i.e. interaction(id1,id2) interaction(id2,id1)

---

**Note:** As with BodyContainer, there is "abstract" class InteractionContainer, and then its concrete implementations. Currently, only InteractionVecMap implementation is used and all the other were removed. Therefore, the abstract InteractionContainer class may disappear in the future, to avoid unnecessary virtual calls.

Further, there is a blueprint for storing interactions inside bodies, as that would give extra advantage of quickly getting all interactions of one particular body (currently, this necessitates loop over all interactions); in that case, InteractionContainer would disappear.

---

**Insert/erase**

Creating new interactions and deleting them is delicate topic, since many eleents of simulation must be synchronized; the exact workflow is described in *Handling interactions*. You will almost certainly never need to insert/delete an interaction manually from the container; if you do, consider designing your code differently.

```
// both insertion and erase are internally protected by a mutex,
// and can be done from parallel sections safely
scene->interactions->insert(shared_ptr<Interaction>(new Interactions(id1,id2)));
scene->interactions->erase(id1,id2);
```

**Iteration**

As with BodyContainer, iteration over interactions should use the `for(const auto& …… : …… )` C++ syntax, also `const shared_ptr<Interaction>&` can be used instead of `auto&`:

```
for(const shared_ptr<Interaction>& i : *scene->interactions){
    if(!i->isReal()) continue;
    /* … */
}
```

> **Warning:** The previously used macro `FOREACH` is now deprecated.

Again, note the usage const reference for `i`. The check `if(!i->isReal())` filters away interactions that exist only *potentially*, i.e. there is only *Bound* overlap of the two bodies, but not (yet) overlap of bodies themselves. The `i->isReal()` function is equivalent to `i->geom && i->phys`. Details are again explained in *Handling interactions*.

In some cases, such as OpenMP-loops requiring integral index (OpenMP >= 3.0 allows parallelization using random-access iterator as well), you need to iterate over interaction indices instead:

```
int nIntr=(int)scene->interactions->size(); // hoist container size
#pragma omp parallel for
for(int j=0; j<nIntr; j++){
    const shared_ptr<Interaction>& i=(*scene->interactions)[j];
    if(!i->isReal()) continue;
    /* … */
}
```

**ForceContainer**

*ForceContainer* holds "generalized forces", i.e. forces, torques, (explicit) dispalcements and rotations for each body.

During each computation step, there are typically 3 phases pertaining to forces:

1. Resetting forces to zero (usually done by the *ForceResetter* engine)

2. Incrementing forces from parallel sections (solving interactions – from *LawFunctor*)

3. Reading absolute force values sequentially for each body: forces applied from different interactions are summed together to give overall force applied on that body (*NewtonIntegrator*, but also various other engine that read forces)

This scenario leads to special design, which allows fast parallel write access:

- each thread has its own storage (zeroed upon request), and only writes to its own storage; this avoids concurrency issues. Each thread identifies itself by the omp_get_thread_num() function provided by the OpenMP runtime.

- before reading absolute values, the container must be synchronized, i.e. values from all threads are summed up and stored separately. This is a relatively slow operation and we provide Force-Container::syncCount that you might check to find cummulative number of synchronizations and compare it against number of steps. Ideally, ForceContainer is only synchronized once at each step.

- the container is resized whenever an element outside the current range is read/written to (the read returns zero in that case); this avoids the necessity of tracking number of bodies, but also is potential danger (such as `scene->forces.getForce(1000000000)`, which will probably exhaust your RAM). Unlike c++, Python does check given id against number of bodies.

```
// resetting forces (inside ForceResetter)
scene->forces.reset()

// in a parallel section
scene->forces.addForce(id,force); // add force

// container is not synced after we wrote to it, sync before reading
scene->forces.sync();
const Vector3r& f=scene->forces.getForce(id);
```

Synchronization is handled automatically if values are read from python:

```
Yade [57]: O.bodies.append(Body())
Out[57]: 3

Yade [58]: O.forces.addF(0,Vector3(1,2,3))

Yade [59]: O.forces.f(0)
Out[59]: Vector3(1,2,3)

Yade [60]: O.forces.f(100)
---------------------------------------------------------------------------
IndexError                                Traceback (most recent call last)
~/yade/lib/x86_64-linux-gnu/yadeflip/py/yade/__init__.py in <module>
----> 1 O.forces.f(100)

IndexError: Body id out of range.
```

### Handling interactions

Creating and removing interactions is a rather delicate topic and number of components must cooperate so that the whole behaves as expected.

Terminologically, we distinguish

**potential interactions,** having neither *geometry* nor *physics*. *Interaction.isReal* can be used to query the status (`Interaction::isReal()` in c++).

**real interactions,** having both *geometry* and *physics*. Below, we shall discuss the possibility of interactions that only have geometry but no physics.

During each step in the simulation, the following operations are performed on interactions in a typical simulation:

1. Collider creates potential interactions based on spatial proximity. Not all pairs of bodies are susceptible of entering interaction; the decision is done in Collider::mayCollide:

   - clumps may not enter interactions (only their members can)

- clump members may not interact if they belong to the same clump

- bitwise AND on both bodies' *masks* must be non-zero (i.e. there must be at least one bit set in common)

2. Collider erases interactions that were requested for being erased (see below).

3. *InteractionLoop* (via *IGeomDispatcher*) calls appropriate *IGeomFunctor* based on *Shape* combination of both bodies, if such functor exists. For real interactions, the functor updates associated *IGeom*. For potential interactions, the functor returns

    **false** if there is no geometrical overlap, and the interaction will stillremain potential-only

    **true** if there is geometrical overlap; the functor will have created an *IGeom* in such case.

    ---

    **Note:** For *real* interactions, the functor *must* return **true**, even if there is no more spatial overlap between bodies. If you wish to delete an interaction without geometrical overlap, you have to do this in the *LawFunctor*.

    This behavior is deliberate, since different *laws* have different requirements, though ideally using relatively small number of generally useful *geometry functors*.

    ---

    ---

    **Note:** If there is no functor suitable to handle given combination of *shapes*, the interaction will be left in potential state, without raising any error.

    ---

4. For real interactions (already existing or just created in last step), *InteractionLoop* (via *IPhysDispatcher*) calls appropriate *IPhysFunctor* based on *Material* combination of both bodies. The functor *must* update (or create, if it doesn't exist yet) associated *IPhys* instance. It is an error if no suitable functor is found, and an exception will be thrown.

5. For real interactions, *InteractionLoop* (via *LawDispatcher*) calls appropriate *LawFunctor* based on combination of *IGeom* and *IPhys* of the interaction. Again, it is an error if no functor capable of handling it is found.

6. *LawDispatcher* takes care of erasing those interactions that are no longer active (such as if bodies get too far apart for non-cohesive laws; or in case of complete damage for damage models). This is triggered by the *LawFunctor* returning false. For this reason it is of upmost importance for the *LawFunctor* to return consistently.

Such interaction will not be deleted immediately, but will be reset to potential state. At the next execution of the collider **InteractionContainer::conditionalyEraseNonReal** will be called, which will completely erase interactions only if the bounding boxes ceased to overlap; the rest will be kept in potential state.

### Creating interactions explicitly

Interactions may still be created explicitly with *utils.createInteraction*, without any spatial requirements. This function searches current engines for dispatchers and uses them. *IGeomFunctor* is called with the **force** parameter, obliging it to return **true** even if there is no spatial overlap.

### Associating Material and State types

Some models keep extra *state* information in the *Body.state* object, therefore requiring strict association of a *Material* with a certain *State* (for instance, *CpmMat* is associated to *CpmState* and this combination is supposed by engines such as *CpmStateUpdater*).

If a *Material* has such a requirement, it must override 2 virtual methods:

1. *Material.newAssocState*, which returns a new *State* object of the corresponding type. The default implementation returns *State* itself.

2. *Material.stateTypeOk*, which checks whether a given *State* object is of the corresponding type (this check is run at the beginning of the simulation for all particles).

In c++, the code looks like this (for *CpmMat*):

```
class CpmMat: public FrictMat {
   public:
      virtual shared_ptr<State> newAssocState() const { return shared_ptr<State>(new CpmState);
↪ }
      virtual bool stateTypeOk(State* s) const { return (bool)dynamic_cast<CpmState*>(s); }
   /* ... */
};
```

This allows one to construct *Body* objects from functions such as *utils.sphere* only by knowing the requires *Material* type, enforcing the expectation of the model implementor.

### 3.1.8 Runtime structure

#### Startup sequence

Yade's main program is python script in core/main/main.py.in; the build system replaces a few `${variables}` in that file before copying it to its install location. It does the following:

1. Process command-line options, set environment variables based on those options.

2. Import main yade module (`import yade`), residing in py/___init___.py.in. This module locates plugins (recursive search for files `lib*.so` in the `lib` installation directory). yade.boot module is used to setup temporary directory, ... and, most importantly, loads plugins.

3. Manage further actions, such as running scripts given at command line, opening *qt.Controller* (if desired), launching the `ipython` prompt.

#### Singletons

There are several "global variables" that are always accessible from c++ code; properly speaking, they are Singletons, classes of which exactly one instance always exists. The interest is to have some general functionality acessible from anywhere in the code, without the necessity of passing pointers to such objects everywhere. The instance is created at startup and can be always retrieved (as non-const reference) using the `instance()` static method (e.g. `Omega::instance().getScene()`).

There are 3 singletons:

**ClassFactory** Registers classes from plugins and able to factor instance of a class given its name as string (the class must derive from `Factorable`). Not exposed to python.

**Omega** Access to simulation(s); deserves separate section due to its importance.

**Logging** Handles logging filters for all named loggers, see *logging verbosity*.

#### Omega

The *Omega* class handles all simulation-related functionality: loading/saving, running, pausing.

In python, the wrapper class to the singleton is instantiated[6] as global variable `O`. For convenience, *Omega* is used as proxy for scene's attribute: although multiple `Scene` objects may be instantiated in c++, it is always the current scene that *Omega* represents.

---

[6] It is understood that instantiating `Omega()` in python only instantiates the wrapper class, not the singleton itself.

The correspondence of data is literal: *Omega.materials* corresponds to `Scene::materials` of the current scene; likewise for *materials*, *bodies*, *interactions*, *tags*, *cell*, *engines*, *initializers*, *miscParams*.

To give an overview of (some) variables:

| Python | c++ |
|--------|-----|
| *Omega.iter* | `Scene::iter` |
| *Omega.dt* | `Scene::dt` |
| *Omega.time* | `Scene::time` |
| *Omega.realtime* | `Omega::getRealTime()` |
| *Omega.stopAtIter* | `Scene::stopAtIter` |

`Omega` in c++ contains pointer to the current scene (`Omega::scene`, retrieved by `Omega::instance().getScene()`). Using *Omega.switchScene*, it is possible to swap this pointer with `Omega::sceneAnother`, a completely independent simulation. This can be useful for example (and this motivated this functionality) if while constructing simulation, another simulation has to be run to dynamically generate (i.e. by running simulation) packing of spheres.

### Engine loop

Running simulation consists in looping over *Engines* and calling them in sequence. This loop is defined in `Scene::moveToNextTimeStep` function in core/Scene.cpp. Before the loop starts, *O.initializers* are called; they are only run once. The engine loop does the following in each iteration over *O.engines*:

1. set `Engine::scene` pointer to point to the current `Scene`.

2. Call `Engine::isActivated()`; if it returns `false`, the engine is skipped.

3. Call `Engine::action()`

4. If *O.timingEnabled*, increment *Engine::execTime* by the difference from the last time reading (either after the previous engine was run, or immediately before the loop started, if this engine comes first). Increment *Engine::execCount* by 1.

After engines are processed, *virtual time* is incremented by *timestep* and *iteration number* is incremented by 1.

### Background execution

The engine loop is (normally) executed in background thread (handled by SimulationFlow class), leaving foreground thread free to manage user interaction or running python script. The background thread is managed by *O.run()* and *O.pause()* commands. Foreground thread can be blocked until the loop finishes using *O.wait()*.

Single iteration can be run without spawning additional thread using *O.step()*.

## 3.1.9 Python framework

### Wrapping c++ classes

Each class deriving from *Serializable* is automatically exposed to python, with access to its (registered) attributes. This is achieved via *YADE_CLASS_BASE_DOC_\* macro family*. All classes registered in class factory are default-constructed in `Omega::buildDynlibDatabase`. Then, each serializable class calls `Serializable::pyRegisterClass` virtual method, which injects the class wrapper into (initially empty) yade.wrapper module. pyRegisterClass is defined by YADE_CLASS_BASE_DOC and knows about class, base class, docstring, attributes, which subsequently all appear in boost::python class definition.

Wrapped classes define special constructor taking keyword arguments corresponding to class attributes; therefore, it is the same to write:

```
Yade [61]: f1=ForceEngine()

Yade [62]: f1.ids=[0,4,5]

Yade [63]: f1.force=Vector3(0,-1,-2)
```

and

```
Yade [64]: f2=ForceEngine(ids=[0,4,5],force=Vector3(0,-1,-2))

Yade [65]: print(f1.dict())
{'force': Vector3(0,-1,-2), 'ids': [0, 4, 5], 'dead': False, 'ompThreads': -1, 'label': ''}

Yade [66]: print(f2.dict())
{'force': Vector3(0,-1,-2), 'ids': [0, 4, 5], 'dead': False, 'ompThreads': -1, 'label': ''}
```

Wrapped classes also inherit from *Serializable* several special virtual methods: *dict()* returning all registered class attributes as dictionary (shown above), *clone()* returning copy of instance (by copying attribute values), *updateAttrs()* and *updateExistingAttrs()* assigning attributes from given dictionary (the former thrown for unknown attribute, the latter doesn't). And *pyDictCustom()* explained also in *preceeding section*.

Read-only property `name` wraps c++ method `getClassName()` returning class name as string. (Since c++ class and the wrapper class always have the same name, getting python type using `__class__` and its property `__name__` will give the same value).

```
Yade [67]: s=Sphere()

Yade [68]: s.__class__.__name__
Out[68]: 'Sphere'
```

### Subclassing c++ types in python

In some (rare) cases, it can be useful to derive new class from wrapped c++ type in pure python. This is done in the *yade.pack module* module: *Predicate* is c++ base class; from this class, several c++ classes are derived (such as *inGtsSurface*), but also python classes (such as the trivial *inSpace* predicate). inSpace derives from python class Predicate; it is, however, not direct wrapper of the c++ `Predicate` class, since virtual methods would not work.

`boost::python` provides special `boost::python::wrapper` template for such cases, where each overridable virtual method has to be declared explicitly, requesting python override of that method, if present. See Overridable virtual functions for more details.

When python code is called from C++, the calling thread must hold the python "Global Interpreter Lock" (GIL). When initalizing the script as well as running one iteration calling `O.step()`, the running thread is the same as python, and no additional code is required. On the other hand, calling python code inside the simulation loop using `O.run()` needs the lock to be acquired by the thread, or a segfault error will occurs. See implementation of *pyGenericPotential* () for a complete exemple.

### Reference counting

Python internally uses reference counting on all its objects, which is not visible to casual user. It has to be handled explicitly if using pure Python/C API with `Py_INCREF` and similar functions.

`boost::python` used in Yade fortunately handles reference counting internally. Additionally, it automatically integrates reference counting for `shared_ptr` and python objects, if class `A` is wrapped as `boost::python::class_<A,shared_ptr<A>>`. Since *all* Yade classes wrapped using *YADE_CLASS_BASE_DOC_* macro family* are wrapped in this way, returning `shared_ptr<…>` objects from is the preferred way of passing objects from c++ to python.

Returning `shared_ptr` is much more efficient, since only one pointer is returned and reference count internally incremented. Modifying the object from python will modify the (same) object in c++ and vice versa. It also makes sure that the c++ object will not be deleted as long as it is used somewhere in python, preventing (important) source of crashes.

#### Custom converters

When an object is passed from c++ to python or vice versa, then either

1. the type is basic type which is transparently passed between c++ and python (int, bool, std::string etc)

2. the type is wrapped by boost::python (such as Yade classes, `Vector3` and so on), in which case wrapped object is returned;[7]

Other classes, including template containers such as `std::vector` must have their custom converters written separately. Some of them are provided in py/wrapper/customConverters.cpp, notably converters between python (homogeneous, i.e. with all elements of the same type) sequences and c++ `std::vector` of corresponding type; look in that source file to add your own converter or for inspiration.

When an object is crossing c++/python boundary, boost::python's global "converters registry" is searched for class that can perform conversion between corresponding c++ and python types. The "converters registry" is common for the whole program instance: there is no need to register converters in each script (by importing `_customConverters`, for instance), as that is done by yade at startup already.

---

**Note:** Custom converters only work for value that are passed by value to python (not "by reference"): some attributes defined using *YADE_CLASS_BASE_DOC_\* macro family* are passed by value, but if you define your own, make sure that you read and understand Why is my automatic to-python conversion not being found?.

In short, the default for `def_readwrite` and `def_readonly` is to return references to underlying c++ objects, which avoids performing conversion on them. For that reason, return value policy must be set to `return_by_value` explicitly, using slighly more complicated `add_property` syntax, as explained at the page referenced.

This deficiency is addressed presently in the file lib/serialization/PyClassCustom.hpp for the `.def_-readonly(…)` function. It can be improved later if the need arises.

---

### 3.1.10 Adding a new python/C++ module

Modules are placed in `py/` directory, the `C++` parts of the modules begin their name with an underscore `_`. The procedure to add a new module is following:

1. Create your new files:

    1. The `yourNewModule.py` file like this.

    2. The `_yourNewModule.cpp` file like this, if part of your module will be written in `C++`.

2. Update the module redirection map in these two places:

    1. `mods` in doc/sphinx/yadeSphinx.py.

    2. `moduleMap` in doc/sphinx/conf.py, if the new module has a `C++` part (this duplication of data will hopefully be soon removed).

---

[7] Wrapped classes are automatically registered when the class wrapper is created. If wrapped class derives from another wrapped class (and if this dependency is declared with the `boost::python::bases` template, which Yade's classes do automatically), parent class must be registered before derived class, however. (This is handled via loop in `Omega::buildDynlibDatabase`, which reiterates over classes, skipping failures, until they all successfully register) Math classes (Vector3, Matrix3, Quaternion) are wrapped in *minieigenHP*. See *high precision documentation* for more details.

3. Add the `C++` file into `py/CMakeLists.txt` like this.

4. Modify the `CMakeLists.txt` but only if the file will depend on cmake compilation variables, eg. like this. The file then needs an additional extension `.in` and be put in two places:

   1. The cmake command to generate the file from `.in` input: like this.

   2. The cmake command to install it: like this.

---

**Hint:** The last step regarding `yourNewModule.py.in` (or `_yourNewModule.cpp.in`) is needed only on very rare occasions, and is included here only for the sake of completeness.

---

### Debugging boundary between python and C++

During normal use all `C++` exceptions are propagated back to `python` interface with full information associated with them. The only situation where this might not be the case is during execution of command `import module` inside a `python` script. It might happen that when importing a new module some cryptic errors occur like: `initialization of module raised unreported exception`. These **unreported exceptions** happen in the situation when the `C++` code executed a `python` code inside it (this is called embedding) and this `python` code threw an exception. The proper way to deal with this situation is to wrap entire module declaration inside a `try {} catch(...) {}` block. It might be possible to deal with specific exceptions also (see here for other example catch blocks), however the general solution is to properly inform `python` that importing this module did not work. In this catch block it is possible to execute `PyErr_Print();` command to see what the problem was and propagate the exception back to python, however during `import module` command only the `SystemError` python exception can get through. Hence the `catch(...)` block after `BOOST_PYTHON_MODULE(_yourNewModule)` should look like this:

```
#include <lib/base/Logging.hpp>

CREATE_CPP_LOCAL_LOGGER("_yourNewModule.cpp");

BOOST_PYTHON_MODULE(_yourNewModule)
try {
        py::def("foo", foo, R"""(
The description of function foo().

:param arg1: description of first argument
:param arg2: description of second argument
:type arg1: type description
:type arg2: type description
:return: return description
:rtype: the return type description

Example usage of foo:

.. ipython::

   In [1]: from yade.yourNewModule import *

   In [1]: foo()

.. note:: Notes, hints and warnings about how to use foo().

        )""");
} catch (...) {
        LOG_FATAL("Importing this module caused an exception and this module is in an↵
↪inconsistent state now.");
```

(continues on next page)

```
        PyErr_Print();
        PyErr_SetString(PyExc_SystemError, __FILE__);
        boost::python::handle_exception();
        throw;
}
```

**Note:** Pay attention to the `_yourNewModule` inside `BOOST_PYTHON_MODULE(…)`, it has to match the file name of the `.cpp` file.

Further reading, about how to work with python exceptions:

1. Example in boost::python reference manual.

2. Example in boost::python tutorial.

3. When PyErr_Print(); is not enough.

### 3.1.11 Maintaining compatibility

In Yade development, we identified compatibility to be very strong desire of users. Compatibility concerns python scripts, *not* simulations saved in XML or old c++ code.

#### Renaming class

Script scripts/rename-class.py should be used to rename class in `c++` code. It takes 2 parameters (old name and new name) and must be run from top-level source directory:

```
$ scripts/rename-class.py OldClassName NewClassName
Replaced 4 occurences, moved 0 files and 0 directories
Update python scripts (if wanted) by running: perl -pi -e 's/\bOldClassName\b/NewClassName/g'␣
↪`ls **/*.py |grep -v py/system.py`
```

This has the following effects:

1. If file or directory has basename `OldClassName` (plus extension), it will be renamed using `bzr`.

2. All occurences of whole word `OldClassName` will be replaced by `NewClassName` in c++ sources.

3. An entry is added to py/system.py, which contains map of deprecated class names. At yade startup, proxy class with `OldClassName` will be created, which issues a `DeprecationWarning` when being instantiated, informing you of the new name you should use; it creates an instance of `NewClassName`, hence not disrupting your script's functioning:

```
Yade [3]: SimpleViscoelasticMat()
/usr/local/lib/yade-trunk/py/yade/__init__.py:1: DeprecationWarning: Class␣
↪`SimpleViscoelasticMat' was renamed to (or replaced by) `ViscElMat', update your code!␣
↪(you can run 'yade --update script.py' to do that automatically)
-> [3]: <ViscElMat instance at 0x2d06770>
```

As you have just been informed, you can run `yade --update` to all old names with their new names in scripts you provide:

```
$ yade-trunk --update script1.py some/where/script2.py
```

This gives you enough freedom to make your class name descriptive and intuitive.

**Renaming class attribute**

Renaming class attribute is handled from c++ code. You have the choice of merely warning at accessing old attribute (giving the new name), or of throwing exception in addition, both with provided explanation. See `deprec` parameter to *YADE_CLASS_BASE_DOC_*_\* macro family* for details.

## 3.2 Yade on GitLab

### 3.2.1 Fast checkout (read-only)

Getting the source code without registering on GitLab can be done via a single command. It will not allow interactions with the remote repository, which you access the read-only way:

```
git clone --recurse-submodules https://gitlab.com/yade-dev/trunk.git
```

### 3.2.2 Branches on GitLab

Most useful commands are listed in the sections below. For more details, see these git guides:

1. ProGit online Book,

2. Guide on setting up git,

3. Git "choose your own adventure",

4. Guide on fixing the conflicts.

**Setup**

1. Register on gitlab.com

2. Add your SSH key to GitLab

3. Set your username and email through terminal

```
git config --global user.name "Firstname Lastname"
git config --global user.email "your_email@youremail.com"
```

You can check these settings with `git config --list`.

4. To fork the repository (optional), click the "Fork" button on the gitlab page, and also fork the YADE data files.

---

**Note:** By default gitlab will try and compile the forked repository, and it will fail if you don't have runners attached to your account. To avoid receiving failure notifications go to repository settings (bottom of left panel->general->permissions) to turn of pipelines.

---

5. Set Up Your Local Repo through terminal. The argument `--recurse-submodules` is to make sure that `./data` directory is filled with the recent data from yade-data (the path is relative to your gitlab profile):

```
git clone --recurse-submodules git@gitlab.com:username/trunk.git
```

This creates a new folder, named trunk, that contains the whole code (make sure username is replaced by your GitLab name). If you already have a cloned yade repository with `./data` directory in it, then you can populate your existing repository using command:

```
git submodule update --init --recursive
```

6. Configure remotes

```
cd to/newly/created/folder
git remote add upstream git@gitlab.com:yade-dev/trunk.git
git remote update
```

Now, your "trunk" folder is linked with two remote repositories both hosted on gitlab.com, the original trunk from yade-dev (called "upstream" after the last command) and the fork which resides in your personal account (called "origin" and always configured by default). Through appropriate commands explained below, you will be able to update your code to include changes commited by others, or to commit yourself changes that others can get.

Holding a fork under personnal account is in fact not strictly necessary. It is recommended, however, and in what follows it is assumed that the above steps have been followed.

### Older versions

In case you want to work with, or compile, an older version of Yade which is not tagged, you can create your own (local) branch of the corresponding daily build. Look here for details.

### Committing and updating

#### Inspecting changes

After changing the source code in the local repository you may start by inspecting them with a few commands. For the "diff" command, it is convenient to copy from the output of "status" instead of typing the path to modified files.

```
git status
git diff path/to/modified/file.cpp
```

#### Pushing changes

Depending on the remote repository you want to push to, follow one of the methods below.

1. Push to yade-dev

Merging changes into yade-dev's master branch cannot be done directly with a push, only by merge request (see below). It is possible however to push changes to a new branch of yade-dev repository for members of that group. It is currently the only way to have merge requests tested by the gitlab CI pipeline before being effectively merged. To push to a new yade-dev/branch:

```
git branch localBranch
git checkout localBranch
git add path/to/new/file.cpp        #Version a newly created file
git commit path/to/new_or_modified/file.cpp -m 'Commit message'   #stage (register) change
↪in the local repository
git pull --rebase upstream master #get updated version of sources from yade-dev repo and
↪apply your commits on the top of them
git push upstream localBranch:newlyCreatedBranch #Push all commits to a new remote branch.
```

The first two lines are optional, if ignored the commits will go the to the default branch, called "master". In the last command **localBranch** is the local branch name on which you were working (possibly **master**) and **newlyCreatedBranch** will be the name of that branch on the remote. Please choose a descriptive name as much as you can (e.g. "fixBug457895").

---

**Note:** If you run into any problems with command `git pull --rebase upstream master`, you *always can revert* or even better fix the conflicts.

---

2. Push to personnal repository

   After previous steps proceed to commit through terminal, "localBranch" should be replaced by a relevant name:

   ```
   git branch localBranch
   git checkout localBranch
   git add path/to/new/file.cpp    #Version a newly created file
   git commit path/to/new_or_modified/file.cpp -m 'Commit message'   #stage (register) change␣
   ↪in the local repository
   git push   #Push all commits to the remote branch
   ```

   The changes will be pushed to your personal fork.

### Updating

You may want to get changes done by others to keep your local and remote repositories synced with the upstream:

```
git pull --rebase upstream master #Pull new updates from the upstream to your branch. Eq. of
↪"bzr update", updating the local branch from the upstream yade-dev/trunk/master
git push  #Merge changes from upstream into your gitlab repo (origin)
```

If you have local uncommited changes this will return an error. A workaround to update while preserving them is to "stash":

```
git stash #backup and hide changes
git pull --rebase upstream master
git push
git stash pop #restore backed up changes
```

### Auto rebase

We promote "rebasing" to avoid confusing logs after each commit/pull/push cycle. It can be convenient to setup automatic rebase, so it does not have to be added everytime in the above commands:

```
git config --global branch.autosetuprebase always
```

Now your file `~/.gitconfig` should include:

```
[branch]
  autosetuprebase = always
```

Check also `.git/config` file in your local trunk folder (rebase = true):

```
[remote "origin"]
  url = git@gitlab.com:yade-dev/trunk.git
  fetch = +refs/heads/*:refs/remotes/origin/*
[branch "master"]
  remote = origin
  merge = refs/heads/master
  rebase = true
```

---

**Pulling a rebased branch**

If someone else rebased on the gitlab server the branch on which you are working on locally, the command `git pull` may complain that the branches have diverged, and refuse to perform operation, in that case this command:

```
git pull --rebase upstream branchName
```

Will match your local branch history with the one present on the gitlab server.

If you are afraid of messing up your local branch you can always make a copy of this branch with command:

```
git branch backupCopyName
```

If you forgot to make that backup-copy and want to go back, then make a copy anyway and go back with this command:

```
git reset --merge ORIG_HEAD
```

The `ORIG_HEAD` backs up the position of `HEAD` before a potentially dangerous operation (merge, rebase, etc.).

A tutorial on fixing the conflicts is a recommended read.

---

**Note:** If you are lost about how to fix your git problems try a git choose your own adventure.

---

### 3.2.3 Merge requests

**Members of yade-dev**

If you have tested your changes and you are ready to merge them into yade-dev's master branch, you'll have to make a "merge request" (MR) from the gitlab.com interface (see the "+" button at the top of the repository webpage). Set source branch and target branch, from yade-dev/trunk/newlyCreatedBranch to yade-dev/trunk/master. The MR will trigger a pipeline which includes compiling, running regression tests, and generating the documentation (the newly built documentation is accessible via settings->pages or by clicking on the "Browse" button in the "Job artifacts" (in the right pane) in the `doc_18_04` build from the pipeline; then navigating to path `Artifacts/install/share/doc`). If the full pipeline succeeds the merge request can be merged into the master branch.

---

**Note:** In case of MR to yade-dev's master from another branch of yade-dev, the pipeline will use group runners attached to yade-dev (the group runners are kindly provided by 3SR, UMS Gricad and Gdańsk University of Technology).

---

**New developers**

Welcome! At start it is very convenient to work on a local fork of YADE in your own gitlab profile. When you are confident that your changes are ready to be merged into official YADE release, please open a Merge Request (MR) in the following way:

1. Make sure that your work is in a separate branch, not in the `master` branch. You can "copy" your branch into another branch with command `git checkout -b myNewFeature`. Please make sure that the amount of changes as compared to the master branch is not large. In case of larger code improvements it is better to split it into several smaller merge requests. This way it will be faster for us to check it and merge.

2. Push your branch to the repository on your gitlab profile with command such as:

```
git push --set-upstream origin myNewFeature
```

3. You should see something like:

```
remote:
remote: To create a merge request for myNewFeature, visit:
remote:   https://gitlab.com/myProfileName/trunk/-/merge_requests/new?merge_request
↪%5Bsource_branch%5D=myNewFeature
remote:
```

4. When you visit the link mentioned above, you will have to select "Change branches" and make sure that correct target branch is selected. Usually that will be `yade-dev/trunk:master`, because this is the official YADE repository.

5. Fill in the title and description then click "Create merge request" at the bottom of the page.

6. After we review the merge request we can click on it to run in our Continuous Integration (CI) pipeline. This pipeline can't start automatically for security reasons. It will be merged after the pipeline checks pass.

Alternatively, create a patch from your commit via:

```
git format-patch origin   #create patch file in current folder)
```

and send to the developers mailing list (yade-dev@lists.launchpad.net) as attachment. In either way, after reviewing your changes they will be added to the main trunk.

When the pull request has been reviewed and accepted, your changes are integrated in the main trunk. Everyone will get them via `git fetch`.

### 3.2.4 Guidelines for pushing

These are general guidelines for pushing to `yade-dev/trunk`.

1. Set autorebase globaly on the computer (only once see above), or at least on current local branch. Non-rebased pull requests will not be accepted on the upstream. This is to keep history linear, and avoid the merge commits.

2. Inspect the diff to make sure you will not commit junk code (typically some "cout<<" left here and there), using in terminal:

```
git diff file1
```

Or using your preferred difftool, such as kdiff3:

```
git difftool -t kdiff3 file1
```

Or, alternatively, any GUI for git: gitg, git-cola...

3. Commit selectively:

```
git commit file1 file2 file3 -m "message" # is good
git commit -a -m "message"                # is bad. It is the best way to commit↵
↪things that should not be commited
```

4. Be sure to work with an up-to-date version launching:

```
git pull --rebase upstream master
```

5. Make sure it compiles and that regression tests pass: try `yade --test` and `yade --check`.

**Thanks a lot for your cooperation to Yade!**

# Chapter 4

# Theoretical background and extensions

## 4.1 DEM formulation

The DEM formulation is presented in earlier chapter 2.1 *DEM formulation* as a common ground for all DEM calculations.

## 4.2 CFD-DEM coupled simulations with Yade and OpenFOAM

The *FoamCoupling* engine provides a framework for Euler-Lagrange fluid-particle simulation with the open source finite volume solver OpenFOAM. The coupling relies on the Message Passing Interface library (MPI), as OpenFOAM is a parallel solver, furthermore communication between the solvers are realised by MPI messages. The *FoamCoupling* engine must be enabled with the ENABLE_MPI flag during compilation:

```
cmake -DCMAKE_INSTALL_PREFIX=/path/to/install /path/to/source -DENABLE_MPI=1
```

Yade sends the particle information (particle position, velocity, etc. ) to all the OpenFOAM processes. Each OpenFOAM process searches the particle in the local mesh, if the particle is found, the hydrodynamic drag force and torque are calculated using the fluid velocity at the particle position (two interpolation methods are available) and the particle velocity. The hydroynamic force is sent to the Yade process and it is added to the force container. The negative of the particle hydrodynamic force (interpolated back to the fluid cell center) is set as source term in the Navier-Stokes equations. The OpenFOAM solver must also be installed to facilitate the MPI connection between Yade and OpenFOAM. Technical details on the coupling methodology can be found in [Kunhappan2017] and [Kunhappan2018].

### 4.2.1 Background

In the standard Euler-Lagrange modelling of particle laden multiphase flows, the particles are treated as point masses. Two approaches are implemented in the present coupling:

1. Point force coupling

2. Volume fraction based force coupling.

In both of the approaches the flow at the particle scale is not resolved and analytical/empirical hydrodynamic force models are used to describe the fluid-particle interactions. For accurate resolution of the particle volume fraction and hydrodynamic forces on the fluid grid the particle size must be smaller than the fluid cell size.

### Point force coupling (*icoFoamYade*)

In the point force coupling, the particles are assumed to be smaller than the smallest fluid length scales, such that the particle Reynolds Number is $Re_p < 1.0$. The particle Reynolds number is defined as the ratio of inertial forces to viscous forces. For a sphere, the associated length-scale is the diameter, therefore:

$$Re_p = \frac{\rho_f |\mathbf{U}_r| d_p}{\mu} \qquad (4.1)$$

where in (4.1) $\rho_f$ is the fluid density, $|\mathbf{U}_r|$ is the norm of the relative velocity between the particle and the fluid, $d_p$ is the particle diameter and $\mu$ the fluid dynamic viscosity. In addition to the Reynolds number, another non-dimensional number that characterizes the particle inertia due to it's mass called Stokes number is defined as:

$$St_k = \frac{\tau_p |\mathbf{U}_f|}{d_p} \qquad (4.2)$$

where in equation (4.2) $\tau_p$ is the particle relaxation time defined as:

$$\tau_p = \frac{\rho_p d_p^2}{18\mu}$$

For $Re_p < 1$ and $St_k < 1$, the hydrodynamic force on the particle can be represented as a point force. This force is calculated using the Stoke's drag force formulation:

$$\mathbf{F}_h = 3\pi\mu d_p (U_f - U_p) \qquad (4.3)$$

The force obtained from (4.3) is applied on the particle and in the fluid side (in the cell where the particle resides), this hydrodynamic force is formulated as a body/volume force:

$$\mathbf{f}_h = \frac{-\mathbf{F}_h}{V_c \rho_f} \qquad (4.4)$$

where in equation (4.4) $V_c$ is the volume of the cell and $\rho_f$ is the fluid density. Hence the Navier-Stokes equations for the combined system is:

$$\frac{\partial \mathbf{U}}{\partial t} + \nabla \cdot (\mathbf{U}\mathbf{U}) = -\frac{\nabla p}{\rho} + \nabla \bar{\bar{\tau}} + \mathbf{f}_h \qquad (4.5)$$

Along with the continuity equation:

$$\nabla \cdot \mathbf{U} = 0 \qquad (4.6)$$

### Volume averaged coupling (*pimpleFoamYade*)

In the volume averaged coupling, the effect of the particle volume fraction is included. The Navier-Stokes equations take the following form:

$$\frac{\partial (\varepsilon_f \mathbf{U}_f)}{\partial t} + \nabla \cdot (\varepsilon_f \mathbf{U}_f \mathbf{U}_f) = -\frac{\nabla p}{\rho} + \varepsilon_f \nabla \bar{\bar{\tau}} - K (U_f - U_p) + \mathbf{S}_u + \varepsilon_f \mathbf{g} \qquad (4.7)$$

Along with the continuity equation:

$$\frac{\partial \varepsilon_f}{\partial t} + \nabla \cdot (\varepsilon_f \mathbf{U}_f) = 0 \tag{4.8}$$

where in equations (4.7) and (4.8) $\varepsilon_f$ is the fluid volume fraction. Note that, we do not solve for $\varepsilon_f$ directly, but obtain it from the local particle volume fraction $\varepsilon_s$, $\varepsilon_f = 1 - \varepsilon_s$. $K$ is the particle drag force parameter, $\mathbf{U}_f$ and $\mathbf{U}_p$ are the fluid and particle velocities respectively. $\mathbf{S}_u$ denotes the explicit source term consisting the effect of other hydrodynamic forces such as the Archimedes/ambient force, added mass force etc. Details on the formulation of these forces are presented in the later parts of this section.

The interpolation and averaging of the Eulerean and Lagrangian quantities are based on a Gaussian envelope $G_\star$. In this method, the the effect of the particle is 'seen' by the neighbouring cells of the cell in which it resides. Let $\mathbf{x}_c$ and $\mathbf{x}_p$ be the fluid cell center and particle position respectively, then the Gaussian filter $G_\star (\mathbf{x}_c - \mathbf{x}_p)$ defined as:

$$G_\star (\mathbf{x}_c - \mathbf{x}_p) = \left(2\pi\sigma^2\right)^{\frac{3}{2}} \exp\left(-\frac{\|\mathbf{x}_c - \mathbf{x}_p\|^2}{2\sigma^2}\right) \tag{4.9}$$

with $\sigma$ being the standard deviation of the filter defined as:

$$\sigma = \delta / \left(2\sqrt{2\ln 2}\right) \tag{4.10}$$

where in equation (4.10) $\delta$ is the cut-off range (at present it's set to $3\Delta x$, with $\Delta x$ being the fluid cell size.) and follows the rule:

$$G_\star (\|\mathbf{x}_c - \mathbf{x}_p\| = \delta/2) = \frac{1}{2} G_\star (\|\mathbf{x}_c - \mathbf{x}_p\| = 0)$$

The particle volume fraction $\varepsilon_{s,c}$ for a fluid cell $c$ is calculated by:

$$\varepsilon_{s,c} = \frac{\sum_{i=1}^{N_p} V_{p,i} G_{\star(i,c)}}{V_c} \tag{4.11}$$

where in (4.11) $N_p$ is the number of particle contributions on the cell $c$, $G_{\star(i,c)}$ is the Gaussian weight obtained from (4.9), $V_{p,i} G_{\star(i,c)}$ forms the individual particle volume contribution. $V_c$ is the fluid cell volume and $\varepsilon_f + \varepsilon_s = 1$

The averaging and interpolation of an Eulerean quantity $\varphi$ from the grid (cells) to the particle position is performed using the following expression:

$$\widetilde{\varphi} = \sum_{i=1}^{N_c} \varphi_i G_{\star(i,p)} \tag{4.12}$$

### Hydrodynamic Force

In equation (4.7) the term $K$ is the drag force parameter. In the present implementation, $K$ is based on the Schiller Naumman drag law, which reads as:

$$K = \frac{3}{4} C_d \frac{\rho_f}{d_p} \left\|\widetilde{\mathbf{U}}_f - \mathbf{U}_p\right\| \varepsilon_f^{-h_{exp}} \tag{4.13}$$

In equation (4.13) $\rho_f$ is the fluid density, $d_p$ the particle diameter, $h_{exp}$ is defined as the 'hindrance coefficient' with the value set as $h_{exp} = 2.65$. The drag force force coefficient $C_d$ is valid for particle Reynolds numbers up to $Re_p < 1000$. The expression for $C_d$ reads as:

$$C_d = \frac{24}{Re_p} \left( 1 + 0.15 Re_p^{0.687} \right) \qquad (4.14)$$

The expression of hydrodynamic drag force on the particle is:

$$\mathbf{F}_{drag} = V_p K (\widetilde{\mathbf{U}}_f - \mathbf{U}_p)$$

In the fluid equations, negative of the drag parameter $(-K)$ is distributed back to the grid based on equation (4.11). Since the drag force includes a non-linear dependency on the fluid velocity $\mathbf{U}_f$, this term is set as an implicit source term in the fluid solver.

The Archimedes/ambient force experienced by the particle is calculated as:

$$\mathbf{F}_{by} = \left( -\widetilde{\nabla p} + \widetilde{\nabla \bar{\bar{\tau}}} \right) V_p \qquad (4.15)$$

where in (4.15), $\widetilde{\nabla p}$ is the averaged pressure gradient at the particle center and $\widetilde{\nabla \bar{\bar{\tau}}}$ is the averaged divergence of the viscous stress at the particle position.

Added mass force:

$$\mathbf{F}_{am} = C_m \left( \frac{D \widetilde{\mathbf{U}}_f}{Dt} - \frac{d \mathbf{U}_p}{dt} \right) V_p \qquad (4.16)$$

where in eaqution (4.16), $\frac{D \widetilde{\mathbf{U}}_f}{Dt}$ is the material derivative of the fluid velocity.

Therefore the net hydrodynamic force on the particle reads as:

$$\mathbf{F}_{hyd} = \mathbf{F}_{drag} + \mathbf{F}_{by} + \mathbf{F}_{am}$$

And on the fluid side the explicit source term $\mathbf{S}_{u,c}$ for a fluid cell c is expressed as :

$$\mathbf{S}_{u,c} = \frac{\sum_{i=1}^{N_p} -\mathbf{F}_{hyd,i} \varepsilon_{s,c} G_{\star(i,c)}}{\rho_f V_c}$$

### 4.2.2 Setting up a case

**In Yade**

Setting a case in the Yade side is fairly straight forward. The python script describing the scene in Yade is based on this method. Make sure the exact wall/periodic boundary conditions are set in Yade as well as in the OpenFOAM. The particles should not leave the fluid domain. In case a particle has 'escaped' the domain, a warning message would be printed/written to the log file and the simulation will break.

The example in examples/openfoam/scriptYade.py demonstrates the coupling. A symbolic link to Yade is created and it is imported in the script. The MPI environment is initialized by calling the initMPI() function before instantiating the coupling engine

```
initMPI()
fluidCoupling = FoamCoupling()
fluidCoupling.getRank()
```

A list of the particle ids and number of particle is passed to the coupling engine

```
sphereIDs = [b.id for b in O.bodies if type(b.shape)==Sphere]
numparts = len(sphereIDs);

fluidCoupling.setNumParticles(numparts)
fluidCoupling.setIdList(sphereIDs)
fluidCoupling.isGaussianInterp = False
```

The type of force/velocity interpolation mode has to be set. For Gaussian envelope interpolation, the *isGaussianInterp* flag has to be set, also the solver *pimpleFoamYade* must be used. The engine is added to the O.engines after the timestepper

```
O.engines = [
ForceResetter(),
...,
GlobalStiffnessTimeStepper,
fluidCoupling ...
newton ]
```

Substepping/data exchange interval is set automatically based on the ratio of timesteps as foamDt/yadeDt (see *exchangeDeltaT* for details).

### In OpenFOAM

There are two solvers available in this git repository. The solver *icoFoamYade* is based on the point force coupling method and the solver *pimpleFoamYade* is based on the volume averaged coupling. They are based on the existing icoFoam and pimpleFoam solvers respectively. Any OpenFOAM supported mesh can be used, for more details on the mesh options and meshing see here. In the present example, the mesh is generated using *blockMesh* utility of OpenFOAM. The case is set up in the usual OpenFOAM way with the directories *0*, *system* and *constant*

```
0/
  U                      ## velocity boundary conditions
  p                      ## pressure boundary conditions
  uSource                ## source term bcs (usually set as calculated).

system/
  controlDict            ## simulation settings : start time, end time, delta T, solution⮠
↪write control etc.
  blockMeshDict          ## mesh setup for using blockMesh utility : define coordinates of⮠
↪geometry and surfaces. (used for simple geometries -> cartesian mesh.)
  decomposeParDict       ## dictionary for setting domain decomposition, (in the present⮠
↪example scotch is used)
  fvSchemes              ## selection of finite volume schemes for calculations of⮠
↪divergence, gradients and interpolations.
  fvSolution             ## linear solver selection, setting of relaxation factors and⮠
↪tolerance criterion,

constant/
  polymesh/              ## mesh information, generated by blockMesh or other mesh utils.
  transportProperties    ## set the fluid and particle properties. (just density of the⮠
↪particle)
```

Note: Always set the timestep less than the particle relaxation time scale, this is not claculated automatically yet! Turbulence modelling based on the RANS equations have not been implemented yet.

However it is possible to use the present formulations for fully resolved turbulent flow simulations via DNS. Dynamic/moving mesh problems are not supported yet. (Let me know if you're interested in implementing any new features.)

To prepare a simulation, follow these steps:

```
blockMesh          ## generate the mesh
decomposePar       ## decompose the mesh
```

Any type of mesh that is supported by OpenFOAM can be used. Dynamic mesh is currently not supported.

**Execution**

The simulation is executed via the following command:

```
mpiexec -n 1 python3 scriptYade.py : -n NUMPROCS icoFoamYade -parallel
```

The video below shows the steps involved in compiling and executing the coupled CFD-DEM simulation

### 4.2.3 Post-Processing

Paraview can be used to visulaize both the Yade solution (use VTKRecorder) and OpenFOAM solution. To visulaize the fluid solution, create an empty file as *name.foam* , open this file in Paraview and in the *properties* section below the pipeline, change "Reconstructed case" to "Decomposed case" , or you can use the reconstructed case itself but after running the *reconstructPar* utility, but this is time consuming.

## 4.3 FEM-DEM hierarchical multiscale modeling with Yade and Escript

Authors: Ning Guo and Jidong Zhao

Institution: Hong Kong University of Science and Technology

Escript download page: https://launchpad.net/escript-finley

mpi4py download page (optional, require MPI): https://bitbucket.org/mpi4py/mpi4py

Tested platforms: Desktop with Ubuntu 10.04, 32 bit; Server with Ubuntu 12.04, 14.04, 64 bit; Cluster with Centos 6.2, 6.5, 64 bit;

### 4.3.1 Introduction

The code is built upon two open source packages: Yade for DEM modules and Escript for FEM modules. It implements the hierarchical multiscale model (FEMxDEM) for simulating the boundary value problem (BVP) of granular media. FEM is used to discretize the problem domain. Each Gauss point of the FEM mesh is embedded a representative volume element (RVE) packing simulated by DEM which returns local material constitutive responses to FEM. Typically, hundreds to thousands of RVEs are involved in a medium-sized problem which is critically time consuming. Hence parallelization is achieved in the code through either multiprocessing on a supercomputer or mpi4py on a HPC cluster (require MPICH or Open MPI). The MPI implementation in the code is quite experimental. The "mpipool.py" is contributed by Lisandro Dalcin, the author of mpi4py package. Please refer to the examples for the usage of the code.

### 4.3.2 Finite element formulation

---

**Note:** This and the following section are a short excerpt from [Guo2014] to provide some theoretical background. Yade users of FEM-DEM coupling are welcome to improve the following two sections.

---

In this coupled FEM/DEM framework on hierarchical multiscale modelling of granular media, the geometric domain $\Omega$ of a given BVP is first discretised into a suitable FEM mesh. After the finite element discretisation, one ends up with the following equation system to be solved,

$$\mathbf{Ku} = \mathbf{f}, \tag{4.17}$$

where $\mathbf{K}$ is the stiffness matrix, $\mathbf{u}$ is the unknown displacement vector at the FEM nodes and $\mathbf{f}$ is the nodal force vector lumped from the applied boundary traction. For a typical linear elastic problem, $\mathbf{K}$ can be formulated from the elastic modulus, and equation (4.17) can be solved directly. Whilst in the case involving nonlinearity such as for granular media where $\mathbf{K}$ depends on state parameters and loading history, Newton–Raphson iterative method needs to be adopted and the stiffness matrix is replaced with the tangent matrix $\mathbf{K_t}$, which is assembled from the tangent operator:

$$\mathbf{K_t} = \int_\Omega \mathbf{B}^\mathsf{T} \mathbf{D} \mathbf{B} \mathrm{d}V, \tag{4.18}$$

where $\mathbf{B}$ is the deformation matrix (i.e. gradient of the shape function), and $\mathbf{D}$ is the matrix form of the rank four tangent operator tensor $\mathbb{D}$. During each Newton–Raphson iteration, both $\mathbf{K_t}$ and internal stress $\boldsymbol{\sigma}$ are updated, and the scheme tries to minimise the residual force $\mathbf{R}$ to find a converged solution:

$$\mathbf{R} = \int_\Omega \mathbf{B}^\mathsf{T} \boldsymbol{\sigma} \mathrm{d}V - \mathbf{f}. \tag{4.19}$$

The tangent operator and the stress tensor at each local Gauss integration point are pivotal variables in the aforementioned calculation and need to be evaluated before each iteration and loading step. A continuum-based conventional FEM usually assumes a constitutive relation for the material and derives the tangent matrix and the stress increment based on this constitutive assumption (e.g. using the elasto-plastic modulus $\mathbf{D^{ep}}$ in equation (4.18) to assemble $\mathbf{K_t}$ and to integrate stress). The coupled FEM/DEM multiscale approach obtains the two quantities from the embedded discrete element assembly at each Gauss point and avoids the needs for phenomenological assumptions.

### 4.3.3 Multiscale solution procedure

The hierarchical multiscale modelling procedure is schematically summarised in the following steps:

1. Discretise the problem domain by suitable FEM mesh and attach each Gauss point with a DEM assembly prepared with suitable initial state.

2. Apply one global loading step, that is, imposed by FEM boundary condition on $\partial\Omega$.

   a) Determine the current tangent operator for each RVE.

   b) Assemble the global tangent matrix using equation (4.18) and obtain a trial solution of displacement $\mathbf{u}$ by solving Equation (4.17) with FEM.

   c) Interpolate the deformation $\nabla\mathbf{u}$ at each Gauss point of the FEM mesh and run the DEM simulation for the corresponding RVE using $\nabla\mathbf{u}$ as the DEM boundary conditions.

   d) Derive the updated total stress for each RVE and use it to evaluate the residual by equation (4.19) for the FEM domain.

   e) Repeat the aforementioned steps from (a) to (d) until convergence is reached and finish the current loading step.

3. Proceed to the next loading step and repeat Step 2.

In interpolating the deformation **u** from the FEM solution for DEM boundary conditions in Step 2(c), we consider both the infinitesimal strain **ε** and rotation **ω**

$$\nabla \mathbf{u} = \underbrace{\frac{1}{2}(\nabla \mathbf{u} + \nabla \mathbf{u}^\mathsf{T})}_{\boldsymbol{\varepsilon}} + \underbrace{\frac{1}{2}(\nabla \mathbf{u} - \nabla \mathbf{u}^\mathsf{T})}_{\boldsymbol{\omega}} \tag{4.20}$$

The corresponding RVE packing will deform according to this prescribed boundary condition.

It is also instructive to add a few remarks on the evolution of stress from the RVE in Step 2(d). In traditional FEM, the stress is updated based on an incremental manner to tackle the nonlinear material response. If small strain is assumed, the incremental stress–strain relation may potentially cause inaccurate numerical results when large deformation occurs in the material, which calls for an alternative formulation for large deformation. This issue indeed can be naturally circumvented in the current hierarchical framework. In our framework, the DEM assembly at each Gauss point will memorise its past state history (e.g. pressure level, void ratio and fabric structure) and will be solved with the current applied boundary condition (including both stretch and rotation) at each loading and iteration step. Towards the end of each loading step, instead of using an incremental stress update scheme, the total true stress (Cauchy stress) is derived directly over the solved DEM assembly through homogenisation and is then returned to the FEM solver for the global solution. In this way, we do not have to resort to the use of other objective stress measures to deal with large deformation problems. However, we note that a proper strain measurement is still required and the FEM mesh should not be severely distorted, otherwise, remeshing of the FEM domain will be required.

More detailed description of the solution procedure can be found in [Guo2013], [Guo2014], [Guo2014b], [Guo2014c], [Guo2015].

### 4.3.4 Work on the YADE side

The version of YADE should be at least rev3682 in which Bruno added the stringToScene function. Before installation, I added some functions to the source code (in "yade" subfolder). But only one function ("Shop::getStressAndTangent" in "./pkg/dem/Shop.cpp") is necessary for the FEMxDEM coupling, which returns the stress tensor and the tangent operator of a discrete packing. The former is homogenized using the Love's formula and the latter is homogenized as the elastic modulus. After installation and we get the executable file: yade-versionNo. We then generate a .py file linked to the executable file by "ln yade-versionNo yadeimport.py". This .py file will serve as a wrapped library of YADE. Later on, we will import all YADE functions into the python script through "from yadeimport import *" (see simDEM.py file).

Open a python terminal. Make sure you can run

```python
import sys
sys.path.append('where you put yadeimport.py')
from yadeimport import *
Omega().load('your initial RVE packing, e.g. 0.yade.gz')
```

If you are successful, you should also be able to run

```python
from simDEM import *
```

### 4.3.5 Work on the Escript side

No particular requirement. But make sure the modules are callable in python, which means the main folder of Escript should be in your PYTHONPATH and LD_LIBRARY_PATH. The modules are wrapped as a class in msFEM*.py.

Open a python terminal. Make sure you can run:

```
from esys.escript import *
from esys.escript.linearPDEs import LinearPDE
from esys.finley import Rectangle
```

(Note: Escript is used for the current implementation. It can be replaced by any other FEM package provided with python bindings, e.g. FEniCS (http://fenicsproject.org). But the interface files "ms-FEM*.py" need to be modified.)

### 4.3.6 Example tests

After Steps 1 & 2, one should be able to run all the scripts for the multiscale analysis. The initial RVE packing (default name "0.yade.gz") should be provided by the user (e.g. using YADE to prepare a consolidated packing), which will be loaded by simDEM.py when the problem is initialized. The sample is initially uniform as long as the same RVE packing is assigned to all the Gauss points in the problem domain. It is also possible for the user to specify different RVEs at different Gauss points to generate an inherently inhomogeneous sample.

While simDEM.py is always required, only one msFEM*.py is needed for a single test. For example, in a 2D (3D) dry test, msFEM2D.py (msFEM3D.py) is needed; similarly for a coupled hydro-mechanical problem (2D only, saturated), msFEMup.py is used which incorporates the u-p formulation. Multiprocessing is used by default. To try MPI parallelization, please set useMPI=True when constructing the problem in the main script. Example tests given in the "example" subfolder are listed below. Note: The initial RVE packing (named 0.yade.gz by default) needs to be generated, e.g. using prepareRVE.py in "example" subfolder for a 2D packing (similarly for 3D).

1. **2D drained biaxial compression test on dry dense sand** (biaxialSmooth.py) *Note*: Test description and result were presented in [Guo2014] and [Guo2014c].

2. **2D passive failure under translational mode of dry sand retained by a rigid and frictionless wall** (retainingSmooth.py) *Note:* Rolling resistance model (CohFrictMat) is used in the RVE packing. Test description and result were presented in [Guo2015].

3. **2D half domain footing settlement problem with mesh generated by Gmsh** (footing.py, footing.msh) *Note:* Rolling resistance model (CohFrictMat) is used in the RVE packing. Six-node triangle element is generated by Gmsh with three Gauss points each. Test description and result were presented in [Guo2015].

4. **3D drained conventional triaxial compression test on dry dense sand using MPI parallelism** (triaxialRough.py) *Note 1:* The simulation is very time consuming. It costs ~4.5 days on one node using multiprocessing (16 processes, 2.0 GHz CPU). When useMPI is switched to True (as in the example script) and four nodes are used (80 processes, 2.2 GHz CPU), the simulation costs less than 24 hours. The speedup is about 4.4 in our test. *Note 2:* When MPI is used, mpi4py is required to be installed. The MPI implementation can be either MPICH or Open MPI. The file "mpipool.py" should also be placed in the main folder. Our test is based on openmpi-1.6.5. This is an on-going work. Test description and result will be presented later.

5. **2D globally undrained biaxial compression test on saturated dense sand with changing permeability using MPI parallelism** (undrained.py) *Note:* This is an on-going work. Test description and result will be presented later.

### 4.3.7 Disclaim

This work extensively utilizes and relies on some third-party packages as mentioned above. Their contributions are acknowledged. Feel free to use and redistribute the code. But there is NO warranty; not even for MERCHANTABILITY or FITNESS FOR A PARTICULAR PURPOSE.

# 4.4 Simulating Acoustic Emissions in Yade

*Suggested citations:*

Caulk, R. (2018), Stochastic Augmentation of the Discrete Element Method for Investigation of Tensile Rupture in Heterogeneous Rock. *Yade Technical Archive.* DOI 10.5281/zenodo.1202039. download full text

Caulk, Robert A. (2020), Modeling acoustic emissions in heterogeneous rocks during tensile fracture with the Discrete Element Method. Open Geomechanics, Volume 2, article no. 2, 19 p. doi : 10.5802/ogeo.5. full text

## 4.4.1 Summary

This document briefly describes the simulation of acoustic emissions (AE) in Yade. Yade's clustered strain energy based AE model follows the methods introduced by [Hazzard2000] and [Hazzard2013]. A validation of Yade's method and a look at the effect of rock heterogeneity on AE during tensile rock failure is discussed in detail in [Caulk2018] and [Caulk2020].

## 4.4.2 Model description

Numerical AE events are simulated by assuming each broken bond (or cluster of broken bonds) represents an event location. Additionally, the associated system strain energy change represents the event magnitude. Once a bond breaks, the strain energies ($E_i$) are summed for all intact bonds within a predefined spatial radius ($\lambda$):

$$E_i = \frac{1}{2}\left(\frac{F_n^2}{k_n} + \frac{F_s^2}{k_s}\right)$$

$$E_o = \sum_i^N E_i$$

where $F_n$, $F_s$ and $k_n$, $k_s$ are the normal and shear force (N) and stiffness (N/m) components of the interaction prior to failure, respectively. Yade's implementation uses the maximum change of strain energy surrounding each broken bond to estimate the moment magnitude of the AE. As soon as the bond breaks, the total strain energy ($E_o = \sum_i^N E_i$) is computed for the radius (set by the user as no. of avg particle diameters, $\lambda$. $E_o$ is used as the reference strain energy to compute $\Delta E = E - E_o$ during subsequent time steps. Finally, $\max(\Delta E)$ is used in the empirical equation derived by [Scholz2003]:

$$M_e = \frac{2}{3}\log \Delta E - 3.2$$

Events are clustered if they occur within spatial and temporal windows of other events, similar to the approach presented by [Hazzard2000] and [Hazzard2013]. The spatial window is simply the user defined $\lambda$ and the temporal window $T_{max}$ is computed as:

$$T_{max} = \text{int}\left(\frac{D_{avg}\lambda}{\max(v_{p1}, v_{p2})\Delta t}\right)$$

where $D_{avg}$ is the average diameter of the particles comprising the failed event (m), $v_{p1}$ and $v_{p2}$ are the P-Wave velocities (m/s) of the particle densities, and $\Delta t$ is the time step of the simulation (seconds/time step). As shown in *fig-cluster*, the final location of a clustered event is simply the average of the clustered event centroids. Here the updated reference strain energy is computed by adding the strain energy of the unique interactions surrounding the new broken bond to the original reference strain energy ($E_o$):

- Original bond breaks, sum strain energy of broken bonds ($N_{orig}$) within spatial window $E_{orig,o} = \sum_{i=1}^{N_{orig}} E_i$

- New broken bond detected within spatial and temporal window of original bond break

- Update reference strain $E_o$ by adding unique bonds ($N_{new}$) within new broken bond spatial window
  $E_{new,o} = E_{orig,o} + \sum_{i=1}^{N_{new}} E_i$

This method maintains a physical reference strain energy for the calculation of $\Delta E = E - E_{new,o}$ and depends strongly on the spatial window size. Ultimately, the clustering increases the number of larger events, which yields more comparable b-values to typical Guttenberg Richter curves [Hazzard2013].

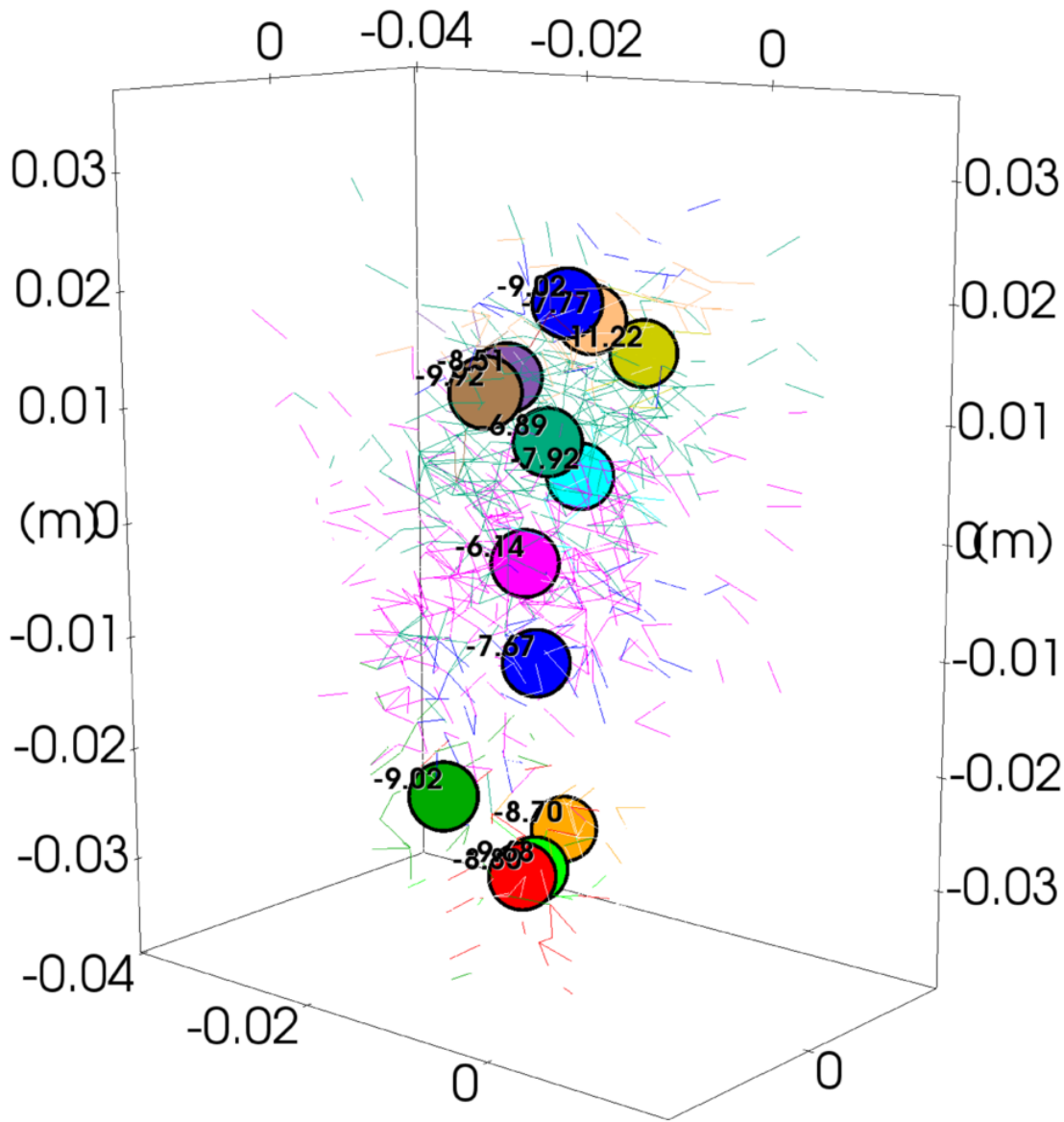

Fig. 1: Example of clustered broken bonds (colored lines) and the final AE events (colored circles) with their event magnitudes.

For a detailed look at the underlying algorithm, please refer to the source code.

### 4.4.3 Activating the algorithm within Yade

The simulation of AE is available as part of Yade's Jointed Cohesive Frictional particle model *(JCFpm)* . As such, your simulation needs to make use of *JCFpmMat* , *JCFpmPhys* , and *Law2_ScGeom_-JCFpmPhys*

Your material assignment and engines list might look something like this:

```
        JCFmat = O.materials.append(JCFpmMat(young=young, cohesion=cohesion,
                density=density, frictionAngle=radians(finalFricDegree),
                tensileStrength=sigmaT, poisson=poisson, label='JCFmat',
                jointNormalStiffness=2.5e6,jointShearStiffness=1e6,jointCohesion=1e6))

        O.engines=[
                ForceResetter(),
                InsertionSortCollider([Bo1_Box_Aabb(),Bo1_Sphere_Aabb
                        ,Bo1_Facet_Aabb()]),
                InteractionLoop(
                        [Ig2_Sphere_Sphere_ScGeom(), Ig2_Facet_Sphere_ScGeom()],
                        [Ip2_FrictMat_FrictMat_FrictPhys(),
                                Ip2_JCFpmMat_JCFpmMat_JCFpmPhys( \
                                        xSectionWeibullScaleParameter=xSectionScale,
                                        xSectionWeibullShapeParameter=xSectionShape,
                                        weibullCutOffMin=weibullCutOffMin,
                                        weibullCutOffMax=weibullCutOffMax)],
                        [Law2_ScGeom_JCFpmPhys_JointedCohesiveFrictionalPM(\
                                recordCracks=True, recordMoments=True,
                                Key=identifier,label='interactionLaw'),
                        Law2_ScGeom_FrictPhys_CundallStrack()]
),

        GlobalStiffnessTimeStepper(),
        VTKRecorder(recorders=['jcfpm','cracks','facets','moments'] \
                        ,Key=identifier,label='vtk'),
                        NewtonIntegrator(damping=0.4)
    ]
```

Most of this simply enables JCFpm as usual, the AE relevant commands are:

```
Law2_ScGeom_JCFpmPhys_JointedCohesiveFrictionalPM(...  recordMoments=True ...)
VTKRecorder(... recorders=[... 'moments' ...])
```

There are some other commands necessary for proper activation and use of the acoustic emissions algorithm:

*clusterMoments*  tells Yade to cluster new broken interactions within the user set spatial radius as described above in the model description. This value is set to True by default.

*momentRadiusFactor*  is λ from the above model description. The momentRadiusFactor changes the number of particle radii beyond the initial interaction that Yade computes the strain energy change. Additionally, Yade uses λ to seek additional broken bonds for clustering. This value is set to 5 by default ( [Hazzard2013] concluded that this value yields accurate strain energy change approximations for the total strain energy change of the system entire system).

*neverErase*  allows old interactions to be stored in memory despite no longer affecting the simulation. This value must be set to True for stable operation of Yade's AE cluster model.

### 4.4.4 Visualizing and post processing acoustic emissions

AE are visualized and post processed in a similar manner to JCFpm cracks. As long as *recordMoments=True*  and *recorder=['moments']* , the simulation will produce timestamped .vtu files for easy Paraview post processing. Within Paraview, the *AE can be filtered according to magnitude, number of constituent interactions, and event time*. *fig-aeexample* shows AE collected during a three point bending test and filtered according to magnitude and time

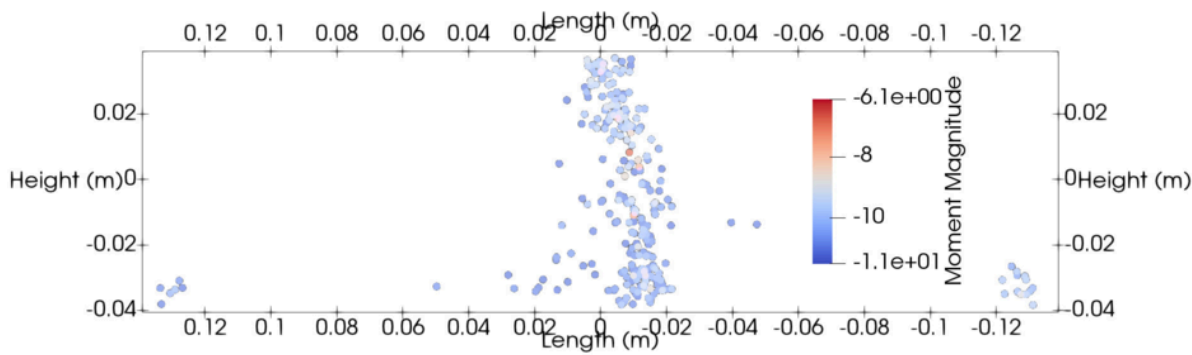

Fig. 2: Example of AE simulated during three point bending test and filtered by magnitude and time.

### 4.4.5 Consideration of rock heterogeneity

[Caulk2018] and [Caulk2020] hypothesize that heterogeneous rock behavior depends on the distribution of interacting grain edge lengths. In support of the hypothesis, [Caulk2018] and [Caulk2020] show how rock heterogeneity can be modeled using cathodoluminescent grain imagery. A Weibull distribution is constructed based on the so called grain edge interaction length distribution. In Yade's *JCFpm* , the Weibull distribution is used to modify the interaction strengths of contacting particles by correcting the interaction area $A_{int}$:

$$A_{int} = \pi(\alpha_w \times \min(R_a, R_b))^2$$

where $\alpha_w$ is the Weibull correction factor, which is distributed as shown in *fig-weibullDist*. The corresponding tensile strength distributions for various Weibull shape parameters are shown in *fig-strengthDist*. Note: a Weibull shape factor of $\infty$ is equivalent to the unaugmented JCFpm model.

In Yade, the application of rock heterogeneity is as simple as passing a Weibull shape parameter to *JCFpmPhys* :

```
Ip2_JCFpmMat_JCFpmMat_JCFpmPhys(
        xSectionWeibullScaleParameter=xSectionScale,
        xSectionWeibullShapeParameter=xSectionShape,
        weibullCutOffMin=weibullCutOffMin,
        weibullCutOffMax=weibullCutOffMax)
```

where the *xSectionWeibullShapeParameter* is the desired Weibull shape parameter. The scale parameter can be assigned in similar fashion. If you want to control the minimum allowable correction factor, you can feed it *weibullCutoffMin* . The maximum correction factor can be controlled in similar fashion.

## 4.5 Using YADE 1D vertical VANS fluid resolution

The goal of the present note is to detail how the DEM-fluid coupling can be used in practice in YADE. It is complementary with the three notes [Maurin2018_VANSbasis], [Maurin2018_VANSfluidResol] and [Maurin2018_VANSvalidations] detailing respectively the theoretical basis of the fluid momentum balance equation, the numerical resolution, and the validation of the code.

All the coupling and the fluid resolution relies only on the engine HydroForceEngine, which use is detailed here. Examples scripts using HydroForceEngine for different purposes can be found in YADE source code in the folder trunk/examples/HydroForceEngine/. In order to get familiar with this engine, it is recommended to read the present note and test/modify the examples scripts.

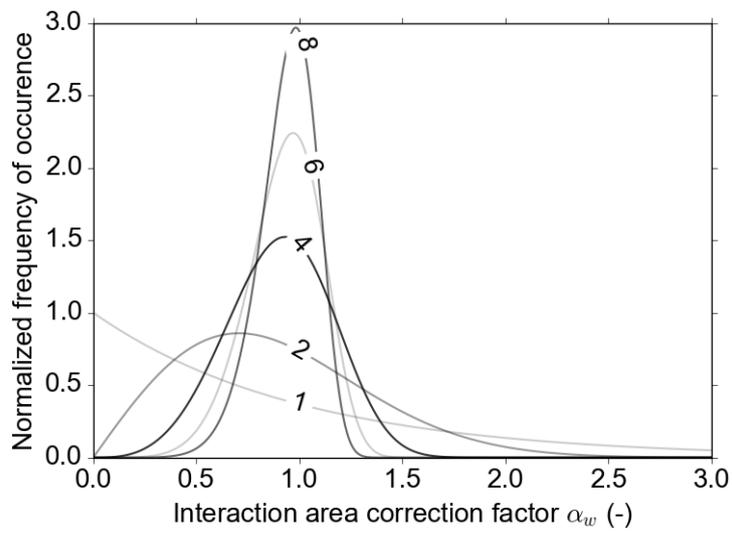

Fig. 3: Weibull distributions for varying shape parameters used to generate $\alpha_w$.

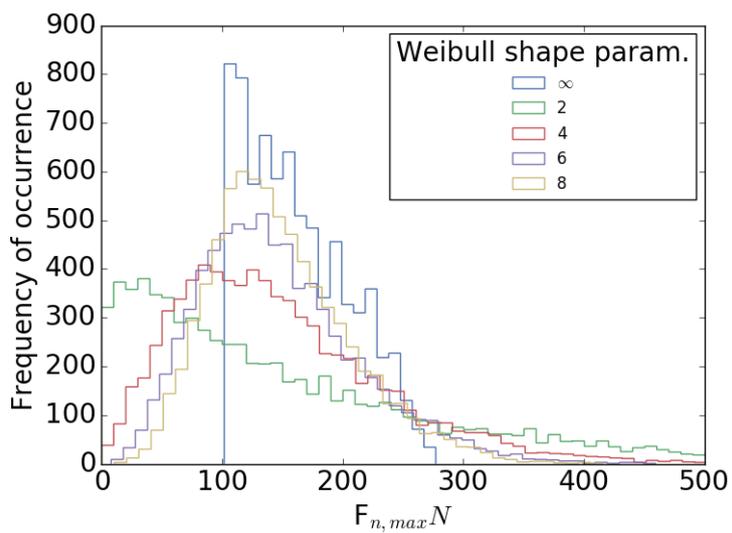

Fig. 4: Maximum DEM particle bond tensile strength distributions for varying Weibull shape parameters.

### 4.5.1 DEM-fluid coupling and fluid resolution in YADE

In YADE, the fluid coupling with the DEM is done through the engine called HydroForceEngine, which is coded in the source in the files trunk/pkg/common/HydroForceEngine.cpp and hpp. HydroForceEngine has three main functions:

- It applies drag and buoyancy to each particle from a 1D vertical fluid velocity profile (HydroForceEngine::action)

- It can evaluates the average drag force, particle velocity and solid volume fraction profiles (HydroForceEngine::averageProfile)

- It can solves the fluid velocity equation detailed in the first section, from given average drag force, particle velocity and solid volume fraction profiles (HydroForceEngine::fluidResolution)

We clearly see the link between the three functions. The idea is to evaluate the average profiles from the DEM, put it as input to the fluid resolution, and apply the fluid forces corresponding to the obtained fluid velocity profile to the particles. In the following, the three points will be detailed separately with precision and imaging with the example scripts available in yade source code at trunk/examples/HydroForceEngine/.

### 4.5.2 Application of drag and buoyancy forces (HydroForceEngine::action)

By default, when adding HydroForceEngine to the list of engine, it applies drag and buoyancy to all the particles which IDs have been passed in argument to HydroForceEngine through the ids variable. This is done for example, in the example script trunk/examples/HydroForceEngine/, in the engine lists:

```
O.engines = [
ForceResetter(),
...
HydroForceEngine(densFluid = densFluidPY,...,ids = idApplyForce),
...
NewtonIntegrator(gravity=gravityVector, label='newtonIntegr')
]
```

where `idApplyForce` corresponds to a list of particle ID to which the hydrodynamic forces should be applied. The expression of the buoyancy and drag force applied to the particles contained in the id list is detailed below.

In case where the fluid is at rest (HydroForceEngine.steadyFlow = False), HydroForceEngine applies buoyancy on a particle p from the fluid density and the acceleration of gravity g as:

$$\mathbf{f}_b^p = -\rho^f V^p \mathbf{g}.$$

Meanwhile, if the fluid flow is steady and turbulent, the buoyancy which is related to the fluid pressure gradient does not have a term in the streamwise direction (see discussion p. 5 of [Maurin2018]). Puting the option HydroForceEngine.steadyFlow to True turns the expression of the buoyancy into:

$$\mathbf{f}_b^p = -\rho^f V^p (\mathbf{g}, \mathbf{e}_x) \mathbf{e}_x.$$

Also, HydroForceEngine applies a drag force to each particles contained in the ids list. This drag force depends on the velocity of the particles and on the fluid velocity, which is defined by a 1D fluid velocity profile, HydroForceEngine.vxFluid. This fluid velocity profile can be evaluated from the fluid model, but can also be imposed by the user and stay constant. From this 1D vertical fluid velocity profile, the drag force applied to particle p reads:

$$\mathbf{f}_D^p = \frac{1}{2} C_d A \rho^f \| u_p^f \mathbf{e}_x - \mathbf{v}^p \| \left( u_p^f \mathbf{e}_x - \mathbf{v}^p \right),$$

where $\mathbf{u}_p^f$ is the fluid velocity at the center of particle p, $\mathbf{v}^p$ is the particle velocity, $\rho^f$ is the fluid density, $A = \pi d^2/4$ is the area of the sphere submitted to the flow, and $C_d$ is the drag coefficient accounts for the effects of particle Reynolds number [Dallavalle1948] and of increased drag due to the presence of other particles (hindrance, [Richardson1954]:

$$C_d = \left(0.44 + \frac{24}{Re_p}\right)(1 - \varphi_p)^{-\gamma} = \left(0.44 + 24 \frac{\nu^f}{\|u_p^f e_x - v^p\|d}\right)(1 - \varphi_p)^{-\gamma}$$

with $\varphi_p$ the solid volume fraction at the center of the particle evaluated from HydroForceEngine.phiPart, and $\gamma$ the Richardson-Zaki exponent, which can be set through the parameter HydroForceEngine.expoRZ (3.1 by default).

HydroForceEngine can also apply a lift force, but this is not done by default (HydroForceEngine.lift = False), and this is not recommended by the author considering the uncertainty on the actual formulation (see discussion p. 6 of [Maurin2015] and [Schmeeckle2007]).

As the fluid velocity profile (HydroForceEngine.vxFluid) and solid volume fraction profile (HydroForceEngine.phiPart) can be imposed by the user, the application of drag and buoyancy to the particles through HydroForceEngine can be done without using the function average-Profile and the fluid resolution. Examples of such use can be found in the source code: trunk/examples/HydroForceEngine/oneWayCouplingfootnote{In this case, we talk about a one-way coupling as the fluid influence the particles but is not influenced back}.

### 4.5.3 Solid phase averaging (HydroForceEngine::averageProfile)

In order to solve the fluid equation, we have seen that it is necessary to compute from the DEM the solid volume fraction, the solid velocity, and the averaged drag profiles. The function HydroForceEngine.averageProfile() has been set up in order to do so. It is designed to evaluate the average profiles over a regular grid, at the position between two mesh nodes. In order to match the fluid velocity profile numerotation, the averaged vector are of size `ndimz + 1` even though the quantities at the top and bottom boundaries are not evaluated and set to zero by defaultfootnote{It is not necessary to evaluate the solid DEM quantities at the boundaries as they are not considered in the fluid resolution, see subsection boundaries of [Maurin2018_VANSfluidResol]}. textcolor{red}{You should do that}

The solid volume fraction profile is evaluated by considering the volume of particles contained in the layer considered. The layer is defined by the mesh step along the wall-normal direction, but extend over the whole length and width of the sample. We perform such an averaging only discretized over the wall-normal direction in order to match the fluid resolution. Meanwhile, this is also physical as, at steady state the problem is unidirectional on average, so that the only variation we should observe in the measured averaged quantities should be along the vertical direction, z. Therefore, the solid volume fraction is evaluated by considering the volume of particles which is contained inside the layer considered $i + 1/2$:

$$\varphi_{i+1/2} = \sum_{p \in [idz;(i+1)dz]} V_{i+1/2}^p;$$

where the sum is over the particles $p$ which have at least a part of their volume inside the layer $i + 1/2$, i.e. in between an elevation of $i * dz$ and $(i + 1) * dz$, and $V_{i+1/2}^p$ is the volume of the particles considered which is contained inside the layer considered. The latter correspond to the integral between two points of a slice of sphere and can be evaluated analytically in cylindrical coordinate. Following this formulation and the formalism of [Jackson2000] with a weighting step function, any particle-associated quantity $K$ can be averaged with the following formulation:

$$\langle K \rangle^p \big|_{i+1/2} = \frac{\sum_{p \in [idz;(i+1)dz]} V_{i+1/2}^p K^p}{\sum_{p \in [idz;(i+1)dz]} V_{i+1/2}^p},$$

Where $K^p$ is the quantity associated with particle $p$, e.g. the particle streamwise velocity. In this case, we can write:

$$\langle v_x \rangle^p \big|_{i+1/2} = \frac{\sum_{p \in [idz;(i+1)dz]} V_{i+1/2}^p v_x^p}{\sum_{p \in [idz;(i+1)dz]} V_{i+1/2}^p},$$

where $v_x^p$ is the velocity of particle $p$. Regarding the evaluation of the average streamwise drag force transmitted by the fluid to the particles, it can be written similarly as:

$$\langle f_{D,x} \rangle^p |_{i+1/2} = \frac{\sum_{p \in [idz;(i+1)dz]} V_{i+1/2}^p f_{D,x}^p}{\sum_{p \in [idz;(i+1)dz]} V_{i+1/2}^p},$$

where $f_{D,x}^p$ is the drag force on particle $p$.

As will be detailed in the next part, these averaged profile can be used for the fluid resolution, but they can also be used for analysis as done for example for bedload transport in [Maurin2015b] [Maurin2018].

### 4.5.4 Fluid resolution\HydroForceEngine::fluidResolution

In order to use the fluid resolution inside the fluid-DEM coupling framework, it is necessary to call the function HydroForceEngine.averageProfile() in order to evaluate the averaged solid volume fraction profile, streamwise velocity and streamwise drag force. The latter is necessary in order to evaluate the terms $\beta$ taken into account in the fluid equation (see [Maurin2018_VANSfluidResol] for details). $\beta$ is defined as:

$$n \langle f_x^f \rangle^p |_{i+1/2} = \beta_{i+1/2} \left( \langle u_x \rangle^f |_{i+1/2} - \langle v_x \rangle^p |_{i+1/2} \right)$$

so that it can be evaluated directly from the averaged drag, particle velocity and the fluid velocity at the last iteration (explicited the term $\beta$ in the fluid resolution):

$$\beta_{i+1/2}^n = \frac{n \langle f_x^f \rangle^p |_{i+1/2}^{n-1}}{\langle u_x \rangle^f |_{i+1/2}^{n-1} - \langle v_x \rangle^p |_{i+1/2}^{n-1}}$$

where the solid variables have been denoted with a superscript $n-1$ as they are known and not re-evaluated at each time stepfootnote{In a way $\beta^n$ should probably be better written as $\beta^{n-1}$}. This terms is called taufsi and is directly evaluated inside the code.

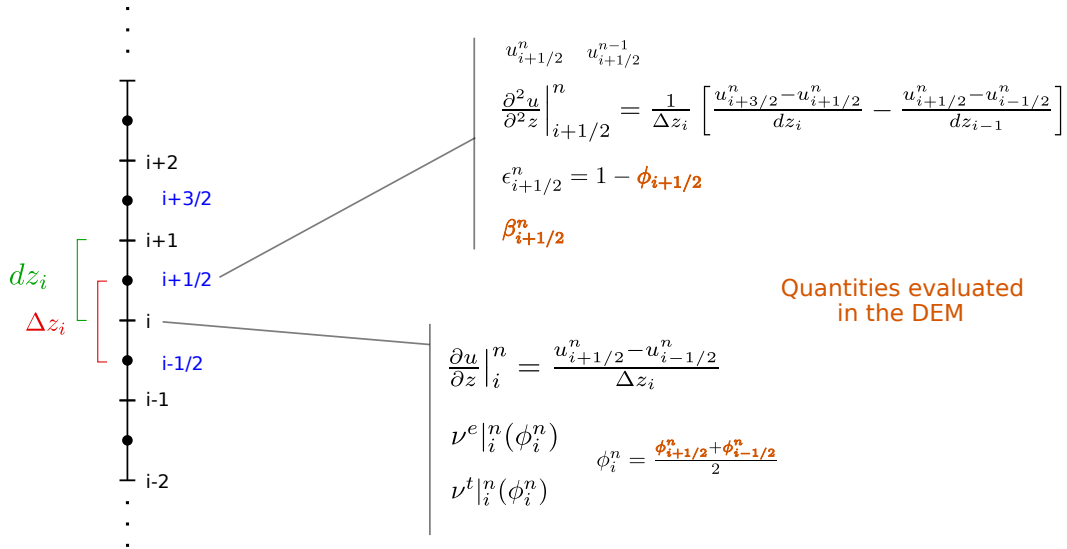

Fig. 5: Schematical picture of the numerical fluid resolution and variables definition with a regular mesh. All the definitions still holds for a mesh with variable spatial step.

All the quantities needed in order to solve the fluid resolution - highlighted in [Maurin2018_VANSfluidResol] and recalled in figure *fig-scheme* - are now explicited. They can be directly evaluated in YADE with the function HydroForceEngine.averageProfile(). From there, the fluid resolution can be performed over a given time $t_{resol}$ with a given time step $\Delta t$ by calling directly the function HydroForceEngine.fluidResolution ($t_{resol}$,$\Delta t$). This will perform the fluid resolution described in [Maurin2018_VANSfluidResol], N = $t_{resol}/\Delta t$ times, with a time step $\Delta t$, considering the vertical profiles of $\beta$, $\langle v_x \rangle$ and $\varphi$ as constant in time. Therefore, one should not only be carefull about the time step, but also about the period of coupling, which should not be too large in order to avoid unphysical behavior in the DEM due to a drastic change of velocity profile not compensated by an increased transmitted drag force.

In the example script in YADE source code, trunk/examples/HydroForceEngine/twoWayCoupling/sedimentTransportExample 1DRANSCoupling.py, the DEM and fluid resolution are coupled with a period of `fluidResolPeriod` = $10^{-2}$s by default, and with a fluid time step of `dtFluid` = $10^{-5}$s. This means that the DEM is let evolved for $10^{-2}$s, and frozen during the fluid resolution which is made over `fluidResolPeriod/dtFluid` = $10^3$ step with $\Delta t = 10^{-5}$. Then, the DEM is let evolved again but with a new fluid velocity profile for $10^{-2}$s, and frozen...etc. This period between two fluid resolution should be tested and taken not too long (see appendix of [Maurin2015b]).

Meanwhile, the fluid resolution can be used in itself, without DEM coupling, in particular to verify the fluid resolution in known cases. This is done in the example folder of YADE source code, trunk/examples/HydroForceEngine/fluidValidation/, where the cases of a poiseuille flow and a log layer have been considered and validated.

## 4.6 Potential Particles and Potential Blocks

The origins of scientific development regarding the algorithms described in this section are traced back to: [Boon2012] (*Potential Blocks* code), [Boon2013b] (*Potential Particles* code) and [Boon2015] (*Block Generation* code).

### 4.6.1 Introduction

This section discusses two codes to simulate (i) non-spherical particles using the concept of the Potential Particles [Houlsby2009], with the solution procedures in [Boon2013] for 3-D and (ii) polyhedral blocks using planar linear inequalities, based on linear programming concepts [Boon2012]. These codes define two shape classes in YADE, namely *PotentialParticle* and *PotentialBlock*. Besides some similarities in syntax, they have distinct differences, concerning morphological characteristics of the particles and the methods used to facilitate contact detection.

The *Potential Particles* code (abbreviated herein as *PP*) is detailed in [Boon2013], where non-spherical particles are assembled as a combination of $2^{nd}$ degree polynomial functions and a fraction of a sphere, while their edges are rounded with a user-defined radius of curvature.

The *Potential Blocks* code (abbreviated herein as *PB*) is used to simulate polyhedral particles with flat surfaces, based on the work of [Boon2012], where a smooth, inner potential particle is used to calculate the contact normal vector. This code is compatible with the *Block Generation* algorithm defined in [Boon2015], in which Potential Blocks can be generated by intersections of original, intact blocks with discontinuity planes.

These two codes are independent, in the sense that either one of them can be compiled/used separately, without enabling the other, while they do not interact with each other (i.e. we cannot establish contact between a PP and a PB). Enabling the PB code causes an automatic compilation of the *Block Generation* algorithm.

### 4.6.2 Potential Particles code (PP)

The concept of *Potential Particles* was introduced and developed by [Houlsby2009]. The problem of contact detection between a pair of potential particles was cast as a constrained optimization problem, where the equations are solved using the Newton-Raphson method in 2-D. In [Boon2013] it was extended to 3-D and more robust solutions were proposed. Many numerical optimization solvers generally cannot cope with discontinuities, ill-conditioned gradients (Jacobians) or curvatures (Hessians), and these obstacles were overcome in [Boon2013], by re-formulating the problem and solving the equations using conic optimization solvers. Previous versions made use of MOSEK (using its academic licence), while currently an in-house code written by [Boon2013] is used to solve the conic optimization problem. A potential particle is defined as in (4.21) [Houlsby2009]:

$$f = (1-k)\left(\sum_{i=1}^{N} \langle a_i x + b_i y + c_i z - d_i \rangle^2 - r^2\right) + k(x^2 + y^2 + z^2 - R^2) \qquad (4.21)$$

where $(a_i, b_i, c_i)$ is the normal vector of the $i^{th}$ plane, defined with respect to the particle's local coordinate system and $d_i$ is the distance of the plane to the local origin. $\langle \rangle$ are Macaulay brackets, i.e., $\langle x \rangle = x$ for $x > 0$; $\langle x \rangle = 0$ for $x \le 0$. The planes are assembled such that their normal vectors point outwards. They are summed quadratically and expanded by a distance $r$, which is also related to the radius of the curvature at the corners. Furthermore, a "shadow" spherical particle is added; $R$ is the radius of the sphere, with $0 < k \le 1$, denoting the fraction of sphericity of the particle. The geometry of some cuboidal potential particles is displayed in Fig. *fig-pp*, for different values of the parameter $k$.

The potential function is normalized for computational reasons in the form (4.22) [Houlsby2009]:

$$f = (1-k)\left(\sum_{i=1}^{N} \frac{\langle a_i x + b_i y + c_i z - d_i \rangle^2}{r^2} - 1\right) + k\left(\frac{x^2 + y^2 + z^2}{R^2} - 1\right) \qquad (4.22)$$

This potential function takes values:

- $f = 0$: on the particle surface
- $f < 0$: inside the particle
- $f > 0$: outside the particle

To ensure numerical stability, it is not advised to use values approaching $k=0$. In particular, the extreme value $k=0$ cannot be used from a theoretical standpoint, since the *Potential Particles* were formulated for strictly convex shapes (curved faces).

### 4.6.3 Potential Blocks code (PB)

The *Potential Blocks* code was developed during the D.Phil. thesis of CW Boon [Boon2013b] and discussed in [Boon2012]. It was developed originally for rock engineering applications, to model polygonal and polyhedral blocks with flat surfaces. The blocks are defined with linear inequalities only and unlike the *PotentialParticle* shape class, no spherical term is considered (so, practically k=0). Although $k$ and $R$ are input parameters of the *PotentialBlock* shape class, their existence during computation is null. In particular, $R$ is used within the source code, denoting a characteristic dimension of the blocks, but does not reflect the radius of a "shadow particle", like it does for the *Potential Particles*. This value of $R$ is used in the *Potential Blocks* code to calculate the initial bi-section step size for line search, to obtain a point on the particle, which in turn is used to calculate the overlap distance during contact.

For a convex particle defined by $N$ planes, the space that it occupies can be defined using the following inequalities (4.23):

$$a_i x + b_i y + c_i z \le d_i, i = 1 : N \qquad (4.23)$$

where $(a_i, b_i, c_i)$ is the unit normal vector of the $i^{th}$ plane, defined with respect to the particle's local coordinate system, and $d_i$ is the distance of the plane to the local origin. According to [Boon2012], an inner, smooth potential particle is used to calculate the contact normal, formulated as in (4.24):

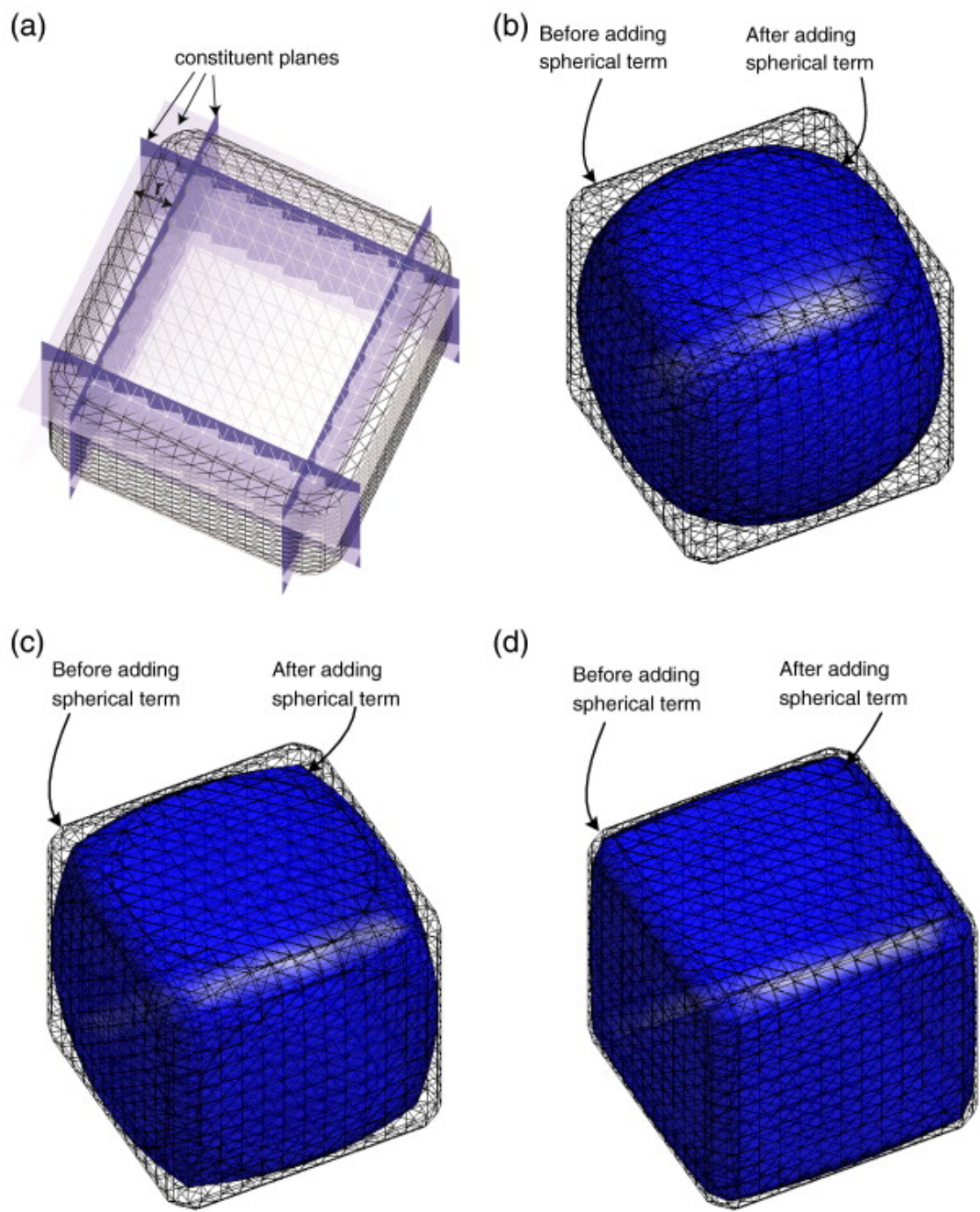

Fig. 6: Construction of potential particles (a) constituent planes are squared and expanded by a constant r. A fraction of sphere is added. Particles with the spherical term are visible in (b) k=0.9, (c) k=0.7, and (d) k=0.4 (after [Boon2013]).

$$f = \sum_{i=1}^{N} \langle a_i x + b_i y + c_i z - d_i + r \rangle^2 \tag{4.24}$$

This potential particle is defined inner by a distance $r$ inside the actual particle, with edges rounded by a radius or curvature $r$, as well (see Fig. *fig-pbInner*).

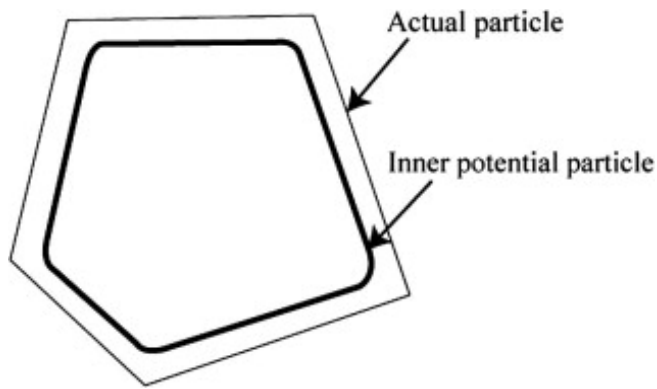

Fig. 7: A potential particle is defined inside the actual particle. The normal vector of the particle at any point can be calculated from the first derivative of the potential particle. (after [Boon2012]).

In YADE, the *Potential Blocks* have a slightly different mathematical expression, since their shape is generated as an assembly of planes as in (4.25):

$$a_i x + b_i y + c_i z - d_i - r = 0, i = 1 : N \tag{4.25}$$

while the inner *Potential Particle* used to calculate the contact normal is defined as in (4.26):

$$f = \sum_{i=1}^{N} \langle a_i x + b_i y + c_i z - d_i \rangle^2. \tag{4.26}$$

Now, the *Potential Block* surface is at a distance of $(d_i + r)$ from the local particle center, while the inner potential particle is at a distance $d$ from the local particle center.

It is worth to emphasize on the fact that the shape of a *Potential Block* is defined using an assembly of planes and not a single, implicit potential function, like we have for the *Potential Particles* code. The inner potential particle in the *Potential Blocks* code is only used to calculate the contact normal.

The problem of establishing intersection between a pair of blocks is cast as a standard linear programming problem of finding a feasible region which satisfies all the linear inequalities defining both blocks. The contact point is calculated as the analytic centre of the feasible region, a well-known concept of interior-point methods in convex optimization calculations. The contact normal is obtained from the gradient of a smooth "potential particle" defined inside the block. The overlap distance is calculated through bi-section searching along the contact normal, within the overlap region.

The linear programming solver for *Potential Blocks* was originally CPLEX, but has been updated to CLP, developed by COIN-OR, since the latter can be downloaded from Ubuntu or Debian's distributions without requiring an academic licence.

### 4.6.4 Engines

The PP and PB codes use their own classes to handle bounding volumes, contact geometry & physics and recording of outputs in vtk format, while they derive the interparticle friction angle from the frictional material class *FrictMat*. The syntax used to invoke these classes is similar, unless if specified otherwise.

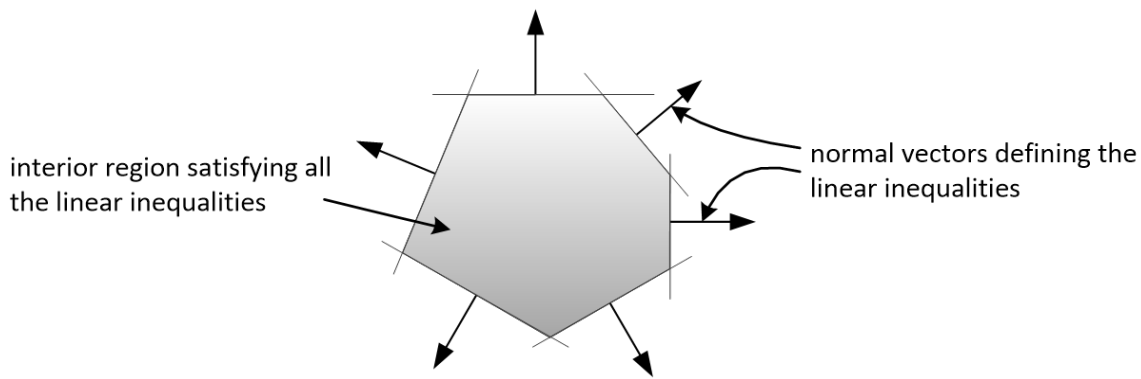

Fig. 8: A potential block. The normal vectors of the faces point outwards (after [Boon2013b]).

| Shape | *PotentialParticle* | *PotentialBlock* |
|---|---|---|
| Material | *FrictMat* | *FrictMat* |
| BoundFunctor | *PotentialParticle2AABB* | *PotentialBlock2AABB* |
| IGeom | *ScGeom* | *ScGeom* |
| IGeomFunctor | *Ig2_PP_PP_ScGeom* | *Ig2_PB_PB_ScGeom* |
| IPhys | *KnKsPhys* | *KnKsPBPhys* |
| IPhysFunctor | *Ip2_FrictMat_FrictMat_KnKsPhys* | *Ip2_FrictMat_FrictMat_KnKsPBPhys* |
| LawFunctor | *Law2_SCG_KnKsPhys_KnKsLaw* | *Law2_SCG_KnKsPBPhys_KnKsPBLaw* |
| VTK Recorder | *PotentialParticleVTKRecorder* | *PotentialBlockVTKRecorder* |

A simple *simulation loop* using the *Potential Blocks* reads as:

```
O.engines=[
        ForceResetter(),
        InsertionSortCollider([PotentialBlock2AABB()], verletDist=0.01),
        InteractionLoop(
                [Ig2_PB_PB_ScGeom(twoDimension=True, unitWidth2D=1.0, calContactArea=True)],
                [Ip2_FrictMat_FrictMat_KnKsPBPhys(kn_i=1e8, ks_i=1e7, Knormal=1e8, Kshear=1e7,
↪viscousDamping=0.2)],
                [Law2_SCG_KnKsPBPhys_KnKsPBLaw(label='law', neverErase=False,
↪allowViscousAttraction=False)]
                ),
        NewtonIntegrator(damping=0.2, exactAsphericalRot=True, gravity=[0,0,-9.81]),
        PotentialBlockVTKRecorder(fileName='./vtk/file_prefix', iterPeriod=1000,
↪twoDimension=True, sampleX=30, sampleY=30, sampleZ=30, maxDimension=0.2, label='vtkRecorder')
]
```

Attention should be given to the *twoDimension* parameter, which defines whether a contact should be handled as 2-D or 3-D.

### 4.6.5 Contact Law

In both codes, the normal force is calculated as:

$$\mathbf{F_n} = \texttt{Knormal} \cdot A_c \cdot u_n \cdot \mathbf{n} \qquad (4.27)$$

where $\texttt{Knormal}$ the normal stiffness coefficient [kN/m³]; $A_c$ the contact area [m²] and $u_n$ the overlap distance. The normal stiffness of each contact [kN/m] is thus $k_n = \texttt{Knormal} \cdot A_c$, where $A_c$ is updated in every timestep.

The shear force is calculated incrementally, using a similar logic. The increment of the shear force vector before slipping of the contact is calculated as:

$$\Delta \mathbf{F_s} = -\mathtt{Kshear} \cdot A_c \cdot \Delta \mathbf{u_s} \tag{4.28}$$

where `Kshear` the shear stiffness coefficient [kN/m$^3$] and $\Delta \mathbf{u_s}$ the current relative shear displacement.

### Contact Area

The contact area is calculated using a heuristic algorithm to detect points on the surface of the overlap volume, searching along the contact shear direction. In essence, it is calculated as the area of a 2D slice of the overlap volume along the shear direction, passing from the contact point. If `twoDimension=True`, the *contactArea* parameter is calculated as:

```
if(twoDimension) { phys->contactArea = phys->jointLength*unitWidth2D;}
```

The *unitWidth2D* parameter is defined by the user (usually equal to 1.0), denoting the out-of-plane width in 2-D simulations. The *contactArea* and *jointLength* parameters are calculated if *calContactArea* =`True`. In the opposite case, they are considered equal to 1.0 and the contact law is degenerated to a linear law with constant stiffness. A minimum value is considered for the *contactArea*, to represent cases where the overlap volume is practically a point.

### Overlap distance

The overlap distance $\mathbf{u_n}$ is calculated using a bracketed bisection search algorithm along the contact normal direction, to find two opposite points on the surface of the overlap region, starting from the contact point. It is stored in the parameter *penetrationDepth*, as the distance between these two opposite points.

## 4.6.6 Shape definition of a PP and a PB

A strong merit of the *Potential Particles* and the *Potential Blocks* codes lies in the fact that the geometric definition of the particle shape and the contact detection problem are resolved using only the equations of the faces of the particles. In this way, using a single data structure, there is no need to store information about the vertices or their connectivity to establish contact, a feature that makes them computationally affordable, while all contacts are handled in the same way (there is no need to distinguish among face-face, face-edge, face-vertex, edge-edge, edge-vertex or vertex-vertex contacts). Due to this, the geometry of a particle is defined in the shape class using the values of the normal vectors of the faces and the distances of the faces from the local origin.

For example, to define a cuboid (6 faces) with rounded edges, an edge length of *D*, centred to its local centroid and aligned to its principal axes, using the *Potential Particles* code, we set:

```
r=D/10.
k=0.3
R=D/2.
b=Body()
b.shape=PotentialParticle( r=r, k=k, R=R,
                           a=[   1.0,   -1.0,    0.0,    0.0,    0.0,    0.0],
                           b=[   0.0,    0.0,    1.0,   -1.0,    0.0,    0.0],
                           c=[   0.0,    0.0,    0.0,    0.0,    1.0,   -1.0],
                           d=[D/2.-r,  D/2.-r,  D/2.-r,  D/2.-r,  D/2.-r,  D/2.-r], ...)
```

The first element of the vector parameters `a`, `b`, `c`, `d` refers to the normal vector of the first plane and its distance from the local origin, the second element to the second plane, and so on.

Using the *Potential Particles* code, this is not a perfect cube, since the particle geometry is defined by a potential function as in (4.22). It is reminded that within this potential function, these planes are summed quadratically, the particle edges are rounded by a radius of curvature $r$ and then the particle faces are curved by the addition of a "shadow" spherical particle with a radius $R$, to a percentage defined by the parameter $k$. A value $r$ is deducted from each element of the vector parameter $d$, to compensate for expanding the potential particle by $r$.

The parameters $a_i, b_i, c_i, d_i$ stated above correspond to the planes used in (4.25):

$$1.0x + 0.0y + 0.0z = D/2 \Leftrightarrow +x = D/2$$
$$-1.0x + 0.0y + 0.0z = D/2 \Leftrightarrow -x = D/2$$
$$0.0x + 1.0y + 0.0z = D/2 \Leftrightarrow +y = D/2$$
$$0.0x - 1.0y + 0.0z = D/2 \Leftrightarrow -y = D/2$$
$$0.0x + 0.0y + 1.0z = D/2 \Leftrightarrow +z = D/2$$
$$0.0x + 0.0y - 1.0z = D/2 \Leftrightarrow -z = D/2$$

To model a cube with an edge of $D$, using the *Potential Blocks* code, we define:

```
r=D/10.
R=D/2.*sqrt(3)
b=Body()
b.shape=PotentialBlock( r=r, R=R,
                        a=[  1.0,    -1.0,     0.0,      0.0,      0.0,      0.0],
                        b=[  0.0,     0.0,     1.0,     -1.0,      0.0,      0.0],
                        c=[  0.0,     0.0,     0.0,      0.0,      1.0,     -1.0],
                        d=[D/2.-r,  D/2.-r,  D/2.-r,   D/2.-r,   D/2.-r,   D/2.-r], ...)
```

Using the *Potential Blocks* code, this particle will have sharp edges and flat faces in what regards its geometry (i.e. the space it occupies), defined by the given planes, while for the calculation of the contact normal, an inner potential particle with rounded edges is used, formulated as in (4.26), located fully inside the actual particle. The distances of the planes from the local origin, stored in the vector parameter $d$, are reduced by $r$ to achieve an exact edge length of $D$, using (4.25). The value of $r$ must be sufficiently small, so that $d_{min} - r > 0$, while it should be sufficiently large, to allow for a proper calculation of the gradient of the inner Potential Particle at the contact point. A recommended value is $r \approx 0.5 * d_{min}$.

To ensure numerical stability, it is advised to normalize the normal vector of each plane, so that $a_i^2 + b_i^2 + c_i^2 = 1$. There is no limit to the number of the particle faces that can be used, a feature that allows the modelling of a variety of convex particle shapes.

In practice, it is usual for the geometry of a particle to be given in terms of vertices & their connectivity (e.g. in the form of a surface mesh, like in .stl files). In such cases, the user can calculate the normal vector of each face, which will give the coefficients $a_i, b_i, c_i$ and using a vertex of each face, then calculate the coefficients $d_i$. A python routine to perform this without any additional effort by the user is currently being developed.

### 4.6.7 Body definition of a PP and a PB

To define a body using the *PotentialParticle* or *PotentialBlock* shape classes, it has to be assembled using the `_commonBodySetup` function, which can be found in the file py/utils.py. For example, to define a *PotentialParticle*:

```
O.materials.append(FrictMat(young=-1,poisson=-1,frictionAngle=radians(0.0),density=2650,label=
↪'frictionless'))

b=Body()
b.shape=PotentialParticle(...)
b.aspherical=True # To be used in conjunction with exactAsphericalRot=True in the
↪NewtonIntegrator
```

```
# V: Volume
# I11, I22, I33: Principal inertias
utils._commonBodySetup(b,V,Vector3(I11,I22,I33), material='frictionless', pos=(0,0,0),␣
↪fixed=False)
b.state.pos=Vector3(xPos,yPos,zPos)
b.state.ori=Quaternion((random.random(),random.random(),random.random()),random.random())
b.shape.volume=V;
O.bodies.append(b)
```

The *PotentialParticle* must be initially defined, so that the local axes coincide with its principal axes, for which the inertia tensor is diagonal. More specifically, the plane coefficients $(a_i, b_i, c_i)$ defining the plane normals must be rotated, so that when the orientation of the particle is zero, the *PotentialParticle* is oriented to its principal axes.

It should be noted that the principal inertia values `I11`, `I22`, `I33` mentioned here are divided with the density of the considered material, since they are multiplied with the density inside the `_commonBodySetup` function. The mass of the particle is calculated within the same function as well, so we do not need to set manually `b.mass=V*density`.

For the *Potential Particles*, the volume and inertia must be calculated manually and assigned to the body as demonstrated above. For the *Potential Blocks*, an automatic calculation has been implemented for the volume and inertia tensor, the user does not have to define the particle to its principal axes, since this is handled automatically within the source code, while if no value is given for the parameter $R$, it is calculated as half the distance of the farthest vertices.

For example, to define a *PotentialBlock*:

```
O.materials.append(FrictMat(young=-1,poisson=-1,frictionAngle=radians(0.0),density=2650,label=
↪'frictionless'))

b=Body()
b.shape=PotentialBlock(R=0.0, ...) #here we set R=0.0 to trigger automatic calculation of R
b.aspherical=True # To be used in conjunction with exactAsphericalRot=True
utils._commonBodySetup(b,b.shape.volume,b.shape.inertia, material='frictionless',␣
↪pos=Vector3(xPos,yPos,zPos), fixed=False)
b.state.ori=b.state.orientation # this will rotate the particle to its initial random system.␣
↪If b.state.ori=Quaternion.Identity, the PB is oriented to its principal axes
O.bodies.append(b)
```

## 4.6.8 Boundary Particles

The PP & PB codes support the definition of *boundary* particles, which interact only with *non-boundary* ones. These particles can have a variety of uses, e.g. to model loading plates acting on a granular sample, while different uses can emerge for different applications. A particle can be set as a boundary one in both codes, using the boolean parameter *isBoundary* inside the shape class.

In the PP code, all particles interact with the same normal and shear contact stiffness *Knormal* and *Kshear*, defined in the *Ip2_FrictMat_FrictMat_KnKsPhys* functor.

The PB code supports the definition of different contact stiffness values for interactions between *boundary* and *non-boundary* or *non-boundary* and *non-boundary* particles. When `isBoundary=False`, the *PotentialBlock* in question is handled to interact with normal and shear stiffness coefficients *Knormal* and *Kshear*, respectively, with other non-boundary particles. When `isBoundary=True`, the *PotentialBlock* in question is handled to interact with normal and shear stiffness coefficients $kn\_i$ and $ks\_i$, respectively, with non-boundary particles.

### 4.6.9 Visualization

Visualization of the *PotentialParticle* and *PotentialBlock* shape classes is offered using the qt environment (OpenGL). Additionally, the *export.VTKExporter.exportPotentialBlocks* function and *PotentialParticleVTKRecorder* and *PotentialBlockVTKRecorder* engines can be used to export geometrical and interaction information of the analyses in vtk format (visualized in Paraview). It should be noted that currently the *PotentialBlockVTKRecorder* records a rounded approximation of the particle, rather than the actual particle with sharp corners and edges.

In the qt environment, the *PotentialParticle* shape class is visualized using the Marching Cubes algorithm, and the level of display accuracy can be determined by the user. This is controlled by the parameters:

```
# Potential Particles
Gl1_PotentialParticle.sizeX=20
Gl1_PotentialParticle.sizeY=20
Gl1_PotentialParticle.sizeZ=20
```

A similar choice exists for output in vtk format, using the *PotentialParticleVTKRecorder* or *PotentialBlockVTKRecorder*, syntaxed as:

```
# Potential Particles
PotentialParticleVTKRecorder(sampleX=30, sampleY=30, sampleZ=30, maxDimension=20)

# Potential Blocks
PotentialBlockVTKRecorder(sampleX=30, sampleY=30, sampleZ=30, maxDimension=20)
```

The parameters sizeX,Y,Z (for OpenGL visualization) and sampleX,Y,Z (for output in vtk format) represent the number of subdivisions of the Aabb of the particle to a grid, which will be used to draw its geometry, in respect to the global axes X, Y, Z. Larger values will result to a more accurate display of the particles' shape, but will slow down the visualization speed in qt and writing speed of the .vtk files and increase the size of the .vtk files. For output in vtk format, users can also define the parameter *maxDimension*, which overrides the selected sampleX,Y,Z values if they are too small, as described below:

if $| \text{xmax} - \text{xmin} | / \text{sampleX} > \text{maxDimension} \Rightarrow \text{sampleX} = | \text{xmax} - \text{xmin} | / \text{maxDimension}$

if $| \text{ymax} - \text{ymin} | / \text{sampleY} > \text{maxDimension} \Rightarrow \text{sampleY} = | \text{ymax} - \text{ymin} | / \text{maxDimension}$

if $| \text{zmax} - \text{zmin} | / \text{sampleZ} > \text{maxDimension} \Rightarrow \text{sampleZ} = | \text{zmax} - \text{zmin} | / \text{maxDimension}$

The *PotentialParticleVTKRecorder* and *PotentialBlockVTKRecorder* also support optionally the recording of the particles' velocities (linear and angular), interaction information (contact point and forces), colors and ids, using:

```
# Potential Particles
PotentialParticleVTKRecorder(..., REC_VELOCITY=True, REC_INTERACTION=True, REC_COLORS=True,␣
↪REC_ID=True)

# Potential Blocks
PotentialBlockVTKRecorder(..., REC_VELOCITY=True, REC_INTERACTION=True, REC_COLORS=True, REC_
↪ID=True)
```

Force chains and other visual outputs are available in qt by default, while they can be extracted in vtk format using the classic *VTKRecorder* or the *export.VTKExporter* class.

A boolean parameter *twoDimension* exists to specify whether the particles will be rendered as 2-D or 3-D in the vtk output:

```
# Potential Particles
PotentialParticleVTKRecorder(..., twoDimension=False)

# Potential Blocks
PotentialBlockVTKRecorder(..., twoDimension=False)
```

This parameter should not be mixed up with the *Ip2_FrictMat_FrictMat_KnKsPBPhys.twoDimension* parameter, which is used to define how the contact forces are calculated, as described in the *Engines* section.

### 4.6.10 Axis-Aligned Bounding Box

The PP & PB codes use their own BoundFunctors, called *PotentialParticle2AABB* and *PotentialBlock2AABB*, respectively, to define the Axis-Aligned Bounding Box of each particle. In both bound functors, a boolean parameter *AabbMinMax* exists, allowing the user to choose between an approximate cubic Aabb or a more accurate one.

In particular, if `AabbMinMax=False`, a cubic Aabb is considered with dimensions `1.0*R`. This is implemented for both the PP and PB codes, even though the *Potential Blocks* do not have a spherical term. In this case, the radius $R$ is used as a reference length, denoting half the diagonal of the cubic Aabb. Usage of this approximate cubic Aabb is not advised in general, since it can increase the number of empty contacts, adding thus to the time needed to facilitate the approximate contact detection, while it relies on the radius $R$, the value of which should enclose the whole particle if this option is activated.

If `AabbMinMax=True`, a more accurate Aabb can be defined. Currently, the initial Aabb of a *PotentialParticle* has to be defined manually by the user, in the particle local coordinate system and for the initial orientation of the particle. To do so, the user has to manually specify the two extreme points of the Aabb: *minAabbRotated*, *maxAabbRotated* inside the shape class. The Aabb for a *PotentialBlock*, on the other hand, is calculated and updated automatically from the vertices of the particle, if the boolean parameter *AabbMinMax* =`True`.

As discussed in the subsection *Visualization*, the dimensions of the Aabb are used as a drawing space in the code implementing rendering of the particles in the qt environment (for the PP code) and for the creation of the output files in vtk format (for both codes). This is achieved by using two auxiliary parameters: *minAabb* and *maxAabb*. For the *Potential Blocks* code only, if these parameters are left unassigned, the drawing space is configured automatically inside the *PotentialBlockVTKRecorder* using the Aabb of the particle. For the particles to be properly rendered as closed surfaces in both qt and vtk outputs using the available codes, we need to define a drawing space slightly larger than the actual one. Here, this drawing space is represented by the Aabb of the particles, and thus the differentiation between the minAabb, maxAabb and minAabbRotated, maxAabbRotated stems from the need to satisfy two conditions: 1. The Aabb used for primary contact detection must be as tight as possible, in order to have the least number of empty contacts and 2. The Aabb used as a rendering space must be slightly larger, in order to have proper rendering. If a dimension of the Aabb used for visualization purposes is defined smaller than the actual one, the faces on that side of the particle are rendered as hollow and only the edges are visualised, a functionality that can be used to e.g. see through boundaries, like demonstrated in the vtk output of the examples/PotentialParticles/cubePPscaled.py example.

To recap, in the *Potential Particles* code, the *minAabbRotated* and *maxAabbRotated* parameters define the initial Aabb used to facilitate primary contact detection, while the *minAabb* and *maxAabb* parameters are used for visualization of the particles in qt and vtk outputs. In the *Potential Blocks* code, the Aabb used to facilitate primary contact detection is calculated automatically from the particles' vertices, which are also used for visualization in qt, while the parameters *minAabb* and *maxAabb* are used for visualization in vtk outputs and can be left unassigned, to trigger an automatic configuration of the drawing space of the particle in the *PotentialBlockVTKRecorder*.

Two brief examples demonstrating the syntax of these features can be found below.

For the *Potential Particles* code:

```
b=Body()
b.shape=PotentialParticle(AabbMinMax=True,
                          minAabbRotated=Vector3(xmin,ymin,zmin),
                          maxAabbRotated=Vector3(xmax,ymax,zmax),
                          minAabb=Vector3(xmin,ymin,zmin),
                          maxAabb=Vector3(xmax,ymax,zmax), ...)
```

For the *Potential Blocks* code:

```
b=Body()
b.shape=PotentialBlock(AabbMinMax=True,
                       minAabb=Vector3(xmin,ymin,zmin),
                       maxAabb=Vector3(xmax,ymax,zmax), ...)
```

### 4.6.11 Block Generation algorithm

The *Potential Blocks* code is compatible with the *Block Generation* algorithm introduced in [Boon2015], which can split particles by their intersection with discontinuity planes, initially developed for the study of rock-masses. This code is hardcoded in YADE in the form of a Preprocessor. Using a single data structure for the definition of the particle shape and the definition of the discontinuities as well, allows the generation of a large number of particles at a reasonable computational cost. The sequential subdivision concept is used along with a linear programming framework. Non-persistent joints can be modelled by introducing more constraints.

An example to demonstrate the usage of this code exists in examples/PotentialBlocks/WedgeYADE.py The discontinuity planes used in this script are included in a csv format in examples/PotentialBlocks/joints/jointC.csv.

The documentation on how to use this code is currently being written.

### 4.6.12 Examples

Examples can be found in the folders examples/PotentialParticles and examples/PotentialBlocks/, where the syntax of the codes is demonstrated.

### 4.6.13 Disclaimer

These codes were developed for academic purposes. Some variables are no longer in use, as the PhD thesis of the original developer spanned over many years, with numerous trials and errors. As this piece of code has many dependencies within the YADE ecosystem, user discretion is advised.

### 4.6.14 References

To acknowledge our scientific contribution, please cite the following:

Potential Blocks

- Boon CW (2013) Distinct Element Modelling of Jointed Rock Masses: Algorithms and Their Verification. D.Phil. Thesis, University of Oxford

- Boon CW, Houlsby GT, Utili S (2012) A new algorithm for contact detection between convex polygonal and polyhedral particles in the discrete element method. Computers and Geotechnics, 44: 73-82

Potential Particles

- Houlsby GT (2009) Potential particles: a method for modelling non-circular particles in DEM. Computers and Geotechnics, 36(6):953-959

- Boon CW, Houlsby GT, Utili S (2013) A new contact detection algorithm for three dimensional non-spherical particles. Powder Technology, S.I. on DEM, 248: 94-102

Block Generation

- Boon CW, Houlsby GT, Utili S (2015) A new rock slicing method based on linear programming. Computers and Geotechnics, 65:12-29

# Chapter 5

# Performance enhancements

## 5.1 Accelerating Yade's FlowEngine with GPU

(Note: we thank Robert Caulk for preparing and sharing this guide)

### 5.1.1 Summary

This document contains instructions for adding Suite Sparse's GPU acceleration to Yade's Pore Finite Volume (PFV) scheme as demonstrated in [Caulk2019]. The guide is intended for intermediate to advanced Yade users. As such, the guide assumes the reader knows how to modify and compile Yade's source files. Readers will find that this guide introduces system requirements, installation of necessary prerequisites, and installation of the modified Yade. Lastly, the document shows the performance enhancement expected by acceleration of the factorization of various model sizes.

### 5.1.2 Hardware, Software, and Model Requirements

- **Hardware:**
    - CUDA-capable GPU with >3 GB memory recommended (64 mb required)
- **Software:**
    - NVIDIA CUDA Toolkit
    - SuiteSparse (CHOLMOD v2.0.0+)
    - Metis (comes with SuiteSparse)
    - CuBlas
    - OpenBlas
    - Lapack
- **Model:**
    - Fluid coupling (Pore Finite Volume aka Yade's "FlowEngine")
    - >10k particles, but likely >30k to see significant speedups
    - Frequent remeshing requirements

### 5.1.3 Install CUDA

The following instructions to install CUDA are a boiled down version of these instructions.

```
lspci | grep -i nvidia #Check your graphics card
# Install kernel headers and development packages
sudo apt-get install linux-headers-$(uname -r)
#Install repository meta-data (see **Note below):
sudo dpkg -i cuda-repo-<distro>_<version>_<architecture>.deb
sudo apt-get update   #update the Apt repository cache
sudo apt-get install cuda #install CUDA
# Add the CUDA library to your path
export PATH=/usr/local/cuda/bin${PATH:+:${PATH}}
export LD_LIBRARY_PATH=/usr/local/cuda/lib64\ ${LD_LIBRARY_PATH:+:${LD_LIBRARY_PATH}}
```

**Note**: use this tool to determine your `<distro>_<version>_<architecture>` values.

Restart your computer.

Verify your CUDA installation by navigating to `/usr/local/cuda/samples` and executing the `make` command. Now you can navigate to `/usr/local/cuda/samples/1_Utilities/deviceQuery/` and execute `./deviceQuery` . Verify the `Result = PASS`.

### 5.1.4 Install OpenBlas, and Lapack

Execute the following command:

```
sudo apt-get install libopenblas-dev liblapack-dev
```

### 5.1.5 Install SuiteSparse

Download the SuiteSparse package and extract the files to `/usr/local/`. Run `make config` and verify `CUDART_LIB` and `CUBLAS_LIB` point to your cuda installed libraries. The typical paths will follow `CUDART_LIB=/usr/local/cuda-x.y/lib64` and `CUBLAS_LIB=/usr/local/cuda-x.y/lib64`. If the paths are blank, you may need to navigate to to `CUDA_PATH` in `/usr/local/SuiteSparse/SuiteSparse_-config/SuiteSparse_config.mk` and modify it manually to point to your cuda installation. Navigate back to the main SuiteSparse folder and execute `make`. SuiteSparse is now compiled and installed on your machine.

Test CHOLMOD's GPU functionality by navigating to `SuiteSparse/CHOLMOD/Demo` and executing `sh gpu.sh`. Note: you will need to download the nd6k.mtx from here and put it in your home directory.

### 5.1.6 Compile Yade

Following the instructions outlined here, run `cmake` with `-DCHOLMOD_GPU=ON` and `-DSUITESPARSEPATH=/usr/local/SuiteSparse` (or your other custom path). Check the output to verify the paths to CHOLMOD (and dependencies such as AMD), SuiteSparse, CuBlas, and Metis are all identified as the paths we created when we installed these packages. Here is an example of the output you need to inspect:

```
-- Found Cholmod in /usr/local/SuiteSparse/lib/libcholmod.so
-- Found OpenBlas in /usr/lib/libopenblas.so
-- Found Metis in /usr/local/SuiteSparse/lib/libmetis.so
-- Found CuBlas in /usr/local/cuda-x.y/libcublas.so
-- Found Lapack in /usr/lib/liblapack.so
```

If you have multiple versions of any of these packages, it is possible the system finds the wrong one. In this case, you will need to either uninstall the old libraries (e.g. `sudo apt-get remove libcholmod` if the other library was installed with apt-get) or edit the paths within `cMake/Find_____.cmake`. If you installed a version of Cuda in a different location than `/usr/local`, you will need to edit `cMake/FindCublas.cmake` to reflect these changes before compilation.

Metis is compiled with SuiteSparse, so the Metis library and Metis include should link to files within `usr/local/SuiteSparse/`. When ready, complete installation with `make -jX install`. Keep in mind that adding `CHOLMOD_GPU` alters `useSolver=4` so to work with the GPU and not the CPU. If you wish to use `useSolver=4` with the CPU without unintended side effects (possible memory leaks), it is recommended to recompile with `CHOLMOD_GPU=OFF`. Of course, `useSolver=3` should always work on the CPU.

### 5.1.7 Controlling the GPU

The GPU accelerated solver can be activated within Yade by setting `flow.useSolver=4`. There are several environment variables that control the allowable memory, allowable GPU matrix size, etc. These are highlighted within the CHOLMOD User Guide, which can be found in `SuiteSparse/CHOLMOD/Doc`. At the minimum, the user needs to set the environment variable by executing `export CHOLMOD_USE_GPU=1`. It is also recommended that you designate half of your available GPU memory with `export CHOLMOD_GPU_MEM_BYTES=3000000000` (for a 6GB graphics card), if you wish to use the `multithread=True` functionality. If you have a multi-gpu setup, you can tell Yade to use one (or both GPUs with SuiteSparse-4.6.0-beta) by executing `export CUDA_VISIBLE_DEVICES=1`, where 1 is the GPU you wish to use.

### 5.1.8 Performance increase

[Catalano2012] demonstrated the performance of DEM+PFV coupling and highlighted its strengths and weaknesses. A significant strength of the DEM+PFV coupling is the asymptotic nature of triangulation costs, volume calculation costs, and force calculation costs ( [Catalano2012], Figure 5.4). In other words, increasing the number of particles beyond ~200k results in negligible additional computational costs. The main weakness of the DEM+PFV coupling is the exponential increase of computational cost of factoring and solving increasingly larger systems of linear equations ( [Catalano2012], Figure 5.7). As shown in Fig. *fig-cpuvsgpu*, the employment of Tesla K20 GPU decreases the time cost of factorization by up to 75% for 2.1 million DOFs and 356k particles.

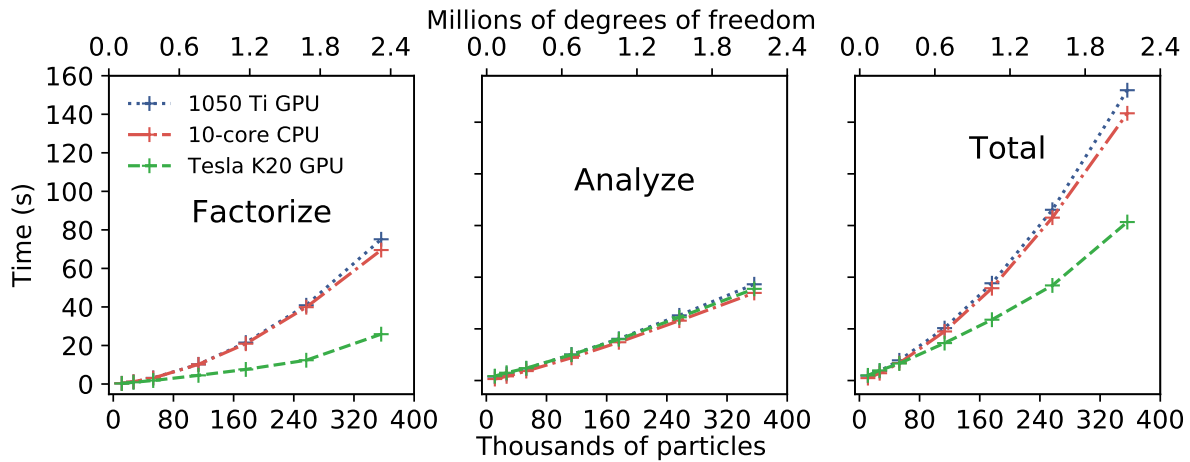

Fig. 1: Time required to factorize and analyze various sized matrices for 10-core CPU, 1050Ti GPU, and Tesla K20 GPU [Caulk2019].

Note: Tesla K20 5GB CPU + 10-core Xeon E5 2.8 GHz CPU

# 5.2 MPI parallelization

The module *mpy* implements parallelization by domain decomposition (distributed memory) using the Message Passing Interface (MPI) implemented by OpenMPI. It aims at exploiting large numbers of compute nodes by running independent instances of Yade on them. The shared memory and the distributed memory approaches are compatible, i.e. it is possible to run hybrid jobs using both, and it may well be the optimal solution in some cases.

Most (initially *all*) calls to OpenMPI library are done in Python using mpi4py. However for the sake of efficiency some critical communications are triggered via python wrappers of C++ functions, wherein messages are produced, sent/received, and processed.

This module development was started in 2018. It received contributions during a HPC hackathon. An extension enables *parallel coupling with OpenFoam*.

---

**Note:** see also *reference documentation of the mpy* module.

---

**Note:** Disclaimer: even though the *yade.mpy* module provides the function *mpirun*, which may seem as a simple replacement for *O.run()*, setting up a simulation with mpy might be deceptively trivial. As of now, it is anticipated that, in general, a simple replacement of "run" by "mpirun" in an arbitrary script will not speedup anything and may even fail miserably (it could be improved in the future). To understand why, and to tackle the problems, basic knowledge of how MPI works will certainly help (specifically mpi4py).

---

## 5.2.1 Concepts

**subdomain**: a (sub)set of bodies attached to one MPI process after domain decomposition - with or without spatial coherence. The corresponding class in Yade is *Subdomain*, a *Shape* instance with helper functions for MPI communications. In some sense *Subdomain* is to subscribed bodies what *Clump* (another *Shape*) is to clump members.

**rank**: subdomain index from 0 to *N*-1 (with *N* the number of mpi processes) to identify subdomains. The rank of the subdomain a body belongs to can be retrieved as *Body.subdomain*. Each subdomain corresponds to an instance of Yade and a specific scene during parallel execution. The rank of the scene is given by *Scene.subdomain*.

**master**: refers to subdomain with *rank* =0. This subdomain does not behave like others. In general master will handle boundary conditions and it will control transitions and termination of the whole simulation. Unlike standard subdomains it may not contain a large number of raw bodies (i.e. not beyond objects bounding the scene such as walls or boxes). In interactive execution master is the process responding to the python prompt.

**splitting and merging**: cutting a master *Scene* into a set of smaller, distributed, scenes is called "splitting". The split is undone by a 'merge', by which all bodies and (optionally) all interactions are sent back to the master thread. Splitting, running, then merging, should leave the scene as if no MPI had been used at all (i.e. as if the same number of iterations had been executed in single-thread). Therefore normal O.run() after that should work as usual.

**intersections**: subsets of bodies in a subdomain intersected by the bounding box of other subdomains (see *fig-subdomains*). *intersection(i,j)* refers to the bodies owned by current (*i*) subdomain and intersecting subdomain *j* (retrieved as *O.\_sceneObj.subD.intersections[j]*); *mirrorIntersection(i,j)* refers to bodies owned by *j* and intersecting current domain (retrieved as *O.\_sceneObj.subD.mirrorIntersections[j]*). The bodies are listed by *Body.id*. By definition *intersection(i,j)=mirrorIntersection(j,i)*.

The intersections and mirror intersections are updated automatically as part of parallel collision detection. They define which body states need to be communicated. The bodies in intersections need to be *sent* to other subdomains (in pratice only updated position and velocity are sent at every iteration), the bodies in mirrorIntersections need to be received from other subdomains.

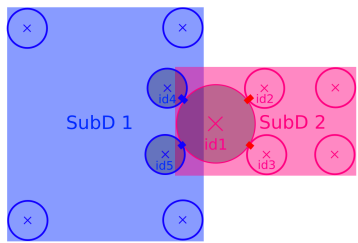

Two overlapping subdomains and their intersections. In this situation we have *SubD1.intersections[SubD2.subdomain]=[id4,id5]* and *SubD1.mirrorIntersections[SubD2.subdomain]=[id1]*, with *SubD1* and *SubD2* instances of *Subdomain*.

### 5.2.2 Walkthrough

For demonstrating the main internal steps in the implemented parallel algorithm let us consider the example script examples/mpi/testMPI_2D.py. Executing this script (interactive or passive mode) with three MPI processes generates the scene as shown in *fig-scene-mpi*. It then executes *mpirun*, which triggers the steps described hereafter.

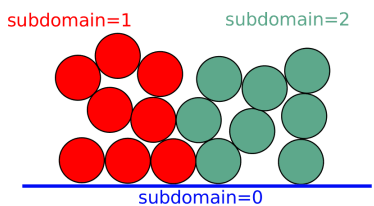

In this scene, we have three MPI processes (three subdomains) and the raw bodies are partitioned among the subdomains/ranks 1 and 2. The master process with subdomain=0 holds the boundary/wall type body. Bodies can be manually assigned or automatically assigned via a domain decomposition algorithm. Details on the dommain decomposition algorithm is presented in the later section of this document.

**Scene splitting** :

In the function *mpy.splitScene*, called at the beginning of mpi execution, specific engines are added silently to the scene in order to handle what will happen next. That very intrusive operation can even change settings of some pre-existing engines, in particular *InsertionSortCollider*, to make them behave with MPI-friendlyness. *InsertionSortCollider.verletDist* is an important factor controlling the efficiency of the simulations. The reason for this will become evident in the later steps.

**Bounds dispatching** : In the next step, the *Body.bound* is dispatched with the *Aabb* extended as shown in figure *fig-regularbounds* (in dotted lines). Note that the *Subdomain Aabb* is obtained from taking the min and max of the owned bodies, see figure *fig-subDBounds* with solid coloured lines for the subdomain *Aabb*. At this time, the min and max of other subdomains are unknown.

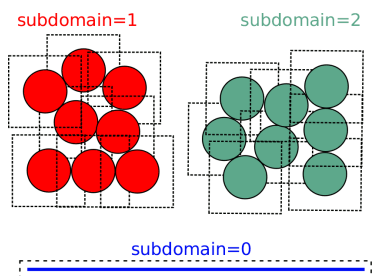

**Update of Domain bounds** : Once the bounds for the regular bodies and the *local subdomain* has been dispatched, information on the other subdomain bounds are obtained via the function

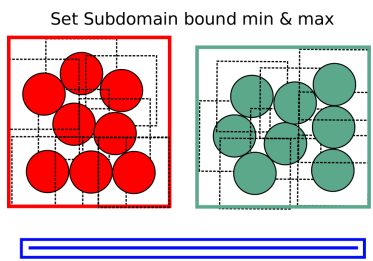

Set Subdomain bound min & max

*mpy.updateDomainBounds*. In this collective communication, each subdomain broadcasts its *Aabb.min* and *Aabb.max* to other subdomains. Figure *fig-subdomain-bounds* shows a schematic in which each subdomain has received the *Aabb.min* and *Aabb.max* of the other subdomains.

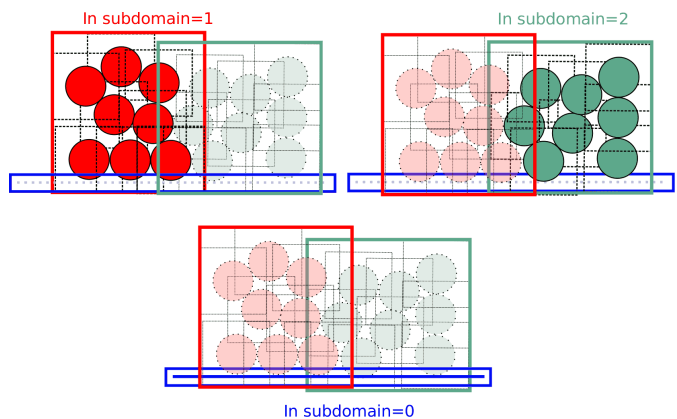

**Parallel Collision detection** :

- Once the *Aabb.min* and *Aabb.max* of the other subdomains are obtained, the collision detection algorithm is used to determine the bodies that have intersections with the remote subdomains. The ids of the identified bodies are then used to build the *Subdomain.intersections* list.

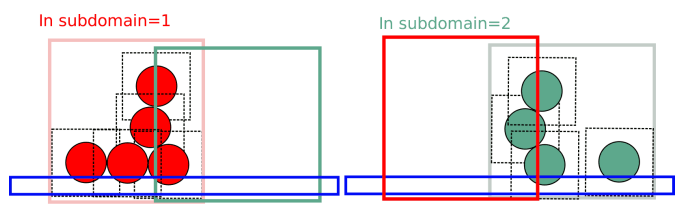

- Next step involves obtaining the ids of the remote bodies intersecting with the current subdomain (*Subdomain.mirrorIntersections*). Each subdomain sends its list of local body intersections to the respective remote subdomains and also receives the list of intersecting ids from the other subdomains. If the remote bodies do not exist within the current subdomain's *BodyContainer*, the subdomain then *requests* these remote bodies from the respective subdomain. A schematic of this operation is shown in figure *fig-mirrorIntersections*, in which subdomain=1 receives three bodies from subdomain=2, and 1 body from subdomain=0. subdomain=2 receives three bodies from subdomain=1. subdomain=0 only sends its bodies and does *not* receive from the worker subdomains. This operation sets the stage for communication of the body states to/from the other subdomains.

**Update states** :

Once the subdomains and the associated intersecting bodies, and remote bodies are identified, *State* of these bodies are sent and received every timestep, by peer-to-peer communications between the interacting subdomains. In the case of an interaction with the master subdomain (subdomain=0), only the total force and torque exerted on master's bodies by a given subdomain are sent. Figure *fig-sendRecvStates*

                                                          

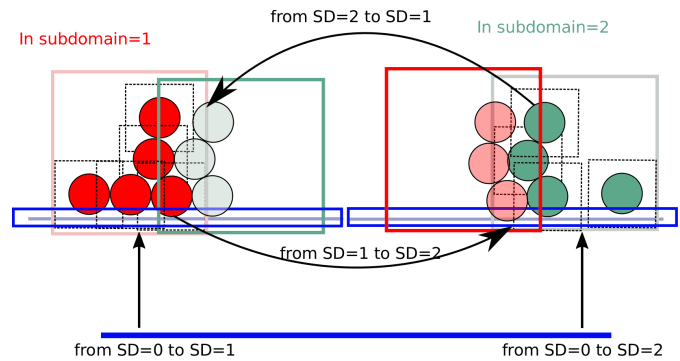

shows a schematic in which the states of the remote bodies between subdomain=1 and subdomain=2 are communicated. Subdomain=0 receives forces and torques from subdomain=1 and subdomain=2.

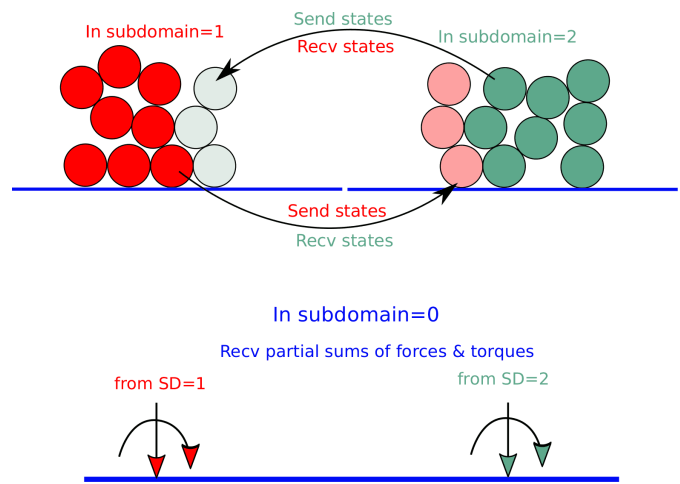

## 5.2.3 MPI initialization and communications

The mpy modules tries to retain one of Yade's most important features: interactive access to the objects of scene (or of multiple scenes in this case), as explained below. Interactive execution does not use the *mpiexec* command of OpenMPI, instead, a pool of workers is spawned by the mpy module after Yade startup. In production one may use passive jobs, and in that case *mpiexec* will preceed the call to Yade.

**Note:** Most examples in this page use 4 mpi processes. It is not a problem, in principle, to run the examples even if the number of available cores is less than 4 (this is called oversubscribing (it may also fail depending on OS and MPI implementation). There is no performance gain to expect from oversubscribing but it is useful for experiments (e.g. for testing the examples in this page on a single-core machine).

### Interactive mode

The interactive mode aims primarily at inspecting the simulation after some MPI execution for debugging. Functions shown here (especially *sendCommand*) may also be usefull in the general case, to achieve advanced tasks such as controlling transitions between phases of a simulation, collecting and processing results.

### Explicit initialization from python prompt

A pool of Yade instances can be spawned with mpy.initialize() as illustrated hereafter. Mind that the next sequences of commands are supposed to be typed directly in the python prompt after starting Yade, it will not give exactly the same result if it is pasted into a script executed by Yade (see the next section on automatic initialization):

```
@suppress
Yade [1]: from yade.utils import *

@suppress
Yade [1]: O.engines=yade.utils.defaultEngines

Yade [2]: wallId=O.bodies.append(box(center=(0,0,0),extents=(2,0,1),fixed=True))

Yade [3]: for x in range(-1,2):
   ...:     O.bodies.append(sphere((x,0.5,0),0.5))
   ...:

Yade [5]: from yade import mpy as mp

@suppress
Yade [5]: mp.COLOR_OUTPUT=False

@doctest
Yade [6]: mp.initialize(4)
Master: I will spawn  3  workers
-> [6]: (0, 4)
```

After mp.initialize(np) the parent instance of Yade takes the role of master process (rank=0). It is the only one executing the commands typed directly in the prompt. The other instances (rank=1 to rank=np-1) are idle and they wait for commands sent from master. Sending commands to the other instances can be done with *mpy.sendCommand()*, which by default returns the result or the list of results. We use that command below to verify that the spawned workers point to different (still empty) scenes:

```
Yade [8]: len(O.bodies)
 -> [8]: 4

Yade [10]: mp.sendCommand(executors="all",command="len(O.bodies)",wait=True) #check content
-> [10]: [4, 0, 0, 0]

Yade [9]: mp.sendCommand(executors="all",command="str(O)") # check scene pointers
 -> [9]:
['<yade.wrapper.Omega object at 0x7f9c0a399490>',
 '<yade.wrapper.Omega object at 0x7f9231213490>',
 '<yade.wrapper.Omega object at 0x7f20086a1490>',
 '<yade.wrapper.Omega object at 0x7f622b47f490>']
```

Sending commands makes it possible to manage all types of message passing using calls to the underlying mpi4py (see mpi4py documentation). Be carefull with sendCommand "blocking" behavior by default. Next example would hang without "wait=False" since both master and worker would be waiting for a message from each other.

```
Yade [3]: mp.sendCommand(executors=1,command="message=comm.recv(source=0); print('received',
↪message)",wait=False)

Yade [4]: mp.comm.send("hello",dest=1)
received hello
```

Every picklable python object (namely, nearly all Yade objects) can be transmitted this way. Remark hereafter the use of *mpy.mprint* (identifies the worker by number and by font colors). Note also that the commands passed via *sendCommand* are executed in the context of the mpy module, for this reason *comm*, *mprint*, *rank* and all objects of the module are accessed without the *mp.* prefix.

```
Yade [3]: mp.sendCommand(executors=1,command="O.bodies.append(comm.recv(source=0))",
↪wait=False) # leaves the worker idle waiting for an argument to append()
```
(continues on next page)

```
Yade [4]: b=Body(shape=Sphere(radius=0.7))  # now create body in the context of master

Yade [5]: mp.comm.send(b,dest=1) # send it to worker 1

Yade [6]: mp.sendCommand(executors="all",command="mprint('received',[b.shape.radius if
↪hasattr(b.shape,'radius') else None for b in O.bodies])")
Master: received [None, 0.5, 0.5, 0.5]
Worker1: received [0.7]
Worker3: received []
Worker2: received []
-> [5]: [None, None, None, None] # printing yields no return value, hence that empty list of
↪returns, "wait=False" argument to sendCommand would suppress it
```

**Explicit initialization from python script**

Though usefull for advanced operations, the function sendCommand() is limited. Basic features of the python language are missing, e.g. function definitions and loops are a problem - in fact every code fragment which can't fit on a single line is. In practice the mpy module provides a mechanism to initialize from a script, where functions and variables will be declared.

Whenever Yade is started with a script as an argument, the script name will be remembered, and if mpy.initialize() is called (by the script itself or interactively in the prompt), all Yade instances will be initialized with that same script. It makes distributing function definitions and simulation parameters trivial (and even distributing scene constructions as seen below).

This behaviour is what happens usually with MPI: all processes execute the same program. It is also what happens with "mpiexec -np N yade ...".

If the first commands above are pasted into a script used to start Yade, all workers insert the same bodies as master (with interactive execution only master was inserting). Here is the script:

```
# script 'test1.py'
wallId=O.bodies.append(box(center=(0,0,0),extents=(2,0,1),fixed=True))
for x in range(-1,2):
        O.bodies.append(sphere((x,0.5,0),0.5))
from yade import mpy as mp
mp.initialize(4)
print( mp.sendCommand(executors="all",command="str(O)",wait=True) )
print( mp.sendCommand(executors="all",command="len(O.bodies)",wait=True) )
```

and the output reads:

```
yade test1.py
...
Running script test1.py
Master: will spawn  3  workers
None
None
None
None
None
None
['<yade.wrapper.Omega object at 0x7feb979403a0>', '<yade.wrapper.Omega object at
↪0x7f5b61ae9440>', '<yade.wrapper.Omega object at 0x7fdd466b8440>', '<yade.wrapper.Omega
↪object at 0x7f8dc7b73440>']
[4, 4, 4, 4]
```

That's because all instances execute the script in the initialize() phase. "None" is printed 2x3 times because the script contains *print( mp.sendCommand(...))* twice, the workers try to execute that too, but for them *sendCommand* returns by default, hence the None.

Though logical, this result is not what we want if we try to split a simulation into pieces. The solution (typical of all mpi programs) is to use the *rank* of the process in conditionals. Different parts of the script can then be executed, differently, by each worker, depending on its rank. In order to produce the same result as before, for instance, the script can be modified as follows:

```python
# script 'test2.py'
from yade import mpy as mp
mp.initialize(4)
if mp.rank==0: # only master
        wallId=O.bodies.append(box(center=(0,0,0),extents=(2,0,1),fixed=True))
        for x in range(-1,2):
        O.bodies.append(sphere((x,0.5,0),0.5))

        print( mp.sendCommand(executors="all",command="str(O)",wait=True) )
        print( mp.sendCommand(executors="all",command="len(O.bodies)",wait=True) )
        print( mp.sendCommand(executors="all",command="str(O)",wait=True) )
```

Resulting in:

```
Running script test2.py
Master: will spawn  3  workers
['<yade.wrapper.Omega object at 0x7f21a8c8d3a0>', '<yade.wrapper.Omega object at
↪0x7f3142e43440>', '<yade.wrapper.Omega object at 0x7fb699b1a440>', '<yade.wrapper.Omega
↪object at 0x7f1e4231e440>']
[4, 0, 0, 0]
```

We could also use *rank* to assign bodies from different regions of space to different workers, as found in example examples/mpi/helloMPI.py, with rank-dependent positions:

```python
# rank is accessed without "mp." prefix as it is interpreted in mpy module's scope
mp.sendCommand(executors=[1,2],command= "ids=O.bodies.append([sphere((xx,1.5+rank,0),0.5) for
↪xx in range(-1,2)])")
```

Keep in mind that the position of the call *mp.initialize(N)* relative to the other commands has no consequence for the execution by the workers (for them initialize() just returns), hence program logic should not rely on it. The workers execute the script from begin to end with the same MPI context, already set when the first line is executed. It can lead to counter intuitive behavior, here is a script:

```python
# testInit.py
# script.py
O.bodies.append([Body() for i in range(100)])

from yade import mpy as mp
mp.mprint("before initialize: rank ", mp.rank,"/", mp.numThreads,"; ",len(O.bodies)," bodies")
mp.initialize(2)
mp.mprint("after initialize: rank ", mp.rank,"/", mp.numThreads,"; ",len(O.bodies)," bodies")
```

and the output:

```
Running script testInit.py
Master: before initialize: rank  0 / 1 ;  100  bodies
Master: will spawn  1  workers
Master: after initialize: rank  0 / 2 ;  100  bodies
Worker1: before initialize: rank  1 / 2 ;  100  bodies
Worker1: after initialize: rank  1 / 2 ;  100  bodies
```

**mpirun (automatic initialization)**

Effectively running a distributed DEM simulation on the basis of the previously described commands would be tedious. The mpy module thus provides the function *mpy.mpirun* to automate most of the steps, as described in *introduction*. Mainly, splitting the scene into subdomains based on rank assigned to bodies and handling collisions between the subdomains as time integration proceeds (includes changing the engine list agressively to make this all happen).

If needed, the first execution of mpirun will call the function initialize(), which can therefore be omitted on the user's side. The subdomains will be merged into a centralized scene on the master process at the end of the iterations depending on the argument *withMerge*.

Here is a concrete example where a floor is assigned to master and multiple groups of spheres are assigned to subdomains:

```python
import os
from yade import mpy as mp

NSTEPS=5000 #turn it >0 to see time iterations, else only initilization
numThreads = 4 # number of threads to be spawned, (in interactive mode).

#materials
young = 5e6
compFricDegree = 0.0
O.materials.append(FrictMat(young=young, poisson=0.5, frictionAngle = radians(compFricDegree),
↪density= 2600, label='sphereMat'))
O.materials.append(FrictMat(young=young*100, poisson = 0.5, frictionAngle = compFricDegree,
↪density =2600, label='wallMat'))

#add spheres

mn,mx=Vector3(0,0,0),Vector3(90,180,90)
pred = pack.inAlignedBox(mn,mx)
O.bodies.append(pack.regularHexa(pred,radius=2.80,gap=0, material='sphereMat'))

#walls (floor)

wallIds=aabbWalls([Vector3(-360,-1,-360),Vector3(360,360,360)],thickness=10.0, material=
↪'wallMat')
O.bodies.append(wallIds)

#engines
O.engines=[
        ForceResetter(),
        InsertionSortCollider([
                Bo1_Sphere_Aabb(),
                Bo1_Box_Aabb()], label = 'collider'), # always add labels.
        InteractionLoop(
                [Ig2_Sphere_Sphere_ScGeom(),Ig2_Box_Sphere_ScGeom()],
                [Ip2_FrictMat_FrictMat_FrictPhys()],
                [Law2_ScGeom_FrictPhys_CundallStrack()],
                label="interactionLoop"
        ),
        GlobalStiffnessTimeStepper(timestepSafetyCoefficient=0.3,  timeStepUpdateInterval=100,
↪parallelMode=True, label = 'timeStepper'),
        NewtonIntegrator(damping=0.1,gravity = (0, -0.1, 0), label='newton'),
        VTKRecorder(fileName='spheres/3d-vtk-', recorders=['spheres', 'intr', 'boxes'],
↪parallelMode=True,iterPeriod=500), #use .pvtu to open spheres, .pvtp for ints, and .vtu for
↪boxes.
]
```

(continues on next page)

```
#set a custom verletDist for efficiency.
collider.verletDist = 1.5

#########  RUN  ##########
# customize mpy
mp.ERASE_REMOTE_MASTER = True    #keep remote bodies in master?
mp.DOMAIN_DECOMPOSITION= True    #automatic splitting/domain decomposition
#mp.mpirun(NSTEPS)               #passive mode run
mp.MERGE_W_INTERACTIONS = False
mp.mpirun(NSTEPS,numThreads,withMerge=True) # interactive run, numThreads is the number of
↪workers to be initialized, see below for withMerge explanation.
mp.mergeScene()  #merge scene after run.
if mp.rank == 0: O.save('mergedScene.yade')

#demonstrate getting stuff from workers, here we get kinetic energy from worker subdomains,
↪notice that the master (mp.rank = 0), uses the sendCommand to tell workers to compute
↪kineticEnergy.
if mp.rank==0:
        print("kinetic energy from workers: "+str(mp.sendCommand([1,2],"kineticEnergy()",
↪True)))
```

The script is then executed:

```
yade script.py
```

For running further timesteps, the mp.mpirun command has to be executed in yade prompt:

```
Yade [0]: mp.mpirun(100,4,withMerge=False) #run for 100 steps and no scene merge.

Yade [1]: mp.sendCommand([1,2],"kineticEnergy()",True) # get kineticEnergy from workers 1 and
↪2.

Yade [2]: mp.mpirun(1,4,withMerge=True) # run for 1 step and merge scene into master. Repeat
↪multiple time to watch evolution in QGL view
```

### Non-interactive execution

Instead of spawning mpi processes after starting Yade, it is possible to run Yade with the classical "mpiexec" from OpenMPI. Importantly, it may be the only method allowed through HPC job submission systems. When using mpiexec there is no interactive shell, or a broken one (which is ok in general in production). The job needs to run (or "*mpirun*") and terminate by itself.

The functions *initialize* and *mpirun* described above handle both interactive and passive executions transparently, and the user scripts should behave the same in both cases. "Should", since what happens behind the scenes is not exactly the same at startup, and it may surface in some occasions (let us know).

Provided that a script calls *mpy.mpirun* with a number of timesteps, the simulation (see e.g. examples/mpi/vtkRecorderExample.py) is executed with the following command:

```
mpiexec -np NUMSUBD+1 yade vtkRecorderExample.py
```

where *NUMSUBD* corresponds to the required number of subdomains.

---

**Note:** Remember that the master process counts one while it does not handle an ordinary subdomain, therefore the number of processes is always *NUMSUBD* +1.

---

### 5.2.4 Splitting

Splitting an initial scene into subdomains and updating the subdomains after particle motion are two critical issues in terms of efficiency. The decomposition can be prescribed on users's side (first section below), but mpy module also provides algorithms for both.

---

**Note:** The mpy module has no requirement in terms of how the subdomains are defined, and using the helper functions described here is not a requirement. Even assigning the bodies randomly from a large cloud to a number of subdomains (such that the subdomains overlap each other and the scene entirely) would work. It would only be suboptimal as the number of interactions between subdomains would increase compared to a proper partition of space.

---

**Split by yourself**

In order to impose a decomposition it is enough to assign *Body.subdomain* a value corresponding to the process rank it should belong to. This can be done either in one centralized scene that is later split, or by inserting the correct subsets of bodies independently in each subdomain (see section on scene construction)

In the example script examples/mpi/testMPI_2D.py the spheres are generated as follows (centralized construction in this example, easily turned into distributed one). For each available worker a bloc of spheres is generated with a different position in space. The spheres in each block are assigned a subdomain rank (and a color for visualisation) so that they will be picked up by the right worker after mpirun().:

```
for sd in range(0,numThreads-1):
        col = next(colorScale)
        ids=[]
        for i in range(N):#(numThreads-1) x N x M spheres, one thread is for master and will↲
↪keep only the wall, others handle spheres
                for j in range(M):
                        id = O.bodies.append(sphere((sd*N+i+j/30.,j,0),0.500,color=col)) #a↲
↪small shift in x-positions of the rows to break symmetry
                        ids.append(id)
                for id in ids: O.bodies[id].subdomain = sd+1
```

**Don't know how to split? Leave it to mpirun**

**Initial split**

mpirun will decide by itself how to distribute the bodies across several subdomains if *DO-MAIN_DECOMPOSITION* =True. In such case the difference between the sequential script and its mpi version is limited to importing mpy and calling mpirun after turning the *DO-MAIN_DECOMPOSITION* flag.

The automatic splitting of bodies to subdomains is based on the Orthogonal Recursive Bisection Algorithm of Berger [Berger1987], and [Fleissner2007]. The partitioning is based on bisecting the space at several *levels*, with the longest axis in each level chosen as the bisection axis. The number of levels is determined as $\texttt{int}(\log_2(N_w))$ with $N_w$ being the number of worker subdomains. A schematic of this decomposition is shown in *fig-bisectionAlgo*, with 4 worker subdomains. At the initial stage (level = 0), we assume that subdomain=1 contains the information of the body positions (and bodies), the longest axis is first determined, this forms the bisecting axis/plane. The list containing the body positions is sorted along the bisecting axis, and the median of this sorted list is determined. The bodies with positions (bisection coordinate) less than the median is coloured with the current subdomain, (SD=1) and the other half is coloured with SD = 2, the subdomain colouring at each level is determined using the following rule:

```
if (subdomain <  1<<level) : this subdomain gets the bodies with position lower than␣
↪the median.
if ((subdomain >  1<<level) and (subdomain <  1<<(level+1) ) ) : this subdomain gets␣
↪the bodies with position greater than median, from subdomain - (1<<level)
```

This process is continued until the number of levels are reached.

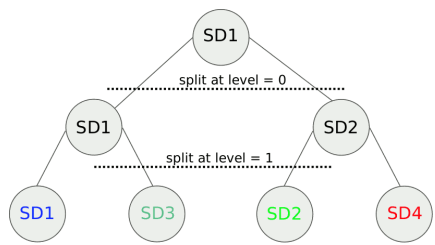

Figure *fig-domainDecompose* shows the resulting partitioning obtained using the ORB algorithm : (a) for 4 subdomains, (b) for 8 subdomains. Odd number of worker subdomains are also supported with the present implementation.

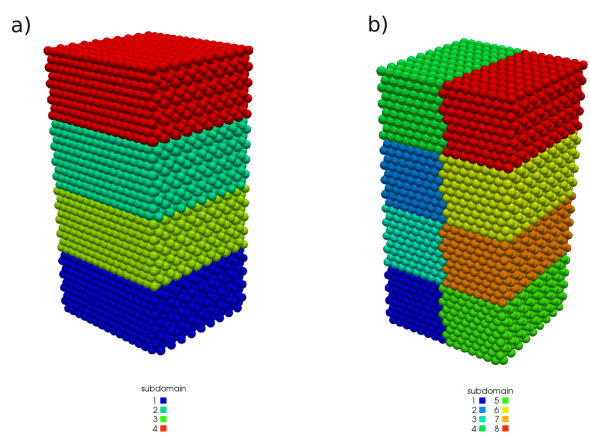

The present implementation can be found in py/bisectionDecomposition.py, and a parallel version can be found here.

---

**Note:** importing py/bisectionDecomposition.py triggers the import of *mpi4py* and ultimately of the MPI library and related environment variables. The *mpy* module may change some mpi settings on import, therefore it is safer to only import *bisectionDecomposition* after some `mpy.initialize()`.

---

### Updating the decomposition (load balancing)

As the bodies move, each subdomain may experience overall distortion and diffusion of bodies leading to an undesirable overlap between multiple subdomains. This subdomain overlap introduces inefficiencies in communications between MPI subdomains, and thus we aim to keep the subdomains as compact as possible by using an algorithm that dynamically reallocates bodies to new subdomains with an objective of minimizing MPI communications. The algorithm exploits *InsertionSortCollider* to reassign bodies efficiently and in synchronicity with collision detection, and it can be activated if *mpy.REALLOCATE_-FREQUENCY* >0.

The algorithm is *not* centralized, which preserves scalability. Additionally, it only engages peer-to-peer communications between MPI workers that share an intersection. The re-assignment depends on a filter for making local decisions. At the moment, there is one filter available called *mpy.medianFilter*. Custom filters can be used instead.

The median filter body re-allocation criterion criterion involves finding the position of a median plane between two subdomains such that after discriminating bodies on the "+" and "-" side of that plane the total number in each subdomain is preserved. It results in the type of split shown in the video hereafter. Even though the median planes seem to rotate rather quickly at some point in this video, there are actually five collision detections between each re-allocation, i.e. thousands of time iterations to effectively rotate the split between two different colors. These progressive rotations are beneficial since the initial split would have resulted in flat discs otherwise.

---

**Note:** This is not a load balancing in the sense of achieving an equal amount of work per core. In fact that sort of balancing is achieved by definition already as soon as each worker is assigned the same amount of bodies (and because a subdomain is really ultimately a list of bodies, not a specific region of space). Instead the objective is to decrease the communication times overall.

---

### Centralized versus distributed scene construction

For the centralized scene construction method, the master process creates all of the bodies of a scene and assigns subdomains to them. As part of mpi initialization some engines will be modified or inserted, then the scene is broadcasted to the workers. Each worker receives the entire scene, identifies its assigned bodies via *Body.subdomain* (if worker's `rank==b.subdomain` the bodies are retained) and erase the others. Such a scene construction was used in the previous example and it is by far the simplest. It makes no real difference with building a scene for non-MPI execution besides calling *mp.mpirun* instead or just *O.run*.

For large number of bodies and processes, though, the centralized scene construction and distribution can consume a significant amount of time. It can also be memory bound since the memory usage is quadratic: suppose N bodies per thread on a 32-core node, centralized construction implies that 32 copies of the entire scene exist simultaneously in memory at some point in time (during the split), i.e. $32^2$N bodies on one single node. For massively parallel applications distributed construction should be prefered.

In distributed mode each worker instantiates its own bodies and insert them in the local *BodyContainer*. Attention need to be paid to properly assign bodies ids since no index should be owned by two different workers initially. Insertion of bodies in *BodyContainer* with imposed ids is done with *BodyContainer.insertAtId*. The distributed mode is activated by setting the `DISTRIBUTED_INSERT` flag ON, the user is in charge of setting up the subdomains and partitioning the bodies. An example of distributed insertion can be found in examples/mpi/parallelBodyInsert3D.py.

The relevant fragment, where the filtering is done by skipping all steps of a loop except the one with proper rank (keep in mind that all workers will run the same loop but they all have a different rank each), reads:

```
#add spheres
subdNo=0
import itertools
_id = 0 #will be used to count total number of bodies regardless of subdomain attribute, so
↪that same ids are not reused for different bodies
for x,y,z in itertools.product(range(int(Nx)),range(int(Ny)),range(int(Nz))):
        subdNo+=1
        if mp.rank!=subdNo: continue
        ids=[]
        for i in range(L):#(numThreads-1) x N x M x L spheres, one thread is for master and
↪will keep only the wall, others handle spheres
                for j in range(M):
                        for k in range(N):
                                dxOndy = 1/5.; dzOndy=1/15.  # shifts in x/y-positions to make
↪columns inclines
                                px= x*L+i+j*dxOndy; pz= z*N+k+j*dzOndy; py = (y*M+j)*(1 -
↪dxOndy**2 -dzOndy**2)**0.5 #so they are always nearly touching initially
```

(continues on next page)

```
                                 id = O.bodies.insertAtId(sphere((px,py,pz),0.500),_
↪id+(N*M*L*(subdNo-1)))
                                 _id+=1
                                 ids.append(id)
        for id in ids: O.bodies[id].subdomain = subdNo

if mp.rank==0: #the wall belongs to master
        WALL_ID=O.bodies.insertAtId(box(center=(Nx*L/2,-0.5,Nz*N/2),extents=(2*Nx*L,0,2*Nz*N),
↪fixed=True),(N*M*L*(numThreads-1)))
```

The bissection algorithm can be used for defining the initial split, in the distributed case too, since it takes a points dataset as input. Provided that all workers work with the same dataset (e.g. the same sequence of a random number generator) they will all reach the same partitioning, and they can instantiate their bodies on this basis.

### 5.2.5 Merging

The possibility of a "merge", shown in the previous example, can be performed using an optional argument of *mpirun* or as a standalone function *mpy.mergeScene* .

If withMerge=True in mpirun then the bodies in master scene are updated to reflect the evolution of their distributed clones. This is done once after finishing the required number of iterations in mpirun. This *merge* operation can include updating interactions. *mpy.mergeScene* does the same within the current iteration. Merging is an expensive task which requires the communication of large messages and, therefore, it should be done purposely and at a reasonable frequency. It can even be the main bottleneck for massively parallel scenes. Nevertheless, it can be useful for debugging with the 3D view, or for various post-processing tasks. The *MERGE_W_INTERACTIONS* provides a full merge, i.e. the interactions in the worker subdomains and between the subdomains are included, otherwise, only the position and states of the bodies are used. Merging with interactions should result in a usual Yade scene, ready for further time-stepping in non-mpi mode or (more useful) for some post-processing. The merge operation is not required for a proper time integration in general.

### 5.2.6 Hints and problems to expect

#### MPI support in engines

For MPI cases, the *parallelMode* flag for *GlobalStiffnessTimeStepper* and *VTKRecorder* have to be turned on. They are the only two engines upgraded with MPI support at the moment.

For other things. Read next section and be careful. If you feel like implementing MPI support for other engines, that would be great, consider using the two available examples as guides. Let us know!

#### Reduction (partial sums)

Quantities such as kinetic energy cannot be obtained for the entire scene just by summing the return value of *kineticEnergy()* from each subdomain. This is because each subdmomain may contain also images of bodies from intersecting subdomains and they may add their velocity, mass, or whatever is summed, to what is returned by each worker. Although some most-used functions of Yade may progressively get mpi support to filter out bodies from remote domains, it is not standard yet and therefore partial sums may need to be implemented on a case-by-case basis, with proper filtering in the user script.

This is just an example of why many things may go wrong if *run* is directly replaced by *mpirun* in a complex script.

**Miscellaneous**

- sendCommand() has a hardcoded latency of 0.001s to not keep all cores 100% busy waiting for a command (with possibly little left to OS). If sendCommand() is used at high frequency in complex algorithms it might be beneficial to decrease that sleep time.

## 5.2.7 Control variables

- VERBOSE_OUTPUT : Details on each *operation/step* (such as *mpy.splitScene*, *mpy.parallelCollide* etc) is printed on the console, useful for debugging purposes

- ACCUMULATE_FORCES : Control force summation on bodies owned by the master.

- ERASE_REMOTE_MASTER : Erase remote bodies in the master subdomain or keep them as unbounded ? Useful for fast merge.

- OPTIMIZE_COM, USE_CPP_MPI : Use optimized communication functions and MPI functions from *Subdomain* class

- YADE_TIMING : Report timing statistics, prints time spent in communications, collision detection and other operations.

- DISTRIBUTED_INSERT : Bodies are created and inserted by each subdomain, used for distributed scene construction.

- DOMAIN_DECOMPOSITION : If true, the bisection decomposition algorithm is used to assign bodies to the workers/subdomains.

- MINIMAL_INTERSECTIONS : Reduces the size of position/velocity communications (at the end of the colliding phase, we can exclude those bodies with no interactions besides body<->subdomain from intersections).

- REALLOCATE_FREQUENCY : if > 0, bodies are migrated between subdomains for efficient load balancing. If =1 realloc. happens each time collider is triggered, else every N collision detection

- REALLOCATE_MINIMAL : Intersections are minimized before reallocations, hence minimizing the number of reallocated bodies

- USE_CPP_REALLOC : Use optimized C++ functions to perform body reallocations

- FLUID_COUPLING : Flag for coupling with OpenFOAM.

## 5.2.8 Benchmark

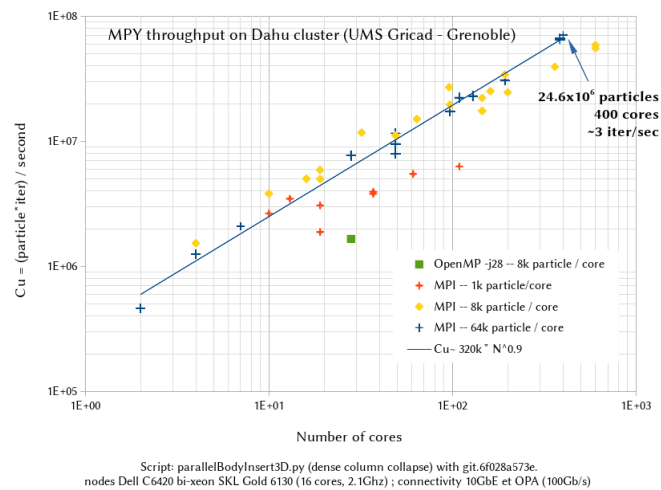

Script: parallelBodyInsert3D.py (dense column collapse) with git.6f028a573e.
nodes Dell C6420 bi-xeon SKL Gold 6130 (16 cores, 2.1Ghz) ; connectivity 10GbE et OPA (100Gb/s)

Comments:

- From 1k particles/core to 8k particles/core there is a clear improvement. Obviously 1k is too small and most of the time is spent in comunications.

- From 8k/core to 64k/core the throughput per core is more or less the same, and the performance is not too far from linear. The data includes elimination of random noise, and overall it is not clear to me which non-linearity comes from the code and which one comes from the hardware.

- Conclusion, if you don't have at least 8k spheres/core (maybe less for more compex shapes) mpi is not your friend. This in line with the estimate of 10k by Dion Weatherley (DEM8+beer)

- It looks like OpenMP sucks, but be aware that the benchmark script is heavily tuned for MPI. It includes huges verletDist and more time wasted on virtual interactions to minimize global updates.

- I believe tuning for OpenMP could make -j26 (or maybe 2xMPIx -j13) on par or faster than 26 MPI threads for less than a million particle. Given the additional difficulty, MPI's niche is for more than a million particles or more than one compute node.

- the nominal per-core throughput is not impressive. On an efficient script my laptop can approach 1e6Cu while we get 0.3e6Cu per core on Dahu. MPI is not to blame here, my laptop would also outperform Dahu on a single core.

# 5.3 Using YADE with cloud computing on Amazon EC2

(Note: we thank Robert Caulk for preparing and sharing this guide)

## 5.3.1 Summary

This guide is intended to help YADE users migrate their simulations to Amazon Web Service (AWS) EC2. Two of the most notable benefits of using scalable cloud computing for YADE include decreased upfront cost and increased productivity. The entire process, from launching an instance, to installing YADE, to running a YADE simulation on the cloud can be executed in under 5 minutes. Once the EC2 instance is running, you can submit YADE scripts the same way you would submit jobs on a local workstation.

## 5.3.2 Launching an EC2 instance

Start by signing into the console on Amazon EC2. This will require an existing or new Amazon account. Once you've signed in, you should find the EC2 console by clicking on 'services' in the upper left hand corner of the AWS homepage. Start by clicking on the `launch an instance` blue button (Fig. *fig-console*). Select the Amazon Machine Image (AMI): `Ubuntu Server 16.04 LTS` (Fig. *fig-ubuntu*).

You will now select the instance type. It is worth looking at the specifications for each of the instances so you can properly select the power you need for you YADE simulation. This document will not go into detail in the selection of size, but you can find plenty of YADE specific performance reports that will help you decide. However, the instance type is an important selection. The `Compute Optimized` instances are necessary for most YADE simulations because they provide access to high performing processors and guaranteed computing power. The C3.2xlarge (Fig. *fig-type*) is equivalent to an 8 core 2.8ghz Xeon E5 with 25 mb of cache, which is likely the best option for medium-large scale YADE simulations.

Before launching, you will be asked to `select an existing key pair or create a new key pair`. Create a new one, download it, and place it in a folder that you know the path to. Modify the permissions on the file by navigating to the same directory in the terminal and typing:

```
chmod 400 KeyPair.pem
```

Now the instance is launched, you will need to connect to it via SSH. On unix systems this is as easy as typing:

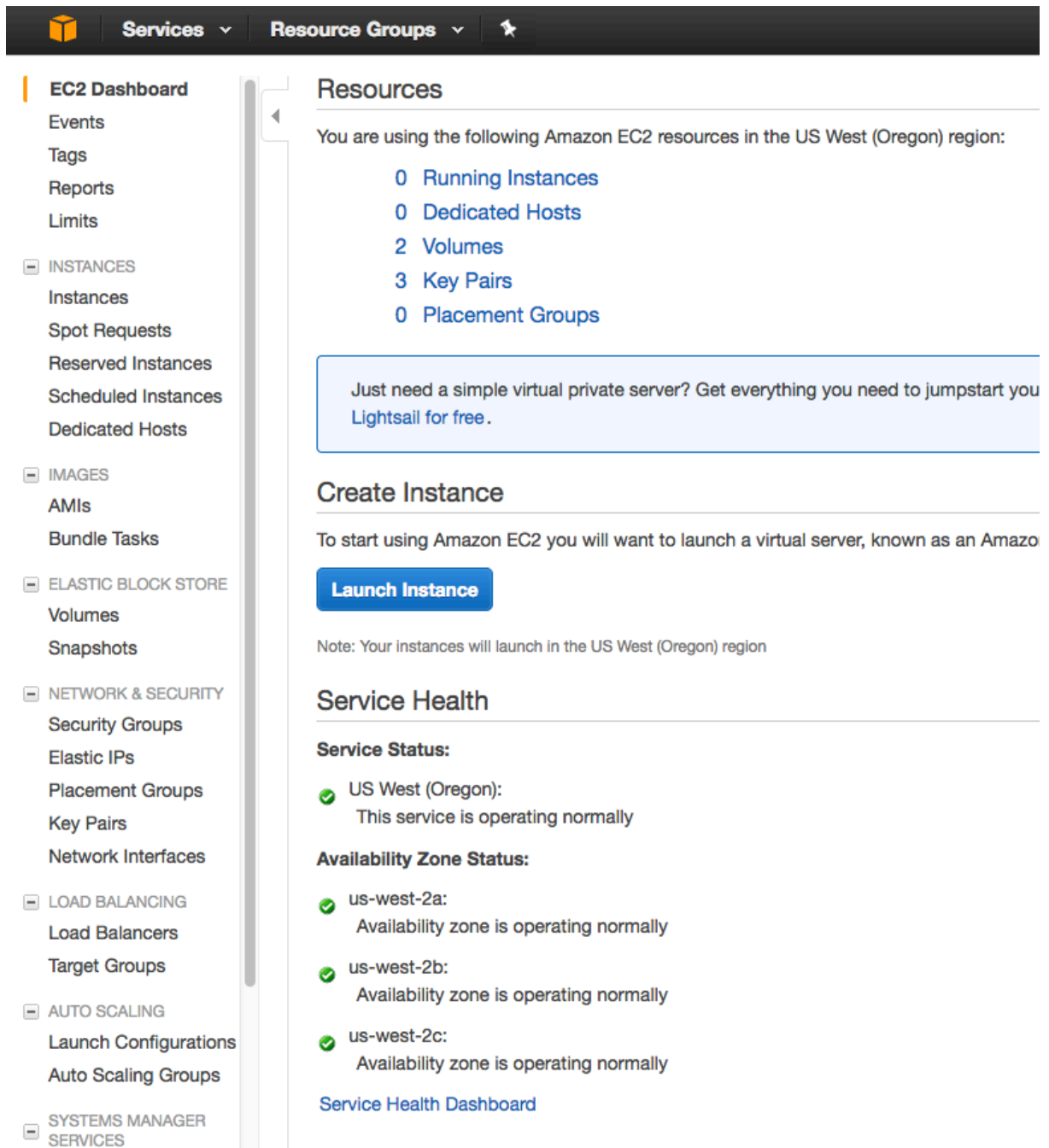

Fig. 2: Amazon Web Services (AWS) Console

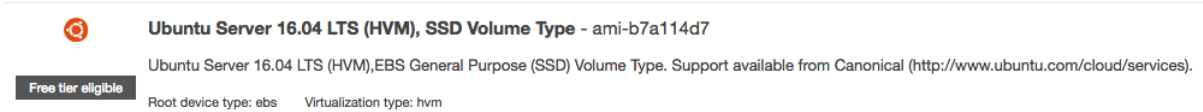

Fig. 3: Select Ubuntu server 16.04 LTS AMI

## C3

**Features:**

- High Frequency Intel Xeon E5-2680 v2 (Ivy Bridge) Processors
- Support for Enhanced Networking
- Support for clustering
- SSD-backed instance storage

| Model | vCPU | Mem (GiB) | SSD Storage (GB) |
|---|---|---|---|
| c3.large | 2 | 3.75 | 2 x 16 |
| c3.xlarge | 4 | 7.5 | 2 x 40 |
| c3.2xlarge | 8 | 15 | 2 x 80 |
| c3.4xlarge | 16 | 30 | 2 x 160 |
| c3.8xlarge | 32 | 60 | 2 x 320 |

**Use Cases**

High performance front-end fleets, web-servers, batch processing, distributed analytics, high performance science and engineering applications, ad serving, MMO gaming, and video-encoding.

Fig. 4: Compute optimized (C3) instance tier

```
ssh -i path/to/KeyPair.pem ubuntu@ec2-XX-XXX-XX-XX.us-west-2.compute.amazon.com
```

into the terminal. There are other options such as using PuTTY, or even a java based terminal on the AWS website. You can find the necessary information by navigating to `Instances` in the left menu of the AWS console. Right click on the instance as shown in Fig. *fig-connect* and click connect.

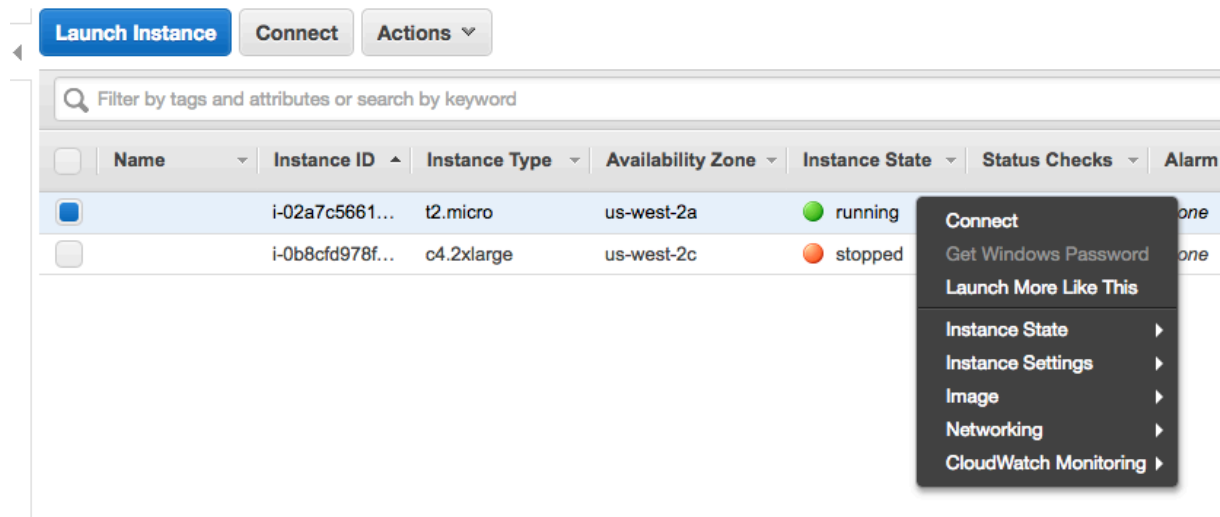

Fig. 5: Connecting to the instance

You will be presented with the public DNS, which should look something like Fig. *fig-dns*.

### 5.3.3 Installing YADE and managing files

After you've connected to the instance through SSH, you will need to install YADE. The following commands should be issued to install yadedaily, python, and some other useful tools:

## 4. Connect to your instance using its Public DNS:

### ec2-35-163-62-84.us-west-2.compute.amazonaws.com

Fig. 6: Public DNS

```
#install yadedaily
sudo bash -c 'echo "deb http://www.yade-dem.org/packages/ xenial/" >> /etc/apt/sources.list'
wget -O - http://www.yade-dem.org/packages/yadedev_pub.gpg | sudo apt-key add -
sudo apt-get update
sudo apt-get install -y yadedaily

# install python
sudo apt-get -y install python
sudo apt-get -y install python-pip python-dev build-essential

# install htop
sudo apt-get -y install htop
```

Note that `..packages/ xenial/` should match the Ubuntu distribution. 16.04 LTS is Xenial, but if you chose to start Ubuntu 14.04, you will need to change 'xenial' to 'trusty'.

Finally, you will need to upload the necessary YADE files. If you have a folder with the contents of your simulation titled `yadeSimulation` you can upload the folder and its contents by issuing the following command:

```
scp -r -i path/to/KeyYADEbox.pem path/to/yadeSimulation ubuntu@ec2-XX-XXX-XX-XX.us-west-2.
↪compute.amazonaws.com:~/yadeSimulation
```

You should now be able to run your simulation by changing to the proper directory and typing:

```
yadedaily nameOfSimulation.py
```

In order to retrieve the output files (folder titled 'out' below) for post processing purposes, you will use the same command that you used to upload the folder, but the remote and local file destinations should be reversed:

```
scp -r -i path/to/KeyYADEbox.pem ubuntu@ec2-XX-XXX-XX-XX.us-west-2.compute.amazonaws.com:~/
↪yadeSimulation/out/ path/to/yadeSimulation/out
```

### 5.3.4 Plotting output in the terminal

One of the main issues encountered with cloud computing is the lack of graphical feedback. There is an easy solution for graphically checking the status of your simulations which makes use of gnuplot's wonderful 'terminal dumb' feature. Any data can be easily plotted by navigating to the subfolder where the simulation is saving its output and typing:

```
gnuplot
set terminal dumb
plot ``data.txt" using 1:2 with lines
```

Where '1:2' refers to the columns in data.txt that you wish to plot against one another. Your output should look something like this:

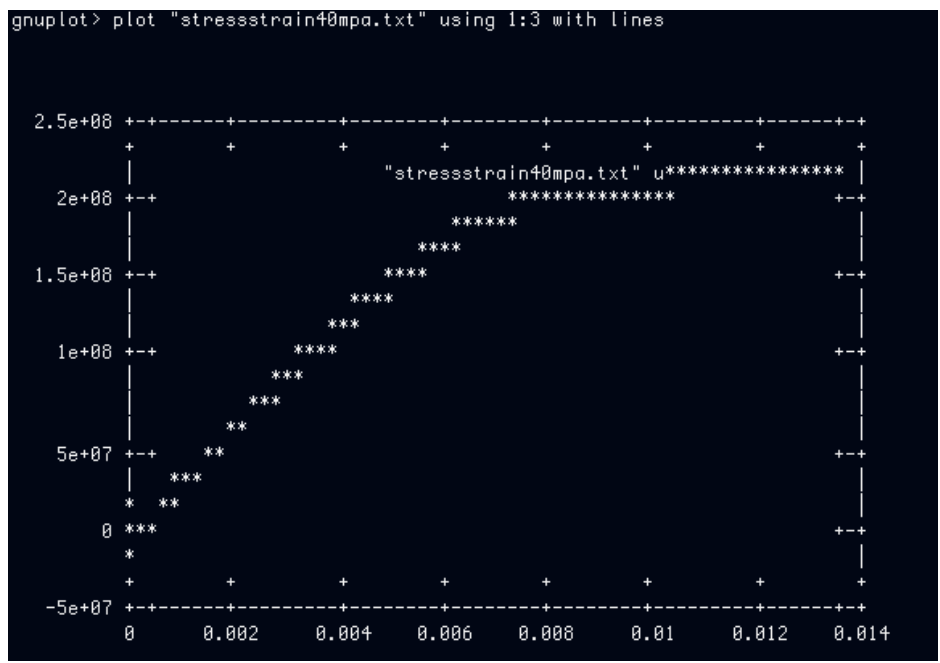

Fig. 7: Gnuplot output

### 5.3.5 Comments

- Amazon AWS allows you to stop your instance and restart it again later with the same files and package installations. If you wish to create several instances that all contain the same installation and file directory you can create a snapshot of your default image which you will be able to use to create various volumes that you can attach to new instances. These actions are all performed very easily and graphically through the EC2 console

- You can use Spot Instances, which are a special type of instance that allow you to bid on unused servers. The price is heavily discounted and worth looking into for any YADE user that wishes to run hundreds of hours of simulations.

- For most simulations, your computational efficiency will decrease if you use above 8 cores per simulation. It is preferable to use yadedaily-batch to distribute your cores accordingly so that you always dedicate 8 cores to each simulation and ensure 100% of the processor is running.

- Create a tmux session to avoid ending YADE simulations upon disconnecting from the server.

```
tmux   # starts a new session
tmux attach -t 0   # attach session 0
tmux kill -t 0   # kill session
## cntrl - b - d to move back to home
## cntrl - b - [ to navigate within the session
```

## 5.4 High precision calculations

Yade supports high and arbitrary precision `Real` type for performing calculations. All tests and checks pass but still the current support is considered experimental. The backend library is boost multiprecision along with corresponding boost math toolkit.

The supported types are following:

| type | bits | decimal places[1] | notes |
|------|------|------------------|-------|
| `float` | 32 | 6 | hardware accelerated (not useful, it is only for testing purposes) |
| `double` | 64 | 15 | hardware accelerated |
| `long double` | 80 | 18 | hardware accelerated |
| `boost float128` | 128 | 33 | depending on processor type it may be hardware accelerated, wrapped by boost |
| `boost mpfr` | Nbit | Nbit/(log(2)/ log(10)) | uses external mpfr library, wrapped by boost |
| `boost cpp_-bin_float` | Nbit | Nbit/(log(2)/ log(10)) | uses boost only, but is slower |

The last two types are arbitrary precision, and their number of bits `Nbit` or decimal places is specified as argument during compilation.

---

**Note:** See file Real.hpp for details. All `Real` types pass the real type concept test from boost concepts. The support for Eigen and CGAL is done with numerical traits.

---

### 5.4.1 Installation

The precompiled Yade package uses `double` type by default. In order to use high precision type Yade has to be compiled and installed from source code by following the regular *installation instructions*. With extra following caveats:

1. Following packages are required to be installed: `python3-mpmath libmpfr-dev libmpfrc++-dev libmpc-dev` (the `mpfr` and `mpc` related packages are necessary only to use `boost::multiprecision::mpfr` type). These packages are already listed in the *default requirements*.

2. A g++ compiler version 9.2.1 or higher is required. It shall be noted that upgrading only the compiler on an existing linux installation (an older one, in which packages for different versions of gcc were not introduced) is difficult and it is not recommended. A simpler solution is to upgrade entire linux installation.

3. During cmake invocation specify:

   1. either number of bits as `REAL_PRECISION_BITS=……`,

   2. or number of requested decimal places as `REAL_DECIMAL_PLACES=……`, but not both

   3. to use MPFR specify `ENABLE_MPFR=ON` (is `OFF` by default).

   The arbitrary precision (`mpfr` or `cpp_bin_float`) types are used only when more than 128 bits or more than 39 decimal places are requested. In such case if `ENABLE_MPFR=OFF` then the slower `cpp_bin_float` type is used. The difference in decimal places between 39 and 33 stems from the fact that 15 bits are used for exponent. Note: a fast quad-double (debian package `libqd-dev`) implementation with 62 decimal places is in the works with boost multiprecision team.

### 5.4.2 Supported modules

During *compilation* several Yade modules can be enabled or disabled by passing an `ENABLE_*` command line argument to cmake. The following table lists which modules are currently working with high precision

---

[1] The amount of decimal places in this table is the amount of places which are completely determined by the binary represenation. *Few additional decimal digits* is necessary to fully reconstruct binary representation. A simple python example to demonstrate this fact: `for a in range(16): print(1./pow(2.,a))`, shows that every binary digit produces "extra" …25 at the end of decimal representation, but these decimal digits are not completely determined by the binary representation, because for example …37 is impossible to obtain there. More binary bits are necessary to represent …37, but the …25 was produced by the last available bit.

---

(those marked with "maybe" were not tested):

| ENABLE_* module name | HP support | cmake default setting | notes |
|---|---|---|---|
| ENABLE_GUI | yes | ON | native support[2] |
| ENABLE_CGAL | yes | ON | native support[2] |
| ENABLE_VTK | yes | ON | supported[3] |
| ENABLE_OPENMP | partial | ON | partial support[4] |
| ENABLE_MPI | maybe | OFF | not tested[5] |
| ENABLE_GTS | yes | ON | supported[6] |
| ENABLE_GL2PS | yes | ON | supported[6] |
| ENABLE_LINSOLV | no | OFF | not supported[7] |
| ENABLE_PARTIALSAT | no | OFF | not supported[7] |
| ENABLE_PFVFLOW | no | OFF | not supported[7] |
| ENABLE_TWOPHASEFLOW | no | OFF | not supported[7] |
| ENABLE_THERMAL | no | OFF | not supported[7] |
| ENABLE_LBMFLOW | yes | ON | supported[6] |
| ENABLE_SPH | maybe | OFF | not tested[8] |
| ENABLE_LIQMIGRATION | maybe | OFF | not tested[8] |
| ENABLE_MASK_ARBITRARY | maybe | OFF | not tested[8] |
| ENABLE_PROFILING | maybe | OFF | not tested[8] |
| ENABLE_POTENTIAL_BLOCKS | no | OFF | not supported[9] |
| ENABLE_POTENTIAL_PARTICLES | yes | ON | supported[10] |
| ENABLE_DEFORM | maybe | OFF | not tested[8] |
| ENABLE_OAR | maybe | OFF | not tested[8] |
| ENABLE_FEMLIKE | yes | ON | supported[6] |
| ENABLE_ASAN | yes | OFF | supported[6] |
| ENABLE_MPFR | yes | OFF | native support[2] |
| ENABLE_LS_DEM | no | ON | not supported[11] |

The unsupported modules are automatically disabled during a high precision `cmake` stage.

## 5.4.3 Double, quadruple and higher precisions

Sometimes a critical section of the calculations in C++ would work better if it was performed in the higher precision to guarantee that it will produce the correct result in the default precision. A simple example is solving a system of linear equations (basically inverting a matrix) where some coefficients are very close to zero. Another example of alleviating such problem is the Kahan summation algorithm.

If *requirements* are satisfied, Yade supports higher precision multipliers in such a way that `RealHP<1>` is the `Real` type described above, and every higher number is a multiplier of the `Real` precision. `RealHP<2>` is double precision of `RealHP<1>`, `RealHP<4>` is quadruple precision and so on. The general formula for

---

[2] This feature is supported natively, which means that specific numerical traits were written for Eigen and for CGAL, as well as GUI and python support was added.

[3] VTK is supported via the compatibility layer which converts all numbers down to `double` type. See *below*.

[4] The OpenMPArrayAccumulator is experimentally supported for `long double` and `float128`. For types `mpfr` and `cpp_bin_float` the single-threaded version of accumulator is used. File lib/base/openmp-accu.hpp needs further testing. If in doubt, compile yade with `ENABLE_OPENMP=OFF`. In all other places OpenMP multithreading should work correctly.

[5] MPI support has not been tested and sending data over network hasn't been tested yet.

[6] The module was tested, the `yade --test` and `yade --check` pass, as well as most of examples are working. But it hasn't been tested extensively for all possible use cases.

[7] Not supported, the code uses external cholmod library which supports only `double` type. To make it work a native Eigen solver for linear equations should be used.

[8] This feature is `OFF` by default, the support of this feature has not been tested.

[9] Potential blocks use external library coinor for linear programming, this library uses `double` type only. To make it work a linear programming routine has to be implemented using Eigen or coinor library should start using C++ templates or a converter/wrapper similar to LAPACK library should be used.

[10] The module is enabled by default, the `yade --test` and `yade --check` pass, as well as most of examples are working. However the calculations are performed at lower `double` precision. A wrapper/converter layer for LAPACK library has been implemented. To make it work with full precision these routines should be reimplemented using Eigen.

[11] Possible future enhancement. See comments there .

---

amount of decimal places is implemented in RealHP.hpp file and the number of decimal places used is simply a multiple N of decimal places in `Real` precision, it is used when native types are not available. The family of available native precision types is listed in the RealHPLadder type list.

All types listed in MathEigenTypes.hpp follow the same naming pattern: `Vector3rHP<1>` is the regular `Vector3r` and `Vector3rHP<N>` for any supported N uses the precision multiplier N. One could then use an Eigen algorithm for solving a system of linear equations with a higher N using `MatrixXrHP<N>` to obtain the result with higher precision. Then continuing calculations in default `Real` precision, after the critical section is done. The same naming convention is used for CGAL types, e.g. `CGAL_AABB_treeHP<N>` which are declared in file AliasCGAL.hpp.

Before we fully move to C++20 standard, one small restriction is in place: the precision multipliers actually supported are determined by these two defines in the RealHPConfig.hpp file:

1. `#define YADE_EIGENCGAL_HP (1)(2)(3)(4)(8)(10)(20)` - the multipliers listed here will work in C++ for `RealHP<N>` in CGAL and Eigen. They are cheap in compilation time, but have to be listed here nonetheless. After we move code to C++20 this define will be removed and all multipliers will be supported via single template constraint. This inconvenience arises from the fact that both CGAL and Eigen libraries offer template specializations only for a *specific* type, not a generalized family of types. Thus this define is used to declare the required template specializations.

---

**Hint:** The highest precision available by default N= (20) corresponds to 300 decimal places when compiling Yade with the default settings, without changing `REAL_DECIMAL_PLACES=……` cmake compilation option.

---

2. `#define YADE_MINIEIGEN_HP (1)(2)` - the precision multipliers listed here are exported to python, they are expensive: each one makes compilation longer by 1 minute. Adding more can be useful only for debugging purposes. The double `RealHP<2>` type is by default listed here to allow exploring the higher precision types from python. Also please note that `mpmath` supports only one precision at a time. Having different `mpmath` variables with different precision is poorly supported, albeit `mpmath` authors promise to improve that in the future. Fortunately this is not a big problem for Yade users because the general goal here is to allow more precise calculations in the critical sections of C++ code, not in python. This problem is partially mitigated by *changing* mpmath precision each time when a `C++   python` conversion occurs. So one should keep in mind that the variable `mpmath.mp.dps` always reflects the precision used by latest conversion performed, even if that conversion took place in GUI (not in the running script). Existing `mpmath` variables are not truncated to lower precision, their extra digits are simply ignored until `mpmath.mp.dps` is increased again, however the truncation might occur during assignment.

On some occasions it is useful to have an intuitive up-conversion between C++ types of different precisions, say for example to add `RealHP<1>` to `RealHP<2>` type. The file UpconversionOfBasicOperatorsHP.hpp serves this purpose. This header is not included by default, because more often than not, adding such two different types will be a mistake (efficiency–wise) and compiler will catch them and complain. After including this header this operation will become possible and the resultant type of such operation will be always the higher precision of the two types used. This file should be included only in `.cpp` files. If it was included in any `.hpp` file then it could pose problems with C++ type safety and will have unexpected consequences. An example usage of this header is in the following test routine.

---

**Warning:** Trying to use N unregistered in `YADE_MINIEIGEN_HP` for a `Vector3rHP<N>` type inside the `YADE_CLASS_BASE_DOC_ATTRS_*` macro to export it to python will not work. Only these N listed in `YADE_MINIEIGEN_HP` will work. However it is safe (and intended) to use these from `YADE_EIGENCGAL_-HP` in the C++ calculations in critical sections of code, without exporting them to python.

---

### 5.4.4 Compatibility

### Python

To declare python variables with `Real` and `RealHP<N>` precision use functions *math.Real(…)*, *math.Real1(…)*, *math.Real2(…)*. Supported are precisions listed in `YADE_MINIEIGEN_HP`, but please note the *mpmath-conversion-restrictions*.

Python has native support for high precision types using `mpmath` package. Old Yade scripts that use *supported modules* can be immediately converted to high precision by switching to `yade.minieigenHP`. In order to do so, the following line:

```
from minieigen import *
```

has to be replaced with:

```
from yade.minieigenHP import *
```

Respectively `import minieigen` has to be replaced with `import yade.minieigenHP as minieigen`, the old name `as minieigen` being used only for the sake of backward compatibility. Then high precision (binary compatible) version of minieigen is used when non `double` type is used as `Real`.

The `RealHP<N>` *higher precision* vectors and matrices can be accessed in python by using the `.HPn` module scope. For example:

```
import yade.minieigenHP as mne
mne.HP2.Vector3(1,2,3) # produces Vector3 using RealHP<2> precision
mne.Vector3(1,2,3)     # without using HPn module scope it defaults to RealHP<1>
```

The respective math functions such as:

```
import yade.math as mth
mth.HP2.sqrt(2) # produces square root of 2 using RealHP<2> precision
mth.sqrt(2)     # without using HPn module scope it defaults to RealHP<1>
```

are supported as well and work by using the respective C++ function calls, which is usually faster than the `mpmath` functions.

> **Warning:** There may be still some parts of python code that were not migrated to high precision and may not work well with `mpmath` module. See *debugging section* for details.

### C++

Before introducing high precision it was assumed that `Real` is actually a POD `double` type. It was possible to use `memset(…)`, `memcpy(…)` and similar functions on `double`. This was not a good approach and even some compiler `#pragma` commands were used to silence the compilation warnings. To make `Real` work with other types, this assumption had to be removed. A single `memcpy(…)` still remains in file openmp-accu.hpp and will have to be removed. In future development such raw memory access functions are to be avoided.

All remaining `double` were replaced with `Real` and any attempts to use `double` type in the code will fail in the gitlab-CI pipeline.

Mathematical functions of all high precision types are wrapped using file MathFunctions.hpp, these are the inline redirections to respective functions of the type that Yade is currently being compiled with. The code will not pass the pipeline checks if `std::` is used. All functions that take `Real` argument should now call these functions in `yade::math::` namespace. Functions which take *only* `Real` arguments may omit `math::` specifier and use ADL instead. Examples:

1. Call to `std::min(a,b)` is replaced with `math::min(a,b)`, because `a` or `b` may be `int` (non `Real`) therefore `math::` is necessary.

2. Call to `std::sqrt(a)` can be replaced with either `sqrt(a)` or `math::sqrt(a)` thanks to ADL, because `a` is always `Real`.

If a new mathematical function is needed it has to be added in the following places:

1. lib/high-precision/MathFunctions.hpp or lib/high-precision/MathComplexFunctions.hpp or lib/high-precision/MathSpecialFunctions.hpp, depending on function type.

2. py/high-precision/_math.cpp, see *math module* for details.

3. py/tests/testMath.py

4. py/tests/testMathHelper.py

The tests for a new function are to be added in py/tests/testMath.py in one of these functions: `oneArgMathCheck(…):`, `twoArgMathCheck(…):`, `threeArgMathCheck(…):`. A table of approximate expected error tolerances in `self.defaultTolerances` is to be supplemented as well. To determine tolerances with better confidence it is recommended to temporarily increase number of tests in the test loop. To determine tolerances for currently implemented functions a `range(1000000)` in the loop was used.

---

**Note:** When passing arguments in `C++` in function calls it is preferred to use `const Real&` rather than to make a copy of the argument as `Real`. The reason is following: in non high-precision regular case both the `double` type and the reference have 8 bytes. However `float128` is 16 bytes large, while its reference is still only 8 bytes. So for regular precision, there is no difference. For all higher precision types it is beneficial to use `const Real&` as the function argument. Also for `const Vector3r&` arguments the speed gain is larger, even without high precision.

---

### String conversions

On the `python` side it is recommended to use *math.Real(…) math.Real1(…)*, or *math.toHP1(…)* to declare `python` variables and *math.radiansHP1(…)* to convert angles to radians using *full Pi precision*.

On the `C++` side it is recommended to use yade::math::toString(…) and yade::math::fromStringReal(…) conversion functions instead of `boost::lexical_cast<std::string>(…)`. The toString and its high precision version toStringHP functions (in file RealIO.hpp) guarantee full precision during conversion. It is important to note that `std::to_string` does not guarantee this and `boost::lexical_cast` does not guarantee this either.

For higher precision types it is possible to control in runtime the precision of `C++` `python` during the `RealHP<N>` string conversion by changing the *math.RealHPConfig.extraStringDigits10* static parameter. Each decimal digit needs $\log_{10}(2) \approx 3.3219$ bits. The `std::numeric_limits<Real>::digits10` provides information about how many decimal digits are completely determined by binary representation, meaning that these digits are absolutely correct. However to convert back to binary more decimal digits are necessary because $\log_2(10) \approx 0.3010299$ decimal digits are used by each bit, and the last digit from `std::numeric_limits<Real>::digits10` is not sufficient. In general 3 or more in *extraStringDigits10* is enough to have an always working number round tripping. However if one wants to only extract results from python, without feeding them back in to continue calculations then a smaller value of *extraStringDigits10* is recommended, like 0 or 1, to avoid a fake sense of having more precision, when it's not there: these extra decimal digits are not correct in decimal sense. They are only there to have working number round tripping. See also a short discussion about this with boost developers. Also see file RealHPConfig.cpp for more details.

---

**Note:** The parameter `extraStringDigits10` does not affect `double` conversions, because `boost::python` uses an internal converter for this particular type. It might be changed in the future if the need arises. E.g. using a class similar to ThinRealWrapper.

---

It is important to note that creating higher types such as `RealHP<2>` from string representation of `RealHP<1>` is ambiguous. Consider following example:

---

```
import yade.math as mth

mth.HP1.getDecomposedReal(1.23)['bits']
Out[2]: '10011101011100001010001111010111000010100011110101110'

mth.HP2.getDecomposedReal('1.23')['bits']   # passing the same arg in decimal format to HP2␣
↪produces nonzero bits after the first 53 bits of HP1
Out[3]:
↪'1001110101110000101000111101011100001010001111010111000010100011110101110000101000111101011100001010001111010111
↪'

mth.HP2.getDecomposedReal(mth.HP1.toHP2(1.23))['bits'] # it is possible to use yade.math.HPn.
↪toHPm(…) conversion, which preserves binary representation
Out[4]:
↪'1001110101110000101000111101011100001010001111010111000000000000000000000000000000000000000000000000000000000000
↪'
```

Which of these two **RealHP<2>** binary representations is more desirable depends on what is needed:

1. The best binary approximation of a `1.23` decimal.

2. Reproducing the 53 binary bits of that number into a higher precision to continue the calculations on **the same** number which was previously in lower precision.

To achieve 1. simply pass the argument `'1.23'` as string. To achieve 2. use *math.HPn.toHPm(…)* or *math.Realn(…)* conversion, which maintains binary fidelity using a single static_cast<RealHP<m>>(…). Similar problem is discussed in mpmath and boost documentation.

The difference between *toHPn* and *Realn* is following: the functions `HPn.toHPm` create a $m \times n$ matrix converting from **RealHP<n>** to **RealHP<m>**. When $n < m$ then extra bits are set to zero (case 2 above, depending on what is required one might say that "precision loss occurs"). The functions *math.Real(…)*, *math.Real1(…)*, *math.Real2(…)* are aliases to the diagonal of this matrix (case 1 above, depending on what is required one might say that "no conversion loss occurs" when using them).

---

**Hint:** All **RealHP<N>** function arguments that are of type higher than `double` can also accept decimal strings. This allows to preserve precision above python default floating point precision.

---

---

**Warning:**  On the contrary all the function arguments that are of type `double` can not accept decimal strings. To mitigate that one can use `toHPn(…)` converters with string arguments.

---

**Hint:**  To make debugging of this problem easier the function *math.toHP1(…)* will raise RuntimeError if the argument is a python float (not a decimal string).

---

---

**Warning:**  I cannot stress this problem enough, please try running `yade --check` (or `yade ./ checkGravityRungeKuttaCashKarp54.py`) in precision different than `double` after changing this line into `g = -9.81`. In this (particular and simple) case the `getCurrentPos()` function fails on the python side because low-precision `g` is multiplied by high-precision `t`.

---

### Complex types

Complex numbers are supported as well.  All standard `C++` functions are available in lib/high-precision/MathComplexFunctions.hpp and also are exported to python in py/high-precision/_math.cpp. There is a cmake compilation option **ENABLE_COMPLEX_MP** which enables using better complex types from

`boost::multiprecision` library for representing `ComplexHP<N>` family of types: `complex128`, `mpc_-complex`, `cpp_complex` and `complex_adaptor`. It is ON by default whenever possible: for boost version >= 1.71. For older boost the `ComplexHP<N>` types are represented by `std::complex<RealHP<N>>` instead, which has larger numerical errors in some mathematical functions.

When using the `ENABLE_COMPLEX_MP=ON` (default) the previously mentioned lib/high-precision/UpconversionOfBasicOperatorsHP.hpp is not functional for complex types, it is a reported problem with the boost library.

When using MPFR type, the `libmpc-dev` package has to be installed (mentioned above).

### Eigen and CGAL

Eigen and CGAL libraries have native high precision support.

- All declarations required by Eigen are provided in files EigenNumTraits.hpp and MathEigen-Types.hpp

- All declarations required by CGAL are provided in files CgalNumTraits.hpp and AliasCGAL.hpp

### VTK

Since VTK is only used to record results for later viewing in other software, such as paraview, the recording of all decimal places does not seem to be necessary (for now). Hence all recording commands in `C++` convert `Real` type down to `double` using `static_cast<double>` command. This has been implemented via classes `vtkPointsReal`, `vtkTransformReal` and `vtkDoubleArrayFromReal` in file VTKCompatibil-ity.hpp. Maybe VTK in the future will support non `double` types. If that will be needed, the interface can be updated there.

### LAPACK

Lapack is an external library which only supports `double` type. Since it is not templatized it is not possible to use it with `Real` type. Current solution is to down-convert arguments to `double` upon calling linear equation solver (and other functions), then convert them back to `Real`. This temporary solution omits all benefits of high precision, so in the future Lapack is to be replaced with Eigen or other templatized libraries which support arbitrary floating point types.

## 5.4.5 Debugging

High precision is still in the experimental stages of implementation. Some errors may occur during use. Not all of these errors are caught by the checks and tests. Following examples may be instructive:

1. Trying to use const references to Vector3r members - a type of problem with results in a segmentation fault during runtime.

2. A part of python code does not cooperate with mpmath - the checks and tests do not cover all lines of the python code (yet), so more errors like this one are expected. The solution is to put the non compliant python functions into py/high-precision/math.py. Then replace original calls to this function with function in `yade.math`, e.g. `numpy.linspace(…)` is replaced with `yade.math.linspace(…)`.

The most flexibility in debugging is with the `long double` type, because special files ThinRealWrap-per.hpp, ThinComplexWrapper.hpp were written for that. They are implemented with boost::operators, using partially ordered field. Note that they do not provide operator++.

A couple of `#defines` were introduced in these two files to help debugging more difficult problems:

1. `YADE_IGNORE_IEEE_INFINITY_NAN` - it can be used to detect all occurrences when `NaN` or `Inf` are used. Also it is recommended to use this define when compiling Yade with `-Ofast` flag, without `-fno-associative-math -fno-finite-math-only -fsigned-zeros`

2. `YADE_WRAPPER_THROW_ON_NAN_INF_REAL`, `YADE_WRAPPER_THROW_ON_NAN_INF_COMPLEX` - can be useful for debugging when calculations go all wrong for unknown reason.

Also refer to *address sanitizer section*, as it is most useful for debugging in many cases.

---

**Hint:** If crash is inside a macro, for example `YADE_CLASS_BASE_DOC_ATTRS_CTOR_PY`, it is useful to know where inside this macro the problem happens. For this purpose it is possible to use `g++` preprocessor to remove the macro and then compile the postprocessed code without the macro. Invoke the preprocessor with some variation of this command:

```
g++ -E -P core/Body.hpp -I ./ -I /usr/include/eigen3 -I /usr/include/python3.7m > /tmp/Body.hpp
```

Maybe use clang-format so that this file is more readable:

```
./scripts/clang-formatter.sh /tmp/Body.hpp
```

Be careful because such files tend to be large and clang-format is slow. So sometimes it is more useful to only use the last part of the file, where the macro was postprocessed. Then replace the macro in the original file in question, and then continue debugging. But this time it will be revealed where inside a macro the problem occurs.

---

**Note:** When *asking questions* about High Precision it is recommended to start the question title with `[RealHP]`.

---

# Chapter 6

# Literature

## 6.1 Yade Technical Archive

### 6.1.1 About

The Yade Technical Archive (YTA) seeks to improve the reproducibility of Yade related publications by clarifying the theory that underlies Yade's opensource code, explaining algorithmic implementations, and providing practical tutorials. In doing so, YTA removes the opacity that commonly exists between readers and computational journal articles, strengthens and improves visibility of existing Yade journal papers, enables academic collaborations, and broadens open access academia.

### 6.1.2 Contribute

**YTA seeks a variety of Yade related materials including, but not limited to:**

- theoretical descriptions of code packages
- user guides and tutorials for code packages
- presentations
- course materials
- supplementary materials for journal articles

### 6.1.3 Contact

If you wish to contribute, please contact rob.caulk@gmail.com. Questions about individual publications are referred to the email address attached to the document description. If you have general questions regarding code, we refer you to our Q&A forum.

### 6.1.4 Archive

Chareyre, Bruno; Caulk, Robert; Chèvremont, William; Guntz, Thomas; Kneib, François; Kunhappen, Deepak; Pourroy, Jean (2019), Calcul distribué MPI pour la dynamique de systèmes particulaires. *Yade Technical Archive.* download full text , watch video summary , read the poster summary

Pirnia, Pouyan; Duhaime Francois; Ethier Yannic; Dubé, Jean-Sébastien (2019), COMSOL-Yade Interface (ICY) instruction guide. *Yade Technical Archive.* download full text, send an email seyed-pouyan.pirnia.1@ens.etsmtl.ca , download helper files

Maurin, Raphael (2018), YADE 1D vertical VANS fluid resolution: Numerical resolution details. *Yade Technical Archive.* download full text, send an email raphael.maurin@imft.fr, follow the tutorial: *Using YADE 1D vertical VANS fluid resolution*

Maurin, Raphael (2018), YADE 1D vertical VANS fluid resolution: Theoretical basis. *Yade Technical Archive.* download full text, send an email raphael.maurin@imft.fr, follow the tutorial: *Using YADE 1D vertical VANS fluid resolution*

Maurin, Raphael (2018), YADE 1D vertical VANS fluid resolution: validations. *Yade Technical Archive.* download full text, send an email raphael.maurin@imft.fr, follow the tutorial: *Using YADE 1D vertical VANS fluid resolution*

Caulk, Robert (2018), Stochastic Augmentation of the Discrete Element Method for Investigation of Tensile Rupture in Heterogeneous Rock. *Yade Technical Archive.* DOI 10.5281/zenodo.1202039. download full text , send an email rob.caulk@gmail.com , follow the tutorial: *Simulating Acoustic Emissions in Yade*

## 6.2 Publications on Yade

Publications on Yade itself or done with Yade are listed on this page.

The first section gives the reference that we kindly ask you to use for citing Yade in publications, as explained in the "Acknowledging Yade" section.

With the increasing rate of publications using Yade it became difficult to list them all, therefore coverage of recent years is only partial. You can help us: if you publish or you know publications related to Yade do not hesitate to add it to this list. If you don't have direct access to the source code, please send the reference (as a bibtex item) to Yade developpers. If a pdf PDF is freely available, add url for direct fulltext downlad. Yade's web server will gladly host such PDF if legally permitted.

---

**Note:** This file is generated from doc/yade-articles.bib, doc/yade-conferences.bib, doc/yade-theses.bib, doc/yade-tech-archive.bib, and doc/citing_yade.bib.

---

### 6.2.1 Citing Yade

Corresponding bibtex entries here. See also "Acknowledging Yade".

### 6.2.2 Journal articles

### 6.2.3 Conference materials and book chapters

### 6.2.4 Master and PhD theses

### 6.2.5 Yade Technical Archive

## 6.3 References

All external articles referenced in Yade documentation.

---

**Note:** This file is generated from doc/references.bib.

---

# Chapter 7

# Indices and tables

- genindex
- modindex
- search

# Python Module Index